\input rztex.qel

\beginOptionen
\Brotschrift{\cmrom 11 pt}
\Schriftarten{\Fraktur}
\Rubriken{\lang}
\Verweise
\Verzeichnisse {\standard}
\endOptionen

\uKopfzeile{}{\Seitennummer}{}
\uKopflinie 0 pt
\Rand = 2.75 cm
\Kolumnenbreite = 16.25 cm
\Kolumnenausgleich

\redefine\tagform#1{\Abschnittswurzel\Zahl\aktAbschnitt -#1}
\redefine\tagstyle#1{\rm(#1)}

% \define\endBeweis{\ \Rahmen .8 em mal .8 em{\ }}

\def\Seq#1#2{ \{ #1 \}_{#2}^{\infty} }
\def\Seqn#1{ \{ #1_n \}_{n=0}^{\infty} }

\def\Ent#1{ [\mkern - 2.5 mu  [ #1 ] \mkern - 2.5 mu ] }

\def\Bar{\, \vert \,}

\def\bigSeq#1#2{ \bigl\{ #1 \bigr\}_{#2}^{\infty} }

\def\bigBar{\, \bigl\vert \bigr. \,}

\def\biggBar{\, \biggl\vert \biggr. \,}

\def\F{\text{I} \mkern - 3.5 mu \text{F} }
\def\D{\text{I} \mkern - 3.5 mu \text{D} }
\def\H{\text{I} \mkern - 3.5 mu \text{H} }

\Titelblatt

\vglue 2.5 cm

{\LARGE
{\bf

\centreline {VERALLGEMEINERTE SUMMATIONSPROZESSE}

\medskip
\smallskip

\centreline {ALS NUMERISCHE HILFSMITTEL}

\medskip
\smallskip

\centreline {F\"UR QUANTENMECHANISCHE UND}
\smallskip

\medskip
\smallskip

\centreline {QUANTENCHEMISCHE RECHNUNGEN}
}

\vskip 4 cm

\centreline {\bf Habilitationsschrift}

\bigskip

\bigskip

\centreline {\bf Naturwissenschaftliche Fakult\"at \Roemisch{4}}
\medskip
\centreline {\bf -- Chemie und Pharmazie --}
\medskip
\centreline {\bf Universit\"at Regensburg}
\medskip
\centreline {\bf 1994}

\vskip 3 cm

\centreline {\bf vorgelegt von}

\bigskip

\bigskip

\vfill

\centreline {\bf Ernst Joachim Weniger}

\bigskip

\bigskip

\centreline {\bf (Korrigierte Version)}

}

\neueSeite
\Titelblatt

\ungeradeSeite
\redefine \Seitennummer {\roemisch \pageno}

\keinTitelblatt
\Pageno = 1

\redefine \Seitennummer {\roemisch \pageno}

\Ueberschrift \large Inhaltsverzeichnis

\medskip

\Eintragi {}{1.}{Einleitung}{1}
\Eintragii {1.}{1.}{Die Rolle der Computer in der theoretischen Chemie}{1}
\Eintragii {1.}{2.}{Die interdisziplin\"are Natur des Fortschritts in der theoretischen Chemie}{2}
\Eintragii {1.}{3.}{Ziel und Inhalt dieser Arbeit}{5}

\Eintragi {}{2.}{Unendliche Reihen und ihre numerische Auswertung}{7}
\Eintragii {2.}{1.}{Konvergenzprobleme bei unendlichen Reihen}{7}
\Eintragii {2.}{2.}{Divergente Reihen}{10}
\Eintragii {2.}{3.}{Das Summationsverfahren von Borel}{13}
\Eintragii {2.}{4.}{Verallgemeinerte Summationsprozesse}{18}

\Eintragi {}{3.}{\"Uber die Konstruktion verallgemeinerter Summationsprozesse}{22}
\Eintragii {3.}{1.}{Konvergenzbeschleunigung und Summation bei alternierenden Reihen}{22}
\Eintragii {3.}{2.}{Die Konstruktion verallgemeinerter Summationsprozesse durch Modellfolgen}{25}
\Eintragii {3.}{3.}{Iterierte Transformationen}{27}

\Eintragi {}{4.}{Pad\'e-Approximationen}{32}
\Eintragii {4.}{1.}{Vorbemerkungen}{32}
\Eintragii {4.}{2.}{Rationale Approximationen f\"ur Potenzreihen}{33}
\Eintragii {4.}{3.}{Pad\'e-Approximationen und Stieltjesreihen}{36}
\Eintragii {4.}{4.}{Die effiziente Berechnung von Pad\'e-Approximationen}{38}
\Eintragii {4.}{5.}{Anwendungen der Pad\'e-Approximationen in den Naturwissenschaften}{47}

\Eintragi {}{5.}{Transformationen mit expliziten Restsummenabsch\"atzungen}{54}
\Eintragii {5.}{1.}{Vorbemerkungen}{54}
\Eintragii {5.}{2.}{Die Levinsche Transformation}{59}
\Eintragii {5.}{3.}{Fakult\"atenreihen}{64}
\Eintragii {5.}{4.}{Darstellung der Korrekturterme durch Fakult\"atenreihen}{67}
\Eintragii {5.}{5.}{Alternative Darstellung der Korrekturterme durch Pochhammersymbole}{70}
\Eintragii {5.}{6.}{Interpolierende Transformationen}{75}
\Eintragii {5.}{7.}{Die Transformation von Potenzreihen}{77}

\Eintragi {}{6.}{Konvergenztheorie verallgemeinerter Summationsprozesse}{88}
\Eintragii {6.}{1.}{Vorbemerkungen}{88}
\Eintragii {6.}{2.}{Modifikationen der Konvergenztheorie von Germain-Bonne}{91}
\Eintragii {6.}{3.}{Transformationen mit expliziten Restsummenabsch\"atzungen}{95}
\Eintragii {6.}{4.}{Eine kritische Bewertung der Konvergenztheorie von Germain-Bonne}{100}
\Eintragii {6.}{5.}{Stieltjesreihen und Stieltjesfunktionen in Summationsprozessen}{101}
\Eintragii {6.}{6.}{Fehlerabsch\"atzungen bei der Transformation von Stieltjesreihen}{105}
\Eintragii {6.}{7.}{Die Summation der Eulerreihe}{109}

\Eintragi {}{7.}{Rationale Approximationen f\"ur die modifizierte Besselfunktion der zweiten Art}{117}
\Eintragii {7.}{1.}{Vorbemerkungen}{117}
\Eintragii {7.}{2.}{Borelsummierbarkeit und Stieltjessummierbarkeit}{121}
\Eintragii {7.}{3.}{Numerische Beispiele}{126}

\Eintragi {}{8.}{Die Berechnung von Mehrzentrenmolek\"ulintegralen exponentialartiger Basisfunktionen mit Hilfe von verallgemeinerten Summationsprozessen}{131}
\Eintragii {8.}{1.}{Molek\"ulrechnungen in der Quantenchemie}{131}
\Eintragii {8.}{2.}{Exponentialartige Basisfunktionen f\"ur Molek\"ulrechnungen}{135}
\Eintragii {8.}{3.}{Linear konvergente Reihenentwicklungen f\"ur Zweizentrenintegrale von $B$-Funktionen}{148}
\Eintragii {8.}{4.}{Logarithmisch konvergente Reihenentwicklungen f\"ur Zweizentrenintegrale von $B$-Funktionen}{157}

\Eintragi {}{9.}{Rationale Approximationen f\"ur Hilfsfunktionen in der Quantenchemie}{166}
\Eintragii {9.}{1.}{Gau{\ss}funktionen als quantenchemische Basisfunktionen}{166}
\Eintragii {9.}{2.}{Mehrzentrenmolek\"ulintegrale von Gau{\ss}funktionen}{170}
\Eintragii {9.}{3.}{Eigenschaften der $F_m$-Funktionen}{173}
\Eintragii {9.}{4.}{Numerische Beispiele}{176}

\Eintragi {}{10.}{Die Summation der St\"orungsreihe f\"ur die Grundzustandsenergie anharmonischer Oszillatoren}{181}
\Eintragii {10.}{1.}{Konvergente und divergente St\"orungsreihen}{181}
\Eintragii {10.}{2.}{Anharmonische Oszillatoren}{184}
\Eintragii {10.}{3.}{Symanzik-Scaling}{189}
\Eintragii {10.}{4.}{Renormierung}{194}
\Eintragii {10.}{5.}{Die Berechnung der St\"orungstheoriekoeffizienten}{202}
\Eintragii {10.}{6.}{Die Divergenz der Levinschen Transformation}{211}
\Eintragii {10.}{7.}{Summationsergebnisse f\"ur die Grundzustandsenergien und die Grenzf\"alle unendlicher Kopplung}{217}
\Eintragii {10.}{8.}{Eine konvergente renormierte St\"orungsreihe f\"ur die Grundzustandsenergie eines anharmonischen Oszillators}{234}

\Eintragi {}{11.}{Konvergenzverbesserung von quantenmechanischen Rechnungen an Polyacetylen mit Hilfe von Extrapolationsverfahren}{249}
\Eintragii {11.}{1.}{Quantenmechanische Rechenverfahren f\"ur quasi-eindimensionale stereo\-regul\"are Polymere}{249}
\Eintragii {11.}{2.}{Polyacetylen als Modellsystem}{254}
\Eintragii {11.}{3.}{Kristallorbital- und Clusterrechnungen an Polyacetylen}{256}
\Eintragii {11.}{4.}{Abbruchfehler bei Kristallorbital- und Clusterrechnungen}{257}
\Eintragii {11.}{5.}{Extrapolationsverfahren}{261}
\Eintragii {11.}{6.}{Extrapolationsergebnisse}{267}

\Eintragi {}{12.}{Zusammenfassung und Ausblicke}{279}

\Eintragi {}{13.}{Literaturverzeichnis}{282}

\Eintrag {\large Danksagungen}{326}

\keinTitelblatt\neueSeite

\redefine \Seitennummer {\Zahl \pageno}
\Pageno = 1

\keinTitelblatt\neueSeite

\beginAbschnittsebene
\aktAbschnitt = 0

\Abschnitt Einleitung

\vskip - 2 \jot

\beginAbschnittsebene

\medskip

\Abschnitt Die Rolle der Computer in der theoretischen Chemie

\smallskip

\aktTag = 0

Nur wenige physikalische Systeme sind so einfach, da{\ss} die entsprechende
zeitunabh\"angige Schr\"odinger\-gleichung {\it exakt\/} gel\"ost werden kann.
Hinzu kommt, da{\ss} einige dieser exakt l\"osbaren Systeme wie beispielsweise
das Teilchen im Potentialkasten, der starre Rotator oder der harmonische
Oszillator ohnehin nur Modellsysteme sind, die in der Natur bestenfalls
n\"aherungsweise realisiert werden. Das einzige Atom, dessen
zeitunabh\"angige Schr\"odingergleichung exakt gel\"ost werden kann, ist das
Wasserstoffatom. Aber schon bei dem einfachsten Molek\"ul, dem
Wasserstoffmolek\"ulion $\rm H_{2}^{+}$, ist eine exakte L\"osung der
Schr\"odingergleichung nur bei Verwendung der Born-Oppenheimer-N\"aherung
[Born und Oppenheimer 1927] m\"oglich [Bates, Ledsham und Stewart 1953;
Bates und Reid 1968]. Bei allen anderen Atomen und Molek\"ulen kann die
Schr\"odingergleich\-ung nur {\it approximativ\/} gel\"ost werden, wobei vor
allem bei komplexeren Atomen und Molek\"ulen zum Teil sehr weitreichende
N\"aherungen notwendig sind.

Die praktische Anwendbarkeit der nichtrelativistischen Quantenmechanik
zur Beschreibung und Analyse der Elektronenstruktur von Atomen und
Molek\"ulen h\"angt also entscheidend davon ab, ob und wie gut es gelingt,
atomare und molekulare Schr\"odingergleichungen mit Hilfe geeigneter
N\"aherungsverfahren numerisch zu l\"osen. Die Situation wird meiner Meinung
nach treffend beschrieben durch ein ber\"uhmtes Zitat von Dirac [1929, S.
714]:

\medskip

\beginSchmaeler
\noindent {\sl The underlying physical laws necessary for the
mathematical theory of a large part of physics and the whole of
chemistry are thus completely known, and the difficulty is only that the
exact application of these laws leads to equations much too complicated
to be soluble. It therefore becomes desirable that approximate practical
methods of applying quantum mechanics should be developed, which can
lead to an explanation of the main features of complex atomic systems
without too much computation.}
\endSchmaeler

\medskip

Die praktische Anwendbarkeit der nichtrelativistischen Quantenmechanik
wird aber auch dadurch erschwert, da{\ss} eine ausreichend genaue
approximative L\"osung atomarer und molekularer Schr\"odinger\-glei\-chungen
keineswegs einfach ist. Schon bei kleinen Atomen oder Molek\"ulen kann der
numerische Aufwand prohibitiv gro{\ss} werden. Deswegen wurde die
Quantenmechanik in ihrer Fr\"uhzeit haupts\"achlich verwendet, um zu einem
{\it qualitativen\/} Verst\"andnis atomarer oder molekularer Ph\"anomene zu
gelangen. Aufgrund der damals sehr begrenzten Rechenkapazit\"at waren {\it
quantitative\/} Aussagen auf der Basis quantenmechanischer Rechnungen
nur in besonders beg\"unstigten Sonderf\"allen m\"oglich. Ein solcher
Sonderfall ist ein freies Atom, das {\it kugelsymmetrisch\/} ist.
Deswegen f\"uhrt das von Hartree [1928] und Fock [1930] eingef\"uhrte
SCF{\footnote[\dagger]{Self-Consistent Field}}-Verfahren nach
Abseparation der Winkelvariablen zu einem System von {\it
eindimensionalen\/} gekoppelten Integrodifferentialgleichungen in der
Radialvariablen $r$. In dem Buch von Hartree [1957] werden die
numerischen Verfahren beschrieben, mit deren Hilfe in der Fr\"uhzeit der
Quantenmechanik SCF-Rechnungen an Atomen unter Verwendung von
Hand\-rechenmaschinen durchgef\"uhrt wurden.

Molek\"ule besitzen -- wenn \"uberhaupt -- eine geringere Symmetrie als
Kugelsymmetrie. Eine Reduktion der molekularen SCF-Gleichungen auf
eindimensionale gekoppelte Integrodifferential\-gleichungen, die ebenso
wie im atomaren Fall numerisch gel\"ost werden k\"onnen, ist deswegen nicht
m\"oglich. Aus diesem Grund sind Rechnungen an Molek\"ulen um
Gr\"o{\ss}en\-ordnungen aufwendiger als analoge Rechnungen an Atomen.
Dementsprechend findet man beispielsweise in den klassischen
Monographien von Pauling und Wilson [1935] oder Hellmann [1937] einen
bereits relativ hoch entwickelten quantenmechanischen Formalismus sowie
zahlreiche Modellrechnungen an einfachen Systemen, aber keine
quantenmechanische Rechnungen an chemisch interessanten Molek\"ulen.
Genaue Rechnungen an gr\"o{\ss}eren Molek\"ulen waren bei den damaligen
numerischen M\"oglichkeiten praktisch nicht durchf\"uhrbar.

Die Situation \"anderte sich radikal, als in den f\"unfziger und sechziger
Jahren programmierbare Rechenanlagen allgemein zug\"anglich wurden. Die
dadurch enorm vergr\"o{\ss}erte Rechenkapazit\"at bewirkte in den
Naturwissenschaften und in der Technik eine Revolution, die immer noch
andauert, und deren gesellschaftliche Konsequenzen auch heute noch nicht
abschlie{\ss}end bewertet werden k\"onnen.

Eine unmittelbare Folge dieser Revolution war, da{\ss} genaue
quantenmechanische Rechnungen an Molek\"ulen praktisch m\"oglich wurden. In
einem autobiographischen Artikel von Clementi [1992] findet man eine
lesenswerte Beschreibung des Wechselspiels der Verbesserung der
Hardware, der Entwicklung von leistungsf\"ahigeren Molek\"ulprogrammen, die
von der verbesserten Hardware profitieren konnten, und der Durchf\"uhrung
quantenmechanischer Rechnungen an immer gr\"o{\ss}eren Molek\"ulen. In diesem
Zusammenhang d\"urfte auch ein \"Ubersichtsartikel von Bolcer und Hermann
[1994] \"uber die Entwicklung der sogenannten {\it computational
chemistry\/} in den Vereinigten Staaten von Interesse sein. Weiterhin
sei noch auf die Proceedings der Boulder Conference on Molecular Quantum
Mechanics verwiesen, die 1960 in Band 32 von Reviews of Modern Physics
ver\"offentlicht wurden. Aus historischer Sicht ist ein Artikel von
Coulson [1960] mit dem Titel {\it Present State of Molecular Structure
Calculations\/} von besonderem Interesse, der ein Abdruck des {\it
after-dinner}-Vortrages am Bankett dieses Kongresses ist.

\medskip

\Abschnitt Die interdisziplin\"are Natur des Fortschritts in der
theoretischen Chemie

\smallskip

\aktTag = 0

Ohne die Entwicklung leistungsf\"ahiger Computer h\"atte es sicherlich keine
moderne theoretische Chemie gegeben, die inzwischen in der Lage ist, auf
viele chemische Fragestellungen verl\"a{\ss}liche Antworten zu geben. Au{\ss}erdem
ist der Fortschritt auf dem Gebiet der Computertechnologie
augenscheinlich noch nicht zum Stillstand gekommen. Man kann also davon
ausgehen, da{\ss} das Leistungsverm\"ogen von Computern in den n\"achsten Jahren
noch deutlich wachsen wird, was sicherlich Folgen f\"ur die theoretische
Chemie haben wird.

Die zentrale Rolle der Computer f\"ur die theoretische Chemie zeigt sich
auch darin, da{\ss} in der englischsprachigen Literatur inzwischen h\"aufig
die Bezeichnung {\it computational chemistry\/} verwendet wird. Es w\"are
aber sicher\-lich falsch, die beeindruckenden Fortschritte der
theoretischen Chemie in den letzten Jahren {\it aus\-schlie{\ss}lich\/} auf
Fortschritte der Computertechnologie zur\"uckf\"uhren zu wollen.

Vor der allgemeinen Verf\"ugbarkeit von Computern wurden Rechnungen
entweder mit Papier und Bleistift oder mit Hilfe von Handrechenmaschinen
durchgef\"uhrt. Numerische Rechnungen waren damals sicherlich ein \"au{\ss}erst
m\"uhsames Gesch\"aft, das man soweit wie m\"oglich zu vermeiden suchte. Die
durch die Entwicklung von Computern ausgel\"oste Revolution bewirkte, da{\ss}
aufwendigere numerische Rechnungen in allen Bereichen der
Naturwissenschaft und der Technik \"uberhaupt erst praktikabel wurden.
Selbstverst\"andlich traten dabei neue, bisher unbekannte numerische und
konzeptionelle Probleme auf. Als Folge der Vergr\"o{\ss}erung der
Rechenkapazit\"at entstand nicht nur ein enormer Bedarf an
leistungsf\"ahigeren numerischen Verfahren, sondern auch an besseren
theoretischen Konzepten. Die Fortschritte der Computertechnologie
befruchteten also nicht nur die numerische Mathematik, die f\"ur unsere
computerorientierte Technologie zu einer Schl\"usseldisziplin wurde,
sondern auch mehr theoretisch orientierte Bereiche der Mathematik, der
Naturwissenschaften und der Technik.

Die Befruchtung der theoretischen Chemie durch Fortschritte in der
Computertechnologie kann am Beispiel der Elektronenkorrelation
demonstriert werden. In der Fr\"uhzeit der computerorientierten
theoretischen Chemie standen nur langsame Rechner mit f\"ur heutige
Verh\"altnisse l\"acherlich kleinen Kernspeichern zur Verf\"ugung. Selbst bei
kleinen Molek\"ulen konnte man damals aus Zeit- und Platzgr\"unden nur
schlechte Basiss\"atze verwenden. Unter diesen Umst\"anden lieferten
SCF-Rechnungen relativ ungenaue Ergebnisse, und man konnte sicher sein,
da{\ss} die bei solchen Rechnungen auftretenden Fehler wesentlich gr\"o{\ss}er
waren als die Korrelationsenergie. Die Gr\"o{\ss}e der Korrelationsenergie und
ihre Ber\"ucksichtigung in quantenmechanischen Rechnungen war damals nur
bei sehr kleinen Systemen wie etwa $\rm He$ oder $\rm H_2$ von
Bedeutung.

Bedingt durch die gro{\ss}en Fortschritte auf dem Gebiet der
Computertechnologie sowie der verwendeten numerischen Verfahren sind
aber seit einiger Zeit sehr aufwendige SCF-Rechnungen mit gro{\ss}en
Basiss\"atzen m\"oglich, die zumindest bei kleineren Systemen extrem genaue
Ergebnisse liefern. Eine Vernachl\"assigung der Elektronenkorrelation ist
unter diesen Umst\"anden nicht l\"anger {\it a priori\/} gerechtfertigt, da
die Korrelationsenergie relativ zu anderen Fehlern nicht mehr unbedingt
klein ist. Dementsprechend ist das Problem der Elektronenkorrelation
inzwischen eines der zentralen Themen der aktuellen Forschung in der
theoretischen Chemie. So spielen in einem Buch von Schaefer [1984],
in dem eine subjektive Auswahl sogenannter {\it landmark papers\/} der
theoretischen Chemie aus den Jahren 1928 - 1983 vorgestellt wird,
Artikel \"uber Korrelation eine ganz wesentliche Rolle.

Die Elektronenkorrelation ist aber auch ein Beispiel daf\"ur, da{\ss} {\it
verschiedene\/} wissenschaftliche Disziplinen sich gegenseitig
befruchten k\"onnen. Viele naturwissenschaftliche und technische Probleme
k\"onnen n\"aherungsweise auf verallgemeinerte Matrixeigenwertprobleme des
Typs
$$
{\bf A} \, {\bf x} \; = \; \lambda \, {\bf B} \, {\bf x}
\tag
$$

\noindent reduziert werden (siehe beispielsweise Cullum und Willoughby
[1985a, S. 18 - 19; 1986]). Dabei sind $\bf A$ und $\bf B$ reelle
symmetrische $n \times n$-Matrizen, $\bf x$ ist ein
$n$-di\-men\-siona\-ler Vektor und $\lambda$ ist der zugeh\"orige
Eigenwert. Aufgrund der gro{\ss}en praktischen Bedeutung verallgemeinerter
Matrix\-eigenwertprobleme gibt es auf diesem Gebiet eine umfangreiche
mathematische Literatur (siehe beispielsweise Garbow, Boyle, Dongarra
und Moler [1977], Golub und Van Loan [1983], Gurlay und Watson [1973],
Parlett [1980], Smith, Boyle, Dongarra, Garbow, Ikebe, Klema und Moler
[1976], Wilkinson [1965] und Wilkinson und Reinsch [1971]). Au{\ss}erdem
findet man in numerischen Programmbibliotheken wie IMSL oder NAG
zahlreiche Programme zur L\"osung verallgemeinerter
Matrixeigenwertprobleme.

Wenn man das Problem der Elektronenkorrelation mit Hilfe der Methode der
sogenannten Konfigurationswechselwirkung{\footnote[\dagger]{Eine
ausf\"uhrliche Beschreibung der Methode der Konfigurationswechselwirkung
zur Beschreibung der Elektronenkorrelation findet man beispielsweise in
einem \"Ubersichtsartikel von Shavitt [1977] oder in Kapitel 6 des Buches
von Harris, Monkhorst und Freeman [1992].}} behandelt, erh\"alt man
ebenfalls ein solches verallgemeinertes Eigenwertproblem, wobei die
Dimensionalit\"at $n$ des Eigenwertproblems durch die L\"ange der
Konfigurationsentwicklung determiniert wird. Allerdings tritt dabei das
Problem auf, da{\ss} die Dimensionalit\"at $n$ des Matrixeigenwertproblems in
der Gr\"o{\ss}enordnung von $10^6$ liegen kann [Davidson 1976, S. 96]. Die
\"ublicherweise verwendeten numerischen Verfahren zur L\"osung des
Eigenwertproblems funktionieren bei derartig gro{\ss}en Matrizen nicht mehr,
da der Rechenzeitbedarf als auch der ben\"otigte Speicher alle Grenzen
sprengen w\"urden (siehe beispielsweise Malmqvist [1992, S. 24]).

Zur Durchf\"uhrung extrem aufwendiger CI{\footnote[\ddagger]{Configuration
Interaction}}-Rechnungen mu{\ss}ten also zuerst spezielle Verfahren zur
L\"osung hochdimensionaler Eigenwertprobleme entwickelt werden [Davidson
1976, S. 103 - 112; Malmqvist 1992, S. 24 - 28; Shavitt 1977, S. 232 -
243]. Inzwischen wird auch von numerischen Mathematikern intensiv \"uber
extrem hochdimensionale Matrixeigenwertprobleme gearbeitet (siehe
beispielsweise Cullum und Willoughby [1985a; 1985b; 1985c; 1986] oder
Golub und Van Loan [1983, Kapitel 9]).

Was die Behandlung der Elektronenkorrelation betrifft, so hat die
theoretische Chemie methodisch sehr viel von der theoretischen Physik
profitiert. Beispielsweise gelang es {\v C}{\' \i}{\v z}ek [1966; 1969],
die aus der statistischen Mechanik stammende [Ursell 1927] und von
Coester [1958] und Coester und K\"ummel [1960] zuerst in der Kernphysik
verwendete Technik der Clusterentwicklungen so umzuformulieren, da{\ss} man
sie zur Behandlung der Elektronenkorrelation in Atomen und Molek\"ulen
verwenden konnte. Die sogenannte {\it Coupled-Cluster}-Methode geh\"ort
heute zu den Standardverfahren der theoretischen Chemie (siehe
beispielsweise Bartlett [1981], Bartlett, Dykstra und Paldus [1984],
Bartlett und Stanton [1994], Bishop [1991], Harris, Monkhorst und
Freeman [1992], K\"ummel, L\"uhrmann und Zabolitzky [1978], Kutzelnigg
[1977], Kvasni\v{c}ka, Laurinc und Biskupi\v{c} [1982] und Paldus
[1992]).

Wenn man eine CI-Rechnung unter Ber\"ucksichtigung einer sehr gro{\ss}en Zahl
von Konfigurationen durchf\"uhrt, wird die Erzeugung der sogenannten
CI-Matrix extrem aufwendig. Paldus [1974] schlug vor, zum Aufbau solcher
CI-Matrizen gruppentheoretische Methoden zu verwenden, die auf der
unit\"aren Symmetrie von Vielelektronenzust\"anden basieren. Die Verwendung
von Liegruppen in der Vielelektronentheorie geh\"ort inzwischen ebenfalls
zu den Standardverfahren der theoretischen Chemie (siehe beispielsweise
Harter und Patterson [1976], Hinze [1981], Matsen und Pauncz [1986],
Paldus [1974; 1976; 1988], Pauncz [1979] und Sutcliffe [1983]). Auch
hier gibt es Vorl\"aufer aus der theoretischen Physik, und zwar vor allem
aus der Elementarteilchentheorie, wo die Verwendung von Liegruppen schon
seit langem \"ublich ist (siehe beispielsweise Georgi [1982], Gibson und
Pollard [1976], Lichtenberg [1978], Lipkin [1967] und Ne'eman [1967]).

Inzwischen spielen Liegruppen auch in anderen Bereichen der
theoretischen Chemie eine erhebliche Rolle, beispielsweise in der
Theorie des quantenmechanischen Drehimpulses [Normand 1980], in der
Spektroskopie [Harter 1993; Judd 1968; 1985; Wybourne 1970] oder zur
Beschreibung der Eigenschaften von Atomen und Molek\"ulen [Adams, {\v
C}{\' \i}{\v z}ek und Paldus 1988; Englefield 1972; Herrick 1983; Judd
1975; Kaplan 1975; Wulfman 1971; 1979].

Eine weitere erfolgreiche Anleihe aus der theoretischen Physik sind
diagrammatische Methoden, die auf Feynman [1949] zur\"uckgehen, und die in
der Vielteilchenst\"orungstheorie von gr\"o{\ss}ter Bedeutung sind [{\v C}{\'
\i}{\v z}ek 1969; Harris, Monkhorst und Freeman 1992; Mattuck 1976;
Paldus und {\v C}{\' \i}{\v z}ek 1975; Wilson 1981; 1984; 1985; 1992a;
1992b; 1992c; 1992d; 1992e]. Inzwischen werden diagrammatische Techniken
aber auch in der Drehimpulsalgebra [Elbaz 1985; El Baz und Castel 1972;
Lindgren und Morrison 1982; Varshalovich, Moskalev und Khersonskii 1988;
Yutsis, Levinson und Vanagas 1962], in der Spektroskopie [Judd 1967] und
in der Gruppentheorie [Stedman 1990] verwendet.

Der Proze{\ss} der Befruchtung der theoretischen Chemie durch benachbarte
Disziplinen ist aber keineswegs abgeschlossen. Beispielsweise wurde in
den letzten Jahren von Herschbach, der 1986 den Nobelpreis f\"ur Chemie
erhalten hatte, die Methode der {\it dimensionalen Skalierung\/}
propagiert, die auf die Methode der dimensionalen Regularisierung von 't
Hooft und Veltman [1972] und 't Hooft [1973] zur\"uckgeht und die bisher
haupts\"achlich in der Quantenfeldtheorie verwendet wurde (siehe
beispielsweise Leibbrandt [1975] oder Narison [1982]). Dabei wird die
zeitunabh\"angige Schr\"odingergleichung in einem $D$-dimensionalen Raum
gel\"ost, wobei die Dimension $D$ so gew\"ahlt wird, da{\ss} die Rechnung
besonders einfach wird (\"ublicherweise $D = 1$ oder $D = \infty$).
Hinterher wird das Ergebnis dann st\"orungstheoretisch oder durch
dimensionale Interpolation auf die physikalische Dimension $D = 3$
transformiert. Eine genauere Beschreibung dieser Methode, weitere
Referenzen und zahlreiche Anwendungsbeispiele aus der theoretischen
Chemie findet man in einem von Herschbach, Avery und Goscinski [1993]
herausgegebenen Buch oder in einem \"Ubersichtsartikel von Herschbach
[1992].

\medskip

\Abschnitt Ziel und Inhalt dieser Arbeit

\smallskip

\aktTag = 0

Die im letzten Unterabschnitt aufgelisteten Beispiele zeigen, da{\ss}
methodische und algorithmische Fortschritte in der theoretischen Chemie
h\"aufig eine Folge der gegenseitigen Befruchtung verschiedener
wissenschaftlicher Disziplinen ist. Das Thema der vorliegenden Arbeit
ist ebenfalls weitgehend interdisziplin\"arer Natur, da mathematische
Techniken, mit deren Hilfe man schlecht konvergierenden oder sogar
divergenten Reihen einen numerischen Wert zuordnen kann, entwickelt und
auf Probleme der Quantenmechanik und der theoretischen Chemie angewendet
werden sollen.

Auf den ersten Blick mag es \"uberraschen, da{\ss} man mathematische
Fragestellungen wie Konvergenzbeschleunigung oder die Summation
divergenter Reihen im Rahmen der theoretischen Chemie behandeln m\"ochte.
Man sollte aber bedenken, da{\ss} die effiziente und genaue Berechnung
unendlicher Reihen auch heute noch ein zentrales Problem der
anwendungsorientierten Mathematik ist, und da{\ss} man keinen Grund zu der
Annahme hat, da{\ss} man in der theoretischen Chemie vor unendlichen Reihen
mit pathologischen Konvergenzeigenschaften verschont sein k\"onnte.

In einem von Roos [1992] herausgegebenen Buch, das die Beitr\"age einer
Sommerschule \"uber moderne Methoden der {\it computational chemistry\/}
enth\"alt, gibt es einen Artikel von Malmqvist [1992] mit dem Titel {\it
Mathematical Tools in Quantum Chemistry\/}. Auf S. 14 - 15 dieses
Artikels wird betont, da{\ss} asymptotische und divergente Reihen in der
quantenmechanischen St\"orungstheorie bei bestimmten Problemtypen
unvermeidlich sind{\footnote[\dagger]{Asymptotische und divergente
Reihen treten nicht nur in der quantenmechanischen St\"orungstheorie auf,
sondern auch in v\"ollig anderen Bereichen der theoretischen Physik und
der theoretischen Chemie: Sie k\"onnen weder bei der im letzten
Unterabschnitt skizzierten Methode der dimensionalen Skalierung [Goodson
und Herschbach 1992; Goodson und L\'{o}pez-Cabrera 1993; Mlodinow und
Shatz 1984; L\'opez-Cabrera, Goodson, Herschbach und Morgan 1992] noch bei
neuronalen Netzen [Albeverio, Tirrozzi und Zegarlinski 1992] vermieden
werden.}}. Die Divergenz einer St\"orungs\-reihe bedeutet nicht, da{\ss} die
Gr\"o{\ss}e, die durch die St\"orungsreihe beschrieben werden soll, undefiniert
ist oder gar divergiert, sondern nur, da{\ss} eine Reihenentwicklung der
betreffenden Gr\"o{\ss}e aus mathematischen Gr\"unden unzul\"assig ist. Trotzdem
ist es mit Hilfe geeigneter Summationsverfahren in vielen F\"allen
m\"oglich, den numerischen Wert der betreffenden Gr\"o{\ss}e aus der divergenten
St\"orungsreihe zu berechnen. Da langsam konvergente oder gar divergente
St\"orungsreihen sehr h\"aufig vorkommen, sind Methoden zur Summation
divergenter Reihen oder zur Verbesserung der Konvergenz langsam
konvergierender Reihen f\"ur die theoretische Chemie \"au{\ss}erst wichtig.

Es ist das Ziel dieser Arbeit, einige neue Verfahren zur
Konvergenzverbesserung und zur Summation divergenter Reihen zu
beschreiben, die vom Autor entwickelt wurden [Weniger 1989; 1991; 1992],
und ihr Leistungsverm\"ogen und ihre N\"utzlichkeit anhand einiger Beispiele
aus der Quantenmechanik und der theoretischen Chemie zu demonstrieren.

Das konventionelle Verfahren zur approximativen Berechnung des Wertes
einer unendlichen Reihe besteht darin, die Terme der Reihe solange
aufzuaddieren, bis Konvergenz erreicht ist. Wie in Abschnitt 2 dieser
Arbeit ausf\"uhrlich besprochen wird, f\"uhrt dieses Verfahren nur dann zum
Ziel, wenn die betreffende unendliche Reihe ausreichend schnell
konvergiert. Es sind aber viele praktisch relevante unendliche Reihen
aus der Quantenmechanik oder der theoretischen Chemie bekannt, die so
langsam konvergieren, da{\ss} eine Aufsummation der Terme praktisch
unm\"oglich ist. Au{\ss}erdem gibt es -- wie schon erw\"ahnt -- viele unendliche
Reihen, die \"uberhaupt nicht konvergieren, die aber eine Funktion
repr\"asentieren, deren Wert man bestimmen m\"ochte.

Es gibt verschiedene Verfahren, mit deren Hilfe man den Wert einer
unendlichen Reihe trotz der oben skizzierten Konvergenzprobleme mit
ausreichender Genauigkeit bestimmen kann. In dieser Arbeit werden dazu
sogenannte {\it verallgemeinerte Summationsprozesse\/} verwendet, deren
Konstruktion in Abschnitt 3 dieser Arbeit besprochen wird.

Ein Beispiel eines solchen verallgemeinerten Summationsprozesses f\"ur
Potenzreihen sind die sogenannten Pad\'e-Approximationen, die in Abschnitt
4 dieser Arbeit behandelt werden. Dabei wird die Folge der Partialsummen
einer Potenzreihe in eine zweifach indizierte Folge von rationalen
Funktionen transformiert. Auf diese Weise kann die Konvergenz vieler
Potenzreihen deutlich verbessert werden, und es ist auch m\"oglich, viele
{\it divergente\/} Potenzreihen zu summieren. Inzwischen werden
Pad\'e-Approximationen in einigen Bereichen der theoretischen Physik
routinem\"a{\ss}ig und mit gro{\ss}em Erfolg verwendet.

Das Schwergewicht dieser Arbeit liegt aber nicht auf den
Pad\'e-Approximationen, sondern auf alternativen Verfahren zur
Konvergenzverbesserung und zur Summation divergenter Reihen [Weniger
1989; 1992], die in Abschnitt 5 dieser Arbeit beschrieben werden und die
zum Teil wesentlich leistungsf\"ahiger sind als Pad\'e-Approximationen.

In Abschnitt 6 dieser Arbeit folgt eine theoretische Analyse der
Konvergenzeigenschaften verallgemeinerter Summationsprozesse, wobei das
Schwergewicht auf den in Abschnitt 5 beschriebenen Transformationen
liegt.

In quantenmechanischen und quantenchemischen Rechnungen ist man immer
wieder mit dem Problem der effizienten und verl\"a{\ss}lichen Berechnung von
speziellen Funktionen und Hilfsfunktionen konfrontiert. Dementsprechend
gibt es eine sehr umfangreiche Literatur \"uber die Berechnung spezieller
Funktionen und Hilfsfunktionen. Au{\ss}erdem enthalten numerische
Programmbibliotheken wie NAG oder IMSL zahlreiche Routinen zur
Berechnung von speziellen Funktionen. Trotzdem sind keineswegs alle
Probleme auf diesem Gebiet befriedigend gel\"ost, und es wird immer noch
sehr intensiv \"uber die Berechnung spezieller Funktionen und
Hilfsfunktionen gearbeitet. In den Abschnitten 7 und 9 dieser Arbeit
wird gezeigt, wie man die modifizierte Besselfunktion der zweiten Art
und die in Mehrzentrenmolek\"ulintegralen von Gau{\ss}funktionen vorkommende
Hilfsfunktion
$$
F_{m} (z) \; = \;
\int\nolimits_{0}^{1} \, u^{2 m} \, \exp \bigl(- z u^2 \bigr) \, \d u
\, , \qquad m \in \N_0, \quad z > 0 \, ,
\tag
$$
mit Hilfe verallgemeinerter Summationsprozesse auf effiziente Weise
berechnen kann.

Exponentialartige Funktionen sind im Prinzip wesentlich besser als
Gau{\ss}funktionen zur Approximation atomarer oder molekularer
Wellenfunktionen geeignet. Bisher haben aber die Probleme mit der
Berechnung ihrer komplizierten Mehrzentrenmolek\"ulintegrale die
routinem\"a{\ss}ige Verwendung exponentialartiger Basisfunktionen in
Molek\"ulrechnungen verhindert. In Abschnitt 8 werden verallgemeinerte
Summationsprozesse zur Auswertung einiger komplizierter und extrem
langsam konvergenter Reihenentwicklungen f\"ur Mehrzentrenmolek\"ulintegrale
exponentialartiger Basisfunktionen verwendet.

Wie schon erw\"ahnt, divergieren viele quantenmechanische St\"orungsreihen.
Besonders gut untersuchte Modellsysteme sind die anharmonischen
Oszillatoren mit einer $\hat{x}^{2 m}$-Anharmonizit\"at, deren
Schr\"odingergleichungen man nicht in geschlossener Form l\"osen kann. In
Abschnitt 10 wird gezeigt, wie man die hochgradig divergente
Rayleigh-Schr\"odingersche St\"orungsreihe f\"ur die Grundzustandsenergie
eines solchen anharmonischen Oszillators mit Hilfe von verallgemeinerten
Summationsprozessen effizient summieren kann. Bemerkenswert ist, da{\ss} es
dabei gelingt, aus den hochgradig divergenten Rayleigh-Schr\"odingerschen
St\"orungsreihen durch Summation {\it transformierte\/} St\"orungsreihen zu
konstruieren, die anscheinend f\"ur alle physikalisch relevanten Werte der
Kopplungskonstante konvergieren.

Zur Zeit wird sowohl experimentell als auch theoretisch sehr intensiv
\"uber quasi-eindimensionale Polymere gearbeitet. In Abschnitt 11 dieser
Arbeit wird am Beispiel des {\it trans}-Polyacetylens demonstriert, da{\ss}
man die Konvergenz sowohl von Kristallorbital- als auch von
Clusterrechnungen an quasi-eindimensionalen Polymeren durch
Extrapolation deutlich verbessern kann.

Diese Arbeit wird abgeschlossen mit einen Ausblick auf verwandte
numerische Methoden, die ebenfalls f\"ur die theoretische Chemie von
Interesse sein d\"urften.

\endAbschnittsebene

\endAbschnittsebene

\keinTitelblatt\neueSeite

\beginAbschnittsebene
\aktAbschnitt = 1

\Abschnitt Unendliche Reihen und ihre numerische Auswertung

\vskip - 2 \jot

\beginAbschnittsebene

\medskip

\Abschnitt Konvergenzprobleme bei unendlichen Reihen

\smallskip

\aktTag = 0

Unendliche Reihen und der damit verbundene Konvergenzbegriff waren von
zentraler Bedeutung bei der Entwicklung der modernen Differential- und
Integralrechnung. Einen guten \"Uberblick \"uber die historische Entwicklung
findet man in Kapitel 20 des sehr lesenswerten Buches {\it Mathematical
Thought from Ancient to Modern Times} von Kline [1972].

Bei allen praktischen Anwendungen der Mathematik und insbesondere bei
der mathematischen Beschreibung naturwissenschaftlicher und technischer
Probleme sind unendliche Reihen von gr\"o{\ss}ter Wichtigkeit. Wenn man
beispielsweise die Integraltafeln von Gradshteyn und Ryzhik [1980]
aufschl\"agt, so stellt man fest, da{\ss} die meisten der dort aufgelisteten
Integrale entweder explizit oder implizit durch unendliche Reihen
dargestellt werden. Unendliche Reihen spielen ebenfalls eine sehr
wesentliche Rolle bei der L\"osung von Differential- und
Integralgleichungen sowie in der quantenmechanischen St\"orungstheorie.
Weiterhin werden unendliche Reihen h\"aufig sowohl zur Definition als
auch zur Berechnung spezieller Funktionen der mathematischen Physik
verwendet.

Dementsprechend umfangreich ist die mathematische Literatur \"uber
unendliche Reihen. Abgesehen davon, da{\ss} unendliche Reihen in jedem
Lehrbuch der Analysis mehr oder weniger ausf\"uhrlich behandelt werden,
gibt es zahlreiche spezielle Monographien \"uber dieses Thema. Beispiele
sind die B\"ucher von Bromwich [1926], D\"orrie [1951], Ferrar [1938],
Fichtenholz [1970a; 1970b; 1970c], Hirschman [1962], Knopp [1964],
Kuntzmann [1971], Lense [1953], Meschkowski [1962; 1963], Rainville
[1967], Ruiz [1993], und Smith [1987]. Au{\ss}erdem gibt es zahlreiche
Formelsammlungen und \"ahnliche B\"ucher, die Material \"uber unendliche
Reihen enthalten (siehe beispielsweise Davis [1962], Hansen [1975], Ross
[1987] und Wheelon [1968]).

Die konventionelle Vorgehensweise zur approximativen Berechnung des
Wertes einer unendli\-chen Reihe besteht darin, da{\ss} man f\"ur $n = 0, 1, 2,
\ldots$ die Partialsummem
$$
s_n \; = \; \sum_{k=0}^{n} \, a_k
\tag
$$
durch Aufaddieren der Terme $a_k$ berechnet. Durch Vergleich der bisher
berechneten Partialsummen versucht man dann, die Gr\"o{\ss}e des
Abbruchfehlers
$$
\sum_{k=n+1}^{\infty} \, a_k
\tag
$$
abzusch\"atzen. Wenn der gesch\"atzte Abbruchfehler der letzten Elemente der
Folge $s_0$, $s_1$, $\ldots$ , $s_n$ noch nicht unterhalb der
vorgegebenen Fehlerschranke liegt, m\"ussen weitere Terme $a_{n+1},
a_{n+2}, \ldots$ aufaddiert werden, bis man schlie{\ss}lich die geforderte
Genauigkeit erreicht hat. Auf diese Weise ist es wenigstens im Prinzip
m\"oglich, den Wert einer konvergenten unendlichen Reihe mit einer
beliebig gro{\ss}en vorgegebenen Genauigkeit zu berechnen. Voraussetzung
daf\"ur ist nat\"urlich, da{\ss} man in der Lage ist, ausreichend viele Terme
$a_m$ mit einer so gro{\ss}en Stellenzahl zu berechnen, da{\ss} eventuell
auftretende numerische Instabilit\"aten keine unl\"osbaren Probleme
aufwerfen.

Ungl\"ucklicherweise sind unendliche Reihen, die bei
naturwissenschaftlichen oder technischen Problemen auftreten, oft so
kompliziert, da{\ss} man nur eine relativ kleine Anzahl von Reihentermen
berechnen kann, was zu einer erheblichen Einschr\"ankung der prinzipiell
erreichbaren Genauigkeit f\"uhrt.

Eine weitere, h\"aufig auftretende Komplikation ist, da{\ss} insbesondere die
Reihenterme $a_m$ mit h\"oheren Indizes $m$ oft nicht mehr genau genug
berechnet werden k\"onnen. Werden solche vergleichsweise ungenauen Terme
zur Berechnung des Wertes einer langsam konvergierenden unendlichen
Reihe verwendet, kann es unter ung\"unstigen Umst\"anden zu einer
katastrophalen Akkumulation von Rundungsfehlern kommen.

Zusammenfassend kann man sagen, da{\ss} der Wert einer unendlichen Reihe
nur dann auf konventionelle Weise durch Aufaddieren der Terme mit
hinreichender Genauigkeit bestimmt werden kann, wenn die Reihe relativ
schnell konvergiert. Ungl\"ucklicherweise trifft dies bei zahlreichen
praktisch relevanten Reihenentwicklungen nicht zu.

Typische Konvergenzprobleme, wie sie oft bei der mathematischen
Beschreibung naturwissenschaftlicher oder technischer Probleme
auftreten, k\"onnen anhand der beiden folgenden einfachen Beispiele
demonstriert werden:

\medskip

\beginEingezogeneBeschreibung \zu \Laenge{{\Roemisch 2}:}

\item {{\Roemisch 1}:} Die geometrische Reihe
$$
\frac {1} {1 - z} \; = \; \sum_{m=0}^{\infty} \, z^m \, .
\tag
$$

\item {{\Roemisch 2}:} Die Reihenentwicklung f\"ur die Riemannsche
Zetafunktion
$$
\zeta (z) \; = \; \sum _{m=0}^{\infty} \; (m+1)^{-z} \, .
\tag
$$

\endEingezogeneBeschreibung

\medskip

W\"urde man den Wert von $1/(1-z)$ mit Hilfe der geometrischen Reihe
berechnen wollen, so w\"are das auf effiziente Weise nur m\"oglich, wenn
$\vert z \vert$ deutlich kleiner als Eins ist. Sobald $\vert z \vert$
sich an Eins ann\"ahert, wird die Konvergenz der geometrischen Reihe trotz
ihrer Einfachheit prohibitiv langsam. Das folgt sofort aus der folgenden
Beziehung f\"ur den Abbruchfehler:
$$
\frac {1} {1 - z} \, - \, \sum_{m=0}^{n} \, z^m \; = \;
\frac {z^{n+1}} {1 - z} \, .
\tag
$$
Da die geometrischen Reihe nur f\"ur $\vert z \vert < 1$ konvergiert, mu{\ss}
man eine {\it analytische Fortsetzung} finden, wenn man den Wert von
$1/(1-z)$ au{\ss}erhalb des Konvergenzkreises mit Hilfe der unendlichen
Reihe (2.1-3) berechnen will. Im Falle der geometrischen Reihe ist das
nat\"urlich eine Trivialit\"at, da die analytische Fortsetzung durch
$1/(1-z)$ gegeben ist.

Die geometrische Reihe dient hier als Modell f\"ur Potenzreihen mit einem
endlichen Konvergenzradius. In der Regel ist es \"au{\ss}erst schwierig, den
Wert einer Potenzreihe in der N\"ahe des Randes des Konvergenzkreises, wo
eine solche Reihe sehr schlecht konvergiert, oder gar au{\ss}erhalb des
Konvergenzkreises, wo sie divergiert und wo man eine analytische
Fortsetzung finden mu{\ss}, mit ausreichender Genauigkeit zu berechnen. Eine
\"Ubersicht \"uber verschiedene Methoden zur analytischen Fortsetzung im
Zusammenhang mit quantenmechanischen Resonanzph\"anomenen findet man in
Abschnitt 1.3 des Buches von Kukulin, Krasnopol'sky, und Hor\'a\v{c}ek
[1989].

In der Literatur \"uber Konvergenzbeschleunigung sagt man, da{\ss} eine Folge
$\Seqn s$ {\it linear} gegen ihren Grenzwert $s$ konvergiert, wenn
$$
\lim_{n \to \infty} \, \frac {s_{n+1} - s} {s_n - s} \; = \;
\rho \, , \qquad 0 < |\rho | < 1 \, ,
\tag
$$
gilt. Mit Hilfe des Cauchyschen Quotientenkriteriums [Knopp 1964, S.
119] kann leicht gezeigt werden, da{\ss} sowohl die Partialsummen der
geometrischen Reihe (2.1-3) als auch die Partialsummen einer Potenzreihe
mit endlichem Konvergenzradius innerhalb ihrer Konvergenzkreise linear
konvergieren.

Wenn eine Potenzreihe einen von Null verschiedenen, aber endlichen
Konvergenzradius hat, so mu{\ss} die Funktion, die durch die Potenzreihe im
Inneren des Konvergenzkreises repr\"asentiert wird, auf dem Rand des
Konvergenzkreises mindestens eine Singularit\"at besitzen. Bei zahlreichen
physikalischen Systemen, die durch Potenzreihen mit endlichen
Konvergenzradien beschrieben werden, liefert die Lage solcher
Singularit\"aten wesentliche Informationen. Ein Beispiel f\"ur ein solches
System sind idealisierte Fl\"ussigkeiten, die laut Voraussetzung aus exakt
scheiben- oder kugelf\"ormigen inkompressiblen Molek\"ulen bestehen, und die
durch Virialentwicklungen f\"ur den Quotienten $p /(\rho k T)$ beschrieben
werden [Baram und Luban 1979]. Dabei sind die Virialentwicklungen
Potenzreihen in der Teilchendichte $\rho = N / V$. Baram und Luban
wollten zeigen, da{\ss} die Virialentwicklung f\"ur eine solche idealisierte
Fl\"ussigkeit eine monoton wachsende Funktion der Teilchenchendichte
$\rho$ repr\"asentiert, die singul\"ar wird, sobald eine Dichte $\rho_c$
erreicht ist, die der dichtesten Packung der idealisierten Molek\"ule
entspricht. Hier tritt allerdings das Problem auf, da{\ss}
Potenzreihenentwicklungen mit endlichem Konvergenzradius normalerweise
mit Ausnahme von endlich vielen Punkten auf dem gesamten Rand des
Konvergenzkreises divergieren [Knopp 1964, S. 415]. Die genaue
Lokalisierung der tats\"achlichen Singularit\"aten der Funktion, die durch
eine solche Virialentwicklung repr\"asentiert wird und die im Prinzip auch
komplexe Singularit\"aten besitzen kann, ist demzufolge nicht m\"oglich,
wenn man nur die unendliche Reihe zur Verf\"ugung hat. Baram und Luban
l\"osten dieses Problem, indem sie aus den Partialsummen ihrer
Virialentwicklungen rationale Approximationen konstruierten und deren
Singularit\"aten bestimmten.

Ein anderes Konvergenzverhalten zeigt die Reihe (2.1-4) f\"ur die
Riemannsche Zetafunktion, die konvergiert, wenn $Re (z) > 1$ gilt. Ist
aber $Re (z)$ nur wenig gr\"o{\ss}er als Eins, konvergiert die Reihe (2.1-4)
so langsam, da{\ss} es auch auf den schnellsten Computern praktisch
unm\"oglich ist, $\zeta (z)$ auf diese Weise zu berechnen. Man kann
zeigen, da{\ss} der Abbruchfehler der Reihe (2.1-4) die folgende
Ordnungsrelation erf\"ullt [Wimp 1981, S. 20]:
$$
\zeta (z) \, - \, \sum _{m=0}^{n-1} \, (m+1)^{-z}
\; = \; O (n^{1 - z}) \, , \qquad n \to \infty \, .
\tag
$$
Mit Hilfe dieser Absch\"atzung folgerten Bender und Orszag [1978, S. 379],
da{\ss} man etwa $10^{20}$ Terme der Reihe (2.1-4) ben\"otigen w\"urde, um
$\zeta (1.1)$ mit einer relativen Genauigkeit von einem Prozent zu
berechnen. Trotzdem ist die Reihe (2.1-4) numerisch nicht nutzlos.
Bender und Orszag [1978] zeigten n\"amlich in Tabelle 8.7 auf S. 380 ihres
Buches, da{\ss} man $\zeta (1.1)$ mit einer Genauigkeit von 26 Stellen
berechnen kann, indem man die ersten 10 Terme der unendlichen Reihe
(2.1-4) in einem speziell auf die Riemannsche Zetafunktion
zugeschnittenen Konvergenzbeschleunigungsverfahren verwendet, welches
bestimmte Eigenschaften der Riemannschen Zetafunktion und der
Reihenentwicklung (2.1-4) geschickt ausn\"utzt.

In der Literatur \"uber Konvergenzbeschleunigung sagt man, da{\ss} eine Folge
$\Seqn s$ {\it logarithmisch} gegen ihren Grenzwert $s$ konvergiert, wenn
$$
\lim_{n \to \infty} \, \frac {s_{n+1} - s} {s_n - s} \; = \; 1
\tag
$$
gilt. Ein Vergleich der Gln. (2.1-7) und (2.1-8) zeigt, da{\ss} die
Partialsummen der Reihe f\"ur die Riemannsche Zetafunktion logarithmisch
konvergieren.

Logarithmisch konvergente Reihen und Folgen treten in der Praxis sehr
h\"aufig auf. Ein Beispiel ist die Berechnung des Bethe-Logarithmus [Bethe
1947] f\"ur den $1s$-Zustand des Wasserstoffatoms mit Hilfe eines
diskreten Basissatzes durch Haywood und Morgan [1985]. Dabei fanden
Haywood und Morgan, da{\ss} auch bei Verwendung von 120 Basisfunktionen nur
eine Genauigkeit von 2 -- 3 Stellen erreicht werden konnte. Eine
genauere Analyse der Abh\"angigkeit des Abbruchfehlers von der Anzahl $N$
der Basisfunktionen ergab, da{\ss} der Abbruchfehler f\"ur gro{\ss}e Werte von $N$
eine Ordnungsabsch\"atzung des Typs $O (1/N)$ erf\"ullt. Diese Absch\"atzung
impliziert nicht nur, da{\ss} die mit $N$ Basisfunktionen erhaltenen
Approximationen logarithmisch gegen den Bethe-Logarithmus konvergieren,
sondern auch, da{\ss} die von Haywood und Morgan verwendeten diskreten
Basisfunktionen eigentlich nicht geeignet sind, um den Bethe-Logarithmus
des Wasserstoffatoms mit ausreichender Genauigkeit zu berechnen. Man
w\"urde n\"amlich etwa $10^{10}$ Basisfunktionen ben\"otigen, um eine
Genauigkeit von 10 Dezimalstellen zu erreichen. Haywood und Morgan
zeigten aber auch, da{\ss} man aus den gleichen Daten, die nur eine
Genauigkeit von 2 -- 3 Stellen lieferten, den Bethe-Logarithmus mit eine
Genauigkeit von 13 Dezimalstellen berechnen kann, wenn man sie als
Eingabedaten f\"ur das Richardsonsche Extrapolationsverfahren [Richardson
1927] verwendet.

\medskip

\Abschnitt Divergente Reihen

\smallskip

\aktTag = 0

Die Unzul\"anglichkeit des konventionellen Verfahrens zur Berechnung des
Wertes einer unendlichen Reihe wird besonders deutlich, wenn man {\it
divergente} Reihen betrachtet. \"Ublicherweise wird eine unendliche Reihe
als divergent bezeichnet, wenn die Folge ihrer Partialsummen nicht gegen
einen festen Wert konvergiert.

Wie im letzten Abschnitt erw\"ahnt wurde, gibt es zahlreiche praktisch
relevante Potenzreihen mit von Null verschiedenen, aber endlichen
Konvergenzradien. Wenn man den Wert einer solchen Potenzreihe au{\ss}erhalb
ihres Konvergenzkreises bestimmen m\"ochte, ist man mit dem Problem
konfrontiert, da{\ss} die Folge der Partialsummen divergiert. Man mu{\ss} also
zuerst eine analytische Fortsetzung der Funktion finden, die im Inneren
des Konvergenzkreises durch die Potenzreihe repr\"asentiert wird.
Zahlreiche Verfahren zur analytischen Fortsetzung von Potenzreihen sind
bekannt. Ausf\"uhrliche Darstellungen findet man beispielsweise in den
B\"uchern von Henrici [1974], Kukulin, Krasnopol'sky und Hor\'a\v{c}ek
[1989] und Markushevich [1977].

Potenzreihen mit von Null verschiedenen, aber endlichen Konvergenzradien
divergieren au{\ss}erhalb ihrer Konvergenzkreise nur relativ schwach. Es
gibt allerdings auch zahlreiche Potenzreihen, deren Koeffizienten so
rasch anwachsen, da{\ss} sie f\"ur jedes von Null verschiedene Argument
divergieren, was bedeutet, da{\ss} sie den Konvergenzradius Null besitzen.
Das in diesem Zusammenhang klassische Beispiel ist das nach Euler
benannte Integral [Hardy 1949, S. 26, Gl. (2.4.4)]
$$
E(z) \; = \;
\int\nolimits^{\infty}_{0} \frac {\e^{-t} \, \d t} {1 + z t } \, .
\tag
$$
Wenn man den Nenner dieses Integrals unter Verwendung von Gl. (2.1-3) in
eine Reihe entwickelt und gliedweise integriert, obwohl die
Reihenentwicklung nur f\"ur $\vert z t \vert < 1$ konvergiert, erh\"alt man
eine divergente Potenzreihe, die sogenannte Eulerreihe [Hardy 1949, S.
26, Gl. (2.4.1)]
$$
E(z) \; \sim \; \sum _{n = 0}^{\infty} \; (-1)^{n} \, n! \, z^n
\; = \; {}_2 F_0 (1,1;- z) \, ,
\qquad z \to 0 \, .
\tag
$$
Offensichtlich divergiert die Eulerreihe (2.2-2) f\"ur jedes Argument $z
\ne 0$. Auf den ersten Blick ist es schwer vorstellbar, da{\ss} derartige
hochgradig divergente Reihen in der angewandten Mathematik oder bei der
mathematischen Naturbeschreibung eine produktive Rolle spielen k\"onnten.
Diese einleuchtende Vorstellung ist aber falsch. Beispielsweise wird in
dem Buch von Bowman und Shenton [1989] gezeigt, da{\ss} hochgradig
divergente Reihen in der Statistik keineswegs un\"ublich sind. Au{\ss}erdem
wird in sp\"ateren Abschnitten noch ausf\"uhrlich diskutiert, da{\ss} divergente
Reihen, die \"ahnlich heftig wie die Eulerreihe oder sogar noch
wesentlich st\"arker divergieren, bei zahlreichen quantenmechanischen
St\"orungsentwicklungen auftreten [Arteca, Fern\'{a}ndez und Castro 1990;
{\v C}{\' \i}{\v z}ek und Vrscay 1982; Kazakov und Shirkov 1980;
Killingbeck 1977; Le Guillou und Zinn-Justin 1990; Simon 1982; 1991;
Wilcox 1966; Zinn-Justin 1981a].

Ein Beispiel sind die anharmonischen Oszillatoren, die durch den
Hamiltonoperator
$$
\hat{H}^{(m)} (\beta) \; = \; \hat{p}^2 + \hat{x}^2 +
\beta \hat{x}^{2m} \, , \qquad m = 2, 3, 4, \ldots \, ,
\tag
$$
beschrieben werden, wobei $\hat{p} = - \i \d / \d x$ der Impulsoperator
ist. In der physikalischen und auch in der physikalisch-chemischen
Literatur werden anharmonische Oszillatoren sehr h\"aufig behandelt, da
sie in vielen Zweigen der Physik und physikalischen Chemie wichtige
Modellsysteme darstellen, die trotz ihrer Einfachheit \"au{\ss}erst
interessante und auch teilweise \"uberaus schwierige mathematische und
konzeptionelle Probleme aufwerfen. Die Rayleigh-Schr\"odingersche
St\"orungsreihe f\"ur die Grundzustandsenergie dieses Hamiltonoperators ist
eine Potenzreihe,
$$
E^{(m)}( \beta ) \; = \;
\sum_{n=0}^{\infty} \, b_{n}^{(m)} \, \beta^{n} \, ,
\tag
$$
die f\"ur jede Kopplungskonstante $\beta \ne 0$ sehr stark divergiert.
Bender und Wu [1969; 1971; 1973] konnten zeigen, da{\ss} die Koeffizienten
$b_{n}^{(m)}$ dieser St\"orungsreihe f\"ur gro{\ss}e Indizes $n$ das folgende
asymptotische Verhalten besitzen:
$$
\beginAligntags
" b_{n}^{(2)} \, " \sim \, " (-1)^{n+1} \,
\frac{(24)^{1/2}}{\pi^{3/2}} \, \Gamma(n+1/2) \, (3/2)^{n} \, ,
\erhoehe\aktTag \\ \tag*{\tagnr a}
" b_{n}^{(3)} \, " \sim \, " (-1)^{n+1} \,
\frac{(128)^{1/2}}{\pi^{2}} \, \Gamma(2 n + 1/2) \,
(16/\pi^{2})^{n} \, ,
\\ \tag*{\tagform\aktTagnr b}
" b_{n}^{(4)} \, " \sim \, " (-1)^{n+1} \,
\left\{
\frac{270 \, [ \Gamma(2/3)]^{3}}{\pi^{5}} \right\}^{1/2} \,
\Gamma (3 n + 1/2) \, (250)^{n} \,
\left\{ \frac{3 \, [ \Gamma (2/3)]^{3}}{4 \pi ^{2}} \right\}^{3n} \, .
\\ \tag*{\tagform\aktTagnr c}
\endAligntags
$$
Aus Gl. (2.2-5a) folgt, da{\ss} der Koeffizient $b_{n}^{(2)}$ der
St\"orungsreihe f\"ur die $\hat{x}^4$-Anharmonizit\"at f\"ur gro{\ss}e $n$ im
wesentlichen wie $n! / n^{1/2}$ w\"achst. Demzufolge divergiert die
St\"orungsreihe (2.2-4) f\"ur $m = 2$ etwas weniger stark als die Eulerreihe
(2.2-2). Im Falle einer $\hat{x}^6$- oder $\hat{x}^8$-Anharmonizit\"at
divergiert die St\"orungsreihe (2.2-4) allerdings wesentlich st\"arker als
die Eulerreihe, da $b_{n}^{(3)}$ und $b_{n}^{(4)}$ f\"ur $n \to \infty$ im
wesentlichen wie $(2 n)! / n^{1/2}$ beziehungsweise $(3 n)! / n^{1/2}$
wachsen. Diese Beispiele zeigen, da{\ss} man ohne die Theorie divergenter
Reihen in der Regel weder die Natur quantenmechanischer
St\"orungsentwicklungen verstehen noch die numerisch nutzbare Information,
die in solchen St\"orungsreihen trotz ihrer Divergenz enthalten ist,
extrahieren kann.

Divergente Reihen und ihre Verwendung waren in der Mathematik lange ein
\"uberaus kontroverses Thema. Einen guten \"Uberblick \"uber die
geschichtliche Entwicklung findet man in Kapitel 47 des Buches von Kline
[1972].

Zu Beginn des achtzehnten Jahrhunderts wurde noch keine genaue
Unterscheidung zwischen konvergenten und divergenten Reihen gemacht. Auf
S. 460 seines Buches bemerkt Kline [1972], da{\ss} Newton, Leibniz, Euler
und sogar noch Lagrange unendliche Reihen im wesentlichen wie endliche
Summen behandelten, und da{\ss} sie die zus\"atzlichen Probleme, die durch die
Erweiterung auf einen unendlichen Summationsproze{\ss} auftreten konnten,
weitgehend ignorierten. Die bei dieser sorglosen Haltung gelegentlich
auftretenden Absurdit\"aten{\footnote[\dagger]{Beispielsweise wurden
offensichtlich absurde Ergebnisse, die mit Hilfe unzul\"assiger
Manipulationen an divergenten Reihen abgeleitet worden waren, von Grandi
in einem Brief an Leibniz als Beweis daf\"ur bezeichnet, da{\ss} Gott in der
Lage sei, die Welt aus dem Nichts zu erschaffen [Heuser 1981, S. 683;
Kline 1972, S. 446].}} erzwangen allerdings im Verlauf des achtzehnten
Jahrhunderts eine Unterscheidung in konvergente und divergente Reihen,
wie sie heute \"ublich ist.

Laut Kline [1972, S. 462 - 463] war sich Euler, der mit divergenten
Reihen auf virtuose, wenn auch nicht immer auf mathematisch ganz
korrekte Weise umzugehen wu{\ss}te, dar\"uber im klaren, da{\ss} man den Wert
einer divergenten Reihe nicht wie bei einer konvergenten Reihe durch
Aufsummieren ihrer Terme bestimmen kann. Euler betonte aber, da{\ss} es in
vielen F\"allen m\"oglich ist, einer divergenten Reihe mit Hilfe eines {\it
verallgemeinerten Summationsprozesses} trotzdem einen Wert zuzuweisen.
Es scheint, da{\ss} Euler eine divergente Reihe immer dann als mathematisch
sinnvolles Objekt gelten lie{\ss}, wenn sie auf nat\"urliche Weise durch
Entwicklung eines mathematisch wohldefinierten Objektes zustandegekommen
war, dem man problemlos einen numerischen Wert zuordnen konnte.
Dementsprechend {\it summierte} Euler die hochgradig divergente
Eulerreihe (2.2-2) dadurch, da{\ss} er ihr den Wert des Integrals (2.2-1)
zuordnete.

Konzeptionell ist damit das Problem der Summation der divergenten
Eulerreihe im Prinzip gel\"ost. Allerdings ist diese Vorgehensweise bei
praktischen Problemen nicht sehr hilfreich, da normalerweise die
Funktion, mit der man die divergente Reihe identifizieren will, nicht
explizit bekannt ist. Ein praktisch n\"utzlicher verallgemeinerter
Summationsproze{\ss} m\"u{\ss}te also in der Lage sein, unter der ausschlie{\ss}lichen
Verwendung der Terme oder der Partialsummen der divergenten Eulerreihe
Approximationen zu liefern, die gegen den Wert des Eulerschen Integrals
konvergieren. Inzwischen sind zahlreiche Verfahren bekannt, die dazu in
der Lage sind, und die auch noch die wesentlich st\"arker divergierenden
St\"orungsreihen f\"ur den anharmonischen Oszillator mit einer $\hat{x}^6$-
oder $\hat{x}^8$-Anharmonizit\"at summieren k\"onnen. In Abschnitt 5 dieser
Zusammenfassung werden einige besonders leistungsf\"ahige, vom Autor
entwickelte Verfahren vorgestellt.

Am Anfang des neunzehnten Jahrhunderts, nach der Kl\"arung des
Konvergenzbegriffes durch Cauchy, wurden divergente Reihen weitgehend
aus der Mathematik verbannt. Abel, der der Sohn eines Pfarrers war,
bezeichnete sie sogar als Erfindungen des Teufels (siehe das Vorwort von
Littlewood in Hardy [1949]). Trotzdem wurden divergente Reihen aufgrund
ihrer unbestreitbaren praktischen N\"utzlichkeit auch weiterhin verwendet,
insbesondere in der Astronomie, wo sie weitgehend unverzichtbar waren.

Ein Beispiel f\"ur eine praktisch \"uberaus n\"utzliche divergente Reihe ist
die Stirlingsche Formel f\"ur den Logarithmus der Gammafunktion [Magnus,
Oberhettinger, und Soni 1966, S. 12], die f\"ur gro{\ss}e Argumente der
Gammafunktion ausgezeichnete Approximationen liefert. Diese Eigenschaft
macht die Stirlingsche Formel sogar in gewisser Weise zur Grundlage der
statistischen Mechanik: Ihre Verwendung beim formalen Aufbau der
statistischen Mechanik macht ganz wesentliche Vereinfachungen m\"oglich,
ohne da{\ss} merkliche Fehler zu bef\"urchten sind.

Inspiriert durch die erfolgreiche Verwendung divergenter Reihen wurden
am Ende des neunzehnten Jahrhunderts neue mathematische Konzepte
entwickelt, n\"amlich die Theorie {\it asymptotischer Reihen} im Sinne
von Poincar\'e [1886] und die Theorie der {\it Summationsverfahren}.

Mit Hilfe der Theorie asymptotischer Potenzreihen konnte man erkl\"aren,
warum divergente Reihen unter bestimmten Bedingungen \"au{\ss}erst genaue
Approximationen liefern k\"onnen. Eine Darstellung dieser auch f\"ur
Konvergenzbeschleunigungs- und Summationsverfahren \"uberaus wich\-tigen
Theorie sowie zahlreiche Anwendungen findet man beispielsweise in den
Monographien von Bleistein und Handelsman [1986], Copson [1965], de
Bruijn [1981], Dingle [1973], Erd\'elyi [1956], Ford [1960], Murray
[1984], Olver [1974], Wasow [1987], und Wong [1989]. Unkonventionelle
Methoden in der Asymptotik werden in dem Buch von van den Berg [1987]
beschrieben. Einen \"Uberblick \"uber neuere Entwicklungen auf dem Gebiet
der Asymptotik findet man in einem \"Ubersichtsartikel von Wimp [1991].

Mit Hilfe der Theorie der Summationsverfahren ist es in vielen F\"allen
m\"oglich, einer unendlichen Reihe, die im Sinne von Cauchy divergiert,
durch einen verallgemeinerten Summationsproze{\ss} einen endlichen Wert
zuzuweisen. Eine \"Ubersicht \"uber die \"altere Theorie dieser
verallgemeinerten Summationsprozesse, die mit den Namen Poisson,
Frobenius, H\"older, Ces\`{a}ro, Laguerre, Stieltjes und Borel verkn\"upft
sind, findet man in den Monographien von Borel [1928], Ford [1960],
Hardy [1949], Peyerimhoff [1969], Powell und Shah [1988], Szasz [1952],
und Van Vleck [1905].

Bei asymptotischen Reihen im Sinne von Poincar\'e gibt es ein sehr
unangenehmes Eindeutigkeitsproblem: Zwar kann man leicht zeigen, da{\ss}
die asymptotische Entwicklung einer Funktion eindeutig ist, wenn sie
existiert. Aber man kann auch zeigen, da{\ss} verschiedene Funktionen die
gleiche asymptotische Reihe besitzen k\"onnen. Wenn man einen
Summationsproze{\ss} auf eine divergente asymptotische Reihe anwendet, ist
man deswegen immer mit der unangenehmen Frage konfrontiert, ob die so
erhaltene Funktion eindeutig bestimmt ist. Wie in Abschnitt
\Roemisch{11} des \"Ubersichtsartikels von Simon [1972] oder in Abschnitt
\Roemisch{12}.4 des Buches von Reed und Simon [1978] diskutiert wird,
konnten aber mit Hilfe des Watsonschen Lemmas [Watson 1912a] und des
Carlemanschen Theorems [Carleman 1926] zus\"atzliche Bedingungen
formuliert werden, die garantieren, da{\ss} man einer divergenten
asymptotischen Reihe wie etwa der Eulerreihe (2.2-2) eindeutig eine
Funktion zuordnen kann.

Die Theorie asymptotischer und divergenter Reihen und ihrer Summation
ist ein mathematisches Forschungsgebiet, welches inzwischen einen hohen
Entwicklungsstand erreicht hat, auf dem aber auch heute noch intensiv
gearbeitet wird [Balser 1992; Balser, Braaksma, Ramis und Sibuya 1991;
Berry 1989; 1991; Berry und Howls 1990; 1991; Boyd 1990a; 1990b;
Braaksma [1992]; Jones 1990; Malgrange und Ramis 1992; Martinet und
Ramis 1991; Olde Daalhuis 1992; 1993; Olver 1990; 1991a; 1991b; Thomann
1990]. Au{\ss}erdem sind divergente Reihen in der Statistik [Bowman und
Shenton 1989] sowie in einigen Bereichen der Quantenmechanik und der
Quantenfeldtheorie unverzichtbar. Trotzdem gibt es zahlreiche
Mathematiker, die divergenten Reihen immer noch sehr kritisch
gegen\"uberstehen, woran die sehr sorglose und mathematisch wenig strenge
Verwendung divergenter Reihen in der Quantenmechanik sicherlich nicht
unschuldig ist. Die Einstellung vieler Mathematiker wird meiner Meinung
nach durch das folgende, aus dem Jahr 1941 stammende Zitat von Haldane
[K\"orner 1988, S. 426] treffend charakterisiert:

\medskip

\beginSchmaeler
\noindent {\sl Cambridge is full of mathematicians who have been so
corrupted by quantum mechanics that they use series which are clearly
divergent, and not even proved to be summable}
\endSchmaeler

\medskip

\Abschnitt Das Summationsverfahren von Borel

\smallskip

\aktTag = 0

Ein vor allem theoretisch \"au{\ss}erst wichtiges Verfahren zur Summation
divergenter Reihen wurde vor mehr als 90 Jahren von Borel [1899]
vorgeschlagen. Mit Hilfe dieses Verfahrens k\"onnen auch hochgradig
divergente Potenzreihen summiert werden, indem man ihnen
Laplaceintegrale zuordnet, die -- wenn sie konvergieren -- analytische
Funktionen des Argumentes der Potenzreihe sind. Voraussetzung f\"ur die
Anwendbarkeit des Borelschen Summationsverfahrens ist, da{\ss} die
Koeffizienten der divergenten Potenzreihe bestimmte globale
Wachstumsbedingungen erf\"ullen.

Die Wirkungsweise des Summationsverfahrens von Borel kann besonders
einfach anhand der Eulerreihe (2.2-2) demonstriert werden. Der
Ausgangspunkt ist die folgende Integraldarstellung f\"ur die Fakult\"at:
$$
n! \; = \; \int\nolimits_{0}^{\infty} \, t^n \, \e^{- t} \, \d t \, .
\tag
$$
Wenn man diese Integraldarstellung in die Eulerreihe (2.2-2) einsetzt,
erh\"alt man nach Vertauschung der Reihenfolge von Summation und
Integration:
$$
\beginAligntags
" E(z) " \; \sim \;
" \sum _{n = 0}^{\infty} \; (-1)^{n} \, z^n \,
\int\nolimits_{0}^{\infty} \, t^n \, \e^{- t} \, \d t
\\ \tag
" " \; = \; " \int\nolimits_{0}^{\infty} \,
\sum _{n = 0}^{\infty} \; (-z t)^{n} \, \, \e^{- t} \, \d t \, .
\\ \tag
\endAligntags
$$
Die geometrische Reihe im Integranden kann gem\"a{\ss} Gl. (2.1-3) geschlossen
aufsummiert werden:
$$
\frac {1} {1 + z t} \; = \; \sum_{n=0}^{\infty} \, (- z t)^n \, .
\tag
$$
Wenn man diese Beziehung in Gl. (2.3-3) einsetzt, sieht man sofort, da{\ss}
die Eulerreihe (2.2-2) offensichtlich durch das nach Euler benannte
Integral (2.2-1) summiert werden kann:
$$
E(z) \; \sim \; \sum _{n = 0}^{\infty} \; (-1)^{n} \, n! \, z^n
\; = \;
\int\nolimits^{\infty}_{0} \frac {\e^{-t} \, \d t} {1 + z t } \, .
\tag
$$
Mit Hilfe der Integraldarstellung (2.3-1) f\"ur die Fakult\"at erh\"alt man
also durch Vertauschung der Reihenfolge von Summation und Integration in
Verbindung mit der analytischen Fortsetzung (2.3-4) aus der formalen
Potenzreihe (2.2-2), die f\"ur jedes Argument $z \ne 0$ {\it divergiert},
eine Darstellung durch ein {\it konvergentes} Laplaceintegral, das
problemlos mit Hilfe von Quadraturverfahren berechnet werden kann.

Diese Vorgehensweise, bei der einer divergenten Potenzreihe ein
konvergentes Integral zugeordnet wird, ist aber, wie Borel [1899]
erkannte, nicht auf die Eulerreihe (2.2-2) beschr\"ankt. Nehmen wir an,
da{\ss} man einer Funktion $f (z)$ die formale Potenzreihe
$$
f (z) \; = \;
\sum_{\nu=0}^{\infty} \> \gamma_{\nu} \, z^{\nu}
\tag
$$
zuordnen kann. Wenn diese Potenzreihe konvergiert, kann man leicht
zeigen, da{\ss} $f (z)$ durch die beiden folgenden Laplaceintegrale
dargestellt werden kann [Whittaker und Watson 1927, S. 141; Hardy 1949,
S. 182, Theorem 122]:
$$
\beginAligntags
" f (z) " \; = \;
" \frac {1} {z} \, \int\nolimits_{0}^{\infty} \,
\exp (- s/z) \, f_B (s) \, \d s
\\ \tag
" " \; = \; "
\int\nolimits_{0}^{\infty} \,
\exp (- t) \, \, f_B (z t) \, \d t \, .
\\ \tag
\endAligntags
$$
Dabei ist $f_B (z)$ die aus der formalen Potenzreihe (2.3-6)
abgeleiteten Borel-Transformierte
$$
f_B (z) \; = \;
\sum_{n=0}^{\infty} \, \frac {\gamma_n} {n!} \, z^n \, .
\tag
$$
Bemerkenswerterweise sind die Integraldarstellungen (2.3-7) und (2.3-8)
f\"ur $f (z)$ aber unter bestimmten Bedingungen auch dann noch g\"ultig,
wenn die formale Potenzreihe in Gl. (2.3-6) divergiert. Nehmen wir dazu
an, da{\ss} die Koeffizienten $\gamma_n$ der formalen Potenzreihe in Gl.
(2.3-6) f\"ur alle $n \in \N_0$ die Ungleichung
$$
\vert \gamma_n \vert \; \le \; C \sigma^n n!
\tag
$$
erf\"ullen, wobei $C$ und $\sigma$ geeignete positive Konstanten sind. Aus
der Ungleichung (2.3-10) folgt, da{\ss} die aus der formalen Potenzreihe
(2.3-6) abgeleitete Borel-Transformierte (2.3-9) in einer Umgebung des
Nullpunktes der komplexen Ebene $\C$ eine analytische Funktion ist, da
die Potenzreihe in Gl. (2.3-9) einen von Null verschiedenen
Konvergenzradius besitzt [Whittaker und Watson 1927, S. 140].

Trotzdem kann man $f_B (z)$ normalerweise nicht direkt in den Integralen
in Gln. (2.3-7) und (2.3-8) verwenden, da man aus der Ungleichung
(2.3-10) nicht schlie{\ss}en kann, da{\ss} die unendliche Reihe in Gl. (2.3-9)
auf der gesamten positiven reellen Halbachse konvergiert. Wenn man aber
eine analytische Fortsetzung der Borel-Transformierten $f_B (z)$ in eine
Umgebung finden kann, welche die gesamte positive reelle Halbachse
enth\"alt und die f\"ur $z \to \infty$ betragsm\"a{\ss}ig nicht schneller w\"achst
als $\exp (z/R)$ mit $R > 0$, dann kann man diese analytische
Fortsetzung in die Laplaceintegrale in Gln. (2.3-7) und (2.3-8)
einsetzen und versuchen, die resultierenden Integrale zu berechnen. Die
Laplaceintegrale in Gln. (2.3-7) und (2.3-8) existieren dann f\"ur alle
$z < R$ [Whittaker und Watson 1927, S. 140], und ihre Berechnung sollte
mit Hilfe von Quadraturverfahren im Prinzip m\"oglich sein.

Man kann explizit zeigen, da{\ss} die divergente Potenzreihe (2.3-6), deren
Koeffizienten die Ungleichung (2.3-10) erf\"ullen, durch die
Laplaceintegrale in Gln. (2.3-7) und (2.3-8) zu $f (z)$ summiert wird.
Eine genaue Diskussion der Voraussetzungen, die dazu erf\"ullt sein
m\"ussen, findet man in Artikeln von Watson [1912a], Nevanlinna [1919] und
Sokal [1980].

Das Summationsverfahren von Borel ist nicht auf divergente Potenzreihen
(2.3-6) beschr\"ankt, deren Koeffizienten $\gamma_n$ gem\"a{\ss} Gl. (2.3-10) im
wesentlichen wie eine Fakult\"at wachsen, sondern es kann leicht so
modifiziert werden, da{\ss} man damit auch wesentlich st\"arker divergente
Potenzreihen summieren kann, deren Koeffizienten $\gamma_n$ im
wesentlichen wie $\Gamma (k n + 1)$ mit $k \ge 1$ wachsen.

Nehmen wir dazu an, da{\ss} die Koeffizienten $\gamma_n$ der formalen
Potenzreihe in Gl. (2.3-6) f\"ur alle $n \in \N_0$ die Ungleichung
$$
\vert \gamma_n \vert \; \le \; C \sigma^n \Gamma (k n + 1) \, ,
\qquad k \ge 1 \, ,
\tag
$$
erf\"ullen, wobei $C$ und $\sigma$ wiederum geeignete positive Konstanten
sind. Wenn man die aus der formalen Potenzreihe in Gl. (2.3-6)
abgeleitete Borel-Transformierte der Ordnung $k$ folgenderma{\ss}en
definiert,
$$
f^{(k)}_B (z) \; = \; \frac {1} {k} \,
\sum_{n=0}^{\infty} \, \frac {\gamma_n} {\Gamma (k n + 1)} \, z^m \, ,
\tag
$$
kann man explizit zeigen, da{\ss} $f (z)$ durch die beiden folgenden
Laplaceintegrale dargestellt werden kann [Graffi, Grecchi und Turchetti
1971, S. 318; Reed und Simon 1978, S. 45, Beispiel 3, und S. 73, Problem
29(b)]:
$$
\beginAligntags
" f (z) " \; = \;
" \frac {1} {z^{1/k}} \, \int\nolimits_{0}^{\infty} \,
\exp \bigl(- (s/z)^{1/k} \bigr) \, f^{(k)}_B (s) \,
s^{- 1 + 1/k} \, \d s
\\ \tag
" " \; = \; " \int\nolimits_{0}^{\infty} \,
\exp \bigl(- t^{1/k} \bigr) \, f^{(k)}_B (z t) \,
t^{- 1 + 1/k} \, \d t \, .
\\ \tag
\endAligntags
$$
Wenn man in diesen Integralen und in Gl. (2.3-12) $k = 1$ setzt, erh\"alt
man die Laplaceintegrale (2.3-7) und (2.3-8) als Spezialf\"alle.

Es gibt Versuche, das Summationsverfahren von Borel so zu erweitern, da{\ss}
damit sogar divergente Reihen summiert werden k\"onnen, deren
Koeffizienten $\gamma_n$ f\"ur $n \to \infty$ im wesentlichen wie $\Gamma
(k n + 1) \exp (\alpha n^2 / 4)$ wachsen, wobei $k$ und $\alpha$
positive Konstanten sind [Grecchi und Maioli 1984a; 1984b]. Potenzreihen
dieses Typs divergieren so stark, da{\ss} sie mit der oben beschriebenen
Variante des Borelschen Summationsverfahrens nicht mehr summiert werden
k\"onnen.

Derartig hochgradig divergente Potenzreihen treten beispielsweise dann
auf, wenn man die Energieeigenwerte von anharmonische Oszillatoren, die
durch einen Hamiltonoperator des folgenden Typs beschrieben werden, mit
Hilfe der St\"orungstheorie berechnen will [Dolgov und Popov 1978; Maioli
1981]:
$$
{\hat H} \; = \; {\hat p^2} + {\hat x^2} +
\beta {\hat x}^m \exp \bigl( \sqrt{\alpha} \hat{x} \bigr) \, ,
\qquad m \in \N_0 \, , \quad \alpha, \beta > 0 \, .
\tag
$$
Da aber anharmonische Oszillatoren mit exponentiellen St\"oroperatoren
bisher nur von geringer praktischer Bedeutung waren und da au{\ss}erdem
nicht klar ist, ob eine Summation von derartig hochgradig divergenten
Potenzreihen tats\"achlich praktisch realisierbar ist, wird die von
Grecchi und Maioli [1984a; 1984b] eingef\"uhrte Variante des
Summationsverfahrens von Borel in dieser Arbeit nicht explizit behandelt.

Wie schon fr\"uher erw\"ahnt wurde, sind etwas weniger stark divergente
Reihen, deren Koeffizienten $\gamma_n$ f\"ur gro{\ss}e $n$ im wesentlichen wie
$n!$ (Eulerreihe, anharmonischer Oszillator mit
$\hat{x}^4$-Anharmonizit\"at) oder $\Gamma (k n + 1)$ mit $k \ge 1$
(anharmonischer Oszillator mit $\hat{x}^6$- bzw.
$\hat{x}^8$-Anharmonizit\"at) wachsen, in der quantenmechanischen
St\"orungstheorie eher die Regel als die Ausnahme [Arteca, Fern\'{a}ndez
und Castro 1990; {\v C}{\' \i}{\v z}ek und Vrscay 1982; Kazakov und
Shirkov 1980; Killingbeck 1977; Le Guillou und Zinn-Justin 1990; Simon
1982; 1991; Wilcox 1966; Zinn-Justin 1981a] Es besteht also ein gro{\ss}er
Bedarf an effizienten Verfahren zur Summation divergenter Reihen dieses
Typs. Dementsprechend umfangreich ist die Literatur \"uber physikalische
Anwendungen des Borelschen Summationsverfahrens.

Avron [1981], Avron, Herbst und Simon [1977] und Le Guillou und
Zinn-Justin [1983] behandelten den Zeemaneffekt des Wasserstoffatoms mit
Hilfe des Summationsverfahrens von Borel. Le Guillou und Zinn-Justin
[1984] behandelten auf diese Weise au{\ss}erdem noch den Zeemaneffekt des
Wasserstoffmolek\"ulions $H_2^{+}$.

Herbst und Simon [1978] bewiesen die Borelsummierbarkeit der
Rayleigh-Schr\"odingerschen St\"orungsreihe f\"ur den Stark-Effekt des
Wasserstoffatoms. Franceschini, Grecchi und Silverstone [1985]
berechneten mit Hilfe des Summationsverfahrens von Borel
Resonanzenergien eines Wasserstoffatoms im elektrischen Feld.

Die Borelsummierbarkeit der $1/R$-Entwicklung des Wasserstoffmolek\"ulions
$H_2^{+}$, wobei $R$ der beiden Wasserstoffkerne ist, wurde von {\v
C}{\' \i}{\v z}ek, Damburg, Graffi, Grecchi, Harrell \Roemisch{2},
Harris, Nakai, Paldus, Propin und Silverstone [1986], von Damburg,
Propin, Graffi, Grecchi, Harrell \Roemisch{2}, {\v C}{\' \i}{\v z}ek,
Paldus und Silverstone [1984] und von Graffi, Grecchi, Harrell
\Roemisch{2} und Silverstone [1985] diskutiert.

Alvarez [1988] verwendete das Borelsche Summationsverfahren zur
Beschreibung von Resonanzzust\"anden anharmonischer Oszillatoren mit
$\hat{x}^3$-Anharmonizit\"at. Das Borelsche Summationsverfahren wurde von
Br\'ezin, Le Guillou und Zinn-Justin [1977], Caswell [1979], Graffi,
Grecchi und Simon [1970], Graffi, Grecchi und Turchetti [1971], Graffi
und Grecchi [1978], Hirsbrunner [1982], Marziani [1984; 1987], Parisi
[1977], Pre\v{s}najder und Kubinec [1991], Seznec und Zinn-Justin [1979]
und Sobelman [1979] zur Summation der divergenten
Rayleigh-Schr\"odingerschen St\"orungsreihen f\"ur die Energieeigenwerte
anharmonischer Oszillatoren mit $\hat{x}^4$-, $\hat{x}^6$- und
$\hat{x}^8$-Anharmoni\-zi\-t\"at verwendet.

Baker und Pirner [1983] verwendeten das Summationsverfahren von Borel
zur Beschreibung der Grundzustandsenergie eines Systems, das aus einer
gro{\ss}en Zahl von Fermionen besteht. Houghton, Reeve und Wallace [1978]
verwendeten das Summationsverfahren von Borel im Zusammenhang mit
kritischen Ph\"anomenen. Popov und Weinberg [1982] summierten mit Hilfe
des Borelschen Summationsverfahrens St\"orungsreihen f\"ur
Hamiltonoperatoren, die zur ph\"anomenologischen Beschreibung von
Quarkzust\"anden verwendet werden.

Silverstone, Nakai und Harris [1985] und Gailitis und Silverstone [1988]
verwendeten das Borelsche Summationsverfahren zur Untersuchung der
Stokes-Diskontinuit\"at asymptotischer Reihen, wie sie in
quantenmechanischen St\"orungstheorie vorkommen. Silverstone [1985; 1986]
verwendete das Summationsverfahren von Borel zur Summation
asymp\-totischer Reihen, die mit Hilfe des
JWKB{\footnote[\dagger]{Jeffreys-Wentzel-Kramers-Brillouin}}-Verfahrens
konstruiert worden waren.

Von besonderer Bedeutung ist das Borelsche Summationsverfahren in der
Quantenfeldtheorie. Zahlreiche Artikel, in denen das Borelsche
Summationsverfahren eine wesentliche Rolle spielt, werden in
\"Ubersichtsartikeln von Zinn-Justin [1981a; 1984] erw\"ahnt oder sind in der
von Le Guillou und Zinn-Justin [1990] herausgegebenen Reprintsammlung
abgedruckt. Au{\ss}erdem gibt es zahlreiche Lehrb\"ucher der
Quantenfeldtheorie, die einen Abschnitt \"uber das Borelsche
Summationsverfahren enthalten [Itzykson und Zuber 1980, Abschnitt 9-4-1;
Negele und Orland 1987, Abschnitt 7.5; Parisi 1988, Abschnitte 5.7 und
5.8; Rivasseau 1991, Abschnitt 1.5; Zinn-Justin 1989, Abschnitte 37.3 -
37.5].

Das Summationsverfahren von Borel ist theoretisch \"au{\ss}erst wichtig, da
man vergleichsweise leicht feststellen kann, ob eine divergente
Potenzreihe \"uberhaupt borelsummierbar sein kann. Die wichtigste
Voraussetzung f\"ur die Anwendbarkeit des Borelschen Summationsverfahrens
ist, da{\ss} die Koeffizienten $\gamma_n$ der formalen Potenzreihe in Gl.
(2.3-6) f\"ur alle $n \in \N_0$ f\"ur ein festes $k \ge 1$ die Ungleichung
(2.3-11) erf\"ullen{\footnote[\ddagger]{Es mu{\ss} aber betont werden, da{\ss} die
Erf\"ullung einer solchen Wachstumsbedingung {\it nicht} ausreicht f\"ur die
Borelsummierbarkeit einer divergenten Reihe. Auf S. 10 von Simon [1982]
wird als Gegenbeispiel eine divergente Reihe angegeben, deren
Koeffizienten die Ungleichung (2.3-10) erf\"ullen, deren
Borel-Transformierte (2.3-9) aber auf der positiven reellen Halbachse
eine Singularit\"at besitzt.}}. Auf diese Weise kann man in vielen F\"allen
zeigen, da{\ss} eine divergente Reihe ein sinnvolles mathematisches Objekt
ist, indem man ihr die Laplaceintegrale in Gln. (2.3-13) und (2.3-14)
zuordnet, die -- wenn sie konvergieren -- analytische Funktionen des
Argumentes der Potenzreihe sind.

Das Summationsverfahren von Borel ist aber nicht nur in theoretischer
Hinsicht \"au{\ss}erst wertvoll, sondern es kann auch -- wie in einigen der
oben genannten Beispiele gezeigt -- zur Berechnung des Wertes einer
Funktion $f (z)$ verwendet werden, die durch eine divergente Potenzreihe
repr\"asentiert wird. Allerdings mu{\ss} man betonen, da{\ss} die Verwendung des
Summationsverfahrens von Borel f\"ur numerische Zwecke in den meisten
F\"allen \"au{\ss}erst schwierig ist. Das Problem besteht nicht darin, da{\ss} man
die Laplaceintegrale in Gln. (2.3-13) und (2.3-14) in den meisten
praktisch relevanten F\"allen mit Hilfe von Quadraturverfahren berechnen
mu{\ss}, sondern darin, da{\ss} man eine analytische Fortsetzung der
Borel-Transformierten (2.3-12) in eine Umgebung der positiven reellen
Halbachse finden mu{\ss}. In vielen F\"allen f\"uhrt diese analytische
Fortsetzung zu \"au{\ss}erst komplizierten numerischen und konzeptionellen
Problemen.

Auf den ersten Blick scheint es m\"oglich, dieses Problem zu vermeiden,
indem man anstelle der Borel-Transformierten (2.3-12) die modifizierte
Borel-Transformierte
$$
f^{(k + 1)}_B (z) \; = \; \frac {1} {k+1} \,
\sum_{n=0}^{\infty} \,
\frac {\gamma_n} {\Gamma ([k + 1] n + 1)} \, z^n
\tag
$$
verwendet. Wenn die Koeffizienten $\gamma_n$ der formalen Potenzreihe
f\"ur alle $n \in \N_0$ die Ungleichung (2.3-11) erf\"ullen, dann
konvergiert die unendliche Reihe f\"ur $f^{(k + 1)}_B (z)$ offensichtlich
gleichm\"a{\ss}ig f\"ur alle $z \in [0, \infty)$ und kann direkt in den
Laplaceintegralen in Gln. (2.3-13) und (2.3-14) verwendet werden,
wenn man dort $k$ durch $k + 1$ ersetzt.

Ungl\"ucklicherweise gibt es bei dieser Vorgehensweise in der Regel
erhebliche numerische Probleme [Zinn-Justin 1981a, S. 157; Marziani
1984, S. 552]. F\"ur gro{\ss}e Argumente $z$ konvergiert die unendliche Reihe
(2.3-16) f\"ur $f^{(k+1)}_B (z)$ sehr schlecht und man w\"urde sehr viele
Terme ben\"otigen, um $f^{(k+1)}_B (z)$ mit ausreichender Genauigkeit
berechnen zu k\"onnen. Bei vielen stark divergenten quantenmechanischen
St\"orungsreihen ist die Berechnung der Koeffizienten mit h\"oheren
Laufindizes aber so aufwendig, da{\ss} man nur eine relativ kleine Anzahl
berechnen kann. In solchen F\"allen kann die Borel-Transformierte
$f^{(k+1)}_B (z)$ durch das Aufaddieren der Terme der unendlichen Reihe
in Gl. (2.3-16) nicht mit ausreichender Genauigkeit berechnet werden.
Au{\ss}erdem sind im Falle gro{\ss}er Argumente $z$ schwerwiegende
Stabilit\"atsprobleme durch Rundungsfehler zu erwarten, wenn die
Koeffizienten $\gamma_n$ der formalen Potenzreihe in Gl. (2.3-6)
alternieren, was bei summierbaren divergenten Potenzreihen normalerweise
der Fall ist.

Ein sehr beliebtes Verfahren zur analytischen Fortsetzung der
Borel-Transformierten (2.3-12) besteht darin, die unendliche Reihe in
Gl. (2.3-12), die normalerweise nur einen endlichen Konvergenzradius
besitzt, durch eine
Pad\'e-Approximation{\footnote[\dagger]{Pad\'e-Approximationen werden in
Abschnitt 4 dieser Arbeit behandelt.}} zu ersetzen, um somit eine
analytische Fortsetzung in eine Umgebung der positiven reellen Halbachse
zu bewirken. Dieses Verfahren, das von Graffi, Grecchi und Simon [1970]
zur Summation quantenmechanischer St\"orungsreihen eingef\"uhrt wurde, wird
in der Literatur \"ublicherweise als die Borel-Pad\'e-Methode bezeichnet.
Allerdings ist auch dieses Verfahren zur analytischen Fortsetzung nicht
frei von Problemen [Graffi, Grecchi und Simon 1970, S. 634; Graffi,
Grecchi und Turchetti 1971, S. 331].

Ein anderes Verfahren zur analytischen Fortsetzung der
Borel-Transformierten (2.3-12) in eine Umgebung der positiven reellen
Halbachsen besteht darin, den Konvergenzradius der Borel-Transformierten
mit Hilfe ordnungsabh\"angiger konformer Abbildungen auszudehnen. Diese
Methode, die zuerst von Le Guillou und Zinn-Justin [1977] verwendet
wurde, wird ausf\"uhrlicher beschrieben von Arteca, Fern\'{a}ndez und
Castro [1990, S. 136 - 138], Itzykson und Zuber [1980, Abschnitt 9-4-1],
Le Guillou und Zinn-Justin [1983], Negele und Orland [1987, Abschnitt
7.5], Parisi [1977; 1988, Abschnitt 5.7], Seznec und Zinn-Justin [1979],
Sobelman [1979] und Zinn-Justin [1981a, Abschnitt 8.2; 1989, Abschnitt
37.5].

Der Nachteil der Methode der ordnungsabh\"angigen konformen Abbildung
besteht darin, da{\ss} man zur Konstruktion der konformen Abbildung gewisse
Informationen \"uber die Lage der Singularit\"aten der Borel-Transformierten
besitzen mu{\ss} [Parisi 1988, S. 90]. Da diese Informationen aber nicht
immer bekannt sind, wird die praktische Anwendbarkeit dieser in vielen
F\"allen sehr leistungsf\"ahigen Methode dadurch erheblich einschr\"ankt.

\medskip

\Abschnitt Verallgemeinerte Summationsprozesse

\smallskip

\aktTag = 0

In den Unterabschnitten 2.1 und 2.2 wurde anhand einiger einfacher
unendlicher Reihen, die entweder nur sehr schlecht oder gar nicht
konvergieren, die Unzul\"anglichkeit des konventionellen Verfahrens zur
Berechnung einer Reihe durch Aufaddieren der Terme demonstriert. H\"atte
man nur dieses Verfahren zur Berechnung des Wertes einer unendlichen
Reihe zur Verf\"ugung, so w\"aren zahlreiche praktisch wichtige
Reihenentwicklungen entweder nur eingeschr\"ankt verwendbar oder sogar
numerisch v\"ollig nutzlos.

Diese Probleme blieben nat\"urlich auch den Begr\"undern der modernen
Differential- und Integralrechnung nicht verborgen. Dementsprechend gab
es schon sehr bald Versuche, die h\"aufig unbefriedigenden
Konvergenzeigenschaften unendlicher Reihen zu verbessern, und die
Theorie der Reihentransformationen ist fast so alt wie die Differential-
und Integralrechnung. Auf S. 249 des Buches von Knopp [1964] wird
erw\"ahnt, da{\ss} das erste Verfahren zur Verbesserung der Konvergenz
unendlicher Reihen 1730 von Stirling in seinem Buch {\it Methodus
Differentialis} [Stirling 1730] beschrieben wurde. Nur 25 Jahre sp\"ater
ver\"offentlichte Euler [1755] eine Reihentransformation, die heute nach
ihm benannt ist.

In rudiment\"arer Form ist die Theorie der Beschleunigungs- und
Extrapolationsverfahren aber noch wesentlich \"alter. In dem einleitenden
Kapitel des Buches von Delahaye [1988] und in einem sp\"ateren Artikel
erw\"ahnte Brezinski [1988; 1989], da{\ss} Extrapolationsverfahren schon 1654
von Huygens und 1674 von Seki Kowa, dem wohl ber\"uhmtesten japanischen
Mathematiker, verwendet wurden, um bessere Approximationen f\"ur $\pi$ zu
erhalten. Huygens verwendete ein lineares Extrapolationsverfahren,
welches ein Spezialfall des Richardsonschen Extrapolationsverfahren
[Richardson 1927] ist, w\"ahrend Seki Kowa den sogenannten
$\Delta^2$-Proze{\ss} verwendete, der \"ublicherweise Aitken [1926]
zugeschrieben wird und der aber gem\"a{\ss} Todd [1962, S. 5] auch schon
Kummer [1837] bekannt war.

Diese fr\"uhen Ideen zur Konvergenzverbesserung oder Summation von
unendlichen Reihen f\"uhrten schlie{\ss}lich dazu, da{\ss} man {\it
verallgemeinerte Summationsprozesse} konstruierte, die den
konventionellen Proze{\ss} der Berechnung des Wertes einer unendlichen Reihe
ersetzten konnten. Beispielsweise werden auf S. 77 - 78 des Buches von
Ford [1960], das urspr\"unglich 1916 ver\"offentlicht wurde, zahlreiche
verallgemeinerte Summationsprozesse aufgelistet, die zur Summation
divergenter Reihen geeignet sind.

Das im letzten Abschnitt beschriebene Summationsverfahren von Borel ist
ein solcher verallgemeinerter Summationsproze{\ss} f\"ur konvergente und
divergente Potenzreihen. Das Problem des Aufsummierens der Terme einer
Potenzreihe wird dabei ersetzt durch das Problem der Berechnung von
Laplaceintegralen. Allerdings ist -- wie im letzten Abschnitt betont --
die praktische Anwendung des Borelschen Summationsverfahrens keineswegs
einfach, da man dabei im Prinzip zwei verschiedene und oft schwierige
Grenzprozessen durchf\"uhren mu{\ss}, n\"amlich die analytische Fortsetzung der
Borel-Transformierten $f_B (z)$ und die Berechnung eines
Laplaceintegrals. In praktischen Rechnungen sind deswegen
verallgemeinerte Summationsprozesse vorzuziehen, bei denen keine
Grenzprozesse durchgef\"uhrt werden m\"ussen und die eine Verbesserung der
Konvergenz oder eine Summation ausschlie{\ss}lich mit Hilfe einer endlichen
Anzahl algebraischer Operationen erreichen. Solche verallgemeinerte
Summationsprozesse sind auch wesentlich besser f\"ur Computer geeignet,
die ja bekanntlich nur die Grundrechenarten beherrschen, wenn man eine
konventionelle Programmiersprache wie FORTRAN 77 verwendet.

Sei ${\cal T}_{\ell}$ ein solcher verallgemeinerter Summationsproze{\ss},
der $\ell + 1$ Partialsummen $s_n, \ldots, s_{n+\ell}$ einer unendlichen
Reihe verwendet, um ein Element einer transformierten Folge $\Seq
{s^{\prime}_n} {n=0}$ mit hoffentlich besseren Konvergenzeigenschaften
zu berechnen:
$$
{\cal T}_{\ell} (s_n, \ldots , s_{n+\ell}) \; = \; s'_n \, ,
\qquad \ell, n \in \N_0 \, .
\tag
$$
In der Literatur \"uber Konvergenzbeschleunigung sagt man, da{\ss} ein
solcher verallgemeinerter Summationsproze{\ss} ${\cal T}_{\ell}$ die
Konvergenz einer Folge $\Seqn s$ verbessert, wenn
$$
\lim_{n \to \infty} \;
\frac { s^{\prime}_n - s }{ s_n - s} \; = \; 0
\tag
$$
gilt. Allerdings ist diese Bedingung allein noch nicht ausreichend, um
einen verallgemeinerten Summationsproze{\ss} ${\cal T}_{\ell}$ praktisch
n\"utzlich zu machen. Wenn wir in Betracht ziehen, da{\ss} im Falle
konvergenter unendlicher Reihen offensichtlich die Beziehung
$$
\sum_{n=0}^{\infty} \; ( \alpha a_n + \beta b_n ) \; = \;
\alpha \sum_{n=0}^{\infty} \; a_n \, + \,
\beta \sum_{n=0}^{\infty} \; b_n \, ,
\qquad \alpha , \beta \in \R \, ,
\tag
$$
erf\"ullt ist, so liegt es nahe zu fordern, da{\ss} ein solcher
verallgemeinerter Summationsproze{\ss} ${\cal T}_{\ell}$ ebenfalls {\it
linear} sein sollte:
$$
\beginAligntags
" {\cal T}_{\ell} (\alpha s_n + \beta t_n, \ldots ,
\alpha s_{n+\ell} + \beta t_{n+\ell}) " \; = \;
" \alpha {\cal T}_{\ell} (s_n, \ldots , s_{n+\ell}) \; + \;
\beta {\cal T}_{\ell} (t_n, \ldots , t_{n+\ell}) \, , \\
" " " \alpha,\beta \in \R \, , \qquad \ell , n \in \N_0 \, .
\\ \tag
\endAligntags
$$
Au{\ss}erdem sollte ein solcher verallgemeinerter Summationsproze{\ss} {\it
regul\"ar} sein, was bedeutet, da{\ss} der Grenzwert einer konvergenten Folge
durch die Transformation nicht ver\"andert wird. Nur dann w\"are
gew\"ahrleistet, da{\ss} ein verallgemeinerter Summationsproze{\ss} im Falle einer
konvergenten Reihe das gleiche Ergebnis liefert wie der konventionelle
Proze{\ss} des Aufaddierens der Terme der Reihe.

Wenn also $\Seqn s$ und $\Seqn t$ zwei Folgen von Partialsummen
unendlicher Reihen gem\"a{\ss} Gl. (2.1-1) sind, die gegen die Grenzwerte $s$
und $t$ konvergieren, dann sollte ein {\it linearer und regul\"arer
verallgemeinerter Summationsproze{\ss}} die folgende Eigenschaft besitzen:
$$
\lim_{n \to \infty} {\cal T}_{\ell}
(\alpha s_n + \beta t_n, \ldots ,
\alpha s_{n+\ell} + \beta t_{n+\ell}) \; = \;
\alpha s \; + \; \beta t \, .
\tag
$$

Es gelang tats\"achlich, verallgemeinerte Summationsprozesse zu
konstruieren, die sowohl linear als auch regul\"ar sind. Es sind dies die
sogenannten {\it regul\"aren Matrixtransformationen}. Sei $\Seqn s$ eine
Folge von Partialsummen einer unendlichen Reihe. Dann erh\"alt man eine
transformierte Folge $\Seq {s^{\prime}_n} {n=0}$, indem man gewichtete
Mittelwerte aus den Elementen der urspr\"unglichen Folge bildet:
$$
s'_n \; = \; \sum _{k = 0}^{n} \; {\mu}_{n k} s_k \, ,
\qquad k, n \in \N_0 \, .
\tag
$$

Der Hauptvorzug dieser Matrixtransformationen, die offensichtlich linear
sind, besteht darin, da{\ss} f\"ur die Gewichte ${\mu}_{n k}$, die eine solche
Transformation definieren, notwendige und hinreichende Bedingungen
formuliert werden konnten, welche die Regularit\"at des Verfahrens
garantieren. Einen guten \"Uberblick \"uber die Theorie regul\"arer
Matrixtransformationen findet man in den B\"uchern von Hardy [1949], Knopp
[1964], Petersen [1966], Peyerimhoff [1969], Powell und Shah [1988], und
Zeller und Beekmann [1970]. Es gibt auch neuere Versuche, die Theorie
der regul\"aren Matrixtransformationen auf der Basis der
Funktionalanalysis darzustellen [Wilansky 1984], oder regul\"are
Matrixtransformationen zur Summation von Fourier- und anderen
Orthogonalreihen zu verwenden [Okuyama 1984].

Die Tatsache, da{\ss} regul\"are Matrixtransformationen bei allen konvergenten
Folgen gefahrlos angewendet werden k\"onnen, da die transformierte Folge
$\Seq {s^{\prime}_n} {n=0}$ gegen den gleichen Grenzwert konvergiert wie
die Ausgangsfolge $\Seqn s$, ist in theoretischer Hinsicht zweifellos
\"au{\ss}erst vorteilhaft: Es ist immer gew\"ahrleistet, da{\ss} man bei
konvergenten Folgen das richtige Ergebnis erh\"alt. In praktischer
Hinsicht ist dieser unbestreitbare theoretische Vorteil aber ein
ernsthafter Nachteil. Diese Feststellung klingt paradox. Man sollte aber
bedenken, da{\ss} man nicht erwarten kann, da{\ss} ein numerisches Verfahren in
einem speziellen Fall besonders gute Ergebnisse liefert, wenn man
gleichzeitig verlangt, da{\ss} dieses Verfahren auch in allen anderen F\"allen
funktionieren soll.

Dementsprechend wurde in den letzten Jahren nur noch relativ wenig \"uber
regul\"are Matrixtransformationen gearbeitet. Der Hauptschwerpunkt der
mathematischen Forschung nicht nur auf dem Gebiet der verallgemeinerten
Summationsprozesse, sondern generell in der Approxima\-tionstheorie,
waren die leistungsf\"ahigeren, aber auch weniger allgemeinen {\it
nichtlinearen} Verfahren (siehe beispielsweise Braess [1986], Cuyt und
Wuytack [1987], Cuyt [1988], Nikishin und Sorokin [1991], und Petrushev
und Popov [1987]), die au{\ss}erdem in der Regel {\it nichtregul\"ar} sind.

Der Unterschied zwischen einem linearen und einem nichtlinearen
verallgemeinerten Summationsproze{\ss} ${\cal T}_{\ell}$ ist, da{\ss} die
Linearit\"atsbedingung (2.4-4) nicht mehr g\"ultig ist. Statt dessen kann
man nur noch fordern, da{\ss} ein nichtlinearer verallgemeinerter
Summationsproze{\ss} ${\cal T}_{\ell}$ {\it trans\-lationsinvariant} ist,
was bedeutet, da{\ss} f\"ur alle zul\"assigen Werte von $\ell, n \in \N_0$ die
Beziehung
$$
{\cal T}_{\ell}
(\alpha s_n + \tau, \ldots , \alpha s_{n+\ell} + \tau) \; = \;
\alpha {\cal T}_{\ell} (s_n, \ldots , s_{n+\ell})
\; + \; \tau
\tag
$$
erf\"ullt ist, wobei $\alpha$ und $\tau$ zwei Konstanten sind. Da
nichtlineare verallgemeinerte Summationsprozesse au{\ss}erdem fast immer
nichtregul\"ar sind, kann man auch nicht erwarten, da{\ss} die Beziehungen
$$
\lim_{n \to \infty} {\cal T}_{\ell} (s_n, \ldots , s_{n+\ell})
\; = \; s
\tag
$$
und
$$
\lim_{\ell \to \infty} {\cal T}_{\ell} (s_n, \ldots , s_{n+\ell})
\; = \; s
\tag
$$
im Falle beliebiger konvergenter Folgen $\Seqn s$ g\"ultig sind.

Nichtlinearit\"at und Nichtregularit\"at sind zweifellos unangenehme
Komplikationen, die man eigentlich instinktiv vermeiden m\"ochte, und die
auch eine theoretische Analyse der Eigenschaften verallgemeinerter
Summationsprozesse stark erschweren. Trotzdem sind sie unverzichtbar,
da die oft beeindruckende Leistungsf\"ahigkeit vieler verallgemeinerter
Summationsprozesse eine direkte Konsequenz ihrer Nichtlinearit\"at und
Nichtregularit\"at ist.

Die intensive Forschung auf dem Gebiet der nichtlinearen
verallgemeinerten Summationsprozesse geht auf Artikel von Shanks [1955]
beziehungsweise Wynn [1956a] zur\"uck. Shanks [1955] gelang es, eine
nichtlineare Transformation zu finden, mit deren Hilfe man
beispielsweise aus den Partialsummen einer Potenzreihe
Pad\'e-Approximationen berechnen kann. Allerdings war die
Reihentransformation von Shanks in ihrer urspr\"unglichen Form nicht sehr
n\"utzlich, da sie als Quotient zweier Determinanten angegeben war, deren
effiziente und verl\"a{\ss}liche Berechnung immer noch ein mehr oder weniger
ungel\"ostes Problem der numerischen Mathematik ist. Wynn [1956a] konnte
nur ein Jahr sp\"ater zeigen, da{\ss} die Reihentransformation von Shanks mit
Hilfe des folgenden zweidimensionalen nichtlinearen Rekursionsschemas,
das heute \"ublicherweise als $\epsilon$-Algorithmus bezeichnet wird, auf
einfache und \"au{\ss}erst effiziente Weise berechnet werden kann:
$$
\beginAligntags
" \epsilon_{-1}^{(n)} " \; = \; " 0 \, ,
\hfill \epsilon_0^{(n)} \, = \, s_n \, ,
\erhoehe\aktTag \\
\tag*{\tagnr a}
" \epsilon_{k+1}^{(n)} " \; = \; " \epsilon_{k-1}^{(n+1)} \, + \,
1 / [\epsilon_{k}^{(n+1)} - \epsilon_{k}^{(n)} ] \, ,
\qquad k,n \in \N_0 \, . \\
\tag*{\tagform\aktTagnr b}
\endAligntags
$$

Die Entdeckung des $\epsilon$-Algorithmus durch Wynn fiel in eine Zeit,
in der -- bedingt durch den raschen Fortschritt der Computertechnologie
-- aufwendigere numerische Rechnungen \"uberhaupt erst m\"oglich wurden,
was zu einem enormen Bedarf an leistungsf\"ahigen numerischen Algorithmen
f\"uhrte. Aufgrund seiner Nichtlinearit\"at ist der $\epsilon$-Algorithmus
ohne Zweifel nicht sehr attraktiv, wenn man nur Papier und Bleistift
zur Verf\"ugung hat. Er ist aber hervorragend geeignet, auf einer
programmierbaren Rechenanlage verwendet zu werden, und er ist heute
einer der ganz wichtigen Algorithmen der numerischen Mathematik, der
nicht nur zur Transformation von unendlichen Reihen, sondern auch in
der numerischen Quadratur (vergleiche zum Beispiel Piessens, de
Doncker-Kapenga, \"Uberhuber und Kahaner [1983] oder Davis und Rabinowitz
[1984]) verwendet wird.

Die beiden Artikel von Shanks [1955] und Wynn [1956a] hatten also eine
enorme Wirkung. Sie initiierten ausgedehnte Forschungen, die ganz
wesentlich zum Verst\"andnis der Eigenschaften des $\epsilon$-Algorithmus
und der eng verwandten Pad\'e-Approximationen beitrugen, und die
schlie{\ss}lich dazu f\"uhrten, da{\ss} Verfahren zur Konvergenzverbesserung und
Summation inzwischen in der theoretischen Physik fast schon routinem\"a{\ss}ig
und in den anderen Naturwissenschaften auch schon sehr h\"aufig verwendet
werden.

Bei diesen Forschungen stellte sich aber auch heraus, da{\ss} der
$\epsilon$-Algorithmus nicht alle Probleme l\"osen kann, die im
Zusammenhang mit schlecht konvergierenden oder divergenten unendlichen
Reihen auftreten. Beispielsweise ist der $\epsilon$-Algorithmus nicht in
der Lage, die Konvergenz der Partialsummen der unendlichen Reihe (2.1-4)
f\"ur die Riemannsche Zetafunktion zu beschleunigen. Deswegen wurde in den
folgenden Jahren ebenfalls sehr intensiv \"uber alternative nichtlineare
Verfahren zur Konvergenzverbesserung oder Summation gearbeitet. Ein
Beispiel ist der von Wynn [1956b] gefundene und sp\"ater von Carstensen
[1990] und Osada [1990a] erweiterte $\rho$-Algorithmus, der zwar im
Falle konvergenter oder divergenter Potenzreihen wirkungslos ist, der
aber logarithmische Konvergenz oft sehr erfolgreich beschleunigen kann.

Die intensive Forschung \"uber Theorie und Anwendungen nichtlinearer
verallgemeinerter Summationsprozesse wird sowohl durch die B\"ucher von
Baker [1975], Baker und Gammel [1970], Baker und Graves-Morris [1981a;
1981b], Brezinski [1977; 1978; 1980a; 1991a, 1991b], Brezinski, Draux,
Magnus, Maroni und Ronveaux [1985], Brezinski und Redivo Zaglia [1991],
Bultheel [1987], Cabannes [1976], Cuyt [1988], Cuyt und Wuytack [1987],
de Bruin und Van Rossum [1981], Delahaye [1988], Draux und van
Ingelandt [1987], Gilewicz [1978], Gilewicz, Pindor und Siemaszko
[1985], Graves-Morris [1973a; 1973b], Graves-Morris, Saff, und Varga
[1984], H{\aa}vie [1989], Saff und Varga [1977], Werner und B\"unger
[1984], Wimp [1981] und Wuytack [1979a] als auch durch die
\"Ubersichtsartikel von Baker [1965; 1972], Brezinski [1985a], Brezinski
und van Iseghem [1991], Gaunt und Guttmann [1974], Gragg [1972],
Guttmann [1989], und Weniger [1989] sowie durch die gro{\ss}e Zahl der
darin enthaltenen Referenzen \"uberzeugend dokumentiert.

\endAbschnittsebene

\endAbschnittsebene

\keinTitelblatt\neueSeite

\beginAbschnittsebene
\aktAbschnitt = 2

\Abschnitt \"Uber die Konstruktion verallgemeinerter Summationsprozesse

\vskip - 2 \jot

\beginAbschnittsebene

\medskip

\Abschnitt Konvergenzbeschleunigung und Summation bei alternierenden
Reihen

\smallskip

\aktTag = 0

Will man die Konvergenz einer unendlichen Reihe mit Hilfe eines
verallgemeinerten Summationsprozesses verbessern oder eine divergente
Reihe summieren, so mu{\ss} man die numerische Information, die in den
Partialsummen $s_0$, $s_1$, $\ldots$ , $s_n$ enthalten ist, auf
effizientere Weise extrahieren als dies durch den konventionellen
Proze{\ss} des Aufaddierens der Terme geschieht.

Wie kann man einen verallgemeinerten Summationsproze{\ss} konstruieren,
der dazu in der Lage ist? Der erste Schritt besteht normalerweise in der
Annahme, da{\ss} die Partialsumme $s_n$ f\"ur jedes $n \in \N_0$
gem\"a{\ss} $$ s_n \; = \; s \, + \, r_n
\tag
$$
in den Grenzwert $s$ und den Summationsrest $r_n$ zerlegt werden kann.

Auf den ersten Blick sieht diese Unterteilung nicht wie eine gro{\ss}e
Errungenschaft aus, denn man ersetzt eine bekannte Gr\"o{\ss}e -- die
Partialsumme $s_n$ -- durch zwei unbekannte Gr\"o{\ss}en $s$ und
$r_n$. Zwar w\"are das Problem der Bestimmung des Grenzwertes $s$ einer
Folge $\Seqn s$ von Partialsummen gel\"ost, wenn man die Summationsreste
$\Seqn r$ kennen w\"urde. Aber mit Ausnahme von einigen wenigen
praktisch irrelevanten Modellproblemen, f\"ur die man eigentlich keine
verallgemeinerten Summationsprozesse ben\"otigt, sind die
Summationsreste $\Seqn r$ unendlicher Reihen nicht in einer numerisch
leicht zug\"anglichen Form bekannt.

Gl\"ucklicherweise ist es aber oft m\"oglich, f\"ur die Summationsreste
Approximationen zu konstruieren. Wenn diese N\"aherungen ausreichend genau
gemacht werden k\"onnen, sollte die anschlie{\ss}ende Elimination der
approximativen Summationsreste entweder eine Verbesserung der Konvergenz
oder eine Summation im Falle einer divergenten Reihe bewirken. Die
Konstruktion von Approximationen f\"ur die Summationsreste $\Seqn r$ einer
Folge $\Seqn s$ von Partialsummen wird dadurch erleichtert, da{\ss} in
vielen F\"allen doch gewisse strukturelle Informationen, die nicht sehr
detailliert sein m\"ussen, \"uber das Verhalten der Summationsreste $\Seqn
r$ als Funktion des Index $n$ vor\-lie\-gen.

Ein verallgemeinerter Summationsproze{\ss}, der auf diese Weise konstruiert
worden ist, wird aus einer Folge $\Seqn s$ von Partialsummen eine neue
Folge $\Seq {s^{\prime}_n} {n=0}$ erzeugen, deren Elemente analog zu Gl.
(3.1-1) f\"ur alle $n \in \N_0$ gem\"a{\ss}
$$
s'_n \; = \; s \, + \, r'_n
\tag
$$
in den Grenzwert $s$ und einen transformierten Summationsrest $r'_n$
zerlegt werden k\"onnen. Dabei werden die transformierten Summationsreste
$\Seq {r^{\prime}_n} {n=0}$ normalerweise f\"ur alle endlichen Indizes $n$
von Null verschieden sein. Das bedeutet, da{\ss} auch die Elemente der
transformierten Folge $\Seq {s^{\prime}_n} {n=0}$ nur Approximationen
f\"ur den Grenzwert $s$ liefern. Aber man sieht sofort, da{\ss} ein
verallgemeinerter Summationsproze{\ss} eine Beschleunigung der Konvergenz
bewirkt, wenn die Folge $\Seq {r^{\prime}_n} {n=0}$ der transformierten
Summationsreste f\"ur $n \to \infty$ schneller gegen Null konvergiert als
die urspr\"ungliche Folge $\Seqn r$. Au{\ss}erdem ist ein verallgemeinerter
Summationsproze{\ss} offensichtlich genau dann in der Lage, eine divergente
Reihe zu summieren, wenn die transformierten Summationsreste f\"ur $n \to
\infty$ gegen Null konvergieren.

Auf diese Weise funktionieren alle in dieser Arbeit verwendeten
verallgemeinerten Summationsprozesse: Man versucht, eine N\"aherung f\"ur
den Summationsrest $r_n$ zu konstruieren, die anschlie{\ss}end entweder
explizit oder implizit aus dem Folgenelement $s_n$ eliminiert wird.

Dieses Konstruktionsprinzip kann am besten durch ein einfaches Beispiel
verdeutlicht werden. Dazu nehmen wir an, da{\ss} die Folgenelemente $s_n$
Partialsummen einer unendlichen Reihe mit reellen und strikt
alternierenden Termen sind,
$$
s_n \; = \; \sum_{k=0}^n \, (-1)^k \, b_k \, .
\tag
$$
Offensichtlich alternieren die Terme dieser Reihe genau dann strikt,
wenn alle $b_n$ mit $n \in \N_0$ das gleiche Vorzeichen besitzen. Der
Summationsrest $r_n$ von $s_n$ ist dann gem\"a{\ss} Gl. (3.1-1) durch
$$
r_n \; = \; - \sum_{k=n+1}^{\infty} \, (-1)^k \, b_k \, .
\tag
$$
gegeben. Nehmen wir nun weiterhin an, da{\ss} alle $b_n$ mit wachsendem $n$
betragsm\"a{\ss}ig streng monoton abnehmen und gegen Null gehen. Damit ist
gew\"ahrleistet, da{\ss} die unendliche Reihe gegen einen Grenzwert $s$
konvergiert. Au{\ss}erdem folgt, da{\ss} die Summationsreste ebenfalls strikt
alternieren, und da{\ss} $r_n$ betragsm\"a{\ss}ig abgesch\"atzt werden kann durch
den ersten Term $b_{n+1}$, der nicht in der Partialsumme (3.1-3)
enthalten ist [Knopp 1964, S. 259]:
$$
\vert r_n \vert < \vert b_{n+1} \vert \, , \qquad n \in \N_0.
\tag
$$
Interessanterweise sind die Summationsreste einer stark divergenten
strikt alternierenden hypergeometrischen Reihe des Typs
$$
{}_2 F_0 (\alpha, \beta, - z) \; = \;
\sum_{m=0}^{\infty} \, \frac {(\alpha)_m \, (\beta)_m} {m!} \, (- z)^m
\, , \qquad \alpha, \beta, z > 0 \, ,
\tag
$$
ebenfalls strikt alternierend und werden analog zu Gl. (3.1-5)
betragsm\"a{\ss}ig vom ersten Term der hypergeometrischen Reihe abgesch\"atzt,
der nicht in der Partialsumme enthalten ist [Carlson 1977, Theorem
5.12-5].

Wenn wir die Konvergenz einer strikt alternierenden Reihe mit den oben
genannten Eigenschaften beschleunigen wollen, so m\"ussen wir einen Weg
finden, um unser Wissen \"uber das Verhalten der Summationsreste (3.1-4)
der alternierenden Reihe auszun\"utzen. Die Absch\"atzung (3.1-5) ist dazu
allein nicht ausreichend und wir ben\"otigen noch eine zus\"atzliche
Annahme \"uber das Verhalten der Summationsreste $r_n$ als Funktion des
Index $n \in \N_0$, um einen verallgemeinerten Summationsproze{\ss} f\"ur
alternierende Reihen mit den gew\"unschten Eigenschaften konstruieren zu
k\"onnen.

Es ist eine relativ nat\"urliche und wenig restriktive Annahme, wenn wir
fordern, da{\ss} der Quotient $r_n / [(-1)^{n+1} b_{n+1}]$ f\"ur alle $n \in
\N_0$ durch eine asymptotische Potenzreihe im Sinne von Poincar\'e in der
Variablen $1/(n + 1)$ dargestellt werden kann:
$$
r_n \; \sim \; (-1)^{n+1} \, b_{n+1} \,
\sum_{j=0}^{\infty} \; c_j \, (n + 1)^{-j} \, , \qquad n \to \infty \, .
\tag
$$
Wenn dieser Ansatz f\"ur den Summationsrest zul\"assig ist, so m\"u{\ss}ten wir im
Prinzip nur noch die unbekannten Koeffizienten $c_j$ in Gl. (3.1-7)
bestimmen. Das Problem der Konvergenzbeschleunigung beziehungsweise der
Summation im Falle einer divergenten Reihe w\"are dann gel\"ost, da wir den
Grenzwert $s$ dann gem\"a{\ss} Gl. (3.1-1) aus den Partialsummen $s_n$
berechnen k\"onnten.

Da aber numerische Algorithmen zwangsl\"aufig endlich sind, ist es nicht
m\"oglich, alle der unendlich vielen unbestimmten Koeffizienten $c_j$ in
Gl. (3.1-7) zu bestimmen. Statt dessen mu{\ss} man sich damit zufrieden
geben, die $k$ f\"uhrenden Koeffizienten $c_0$, $c_1$, $\ldots$ ,
$c_{k-1}$ aus Gl. (3.1-7) zu bestimmen. Dazu nimmt man an, da{\ss} der
Quotient $r_n / [(-1)^{n+1} b_{n+1}]$ f\"ur alle $n \in \N_0$ durch ein
Polynom vom Grade $k-1$ in der Variablen $1/(n + 1)$ dargestellt werden
kann:
$$
r_n \; = \; (-1)^{n+1} \, b_{n+1} \,
\sum_{j=0}^{k-1} \; c_j \, (n + 1)^{-j} \, .
\tag
$$
Mit Hilfe dieser Annahme f\"ur den Summationsrest kann ein
verallgemeinerter Summationsproze{\ss} konstruiert werden, der -- wie die
Erfahrung und auch einige theoretische Untersuchungen zeigen -- die
Konvergenz vieler alternierender Reihen verbessert und der auch
divergente alternierende Reihen vom Typ der nichtabbrechenden
hypergeometrischen Reihe ${}_2 F_0$ in Gl. (3.1-6) summieren kann.
Allerdings f\"uhrt die oben erw\"ahnte notwendige Beschr\"ankung auf $k$
unbestimmte Koeffizienten $c_0$, $c_1$, $\ldots$ , $c_{k-1}$ in Gl.
(3.1-8) dazu, da{\ss} ein solcher verallgemeinerter Summationsproze{\ss} f\"ur
endliche $k \in \N$ in der Regel nicht den exakten Wert einer
alternierenden Reihe liefert, sondern nur eine Approximation.

Wie aus Gln. (3.1-1) und (3.1-8) sofort folgt, wird also angenommen, da{\ss}
die in Gl. (3.1-3) definierte Partialsumme $s_n$ f\"ur alle $n \in \N_0$
durch den folgenden endlichen Ausdruck dargestellt werden kann:
$$
s_n \; = \; s \> + \> (-1)^{n+1} \, b_{n+1} \,
\sum_{j=0}^{k-1} \, c_j \, (n + 1)^{-j} \, .
\tag
$$
Dieser Ausdruck enth\"alt $k+1$ Unbekannte, den Grenzwert $s$ und die $k$
Koeffizienten $c_0$, $\ldots$ , $c_{k-1}$. Da alle Unbekannten in Gl.
(3.1-9) {\it linear} vorkommen, ist ihre Bestimmung im Prinzip
unproblematisch. Man ben\"otigt nur die numerischen Werte von $k+1$
Folgenelementen $s_n$, $s_{n+1}$, $\ldots$ , $s_{n+k}$, um den
Grenzwert $s$ der Modellfolge (3.1-9) zu bestimmen. Wenn man dazu die
Cramersche Regel verwendet, erh\"alt man f\"ur den verallgemeinerten
Summationsproze{\ss} $T_k^{(n)} (s_n)$, der den Grenzwert $s$ der
Modellfolge (3.1-9) exakt bestimmen kann, eine Darstellung als Quotient
zweier Determinanten:

\smallskip

$$
T_k^{(n)} (s_n) \; = \; \frac
{
\vmatrix
{
s_n " \ldots " s_{n+k} \\ [1\jot]
(-1)^{n+1} b_{n+1} " \ldots " (-1)^{n+k+1} b_{n+k+1} \\
\vdots " \ddots " \vdots \\ [1\jot]
{\displaystyle \frac {(-1)^{n+1} b_{n+1} } {(n+1)^{k-1}} } " \ldots "
{\displaystyle \frac {(-1)^{n+k+1} b_{n+k+1}} {(n+k+1)^{k-1}}}
}
}
{
\vmatrix
{
1 " \ldots " 1      \\ [1\jot]
(-1)^{n+1} b_{n+1} " \ldots " (-1)^{n+k+1} b_{n+k+1} \\
\vdots " \ddots " \vdots \\ [1\jot]
{\displaystyle \frac {(-1)^{n+1} b_{n+1} } {(n+1)^{k-1}} } " \ldots "
{\displaystyle \frac {(-1)^{n+k+1} b_{n+k+1}} {(n+k+1)^{k-1}} }
}
} \, .
\tag
$$

\smallskip

Allerdings ist diese Darstellung des verallgemeinerten
Summationsprozesses $T_k^{(n)} (s_n)$ numerisch nicht sehr hilfreich, da
die effiziente und verl\"a{\ss}liche Berechnung von Determinanten ein immer
noch nicht befriedigend gel\"ostes Problem der numerischen Mathematik ist.
Um eine einfache explizite Darstellung f\"ur $T_k^{(n)} (s_n)$ zu
konstruieren, schreiben wir die Modellfolge (3.1-9) auf folgende Weise
um:
$$
(n+1)^{k-1} \, \frac {s_n - s} {(-1)^{n+1} \, b_{n+1}} \; = \;
\sum_{j=0}^{k-1} \, c_j \, (n+1)^{k-j-1}
\, , \qquad n \in \N_0 \, .
\tag
$$
Die rechte Seite dieser Beziehung ist ein Polynom vom Grade $k - 1$ in
$n$. Bekanntlich [Milne-Thomson 1981, S. 29] wird ein solches Polynom
in $n$ annihiliert durch Anwendung der $k$-ten Potenz des
Differenzenoperators $\Delta$, der auf folgende Weise definiert ist:
$$
\beginAligntags
" \Delta^0 f (n) \; " = \; " f (n) \, ,
\erhoehe\aktTag \\ \tag*{\tagnr a}
" \Delta f (n) \; " = \; " f (n+1) \, - \, f (n) \, ,
\\ \tag*{\tagform\aktTagnr b}
" \Delta^k f (n) \; " = \; " \Delta \, [ \Delta^{k-1} f (n)] \, ,
\qquad k \ge 1 \, .
\\ \tag*{\tagform\aktTagnr c}
\endAligntags
$$
Hierbei ist $f$ eine Funktion, die f\"ur alle positiven und negativen
ganzen Zahlen $n$ definiert ist und von der angenommen wird, da{\ss} sie f\"ur
alle negativen Werte von $n$ Null ist.

Da $\Delta^k$ linear ist, erh\"alt man aus Gl. (3.1-11) sofort die
folgende Darstellung:
$$
T_{k}^{(n)} (s_n) \; = \; \frac
{
\Delta^k \, \bigl\{ (n + 1)^{k-1} \>
s_n / (-1)^{n+1} b_{n+1} \bigr\}
}
{
\Delta^k \, \bigl\{ (n + 1)^{k-1}
/ (-1)^{n+1} b_{n+1} \bigr\}
}
\; , \qquad k, n \in \N_0 \, .
\tag
$$
Wenn wir jetzt noch die folgende Beziehung verwenden [Milne-Thomson
1981, S. 33],
$$
\Delta^k f(n) \; = \; (-1)^k \, \sum_{j=0}^k (-1)^j \binom k j
f(n+j) \, , \qquad k \in \N_0 \, ,
\tag
$$
erhalten wir f\"ur den verallgemeinerten Summationsproze{\ss} $T_{k}^{(n)}
(s_n)$ eine explizite Darstellung als Quotient zweier endlicher Summen:
$$
T_{k}^{(n)} (s_n) \; = \;
\frac
{\displaystyle
\sum_{j=0}^{k} \; \binom {k} {j} \;
(n + j + 1)^{k-1} \; \frac {s_{n+j}} {b_{n+j+1}}
}
{\displaystyle
\sum_{j=0}^{k} \; \; \binom {k} {j} \;
(n + j + 1)^{k-1} \; \frac {1} {b_{n+j+1}}
}
\; , \qquad k, n \in \N_0 \; .
\tag
$$
Dieser verallgemeinerte Summationsproze{\ss} $T_{k}^{(n)} (s_n)$ ist in der
Lage, sowohl die Konvergenz alternierender Reihen betr\"achtlich zu
verbessern als auch zahlreiche alternierende divergente Reihen vom Typ
der divergenten hypergeometrischen Reihe ${}_2 F_0$ in Gl. (3.1-6)
effizient zu summieren. Wie sp\"ater gezeigt wird, ist $T_{k}^{(n)} (s_n)$
n\"amlich ein Spezialfall einer allgemeineren Klasse nichtlinearer
Transformationen, die von Levin [1973] eingef\"uhrt worden war.
Ausgedehnte numerische Untersuchungen, die von Smith und Ford [1979;
1982] und vom Autor [Weniger 1989, Abschnitte 13 und 14] durchgef\"uhrt
wurden, zeigen, da{\ss} die Levinsche Transformation, die in Abschnitt 5
dieser Arbeit eingehender behandelt wird, zu den leistungsf\"ahigsten und
vielseitigsten Verfahren zur Konvergenzverbesserung bzw. zur Summation
von unendlichen Reihen geh\"ort, die zur Zeit bekannt sind.

\medskip

\Abschnitt Die Konstruktion verallgemeinerter Summationsprozesse durch
Modellfolgen

\smallskip

\aktTag = 0

Im letzten Abschnitt wurde ein sehr leistungsf\"ahiger verallgemeinerter
Summationsproze{\ss} $T_{k}^{(n)} (s_n)$ f\"ur konvergente und divergente
alternierende Reihen konstruiert, indem man annahm, da{\ss} der
Summationsrest $r_n$ einer alternierenden Reihe dargestellt werden kann
durch eine obere Schranke gem\"a{\ss} Gl. (3.1-5) multipliziert mit den
f\"uhrenden Termen einer asymptotischen Potenzreihe im Sinne von Poincar\'e
in der Variablen $1/(n+1)$. Allerdings kann man nicht erwarten, da{\ss} der
Ansatz (3.1-8) auch bei einer logarithmisch konvergenten Reihe zum Ziel
f\"uhren w\"urde, da in einem solchen Fall die Absch\"atzung (3.1-5) nicht
gelten w\"urde.

Wenn eine unendliche Reihe, deren Konvergenz verbessert werden soll,
nicht alterniert, sondern einen anderen Konvergenztyp besitzt, m\"ussen
die Annahmen, die dem verallgemeinerten Summationsproze{\ss} $T_{k}^{(n)}
(s_n)$ zugrundeliegen, entsprechend modifiziert werden. Das geschieht am
einfachsten dadurch, da{\ss} man die Absch\"atzung (3.1-5) f\"ur den
Summationsrest alternierender Reihen durch Absch\"atzungen f\"ur die
Summationsreste unendlicher Reihen anderen Typs ersetzt. Man erh\"alt auf
diese Weise anstelle der Modellfolge (3.1-9) andere Modellfolgen, die
zur Approximation der Partialsummen von unendlichen Reihen anderen Typs
geeignet sein sollten.

Dieses Konstruktionsprinzip f\"ur verallgemeinerte Summationsprozesse auf
der Basis von Modellfolgen kann auf einfache Weise noch weiter
verallgemeinert werden, indem man Modellfolgen des folgenden Typs
betrachtet, die ebenfalls aus endlich vielen Termen bestehen und die
Modellfolge (3.1-9) als Spezialfall enthalten:
$$
s_n \; = \; s \, + \, \sum_{j=0}^{k-1} \> c_j \, f_j (n) \, ,
\qquad k, n \in \N_0 \, .
\tag
$$
Die Koeffizienten $c_j$ dieser Modellfolge sind wie in Gl. (3.1-9)
unbestimmt. Was die Funktionen $f_j (n)$ mit $j, n \in \N_0$ betrifft,
so wird angenommen, da{\ss} sie von Null verschiedene bekannte Funktionen
von $n$ sind, die ansonsten im Prinzip beliebig sind. Der Ansatz
(3.2-1), der gem\"a{\ss} Sidi [1988a, S. 238] zuerst von Hart, Cheney, Lawson,
Maehly, Mesztenyi, Rice, Thacher und Witzgall [1968, S. 39] verwendet
wurde, ist zur Beschreibung sowohl konvergenter als auch divergenter
Folgen von Partialsummen geeignet, je nach dem, ob die Funktionen
$\{ f_j (n) \}_{j=0}^{\infty}$ f\"ur $n \to \infty$ konvergieren oder
divergieren.

Allerdings sollten die Elemente der Modellfolge (3.2-1) in der Lage
sein, gute N\"aherungen f\"ur Partialsummen zu liefern, die bei praktischen
Problemen auftreten. Au{\ss}erdem sollte die G\"ute der N\"aherung zunehmen,
wenn man die Summationsgrenze $k$ in Gl. (3.2-1) vergr\"o{\ss}ert. Demzufolge
ist es sicherlich sinnvoll zu fordern, da{\ss} die Funktionen $\{ f_j (n)
\}_{j=0}^{\infty}$ eine asymptotische Folge bilden sollten, d.~h.,
$$
f_{j+1} (n) \; = \; o \bigl(f_j (n) \bigr) \, ,
\qquad n \to \infty \, .
\tag
$$

Brezinski [1980b, S. 176] zeigte, da{\ss} man zahlreiche sehr
leistungsf\"ahige verallgemeinerte Summationsprozesse erh\"alt, wenn man die
unbestimmten Funktionen $f_j (n)$ in Gl. (3.2-1) auf geeignete Weise
spezialisiert. Beispiele sind die Transformation von Shanks [1955], das
Richardsonsche Extrapolationsverfahren [1927], der Wynnsche
$\rho$-Algorithmus [1956b] oder die Reihentransformation von Levin
[1973] und ihre Verallgemeinerung durch Sidi [1982]. Einige vom Autor
gefundene Reihentransformationen [Weniger 1989, Abschnitte 8 und 9], die
eng mit der Reihentransformation von Levin [1973] verwandt sind und die
in Abschnitt 5 dieser Arbeit behandelt werden, wurden ebenfalls auf der
Basis von Modellfolgen konstruiert, die Spezialf\"alle der allgemeinen
Modellfolge (3.2-1) sind.

Es ist es m\"oglich, einen verallgemeinerten Summationsproze{\ss} $E_k (s_n)$
zu konstruieren, der den exakten Grenzwert $s$ der Modellfolge (3.2-1)
sogar dann bestimmen kann, wenn nur die numerischen Werte der
unspezifizierten Funktionen $f_j (n)$ bekannt sind. Die exakte
funktionale Form der $f_j (n)$ mu{\ss} dazu nicht bekannt sein. Da alle
$k+1$ Unbekannte in Gl. (3.2-1) -- der Grenzwert $s$ und die $k$
Koeffizienten $c_0$, $c_1$, $\ldots$ , $c_{k-1}$ -- linear vorkommen,
folgt aus der Cramerschen Regel, da{\ss} $E_k (s_n)$ durch den folgenden
Quotienten zweier Determinanten definiert werden kann:

\smallskip

$$
E_k (s_n) \; = \; \frac
{
\vmatrix{
s_n " \ldots " s_{n+k} \\ [1\jot]
f_0 (n) " \ldots " f_0 (n+k) \\
\vdots " \ddots " \vdots \\
f_{k-1} (n) " \ldots " f_{k-1} (n+k) }
}
{
\vmatrix{
1 " \ldots " 1       \\ [1\jot]
f_0 (n) " \ldots " f_0 (n+k) \\
\vdots " \ddots " \vdots \\
f_{k-1} (n) " \ldots " f_{k-1} (n+k) }
} \, .
\tag
$$

\smallskip

Wenn also die numerischen Werte der Funktionen $f_i (n + j)$ mit $0 \le
i \le k - 1$ und $0 \le j \le k$ und der $k+1$ Folgenelemente $s_n$,
$s_{n+1}$, $\ldots$ , $s_{n+k}$ bekannt sind, liefert Gl. (3.2-3) den
Grenzwert $s$ der Modellfolge (3.2-1):
$$
E_k (s_n) \; = \; s \, , \qquad k, n \in \N_0 \, .
\tag
$$

Die Definition eines verallgemeinerten Summationsprozesses als Quotient
von Determinanten ist aber nicht sehr zweckm\"a{\ss}ig, da -- wie schon
mehrfach erw\"ahnt -- die effiziente und verl\"a{\ss}liche Berechnung von
Determinanten ein immer noch nicht befriedigend gel\"ostes Problem der
numerischen Mathematik ist. Der verallgemeinerte Summationsproze{\ss} $E_k
(s_n)$, der eine Verallgemeinerung des Richardsonschen
Extrapolationsverfahrens [1927] darstellt, ist nur dann praktisch
n\"utzlich, wenn andere Verfahren zu seiner Berechnung zur Verf\"ugung
stehen, beispielsweise ein nicht zu komplizierter expliziter Ausdruck
oder noch besser ein einfaches Rekursionsschema.

Es scheint, da{\ss} Schneider [1975] als erster ein allerdings sehr
kompliziertes Rekursionsschema fand, mit dessen Hilfe der
verallgemeinerte Summationsproze{\ss} $E_k (s_n)$ berechnet werden kann.
Allerdings wurde dieser Artikel offensichtlich \"ubersehen und H{\aa}vie
[1979] und Brezinski [1980b] leiteten unabh\"angig voneinander und unter
Verwendung verschiedener Techniken das von Schneider [1975] gefundene
Rekursionsschema erneut ab. Einen \"Uberblick \"uber die verschiedenen
mathematischen Techniken, die von den oben genannten Autoren verwendet
wurden, findet man in einem Artikel von Brezinski [1989]. Die
Konvergenzeigenschaften des verallgemeinerten Summationsprozesses $E_k
(s_n)$ in Konvergenzbeschleunigungsprozessen wurden von Brezinski
[1980b] analysiert. Au{\ss}erdem ver\"offentlichte Brezinski [1982] ein
FORTRAN-Programm f\"ur das oben erw\"ahnte Rekursionsschema. Vor kurzem
wurde auch eine Implementation des Rekursionsschemas f\"ur $E_k (s_n)$ in
MAPLE [Char, Geddes, Gonnet, Leong, Monagan und Watt 1991a] von
Grotendorst [1990] ver\"offentlicht.

Ford und Sidi [1987] gelang es sp\"ater, ein effizienteres
Rekursionsschema zu konstruieren, das den gleichen verallgemeinerten
Summationsproze{\ss} mit einer geringeren Anzahl von Rechenoperationen
berechnet als das von Schneider [1975], H{\aa}vie [1979] und Brezinski
[1980b] unabh\"angig voneinander abgeleitete Rekursionsschema. Allerdings
ist auch das Rekursionsschema von Ford und Sidi [1987] immer noch
deutlich weniger effizient als die vergleichsweise einfachen
Rekursionsschemata der oben erw\"ahnten Transformationen, die man erh\"alt,
wenn man die unspezifizierten Funktionen $f_j (n)$ in Gl. (3.2-1) auf
geeignete Weise spezialisiert.

Eine weitere Komplikation ist, da{\ss} es aufgrund der Allgemeinheit des
verallgemeinerten Summationsprozesses $E_k (s_n)$ mit unspezifizierten
Funktionen $f_j (n)$ vergleichsweise schwierig ist, seine Eigenschaften
in Konvergenzbeschleunigungs- oder Summationsprozessen theoretisch zu
analysieren. Deswegen ist es normalerweise einfacher und auch
effizienter, nicht die ganz allgemeine Transformation $E_k (s_n)$ zu
verwenden, sondern die oben erw\"ahnten Spezialf\"alle, die man durch
explizite Wahl der Funktionen $f_j (n)$ in Gl. (3.2-1) erh\"alt.

Au{\ss}erdem sollte man bedenken, da{\ss} es f\"ur die erfolgreiche praktische
Anwendung eines verallgemeinerten Summationsprozesses, der auf der ganz
allgemeinen Modellfolge (3.2-1) basiert, nicht so wichtig ist zu wissen,
da{\ss} man ihn auch f\"ur unspezifizierte Funktionen $f_j (n)$ rekursiv
berechnen kann. Vielmehr mu{\ss} man Funktionen $f_j (n)$ finden, die zur
Behandlung des vorliegendem Problems besonders gut geeignet sind.

\medskip

\Abschnitt Iterierte Transformationen

\smallskip

\aktTag = 0

Ausgehend entweder von der allgemeinen Transformation $E_k (s_n)$, Gl.
(3.2-3), oder von einem der zahlreichen Spezialf\"alle, die im letzten
Abschnitt erw\"ahnt wurden, k\"onnen verallgemeinerte Summationsprozesse
konstruiert werden, die nicht auf Modellfolgen des Typs von Gl. (3.2-1)
basieren. Nehmen wir dazu an, da{\ss} ein verallgemeinerter
Summationsproze{\ss} ${\cal T}_k^{(n)}$ mit einer festen
Transformationsordnung $k \in \N_0$ eine Folge $\Seqn s$ von
Partialsummen in eine Folge $\Seqn {s'}$ mit
$$
s'_n \; = \; {\cal T}_k^{(n)} \, , \qquad n \in \N_0 \, ,
\tag
$$
transformiert. Nehmen wir au{\ss}erdem an, da{\ss} f\"ur ein $\kappa \in \N$,
das normalerweise eine relativ kleine ganze Zahl ist, ${\cal
T}_{\kappa}^{(n)}$ explizit angegeben werden kann als Funktion der
Folgenelemente $s_n$, $s_{n+1}$, $\ldots$ , $s_{n+ \lambda}$,
$$
{\cal T}_{\kappa}^{(n)} \; = \;
F (s_n, s_{n+1}, \ldots, s_{n+\lambda}) \, .
\tag
$$

Dieser Proze{\ss} der Berechnung transformierter Gr\"o{\ss}en kann iteriert
werden. Das bedeutet, da{\ss} die Elemente der transformierten Folge (3.3-1)
als Eingabedaten f\"ur den verallgemeinerten Summationsproze{\ss} (3.3-2)
verwendet werden. Dabei werden in Gl. (3.3-2) die Folgenelemente $s_n$,
$s_{n+1}$, $\ldots$ , $s_{n+ \lambda}$ durch die transformierten
Folgenelemente $s'_n$, $s'_{n+1}$, $\ldots$ , $s'_{n+\lambda}$ ersetzt.

Dieser Vorgang kann im Prinzip beliebig oft wiederholt werden. Man
erh\"alt auf diese Weise einen neuen verallgemeinerten Summationsproze{\ss}
${\mit \Theta}_k^{(n)}$. Dazu definieren wir
$$
{\mit \Theta}_0^{(n)} \; = \; s_n \, , \qquad n \in \N_0 \, ,
\tag
$$
und schreiben Gl. (3.3-2) auf folgende Weise um:
$$
{\mit \Theta}_1^{(n)} \; = \;
F \left( {\mit \Theta}_0^{(n)}, {\mit \Theta}_0^{(n+1)},
\ldots, {\mit \Theta}_0^{(n + \lambda)} \right)
\, , \qquad n \in \N_0 \, .
\tag
$$
Durch Iteration dieser Beziehung erh\"alt man ein Rekursionsschema, mit
dessen Hilfe der verallgemeinerte Summationsproze{\ss} ${\mit
\Theta}_{k+1}^{(n)}$ f\"ur $k \ge 1$ berechnet werden kann:
$$
{\mit \Theta}_{k+1}^{(n)} \; = \; F \left(
{\mit \Theta}_k^{(n)}, {\mit \Theta}_k^{(n+1)}, \ldots,
{\mit \Theta}_k^{(n + \lambda)} \right) \, ,
\qquad k, n \in \N_0 \, .
\tag
$$

Der wahrscheinlich am besten bekannte verallgemeinerte
Summationsproze{\ss}, der auf einfache Weise iteriert werden kann, ist der
Aitkensche $\Delta^2$-Proze{\ss} [Aitken 1926],
$$
{\cal A}_1^{(n)} \; = \; s_n \; - \; \frac
{ [ \Delta s_n ]^2 } { \Delta^2 s_n } \, ,
\qquad n \in\N_0 \, ,
\tag
$$
der per Konstruktion in der Lage ist, den Grenzwert von Folgen des Typs
$$
s_n \; = \; s \, + \, c \lambda^n \; , \qquad c \ne 0 \, , \quad
\lambda \ne 1 \, , \quad n \in \N_0 \, ,
\tag
$$
aus drei aufeinanderfolgenden Folgenelementen $s_n$, $s_{n+1}$ und
$s_{n+2}$ exakt zu bestimmen. Dieser $\Delta^2$-Proze{\ss} kann
offensichtlich auf folgende Weise iteriert werden:
$$
\beginAligntags
" {\cal A}_0^{(n)} \, " = " \, s_n \; , \hfill
\erhoehe\aktTag
\\ \tag*{\tagnr a}
" {\cal A}_{k+1}^{(n)} \, " = " \, {\cal A}_{k}^{(n)} \, - \,
\frac {\bigl[\Delta {\cal A}_{k}^{(n)}\bigr]^2}
{\Delta^2 {\cal A}_{k}^{(n)}} \; , \qquad k,n \in \N_0 \, .
\\ \tag*{\tagform\aktTagnr b}
\endAligntags
$$
In dieser als auch in allen anderen Arbeiten des Autors wird immer die
Konvention verwendet, da{\ss} bei zweifach indizierten Gr\"o{\ss}en der in Gl.
(3.1-12) definierte Differenzenoperator $\Delta$ immer nur auf den
oberen Index $n$ und nicht auf den unteren Index $k$ wirkt. Es gilt also
in Gl. (3.3-8):
$$
\Delta {\cal A}_{k}^{(n)} \; = \;
{\cal A}_{k}^{(n+1)} \, - \, {\cal A}_{k}^{(n)} \, ,
\qquad n \in \N_0 \, .
\tag
$$

Der Aitkensche $\Delta^2$-Proze{\ss} (3.3-6) kann auch als Spezialfall
des Wynnschen $\epsilon$-Algorithmus (2.4-10) aufgefa{\ss}t werden. Setzt
man n\"amlich in Gl. (2.4-10b) $k = 1$, so ergibt ein Vergleich mit Gl.
(3.3-6):
$$
{\cal A}_1^{(n)} \; = \; \epsilon_2^{(n)} \, , \qquad n \in \N_0 \, .
\tag
$$
Offensichtlich sind der Wynnsche $\epsilon$-Algorithmus (2.4-10) und
der iterierte Aitkensche $\Delta^2$-Proze{\ss} (3.3-8) verschiedene
Verallgemeinerungen des Aitkenschen $\Delta^2$-Prozesses (3.3-6). Es ist
deswegen auch nicht \"uberraschend, da{\ss} der $\epsilon$-Algorithmus und der
iterierte $\Delta^2$-Proze{\ss} aufgrund ihrer gemeinsamen Abstammung
\"ahnliche Eigenschaften in Konvergenzbeschleunigungs- und
Summationsprozessen aufweisen. Beide Transformationen sind
leistungsf\"ahige Beschleuniger im Falle linearer Konvergenz, und sie sind
auch in der Lage, zahlreiche divergente alternierende Reihen zu
summieren. Die Hauptschw\"ache beider Transformationen ist, da{\ss} sie nicht
in der Lage sind, logarithmische Konvergenz zu beschleunigen.

Trotz ihrer unbestreitbaren \"Ahnlichkeiten unterscheiden sich der
$\epsilon$-Algorithmus und der iterierte $\Delta^2$-Proze{\ss} bei
speziellen Problemen aber oft sehr deutlich [Brezinski und Lembarki
1986; Drummond 1981, Tabelle 3; Weniger 1989, Tabelle 13-1; Weniger und
{\v C}{\'\i}{\v z}ek 1990, Tabelle 1]. Allerdings ist ein derartig
unterschiedliches Verhalten eng verwandter verallgemeinerter
Summationsprozesse keineswegs un\"ublich [Weniger 1991].

Durch das Iterieren expliziter Ausdr\"ucke vom Typ von Gl. (3.3-4) k\"onnen
viele leistungsf\"ahige verallgemeinerte Summationsprozesse konstruiert
werden [Bhowmick, Bhattacharya und Roy 1989; Homeier 1993; Weniger
1991]. Allerdings gibt es dabei manchmal erhebliche Probleme, die
leistungsf\"ahigste Iteration zu finden. Der Schritt von Gl. (3.3-4) zu
Gl. (3.3-5) beinhaltet oft ein erhebliches Ma{\ss} an Willk\"ur und wird
normalerweise nicht so offensichtlich und einleuchtend sein wie im Falle
des Aitkenschen $\Delta^2$-Prozesses (3.3-6), der in dem
Rekursionsschema (3.3-8) eine sehr nat\"urliche Iteration besitzt. In
diesem Zusammenhang sollte man auch bedenken, da{\ss} der Wynnsche
$\epsilon$-Algorithmus (2.4-10) aufgrund von Gl. (3.3-10) ebenfalls als
eine, wenn auch weniger naheliegende Iteration des Aitkenschen
$\Delta^2$-Prozesses (3.3-6) interpretiert werden kann.

Diese Nichteindeutigkeitsprobleme bei der Konstruktion iterierter
Transformationen sind besonders offensichtlich, wenn die rechte Seite
von Gl. (3.3-2) explizit von $n$ abh\"angt und nicht nur implizit \"uber
die Eingabedaten $s_n$, $s_{n+1}$, $\ldots$ , $s_{n + \lambda}$. In
einem solchen Fall gibt es normalerweise keine eindeutig bestimmte
Verallgemeinerung der in Gln. (3.3-2) und (3.3-4) vorkommenden Funktion
$F$ f\"ur h\"ohere Transformationsordnungen $k$. Man mu{\ss} dann zus\"atzlich
noch in Betracht ziehen, da{\ss} $F$ f\"ur h\"ohere Werte von $k$ ebenfalls
explizit von $k$ abh\"angen kann. Anstelle von Gl. (3.3-5) mu{\ss} man dann
mit Rekursionsbeziehungen des folgenden Typs rechnen:
$$
{\mit \Theta}_{k+1}^{(n)} \; = \; F_k^{(n)} \left(
{\mit \Theta}_k^{(n)}, {\mit \Theta}_k^{(n+1)}, \ldots,
{\mit \Theta}_k^{(n + \lambda)} \right) \, ,
\qquad k, n \in \N_0 \, .
\tag
$$

Ungl\"ucklicherweise gibt es keine allgemeing\"ultige Regel, mit deren Hilfe
man vorhersagen kann, welche der zahlreichen zul\"assigen
Verallgemeinerungen $F_k^{(n)}$ einer vorgegebenen explizit
$n$-abh\"angigen Funktion $F_0^{(n)}$ die besten Resultate in
Konvergenzbeschleunigungs- und Summationsprozessen liefern wird. Das
Nichteindeutigkeitsproblem wird dadurch versch\"arft, da{\ss} verschiedene
Iterationen $F_k^{(n)}$ der gleichen Funktion $F_0^{(n)}$ ganz
unterschiedliche Eigenschaften haben k\"onnen [Weniger 1991].

Ein Beispiel f\"ur eine Transformation, die explizit von $n$ abh\"angt, ist
der Wynnsche $\rho$-Algorithmus [Wynn 1956b]:
$$
\beginAligntags
" \rho_{-1}^{(n)} " \; = \; " 0 \, ,
\hfill \rho_0^{(n)} \; = \; s_n \, ,
\erhoehe\aktTag
\\ \tag*{\tagnr a}
" \rho_{k+1}^{(n)} " \; = \; " \rho_{k-1}^{(n+1)} \, + \,
\frac
{x_{n+k+1} - x_n} {\rho_{k}^{(n+1)} - \rho_{k}^{(n)}} \, ,
\qquad k,n \in \N_0 \, .
\\ \tag*{\tagform\aktTagnr b}
\endAligntags
$$

Der Wynnsche $\rho$-Algorithmus ist formal fast identisch mit dem
Wynnschen $\epsilon$-Algorithmus, Gl. (2.4-10). Der einzige Unterschied
besteht darin, da{\ss} in Gl. (3.3-12) au{\ss}erdem noch Interpolationspunkte
$\Seqn x$ vorkommen, von denen angenommen wird, da{\ss} sie f\"ur alle $n \in
\N_0$ positiv und voneinander verschieden sind und da{\ss} sie au{\ss}erdem mit
zunehmendem $n$ unbeschr\"ankt wachsen,
$$
\beginMultiline
0 < x_0 < x_1 < x_2 < \cdots < x_m < x_{m+1} < \cdots \, ,
\erhoehe\aktTag \\ \tag*{\tagnr a}
\lim_{n \to \infty} \; x_n \; = \; \infty \, .
\\ \tag*{\tagform\aktTagnr b}
\endMultiline
$$
Die folgenden Interpolationspunkte werden beim Wynnschen
$\rho$-Algorithmus am h\"aufigsten verwendet:
$$
x_n = n + \beta \, , \qquad n \in \N_0 \, , \quad \beta \in \R_{+} \, .
\tag
$$
Wenn man diese Interpolationspunkte in Gl. (3.3-12) verwendet, erh\"alt
man die Standardform des Wynnschen $\rho$-Algorithmus:
$$
\beginAligntags
" \rho_{-1}^{(n)} " \; = \; " 0 \, ,
\hfill \rho_0^{(n)} \; = \; s_n \, ,
\erhoehe\aktTag \\ \tag*{\tagnr a}
" \rho_{k+1}^{(n)} " \; = \; " \rho_{k-1}^{(n+1)} \, + \,
\frac {k+1} {\rho_{k}^{(n+1)} - \rho_{k}^{(n)} }
\, , \qquad k,n \in \N_0 \, .
\\ \tag*{\tagform\aktTagnr b}
\endAligntags
$$
Bemerkenswerterweise h\"angt Gl. (3.3-15b) im Gegensatz zu Gl. (3.3-12b)
nicht mehr explizit von $n$ ab, sondern nur noch implizit \"uber die
Elemente $\rho_{k-1}^{(n+1)}$, $\rho_{k}^{(n+1)}$ und $\rho_{k}^{(n)}$.

Da der $\epsilon$-Algorithmus und der $\rho$-Algorithmus formal fast
identisch sind, kann man bei der Kon\-struktion einer iterierten
Transformation, die auf dem $\rho$-Algorithmus basiert, ebenso verfahren
wie im Falle des Aitkenschen $\Delta^2$-Prozesses (3.3-6), der gem\"a{\ss}
Gl. (3.3-10) mit $\epsilon_2^{(n)}$ identisch ist. Man dr\"uckt also
zuerst $\rho_2^{(n)}$ durch die Partialsummen $s_n$, $s_{n+1}$ und
$s_{n+2}$ und die Interpolationspunkte $x_n$, $x_{n+1}$ und $x_{n+2}$
aus. Der resultierende Ausdruck f\"ur $\rho_2^{(n)}$ kann dann iteriert
werden.

Wenn wir in Gl. (3.3-12) $k = 1$ setzen, erhalten wir den folgenden
Ausdruck f\"ur das $\rho$-Analogon des Aitkenschen $\Delta^2$-Prozesses:
$$
\rho_2^{(n)} \; = \; s_{n+1} \, + \, \frac
{(x_{n+2} - x_n) [ \Delta s_{n+1} ] [ \Delta s_n ] }
{ [ \Delta x_{n+1} ] [ \Delta s_{n} ] -
[ \Delta x_n ] [ \Delta s_{n+1} ] }
\, , \qquad n \in \N_0 \, .
\tag
$$

Dieser Ausdruck, der eine Art gewichteter $\Delta^2$-Proze{\ss} ist und der
explizit von $n$ abh\"angt, kann auf verschiedene Weise iteriert werden.
Das Problem ist, da{\ss} es nicht {\it a priori} klar ist, wie die Indizes
der Interpolationspunkte $\Seqn x$ in einer Iteration von Gl. (3.3-16)
gew\"ahlt werden sollen. Wenn man aber in Betracht zieht, da{\ss} in Gl.
(3.3-12b) die Differenzen der Indizes der Interpolationspunkte
$x_{n+k+1}$ und $x_n$ mit zunehmendem $k$ linear wachsen, und wenn man
au{\ss}erdem in Betracht zieht, da{\ss} ein Rekursionsschritt einer Iteration
von Gl. (3.3-16) zwei Rekursionschritten von Gl. (3.3-12b) entspricht,
so scheint das folgende nichtlineare Rekursionsschema [Weniger 1989,
Abschnitt 6.3] eine relativ naheliegende Iteration der elementaren
Transformation (3.3-16) zu sein:
$$
\beginAligntags
" {\cal W}_{0}^{(n)} \; " = \; " s_n \, ,
\hfill n \in \N_0 \, , \erhoehe\aktTag \\ \tag*{\tagnr a}
" {\cal W}_{k+1}^{(n)} \; " = \; " {\cal W}_{k}^{(n+1)} \,
+ \frac
{ (x_{n + 2 k + 2} - x_n ) \bigl[ \Delta {\cal W}_k^{(n+1)}
\bigr] \bigl[ \Delta {\cal W}_k^{(n)} \bigr] }
{(x_{n + 2 k + 2} - x_{n+1}) \bigl[ \Delta {\cal W}_k^{(n)} \bigr] -
(x_{n + 2 k + 1} - x_n) \bigl[ \Delta {\cal W}_k^{(n+1)} \bigr] } \, ,
\\
" " " k, n \in \N_0 \, .
\\ \tag*{\tagform\aktTagnr b}
\endAligntags
$$
Wenn wir die Interpolationspunkte in diesem Rekursionsschema gem\"a{\ss} Gl.
(3.3-14) w\"ahlen, erhalten wir die Standardform von ${\cal W}_k^{(n)}$
[Weniger 1989, Abschnitt 6.3]:
$$
\beginAligntags
" {\cal W}_{0}^{(n)} \; " = \; " s_n \, ,
\hfill n \in \N_0 \, , \erhoehe\aktTag \\ \tag*{\tagnr a}
" {\cal W}_{k+1}^{(n)} \; " = \; " {\cal W}_{k}^{(n+1)} \, -
\frac
{ (2 k + 2) \bigl[ \Delta {\cal W}_k^{(n+1)} \bigr]
\bigl[ \Delta {\cal W}_k^{(n)} \bigr] }
{ (2 k + 1) \Delta^2 {\cal W}_k^{(n)} }
\, , \hfill \qquad k,n \in \N_0 \, . \\
\tag*{\tagform\aktTagnr b}
\endAligntags
$$

Allerdings ist das nicht die einzige M\"oglichkeit, die elementare
Transformation (3.3-16) zu iterieren. Man kann beispielsweise auch
fordern, da{\ss} die Indizes der Interpolationspunkte sich w\"ahrend der
Rekursion \"uberhaupt nicht \"andern sollen. Damit erh\"alt man die folgende
Transformation [Weniger 1991, Gl. (2.14)]:
$$
\beginAligntags
" {\mit B}_{0}^{(n)} \; " = \; " s_n \, ,
\hfill n \in \N_0 \, , \erhoehe\aktTag \\ \tag*{\tagnr a}
" {\mit B}_{k+1}^{(n)} \; " = \; " {\mit B}_{k}^{(n+1)} \,
+ \frac
{ (x_{n + 2} - x_n ) \bigl[ \Delta {\mit B}_k^{(n+1)}
\bigr] \bigl[ \Delta {\mit B}_k^{(n)} \bigr] }
{(x_{n + 2} - x_{n+1}) \bigl[ \Delta {\mit B}_k^{(n)} \bigr] -
(x_{n + 1} - x_n) \bigl[ \Delta {\mit B}_k^{(n+1)} \bigr] } \, ,
\\
" " " k,n \in \N_0 \, .
\\ \tag*{\tagform\aktTagnr b}
\endAligntags
$$
Wenn wir die Interpolationspunkte in diesem Rekursionsschema gem\"a{\ss} Gl.
(3.3-14) w\"ahlen, erhalten wir die Standardform dieser Transformation:
$$
\beginAligntags
" {\mit B}_{0}^{(n)} \; " = \; " s_n \, ,
\hfill n \in \N_0 \, , \erhoehe\aktTag \\ \tag*{\tagnr a}
" {\mit B}_{k+1}^{(n)} \; " = \; " {\mit B}_{k}^{(n+1)} \, -
\frac
{ 2\, \bigl[ \Delta {\mit B}_k^{(n+1)} \bigr]
\bigl[ \Delta {\mit B}_k^{(n)} \bigr] }
{ \Delta^2 {\mit B}_k^{(n)} }
\, , \hfill \qquad k,n \in \N_0 \, . \\
\tag*{\tagform\aktTagnr b}
\endAligntags
$$
Diese Standardform von ${\mit B}_k^{(n)}$ war von Bhowmick, Bhattacharya
und Roy [1989] abgeleitet worden. Bei ihrer Ableitung gingen Bhowmick,
Bhattacharya und Roy allerdings nicht von der allgemeinen Form von
$\rho_2^{(n)}$ gem\"a{\ss} Gl. (3.3-16) aus, sondern von der Standardform
dieser Transformation,
$$
\rho_2^{(n)} \; = \; s_{n+1} \, - \, \frac
{2 [ \Delta s_{n+1} ] [ \Delta s_n ] }
{ \Delta^2 s_{n} }
\, , \qquad n \in \N_0 \, .
\tag
$$
Der Unterschied zwischen den Gln. (3.3-16) und (3.3-21) ist, da{\ss} Gl.
(3.3-21) nicht explizit von $n$ abh\"angt, sondern nur implizit \"uber die
Partialsummen $s_n$, $s_{n+1}$, und $s_{n+2}$, wogegen in Gl. (3.3-16)
auch noch die Interpolationspunkte $x_n$, $x_{n+1}$, und $x_{n+2}$
vorkommen.

Numerische Tests [Weniger 1991, Tabelle 1] ergaben, da{\ss} die
Transformation ${\mit B}_k^{(n)}$ weit weniger leistungsf\"ahig ist als
etwa die eng verwandte Transformation ${\cal W}_k^{(n)}$, die ebenfalls
durch Iteration von Gl. (3.3-16) abgeleitet wurde. Dieses und andere
Beispiele [Weniger 1991] zeigen, da{\ss} man bei der Konstruktion iterierter
Transformationen sehr vorsichtig sein mu{\ss}, um aus der Vielzahl der
m\"oglichen Iterationen einer elementaren Transformation die wirklich
leistungsf\"ahigen Varianten zu finden.

\endAbschnittsebene

\endAbschnittsebene

\keinTitelblatt\neueSeite

\beginAbschnittsebene
\aktAbschnitt = 3

\Abschnitt Pad\'e-Approximationen

\vskip - 2 \jot

\beginAbschnittsebene

\medskip

\Abschnitt Vorbemerkungen

\smallskip

\aktTag = 0

Wie in Abschnitt 4.5 ausf\"uhrlich diskutiert wird, sind
Pad\'e-Approximationen zweifellos das am weitesten verbreitete Verfahren
zur Verbesserung der Konvergenz unendlicher Reihen und zur Summation
divergenter Reihen. In einigen Bereichen der theoretischen Physik werden
Pad\'e-Approximationen inzwischen quasi routinem\"a{\ss}ig und mit
gro{\ss}em Erfolg verwendet.

Andere verallgemeinerte Summationsprozesse werden in den
Naturwissenschaften und in der Technik bisher noch vergleichsweise wenig
verwendet. Das liegt zu einem gro{\ss}en Teil daran, da{\ss} die meisten
dieser Verfahren neueren Datums sind und demzufolge bisher fast
ausschlie{\ss}lich Spezialisten bekannt sind. Ausnahmen sind
beispielsweise der Aitkensche $\Delta^2$-Proze{\ss}, Gl. (3.3-6), oder
das ebenfalls sehr alte Richardsonsche Extrapolationsverfahren
[Richardson 1927], das h\"aufig verwendet wird, um logarithmische
Konvergenz zu beschleunigen.

Da Pad\'e-Approximationen einen {\it de facto} Standard darstellen, an dem
sich alle anderen Verfahren zu messen haben, sollen sie hier
ausf\"uhrlicher dargestellt werden, obwohl das Schwergewicht der Forschung
des Autors nicht auf dem Gebiet der Pad\'e-Approximationen liegt.

Pad\'e-Approximationen und ihre Eigenschaften sind in der mathematischen
Literatur sehr ausf\"uhrlich behandelt worden. Die \"altere Geschichte der
Pad\'e-Approximationen bis zum Jahre 1939 wird in einer Monographie von
Brezinski [1991a] beschrieben. Die mathematischen Eigen\-schaften der
Pad\'e-Approximationen und zahlreiche Anwendungen findet man in B\"uchern
von Baker [1975; 1990], Baker und Gammel [1970], Baker und Graves-Morris
[1981a; 1981b], Brezinski [1977; 1978; 1980a; 1991b], Brezinski, Draux,
Magnus, Maroni und Ronveaux [1985], Brezinski und Redivo Zaglia [1991],
Bultheel [1987], Cabannes [1976], Cuyt [1984; 1988], Cuyt und Wuytack
[1987], de Bruin und Van Rossum [1981], Draux und van Ingelandt [1987],
Gilewicz [1978], Gilewicz, Pindor und Siemaszko [1985], Graves-Morris
[1973a; 1973b], Graves-Morris, Saff und Varga [1984], Saff und Varga
[1977], Werner und B\"unger [1984], Wimp [1981] und Wuytack [1979a] sowie
in \"Ubersichtsartikeln von Baker [1965; 1972], Basdevant [1972],
Brezinski und van Iseghem [1991], Gragg [1972] und Zinn-Justin [1971].
Au{\ss}erdem sei hier noch auf umfangreiche, von Brezinski [1976; 1991c]
herausgegebene Bibliographien \"uber Pad\'e-Approximationen und verwandte
Probleme hingewiesen.

Wenn man in Kettenbr\"uchen des Typs
$$
{\cal C} (z) \; = \;
c_0 \, + \,
{c_1 z \over\displaystyle 1 +
{\strut c_2 z \over\displaystyle 1 +
{\strut c_3 z \over\displaystyle 1 +
{\strut c_4 z \over\displaystyle 1 + \ldots }}}}
\tag
$$
mit $c_1, c_2, c_3 \ldots \neq 0$ willk\"urlich ein $c_j$ mit $j \in \N$
gleich Null setzt, so erh\"alt man einen sogenannten N\"aherungsbruch. Diese
N\"aherungsbr\"uche sind rationale Funktionen in $z$, die spezielle
Pad\'e-Approximationen sind [Baker und Graves-Morris 1981a, S. 109,
Theorem 4.2.1]. Demzufolge werden Pad\'e-Approximationen auch in B\"uchern
\"uber Kettenbr\"uche behandelt, beispielsweise in den Monographien von
Bowman und Shenton [1989], Jones und Thron [1980], Lorentzen und
Waadeland [1992], Perron [1957] und Wall [1973].

In diesem Zusammenhang sei noch erw\"ahnt, da{\ss} man die numerischen
Eigenschaften mancher unendlicher Reihen auch dadurch verbessern kann,
indem man sie in Kettenbr\"uche transformiert. Beispiele daf\"ur findet man
zus\"atzlich zu den oben genannten Monographien \"uber Kettenbr\"uche auch in
B\"uchern von Borel [1928], Cuyt und Wuytack [1987], Henrici [1974],
Khovanskii [1963] und van der Laan und Temme [1980] oder auch in
Artikeln von Baltus und Jones [1989], H\"anggi, Roesel und Trautmann
[1980], Jones und Thron [1985, 1988], Lorentzen [1992] und Meyer [1985;
1986].

Aufgrund dieser Tatsache ist es nicht verwunderlich, da{\ss} Kettenbr\"uche
nicht nur in der Mathematik, sondern auch h\"aufig in den
Naturwissenschaften verwendet worden sind. Die Berechnung der
Coulombfunktionen $F_{\lambda} (\eta, x)$ und $G_{\lambda} (\eta, x)$
mit Hilfe von Kettenbr\"uchen wird von Barnett [1982] beschrieben.
Anwendungen von Kettenbr\"uchen in der Atomphysik werden von
Hor\'{a}\v{c}ek und Sasakawa [1983] und von Swain [1986] beschrieben.
Peratt [1984] verwendete Kettenbr\"uche in der Plasmaphysik. Zahlreiche
festk\"orperphysikalische Anwendungen findet man in dem Buch von Pettifor
und Weaire [1985] oder in Artikeln von Grosso und Pastori Parravicini
[1985a; 1985b], Lackner [1987], Sherman [1987], Wheeler [1984] und
Yoshino [1988]. Gerck und d'Olivera [1980] verwendeten Kettenbr\"uche zur
Berechnung der Eigenwerte tridiagonaler Matrizen, die man erh\"alt, wenn
man eine radiale Schr\"odingergleichung durch ein Differenzenschema
ersetzt. Masson verwendete Kettenbr\"uche zur Berechnung der Eigenwerte
rotierender harmonischer Oszillatoren [Masson 1983] und zur Beschreibung
der Wechselwirkung von Bosonen [Masson 1986]. Swain [1976] verwendete
Kettenbr\"uche zur L\"osung linearer Gleichungssysteme.

Au{\ss}erdem gibt es zahlreiche Arbeiten \"uber die Verwendungen von
Kettenbr\"uchen in der quantenmechanischen St\"orungstheorie. Feenberg
[1958] zeigte, da{\ss} Kettenbruchdarstellungen von St\"orungsreihen
Invarianzeigenschaften bez\"uglich bestimmter Skalentransformationen
aufweisen. Goscinski [1967] bestimmte obere und untere Schranken in der
Brillouin-Wignerschen St\"orungs\-theorie mit Hilfe von Kettenbr\"uchen.
Jackson und Swain [1981] verglichen die Verwendung von
Projektionsoperatoren mit der Darstellung der St\"orungsentwicklungen
durch Kettenbr\"uche. Schlie{\ss}lich wurden Kettenbr\"uche in Artikeln von Reid
[1967], {\v C}{\' \i}{\v z}ek und Vrscay [1982; 1984; 1985], Vrscay
[1985; 1986], Vrscay und {\v C}{\' \i}{\v z}ek [1986] und Vrscay und
Handy [1989] zur Summation divergenter quantenmechanischer
St\"orungsreihen verwendet.

\medskip

\Abschnitt Rationale Approximationen f\"ur Potenzreihen

\smallskip

\aktTag = 0

Nehmen wir an, da{\ss} man einer Funktion $f (z)$ die folgende formale
Potenzreihe zuordnen kann:
$$
f (z) \; = \;
\sum_{\nu=0}^{\infty} \> \gamma_{\nu} \, z^{\nu} \, ,
\qquad \gamma_0 \ne 0 \, .
\tag
$$
Die Pad\'e-Approximation von $f (z)$ ist der Quotient zweier Polynome
$P_{\ell} (z)$ und $Q_m (z)$ mit den Graden $\ell$ und $m$:
$$
[ \, \ell \, / \, m \, ]_f \, (z) \; = \;
P_{\ell} (z) \, / \, Q_m (z) \, .
\tag
$$
Die Polynome $P_{\ell} (z)$ und $Q_m (z)$, die als teilerfremd
vorausgesetzt werden, sind so zu w\"ahlen, da{\ss} die Taylorentwicklung der
Pad\'e-Approximation (4.2-2) soweit wie m\"oglich mit der formalen
Potenzreihe (4.2-1) f\"ur $f (z)$ \"ubereinstimmt,
$$
f (z) \, - \, P_{\ell} (z) \, / \, Q_m (z) \; = \;
O (z^{\ell + m +1}) \, , \qquad z \to 0 \, .
\tag
$$

Der Wert einer rationalen Funktion \"andert sich nicht, wenn man sowohl
den Z\"ahler als auch den Nenner mit einem gemeinsamen Faktor
multipliziert. Demzufolge sind nur $\ell + m + 1$ der $\ell + m + 2$
Polynomkoeffizienten in Gl. (4.2-2) unabh\"angig. \"Ublicherweise normiert
man die Koeffizienten der Polynome $P_{\ell} (z)$ und $Q_m (z)$ auf
solche Weise, da{\ss} der konstante Term von $Q_m (z)$ gleich Eins gesetzt
wird.

Man ben\"otigt also nur $\ell + m + 1$ Gleichungen, um die Koeffizienten
der Polynome
$$
P_{\ell} (z) \; = \;
p_0 \, + \, p_1 z \, + \, \ldots \, + \, p_{\ell} z^{\ell}
\tag
$$
und
$$
Q_m (z) \; = \;
1 \, + \, q_1 z \, + \, \ldots \, + \, q_m z^m
\tag
$$
so zu bestimmen, da{\ss} Gl. (4.2-3) erf\"ullt ist. Wenn man die
Bestimmungsgleichung (4.2-3) auf folgende Weise umschreibt,
$$
Q_m (z) \, f (z) \, - \, P_{\ell} (z) \; = \;
O(z^{\ell + m +1}) \, , \qquad z \to 0 \, ,
\tag
$$
und mit den Gln. (4.2-1), (4.2-4) und (4.2-5) kombiniert, so folgt, da{\ss}
die Koeffizienten $p_{\lambda}$ und $q_{\mu}$ der Polynome $P_{\ell}
(z)$ und $Q_m (z)$ das folgende Gleichungssystem erf\"ullen m\"ussen [Baker
1975, S. 6]:
$$
\beginAligntags
\gamma_0 \hfill \; " = \; " p_0 \, , \hfill \\
\gamma_1 \, + \, \gamma_0 q_1 \hfill \; " = \; " p_1 \, , \hfill \\
\gamma_2 \, + \, \gamma_1 q_1 \, + \, \gamma_0 q_2 \, , \hfill
\; " = \; " p_2 \, , \hfill \\
\; \vdots \hfill " " \vdots \hfill \\
\gamma_{\ell} \, + \, \gamma_{\ell-1} q_1 \, + \, \ldots \, +
\, \gamma_0 q_{\ell}
\hfill \; " = \; " p_{\ell} \, , \hfill \\ \tag
\gamma_{\ell+1} \, + \, \gamma_{\ell} q_1 \, + \, \ldots \, +
\, \gamma_{\ell-m+1} q_m \hfill \; " = \; " 0 \, , \hfill \\
\; \vdots \hfill " " \vdots \hfill \\
\gamma_{\ell+m} \, + \, \gamma_{\ell+m-1} q_1 \, + \, \ldots \, +
+ \, \gamma_{\ell} q_m \hfill \; " = \; " 0 \, . \hfill \\
\endAligntags
$$
Bei diesem Gleichungssystem wird angenommen, da{\ss} f\"ur $n < 0$
$$
\gamma_n \; = \; 0
\tag
$$
und f\"ur $j > m$
$$
q_j \; = \; 0
\tag
$$
gelten soll.

Wenn das Gleichungssystem (4.2-7) f\"ur die Polynomkoeffizienten der
Pad\'e-Approximation $[ \ell / m ]_f (z)$ der formalen Potenzreihe (4.2-1)
eine L\"osung besitzt, so ist die L\"osung durch den folgenden Quotienten
zweier Determinanten gegeben [Baker 1975, Gl. (1.27)]:
$$
[ \, \ell \, / \, m \, ]_f \, (z) \; = \; \frac
{
\vmatrix
{ \displaystyle
\gamma_{\ell-m+1} " \gamma_{\ell-m+2} " \ldots " \gamma_{\ell}
\\ [1\jot]
\gamma_{\ell-m+2} " \gamma_{\ell-m+3} " \ldots " \gamma_{\ell+1}
\\ [1\jot]
\vdots " \ddots " \vdots \\ [1\jot]
\gamma_{\ell} " \gamma_{\ell+1} " \ldots " \gamma_{\ell+m}
\\ [1\jot]
{\displaystyle \sum_{j=m}^{\ell} \gamma_{j-m} z^j} "
{\displaystyle \sum_{j=m-1}^{\ell} \gamma_{j-m+1} z^j} " \ldots "
{\displaystyle \sum_{j=0}^{\ell} \gamma_{j} z^j} \\ [1\jot]
}
}
{
\vmatrix
{
\gamma_{\ell-m+1} " \gamma_{\ell-m+2} " \ldots " \gamma_{\ell}
\\ [1\jot]
\gamma_{\ell-m+2} " \gamma_{\ell-m+3} " \ldots " \gamma_{\ell+1}
\\ [1\jot]
\vdots " \ddots " \vdots \\ [1\jot]
\gamma_{\ell} " \gamma_{\ell+1} " \ldots " \gamma_{\ell+m}
\\ [1\jot]
z^m " z^{m-1} " \ldots " 1
}
} \, .
\tag
$$

Diese Determinantendarstellung zeigt, da{\ss} man die Koeffizienten
$\gamma_0$, $\gamma_1$, $\ldots$ , $\gamma_{\ell+m}$ der Potenzreihe
(4.2-1) zur Konstruktion der Pad\'e-Approxi\-ma\-tion $[ \ell / m ]_f (z)$
ben\"otigt. Die Ordnungsrelation (4.2-3) impliziert also, da{\ss} eine
Pad\'e-Approximation $[ \ell / m ]_f (z)$ {\it alle} $\ell + m + 1$
Koeffizienten der formalen Potenzreihe (4.2-1) {\it exakt} reproduziert,
die zu ihrer Konstruktion verwendet wurden.

Was hat man gewonnen, wenn man eine Folge von Partialsummen
$$
f_n (z) \; = \;
\sum_{\nu=0}^n \> \gamma_{\nu} \, z^{\nu}
\tag
$$
der Potenzreihe (4.2-1) in eine Folge von Pad\'e-Approximationen
transformiert? Nehmen wir an, da{\ss} die Potenzreihe f\"ur $f (z)$ einen von
Null verschiedenen, aber endlichen Konvergenzradius $0 < R < \infty$
besitzt. Das bedeutet, da{\ss} die Funktion $f (z)$ irgendwo auf dem
Konvergenzkreis $\vert z \vert = R$ eine Singularit\"at $z_0$ besitzt.
Au{\ss}erdem kann man $f (z)$ nur dann durch Aufaddieren der Terme der
Potenzreihe (4.2-1) berechnen, wenn $\vert z \vert < R$ gilt.

Die Singularit\"at $z_0$, die irgendwo auf dem Konvergenzkreis $\vert z
\vert = R$ liegt, beeinflu{\ss}t also das numerische Verhalten der
Potenzreihe (4.2-1) in der gesamten komplexen Ebene $\C$. In dieser
Beziehung besitzen rationale Funktionen, d. h. Quotienten zweier
Polynome, wesentlich g\"unstigere Eigenschaften. Zwar werden rationale
Funktionen an den Nullstellen des Nennerpolynoms singul\"ar, ansonsten
kann man ihren Wert aber \"uberall in der komplexen Ebene $\C$ problemlos
berechnen. Pad\'e-Approximationen sind also prinzipiell dazu geeignet,
einer Potenzreihe auch au{\ss}erhalb ihres Konvergenzkreises einen Wert
zuzuordnen.

Wenn Folgen von Pad\'e-Approximationen auch au{\ss}erhalb des
Konvergenzkreises einer Potenzreihe ausreichend schnell konvergieren,
dann sind Pad\'e-Approximationen tat\-s\"achlich ein praktisch n\"utzliches
Verfahren zur analytischen Fortsetzung dieser Potenzreihe mit endlichem
Konvergenzradius $0 < R < \infty$.

Eine Pad\'e-Approximation $[ \ell / m ]_f (z)$ ben\"utzt ebenso wie die
Partialsummen $f_0 (z)$, $f_1 (z)$, $\ldots$ , $f_{\ell+m} (z)$ der
Potenzreihe (4.2-1) nur die numerische Information, die in den
Koeffizienten $\gamma_0$, $\gamma_1$, $\ldots$ , $\gamma_{\ell+m}$
enthalten ist. Die in diesem Zusammenhang zentrale Frage lautet also:
Warum konvergieren Pad\'e-Approximationen h\"aufig wesentlich besser als
die Partialsummen einer unendlichen Reihe, aus denen sie konstruiert
wurden?

Wenn man eine Pad\'e-Approximation $[ \ell / m ]_f (z)$ in eine
Taylorreihe um den Nullpunkt entwickelt, so werden aufgrund von Gl.
(4.2-3) die ersten $\ell+m+1$ Koeffizienten $\gamma_0$, $\gamma_1$,
$\ldots$ , $\gamma_{\ell+m}$ exakt reproduziert. Man erh\"alt also die
folgende Reihenentwicklung:
$$
\beginAligntags
" [ \, \ell \, / \, m \, ]_f \, (z) \; " = \;
\sum_{\nu=0}^{\ell+m} \, \gamma_{\nu} \, z^{\nu} \, + \, "
\sum_{\nu=\ell+m+1}^{\infty} \, \Gamma_{\nu}^{(\ell, m)} \, z^{\nu}
\hfill \\ \tag
" " = \; f_{\ell+m} (z) \hfill \, + \, "
\sum_{\nu=\ell+m+1}^{\infty} \, \Gamma_{\nu}^{(\ell, m)} \, z^{\nu}
\, . \hfill \\ \tag
\endAligntags
$$

Diese Beziehung zeigt, da{\ss} die Pad\'e-Approximation $[ \ell / m ]_f (z)$
geschrieben werden kann als die Partialsumme $f_{\ell+m} (z)$ mit dem
h\"ochsten Index $\ell+m$, der bei der Konstruktion der
Pad\'e-Approximation verwendet wurde, plus einer zus\"atzlichen Potenzreihe
mit den Koeffizienten $\Gamma_{\nu}^{(\ell, m)}$ mit $\nu > \ell + m$.
Demzufolge mu{\ss} die numerische \"Uberlegenheit der Pad\'e-Approximation
$[ \ell / m ]_f (z)$ gegen\"uber den Partialsummen an der zus\"atzlichen
Potenzreihe liegen. Zwar gibt es im allgemeinen keinen Grund zu der
Annahme, da{\ss} etwa
$$
\gamma_{\nu} \; = \; \Gamma_{\nu}^{(\ell, m)} \, ,
\qquad \nu > \ell + m \, ,
\tag
$$
f\"ur endliche Werte von $\ell$ und $m$ gelten k\"onnte. Wenn
Pad\'e-Approximationen aber in der Lage sind, den Wert der Funktion $f
(z)$ beliebig genau zu approximieren und au{\ss}erdem noch g\"unstigere
numerische Eigenschaften besitzen als die Partialsummen, aus denen sie
konstruiert wurden, dann mu{\ss} die Reihe
$$
\sum_{\nu=\ell+m+1}^{\infty} \, \Gamma_{\nu}^{(\ell, m)} \, z^{\nu} \, ,
\tag
$$
die zus\"atzlich noch in der Pad\'e-Approximation $[ \ell / m ]_f (z)$
enthalten ist, f\"ur wachsendes $\ell$ und $m$ gegen den exakten
Abbruchfehler
$$
\sum_{\nu=\ell+m+1}^{\infty} \, \gamma_{\nu} \, z^{\nu}
\tag
$$
der Partialsumme $f_{\ell+m} (z)$ konvergieren. Das bedeutet dann
nat\"urlich auch, da{\ss} die Koeffizienten $\Gamma_{\nu}^{(\ell, m)}$ f\"ur
wachsendes $\ell$ und $m$ gegen die entsprechenden Koeffizienten
$\gamma_{\nu}$ der Potenzreihe (4.2-1) konvergieren, die nicht in der
Partialsumme $f_{\ell+m} (z)$ enthalten sind.

Ideen dieser Art wurden schon von Sidi und Levin [1983] und von
Brezinski [1985b] in Artikeln \"uber verallgemeinerte Summationsprozesse,
die Partialsummen von Potenzreihen in rationale Funktionen
transformieren, behandelt. In diesen Artikeln wurden die F\"ahigkeiten
verallgemeinerter Summationsprozesse diskutiert, Approximationen f\"ur
die h\"oheren Taylorkoeffizienten zu liefern, die nicht zur Konstruktion
der entsprechenden rationalen Funktionen verwendet wurden.

\medskip

\Abschnitt Pad\'e-Approximationen und Stieltjesreihen

\smallskip

\aktTag = 0

Im letzten Unterabschnitt wurde der Versuch unternommen, plausibel zu
machen, warum Pad\'e-Approximationen \"uberhaupt die Konvergenz einer
Potenzreihe vom Typ von Gl. (4.2-1) verbessern k\"onnen. Dabei wurde
argumentiert, da{\ss} die Pad\'e-Approximation $[ \ell / m ]_f (z)$ einer
Funktion $f (z)$ gem\"a{\ss} Gl. (4.2-13) in die Partialsumme $f_{\ell+m} (z)$
und eine zus\"atzliche Potenzreihe (4.2-15) aufgespalten werden kann,
wobei die zus\"atzliche Potenzreihe f\"ur die g\"unstigeren numerischen
Eigenschaften der Pad\'e-Approximation verantwortlich ist, da sie eine
Approximation f\"ur den exakten Abbruchfehler (4.2-16) der Partialsumme
$f_{\ell+m} (z)$ liefern kann.

Allerdings ist dieser Erkl\"arungsversuch mathematisch unbefriedigend, da
keinerlei Aussagen gemacht werden k\"onnen, welche Voraussetzungen
entweder die Funktion $f (z)$ oder die zugeh\"orige formale Potenzreihe
(4.2-1) erf\"ullen mu{\ss}, damit die zus\"atzliche Potenzreihe (4.2-15) gegen
den exakten Abbruchfehler (4.2-16) der Partialsumme $f_{\ell+m} (z)$
konvergiert. Weiterhin ist dieser simple Erkl\"arungsversuch nicht in der
Lage, plausibel zu machen, warum Pad\'e-Approximationen divergente
Potenzreihen summieren k\"onnen. Man ben\"otigt also eine wesentlich
aufwendigere mathematische Theorie, um die Konvergenzeigenschaften der
Pad\'e-Approximationen ausreichend verstehen zu k\"onnen.

Ideal w\"are in diesem Zusammenhang eine mathematische Theorie, mit deren
Hilfe man vorhersagen kann, da{\ss} eine Folge von Pad\'e-Approximationen dann
und nur dann gegen $f (z)$ konvergiert, wenn diese Funktion bestimmte,
leicht \"uberpr\"ufbare Eigenschaften besitzt. Zwar gibt es eine
hoch\-entwickelte Konvergenztheorie der Pad\'e-Approximationen [Baker
1975, Abschnitte 10 - 14; 1990, Abschnitt 17; Baker und Graves-Morris
1981a, Abschnitt 6], aber eine derartig bequeme und leistungsf\"ahige
mathematische Theorie wie die oben skizzierte gibt es bisher nicht, und
m\"oglicherweise wird es sie auch nie geben [Baker 1965, S. 3].

Es gibt allerdings eine bestimmte Klasse von Funktionen, die durch
konvergente oder divergente Potenzreihen vom Typ von Gl. (4.2-1)
dargestellt werden k\"onnen, bei denen Fragestellungen dieser Art
weitgehend beantwortet werden k\"onnen. Wenn man beispielsweise zeigen
kann, da{\ss} die Koeffizienten $\gamma_{\nu}$ der formalen Potenzreihe
(4.2-1) f\"ur alle endlichen Werte von $\nu$ endlich sind und au{\ss}erdem f\"ur
alle $\nu \in \N_0$ die Beziehung
$$
\gamma_{\nu} \; = \; (-1)^{\nu} \, \mu_{\nu}
\tag
$$
mit
$$
\mu_{\nu} \; = \; \int\nolimits_{0}^{\infty} \, t^{\nu} \d \psi (t)
\tag
$$
erf\"ullen, wobei $\psi (t)$ ein eindeutig bestimmtes positives Ma{\ss} auf $0
\le t < \infty$ ist und dort unendlich viele verschiedene Werte annimmt
[Baker und Graves-Morris 1981a, S. 159], dann ist $f (z)$ eine {\it
Stieltjesfunktion} [Stieltjes 1894], welche die folgende Darstellung als
{\it Stieltjesintegral} besitzt:
$$
f(z) \; = \; \int\nolimits_{0}^{\infty} \,
\frac {\d \psi (t)} {1 + z t} \, ,
\qquad \vert \arg (z) \vert < \pi \, .
\tag
$$
Eine \"Ubersicht \"uber die mathematischen Eigenschaften solcher
Stieltjesintegrale findet man beispielsweise in einem Artikel von Widder
[1938]. Stieltjesfunktionen und die zugeh\"origen Stieltjesreihen
$$
f(z) \; = \; \sum_{m=0}^{\infty} \; (-1)^m \, \mu_m \, z^m \, ,
\tag
$$
spielen eine sehr gro{\ss}e Rolle in der Theorie divergenter Reihen, da sie
eine vergleichsweise hochentwickelte Konvergenztheorie besitzen [Baker
1975; 1990; Baker und Graves-Morris 1981a; Borel 1928; Bowman und
Shenton 1989; Perron 1957; Wall 1973]. Offensichtlich ist das nach Euler
benannte Integral (2.2-1) ein Stieltjesintegral, und die Eulerreihe
(2.2-2) ist die zugeh\"orige Stieltjesreihe.

Wenn man das zu $f (z)$ geh\"orende positve Ma{\ss} $\psi (t)$ aus den
Momenten $\Seqn {\mu}$ berechnen kann, besitzt man sofort eine
Darstellung der Stieltjesfunktion $f (z)$ durch das Stieltjesintegral
(4.3-3), die auch im Falle der Divergenz der Stieltjesreihe (4.3-3) zur
Berechnung von $f (z)$ verwendet werden kann. Die L\"osung solcher
Momentenprobleme ist ein altes Problem der Mathematik [Shohat und
Tamarkin 1950], und es sind auch schon seit langem notwendige und
hinreichende Bedingungen bekannt, die garantieren, da{\ss} eine solche
Stieltjesreihe durch Pad\'e-Approximationen beliebig genau approximiert
werden kann. Ein Beispiel ist die Bedingung, da{\ss} alle aus den Momenten
$\Seqn {\mu}$ gebildeten Hankeldeterminanten strikt positiv sein m\"ussen
[Baker und Graves-Morris 1981a, Theorem 5.1.2]. Allerdings ist die
Anwendung dieser und \"ahnlicher Bedingungen bei praktischen Problemen
etwa aus der St\"orungstheorie keineswegs einfach, da dazu alle Momente
$\Seqn {\mu}$ bekannt sein m\"ussen.

Es gibt aber eine einfache {\it hinreichende} Wachstumsbedingung f\"ur die
Stieltjesmomente $\Seqn {\mu}$, die sogenannte {\it Carlemanbedingung},
die vergleichsweise einfach angewendet werden kann. Wenn die
Stieltjes-Momente $\Seqn {\mu}$ die Beziehung
$$
\sum_{k = 0}^{\infty} \, \mu_k^{- 1/(2 k)} \; = \; \infty
\tag
$$
erf\"ullen, dann besitzt das entsprechende Momentenproblem eine eindeutige
L\"osung, und die Folge $[n + j / n]$ der Pad\'e-Approximationen konvergiert
f\"ur $n \to \infty$ und f\"ur festes $j \ge - 1$ gegen den Wert der
entsprechenden Stieltjesfunktion [Henrici 1977, Theorem 12.11f und
Corollar 12.11h]. Man kann zeigen, da{\ss} die Carlemanbedingung erf\"ullt
ist, wenn die Momente $\mu_n$ f\"ur $n \to \infty$ nicht schneller wachsen
als $C^{n + 1} (2 n)!$, wobei $C$ eine geeignete positive Konstante ist
[Simon 1982, Theorem 1.3]. Unter diesen Voraussetzungen kann man auch
zeigen, da{\ss} die Pad\'e-Approximationen $[ n / n ]$ und $[n / n + 1 ]$ im
Falle einer Stieltjesreihe mit positivem Argument f\"ur $n \to \infty$
eine konvergente Folge von oberen und unteren Schranken liefern [Baker
und Graves-Morris 1981a, Theorem 5.2.2].

Aus der Carlemanbedingung (4.3-5) folgt beispielsweise, da{\ss} die
Eulerreihe (2.2-2) durch die Pad\'e-Approximationen $[n + j / n]$ f\"ur
festes $j \ge - 1$ summiert wird, und da{\ss} diese Folge von
Pad\'e-Approximationen f\"ur $n \to \infty$ gegen das nach Euler benannte
Integral (2.2-1) konvergiert.

Die Tatsache, da{\ss} die Carlemanbedingung (4.3-5) nur dann erf\"ullt ist,
wenn die Momente f\"ur $n \to \infty$ nicht schneller wachsen als $C^{n +
1} (2 n)!$, wobei $C$ eine geeignete positive Konstante ist [Simon 1982,
Theorem 1.3], hat -- wie sp\"ater noch ausf\"uhrlicher diskutiert wird --
weitreichende Konsequenzen f\"ur die praktische Anwendbarkeit von
Pad\'e-Approximationen zur Summation von extrem stark divergenten Reihen,
die in der quantenmechanischen St\"orungstheorie vorkommen.

Wenn man die Beziehung
$$
\sum_{m=0}^{n} \, (- z t)^m \; = \;
\frac {1 - (- z t)^{n+1}} {1 + z t}
\tag
$$
in Gl. (4.3-3) verwendet und die Momente $\mu_m$ gem\"a{\ss} Gl. (4.3-2)
berechnet, folgt sofort, da{\ss} der Abbruchfehler einer Stieltjesreihe auf
folgende Weise durch ein Stieltjesintegral dargestellt werden kann:
$$
f (z) \; = \; \sum_{m=0}^{n} \, (-1)^m \, \mu_m \, z^m \; + \;
(- z)^{n+1} \,
\int\nolimits_{0}^{\infty} \,
\frac {t^{n+1} \, \d \psi (t)} {1 + z t}
\, , \qquad \vert \arg (z) \vert < \pi \, .
\tag
$$
Ob eine Stieltjesreihe konvergiert oder divergiert, h\"angt von dem
Verhalten des Integrals auf der rechten Seite von Gl. (4.3-7) f\"ur $n \to
\infty$ ab. Wenn das Argument $z$ der Stieltjesfunktion $f (z)$ positiv
ist, dann kann der Abbruchfehler offensichtlich durch den ersten Term
der Stieltjesreihe abgesch\"atzt werden, der nicht in der Partialsumme
enthalten ist:
$$
z^{n+1} \, \int\nolimits_{0}^{\infty} \,
\frac {t^{n+1} \, \d \psi (t)} {1 + z t} \; \le \;
\mu_{n + 1} \, z^{n+1}
\; = \; (- 1)^{n + 1} \, \gamma_{n + 1} \, z^{n+1} \, ,
\qquad z \ge 0 \, .
\tag
$$
Diese Absch\"atzung wird sp\"ater noch eine wichtige Rolle spielen, und zwar
im Zusammenhang mit anderen verallgemeinerten Summationsprozessen, die
in Abschnitt 5 behandelt werden und die im Gegensatz zu
Pad\'e-Approximationen zus\"atzliche Informationen \"uber das Verhalten des
Abbruchfehlers einer unendlichen Reihe als Funktion des Summationsindex
$n$ nutzbringend verwerten k\"onnen.

\medskip

\Abschnitt Die effiziente Berechnung von Pad\'e-Approximationen

\smallskip

\aktTag = 0

Die Determinantendarstellung (4.2-10) f\"ur die Pad\'e-Approximation $[ \ell
/ m ]_f (z)$ ist numerisch nicht sehr n\"utzlich, da -- wie schon h\"aufiger
erw\"ahnt -- die effiziente und verl\"a{\ss}liche Berechnung von Determinanten
ein immer noch nicht befriedigend gel\"ostes Problem der numerischen
Mathematik ist. Gl\"ucklicherweise gibt es zahlreiche andere wesentlich
leistungsf\"ahigere Verfahren zur Berechnung von Pad\'e-Approximationen.
Einen \"Uberblick \"uber die St\"arken und Schw\"achen der verschiedenen
Verfahren findet man in einer kommentierten Bibliographie von Wuytack
[1979b], in Abschnitt \Roemisch{2}.3 des Buches von Cuyt und Wuytack
[1987], oder in einem Artikel von Cabay und Kossowski [1990].

Im Rahmen der Forschungen des Autors wurden immer nur die numerischen
Werte von Pad\'e-Approximationen ben\"otigt, und es gab nie einen Grund,
die Koeffizienten $p_{\lambda}$ und $q_{\mu}$ der Z\"ahler- und
Nennerpolynome $P_{\ell} (z)$ und $Q_m (z)$ explizit zu bestimmen.
Deswegen wurden Pad\'e-Approximationen in allen F\"allen mit Hilfe des
Wynnschen $\epsilon$-Algorithmus, Gl. (2.4-10), berechnet. Wynn [1956a]
konnte n\"amlich zeigen, da{\ss} die Elemente der $\epsilon$-Tafel mit
geraden unteren Indizes Pad\'e-Approximationen der Potenzreihe (4.2-1)
ergeben,
$$
\epsilon_{2 k}^{(n)} \; = \;
[ \, n \, + \, k \, / \, k \, ]_f \, (z) \, ,
\qquad k,n \in \N_0 \, ,
\tag
$$
wenn man die Partialsummen (4.2-11) gem\"a{\ss}
$$
s_n \; = \; f_n (z) \, , \qquad n \in \N_0 \, ,
\tag
$$
als Eingabedaten in Gl. (2.4-10a) verwendet. Die Elemente der
$\epsilon$-Tafel mit ungeraden unteren Indizes sind nur Hilfsgr\"o{\ss}en, die
divergieren, wenn die Pad\'e-Approximationen $[ n + k / k ]_f \, (z)$
gegen $f (z)$ konvergieren.

Der Wynnsche $\epsilon$-Algorithmus (2.4-10) kann trotz seiner
Nichtlinearit\"at und seiner zweidimensionalen Struktur sehr leicht in
ein Programm \"ubersetzt werden, das bequem in Beschleunigungs- und
Summationsverfahren verwendet werden kann. Beim Entwurf eines solchen
Programms sollte man allerdings bedenken, da{\ss} man normalerweise nicht
vorhersagen kann, wieviele Eingabedaten ben\"otigt werden, um eine
bestimmte Genauigkeit zu erreichen. Ein solches Programm sollte also
{\it inputgesteuert} sein. Das bedeutet, da{\ss} man die Inputdaten $s_0$,
$s_1$, $\ldots$ , $s_m$, $\ldots$ f\"ur Gl. (2.4-10a) nacheinander
einlesen sollte, wobei mit $s_0$ begonnen wird. Nach dem Einlesen eines
neuen Startwertes $s_m$ sollten so viele neue Elemente
$\epsilon_k^{(n)}$ berechnet werden, wie es die Rekursion (2.4-10b)
erlaubt. Dasjenige neue Element $\epsilon_k^{(n)}$ mit dem gr\"o{\ss}ten
geraden unteren Index $k = 2 \kappa$ sollte als Approximation des
Grenzwertes $s$ der Folge $\Seqn s$ der Eingabedaten verwendet und auf
Konvergenz \"uberpr\"uft werden.

Die bei der praktischen Verwendung des $\epsilon$-Algorithmus
auftretenden programmtechnischen Probleme k\"onnen verdeutlicht werden,
indem man die Elemente $\epsilon_k^{(n)}$ der $\epsilon$-Tafel auf
solche Weise in einem rechteckigen Schema anordnet, da{\ss} der obere Index
$n$ die Zeile und der untere Index $k$ die Spalte einer
zweidimensionalen unendlichen Matrix bezeichnet:

$$
\matrix{
\epsilon_0^{(0)} " \epsilon_1^{(0)} " \epsilon_2^{(0)} " \ldots "
\epsilon_n^{(0)} " \ldots \\
\epsilon_0^{(1)} " \epsilon_1^{(1)} " \epsilon_2^{(1)} " \ldots "
\epsilon_n^{(1)} " \ldots \\
\epsilon_0^{(2)} " \epsilon_1^{(2)} " \epsilon_2^{(2)} " \ldots "
\epsilon_n^{(2)} " \ldots \\
\epsilon_0^{(3)} " \epsilon_1^{(3)} " \epsilon_2^{(3)} " \ldots "
\epsilon_n^{(3)} " \ldots \\
\vdots " \vdots " \vdots " \ddots " \vdots " \ddots \\
\epsilon_0^{(n)} " \epsilon_1^{(n)} " \epsilon_2^{(n)} " \ldots "
\epsilon_n^{(n)} " \ldots \\
\vdots " \vdots " \vdots " \ddots " \vdots  " \ddots }
\tag
$$

Die Gr\"o{\ss}en in der ersten Spalte dieser Matrix sind die Startwerte
$\epsilon_0^{(n)} = s_n$ der Rekursion gem\"a{\ss} Gl. (2.4-10a). Die
restlichen Elemente dieser Matrix k\"onnen mit Hilfe von Gl. (2.4-10b)
berechnet werden. Diese nichtlineare Viertermrekursion verkn\"upft die
folgenden Elemente, die in der $\epsilon$-Tafel (4.4-3) an den Ecken
eines Rhombus liegen:
$$
\beginMatrix
\beginFormat &\Formel\links \endFormat
\+" " \epsilon_{k}^{(n)} \qquad " \epsilon_{k+1}^{(n)} "
\\ [1\jot]
\+" \epsilon_{k-1}^{(n+1)} \qquad " \epsilon_{k}^{(n+1)} \qquad " " \\
\endMatrix
\tag
$$

\medskip

Wenn man die Partialsummen $s_0$, $s_1$, $\ldots$ , $s_m$ als
Startwerte gem\"a{\ss} Gl. (2.4-10a) verwendet, so kann man mit Hilfe der
Rekursionsformel (2.4-10b) alle Elemente $\epsilon_j^{(\mu-j)}$ mit $0
\le \mu \le m$ und $0 \le j \le \mu$ berechnen. Diese Elemente bilden
offensichtlich ein gleichseitiges Dreieck in der linken oberen Ecke der
$\epsilon$-Tafel (4.4-3). Wenn man die n\"achste Partialsumme $s_{m+1}$
ebenfalls als Startwert gem\"a{\ss} Gl. (2.4-10a) verwendet, so wird dieses
Dreieck durch die Gegendiagonale $\epsilon_j^{(m-j+1)}$ mit $0 \le j
\le m+1$ vergr\"o{\ss}ert. Diese geometrischen Zusammenh\"ange werden
deutlicher, wenn man den Wynnschen $\epsilon$-Algorithmus, Gl.
(2.4-10), auf folgende Weise umformuliert:
$$
\beginAligntags
" \epsilon_0^{(n)} " \; = \; " s_n \, , \qquad n \ge 0 \, ,
\erhoehe\aktTag \\ \tag*{\tagnr a}
" \epsilon_1^{(n-1)} " \; = \; " 1 / [ s_{n} - s_{n-1} ] \, ,
\qquad n \ge 1 \, , \\ \tag*{\tagform\aktTagnr b}
" \epsilon_{j}^{(n-j)} " \; = \; " \epsilon_{j-2}^{(n-j+1)} \, + \,
1 / [\epsilon_{j-1}^{(n-j+1)} - \epsilon_{j-1}^{(n-j)}] \, ,
\quad n \ge 2 \, , \quad 2 \le j \le n \, .
\\ \tag*{\tagform\aktTagnr c}
\endAligntags
$$

Da nur die Elemente der $\epsilon$-Tafel mit geraden unteren Indizes
Pad\'e-Approximationen gem\"a{\ss} Gl. (4.4-1) produzieren, mu{\ss} man die N\"aherung
zum Grenzwert der zu transformierenden Folge $\Seqn s$ unterschiedlich
w\"ahlen, je nachdem ob der Index der letzten Partialsumme $s_m$, die in
der Rekursion verwendet wurde, gerade oder ungerade ist. Wenn $m$ gerade
ist, $m = 2 \mu$, w\"ahlen wir die Approximation auf folgende Weise,
$$
\left\{s_0, s_1, \ldots , s_{2 \mu} \right\} \; \to \;
\epsilon_{2 \mu}^{(0)} \, .
\tag
$$
Wenn $m$ dagegen ungerade ist, $m = 2 \mu + 1$, verwenden wir die
Transformation
$$
\left\{s_1, s_2, \ldots , s_{2 \mu + 1} \right\} \; \to \;
\epsilon_{2 \mu}^{(1)} \, .
\tag
$$
Mit Hilfe der Notation $\Ent x$ f\"ur den ganzzahligen Anteil von $x$,
der die gr\"o{\ss}te ganze Zahl $\nu$ ist, welche die Beziehung $\nu \le x$
erf\"ullt, kann man die beiden obigen Beziehungen zu einer einzigen
Gleichung zusammenfassen:
$$
\left\{ s_{m - 2 \Ent {m/2}}, s_{m - 2 \Ent {m/2} + 1}, \ldots , s_m
\right\} \; \to \;
\epsilon_{2 \Ent {m/2}}^{(m - 2 \Ent {m/2})} \, .
\tag
$$
Wenn die Eingabedaten in Gl. (2.4-10a) Partialsummen einer Potenzreihe
sind, so folgt aus Gl. (4.4-1), da{\ss} die Approximationen (4.4-8) f\"ur
wachsendes $m$ eine treppenartige Folge von Elementen der Pad\'e-Tafel
ergeben:
$$
[0/0] \; , \; [1/0] \; , \; [1/1] \; , \ldots , \; [\nu / \nu] \; , \;
[\nu + 1/ \nu] \; , \; [\nu +1/ \nu +1] \; , \ldots \; .
\tag
$$

Wenn man die Partialsumme einer formalen Potenzreihe (4.2-1) in
rationale Funktionen verwandelt, so ist es naheliegend, die Grade der
Z\"ahler- und Nennerpolynome so zu w\"ahlen, da{\ss} sie gleich sind oder --
wenn das nicht m\"oglich ist -- da{\ss} sie sich so wenig wie m\"oglich
unterscheiden. In manchen F\"allen kann man explizit zeigen, da{\ss} eine
solche Wahl der Z\"ahler- und Nennerpolynome tats\"achlich optimal ist.

Wenn $f (z)$ beispielsweise eine Stieltjesfunktion gem\"a{\ss} Gl. (4.3-3)
ist, dann ergeben die diagonalen Pad\'e-Approximationen $[ n / n ]$ laut
Wynn [1968] die genauesten Approximationen, die man unter Verwendung der
Partialsummen $f_0 (z)$, $f_1 (z)$, $\cdots$, $f_{2 n} (z)$ erhalten
kann. Wenn man dagegen die Partialsummen $f_0 (z)$, $f_1 (z)$, $\cdots$,
$f_{2 n + 1} (z)$ verwendet, dann liefern entweder die
Pad\'e-Approximationen $[ n + 1 / n ]$ oder $[ n / n + 1 ]$ die
genauesten Approximationen. Wir sehen also, da{\ss} unter diesen
Voraus\-setzungen die treppenartige Folge (4.4-9) von
Pad\'e-Approximationen die Information, die in den Reihenkoeffizienten
$\gamma_{\nu}$ enthalten ist, optimal ausn\"utzt.

Die rhombusartige Struktur (4.4-4) der Viertermrekursion (2.4-10b)
suggeriert, da{\ss} ein Programm f\"ur den $\epsilon$-Algorithmus entweder
ein zweidimensionales Feld oder wenigstens zwei eindimensionale Felder
ben\"otigen w\"urde. Wynn [1965] konnte aber zeigen, da{\ss} ein einziges
eindimensionales Feld ausreicht.

Das Wynnsche Verfahren, das in der Literatur \"ublicherweise {\it moving
lozenge technique} genannt wird, basiert auf der Beobachtung, da{\ss} zur
Berechnung eines neuen Elementes $\epsilon_j^{(m-j+1)}$ nur die beiden
benachbarten Elemente $\epsilon_{j-1}^{(m-j+1)}$ und
$\epsilon_{j-2}^{(m-j+2)}$ aus der gesamten dar\"uberliegenden
Gegendiagonalen $\epsilon_{\mu}^{(m-\mu)}$ mit $0 \le \mu \le m$
ben\"otigt werden. Bei der Wynnschen {\it moving lozenge technique}
werden diese Gr\"o{\ss}en in Hilfsvariablen abgespeichert, w\"ahrend die
Rekursion (4.4-5) die aktuelle Gegendiagonale $\epsilon_j^{(m-j+1)}$
mit $0 \le j \le m+1$ durchl\"auft und die beim vorherigen Durchlauf
abgespeicherten Gr\"o{\ss}en $\epsilon_{\mu}^{(m - \mu)}$ mit $0 \le \mu \le
m$ \"uberschreibt. Eine gute Beschreibung der Wynnschen {\it moving
lozenge technique} findet man in Abschnitt 4.2.1.2. des Buches von
Brezinski [1978].

Wynn [1965] verwendete f\"ur die Rekursion ein eindimensionales Feld
${\cal E}$, in dem er die Elemente der aktuellen Gegendiagonale der
$\epsilon$-Tafel auf solche Weise abspeicherte, da{\ss} der Index des
Feldelementes mit dem unteren Index des entsprechenden Elementes der
$\epsilon$-Tafel \"ubereinstimmte,
$$
\epsilon_j^{(m-j)} \; \to \; {\cal E}(j) \, ,
\qquad m \ge 0 \, , \quad 0 \le j \le m \, .
\tag
$$
Wenn diese Konvention verwendet wird, ben\"otigt man drei Hilfsvariablen
bei der Rekursion. Die Wynnsche {\it moving lozenge technique} kann
aber noch verbessert werden, indem man die Elemente der aktuellen
Gegendiagonale so in einem eindimensionalen Feld $E$ abspeichert, da{\ss}
der obere Index des entsprechenden Elementes der $\epsilon$-Tafel mit
dem Index des Feldelementes \"ubereinstimmt [Weniger 1989, Abschnitt 4.3],
$$
\epsilon_j^{(m-j)} \; \to \; E(m-j) \, ,
\qquad m \ge 0 \, , \quad 0 \le j \le m \, .
\tag
$$
Auf diese Weise ben\"otigt man nur zwei Hilfsvariablen. Au{\ss}erdem wird das
resultierende Programm wesentlich einfacher.

Unter Verwendung der Konvention (4.4-11) kann das Rekursionsschema
(4.4-5) auf folgende Weise in ein Rekursionsschema f\"ur die Elemente des
eindimensionalen Feldes $E$ umformuliert werden:
$$
\beginAligntags
" E(n) " \, \gets \, " s_n \, , \qquad n \ge 0 \, ,
\erhoehe\aktTag \\ \tag*{\tagnr a}
" E(n-1) " \, \gets \, " 1 / [ E(n) - E'(n-1) ] \, ,
\qquad n \ge 1 \, , \\ \tag*{\tagform\aktTagnr b}
" E(n-j) " \, \gets \, " E'(n-j+1) \, + \, 1 / [ E(n-j+1) - E'(n-j) ]
\, , \hfill \\
" " " n \ge 2 \, , \qquad 2 \le j \le n \, .
\\ \tag*{\tagform\aktTagnr c}
\endAligntags
$$
Die indizierten Feldelemente $E'(n-j)$ und $E'(n-j+1)$ m\"ussen in
Hilfsvariablen gespeichert werden, da sie bei der Berechnung der
aktuellen Gegendiagonale $\epsilon_j^{(n-j)}$ mit $0 \le j \le n$
\"uberschrieben werden.

Das folgende FORTRAN-Programm EPSAL [Weniger 1989, Abschnitt 4.3]
zeigt, da{\ss} der Wynnsche $\epsilon$-Algorithmus tats\"achlich auf sehr
einfache Weise programmiert werden kann, wenn man das Rekursionsschema
(4.4-12) verwendet, das auf der Modifikation (4.4-11) der Wynnschen
{\it moving lozenge technique} [Wynn 1965] basiert.

\bigskip

\listenvon{epsal.for}

\medskip

Der Wynnsche $\epsilon$-Algorithmus, Gl. (2.4-10), und ebenso das daraus
abgeleitete Rekursionsschema (4.4-12) kann nicht zur Berechnung von
Pad\'e-Approximationen verwendet werden, wenn benachbarte Elemente
$\epsilon_k^{(n+1)}$ und $\epsilon_k^{(n)}$ entweder identisch sind oder
wenn sie sich so wenig unterscheiden, da{\ss} der Nenner
$1/[\epsilon_k^{(n+1)} - \epsilon_k^{(n)}]$ bei Verwendung einer
Gleitkommaarithmetik OVERFLOW ergibt. Wynn [1965] gab spezielle Regeln
f\"ur den $\epsilon$-Algorithmus an, mit deren Hilfe man
Pad\'e-Approximationen auch dann berechnen kann, wenn solche
Singularit\"aten auftreten. FORTRAN-Programme, die diese speziellen Regeln
verwenden, wurden von Brezinski und Redivo Zaglia [1991, S. 93]
beschrieben. Die laut Brezinski und Redivo Zaglia [1991, S. 94]
leistungsf\"ahigsten Programme zur Berechnung von Pad\'e-Approximationen
beim Auftreten von Singularit\"aten sind in den Anh\"angen des Buches von
Draux und van Ingelandt [1987] abgedruckt.

Die Subroutine EPSAL ist gesichert gegen OVERFLOW aufgrund von exakter
oder approximativer Identit\"at zweier benachbarter Elemente der
$\epsilon$-Tafel. Bei einer gr\"o{\ss}eren Zahl von Singularit\"aten sollte man
dieses Programm aber nicht mehr verwenden. Allerdings treten solche
Pathologien in naturwissenschaftlichen und technischen Anwendungen
normalerweise nicht auf, und es ist in den meisten F\"allen m\"oglich, den
Wynnschen $\epsilon$-Algorithmus, Gl. (2.4-10), und damit die
Subroutine EPSAL, direkt zu verwenden.

Wenn man den $\epsilon$-Algorithmus unter Verwendung des
Rekursionsschemas (4.4-12) programmiert, kann man die numerischen Werte
der Pad\'e-Approximationen (4.4-9) nicht nur mit einer sehr geringen
Anzahl von Rechenoperationen, sondern auch mit einem minimalen
Speicherbedarf berechnen. Eine solche Optimierung des Speicherbedarfes
ist an sich v\"ollig \"uberfl\"ussig, wenn man die Berechnung der
Pad\'e-Approximationen in einer konventionellen Programmiersprache wie
etwa FOR\-TRAN 77 unter Verwendung einer Gleitkomma\-arithmetik mit
einer festen Stellenzahl durchf\"uhrt. Die maximal erreichbare
Komplexit\"at einer Transformation von Partialsummen einer Potenzreihe in
Pad\'e-Approximationen wird normalerweise durch die Anzahl der zur
Verf\"ugung stehenden Terme und leider auch manchmal durch numerische
Instabilit\"aten beschr\"ankt. Der Speicherbedarf der $\epsilon$-Tafel
(4.4-3) ist bei allen praktisch relevanten Problemen vernachl\"assigbar
klein.

Bis vor relativ kurzer Zeit gab es keine praktikable Alternative zu
numerisch orientierten Programmiersprachen wie FORTRAN 77 oder ALGOL. In
den letzten Jahren sind aber Compu\-ter\-algebrasysteme wie REDUCE,
MACSYMA, MAPLE oder MATHEMATICA auf einer gro{\ss}en Anzahl von Rechnertypen
verf\"ugbar geworden. In einem von Grossman [1989] herausgegebenen Buch
werden Anwendungsm\"oglichkeiten solcher Compu\-ter\-algebrasysteme in den
Naturwissenschaften diskutiert. Einen \"Uberblick \"uber die zur Zeit
erh\"altlichen Computeralgebrasysteme sowie eine kritische Diskussion
ihrer St\"arken und Schw\"achen findet man in dem Buch von Harper, Wooff und
Hodgkinson [1991]. Der aktuelle Stand der Forschung wird in den von
Boyle und Caviness [1990] herausgegebenen Proceedings eines Workshops
\"uber symbolisches Rechnen beschrieben. Dort oder in dem von Maple
Waterloo Software herausgegebenen Maple Technical Newsletter findet man
auch zahlreiche naturwissenschaftliche und technische Anwendungen
solcher Computeralgebrasysteme. Bisher haben Anwendungen aus der
Mathematik oder der theoretischen Physik dominiert. So wurden die
St\"arken und Schw\"achen verschiedener Computeralgebrasysteme bei der
Behandlung typischer physikalischer Probleme von Cook, Dubisch, Sowell,
Tam und Donnelly [1992a; 1992b] untersucht, und Scott, Moore, Fee,
Monagan, Labahn und Geddes [1990] diskutierten, wie man symbolische
Manipulationen zur L\"osung st\"orungstheoretischer Probleme aus der
Quantenmechanik verwenden kann. Inzwischen gibt es aber schon zahlreiche
chemische Anwendungen. Ein Beispiel ist der Artikel von Holmes und Bell
[1991], in dem Probleme aus der Reaktionskinetik mit Hilfe von
symbolischen Manipulationen behandelt werden.

Wenn man ein solches Computeralgebrasystem anstelle einer
konventionellen problemorientierten Programmiersprache wie FORTRAN 77
im Zusammenhang mit verallgemeinerten Summationsprozessen verwendet,
werfen numerischen Instabilit\"aten praktisch nie un\"uberwindliche
Probleme auf, da diese Systeme \"ublicherweise sowohl eine exakte
rationale Arithmetik als auch eine Gleitkomma\-arithmetik mit einer im
Prinzip unbeschr\"ankten und frei w\"ahlbaren Genauigkeit besitzen. Wenn
man aber exakt rational oder reell mit einer sehr hohen Genauigkeit
rechnet, kann der Speicherbedarf eines solchen Systems so gro{\ss} werden,
da{\ss} er der komplexit\"atsbestimmende Faktor wird.

Diese Erfahrung blieb dem Autor auch nicht erspart. Bei einigen sehr
aufwendigen Rechnungen in MAPLE [Char, Geddes, Gonnet, Leong, Monagan
und Watt 1991a], die an der Universit\"at von Waterloo durchgef\"uhrt
wurden, konnte der 64 MB gro{\ss}e Hauptspeicher einer Silicon Graphics
4D-340~S mit Vierfachprozessor problemlos gef\"ullt oder sogar \"uberf\"ullt
werden. Speicherprobleme traten sowohl bei der exakt rationalen
Berechnung der Koeffizienten von St\"orungsreihen f\"ur anharmonische
Oszillatoren [Weniger, {\v C}{\'\i}{\v z}ek, und Vinette 1991; 1993] als
auch bei eindimensionalen Diffusions- und W\"armeleitungsgleichungen mit
nichtlinearen Randbedingungen auf, bei denen es aufgrund von
Instabilit\"aten notwendig war, mit einer Genauigkeit von 170
Dezimalstellen zu rechnen [{\v C}{\'\i}{\v z}ek, Vinette und Weniger
1991, Abschnitt 4]. In beiden F\"allen erwies sich der vorhandene Speicher
als der komplexit\"atsbestimmende Faktor, was letztlich dazu f\"uhrte, da{\ss}
nicht alle Rechnungen so durchgef\"uhrt werden konnten, wie es eigentlich
geplant war.

Wenn man also in der Lage sein will, Summations- und
Konvergenzbeschleunigungsprozesse entweder exakt rational oder reell mit
hoher Genauigkeit in MAPLE oder in einer anderen Sprache f\"ur formale
Manipulationen durchzuf\"uhren, dann ist es wichtig, alle
verallgemeinerten Summationsprozesse so zu programmieren, da{\ss} der
Speicherbedarf minimal wird. Das gelang auch in allen F\"allen durch die
Entwicklung von Techniken, die Verallgemeinerungen der Wynnschen {\it
moving lozenge technique} [Wynn 1965] sind (siehe Abschnitte 5.2, 6.2,
6.3, 7.5, 10.2 und 10.3 von Weniger [1989]). Hier sei noch erw\"ahnt, da{\ss}
schon seit relativ langer Zeit spezielle Algorithmen f\"ur die symbolische
Berechnung von Pad\'e-Approximationen bekannt sind [Geddes 1979].

Der algorithmisch komplizierteste verallgemeinerte Summationsproze{\ss},
mit dem der Autor bisher konfrontiert war, ist der durch das folgende
Rekursionsschema definierte und durch Modifikation des Wynnschen
$\epsilon$-Algorithmus abgeleitete $\theta$-Algorithmus von Brezinski
[1971]:
$$
\beginAligntags
" \theta_{-1}^{(n)} \; " = \; " 0 \, , \hfill
" \theta_0^{(n)} \; = \; s_n \, ,
\erhoehe\aktTag \\ \tag*{\tagnr a}
" \theta_{2 k + 1}^{(n)} \; " = \; " \theta_{2 k-1}^{(n+1)}
\, + \, 1 / [\Delta \theta_{2 k}^{(n)}] \, , \hfill "
\\ \tag*{\tagform\aktTagnr b}
" \theta_{2 k+2}^{(n)} \; " = \; " \theta_{2 k}^{(n+1)} \, +
\, \frac
{[\Delta \theta_{2 k}^{(n+1)}] \,
[\Delta \theta_{2 k + 1}^{(n+1)}]}
{\Delta^2 \theta_{2 k+1}^{(n)}}
\, , \qquad " k,n \in \N_0 \, .
\\ \tag*{\tagform\aktTagnr c}
\endAligntags
$$
Numerische Studien [Smith und Ford 1979; 1982; Weniger 1989, Abschnitte
13 und 14] ergaben, da{\ss} der $\theta$-Algorithmus eine sehr
leistungsf\"ahige und auch sehr vielseitige Transformation ist, die
sowohl zur Beschleunigung linearer und logarithmischer Konvergenz als
auch zur Summation divergenter alternierender Reihen geeignet ist.

Wenn man die Elemente $\theta_k^{(n)}$ analog zur $\epsilon$-Tafel
(4.4-3) auf solche Weise in einem rechteckigen Schema anordnet, da{\ss} der
obere Index $n$ die Zeile und der untere Index $k$ die Spalte einer
zweidimensionalen unendlichen Matrix bezeichnet, dann verkn\"upft die
nichtlineare Viertermrekursion (4.4-13b) die folgenden Elemente, die in
der $\theta$-Tafel an den Ecken eines Rhombus liegen:
$$
\beginMatrix
\beginFormat &\Formel\links \endFormat
\+" " \theta_{2 k}^{(n)} \qquad " \theta_{2 k+1}^{(n)} "
\\ [1\jot]
\+" \theta_{2 k-1}^{(n+1)} \qquad " \theta_{2 k}^{(n+1)} \qquad " " \\
\endMatrix
\tag
$$
Analog kann die nichtlineare Sechstermrekursion (4.4-13c) auf folgende
Weise geometrisch verdeutlicht werden:
$$
\beginMatrix
\beginFormat &\Formel\links \endFormat
\+" " \theta_{2 k+1}^{(n)} \qquad " \theta_{2 k+2}^{(n)} "
\\ [1\jot]
\+" \theta_{2 k}^{(n+1)} \qquad
" \theta_{2 k+1}^{(n+1)} \qquad " " \\ [1\jot]
\+" \theta_{2 k}^{(n+2)} \qquad
" \theta_{2 k+1}^{(n+2)} \qquad " " \\
\endMatrix
\tag
$$

Die nichtlineare Viertermrekursion (4.4-13b) und damit der Rhombus
(4.4-14) ist strukturell identisch mit der Rekursionsformel (2.4-10b)
des $\epsilon$-Algorithmus und dem zugeh\"origen Rhombus (4.4-4), aber die
nichtlineare Sechstermrekursion (4.4-13c) ist wesentlich komplizierter
als Gl. (2.4-10b). Demzufolge kann man nicht erwarten, da{\ss} ein
speicheroptimierendes Programm f\"ur den $\theta$-Algorithmus \"ahnlich
einfach sein wird wie das FORTRAN-Programm EPSAL, das auf dem
Rekursionsschema (4.4-12) basiert. Trotzdem ist es m\"oglich, den
$\theta$-Algorithmus speicheroptimierend so zu programmieren, da{\ss} nur
zwei eindimensionale Felder $A$ und $B$ ben\"otigt werden.

Aus den bildlichen Darstellungen (4.4-14) und (4.4-15) folgt, da{\ss} die
Rekursionen (4.4-13b) und (4.4-13c) einen relativ komplizierten Pfad in
der $\theta$-Tafel verfolgen m\"ussen. Nehmen wir also an, da{\ss} die
Eingabedaten $s_0, s_1, \ldots , s_{n-1}$ eingelesen worden sind, und
da{\ss} so viele Elemente der $\theta$-Tafel berechnet worden sind, wie es
aufgrund des Rekursionsschemas (4.4-13) m\"oglich ist. Wenn man nun auch
den n\"achsten Startwert $s_n$ in dem Rekursionsschema (4.4-13) verwendet,
dann kann man auch noch die Elemente $\theta_j^{(n - \Ent {3 j / 2})}$
mit $0 \le j \le \Ent {(2 n + 1)/3}$ berechnen, wobei $\Ent {x}$ wieder
der ganzzahlige Anteil von $x$ ist.

In diesem Zusammenhang ist es empfehlenswert, das Rekursionsschema
(4.4-13) auf folgende Weise umzuschreiben:
$$
\beginAligntags
" \theta_0^{(n)} \; " = \; " s_n \, , \qquad n \ge 0 \, , "
\erhoehe\aktTag \\ \tag*{\tagnr a}
" \theta_1^{(n-1)} \; " = \; "
1/[\theta_0^{(n)} - \theta_0^{(n-1)}]
\, , \qquad n \ge 1 \, , "
\\ \tag*{\tagform\aktTagnr b}
" \theta_{2 j}^{(n-3 j)} \; " = \; " \theta_{2 j -2}^{(n-3 j+1)}
\, + \, \frac
{ [\Delta \theta_{2 j - 2}^{(n-3 j+1)}] \,
[\Delta \theta_{2 j - 1}^{(n-3 j+1)}] }
{ \Delta^2 \theta_{2 j - 1}^{(n-3 j)} }
\, , \qquad n \ge 3 \, , \; 1 \le j \le \, \Ent {n/3} \, , " \qquad
\\ \tag*{\tagform\aktTagnr c}
" \theta_{2 j + 1}^{(n-3 j-1)} \; " = \;
" \theta_{2 j-1}^{(n-3 j)}
\, + \, 1 / [\Delta \theta_{2 j}^{(n-3 j-1)}] \, , \qquad
n \ge 4 \, , \; j \le \Ent {(n-1)/3} \, . "
\\ \tag*{\tagform\aktTagnr d}
\endAligntags
$$

Wenn der Index des letzten Startwertes $s_n$, der in der Rekursion
(4.4-13) verwendet wurde, gerade ist, $n = 2 m$, verwenden wir die
Konvention, da{\ss} die neu zu berechnenden Elemente gem\"a{\ss} der Regel
$$
\theta_j^{(2 m - \Ent {3 j / 2})} \; \to \; A (j) \, ,
\qquad 0 \le j \le \Ent {(4 m + 1) / 3} \, ,
\tag
$$
in dem Feld $A$ abgespeichert werden. Wenn dagegen der Index $n$ des
letzten Startwertes $s_n$ ungerade ist, $n = 2 m + 1$, werden die neu
zu berechnenden Elemente gem\"a{\ss} der Regel
$$
\theta_j^{(2 m - \Ent {3 j / 2} + 1)} \; \to \; B (j) \, ,
\qquad 0 \le j \le \Ent {(4 m + 3) / 3} \, ,
\tag
$$
in dem Feld $B$ abgespeichert. Das Rekursionsschema (4.4-16) kann dann
unter Verwendung dieser Konventionen in Relationen zwischen Elementen
der Felder $A$ und $B$ \"ubersetzt werden.

Wenn der Index $n$ des letzten Startwertes $s_n$, der in der Rekursion
verwendet wurde, gerade ist, $n = 2 m$, enth\"alt das Feld $B$ die
Elemente $\theta_j^{(2 m -\Ent {3 j/2}-1)}$ mit $0 \le j \le \Ent {(4 m
- 1)/3}$, w\"ahrend in $A$ die Elemente $\theta_j^{(2 m -\Ent {3
j/2}-2)}$ mit $0 \le j \le \Ent {(4 m - 3)/3}$ gespeichert sind. Unter
Verwendung der Konventionen (4.4-17) und (4.4-18) erh\"alt man dann f\"ur
Gl. (4.4-16):
$$
\beginAligntags
" A(0) \; " \gets \; " s_{2 m} \, , \qquad m \in \N_0 \, ,
\erhoehe\aktTag \\ \tag*{\tagnr a}
" A(1) \; " \gets \; " 1 / [ A(0) - B(0)] \, ,
\\ \tag*{\tagform\aktTagnr b}
" A(2 j) \; " \gets \; " A'(2 j - 2) \, + \, \frac
{[B(2 j - 2) - A'(2 j - 2)] \, [A(2 j -1) - B(2 j - 1)]}
{A(2 j - 1) - 2 B(2 j -1) + A'(2 j - 1)} \, , \\
" j \le \Ent {2 m / 3} \, , \span\omit \span\omit
\\ \tag*{\tagform\aktTagnr c}
" A(2 j + 1) \; " \gets \; " A'(2 j - 1) \, + \,
1/[B(2 j) - A(2 j)] \, ,
\qquad j \le \Ent {(2 m - 1) / 3} \, .
\\ \tag*{\tagform\aktTagnr d}
\endAligntags
$$
Die indizierten Feldelemente $A'(2 j - 2)$ und $A'(2 j - 1)$ beziehen
sich auf die Belegung von $A$ nach der Berechnung der Elemente
$\theta_j^{(2 m -\Ent {3 j/2}-2)}$ mit $0 \le j \le \Ent {(4 m -
3)/3}$. Da diese Elemente w\"ahrend der Rekursion \"uberschrieben werden,
m\"ussen sie in Hilfsvariablen gespeichert werden.

Nehmen wir nun an, da{\ss} der Index $n$ des letzten Startwertes $s_n$, der
in der Rekursion verwendet wurde, ungerade ist, $n = 2 m + 1$. Dann
enth\"alt das Feld $A$ die Elemente $\theta_j^{(2 m -\Ent {3 j/2})}$ mit
$0 \le j \le \Ent {(4 m + 1)/3}$, und $B$ enth\"alt die Elemente
$\theta_j^{(2 m -\Ent {3 j/2}-1)}$ mit $0 \le j \le \Ent {(4 m -
1)/3}$. Unter Verwendung der Konventionen (4.4-17) und (4.4-18) erh\"alt
man dann f\"ur Gl. (4.4-16):
$$
\beginAligntags
" B(0) \; " \gets \; " s_{2 m + 1} \, , \qquad m \in \N_0 \, ,
\erhoehe\aktTag \\ \tag*{\tagnr a}
" B(1) \; " \gets \; " 1 / [ B(0) - A(0)] \, ,
\\ \tag*{\tagform\aktTagnr b}
" B(2 j) " \; \gets \; " B'(2 j - 2) \, + \, \frac
{[A(2 j - 2) - B'(2 j - 2)] \, [B(2 j -1) - A(2 j - 1)]}
{B(2 j - 1) - 2 A(2 j -1) + B'(2 j - 1)} \, , \\
" j \le \Ent {(2 m + 1) / 3} \, , \span\omit \span\omit
\\ \tag*{\tagform\aktTagnr c}
" B(2 j + 1) \; " \gets \; " B'(2 j - 1) \, + \,
1/[A(2 j) - B(2 j)] \, ,
\qquad j \le \Ent { 2 m / 3} \, .
\\ \tag*{\tagform\aktTagnr d}
\endAligntags
$$
Die indizierten Feldelemente $B'(2 j - 2)$ und $B'(2 j - 1)$ beziehen
sich auf die Belegung von $B$ nach der Berechnung der Elemente
$\theta_j^{(2 m -\Ent {3 j/2}-1)}$ mit $0 \le j \le \Ent {(4 m -
1)/3}$. Da diese Elemente w\"ahrend der Rekursion \"uberschrieben werden,
m\"ussen sie in Hilfsvariablen gespeichert werden.

Das folgende speicheroptimierende MAPLE-Programm f\"ur den
$\theta$-Algorithmus, das auf den Rekursionen (4.4-19) und (4.4-20)
basiert, ist eine direkte \"Ubersetzung des in Abschnitt 10.2 von Weniger
[1989] abgedruckten FORTRAN-Programms THETA.

\medskip

\listenvon{theta.mpl}

\smallskip

\Abschnitt Anwendungen der Pad\'e-Approximationen in den
Naturwissenschaften

\smallskip

\aktTag = 0

Wie sp\"ater noch ausf\"uhrlicher diskutiert werden wird, sind in der
quantenmechanischen St\"orungstheorie Probleme mit schlechter Konvergenz
oder gar Divergenz der St\"orungsreihen eher die Regel als die Ausnahme
[Arteca, Fern\'{a}ndez und Castro 1990; {\v C}{\' \i}{\v z}ek und Vrscay
1982; Kazakov und Shirkov 1980; Killingbeck 1977; Le Guillou und
Zinn-Justin 1990; Simon 1982; 1991; Wilcox 1966; Zinn-Justin 1981a].
Dementsprechend umfangreich ist dann auch die Literatur \"uber die
Verwendung von Pad\'e-Approximationen in der quantenmechanischen
St\"orungstheorie. Beispielsweise verwendete Adams [1988]
Pad\'e-Approximationen zur st\"orungstheoretischen Behandlung eines
zweidimensionalen Wasserstoffatoms in einem \"au{\ss}eren Magnetfeld. Amos
[1978], Baker und Chisholm [1966] und Br\"andas und Goscinski [1970]
diskutierten die Summation divergenter St\"orungsreihen f\"ur Energien und
Wellenfunktionen mit Hilfe von Pad\'e-Approximationen. Belki\'{c} [1989]
kombinierte Pad\'e-Approximationen mit anderen nichtlinearen
Transformationen, um die divergente St\"orungsreihe f\"ur den quadratischen
Zeemaneffekt zu summieren. Bhattacharyya [1982; 1987] verwendete sowohl
eine spezielle Variante der Euler-Transformation als auch
Pad\'e-Approximationen zur Summation divergenter St\"orungsreihen. Au{\ss}erdem
verwendete Bhattacharyya [1989a] Pad\'e-Approximationen, um Aussagen \"uber
das asymptotische Verhalten bestimmter Eigenschaften anharmonischer
Oszillatoren machen zu k\"onnen, die durch divergente St\"orungs\-reihen
definiert sind. Khalil [1992] verwendete Pad\'e-Approximationen, um obere
und untere Schranken f\"ur Grundzustandseigenwerte mit Hilfe der
Rayleigh-Schr\"odingerschen und der Brillouin-Wignerschen St\"orungstheorie
zu bestimmen. Langhoff und Karplus [1970] berechneten atomare
Polarisierbarkeiten mit Hilfe von Pad\'e-Approxi\-mationen. Reinhardt
[1982] summierte den Real- und Imagin\"arteil atomarer Stark-Eigenwerte
mit Hilfe von Pad\'e-Approximatio\-nen. Iafrate und Mendelsohn [1970]
verwendeten Pad\'e-Approxima\-tionen f\"ur St\"orungsreihen bei
Ein\-elektronenatomen mit verallgemeinerten Zentralpotentialen. Wilson,
Silver, und Farrel [1977] zeigten, da{\ss} die speziellen
Pad\'e-Approximationen $[N+1/N]$ Invarianzeigenschaften in der
Rayleigh-Schr\"odingerschen St\"orungstheorie besitzen. Goodson und
Herschbach [1992] und Goodson und L\'{o}pez-Cabrera [1993] verwendeten
Pad\'e-Approximationen, um divergente St\"orungsreihen, die bei der
Behandlung von Coulomb-Systemen mit Hilfe dimensionaler Regularisierung
auftreten, zu summieren. Witwit und Killingbeck [1992] verwendeten
Pad\'e-Approximationen zur Summation divergenter St\"orungsreihen f\"ur die
Energieeigenwerte von eindimensionalen Schr\"o\-dinger\-gleichungen mit
rationalen Potentialen.

Besonders intensiv untersuchte quantenmechanische Modellsysteme sind die
in Gl. (2.2-3) definierten anharmonischen Oszillatoren mit ${\hat{x}}^{2
m}$-Anharmonizit\"at, deren St\"orungsreihen f\"ur die Energieeigenwerte
bekanntlich hochgradig divergent sind [Bender und Wu 1969; 1971; 1973].
Loeffel, Martin, Simon und Wightman [1969] konnten zeigen, da{\ss} die
diagonalen Pad\'e-Approximatio\-nen der Eigenwerte des anharmonischen
Oszillators mit einer ${\hat{x}}^4$-Anharmonizit\"at konvergieren. Graffi,
Grecchi und Simon [1970] verglichen die Konvergenz der Pad\'e-Summation
und der Borel-Pad\'e-Summation der divergenten St\"orungsreihe f\"ur die
Grundzustandsenergie des anharmonischen Oszillators mit einer
${\hat{x}}^4$-Anharmonizit\"at. Loeffel und Martin [1972] studierten die
analytische Struktur der Energieniveaus des anharmonischen Oszillators
mit einer ${\hat{x}}^4$-Anharmonizit\"at und die Konvergenz der
Pad\'e-Approximationen. Marziani [1984] verwendete eine Kombination des
Borelschen Summationsverfahrens und der konfluenten Form des Wynnschen
$\epsilon$-Algorithmus zur Summation der divergenten St\"orungsreihen f\"ur
Energie\-eigenwerte anharmonischer Oszillatoren mit ${\hat{x}}^4$-,
${\hat{x}}^6$- und ${\hat{x}}^8$-Anharmonizit\"aten. In Abschnitt
\Roemisch{4} seines langen Artikels \"uber die analytischen Eigenschaften
der Energie\-eigenwerte des anharmonischen Oszillators mit
${\hat{x}}^4$-Anharmonizit\"at bewies Simon [1970] einige S\"atze \"uber die
Konvergenz von Pad\'e-Approximationen und illuminierte seine theoretischen
Schlu{\ss}folgerungen durch numerische Beispiele. Bansal, Srivastava und
Vishwamittar [1991; 1992] und Srivastava und Vishwamittar [1991]
verwendeten Pad\'e-Approximationen, um renormierte St\"orungsreihen f\"ur die
Energieeigenwerte anharmonischer Oszillatoren mit den Potentialen $V (x)
= m (\frac {1}{2} \omega^2 x^2 + \frac {1}{3} \alpha x^3 + \frac {1}{6}
\beta x^6)$ und $V (x) = m (\frac {1}{2} \omega^2 x^2 + \frac {1}{3}
\alpha x^3 + \frac {1}{4} \beta x^4)$ zu summieren. Guardiola und Ros
[1992] und Guardiola, Sol\'{\i}s und Ros [1992] verwendeten
Pad\'e-Approximationen, um die renormierte St\"orungsreihen f\"ur die
Energieeigenwerte anharmonischer Oszillatoren mit den Potentialen $V = b
/ x^4 + c / x^6$ und $V = \lambda x^{2 N}$ zu berechnen. Chandra und
Bhattacharyya [1993] untersuchten, wie man die Energieeigenwerte
anharmonischer Oszillatoren auf effiziente Weise im Falle gro{\ss}er
Kopplungskonstanten mit Hilfe von Pad\'e-Approximationen berechnen kann.

Fried und Ezra [1989] verwendeten Pad\'e-Approximationen im Zusammenhang
mit der st\"orungs\-theoretischen Behandlungen von beinahe resonanten
gekoppelten Molek\"ulschwingungen. Stevens, Kinsey und Johnson [1992]
verwendeten Pad\'e-Approximationen und andere rationale Approximationen
zur analytischen Darstellung von Potentialfl\"achen zwei- und mehratomiger
Molek\"ule im Rahmen der Born-Oppenheimer-N\"aherung. {\v C}{\'\i}{\v z}ek,
{\v S}pirko und Bludsk\'{y} [1993] verwendeten Pad\'{e}-Approximationen
zur Summation divergenter St\"orungsreihen, die bei der
quantenmechanischen Beschreibung von Schwingungsspektren mehratomiger
Molek\"ule auftreten. Znojil verwendete Pad\'e-Approximationen zur
Darstellung von Potentialen bei der st\"orungstheoretischen Behandlung
quantenmechanischer St\"orungsreihen [Znojil 1988; 1991] oder zur
Summation verschiedener St\"orungsreihen bei dreidimensionalen isotropen
anharmonischen Oszillatoren [Znojil 1993].

Pad\'e-Approximationen wurden auch in der Streutheorie verwendet. Coleman
[1976] verwendete den $\epsilon$-Algorithmus zur L\"osung von
Integrodifferentialgleichungen, die bei Streuproblemen auftreten, und
Johnson und Reinhardt [1984] zeigten, da{\ss} divergente $L^2$-Entwicklungen
von $t$-Matrix-Amplituden Pad\'e-summierbar sind. Auf S. 327 des Buches
von Newton [1982] findet man zahlreiche Referenzen \"uber die Verwendung
von Pad\'e-Approximationen zur Summation von Partialwellenentwicklungen.

In dem Buch von Kukulin, Krasnopol'sky und Hor\'a\v{c}ek [1989] wurden
Pad\'e-Approximationen zur Beschreibung von Resonanzzust\"anden in der
Atom-, Molek\"ul- und Kernphysik verwendet. In dem \"Ubersichtsartikel von
Zinn-Justin [1971] wurden Anwendungen von Pad\'e-Approximationen in der
Streutheorie, in der Quantenfeldtheorie und in der
Elementarteilchenphysik diskutiert.

Eine sehr wichtige Rolle spielen Pad\'e-Approximationen auch in der
Theorie kritischer Ph\"anome\-ne. Hierzu sei auf die vor kurzem
erschienene Monographie von Baker [1990] und die zahlreichen darin
enthaltenen Referenzen verwiesen. \"Ahnliche Probleme behandelt auch ein
\"Ubersichtsartikel von Navarro [1990] \"uber die magnetischen
Eigenschaften von \"Ubergangs\-metall\-verbindungen, die eine im
wesentlichen zweidimensionale Schichtstruktur besitzen.
Pad\'e-Approximationen spielen auch in der Perkolation eine erhebliche
Rolle [Onody und Neves 1992; Stauffer und Aharony 1992].

Einen unorthodoxen Ansatz zur analytischen L\"osung eindimensionaler
Schr\"odingergleichungen verwendeten Fern\'{a}ndez, Ma und Tipping [1989]
und Fern\'{a}ndez [1992b]. Sie transformierten Schr\"odingergleichungen
f\"ur anharmonische Oszillatoren und gest\"orte Coulomb-Potentiale in die
entsprechenden Riccatigleichungen. Als L\"osungen der Riccatigleichungen
erhielten sie Potenzreihen f\"ur die logarithmische Ableitungen der
Wellenfunktionen, die dann zur Verbesserung der Konvergenzeigenschaften
in Pad\'e-Approximationen transformiert wurde.

Pad\'e-Approximationen wurden inzwischen auch bei
Dichtefunktionalrechnungen verwendet. DePristo und Kress [1987]
verwendete sie zur Konstruktion von Funktionalen der kinetischen
Energie, und Cedillo, Robles, und G\'{a}zquez [1988] konstruierten auf
diese Weise neue nichtlokale Austauschfunktionale.

Cioslowski [1988a; 1988b] verwendete Pad\'e-Approximationen zur Berechnung
der Hartree-Fock-Austauschenergie und zur Darstellung der spektralen
Dichteverteilung in der chemischen Graphentheorie. Au{\ss}erdem verwendeten
Cioslowski [1988c; 1990a; 1993a] und Cioslowski und Bessis [1988]
Pad\'e-Approximationen, um Gesamtenergien und Eigenschaften von
eindimensionalen Polymeren aus Rechnungen an endlichen Molek\"ulketten zu
bestimmen. Sarkar, Bhattacharyya und Bhattacharyya [1989] empfahlen
Pad\'e-Approximationen als Hilfsmittel bei Konvergenzproblemen von SCF-
und MCSCF-Rechnungen.

Neben diesen quantenmechanischen Beispielen gibt es aber auch zahlreiche
Anwendungen von Pad\'e-Approxi\-mationen in mehr klassisch orientierten
Bereichen der Naturwissenschaften und in der Technik. Im zweiten Teil
des Buches von Cabannes [1976] findet man zahlreiche Artikel \"uber
Anwendungen von Pad\'e-Approximationen in der Str\"omungsmechanik. Cohen
[1991] verwendete Pad\'e-Approximationen zur Beschreibung elastischer
Eigenschaften von Polymeren, und {\v C}{\'\i}{\v z}ek, Paldus, Ramgulam
und Vinette [1987] berechneten damit polarographische Str\"ome. Rogers und
Ames [1989, Abschnitt 2.10] verwendeten Pad\'e-Approximationen, um die
Konvergenzeigenschaften von Reihenentwicklungen bei nichtlinearen
Randwertproblemen mit sph\"arischer oder zylindrischer Symmetrie zu
verbessern. In den B\"uchern von Gray und Gubbins [1984] und von Hansen
und McDonald [1986] werden Anwendungen der Pad\'e-Approximationen in der
Theorie der Fl\"ussigkeiten beschrieben. Janse van Rensburg [1993]
verwendete Pad\'e-Approximationen, um Virialkoeffizienten inerter
Fl\"ussigkeiten aus exakt kugelf\"ormigen und exakt scheibenf\"ormigen
Molek\"ulen zu berechnen. Schlie{\ss}lich sei noch Basdevant [1972, s. 301]
erw\"ahnt, der demonstrierte, da{\ss} die Zustandsgleichung realer Gase nach
van der Waals auch als $[1/1]$-Pad\'e-Approximation einer
Virialentwicklung interpretiert werden kann, die eine Potenzreihe in der
Teilchendichte $\rho = N/V$ ist.

Diese unvollst\"andige Liste von naturwissenschaftlichen und technischen
Anwendungen zeigt, da{\ss} Pad\'e-Approximationen inzwischen zu den ganz
wichtigen numerischen Hilfsmitteln eines theoretisch arbeitenden
Naturwissenschaftlers geh\"oren (weitere Anwendungsbeispiele findet man in
den B\"uchern von Baker [1975] und Baker und Graves-Morris [1981a; 1981b]
oder in den von Brezinski [1976; 1991c] herausgegebenen Bibliographien
\"uber Pad\'e-Approximationen). Das Schwergewicht der Anwendungen lag bisher
in der theoretischen Physik. Es ist aber damit zu rechnen, da{\ss}
Pad\'e-Approximationen und verwandte Verfahren in der Zukunft vermehrt
auch in anderen Bereichen der Naturwissenschaften und in der Technik
Verwendung finden werden.

Trotz der unbestreitbaren Erfolge der Pad\'e-Approximationen sind diese
aber kein Allheilmittel bei Konvergenzproblemen, und es gibt zahlreiche
Probleme, bei denen Pad\'e-Approximationen entweder nicht leistungsf\"ahig
genug oder bei denen sie \"uberhaupt nicht anwendbar sind.

So sind Pad\'e-Approximationen offensichtlich nicht leistungsf\"ahig genug,
um einige besonders stark divergente St\"orungsreihen zu summieren. Ein
Beispiel ist die St\"orungsreihe (2.2-4) f\"ur die Grundzustandsenergie des
anharmonischen Oszillators mit einer ${\hat{x}}^8$-Anharmonizit\"at. Die
Koeffizienten $b_n^{(4)}$ dieser St\"orungsreihe wachsen gem\"a{\ss} Gl.
(2.2-5c) im wesentlichen wie $(3 n)!/n^{1/2}$ f\"ur $n \to \infty$. Daraus
folgt, da{\ss} die Carlemanbedingung (4.3-5), die nur eine {\it notwendige}
Bedingung ist, nicht erf\"ullt ist [Simon 1982, Theorem 1.3]. Es ist also
nicht klar, ob Pad\'e-Approximationen in der Lage sind, die St\"orungsreihe
(2.2-4) im Falle der ${\hat{x}}^8$-Anharmonizit\"at zu summieren. Als
Graffi, Grecchi und Turchetti [1971] versuchten, diese St\"orungsreihe zu
summieren, kamen sie aufgrund ihrer numerischen Ergebnisse zu dem
Schlu{\ss}, da{\ss} Pad\'e-Approximationen mit gro{\ss}er Wahrscheinlichkeit nicht in
der Lage sind, diese St\"orungsreihe zu summieren. Sp\"ater gelang es Graffi
und Grecchi [1978] explizit zu zeigen, da{\ss} der Stieltjes-Kettenbruch,
der einer St\"orungsreihe f\"ur den Energieeigenwert des anharmonischen
Oszillators mit einer ${\hat{x}}^{2 m}$-Anharmonizit\"at zugeordnet werden
kann, f\"ur $m > 2$ nicht konvergiert, was impliziert, da{\ss}
Pad\'e-Approximationen f\"ur $m > 2$ ebenfalls nicht konvergieren. Wie
sp\"ater noch ausf\"uhrlich behandelt wird, sind Pad\'e-Approximationen aber
auch bei anderen hochgradig divergenten Reihen weniger effizient als die
verallgemeinerten Summationsprozesse, die in Abschnitt 5 dieser Arbeit
besprochen werden.

Es scheint, da{\ss} die relative Ineffizienz der Pad\'e-Approximationen bei
hochgradig divergenten Reihen konstruktionsbedingt ist. Der Wynnsche
$\epsilon$-Algorithmus (2.4-10), der im Falle einer Potenzreihe
Pad\'e-Approximationen gem\"a{\ss} Gl. (4.4-1) produziert, ist exakt f\"ur
Modellfolgen des Typs [Shanks 1955; Wynn 1956a]
$$
s_n \; = \; s \, + \, \sum_{j=0}^{k-1} \, c_j \, \Delta s_{n+j} \, ,
\qquad n \in \N_0 \, .
\tag
$$
Wenn die Folgenelemente $s_n$ Partialsummen einer unendlichen Reihe
sind,
$$
s_n \; = \; \sum_{\nu = 0}^{n} \, a_{\nu} \, ,
\qquad n \in \N_0 \, ,
\tag
$$
kann die Modellfolge (4.5-1) auch auf folgende Weise geschrieben
werden:
$$
s_n \; = \; s \, + \, \sum_{j=0}^{k-1} \, c_j \, a_{n+j+1} \, ,
\qquad n \in \N_0 \, .
\tag
$$

Beim $\epsilon$-Algorithmus wird also implizit angenommen, da{\ss} der in
Gl. (3.1-1) definierte Abbruchfehler $r_n$ durch eine Linearkombination
aus den n\"achsten $k$ Termen $a_{n+1}$, $a_{n+2}$, $\ldots$ , $a_{n+k}$,
die nicht in der Partialsumme $s_n$ enthalten sind, approximiert werden
kann. Wenn die Reihe, die transformiert werden soll, eine Potenzreihe
mit endlichem Konvergenzradius ist, sollte diese Annahme wenigstens im
Inneren des Konvergenzkreises vern\"unftig sein und ausreichend genaue
Approximationen f\"ur die Abbruchfehler liefern. Tats\"achlich kann man
mit Hilfe von Pad\'e-Approximationen im Falle konvergenter Potenzreihen
oft sehr befriedigende Ergebnisse erhalten.

Im Falle hochgradig divergenter Reihen wie beispielsweise der Eulerreihe
(2.2-2) oder der St\"orungsreihe (2.2-4) f\"ur die Grundzustandsenergie
anharmonischer Oszillatoren (2.2-3) mit Anharmonizit\"at ${\hat{x}}^{2 m}$
ist die Modellfolge (4.5-1) aber nicht in der Lage, ausreichend gute
Approximationen f\"ur die tats\"achlichen Abbruchfehler zu liefern. So folgt
aus Gln. (4.3-7) und (4.3-8), da{\ss} der Abbruchfehler $r_n$ der Eulerreihe
f\"ur $z \ge 0$ betragsm\"a{\ss}ig durch den Absolutbetrag $(n+1)! z^{n+1}$ des
ersten Termes, der nicht in der Partialsumme enthalten war, abgesch\"atzt
werden kann. Wenn man aber versucht, die Eulerreihe mit Hilfe des
$\epsilon$-Algorithmus (2.4-10) zu summieren, dann folgt aus der
Modellfolge (4.5-1), da{\ss} stillschweigend vorausgesetzt wird, da{\ss} der
Abbruchfehler der Eulerreihe f\"ur $n \to \infty$ betragsm\"a{\ss}ig von der
Ordnung $O \bigl((n+k)! z^{n+k}\bigr)$ ist. F\"ur gro{\ss}e
Transformationsordnungen $k \in \N$ ist das nat\"urlich eine sehr
schlechte Absch\"atzung des tats\"achlichen Abbruchfehlers, und der
$\epsilon$-Algorithmus liefert im Falle der Eulerreihe (2.2-4) auch nur
relativ schlechte Ergebnisse [Weniger 1989, Tabelle 13.1].

In Abschnitt 5 werden einige verallgemeinerte Summationsprozesse
vorgestellt, die im Gegensatz zum $\epsilon$-Algorithmus von der
Zusatzinformation profitieren k\"onnen, da{\ss} der Abbruchfehler $r_n$ einer
Stieltjesreihe im Falle eines positiven Argumentes gem\"a{\ss} Gl. (4.3-8)
betragsm\"a{\ss}ig durch den ersten Term abgesch\"atzt werden kann, der nicht in
der Partialsumme enthalten ist. Wie in einigen Arbeiten gezeigt werden
konnte [Weniger und Steinborn 1989a; Weniger 1989; 1990; 1992; Weniger
und {\v C}{\'\i}{\v z}ek 1990; Weniger, {\v C}{\'\i}{\v z}ek und Vinette
1991; 1993; {\v C}{\'\i}{\v z}ek, Vinette und Weniger 1991; 1993;
Grotendorst 1991], sind diese verallgemeinerten Summationsprozesse zur
Summation hochgradig divergenter Reihen wesentlich besser als
Pad\'e-Approximationen geeignet.

Ein Problemkreis, bei denen Pad\'e-Approximationen \"uberhaupt nicht helfen
k\"onnen, sind logarithmisch konvergente Reihen, deren Partialsummen Gl.
(2.1-8) erf\"ullen. Beispielsweise wendete Wynn [1966] den
$\epsilon$-Algorithmus (2.4-10) auf die Elemente der Modellfolge
$$
s_n \; \sim \; s \, + \,
\sum_{j=0}^{\infty} \, c_j \, / \, (n + \beta)^{j+1} \, ,
\qquad \beta \in \R_{+} \, , \quad n \to \infty \, ,
\tag
$$
an, die offensichtlich logarithmisch konvergiert. F\"ur festes $k$ und
unter der Annahme, da{\ss} $c_0 \ne 0$ gilt, gelang es Wynn [1966] zu
zeigen, da{\ss} die Anwendung des $\epsilon$-Algorithmus auf die Modellfolge
(4.5-4) Approximationen ergibt, welche die folgende asymptotische
Absch\"atzung erf\"ullen:
$$
\epsilon_{2 k}^{(n)} \, \sim \, s \, + \, \frac
{c_0} {(k+1) (n + \beta )} \, , \qquad n \to \infty \, .
\tag
$$
Ein Vergleich dieser Beziehung mit Gln. (2.4-2) zeigt, da{\ss} der
$\epsilon$-Algorithmus offensichtlich nicht in der Lage ist, die
Konvergenz der Modellfolge (4.5-4) zu verbessern.

Auch bei anderen logarithmisch konvergenten Folgen und Reihen ist der
$\epsilon$-Algorithmus offensichtlich nicht in der Lage, eine
Konvergenzverbesserung zu erreichen. Ein sehr beliebtes Testbeispiel in
der Literatur \"uber Konvergenzbeschleunigung ist die Reihenentwicklung
(2.1-4) f\"ur die Riemannsche Zetafunktion. Auf S. 20 des Buches von
Wimp [1981] findet man die folgende asymptotische Absch\"atzung f\"ur den
Abbruchfehler der f\"ur ihre schlechte Konvergenz ber\"uchtigten
Reihenentwicklung (2.1-4),
$$
\zeta (z) \, - \, \sum_{j=0}^{n-1} \, (j+1)^{- z} \; \sim \;
n^{1-z} \, \left\{ \alpha_0 \, + \, \frac {\alpha_1} {n} \, + \,
\frac {\alpha_2} {n^2} \, + \, \ldots \; \right\}
\, , \qquad n \to \infty \, ,
\tag
$$
mit
$$
\beginAligntags
" \alpha_0 \; " = \; " 1 / (z - 1) \, ,
\erhoehe\aktTag \\ \tag*{\tagnr a}
" \alpha_1 \; " = \; " - 1 / 2 \, ,
\\ \tag*{\tagform\aktTagnr b}
" \alpha_{2 k + 1} \; " = \; " 0 \, , " " \qquad k \ge 1 \, ,
\\ \tag*{\tagform\aktTagnr c}
" \alpha_{2 k} \; " = \; " \frac {(z)_{2 k - 1}} {(2 k)!} B_{2 k} \, .
" " \qquad k \ge 1 \, .
\\ \tag*{\tagform\aktTagnr d}
\endAligntags
$$
Das in Gl. (4.5-7d) vorkommende Symbol $B_{2 k}$ ist eine
Bernoulli-Zahl, die beispielsweise in Gl. (1.08) auf S. 282 des Buches
von Olver [1974] definiert ist.

Weil $\zeta (2 m)$ f\"ur alle $m \in \N$ in geschlossener Form
ausgedr\"uckt werden kann [Magnus, Oberhettinger und Soni 1966, S. 19],
ist die unendliche Reihe f\"ur
$$
\zeta (2) \; = \; \pi^2 / 6
\; = \; \sum_{m=0}^{\infty} \, (m+1)^{- 2}
\tag
$$
in der Literatur \"uber Konvergenzbeschleunigung das am h\"aufigsten
verwendete Testproblem, um festzustellen, ob ein verallgemeinerter
Summationsproze{\ss} \"uberhaupt in der Lage ist, logarithmische Konvergenz
zu beschleunigen.

Aus Gln. (4.5-6) und (4.5-7) folgt, da{\ss} der Summationsrest der
Partialsumme
$$
s_n \; = \; \sum_{m=0}^n \, (m+1)^{- 2}
\, , \qquad n \in \N_0 \, ,
\tag
$$
der unendlichen Reihe f\"ur $\zeta (2)$ eine Potenzreihe in der Variablen
$(n + 1)^{- 1}$ ist. Das bedeutet, da{\ss} die Partialsummen (4.5-9) vom
Typ der Modellfolge (4.5-4) sind. Aufgrund der Absch\"atzung (4.5-5) ist
demzufolge nicht zu erwarten, da{\ss} der $\epsilon$-Algorithmus die
Konvergenz der unendlichen Reihe f\"ur $\zeta (2)$ beschleunigen kann.

In Tabelle 4-1 wird sowohl der Wynnsche $\epsilon$-Algorithmus, Gl.
(2.4-10), als auch die Standardform des Wynnsche $\rho$-Algorithmus,
Gl. (3.3-15), auf die Partialsummen (4.5-9) der unendlichen Reihe f\"ur
$\zeta (2)$ angewendet.

Die Approximationen zum Grenzwert der Folge der Partialsummen (4.5-9)
wurden im Falle des $\epsilon$-Algorithmus gem\"a{\ss} Gl. (4.4-8) gew\"ahlt.
Auch bei der Standardform des $\rho$-Algorithmus mu{\ss} man die N\"aherung
zum Grenzwert der zu transformierenden Folge $\Seqn s$ unterschiedlich
w\"ahlen, je nachdem ob der Index der letzten Partialsumme $s_m$, die in
der Rekursion verwendet wurde, gerade oder ungerade ist. Mit Hilfe der
Notation $\Ent x$ f\"ur den ganzzahligen Anteil von $x$, der die gr\"o{\ss}te
ganze Zahl $\nu$ ist, welche die Beziehung $\nu \le x$ erf\"ullt, kann man
die Approximation zum Grenzwert $s$ auf folgende Weise schreiben:
$$
\left\{ s_{m - 2 \Ent {m/2}}, s_{m - 2 \Ent {m/2} + 1}, \ldots , s_m
\right\} \; \to \;
\rho_{2 \Ent {m/2}}^{(m - 2 \Ent {m/2})} \, .
\tag
$$

\beginFloat

\medskip

\beginTabelle [to 12 cm]
\beginFormat \rechts " \mitte " \mitte " \mitte
\endFormat
\+ " \links {\bf Tabelle 4-1} \@ \@ \@ " \\
\+ " \links {Beschleunigung der Konvergenz der Reihe (4.5-8) f\"ur
$\zeta (2) = \pi^2 / 6$}
\@ \@ \@ " \\
\- " \- " \- " \- " \- " \\ \sstrut {} {1.5 \jot} {1.5 \jot}
\+ " \rechts {$n$} " Partialsumme $s_n$
" $\epsilon_{2 \Ent {n/2}}^{(n - 2 \Ent {n/2})}$
" $\rho_{2 \Ent {n/2}}^{(n - 2 \Ent {n/2})}$
" \\ \sstrut {} {1 \jot} {1.7 \jot}
\+ " " Gl. (4.5-9) " Gl. (2.4-10) " Gl. (3.3-15)
" \\ \sstrut {} {1.7 \jot} {1 \jot}
\- " \- " \- " \- " \- " \\ \sstrut {} {1 \jot} {1 \jot}
\+ "  0 "  1.0000000000000 "  1.0000000000000 "  1.0000000000000 " \\
\+ "  1 "  1.2500000000000 "  1.2500000000000 "  1.2500000000000 " \\
\+ "  2 "  1.3611111111111 "  1.4500000000000 "  1.6500000000000 " \\
\+ "  3 "  1.4236111111111 "  1.5039682539683 "  1.6468253968254 " \\
\+ "  4 "  1.4636111111111 "  1.5516174402250 "  1.6448948948949 " \\
\+ "  5 "  1.4913888888889 "  1.5717673885475 "  1.6449225865209 " \\
\+ "  6 "  1.5117970521542 "  1.5903054136156 "  1.6449343761237 " \\
\+ "  7 "  1.5274220521542 "  1.5999841551511 "  1.6449341453753 " \\
\+ "  8 "  1.5397677311665 "  1.6090869062812 "  1.6449340643806 " \\
\+ "  9 "  1.5497677311665 "  1.6144742952375 "  1.6449340662783 " \\
\+ " 10 "  1.5580321939765 "  1.6196099135256 "  1.6449340668680 " \\
\+ " 11 "  1.5649766384209 "  1.6229152921442 "  1.6449340668525 " \\
\+ " 12 "  1.5708937981842 "  1.6260947324348 "  1.6449340668481 " \\
\+ " 13 "  1.5759958390005 "  1.6282682531047 "  1.6449340668482 " \\
\+ " 14 "  1.5804402834450 "  1.6303723744034 "  1.6449340668482 " \\
\- " \- " \- " \- " \- " \\ \sstrut {} {1 \jot} {1 \jot}
\+ " \links {$\pi^2 /6$} " 1.6449340668482 " 1.6449340668482 "
1.6449340668482 " \\
\- " \- " \- " \- " \- " \\ \sstrut {} {1 \jot} {1 \jot}

\endTabelle

\medskip

\endFloat

Die Ergebnisse in Tabelle 4-1 zeigen, da{\ss} der $\epsilon$-Algorithmus
offensichtlich nicht in der Lage ist, die Konvergenz der unendlichen
Reihe f\"ur $\zeta (2) = \pi^2 / 6$ zu beschleunigen, wohingegen der eng
verwandte $\rho$-Algorithmus sehr gute Ergebnisse liefert.

Die Unf\"ahigkeit, logarithmische Konvergenz zu verbessern, ist eine der
Hauptschw\"achen des ansonsten sehr leistungsf\"ahigen
$\epsilon$-Algorithmus (2.4-10). Da logarithmische Konvergenz bei
praktischen Problemen aber sehr h\"aufig
vorkommt{\footnote[\dagger]{Logarithmisch konvergente Folgen und Reihen
und ihre effiziente Berechnung mit Hilfe verallgemeinerter
Summationsprozesse waren das zentrale numerische Problem in einigen
Artikeln des Autors, in denen so unterschiedliche Themen behandelt wurden
wie quantenmechanische Rechnungen an {\it quasi\/}-eindimensionalen
organischen Polymeren [Cioslowski und Weniger 1993; Weniger und Liegener
1990] oder die Berechnung von Mehrzentrenmolek\"ulintegrale
exponentialartiger Basisfunktionen durch unendliche Reihen [Grotendorst,
Weniger und Steinborn 1986; Steinborn und Weniger 1990; Weniger,
Grotendorst und Steinborn 1986a]}}, wurde schon sehr bald intensiv \"uber
alternative verallgemeinerte Summationsprozesse gearbeitet, die
logarithmische Konvergenz beschleunigen k\"onnen [Lubkin 1952; Salzer
1954; Salzer 1956; Salzer und Kimbro 1961; Wynn 1956b].

In B\"uchern von Brezinski [1977; 1978], Brezinski und Redivo Zaglia
[1991], Delahaye [1988] und Wimp [1981] oder in Artikeln von Brezinski,
Delahaye und Germain-Bonne [1983], Delahaye und Germain-Bonne [1980;
1982], Kowalewski [1981] und Osada [1990b] wurde aber gezeigt, da{\ss} die
Verbesserung logarithmischer Konvergenz ein \"uberaus schwieriges Problem
ist, und zwar sowohl theoretisch als auch praktisch. Insbesondere
konnten Delahaye und Germain-Bonne [1980; 1982] zeigen, da{\ss} es keine
Transformation geben kann, die in der Lage ist, die Konvergenz {\it
aller} logarithmisch konvergenten Folgen zu verbessern. In dieser
Beziehung unterscheidet sich logarithmische Konvergenz ganz erheblich
von linearer Konvergenz. Wie in Abschnitt 6 besprochen wird, konnte
Germain-Bonne [1973] f\"ur eine bestimmte Klasse von verallgemeinerten
Summationsprozessen notwendige und hinreichende Bedingungen formulieren,
die ein solcher verallgemeinerter Summationsproze{\ss} erf\"ullen mu{\ss}, um {\it
jede} linear konvergente Folge beschleunigen zu k\"onnen. Eine Erweiterung
der Theorie von Germain-Bonne auf eine gr\"o{\ss}ere Klasse von
verallgemeinerten Summationsprozessen [Weniger 1989, Abschnitt 12] wird
in den Abschnitten 6.2 und 6.3 dieser Arbeit beschrieben.

Trotz der oben genannten Probleme gab es in den letzten Jahren einige
erhebliche Fortschritte. Eine gro{\ss}e Anzahl von neuen verallgemeinerten
Summationsprozessen, die in der Lage sind, die Konvergenz zahlreicher
logarithmisch konvergierender Folgen und Reihen auf effiziente Weise zu
beschleunigen, und eine Diskussion ihrer theoretischen Eigenschaften
findet man in Artikeln von Bhattacharya, Roy und Bhowmick [1989],
Bj{\o}rstad, Dahlquist und Grosse [1981], Bhowmick, Bhattacharya und Roy
[1989], Brezinski [1971], Carstensen [1990], Cordellier [1979], Drummond
[1976; 1984], Homeier [1993], Levin [1973], Matos [1989; 1990a; 1990b],
Osada [1990a], Sablonniere [1991; 1992], Sedogbo [1990], Weniger [1989;
1991], Weniger und Liegener [1990], und Wimp [1972].

\endAbschnittsebene

\endAbschnittsebene

\keinTitelblatt\neueSeite

\beginAbschnittsebene
\aktAbschnitt = 4

\Abschnitt Transformationen mit expliziten Restsummenabsch\"atzungen

\vskip - 2 \jot

\beginAbschnittsebene

\medskip

\Abschnitt Vorbemerkungen

\smallskip

\aktTag = 0

Wie in Abschnitt 3.1 ausf\"uhrlich diskutiert wurde, versuchen alle in
dieser Arbeit behandelten verallgemeinerten Summationsprozesse, die
Elemente einer Folge $\Seqn s$ von Partialsummen in eine neue Folge
$\Seqn {s^{\prime}}$ mit g\"unstigeren Konvergenzeigenschaften zu
transformieren. Der erste Schritt besteht in der Annahme, da{\ss} eine
Partialsumme $s_n$ f\"ur jedes $n \in \N_0$ gem\"a{\ss}
$$
s_n \; = \; s \, + \, r_n
\tag
$$
in den Grenzwert $s$ und einen Summationsrest $r_n$ zerlegt werden
kann. Verallgemeinerte Summationsprozesse unterscheiden sich bez\"uglich
der Annahmen, die sie entweder explizit oder implizit \"uber das
Verhalten der Summationsreste $r_n$ als Funktion des Index $n$ machen.
Die unterschiedlichen Annahmen f\"uhren zu unterschiedlichen Strategien,
um die Summationsreste aus den Folgenelementen zu eliminieren.
Allerdings wird auf diese Weise der Grenzwert $s$ einer Folge in der
Regel nicht exakt, sondern nur approximativ bestimmt, da aus den
Eingabedaten $\Seqn s$ eine transformierte Folge $\Seqn {s^{\prime}}$
erzeugt wird, deren Elemente sich ebenfalls f\"ur alle $n \in \N_0$ gem\"a{\ss}
$$
s'_n \; = \; s \, + \, r'_n
\tag
$$
in den Grenzwert $s$ und einen im allgemeinen von Null verschiedenen
transformierten Summationsrest $r'_n$ zerlegen lassen. Wenn das
Verfahren erfolgreich war, werden die transformierten Summationsreste
$\Seqn {r^{\prime}}$ f\"ur $n \to \infty$ aber wesentlich schneller gegen
Null konvergieren als die urspr\"unglichen Summationsreste $\Seqn r$.

Bisher wurden nur verallgemeinerte Summationsprozesse behandelt, die,
obwohl sie zwar explizit oder implizit gewisse Annahmen \"uber das
Verhalten der Summationsreste machen, als Eingabedaten ausschlie{\ss}lich
Elemente einer Folge von Partialsummen $\Seqn s$ verwenden. Weitere
Informationen \"uber das Verhalten der Summationsreste werden nicht zur
Konstruktion der Elemente $s'_n$ der transformierten Folge verwendet.
Auf den ersten Blick scheint das eine sehr vorteilhafte Eigenschaft zu
sein, die eine praktische Anwendung erleichtert. In bestimmten
Situationen kann sich dieser scheinbar offensichtliche Vorteil aber in
einen gravierenden Nachteil verwandeln. So gibt es zahlreiche praktisch
relevante divergente Reihen, die so stark divergieren, da{\ss} sie von
verallgemeinerten Summationsprozessen, die solche Zusatzinformationen
nicht verwerten, entweder nur relativ schlecht oder \"uberhaupt nicht
summiert werden k\"onnen.

Beispiele sind die in Gl.~(4.3-4) definierten Stieltjesreihen, die --
wie etwa in Gl.~(2.2-5) gezeigt wurde -- in manchen F\"allen extrem stark
divergieren. Die Summationsreste solcher Stieltjesreihen k\"onnen im
Falle eines positiven Argumentes gem\"a{\ss} Gl.~(4.3-8) betragsm\"a{\ss}ig durch
den ersten Term abgesch\"atzt werden, der nicht in der Partialsumme
enthalten ist. In Abschnitt 4.4 wurde darauf hingewiesen, da{\ss} der
Wynnsche $\epsilon$-Algorithmus, Gl.~(2.4-10), nicht in der Lage ist,
von einer solchen zus\"atzlichen Information zu profitieren, was letztlich
erkl\"art, warum der $\epsilon$-Algorithmus bei hochgradig divergenten
Reihen wie etwa der Eulerreihe (2.2-2) oder der St\"orungsreihe (2.2-4)
f\"ur die Grundzustandsenergie des anharmonischen Oszillators (2.2-3) mit
einer ${\hat{x}}^{2 m}$-Anharmonizit\"at ($m = 2, 3, 4$) nur noch
vergleichsweise wenig bewirkt.

In diesem Abschnitt werden verallgemeinerte Summationsprozesse
behandelt, die solche Zusatzinformationen nutzbringend verwenden k\"onnen,
und die -- wie in einigen Artikeln gezeigt wurde -- im Falle hochgradig
divergenter Reihen den Wynnschen $\epsilon$-Algorithmus oder auch andere
verallgemeinerte Summationsprozesse ganz klar deklassieren [Weniger und
Steinborn 1989a; Weniger 1989; 1990; 1992; Weniger und {\v C}{\'\i}{\v
z}ek 1990; Weniger, {\v C}{\'\i}{\v z}ek und Vinette 1991; 1993; {\v
C}{\'\i}{\v z}ek, Vinette und Weniger 1991; 1993; Grotendorst 1991].

Es ist die typische Eigenschaft der in diesem Abschnitt behandelten
verallgemeinerten Summationsprozesse, da{\ss} sie als Eingabedaten nicht nur
die Elemente einer Folge $\Seqn s$ von Partialsummen verwenden, sondern
auch die Elemente einer Folge $\Seqn \omega$ von expliziten {\it
Restsummenabsch\"atzungen}. Wenn man Restsummenabsch\"atzungen finden kann,
die gute Approximationen f\"ur die tats\"achlichen Summationsreste liefern,
erh\"alt man verallgemeinerte Summationsprozesse, die -- wie zahlreiche
Beispiele zeigen -- extrem leistungsf\"ahig sein k\"onnen.

Alle verallgemeinerten Summationsprozesse, die in diesem Abschnitt
behandelten werden, basieren auf Modellfolgen, deren Summationsreste
$r_n$ f\"ur alle $n \in \N_0$ als Produkt einer Restsummenabsch\"atzung
$\omega_n$ und eines Korrekturterms $z_n$ dargestellt werden,
$$
s_n \; = \; s \, + \, \omega_n \, z_n \, , \qquad n \in \N_0 \, .
\tag
$$
Wenn es m\"oglich w\"are, sowohl die Restsummenabsch\"atzungen $\Seqn \omega$
als auch die Korrektur\-terme $\Seqn z$ so zu w\"ahlen, da{\ss} die Produkte
$\{ \omega_n z_n \}_{n=0}^{\infty}$ die Summationsreste $\Seqn r$ einer
gegebenen Folge von Partialsummen $\Seqn s$ beliebig genau approximieren
k\"onnen, w\"are das Problem der Konvergenzbeschleunigung und der Summation
gel\"ost. Da die Summa\-tionsreste bei praktisch relevanten Problemen aber
entweder numerisch nicht leicht zug\"anglich oder unbekannt sind, mu{\ss} man
sich in der Praxis damit zufrieden geben, sowohl die
Restsummenabsch\"atzungen $\Seqn \omega$ als auch die Korrekturterme
$\Seqn z$ aufgrund mehr oder weniger plausibler Annahmen zu w\"ahlen. Ob
diese Annahmen f\"ur eine gegebene Folge von Partialsummen berechtigt
waren oder nicht, kann man dann allerdings nur empirisch \"uberpr\"ufen,
n\"amlich anhand des Erfolges oder Mi{\ss}erfolges eines
Konvergenzbeschleunigungs- und Summationsverfahrens.

Wie sp\"ater noch ausf\"uhrlicher diskutiert wird, soll die
Restsummenabsch\"atzung $\omega_n$ so gew\"ahlt werden, da{\ss} der Quotient
$[s_n - s]/\omega_n$ f\"ur alle $n \in \N_0$ nur relativ schwach von $n$
abh\"angt, und f\"ur $n \to \infty$ gegen eine von Null verschiedene
Konstante konvergiert:
$$
s_n - s \; = \; \omega_n \bigl[ c + O (n^{- 1}) \bigr] \, .
\tag
$$
In praktischen Anwendungen ist es sicherlich sinnvoll, zus\"atzlich noch
zu fordern, da{\ss} die $\omega_n$ f\"ur alle endlichen Werte von $n$
ungleich Null sind und da{\ss} aus $m \ne n$ auch $\omega_m \ne \omega_n$
folgt. Ansonsten sind die Restsummenabsch\"atzungen $\omega_n$ weitgehend
beliebige Funktionen des Index $n$. Die Folge $\Seqn \omega$ kann also
sowohl konvergieren als auch divergieren. Demzufolge sind
verallgemeinerte Summationsprozesse, die auf Modellfolgen vom Typ von
Gl. (5.1-3) basieren, prinzipiell sowohl zur Konvergenzbeschleunigung
als auch zur Summation geeignet.

Es ist unmittelbar einleuchtend, da{\ss} ein verallgemeinerter
Summationsproze{\ss}, der f\"ur die Modellfolge (5.1-3) exakt ist, nur dann
gute Ergebnisse liefert, wenn man f\"ur eine gegebene Folge $\Seqn s$ von
Partialsummen eine Folge $\Seqn \omega$ von Restsummenabsch\"atzungen
finden kann, die schon f\"ur relativ kleine Werte von $n$ die
asymptotische Bedingung (5.1-4) ausreichend gut erf\"ullt. Demzufolge ist
die Wahl der Restsummenabsch\"atzungen f\"ur den Erfolg oder Mi{\ss}erfolg eines
Konvergenzbeschleunigungs- oder Summationsverfahrens von gr\"o{\ss}ter
Bedeutung. Aber auch die Wahl der Korrekturterme $\Seqn z$ spielt -- wie
sp\"ater gezeigt wird -- f\"ur die Eigenschaften der resultierenden
verallgemeinerten Summationsprozesse eine ganz entscheidende Rolle.

Sei ${\hat T}$ ein Operator, der auf Elemente von Folgen wirkt. Ein
solcher Operator wird {\it linear} genannt, wenn
$$
{\hat T} (a \xi_n + b \eta_n) \; = \;
a \, {\hat T} (\xi_n) \, + \, b \, {\hat T} (\eta_n)
\tag
$$
f\"ur beliebige Folgen $\Seqn \xi$ und $\Seqn \eta$ und f\"ur beliebige
reelle Konstanten $a$ und $b$ gilt. Wenn ein solcher linearer Operator
${\hat T}$ die Korrekturterme $z_n$ {\it annihiliert},
$$
{\hat T} (z_n) \; = \; 0 \, , \qquad n \in \N_0 \, ,
\tag
$$
kann man sofort einen verallgemeinerten Summationsproze{\ss} angeben, der
f\"ur die Modellfolge (5.1-3) exakt ist. Man mu{\ss} nur den
Annihilationsoperator ${\hat T}$ auf den Quotienten $[s_n - s] /
\omega_n$ anwenden. Da ${\hat T}$ laut Voraussetzung linear ist, erh\"alt
man [Weniger 1989, Abschnitt 3.2]{\footnote[\dagger]{Die hier skizzierte
Technik, verallgemeinerte Summationsprozesse mit Hilfe von
Annihilationsoperatoren zu konstruieren, wurde inzwischen von Brezinski
und Matos [1993] und Brezinski und Redivo Zaglia [1993a; 1993b]
aufgegriffen. Sie konnten zeigen, da{\ss} man alle bekannten
Extrapolationsverfahren durch Annihilationsoperatoren darstellen kann,
und da{\ss} man auf diese Weise auch zus\"atzliche Einblicke in die
Eigenschaften der Extrapolationsverfahren gewinnen kann.}}:
$$
{\cal T} (s_n, \omega_n) \; = \; \frac { {\hat T} (s_n / \omega_n ) }
{ {\hat T} (1 / \omega_n ) } \; = \; s \, ,
\qquad n \in \N_0 \, .
\tag
$$

Wenn man verallgemeinerte Summationsprozesse konstruieren will, die f\"ur
Modellfolgen vom Typ von Gl. (5.1-3) exakt sind, so mu{\ss} man Folgen
$\Seqn z$ von Korrekturtermen finden, die einerseits schon f\"ur kleine
Werte von $n$ gute Approximationen f\"ur Summationsreste, die bei
praktischen Problemen vorkommen, produzieren k\"onnen, und die
andererseits ausreichend einfache Annihilationsoperatoren ${\hat T}$
besitzen. Nur dann kann man hoffen, leistungsf\"ahige verallgemeinerte
Summationsprozesse ${\cal T} (s_n, \omega_n)$ zu erhalten, die au{\ss}erdem
noch einfach angewendet werden k\"onnen. Ideal w\"are eine nicht zu
komplizierte explizite Darstellung f\"ur ${\cal T} (s_n, \omega_n)$ und
au{\ss}erdem noch ein einfaches und numerisch stabiles Rekursionsschema.

In diesem Abschnitt werden ausschlie{\ss}lich Korrekturterme $z_n$
betrachtet, die nach Multiplikation mit einer geeigneten Funktion $w_k
(n)$ ein Polynom $P_{k - 1} (n)$ vom Grade $k - 1$ in $n$ ergeben. Die
Multiplikation der Modellfolge (5.1-3) mit $w_k (n)$ ergibt also einen
Ausdruck des folgenden Typs:
$$
w_k (n) \, \frac {s_n - s} {\omega_n}
\; = \; \, P_{k - 1} (n) \, , \qquad k, n \in \N_0 \, .
\tag
$$
Man kann jetzt ausn\"utzen, da{\ss} ein Polynom vom Grade $k - 1$ in $n$
durch Anwendung der $k$-ten Potenz des Differenzenoperators $\Delta$
annihiliert wird [Milne-Thomson 1981, S. 29]. Alle verallgemeinerten
Summationsprozesse dieses Abschnitts sind also Ausdr\"ucke des Typs
$$
{\cal T}_k^{(n)} \bigl( w_k (n); s_n, \omega_n \bigr)
\; = \; \frac
{\Delta^k \[ w_k (n) s_n / \omega_n \]}
{\Delta^k \[ w_k (n) / \omega_n \]}
\, , \qquad k, n \in \N_0 \, ,
\tag
$$
die sich nur in den bisher unspezifizierten Funktionen $w_k (n)$
unterscheiden. Ein Vergleich der Gln. (5.1-7) und (5.1-9) zeigt, da{\ss} in
Gl. (5.1-9) der {\it gewichtete Differenzenoperator} $\Delta^k w_k (n)$
als Annihilationsoperator ${\hat T}$ fungiert.

In den n\"achsten Unterabschnitten werden verschiedene Folgen $\Seqn z$
von Korrekturtermen und die daraus resultierenden verallgemeinerten
Summationsprozesse diskutiert. Dabei ergeben unterschiedliche Folgen von
Korrekturtermen verallgemeinerte Summationsprozesse mit ganz
unterschiedlichen numerischen Eigenschaften [Weniger und Steinborn
1989a; Weniger 1989; 1990; 1992; Weniger und {\v C}{\'\i}{\v z}ek 1990;
Weniger, {\v C}{\'\i}{\v z}ek und Vinette 1991; 1993; {\v C}{\'\i}{\v
z}ek, Vinette und Weniger 1991; 1993; Grotendorst 1991].

Man kann auch von Korrekturtermen $z_n$ ausgehen, die {\it nicht} nach
Multiplikation mit einer geeigneten Gr\"o{\ss}e in Polynome in $n$
transformiert werden k\"onnen und die demzufolge auch nicht durch
gewichtete Differenzenoperatoren annihiliert werden k\"onnen. Sidi [1982]
verwendete beispielsweise die Modellfolge
$$
s_n \; = \; s \, + \, \omega_n \, \sum_{j=0}^{k-1} \, c_j x_n^j \, ,
\qquad k, n \in \N_0 \, ,
\tag
$$
zur Konstruktion eines verallgemeinerten Summationsprozesses, wobei
$\Seqn x$ eine Folge von Interpolationspunkten ist, deren Elemente die
folgenden Bedingungen erf\"ullen:
$$
\beginMultiline
x_0 > x_1 > x_2 > \cdots > x_m > x_{m+1} > \cdots > 0 \, ,
\erhoehe\aktTag \\ \tag*{\tagnr a}
\lim_{n \to \infty} \; x_n \; = \; 0 \, .
\\ \tag*{\tagform\aktTagnr b}
\endMultiline
$$
Wenn man die Modellfolge (5.1-10) auf folgende Weise umformuliert,
$$
[s_n - s] / \omega_n \; = \; \sum_{j=0}^{k-1} \, c_j \, x_n^j \, ,
\tag
$$
erh\"alt man einen Ausdruck, dessen rechte Seite ein Polynom vom Grade
$k-1$ in der Variablen $x_n$ ist. Man kann also den Grenzwert $s$ der
Modellfolge (5.1-10) gem\"a{\ss} Gl. (5.1-7) bestimmen, wenn man einen
linearen Operator ${\hat T}$ finden kann, der Polynome in der Variablen
$x_n$ annihiliert.

Bekanntlich kann die Annihilation des Polynoms auf der rechten Seite
von Gl. (5.1-12) mit Hilfe von {\it dividierten Differenzen} erreicht
werden, die beispielsweise in der Newtonschen Interpolationsformel
vorkommen, und die in jedem Buch der numerischen Mathematik oder auch
in B\"uchern \"uber Differenzenrechnung wie N\"orlund [1954] oder
Milne-Thomson [1981] behandelt werden.

Sei $\Seqn x$ eine Folge von Interpolationspunkten, deren Elemente Gl.
(5.1-11) erf\"ullen. Die dividierten Differenzen der Ordnungen $0, 1,
\ldots , k, k+1, \ldots $ einer gegebenen Funktion $f$ werden
\"ublicherweise durch das Rekursionsschema
$$
\beginAligntags
" f [x_n] \> " = " \> f (x_n)\; ,
\hfill \erhoehe\aktTag \\ \tag*{\tagnr a}
" f [x_n, \ldots , x_{n+k+1}] \> " = " \> \frac
{f [x_{n+1}, \ldots , x_{n+k+1}] - f [x_n, \ldots , x_{n+k}] }
{ x_{n+k+1} - x_n } \, , \qquad k, n \in \N_0 \, ,
\\ \tag*{\tagform\aktTagnr b}
\endAligntags
$$
definiert [Powell 1981, Gl. (5.14)]. Die dividierten Differenzen $f
[x_n, \ldots , x_{n+k}]$ besitzen auch eine geschlossene Darstellung
als Summe [Powell 1981, Gl. (5.2)]:
$$
f [x_n, \ldots , x_{n+k}] \; = \; \sum_{j=0}^k \, f (x_{n+j}) \,
\prod^k \Sb{i=0 \\ i\ne j} \, \frac {1} {x_{n+j} - x_{n+i} } \, ,
\qquad k, n \in \N_0 \, .
\tag
$$
Aus Gln. (5.1-13) und (5.1-14) folgt, da{\ss} die dividierten Differenzen
$f [x_n, \ldots , x_{n+k}]$ lineare Funktionen der Eingabedaten $f
(x_n)$, $\ldots$ , $f (x_{n+k})$ sind. Wenn $P_m (x)$ ein Polynom vom
Grade $m$ in $x$ ist, $$
P_m (x) \; = \; c_0 + c_1 x + c_2 x^2 + \cdots + c_m x^m \, ,
\tag
$$
dann kann man zeigen, da{\ss} alle dividierten Differenzen mit $k > m$
dieses Polynom annihilieren [N\"orlund 1954, S. 9]:
$$
P_m [x_n, \ldots , x_{n+k}] \, = \, 0 \, , \qquad k > m \, .
\tag
$$

Man mu{\ss} jetzt nur noch annehmen, da{\ss} zwei Funktionen ${\cal S} (x)$ und
${\it \Omega} (x)$ einer kontinuierlichen Variable $x$ existieren, die
an den Interpolationspunkten $x_n$ mit $s_n$ beziehungsweise $\omega_n$
\"ubereinstimmen,
$$
\beginAligntags
" {\cal S} (x_n) \; " = \; " s_n \, ,
\erhoehe\aktTag \\ \tag*{\tagnr a}
" {\it \Omega} (x_n) \; " = \; " \omega_n \, ,
\\ \tag*{\tagform\aktTagnr b}
\endAligntags
$$
und die au{\ss}erdem noch die Beziehung
$$
[ {\cal S} (x) - s ] / {\it \Omega} (x) \; = \;
\sum_{j=0}^{k-1} \, c_j x^j
\tag
$$
erf\"ullen, um einen verallgemeinerten Summationsproze{\ss} konstruieren zu
k\"onnen, der f\"ur die Elemente der Modellfolge (5.1-10) exakt ist. Sidi
[1982] definierte den verallgemeinerten Summationsproze{\ss} ${\cal
R}_k^{(n)} (s_n, \omega_n, x_n)$, den er als Verallgemeinerung des
Richardsonschen Extrapolationsprozesses [Richardson 1927] bezeichnete
und der f\"ur die Modellfolge (5.1-10) exakt ist, indem er dividierte
Differenzen der Funktionen ${\cal S} (x) / {\it \Omega} (x)$ und $1 /
{\it \Omega} (x)$ an den Interpo\-la\-tions\-punkten $x_n, \ldots ,
x_{n+k}$ bildete:
$$
{\cal R}_k^{(n)} (s_n, \omega_n, x_n) \; = \; \frac
{\{ {\cal S} (x) / {\it \Omega} (x) \} [x_n, \ldots , x_{n+k}] }
{\{ 1 / {\it \Omega} (x) \} [x_n, \ldots , x_{n+k}] }
\, , \qquad k, n \in \N_0 \, .
\tag
$$
Der in Gl. (5.1-7) definierte allgemeine Annihilationsoperator ${\hat
T}$ ist in diesem Fall die dividierte Differenz an den
Interpolationspunkten $x_n$, $x_{n+1}$, $\ldots$ , $x_{n+k}$.

Aus Gl.~(5.1-13) folgt, da{\ss} sowohl der Z\"ahler als auch der Nenner
des verallgemeinerten Summationsprozesses (5.1-19) mit Hilfe des
Rekursionsschemas
$$
R_{k+1}^{(n)} \, = \, \frac
{ R_{k}^{(n+1)} - R_{k}^{(n)} }
{ x_{n+k+1} - x_n } \; , \qquad k, n \in \N_0 \, ,
\tag
$$
berechnet werden kann [Weniger 1989, Abschnitt 7.4]. Wenn man die
Startwerte
$$
R_0^{(n)} \; = \; s_n / \omega_n \, , \qquad n \in \N_0 \, ,
\tag
$$
verwendet, erh\"alt man den Z\"ahler der Transformation ${\cal R}_k^{(n)}
(s_n, \omega_n, x_n)$, w\"ahrend die Startwerte
$$
R_0^{(n)} \; = \; 1 / \omega_n \, , \qquad n \in \N_0 \, ,
\tag
$$
den Nenner dieser Transformation ergeben.

Der verallgemeinerte Summationsproze{\ss} ${\cal R}_k^{(n)} (s_n, \omega_n,
x_n)$, Gl. (5.1-19), der weitgehend beliebige Interpolationspunkte
$\Seqn x$ zul\"a{\ss}t, ist allgemeiner als die verallgemeinerten
Summationsprozesse vom Typ von Gl. (5.1-9). Trotzdem werden in dieser
Arbeit ausschlie{\ss}lich die spezielleren verallgemeinerten
Summationsprozesse vom Typ von Gl. (5.1-9) behandelt.

Der Grund ist, da{\ss} es bei praktischen Anwendungen fast nicht m\"oglich
ist, die zus\"atzlichen M\"oglichkeiten, die der verallgemeinerte
Summationsproze{\ss} ${\cal R}_k^{(n)} (s_n, \omega_n, x_n)$ verglichen mit
Transformationen von Typ von Gl. (5.1-9) bietet, auszun\"utzen. Es gibt
bisher keine mathematische Theorie, die angibt, wie man die
Interpolationspunkte $\Seqn x$ als Funktion des Index $n$ w\"ahlen sollte,
um bei gegebenen Eingabedaten $\Seqn s$ und Restsummenabsch\"atzungen
$\Seqn \omega$ m\"oglichst gute Ergebnisse erhalten zu k\"onnen. Bisher
wurden bei praktischen Anwendungen haupts\"achlich die Interpolationspunkte
$$
x_n \; = \; 1/(n + \zeta) \, ,
\qquad n \in \N_0 \, , \quad \zeta \, > \, 0 \, ,
\tag
$$
verwendet. Diese Wahl ergibt aber -- wie im n\"achsten Unterabschnitt
gezeigt wird -- die sehr leistungsf\"ahige Levinsche Transformation
[Levin 1973], die demzufolge ein Spezialfall der von Sidi [1982]
eingef\"uhrten Verallgemeinerung ${\cal R}_k^{(n)} (s_n, \omega_n, x_n)$
des Richardsonschen Extrapolationsprozesses [Richardson 1927] ist.

Aufgrund der oben erw\"ahnten Probleme mit der Wahl der
Interpolationspunkte wurde bei praktischen Anwendungen vom Autor bisher
nie ${\cal R}_k^{(n)} (s_n, \omega_n, x_n)$ verwendet, sondern
ausschlie{\ss}lich die im n\"achsten Abschnitt besprochene Levinsche
Transformation.

\medskip

\Abschnitt Die Levinsche Transformation

\smallskip

\aktTag = 0

Die Idee, explizite Restsummenabsch\"atzungen $\Seqn \omega$ in
verallgemeinerten Summationspro\-zessen zu verwenden, geht auf Levin
[1973] zur\"uck. Sein Ausgangspunkt war die Modellfolge
$$
s_n \; = \; s \, + \, \omega_n \,
\sum_{j=0}^{k-1} \, c_j / (n + \zeta)^j \, ,
\qquad k, n \in \N_0 \, ,
\tag
$$
die ein Spezialfall der von Sidi [1982] eingef\"uhrten Modellfolge
(5.1-10) ist. Ein Vergleich der Gln. (5.1-3) und (5.2-1) zeigt, da{\ss}
Levin [1973] die Korrekturterme $z_n$ durch Polynome vom Grade $k - 1$
in der Variablen $1/(n + \zeta)$ mit $\zeta > 0$ darstellte. Das ist
eine relativ naheliegende Annahme. Die Modellfolge (5.2-1) erh\"alt man
n\"amlich, wenn man asymptotische Potenzreihen im Sinne von Poincar\'e des
Typs
$$
s_n \; \sim \; s \, + \,
\omega_n \, \sum_{j=0}^{\infty} \, c_j / (n + \zeta)^j \, ,
\qquad n \to \infty \, ,
\tag
$$
nach $k$ Termen abbricht. Offensichtlich erf\"ullt eine solche
asymptotische Potenzreihe automatisch die asymptotische Bedingung
(5.1-4).

Die Modellfolge (5.2-1) ist auch ein Spezialfall der Modellfolge
(3.2-1), die den allgemeinen Extrapolationsproze{\ss} $E_k (s_n)$, Gl.
(3.2-3), ergibt. Sie enth\"alt ebenfalls $k + 1$ Unbekannte, die alle
linear vorkommen -- n\"amlich den Grenzwert $s$ und die $k$ Koeffizienten
$c_0$, $c_1$, $\ldots$ , $c_{k - 1}$. Nehmen wir an, da{\ss} die numerischen
Werte von $k + 1$ Folgenelementen $s_n$, $s_{n+1}$, $\ldots$ , $s_{n+k}$
und $k + 1$ Restsummenabsch\"atzungen $\omega_n$, $\omega_{n+1}$, $\ldots$
, $\omega_{n+k}$ bekannt sind. Dann kann man die Levinsche
Transformation ${\cal L}_k^{(n)} (\zeta, s_n, \omega_n)$, die exakt ist
f\"ur Elemente der Modellfolge (5.2-1), analog zu Gl. (3.2-3) unter
Verwendung der Cramerschen Regel als Quotient zweier Determinanten
definieren:

\smallskip

$$
{\cal L}_k^{(n)} (\zeta , s_n , \omega_n) \; = \; \frac
{
\vmatrix
{
s_n " \ldots " s_{n+k} \\ [1\jot]
\omega_n " \ldots " \omega_{n+k} \\
\vdots " \ddots " \vdots \\ [1\jot]
\omega_n / (\zeta + n)^{k-1} " \ldots "
\omega_{n+k} / (\zeta + n + k)^{k-1}
}
}
{
\vmatrix
{
1 " \ldots " 1       \\ [1\jot]
\omega_n " \ldots " \omega_{n+k} \\
\vdots " \ddots " \vdots \\ [1\jot]
\omega_n / (\zeta + n)^{k-1} " \ldots "
\omega_{n+k} / (\zeta + n + k)^{k-1}
}
} .
\tag
$$

\smallskip

Allerdings ist -- wie schon mehrfach erw\"ahnt -- die Darstellung eines
verallgemeinerten Summationsprozesses durch Determinanten f\"ur
praktische Zwecke nicht sehr n\"utzlich. Man kann aber leicht explizite
Darstellungen der Levinschen Transformation ${\cal L}_k^{(n)} (\zeta,
s_n , \omega_n)$ als Quotient zweier endlicher Summen finden. Die
urspr\"ungliche Ableitung von Levin [1973] basiert darauf, da{\ss} die
Determinanten in Gl. (5.2-3) durch Vandermondesche Determinanten
ausgedr\"uckt werden k\"onnen, f\"ur die explizite Ausdr\"ucke bekannt sind.
F\"ur unsere Zwecke ist aber eine alternative Ableitung vorzuziehen, die
auf Sidi [1979] zur\"uckgeht und die bestimmte Eigenschaften des in Gl.
(3.1-12) definierten Differenzenoperators $\Delta$ ausn\"utzt. Dazu
schreiben wir Gl. (5.2-1) auf folgende Weise um:
$$
(n + \zeta)^{k-1} [s_n - s] / \omega_n \; = \;
\sum_{j=0}^{k-1} \, c_j \, (n + \zeta)^{k-j-1} \, .
\tag
$$
Der Ausdruck auf der rechten Seite ist ein Polynom vom Grade $k - 1$ in
$n$. Bekanntlich wird ein solches Polynom durch $k$-fache Anwendung des
Differenzenoperators $\Delta$ annihiliert [Milne-Thomson 1981, S. 29].
Da der Differenzenoperator $\Delta^k$ linear ist, erhalten wir aus Gl.
(5.2-4) die folgende Darstellung f\"ur die Levinsche Transformation:
$$
{\cal L}_k^{(n)} (\zeta , s_n , \omega_n) \; = \; \frac
{ \Delta^k \, \{ (n + \zeta)^{k-1} \> s_n / \omega_n\} }
{ \Delta^k \, \{ (n + \zeta)^{k-1} / \omega_n \} }
\, , \qquad k, n \in \N_0 \, .
\tag
$$
Der in Gl. (5.1-7) definierte allgemeine Annihilationsoperator ${\hat
T}$ ist im Falle der Levinschen Transformation also der gewichtete
Differenzenoperator $\Delta^k (n + \zeta)^{k-1}$. Mit Hilfe von Gl.
(3.1-14) k\"onnen die Differenzenoperatoren in Gl. (5.2-5) in
geschlossener Form ausgedr\"uckt werden. Man erh\"alt auf diese Weise die
folgende Darstellung der Levinschen Transformation als Quotient zweier
endlicher Summen [Sidi 1979; Weniger 1989, Abschnitt 7.2]:
$$
{\cal L}_{k}^{(n)} (\zeta , s_n, \omega_n) \; = \;
\frac
{\displaystyle
\sum_{j=0}^{k} \; ( - 1)^{j} \; \binom {k} {j} \;
\frac {(\zeta + n +j )^{k-1}} {(\zeta + n + k )^{k-1}} \;
\frac {s_{n+j}} {\omega_{n+j}} }
{\displaystyle
\sum_{j=0}^{k} \; ( - 1)^{j} \; \binom {k} {j} \;
\frac {(\zeta + n +j )^{k-1}} {(\zeta + n + k )^{k-1}} \;
\frac {1} {\omega_{n+j}} }
\; , \qquad k, n \in \N_0 \, .
\tag
$$
Der gemeinsame Faktor $(\zeta + n + k )^{k-1}$ im Z\"ahler und Nenner von
Gl. (5.2-6) soll den Betrag der Terme in den Summen verringern. Auf
diese Weise wird die Gefahr von OVERFLOW verringert, was sonst bei
h\"oheren Transformationsordnungen $k$ nicht auszuschlie{\ss}en w\"are.

Die beiden Summen in Gl. (5.2-6) k\"onnen auch mit Hilfe der folgenden
Dreitermrekursion berechnet werden [Longman 1981; Fessler, Ford und
Smith 1983a; Weniger 1989, Abschnitt 7.2; Brezinski und Redivo Zaglia
1991, Abschnitt 2.7]:
$$
L_{k+1}^{(n)} (\zeta) \; = \; L_k^{(n+1)} (\zeta) \, - \, \frac
{ (\zeta + n) (\zeta + n + k)^{k-1} } { (\zeta + n + k + 1)^k } \>
L_k^{(n)} (\zeta) \, , \qquad k, n \in \N_0 \, .
\tag
$$
Wenn man die Startwerte
$$
L_0^{(n)} (\zeta) \; = \; s_n / \omega_n \, ,
\qquad n \in \N_0 \, ,
\tag
$$
verwendet, ergibt die Dreitermrekursion (5.2-7) den Z\"ahler der
Levinschen Transformation, w\"ahrend die Startwerte
$$
L_0^{(n)} (\zeta) \; = \; 1 / \omega_n \, ,
\qquad n \in \N_0 \, ,
\tag
$$
den Nenner der Levinschen Transformation ergeben.

Um die Levinsche Transformation ${\cal L}_k^{(n)} (\zeta , s_n ,
\omega_n)$ in Konvergenzbeschleunigungs- und Summationsverfahren
verwenden zu k\"onnen, mu{\ss} man erst f\"ur eine gegebene Folge von
Partialsummen
$$
s_n \; = \; \sum_{\nu = 0}^n a_{\nu} \, , \qquad n \in \N_0 \, ,
\tag
$$
eine Folge $\Seqn \omega$ von Restsummenabsch\"atzungen finden, deren
Elemente Gl. (5.1-4) erf\"ullen. Allerdings legt die asymptotische
Bedingung (5.1-4) die Restsummenabsch\"atzungen nicht eindeutig fest.
Deswegen ist es wenigstens im Prinzip immer m\"oglich, f\"ur eine gegebene
Folge $\Seqn s$ von Partialsummen mehrere verschiedene Folgen von
Restsummenabsch\"atzungen zu finden, die alle Gl. (5.1-4) erf\"ullen, die
aber in Konvergenzbeschleunigungs- und Summationsprozessen
unterschiedlich wirksam sein k\"onnen [Weniger 1989, Tabelle 14-6;
Steinborn und Weniger 1990, Tabelle 1].

Manchmal ist es m\"oglich, explizite Ausdr\"ucke f\"ur
Restsummenabsch\"atzungen zu finden, die numerisch au{\ss}erdem noch relativ
leicht zug\"anglich sind. Wenn beispielsweise die Terme $a_{\nu}$ einer
Reihe explizit bekannt sind und eine ausreichend einfache Struktur
besitzen, ist es unter Umst\"anden m\"oglich, f\"ur die Summationsreste
$$
r_n \; = \; - \sum_{k=n+1}^{\infty} \, a_k \, , \qquad n \in \N_0 \, ,
\tag
$$
eine Integraldarstellung zu finden, f\"ur die man dann eventuell auch
noch einen einfachen expliziten N\"aherungsausdruck ableiten kann. Wenn
man einen solchen {\it expliziten} Ausdruck f\"ur $\omega_n$ in Gl.
(5.2-6) verwendet, ist die Levinsche Transformation ${\cal L}_k^{(n)}
(\zeta , s_n , \omega_n)$ {\it linear} in den Eingabedaten $s_n$,
$s_{n+1}$, $\ldots$ , $s_{n+k}$.

In den meisten F\"allen wird es aber nicht m\"oglich sein, solche expliziten
Restsummenabsch\"atzungen zu finden, da nicht genug \"uber das Verhalten der
Restsummen $\Seqn r$ als Funktion des Index $n$ bekannt ist. Bei
praktischen Problemen kennt man normalerweise nur die numerischen Werte
einer relativ kleinen Zahl von Eingabedaten $s_m$, $s_{m+1}$, $\ldots$ ,
$s_{m + \ell}$. Man mu{\ss} also Mittel und Wege finden, wie man die
ben\"otigten Restsummenabsch\"atzungen direkt aus den numerischen Werten der
wenigen vorhandenen Eingabedaten bestimmt.

Wenn man in Gl. (5.2-6) Restsummenabsch\"atzungen $\omega_n$ verwendet,
die aus den Partialsum\-men $s_n$, $s_{n+1}$, $\ldots$ , $s_{n+k}$
berechnet wurden, ist die Levinsche Transformation ${\cal L}_k^{(n)}
(\zeta , s_n , \omega_n)$ offensichtlich {\it nichtlinear} in den
Eingabedaten $s_n$, $s_{n+1}$, $\ldots$ , $s_{n+k}$.

Mit Hilfe einfacher heuristischer \"Uberlegungen gelang es Levin [1973],
f\"ur Folgen $\Seqn s$ von Partialsummen Restsummenabsch\"atzungen $\Seqn
\omega$ zu finden, die trotz ihrer Einfachheit in vielen F\"allen
erstaunlich gut funktionieren.

Im Falle logarithmischer Konvergenz, d. h., wenn die Eingabedaten Gl.
(2.1-8) erf\"ullen, schlug Levin [1973] die Restsummenabsch\"atzung
$$
\omega_n \; = \; (\zeta + n) a_n \; = \; (\zeta + n) \Delta s_{n-1}
\, , \qquad n \in \N_0 \, ,
\tag
$$
vor, deren Verwendung in Gl. (5.2-6) die Levinsche $u$-Transformation
ergibt:
$$
u_{k}^{(n)} (\zeta, s_n) \; = \;
\frac
{\displaystyle
\sum_{j=0}^{k} \; ( - 1)^{j} \; \binom {k} {j} \;
\frac {(\zeta + n +j )^{k-2}} {(\zeta + n + k )^{k-1}} \;
\frac {s_{n+j}} {a_{n+j}}
}
{\displaystyle
\sum_{j=0}^{k} \; ( - 1)^{j} \; \binom {k} {j} \;
\frac {(\zeta + n +j )^{k-2}} {(\zeta + n + k )^{k-1}} \;
\frac {1} {a_{n+j}}
} \; .
\tag
$$
Die Restsummenabsch\"atzung (5.2-12) war urspr\"unglich von Levin [1973] f\"ur
die Beschleunigung der Konvergenz extrem langsam konvergierender
monotoner Reihen vom Typ der Reihenentwicklung (2.1-4) f\"ur die
Riemannsche Zetafunktion konzipiert worden. Trotzdem ist die
$u$-Transformation in der Lage, lineare Konvergenz zu beschleunigen und
auch zahlreiche alternierende divergente Reihen effizient zu summieren.
Laut Smith und Ford [1979; 1982] geh\"ort die $u$-Transformation zu den
leistungsf\"ahigsten und vielseitigsten verallgemeinerten
Summationsprozessen, die zur Zeit bekannt sind.

Im Falle alternierender Reihen schlug Levin [1973] die
Restsummenabsch\"atzung
$$
\omega_n \; = \; a_n \; = \; \Delta s_{n-1}
\, , \qquad n \in \N_0 \, ,
\tag
$$
vor, deren Verwendung in Gl. (5.2-6) die Levinsche $t$-Transformation
ergibt:
$$
t_{k}^{(n)} (\zeta, s_n) \; = \;
\frac
{\displaystyle
\sum_{j=0}^{k} \; ( - 1)^{j} \; \binom {k} {j} \;
\frac {(\zeta + n +j )^{k-1}} {(\zeta + n + k )^{k-1}} \;
\frac {s_{n+j}} {a_{n+j}}
}
{\displaystyle
\sum_{j=0}^{k} \; ( - 1)^{j} \; \binom {k} {j} \;
\frac {(\zeta + n +j )^{k-1}} {(\zeta + n + k )^{k-1}} \;
\frac {1} {a_{n+j}}
} \; .
\tag
$$
Die $t$-Transformation ist eine leistungsf\"ahige Transformation im Falle
linearer Konvergenz, und sie ist ebenfalls in der Lage, zahlreiche
alternierende divergente Reihen effizient zu summieren. Allerdings
versagt sie v\"ollig bei logarithmischer Konvergenz.

Da der Summationsrest $r_n$ einer konvergenten Reihe mit strikt
alternierenden Termen $a_{\nu}$ betragsm\"a{\ss}ig abgesch\"atzt werden kann
durch den ersten Term $a_{n+1}$, der nicht in der Partialsumme $s_n$
enthalten ist, und da die Summationsreste $r_n$ ebenfalls strikt
alternieren [Knopp 1964, S. 259], ist die Restsummenabsch\"atzung
$$
\omega_n \; = \; a_{n+1} \; = \; \Delta s_n
\, , \qquad n \in \N_0 \, ,
\tag
$$
laut Smith und Ford [1979] die nat\"urlichste {\it einfache}
Restsummenabsch\"atzung f\"ur konvergente alternierende Reihen. Diese
Restsummenabsch\"atzung, die auch mit der Absch\"atzung (4.3-8) f\"ur den
Abbruchfehler einer Stieltjesreihe \"ubereinstimmt, ist ebenfalls die
beste einfache Restsummenabsch\"atzungen f\"ur nichtabbrechende, d. h.
divergente hypergeometrische Reihen des Typs
$$
{}_2 F_0 (\alpha, \beta, - z) \; = \;
\sum_{m=0}^{\infty} \, \frac {(\alpha)_m \, (\beta)_m} {m!} \, (- z)^m
\, , \qquad \alpha, \beta, z > 0 \, ,
\tag
$$
da ihr Abbruchfehler ebenfalls ein strikt alternierendes Vorzeichen
besitzt und betragsm\"a{\ss}ig vom ersten Term der hypergeometrischen Reihe,
der nicht in der Partialsumme enthalten ist, abgesch\"atzt werden kann
[Carlson 1977, Theorem 5.12-5].

Wenn man die Restsummenabsch\"atzung (5.2-16) in Gl. (5.2-6) verwendet,
erh\"alt man die folgende, von Smith und Ford [1979] eingef\"uhrte
Modifikation der Levinschen $t$-Transformation:
$$
d_{k}^{(n)} (\zeta, s_n) \; = \;
\frac
{\displaystyle
\sum_{j=0}^{k} \; ( - 1)^{j} \; \binom {k} {j} \;
\frac {(\zeta + n +j )^{k-1}} {(\zeta + n + k )^{k-1}} \;
\frac {s_{n+j}} {a_{n+j+1}}
}
{\displaystyle
\sum_{j=0}^{k} \; ( - 1)^{j} \; \binom {k} {j} \;
\frac {(\zeta + n +j )^{k-1}} {(\zeta + n + k )^{k-1}} \;
\frac {1} {a_{n+j+1}}
} \; .
\tag
$$
Die $d$-Transformation hat \"ahnliche Eigenschaften wie die
$t$-Transformation: Sie versagt zwar v\"ollig bei logarithmischer
Konvergenz, bei linearer Konvergenz und im Falle alternierender
divergenter Reihen liefert sie aber h\"aufig sehr gute Ergebnisse.

Ein Vergleich der Gln. (3.1-15) und (5.2-18) zeigt, da{\ss} man die in
Abschnitt 3.1 zur Beschleunigung der Konvergenz strikt alternierender
Reihen konstruierte Transformation $T_k^{(n)}$ erh\"alt, wenn man in Gl.
(5.2-18) $\zeta = 1$ und $a_n = (- 1)^n b_n$ setzt.

Die dritte einfache Restsummenabsch\"atzung, die Levin [1973] vorschlug,
basiert auf dem Aitkenschen $\Delta^2$-Proze{\ss}, Gl. (3.3-6):
$$
\omega_n \; = \; \frac {a_n a_{n+1}} {a_n - a_{n+1}}
\; = \; \frac
{\Delta s_{n-1} \Delta s_n} {\Delta s_{n-1} - \Delta s_n}
\, , \qquad n \in \N_0 \, .
\tag
$$
Wenn man diese Restsummenabsch\"atzung in Gl. (5.2-6) verwendet, erh\"alt
man die Levinsche $v$-Transformation:
$$
v_{k}^{(n)} (\zeta, s_n) \; = \;
\frac
{\displaystyle
\sum_{j=0}^{k} \; ( - 1)^{j} \; \binom {k} {j} \;
\frac {(\zeta + n +j )^{k-1}} {(\zeta + n + k )^{k-1}} \;
\frac {a_{n+j} - a_{n+j+1}} {a_{n+j} a_{n+j+1}} s_{n+j}
}
{\displaystyle
\sum_{j=0}^{k} \; ( - 1)^{j} \; \binom {k} {j} \;
\frac {(\zeta + n +j )^{k-1}} {(\zeta + n + k )^{k-1}} \;
\frac {a_{n+j} - a_{n+j+1}} {a_{n+j} a_{n+j+1}}
} \; .
\tag
$$
Die $v$-Transformation ist \"ahnlich leistungsf\"ahig und vielseitig wie
die $u$-Transformation, da sie sowohl zur Beschleunigung linearer und
logarithmischer Konvergenz geeignet ist als auch zur Summation
divergenter alternierender Reihen.

Weitere Spezialf\"alle der Levinschen Transformation ${\cal L}_k^{(n)}
(\zeta , s_n , \omega_n)$, die auf anderen Restsummenabsch\"atzungen
basieren, sind in Abschnitt 7 von Weniger [1989] beschrieben.

Die Levinschen Restsummenabsch\"atzungen (5.2-12), (5.2-14) und (5.2-19)
sowie die von Smith und Ford [1979] eingef\"uhrte Restsummenabsch\"atzung
(5.2-16) wurden auf der Basis einfacher heuristischer \"Uberlegungen
gefunden. Die oben erw\"ahnten Restsummenabsch\"atzungen k\"onnen zum Teil
aber auch durch asymptotische Absch\"atzungen zus\"atzlich motiviert
werden. Wenn beispielsweise die Terme $a_n$ einer Reihe in Falle
gro{\ss}er Indizes $n$ die Beziehung
$$
a_n \; \sim \; \lambda^n n^{\Theta} \left\{ \alpha_0 +
\frac {\alpha_1}{n} + \frac {\alpha_2}{n^2} + \cdots \; \right\}
\, , \qquad n \to \infty \, ,
\tag
$$
mit $\alpha_0 \ne 0$ erf\"ullen, dann gibt es Konstanten $\zeta_j$ und
$\gamma_j$, so da{\ss} der Summationsrest $r_n$ der Partialsumme $s_n$ dieser
Reihe die Beziehung
$$
r_n \; \sim \; \frac {\lambda^{n+1} n^{\Theta}} {\lambda - 1}
\left\{ \alpha_0 + \frac {\zeta_1}{n} + \frac {\zeta_2}{n^2}
+ \cdots \; \right\} \, , \qquad n \to \infty \, ,
\tag
$$
erf\"ullt, wenn $| \lambda | < 1$ gilt, und
$$
r_n \; \sim \; - \frac {n^{\Theta + 1}} {\Theta + 1}
\left\{ \alpha_0 + \frac {\gamma_1}{n} + \frac {\gamma_2}{n^2} + \cdots
\; \right\} \, , \qquad n \to \infty \, ,
\tag
$$
wenn $\lambda = 1$ und $Re(\Theta) < - 1$ gilt [Wimp 1981, S. 19].

Wenn wir annehmen, da{\ss} Gl. (5.2-21) gilt, dann entspricht die f\"uhrende
Ordnung der Restsummenabsch\"atzung (5.2-16), die die $d$-Transformation
ergibt, im Falle linearer Konvergenz ($| \lambda | < 1$) der f\"uhrenden
Ordnung in Gl. (5.2-22). Im Falle logarithmischer Konvergenz ($\lambda =
1$ und $Re (\Theta) < - 1$) entspricht die f\"uhrende Ordnung in Gl.
(5.2-23) der f\"uhrenden Ordnung der Restsummenabsch\"atzung (5.2-12), die
die $u$-Transformation ergibt.

FORTRAN-Programme f\"ur die allgemeine Levinsche Transformation ${\cal
L}_k^{(n)} (\zeta, s_n, \omega_n)$ findet man in Abschnitt 7.5 von
Weniger [1989], und auf der zu dem Buch von Brezinski und Redivo Zaglia
[1991] geh\"orenden DOS-Diskette. Au{\ss}erdem wurde ein FORTRAN-Programm f\"ur
die $u$-Transformation von Fessler, Ford und Smith [1983a; 1983b]
ver\"offentlicht. MAPLE-Programme f\"ur die $u$-, $t$-, $d$- und
$v$-Transformation wurden von Grotendorst [1989; 1991] ver\"offentlicht.

Normalerweise gibt die Transformation ${\cal L}_k^{(n)} (\zeta, s_n,
\omega_n)$ mit dem maximalen unteren Index $k$ und dem minimalen oberen
Index $n$ in Konvergenzbeschleunigungs- und Summationsverfahren das
beste Ergebnis. Als N\"aherung zum Grenzwert $s$ der zu transformierenden
Folge $\Seqn s$ ver\-wen\-den wir deswegen [Weniger 1989, Gl. (7.5-4)]:
$$
\{ s_0, \omega_0; s_1, \omega_1; \ldots ; s_m, \omega_m \} \; \to \;
{\cal L}_m^{(0)} (\zeta, s_0, \omega_0) \, ,
\qquad m \in \N_0 \, .
\tag
$$
Der Grenzwert $s$ der zu transformierenden Folge $\Seqn s$ wird also
durch die Folge
$$
{\cal L}_0^{(0)} (\zeta, s_0, \omega_0), \;
{\cal L}_1^{(0)} (\zeta, s_0, \omega_0), \; \ldots , \;
{\cal L}_m^{(0)} (\zeta, s_0, \omega_0), \; \ldots \; ,
\tag
$$
von Transformationen approximiert.

\medskip

\Abschnitt Fakult\"atenreihen

\smallskip

\aktTag = 0

Im letzten Abschnitt wurde gezeigt, da{\ss} man die Levinsche Transformation
${\cal L}_k^{(n)} (\zeta, s_n, \omega_n)$ erh\"alt, wenn man den
Korrekturterm $z_n$ -- oder \"aquivalenterweise den Quotienten $[s_n - s]
/ \omega_n$ -- in der Modellfolge (5.1-3) durch ein Polynom vom Grade $k
- 1$ in der Variablen $1/(n + \zeta)$ darstellt. Im Prinzip entspricht
dieser Ansatz der Annahme, da{\ss} der Quotient $[s_n - s]/\omega_n$ durch
eine asymptotische Potenzreihe im Sinne von Poincar\'e gem\"a{\ss} Gl. (5.2-2)
dargestellt werden kann. Die Levinsche Transformation versucht dann, die
f\"uhrenden $k$ Terme dieser Potenzreihe zu eliminieren.

Ein relativ naheliegender Ansatz, der zu anderen verallgemeinerten
Summationsprozessen f\"uhrt, besteht darin, die Folge $\Seq {(n+\zeta)^{-
j}} {j=0}$ der Potenzen in der asymptotischen Reihe (5.2-2) durch eine
alternative Folge $\Seq {\phi_j (n)} {j=0}$ zu ersetzen. Man m\"u{\ss}te dann
versuchen, eine Transformation zu konstruieren, die f\"ur die Modellfolge
$$
s_n \; = \; s \, + \, \omega_n \,
\sum_{j=0}^{k-1} \, c_j \, \phi_j (n) \, ,
\qquad k, n \in \N_0 \, ,
\tag
$$
exakt ist, die ebenfalls ein Spezialfall der sehr allgemeinen
Modellfolge (3.2-1) ist. Wenn man in der Determinantendarstellung
(3.2-3) f\"ur den allgemeinen Extrapolationsproze{\ss} $E_k (s_n)$ $f_j (n)
= \omega_n \phi_j (n)$ setzen w\"urde, erhielte man sofort eine
Determinantendarstellung f\"ur einen verallgemeinerten Summationsproze{\ss},
der f\"ur die Modellfolge (5.3-1) exakt w\"are.

Im Prinzip k\"onnte man in Gl. (5.3-1) jede Funktionenfolge $\Seq {\phi_j
(n)} {j=0}$ verwenden, deren Elemente die Beziehungen
$$
\beginAligntags
" \phi_0 (n) \; " = \; " 1 \, , " " n \in \N_0 \, ,
\erhoehe\aktTag \\ \tag*{\tagnr a}
" \phi_{j+1} (n) \; " = \; " o ( \phi_j (n) ) \, ,
\qquad " " j \in \N_0 \, , \quad n \to \infty \, ,
\\ \tag*{\tagform\aktTagnr b}
\endAligntags
$$
erf\"ullen, was bedeutet, da{\ss} $\Seq {\phi_j (n)} {j=0}$ eine asymptotische
Folge f\"ur $n \to \infty$ sein mu{\ss}. Allerdings sind solche
Minimalforderungen nicht ausreichend, um auf der Basis von Gl. (5.1-7)
einen leistungsf\"ahigen und benutzerfreundlichen verallgemeinerten
Summationsproze{\ss} konstruieren zu k\"onnen. Man mu{\ss} einen
Annihilationsoperator finden, der die Summe auf der rechten Seite von
Gl. (5.3-1) auf ausreichend einfache Weise annihilieren kann, was die
Zahl der potentiell verwendbaren Funktionenfolgen $\Seq {\phi_j (n)}
{j=0}$ stark einschr\"ankt. Au{\ss}erdem sollte ein neuer verallgemeinerter
Summationsproze{\ss} \"ahnlich leistungsf\"ahig sein wie die Levinsche
Transformation. Da die effiziente und verl\"a{\ss}liche Berechnung gro{\ss}er
Determinanten in der Regel nicht m\"oglich ist, w\"are ein einfaches
Rekursionsschema vom Typ von Gl. (5.2-7), mit dessen Hilfe der neue
verallgemeinerte Summationsproze{\ss} berechnet werden kann, sehr
erstrebenswert. Ein expliziter Ausdruck vom Typ von Gl. (5.2-6) w\"are
ebenfalls sehr hilfreich, da man auf diese Weise bessere Chancen h\"atte,
den Mechanismus als auch die Grenzen des auf der Basis der Modellfolge
(5.3-1) konstruierten neuen verallgemeinerten Summationsprozesses zu
verstehen.

Es ist nicht leicht, eine Funktionenfolge $\Seq {\phi_j (n)} {j=0}$ zu
finden, die zu einem neuen verallgemeinerten Summationsproze{\ss} f\"uhrt, der
die oben genannten Eigenschaften besitzt. Wenn man aber annimmt, da{\ss} der
Quotient $[s_n - s] / \omega_n$ durch eine {\it Fakult\"atenreihe}
dargestellt werden kann, erh\"alt man einen verallgemeinerten
Summationsproze{\ss} mit den gew\"unschten Eigenschaften [Weniger 1989,
Abschnitt 8].

Sei $\Omega (z)$ eine Funktion, die f\"ur $| z | \to \infty$
verschwindet. Eine Fakult\"atenreihe f\"ur $\Omega (z)$ ist eine
Entwicklung des Typs
$$
\Omega (z) \; = \; \frac {c_0} {z} \, + \, \frac {c_1} {z (z+1)}
\, + \, \frac {c_2} {z (z+1) (z+2)} \, + \, \cdots \; = \;
\sum_{\nu=0}^{\infty} \frac {c_{\nu}} {(z)_{\nu+1}} \, .
\tag
$$
Hierbei ist $(z)_{\nu+1}$ ein Pochhammersymbol, das \"ublicherweise als
Quotient zweier Gammafunktionen definiert wird [Magnus, Oberhettinger
und Soni 1966, S. 3],
$$
(z)_{\nu+1} \; = \; \Gamma (z+\nu+1) / \Gamma (z) \; = \;
z (z+1) \ldots (z+\nu) \, , \qquad \nu \in \N_0 \, .
\tag
$$

Fakult\"atenreihen haben eine lange Tradition in der Mathematik. Ein
gro{\ss}er Teil des Buches von Stirling [1730] besch\"aftigt sich mit
Fakult\"atenreihen. Im neunzehnten Jahrhundert wurde die Theorie der
Fakult\"atenreihen weiterentwickelt und um die Jahrhundertwende zu einem
gewissen Abschlu{\ss} gebracht. Eine relativ vollst\"andige Darstellung der
\"alteren Literatur \"uber Fakult\"atenreihen findet man in B\"uchern von
Nielsen [1965] und N\"orlund [1926; 1954]. In diesen B\"uchern findet man
auch eine ausf\"uhrliche Behandlung der grundlegenden mathematischen
Eigenschaften der Fakult\"atenreihen.

Fakult\"atenreihen haben eine bemerkenswerte Eigenschaft, die sich auch
bei theoretischen Untersuchungen der Konvergenzeigenschaften einiger
verallgemeinerter Summationsprozesse als sehr n\"utzlich erwies [Weniger
1989, Abschnitt 13]: Es ist extrem einfach, h\"ohere Potenzen des
Differenzenoperators $\Delta$ auf Fakult\"atenreihen anzuwenden.
Demzufolge spielen Fakult\"atenreihen in der Theorie der
Differenzengleichungen eine \"ahnliche Rolle wie die Potenzreihen in der
Theorie der Differentialgleichungen. Das erkl\"art, warum Fakult\"atenreihen
in den klassischen B\"uchern \"uber Differenzengleichungen [Meschkowski
1959; Milne-Thomson 1981; N\"orlund 1926; 1954] relativ ausf\"uhrlich
behandelt werden. Einen kurzen Abschnitt \"uber Fakult\"atenreihen findet
man auch in dem Buch von L\"osch und Schoblik [1951, Abschnitt 3.2].

Im Zusammenhang mit der Summation divergenter Reihen sind sowohl die
B\"ucher von Borel [1928] \"uber divergente Reihen und von Doetsch [1955]
\"uber Laplace-Transformationen als auch ein \"Ubersichtsartikel von Thomann
[1990] von besonderem Interesse, da in ihnen der Zusammenhang zwischen
Fakult\"atenreihen und Summierbarkeit behandelt wird.

In den letzten Jahren hat aber das Interesse der Mathematiker an
Fakult\"atenreihen anscheinend stark nachgelassen, was aus der relativ
geringen Anzahl von neueren Referenzen geschlossen werden kann. Eine
l\"obliche Ausnahme ist das Buch von Wasow [1987], das ein Kapitel \"uber
Fakult\"atenreihen enth\"alt. Au{\ss}erdem gibt es noch einige neuere Artikel:
Ramis und Thomann [1981] behandelten numerische Anwendungen von
Fakult\"atenreihen, Iseki und Iseki [1980] diskutierten asymptotische
Absch\"atzungen f\"ur den Abbruchfehler von Fakult\"atenreihen, und Dunster
und Lutz [1991] versuchten, Besselfunktionen mit Hilfe von
Fakult\"atenreihen zu berechnen.

Eine Fakult\"atenreihe ist immer auch eine asymptotische Reihe f\"ur $z \to
\infty$. Aus Gl. (5.3-3) und aus der Beziehung $(z)_n - z^n = O
(z^{n-1})$, $z \to \infty$, folgt
$$
\Omega (z) \, - \,
\sum_{\nu=0}^{n-1} \frac {c_{\nu}} {(z)_{\nu+1}}
\; = \; O (z^{- n - 1}) \, , \qquad z \to \infty \, .
\tag
$$
Diese Beziehung impliziert, da{\ss} $\Seq {1/(z)_n} {n=0}$ eine
asymptotische Folge f\"ur $z \to \infty$ ist. Trotzdem unterscheiden sich
Fakult\"atenreihen und asymptotische Potenzreihen in $1/z$ ganz erheblich
bez\"uglich ihrer numerischen Eigenschaften. Das ist eine direkte Folge
der Tatsache, da{\ss} das Argument $z$ einer Fakult\"atenreihe in einem
Pochhammersymbol und nicht als Potenz vorkommt.

Eine Potenzreihe konvergiert im Inneren eines Kreises, der
g\"unstigstenfalls mit der gesamten komplexen Ebene $\C$ \"ubereinstimmt
und der ung\"unstigstenfalls aus einem einzigen Punkt besteht. Wenn eine
Fakult\"atenreihe aber \"uberhaupt konvergiert, dann konvergiert sie nach
Landau [1906] in einer Halbebene. Wenn eine Fakult\"atenreihe also f\"ur
einen Punkt $z_0 \in \C$ konvergiert, dann konvergiert sie mit der
m\"oglichen Ausnahme der Punkte $z = 0, -1, -2, \ldots$ f\"ur alle $z \in
\C$ mit $Re(z) > Re(z_0)$.

Die unterschiedlichen Konvergenzeigenschaften von Potenzreihen und
Fakult\"atenreihen kann man anhand der Reihen
$$
\frac {1} {x} \, - \, \frac {1} {x^2} \, + \, \frac
{2} {x^3} \, - \, \frac {6} {x^4} \, + \cdots \; = \;
\sum_{m=0}^{\infty} \frac {(-1)^m m!} {x^{m+1}}
\tag
$$
und
$$
\frac {1} {x} \, - \, \frac {1} {(x)_2} \, + \, \frac
{2} {(x)_3} \, - \, \frac {6} {(x)_4} \, + \cdots
\; = \; \sum_{m=0}^{\infty} \frac {(-1)^m m!} {(x)_{m+1}}
\tag
$$
demonstrieren, die beide die gleichen numerischen Koeffizienten $c_m =
(-1)^m m!$ besitzen. Die Potenzreihe (5.3-6) divergiert f\"ur alle $\vert
x \vert < \infty$, wogegen die Fakult\"atenreihe (5.3-7) f\"ur alle $x > 0$
konvergiert.

Aufgrund der unterschiedlichen Konvergenzeigenschaften von
Fakult\"atenreihen und Potenzreihen kann es vorkommen, da{\ss} eine Funktion
$\Omega (z)$, die eine Darstellung durch eine {\it divergente}
asymptotische Potenzreihe besitzt,
$$
\Omega (z) \; \sim \; \frac {c'_0}{z} \, + \, \frac {c'_1}{z^2}
\, + \, \frac {c'_3}{z^3} \, + \, \cdots \; ,
\qquad z \to \infty \, ,
\tag
$$
auch eine Darstellung durch eine {\it konvergente} Fakult\"atenreihe
gem\"a{\ss} Gl. (5.3-3) besitzt. Ein Beispiel ist die unvollst\"andige
Gammafunktion
$$
\Gamma (a, x) \; = \;
\int\nolimits_{x}^{\infty} \, \e^{- t} \, t^{a-1} \d t \, .
\tag
$$
Sie besitzt sowohl eine Darstellung durch eine {\it divergente}
asymptotische Potenzreihe [Erd\'elyi, Magnus, Oberhettinger und Tricomi
1953, S. 135],
$$
\Gamma (a, x) \; = \;
x^{a - 1} \, \e^{- x} \, {}_2 F_0 (1, 1 - a; - 1/x)
\tag
$$
als auch eine Darstellung durch eine {\it konvergente} Fakult\"atenreihe
[Erd\'elyi, Magnus, Oberhettinger und Tricomi 1953, S. 139],
$$
\beginAligntags
" \Gamma (a, x) \; " = \;
" x^a \, \e^{- x} \,
\sum_{n=0}^{\infty} \, \frac {c_n (a)} {(x)_{n+1}} \, , \\
" c_n (a) \; " = \; " \frac {(- 1)^n n!} {\Gamma (1 - a)} \,
\int\nolimits_{0}^{\infty} \, \e^{- t} \, t^{- a} \,
\binom {t} {n} \d t \, .
\\ \tag
\endAligntags
$$

Die algebraischen Prozesse, mit deren Hilfe man Potenzreihen und
Fakult\"atenreihen ineinander \"uberf\"uhren kann, wurden schon in dem Buch
von Stirling [1730] beschrieben. Eine etwas modernere Beschreibung der
Stirlingschen Methode findet man in dem urspr\"unglich aus dem Jahre 1906
stammenden Buch von Nielsen [1965, S. 272 - 282]. Au{\ss}erdem gibt es noch
einen langen Artikel von Watson [1912b], in dem die bei der
Transformation einer asymptotischen Reihe in eine konvergente
Fakult\"atenreihe auftretenden mathematischen Probleme ausf\"uhrlich
behandelt werden.

\medskip

\Abschnitt Darstellung der Korrekturterme durch Fakult\"atenreihen

\smallskip

\aktTag = 0

In diesem Abschnitt wollen wir annehmen, da{\ss} der Korrekturterm $z_n$ in
Gl. (5.1-3) durch eine abgebrochene Fakult\"atenreihe dargestellt werden
kann. Unser Ausgangspunkt ist also die Modellfolge
$$
s_n \; = \; s \, + \, \omega_n \,
\sum_{j=0}^{k-1} \, c_j / (n + \zeta)_j
\, , \qquad k, n \in \N_0 \, ,
\tag
$$
die formal fast v\"ollig identisch ist mit der Modellfolge (5.2-1), die
die Levinsche Transformation ${\cal L}_k^{(n)} (\zeta, s_n, \omega_n)$
ergibt. Der einzige Unterschied ist, da{\ss} die Potenzen $(n+\zeta)^j$ in
Gl. (5.2-1) durch Pochhammersymbole $(n+\zeta)_j$ ersetzt wurden.

Was die Folge $\Seqn \omega$ der Restsummenabsch\"atzungen betrifft, so
machen wir die gleichen Annahmen wie im Falle der Levinschen
Transformation. Wir gehen also davon aus, da{\ss} die
Restsummenabsch\"atzungen $\Seqn \omega$ weitgehend beliebige Funktionen
des Index $n$ sind, die sowohl divergieren als auch konvergieren
k\"onnen. Analog zur Levinschen Transformation nehmen wir ebenfalls an,
da{\ss} die $\omega_n$ f\"ur alle endlichen Werte von $n$ ungleich Null sind
und da{\ss} aus $m \ne n$ auch $\omega_m \ne \omega_n$ folgt.

Weiterhin nehmen wir an, da{\ss} in Gl. (5.4-1) $\zeta > 0$ gilt. Dadurch
ist garantiert, da{\ss} die Pochhammersymbole $(n+\zeta)_j$ in Gl. (5.4-1)
immer von Null verschieden sind. Das w\"are an sich schon gew\"ahrleistet,
wenn $\zeta$ keine negative ganze Zahl oder Null ist. Da die
Modellfolge (5.4-1) aber als endliche Approximation f\"ur unendliche
Fakult\"atenreihen des Typs
$$
s_n \; \sim \; s \, + \, \omega_n \,
\sum_{j=0}^{\infty} \, c_j / (n + \zeta)_j
\, , \qquad n \to \infty \, ,
\tag
$$
\"uber einen weiten Bereich von $n$-Werten dienen soll, darf sich das
Vorzeichenmuster der Terme der Fakult\"atenreihe nicht \"andern, wenn $n$
gr\"o{\ss}er wird. Negative Werte von $\zeta$ m\"ussen demzufolge ausgeschlossen
werden, da sie zu unterschiedlichen Vorzeichen der Terme der
Fakult\"atenreihe f\"uhren, je nachdem ob $n + \zeta < 0$ oder $n + \zeta >
0$ gilt. Demzufolge ist in Gl. (5.4-1) die Einschr\"ankung $\zeta > 0$
notwendig. Ansonsten ist $\zeta$ aber im Prinzip beliebig. In fast allen
praktischen Anwendungen des Autors wurde bisher $\zeta = 1$ verwendet.

Die Modellfolge (5.4-1) ist ein Spezialfall der Modellfolge (3.2-1). Sie
enth\"alt ebenfalls $k + 1$ Unbekannte, die alle linear vorkommen --
n\"amlich den Grenzwert $s$ und die $k$ Koeffizienten $c_0$, $c_1$,
$\ldots$ , $c_{k - 1}$. Wenn also die numerischen Werte von $k + 1$
Folgenelementen $s_n$, $s_{n+1}$, $\ldots$ , $s_{n+k}$ und $k + 1$
Restsummenabsch\"atzungen $\omega_n$, $\omega_{n+1}$, $\ldots$ ,
$\omega_{n+k}$ bekannt sind, kann man unter Verwendung der Cramerschen
Regel den verallgemeinerten Summationsproze{\ss} ${\cal S}_k^{(n)} (\zeta,
s_n, \omega_n)$, der f\"ur Elemente der Modellfolge (5.4-1) exakt ist, als
Quotient zweier Determinanten definieren [Weniger 1989, Abschnitt 8.2]:

\smallskip

$$
{\cal S}_k^{(n)} (\zeta , s_n , \omega_n) \; = \; \frac
{
\vmatrix
{
s_n " \ldots " s_{n+k} \\ [1\jot]
\omega_n " \ldots " \omega_{n+k} \\
\vdots " \ddots " \vdots \\ [1\jot]
\omega_n / (\zeta + n)_{k-1} " \ldots "
\omega_{n+k} / (\zeta + n + k)_{k-1}
}
}
{
\vmatrix
{
1 " \ldots " 1       \\ [1\jot]
\omega_n " \ldots " \omega_{n+k} \\
\vdots " \ddots " \vdots \\ [1\jot]
\omega_n / (\zeta + n)_{k-1} " \ldots "
\omega_{n+k} / (\zeta + n + k)_{k-1}
}
} .
\tag
$$

\smallskip

Wie im Falle der Levinschen Transformation h\"atte man auch f\"ur diese
Transformation gerne eine explizite Darstellung. Dazu wird die
Modellfolge (5.4-1) auf folgende Weise umformuliert:
$$
(\zeta + n)_{k-1} [s_n - s] / \omega_n \; = \;
\sum_{j=0}^{k-1} \, c_j \> (\zeta + n + j)_{k-j-1} \, .
\tag
$$
Da ein Pochhammersymbol $(a)_m$ ein Polynom $m$-ten Grades in $a$ ist,
ist der Ausdruck auf der rechten Seite ein Polynom vom Grade $k - 1$ in
$n$. Bekanntlich wird ein solches Polynom durch $k$-fache Anwendung des
Differenzenoperators $\Delta$ annihiliert [Milne-Thomson 1981, S. 29].
Da der Differenzenoperator $\Delta^k$ linear ist, erhalten wir aus Gl.
(5.4-4) die folgende Darstellung [Weniger 1989, Abschnitt 8.2;
Brezinski und Redivo Zaglia 1991, Abschnitt 2.7]:
$$
{\cal S}_k^{(n)} (\zeta , s_n , \omega_n) \; = \; \frac
{\Delta^k \, \{(\zeta + n)_{k-1} \> s_n / \omega_n\} }
{\Delta^k \, \{(\zeta + n)_{k-1} / \omega_n\} } \, .
\tag
$$
Ein Vergleich mit Gl. (5.1-7) zeigt, da{\ss} der allgemeine
Annihilationsoperator ${\hat T}$ f\"ur den Korrekturterm $z_n$ in diesem
Falle der gewichtete Differenzenoperator $\Delta^k (n + \zeta)_{k-1}$
ist. Mit Hilfe von Gl. (3.1-14) k\"onnen die Differenzenoperatoren in Gl.
(5.4-5) in geschlossener Form ausgedr\"uckt werden. Man erh\"alt auf diese
Weise f\"ur die Transformation ${\cal S}_k^{(n)} (\zeta , s_n ,
\omega_n)$ die folgende Darstellung als Quotient zweier endlicher
Summen [Weniger 1989, Abschnitt 8.2]:
$$
{\cal S}_{k}^{(n)} (\zeta, s_n, \omega_n) \; = \;
\frac
{\displaystyle
\sum_{j=0}^{k} \; ( - 1)^{j} \; \binom {k} {j} \;
\frac {(\zeta + n +j )_{k-1}} {(\zeta + n + k )_{k-1}} \;
\frac {s_{n+j}} {\omega_{n+j}} }
{\displaystyle
\sum_{j=0}^{k} \; ( - 1)^{j} \; \binom {k} {j} \;
\frac {(\zeta + n +j )_{k-1}} {(\zeta + n + k )_{k-1}} \;
\frac {1} {\omega_{n+j}} }
\; , \qquad k, n \in \N_0 \, .
\tag
$$
Der gemeinsame Faktor $(\zeta + n + k )_{k-1}$ im Z\"ahler und Nenner von
Gl. (5.4-6) soll den Betrag der Terme in den Summen verringern, um auf
diese Weise OVERFLOW weniger wahrscheinlich zu machen.

Die Transformation (5.4-6) wurde schon von Sidi [1981] verwendet, um
explizite Ausdr\"ucke f\"ur Pad\'e-Approximationen f\"ur spezielle
hypergeometrische Reihen abzuleiten. Es scheint aber, da{\ss} Sidi die
Transformation ${\cal S}_{k}^{(n)} (\zeta , s_n, \omega_n)$ nicht als
verallgemeinerten Summationsproze{\ss} verwendet hat. Das ist -- wie sp\"ater
noch ausf\"uhrlich gezeigt wird -- sicherlich ein Vers\"aumnis, da diese
Transformation vor allem bei divergenten alternierenden Reihen
exzellente Ergebnisse liefert [Weniger und Steinborn 1989a; Weniger
1989; 1990; 1992; Weniger und {\v C}{\'\i}{\v z}ek 1990; Weniger, {\v
C}{\'\i}{\v z}ek und Vinette 1991; 1993; {\v C}{\'\i}{\v z}ek, Vinette
und Weniger 1991; 1993; Grotendorst 1991]. Allerdings wurde ${\cal
S}_{k}^{(n)} (\zeta , s_n, \omega_n)$ sp\"ater von Shelef [1987] im Rahmen
ihrer von Sidi betreuten Master Thesis f\"ur die numerische Inversion von
Laplacetransformationen verwendet.

Die beiden Summen in Gl. (5.4-6) k\"onnen auch mit Hilfe der folgenden
Dreitermrekursion berechnet werden [Weniger 1989, Abschnitt 8.3;
Brezinski und Redivo Zaglia 1991, Abschnitt 2.7]:
$$
S_{k+1}^{(n)} (\zeta) \; = \; S_k^{(n+1)} (\zeta)
\, - \, \frac
{ (\zeta + n + k ) (\zeta + n + k - 1) }
{ (\zeta + n + 2 k ) (\zeta + n + 2 k - 1) } \>
S_k^{(n)} (\zeta) \, , \qquad k, n \ge 0 \, .
\tag
$$
Wenn man die Startwerte
$$
S_0^{(n)} (\zeta) \; = \; s_n / \omega_n \, ,
\qquad n \in \N_0 \, ,
\tag
$$
verwendet, ergibt die Dreitermrekursion (5.4-7) den Z\"ahler von Gl.
(5.4-6), w\"ahrend die Startwerte
$$
S_0^{(n)} (\zeta) \; = \; 1 / \omega_n \, ,
\qquad n \in \N_0 \, ,
\tag
$$
den Nenner von Gl. (5.4-6) ergeben.

Man mu{\ss} noch diskutieren, wie man die Restsummenabsch\"atzungen $\Seqn
\omega$ in Falle der Transformation ${\cal S}_{k}^{(n)} (\zeta, s_n,
\omega_n)$ w\"ahlen soll. In Unterabschnitt 5.1 wurde argumentiert, da{\ss}
die Restsummenabsch\"atzungen so gew\"ahlt werden sollen, da{\ss} $\omega_n$
proportional zum dominanten Term einer asymptotischen Entwicklung des
Summationsrestes $r_n$ f\"ur $n \to \infty$ ist,
$$
r_n \; = \; s_n - s \; = \; \omega_n [ c + O(n^{-1}) ] \, ,
\qquad n \to \infty \, .
\tag
$$
Da der dominante Term sich nicht \"andert, wenn man eine asymptotische
Potenzreihe f\"ur den Quotienten $[s_n - s] / \omega_n$ in eine
Fakult\"atenreihe transformiert, sollte man im Falle der Transformation
${\cal S}_{k}^{(n)} (\zeta, s_n, \omega_n)$ die gleichen
Restsummenabsch\"atzungen verwenden k\"onnen wie im Falle der Levinschen
Transformation.

Wenn wir also die Restsummenabsch\"atzung (5.2-12) in Gl. (5.4-6)
verwenden, erhalten wir ein Analogon der in Gl. (5.2-13) definierten
Levinschen $u$-Transformation [Weniger 1989, Gl. (8.4-2)]:
$$
y_{k}^{(n)} (\zeta, s_n) \; = \;
\frac
{\displaystyle
\sum_{j=0}^{k} \; ( - 1)^{j} \; \binom {k} {j} \;
\frac {(\zeta + n +j + 1 )_{k-2}} {(\zeta + n + k )_{k-1}} \;
\frac {s_{n+j}} {a_{n+j}} }
{\displaystyle
\sum_{j=0}^{k} \; ( - 1)^{j} \; \binom {k} {j} \;
\frac {(\zeta + n +j + 1 )_{k-2}} {(\zeta + n + k )_{k-1}} \;
\frac {1} {a_{n+j}} }
\; .
\tag
$$
Ebenso ergibt die Verwendung der Restsummenabsch\"atzung (5.2-14) in Gl.
(5.4-6) ein Analogon der Levinschen $t$-Transformation [Weniger 1989,
Gl. (8.4-3)],
$$
{\tau}_{k}^{(n)} (\zeta, s_n) \; = \;
\frac
{\displaystyle
\sum_{j=0}^{k} \; ( - 1)^{j} \; \binom {k} {j} \;
\frac {(\zeta + n +j )_{k-1}} {(\zeta + n + k )_{k-1}} \;
\frac {s_{n+j}} {a_{n+j}} }
{\displaystyle
\sum_{j=0}^{k} \; ( - 1)^{j} \; \binom {k} {j} \;
\frac {(\zeta + n +j )_{k-1}} {(\zeta + n + k )_{k-1}} \;
\frac {1} {a_{n+j}} }
\; ,
\tag
$$
und die Restsummenabsch\"atzung (5.2-16) ergibt ein Analogon der
urspr\"unglich von Smith und Ford [1979] eingef\"uhrten $d$-Transformation
[Weniger 1989, Gl. (8.4-4)]:
$$
{\delta}_{k}^{(n)} (\zeta, s_n) \; = \;
\frac
{\displaystyle
\sum_{j=0}^{k} \; ( - 1)^{j} \; \binom {k} {j} \;
\frac {(\zeta + n +j )_{k-1}} {(\zeta + n + k )_{k-1}} \;
\frac {s_{n+j}} {a_{n+j+1}} }
{\displaystyle
\sum_{j=0}^{k} \; ( - 1)^{j} \; \binom {k} {j} \;
\frac {(\zeta + n +j )_{k-1}} {(\zeta + n + k )_{k-1}} \;
\frac {1} {a_{n+j+1}} }
\; .
\tag
$$
Schlie{\ss}lich kann man in Gl. (5.4-6) auch die Restsummenabsch\"atzung
(5.2-19) verwenden. Man erh\"alt dann ein Analogon der Levinschen
$v$-Transformation [Weniger 1989, Gl. (8.4-5)]:
$$
{\phi}_{k}^{(n)} (\zeta, s_n) \; = \;
\frac
{\displaystyle
\sum_{j=0}^{k} \; ( - 1)^{j} \; \binom {k} {j} \;
\frac {(\zeta + n +j )_{k-1}} {(\zeta + n + k )_{k-1}} \;
\frac {a_{n+j} - a_{n+j+1}} {a_{n+j} a_{n+j+1}} s_{n+j} }
{\displaystyle
\sum_{j=0}^{k} \; ( - 1)^{j} \; \binom {k} {j} \;
\frac {(\zeta + n +j )_{k-1}} {(\zeta + n + k )_{k-1}} \;
\frac {a_{n+j} - a_{n+j+1}} {a_{n+j} a_{n+j+1}} }
\; .
\tag
$$

Weitere Spezialf\"alle der Transformation ${\cal S}_{k}^{(n)} (\zeta,
s_n, \omega_n)$, die auf alternativen Restsummenabsch\"atzungen basieren,
findet man in Abschnitt 8 von Weniger [1989].

Numerische Tests ergaben, da{\ss} ${\cal S}_{k}^{(n)} (\zeta, s_n,
\omega_n)$ und seine dazu im Prinzip geeigneten Varianten $y_{k}^{(n)}
(\zeta, s_n)$ und ${\phi}_{k}^{(n)} (\zeta, s_n)$ logarithmische
Konvergenz in den meisten F\"allen deutlich weniger gut beschleunigen als
die in Abschnitt 5.2 besprochene Varianten $u_{k}^{(n)} (\zeta, s_n)$
und $v_{k}^{(n)} (\zeta, s_n)$ der Levinschen Transformation [Weniger
1989, Abschnitt 14]. Was lineare Konvergenz betrifft, so sind analoge
Varianten von ${\cal S}_{k}^{(n)} (\zeta, s_n, \omega_n)$ und ${\cal
L}_{k}^{(n)} (\zeta, s_n, \omega_n)$ anscheinend in etwa gleich
leistungsf\"ahig [Weniger 1989; Weniger und Steinborn 1989a].

Die wirkliche St\"arke der neuen Transformation ${\cal S}_{k}^{(n)}
(\zeta, s_n, \omega_n)$ scheint aber die Summation hochgradig
divergenter alternierender Reihen zu sein. In allen bisher
durchgef\"uhrten numerischen Tests bei hochgradig divergenten Reihen
deklassierten sowohl ${\cal S}_{k}^{(n)} (\zeta, s_n, \omega_n)$ als
auch die Levinsche Transformation ${\cal L}_{k}^{(n)} (\zeta, s_n,
\omega_n)$ Pad\'e-Approximationen ganz klar. Au{\ss}erdem waren die Varianten
von ${\cal S}_{k}^{(n)} (\zeta, s_n, \omega_n)$ in allen bisher
bekannten Beispielen mindestens so effizient wie die analogen Varianten
der Levinschen Transformation ${\cal L}_{k}^{(n)} (\zeta, s_n,
\omega_n)$, und in manchen F\"allen war ${\cal S}_{k}^{(n)} (\zeta, s_n,
\omega_n)$ sogar deutlich leistungsf\"ahiger als die Levinsche
Transformation [Weniger und Steinborn 1989a; Weniger 1989; 1990; 1992;
Weniger und {\v C}{\'\i}{\v z}ek 1990; Weniger, {\v C}{\'\i}{\v z}ek und
Vinette 1991; 1993; {\v C}{\'\i}{\v z}ek, Vinette und Weniger 1991;
1993; Grotendorst 1991].

MAPLE-Programme f\"ur die oben definierten $y$-, $\tau$-, $\delta$- und
$\phi$-Varianten des neuen verallgemeinerten Summationsprozesses ${\cal
S}_{k}^{(n)} (\zeta, s_n, \omega_n)$ wurden von Grotendorst [1991]
ver\"offentlicht.

Ebenso wie bei der Levinschen Transformation, Gl. (5.2-6), sollten auch
die Transformatio\-nen ${\cal S}_k^{(n)} (\zeta, s_n, \omega_n)$ mit
maximalen unteren Indizes $k$ und minimalen oberen Indizes $n$ in
Konvergenz\-beschleunigungs- und Summationsverfahren die besten
Ergebnisse liefern. Als N\"aherung zum Grenzwert $s$ der zu
transformierenden Folge $\Seqn s$ verwenden wir deswegen [Weniger 1989,
Gl. (8.3-12)]:
$$
\{ s_0, \omega_0; s_1, \omega_1; \ldots ; s_m, \omega_m \} \; \to \;
{\cal S}_m^{(0)} (\zeta, s_0, \omega_0) \, ,
\qquad m \in \N_0 \, .
\tag
$$
Der Grenzwert $s$ der zu transformierenden Folge $\Seqn s$ wird also
durch die Transformationen
$$
{\cal S}_0^{(0)} (\zeta, s_0, \omega_0), \;
{\cal S}_1^{(0)} (\zeta, s_0, \omega_0), \; \ldots , \;
{\cal S}_m^{(0)} (\zeta, s_0, \omega_0), \; \ldots \; ,
\tag
$$
approximiert.

\medskip

\Abschnitt Alternative Darstellung der Korrekturterme durch
Pochhammersymbole

\smallskip

\aktTag = 0

Im letzten Unterabschnitt wurde gezeigt, da{\ss} ein neuer verallgemeinerter
Summationsproze{\ss} ${\cal S}_{k}^{(n)} (\zeta, s_n, \omega_n)$
konstruiert werden kann, der \"ahnliche Eigenschaften wie die Levinsche
Transformation ${\cal L}_{k}^{(n)} (\zeta, s_n, \omega_n)$ besitzt,
indem man annimmt, da{\ss} der Korrekturterm $z_n$ in Gl. (5.1-3) durch eine
abgebrochene Fakult\"atenreihe dargestellt werden kann. Im Prinzip
bedeutet dieser Ansatz, da{\ss} man nicht wie bei der Levinschen
Transformation von der Existenz einer asymptotischen Potenzreihe in der
Variablen $1/(n + \zeta)$ f\"ur den Quotienten $[s_n - s] / \omega_n$
gem\"a{\ss} Gl. (5.2-2) ausgeht. Statt dessen nimmt man an, da{\ss} der Quotient
$[s_n - s] / \omega_n$ f\"ur alle $n \in \N_0$ gem\"a{\ss} Gl. (5.4-2) in eine
Fakult\"atenreihe mit den Pochhammersymbolen $(n + \zeta)_j$ entwickelt
werden kann.

Allerdings sind die M\"oglichkeiten, alternative verallgemeinerte
Summationsprozesse mit \"ahn\-lichen Eigenschaften wie die Levinsche
Transformation zu konstruieren, damit noch l\"angst nicht ersch\"opft. In
den letzten Jahren sind im Zusammenhang mit quantenmechanischen
St\"orungsent\-wicklungen hoher Ordnung einige Artikel erschienen, in
denen {\it asymptotische Approximationen} der folgenden Art verwendet
wurden [Alvarez 1988, S. 4079, Gl. (4), S. 4081, Gl. (24); {\v C}{\'
\i}{\v z}ek, Damburg, Graffi, Grecchi, Harrell \Roemisch{2}, Harris,
Nakai, Paldus, Propin und Silverstone 1986, S. 13, Gl. (2), S. 15, Gln.
(28) und (29), S. 36, Gl. (229), S. 37, Tabelle \Roemisch{4}, S. 38-39,
Gl. (232), S. 43, Gl. (236), S. 45, Gl. (238); Silverstone 1990, S.
298, Gl. (15); Silverstone, Adams, {\v C}{\' \i}{\v z}ek und Otto 1979,
S. 1499, Gl. (5); Silverstone, Harris, {\v C}{\' \i}{\v z}ek und Paldus
1985, S. 1966, Gl. (1), S. 1977, Gl. (69), S. 1979, Gl. (71); Voros
1983, Appendix B]:
$$
\beginAligntags
" {\tilde f}_n (z) \; " \sim \;
" \frac {c_0} {z} \, + \, \frac {c_1} {z (z-1)}
\, + \, \frac {c_2} {z (z-1) (z-2)} \, + \, \cdots \, + \,
\frac {c_n} {z (z-1) (z-2) \ldots (z-n)} \\
" " = \; " \sum_{\nu=0}^n \> (-1)^{\nu+1} \;
\frac {c_{\nu}} {(-z)_{\nu+1} }
\, , \qquad z \to \infty \, .
\\ \tag
\endAligntags
$$
Auf den ersten Blick sieht eine solche Summe wie eine nach endlich
vielen Termen abgebrochene Fakult\"atenreihe aus, da das Argument $z$
ebenfalls in einem Pochhammersymbol vorkommt. Die Tatsache, da{\ss} in Gl.
(5.5-1) keine Pochhammersymbole des Typs $(z)_{\nu + 1}$ wie in Gl.
(5.3-3) sondern Pochhammersymbole des Typs
$$
\beginAligntags
" (-z)_{\nu+1} \;" = \; " (-z)(-z+1) \ldots (-z+\nu) \\
" "= \; " (-1)^{\nu+1} z (z-1) \ldots (z-\nu) \, ,
\qquad \nu \in \N_0 \, ,
\\ \tag
\endAligntags
$$
vorkommen, hat einige sehr weitreichende Konsequenzen. Wenn eine
Funktion $\Omega (z)$ durch eine konvergente oder asymptotische
Fakult\"atenreihe gem\"a{\ss} Gl. (5.3-3) dargestellt werden kann, dann ist der
Abbruchfehler einer Fakult\"atenreihe nach $n$ Termen f\"ur $z \to \infty$
immer von der Ordnung $O (z^{- n - 1})$. Analoges gilt auch, wenn
$\Omega (z)$ eine asymptotische Potenzreihe in $1/z$ gem\"a{\ss} Gl. (5.3-8)
besitzt. Der Abbruchfehler einer solchen asymptotischen Potenzreihe
nach $n$ Termen erf\"ullt automatisch die Ordnungsrelation
$$
\Omega (z) \, - \,
\sum_{\nu=0}^{n-1} \frac {c'_{\nu}} {z^{\nu + 1}}
\; = \; O (z^{- n - 1}) \, , \qquad z \to \infty \, .
\tag
$$
Wichtig ist, da{\ss} die Zahl $n$ der Terme in Gln. (5.3-5) und (5.5-3)
beliebig ist. Im Falle der asymptotischen Approximation (5.5-1) kann die
Zahl der Terme aber nicht beliebig gro{\ss} werden. Wenn beispielsweise $z$
eine positive reelle Zahl ist, ist die endliche Summe in Gl. (5.5-1) nur
dann sinnvoll als asymptotische Approximation zu verwenden, wenn $z > n$
gilt. Falls $n > z$ gilt, werden entweder die sp\"ateren Terme der Summe
irregul\"are Vorzeichenmuster aufweisen, oder -- wenn $z$ eine positive
ganze Zahl sein sollte -- werden sogar einige Pochhammersymbole Null
sein. Demzufolge kann die Summationsgrenze $n$ in Gl. (5.5-1) f\"ur einen
festen Wert des Argumentes $z$ nicht beliebig gro{\ss} werden. Das bedeutet
auch, da{\ss} man eine asymptotische Approximation vom Typ von Gl. (5.5-1)
nicht dadurch erhalten kann, indem man eine asymptotische Reihe nach
endlich vielen Termen abbricht. Ein solcher Ausdruck kann also
ausschlie{\ss}lich als eine asymptotische Approximation interpretiert
werden, die nur aus einer endlichen Anzahl von Termen bestehen kann. Ein
weiteres Problem von solchen asymptotischen Approximationen ist, da{\ss} man
ohne Zusatzinformationen keine Aussage \"uber das Verhalten des
Approximationsfehlers f\"ur $z \to \infty$ machen kann.

Asymptotische Approximationen von der Art von Gl. (5.5-1) sind --
soweit es dem Autor bekannt ist -- bisher noch nicht in der
mathematischen Literatur behandelt worden. Demzufolge ist das in diesem
Unterabschnitt behandelte Material in gewisser Weise eher
experimenteller Natur und mathematisch weniger gut fundiert, als es
sonst der Fall ist.

Man mu{\ss} allerdings betonen, da{\ss} diese Einw\"ande keineswegs ausschlie{\ss}en,
da{\ss} endliche Summen vom Typ von Gl. (5.5-1) ausgezeichnete
Approximationen liefern k\"onnen, wenn man $z$ und $n$ geeigneten
Einschr\"ankungen unterwirft. Wir werden sp\"ater sehen, da{\ss} man auf der
Basis solcher asymptotischer Approximationen verallgemeinerte
Summationsprozesse konstruieren kann, die in manchen F\"allen ganz
vorz\"ugliche Ergebnisse liefern.

Der Ausgangspunkt f\"ur die Konstruktion eines neuen verallgemeinerten
Summationsprozesses ist die Modellfolge
$$
s_n \; = \; s \, + \, \omega_n \,
\sum_{j=0}^{k-1} \, c_j / (- \xi - n)_j \, , \qquad k,n \in \N_0 \, ,
\tag
$$
die offensichtlich f\"ur den Quotienten $[s_n - s]/\omega_n$ asymptotische
Approximationen vom Typ von Gl. (5.5-1) liefert, und die formal fast
v\"ollig identisch ist mit der Modellfolge (5.4-1). Der einzige
Unterschied ist, da{\ss} die Pochhammersymbole $(n + \zeta)_j$ in Gl.
(5.4-1) durch Pochhammersymbole $(- \xi - n)_j$ ersetzt sind. Was die
Folge $\Seqn \omega$ der Restsummenabsch\"atzungen betrifft, so machen wir
wieder die \"ublichen Annahmen. Wir gehen also davon aus, da{\ss} die
Restsummenabsch\"atzungen $\Seqn \omega$ weitgehend beliebige Funktionen
des Index $n$ sind, die f\"ur $n \to \infty$ sowohl divergieren als auch
konvergieren k\"onnen. Au{\ss}erdem nehmen wir an, da{\ss} die $\omega_n$ f\"ur alle
endlichen Werte von $n$ ungleich Null sind und da{\ss} aus $m \ne n$ auch
$\omega_m \ne \omega_n$ folgt.

Der Parameter $\xi$ in Gl. (5.5-4) mu{\ss} so gew\"ahlt werden, da{\ss} die
Pochhammersymbole $(- \xi - n)_j$ f\"ur alle zul\"assigen Werte von $n$ und
$j$ immer ungleich Null sind. Au{\ss}erdem darf das Vorzeichenmuster in Gl.
(5.5-4) nicht zerst\"ort werden, wenn $n$ und $k$ gr\"o{\ss}er werden. Diese
zwei Forderungen sind sicherlich erf\"ullt, wenn $\xi$ eine positive
reelle Zahl ist, die au{\ss}erdem noch die Ungleichung $\xi \ge k - 1$
erf\"ullt.

Die Modellfolge (5.5-4), die ein Spezialfall der Modellfolge (3.2-1)
ist, enth\"alt $k+1$ Unbekannte, die alle linear vorkommen -- n\"amlich den
Grenzwert $s$ und die $k$ Koeffizienten $c_0$, $c_1$, $\ldots$ ,
$c_{k-1}$. Wenn also die numerischen Werte von $k+1$ Folgenelemente
$s_n$, $s_{n+1}$, $\ldots$ , $s_{n+k}$ und $k + 1$
Restsummenabsch\"atzungen $\omega_n$, $\omega_{n+1}$, $\ldots$ ,
$\omega_{n+k}$ bekannt sind, kann man mit Hilfe der Cramerschen Regel
den verallgemeinerten Summationsproze{\ss} ${\cal M}_k^{(n)} (\xi, s_n,
\omega_n)$, der exakt ist f\"ur die Elemente der Modellfolge (5.5-4), als
Quotient zweier Determinanten definieren [Weniger 1989, Abschnitt 9.2]:

\smallskip

$$
{\cal M}_k^{(n)} (\xi , s_n , \omega_n) \; = \; \frac
{
\vmatrix
{
s_n " \ldots " s_{n+k} \\ [1\jot]
\omega_n " \ldots " \omega_{n+k} \\
\vdots " \ddots " \vdots \\ [1\jot]
\omega_n / (- \xi - n)_{k-1} " \ldots "
\omega_{n+k} / (- \xi - n - k)_{k-1}
}
}
{
\vmatrix
{
1 " \ldots " 1       \\ [1\jot]
\omega_n " \ldots " \omega_{n+k} \\
\vdots " \ddots " \vdots \\ [1\jot]
\omega_n / (- \xi - n)_{k-1} " \ldots "
\omega_{n+k} / (- \xi - n - k)_{k-1}
}
} .
\tag
$$

\smallskip

Man h\"atte nat\"urlich auch hier gerne wieder eine explizite Darstellung.
Dazu wird die Modellfolge (5.5-4) auf folgende Weise umformuliert:
$$
(- \xi - n)_{k-1} [s_n - s] / \omega_n \; = \;
\sum_{j=0}^{k-1} \, c_j \> (- \xi - n + j)_{k-j-1} \, .
\tag
$$
Der Ausdruck auf der rechten Seite ist ein Polynom vom Grade $k - 1$ in
$n$. Bekanntlich wird ein solches Polynom durch $k$-fache Anwendung des
Differenzenoperators $\Delta$ annihiliert [Milne-Thomson 1981, S. 29].
Da der Differenzenoperator $\Delta^k$ linear ist, erhalten wir aus Gl.
(5.5-6) die folgende Darstellung [Weniger 1989, Abschnitt 9.2; Brezinski
und Redivo Zaglia 1991, Abschnitt 2.7]:
$$
{\cal M}_k^{(n)} (\xi, s_n, \omega_n) \; = \; \frac
{\Delta^k \, \{(- \xi - n)_{k-1} \> s_n / \omega_n\} }
{\Delta^k \, \{(- \xi - n)_{k-1} / \omega_n\} } \, .
\tag
$$
Ein Vergleich mit Gl. (5.1-7) zeigt, da{\ss} der allgemeine
Annihilationsoperator ${\hat T}$ f\"ur den Korrekturterm $z_n$ in diesem
Falle der gewichtete Differenzenoperator $\Delta^k (- \xi - n)_{k-1}$
ist. Mit Hilfe von Gl. (3.1-14) k\"onnen die Differenzenoperatoren in Gl.
(5.5-7) in geschlossener Form ausgedr\"uckt werden. Man erh\"alt auf diese
Weise f\"ur die Transformation ${\cal M}_k^{(n)} (\xi, s_n, \omega_n)$
die folgende Darstellung als Quotient zweier endlicher Summen [Weniger
1989, Abschnitt 9.2]:
$$
{\cal M}_{k}^{(n)} (\xi ,s_n, \omega_n) \; = \;
\frac
{\displaystyle
\sum_{j=0}^{k} \; ( - 1)^{j} \; \binom {k} {j} \;
\frac {(- \xi - n - j )_{k-1}} {(- \xi - n - k )_{k-1}} \;
\frac {s_{n+j}} {\omega_{n+j}} }
{\displaystyle
\sum_{j=0}^{k} \; ( - 1)^{j} \; \binom {k} {j} \;
\frac {(- \xi - n - j )_{k-1}} {(- \xi - n - k )_{k-1}} \;
\frac {1} {\omega_{n+j}} } \; , \qquad k,n \in \N_0 \, .
\tag
$$
Der gemeinsame Faktor $(- \xi - n - k )_{k-1}$ im Z\"ahler und Nenner von
Gl. (5.5-8) soll den Betrag der Terme in den Summen verringern, um
OVERFLOW weniger wahrscheinlich zu machen.

Die beiden Summen in Gl.~(5.5-8) k\"onnen auch mit Hilfe der folgenden
Dreitermrekursion berechnet werden [Weniger 1989, Abschnitt 9.3;
Brezinski und Redivo Zaglia 1991, Abschnitt 2.7]:
$$
M_{k+1}^{(n)} (\xi) \; = \; M_k^{(n+1)} (\xi)
\, - \, \frac
{ \xi + n - k + 1 } { \xi + n + k + 1 } \>
M_k^{(n)} (\xi) \, , \qquad k, n \ge 0 \, .
\tag
$$
Wenn man die Startwerte
$$
M_0^{(n)} (\xi) \; = \; s_n / \omega_n \, ,
\qquad n \in \N_0 \, ,
\tag
$$
verwendet, ergibt die Dreitermrekursion (5.5-9) den Z\"ahler von Gl.
(5.5-8), w\"ahrend die Startwerte
$$
M_0^{(n)} (\xi) \; = \; 1 / \omega_n \, ,
\qquad n \in \N_0 \, ,
\tag
$$
den Nenner von Gl. (5.5-8) ergeben.

Man mu{\ss} noch diskutieren, wie man die Restsummenabsch\"atzungen $\Seqn
\omega$ im Falle der Transformation ${\cal M}_{k}^{(n)} (\xi, s_n,
\omega_n)$ w\"ahlen soll. Die einfachste Vorgehensweise besteht wiederum
darin, im wesentlichen die gleichen einfachen Restsummenabsch\"atzungen
zu verwenden wie im Falle der Levinschen Transformation. Wenn wir also
in Gl. (5.5-8) die Restsummenabsch\"atzung
$$
\omega_n \; = \; (- \xi - n) a_n
\tag
$$
verwenden, erhalten wir ein Analogon der Levinschen $u$-Transformation
[Weniger 1989, Gl. (9.4-2)]:
$$
Y_{k}^{(n)} (\xi, s_n) \; = \;
\frac
{\displaystyle
\sum_{j=0}^{k} \; ( - 1)^{j} \; \binom {k} {j} \;
\frac
{(- \xi - n - j + 1)_{k-2}} {(- \xi - n - k )_{k-1}} \;
\frac {s_{n+j}} {a_{n+j}} }
{\displaystyle
\sum_{j=0}^{k} \; ( - 1)^{j} \; \binom {k} {j} \;
\frac
{(- \xi - n - j + 1)_{k-2}} {(- \xi - n - k )_{k-1}} \;
\frac {1} {a_{n+j}} }
\; .
\tag
$$
Ebenso ergibt die Verwendung der Restsummenabsch\"atzung (5.2-14) in Gl.
(5.5-8) ein Analogon der Levinschen $t$-Transformation [Weniger 1989,
Gl. (9.4-3)],
$$
T_{k}^{(n)} (\xi, s_n) \; = \;
\frac
{\displaystyle
\sum_{j=0}^{k} \; ( - 1)^{j} \; \binom {k} {j} \;
\frac
{(- \xi - n - j )_{k-1}} {(- \xi - n - k )_{k-1}} \;
\frac {s_{n+j}} {a_{n+j}} }
{\displaystyle
\sum_{j=0}^{k} \; ( - 1)^{j} \; \binom {k} {j} \;
\frac
{(- \xi - n - j )_{k-1}} {(- \xi - n - k )_{k-1}} \;
\frac {1} {a_{n+j}} }
\; .
\tag
$$
und die Restsummenabsch\"atzung (5.2-16) ergibt ein Analogon der
urspr\"unglich von Smith und Ford [1979] eingef\"uhrten $d$-Transformation
[Weniger 1989, Gl. (9.4-4)]:
$$
{\Delta}_{k}^{(n)} (\xi, s_n) \; = \;
\frac
{\displaystyle
\sum_{j=0}^{k} \; ( - 1)^{j} \; \binom {k} {j} \;
\frac
{(- \xi - n - j )_{k-1}} {(- \xi - n - k )_{k-1}} \;
\frac {s_{n+j}} {a_{n+j+1}} }
{\displaystyle
\sum_{j=0}^{k} \; ( - 1)^{j} \; \binom {k} {j} \;
\frac
{(- \xi - n - j )_{k-1}} {(- \xi - n - k )_{k-1}} \;
\frac {1} {a_{n+j+1}} }
\; .
\tag
$$
Schlie{\ss}lich kann man noch die Restsummenabsch\"atzung (5.2-19) in Gl.
(5.5-8) verwenden. Man erh\"alt dann ein Analogon der Levinschen
$v$-Transformation [Weniger 1989, Gl. (9.4-5)]:
$$
{\Phi}_{k}^{(n)} (\xi, s_n) \; = \;
\frac
{\displaystyle
\sum_{j=0}^{k} \; ( - 1)^{j} \; \binom {k} {j} \;
\frac
{(- \xi - n - j )_{k-1}} {(- \xi - n - k )_{k-1}} \;
\frac {a_{n+j} - a_{n+j+1}} {a_{n+j} a_{n+j+1}} s_{n+j} }
{\displaystyle
\sum_{j=0}^{k} \; ( - 1)^{j} \; \binom {k} {j} \;
\frac
{(- \xi - n - j )_{k-1}} {(- \xi - n - k )_{k-1}} \;
\frac {a_{n+j} - a_{n+j+1}} {a_{n+j} a_{n+j+1}} }
\; .
\tag
$$

Weitere Spezialf\"alle der Transformation ${\cal M}_{k}^{(n)} (\zeta,
s_n, \omega_n)$, die auf alternativen Restsummenabsch\"atzungen basieren,
findet man in Abschnitt 9 von Weniger [1989].

Numerische Tests ergaben, da{\ss} ${\cal S}_{k}^{(n)} (\zeta, s_n,
\omega_n)$ und ${\cal M}_{k}^{(n)} (\xi, s_n, \omega_n)$ \"ahnliche
Eigenschaften besitzen. So beschleunigen die Varianten $Y_{k}^{(n)}
(\xi, s_n)$ und ${\Phi}_{k}^{(n)} (\xi, s_n)$, die dazu im Prinzip
geeignet sein sollten, logarithmische Konvergenz in den meisten F\"allen
deutlich weniger gut als die in Abschnitt 5.2 besprochenen analogen
Varianten $u_{k}^{(n)} (\zeta, s_n)$ und $v_{k}^{(n)} (\zeta, s_n)$ der
Levinschen Transformation [Weniger 1989, Abschnitt 14]. Was lineare
Konvergenz betrifft, so sind analoge Varianten von ${\cal M}_{k}^{(n)}
(\xi, s_n, \omega_n)$, ${\cal S}_{k}^{(n)} (\zeta, s_n, \omega_n)$ und
${\cal L}_{k}^{(n)} (\zeta, s_n, \omega_n)$ in den meisten F\"allen in
etwa gleich leistungsf\"ahig [Weniger 1989; Weniger und Steinborn 1989a].
Auch im Falle hochgradig divergenter Reihen liefert ${\cal M}_{k}^{(n)}
(\xi, s_n, \omega_n)$ und seine Varianten zum Teil sehr gute Ergebnisse
[Weniger und Steinborn 1989a; Weniger 1989; Weniger und {\v C}{\'\i}{\v
z}ek 1990].

Trotzdem ist die Verwendung der Transformation
${\cal M}_{k}^{(n)} (\xi, s_n, \omega_n)$ und seiner Varianten in
Konvergenzbeschleunigungs- und Summationsprozessen weniger bequem als
die Verwendung der in Abschnitt 5.4 besprochenen Transformation
${\cal S}_{k}^{(n)} (\zeta, s_n, \omega_n)$. Der Grund ist, da{\ss} man den
Parameter $\xi$ in ${\cal M}_{k}^{(n)} (\xi, s_n, \omega_n)$ so w\"ahlen
mu{\ss}, da{\ss} die Ungleichung $\xi \ge k - 1$ erf\"ullt ist. Man mu{\ss} sich also
von Anfang an auf eine gr\"o{\ss}te zul\"assige Transformationsordnung $k_{max}$
festlegen, was sicherlich unpraktisch ist.

Ebenso wie bei den Transformationen (5.2-6) und (5.4-6) sollten auch die
Transformationen ${\cal M}_k^{(n)} (\xi, s_n, \omega_n)$ mit maximalen
unteren Indizes $k$ und minimalen oberen Indizes $n$ in
Konvergenzbeschleunigungs- und Summationsverfahren die besten Ergebnisse
liefern. Als N\"aher\-ung zum Grenzwert $s$ der zu transformierenden
Folge $\Seqn s$ verwenden wir deswegen [Weniger 1989, Gl. (9.3-12)]:
$$
\{ s_0, \omega_0; s_1, \omega_1; \ldots ; s_m, \omega_m \} \; \to \;
{\cal M}_m^{(0)} (\xi, s_0, \omega_0) \, ,
\qquad m \in \N_0 \, .
\tag
$$
Der Grenzwert $s$ der zu transformierenden Folge $\Seqn s$ wird also
durch die Transformationen
$$
{\cal M}_0^{(0)} (\xi, s_0, \omega_0), \;
{\cal M}_1^{(0)} (\xi, s_0, \omega_0), \; \ldots , \;
{\cal M}_m^{(0)} (\xi, s_0, \omega_0), \; \ldots \; ,
\tag
$$
approximiert.

\medskip

\Abschnitt Interpolierende Transformationen

\smallskip

\aktTag = 0

Wenn man in der allgemeinen Transformation ${\cal T}_k^{(n)} \bigl( w_k
(n); s_n, \omega_n \bigr)$, Gl. (5.1-9), den Effekt der
Differenzenoperatoren mit Hilfe von Gl. (3.1-14) in geschlossener Form
ausdr\"uckt, erh\"alt man den folgenden Ausdruck:
$$
{\cal T}_k^{(n)} \bigl( w_k (n); s_n, \omega_n \bigr)
\; = \; \frac
{\displaystyle
\sum_{j=0}^{k} \; ( - 1)^{j} \; \binom {k} {j} \;
w_k (n+j) \; \frac {s_{n+j}} {\omega_{n+j}} }
{\displaystyle
\sum_{j=0}^{k} \; ( - 1)^{j} \; \binom {k} {j} \;
w_k (n+j) \; \frac {1} {\omega_{n+j}} }
\; , \qquad k,n \in \N_0 \, .
\tag
$$
Wie man sieht, ist dieser verallgemeinerte Summationsproze{\ss} ein
gewichteter Mittelwert der $k + 1$ Quotienten $s_n/\omega_n$,
$s_{n+1}/\omega_{n+1}$, $\ldots$ , $s_{n+k}/\omega_{n+k}$. Sowohl die
Levinsche Transformation ${\cal L}_{k}^{(n)} (\zeta, s_n, \omega_n)$,
Gl. (5.2-6), als auch ${\cal S}_{k}^{(n)} (\zeta, s_n, \omega_n)$, Gl.
(5.4-6), und ${\cal M}_{k}^{(n)} (\xi, s_n, \omega_n)$, Gl. (5.5-8),
sind Spezialf\"alle der Transformation (5.6-1), da sie ebenfalls
gewichtete Mittelwerte der Quotienten $s_n/\omega_n$,
$s_{n+1}/\omega_{n+1}$, $\ldots$ , $s_{n+k}/\omega_{n+k}$ sind, die sich
nur bez\"uglich der Gewichtsfaktoren $w_k (n)$ unterscheiden. Im Falle der
Levinschen Transformation gilt $w_k (n) = (n + \zeta)^{k - 1}$, wogegen
$w_k (n) = (n + \zeta)_{k - 1}$ beziehungsweise $w_k (n) =
(- n - \xi)_{k - 1}$ bei den Transformationen ${\cal S}_{k}^{(n)}
(\zeta, s_n, \omega_n)$ und ${\cal M}_{k}^{(n)} (\xi, s_n, \omega_n)$
gilt. Wichtig ist in diesem Zusammenhang, da{\ss} Pochhammersymbole $(n + j
+ \zeta)_{k - 1}$ und $(- n - j - \xi)_{k - 1}$ f\"ur zunehmendes $n$ und
festes $k$ schneller wachsen als Potenzen $(n + j + \zeta)^{k - 1}$.
Demzufolge erhalten die Quotienten $s_{n+j}/\omega_{n+j}$ mit gr\"o{\ss}eren
Werten von $j$ in ${\cal S}_{k}^{(n)} (\zeta, s_n, \omega_n)$ und ${\cal
M}_{k}^{(n)} (\xi, s_n, \omega_n)$ ein gr\"o{\ss}eres Gewicht als in der
Levinschen Transformation ${\cal L}_{k}^{(n)} (\zeta, s_n, \omega_n)$.

Wie sp\"ater noch ausf\"uhrlicher diskutiert wird, unterscheiden sich die
numerischen Eigenschaften dieser drei Transformationen ${\cal
L}_{k}^{(n)} (\zeta, s_n, \omega_n)$, ${\cal S}_{k}^{(n)} (\zeta, s_n,
\omega_n)$ und ${\cal M}_{k}^{(n)} (\xi, s_n, \omega_n)$ zum Teil ganz
erheblich [Weniger 1989; 1990; 1992; Weniger und Steinborn 1989a;
Weniger und {\v C}{\'\i}{\v z}ek 1990; Weniger, {\v C}{\'\i}{\v z}ek und
Vinette 1991; 1993; Grotendorst 1991]. Da diese Transformationen sich
nur bez\"uglich der Gewichte $w_k (n + j)$ der Quotienten
$s_{n+j}/\omega_{n+j}$ in den Summen in Gln. (5.2-6), (5.4-6) und
(5.5-8) unterscheiden, sollte es sicherlich interessant sein, genauer zu
untersuchen, ob die unterschiedlichen Eigenschaften dieser
Transformationen in Konvergenzbeschleunigungs- und Summationsprozessen
auf diese Weise erkl\"art werden k\"onnen.

Man kann leicht einen verallgemeinerten Summationsproze{\ss} konstruieren,
der in Abh\"angigkeit von einem kontinuierlichen Parameter sowohl die
Levinsche Transformation ${\cal L}_{k}^{(n)} (\zeta, s_n, \omega_n)$ als
auch ${\cal S}_{k}^{(n)} (\zeta, s_n, \omega_n)$ reproduzieren kann.
Dazu mu{\ss} man nur die Beziehung
$$
\lim_{\alpha \to \infty} \, \frac {(\alpha n)_m} {\alpha^m}
\; = \; n^m
\tag
$$
ausn\"utzen. Unser Ausgangspunkt ist also die Modellfolge [Weniger 1992,
Gl. (3.1)]
$$
s_n \; = \; s \, + \, \omega_n \,
\sum_{j=0}^{k-1} \, c_j / (\alpha [n + \zeta])_j \, ,
\qquad k, n \in \N_0 \, , \quad \alpha, \zeta > 0 \, .
\tag
$$
Die Modellfolge (5.6-3), die eine Verallgemeinerung der Modellfolge
(5.4-1) ist, ist ebenfalls ein Spezialfall der Modellfolge (3.2-1). Sie
enth\"alt $k + 1$ Unbekannte, die alle linear vorkommen -- n\"amlich den
Grenzwert $s$ und die $k$ Koeffizienten $c_0$, $c_1$, $\ldots$ , $c_{k -
1}$. Wenn man die numerischen Werte von $k + 1$ Folgenelemente $s_n$,
$s_{n+1}$, $\ldots$ , $s_{n+k}$ und $k + 1$ Restsummenabsch\"atzungen
$\omega_n$, $\omega_{n+1}$, $\ldots$ , $\omega_{n+k}$ kennt, kann man
unter Verwendung der Cramerschen Regel den verallgemeinerten
Summationsproze{\ss} ${\cal C}_{k}^{(n)} (\alpha, \zeta, s_n, \omega_n)$,
der exakt ist f\"ur die Elemente der Modellfolge (5.6-3), als Quotient
zweier Determinanten definieren:

\smallskip

$$
{\cal C}_{k}^{(n)} (\alpha, \zeta, s_n, \omega_n) \; = \; \frac
{
\vmatrix
{
s_n " \ldots " s_{n+k} \\ [1\jot]
\omega_n " \ldots " \omega_{n+k} \\
\vdots " \ddots " \vdots \\ [1\jot]
\omega_n / (\alpha [\zeta + n])_{k-1} " \ldots "
\omega_{n+k} / (\alpha [\zeta + n + k])_{k-1}
}
}
{
\vmatrix
{
1 " \ldots " 1       \\ [1\jot]
\omega_n " \ldots " \omega_{n+k} \\
\vdots " \ddots " \vdots \\ [1\jot]
\omega_n / (\alpha [\zeta + n])_{k-1} " \ldots "
\omega_{n+k} / (\alpha [\zeta + n + k])_{k-1}
}
} .
\tag
$$

\smallskip

Eine Darstellung dieser Transformation als Quotient zweier endlicher
Summen kann leicht abgeleitet werden. Dazu schreiben wir die Modellfolge
(5.6-3) auf folgende Weise um:
$$
(\alpha [\zeta + n])_{k-1} [s_n - s] / \omega_n \; = \;
\sum_{j=0}^{k-1} \, c_j \> (\alpha [\zeta + n + j])_{k-j-1} \, .
\tag
$$
Die rechte Seite von Gl. (5.6-5) ist ein Polynom vom Grade $k - 1$ in
$n$, das durch Anwendung der $k$-ten Potenz des Differenzenoperators
$\Delta$ annihiliert wird [Milne-Thomson 1981, S. 29]. Da der
Differenzenoperator $\Delta^k$ linear ist, erh\"alt man aus Gl. (5.6-5)
die folgende Darstellung:
$$
{\cal C}_{k}^{(n)} (\alpha, \zeta, s_n, \omega_n) \; = \; \frac
{\Delta^k \, \{(\alpha [\zeta + n])_{k-1} \> s_n / \omega_n\} }
{\Delta^k \, \{(\alpha [\zeta + n])_{k-1} / \omega_n\} } \, .
\tag
$$
Ein Vergleich mit Gl. (5.1-7) zeigt, da{\ss} der allgemeine
Annihilationsoperator ${\hat T}$ f\"ur den Korrekturterm $z_n$ in diesem
Falle der gewichtete Differenzenoperator $\Delta^k (\alpha [n +
\zeta])_{k-1}$ ist. Die Differenzenoperatoren in Gl. (5.6-6) k\"onnen mit
Hilfe von Gl. (3.1-14) in geschlossener Form ausgedr\"uckt werden. Man
erh\"alt auf diese Weise f\"ur die Transformation ${\cal C}_{k}^{(n)}
(\alpha, \zeta, s_n, \omega_n)$ eine Darstellung als Quotient zweier
endlicher Summen [Weniger 1992, Gl. (3.2)]:
$$
{\cal C}_{k}^{(n)} (\alpha, \zeta, s_n, \omega_n) \; = \;
\frac
{\displaystyle
\sum_{j=0}^{k} \; ( - 1)^{j} \; \binom {k} {j} \;
\frac {(\alpha [\zeta + n +j])_{k-1}}
{(\alpha [\zeta + n + k])_{k-1}} \;
\frac {s_{n+j}} {\omega_{n+j}} }
{\displaystyle
\sum_{j=0}^{k} \; ( - 1)^{j} \; \binom {k} {j} \;
\frac {(\alpha [\zeta + n +j])_{k-1}}
{(\alpha [\zeta + n + k])_{k-1}} \;
\frac {1} {\omega_{n+j}} } \; ,
\qquad k, n \in \N_0 \, .
\tag
$$
F\"ur $\alpha = 1$ ist ${\cal C}_{k}^{(n)} (\alpha, \zeta, s_n,
\omega_n)$ offensichtlich identisch mit ${\cal S}_{k}^{(n)} (\zeta,
s_n, \omega_n)$, und f\"ur $\alpha \to \infty$ geht ${\cal C}_{k}^{(n)}
(\alpha, \zeta, s_n, \omega_n)$ aufgrund von Gl. (5.6-2) in die
Levinsche Transformation ${\cal L}_{k}^{(n)} (\zeta, s_n, \omega_n)$
\"uber.

Die beiden Summen in Gl. (5.6-7) k\"onnen auch mit Hilfe der folgenden
Dreitermrekursion berechnet werden [Weniger 1992, Gl. (3.3)]:
$$
\beginAligntags
C_{k+1}^{(n)} (\alpha, \zeta) \; = \; " C_k^{(n+1)} (\alpha, \zeta)
\, - \, (\alpha [\zeta + n] + k - 2) \,
\frac
{(\alpha [n + \zeta + k - 1])_{k-2}}
{(\alpha [n + \zeta + k])_{k-1}}
\>
C_k^{(n)} (\alpha, \zeta) \, , \\
" k, n \ge 0 \, , \qquad \alpha, \zeta > 0 \, .
\\ \tag
\endAligntags
$$
Wenn man die Startwerte
$$
C_0^{(n)} (\alpha, \zeta) \; = \; s_n / \omega_n \, ,
\qquad n \in \N_0 \, ,
\tag
$$
verwendet, ergibt die Dreitermrekursion (5.6-8) den Z\"ahler von Gl.
(5.6-7), und die Startwerte
$$
C_0^{(n)} (\alpha, \zeta) \; = \; 1 / \omega_n \, ,
\qquad n \in \N_0 \, ,
\tag
$$
ergeben den Nenner von Gl. (5.6-7).

Die Approximationen zum Grenzwert $s$ der zu transformierenden Folge
werden im Prinzip auf gleiche Weise gew\"ahlt wie im Fall der
Transformation ${\cal S}_{k}^{(n)} (\zeta, s_n, \omega_n)$, Gl. (9.4-6):
$$
\{ s_0, \omega_0; s_1, \omega_1; \ldots ; s_m, \omega_m \} \; \to \;
{\cal C}_m^{(0)} (\alpha, \zeta, s_0, \omega_0) \, ,
\qquad m \in \N_0 \, .
\tag
$$
Der Grenzwert $s$ der zu transformierenden Folge $\Seqn s$ wird also
durch die Transformationen
$$
{\cal C}_0^{(0)} (\alpha, \zeta, s_0, \omega_0), \;
{\cal C}_1^{(0)} (\alpha, \zeta, s_0, \omega_0), \; \ldots , \;
{\cal C}_m^{(0)} (\alpha, \zeta, s_0, \omega_0), \; \ldots \; ,
\tag
$$
approximiert.

\medskip

\Abschnitt Die Transformation von Potenzreihen

\smallskip

\aktTag = 0

Potenzreihen spielen sowohl in der Analysis als auch in
naturwissenschaftlichen und technischen Anwendungen eine besondere
Rolle. Aus diesem Grund werden in diesem Unterabschnitt die
Eigenschaften derjenigen rationalen Approximationen diskutiert, die man
erh\"alt, wenn man die ver\-allgemeinerten Summationsprozessen ${\cal
L}_{k}^{(n)} (\zeta, s_n, \omega_n)$, Gl. (5.2-6), ${\cal S}_{k}^{(n)}
(\zeta, s_n, \omega_n)$, Gl. (5.4-6), und ${\cal M}_{k}^{(n)} (\xi, s_n,
\omega_n)$, Gl. (5.5-8), zur Konvergenzverbesserung oder zur Summation
von Potenz\-reihen verwendet. Nehmen wir also an, da{\ss} man einer Funktion
$f (z)$ eine formale Potenzreihe

$$
f (z) \; = \;
\sum_{\nu=0}^{\infty} \> \gamma_{\nu} \, z^{\nu} \,
\tag
$$
zuordnen kann, die entweder konvergiert oder divergiert. Au{\ss}erdem
nehmen wir an, da{\ss} alle Reihenkoeffizienten $\gamma_n$ von Null
verschieden sind. Wir wollen die Partialsummen
$$
f_n (z) \; = \;
\sum_{\nu = 0}^{n} \, \gamma_{\nu} \, z^{\nu}
\tag
$$
der Potenzreihe (5.7-1) f\"ur $f (z)$ als Eingabedaten in den
verallgemeinerten Summationsprozessen ${\cal L}_{k}^{(n)} (\zeta, s_n,
\omega_n)$, ${\cal S}_{k}^{(n)} (\zeta, s_n, \omega_n)$ und ${\cal
M}_{k}^{(n)} (\xi, s_n, \omega_n)$ verwenden. Die einfachste
Restsummenabsch\"atzung f\"ur die Partialsumme $f_n (z)$ der Potenzreihe
(5.7-1) ist der erste Term $\gamma_{n + 1} z^{n + 1}$ der Reihe, der
nicht in $f_n (z)$ enthalten ist. Diese Restsummenabsch\"atzung entspricht
der von Smith und Ford [1979] eingef\"uhrten Restsummenabsch\"atzung
(5.2-16), welche die Transformationen $d_{k}^{(n)} (\zeta, s_n)$, Gl.
(5.2-18), $\delta_{k}^{(n)} (\zeta, s_n)$, Gl. (5.4-13), und
$\Delta_{k}^{(n)} (\xi, s_n)$, Gl. (5.5-15), ergibt.

Wenn man in Gln. (5.2-6), (5.4-6) und (5.5-8) als Eingabedaten die
Partialsummen (5.7-2) und als Restsummenabsch\"atzung $\omega_n =
\gamma_{n + 1} z^{n + 1}$ verwendet, erh\"alt man die folgenden
rationalen Funktionen:
$$
\beginAligntags
" d_k^{(n)} \bigl(\zeta, f_n (z) \bigr) \; " = \; "
\frac
{\displaystyle
\sum_{j=0}^{k} \; ( - 1)^{j} \; \binom {k} {j} \;
\frac
{(\zeta + n +j )^{k-1}} {(\zeta + n + k )^{k-1}} \;
\frac {z^{k - j} f_{n+j} (z)} {\gamma_{n+j+1}} }
{\displaystyle
\sum_{j=0}^{k} \; ( - 1)^{j} \; \binom {k} {j} \;
\frac
{(\zeta + n +j )^{k-1}} {(\zeta + n + k )^{k-1}} \;
\frac {z^{k - j}} {\gamma_{n+j+1}} }
\; ,
\\ \tag
" {\delta}_k^{(n)} \bigl(\zeta, f_n (z) \bigr) \; " = \; "
\frac
{\displaystyle
\sum_{j=0}^{k} \; ( - 1)^{j} \; \binom {k} {j} \;
\frac
{(\zeta + n +j )_{k-1}} {(\zeta + n + k )_{k-1}} \;
\frac {z^{k - j} f_{n+j} (z)} {\gamma_{n+j+1}} }
{\displaystyle
\sum_{j=0}^{k} \; ( - 1)^{j} \; \binom {k} {j} \;
\frac
{(\zeta + n +j )_{k-1}} {(\zeta + n + k )_{k-1}} \;
\frac {z^{k - j}} {\gamma_{n+j+1}} }
\; ,
\\ \tag
" {\Delta}_k^{(n)} \bigl(\xi, f_n (z) \bigr) \; " = \; "
\frac
{\displaystyle
\sum_{j=0}^{k} \; ( - 1)^{j} \; \binom {k} {j} \;
\frac
{( - \xi - n - j )_{k-1}} {( - \xi - n - k )_{k-1}} \;
\frac {z^{k - j} f_{n+j} (z)} {\gamma_{n+j+1}} }
{\displaystyle
\sum_{j=0}^{k} \; ( - 1)^{j} \; \binom {k} {j} \;
\frac
{( - \xi - n - j )_{k-1}} {( - \xi - n - k )_{k-1}} \;
\frac {z^{k - j}} {\gamma_{n+j+1}} }
\; .
\\ \tag
\endAligntags
$$
Die Z\"ahler dieser Ausdr\"ucke sind offensichtlich Polynome vom Grade $k +
n$ in $z$, und die Nenner sind Polynome vom Grade $k$ in $z$. In dieser
Hinsicht sind $d_k^{(n)} \bigl(\zeta, f_n (z) \bigr)$,
${\delta}_k^{(n)} \bigl(\zeta, f_n (z) \bigr)$ und ${\Delta}_k^{(n)}
\bigl(\xi, f_n (z) \bigr)$ vergleichbar mit den Pad\'e-Approximationen
$\epsilon_{2 k}^{(n)} = [ k + n / k ]$, die ebenfalls Quotienten von
Polynomen der Grade $k + n$ beziehungsweise $k$ in $z$ sind. Allerdings
ben\"otigt man zur Berechnung der rationalen Funktionen $d_k^{(n)}
\bigl(\zeta, f_n (z) \bigr)$, ${\delta}_k^{(n)} \bigl(\zeta, f_n (z)
\bigr)$ und ${\Delta}_k^{(n)} \bigl(\xi, f_n (z) \bigr)$ nur die
numerischen Werte der $k + 2$ Partialsummen $f_n (z)$, $f_{n+1} (z)$,
$\ldots$ , $f_{n+k+1} (z)$, wogegen man zur Berechnung der
Pad\'e-Approximation $[ k + n / k ]$ die numerischen Werte der $2 k + 1$
Partialsummen $f_n (z)$, $f_{n+1} (z)$, $\ldots$ , $f_{n+2 k} (z)$
ben\"otigt.

Die Transformationen $d_k^{(n)} \bigl(\zeta, f_n (z) \bigr)$,
${\delta}_k^{(n)} \bigl(\zeta, f_n (z) \bigr)$ und ${\Delta}_k^{(n)}
\bigl(\xi, f_n (z) \bigr)$ sind also in gewisser Weise effizienter als
Pad\'e-Approximationen, da sie bei einer gleichen Zahl von Eingabedaten
Quotienten von Polynomen mit gr\"o{\ss}eren Polynomgraden erzeugen. Man mu{\ss}
aber ber\"ucksichtigen, da{\ss} bei einer Pad\'e-Approximation $[ k + n / k ]$
von den $2 k + n + 2$ Polynomkoeffizienten $2 k + n + 1$ unabh\"angig
sind. Dagegen sind von den $2 k + n + 2$ Polynomkoeffizienten in den
Z\"ahler- und Nennersummen in Gln. (5.7-3) - (5.7-5) nur $k + 2$
unabh\"angig.

Man kann beim heutigen Stand des Wissens keine allgemeinen Aussage
dar\"uber machen, ob dieses hohe Ma{\ss} an Abh\"angigkeit der
Polynomkoeffizienten die numerischen Eigenschaften der Transformationen
$d_k^{(n)} \bigl(\zeta, f_n (z) \bigr)$, ${\delta}_k^{(n)} \bigl(\zeta,
f_n (z) \bigr)$ und ${\Delta}_k^{(n)} \bigl(\xi, f_n (z) \bigr)$ in
Konvergenzbeschleunigungs- und Summationsverfahren positiv oder negativ
beeinflu{\ss}t. Man kann aber relativ leicht die folgenden asymptotischen
Absch\"atzungen f\"ur $z \to 0$ beweisen:
$$
\beginAligntags
" f (z) \, - \, d_k^{(n)} \bigl(\zeta, f_n (z) \bigr)
\; " = \; " O (z^{k + n + 2}) \, ,
\\ \tag
" f (z) \, - \, {\delta}_k^{(n)} \bigl(\zeta, f_n (z) \bigr)
\; " = \; " O (z^{k + n + 2}) \, ,
\\ \tag
" f (z) \, - \, {\Delta}_k^{(n)} \bigl(\xi, f_n (z) \bigr)
\; " = \; " O (z^{k + n + 2}) \, .
\\ \tag
\endAligntags
$$
Diese Ordnungsabsch\"atzungen sind formal fast identisch mit der
Ordnungsabsch\"atzung (4.2-3), die man -- wie in Abschnitt 4.2 skizziert
wurde -- zur Konstruktion von Pad\'e-Approximationen verwenden kann.

Zum Beweis der asymptotischen Absch\"atzungen (5.7-6) - (5.7-8) f\"uhrt man
den Abbruchfehler $r_n (z)$ der Potenzreihe (5.7-1) f\"ur $f (z)$ gem\"a{\ss}
$$
r_n (z) \; = \; f_n (z) \, - \, f (z) \; = \; -
\sum_{\nu = 0}^{\infty} \, \gamma_{n + \nu + 1} \, z^{n + \nu + 1}
\tag
$$
ein. Wenn man jetzt noch ausn\"utzt, da{\ss} die Transformationen $d_k^{(n)}
\bigl(\zeta, f_n (z) \bigr)$, ${\delta}_k^{(n)} \bigl(\zeta, f_n (z)
\bigr)$ und ${\Delta}_k^{(n)} \bigl(\xi, f_n (z) \bigr)$ bez\"uglich der
$k + 1$ Partialsummen $f_n (z)$, $f_{n+1} (z)$, $\ldots$ , $f_{n+k}
(z)$ translationsinvariant sind, erh\"alt man:
$$
\beginAligntags
" f (z) \, - \, d_k^{(n)} \bigl(\zeta, f_n (z) \bigr) \\
" \qquad \; = \; - \; z^{k + n + 1} \;
\frac
{ \displaystyle
\sum_{j=0}^{k} \; ( - 1)^{j} \; \binom {k} {j} \;
\frac
{(\zeta + n +j )^{k-1}} {(\zeta + n + k )^{k-1}} \;
\frac {r_{n+j} (z)} {\gamma_{n+j+1} z^{n+j+1}} }
{\displaystyle
\sum_{j=0}^{k} \; ( - 1)^{j} \; \binom {k} {j} \;
\frac
{(\zeta + n +j )^{k-1}} {(\zeta + n + k )^{k-1}} \;
\frac {z^{k - j}} {\gamma_{n+j+1}} }
\; ,
\\ \tag
" f (z) \, - \, \delta_k^{(n)} \bigl(\zeta, f_n (z) \bigr) \\
" \qquad \; = \; - \; z^{k + n + 1} \;
\frac
{ \displaystyle
\sum_{j=0}^{k} \; ( - 1)^{j} \; \binom {k} {j} \;
\frac
{(\zeta + n +j )_{k-1}} {(\zeta + n + k )_{k-1}} \;
\frac {r_{n+j} (z)} {\gamma_{n+j+1} z^{n+j+1}} }
{\displaystyle
\sum_{j=0}^{k} \; ( - 1)^{j} \; \binom {k} {j} \;
\frac
{(\zeta + n +j )_{k-1}} {(\zeta + n + k )_{k-1}} \;
\frac {z^{k - j}} {\gamma_{n+j+1}} }
\; ,
\\ \tag
" f (z) \, - \, \Delta_k^{(n)} \bigl(\xi, f_n (z) \bigr) \\
" \qquad \; = \; - \; z^{k + n + 1} \;
\frac
{ \displaystyle
\sum_{j=0}^{k} \; ( - 1)^{j} \; \binom {k} {j} \;
\frac
{( - \xi - n - j )_{k-1}} {( - \xi - n - k )_{k-1}} \;
\frac {r_{n+j} (z)} {\gamma_{n+j+1} z^{n+j+1}} }
{\displaystyle
\sum_{j=0}^{k} \; ( - 1)^{j} \; \binom {k} {j} \;
\frac
{( - \xi - n - j )_{k-1}} {( - \xi - n - k )_{k-1}} \;
\frac {z^{k - j}} {\gamma_{n+j+1}} }
\; .
\\ \tag
\endAligntags
$$

Da die Nenner in Gln. (5.7-10) - (5.7-12) Polynome von Grade $k$ in $z$
sind, deren konstante Terme laut Voraussetzung ungleich Null sind, mu{\ss}
man nur noch zeigen, da{\ss} die Z\"ahlersummen mindestens von der Ordnung $O
(z)$ f\"ur $z \to 0$ sind, um die Absch\"atzungen (5.7-6) - (5.7-8) zu
beweisen. Aus Gl. (5.7-9) folgt:
$$
\beginAligntags
" \frac {r_{n+j} (z)} {\gamma_{n+j+1} z^{n+j+1}} \; " = \;
" - 1 - \rho_{n+j} (z) \, ,
\erhoehe\aktTag \\ \tag*{\tagnr a} \openup 1 \jot
" \rho_{n+j} (z) \; " = \; " \sum_{\nu=0}^{\infty} \,
\frac {\gamma_{n+j+\nu+2}} {\gamma_{n+j+1}} z^{\nu+1} \, .
\\ \tag*{\tagform\aktTagnr b}
\endAligntags
$$
Aus Gl. (5.7-13b) folgt $\rho_{n + j} (z) = O (z)$ f\"ur $z \to 0$. Im
n\"achsten Schritt eliminieren wir aus Gln. (5.7-10) - (5.7-12) die
gemeinsamen Faktoren $(\zeta + n + k )^{k-1}$, $(\zeta + n + k )_{k-1}$
und $( - \xi - n - k )_{k-1}$ und schreiben die resultierenden
Z\"ahlersummen der rationalen Funktionen mit Hilfe von Gln. (3.1-14) und
(5.7-13) auf folgende Weise um:
$$
\beginAligntags
" \sum_{j=0}^{k} \; ( - 1)^{j} \; \binom {k} {j} \,
(\zeta + n +j )^{k-1} \,
\frac {r_{n+j} (z)} {\gamma_{n+j+1} z^{n+j+1}} \\
" \qquad \; = \; (- 1)^{k+1} \,
\bigl[ \Delta^k \, (\zeta + n)^{k-1} \bigr]
\; + \; (- 1)^{k+1} \,
\bigl[ \Delta^k \, (\zeta + n)^{k-1} \, \rho_n (z) \bigr] \, ,
\\ \tag
" \sum_{j=0}^{k} \; ( - 1)^{j} \; \binom {k} {j} \,
(\zeta + n +j )_{k-1} \,
\frac {r_{n+j} (z)} {\gamma_{n+j+1} z^{n+j+1}} \\
" \qquad \; = \; (- 1)^{k+1} \,
\bigl[ \Delta^k \, (\zeta + n)_{k-1} \bigr]
\; + \; (- 1)^{k+1} \,
\bigl[ \Delta^k \, (\zeta + n)_{k-1} \, \rho_n (z) \bigr] \, ,
\\ \tag
" \sum_{j=0}^{k} \; ( - 1)^{j} \; \binom {k} {j} \,
(- \xi - n - j)_{k-1} \,
\frac {r_{n+j} (z)} {\gamma_{n+j+1} z^{n+j+1}} \\
" \qquad \; = \; (- 1)^{k+1} \,
\bigl[ \Delta^k \, (- \xi - n)_{k-1} \bigr]
\; + \; (- 1)^{k+1} \,
\bigl[ \Delta^k \, (- \xi - n)_{k-1} \, \rho_n (z) \bigr] \, .
\\ \tag
\endAligntags
$$

Da die $k$-te Potenz des Differenzenoperators $\Delta$ Polynome vom
Grade $k - 1$ in $n$ annihiliert [Milne-Thomson 1981, S. 29], bleiben
auf den rechten Seiten dieser Beziehungen nur die $k$-ten gewichteten
Differenzen von $\rho_n (z)$ \"ubrig. Da $\rho_n (z) = O (z)$ f\"ur $z \to
0$ gilt, sind auch die gewichteten $k$-ten Differenzen von $\rho_n (z)$
mindestens von der Ordnung $O (z)$ f\"ur $z \to 0$, was die Absch\"atzungen
(5.7-6) - (5.7-8) beweist.

Die Absch\"atzungen (5.7-6) - (5.7-8) implizieren, da{\ss}
Taylorentwicklungen der rationalen Funktionen $d_k^{(n)} \bigl(\zeta,
f_n (z) \bigr)$, ${\delta}_k^{(n)} \bigl(\zeta, f_n (z) \bigr)$ und
${\Delta}_k^{(n)} \bigl(\xi, f_n (z) \bigr)$ {\it alle} Koeffizienten
$\gamma_0$, $\gamma_1$, $\ldots$ , $\gamma_{n + k + 1}$ der unendlichen
Reihe (5.7-1) exakt reproduzieren, die zur Konstruktion der rationalen
Funktionen verwendet wurden.

Die Restsummenabsch\"atzung $\omega_n = \gamma_n z^n$ ist ebenfalls zur
Verbesserung der Konvergenzeigenschaften einer Potenzreihe oder zur
Summation einer divergenten Potenzreihe geeignet. Sie entspricht der von
Levin [1973] eingef\"uhrten Restsummenabsch\"atzung (5.2-14), welche die
Transformationen $t_{k}^{(n)} (\zeta, s_n)$, Gl. (5.2-15),
$\tau_{k}^{(n)} (\zeta, s_n)$, Gl. (5.4-12), und $T_{k}^{(n)} (\xi,
s_n)$, Gl. (5.5-14), ergibt.

Wenn man in Gln. (5.2-6), (5.4-6) und (5.5-8) als Eingabedaten die
Partialsummen (5.7-2) und die Restsummenabsch\"atzung $\omega_n =
\gamma_n z^n$ verwendet, erh\"alt man die folgenden rationalen Funktionen:
$$
\beginAligntags
" t_k^{(n)} \bigl(\zeta, f_n (z) \bigr) \; " = \; "
\frac
{\displaystyle
\sum_{j=0}^{k} \; ( - 1)^{j} \; \binom {k} {j} \;
\frac
{(\zeta + n +j )^{k-1}} {(\zeta + n + k )^{k-1}} \;
\frac {z^{k - j} f_{n+j} (z)} {\gamma_{n+j}} }
{\displaystyle
\sum_{j=0}^{k} \; ( - 1)^{j} \; \binom {k} {j} \;
\frac
{(\zeta + n +j )^{k-1}} {(\zeta + n + k )^{k-1}} \;
\frac {z^{k - j}} {\gamma_{n+j}} }
\; ,
\\ \tag
" {\tau}_k^{(n)} \bigl(\zeta, f_n (z) \bigr) \; " = \; "
\frac
{\displaystyle
\sum_{j=0}^{k} \; ( - 1)^{j} \; \binom {k} {j} \;
\frac
{(\zeta + n +j )_{k-1}} {(\zeta + n + k )_{k-1}} \;
\frac {z^{k - j} f_{n+j} (z)} {\gamma_{n+j}} }
{\displaystyle
\sum_{j=0}^{k} \; ( - 1)^{j} \; \binom {k} {j} \;
\frac
{(\zeta + n +j )_{k-1}} {(\zeta + n + k )_{k-1}} \;
\frac {z^{k - j}} {\gamma_{n+j}} }
\; ,
\\ \tag
" T_k^{(n)} \bigl(\xi, f_n (z) \bigr) \; " = \; "
\frac
{\displaystyle
\sum_{j=0}^{k} \; ( - 1)^{j} \; \binom {k} {j} \;
\frac
{( - \xi - n - j )_{k-1}} {( - \xi - n - k )_{k-1}} \;
\frac {z^{k - j} f_{n+j} (z)} {\gamma_{n+j}} }
{\displaystyle
\sum_{j=0}^{k} \; ( - 1)^{j} \; \binom {k} {j} \;
\frac
{( - \xi - n - j )_{k-1}} {( - \xi - n - k )_{k-1}} \;
\frac {z^{k - j}} {\gamma_{n+j}} }
\; .
\\ \tag
\endAligntags
$$
Die Z\"ahler dieser Ausdr\"ucke sind Polynome vom Grade $k + n$ in $z$, und
die Nenner sind Polynome vom Grade $k$ in $z$. In dieser Hinsicht sind
sowohl $t_k^{(n)} \bigl(\zeta, f_n (z) \bigr)$, ${\tau}_k^{(n)}
\bigl(\zeta, f_n (z) \bigr)$ und $T_k^{(n)} \bigl(\xi, f_n (z) \bigr)$,
die auf der Restsummenabsch\"atzung $\omega_n = \gamma_n z^n$ basieren,
als auch $d_k^{(n)} \bigl(\zeta, f_n (z) \bigr)$, ${\delta}_k^{(n)}
\bigl(\zeta, f_n (z) \bigr)$ und ${\Delta}_k^{(n)} \bigl(\xi, f_n (z)
\bigr)$, die auf der Restsummenabsch\"atzung $\omega_n = \gamma_{n+1}
z^{n+1}$ basieren, als auch die Pad\'e-Approximationen $\epsilon_{2
k}^{(n)} = [ k + n / k ]$, die ebenfalls Quotienten von Polynomen der
Grade $k + n$ beziehungsweise $k$ in $z$ sind, \"aquivalent. Allerdings
ben\"otigt man zur Berechnung der rationalen Funktionen $t_k^{(n)}
\bigl(\zeta, f_n (z) \bigr)$, ${\tau}_k^{(n)} \bigl(\zeta, f_n (z)
\bigr)$ und $T_k^{(n)} \bigl(\xi, f_n (z) \bigr)$ nur die numerischen
Werte der $k + 1$ Partialsummen $f_n (z)$, $f_{n+1} (z)$, $\ldots$ ,
$f_{n+k} (z)$, wogegen man zur Berechnung der Pad\'e-Approximation $[ k +
n / k ]$ die numerischen Werte der $2 k + 1$ Partialsummen $f_n (z)$,
$f_{n+1} (z)$, $\ldots$ , $f_{n+2 k} (z)$ ben\"otigt.

Man kann relativ leicht die folgenden asymptotischen Absch\"atzungen
f\"ur $z \to 0$ beweisen:
$$
\beginAligntags
" f (z) \, - \, t_k^{(n)} \bigl(\zeta, f_n (z) \bigr)
\; " = \; " O (z^{k + n + 1}) \, ,
\\ \tag
" f (z) \, - \, {\tau}_k^{(n)} \bigl(\zeta, f_n (z) \bigr)
\; " = \; " O (z^{k + n + 1}) \, ,
\\ \tag
" f (z) \, - \, T_k^{(n)} \bigl(\xi, f_n (z) \bigr)
\; " = \; " O (z^{k + n + 1}) \, .
\\ \tag
\endAligntags
$$
Diese Absch\"atzungen zeigen, da{\ss} Taylorentwicklungen der rationalen
Funktionen $t_k^{(n)} \bigl(\zeta, f_n (z) \bigr)$, ${\tau}_k^{(n)}
\bigl(\zeta, f_n (z) \bigr)$ und $T_k^{(n)} \bigl(\xi, f_n (z) \bigr)$
ebenfalls {\it alle} Koeffizienten $\gamma_0$, $\gamma_1$, $\ldots$ ,
$\gamma_{n + k}$ der unendlichen Reihe (5.7-1) exakt reproduzieren, die
zur Konstruktion der rationalen Funktionen verwendet wurden.

Zum Beweis der Absch\"atzungen (5.7-20) - (5.7-22) verwenden wir Gl.
(5.7-9) und n\"utzen aus, da{\ss} die Transformationen $t_k^{(n)}
\bigl(\zeta, f_n (z) \bigr)$, ${\tau}_k^{(n)} \bigl(\zeta, f_n (z)
\bigr)$ und $T_k^{(n)} \bigl(\xi, f_n (z) \bigr)$ bez\"uglich der $k + 1$
Partialsummen $f_n (z)$, $f_{n+1} (z)$, $\ldots$ , $f_{n+k} (z)$
translationsinvariant sind. Damit erhalten wir:
$$
\beginAligntags
" f (z) \, - \, t_k^{(n)} \bigl(\zeta, f_n (z) \bigr) \\
" \qquad \; = \; - \; z^{k + n} \;
\frac
{ \displaystyle
\sum_{j=0}^{k} \; ( - 1)^{j} \; \binom {k} {j} \;
\frac
{(\zeta + n +j )^{k-1}} {(\zeta + n + k )^{k-1}} \;
\frac {r_{n+j} (z)} {\gamma_{n+j} z^{n+j}} }
{\displaystyle
\sum_{j=0}^{k} \; ( - 1)^{j} \; \binom {k} {j} \;
\frac
{(\zeta + n +j )^{k-1}} {(\zeta + n + k )^{k-1}} \;
\frac {z^{k - j}} {\gamma_{n+j}} }
\; ,
\\ \tag
" f (z) \, - \, \tau_k^{(n)} \bigl(\zeta, f_n (z) \bigr) \\
" \qquad \; = \; - \; z^{k + n} \;
\frac
{ \displaystyle
\sum_{j=0}^{k} \; ( - 1)^{j} \; \binom {k} {j} \;
\frac
{(\zeta + n +j )_{k-1}} {(\zeta + n + k )_{k-1}} \;
\frac {r_{n+j} (z)} {\gamma_{n+j} z^{n+j}} }
{\displaystyle
\sum_{j=0}^{k} \; ( - 1)^{j} \; \binom {k} {j} \;
\frac
{(\zeta + n +j )_{k-1}} {(\zeta + n + k )_{k-1}} \;
\frac {z^{k - j}} {\gamma_{n+j}} }
\; ,
\\ \tag
" f (z) \, - \, T_k^{(n)} \bigl(\xi, f_n (z) \bigr) \\
" \qquad \; = \; - \; z^{k + n} \;
\frac
{ \displaystyle
\sum_{j=0}^{k} \; ( - 1)^{j} \; \binom {k} {j} \;
\frac
{( - \xi - n - j )_{k-1}} {( - \xi - n - k )_{k-1}} \;
\frac {r_{n+j} (z)} {\gamma_{n+j} z^{n+j}} }
{\displaystyle
\sum_{j=0}^{k} \; ( - 1)^{j} \; \binom {k} {j} \;
\frac
{( - \xi - n - j )_{k-1}} {( - \xi - n - k )_{k-1}} \;
\frac {z^{k - j}} {\gamma_{n+j}} }
\; .
\\ \tag
\endAligntags
$$

Da die Nenner in Gln. (5.7-23) - (5.7-25) Polynome von Grade $k$ in $z$
sind, deren konstante Terme laut Voraussetzung ungleich Null sind, mu{\ss}
man nur noch zeigen, da{\ss} die Z\"ahlersummen mindestens von der Ordnung $O
(z)$ f\"ur $z \to 0$ sind, um die Absch\"atzungen (5.7-20) - (5.7-22) zu
beweisen.

Zum Beweis der Absch\"atzungen (5.7-20) - (5.7-22) eliminieren wir aus
Gln. (5.7-23) - (5.7-25) die gemeinsamen Faktoren $(\zeta + n +
k)^{k-1}$, $(\zeta + n + k)_{k-1}$ und $( - \xi - n - k)_{k-1}$ und
schreiben die resultierenden Z\"ahlersummen der rationalen Funktionen mit
Hilfe von Gl. (3.1-14) auf folgende Weise um:
$$
\beginAligntags
" \sum_{j=0}^{k} \; ( - 1)^{j} \; \binom {k} {j} \,
(\zeta + n +j )^{k-1} \,
\frac {r_{n+j} (z)} {\gamma_{n+j} z^{n+j}} \\
" \qquad \; = \;
(- 1)^{k+1} \, \left[ \Delta^k \, (\zeta + n)^{k-1} \,
\frac {r_n (z)} {\gamma_n z^n} \right] \, ,
\\ \tag
" \sum_{j=0}^{k} \; ( - 1)^{j} \; \binom {k} {j} \,
(\zeta + n +j )_{k-1} \,
\frac {r_{n+j} (z)} {\gamma_{n+j} z^{n+j}} \\
" \qquad \; = \;
(- 1)^{k+1} \,
\left[ \Delta^k \, (\zeta + n)_{k-1} \,
\frac {r_n (z)} {\gamma_n z^n} \right] \, ,
\\ \tag
" \sum_{j=0}^{k} \; ( - 1)^{j} \; \binom {k} {j} \,
(- \xi - n - j)_{k-1} \,
\frac {r_{n+j} (z)} {\gamma_{n+j} z^{n+j}} \\
" \qquad \; = \;
(- 1)^{k+1} \,
\left[ \Delta^k \, (- \xi - n)_{k-1} \,
\frac {r_n (z)} {\gamma_n z^n} \right] \, .
\\ \tag
\endAligntags
$$
Aus Gl. (5.7-9) folgt
$$
\frac {r_n (z)} {\gamma_n z^n}
\; = \; - \, \sum_{\nu=0}^{\infty} \,
\frac {\gamma_{n+\nu+1}} {\gamma_n} z^{\nu+1}
\; = \; O (z) \, , \qquad z \to 0 \, .
\tag
$$
Aus dieser Beziehung folgt, da{\ss} die $k$-ten gewichteten Differenzen in
Gln. (5.7-26) - (5.7-28) mindestens von der Ordnung $O (z)$ f\"ur $z \to
0$ sind, was die Absch\"atzungen (5.7-20) - (5.7-22) beweist.

Obwohl die Restsummenabsch\"atzungen $\omega_n = (\zeta + n) \gamma_n z^n$
beziehungsweise $\omega_n = (- \xi - n) \gamma_n z^n$ urspr\"unglich zur
Beschleunigung von logarithmischer Konvergenz eingef\"uhrt wurden, sind
sie auch zur Verbesserung der Konvergenz einer Potenzreihe oder zur
Summation divergenter Potenzreihen geeignet. Sie entsprechen den
Restsummenabsch\"atzungen (5.2-12) beziehungsweise (5.5-12), welche die
Transformationen $u_{k}^{(n)} (\zeta, s_n)$, Gl. (5.2-13), $y_{k}^{(n)}
(\zeta, s_n)$, Gl. (5.4-11), und $Y_{k}^{(n)} (\xi, s_n)$, Gl. (5.5-13),
ergeben.

Wenn man in Gln. (5.2-6), (5.4-6) und (5.5-8) als Eingabedaten die
Partialsummen (5.7-2) und in Gln. (5.2-6) und (5.4-6) die
Restsummenabsch\"atzung $\omega_n = (\zeta + n) \gamma_n z^n$
beziehungsweise in Gl. (5.5-8) die Restsummenabsch\"atzung $\omega_n =
(- \xi - n) \gamma_n z^n$ verwendet, erh\"alt man die folgenden
rationalen Funktionen:
$$
\beginAligntags
" u_k^{(n)} \bigl(\zeta, f_n (z) \bigr) \; " = \; "
\frac
{\displaystyle
\sum_{j=0}^{k} \; ( - 1)^{j} \; \binom {k} {j} \;
\frac
{(\zeta + n +j )^{k-2}} {(\zeta + n + k )^{k-1}} \;
\frac {z^{k - j} f_{n+j} (z)} {\gamma_{n+j}} }
{\displaystyle
\sum_{j=0}^{k} \; ( - 1)^{j} \; \binom {k} {j} \;
\frac
{(\zeta + n +j )^{k-2}} {(\zeta + n + k )^{k-1}} \;
\frac {z^{k - j}} {\gamma_{n+j}} }
\; ,
\\ \tag
" y_k^{(n)} \bigl(\zeta, f_n (z) \bigr) \; " = \; "
\frac
{\displaystyle
\sum_{j=0}^{k} \; ( - 1)^{j} \; \binom {k} {j} \;
\frac
{(\zeta + n + j + 1)_{k-2}} {(\zeta + n + k )_{k-1}} \;
\frac {z^{k - j} f_{n+j} (z)} {\gamma_{n+j}} }
{\displaystyle
\sum_{j=0}^{k} \; ( - 1)^{j} \; \binom {k} {j} \;
\frac
{(\zeta + n + j + 1)_{k-2}} {(\zeta + n + k )_{k-1}} \;
\frac {z^{k - j}} {\gamma_{n+j}} }
\; ,
\\ \tag
" Y_k^{(n)} \bigl(\xi, f_n (z) \bigr) \; " = \; "
\frac
{\displaystyle
\sum_{j=0}^{k} \; ( - 1)^{j} \; \binom {k} {j} \;
\frac
{( - \xi - n - j + 1)_{k-2}} {( - \xi - n - k )_{k-1}} \;
\frac {z^{k - j} f_{n+j} (z)} {\gamma_{n+j}} }
{\displaystyle
\sum_{j=0}^{k} \; ( - 1)^{j} \; \binom {k} {j} \;
\frac
{( - \xi - n - j + 1)_{k-2}} {( - \xi - n - k )_{k-1}} \;
\frac {z^{k - j}} {\gamma_{n+j}} }
\; .
\\ \tag
\endAligntags
$$
Die Z\"ahler dieser Ausdr\"ucke sind wiederum Polynome vom Grade $k + n$ in
$z$, und die Nenner sind Polynome vom Grade $k$ in $z$. In dieser
Hinsicht sind $u_k^{(n)} \bigl(\zeta, f_n (z) \bigr)$, $y_k^{(n)}
\bigl(\zeta, f_n (z) \bigr)$ und $Y_k^{(n)} \bigl(\xi, f_n (z) \bigr)$
\"aquivalent zu den bisher in diesem Unterabschnitt besprochenen
Varianten von ${\cal L}_{k}^{(n)} (\zeta, s_n, \omega_n)$, Gl. (5.2-6),
${\cal S}_{k}^{(n)} (\zeta, s_n, \omega_n)$, Gl. (5.4-6), und ${\cal
M}_{k}^{(n)} (\xi, s_n, \omega_n)$, Gl. (5.5-8), als auch zu den
Pad\'e-Approximationen $\epsilon_{2 k}^{(n)} = [ k + n / k ]$, die
ebenfalls Quotienten von Polynomen der Grade $k + n$ beziehungsweise
$k$ in $z$ sind. Allerdings ben\"otigt man zur Berechnung der rationalen
Funktionen $u_k^{(n)} \bigl(\zeta, f_n (z) \bigr)$, $y_k^{(n)}
\bigl(\zeta, f_n (z) \bigr)$ und $Y_k^{(n)} \bigl(\xi, f_n (z) \bigr)$
nur die numerischen Werte der $k + 1$ Partialsummen $f_n (z)$, $f_{n+1}
(z)$, $\ldots$ , $f_{n+k} (z)$, wogegen man zur Berechnung der
Pad\'e-Approximation $[ k + n / k ]$ die numerischen Werte der $2 k + 1$
Partialsummen $f_n (z)$, $f_{n+1} (z)$, $\ldots$ , $f_{n+2 k} (z)$
ben\"otigt.

Die folgenden asymptotischen Absch\"atzungen f\"ur $z \to 0$ k\"onnen
ebenfalls relativ leicht bewiesen werden:
$$
\beginAligntags
" f (z) \, - \, u_k^{(n)} \bigl(\zeta, f_n (z) \bigr)
\; " = \; " O (z^{k + n + 1}) \, ,
\\ \tag
" f (z) \, - \, y_k^{(n)} \bigl(\zeta, f_n (z) \bigr)
\; " = \; " O (z^{k + n + 1}) \, ,
\\ \tag
" f (z) \, - \, Y_k^{(n)} \bigl(\xi, f_n (z) \bigr)
\; " = \; " O (z^{k + n + 1}) \, .
\\ \tag
\endAligntags
$$
Diese Absch\"atzungen zeigen, da{\ss} Taylorentwicklungen der rationalen
Funktionen $u_k^{(n)} \bigl(\zeta, f_n (z) \bigr)$, $y_k^{(n)}
\bigl(\zeta, f_n (z) \bigr)$ und $Y_k^{(n)} \bigl(\xi, f_n (z) \bigr)$
ebenfalls {\it alle} Koeffizienten $\gamma_0$, $\gamma_1$, $\ldots$ ,
$\gamma_{n + k}$ der unendlichen Reihe (5.7-1) exakt reproduzieren, die
zur Konstruktion der rationalen Funktionen verwendet wurden.

Zum Beweis der Absch\"atzungen (5.7-33) - (5.7-35) verwenden wir wiederum
Gl. (5.7-9) und n\"utzen aus, da{\ss} die Transformationen $u_k^{(n)}
\bigl(\zeta, f_n (z) \bigr)$, $y_k^{(n)} \bigl(\zeta, f_n (z) \bigr)$
und $Y_k^{(n)} \bigl(\xi, f_n (z) \bigr)$ bez\"uglich der $k + 1$
Partialsummen $f_n (z)$, $f_{n+1} (z)$, $\ldots$ , $f_{n+k} (z)$
translationsinvariant sind. Damit erhalten wir:
$$
\beginAligntags
" f (z) \, - \, u_k^{(n)} \bigl(\zeta, f_n (z) \bigr) \\
" \qquad \; = \; - \; z^{k + n} \;
\frac
{ \displaystyle
\sum_{j=0}^{k} \; ( - 1)^{j} \; \binom {k} {j} \;
\frac
{(\zeta + n +j )^{k-2}} {(\zeta + n + k )^{k-1}} \;
\frac {r_{n+j} (z)} {\gamma_{n+j} z^{n+j}} }
{\displaystyle
\sum_{j=0}^{k} \; ( - 1)^{j} \; \binom {k} {j} \;
\frac
{(\zeta + n +j )^{k-2}} {(\zeta + n + k )^{k-1}} \;
\frac {z^{k - j}} {\gamma_{n+j}} }
\; ,
\\ \tag
" f (z) \, - \, y_k^{(n)} \bigl(\zeta, f_n (z) \bigr) \\
" \qquad \; = \; - \; z^{k + n} \;
\frac
{ \displaystyle
\sum_{j=0}^{k} \; ( - 1)^{j} \; \binom {k} {j} \;
\frac
{(\zeta + n + j + 1)_{k-2}} {(\zeta + n + k )_{k-1}} \;
\frac {r_{n+j} (z)} {\gamma_{n+j} z^{n+j}} }
{\displaystyle
\sum_{j=0}^{k} \; ( - 1)^{j} \; \binom {k} {j} \;
\frac
{(\zeta + n + j +1)_{k-2}} {(\zeta + n + k )_{k-1}} \;
\frac {z^{k - j}} {\gamma_{n+j}} }
\; ,
\\ \tag
" f (z) \, - \, Y_k^{(n)} \bigl(\xi, f_n (z) \bigr) \\
" \qquad \; = \; - \; z^{k + n} \;
\frac
{ \displaystyle
\sum_{j=0}^{k} \; ( - 1)^{j} \; \binom {k} {j} \;
\frac
{( - \xi - n - j + 1)_{k-2}} {( - \xi - n - k )_{k-1}} \;
\frac {r_{n+j} (z)} {\gamma_{n+j} z^{n+j}} }
{\displaystyle
\sum_{j=0}^{k} \; ( - 1)^{j} \; \binom {k} {j} \;
\frac
{( - \xi - n - j + 1)_{k-2}} {( - \xi - n - k )_{k-1}} \;
\frac {z^{k - j}} {\gamma_{n+j}} }
\; .
\\ \tag
\endAligntags
$$

Da die Nenner in Gln. (5.7-36) - (5.7-38) Polynome von Grade $k$ in $z$
sind, deren konstante Terme laut Voraussetzung ungleich Null sind, mu{\ss}
man nur noch zeigen, da{\ss} die Z\"ahlersummen mindestens von der Ordnung $O
(z)$ f\"ur $z \to 0$ sind, um die Absch\"atzungen (5.7-33) - (5.7-35) zu
beweisen.

Zum Beweis der Absch\"atzungen (5.7-33) - (5.7-35) eliminieren wir aus
Gln. (5.7-36) - (5.7-38) die gemeinsamen Faktoren $(\zeta + n +
k)^{k-1}$, $(\zeta + n + k)_{k-1}$ und $( - \xi - n - k)_{k-1}$, und
schreiben die resultierenden Z\"ahlersummen der rationalen Funktionen mit
Hilfe von Gl. (3.1-14) auf folgende Weise um:
$$
\beginAligntags
" \sum_{j=0}^{k} \; ( - 1)^{j} \; \binom {k} {j} \,
(\zeta + n +j )^{k-2} \,
\frac {r_{n+j} (z)} {\gamma_{n+j} z^{n+j}} \\
" \qquad \; = \;
(- 1)^{k+1} \, \left[ \Delta^k \, (\zeta + n)^{k-2} \,
\frac {r_n (z)} {\gamma_n z^n} \right] \, ,
\\ \tag
" \sum_{j=0}^{k} \; ( - 1)^{j} \; \binom {k} {j} \,
(\zeta + n + j + 1)_{k-2} \,
\frac {r_{n+j} (z)} {\gamma_{n+j} z^{n+j}} \\
" \qquad \; = \;
(- 1)^{k+1} \,
\left[ \Delta^k \, (\zeta + n + 1)_{k-2} \,
\frac {r_n (z)} {\gamma_n z^n} \right] \, ,
\\ \tag
" \sum_{j=0}^{k} \; ( - 1)^{j} \; \binom {k} {j} \,
(- \xi - n - j + 1)_{k-2} \,
\frac {r_{n+j} (z)} {\gamma_{n+j} z^{n+j}} \\
" \qquad \; = \;
(- 1)^{k+1} \,
\left[ \Delta^k \, (- \xi - n + 1)_{k-2} \,
\frac {r_n (z)} {\gamma_n z^n} \right] \, .
\\ \tag
\endAligntags
$$
Aus Gl. (5.7-29) folgt, da{\ss} auch die $k$-ten gewichteten Differenzen in
Gln. (5.7-39) - (5.7-41) mindestens von der Ordnung $O (z)$ f\"ur $z \to
0$ sind, was die Absch\"atzungen (5.7-33) - (5.7-35) beweist.

Man kann auch die Restsummenabsch\"atzung $\omega_n = \gamma_n
\gamma_{n+1} z^{n+1} / [\gamma_n - z \gamma_{n+1}]$, die auf dem
Aitkenschen $\Delta^2$-Proze{\ss}, Gl. (3.3-6), basiert, zur Verbesserung
der Konvergenzeigenschaften einer Potenzreihe oder zur Summation einer
divergenten Potenzreihe verwenden. Sie entspricht der von Levin [1973]
eingef\"uhrten Restsummenabsch\"atzung (5.2-19), welche die Transformationen
$v_{k}^{(n)} (\zeta, s_n)$, Gl. (5.2-20), $\phi_{k}^{(n)} (\zeta, s_n)$,
Gl. (5.4-14), und $\Phi_{k}^{(n)} (\xi, s_n)$, Gl. (5.5-16), ergibt.

Wenn man in Gln. (5.2-6), (5.4-6) und (5.5-8) als Eingabedaten die
Partialsummen (5.7-2) und die Restsummenabsch\"atzung $\omega_n =
\gamma_n \gamma_{n+1} z^{n+1} / [\gamma_n - z \gamma_{n+1}]$ verwendet,
erh\"alt man die folgenden rationalen Funktionen:
$$
\beginAligntags
" v_k^{(n)} \bigl(\zeta, f_n (z) \bigr) \; " = \; "
\frac
{\displaystyle
\sum_{j=0}^{k} \; ( - 1)^{j} \; \binom {k} {j} \;
\frac
{(\zeta + n +j )^{k-1}} {(\zeta + n + k )^{k-1}} \;
\frac {z^{k - j} (\gamma_{n+j} - z \gamma_{n+j+1}) f_{n+j} (z)}
{\gamma_{n+j} \gamma_{n+j+1}} }
{\displaystyle
\sum_{j=0}^{k} \; ( - 1)^{j} \; \binom {k} {j} \;
\frac
{(\zeta + n +j )^{k-1}} {(\zeta + n + k )^{k-1}} \;
\frac {z^{k - j} (\gamma_{n+j} - z \gamma_{n+j+1})}
{\gamma_{n+j} \gamma_{n+j+1}} }
\; ,
\\ \tag
" \phi_k^{(n)} \bigl(\zeta, f_n (z) \bigr) \; " = \; "
\frac
{\displaystyle
\sum_{j=0}^{k} \; ( - 1)^{j} \; \binom {k} {j} \;
\frac
{(\zeta + n +j )_{k-1}} {(\zeta + n + k )_{k-1}} \;
\frac {z^{k - j} (\gamma_{n+j} - z \gamma_{n+j+1}) f_{n+j} (z)}
{\gamma_{n+j} \gamma_{n+j+1}} }
{\displaystyle
\sum_{j=0}^{k} \; ( - 1)^{j} \; \binom {k} {j} \;
\frac
{(\zeta + n +j )_{k-1}} {(\zeta + n + k )_{k-1}} \;
\frac {z^{k - j} (\gamma_{n+j} - z \gamma_{n+j+1})}
{\gamma_{n+j} \gamma_{n+j+1}} }
\; ,
\\ \tag
" \Phi_k^{(n)} \bigl(\xi, f_n (z) \bigr) \; " = \; "
\frac
{\displaystyle
\sum_{j=0}^{k} \; ( - 1)^{j} \; \binom {k} {j} \;
\frac
{( - \xi - n - j )_{k-1}} {( - \xi - n - k )_{k-1}} \;
\frac {z^{k - j} (\gamma_{n+j} - z \gamma_{n+j+1}) f_{n+j} (z)}
{\gamma_{n+j} \gamma_{n+j+1}} }
{\displaystyle
\sum_{j=0}^{k} \; ( - 1)^{j} \; \binom {k} {j} \;
\frac
{( - \xi - n - j )_{k-1}} {( - \xi - n - k )_{k-1}} \;
\frac {z^{k - j} (\gamma_{n+j} - z \gamma_{n+j+1})}
{\gamma_{n+j} \gamma_{n+j+1}} }
\; . \quad
\\ \tag
\endAligntags
$$
Die Z\"ahler dieser Ausdr\"ucke sind Polynome vom Grade $k + n + 1$ in $z$,
und die Nenner sind Polynome vom Grade $k + 1$ in $z$. In dieser
Hinsicht unterscheiden sich $v_k^{(n)} \bigl(\zeta, f_n (z) \bigr)$,
$\phi_k^{(n)} \bigl(\zeta, f_n (z) \bigr)$ und $\Phi_k^{(n)} \bigl(\xi,
f_n (z) \bigr)$ von den bisher in diesem Unterabschnitt besprochenen
Varianten von ${\cal L}_{k}^{(n)} (\zeta, s_n, \omega_n)$, Gl. (5.2-6),
${\cal S}_{k}^{(n)} (\zeta, s_n, \omega_n)$, Gl. (5.4-6), und ${\cal
M}_{k}^{(n)} (\xi, s_n, \omega_n)$, Gl. (5.5-8), die allesamt
Quotienten von Polynomen der Grade $k + n$ beziehungsweise $k$ in $z$
sind. Zur Berechnung von $v_k^{(n)} \bigl(\zeta, f_n (z) \bigr)$,
$\phi_k^{(n)} \bigl(\zeta, f_n (z) \bigr)$, und $\Phi_k^{(n)}
\bigl(\xi, f_n (z) \bigr)$ ben\"otigt man die numerischen Werte der $k +
2$ Partialsummen $f_n (z)$, $f_{n+1} (z)$, $\ldots$ , $f_{n+k+1} (z)$.

Man kann relativ leicht die folgenden asymptotischen Absch\"atzungen
f\"ur $z \to 0$ beweisen:
$$
\beginAligntags
" f (z) \, - \, v_k^{(n)} \bigl(\zeta, f_n (z) \bigr)
\; " = \; " O (z^{k + n + 2}) \, ,
\\ \tag
" f (z) \, - \, \phi_k^{(n)} \bigl(\zeta, f_n (z) \bigr)
\; " = \; " O (z^{k + n + 2}) \, ,
\\ \tag
" f (z) \, - \, \Phi_k^{(n)} \bigl(\xi, f_n (z) \bigr)
\; " = \; " O (z^{k + n + 2}) \, .
\\ \tag
\endAligntags
$$
Diese Absch\"atzungen zeigen, da{\ss} Taylorentwicklungen der rationalen
Funktionen $v_k^{(n)} \bigl(\zeta, f_n (z) \bigr)$, $\phi_k^{(n)}
\bigl(\zeta, f_n (z) \bigr)$ und $\Phi_k^{(n)} \bigl(\xi, f_n (z)
\bigr)$ wiederum alle $k+n+2$ Koeffizienten $\gamma_0$, $\gamma_1$,
$\ldots$ , $\gamma_{n + k + 1}$ der unendlichen Reihe (5.7-1) exakt
reproduzieren, die zur Konstruktion der rationalen Funktionen verwendet
wurden.

Zum Beweis der Absch\"atzungen (5.7-45) - (5.7-47) verwenden wir Gl.
(5.7-9) und n\"utzen aus, da{\ss} die Transformationen $v_k^{(n)}
\bigl(\zeta, f_n (z) \bigr)$, $\phi_k^{(n)} \bigl(\zeta, f_n (z)
\bigr)$ und $\Phi_k^{(n)} \bigl(\xi, f_n (z) \bigr)$ bez\"uglich der $k +
1$ Partialsummen $f_n (z)$, $f_{n+1} (z)$, $\ldots$ , $f_{n+k} (z)$
translationsinvariant sind. Damit erhalten wir:
$$
\beginAligntags
" f (z) \, - \, v_k^{(n)} \bigl(\zeta, f_n (z) \bigr) \\
" \qquad \; = \; - \; z^{k + n + 1} \;
\frac
{ \displaystyle
\sum_{j=0}^{k} \; ( - 1)^{j} \; \binom {k} {j} \;
\frac
{(\zeta + n +j )^{k-1}} {(\zeta + n + k )^{k-1}} \;
\frac {(\gamma_{n+j} - z \gamma_{n+j+1}) \, r_{n+j} (z)}
{\gamma_{n+j} \gamma_{n+j+1} z^{n+j+1}} }
{\displaystyle
\sum_{j=0}^{k} \; ( - 1)^{j} \; \binom {k} {j} \;
\frac
{(\zeta + n +j )^{k-1}} {(\zeta + n + k )^{k-1}} \;
\frac {z^{k - j} (\gamma_{n+j} - z \gamma_{n+j+1})}
{\gamma_{n+j} \gamma_{n+j+1}} }
\; ,
\\ \tag
" f (z) \, - \, \phi^{(n)} \bigl(\zeta, f_n (z) \bigr) \\
" \qquad \; = \; - \; z^{k + n + 1} \;
\frac
{ \displaystyle
\sum_{j=0}^{k} \; ( - 1)^{j} \; \binom {k} {j} \;
\frac
{(\zeta + n +j )_{k-1}} {(\zeta + n + k )_{k-1}} \;
\frac {(\gamma_{n+j} - z \gamma_{n+j+1}) \, r_{n+j} (z)}
{\gamma_{n+j} \gamma_{n+j+1} z^{n+j+1}} }
{\displaystyle
\sum_{j=0}^{k} \; ( - 1)^{j} \; \binom {k} {j} \;
\frac
{(\zeta + n +j )_{k-1}} {(\zeta + n + k )_{k-1}} \;
\frac {z^{k - j} (\gamma_{n+j} - z \gamma_{n+j+1})}
{\gamma_{n+j} \gamma_{n+j+1}} }
\; ,
\\ \tag
" f (z) \, - \, \Phi_k^{(n)} \bigl(\xi, f_n (z) \bigr) \\
" \qquad \; = \; - \; z^{k + n + 1} \;
\frac
{ \displaystyle
\sum_{j=0}^{k} \; ( - 1)^{j} \; \binom {k} {j} \;
\frac
{( - \xi - n - j )_{k-1}} {( - \xi - n - k )_{k-1}} \;
\frac {(\gamma_{n+j} - z \gamma_{n+j+1}) \, r_{n+j} (z)}
{\gamma_{n+j} \gamma_{n+j+1} z^{n+j+1}} }
{\displaystyle
\sum_{j=0}^{k} \; ( - 1)^{j} \; \binom {k} {j} \;
\frac
{( - \xi - n - j )_{k-1}} {( - \xi - n - k )_{k-1}} \;
\frac {z^{k - j} (\gamma_{n+j} - z \gamma_{n+j+1})}
{\gamma_{n+j} \gamma_{n+j+1}} }
\; .
\\ \tag
\endAligntags
$$

Da die Nenner in Gln. (5.7-48) - (5.7-50) Polynome von Grade $k + 1$ in
$z$ sind, deren konstante Terme laut Voraussetzung ungleich Null sind,
mu{\ss} man nur noch zeigen, da{\ss} die Z\"ahlersummen mindestens von der
Ordnung $O (z)$ f\"ur $z \to 0$ sind, um die Absch\"atzungen (5.7-45) -
(5.7-47) zu beweisen.

Zum Beweis der Absch\"atzungen (5.7-45) - (5.7-47) eliminieren wir aus
Gln. (5.7-48) - (5.7-50) die gemeinsamen Faktoren $(\zeta + n +
k)^{k-1}$, $(\zeta + n + k)_{k-1}$ und $( - \xi - n - k)_{k-1}$ und
schreiben die resultierenden Z\"ahlersummen der rationalen Funktionen mit
Hilfe von Gl. (3.1-14) auf folgende Weise um:
$$
\beginAligntags
" \sum_{j=0}^{k} \; ( - 1)^{j} \; \binom {k} {j} \,
(\zeta + n +j )^{k-1} \,
\frac {(\gamma_{n+j} - z \gamma_{n+j+1}) \, r_{n+j} (z)}
{\gamma_{n+j} \gamma_{n+j+1} z^{n+j+1}} \\
" \qquad \; = \;
(- 1)^{k+1} \, \left[ \Delta^k \, (\zeta + n)^{k-1} \,
\frac {(\gamma_{n} - z \gamma_{n+1}) \, r_{n} (z)}
{\gamma_{n} \gamma_{n+1} z^{n+1}} \right] \, ,
\\ \tag
" \sum_{j=0}^{k} \; ( - 1)^{j} \; \binom {k} {j} \,
(\zeta + n +j )_{k-1} \,
\frac {(\gamma_{n+j} - z \gamma_{n+j+1}) \, r_{n+j} (z)}
{\gamma_{n+j} \gamma_{n+j+1} z^{n+j+1}} \\
" \qquad \; = \;
(- 1)^{k+1} \,
\left[ \Delta^k \, (\zeta + n)_{k-1} \,
\frac {(\gamma_{n} - z \gamma_{n+1}) \, r_{n} (z)}
{\gamma_{n} \gamma_{n+1} z^{n+1}} \right] \, ,
\\ \tag
" \sum_{j=0}^{k} \; ( - 1)^{j} \; \binom {k} {j} \,
(- \xi - n - j)_{k-1} \,
\frac {(\gamma_{n+j} - z \gamma_{n+j+1}) \, r_{n+j} (z)}
{\gamma_{n+j} \gamma_{n+j+1} z^{n+j+1}} \\
" \qquad \; = \;
(- 1)^{k+1} \,
\left[ \Delta^k \, (- \xi - n)_{k-1} \,
\frac {(\gamma_{n} - z \gamma_{n+1}) \, r_{n} (z)}
{\gamma_{n} \gamma_{n+1} z^{n+1}} \right] \, .
\\ \tag
\endAligntags
$$
Mit Hilfe von Gl. (5.7-9) erhalten wir die folgenden Absch\"atzungen f\"ur
$z \to 0$:
$$
\beginAligntags
" \frac {\gamma_n r_n (z)}
{\gamma_n \gamma_{n+1} z^{n+1}}
\; " = \;
" - 1 \, - \, \sum_{\nu=0}^{\infty} \,
\frac {\gamma_{n+\nu+2}} {\gamma_{n+1}} z^{\nu+1}
\; " = \; " - 1 \, + \, O (z) \, ,
\\ \tag
" \frac {z \gamma_{n+1} r_n (z)}
{\gamma_n \gamma_{n+1} z^{n+1}}
\; " = \;
" - z \, \sum_{\nu=0}^{\infty} \,
\frac {\gamma_{n+\nu+1}} {\gamma_n} z^{\nu}
" \; = \; " O (z) \, .
\\ \tag
\endAligntags
$$
Aus Gl. (5.7-54) folgt, da{\ss} die $z$-unabh\"angigen Anteile in Gln.
(5.7-51) - (5.7-53) Polynome vom Grade $k - 1$ in $n$ sind, die durch
$k$-fache Anwendung des Differenzenoperators $\Delta$ annihiliert
werden [Milne-Thomson 1981, S. 29]. Die Absch\"atzungen (5.7-54) und
(5.7-55) implizieren also, da{\ss} die nichtverschwindenden $k$-ten
gewichteten Differenzen in Gln. (5.7-51) - (5.7-53) mindestens von der
Ordnung $O (z)$ f\"ur $z \to 0$ sind, was die Absch\"atzungen (5.7-45) -
(5.7-47) beweist.

Man sollte hier noch erw\"ahnen, da{\ss} es gewisse Parallelen gibt zwischen
der Theorie der {\it Konvergenzfaktoren} [Airey 1937; Dingle 1973,
Kapitel 21 - 26; Miller 1952; Neuhaus und Schottlaender 1975] und den
verallgemeinerten Summationsprozessen $d_{k}^{(n)} (\zeta, s_n)$, Gl.
(5.2-18), $\delta_{k}^{(n)} (\zeta, s_n)$, Gl. (5.4-13), und
$\Delta_{k}^{(n)} (\xi, s_n)$, Gl. (5.5-15). In der Theorie der
Konvergenzfaktoren wird der Abbruchfehler
$$
r_n \; = \; - \sum_{k=n+1}^{\infty} \, a_k
\tag
$$
der Partialsumme $s_n$ einer divergenten Reihe dargestellt als der
erste Term $a_{n+1}$, der nicht in der Partialsumme $s_n$ enthalten
ist, multipliziert mit einem Konvergenzfaktor $\psi_n$. Dieser
Konvergenzfaktor sollte nat\"urlich so gew\"ahlt werden, da{\ss}
$$
s_n \; = \; s \, + \, a_{n+1} \psi_n
\tag
$$
exakt gilt. Diese Beziehung ist aber formal identisch mit Gl. (5.1-3),
wenn man $\omega_n = a_{n+1}$ und $z_n = \psi_n$ setzt.

Die verallgemeinerten Summationsprozesse $d_{k}^{(n)} (\zeta, s_n)$,
$\delta_{k}^{(n)} (\zeta, s_n)$ und $\Delta_{k}^{(n)} (\xi, s_n)$, die
auf der Restsummenabsch\"atzung (5.2-16) basieren, versuchen also, den in
Gl. (5.7-57) vorkommenden Konvergenzfaktor $\psi_n$ ausschlie{\ss}lich mit
Hilfe numerischer Informationen und gewisser Annahmen \"uber seine
funktionale Form approximativ zu bestimmen.

Im Falle des verallgemeinerten Summationsprozesses $d_{k}^{(n)} (\zeta,
s_n)$ wird angenommen, da{\ss} der Konvergenzfaktor durch eine abgebrochene
asymptotische Potenzreihe in $1/(n + \zeta)$ mit unbestimmten
Koeffizienten $c_0$, $c_1$, $\ldots$ , $c_{k-1}$ approximiert werden
kann. Im Falle des verallgemeinerten Summationsprozesses
$\delta_{k}^{(n)} (\zeta, s_n)$ wird angenommen, da{\ss} der
Konvergenzfaktor durch eine abgebrochene Fakult\"atenreihe mit den
Pochhammersymbolen $(n + \zeta)_j$ und den unbestimmten Koeffizienten
$c_0$, $c_1$, $\ldots$ , $c_{k-1}$ approximiert werden kann.
Schlie{\ss}lich wird im Falle des verallgemeinerten Summationsprozesses
$\Delta_{k}^{(n)} (\xi, s_n)$ angenommen, da{\ss} der Konvergenzfaktor
durch eine asymptotische N\"aherung vom Typ von Gl. (5.5-1) mit den
Pochhammersymbolen $(- n + \xi)_j$ und den unbestimmten Koeffizienten
$c_0$, $c_1$, $\ldots$ , $c_{k-1}$ approximiert werden kann.

Der Erfolg eines solchen Ansatzes zur Summation einer divergenten Reihe
wird nat\"urlich ganz entscheidend davon abh\"angen, ob und wie gut man mit
Hilfe der obengenannten Annahmen die Konvergenzfaktoren der divergenten
Reihe approximieren kann. In allen drei F\"allen m\"ussen die unbestimmten
Koeffizienten $c_0$, $c_1$, $\ldots$ , $c_{k-1}$, deren Existenz
vorausgesetzt wird, auf rein numerische Weise bestimmt werden.

\endAbschnittsebene

\endAbschnittsebene

\keinTitelblatt\neueSeite

\beginAbschnittsebene
\aktAbschnitt = 5

\Abschnitt Konvergenztheorie verallgemeinerter Summationsprozesse

\vskip - 2 \jot

\beginAbschnittsebene

\medskip

\Abschnitt Vorbemerkungen

\smallskip

\aktTag = 0

Bei der mathematischen Behandlung naturwissenschaftlicher und
technischer Problemen ist man oft mit Situationen der folgenden Art
konfrontiert: Aufgrund des damit verbundenen numerischen Aufwandes ist
es nicht m\"oglich, so viele Partialsummen $s_0$, $s_1$, $\ldots$ ,
$s_{\ell}$ einer langsam konvergierenden unendlichen Reihe zu berechnen,
da{\ss} ihr Wert $s$ mit der gew\"unschten Genauigkeit bestimmt werden
kann.  Weitere, h\"aufig auftretende Komplikationen sind, da{\ss} man
den Wert einer Potenzreihe mit endlichem Konvergenzradius am Rand oder
au{\ss}erhalb ihres Konvergenzkreises berechnen mu{\ss}, oder da{\ss}
der Konvergenzradius einer Potenzreihe Null ist.

Wenn die Folge $\Seqn s$ der Partialsummen langsam gegen einen
Grenzwert $s$ konvergiert, kann man versuchen, die Konvergenz der
unendlichen Reihe mit Hilfe eines verallgemeinerten Summationsprozesses
$G_k^{(n)}$, der $k + 1$ Folgenelemente $s_n$, $s_{n+1}$, $\ldots$ ,
$s_{n+k}$ als Eingabedaten verwendet, zu verbessern. Allerdings stellt
sich dabei die Frage, ob $G_k^{(n)}$ f\"ur diese Reihe {\it regul\"ar} ist.
Man m\"ochte also wissen, ob die Transformationen
$$
s'_n \; = \; G_k^{(n)} (s_n, s_{n+1}, \ldots , s_{n+k})
\tag
$$
f\"ur $n \to \infty$ tats\"achlich gegen den gleichen Grenzwert $s$
konvergieren wie die Folge $\Seqn s$ der Partialsummen.

Weiterhin stellt sich in diesem Zusammenhang immer die Frage, welche
Eigenschaften eine langsam konvergierende Folge $\Seqn s$ haben mu{\ss},
damit ein gegebener verallgemeinerter Summationsproze{\ss} $G_k^{(n)}$ ihre
Konvergenz \"uberhaupt verbessern kann. In der Literatur \"uber
Konvergenzbeschleunigung sagt man, da{\ss} $G_k^{(n)}$ die Konvergenz einer
Folge $\Seqn s$ beschleunigt, wenn
$$
\lim_{n \to \infty} \;
\frac
{G_k^{(n)} (s_n, s_{n+1}, \ldots , s_{n+k}) - s}
{ s_n - s}
\; = \; 0
\tag
$$
f\"ur festes $k \in \N$ gilt.

Was die Summation divergenter Reihen betrifft, so w\"aren allgemeine
Kriterien, mit deren Hilfe man vorhersagen k\"onnte, ob eine divergente
Folge $\Seqn s$ \"uberhaupt von einem gegebenen verallgemeinerten
Summationsproze{\ss} $G_k^{(n)}$ gegen einen verallgemeinerten Grenzwert
$s$ summiert werden kann und ob dieser verallgemeinerte Grenzwert $s$
durch $G_k^{(n)}$ eindeutig bestimmt ist, ohne Zweifel \"uberaus wertvoll.

Zus\"atzlich zu den Antworten auf solche Fragen w\"are man auch sehr an
Fehlerabsch\"atzungen interessiert, die Aussagen dar\"uber machen, wie gut
ein verallgemeinerter Summationsproze{\ss} $G_k^{(n)}$ den
(verallgemeinerten) Grenzwert $s$ einer konvergenten oder divergenten
Folge $\Seqn s$ in Abh\"angigkeit von $k$ und $n$ approximieren kann.

Jede sinnvolle Konvergenztheorie verallgemeinerter Summationsprozesse
mu{\ss} \"uber diese fundamentalen Fragen Aussagen machen. Eine Beantwortung
der oben genannten Fragen scheint sehr schwierig zu sein, da bei
praktischen Problemen normalerweise nur die numerischen Werte einer
relativ kleinen Zahl von Folgenelementen bekannt und die
(verallgemeinerten) Grenzwerte unbekannt sind. Trotzdem ist es aber oft
m\"oglich, gewisse Aussagen allgemeinerer Natur zu machen, wenn die
Folgen, die transformiert werden sollen, geeignete Bedingungen erf\"ullen.

Bei Pad\'e-Approximationen gibt es eine hochentwickelte Konvergenztheorie,
die in den ent\-sprechenden Kapiteln der B\"ucher von Baker [1975; 1990],
Baker und Graves-Morris [1981a; 1981b], Cuyt und Wuytack [1987], Perron
[1957], und Wall [1973] oder in dem \"Ubersichtsartikel von Brezinski und
van Iseghem [1991] beschrieben wird. Dort findet man auch zahlreiche
andere Referenzen. Dagegen ist die Konvergenztheorie anderer
verallgemeinerter Summationsprozesse, die in der Regel wesentlich j\"unger
sind als Pad\'e-Approximationen, in den meisten F\"allen noch nicht
ausreichend entwickelt, und die theoretischen Eigenschaften dieser
verallgemeinerten Summationsprozesse sind bisher nur unzureichend
verstanden.

Es ist nicht das Ziel dieses Abschnittes, einen ann\"ahernd vollst\"andigen
\"Uberblick \"uber den augen\-blicklichen Stand der Konvergenztheorie
nichtlinearer verallgemeinerter Summationspro\-zesse zu geben. Dazu sei
auf B\"ucher von Brezinski [1977], Brezinski und Redivo Zaglia [1991],
Delahaye [1988], und Wimp [1981] verwiesen. Das Schwergewicht dieses
Abschnitts liegt auf eigenen Arbeiten \"uber Konvergenzeigenschaften
nichtlinearer verallgemeinerter Summationsprozesse mit besonderer
Betonung der in Abschnitt 5 behandelten Verfahren, die wie die Levinsche
Transformation ${\cal L}_k^{(n)} (\zeta , s_n, \omega_n)$, Gl. (5.2-6),
explizite Restsummenabsch\"atzungen verwenden.

Der erste Schritt einer theoretischen Analyse verallgemeinerter
Summationsprozesse besteht darin, da{\ss} man sich \"uber die {\it
wesentlichen} Eigenschaften derjenigen verallgemeinerter
Summationsprozesse klar wird, die man analysieren will. So besitzen
{\it lineare} Transformationen Eigenschaften, die vom theoretischen
Standpunkt besonders g\"unstig sind. Ein verallgemeinerte
Summationsproze{\ss} ${\bf L}_k^{(n)}$, der auf $k + 1$ Eingabedaten wirkt,
wird linear genannt, wenn
$$
\beginAligntags
" {\bf L}_k^{(n)}
(\alpha s_n + \beta t_n, \alpha s_{n+1} + \beta t_{n+1},
\ldots , \alpha s_{n+k} + \beta t_{n+k}) \\
" \qquad \; = \; \alpha \,
{\bf L}_k^{(n)} (s_n, s_{n+1}, \ldots , s_{n+k}) \, + \,
\beta \, {\bf L}_k^{(n)} (t_n, t_{n+1}, \ldots , t_{n+k})
\\ \tag
\endAligntags
$$
f\"ur beliebige reelle Konstanten $\alpha$ und $\beta$ und f\"ur beliebige
Folgen $\Seqn s$ und $\Seqn t$ gilt.

Wie schon in Abschnitt 2.3 erw\"ahnt, besitzen diejenigen lineare
Transformationen, welche die Elemente einer transformierten Folge $\Seqn
{s'}$ gem\"a{\ss} Gl. (2.4-6) als gewichtete Mittelwerte aus den Eingabedaten
$\Seqn s$ berechnen, eine hochentwickelte Konvergenztheorie [Hardy
1949; Knopp 1964; Petersen 1966; Peyerimhoff 1969; Powell und Shah
1988; Zeller und Beekmann 1970]. Beispielsweise k\"onnen relativ leicht
notwendige und hinreichende Bedingungen formuliert werden, die
garantieren, da{\ss} eine transformierte Folge $\Seqn {s'}$ gegen den
gleichen Grenzwert $s$ konvergiert wie die urspr\"ungliche Folge $\Seqn
s$.

Allerdings sind lineare Transformationen h\"aufig nicht besonders
leistungsf\"ahig, was die praktische N\"utzlichkeit solcher Transformationen
in Konvergenzbeschleunigungs- und Summationsprozessen trotz ihrer
unbestreitbaren theoretischen Vorz\"uge erheblich einschr\"ankt. In den
letzten Jahren wurde deswegen nur noch relativ wenig \"uber lineare
verallgemeinerte Summationsprozesse gearbeitet. Der Schwerpunkt der
aktuellen mathematischen Forschung auf diesem Gebiet und auch des Autors
sind die in der Regel leistungsf\"ahigeren und theoretisch allerdings auch
schwierigeren {\it nichtlinearen} verallgemeinerten Summationsprozesse.
Wenn $G_k^{(n)}$ ein solcher nichtlinearer verallgemeinerter
Summationsproze{\ss} ist, der auf $k + 1$ Eingabedaten wirkt, dann wird es,
wenn $\alpha$ und $\beta$ beliebige reelle Konstanten und $\Seqn s$ und
$\Seqn t$ beliebige Folgen sind, im allgemeinen keine Beziehung vom Typ
von Gl. (6.1-3) geben, sondern es wird normalerweise
$$
\beginAligntags
" G_k^{(n)}
(\alpha s_n + \beta t_n, \alpha s_{n+1} + \beta t_{n+1},
\ldots , \alpha s_{n+k} + \beta t_{n+k}) \\
" \qquad \; \ne \; \alpha \,
G_k^{(n)} (s_n, s_{n+1}, \ldots , s_{n+k}) \, + \,
\beta \, G_k^{(n)} (t_n, t_{n+1}, \ldots , t_{n+k})
\\ \tag
\endAligntags
$$
gelten. Da nichtlineare verallgemeinerte Summationsprozesse $G_k^{(n)}$
in der Regel {\it nichtregul\"ar} sind, kann man au{\ss}erdem nicht erwarten,
da{\ss} die Beziehungen
$$
\lim_{n \to \infty} G_k^{(n)} (s_n, s_{n+1}, \ldots , s_{n+k})
\; = \; s \, , \qquad k \in \N \, ,
\tag
$$
und
$$
\lim_{k \to \infty} G_k^{(n)} (s_n, s_{n+1}, \ldots , s_{n+k})
\; = \; s \, , \qquad n \in \N_0 \, ,
\tag
$$
f\"ur beliebige konvergente Folgen $\Seqn s$ notwendigerweise erf\"ullt
sind.

Nichtlinearit\"at und Nichtregularit\"at sind unangenehme Komplikationen,
die eine theoretische Analyse der Eigenschaften verallgemeinerter
Summationsprozesse erheblich erschweren, und die man instinktiv lieber
vermeiden w\"urde. Trotzdem sind Nichtlinearit\"at und Nichtregularit\"at in
der Praxis {\it unverzichtbar}, da die oft beeindruckenden Erfolge
verallgemeinerter Summationsprozesse eine direkte Konsequenz ihrer
Nichtlinearit\"at und Nichtregularit\"at sind.

Man mu{\ss} jetzt versuchen, eine allgemeine Eigenschaft aller in dieser
Arbeit vorkommenden verallgemeinerten Summationsprozesse $G_k^{(n)}$ zu
finden, die -- obwohl weniger restriktiv als Linearit\"at -- trotzdem eine
theoretische Analyse der Konvergenzeigenschaften erm\"oglicht. Diese
Eigenschaft ist die {\it Translationsinvarianz}. Wenn $G_k^{(n)}$ ein
translationsinvarianter verallgemeinerter Summationsproze{\ss} ist, der auf
$k + 1$ Folgenelemente $s_n$, $s_{n+1}$, $\ldots$ , $s_{n+k}$ wirkt, und
wenn $\alpha \ne 0$ und $t$ zwei reelle Zahlen sind, so gilt:
$$
G_k^{(n)} (\alpha s_n + t, \alpha s_{n+1} + t, \ldots ,
\alpha s_{n+k} + t) \; = \;
\alpha \, G_k^{(n)} (s_n, s_{n+1}, \ldots , s_{n+k}) \, + \, t \, .
\tag
$$
Ein Vergleich mit Gl.~(6.1-3) zeigt, da{\ss} ein linearer verallgemeinerter
Summationsproze{\ss} auto\-matisch translationsinvariant ist. In Abschnitt
1.4 von Brezinski und Redivo Zaglia [1991] werden Eigenschaften
translationsinvarianter verallgemeinerter Summationsprozesse ausf\"uhrlich
diskutiert.

Mit Hilfe der Translationsinvarianz ist eine theoretische Analyse der
Konvergenzeigenschaften zahlreicher nichtlinearer verallgemeinerter
Summationsprozesse m\"oglich. Sei $\Seqn s$ eine Folge, die entweder
gegen den Grenzwert $s$ konvergiert oder die gegen den
verallgemeinerten Grenzwert $s$ summiert werden kann. Dann erf\"ullt ein
translationsinvarianter verallgemeinerter Summationsproze{\ss}
$G_k^{(n)}$, der auf $k + 1$ Eingabedaten wirkt, die folgende Beziehung:
$$
G_k^{(n)} (s_n, s_{n+1}, \ldots , s_{n+k}) \; = \; s \, + \,
G_k^{(n)} (s_n - s, s_{n+1} - s, \ldots , s_{n+k} - s) \, .
\tag
$$
Der Ausdruck auf der rechten Seite von Gl. (6.1-8) ist ein Ma{\ss} daf\"ur,
wie gut ein verallgemeinerter Summationsproze{\ss} $G_k^{(n)}$ den
(verallgemeinerten) Grenzwert $s$ einer Folge $\Seqn s$ approximiert.

Wenn man in der Lage ist, die Gr\"o{\ss}e des Transformationsfehlers in Gl.
(6.1-8) in Abh\"angigkeit von $k$ und $n$ abzusch\"atzen, kann man
Konvergenzbeschleunigungs- und Summationsprozesse theoretisch
analysieren, ohne den (verallgemeinerten) Grenzwert $s$ kennen zu
m\"ussen. Allerdings ben\"otigt man dazu noch eine etwas genauere
Terminologie. So wird in der mathematischen Literatur (vergleiche
beispielsweise Wimp [1981, S. 3]) eine Folge ${\cal W} = \{ (n_j, k_j)
\}_{j=0}^{\infty}$ von geordneten Paaren ganzer Zahlen $n_j, k_j \in
\N_0$ {\it Weg} genannt, wenn die Folge die Anfangswerte $n_0 = k_0 = 0$
besitzt, und wenn f\"ur alle $j\in \N_0$ sowohl $n_{j+1} \ge n_j$ als auch
$k_{j+1} \ge k_j$ gilt. Au{\ss}erdem mu{\ss} f\"ur jedes $j\in \N_0$ mindestens
eine der beiden Beziehungen $n_{j+1} = n_j + 1$ und $k_{j+1} = k_j + 1$
erf\"ullt sein. Offensichtlich impliziert $j \to \infty$ immer $n_j + k_j
\to \infty$.

Wege, bei denen $k_j$ f\"ur $j \to \infty$ konstant wird, werden {\it
vertikal} genannt, und Wege, bei denen $n_j$ f\"ur $j \to \infty$
konstant wird, werden {\it horizontal} genannt.

Der folgende Satz, der leicht mit Hilfe von Gl. (6.1-8) bewiesen werden
kann, ist die Grundlage einer theoretischen Analyse von
Summationsprozessen [Weniger 1989, Theorem 13-3]:

\medskip

\beginEinzug \sl \parindent = 0 pt

\Auszug {\bf Satz 6-1:} Ein verallgemeinerter Summationsproze{\ss}
$G_k^{(n)} (s_n, s_{n+1}, \ldots, s_{n+k})$ mit $k, n \in \N_0$ sei
translationsinvariant gem\"a{\ss} Gl. (6.1-7). Dann ist
$$
\lim_{j \to \infty} \; G_{k_j}^{({n_j})}
(s_{n_j} - s, s_{n_j + 1} - s, \ldots, s_{n_j + k_j} - s) \; = \; 0
\tag
$$
eine notwendige und hinreichende Bedingung, da{\ss} dieser verallgemeinerte
Summationsproze{\ss} auf einem Weg ${\cal W} = \{ (n_j, k_j)
\}_{j=0}^{\infty}$ eine divergente Folge $\Seqn s$ zu $s$ summieren
kann.

\endEinzug

\medskip

Mit Hilfe eines typischen $2 \epsilon$-Beweises kann man leicht zeigen,
da{\ss} der verallgemeinerte Grenzwert $s$ einer divergenten Folge
$\Seqn s$ auf einen bestimmten Weg ${\cal W} =
\{ (n_j, k_j) \}_{j=0}^{\infty}$ eindeutig bestimmt ist. Man kann aber
keine Aussagen machen, ob ein verallgemeinerter Summationsproze{\ss}
$G_k^{(n)}$ auch auf verschiedenen Wegen einer divergenten Folge $\Seqn
s$ einen eindeutig bestimmten verallgemeinerten Grenzwert $s$ zuordnet.
Man kann also nicht {\it a priori} ausschlie{\ss}en, da{\ss} ein
verallgemeinerter Summationsproze{\ss} $G_k^{(n)}$ einer divergenten Folge
$\Seqn s$ auf einem Weg ${\cal W}$ einen verallgemeinerten Grenzwert $s$
und auf einem anderen Weg ${\cal W}'$ einen anderen verallgemeinerten
Grenzwert $s' \ne s$ zuordnet. Bei Summationsprozessen wird man in der
Praxis aber immer horizontale Wege ${\cal W} = \{ (n_j, k_j)
\}_{j=0}^{\infty}$ verwenden, bei denen $n_j$ f\"ur $j \to \infty$
konstant wird.

Satz 6-1 kann nat\"urlich leicht so umformuliert werden, da{\ss} er auch zur
theoretischen Analyse von Konvergenzbeschleunigungsprozessen verwendet
werden kann [Weniger 1989, Theorem 13-4]:

\medskip

\beginEinzug \sl \parindent = 0 pt

\Auszug {\bf Satz 6-2:} Ein verallgemeinerter Summationsproze{\ss}
$G_k^{(n)} (s_n, s_{n+1}, \ldots, s_{n+k})$ mit $k, n \in \N_0$ sei
translationsinvariant gem\"a{\ss} Gl. (6.1-7), und $\Seqn s$ sei eine Folge,
die gegen einen Grenzwert $s$ konvergiert. Dann ist
$$
\lim_{j \to \infty} \; G_{k_j}^{({n_j})}
(s_{n_j} - s, s_{n_j + 1} - s, \ldots, s_{n_j + k_j} - s) \; = \; 0
\tag
$$
eine notwendige und hinreichende Bedingung, da{\ss} dieser verallgemeinerte
Summationsproze{\ss} $G_k^{(n)}$ auf einem Weg ${\cal W} = \{ (n_j, k_j)
\}_{j=0}^{\infty}$ gegen den gleichen Grenzwert $s$ wie die Folge $\Seqn
s$ der Eingabedaten konvergiert,
$$
\lim_{j \to \infty} \;
G_{k_j}^{({n_j})} (s_{n_j}, s_{n_j + 1}, \ldots, s_{n_j + k_j})
\; = \; s \, .
\tag
$$

\endEinzug

\medskip

\Abschnitt Modifikationen der Konvergenztheorie von Germain-Bonne

\smallskip

\aktTag = 0

Der erste erfolgreiche Versuch, eine allgemeine Theorie der Regularit\"at
verallgemeinerter Summationsprozesse und ihrer F\"ahigkeit zur
Konvergenzbeschleunigung zu entwickeln, geht auf Germain-Bonne [1973]
zur\"uck. Germain-Bonne interpretierte verallgemeinerte Summationsprozesse
$G_k$ mit $k \in \N_0$ als Funktionen, die auf $(k + 1)$-dimensionalen
Vektoren ${\bf x} = (x_1, x_2, \ldots, x_{k+1})$ definiert sind. Wenn
die Folgen $\Seqn s$, die transformiert werden sollen, reell sind, so
sind die von Germain-Bonne behandelten verallgemeinerten
Summationsprozesse also Funktionen des Typs $G_k : \R^{k+1} \to \R$.

Weiterhin nahm Germain-Bonne an, da{\ss} verallgemeinerte Summationsprozesse
$G_k$ einige sehr allgemeine und wenig einschr\"ankende Eigenschaften wie
{\it Stetigkeit\/}, {\it Homogenit\"at\/} und {\it Translativit\"at\/}
zumindest f\"ur geeignete Teilmengen des $\R^{k+1}$ besitzen. Auf der
Basis dieser sehr allgemeinen Postulate gelang es Germain-Bonne,
Bedingungen zu formulieren, welche die Regularit\"at eines
verallgemeinerten Summationsprozesses garantieren. Au{\ss}erdem gelang es
Germain-Bonne, ein allgemeines Kriterium zu formulieren, mit dessen
Hilfe man entscheiden kann, ob ein verallgemeinerter Summationsproze{\ss}
$G_k$ lineare Konvergenz beschleunigen kann. Eine gute Behandlung der
Theorie von Germain-Bonne [1973] findet man in dem Buch von Wimp [1981,
S. 101 - 105].

Allerdings ist die urspr\"ungliche Variante der Theorie von Germain-Bonne
in vielen F\"allen nicht anwendbar. Der Grund daf\"ur ist, da{\ss} Germain-Bonne
ausschlie{\ss}lich verallgemeinerte Summationsprozesse $G_k$ behandelt, die
nur {\it implizit} \"uber die $k+1$ Eingabedaten $s_n, s_{n+1}, \ldots,
s_{n+k}$ von $n$ abh\"angen, aber nicht {\it explizit}. Demzufolge kann
die urspr\"ungliche Variante der Theorie von Germain-Bonne nur bei
verallgemeinerten Summationsprozessen wie dem Wynnschen
$\epsilon$-Algorithmus, Gl. (2.4-10), dem iterierten Aitkenschen
$\Delta^2$-Proze{\ss}, Gl. (3.3-8), oder dem Brezinskischen
$\theta$-Algorithmus, Gl. (4.4-13), deren Rekursionsschemata nicht
explizit von $n$ abh\"angen, verwendet werden.

Es gibt aber zahlreiche verallgemeinerte Summationsprozesse, die eng
verwandt sind mit dem iterierten Aitkenschen $\Delta^2$-Proze{\ss}, Gl.
(3.3-8), deren Rekursionsschemata aber {\it explizit} von $n$ abh\"angen
[Weniger 1989, Abschnitt 11.2; Weniger 1991]. Andere verallgemeinerte
Summationsprozesse, die explizit von $n$ abh\"angen, sind die Varianten
von ${\cal L}_{k}^{(n)} (\zeta, s_n, \omega_n)$, Gl. (5.2-6), ${\cal
S}_{k}^{(n)} (\zeta, s_n, \omega_n)$, Gl. (5.4-6), und ${\cal
M}_{k}^{(n)} (\xi, s_n, \omega_n)$, Gl. (5.5-8), die man erh\"alt, wenn
man die Restsummenabsch\"atzungen (5.2-12), (5.2-14), (5.2-16), (5.2-19)
beziehungsweise (5.5-12) verwendet. Demzufolge kann die urspr\"ungliche
Variante der Theorie von Germain-Bonne nicht zur theoretischen Analyse
dieser Transformationen verwendet werden. In Abschnitt 12.1 von Weniger
[1989] wurde gezeigt, da{\ss} man diese Theorie aber leicht so modifizieren
kann, da{\ss} sie auch f\"ur verallgemeinerte Summationsprozesse anwendbar
wird, deren Rekursionsschemata explizit von $n$ abh\"angen.

In diesem Abschnitt soll diese Variante der Theorie von Germain-Bonne
[1973] beschrieben werden. Dabei werden haupts\"achlich S\"atze angegeben
und ihre Implikationen kurz diskutiert. Die Beweise dieser S\"atze und
ausf\"uhrlichere Diskussionen findet man in Abschnitt 12.1 von Weniger
[1989].

Um bei der formalen Theorie der Konvergenzbeschleunigung von
Germain-Bonne [1973] Aussagen \"uber die Regularit\"at und das
Beschleunigungsverm\"ogen verallgemeinerter Summationsprozesse machen zu
k\"onnen, ist es notwendig, den Definitionsbereich der verallgemeinerten
Summationsprozesse auf geeignete Weise einzuschr\"anken. In diesem
Zusammenhang ist es vorteilhaft, zuerst die folgenden Mengensymbole zu
definieren:

\medskip

\beginBeschreibung \zu \Laenge{$\H^n$ :} {\sl

\item{$\F^n$ \hfill :} Menge der Vektoren $(x_1, \ldots , x_n) \in
\R^n$ mit von Null verschiedenen Komponenten, d.~h., es gilt $x_j \neq
0$ f\"ur alle $j = 1, 2, \ldots , n$.

\item{$\D^n$ \hfill :} Menge der Vektoren $(x_1, \ldots , x_n) \in
\R^n$ mit untereinander verschiedenen Komponenten, d.~h., $i \neq j$
impliziert $x_i \neq x_j$ f\"ur alle $i,j = 1,2,\ldots , n$.

\item{$\H^n$ \hfill :} Durchschnitt von $\F^n$ und $\D^n$, d.~h., die
Menge der Vektoren $(x_1, \ldots , x_n) \in \R^n$ mit von Null
verschiedenen und untereinander verschiedenen Komponenten.}

\endBeschreibung

\medskip

Um die Theorie von Germain-Bonne [1973] so zu modifizieren, da{\ss} sie
auch auf verallgemeinerte Summationsprozesse $G_k^{(n)}$ angewendet
werden kann, die explizit von $n$ abh\"angen, nehmen wir an, da{\ss} ein
verallgemeinerter Summationsproze{\ss} $G_k^{(n)}$ f\"ur festes $k \in \N_0$
eine Funktion ist, die f\"ur Vektoren ${\bf x} = (x_1, x_2, \ldots,
x_{k+1}) \in \R^{k+1}$ definiert ist, und die explizit von $n \in \N_0$
abh\"angen kann. Au{\ss}erdem nehmen wir noch an, da{\ss} ein solcher
verallgemeinerter Summationsproze{\ss} $G_k^{(n)} : \R^{k+1} \to \R$ f\"ur
festes $k \in \N_0$ und f\"ur alle $n \in \N_0$ die folgenden
Eigenschaften besitzt [Weniger 1989, Abschnitt 12.1]:

\medskip

\beginBeschreibung \zu \Laenge{(H-0):} \sl

\item {(H-0):} $G_k^{(n)}$ ist definiert und stetig auf einer
Untermenge ${\bf X}^{(n)}$ des $\R^{k+1}$.

\item {(H-1):} $G_k^{(n)}$ ist eine homogene Funktion vom Grade Eins.
Demzufolge erf\"ullt $G_k^{(n)}$ f\"ur beliebige Vektoren ${\bf x} \in {\bf
X}^{(n)}$ und f\"ur alle $\lambda \in \R$, bei denen auch $\lambda {\bf
x}$ immer noch ein Element von ${\bf X}^{(n)}$ ist, die folgende
Beziehung:
$$
G_k^{(n)} (\lambda x_1, \lambda x_2, \ldots, \lambda x_{k+1})
\; = \; \lambda \> G_k^{(n)} (x_1, x_2, \ldots, x_{k+1}) \, .
\tag
$$
\item {(H-2):} $G_k^{(n)}$ ist translationsinvariant. F\"ur beliebige $t
\in \R$ und f\"ur beliebige Vektoren ${\bf x} \in {\bf X}^{(n)}$ gilt
also:
$$
G_k^{(n)} (x_1 + t, x_2 + t, \ldots, x_{k+1} + t)
\; = \; G_k^{(n)} (x_1, x_2, \ldots, x_{k+1}) \; + \; t \, .
\tag
$$
\item {(H-3):} Es existiert eine Teilmenge ${\bf X}^{(\infty)}$ des
$\R^{k+1}$, so da{\ss} der durch den Grenzproze{\ss} $n \to \infty$ definierte
verallgemeinerte Summationsproze{\ss}
$$
G_k^{(\infty)} (x_1, x_2, \ldots, x_{k+1}) \; = \;
\lim_{n \to \infty} \> G_k^{(n)} (x_1, x_2, \ldots, x_{k+1})
\tag
$$
\item {} f\"ur jeden Vektor ${\bf x}$ $=$ $(x_1, x_2, \ldots, x_{k+1})
\in {\bf X}^{(\infty)}$ eindeutig definiert und stetig ist. Au{\ss}erdem
wird angenommen, da{\ss} der Grenzwert $G_k^{(\infty)}$ ebenfalls homogen
und translationsinvariant gem\"a{\ss} (H-1) und (H-2) ist.

\endBeschreibung

\medskip

Die ersten drei Postulate (H-0) -- (H-2) sind im wesentlichen identisch
mit analogen Postulaten von Germain-Bonne [1973]. Der Hauptunterschied
besteht darin, da{\ss} hier der verallgemeinerte Summationsproze{\ss}
$G_k^{(n)}$ explizit von $n$ abh\"angen kann und nicht nur implizit \"uber
die $k+1$ Eingabedaten $s_n$, $s_{n+1}$, $\ldots$, $s_{n+k}$. Da einige
wesentliche Resultate der Theorie von Germain-Bonne auf dem Verhalten
der nicht explizit $n$-abh\"angigen Transformation $G_k (s_n, s_{n+1},
\ldots, s_{n+k})$ f\"ur $n \to \infty$ basieren, mu{\ss} man bei explizit
$n$-abh\"angigen verallgemeinerten Summationsprozessen zus\"atzlich
fordern, da{\ss} Postulat (H-3) gilt. In Abschnitt 12.1 von Weniger [1989]
findet man eine detailliertere Diskussion der Postulate (H-0) -- (H-3).
Dort ist auch das folgende Satz bewiesen [Weniger 1989, Theorem 12-1]:

\medskip

\beginEinzug \sl \parindent = 0 pt

\Auszug {\bf Satz 6-3:} Wenn ein verallgemeinerter Summationsproze{\ss}
$G_k^{(n)}$ f\"ur alle $n \in \N_0$ eine stetige Funktion auf $\R^{k+1}$
ist, und wenn der Grenzwert $G_k^{(\infty)}$ ebenfalls auf $\R^{k+1}$
stetig ist und au{\ss}erdem (H-1) und (H-2) erf\"ullt, dann ist $G_k^{(n)}$
f\"ur jede konvergente Folge $\Seqn s$ regul\"ar.

\endEinzug

\medskip

Allerdings ist Satz 6-3 bei nichtlinearen verallgemeinerten
Summationsprozessen, die ja normalerweise nichtregul\"ar sind, nicht
besonders hilfreich. Eine nichtlineare Transformation $G_k^{(n)}$ ist
normalerweise eine rationale Funktion der $k+1$ Eingabedaten $s_n,
s_{n+1}, \ldots, s_{n+k}$. Da rationale Funktionen Pole haben, kann man
nicht erwarten, da{\ss} nichtlineare verallgemeinerte Summationsprozesse auf
$\R^{k+1}$ stetig sind. Dementsprechend kann man bei einer nichtlinearen
Transformation auch nicht erwarten, da{\ss} die Konvergenz einer beliebigen
Folge $\Seqn s$ die Konvergenz der transformierten Folge $\Seqn {s'}$
impliziert, und wenn die transformierte Folge doch konvergiert, dann
kann man nicht erwarten, da{\ss} $\Seqn s$ und $\Seqn {s'}$ notwendigerweise
gegen den gleichen Grenzwert konvergieren. Der folgende Satz [Weniger
1989, Theorem 12-2] zeigt aber, da{\ss} verallgemeinerte Summationsprozesse
$G_k^{(n)}$ dennoch gewisse Stetigkeitseigenschaften besitzen. Dazu mu{\ss}
man aber die Menge der zul\"assigen Eingabedaten auf geeignete Weise
einschr\"anken [Weniger 1989, Theorem 12-2].

\medskip

\beginEinzug \sl \parindent = 0 pt

\Auszug {\bf Satz 6-4:} Sei $\D^{k+1}$ die Menge der Vektoren ${\bf x}
\in \R^{k+1}$ mit untereinander verschiedenen Komponenten. Jeder
verallgemeinerte Summationsproze{\ss} $G_k^{(n)}$, der f\"ur alle Vektoren
${\bf x}\in \D^{k+1}$ eine stetige Funktion ist und au{\ss}erdem (H-1) und
(H-2) erf\"ullt, kann auf folgende Weise dargestellt werden:
$$
G_k^{(n)} (x_1, x_2, \ldots, x_{k+1}) \; = \;
x_1 \, + \, (x_2 - x_1) \>
g_k^{(n)} \biggl( \frac {x_3 - x_2}{x_2 - x_1}, \ldots ,
\frac {x_{k+1} - x_k}{x_k - x_{k-1}} \biggr) \, .
\tag
$$
Dabei ist der assoziierte verallgemeinerte Summationsproze{\ss}
$g_k^{(n)}$, der auf einer Teilmenge von $\F^{k-1}$ (der Menge der
Vektoren ${\bf y} \in \R^{k-1}$ mit von Null verschiedenen Komponenten)
definiert und stetig ist, gegeben durch
$$
\beginAligntags
" g_k^{(n)} \biggl( \frac {x_3 - x_2}{x_2 - x_1}, \ldots ,
\frac {x_{k+1} - x_k}{x_k - x_{k-1}} \biggr) \\
" \qquad \; = \;
G_k^{(n)} \biggl( 0, 1, 1 + \frac {x_3 - x_2}{x_2 - x_1},
\ldots , \sum_{j=0}^{k-1} \> \prod_{i=0}^{j-1}
\frac {x_{i+3} - x_{i+2}}{x_{i+2} - x_{i+1}} \biggr) \, .
\\ \tag
\endAligntags
$$

\endEinzug

\medskip

Mit Hilfe dieses Satzes kann man diejenigen konvergenten Folgen
$\Seqn s$ charakterisieren, f\"ur die ein verallgemeinerter
Summationsproze{\ss} $G_k^{(n)}$, der (H-0) -- (H-3) erf\"ullt, {\it
regul\"ar} ist [Weniger 1989, Theorem 12-3].

\medskip

\beginEinzug \sl \parindent = 0 pt

\Auszug {\bf Satz 6-5:} Sei $G_k^{(n)}$ ein verallgemeinerter
Summationsproze{\ss}, der die Bedingungen von Satz 6-4 erf\"ullt. Es
existiert also gem\"a{\ss} Gln. (6.2-4) und (6.2-5) ein assoziierter
verallgemeinerter Summationsproze{\ss} $g_k^{(n)}$, der definiert und
stetig ist auf derjenigen Teilmenge der Vektoren ${\bf y} = (y_1, y_2,
\ldots, y_{k-1}) \in \F^{k-1}$, die durch die Vektoren ${\bf x} = (x_1,
x_2, \ldots, x_{k+1}) \in \D^{k+1}$ gem\"a{\ss}
$$
y_{\mu} \; = \; (x_{\mu+2} - x_{\mu+1}) \, / \,
(x_{\mu+1} - x_{\mu}) \, , \qquad 1 \le \mu \le k - 1 \, ,
\tag
$$
erzeugt werden. Wenn der Grenzwert
$$
g_k^{(\infty)} (y_1, y_2, \ldots , y_{k-1}) \; = \;
\lim_{n \to \infty} g_k^{(n)} (y_1, y_2, \ldots , y_{k-1})
\tag
$$
des assoziierten verallgemeinerten Summationsprozesses $g_k^{(n)}$ auf
der gleichen Teilmenge von $\F^{k-1}$ definiert und stetig ist wie
$g_k^{(n)}$, dann ist $G_k^{(n)}$ f\"ur jede konvergente Folge $\Seqn s$
regul\"ar, welche die folgenden Eigenschaften besitzt:

\beginBeschreibung \zu \Laenge{(ii):}

\item{(i) \hfill :} F\"ur hinreichend gro{\ss}e Werte von $n \in \N_0$ sind
die Folgenelemente $s_n$, $s_{n+1}$, $s_{n+2}$, $\ldots$ alle
untereinander verschieden.
\item{(ii):} F\"ur hinreichend gro{\ss}e Werte von $n \in \N_0$ erf\"ullen die
Quotienten $\Delta s_{n+1} / \Delta s_n$ die Ungleichung
$$
c \le \vert \Delta s_{n+1} / \Delta s_n \vert \le c^{\prime} \, ,
\qquad 0 < c < c^{\prime} < \infty \, .
\tag
$$

\endBeschreibung

\endEinzug

\medskip

Mit Hilfe von Satz 6-5 kann man Bedingungen formulieren, die
garantieren, da{\ss} ein verallgemeinerter Summationsproze{\ss} $G_k^{(n)}$,
der explizit von $n$ abh\"angt, lineare Konvergenz beschleunigt [Weniger
1989, Theorem 12-4]:

\medskip

\beginEinzug \sl \parindent = 0 pt

\Auszug {\bf Satz 6-6:} $\Seqn s$ sei eine Folge, die linear gegen
ihren Grenzwert $s$ konvergiert, d.~h.,
$$
\lim_{n \to \infty} \> \frac {s_{n+1} - s}{s_n - s} \; = \; \rho
\, , \qquad 0 < |\rho| < 1 \, .
\tag
$$
Eine notwendige und hinreichende Bedingung, da{\ss} ein verallgemeinerter
Summationsproze{\ss} $G_k^{(n)} : \R^{k+1} \to \R$ die Konvergenz von $\Seqn
s$ beschleunigt, ist, da{\ss} der in Gl. (6.2-5) definierte assoziierte
verallgemeinerte Summationsproze{\ss} $g_k^{(n)}$ die folgende Beziehung
erf\"ullt:
$$
\lim_{n \to \infty} \>
g_k^{(n)} (\rho_n, \rho_{n+1}, \ldots, \rho_{n+k-2}) \; = \;
g_k^{(\infty)} (\rho, \rho, \ldots, \rho) \; = \;
\frac {1} {1 - \rho} \, .
\tag
$$
Dabei ist $\Seqn {\rho}$ eine beliebige Folge, die gegen $\rho$
konvergiert. Die notwendige und hinreichende Bedingung (6.2-10) kann
auch auf folgende Weise formuliert werden:
$$
\beginAligntags
" \lim_{n \to \infty} G_k^{(n)} \biggl(0, 1, 1+\rho_n , \ldots,
\sum_{j=0}^{k-1} \, \prod_{i=0}^{j-1} \, \rho_{n+i} \biggr) \\
" \qquad \; = \; G_k^{(\infty)} \biggl(0, 1, 1+\rho, \ldots,
\sum_{j=0}^{k-1}\rho^j \biggr) \; = \; \frac {1} {1 - \rho} \, ,
\qquad 0 < \vert \rho \vert < 1 \, .
\\ \tag
\endAligntags
$$

\endEinzug

\medskip

Die Folge $0, 1, 1 + \rho, 1 + \rho^2, \ldots$ , auf die
$G_k^{(\infty)}$ in Gl. (6.2-11) wirkt, ist abgesehen vom ersten Element
und von einer Indexverschiebung identisch mit den Partialsummen der
geometrischen Reihe. Man erkennt dies sofort, wenn man die Elemente
dieser Folge von Partialsummen auf folgende Weise schreibt:
$$
\sigma_n (\rho) \; = \; \frac {1 - \rho^n} {1 - \rho} \; = \;
\sum_{\nu=0}^{n-1} \, \rho^{\nu} \, , \qquad 0 < |\rho| < 1 \, ,
\quad n \in \N_0 \, .
\tag
$$
Diese indexverschobene Folge $\Seq {\sigma_n (\rho)} {n=0}$ konvergiert
f\"ur $n \to \infty$ offensichtlich gegen $1 / (1 - \rho)$. Aus Satz 6-6
folgt also, da{\ss} ein verallgemeinerter Summationsproze{\ss} $G_k^{(n)}$ nur
dann lineare Konvergenz beschleunigen kann, wenn $G_k^{(\infty)}$ exakt
ist f\"ur die indexverschobene Folge $\Seq {\sigma_n (\rho)} {n=0}$ der
Partialsummen der geometrischen Reihe. Die Exaktheit eines
verallgemeinerten Summationsprozesses $G_k^{(n)}$ f\"ur die Partialsummen
der geometrischen Reihe und die F\"ahigkeit, lineare Konvergenz zu
beschleunigen, sind also eng miteinander verkn\"upft sind.

Mit Hilfe der in diesem Abschnitt angegebenen S\"atze kann die Regularit\"at
und das Beschleunigungsverm\"ogen zahlreicher verallgemeinerter
Summationsprozesse analysiert werden. So wird beispielsweise in
Theoremen 12.9 und 12.10 von Weniger [1989] gezeigt, da{\ss} $u_k^{(n)}
(\zeta, s_n)$, Gl. (5.2-13), $t_k^{(n)} (\zeta, s_n)$, Gl. (5.2-15),
$d_k^{(n)} (\zeta, s_n)$, Gl. (5.2-18), $v_k^{(n)} (\zeta, s_n)$, Gl.
(5.2-11), $y_k^{(n)} (\zeta, s_n)$, Gl. (5.4-11), $\tau_k^{(n)} (\zeta,
s_n)$, Gl. (5.4-12),$\delta_k^{(n)} (\zeta, s_n)$, Gl. (5.4-13),
$\phi_k^{(n)} (\zeta, s_n)$, Gl. (5.4-14), $Y_k^{(n)} (\xi, s_n)$, Gl.
(9.4-13), und $T_k^{(n)} (\xi, s_n)$, Gl. (9.4-14), $\Delta_k^{(n)}
(\xi, s_n)$, Gl. (5.5-15), und $\Phi_k^{(n)} (\xi, s_n)$, Gl. (5.5-16),
exakt sind f\"ur die geometrische Reihe (2.1-3) und auch lineare
Konvergenz beschleunigen.

\medskip

\Abschnitt Transformationen mit expliziten Restsummenabsch\"atzungen

\smallskip

\aktTag = 0

Die Restsummenabsch\"atzungen $\Seqn \omega$ spielen eine ganz
entscheidende Rolle f\"ur eine erfolgreiche Verwendung der
verallgemeinerten Summationsprozesse ${\cal L}_{k}^{(n)} (\zeta, s_n,
\omega_n)$, Gl. (5.2-6), ${\cal S}_{k}^{(n)} (\zeta, s_n, \omega_n)$,
Gl. (5.4-6), und ${\cal M}_{k}^{(n)} (\xi, s_n, \omega_n)$, Gl. (5.5-8),
in Konvergenzbeschleunigungs- und Summationsverfahren. Es ist aus
zahlreichen Rechnungen bekannt, da{\ss} die Restsummenabsch\"atzungen
(5.2-12), (5.2-14), (5.2-16), (5.2-19) beziehungsweise (5.5-12) trotz
ihrer Einfachheit oft erstaunlich gute Ergebnisse liefern. Trotzdem kann
man aber nicht erwarten, da{\ss} diese einfachen Restsummenabsch\"atzungen
immer befriedigende Ergebnisse liefern werden, und es gibt Situationen,
in denen die Verwendung alternativer Restsummenabsch\"atzungen $\Seqn
\omega$ vorzuziehen w\"are.

So ist es manchmal m\"oglich, numerisch leicht zug\"angliche Approximationen
der exakten Summationsreste $\Seqn r$ einer unendlichen Reihe zu finden
und diese mit guten Ergebnissen in Konvergenzbeschleunigungsprozessen zu
verwenden [Weniger 1989, Tabelle 14-6; Steinborn und Weniger 1990,
Tabelle 1]. In einem solchen Fall w\"urde man gerne vorher wissen, welche
Eigenschaften die beiden Folgen $\Seqn s$ und $\Seqn \omega$ haben
m\"ussen, damit ein solcher verallgemeinerter Summationsproze{\ss} regul\"ar
ist, und unter welchen Voraussetzungen die Konvergenz einer gegebenen
Folge $\Seqn s$ von Partialsummen beschleunigt wird.

Wenn man solche explizite Restsummenabsch\"atzungen $\Seqn \omega$
verwendet, die nicht aus den Partialsummen $\Seqn s$ berechnet wurden,
sind die verallgemeinerten Summationsprozesse ${\cal L}_{k}^{(n)}
(\zeta, s_n, \omega_n)$, ${\cal S}_{k}^{(n)} (\zeta, s_n, \omega_n)$ und
${\cal M}_{k}^{(n)} (\xi, s_n, \omega_n)$ Funktionen von $2 k + 2$
Eingabedaten, n\"amlich von $k + 1$ Partialsummen $s_n$, $s_{n+1}$,
$\ldots$ , $s_{n+k}$ und von $k + 1$ Restsummenabsch\"atzungen $\omega_n$,
$\omega_{n+1}$, $\ldots$ , $\omega_{n+k}$. Man kann also weder die
urspr\"ungliche Version der Konvergenztheorie von Germain-Bonne [1973]
noch die im letzten Unterabschnitt beschriebene Modifikation der
Germain-Bonneschen Theorie zur Beantwortung dieser Fragen verwenden. In
Abschnitt 12.3 von Weniger [1989] wurde aber gezeigt, da{\ss} diese Fragen
mit Hilfe einer geeigneten Modifikation der Theorie von Germain-Bonne
zumindest partiell beantwortet werden k\"onnen.

Die verallgemeinerten Summationsprozesse ${\cal L}_{k}^{(n)} (\zeta,
s_n, \omega_n)$, Gl. (5.2-6), ${\cal S}_{k}^{(n)} (\zeta, s_n,
\omega_n)$, Gl. (5.4-6), und ${\cal M}_{k}^{(n)} (\xi, s_n, \omega_n)$,
Gl. (5.5-8), sind alle vom Typ von Gl. (5.6-1). Wenn wir also annehmen,
da{\ss} die Restsummenabsch\"atzungen $\Seqn \omega$ nicht explizit von den
Eingabedaten $\Seqn s$ abh\"angen, sind die oben genannten
Transformationen {\it lineare} Funktionen der $k + 1$ Partialsummen
$s_n$, $s_{n+1}$, $\ldots$ , $s_{n+k}$. Au{\ss}erdem sind sie definiert und
exakt f\"ur konstante Folgen. Aus der Linearit\"at folgt, da{\ss} diese
Transformationen {\it stetige} Funktionen ihrer ersten $k + 1$
Variablen sind, wenn die Elemente der Folge $\Seqn s$ f\"ur alle $n \in
\N_0$ beschr\"ankt sind.

Die Stetigkeit der oben genannten Transformationen bez\"uglich ihrer
zweiten $k + 1$ Variablen $\omega_n$, $\omega_{n+1}$, $\ldots$ ,
$\omega_{n+k}$ ist ein wesentlich schwierigeres Problem. Da die
Restsummenabsch\"atzungen immer in Nennern vorkommen, d\"urfen die Elemente
von $\Seqn \omega$ f\"ur endliche Werte von $n$ nie Null sein. Da die oben
genannten Transformationen vom Typ von Gl.~(5.6-1) und damit auto\-matisch
auch vom Typ von Gl. (5.1-9) sind, mu{\ss} man fordern, da{\ss} der Z\"ahler einer
solchen Transformation, der die $k$-te gewichtete Differenz von
$1/\omega_n$ ist, nicht verschwindet. Eine notwendige, aber
ungl\"ucklicherweise nicht hinreichende Bedingung f\"ur das
Nichtverschwinden des Nenners einer Transformation vom Typ von Gl.
(5.1-9) ist, da{\ss} die Restsummenabsch\"atzungen $\Seqn \omega$ f\"ur jeden
endlichen Wert von $n$ ungleich Null und untereinander verschieden sind.

In diesem Unterabschnitt ist $\Gamma_k^{(n)}$ ein verallgemeinerter
Summationsproze{\ss}, der auf $k+1$ Elemente einer konvergenten Folge
$\Seqn s$ wirkt, und der ebenfalls $k+1$ Elemente einer Folge $\Seqn
\omega$ von Restsummenabsch\"atzungen ben\"otigt. Der obere Index $n$ gibt
dabei an, da{\ss} $\Gamma_k^{(n)}$ explizit von $n$ abh\"angen kann.

F\"ur festes $k \in \N_0$ ist ein verallgemeinerter Summationsproze{\ss}
$\Gamma_k^{(n)}$ also eine Funktion, die explizit von $n \in \N_0$
abh\"angen kann, und die definiert ist f\"ur Vektoren ${\bf x} = (x_1, x_2,
\ldots, x_{k+1}) \in \R^{k+1}$ und ${\bf z} = (z_1, z_2, \ldots,
z_{k+1}) \in \H^{k+1}$. Au{\ss}erdem nehmen wir an, da{\ss} ein solcher
verallgemeinerter Summationsproze{\ss} $\Gamma_k^{(n)} : \R^{k+1} \times
\H^{k+1} \to \R$ f\"ur festes $k \in \N_0$ und f\"ur alle $n \in \N_0$ noch
die folgenden Eigenschaften besitzt:

\medskip

\beginBeschreibung \zu \Laenge{(A-0):} \sl

\item {(A-0):} $\Gamma_k^{(n)}$ ist definiert und stetig auf einer
Teilmenge von $\R^{k+1} \times \H^{k+1}$.
\item {(A-1):} $\Gamma_k^{(n)}$ ist eine homogene Funktion vom Grade
Eins der ersten $k+1$ Variablen und eine homogene Funktion vom Grade
Null der zweiten $k+1$ Variablen. Es gilt also f\"ur alle Vektoren ${\bf
x} \in \R^{k+1}$ und ${\bf z} \in \H^{k+1}$, f\"ur die $\Gamma_k^{(n)}$
definiert und stetig ist, und f\"ur alle $\lambda, \mu \in \R$ mit $\mu
\neq 0$:
$$
\beginAligntags
" \Gamma_k^{(n)}
(\lambda x_1, \lambda x_2, \ldots, \lambda x_{k+1} \Bar
z_1, z_2, \ldots, z_{k+1}) \\
" \qquad \; = \; \lambda \>
\Gamma_k^{(n)} (x_1, x_2, \ldots, x_{k+1} \Bar
z_1, z_2, \ldots, z_{k+1}) \, ,
\erhoehe\aktTag \\ \tag*{\tagnr a}
" \Gamma_k^{(n)}
(x_1, x_2, \ldots, x_{k+1} \Bar
\mu z_1, \mu z_2, \ldots, \mu z_{k+1}) \\
" \qquad \; = \;
\Gamma_k^{(n)} (x_1, x_2, \ldots, x_{k+1} \Bar
z_1, z_2, \ldots, z_{k+1}) \, .
\\ \tag*{\tagform\aktTagnr b}
\endAligntags
$$
\item {(A-2):} $\Gamma_k^{(n)}$ ist linear in den ersten $k+1$
Variablen. F\"ur alle Vektoren ${\bf x}, {\bf y} \in \R^{k+1}$ und ${\bf
z} \in \H^{k+1}$, f\"ur die $\Gamma_k^{(n)}$ definiert und stetig ist,
gilt also:
$$
\beginAligntags
" \Gamma_k^{(n)} (x_1 + y_1, x_2 + y_2, \ldots, x_{k+1} + y_{k+1}
\Bar z_1, z_2, \ldots, z_{k+1}) \\
" \qquad \; = \; \Gamma_k^{(n)} (x_1, x_2, \ldots, x_{k+1} \Bar
z_1, z_2, \ldots, z_{k+1}) \\
" \qquad \; + \;
\Gamma_k^{(n)} (y_1, y_2, \ldots, y_{k+1} \Bar
z_1, z_2, \ldots, z_{k+1}) \, . \qquad
\\ \tag
\endAligntags
$$
\item {(A-3):} Wenn ${\bf c} = (c, c, \ldots, c) \in \R^{k+1}$ ein
Vektor mit konstanten Komponenten ist, und wenn ${\bf z}$ zu der
Teilmenge von $\H^{k+1}$ geh\"ort, f\"ur die $\Gamma_k^{(n)}$ definiert und
stetig ist, dann ist $\Gamma_k^{(n)}$ exakt, d.~h.,
$$
\Gamma_k^{(n)} (c, c, \ldots, c \Bar z_1, z_2, \ldots, z_{k+1})
\; = \; c \, .
\tag
$$
\item {(A-4):} Es existiert eine Teilmenge von $\H^{k+1}$, so da{\ss} f\"ur
alle beschr\"ankten Vektoren ${\bf x} \in \R^{k+1}$ und f\"ur alle Vektoren
${\bf z}$, die zu dieser Teilmenge geh\"oren, der durch den Grenzwert $n
\to \infty$ definierte verallgemeinerte Summationsproze{\ss}
$$
\beginAligntags
" \Gamma_k^{(\infty)} (x_1, x_2, \ldots, x_{k+1} \Bar
z_1, z_2, \ldots, z_{k+1}) \\
" \qquad \; = \; \lim_{n \to \infty} \>
\Gamma_k^{(n)} (x_1, x_2, \ldots, x_{k+1} \Bar
z_1, z_2, \ldots, z_{k+1})
\\ \tag
\endAligntags
$$
\item {} eindeutig definiert und stetig ist auf dieser Teilmenge von
$\R^{k+1} \times \H^{k+1}$. Au{\ss}erdem wird angenommen, da{\ss}
$\Gamma_k^{(\infty)}$ ebenfalls homogen und linear gem\"a{\ss} (A-1) und
(A-2) ist.

\endBeschreibung

\medskip

Mit Hilfe der Annahmen (A-0) - (A-4) kann man auch bei verallgemeinerten
Summationsprozesses, die wie die Levinsche Transformation ${\cal
L}_k^{(n)} (\zeta , s_n, \omega_n)$, Gl. (5.2-6), explizite
Restsummenabsch\"atzungen verwenden, Regularit\"atseigenschaften und die
F\"ahigkeit, lineare Konvergenz zu beschleunigen, theoretisch analysieren.

\"Ahnlich wie im vorherigen Unterabschnitt werden auch hier haupts\"achlich
S\"atze angegeben und ihre Implikationen kurz diskutiert. Die Beweise
dieser S\"atze und ausf\"uhrlichere Diskussionen findet man in Abschnitt
12.3 von Weniger [1989].

Satz 6-4, in dem ein zu $G_k^{(n)}$ assoziierter verallgemeinerter
Summationsproze{\ss} $g_k^{(n)}$ gem\"a{\ss} Gln. (6.2-4) und (6.2-5) eingef\"uhrt
wurde, war ein wesentlicher Zwischenschritt bei der Untersuchung der
Regularit\"atseigenschaften von verallgemeinerten Summationsprozessen
$G_k^{(n)} : \R^{k+1} \to \R$ und ihrer F\"ahigkeit, lineare Konvergenz zu
beschleunigen. Auch bei verallgemeinerten Summationsprozessen
$\Gamma_k^{(n)} : \R^{k+1} \times \H^{k+1} \to \R$ kann ein analoges
Resultat abgeleitet werden [Weniger 1989, Theorem 12-12]:

\medskip

\beginEinzug \sl \parindent = 0 pt

\Auszug {\bf Satz 6-7:} Jeder verallgemeinerte Summationsproze{\ss}
$\Gamma_k^{(n)}$, der f\"ur alle Vektoren ${\bf x}\in \R^{k+1}$ und f\"ur
alle ${\bf z}$, die zu einer geeigneten Teilmenge von $\H^{k+1}$
geh\"oren, definiert und stetig ist und der au{\ss}erdem die Voraussetzungen
(A-0) - (A-4) erf\"ullt, kann auf folgende Weise geschrieben werden:
$$
\beginAligntags
" \Gamma_k^{(n)} (x_1, x_2, \ldots, x_{k+1} \Bar
z_1, z_2, \ldots, z_{k+1}) \\
" \qquad \; = \; x_1 \, + \, z_1 \>
\gamma_k^{(n)} \biggl(\frac {x_2 - x_1} {z_1}, \ldots ,
\frac {x_{k+1} - x_k} {z_k} \biggBar \frac {z_2} {z_1},
\ldots, \frac {z_{k+1}} {z_k} \biggr) \, .
\\ \tag
\endAligntags
$$
Dabei ist der assoziierte verallgemeinerte Summationsproze{\ss}
$\gamma_k^{(n)}$, der definiert und stetig ist auf einer geeigneten
Teilmenge von $\R^k \times \F^k$, gegeben durch
$$
\beginAligntags
" \gamma_k^{(n)} \biggl(\frac {x_2 - x_1} {z_1}, \ldots ,
\frac {x_{k+1} - x_k} {z_k} \biggBar \frac {z_2} {z_1},
\ldots, \frac {z_{k+1}} {z_k} \biggr) \\
" \qquad \; = \;
\Gamma_k^{(n)} \biggl( 0, \frac {x_2 - x_1}{z_1}, \ldots ,
\sum_{j=0}^{k-1} \, \frac {x_{j+2} - x_{j+1}} {z_{j+1}} \,
\prod_{i=0}^{j-1} \, \frac {z_{i+2}}{z_{i+1}} \biggBar
1, \frac {z_2}{z_1}, \ldots,
\prod_{j=0}^{k-1} \, \frac {z_{j+2}}{z_{j+1}} \biggr) \, .
\\ \tag
\endAligntags
$$

\endEinzug

\medskip

Mit Hilfe von Satz 6-7 k\"onnen Bedingungen angegeben werden, die die
Folgen $\Seqn s$ und $\Seqn \omega$ erf\"ullen m\"ussen, damit ein
verallgemeinerter Summationsproze{\ss} $\Gamma_k^{(n)}$, der die
Eigenschaften (A-0) -- (A-4) besitzt, regul\"ar ist [Weniger 1989,
Theorem 12-13].

\medskip

\beginEinzug \sl \parindent = 0 pt

\Auszug {\bf Satz 6-8:} $\Gamma_k^{(n)}$ sei ein verallgemeinerter
Summationsproze{\ss}, der die Voraussetzungen von Satz 6-7 erf\"ullt. Es
existiert also gem\"a{\ss} Gln. (6.3-5) und (6.3-6) ein assoziierter
verallgemeinerter Summationsproze{\ss} $\gamma_k^{(n)}$, der stetig ist
auf einer geeigneten Teilmenge von $\R^k \times \F^k$. Nehmen wir
au{\ss}erdem an, da{\ss} eine Folge $\Seqn s$ gegen einen Grenzwert $s$
konvergiert, und da{\ss} die Elemente der Folge $\Seqn \omega$ der
Restsummenabsch\"atzungen, obwohl sie f\"ur alle endlichen Werte von $n$
von Null verschieden sind, f\"ur $n \to \infty$ gegen Null konvergieren.
Dann ist $\Gamma_k^{(n)}$ regul\"ar, wenn die Elemente von $\Seqn s$ und
$\Seqn \omega$ die folgenden Bedingungen erf\"ullen:

\beginBeschreibung \zu \Laenge{(ii):}

\item{(i) \hfill :} F\"ur hinreichend gro{\ss}e Werte von $n \in \N_0$ sind
die Quotienten $\Delta s_n / \omega_n$ beschr\"ankt, d.~h.,
$$
\vert \Delta s_n / \omega_n \vert \le c \, ,
\qquad 0 \le c < \infty \, .
\tag
$$
\item{(ii):} F\"ur alle beschr\"ankten Vektoren ${\bf y} = (y_1, y_2,
\ldots, y_k) \in \R^k$ ist der assoziierte verallgemeinerte
Summationsproze{\ss} $\gamma_k^{(n)}$ beschr\"ankt f\"ur $n \to \infty$:
$$
\qquad \qquad \lim_{n \to \infty} \> \biggl| \gamma_k^{(n)}
\bigl( y_1, \ldots, y_k \bigBar \omega_{n+1} / \omega_n,
\ldots, \omega_{n+k} / \omega_{n+k-1} \bigr) \biggr|
\; \le \; M \, , \qquad 0 < M < \infty \, .
\tag
$$

\endBeschreibung

\endEinzug

\medskip

Ein Vergleich der S\"atze 6-5 und 6-8 zeigt, da{\ss} hier nur relativ
schwache Voraussetzungen an eine konvergente Folge $\Seqn s$ gestellt
werden m\"ussen, um die Regularit\"at von $\Gamma_k^{(n)}$ zu garantieren.
Die Voraussetzungen, welche die Folge $\Seqn \omega$ zu erf\"ullen hat,
sind aber sehr restriktiv.

Der n\"achste Satz gibt an, unter welchen Voraussetzung ein
verallgemeinerter Summationsproze{\ss} $\Gamma_k^{(n)}$ lineare Konvergenz
beschleunigen kann [Weniger 1989, Theorem 12-14]. Er ist fast v\"ollig
identisch mit dem analogen Satz 6-6, in dem verallgemeinerte
Summationsprozesse $G_k^{(n)}$ behandelt werden. In beiden F\"allen m\"ussen
die Grenzwerte $\Gamma_k^{(\infty)}$ und $G_k^{(\infty)}$ exakt sein f\"ur
die indexverschobene Folge (6.2-12) von Partialsummen der geometrischen
Reihe.

\medskip

\beginEinzug \sl \parindent = 0 pt

\Auszug {\bf Satz 6-9:} Nehmen wir an, da{\ss} die Elemente der Folgen
$\Seqn s$ und $\Seqn \omega$ die Bedingungen
$$
\beginAligntags
" \text{(i)} " : \qquad
" \lim_{n \to \infty} \> s_n \; = \; s \, , \hfill
\\ \tag
" \text{(ii)} " : \qquad " \lim_{n \to \infty} \>
\frac {s_n - s} {\omega_n} \; = \; c \, ,
\qquad c \ne 0 \, , \hfill
\\ \tag
" \text{(iii)} " : \qquad " \lim_{n \to \infty} \>
\frac {\omega_{n+1}} {\omega_n} \; = \; \rho \, ,
\qquad 0 < \vert \rho \vert < 1 \, , \hfill
\\ \tag
\endAligntags
$$
erf\"ullen. Eine notwendige und hinreichende Bedingung, da{\ss} ein
verallgemeinerter Summationsproze{\ss} $\Gamma_k^{(n)}$ die Konvergenz von
$\Seqn s$ beschleunigt, ist, da{\ss} der assoziierte verallgemeinerte
Summationsproze{\ss} $\gamma_k^{(n)}$ die folgende Bedingung erf\"ullt:
$$
\beginAligntags
" \lim_{n \to \infty} \>
\gamma_k^{(n)} (y_n, y_{n+1}, \ldots, y_{n+k-1} \Bar
z_n, z_{n+1}, \ldots, z_{n+k-1}) \\
" \qquad \; = \; \gamma_k^{(\infty)} (y, y, \ldots, y \Bar
z, z, \ldots, z) \; = \;
\frac {y} {1 - z} \, .
\\ \tag
\endAligntags
$$
Dabei sind $\Seqn y$ und $\Seqn z$ im Prinzip beliebige Folgen, die
gegen $y$ beziehungsweise $z$ konvergieren. Die notwendige und
hinreichende Bedingung (6.3-12) kann auch auf folgende Weise formuliert
werden:
$$
\beginAligntags
" \lim_{n \to \infty} \> \Gamma_k^{(n)} \biggl(0, y_n, \ldots,
\, \sum_{j=0}^{k-1} y_{n+j} \, \prod_{i=0}^{j-1} \, z_{n+i} \biggBar
1, z_n, \ldots, \prod_{j=0}^{k-1} \, z_{n+j} \biggr) \\
" \qquad \; = \; \Gamma_k^{(\infty)} \biggl(0, y, \ldots,
y \, \sum_{j=0}^{k-1}\, z^j \biggBar 1, z, \ldots, z^k \biggr)
\; = \; \frac {y} {1 - z} \, ,
\qquad 0 < \vert z \vert < 1 \, .
\\ \tag
\endAligntags
$$

\endEinzug

\medskip

Man sollte hier noch anmerken, da{\ss} alle Restsummenabsch\"atzungen, die
Gl.~(5.1-4) erf\"ullen, automatisch auch Bedingung (ii) erf\"ullen. Das ist
eine weitere Best\"atigung daf\"ur, da{\ss} man die Restsummenabsch\"atzungen so
w\"ahlen sollte, da{\ss} $\omega_n$ proportional zu dem f\"uhrenden Term der
asymptotischen Entwicklung von $s_n - s$ f\"ur $n \to \infty$ ist.

Die ersten $k+1$ Argumente von $\Gamma_k^{(\infty)}$ in Gl. (6.3-13)
sind bis auf den multiplikativen Faktor $y$ identisch mit den Elementen
der Folge $\Seq {\sigma_n (\rho)} {n=0}$, die in Gl. (6.2-12) definiert
wurden. Man kann also die rechte Seite von Gl. (6.3-13) auch auf
folgende Weise schreiben:
$$
\Gamma_k^{(\infty)} \biggl( 0, y, \ldots, y \frac{1-z^k} {1-z}
\biggBar 1,z, \ldots, z^k \biggr) \; = \;
\frac {y}{1 - z} \, , \qquad 0 < | z | < 1 \, .
\tag
$$

Mit Hilfe des n\"achsten Satzes kann man beweisen, da{\ss} die
verallgemeinerten Summationsprozesse ${\cal L}_{k}^{(n)} (\zeta, s_n,
\omega_n)$, Gl. (5.2-6), ${\cal S}_{k}^{(n)} (\zeta, s_n, \omega_n)$,
Gl. (5.4-6), und ${\cal M}_{k}^{(n)} (\xi, s_n, \omega_n)$, Gl. (5.5-8),
lineare Konvergenz beschleunigen, wenn man die Restsummenabsch\"atzungen
so w\"ahlt, da{\ss} $\omega_n$ proportional zu dem f\"uhrenden Term der
asymptotischen Entwicklung von $s_n - s$ f\"ur $n \to \infty$ ist [Weniger
1989, Theorem 12-15].

\medskip

\beginEinzug \sl \parindent = 0 pt

\Auszug {\bf Satz 6-10:} Wir nehmen an, da{\ss} ein verallgemeinerter
Summationsproze{\ss} $T_k^{(n)} (s_n, \omega_n)$ auf folgende Weise
geschrieben werden kann:
$$
T_k^{(n)} (s_n, \omega_n) \; = \; \frac
{\displaystyle
\sum_{j=0}^{k} \; ( - 1)^{j} \; \binom {k} {j} \;
f_k (n+j) \; \frac {s_{n+j}} {\omega_{n+j}} }
{\displaystyle
\sum_{j=0}^{k} \; ( - 1)^{j} \; \binom {k} {j} \;
f_k (n+j) \; \frac {1} {\omega_{n+j}} }
\; , \qquad k,n \in \N_0 \, .
\tag
$$
Wenn die Folgen $\Seqn s$ und $\Seqn \omega$ so wie in Satz 6-9 gew\"ahlt
sind und wenn die Koeffizienten $f_k (n)$ au{\ss}erdem die Bedingung
$$
\lim_{n \to \infty} \; f_k (n) \; = \; 1 \, ,
\qquad k \in \N_0 \, ,
\tag
$$
erf\"ullen, dann beschleunigt $T_k^{(n)} (s_n, \omega_n)$ die Konvergenz
von $\Seqn s$ f\"ur $k \ge 1$.

\endEinzug

\medskip

Mit Hilfe des n\"achsten Satzes kann man beweisen, da{\ss} die Varianten der
verallgemeinerten Summationsprozesse ${\cal L}_{k}^{(n)} (\zeta, s_n,
\omega_n)$, Gl. (5.2-6), ${\cal S}_{k}^{(n)} (\zeta, s_n, \omega_n)$,
Gl. (5.4-6), und ${\cal M}_{k}^{(n)} (\xi, s_n, \omega_n)$, Gl. (5.5-8),
die auf den Restsummenabsch\"atzungen (5.2-12), (5.2-14), (5.2-16)
beziehungsweise (5.5-12) basieren, exakt sind f\"ur die geometrische Reihe
[Weniger 1989, Theorem 12-16].

\medskip

\beginEinzug \sl \parindent = 0 pt

\Auszug {\bf Satz 6-11:} Wir nehmen an, da{\ss} ein verallgemeinerter
Summationsproze{\ss} $T_k^{(n)} (s_n, \omega_n)$ auf folgende Weise
geschrieben werden kann:
$$
T_k^{(n)} (s_n, \omega_n) \; = \; \frac
{\displaystyle
\sum_{j=0}^{k} \; ( - 1)^{j} \; \binom {k} {j} \;
\phi_{k-1} (n+j) \; \frac {s_{n+j}} {\omega_{n+j}} }
{\displaystyle
\sum_{j=0}^{k} \; ( - 1)^{j} \; \binom {k} {j} \;
\phi_{k-1} (n+j) \; \frac {1} {\omega_{n+j}} }
\; , \qquad k,n \in \N_0 \, .
\tag
$$
Wenn die Koeffizienten $\phi_{k-1} (n)$ f\"ur hinreichend gro{\ss}e Werte von
$k \in \N$ Polynome vom Grade $\le k - 1$ in $n$ sind, dann ist
$T_k^{(n)} (s_n, \omega_n)$ f\"ur hinreichend gro{\ss}e Werte von $k \in \N$
exakt f\"ur die Partialsummen der geometrischen Reihe (2.1-3), wenn die
Restsummenabsch\"atzungen gem\"a{\ss} $\omega_n = z^{n + \alpha}$ mit $\alpha
\in \R$ gew\"ahlt werden.

\endEinzug

\medskip

\Abschnitt Eine kritische Bewertung der Konvergenztheorie von
Germain-Bonne

\smallskip

\aktTag = 0

Mit Hilfe von entweder der urspr\"unglichen Version der Konvergenztheorie
von Germain-Bonne [1973] oder den in den beiden letzten Unterabschnitten
vorgestellten Modifikationen kann entschieden werden, ob ein
verallgemeinerter Summationsproze{\ss} regul\"ar ist, d.~h., ob die
transformierte Folge $\Seqn {s'}$ gegen den gleichen Grenzwert
konvergiert wie die urspr\"ungliche Folge $\Seqn s$. Au{\ss}erdem konnten
notwendige und hinreichende Bedingungen formuliert werden, mit deren
Hilfe man vorhersagen kann, ob ein verallgemeinerter Summationsproze{\ss}
in der Lage ist, lineare Konvergenz zu beschleunigen.

In theoretischer Hinsicht sind dies bemerkenswerte Errungenschaften,
vor allem in Anbetracht der Tatsache, da{\ss} die verallgemeinerten
Summationsprozesse nur einige wenig restriktive Eigenschaften wie
Stetigkeit, Homogenit\"at und Translativit\"at besitzen m\"ussen, um die
Theorie von Germain-Bonne anwendbar zu machen. Auch die Folgen $\Seqn
s$, die transformiert werden sollen, m\"ussen keine besonderen
Eigenschaften besitzen. In den meisten F\"allen ist es ausreichend, da{\ss}
die Folgen, die transformiert werden sollen, konvergieren, oder -- wenn
die Beschleunigung linearer Konvergenz analysiert werden soll -- da{\ss}
die Folgen linear konvergieren.

Man sollte aber nicht \"ubersehen, da{\ss} die Konvergenztheorie von
Germain-Bonne [1973] trotz ihrer unbestreitbaren mathematischen Eleganz
einige schwerwiegende M\"angel aufweist, die ihre praktische N\"utzlichkeit
erheblich verringert. So ist die sehr allgemeine Natur der Theorie von
Germain-Bonne unbestreitbar ein Vorteil, da es auf diese Weise m\"oglich
ist, die Eigenschaften einer gro{\ss}en Klasse von verallgemeinerten
Summationsprozessen theoretisch zu analysieren. Andererseits impliziert
diese Allgemeinheit aber auch, da{\ss} die Ergebnisse dieser Theorie nicht
so spezifisch und detailliert sein k\"onnen, wie man es in den meisten
F\"allen gerne h\"atte.

Die Konvergenztheorie von Germain-Bonne kann nur Aussagen dar\"uber
machen, ob ein verallgemeinerter Summationsproze{\ss} regul\"ar ist oder ob
er lineare Konvergenz beschleunigen kann. In praktischer Hinsicht ist
eine solche Aussage aber genauso n\"utzlich -- oder nutzlos -- wie die
Aussage, da{\ss} eine bestimmte Reihe konvergiert.

Die Konvergenztheorie von Germain-Bonne ist \"ahnlich wie
Konvergenzbeweise bei unendlichen Reihen asymptotischer Natur, da nur
das Verhalten der Folgenelemente $s_n$ und der Transformationen
$G_k^{(n)}$ oder $\Gamma_k^{(n)}$ mit gro{\ss}en Werten von $n$ eine Rolle
spielen. Diese asymptotische Natur der Theorie von Germain-Bonne ist
essentiell, da nur dadurch die technischen Probleme so stark vereinfacht
werden k\"onnen, da{\ss} eine theoretische Analyse \"uberhaupt m\"oglich wird.
Andererseits bezieht sich die Theorie von Germain-Bonne auf eine
Situation (Folgenelemente $s_n$ oder Transformationen $s^{\prime}_n$ mit
gro{\ss}en Indizes $n$), die man durch die Verwendung verallgemeinerter
Summationsverfahren eigentlich vermeiden m\"ochte. Au{\ss}erdem sollte man
bedenken, da{\ss} ein numerisches Verfahren nicht notwendigerweise f\"ur
mittelgro{\ss}e oder gar kleine Werte von $n$ besonders leistungsf\"ahig sein
mu{\ss}, auch wenn es gew\"ahrleistet ist, da{\ss} dieses Verfahren f\"ur den
Grenzfall $n \to \infty$ sehr gute Ergebnisse liefert.

Bei praktischen Rechnungen wird normalerweise nur eine relativ kleine
Zahl von Eingabedaten $s_n$, $s_{n+1}$, $\ldots$ , $s_{n+k}$ bekannt
sein, und man m\"ochte wissen, wie man die Information, die in diesen
Daten enthalten ist, auf optimale Weise extrahieren kann. In dieser
Hinsicht kann die Theorie von Germain-Bonne \"uberhaupt nicht helfen, da
sie nicht unterscheiden kann zwischen verschiedenen verallgemeinerten
Summationsprozessen, die alle lineare Konvergenz beschleunigen k\"onnen.

Ein weiterer Einwand ist, da{\ss} die Theorie von Germain-Bonne im
wesentlichen eine erfolgreiche Theorie der Beschleunigung linearer
Konvergenz ist und bei der Beschleunigung logarithmischer Konvergenz
nicht angewendet werden kann. Das ist ohne Zweifel sehr bedauerlich, da
die Beschleunigung logarithmischer Konvergenz ein wesentlich
unangenehmeres Problem ist als die Beschleunigung linearer Konvergenz --
und zwar sowohl in theoretischer als auch in praktischer Hinsicht.

Ein anderes Problem von gro{\ss}er praktischer Wichtigkeit ist die Summation
divergenter Reihen durch verallgemeinerte Summationsprozesse. Eine
Theorie, bei der Grenz\"uberg\"ange $n \to \infty$ durchgef\"uhrt werden
m\"ussen, ist bei divergenten Reihen sinnlos. Stattdessen sollte eine
Konvergenztheorie Aussagen dar\"uber machen, ob und wie gut die
Transformationen $G_k^{(n)}$ oder $\Gamma_k^{(n)}$ den verallgemeinerten
Grenzwert $s$ einer divergenten Folge $\Seqn s$ approximieren k\"onnen,
und wie die Qualit\"at der Approximation sich \"andert, wenn man f\"ur festes
$n$ die Transformationsordnung $k$ vergr\"o{\ss}ert.

Man sieht also, da{\ss} die mathematisch sehr elegante Konvergenztheorie von
Germain-Bonne bei einigen praktisch sehr relevanten Problemen \"uberhaupt
nicht anwendbar ist und durch alternative theoretische Ans\"atze erg\"anzt
werden mu{\ss}. Es ist aber nicht wahrscheinlich, da{\ss} man die
Konvergenztheorie von Germain-Bonne verbessern kann ohne wesentlich
detailliertere Annahmen zu machen sowohl \"uber die verallgemeinerten
Summationsprozesse, deren Eigenschaften analysiert werden sollen, als
auch \"uber die Folgen, die transformiert werden sollen.

\medskip

\Abschnitt Stieltjesreihen und Stieltjesfunktionen in
Summationsprozessen

\smallskip

\aktTag = 0

In diesem Unterabschnitt wird die Summation hochgradig divergenter
asymptotischer Potenzreihen
$$
f (z) \; \sim \; \sum_{m=0}^{\infty} \, \gamma_m z^m \, ,
\qquad z \to 0 \, ,
\tag
$$
behandelt, wie sie beispielsweise in der quantenmechanischen
St\"orungstheorie vorkommen.

Bei der Summation asymptotischer Reihen gibt es ein sehr unangenehmes
Eindeutigkeitsproblem. Wenn eine gegebene Funktion $f (z)$ bez\"uglich
einer asymptotischen Folge eine asymptotische Reihe besitzt, dann ist
diese {\it eindeutig} bestimmt. Die Umkehrung dieser Aussage ist aber
nicht richtig, d.~h., verschiedene Funktionen k\"onnen die gleiche
asymptotische Reihe besitzen. Bei der theoretischen Analyse von
Summationsprozessen, die ja versuchen, einer divergenten asymptotischen
Reihe einen endlichen Wert zuzuordnen, ist diese Nichteindeutigkeit eine
sehr unangenehme Komplikation. Demzufolge mu{\ss} man die Menge der
zul\"assigen asymptotischen Reihen auf geeignete Weise einschr\"anken, um
dieses Problem vermeiden zu k\"onnen.

Ein weiteres Problem ist, da{\ss} man nicht jede divergente Reihe summieren
kann. So scheint es, da{\ss} man beispielsweise der formalen Potenzreihe
[Baker 1990, Gl. (13.46)]
$$
f (z) \; = \; 1 \, + \, 1! z \, + \, 2! z^2 \, + \, 4! z^4
\, + \, 8! z^8 \, + \, 16! z^{16} \, + \, \ldots
\, .
\tag
$$
nur f\"ur $z = 0$ einen endlichen Wert zuordnen kann.

Die oben erw\"ahnten Nichteindeutigkeits- und Existenzprobleme k\"onnen
weitgehend vermieden werden, wenn man sich bei theoretischen
Betrachtungen auf die schon in Abschnitt 4.3 erw\"ahnten Stieltjesreihen
beschr\"ankt, die unter den divergenten Reihen eine besondere Position
einnehmen, da sie eine hochentwickelte Darstellungs- und
Konvergenztheorie besitzen [Baker 1975; Baker und Graves-Morris 1981a;
Borel 1928; Bowman und Shenton 1989; Perron 1957; Wall 1973]. Man kann
beispielsweise zeigen, da{\ss} Pad\'e-Approximationen auch hochgradig
divergente Stieltjesreihen summieren k\"onnen, wenn die Koeffizienten
$\gamma_n$ dieser Reihe mit wachsendem $n$ betragsm\"a{\ss}ig nicht schneller
wachsen als $C^{n+1} (2n)!$, wobei $C$ eine positive Konstante ist
[Simon 1982, Theorem 1.3]. Das impliziert beispielsweise, da{\ss} die
hochgradig divergente Eulerreihe (2.2-2), die eine Stieltjesreihe ist,
durch Pad\'e-Approximationen summiert werden kann.

Wie sp\"ater noch ausf\"uhrlicher diskutiert wird, sind die St\"orungsreihen
(2.2-4) f\"ur die Grundzustandsenergien der anharmonischen Oszillatoren
(2.2-3) mit $\hat{x}^4$- beziehungsweise
$\hat{x}^6$-Anhar\-moni\-zi\-t\"at, abgesehen vom ersten Term, ebenfalls
Stieltjesreihen. Aus den asymptotischen Absch\"atzungen (2.2-5a) und
(2.2-5b) f\"ur die Koeffizienten $b_n^{(2)}$ und $b_n^{(3)}$ folgt dann,
da{\ss} Pad\'e-Approximationen diese divergenten St\"orungsreihen summieren
k\"onnen.

Allerdings sind Pad\'e-Approximationen trotz ihrer hochentwickelten
Konvergenztheorie nicht besonders gut zur Summation hochgradig
divergenter Stieltjesreihen geeignet. Wir werden sp\"ater sehen, da{\ss}
${\cal L}_{k}^{(n)} (\zeta, s_n, \omega_n)$, Gl. (5.2-6), ${\cal
S}_{k}^{(n)} (\zeta, s_n, \omega_n)$, Gl. (5.4-6), und ${\cal
M}_{k}^{(n)} (\xi, s_n, \omega_n)$, Gl. (5.5-8), in dieser Hinsicht
wesentlich leistungsf\"ahiger sind als Pad\'e-Approximationen oder der von
Brezinski [1971] eingef\"uhrte $\theta$-Algorithmus, Gl. (4.4-13). Das ist
wahrscheinlich eine direkte Konsequenz der Tatsache, da{\ss} die oben
erw\"ahnten verallgemeinerten Summationsprozesse explizite
Restsummenabsch\"atzungen $\Seqn \omega$ verwenden, und da{\ss} man au{\ss}erdem
bei Stieltjesreihen sehr leicht Restsummenabsch\"atzungen finden kann, die
trotz ihrer Einfachheit sehr gute obere Schranken f\"ur die Abbruchfehler
liefern.

Obwohl Stieltjesreihen und Stieltjesfunktionen schon in Abschnitt 4.3
eingef\"uhrt und ihre Eigenschaften besprochen wurden, sollen in diesem
Unterabschnitt alle wesentlichen Eigenschaften noch einmal im
Zusammenhang mit der Summation divergenter Reihen durch verallgemeinerte
Summationsprozesse besprochen werden.

Eine Funktion $f(z)$ mit $z \in \C$ wird {\it Stieltjesfunktion}
genannt, wenn man sie durch ein {\it Stieltjesintegral}
$$
f(z) \; = \; \int\nolimits_{0}^{\infty} \,
\frac {\d \psi (t)} {1 + z t} \, ,
\qquad \vert \arg (z) \vert < \pi \, ,
\tag
$$
darstellen kann. Dabei ist $\psi (t)$ ein positives Ma{\ss} auf $0 \le t <
\infty$, das dort unendlich viele verschiedene Werte annimmt [Baker und
Graves-Morris 1981a, S. 159], und das f\"ur alle $m \in \N_0$ {\it
endliche\/} und {\it positive\/} Momente $\mu_m$ besitzt, die durch
$$
\mu_m \; = \; \int\nolimits_{0}^{\infty} \, t^m \d \psi (t)
\, , \qquad m \in \N_0 \, ,
\tag
$$
definiert sind. Eine konvergente oder divergente formale
Reihenentwicklung
$$
f(z) \; = \; \sum_{m=0}^{\infty} \; (-1)^m \, \mu_m \, z^m
\tag
$$
wird {\it Stieltjesreihe} genannt, wenn die Koeffizienten $\mu_m$
Momente eines positives Ma{\ss}es $\psi (t)$ auf $0 \le t < \infty$ gem\"a{\ss}
Gl. (6.5-4) sind, d.~h., wenn
$$
f(z) \; = \; \sum_{m=0}^{\infty} \; (-1)^m \, z^m \,
\int\nolimits_{0}^{\infty} \, t^m \d \psi (t)
\tag
$$
gilt. Auf diese Weise ist gew\"ahrleistet, da{\ss} zu jeder Stieltjesreihe
mindestens eine Stieltjesfunktion existiert.

Da Stieltjes-Momente $\mu_n$ laut Voraussetzung immer positiv sind,
folgt aus Gl. (6.5-5), da{\ss} eine Stieltjesreihe strikt alternierende
Terme besitzt, wenn das Argument $z$ positiv ist. Demzufolge kann die
formale Potenzreihe (6.5-2), der man nur f\"ur $z = 0$ einen endlichen
Wert zuordnen kann, keine Stieltjesreihe sein. Dagegen ist das nach
Euler benannte Integral (2.2-1) offensichtlich eine Stieltjesfunktion,
und die zugeh\"orige asymptotische Potenzreihe (2.2-2), die sogenannte
Eulerreihe, ist eine Stieltjesreihe.

Wenn man die Beziehung
$$
\sum_{m=0}^{n} \, (- z t)^m \; = \;
\frac {1 - (- z t)^{n+1}} {1 + z t}
\tag
$$
in Gl. (6.5-3) verwendet und die Momente $\mu_m$ gem\"a{\ss} Gl. (6.5-4)
berechnet, erkennt man, da{\ss} eine Stieltjesfunktion $f (z)$ auf folgende
Weise dargestellt werden kann:
$$
f (z) \; = \; \sum_{m=0}^{n} \, (-1)^m \, \mu_m \, z^m \; + \;
(- z)^{n+1} \,
\int\nolimits_{0}^{\infty} \,
\frac {t^{n+1} \, \d \psi (t)} {1 + z t}
\, , \qquad \vert \arg (z) \vert < \pi \, .
\tag
$$
Ob die Folge $\Seq {f_n (z)} {n=0}$ der Partialsummen
$$
f_n (z) \; = \; \sum_{m=0}^{n} \, (-1)^m \, \mu_m \, z^m
\tag
$$
gegen $f (z)$ konvergiert oder ob sie divergiert, h\"angt von dem
Verhalten des Integrals auf der rechten Seite von Gl. (6.5-8) f\"ur $n \to
\infty$ ab. Der n\"achste Satz zeigt, da{\ss} der Abbruchfehler einer
Stieltjesreihe abgesch\"atzt werden kann durch den ersten Term der
Potenzreihe (6.5-5), der nicht in der Partialsumme auf der rechten Seite
von Gl. (6.5-8) enthalten ist [Weniger 1989, Theorem 13-2].

\medskip

\beginEinzug \sl \parindent = 0 pt

\Auszug {\bf Satz 6-12:} Der Abbruchfehler in Gl. (6.5-8),
$$
R_n (z) \; = \; (-z)^{n+1} \, \int\nolimits_{0}^{\infty} \,
\frac {t^{n+1} \, \d \psi (t)} {1 + z t} \, ,
\tag
$$
erf\"ullt in Abh\"angigkeit von $\theta = \arg (z)$ die folgenden
Ungleichungen:
$$
\beginAligntags
" \vert R_n (z) \vert \; \le \;
\mu_{n+1} \, \vert z^{n+1} \vert \, ,
\qquad " \vert \theta \vert \le \pi /2 \, ,
\erhoehe\aktTag \\ \tag*{\tagnr a}
" \vert R_n (z) \vert \; \le \;
\mu_{n+1} \, \vert z^{n+1} \Funk{cosec} \, \theta \vert \, ,
\qquad " \pi / 2 < \vert \theta \vert < \pi \, .
\\ \tag*{\tagform\aktTagnr b}
\endAligntags
$$

\endEinzug

\medskip

Aus Satz 6-12 folgt auch, da{\ss} man einer Stieltjesfunktion (6.5-3) in
einem Sektor $\vert \arg (z) \vert < \theta$ mit $\theta < \pi$ eine
asymptotische Potenzreihe zuordnen kann, und da{\ss} diese asymptotische
Reihe die zugeh\"orige Stieltjesreihe (6.5-5) ist [Simon 1972, S. 398].
Man kann au{\ss}erdem beweisen, da{\ss} man jeder Stieltjesreihe mindestens
eine Stieltjesfunktion zuordnen kann. Um diese f\"ur Summationsprozesse
unangenehme Nichteindeutigkeit ausschlie{\ss}en zu k\"onnen, ben\"otigt man
noch ein zus\"atzliches Kriterium, mit dessen Hilfe man garantieren kann,
da{\ss} einer Stieltjesreihe auf eindeutige Weise genau eine
Stieltjesfunktion zugeordnet werden kann. Wenn die Stieltjesreihe
konvergieren w\"urde, g\"abe es dieses Problem nicht, da konvergente
Potenzreihen auf eindeutige Weise eine analytische Funktion
beschreiben. Man ben\"otigt also eine Bedingung, die st\"arker ist als die
Existenz einer asymptotischen Entwicklung vom Typ von Gl. (6.5-3), und
die schw\"acher ist als die Forderung, da{\ss} eine solche Reihenentwicklung
konvergent und damit eindeutig ist.

Man kann tats\"achlich ein solches Kriterium finden. Auf der Basis des
Theorems von Carleman [Carleman 1926, Abschnitt \Roemisch{5}; Reed und
Simon 1978, Theoreme XII.17 und XII.18; Baker 1990, S. 223] k\"onnen
notwendige und hinreichende Bedingungen formuliert werden, die
garantieren, da{\ss} eine umkehrbar eindeutige Zuordnung zwischen einer
Stieltjesfunktion und der zugeh\"origen asymptotischen Potenzreihe
existiert.

Man sagt, da{\ss} eine Stieltjesfunktion $f (z)$, die in einem geeigneten
Sektor ${\cal S}$ der komplexen Ebene analytisch ist, eine {\it starke
asymptotische Bedingung} erf\"ullt, und die zugeh\"orige Stieltjesreihe wird
als {\it starke asymptotische Reihe} bezeichnet, wenn man geeignete
positive Konstanten $A$ und $\chi$ finden kann, so da{\ss}
$$
\biggl\vert f (z) \; - \;
\sum_{m=0}^n \, (-1)^m \mu_m z^m \biggr\vert
\; \le \; A \chi^{n+1} (n+1)! \, {\vert z \vert}^{n+1}
\tag
$$
f\"ur alle $n \in \N_0$ und f\"ur alle $z \in {\cal S}$ gilt.

Die G\"ultigkeit einer solchen starken asymptotischen Bedingung
impliziert, da{\ss} eine Stieltjesfunktion $f (z)$ eindeutig bestimmt ist
durch die zugeh\"orige asymptotische Reihe [Reed und Simon 1978, S. 40].
Offensichtlich kann eine Stieltjesfunktion $f (z)$ eine solche starke
asymptotische Bedingung nur dann erf\"ullen, wenn die Stieltjes-Momente
$\mu_n$, die durch Gln. (6.5-5) und (6.5-6) definiert sind, f\"ur alle $n
\in \N_0$ die folgende Bedingung erf\"ullen [Reed und Simon 1978, S. 43]:
$$
\mu_n \; \le \; A \, \chi^n \, n! \, .
\tag
$$
Die Momente der Eulerreihe (2.2-2) erf\"ullen offensichtlich diese
Bedingung. Daraus folgt, da{\ss} das nach Euler benannte Integral (2.2-1)
eindeutig bestimmt ist durch die zugeh\"orige asymptotische Reihe (2.2-2).
Auf gleiche Weise folgt aus Gl. (2.2-5a), da{\ss} die St\"orungsreihe (2.2-4)
f\"ur die Grundzustandsenergie des anharmonischen Oszillators (2.2-3) mit
$\hat{x}^4$-Anharmonizit\"at, die ja mit Ausnahme des ersten Terms eine
Stieltjesreihe ist, eine starke asymptotische Reihe ist. Das bedeutet,
da{\ss} die Grundzustandsenergie des anharmonischen Oszillators mit
$\hat{x}^4$-Anharmonizit\"at auf eindeutige Weise durch die divergente
St\"orungsreihe (2.2-4) bestimmt ist [Reed und Simon 1978, S. 41].

In der quantenmechanischen St\"orungstheorie gibt es aber auch
Stieltjesreihen, deren Momente $\mu_n$ wie $(k n)!$ mit $k > 1$ f\"ur $n
\to \infty$ wachsen. So folgt beispielsweise aus den asymptotischen
Absch\"atzungen (2.2-5b) und (2.2-5c), da{\ss} die Koeffizienten $b_n^{(m)}$
der St\"orungsreihen (2.2-4) f\"ur die Grundzustandsenergien der
anharmonischen Oszillatoren (2.2-3) mit $\hat{x}^6$- beziehungsweise
$\hat{x}^8$-Anharmonizit\"at im wesentlichen wie $(2 n)! / n^{1/2}$
beziehungsweise $(3 n)! / n^{1/2}$ f\"ur $n \to \infty$ wachsen. Diese
Reihen k\"onnen nat\"urlich keine starke asymptotische Bedingung erf\"ullen,
da ihre Koeffizienten schneller als $\chi^n n!$ wachsen. Man kann aber
zeigen, da{\ss} eine Funktion $f (z)$, die in einem entsprechenden Sektor
$\cal{S}$ der komplexen Ebene analytisch ist, ebenfalls eindeutig durch
ihre zugeh\"orige asymptotische Reihe bestimmt ist, wenn $f (z)$ eine {\it
modifizierte starke asymptotische Bedingung der Ordnung $k$} erf\"ullt,
und wenn ihre zugeh\"orige asymptotische Reihe eine {\it modifizierte
starke asymptotische Reihe der Ordnung $k$} ist [Reed und Simon 1978,
S. 43]. Das bedeutet, da{\ss} man geeignete positive Konstanten $A$ und
$\chi$ finden kann, so da{\ss}
$$
\biggl\vert f (z) \; - \;
\sum_{m=0}^n \, (-1)^m \mu_m z^m \biggr\vert \; \le \;
A \chi^{n+1} [k (n+1)]! \, {\vert z \vert}^{n+1}
\tag
$$
f\"ur alle $n \in \N_0$ und f\"ur alle $z \in {\cal S}$ gilt.

Offensichtlich kann eine Stieltjesfunktion $f (z)$ eine solche
modifizierte starke asymptotische Bedingung der Ordnung $k$ nur dann
erf\"ullen, wenn die Stieltjes-Momente $\mu_n$ f\"ur alle $n \in \N_0$ die
folgende Bedingung erf\"ullen [Simon 1972, S. 406]:
$$
\mu_n \; \le \; A \, \chi^n \, (k n)! \, .
\tag
$$

Die Absch\"atzungen (6.5-11a) und (6.5-11b) f\"ur den Abbruchfehler $R_n
(z)$ in Satz 6-12 sind von erheblicher Bedeutung f\"ur
Konvergenzbeschleunigungs- und Summationsprozesse, da man mit ihrer
Hilfe einfache und praktikable Restsummenabsch\"atzungen $\Seqn \omega$
f\"ur Stieltjesreihen finden kann. Aufgrund der Tatsache, da{\ss} die
verallgemeinerten Summationsprozesse ${\cal L}_{k}^{(n)} (\zeta, s_n,
\omega_n)$, Gl. (5.2-6), ${\cal S}_{k}^{(n)} (\zeta, s_n, \omega_n)$,
Gl. (5.4-6), und ${\cal M}_{k}^{(n)} (\xi, s_n, \omega_n)$, Gl. (5.5-8),
homogene Funktionen vom Grade Null ihrer zweiten $k + 1$ Eingabedaten
$\omega_n$, $\omega_{n+1}$, $\ldots$ , $\omega_{n+k}$ sind, k\"onnen alle
multiplikative Faktoren, die unabh\"angig von $n$ sind, in den
Restsummenabsch\"atzungen weggelassen werden. Man mu{\ss} also die beiden
F\"alle $\vert \theta \vert \le \pi$ und $\pi / 2 < \vert \theta \vert <
\pi$ in Satz 6-12 nicht unterscheiden. In jedem Sektor $\vert \arg (z)
\vert < \theta$ mit $\theta < \pi$ kann man also den Abbruchfehler $R_n
(z)$ einer Stieltjesreihe durch den ersten Term absch\"atzen, der nicht in
der Partialsummen enthalten ist:
$$
\omega_n \; = \; (-1)^{n+1} \, \mu_{n+1} \, z^{n+1} \, ,
\qquad n \in \N_0 \, .
\tag
$$
Diese Restsummenabsch\"atzung f\"ur Stieltjesreihen ist aber ein
Spezialfall der von Smith und Ford [1979] eingef\"uhrten
Restsummenabsch\"atzung (5.2-16). Wenn man also eine divergente
Stieltjesreihe mit Hilfe geeigneter Varianten der verallgemeinerten
Summationsprozesse ${\cal L}_k^{(n)} (\zeta, s_n, \omega_n)$, ${\cal
S}_k^{(n)} (\zeta, s_n, \omega_n)$, und ${\cal M}_k^{(n)} (\xi, s_n,
\omega_n)$ summieren will, dann ist es aufgrund von Satz 6-12
naheliegend, die verallgemeinerten Summationsprozesse $d_k^{(n)}
(\zeta, s_n)$, Gl. (5.2-18), $\delta_k^{(n)} (\zeta, s_n)$, Gl.
(5.4-13), und $\Delta_k^{(n)} (\xi, s_n)$, Gl. (5.5-15), zu verwenden,
die alle auf der Restsummenabsch\"atzung (5.2-16) basieren.

\medskip

\Abschnitt Fehlerabsch\"atzungen bei der Transformation von
Stieltjesreihen

\smallskip

\aktTag = 0

In diesem Abschnitt wird versucht, die Anwendung verallgemeinerter
Summationsprozesse auf konvergente oder divergente Stieltjesreihen
theoretisch zu analysieren. Die so erhaltenen Ergebnisse sind aber auch
im Zusammenhang mit Konvergenzbeschleunigungs- und Summationsprozessen
bei alternierenden Reihen, die keine Stieltjesreihen sind, von
Interesse. In vielen F\"allen k\"onnen die in diesem Unterabschnitt
vorgestellten Ergebnisse ohne wesentliche \"Anderungen auch auf andere
alternierende Reihen \"ubertragen werden.

Bisher wurden haupts\"achlich theoretische Arbeiten ver\"offentlicht, in
denen die Darstellung konvergenter oder divergenter Stieltjesreihen
durch Pad\'e-Approximationen analysiert wurde [Allen, Chui, Madych,
Narcowich und Smith 1975; Baker 1969; Common 1968; Graves-Morris 1981;
Karls\-son und von Sydow 1976; Wynn 1968]. Dann gibt es noch einen
Artikel von Common [1968] \"uber die Darstellung von Stieltjesreihen durch
Verallgemeinerungen von Pad\'e-Approximationen. Bisher gibt es aber kaum
theoretische Artikel, in denen die Anwendung anderer verallgemeinerter
Summationsprozesse auf Stieltjesreihen behandelt wird. Das ist an sich
nicht \"uberraschend, weil viele verallgemeinerte Summationsprozesse nur
durch relativ komplizierte Rekursionsschemata definiert sind, wobei
ansonsten nur wenig \"uber ihre theoretischen Eigenschaft bekannt ist.

Ein gutes Beispiel ist der von Brezinski [1971] eingef\"uhrte
$\theta$-Algorithmus, der durch das \"au{\ss}erst komplizierte
Rekursionsschema (4.4-13) definiert ist. Numerische Studien [Smith und
Ford 1979; 1982; Weniger 1989, Abschnitte 13 und 14] ergaben, da{\ss} der
$\theta$-Algorithmus eine sehr leistungsf\"ahige und auch sehr vielseitige
Transformation ist, die sowohl zur Beschleunigung linearer und
logarithmischer Konvergenz als auch zur Summation divergenter
alternierender Reihen geeignet ist. Trotzdem sind die theoretischen
Eigenschaften des $\theta$-Algorithmus bisher weitgehend unverstanden.
So konnten Brezinski und Walz [1991, Theorem 5.3] erst k\"urzlich eine
Determinantendarstellung f\"ur $\theta_{2 k}^{(n)}$ abzuleiten.

Beim augenblicklichen Stand des Wissens erscheint also eine
detailliertere theoretische Analyse der Eigenschaften der meisten
verallgemeinerten Summationsverfahren in Konvergenzbeschleuni\-gungs- und
Summationsprozessen aufgrund der oben beschriebenen Schwierigkeiten mehr
oder weniger unm\"oglich zu sein.

Bei den in Abschnitt 5 behandelten verallgemeinerten Summationsprozessen
${\cal L}_{k}^{(n)} (\zeta, s_n, \omega_n)$, Gl. (5.2-6), ${\cal
S}_{k}^{(n)} (\zeta, s_n, \omega_n)$, Gl. (5.4-6), und ${\cal
M}_{k}^{(n)} (\xi, s_n, \omega_n)$, Gl. (5.5-8), die ja einfache
explizite Darstellungen besitzen, ist eine theoretische Analyse aber
vergleichsweise einfach, wenn die Partialsummen $\Seqn s$ und die
Restsummenabsch\"atzungen $\Seqn \omega$ geeigneten Einschr\"ankungen
unterworfen werden. So gelang es Sidi [1979; 1980; 1990], zahlreiche
Konvergenzabsch\"atzungen f\"ur die $u$- und $t$-Transformation von Levin,
Gln. (5.2-13) und (5.2-15), abzuleiten. Sidi [1986b] analysierte
ebenfalls die Summation hochgradig divergenter Potenzreihen, deren
Koeffizienten bestimmte Eigenschaften besitzen, mit Hilfe der $u$- und
$t$-Transformationen. Er konnte zeigen, da{\ss} die von ihm betrachteten
divergenten Reihen asymptotische Entwicklungen borelartiger Integrale
sind. Auf der Basis seiner Untersuchungen kam Sidi [1986b] zu der
Schlu{\ss}folgerung, da{\ss} die durch die Anwendung der $u$- beziehungsweise
$t$-Transformation erhaltenen rationalen Funktionen im Falle der von ihm
betrachteten divergenten Reihen wahrscheinlich gegen diese borelartigen
Integrale konvergieren, ohne diese Vermutung aber explizit beweisen zu
k\"onnen.

Die Konvergenzuntersuchungen von Sidi [1979; 1980; 1990] basieren auf
der Tatsache, da{\ss} die Levinsche Transformation ${\cal L}_{k}^{(n)}
(\zeta, s_n, \omega_n)$ gem\"a{\ss} Gl. (5.2-5) eine Transformation vom Typ
von Gl. (5.1-9) ist,
$$
{\cal T}_k^{(n)} \bigl( P_{k - 1} (n); s_n, \omega_n \bigr)
\; = \; \frac
{\Delta^k \[ P_{k - 1} (n) s_n / \omega_n \]}
{\Delta^k \[ P_{k - 1} (n) / \omega_n \]} \, , \qquad k, n \in \N_0 \, ,
\tag
$$
wobei $P_{k - 1} (n)$ ein Polynom vom Grade $k - 1$ in $n$ ist. Da eine
solche Transformation trans\-lationsinvariant ist, kann man sie gem\"a{\ss}
Gl. (6.1-8) auf folgende Weise ausdr\"ucken,
$$
{\cal T}_k^{(n)} \bigl( P_{k - 1} (n); s_n, \omega_n \bigr)
\; = \; s \, + \, \frac
{\Delta^k \[ P_{k - 1} (n) [s_n - s] / \omega_n \]}
{\Delta^k \[ P_{k - 1} (n) / \omega_n \]} \, , \qquad k, n \in \N_0 \, ,
\tag
$$
wobei $s$ der (verallgemeinerte) Grenzwert der zu transformierenden
Folge $\Seqn s$ ist.

Anhand von Gl. (6.6-2) kann man erkennen, wie solche verallgemeinerte
Summationsprozesse wirken: F\"ur eine gegebene Folge $\Seqn s$ von
Partialsummen mu{\ss} man die Restsummenabsch\"atzungen $\Seqn \omega$ so
w\"ahlen, da{\ss} $[s_n - s] / \omega_n$ weniger stark von $n$ abh\"angt als $1
/ \omega_n$. Der gewichtete Differenzenoperator $\Delta^k P_{k - 1} (n)$
wird dann den Z\"ahler des Quotienten in Gl. (6.6-2) schneller
annihilieren als den Nenner, was dazu f\"uhrt, da{\ss} der Quotient in Gl.
(6.6-2) gegen Null und ${\cal T}_k^{(n)} \bigl( P_{k - 1} (n); s_n,
\omega_n \bigr)$ gegen den (verallgemeinerten) Grenzwert $s$
konvergiert. Optimal w\"are es, wenn man die Restsummenabsch\"atzungen so
w\"ahlen k\"onnte, da{\ss}
$$
\omega_n \; = \; c \, (s_n - s) \, , \qquad c \ne 0 \, ,
\tag
$$
f\"ur alle $n \in \N_0$ gilt. Der Z\"ahler des Quotienten auf der rechten
Seite von Gl. (6.6-2) w\"are dann die $k$-te Differenz eines Polynoms vom
Grade $k - 1$ in $n$, die bekanntlich f\"ur $k \ge 1$ Null ist
[Milne-Thomson 1981, S. 29]. Da wir immer stillschweigend davon
ausgehen, da{\ss} ${\Delta}^k [ P_{k - 1} (n) / \omega_n ] \ne 0$ gilt, w\"are
die Transformation ${\cal T}_k^{(n)} \bigl( P_{k - 1} (n); s_n, \omega_n
\bigr)$ dann f\"ur die Folge $\Seqn s$ f\"ur $k \ge 1$ exakt.

Bei praktisch relevanten Problemen kann man aber leider nicht erwarten,
da{\ss} man eine solche Folge von Restsummenabsch\"atzungen finden kann.
Realistischerweise kann man bestenfalls hoffen, die
Restsummenabsch\"atzungen so w\"ahlen zu k\"onnen, da{\ss}
$$
s_n - s \; = \; \omega_n \bigl[ c + O (n^{- 1}) \bigr]
\tag
$$
f\"ur alle $n \in \N_0$ gilt.

Die Beweistechniken, die Sidi [1979; 1980; 1990] zur Untersuchung
theoretischer Eigenschaften von Varianten der Levinschen Transformation
${\cal L}_{k}^{(n)} (\zeta, s_n, \omega_n)$, Gl. (5.2-6), verwendete,
k\"onnen leicht so modifiziert werden, da{\ss} man sie auch bei den eng
verwandten verallgemeinerten Summationsprozessen ${\cal S}_{k}^{(n)}
(\zeta, s_n, \omega_n)$, Gl. (5.4-6), und ${\cal M}_{k}^{(n)} (\xi, s_n,
\omega_n)$, Gl. (5.5-8), erfolgreich anwenden kann [Weniger 1989,
Abschnitt 13]. Das Schwergewicht dieses Unterabschnitt liegt also auf
der theoretischen Analyse der Transformation von konvergenten oder
divergenten Stieltjesreihen durch den verallgemeinerten Summationsproze{\ss}
${\cal S}_{k}^{(n)} (\zeta, s_n, \omega_n)$, der -- wie in sp\"ateren
Abschnitten noch ausf\"uhrlicher diskutiert wird -- bei hochgradig
divergenten alternierenden Reihen besonders wirksam ist und der einer
theoretischen Analyse seiner Konvergenzeigenschaften auch
vergleichsweise leicht zug\"anglich ist. \"Ahnlich wie in den
Unterabschnitten 6.2 und 6.3 werden auch hier haupts\"achlich S\"atze
angegeben und ihre Implikationen kurz diskutiert. Die Beweise dieser
S\"atze und ausf\"uhrlichere Diskussionen findet man in Abschnitt 13 von
Weniger [1989].

Die in diesem Abschnitt besprochene Analyse der Konvergenz von
verallgemeinerten Summationsprozessen basiert auf den S\"atzen 6-1 und
6-2. Dabei wird im Prinzip immer versucht, die Gr\"o{\ss}e des
Transformationsfehlers $G_k^{(n)} (s_n - s, s_{n+1} - s, \ldots, s_{n+k}
- s)$ in Abh\"angigkeit von $k$ und $n$ abzusch\"atzen. Dazu wird
angenommen, da{\ss} die Folgen $\Seqn s$ und $\Seqn \omega$ die folgenden
Eigenschaften besitzen:

\medskip

\beginBeschreibung \zu \Laenge{(S-0):} \sl

\item{(S-0):} Die Elemente der Folge $\Seqn s$ sind die Partialsummen
einer unendlichen Reihe, die entweder gegen den Grenzwert $s$
konvergiert oder -- falls sie divergiert -- gegen den verallgemeinerten
Grenzwert $s$ summiert werden kann.
\item{(S-1):} Die Elemente der Folge $\Seqn \omega$ der
Restsummenabsch\"atzungen von $\Seqn s$ besitzen strikt alternierende
Vorzeichen.
\item{(S-2):} F\"ur alle $n \in \N_0$ besitzt der Quotient $(s_n - s) /
\omega_n$ eine Darstellung durch eine Fakult\"atenreihe,
$$
\frac{s_n - s} {\omega_n} \; = \; \sum_{j=0}^{\infty}
\frac {c_j} {(\zeta + n)_j} \, ,
\qquad \zeta \in \R_{+} \, , \quad n \in \N_0 \, .
\tag
$$

\endBeschreibung

\medskip

Auf der Basis dieser Annahmen kann man sowohl die Summation divergenter
Stieltjesreihen als auch die Beschleunigung der Konvergenz bestimmter
alternierender Reihen analysieren.

Wenn man die Restsummenabsch\"atzungen $\Seqn \omega$ einer Stieltjesreihe
gem\"a{\ss} Gl. (6.5-16) w\"ahlt, dann folgt aus der Positivit\"at der
Stieltjesmomente $\Seqn \mu$, da{\ss} man aufgrund von Postu\-lat (S-1) nur
Potenzreihen mit positivem Argument $z$ betrachten kann. W\"urde man f\"ur
$z$ beliebige komplexe Werte zulassen, w\"are nicht mehr garantiert, da{\ss}
die Restsummenabsch\"atzungen (6.5-16) strikt alternierende Vorzeichen
besitzen.

Die Forderung, da{\ss} $(s_n - s)/\omega_n$ gem\"a{\ss} Gl. (6.6-5) durch eine
Fakult\"atenreihe dargestellt werden kann, mag auf den ersten Blick wie
eine schwerwiegende Einschr\"ankung der Allgemeinheit wirken. Diese
Forderung ist aber nicht unbedingt restriktiver als die scheinbar
naheliegendere Forderung, da{\ss} $(s_n - s)/\omega_n$ durch eine
Potenzreihe in inversen Potenzen von $\zeta+n$ dargestellt werden kann:
$$
\frac{s_n - s} {\omega_n} \; = \; \sum_{j=0}^{\infty}
\frac {c_j^{\prime}} {(\zeta + n)^j} \, ,
\qquad \zeta \in \R_{+} \, , \quad n \in \N_0 \, .
\tag
$$
Auf S. 272 - 282 des Buchs von Nielsen [1965] werden die algebraischen
Prozesse beschrieben, mit deren Hilfe man Potenzreihen und
Fakult\"atenreihen ineinander \"uberf\"uhren kann.

Mit Hilfe der Annahmen (S-0) - (S-2) k\"onnen quantitative Absch\"atzungen
des Transformationsfehlers von ${\cal S}_k^{(n)} (\zeta, s_n,
\omega_n)$, Gl. (5.4-6), in Konvergenzbeschleunigungs- und
Summationsverfahren abgeleitet werden [Weniger 1989, Theorem 13-5].

\medskip

\beginEinzug \sl \parindent = 0 pt

\Auszug {\bf Satz 6-13:} Nehmen wir an, da{\ss} die Folgen $\Seqn s$ und
$\Seqn \omega$ die Voraussetzungen (S-0) - (S-2) erf\"ullen, und da{\ss} der
verallgemeinerte Summationsproze{\ss} ${\cal S}_k^{(n)} (\zeta, s_n,
\omega_n)$, Gl. (5.4-6), zur Transformation der Partialsummen $\Seqn s$
verwendet wird. Dann erhalten wir f\"ur festes $k \in \N$ und f\"ur alle $n
\in \N_0$ die folgende Absch\"atzung f\"ur den Transformationsfehler:
$$
\left\vert
{\cal S}_k^{(n)} (\zeta, s_n, \omega_n) \, - \, s
\right\vert
\; \le \;
\left\vert
\frac {\omega_n} {(\zeta + n)_{2 k} } \, \sum_{j=0}^{\infty} \,
\frac {c_{k + j} \, (j+1)_k } {(\zeta + n + 2 k)_j }
\right\vert \, .
\tag
$$
Das impliziert f\"ur festes $k \in \N$ und f\"ur gro{\ss}e Werte von $n$ die
folgende Ordnungsabsch\"atzung:
$$
\frac
{{\cal S}_k^{(n)} (\zeta, s_n, \omega_n) \, - \, s} {s_n - s}
\; = \; O (n^{- 2 k}) \, , \qquad n \to \infty \, .
\tag
$$

\endEinzug

\medskip

Es ist eine typische Eigenschaft der Fehlerabsch\"atzung (6.6-7) und auch
anderer Fehlerabsch\"atzungen, die sp\"ater behandelt werden, da{\ss} sie direkt
proportional zu $\omega_n$ ist. Demzufolge mu{\ss} man keine Unterscheidung
zwischen konvergenten und divergenten Reihen machen. Aus der
Fehlerabsch\"atzung (6.6-7) folgt auch, da{\ss} ${\cal S}_k^{(n)} (\zeta, s_n,
\omega_n)$ eine divergente Reihe, die die Voraussetzungen (S-0) - (S-2)
erf\"ullt, auf einem horizontalen Weg ${\cal W}$ summieren kann, wenn die
Koeffizienten $c_j$ der Fakult\"atenreihe (6.6-5) nicht zu schnell f\"ur $j
\to \infty$ wachsen.

Es w\"are sicherlich interessant, auf analoge Weise Absch\"atzungen des
Transformationsfehlers auch im Falle des verallgemeinerten
Summationsprozesses ${\cal M}_k^{(n)} (\xi, s_n, \omega_n)$, Gl.
(5.5-8), abzuleiten. Wenn wir diese Analyse auf der Basis der
Voraussetzungen (S-0) - (S-2) durchf\"uhren w\"urden, h\"atten wir aber mit
erheblichen technischen Problemen zu k\"ampfen, und wir w\"urden letztlich
nur Ausdr\"ucke ableiten, die so kompliziert sind, da{\ss} sie nur wenig zu
unserem Verst\"andnis beitragen k\"onnten. Der Grund f\"ur diese technischen
Probleme ist, da{\ss} wir im Falle beliebiger Werte von $\zeta$ und $\xi$
das Leibnizsche Theorem f\"ur endliche Differenzen ben\"otigen w\"urden
[Milne-Thomson 1981, S. 35],
$$
\Delta^k \, [ f (n) g (n) ] \; = \; \sum_{j=0}^k \,
\binom {k} {j} \, [\Delta^j \, f (n)] \, [\Delta^{k-j} g(n)] \, .
\tag
$$

Viel aufschlu{\ss}reicher und hilfreicher ist dagegen die folgende
Beobachtung [Weniger 1989, Theorem 13-8]:

\medskip

\beginEinzug \sl \parindent = 0 pt

\Auszug {\bf Satz 6-14:} Nehmen wir an, da{\ss} $\xi = \zeta + k - 2$
gilt. Dann folgt f\"ur alle $n \in \N_0$:
$$
{\cal M}_k^{(n)} (\xi, s_n, \omega_n) \; = \;
{\cal S}_k^{(n)} (\zeta, s_n, \omega_n) \, .
\tag
$$

\endEinzug

\medskip

Man darf diesen Satz nicht \"uberinterpretieren. Satz 6-14 impliziert
nicht, da{\ss} die beiden Strings ${\cal M}_j^{(n)} (\xi, s_n, \omega_n)$
und ${\cal S}_j^{(n)} (\zeta, s_n, \omega_n)$ mit $0 \le j \le k$
identisch sind, wenn $\xi = \zeta + k -2$ gilt. Nur die letzten Elemente
dieser Strings sind gem\"a{\ss} Gl. (6.6-10) identisch, nicht aber die
anderen. Die Transformationen ${\cal S}_j^{(n)} (\zeta, s_n, \omega_n)$,
Gl. (5.4-6), und ${\cal M}_k^{(n)} (\xi, s_n, \omega_n)$, Gl. (5.5-8),
k\"onnen also durchaus verschiedene Eigenschaften in
Konvergenzbeschleunigungs- und Summationsverfahren aufweisen.

In Falle der Levinschen Transformation ${\cal L}_k^{(n)} (\zeta, s_n,
\omega_n)$, Gl. (5.2-6), w\"urde eine Analyse des Transformationsfehlers
in Stil von Satz 6-13 ebenfalls zu vergleichsweise komplizierten
Ausdr\"ucken f\"uhren. Diese Schwierigkeiten sind eine direkte Konsequenz
der Tatsache, da{\ss} die Modellfolge (5.2-1), welche der Ausgangspunkt der
Konstruktion der Levinschen Transformation war, identisch ist mit den
ersten $k$ Termen der asymptotischen Potenzreihe (5.2-2), und da{\ss} in der
Differenzenrechnung Pochhammersymbole und nicht Potenzen die einfachsten
Funktionen sind. Man kann aber wenigstens eine Ordnungsabsch\"atzung f\"ur
den Transformationsfehler der Levinschen Transformation auf einfache
Weise erhalten [Weniger 1989, Theorem 13-9:]:

\medskip

\beginEinzug \sl \parindent = 0 pt

\Auszug {\bf Satz 6-15:} Nehmen wir an, da{\ss} die Folgen $\Seqn s$ und
$\Seqn \omega$ die Annahmen (S-0) und (S-1) und Gl. (6.6-6) erf\"ullen, und
da{\ss} ${\cal L}_k^{(n)} (\zeta, s_n, \omega_n)$, Gl. (5.2-6), f\"ur die
Transformation der Partialsummen $\Seqn s$ verwendet wird. Dann erhalten
wir f\"ur gro{\ss}e Werte von $n$ und f\"ur festes $k \in \N$ die folgende
Ordnungsabsch\"atzung f\"ur den Transformationsfehler:
$$
\frac
{{\cal L}_k^{(n)} (\zeta, s_n, \omega_n) \, - \, s} {s_n - s}
\; = \; O (n^{- 2 k}) \, , \qquad n \to \infty \, .
\tag
$$

\endEinzug

\medskip

Ein Vergleich der Ordnungsabsch\"atzungen (6.6-8) und (6.6-11) zeigt, da{\ss}
die Levinsche Transformation ${\cal L}_k^{(n)} (\zeta, s_n, \omega_n)$,
Gl. (5.2-6), und ${\cal S}_k^{(n)} (\zeta, s_n, \omega_n)$, Gl. (5.4-6),
auf einem vertikalen Weg in etwa gleich leistungsf\"ahig sein sollten. Wir
werden sp\"ater sehen, da{\ss} diese Schlu{\ss}folgerung in vielen F\"allen in etwa
richtig ist. Es gibt aber gen\"ugend Beispiele, bei denen ${\cal
S}_k^{(n)} (\zeta, s_n, \omega_n)$ zum Teil wesentlich bessere
Ergebnisse liefert als ${\cal L}_k^{(n)} (\zeta, s_n, \omega_n)$
[Weniger und Steinborn 1989a; Weniger 1989; 1990; 1992; Weniger und {\v
C}{\'\i}{\v z}ek 1990; Weniger, {\v C}{\'\i}{\v z}ek und Vinette 1991;
1993; {\v C}{\'\i}{\v z}ek, Vinette und Weniger 1991; 1993; Grotendorst
1991]. Eine genauere Analyse der Transformationsfehler ist auf der Basis
der Annahmen (S-0) - (S-2) beziehungsweise Gl. (6.6-6) nicht m\"oglich.
Eine Verfeinerung der Absch\"atzungen w\"urde wesentlich detailliertere
Informationen \"uber die Partialsummen $\Seqn s$ als auch \"uber die
Restsummenabsch\"atzungen $\Seqn \omega$ voraussetzen.

Die in diesem Unterabschnitt durchgef\"uhrte Analyse der
Transformationsfehler ist auf konvergente oder divergente Folgen $\Seqn
s$ mit strikt alternierenden Restsummenabsch\"atzungen $\Seqn \omega$
beschr\"ankt. Diese Einschr\"ankung ist essentiell f\"ur den Beweis der S\"atze
6-13 und 6-15. Wenn man den Transformationsfehler von Folgen mit
nichtalternierenden Restsummenabsch\"atzungen auf diese Weise analysieren
will, mu{\ss} man zus\"atzliche Annahmen \"uber das Verhalten der
Restsummenabsch\"atzungen $\Seqn \omega$ in Abh\"angigkeit vom Index $n$
machen. Beispielsweise k\"onnte man im Falle logarithmischer Konvergenz
die folgende Annahme machen [Weniger 1989, Gl. (13.2-34)]:
$$
\frac {(\zeta+n)_{k-1}}{\omega_n} \; = \;
\sum_{j=0}^{\infty} \, c_j \,
\frac {\Gamma (\zeta + n + k - 1)} {\Gamma (\delta + n + j)}
\, , \qquad \zeta, \delta \in \R_{+} \, ,
\quad n \in \N_0 \, .
\tag
$$
Mit Hilfe \"ahnlicher Annahmen gelang es Sidi [1979; 1980], Absch\"atzungen
f\"ur Transformationsfehler der Levinschen Transformation zu konstruieren.

\medskip

\Abschnitt Die Summation der Eulerreihe

\smallskip

\aktTag = 0

In Abschnitt 4.5 wurde anhand zahlreicher Beispiele gezeigt, da{\ss}
Pad\'e-Approximationen inzwischen einen {\it de facto} Standard
darstellen, an dem sich alle anderen Verfahren zu messen haben. Es w\"are
deswegen sicherlich interessant, Pad\'e-Approximationen mit den in
Abschnitt 5 dieser Arbeit beschriebenen verallgemeinerten
Summationsprozessen genauer zu vergleichen, welche als Eingabedaten
nicht nur die Elemente einer Folge $\Seqn s$ von Partialsummen
verwenden, sondern zus\"atzlich noch die Elemente einer Folge $\Seqn
\omega$ von expliziten Restsummenabsch\"atzungen.

Ungl\"ucklicherweise sind die theoretischen Fehlerabsch\"atzungen f\"ur die
Transformation von Stieltjesreihen durch Pad\'e-Approximationen, die aus
der Literatur bekannt sind [Allen, Chui, Madych, Narcowich und Smith
1975; Baker 1969; Common 1968; Graves-Morris 1981; Karlsson und von
Sydow 1976; Wynn 1968] nicht direkt vergleichbar mit den
Fehlerabsch\"atzungen in den S\"atzen 6-13 und 6-14, wo der
Transformationsfehler immer proportional zur Restsummenabsch\"atzung
$\omega_n$ ist. Zum Beispiel wird der Transformationsfehler in Artikeln
von Allen, Chui, Madych, Narcowich und Smith [1975] und Karlsson und von
Sydow [1976] durch Polynome ausgedr\"uckt, die orthogonal sind bez\"uglich
des im Stieltjesintegral (6.5-3) vorkommenden Ma{\ss}es $\psi (t)$, das bei
praktischen Anwendungen normalerweise nicht bekannt ist.

Die verallgemeinerten Summationsprozesse ${\cal L}_k^{(n)} (\zeta, s_n,
\omega_n)$, Gl. (5.2-6), und ${\cal S}_k^{(n)} (\zeta, s_n, \omega_n)$,
Gl. (5.4-6), sind ebenso wie Pad\'e-Approximationen zur Summation
divergenter Stieltjesreihen geeignet. Um das Leistungsverm\"ogen dieser
verschiedenen Verfahren auf der Basis theoretischer Fehlerabsch\"atzungen
vergleichen zu k\"onnen, w\"urde man also zuerst ein Analogon von Satz 6-13
f\"ur Pad\'e-Approximationen ableiten m\"ussen. Ungl\"ucklicherweise ist bisher
noch kein Theorem f\"ur Pad\'e-Approximationen bekannt, in dem der
Transformationsfehler durch eine Reihenentwicklung vom Typ von Gl.
(6.6-7) dargestellt wird.

Es gibt eine bemerkenswerte Ausnahme: Im Falle der Eulerreihe (2.2-2)
konnte Sidi [1981] zeigen, da{\ss} ihre Pad\'e-Approximationen in
geschlossener Form ausgedr\"uckt werden k\"onnen mit Hilfe des
verallgemeinerten Summationsprozesses von Drummond [1972], der
folgenderma{\ss}en definiert ist [Weniger 1989, Gln. (9.5-3) und (9.5-4)]:
$$
{\cal D}_{k}^{(n)} (s_n, \omega_n) \; = \; \frac
{\Delta^k \, \{ s_n / \omega_n\} }
{\Delta^k \, \{ 1 / \omega_n\} }
\; = \; \frac
{\displaystyle
\sum_{j=0}^{k} \; ( - 1)^{j} \; \binom {k} {j} \;
\frac {s_{n+j}} {\omega_{n+j}} }
{\displaystyle
\sum_{j=0}^{k} \; ( - 1)^{j} \;
\binom {k} {j} \; \frac {1} {\omega_{n+j}} } \; ,
\qquad k,n \in \N_0 \, .
\tag
$$
Bei der Konstruktion der expliziten Darstellung der Pad\'e-Approximationen
der Eulerreihe ging Sidi [1981] von der bekannten Tatsache aus, da{\ss} die
Pad\'e-Approximationen $[n + k / k]$ einer formalen Potenzreihe gem\"a{\ss} Gl.
(4.4-1) mit Hilfe des Wynnschen $\epsilon$-Algorithmus, Gl. (2.4-10),
berechnet werden k\"onnen und da{\ss} der $\epsilon$-Algorithmus exakt ist f\"ur
die Modellfolge (4.5-1). Im Fall der Eulerreihe (2.2-2) gilt
$$
s_n \; = \; \sum_{\nu=0}^{n} \, (-1)^{\nu} \, {\nu}! \, z^{\nu} \, .
\tag
$$
Wenn wir diese Beziehung in der Modellfolge (4.5-1) verwenden,
erhalten wir
$$
s_n \; = \; s \, + \, \sum_{j=0}^{k-1} \,
c_j \, (-1)^{n+j+1} \, (n+j+1)! \, z^{n+j+1} \, .
\tag
$$
Diese Modellfolge kann auf folgende Weise umgeschrieben werden:
$$
\frac {s_n - s} {(-1)^{n+1} (n+1)! z^{n+1} } \; = \;
\sum_{j=0}^{k-1} \, c_j \, (-1)^j \, z^j \, (n + 2)_j \, .
\tag
$$
Die Summe auf der rechten Seite ist ein Polynom vom Grade $k-1$ in $n$,
das durch den Differen\-zen\-operator $\Delta^k$ annihiliert wird
[Milne-Thomson 1981, S. 29]. Die Drummondsche Transformation ist exakt
f\"ur die Modellfolge [Weniger 1989, Gl. (9.5-2)]
$$
[s_n - s] / \omega_n \; = \; P_{k-1} (n) \, ,
\qquad k,n \in \N_0 \, ,
\tag
$$
wobei $P_{k-1} (n)$ ein Polynom vom Grade $k-1$ in $n$ ist. Aus Gln.
(6.7-4) und (6.7-5) folgt, da{\ss} die Pad\'e-Approximationen $[n + k / k]$
der Eulerreihe (2.2-2) die folgende Darstellung durch die Drummondsche
Transformation (6.7-1) besitzen, wobei $s_n$ die Partialsumme (6.7-2)
ist:
$$
[n + k / k] \; = \; \epsilon_{2 k}^{(n)}
\; = \; {\cal D}_k^{(n)} (s_n, \Delta s_n) \, ,
\qquad k,n \in \N_0 \, .
\tag
$$

F\"ur die Drummondsche Transformation (6.7-1) kann man aber eine
theoretische Analyse des Transformationsfehlers im Stil von Satz 6-13
durchf\"uhren [Weniger 1989, Theorem 13-7]:

\medskip

\beginEinzug \sl \parindent = 0 pt

\Auszug {\bf Theorem 6-16:} Nehmen wir an, da{\ss} die Folgen
$\Seqn s$ und $\Seqn \omega$ die Bedingungen (S-0) - (S-2) erf\"ullen, und
da{\ss} die Drummondsche Transformation ${\cal D}_k^{(n)} (s_n, \omega_n)$,
Gl. (6.7-1), f\"ur die Transformation von $\Seqn s$ verwendet wird. Dann
erhalten wir f\"ur festes $k \in \N$ und f\"ur alle $n \in \N_0$ die
folgende Absch\"atzung f\"ur den Transformationsfehler:
$$
\left\vert
{\cal D}_k^{(n)} (\zeta, s_n, \omega_n) \, - \, s
\right\vert
\; \le \;
\left\vert
\frac {\omega_n} {(\zeta + n)_{k + 1} } \, \sum_{j=0}^{\infty} \,
\frac {c_{j + 1} \, (j+1)_k } {(\zeta + n + k + 1)_j }
\right\vert \, .
\tag
$$
Das impliziert f\"ur festes $k \in \N$ und f\"ur gro{\ss}e Werte von $n$ die
folgende Ordnungsabsch\"atzung:
$$
\frac
{{\cal D}_k^{(n)} (s_n, \omega_n) \, - \, s} {s_n - s}
\; = \; O (n^{- k - 1}) \, , \qquad n \to \infty \, .
\tag
$$

\endEinzug

\medskip

Ein Vergleich dieses Satzes mit den S\"atzen 6-13 und 6-15 zeigt, da{\ss} die
Drummondsche Transformation ${\cal D}_k^{(n)} (s_n, \omega_n)$, Gl.
(6.7-1), vor allem f\"ur gr\"o{\ss}ere Werte der Transformationsordnung $k$
deutlich weniger effizient sein sollte als die Levinsche Transformation
${\cal L}_k^{(n)} (\zeta, s_n, \omega_n)$, Gl. (5.2-6), oder ${\cal
S}_k^{(n)} (\zeta, s_n, \omega_n)$, Gl. (5.4-6).

Aus Gl. (6.7-6) folgt au{\ss}erdem, da{\ss} die Drummondsche Transformation
(6.7-1) im Falle der Eulerreihe (2.2-2) wesentlich effizienter ist als
Pad\'e-Approximationen. Aus dem Rekursionsschema (2.4-10) folgt, da{\ss} man
zur Berechnung der diagonalen Pad\'e-Approximationen $[n / n]$, die
Quotienten zweier Polynome $P_n (z)$ und $Q_n (z)$ vom Grade $n$ in $z$
sind, die Partialsummen $s_0$, $s_1$, $\ldots$, $s_{2 n}$ der Eulerreihe
ben\"otigt. Wenn man dagegen die diagonalen Pad\'e-Approximationen $[n / n]$
gem\"a{\ss} Gl. (6.7-6) \"uber die Drummondsche Transformation ${\cal D}_n^{(0)}
(s_0,\Delta s_0)$ berechnet, deren Z\"ahler und Nenner auch rekursiv
berechnet werden k\"onnen [Weniger 1989, Abschnitt 9.5], dann ben\"otigt man
nur die Partialsummen $s_0$, $s_1$, $\ldots$, $s_{n + 1}$. Im Falle der
Eulerreihe ist die Drummondsche Transformation demzufolge in etwa
doppelt so effizient wie der Wynnsche $\epsilon$-Algorithmus.

Es ist eine naheliegende Idee, Satz 6-16 f\"ur eine theoretische Analyse
des Transformationsfehlers bei der Summation der Eulerreihe durch
Pad\'e-Approximationen zu verwenden. Die Bedingungen (S-0) und (S-1) sind
offensichtlich erf\"ullt. Ungl\"ucklicherweise ist aber nicht klar, ob und
wie eine Folge $\Seqn \omega$ von Restsummenabsch\"atzungen f\"ur die
Eulerreihe gefunden werden kann, deren Elemente die Bedingung (S-2)
erf\"ullt. Bei Stieltjesreihen ist die Restsummenabsch\"atzung (6.5-16)
besonders naheliegend, da sie gem\"a{\ss} Satz 6-12 eine obere Schranke f\"ur
den Abbruchfehler (6.5-10) liefert. Im Falle der Eulerreihe m\"u{\ss}ten die
Restsummenabsch\"atzungen also folgenderma{\ss}en gew\"ahlt werden:
$$
\omega_n \; = \; (-1)^{n+1} \, (n+1)! \, z^{n+1} \, ,
\qquad n \in \N_0 \, .
\tag
$$
Leider konnte nicht explizit bewiesen werden, da{\ss} der Quotient $[s_n -
s] / \omega_n$ in Falle der Eulerreihe eine Darstellung durch eine
Fakult\"atenreihe gem\"a{\ss} Gl. (6.6-5) besitzt. Man kann also nur numerisch
untersuchen, ob die Absch\"atzung des Transformationsfehlers in Satz 6-16
eine angemessene Beschreibung der Summation der Eulerreihe (2.2-2) durch
Pad\'e-Approximationen liefert.

Eng verwandt mit dem nach Euler benannten Integral (2.2-1) ist das
sogenannte Exponential\-integral [Magnus, Oberhettinger und Soni 1966, S.
342]
$$
E_1 (z) \; = \; \int\nolimits_{z}^{\infty} \,
\frac {\e^{- x} } {x} \d x \, .
\tag
$$
Diese Beziehung kann leicht auf folgende Weise umformuliert werden
[Magnus, Oberhettinger und Soni 1966, S. 344]:
$$
z \, \e^z \, E_1 (z) \; = \;
\int\nolimits_{0}^{\infty} \,
\frac {\e^{- t} \d t} {1 + t / z} \, .
\tag
$$
Ein Vergleich dieses Integrals mit dem nach Euler benannten Integral
(2.2-1) zeigt, da{\ss} man ihm die Eulerreihe mit dem Argument $1/z$ als
asymptotische Potenzreihe f\"ur $z \to \infty$ zuordnen kann:
$$
z \, \e^z \, E_1 (z) \; \sim \;
\sum_{m=0}^{\infty} \, (-1)^m \, m! \, z^{- m} \; = \;
{}_2 F_0 (1,1; - 1/z) \, , \qquad z \to \infty \, .
\tag
$$
Der Konvergenzradius der hypergeometrischen Reihe ${}_2 F_0$ ist
offensichtlich gleich Null. Das bedeutet, da{\ss} diese Reihe f\"ur alle
endlichen Werte von $z$ sehr stark divergiert. Da aber alternative
Berechnungsm\"oglichkeiten f\"ur das Exponential\-integral $E_1 (z)$ mit $z
\in \R_{+}$ bekannt sind, ist Gl. (6.7-12) sehr gut geeignet, um das
F\"ahigkeit eines verallgemeinerten Summationsprozesses zu testen, auch
hochgradig divergente Reihen zu summieren. Die Vergleichswerte f\"ur das
Exponential\-integral $E_1 (z)$ werden hier mit Hilfe der FORTRAN
FUNCTION S13AAF der NAG Library berechnet. Dieses Programm berechnet
eine Approximation des Exponential\-integrals in DOUBLE PRECISION (15 -
16 Dezimalstellen) mit Hilfe geeigneter Tschebyscheffentwicklungen.

\beginFloat

\medskip

\beginTabelle [to \kolumnenbreite]
\beginFormat \rechts " \rechts " \mitte " \mitte " \mitte
\endFormat
\+ " \links {\bf Tabelle 6-1} \@ \@ \@ \@ " \\
\+ " \links {Summation der asymptotischen Reihe ${}_2 F_0 (1,1; -1/z)
\; = \; z \, \e^z \, E_1 (z)$ f\"ur $z \; = \; 3$}
\@ \@ \@ \@ " \\
\- " \- " \- " \- " \- " \- " \\ \sstrut {} {1.5 \jot} {1.5 \jot}
\+ " \rechts {$n$} " \mitte {Partialsumme $s_n$}
" ${\cal A}_{\Ent {n/2}}^{(n - 2 \Ent {n/2})}$
" ${\cal D}_n^{(0)} (s_0, \Delta s_0)$
" $\epsilon_{2 \Ent {n/2}}^{(n - 2 \Ent {n/2})}$ " \\
\+ " " \mitte {Gl. (6.7-13)} " Gl. (3.3-8) " Gl. (6.7-1) " Gl.
(2.4-10) " \\
\- " \- " \- " \- " \- " \- " \\ \sstrut {} {1 \jot} {1 \jot}
\+ "  10 "  $ 0.4831550069 \times 10^{02}$  "  0.78625130019479 "
0.78625125348502 "  0.78626367674141 " \\
\+ "  11 " $ -0.1770160037 \times 10^{03}$  "  0.78625114835779 "
0.78625123263883 "  0.78624220653206 " \\
\+ "  12 "  $ 0.7243100137 \times 10^{03}$  "  0.78625122394910 "
0.78625122525386 "  0.78625447790898 " \\
\+ "  13 " $-0.3181436062 \times 10^{04}$  "  0.78625121766831 "
0.78625122252501 "  0.78624881508686 " \\
\+ "  14 " $ 0.1504537896 \times 10^{05}$  "  0.78625122089403 "
0.78625122147819 "  0.78625215335611 " \\
\+ "  15 " $-0.7608869613 \times 10^{05}$  "  0.78625122063943 "
0.78625122106292 "  0.78625052018310 " \\
\+ "  16 " $ 0.4099597043 \times 10^{06}$  "  0.78625122077179 "
0.78625122089311 "  0.78625150842397 " \\
\+ "  17 " $-0.2344314565 \times 10^{07}$  "  0.78625122076057 "
0.78625122082175 "  0.78625100153477 " \\
\+ "  18 " $ 0.1418133105 \times 10^{08}$  "  0.78625122076626 "
0.78625122079099 "  0.78625131522011 " \\
\+ "  19 " $-0.9048109119 \times 10^{08}$  "  0.78625122076568 "
0.78625122077742 "  0.78625114787954 " \\
\+ "  20 " $ 0.6072683904 \times 10^{09}$  "  0.78625122076597 "
0.78625122077131 "  0.78625125348502 " \\
\+ "  21 " $-0.4276977981 \times 10^{10}$  "  0.78625122076594 "
0.78625122076850 "  0.78625119524201 " \\
\+ "  22 " $ 0.3154082874 \times 10^{11}$  "  0.78625122076595 "
0.78625122076718 "  0.78625123263883 " \\
\+ "  23 " $-0.2430623561 \times 10^{12}$  "  0.78625122076596 "
0.78625122076656 "  0.78625121141456 " \\
\+ "  24 " $ 0.1953763123 \times 10^{13}$  "  0.78625122076595 "
0.78625122076626 "  0.78625122525386 " \\
\+ "  25 " $-0.1635311587 \times 10^{14}$  "  0.78625122076596 "
0.78625122076611 "  0.78625121720071 " \\
\+ "  26 " $ 0.1423065021 \times 10^{15}$  "  0.78625122076596 "
0.78625122076603 "  0.78625122252501 " \\
\+ "  27 " $-0.1285630059 \times 10^{16}$  "  0.78625122076596 "
0.78625122076600 "  0.78625121935772 " \\
\+ "  28 " $ 0.1204177785 \times 10^{17}$  "  0.78625122076596 "
0.78625122076598 "  0.78625122147819 " \\
\+ "  29 " $-0.1167898319 \times 10^{18}$  "  0.78625122076596 "
0.78625122076597 "  0.78625122019177 " \\
\+ "  30 " $ 0.1171526266 \times 10^{19}$  "  0.78625122076596 "
0.78625122076596 "  0.78625122106292 " \\
\- " \- " \- " \- " \- " \- " \\ \sstrut {} {1 \jot} {1 \jot}
\+ " \links {NAG FUNCTION S13AAF} \@        "  0.78625122076594 "
0.78625122076594 "  0.78625122076594 " \\
\- " \- " \- " \- " \- " \- " \\ \sstrut {} {1 \jot} {1 \jot}
\endTabelle

\medskip

\endFloat

In Tabelle 6-1 wird der iterierte Aitkensche $\Delta^2$-Proze{\ss}, Gl.
(3.3-8), die Drummondsche Transformation, Gl. (6.7-1), und der Wynnsche
$\epsilon$-Algorithmus, Gl. (2.4-10), auf die Partialsummen
$$
s_n \; = \; \sum_{m=0}^n \, (-1)^m \, m! \; z^{- m} \, ,
\qquad n \in \N_0 \, ,
\tag
$$
der divergenten Reihe ${}_2 F_0$ in Gl. (6.7-12) mit $z = 3$ angewendet.
In Tabelle 6-1 verwendet die Drummondsche Transformation die
Restsummenabsch\"atzung (6.5-16). Im Falle der hypergeometrischen Reihe in
Gl. (6.7-12) bedeutet dies
$$
\omega_n \; = \; \Delta s_n \; = \;
(-1)^{n+1} \, (n+1)! \, z^{- n - 1} \, , \qquad n \in \N_0 \, .
\tag
$$

Im Falle des Aitkenschen iterierten $\Delta^2$-Prozesses, Gl. (3.3-8),
werden die Approximationen zum Grenzwert $s$ der zu transformierenden
Folge $\Seqn s$ folgenderma{\ss}en gew\"ahlt [Weniger 1989, Gl. (5.2-6)]:
$$
\left\{ s_{m - 2 \Ent {m/2}}, s_{m - 2 \Ent {m/2} + 1},
\ldots , s_m \right\} \, \to \;
{\cal A}_{\Ent {m/2}}^{(m - 2 \Ent {m/2})} \, .
\tag
$$
Dabei wird die Notation $\Ent x$ f\"ur den ganzzahligen Anteil von $x$
verwendet, der die gr\"o{\ss}te nat\"urliche Zahl $\nu$ ist, welche die
Ungleichung $\nu \le x$ erf\"ullt.

Im Falle der Drummondschen Transformation, Gl. (6.7-1), werden die
Approximationen zum Grenzwert $s$ der zu transformierenden Folge $\Seqn
s$ folgenderma{\ss}en gew\"ahlt [Weniger 1989, Gl. (9.5-8)]:
$$
\{ s_0, \omega_0; s_1, \omega_1; \ldots ; s_m, \omega_m \} \; \to \;
{\cal D}_m^{(0)} (s_0, \omega_0) \, ,
\qquad m \in \N_0 \, .
\tag
$$

Die Partialsummen und die Transformationen in Tabelle 6-1 wurden in
QUADRUPLE PRECISION (31 - 32 Dezimalstellen) berechnet. Um die
numerische Stabilit\"at dieser Summationen \"uberpr\"ufen zu k\"onnen, wurden
diese Rechnungen in DOUBLE PRECISION (15 - 16 Dezimalstellen)
wiederholt. Dabei ergab sich eine \"Ubereinstimmung auf mindestens 12
Dezimalstellen.

Ein Vergleich der drei verallgemeinerten Summationsprozesse in Tabelle
6-1 ist aufgrund ihrer verwandtschaftlichen Beziehungen von Interesse.
Sowohl der Aitkensche iterierte $\Delta^2$-Proze{\ss} als auch der Wynnsche
$\epsilon$-Algorithmus sind Iterationen des Aitkenschen
$\Delta^2$-Prozesses, Gl. (3.3-6), und man m\"ochte nat\"urlich wissen,
welche dieser beiden Verallgemeinerungen die besseren Ergebnisse
liefert. Da die hypergeometrische Reihe in Gl. (6.7-12) die Eulerreihe
mit dem Argument $1/z$ ist, kann man die G\"ultigkeit von Gl. (6.7-6)
\"uberpr\"ufen, indem man die Ergebnisse des $\epsilon$-Algorithmus und der
Drummondschen Transformation vergleicht.

Der klare Sieger in Tabelle 6-1 ist der Aitkensche iterierte
$\Delta^2$-Proze{\ss}, der f\"ur $n = 23$ eine Genauigkeit von 14
Dezimalstellen liefert{\footnote[\dagger]{Die letzte Stelle, die die NAG
FUNCTION S13AAF in Tabelle 6-1 liefert, ist falsch, und das Resultat des
iterierten $\Delta^2$-Prozesses ist korrekt.}}. Deutlich weniger
effizient ist die Drummondsche Transformation, die f\"ur $n = 30$ eine
Genauigkeit von 14 Dezimalstellen liefert, und der klare Verlierer ist
der $\epsilon$-Algorithmus. Die Ergebnisse in Tabelle 6-1 zeigen
au{\ss}erdem, da{\ss} die Pad\'e-Approximationen der Eulerreihe tats\"achlich durch
die Drummondsche Transformation gem\"a{\ss} Gl. (6.7-6) berechnet werden
k\"onnen, denn wir beobachten
$$
{\cal D}_n^{(0)} (s_0, \Delta s_0) \; = \;
\epsilon_{2 n}^{(0)} \, .
\tag
$$
Da ${\cal D}_{30}^{(0)} (s_0, \Delta s_0)$ eine Genauigkeit von
14 Dezimalstellen liefert, folgt aus dieser Beziehung, da{\ss} der
$\epsilon$-Algorithmus die Partialsummen $s_0$, $s_1$, $\ldots$
, $s_{60}$ der asymptotischen Reihe (6.7-12) ben\"otigen w\"urde, um die
gleiche Genauigkeit zu produzieren.

In Tabelle 6-2 werden die verallgemeinerten Summationsprozesse
$d_n^{(0)} (\zeta, s_0)$, Gl. (5.2-18), und ${\delta}_n^{(0)} (\zeta,
s_0)$, Gl. (5.4-13), mit $\zeta = 1$ und ${\Delta}_n^{(0)} (\xi, s_0)$,
Gl. (5.5-15), mit $\xi = 17$ zur Summation der divergenten Reihe in Gl.
(6.7-12) verwendet. Die drei verallgemeinerten Summationsprozesse
verwenden die gleiche Restsummenabsch\"atzung (6.7-14) wie ${\cal
D}_n^{(0)} (s_0, \Delta s_0)$ in Tabelle 6-1.

Tabelle 6-2 wurde ebenfalls in QUADRUPLE PRECISION berechnet. Eine
Wiederholung dieser Rechnungen in DOUBLE PRECISION ergab, da{\ss} bei den
Transformationen ${\delta}_n^{(0)} (1, s_0)$ und ${\Delta}_n^{(0)} (17,
s_0)$ alle 14 Stellen mit den vierfach genauen Ergebnissen
\"ubereinstimmten. Nur bei der Levinschen Transformation $d_n^{(0)} (1,
s_0)$ geschah es manchmal, da{\ss} die letzte ausgegebene Stelle des doppelt
genauen Ergebnisses falsch war.

\beginFloat

\medskip

\beginTabelle [to \kolumnenbreite]
\beginFormat \rechts " \rechts " \mitte " \mitte " \mitte
\endFormat
\+ " \links {\bf Tabelle 6-2} \@ \@ \@ \@ " \\
\+ " \links {Summation der asymptotischen Reihe ${}_2 F_0 (1,1; -1/z)
\; = \; z \, \e^z \, E_1 (z)$ f\"ur $z \; = \; 3$}
\@ \@ \@ \@ " \\
\- " \- " \- " \- " \- " \- " \\ \sstrut {} {1 \jot} {1 \jot}
\+ " \rechts {$n$} " \mitte {Partialsumme $s_n$}
" $d_n^{(0)} (1, s_0)$ " ${\delta}_n^{(0)} (1, s_0)$
" ${\Delta}_n^{(0)} (17, s_0)$ " \\
\+ " " \mitte {Gl. (6.7-13)} " Gl. (5.2-18) " Gl. (5.4-13) " Gl.
(5.5-15)" \\
\- " \- " \- " \- " \- " \- " \\ \sstrut {} {1 \jot} {1 \jot}
\+ "    3 " $  0.6666666667 \times 10^{00}$  "  0.78709677419355 "
0.78672985781991 "  0.78633660627852 " \\
\+ "    4 " $  0.9629629630 \times 10^{00}$  "  0.78607714016933 "
0.78622197922362 "  0.78625813355638 " \\
\+ "    5 " $  0.4691358025 \times 10^{00}$  "  0.78628225839245 "
0.78625036724446 "  0.78625167667778 " \\
\+ "    6 " $  0.1456790123 \times 10^{01}$  "  0.78624675493384 "
0.78625141640628 "  0.78625123654802 " \\
\+ "    7 " $ -0.8477366255 \times 10^{00}$  "  0.78625162955159 "
0.78625123162756 "  0.78625121997903 " \\
\+ "    8 " $  0.5297668038 \times 10^{01}$  "  0.78625123599599 "
0.78625121903376 "  0.78625122068020 " \\
\+ "    9 " $ -0.1313854595 \times 10^{02}$  "  0.78625120523222 "
0.78625122051031 "  0.78625122077447 " \\
\+ "   10 " $  0.4831550069 \times 10^{02}$  "  0.78625122396512 "
0.78625122077239 "  0.78625122076641 " \\
\+ "   11 " $ -0.1770160037 \times 10^{03}$  "  0.78625122056582 "
0.78625122077131 "  0.78625122076576 " \\
\+ "   12 " $  0.7243100137 \times 10^{03}$  "  0.78625122068924 "
0.78625122076646 "  0.78625122076598 " \\
\+ "   13 " $ -0.3181436062 \times 10^{04}$  "  0.78625122079175 "
0.78625122076590 "  0.78625122076596 " \\
\+ "   14 " $  0.1504537896 \times 10^{05}$  "  0.78625122076354 "
0.78625122076593 "  0.78625122076595 " \\
\+ "   15 " $ -0.7608869613 \times 10^{05}$  "  0.78625122076528 "
0.78625122076595 "  0.78625122076596 " \\
\+ "   16 " $  0.4099597043 \times 10^{06}$  "  0.78625122076622 "
0.78625122076596 "  0.78625122076596 " \\
\+ "   17 " $ -0.2344314565 \times 10^{07}$  "  0.78625122076593 "
0.78625122076596 "  0.78625122076596 " \\
\+ "   18 " $  0.1418133105 \times 10^{08}$  "  0.78625122076595 "
0.78625122076596 "  0.78625122076596 " \\
\- " \- " \- " \- " \- " \- " \\ \sstrut {} {1 \jot} {1 \jot}
\+ " \links {NAG FUNCTION S13AAF} \@   "  0.78625122076594 "
0.78625122076594 "  0.78625122076594 " \\
\- " \- " \- " \- " \- " \- " \\ \sstrut {} {1 \jot} {1 \jot}
\endTabelle

\medskip

\endFloat

Satz 6-16 impliziert, da{\ss} die Drummondsche Transformation ${\cal
D}_k^{(n)} (s_n, \omega_n)$, Gl. (6.7-1), deutlich weniger
leistungsf\"ahig sein sollte als die verallgemeinerten Summationsprozesse
${\cal L}_k^{(n)} (\zeta, s_n, \omega_n)$, Gl. (5.2-6), ${\cal
S}_k^{(n)} (\zeta, s_n, \omega_n)$, Gl. (5.4-6), und ${\cal M}_k^{(n)}
(\xi, s_n, \omega_n)$, Gl. (5.5-8). Die Ergebnisse in Tabellen 6-1 und
6-2 best\"atigen dies. Selbst $d_n^{(0)} (\zeta, s_0)$, das in Tabelle 6-2
etwas weniger gute Ergebnisse produzierte als die beiden anderen
Transformationen, ist deutlich leistungsf\"ahiger als die
verallgemeinerten Summationsprozesse in Tabelle 6-1, und sowohl
$\Delta_{n}^{(0)} (\xi, s_0)$ als auch $\delta_{n}^{(0)} (\zeta, s_0)$
sind in etwa doppelt so leistungsf\"ahig wie ${\cal D}_n^{(0)} (s_0,
\Delta s_0)$. Diese Beobachtung ist in \"Ubereinstimmung mit den
theoretischen Voraussagen in den S\"atzen 6-13, 6-14 und 6-16.

In Anbetracht der langsamen Konvergenz scheint eine Summation der
divergenten Reihe in Gl. (6.7-12) durch Pad\'e-Approximationen nicht
praktikabel zu sein, wenn das Argument der hypergeometrischen Reihe
deutlich kleiner ist als $z = 3$ wie in Tabelle 6-1. Wenn man aber
geeignete Varianten der verallgemeinerten Summationsprozesse ${\cal
L}_k^{(n)} (\zeta, s_n, \omega_n)$, Gl. (5.2-6), ${\cal S}_k^{(n)}
(\zeta, s_n, \omega_n)$, Gl. (5.4-6), und ${\cal M}_k^{(n)} (\xi, s_n,
\omega_n)$, Gl. (5.5-8), verwendet, kann die divergente Reihe in Gl.
(6.7-12) auch im Falle relativ kleiner Argumente mit vertretbarem
Aufwand summiert werden. Tabelle 6-3 zeigt, da{\ss} die verallgemeinerten
Summationsprozesse $d_n^{(0)} (\zeta, s_0)$, Gl. (5.2-18), und
${\delta}_n^{(0)} (\zeta, s_0)$, Gl. (5.4-13), mit $\zeta = 1$ und
${\Delta}_n^{(0)} (\xi, s_0)$, Gl. (5.5-15), mit $\xi = 29$ die
divergente asymptotische Reihe ${}_2 F_0$ in Gl. (6.7-12) selbst dann
noch mit einer Genauigkeit von 14 Dezimalstellen summieren k\"onnen, wenn
das Argument dieser asymptotischen Potenzreihe so klein wie $z = 1/2$
ist.

\beginFloat

\medskip

\beginTabelle [to \kolumnenbreite]
\beginFormat \rechts " \rechts " \mitte " \mitte " \mitte
\endFormat
\+ " \links {\bf Tabelle 6-3} \@ \@ \@ \@ " \\
\+ " \links {Summation der asymptotischen Reihe ${}_2 F_0 (1,1; -1/z)
\; = \; z \, \e^z \, E_1 (z)$ f\"ur $z \; = \; 1/2$}
\@ \@ \@ \@ " \\
\- " \- " \- " \- " \- " \- " \\ \sstrut {} {1 \jot} {1 \jot}
\+ " \rechts {$n$} " \mitte {Partialsumme $s_n$}
" $d_n^{(0)} (1, s_0)$ " ${\delta}_n^{(0)} (1, s_0)$
" ${\Delta}_n^{(0)} (29, s_0)$ " \\
\+ " " \mitte {Gl. (6.7-13)} " Gl. (5.2-18) " Gl. (5.4-13) " Gl.
(5.5-15)" \\
\- " \- " \- " \- " \- " \- " \\ \sstrut {} {1 \jot} {1 \jot}
\+ "   15 " $ -0.4147067254 \times 10^{17}$  "  0.46145531715043 "
0.46145531958535 "  0.46145595366489 " \\
\+ "   16 " $  0.1329725286 \times 10^{19}$  "  0.46145530923846 "
0.46145531701552 "  0.46145551453546 " \\
\+ "   17 " $ -0.4529093729 \times 10^{20}$  "  0.46145531613431 "
0.46145531625982 "  0.46145536941468 " \\
\+ "   18 " $  0.1633052915 \times 10^{22}$  "  0.46145531735759 "
0.46145531613493 "  0.46145532757622 " \\
\+ "   19 " $ -0.6214401349 \times 10^{23}$  "  0.46145531627646 "
0.46145531616450 "  0.46145531778365 " \\
\+ "   20 " $  0.2488938643 \times 10^{25}$  "  0.46145531605612 "
0.46145531620445 "  0.46145531622965 " \\
\+ "   21 " $ -0.1046565329 \times 10^{27}$  "  0.46145531622971 "
0.46145531622787 "  0.46145531618769 " \\
\+ "   22 " $  0.4609744216 \times 10^{28}$  "  0.46145531627375 "
0.46145531623807 "  0.46145531623838 " \\
\+ "   23 " $ -0.2122526902 \times 10^{30}$  "  0.46145531624564 "
0.46145531624153 "  0.46145531624494 " \\
\+ "   24 " $  0.1019714416 \times 10^{32}$  "  0.46145531623631 "
0.46145531624231 "  0.46145531624191 " \\
\+ "   25 " $ -0.5102726985 \times 10^{33}$  "  0.46145531624080 "
0.46145531624227 "  0.46145531624156 " \\
\+ "   26 " $  0.2655415912 \times 10^{35}$  "  0.46145531624283 "
0.46145531624210 "  0.46145531624194 " \\
\+ "   27 " $ -0.1434925159 \times 10^{37}$  "  0.46145531624214 "
0.46145531624197 "  0.46145531624189 " \\
\+ "   28 " $  0.8040791666 \times 10^{38}$  "  0.46145531624170 "
0.46145531624191 "  0.46145531624184 " \\
\+ "   29 " $ -0.4666476909 \times 10^{40}$  "  0.46145531624180 "
0.46145531624188 "  0.46145531624188 " \\
\+ "   30 " $  0.2801466126 \times 10^{42}$  "  0.46145531624189 "
0.46145531624187 "  0.46145531624187 " \\
\- " \- " \- " \- " \- " \- " \\ \sstrut {} {1 \jot} {1 \jot}
\+ " \links {NAG FUNCTION S13AAF} \@  "  0.46145531624187 "
0.46145531624187 "  0.46145531624187 " \\
\- " \- " \- " \- " \- " \- " \\ \sstrut {} {1 \jot} {1 \jot}
\endTabelle

\medskip

\endFloat

In Tabelle 6-3 war es essentiell, QUADRUPLE PRECISION zu verwenden. In
DOUBLE PRECISION beobachtet man in Abh\"angigkeit von der
Transformationsordnung den Verlust zahlreicher Stellen durch
Rundungsfehler. Die h\"ochste Genauigkeit in DOUBLE PRECISION liefert
${\Delta}_n^{(0)} (\xi, s_0)$ f\"ur $n = 20$ (10 Dezimalstellen). F\"ur
gr\"o{\ss}ere Werte der Transformationsordnung $n$ nimmt die Genauigkeit der
Summationsergebnisse rasch ab. Beispielsweise produziert die Levinsche
Transformation $d_{30}^{(0)} (\zeta, s_0)$ schon v\"ollig unsinnige
Resultate, und auch ${\delta}_{30}^{(0)} (\zeta, s_0)$ und
${\Delta}_{30}^{(0)} (\xi, s_0)$ erreichen nur noch eine Genauigkeit von
3 Dezimalstellen.

Wenn man die anderen Varianten der verallgemeinerten Summationsprozesse
${\cal L}_k^{(n)} (\zeta, s_n, \omega_n)$, Gl. (5.2-6), ${\cal
S}_k^{(n)} (\zeta, s_n, \omega_n)$, Gl. (5.4-6), und ${\cal M}_k^{(n)}
(\xi, s_n, \omega_n)$, Gl. (5.5-8), die auf den Restsummenabsch\"atzungen
(5.2-12), (5.2-14), (5.2-16), (5.2-19) beziehungsweise (5.5-12)
basieren, zur Summation der divergenten Reihe ${}_2 F_0$ in Gl. (6.7-12)
verwendet, findet man, da{\ss} diese verallgemeinerten Summationsprozesse
\"ahnlich wirkungsvoll sind wie $d_n^{(0)} (\zeta, s_0)$,
${\delta}_n^{(0)} (\zeta, s_0)$ oder ${\Delta}_n^{(0)} (\xi, s_0)$, die
in Tabellen 6-2 und 6-3 verwendet wurden.

Es ist eine bemerkenswerte Beobachtung, da{\ss} Pad\'e-Approximationen, selbst
wenn sie mit Hilfe der Drummondschen Transformation gem\"a{\ss} Gl. (6.7-6)
und nicht mit Hilfe des Wynnschen $\epsilon$-Algorithmus berechnet
werden, den verallgemeinerten Summationsprozessen $d_n^{(0)} (\zeta,
s_0)$, Gl. (5.2-18), ${\delta}_n^{(0)} (\zeta, s_0)$, Gl. (5.4-13), oder
${\Delta}_n^{(0)} (\xi, s_0)$, Gl. (5.5-15), hoffnungslos unterlegen
sind, was die Summation der hochgradig divergenten Eulerreihe betrifft.
Diese Unterlegenheit der Pad\'e-Approximationen in Summationsprozessen ist
wahrscheinlich konstruktionsbedingt. Die Pad\'e-Approximationen $[n +k /
k]$ f\"ur die Eulerreihe k\"onnen auf der Basis der Modellfolge (6.7-3)
konstruiert werden. Der Summationsrest dieser Modellfolge ist
betragsm\"a{\ss}ig von der Ordnung $O (z^{n+k} n^{n+k})$ f\"ur $n \to \infty$.
Aus Satz 6-12 folgt aber, da{\ss} der tats\"achliche Summationsrest der
Eulerreihe f\"ur alle $z \in \R_{+}$ betragsm\"a{\ss}ig abgesch\"atzt werden kann
durch $(n+1)! z^{n+1}$. Diese Absch\"atzung ist aber nur von der Ordnung
$O (z^{n+1} n^{n+1})$ f\"ur $n \to \infty$.

Die Restsummenabsch\"atzung, die {\it implizit} in der Modellfolge f\"ur die
Pad\'e-Approximationen $[n +k / k]$ enthalten ist, ist sowohl im Falle der
Eulerreihe als auch bei anderen hochgradig divergenten Reihen
unrealistisch gro{\ss}. Demzufolge kann man auch nicht erwarten, da{\ss}
Pad\'e-Approximationen bei der Summation hochgradig divergenter Reihen
\"ahnlich gute Ergebnisse liefern wie verallgemeinerte Summationsprozesse,
die wesentlich bessere {\it explizite} Restsummenabsch\"atzungen
verwenden. Zahlreiche Rechnungen demonstrieren, da{\ss} die Unterlegenheit
der Pad\'e-Approximationen in Summationsprozessen offensichtlich
allgemeinerer Natur ist [Weniger und Steinborn 1989a; Weniger 1989;
1990; 1992; Weniger und {\v C}{\'\i}{\v z}ek 1990; Weniger, {\v
C}{\'\i}{\v z}ek und Vinette 1991; 1993; {\v C}{\'\i}{\v z}ek, Vinette
und Weniger 1991; 1993; Grotendorst 1991].

In den Abschnitten 10 und 11 von Weniger [1989] oder in Weniger [1991]
werden noch zahlreiche andere verallgemeinerte Summationsprozesse
beschrieben, die zur Summation der Eulerreihe geeignet sind. Numerische
Tests ergaben aber, da{\ss} diese verallgemeinerten Summationsprozesse, die
\"ahnlich wie der Wynnsche $\epsilon$-Algorithmus nur die Partialsummen
einer unendlichen Reihe als Eingabedaten verwenden, die divergente Reihe
${}_2 F_0$ in Gl. (6.7-12) deutlich weniger effizient summieren als die
oben erw\"ahnten verallgemeinerten Summationsprozesse, welche die
zus\"atzlichen Informationen, die in realistischen expliziten
Restsummenabsch\"atzungen enthalten sind, nutzbringend verwerten k\"onnen
[Weniger 1989, S. 330].

\endAbschnittsebene

\endAbschnittsebene

\keinTitelblatt\neueSeite

\beginAbschnittsebene
\aktAbschnitt = 6

\Abschnitt Rationale Approximationen f\"ur die modifizierte
Besselfunktion der zweiten Art

\vskip - 2 \jot

\beginAbschnittsebene

\medskip

\Abschnitt Vorbemerkungen

\smallskip

\aktTag = 0

Die effiziente und verl\"a{\ss}liche Berechnung spezieller Funktionen
ist f\"ur einen numerisch arbeitenden Naturwissenschaftler ein
h\"aufiges und keineswegs immer leicht zu l\"osendes
Problem. Dementsprechend umfangreich ist die Literatur auf diesem
Gebiet. Eine Beschreibung verschiedener Verfahren zur Berechnung
spezieller Funktionen findet man in einem
\"Ubersichtsartikel von Gautschi [1975]. Dar\"uber hinaus gibt es
Monographien, die sich vorwiegend oder sogar ausschlie{\ss}lich mit der
Berechnung spezieller Funktionen besch\"aftigen [Luke 1969a; 1969b; 1975;
1977; van der Laan und Temme 1980].

Au{\ss}erdem enthalten numerische Programmbibliotheken wie etwa die NAG
Library Programme f\"ur die wichtigsten speziellen Funktionen. Allerdings
sind diese Programmbibliotheken nicht vollst\"andig, und es gibt
zahlreicher spezielle Funktionen, deren effiziente und verl\"a{\ss}liche
Berechnung immer noch nicht auf v\"ollig befriedigende Weise m\"oglich ist.

Da f\"ur die Mehrheit der speziellen Funktionen Darstellungen durch
konvergente oder divergente Reihen bekannt sind, ist es naheliegend,
verallgemeinerte Summationsprozesse zur Berechnung von speziellen
Funktionen zu verwenden. Pad\'e-Approximationen und Kettenbr\"uche geh\"oren
inzwischen zu den Standardverfahren und werden erfolgreich zur
Berechnung von vielen speziellen Funktionen verwendet [Gautschi 1975,
Abschnitt 1.4; Luke 1975; 1977; van der Laan und Temme 1980, Abschnitt
\Roemisch{2}.4]. In diesem Abschnitt soll aber am Beispiel der {\it
modifizierten Besselfunktion der zweiten Art} gezeigt werde, da{\ss} die in
Abschnitt 5 dieser Arbeit beschriebenen verallgemeinerten
Summationsprozesse, die als Eingabedaten nicht nur die Partialsummen
$\Seqn s$ einer unendlichen Reihe, sondern auch explizite
Restsummenabsch\"atzungen $\Seqn \omega$ verwenden, wesentlich bessere
Ergebnisse liefern k\"onnen als Pad\'e-Approximationen und Kettenbr\"uche
[Weniger und {\v C}{\'\i}{\v z}ek 1990].

Die modifizierte Besselfunktion der zweiten Art wurde deswegen als
Demonstrationsobjekt gew\"ahlt, weil sie ohne Zweifel zu den wichtigeren
speziellen Funktionen geh\"ort, weil bei ihrer Berechnung interessante
praktische und theoretische Probleme auftreten, und weil sie indirekt
beim wissenschaftlichen Werdegang des Autors eine wesentliche Rolle
gespielt hat.

Die Besselfunktionen sind L\"osungen der {\it Besselschen
Differentialgleichung} [Magnus, Oberhettinger und Soni 1966, S. 65]:
$$
\Bigl\{ z^2 \frac {d^2}{d z^2} \, + \, z \frac {d}{d z} \, + \,
( z^2 \, - \, \nu^2 ) \Bigr\} \; u (z) \; = \; 0 \, ,
\qquad \nu, z \in \C \, .
\tag
$$
Besselfunktionen geh\"oren zu den am meisten verwendeten speziellen
Funktionen \"uberhaupt. Eine unvoll\-st\"andige Liste von Anwendungen der
Besselfunktionen in verschiedenen Bereichen der Mathematik, der
Naturwissenschaften und der Technik findet man auf S. 98 des Buches von
Lebedev [1972] oder in den Abschnitten \Roemisch{2} und \Roemisch{7} des
Buches von Bowman [1958]. Einen \"Uberblick \"uber die geschichtliche
Entwicklung der Besselfunktionen findet man in Abschnitt \Roemisch{1}
des Buches von Watson [1966].

Wenn man in der Besselschen Differentialgleichung (7.1-1) $z$ durch
$\i \, z$ ersetzt, erh\"alt man die {\it modifizierte Besselsche
Differentialgleichung} [Magnus, Oberhettinger und Soni 1966, S. 66]:
$$
\Bigl\{ z^2 \frac {d^2}{d z^2} \, + \, z \frac {d}{d z} \, - \,
( z^2 \, + \, \nu^2 ) \Bigr\} \; u (z) \; = \; 0 \, ,
\qquad \nu, z \in \C \, .
\tag
$$
Die erste L\"osung der Besselschen modifizierten Differentialgleichung,
die \"ublicherweise {\it modifizierte Besselfunktion der ersten Art}
genannt wird, kann durch die folgende Reihenentwicklung definiert werden
[Magnus, Oberhettinger und Soni 1966, S. 66]:
$$
I_{\nu} (z) \; = \; \sum_{m=0}^{\infty} \,
\frac {(z/2)^{\nu + 2 m}} {m! \, \Gamma (\nu + m + 1)}
\; = \; \frac {(z/2)^{\nu}} {\Gamma (\nu + 1)} \,
{}_0 F_1 (\nu + 1; z^2/4) \, ,
\qquad \nu, z \in \C \, .
\tag
$$
Die Potenzreihe (7.7-3) f\"ur $I_{\nu} (z)$ konvergiert absolut und
gleichm\"a{\ss}ig auf jeder kompakten Teilmenge von $\C$. F\"ur $\nu, z > 0$
sind alle Terme dieser Reihe positiv, und der Verlust signifikanter
Stellen durch Ausl\"oschung ist nicht zu bef\"urchten. F\"ur $\nu > 0$ kann
man zeigen, da{\ss} die Terme dieser Reihe zumindest f\"ur hinreichend gro{\ss}e
Laufindizes $m$ durch die Terme der Potenzreihe f\"ur $\exp(z/2)$
majorisiert werden. Daraus folgt, da{\ss} $I_{\nu} (z)$ f\"ur nicht zu gro{\ss}e
Argumente $z$ effizient und zuverl\"assig mit Hilfe der Potenzreihe
(7.1-3) berechnet werden kann. Wenn das Argument $z$ aber gro{\ss} ist, oder
wenn man gleichzeitig ganze Strings $I_{\nu + n} (z)$ mit $n = 0, 1, 2,
\ldots$ ben\"otigt, ist es vorteilhaft, die modifizierten
Besselfunktionen mit Hilfe des ber\"uhmten Miller-Algorithmus rekursiv zu
berechnen. Gute Beschreibungen des Miller-Algorithmus findet man in
einem \"Ubersichtsartikel von Gautschi [1967] oder in einem Buch von Wimp
[1984].

Da die Ordnung $\nu$ in der modifizierten Besselschen
Differentialgleichung (7.1-2) quadratisch vorkommt, ist nicht nur
$I_{\nu} (z)$ eine L\"osung dieser Differentialgleichung, sondern auch die
Funktion $I_{- \nu} (z)$. Demzufolge scheint es naheliegend zu sein,
$I_{- \nu} (z)$ als {\it modifizierte Besselfunktion der zweiten Art}
zu bezeichnen. Wenn die Ordnung $\nu$ aber eine positive oder negative
ganze Zahl ist, dann sind die beiden L\"osungen $I_{\nu} (z)$ und
$I_{- \nu} (z)$ nicht mehr linear unabh\"angig, und es gilt [Magnus,
Oberhettinger und Soni 1966, S. 70]
$$
I_n (z) \; = \; I_{- n} (z) \, , \qquad n \in \N_0 \, .
\tag
$$
Man kann aber eine zweite L\"osung der modifizierten Besselschen
Differentialgleichung (7.1-2) konstruieren, die f\"ur alle $\nu \in \C$
von der ersten L\"osung $I_{\nu} (z)$ linear unabh\"angig ist, und die
deswegen \"ublicherweise {\it modifizierte Besselfunktion der zweiten
Art} genannt wird{\footnote[\dagger]{In der mathematischen Literatur
werden au{\ss}erdem noch die folgenden Namen f\"ur $K_{\nu} (z)$ verwendet:
Bassetfunktion, Macdonaldfunktion, modifizierte Besselfunktion der
dritten Art, hyperbolische Besselfunktion der zweiten (dritten) Art,
modifizierte Hankelfunktion der zweiten (dritten) Art.}} [Magnus,
Oberhettinger und Soni 1966, S. 66]:
$$
K_{\nu} (z) \; = \; \frac {\pi} {2 \sin (\pi \nu)} \,
\bigl\{ I_{- \nu} (z) \, - \, I_{\nu} (z) \bigr\} \, ,
\qquad \nu, z \in \C \, .
\tag
$$
Aus dieser Beziehung folgt sofort, da{\ss} $K_{\nu} (z)$ nicht vom
Vorzeichen von $\nu$ abh\"angt,
$$
K_{\nu} (z) \; = \; K_{- \nu} (z) \, .
\tag
$$

Wie schon erw\"ahnt, wirft die effiziente und verl\"a{\ss}liche Berechnung von
$I_{\nu} (z)$ keine besonderen Probleme auf. Demzufolge liegt es nahe,
$K_{\nu} (z)$ direkt durch Gl. (7.1-5) als Differenz zweier
$I$-Funktionen zu berechnen. Ungl\"ucklicherweise kann Gl. (7.1-5) in
vielen F\"allen nicht verwendet werden. Wenn $\nu$ eine ganze Zahl ist,
$\nu = n$ mit $n \in \N_0$, dann ist die rechte Seite von Gl. (7.1-5)
aufgrund von Gl. (7.1-4) undefiniert. Man mu{\ss} dann einen Grenzproze{\ss}
$\nu \to n$ durchf\"uhren, der den folgenden Ausdruck f\"ur die modifizierte
Besselfunktion mit ganzzahliger Ordnung $n$ ergibt [Abramowitz und
Stegun 1972, S. 375]:
$$
\beginAligntags
K_n (z) \; " = \; " " \frac {(2/z)^n} {2} \, \sum_{k=0}^{n-1} \,
\frac {(n-k-1)!}{k!} \, (- z^2/4)^k \, + \,
(-1)^{n+1} \ln (z/2) \, I_n (z) \\
" + "" \frac {(- z/2)^n} {2} \, \sum_{k=0}^{\infty} \,
\bigl\{ \psi (k+1) \, + \, \psi (n+k+1) \bigr\} \,
\frac {(z/2)^{2 k}}{k! \, (n+k)!} \, . \\
\tag
\endAligntags
$$
Wenn $n = 0$ gilt, ist die erste Summe in Gl. (7.1-7) eine leere Summe,
die Null ist.

In der Praxis ist die Berechnung der modifizierten Besselfunktionen
$K_n (z)$ mit ganzzahliger Ordnung $n$ unproblematisch, da
leistungsf\"ahige Programme in numerischen Programmbibliotheken vorhanden
sind. Beispiele sind die Programme S18CCF und S18CDF der NAG Library,
die $\e^x K_0 (x)$ beziehungsweise $\e^x K_1 (x)$ f\"ur $x > 0$ mit Hilfe
geeigneter Tschebyscheffentwicklungen berechnen. Die anderen Funktionen
$K_n (x)$ mit $n \ge 2$ k\"onnen bequem mit Hilfe der Dreitermrekursion
[Magnus, Oberhettinger und Soni 1966, S. 67]
$$
K_{\nu + 1} (z) \; = \; (2 \nu / z) \, K_{\nu} (z)
\, + \, K_{\nu - 1} (z)
\tag
$$
berechnet werden, die bekanntlich {\it aufw\"artsstabil} ist.

Wenn die Ordnung $\nu$ sich nur wenig von einer ganzen Zahl
unterscheidet, ist es nicht empfehlenswert, $K_{\nu} (z)$ mit Hilfe von
Gl. (7.1-5) zu berechnen, da dann Stellenverlust zu erwarten ist. In
einem solchen Fall sollte man $K_{\nu} (z)$ laut Campbell [1981] und
Thompson und Barnett [1987] mit Hilfe der folgenden, von Goldstein und
Thaler [1959] abgeleiteten Neumannreihe berechnen:
$$
K_{\nu} (z) \; = \;
\sum_{m=0}^{\infty} \, \delta_m \, I_{\nu + 2 m} (z) \, .
\tag
$$
F\"ur die beiden ersten Koeffizienten $\delta_0$ und $\delta_1$ der
Neumannreihe (7.1-9) sind explizite Ausdr\"ucke bekannt. Die h\"oheren
Koeffizienten $\delta_m$ mit $m \ge 2$ sind dagegen rekursiv definiert
[Goldstein und Thaler 1959, S. 106 - 107]:
$$
\beginAligntags
" \delta_0 \; " = \; " - \, \frac {1} {2 \nu} \,
\left[ \frac {\pi \nu} {\sin (\pi \nu)} \, - \,
{\left( \frac {2} {x} \right)}^2 \, \bigl[ \Gamma (\nu + 1) \bigr]^2
\right] \, ,
\erhoehe\aktTag \\ \tag*{\tagnr a}
" \delta_1 \; " = \; " {\left( \frac {2} {x} \right)}^2 \,
\frac {2 + \nu} {1 - \nu} \,
\bigl[ \Gamma (\nu + 1) \bigr]^2 \, ,
\\ \tag*{\tagform\aktTagnr b}
" \delta_{m+1} \; " = \; " \frac {m (\nu + 2 m - 2) (m - \nu)}
{(\nu + 2 m) (2 \nu + m - 1) (\nu + m - 1)}
\; \delta_m \, , \qquad m \ge 1 \, .
\\ \tag*{\tagform\aktTagnr c}
\endAligntags
$$
Da $I_{\nu} (z)$ f\"ur $\nu, z > 0$ eine strikt monoton abnehmende
Funktion der Ordnung $\nu$ ist [Magnus, Oberhettinger und Soni 1966, S.
151], sollte die Neumannreihe (7.1-9) normalerweise schnell genug
konvergieren, um praktisch n\"utzlich zu sein.

Ungl\"ucklicherweise gibt es neben Ordnungen, die sich nur wenig von einer
ganzen Zahl unterscheiden, auch noch andere Komplikationen, welche die
praktische Verwendbarkeit der Darstellung (7.1-5) zur Berechnung von
$K_{\nu} (z)$ einschr\"anken. Selbst wenn die Ordnung $\nu$ sich deutlich
von der n\"achsten ganzen Zahl unterscheidet, kann die Differenz $I_{-
\nu} (z) - I_{\nu} (z)$ in Gl. (7.1-5) nur dann mit ausreichender
Genauigkeit berechnet werden, wenn $z$ relativ klein ist. Das folgt
sofort aus den asymptotischen Entwicklungen der modifizierten
Besselfunktionen $K_{\nu} (z)$ und $I_{\nu} (z)$ f\"ur $\vert z \vert \to
\infty$ [Watson 1966, S. 202 - 203]:
$$
\beginAligntags
" K_{\nu} (z) \; " \sim \;
" [\pi / (2 z)]^{1/2} \, \e^{- z} \,
{}_2 F_0 \bigl( 1/2 + \nu, 1/2 - \nu; - 1/(2 z) \bigr) \, ,
\\ \tag
" I_{\nu} (z) \; " \sim \;
" \frac {\e^z} {(2 \pi z)^{1/2}} \,
{}_2 F_0 \bigl( 1/2 + \nu, 1/2 - \nu; 1/(2 z) \bigr) \, \\
" " + \,
" \frac {\e^{- z \pm \i \pi (\nu + 1/2)}} {(2 \pi z)^{1/2}} \,
{}_2 F_0 \bigl( 1/2 + \nu, 1/2 - \nu; - 1/(2 z) \bigr) \, . \\
\tag
\endAligntags
$$
Die asymptotische Entwicklung f\"ur $K_{\nu} (z)$ ist f\"ur $\vert \arg (z)
\vert < 3 \pi / 2$ g\"ultig. In der analogen asymp\-totischen Entwicklung
f\"ur $I_{\nu} (z)$ mu{\ss} man in Abh\"angigkeit von $\arg (z)$ zwei F\"alle
unterscheiden: In der zweiten Exponentialfunktion ist f\"ur $- \pi / 2 <
\arg (z) < 3 \pi / 2$ das positive Vorzeichen zu verwenden, und das
negative f\"ur $- 3 \pi / 2 < \arg (z) < \pi / 2$.

Wenn die Ordnung $\nu$ halbzahlig ist, $\nu = \pm (n + 1/2)$ und $n \in
\N_0$, dann brechen die beiden asymptotischen Reihen (7.1-11) und
(7.1-12) nach einer endlichen Anzahl von Termen ab und ergeben die
expliziten Ausdr\"ucke f\"ur die modifizierten Besselfunktionen $I_{\pm (n
+ 1/2)} (z)$ und $K_{n + 1/2} (z)$ [Watson 1966, S. 80]:
$$
\beginAligntags
" I_{n + 1/2} (z) \; = \; [2 \pi z]^{- 1/2} \, \\
" \qquad \times \left\{
e^z \, \sum_{\nu = 0}^{n} \,
\frac {(-1)^{\nu} (n + \nu)!} {{\nu}! (n - \nu)! (2 z)^{\nu}} \, + \,
(-1)^{n+1} \, e^{- z} \, \sum_{\nu = 0}^{n} \,
\frac {(n + \nu)!} {{\nu}! (n - \nu)! (2 z)^{\nu}}
\right\} \, , \qquad
\\ \tag
" I_{- n - 1/2} (z) \; = \; [2 \pi z]^{- 1/2} \, \\
" \qquad \times \left\{
e^z \, \sum_{\nu = 0}^{n} \,
\frac {(-1)^{\nu} (n + \nu)!} {{\nu}! (n - \nu)! (2 z)^{\nu}} \, + \,
(-1)^n \, e^{- z} \, \sum_{\nu = 0}^{n} \,
\frac {(n + \nu)!} {{\nu}! (n - \nu)! (2 z)^{\nu}}
\right\} \, ,
\\ \tag
" K_{n + 1/2} (z) \; = \; [\pi/(2 z)]^{1/2} \,
e^{- z} \, \sum_{\nu = 0}^{n} \,
\frac {(n + \nu)!} {{\nu}! (n - \nu)! (2 z)^{\nu}} \, .
\\ \tag
\endAligntags
$$
Wenn die Ordnung $\nu$ einer $K$-Funktion halbzahlig ist, $\nu = \pm (n
+ 1/2)$ und $n \in \N_0$, dann kann man sie problemlos mit Hilfe der
Rekursionsformel (7.1-8) und den Startwerten
$$
K_{\pm 1/2} (z) \; = \; [\pi/(2 z)]^{1/2} \, e^{- z}
\tag
$$
berechnen. Die speziellen Techniken zur Berechnung von $K$-Funktionen,
die in diesem Abschnitt noch beschrieben werden, sind im Falle
halbzahliger Ordnungen also \"uberfl\"ussig.

Aus den beiden asymptotischen Reihen (7.1-11) und (7.1-12) folgt, da{\ss}
$K_{\nu} (z)$ f\"ur gro{\ss}e Werte von $z$ in etwa wie $\exp (- z)$ f\"allt,
wogegen $I_{\pm \nu} (z)$ in etwa wie $\exp (z)$ w\"achst. In Gl. (7.1-5)
wird also eine exponentiell fallende Funktion als Differenz zweier
exponentiell wachsender Funktionen dargestellt. F\"ur gr\"o{\ss}ere Werte von
$z$ f\"uhrt das unweigerlich zu Stellenverlust durch Ausl\"oschung und
letztlich zu v\"ollig unsinnigen Ergebnissen. $K_{\nu} (z)$ kann also nur
dann ausreichend genau mit Hilfe von Gl. (7.1-5) berechnet werden, wenn
$z$ vergleichsweise klein ist.

Wenn $z$ sehr gro{\ss} ist, kann man $K_{\nu} (z)$ auf einfache Weise
berechnen, indem man die divergente Reihe ${}_2 F_0$ in Gl. (7.1-11) auf
optimale Weise abbricht. Diese Vorgehensweise, die beispielsweise in
Abschnitt 14 von Olver [1974] beschrieben ist, liefert aber nur dann
ausreichend genaue Approximationen, wenn $z$ sehr gro{\ss} ist. Wenn $z$ nur
mittelgro{\ss} ist, ergibt das optimale Abbrechen der divergenten Reihe
Approximationen, die in den meisten F\"allen nicht genau genug sind. Man
ben\"otigt also alternative Algorithmen, um die modifizierte
Besselfunktion $K_{\nu} (z)$ auch im problematischen Bereich
mittelgro{\ss}er Argumente, in dem weder Gl. (7.1-5) noch die asymptotische
Reihe (7.1-11) verwendet werden k\"onnen, berechnen zu k\"onnen.

Es gibt zahlreiche Arbeiten \"uber alternative Algorithmen zur Berechnung
von $K_{\nu} (z)$, die nicht auf Gl. (7.1-5) basieren. Hitotumatu [1967]
berechnete Quotienten $K_{\nu+1} (z)/K_{\nu} (z)$ von modifizierten
Besselfunktionen mit Hilfe von Kettenbr\"uchen. Luke [1971] konstruierte
f\"ur $K_{\nu} (z)$ abgebrochene doppelte Tschebyscheffentwicklungen in
den Variablen $\nu$ und $z$. Kostroun [1980] approximierte eine bekannte
Integraldarstellung f\"ur $K_{\nu} (z)$ durch eine unendliche Reihe. Dann
gibt es noch einen Algorithmus von Temme [1975], der auf der folgenden
Darstellung von $K_{\nu} (z)$ durch eine Kummerfunktion basiert [Magnus,
Oberhettinger und Soni 1966, S. 283]:
$$
K_{\nu} (z) \; = \; \pi^{1/2} \, (2 z)^{\nu} \, \e^{- z} \,
U (\nu + 1/2, 2 \nu + 1, 2 z) \, .
\tag
$$
Temme [1975] zeigte, da{\ss} die Kummerfunktion in Gl. (7.1-17) mit Hilfe
des Miller-Algorithmus rekursiv berechnet werden kann. Campbell [1980]
konnte den Algorithmus von Temme noch verbessern. Diese Modifikation des
Temmeschen Algorithmus ist die Basis von Programmen von Campbell [1981;
1982] und von Thompson und Barnett [1987]. Diese Programme k\"onnen
$K_{\nu} (z)$ f\"ur beliebige $\nu \in \R$ und $z \in \C$ berechnen.
Dunster und Lutz [1991] berechneten Besselfunktionen mit Hilfe von
Fakult\"atenreihen, und Hautot [1082] konstruierte verallgemeinerte
Pad\'e-Approximationen f\"ur $K_{\nu} (z)$.

In diesem Abschnitt wollen wir aber anders vorgehen: Wir wollen
versuchen, $K_{\nu} (z)$ in dem problematischen Bereich mittelgro{\ss}er
Argumente durch rationale Funktionen zu approximieren, die durch
verallgemeinerte Summationsprozesse entstanden sind. Der Ausgangspunkt
f\"ur die Konstruktion verschiedener rationaler Approximationen ist die
divergente hypergeometrische Reihe ${}_2 F_0$ in Gl. (7.1-11). Das
Schwergewicht liegt dabei nicht wie sonst in der Literatur auf den
Pad\'e-Approximationen, sondern auf den bei solchen Problemen deutlich
leistungsf\"ahigeren verallgemeinerten Summationsprozessen, die in
Abschnitt 5 beschrieben wurden.

\medskip

\Abschnitt Borelsummierbarkeit und Stieltjessummierbarkeit

\smallskip

\aktTag = 0

In diesem Unterabschnitt soll untersucht werden, ob die divergente
asymptotische Reihe ${}_2 F_0$ in Gl. (7.1-11) borelsummierbar ist, und
ob sie eine Stieltjesreihe ist.

Aus der folgenden asymptotischen Beziehung f\"ur den Quotienten zweier
Gammafunktionen [Magnus, Oberhettinger und Soni 1966, S. 12],
$$
\frac {\Gamma (z + \alpha)} {\Gamma (z + \beta)} \; = \;
z^{\alpha - \beta} \bigl\{ 1 + O (1/z) \bigr\} \, ,
\qquad \vert \arg (z) \vert \, < \, \pi \, ,
\quad \vert z \vert \, \to \, \infty \, ,
\tag
$$
erhalten wir eine asymptotische Absch\"atzung f\"ur die Koeffizienten der
divergenten Reihe ${}_2 F_0$ in Gl. (7.1-11):
$$
\frac {(1/2+\nu)_n \, (1/2-\nu)_n} {n!} \; = \;
\frac {(n-1)!} {\Gamma (1/2+\nu) \Gamma (1/2-\nu)}
\bigl\{ 1 + O (1/n) \bigr\} \, , \qquad n \to \infty \, .
\tag
$$
Offensichtlich erf\"ullen die Koeffizienten der divergenten Reihe ${}_2
F_0$ in Gl. (7.1-11) eine Ungleichung vom Typ von Gl. (2.3-10). Daraus
folgt, da{\ss} die Carlemanbedingung (4.3-5) erf\"ullt ist, was impliziert,
da{\ss} die divergente asymptotische Reihe ${}_2 F_0$ in Gl. (7.1-11) durch
die Pad\'e-Approximationen $[n + k / n]$ mit $k \ge - 1$ und $n \to
\infty$ auf kompakten Teilmengen der geschlitzten komplexen Ebene $\C
\setminus (- \infty, 0]$ summiert wird. Au{\ss}erdem folgt aus Gl. (7.2-2),
da{\ss} die divergente asymptotische Reihe ${}_2 F_0$ in Gl. (7.1-11) eine
starke asymptotische Reihe der Ordnung Eins ist, was impliziert, da{\ss}
$K_{\nu} (z)$ durch die divergente Reihe ${}_2 F_0$ in Gl. (7.1-11)
eindeutig bestimmt ist und da{\ss} die Borel-Transformierte der divergenten
Reihe ${}_2 F_0$ die hypergeometrische Reihe
$$
{}_2 F_1 \bigl(1/2 + \nu, 1/2 - \nu; 1; - t/(2 z) \bigr) \; = \;
\sum_{k=0}^{\infty} \, \frac {(1/2 - \nu)_k (1/2 + \nu)_k} {(k!)^2} \,
\bigl( - t/(2 z) \bigr)^k
\tag
$$
ist. Diese Reihe konvergiert f\"ur $\vert t \vert < \vert 2 z \vert$. Sie
kann aber mit Hilfe von Integraldarstellungen in eine Umgebung der
positiven reellen Halbachse analytisch fortgesetzt werden [Slater 1966,
S. 19 - 26]. Um die Borelsummierbarkeit der divergenten Reihe ${}_2 F_0$
in Gl. (7.1-11) f\"ur die modifizierte Besselfunktion $K_{\nu} (z)$
explizit zu zeigen, mu{\ss} man also beweisen, da{\ss} die Beziehung
$$
(2 z / \pi)^{1/2} \, \e^z \, K_{\nu} (z) \; = \;
\int\nolimits_{0}^{\infty} \, \e^{- t}
{}_2 F_1 \bigl(1/2+\nu,1/2-\nu;1;- t/(2 z) \bigr) \d t \, ,
\qquad \vert \arg (z) \vert \, < \, \pi \, ,
\tag
$$
erf\"ullt ist. Das kann relativ leicht mit Hilfe einer bekannten
Darstellung der Whittakerfunktion durch ein Laplaceintegral geschehen
[Buchholz 1969, S. 78, Gl. (30)],
$$
\beginAligntags
" W_{\kappa, \mu/2} (z) " \; = \;
" \frac {z^{\kappa} \e^{- z/2}} {\Gamma (\alpha )} \,
\int\nolimits_{0}^{\infty} \, \e^{- t} t^{\alpha - 1}
{}_2 F_1 \bigl( (1+\mu)/2 - \kappa,
(1-\mu)/2 - \kappa; \alpha ; - t/z \bigr) \d t \, , \\
" " " Re (\alpha) \, > \, 0 \, ,
\qquad \vert \arg (z) \vert \, < \, \pi \, ,
\\ \tag
\endAligntags
$$
die Gl. (7.2-4) als Spezialfall enth\"alt.

Um zu zeigen, da{\ss} Gl. (7.2-4) ein Spezialfall von Gl. (7.2-5) ist, mu{\ss}
man in Gl. (7.2-5) nur $\alpha = 1$ und $\kappa = 0$ setzen und die
Tatsache ausn\"utzen, da{\ss} die modifizierte Besselfunktion der zweiten Art
ein Spezialfall der Whittakerfunktion ist [Buchholz 1969, S. 208, Gl.
(2c)]:
$$
W_{0,\mu/2} (z) \; = \; (z/2)^{1/2} \, K_{\mu/2} (z/2) \, .
\tag
$$

Silverstone, Nakai und Harris [1985, S. 1343] haben darauf hingewiesen,
da{\ss} der Ausdruck f\"ur die Laplaceintegrale in Gln. (7.2-4) und (7.2-5)
in vielen Integraltafeln und mathematischen Formelsammlungen falsch ist.
Eine einfache Ableitung von Gl. (7.2-5) auf der Basis des
Faltungstheorems f\"ur Laplace-Transformationen wurde von Gailitis und
Silverstone [1988, S. 112] ver\"offentlicht.

Im n\"achsten Schritt soll gezeigt werden, da{\ss} die divergente Reihe ${}_2
F_0$ in Gl. (7.1-11) eine Stieltjesreihe ist. Stieltjesreihen und
Stieltjesfunktionen wurden schon in Abschnitten 4.3 und 6.5 behandelt.
Dort wurden Stieltjesreihen betrachtet, die asymptotische Potenzreihen
f\"ur $z \to 0$ sind. Die divergente Reihe ${}_2 F_0$ in Gl. (7.1-11) ist
aber eine asymptotische Reihe f\"ur $\vert z \vert \to \infty$ in Potenzen
von $1/z$. Deswegen k\"onnen die in Abschnitten 4.3 und 6.5 angegebenen
Beziehungen nicht direkt \"ubernommen werden und m\"ussen zuerst auf
entsprechende Weise modifiziert werden.

In Abschnitten 4.3 und 6.5 wurden Stieltjesfunktionen auf folgende
Weise definiert,
$$
f (z) \; = \; \int\nolimits_{0}^{\infty} \,
\frac {\d \psi (t)} {1 + z t} \, ,
\qquad \vert \arg (z) \vert < \pi \, ,
\tag
$$
wobei $\psi (t)$ ein positives Ma{\ss} auf $0 \le t < \infty$ ist, das dort
unendlich viele verschiedene Werte annimmt [Baker und Graves-Morris
1981a, S. 159]. Einer solchen Stieltjesfunktion $f (z)$ kann man eine
asymptotische Potenzreihe zuordnen,
$$
f(z) \; \sim \; \sum_{m=0}^{\infty} \; (-1)^m \, \mu_m \, z^m \, ,
\qquad z \to 0 \, .
\tag
$$
Da die Koeffizienten $\mu_n$ f\"ur alle $n \in \N_0$ Momente eines
positiven Ma{\ss}es $\psi (t)$ sind,
$$
\mu_n \; = \; \int\nolimits_{0}^{\infty} \, t^n \d \psi (t) \, ,
\tag
$$
ist die formale Potenzreihe (7.2-8) eine Stieltjesreihe.

Hier dagegen ist es sinnvoller, Stieltjesfunktionen $F (z)$ zu
betrachten, die folgenderma{\ss}en definiert sind:
$$
F (z) \; = \; \frac {1} {z} f (1/z) \, .
\tag
$$
Offensichtlich gilt dann:
$$
F (z) \; = \; \int\nolimits_{0}^{\infty} \,
\frac {\d \psi (t)} {z + t} \, ,
\qquad \vert \arg (z) \vert < \pi \, .
\tag
$$
Man kann zeigen, da{\ss} eine solche Stieltjesfunktion $F (z)$ f\"ur alle $z
> 0$ positiv und f\"ur alle $z \in \C \setminus (- \infty, 0]$ eine
analytische Funktion ist [Henrici 1977, S. 581].

Wenn man die Beziehung
$$
\frac {1} {z + t} \; = \;
\sum_{k=0}^{n} \, \frac {(- t)^k} {z^{k+1}} \, + \,
\frac {(- t)^{n+1}} {z^{n+1}} \, \frac {1} {z + t} \, ,
\tag
$$
in dem Stieltjesintegral (7.2-11) verwendet, sieht man, da{\ss} eine solche
Stieltjesfunktion $F (z)$ f\"ur alle $n \in \N_0$ durch die ersten $n+1$
Terme einer Potenzreihe in $1/z$ und ein Integral, das den
Abbruchfehler repr\"asentiert, dargestellt werden kann:
$$
F (z) \; = \;
\sum_{k=0}^{n} \, \frac {(- 1)^k \mu_k} {z^{k+1}} \, + \,
\frac {(- 1)^{n+1}} {z^{n+1}}
\int\nolimits_{0}^{\infty} \frac {t^{n+1} \d \psi (t)} {z + t} \, ,
\qquad n \in \N_0 \, .
\tag
$$
Ein Vergleich mit Satz 6-12 zeigt, da{\ss} der Abbruchfehler
$$
R_n (z) \; = \; \frac {(- 1)^{n+1}} {z^{n+1}}
\int\nolimits_{0}^{\infty} \frac {t^{n+1} \d \psi (t)} {z + t}
\tag
$$
in Abh\"angigkeit von $\theta = \arg (z)$ die folgende Ungleichung
erf\"ullt:
$$
\vert R_n (z) \vert \; \le \;
\cases
{ \mu_{n+1} \, / \, \vert z^{n+2} \vert \, ,
\qquad \quad @ $\vert \theta \vert \le \pi /2$ \, ,
\\
\\
\mu_{n+1} \, / \, \vert z^{n+2} \sin (\theta) \vert \, ,
\qquad \quad @ $\pi / 2 < \vert \theta \vert < \pi$ \, .
}
\tag
$$
Aus dieser Ungleichung f\"ur den Abbruchfehler in Gl. (7.2-13) folgt, da{\ss}
eine Stieltjesfunktion $F (z)$ eine asymptotische Reihe im Sinne von
Poincar\'e in Potenzen von $1/z$ besitzt,
$$ F (z) \; \sim \;
\sum_{k=0}^{\infty} \, \frac {(- 1)^k \mu_k} {z^{k+1}} \, ,
\qquad \vert z \vert \to \infty \, ,
\tag
$$
die f\"ur alle $z \in \C \setminus (- \infty, 0]$ g\"ultig ist [Henrici
1977, S. 584].

In Abschnitt 4.3 wurde schon erw\"ahnt, da{\ss} das Problem der Summation
einer divergenten Stieltjesreihe im Prinzip gel\"ost ist, wenn man aus
den Momenten $\Seqn {\mu}$ das entsprechende positive Ma{\ss} $\psi (t)$
eindeutig berechnen kann. Die Stieltjesfunktion kann dann im Prinzip
mit Hilfe des entsprechenden Stieltjesintegrals berechnet werden, und
es spielt keine Rolle, ob die Stieltjesreihe konvergiert oder
divergiert.

Nehmen wir nun an, da{\ss} eine divergente asymptotische Reihe
$$
\Phi (z) \; \sim \; \sum_{k=0}^{\infty} \, \frac {a_k} {z^{k+1}} \, ,
\qquad \vert z \vert \to \infty \, ,
\tag
$$
gegeben ist. Wenn man zeigen kann, da{\ss} die Koeffizienten $a_k$ in Gl.
(7.2-17) f\"ur alle $k \in \N_0$ die Beziehung
$$
a_k \; = \; (- 1)^k {\tilde \mu}_k
\tag
$$
erf\"ullen, wobei die Koeffizienten ${\tilde \mu}_k$ f\"ur alle $k
\in \N_0$ die positiven und endlichen Momente einer reellen,
nichtabnehmenden und beschr\"ankten Funktion ${\tilde \psi} (t)$ sind,
$$
{\tilde \mu}_k \; = \;
\int\nolimits_{0}^{\infty} \, t^k \, \d {\tilde \psi} (t) \, ,
\tag
$$
die auf $[0, \infty)$ definiert ist und die dort unendlich viele
verschiedene Werte annimmt [Baker und Graves-Morris 1981a, S. 159], dann
folgt, da{\ss} die asymptotische Reihe (7.2-17) eine Stieltjesreihe ist,
und da{\ss} sie f\"ur alle $z \in \C \setminus (- \infty, 0]$ mit dem
folgenden Stieltjesintegral identifiziert werden kann:
$$
\Phi (z) \; = \;
\int\nolimits_{0}^{\infty} \frac {\d {\tilde \psi} (t)} {z + t} \, .
\tag
$$

Wenn wir also zeigen wollen, da{\ss} die divergente asymptotische Reihe f\"ur
die modifizierte Besselfunktion der zweiten Art,
$$
\beginAligntags
" [2 / (\pi z)]^{1/2} \, \e^z \, K_{\nu} (z) \;
" " \; \sim \; z^{- 1} \, {}_2 F_0 \bigl(1/2 + \nu, 1/2 - \nu; - 1/(2
z) \bigr) \\
" " " \; = \; \sum_{n=0}^{\infty} \,
\frac {(1/2+\nu)_n \, (1/2-\nu)_n} {n! \, 2^n}
\frac {(- 1)^n} {z^{n+1}} \, , \\
\tag
\endAligntags
$$
durch ein Stieltjesintegral vom Typ von Gl. (7.2-20) summiert werden
kann, dann m\"ussen wir eine reelle, nichtabnehmende und beschr\"ankte
Funktion $\psi_{\nu} (t)$ finden, die auf $[0, \infty)$ definiert ist,
die dort unendlich viele verschiedene Werte annimmt [Baker und
Graves-Morris 1981a, S. 159], und die f\"ur alle $n \in \N_0$ die
positiven und endlichen Momente
$$
\mu_n \; = \; \frac {(1/2+\nu)_n \, (1/2-\nu)_n} {n! \, 2^n}
\; = \; \int\nolimits_{0}^{\infty} \, t^n \, \d \psi_{\nu} (t) \, ,
\tag
$$
besitzt.

Die Forderung, da{\ss} die Stieltjesmomente (7.2-9) reell sein m\"ussen,
impliziert, da{\ss} die Ordnung $\nu$ der modifizierten Besselfunktion
$K_{\nu} (z)$ ebenfalls reell sein mu{\ss}. Aufgrund der Symmetriebeziehung
(7.1-6) k\"onnen wir ohne Einschr\"ankung der Allgemeinheit von nun an
annehmen, da{\ss} die Ordnung $\nu$ positiv ist.

Wenn die Ordnung der modifizierten Besselfunktion $K_{\nu} (z)$
halbzahlig ist, $\nu = n + 1/2$ mit $n \in \N_0$, dann bricht die
hypergeometrische Reihe ${}_2 F_0$ in Gl. (7.1-11) nach einer endlichen
Anzahl von Termen ab und man erh\"alt Gl. (7.1-15). Diese speziellen
$K$-Funktionen k\"onnen problemlos mit Hilfe der Rekursionsformel (7.1-8)
und den Startwerten (7.1-16) berechnet werden.

In folgenden Text beschr\"ankt man sich folglich auf Ordnungen $\nu$, die
{\it positiv} und {\it nicht halbzahlig} sind, d. h., es soll $\nu \ge
0$ und $\nu \ne n + 1/2$ mit $n \in \N_0$ gelten. Der Ausgangspunkt
unserer Analyse ist das Laplaceintegral [Gradshteyn und Ryzhik 1980, S.
712, Gl. (6.631.3)]
$$
\beginAligntags
" \int\nolimits_{0}^{\infty} \, x^{\mu - 1} \, \e^{- \alpha x}
\, K_{\nu} (\beta x) \, \d x \\
" = \;
\frac {\pi^{1/2} (2 \beta)^{\nu}} {(\alpha + \beta)^{\mu + \nu}}
\; \frac {\Gamma (\mu + \nu) \, \Gamma (\mu - \nu)}
{\Gamma (\mu + 1/2)} \;
{}_2 F_1 \bigl( \mu + \nu, \nu + 1/2; \mu + 1/2;
(\alpha - \beta)/(\alpha + \beta) \bigr) \, , \\
" \qquad Re (\mu \pm \nu) \, > \, 0 \, ,
\qquad Re (\alpha + \beta) \, > \, 0 \, . \\
\tag
\endAligntags
$$
In diesem Integral setzen wir jetzt $\mu = n + 1/2$ mit $n \in \N_0$,
und $\alpha = \beta = 1$. Dann bricht die hypergeometrische Reihe ${}_2
F_1$ nach dem ersten Term ab und wir erhalten:
$$
\beginAligntags
" \int\nolimits_{0}^{\infty} \, x^{n - 1/2} \, \e^{- x}
\, K_{\nu} (x) \, \d x " \; = \; " (\pi/2)^{1/2} \;
\frac {\Gamma (1/2 + \nu + n) \, \Gamma (1/2 - \nu + n)}
{n! \, 2^n} \, , \\
" " " Re (\pm \nu) \, < \, n + 1/2 \, , \quad n \in \N_0 \, . \\
\tag
\endAligntags
$$
Wenn wir annehmen, da{\ss} $0 \le \nu < 1/2$ gilt, dann existiert das
Integral in Gl. (7.2-24) f\"ur alle $n \in \N_0$, und die rechte Seite
von Gl. (7.2-24) ist positiv. Ein geeignetes reelles, nichtabnehmendes
und beschr\"anktes Ma{\ss} $\psi_{\nu} (t)$, das auf $[0, \infty)$ definiert
ist und dort unendlich viele verschiedene Werte annimmt, und das f\"ur
alle $n \in \N_0$ Gl. (7.2-22) erf\"ullt, ist durch das Integral
$$
\psi_{\nu} (t) \; = \;
\frac {(2 / \pi)^{1/2}} {\Gamma (1/2 + \nu) \Gamma (1/2 - \nu)} \,
\int\nolimits_{0}^{t} \, s^{- 1/2} \e^{- s} K_{\nu} (s) \d s \, ,
\qquad 0 \, \le \, \nu \, < \, 1/2 \, ,
\tag
$$
gegeben. Aus den Gln.~(7.2-11), (7.2-22), (7.2-24) und (7.2-25) folgt
also, da{\ss} die modifizierte Besselfunktion der zweiten Art eine
Darstellung durch ein Stieltjesintegral besitzt:
$$
\beginAligntags
" K_{\nu} (z) " \; = \; "
\frac {z^{1/2} \e^{- z}} {\Gamma (1/2 + \nu) \Gamma (1/2 - \nu)}
\, \int\nolimits_{0}^{\infty} \, \frac {t^{- 1/2}} {z + t} \,
\e^{- t} K_{\nu} (t) \d t \, , \\
" " " 0 < \nu \, < \, 1/2 \, ,
\qquad \vert \arg (z) \vert \, < \, \pi \, .
\\ \tag
\endAligntags
$$
Diese Beziehung ist ein Spezialfall einer allgemeineren Beziehung f\"ur
Whittakerfunktionen, die von Gargantini und Henrici [1967, Gl. (9)]
abgeleitet wurde.

Nehmen wir nun an, da{\ss} $\nu > 1/2$ und $\nu \ne n + 1/2$ mit $n \in
\N_0$ gilt. Au{\ss}erdem verwenden wir wieder die Notation $\Ent {x}$ f\"ur
den ganzzahligen Anteil von $x$, der die gr\"o{\ss}te ganze Zahl $m$ ist,
welche die Ungleichung $m \le x$ erf\"ullt. Dann besitzen die Momente
$\mu_n$ in Gl. (7.2-22) strikt alternierende Vorzeichen f\"ur alle Indizes
$n$, welche die Ungleichung $n \le n'$ mit $n' = \Ent{\nu + 1/2}$
erf\"ullen. Dagegen haben die Momente $\mu_n$ das gleiche Vorzeichen f\"ur
alle Indizes $n \ge n'$. Au{\ss}erdem existiert das Integral in Gl. (7.2-24)
f\"ur alle $n \ge n'$, und es divergiert f\"ur $n < n'$. Diese Beobachtungen
legen nahe, die hypergeometrische Reihe ${}_2 F_0$ in Gl. (7.2-21) auf
folgende Weise umzuschreiben:
$$
\beginAligntags
" \frac {1} {z} \;
{}_2 F_0 \bigl( 1/2 + \nu, 1/2 - \nu; -1/(2 z) \bigr) \; = \;
\sum_{n=0}^{n' - 1} \, \frac
{(1/2+\nu)_n \, (1/2-\nu)_n} {n! \, 2^n} \,
\frac {(- 1)^n} {z^{n+1}} \\
" + \, \frac
{(1/2+\nu)_{n'} \, (1/2-\nu)_{n'}} {(n')! \, 2^{n'}} \,
\frac {(- 1)^{n'}} {z^{n'}} \,
\sum_{m=0}^{\infty} \,
\frac {(1/2+\nu+n')_m \, (1/2-\nu+n')_m} {(n'+1)_m \, 2^m} \,
\frac {(- 1)^m} {z^{m+1}} \, , \\
" \qquad \nu > 1/2 \, ,
\qquad \nu \ne n + 1/2 \, , \quad n \in \N_0 \, ,
\qquad n' = \Ent {\nu + 1/2} \, . \\
\tag
\endAligntags
$$

Wir wollen jetzt zeigen, da{\ss} die unendliche Reihe auf der rechten Seite
von Gl.~(7.2-27) eine Stieltjesreihe ist und wir wollen sie mit einem
Stieltjesintegral identifizieren. Aus Gln.~(7.2-24) und (7.2-25) folgt,
da{\ss} ein geeignetes reelles, nichtabnehmendes und beschr\"anktes Ma{\ss}
$\psi'_{\nu} (t)$, das auf $[0, \infty)$ definiert ist und dort
unendlich viele verschiedene Werte annimmt, und das f\"ur alle $n \in
\N_0$ die reellen, positiven und endlichen Momente
$$
\mu'_m \; = \;
\frac {(1/2+\nu+n')_m \, (1/2-\nu+n')_m} {(n'+1)_m \, 2^m} \;
= \; \int\nolimits_{0}^{\infty} \, t^m \, \d \psi'_{\nu} (t)
\tag
$$
besitzt, durch das folgende Integral gegeben ist:
$$
\beginAligntags
" \psi'_{\nu} (t) " \; = \; "
\frac {(2 / \pi)^{1/2} \, (n')! \, 2^{n'}}
{\Gamma (1/2 + \nu + n') \, \Gamma (1/2 - \nu + n')} \,
\int\nolimits_{0}^{t} \, s^{n'- 1/2} \,
\e^{- s} \, K_{\nu} (s) \, \d s \, , \\
" " " \nu > 1/2 \, ,
\qquad \nu \ne n + 1/2 \, , \quad n \in \N_0 \, ,
\qquad n' = \Ent {\nu + 1/2} \, . \\
\tag
\endAligntags
$$
Es folgt also aus Gln.~(7.2-11), (7.2-27), (7.2-28) und (7.2-29), da{\ss}
$K_{\nu} (z)$ f\"ur $\nu > 1/2$ die folgende Darstellung durch ein
Stieltjesintegral besitzt [Weniger und {\v C}{\'\i}{\v z}ek 1990, Gl.
(2.25)]:
$$
\beginAligntags
" K_{\nu} (z) \, - \, (\pi/2)^{1/2} \, z^{1/2} \, \e^{- z} \,
\sum_{m=0}^{n' - 1} \, \frac
{(1/2+\nu)_m \, (1/2-\nu)_m} {m! \, 2^m} \,
\frac {(- 1)^m} {z^{m+1}} \\
" = \; \frac {z^{1/2} \, \e^{- z}}
{\Gamma (1/2 + \nu) \, \Gamma (1/2 - \nu)} \,
\frac {(- 1)^{n'}} {z^{n'}} \,
\int\nolimits_{0}^{\infty} \frac {t^{n'- 1/2}} {z + t} \,
\e^{- t} \, K_{\nu} (t) \, \d t \, , \\
" \qquad \nu > 1/2 \, ,
\qquad \nu \ne n + 1/2 \, , \quad n \in \N_0 \, ,
\qquad n' = \Ent {\nu + 1/2} \, ,
\qquad \vert \arg (z) \vert \, < \, \pi \, . \qquad \\
\tag
\endAligntags
$$

Die meisten Einschr\"ankungen in Gl. (7.2-30) sind \"uberfl\"ussig und man
kann zeigen, da{\ss} Gl. (7.2-30) f\"ur alle $\nu \in \R$ g\"ultig ist. Aufgrund
von Gl. (7.1-6) mu{\ss} man nur die G\"ultigkeit von Gl. (7.2-30) f\"ur $\nu \ge
0$ beweisen. Wenn beispielsweise $0 \le \nu < 1/2$ gilt, folgt $n' = 0$
und die Summe auf der linken Seite von Gl. (7.2-30) ist eine leere Summe
und demzufolge Null, und Gl. (7.2-30) vereinfacht sich dann zu Gl.
(7.2-26). Au{\ss}erdem vereinfachen sich dann Gl. (7.2-27) und Gl. (7.2-29)
zu Gl. (7.1-11) beziehungsweise Gl. (7.2-25). Wenn wir $n'$ in Gl.
(7.2-30) gem\"a{\ss} der Regel $n' \; = \; \Ent {\vert \nu \vert + 1/2}$
w\"ahlen, folgt, da{\ss} die Darstellung von $K_{\nu} (z)$ durch ein
Stieltjesintegral f\"ur alle reellen Ordnungen $\nu$ g\"ultig ist, die
nicht halbzahlig sind, d. h., wenn $\nu \in \R$ und $\nu \ne \pm (n +
1/2)$ mit $n \in \N_0$ gilt. Gl. (7.2-30) ist aber auch dann noch
g\"ultig, wenn die Ordnung $\nu$ halbzahlig ist. Dann verschwindet n\"amlich
die rechte Seite von Gl. (7.2-30), da eine der beiden Gammafunktionen
singul\"ar wird, und wir erhalten die explizite Darstellung der
modifizierten Besselfunktion mit halbzahliger Ordnung, Gl. (7.1-15).

\medskip

\Abschnitt Numerische Beispiele

\smallskip

\aktTag = 0

Im letzten Unterabschnitt wurde gezeigt, da{\ss} die divergente
hypergeometrische Reihe ${}_2 F_0$ in Gl. (7.1-11) f\"ur $K_{\nu} (z)$
eine starke asymptotische Reihe der Ordnung Eins ist und da{\ss} sie gem\"a{\ss}
Gl. (7.2-4) borelsummierbar ist. Da $K_{\nu} (z)$ gem\"a{\ss} Gl. (7.2-30) auf
eindeutige Weise durch ein Stieltjesintegral dargestellt werden kann,
ist die divergente Reihe in Gl. (7.1-11) auch eine Stieltjesreihe.

In theoretischer Hinsicht sind damit im Prinzip alle Probleme, die im
Zusammenhang mit der Summation der divergenten hypergeometrischen Reihe
${}_2 F_0$ in Gl. (7.1-11) auftreten, gel\"ost, da man der divergenten
Reihe f\"ur alle $z \in \C \setminus (- \infty, 0]$ auf eindeutige Weise
die modifizierte Besselfunktion $K_{\nu} (z)$ zuordnen kann. In
praktischer Hinsicht ist aber weder das Laplaceintegral in Gl. (7.2-4)
noch das Stieltjesintegral in Gl. (7.2-30) zur Berechnung der
modifizierten Besselfunktion $K_{\nu} (z)$ geeignet.

Das Laplaceintegral in Gl. (7.2-4) kann zwar im Prinzip zur Berechnung
der modifizierten Besselfunktion $K_{\nu} (z)$ verwendet werden. Da man
aber eine analytische Fortsetzung der hypergeometrischen Reihe ${}_2
F_1$ in Gl. (7.2-3) in eine Umgebung der positiv reellen Halbachse
ben\"otigt, w\"are eine numerische Auswertung des Laplaceintegrals in Gl.
(7.2-4) \"au{\ss}erst schwierig und wahrscheinlich auch sehr aufwendig.

Die Darstellung der modifizierten Besselfunktion $K_{\nu}$ durch ein
Stieltjesintegral gem\"a{\ss} Gl. (7.2-30) ist f\"ur numerische Zwecke v\"ollig
unbrauchbar, da das Stieltjesintegral $K_{\nu}$ ebenfalls im
Integranden enth\"alt.

In diesem Unterabschnitt soll gezeigt werden, da{\ss} man $K_{\nu} (z)$ auf
effiziente Weise im problematischen Bereich mittelgro{\ss}er Argumente $z$
berechnen kann, wenn man geeignete verallgemeinerte Summationsprozesse
auf die Partialsummen
$$
s_n \; = \; \sum_{m=0}^{n} \,
\frac {(1/2+\nu)_m \, (1/2-\nu)_m} {m!} \;
\frac {(-1)^m} {(2 z)^m} \, , \qquad n \in \N_0 \, ,
\tag
$$
der hypergeometrischen Reihe
$$
{}_2 F_0 \bigl( 1/2 + \nu, 1/2 - \nu; - 1/(2 z)\bigr)
\; = \; (2 z / \pi)^{1/2} \, \e^z \, K_{\nu} (z)
\tag
$$
anwendet.

In Tabelle 7-1 wird der Wynnsche $\epsilon$-Algorithmus, Gl. (2.4-10),
der iterierte Aitkensche $\Delta^2$-Proze{\ss}, Gl. (3.3-8), und der
Brezinskische $\theta$-Algorithmus, Gl. (4.4-13), auf die Partialsummen
(7.3-1) der hypergeometrischen Reihe (7.3-2) f\"ur $\nu = 0$ und $z = 5/2$
angewendet. Die Vergleichswerte f\"ur $K_0 (z)$ wurden mit der FORTRAN
FUNCTION S18CCF der NAG Library berechnet. Dieses Programm berechnet
eine Approximation f\"ur $\e^z \, K_0 (z)$ in DOUBLE PRECISION (15 - 16
Dezimalstellen) mit Hilfe geeigneter Tschebyscheffentwicklungen.

\beginFloat

\medskip

\beginTabelle [to \kolumnenbreite]
\beginFormat \rechts " \rechts " \mitte " \mitte " \mitte
\endFormat
\+ " \links {\bf Tabelle 7-1} \@ \@ \@ \@ " \\
\+ " \links {Summation der asymptotischen Reihe} \@ \@ \@ \@ " \\
\+ " \links {${}_2 F_0 \bigl( 1/2, 1/2; - 1/(2 z)\bigr)
\; = \; (2 z / \pi)^{1/2} \, \e^z \, K_0 (z)$
f\"ur $z \; = \; 5/2$} \@ \@ \@ \@ " \\
\- " \- " \- " \- " \- " \- " \\ \sstrut {} {1.5 \jot} {1.5 \jot}
\+ " \rechts {$n$} " \mitte {Partialsumme $s_n$}
" $\epsilon_{2 \Ent {n/2}}^{(n - 2 \Ent {n/2})}$
" ${\cal A}_{\Ent {n/2}}^{(n - 2 \Ent {n/2})}$
" $\theta_{2 \Ent {n/3}}^{(n - 3 \Ent {n/3})}$ " \\
\+ " " \mitte {Gl. (7.3-1)} " Gl. (2.4-10) " Gl. (3.3-8)
" Gl. (4.4-13) " \\
\- " \- " \- " \- " \- " \- " \\ \sstrut {} {1 \jot} {1 \jot}
\+ " $ 5$ " $ 0.9571080078 \times 10^{00}$ " $0.95820166036768$ "
$0.95821205820847$ " $0.95823017369329$ " \\
\+ " $ 6$ " $ 0.9594529736 \times 10^{00}$ " $0.95822587523910$ "
$0.95822158747979$ " $0.95822101004996$ " \\
\+ " $ 7$ " $ 0.9566222649 \times 10^{00}$ " $0.95821903805091$ "
$0.95822073340076$ " $0.95822097917668$ " \\
\+ " $ 8$ " $ 0.9606029491 \times 10^{00}$ " $0.95822158404933$ "
$0.95822101508423$ " $0.95822101830371$ " \\
\+ " $ 9$ " $ 0.9542117395 \times 10^{00}$ " $0.95822074460113$ "
$0.95822099379856$ " $0.95822099880151$ " \\
\+ " $10$ " $ 0.9657478728 \times 10^{00}$ " $0.95822108688400$ "
$0.95822100159248$ " $0.95822100146794$ " \\
\+ " $11$ " $ 0.9426231692 \times 10^{00}$ " $0.95822096098084$ "
$0.95822100111741$ " $0.95822100129348$ " \\
\+ " $12$ " $ 0.9935938700 \times 10^{00}$ " $0.95822101599077$ "
$0.95822100131359$ " $0.95822100130552$ " \\
\+ " $13$ " $ 0.8710681470 \times 10^{00}$ " $0.95822099401520$ "
$0.95822100130802$ " $0.95822100129665$ " \\
\+ " $14$ " $ 0.1190072618 \times 10^{01}$ " $0.95822100415929$ "
$0.95822100131131$ " $0.95822100133621$ " \\
\+ " $15$ " $ 0.2957967502 \times 10^{00}$ " $0.95822099983413$ "
$0.95822100131230$ " $0.95822100130860$ " \\
\+ " $16$ " $ 0.2981418967 \times 10^{01}$ " $0.95822100192275$ "
$0.95822100131242$ " $0.95822100131067$ " \\
\+ " $17$ " $-0.5620471016 \times 10^{01}$ " $0.95822100098366$ "
$0.95822100131234$ " $0.95822100131230$ " \\
\+ " $18$ " $ 0.2364984906 \times 10^{02}$ " $0.95822100145466$ "
$0.95822100131232$ " $0.95822100131231$ " \\
\+ " $19$ " $-0.8180033038 \times 10^{02}$ " $0.95822100123328$ "
$0.95822100131232$ " $0.95822100131232$ " \\
\+ " $20$ " $ 0.3191739770 \times 10^{03}$ " $0.95822100134797$ "
$0.95822100131232$ " $0.95822100131233$ " \\
\- " \- " \- " \- " \- " \- " \\ \sstrut {} {1 \jot} {1 \jot}
\+ " \links {NAG FUNCTION S18CCF} \@ " $0.95822100131232$ "
$0.95822100131232$ " $0.95822100131232$ " \\
\- " \- " \- " \- " \- " \- " \\ \sstrut {} {1 \jot} {1 \jot}
\endTabelle

\medskip

\endFloat

Im Falle des $\theta$-Algorithmus, Gl. (4.4-13), werden die
Approximationen zum Grenzwert $s$ der zu transformierenden Folge $\Seqn
s$ folgenderma{\ss}en gew\"ahlt [Weniger 1989, Gl. (10.2-8)]:
$$
\left\{ s_{m - 3 \Ent {m/3}}, s_{m - 3 \Ent {m/3} + 1}, \ldots , s_m
\right\} \; \to \;
\theta_{2 \Ent {m/3}}^{(m - 3 \Ent {m/3})} \, .
\tag
$$
Dabei wird die Notation $\Ent x$ f\"ur den ganzzahligen Anteil von $x$
verwendet, der die gr\"o{\ss}te nat\"urliche Zahl $\nu$ ist, welche die
Ungleichung $\nu \le x$ erf\"ullt.

Die Partialsummen und die Transformationen in Tabelle 7-1 wurden in
QUADRUPLE PRECISION (31 - 32 Dezimalstellen) berechnet. Um die
numerische Stabilit\"at dieser Summationen \"uberpr\"ufen zu k\"onnen, wurden
die Rechnungen in DOUBLE PRECISION (15 - 16 Dezimalstellen) wiederholt.
Dabei ergab sich eine \"Ubereinstimmung aller ausgegebenen 14
Dezimalstellen.

Der knappe Gewinner in Tabelle 7-1 ist der iterierte Aitkensche
$\Delta^2$-Proze{\ss}, Gl. (3.3-8), der die Partialsummen $s_0$, $s_1$,
$\ldots$ , $s_{18}$ ben\"otigt, um eine Genauigkeit von 14 Dezimalstellen
zu produzieren, gefolgt von dem Brezinskischen $\theta$-Algorithmus, Gl.
(4.4-13), und der klare Verlierer ist der Wynnsche
$\epsilon$-Algorithmus, Gl. (2.4-10), der aus den Partialsummen $s_0$,
$s_1$, $\ldots$ , $s_{20}$ nur eine Genauigkeit von 10 Dezimalstellen
extrahieren kann.

Die Ergebnisse in Tabelle 7-1 sind typisch f\"ur die F\"ahigkeit der dort
vorkommenden verallgemeinerten Summationsprozesse, die divergente Reihe
(7.3-2) zu summieren. F\"ur ein kleines Argument der modifizierten
Besselfunktion $K_{\nu} (z)$ divergiert die hypergeometrische Reihe
st\"arker als f\"ur ein gro{\ss}es Argument. Demzufolge ist es auch wesentlich
schwerer, die hypergeometrische Reihe f\"ur kleine als f\"ur gro{\ss}e Argumente
zu summieren. Beispielsweise extrahiert der iterierte Aitkensche
$\Delta^2$-Proze{\ss}, Gl. (3.3-8), f\"ur $\nu = 0$ und $z = 1$ aus den
Partialsummen $s_0$, $s_1$, $\ldots$ , $s_{20}$ eine Genauigkeit von 10
Stellen, wogegen f\"ur $z = 4$ nur die Partialsummen $s_0$, $s_1$,
$\ldots$ , $s_{15}$ ben\"otigt werden, um eine Genauigkeit von 14 Stellen
zu erreichen. In allen betrachteten F\"allen war aber das
Leistungsverm\"ogen der betrachteten verallgemeinerten Summationsprozesse
so wie in Tabelle 7-1. Der iterierte Aitkensche $\Delta^2$-Proze{\ss}
lieferte also immer geringf\"ugig bessere Ergebnisse als der Brezinskische
$\theta$-Algorithmus, und der klare Verlierer war immer der Wynnsche
$\epsilon$-Algorithmus. Bei diesen numerischen Tests stellte sich auch
heraus, da{\ss} die Konvergenzgeschwindigkeit der Summationsverfahren nur
unwesentlich von der Ordnung $\nu$ der modifizierten Besselfunktion
$K_{\nu} (z)$ abh\"angt.

Die verallgemeinerten Summationsprozesse in Tabelle 7-1 verwenden als
Eingabedaten nur die Elemente einer Folge $\Seqn s$ von Partialsummen
einer konvergenten oder divergenten unendlichen Reihe. Keine weitere
Information \"uber das Verhalten der Partialsummen $s_n$ als Funktion des
Index $n$ wird zur Konstruktion der verallgemeinerten Summationsprozesse
ben\"otigt. In vielen F\"allen, in denen nur wenig \"uber das
Konvergenzverhalten der Partialsummen bekannt ist, ist dies sicherlich
sehr vorteilhaft. Bei einer Reihe, die so stark divergiert wie die
divergente hypergeometrische Reihe (7.3-2), ist es aber vorteilhaft,
verallgemeinerte Summationsprozesse zu verwenden, die in der Lage sind,
zus\"atzliche Informationen \"uber das Verhalten der Eingabedaten wie etwa
die Fehlerabsch\"atzung (7.2-15) nutzbringend zu verwenden.

In Tabelle 7-2 werden die verallgemeinerten Summationsprozesse
$d_n^{(0)} (\zeta, s_0)$, Gl. (5.2-18), und ${\delta}_n^{(0)} (\zeta,
s_0)$, Gl. (5.4-13), mit $\zeta = 1$ und ${\Delta}_n^{(0)} (\xi, s_0)$,
Gl. (5.5-15), mit $\xi = 14$ auf die gleiche Folge von Partialsummen
(7.3-1) der hypergeometrischen Reihe (7.3-2) mit $\nu = 0$ und $z = 5/2$
angewendet wie in Tabelle 7-1.

\beginFloat

\medskip

\beginTabelle [to \kolumnenbreite]
\beginFormat \rechts " \rechts " \mitte " \mitte " \mitte
\endFormat
\+ " \links {\bf Tabelle 7-2} \@ \@ \@ \@ " \\
\+ " \links {Summation der asymptotischen Reihe} \@ \@ \@ \@ " \\
\+ " \links {${}_2 F_0 \bigl( 1/2, 1/2; - 1/(2 z)\bigr)
\; = \; (2 z / \pi)^{1/2} \, \e^z \, K_0 (z)$
f\"ur $z \; = \; 5/2$} \@ \@ \@ \@ " \\
\- " \- " \- " \- " \- " \- " \\ \sstrut {} {1.5 \jot} {1.5 \jot}
\+ " \rechts {$n$} " \mitte {Partialsumme $s_n$}
" $d_n^{(0)} (1, s_0)$
" ${\delta}_n^{(0)} (1, s_0)$
" ${\Delta}_n^{(0)} (14, s_0)$ " \\
\+ " " \mitte {Gl. (7.3-1)} " Gl. (5.2-18) " Gl. (5.4-13)
" Gl. (5.5-15) " \\
\- " \- " \- " \- " \- " \- " \\ \sstrut {} {1 \jot} {1 \jot}
\+ " $ 0$ " $ 0.1000000000 \times 10^{01}$ " $1.00000000000000$ "
$1.00000000000000$ " $1.00000000000000$ " \\
\+ " $ 1$ " $ 0.9500000000 \times 10^{00}$ " $0.95918367346939$ "
$0.95918367346939$ " $0.95918367346939$ " \\
\+ " $ 2$ " $ 0.9612500000 \times 10^{00}$ " $0.95807439824945$ "
$0.95807439824945$ " $0.95823899371069$ " \\
\+ " $ 3$ " $ 0.9565625000 \times 10^{00}$ " $0.95824282725151$ "
$0.95823495012615$ " $0.95822045436674$ " \\
\+ " $ 4$ " $ 0.9594335937 \times 10^{00}$ " $0.95821784425945$ "
$0.95822015735496$ " $0.95822082475511$ " \\
\+ " $ 5$ " $ 0.9571080078 \times 10^{00}$ " $0.95822143450803$ "
$0.95822102000524$ " $0.95822097156909$ " \\
\+ " $ 6$ " $ 0.9594529736 \times 10^{00}$ " $0.95822094720605$ "
$0.95822100308861$ " $0.95822099647917$ " \\
\+ " $ 7$ " $ 0.9566222649 \times 10^{00}$ " $0.95822100697985$ "
$0.95822100119257$ " $0.95822100052914$ " \\
\+ " $ 8$ " $ 0.9606029491 \times 10^{00}$ " $0.95822100092938$ "
$0.95822100130426$ " $0.95822100119216$ " \\
\+ " $ 9$ " $ 0.9542117395 \times 10^{00}$ " $0.95822100129607$ "
$0.95822100131307$ " $0.95822100129572$ " \\
\+ " $10$ " $ 0.9657478728 \times 10^{00}$ " $0.95822100132346$ "
$0.95822100131240$ " $0.95822100131034$ " \\
\+ " $11$ " $ 0.9426231692 \times 10^{00}$ " $0.95822100131021$ "
$0.95822100131232$ " $0.95822100131213$ " \\
\+ " $12$ " $ 0.9935938700 \times 10^{00}$ " $0.95822100131253$ "
$0.95822100131232$ " $0.95822100131231$ " \\
\+ " $13$ " $ 0.8710681470 \times 10^{00}$ " $0.95822100131233$ "
$0.95822100131232$ " $0.95822100131232$ " \\
\+ " $14$ " $ 0.1190072618 \times 10^{01}$ " $0.95822100131232$ "
$0.95822100131232$ " $0.95822100131232$ " \\
\+ " $15$ " $ 0.2957967502 \times 10^{00}$ " $0.95822100131233$ "
$0.95822100131232$ " $0.95822100131232$ " \\
\- " \- " \- " \- " \- " \- " \\ \sstrut {} {1 \jot} {1 \jot}
\+ " \links {NAG FUNCTION S18CCF} \@ " $0.95822100131232$ "
$0.95822100131232$ " $0.95822100131232$ " \\
\- " \- " \- " \- " \- " \- " \\ \sstrut {} {1 \jot} {1 \jot}
\endTabelle

\medskip

\endFloat

Die Partialsummen und Transformationen in Tabelle 7-2 wurden
in QUADRUPLE PRECISION berechnet. Eine Wiederholung dieser Rechnungen
in DOUBLE PRECISION ergab v\"ollige \"Ubereinstimmung aller ausgegebenen
14 Dezimalstellen.

Der klare Gewinner in Tabelle 7-2 ist $\delta_k^{(n)} (\zeta, s_n)$, Gl.
(5.4-13), gefolgt von $\Delta_k^{(n)} (\xi, s_n)$, Gl. (5.5-15). Aber
auch die in Tabelle 7-2 am wenigsten leistungsf\"ahige Transformation
$d_k^{(n)} (\zeta, s_n)$, Gl. (5.2-18), ist immer noch deutlich
leistungsf\"ahiger als der iterierte Aitkensche $\Delta^2$-Proze{\ss}, Gl.
(3.3-8), der in Tabelle 7-1 die besten Ergebnisse lieferte. Wir sehen
also, da{\ss} man im Falle der divergenten hypergeometrischen Reihe (7.3-2)
\"ahnliche Ergebnisse erh\"alt wie im Falle der divergenten Reihe (6.7-12)
f\"ur das Exponentialintegral. Verallgemeinerte Summationsprozesse, die
explizite Restsummenabsch\"atzungen $\Seqn \omega$ verwenden, sind
offensichtlich auch hier wesentlich leistungsf\"ahiger als verallgemeinere
Summationsprozesse, die nur die Partialsummen $\Seqn s$ einer
unendlichen Reihe als Eingabedaten verwenden.

\"Ahnlich wie beim Exponentialintegral $E_1 (z)$, Gl. (6.7-10), dessen
Berechnung durch verallgemeinerte Summationsprozesse in Abschnitt 6.7
behandelt wurde, sind Pad\'e-Approximationen auch im Falle der
modifizierten Besselfunktion $K_{\nu} (z)$ deutlich weniger
leistungsf\"ahig als die anderen verallgemeinerten Summationsprozesse. Die
Unterlegenheit der Pad\'e-Approximationen ist in Tabelle 7-1 aber weniger
eklatant als in Tabelle 6-1.

\beginFloat

\medskip

\beginTabelle [to \kolumnenbreite]
\beginFormat \rechts " \rechts " \mitte " \mitte " \mitte
\endFormat
\+ " \links {\bf Tabelle 7-3} \@ \@ \@ \@ " \\
\+ " \links {Summation der asymptotischen Reihe} \@ \@ \@ \@ " \\
\+ " \links {${}_2 F_0 \bigl( 3/2, - 1/2; - 1/(2 z)\bigr)
\; = \; (2 z / \pi)^{1/2} \, \e^z \, K_1 (z)$
f\"ur $z \; = \; 3/10$} \@ \@ \@ \@ " \\
\- " \- " \- " \- " \- " \- " \\ \sstrut {} {1.5 \jot} {1.5 \jot}
\+ " \rechts {$n$} " \mitte {Partialsumme $s_n$}
" $d_n^{(0)} (1, s_0)$
" ${\delta}_n^{(0)} (1, s_0)$
" ${\Delta}_n^{(0)} (27, s_0)$ " \\
\+ " " \mitte {Gl. (7.3-1)} " Gl. (5.2-18) " Gl. (5.4-13)
" Gl. (5.5-15)" \\
\- " \- " \- " \- " \- " \- " \\ \sstrut {} {1 \jot} {1 \jot}
\+ " $13$ " $ 0.1177864357 \times 10^{12}$ " $1.80277387044868$ "
$1.80277385999343$ " $1.80275692790040$ " \\
\+ " $14$ " $-0.2550875490 \times 10^{13}$ " $1.80277386153044$ "
$1.80277385596725$ " $1.80276584849037$ " \\
\+ " $15$ " $ 0.5949551429 \times 10^{14}$ " $1.80277385343933$ "
$1.80277385530847$ " $1.80277015760837$ " \\
\+ " $16$ " $-0.1486816856 \times 10^{16}$ " $1.80277385488427$ "
$1.80277385539919$ " $1.80277219785268$ " \\
\+ " $17$ " $ 0.3963448025 \times 10^{17}$ " $1.80277385596071$ "
$1.80277385553283$ " $1.80277313939011$ " \\
\+ " $18$ " $-0.1122613292 \times 10^{19}$ " $1.80277385574008$ "
$1.80277385560233$ " $1.80277356003622$ " \\
\+ " $19$ " $ 0.3366835621 \times 10^{20}$ " $1.80277385558458$ "
$1.80277385562788$ " $1.80277374042823$ " \\
\+ " $20$ " $-0.1065871242 \times 10^{22}$ " $1.80277385561735$ "
$1.80277385563459$ " $1.80277381386546$ " \\
\+ " $21$ " $ 0.3551999993 \times 10^{23}$ " $1.80277385564154$ "
$1.80277385563532$ " $1.80277384182829$ " \\
\+ " $22$ " $-0.1242906748 \times 10^{25}$ " $1.80277385563681$ "
$1.80277385563480$ " $1.80277385158842$ " \\
\+ " $23$ " $ 0.4556326097 \times 10^{26}$ " $1.80277385563281$ "
$1.80277385563434$ " $1.80277385462456$ " \\
\+ " $24$ " $-0.1746235347 \times 10^{28}$ " $1.80277385563345$ "
$1.80277385563409$ " $1.80277385543265$ " \\
\+ " $25$ " $ 0.6983611904 \times 10^{29}$ " $1.80277385563415$ "
$1.80277385563400$ " $1.80277385560535$ " \\
\+ " $26$ " $-0.2909320521 \times 10^{31}$ " $1.80277385563407$ "
$1.80277385563397$ " $1.80277385563179$ " \\
\+ " $27$ " $ 0.1260495433 \times 10^{33}$ " $1.80277385563395$ "
$1.80277385563397$ " $1.80277385563398$ " \\
\+ " $28$ " $-0.5671342237 \times 10^{34}$ " $1.80277385563395$ "
$1.80277385563397$ " $1.80277385563397$ " \\
\- " \- " \- " \- " \- " \- " \\ \sstrut {} {1 \jot} {1 \jot}
\+ " \links {NAG FUNCTION S18CDF} \@ " $1.80277385563397$ "
$1.80277385563397$ " $1.80277385563397$ "
\\ \- " \- " \- " \- " \- " \- " \\ \sstrut {} {1 \jot} {1 \jot}
\endTabelle

\medskip

\endFloat

Aufgrund der langsamen Konvergenz der Pad\'e-Approximationen ist es
sicherlich nicht sinnvoll, die divergente Reihe (7.3-2) im Falle kleiner
Argumente durch Pad\'e-Approximationen summieren zu wollen. Wenn man aber
geeignete Varianten der verallgemeinerten Summationsprozesse ${\cal
L}_k^{(n)} (\zeta, s_n, \omega_n)$, Gl. (5.2-6), ${\cal S}_k^{(n)}
(\zeta, s_n, \omega_n)$, Gl. (5.4-6), und ${\cal M}_k^{(n)} (\xi, s_n,
\omega_n)$, Gl. (5.5-8), verwendet, kann diese divergente Reihe auch im
Falle relativ kleiner Argumente mit vertretbarem Aufwand summiert
werden. In Tabelle 7-3 werden die Partialsummen (7.3-1) der
hypergeometrischen Reihe (7.3-2) mit $\nu = 1$ und $z = 3/10$ durch die
verallgemeinerten Summationsprozesse $d_n^{(0)} (\zeta, s_0)$, Gl.
(5.2-18), und ${\delta}_n^{(0)} (\zeta, s_0)$, Gl. (5.4-13), mit $\zeta
= 1$ und ${\Delta}_n^{(0)} (\xi, s_0)$, Gl. (5.5-15), mit $\xi = 27$
summiert. Die Ergebnisse zeigen, da{\ss} man die Partialsummen (7.3-1) der
divergente asymptotische Reihe ${}_2 F_0$ in Gl. (7.3-2) selbst dann
noch mit einer Genauigkeit von 14 Dezimalstellen summieren kann, wenn
das Argument dieser asymptotischen Potenzreihe so klein wie $z = 3/10$
ist. Die Vergleichswerte f\"ur $K_1 (z)$ wurden mit der FORTRAN FUNCTION
S18CCF der NAG Library berechnet. Dieses Programm berechnet eine
Approximation f\"ur $\e^z \, K_1 (z)$ in DOUBLE PRECISION mit Hilfe
geeigneter Tschebyscheffentwicklungen.

Aufgrund der sehr schnellen Divergenz der Partialsummen (7.3-1) f\"ur $z =
3/10$ ist es in Tabelle 7-3 n\"otig, QUADRUPLE PRECISION zu verwenden.
Eine Wiederholung dieser Rechnungen in DOUBLE PRECISION ergab einen
schweren Verlust signifikanter Stellen. Die besten Ergebnisse in DOUBLE
PRECISION lieferte $\delta_n^{(0)} (\zeta, s_n)$, Gl. (5.4-13), das eine
Genauigkeit von ann\"ahernd 11 Stellen f\"ur $n = 18, 19$ ergab. F\"ur gr\"o{\ss}ere
Werte der Transformationsordnung $n$ nahm die Genauigkeit der
Summationsergebnisse wieder ab.

Die Ergebnisse in Tabellen 7-2 und 7-3 sind typisch f\"ur die F\"ahigkeit
der verallgemeinerten Summationsprozesse ${\cal L}_k^{(n)} (\zeta, s_n,
\omega_n)$, Gl. (5.2-6), ${\cal S}_k^{(n)} (\zeta, s_n, \omega_n)$, Gl.
(5.4-6), und ${\cal M}_k^{(n)} (\xi, s_n, \omega_n)$, Gl. (5.5-8), und
ihrer Varianten, die divergente hypergeometrische Reihe (7.3-2) zu
summieren. Ausgiebige Untersuchungen zeigten, da{\ss} ${\cal S}_k^{(n)}
(\zeta, s_n, \omega_n)$ immer die besten Ergebnisse lieferte. In den
meisten F\"allen war ${\cal M}_k^{(n)} (\xi, s_n, \omega_n)$ etwas
leistungsf\"ahiger als die Levinsche Transformation ${\cal L}_k^{(n)}
(\zeta, s_n, \omega_n)$. Da der Parameter $\xi$ aber die Ungleichung
$\xi \ge k - 1$ zu erf\"ullen hat, mu{\ss} man sich bei Verwendung von ${\cal
M}_k^{(n)} (\xi, s_n, \omega_n)$ und seine Varianten auf eine maximale
Transformationsordnung $k_{max}$ festlegen, was sicherlich unbequem ist.

Bei diesen Untersuchungen wurde au{\ss}erdem beobachtet, da{\ss} die
Konvergenzgeschwindigkeit der Summationsverfahren nur unwesentlich von
der Ordnung $\nu$ der modifizierten Besselfunktion $K_{\nu} (z)$
abh\"angt.

Die hier vorgestellten Ergebnisse zeigen, da{\ss} verallgemeinerte
Summationsprozesse bei der Berechnung spezieller Funktionen sehr
n\"utzlich sein k\"onnen. Es w\"are sicherlich interessant, genauer zu
untersuchen, welche spezielle Funktionen \"ahnlich gut wie die
modifizierte Besselfunktion $K_{\nu} (z)$ mit Hilfe von
verallgemeinerten Summationsprozessen berechnet werden k\"onnen.

\endAbschnittsebene

\endAbschnittsebene

\keinTitelblatt\neueSeite

\beginAbschnittsebene
\aktAbschnitt = 7

%%%%%%%%%%%%%%%%%%%%%%%%%%%%%%%%%%%%%%%%%%%%%%%%%%%%%%%%%%%%%%%%%%%%%%%%

\def\abs#1{ \vert #1 \vert }

\define \gaunt#1#2#3#4#5#6
{ \langle #1 \, #2 \vert #3 \, #4 \vert #5 \, #6 \rangle }

\define \gntcf#1#2#3#4#5#6
{ \langle {\ell}_{#1} \, #2 \vert {\ell}_{#3} \, #4
\vert {\ell}_{#5} \, #6 \rangle }

\define\gntsm{{}\,\Stapel{{}_{\ell_{max}} \\ \sum{}^{(2)} \\
{}^{\ell = \ell_{min}}} \, {}}

%%%%%%%%%%%%%%%%%%%%%%%%%%%%%%%%%%%%%%%%%%%%%%%%%%%%%%%%%%%%%%%%%%%%%%%%

\Abschnitt Die Berechnung von Mehrzentrenmolek\"ulintegralen
exponentialartiger Basisfunktionen mit Hilfe von verallgemeinerten
Summationsprozessen

\vskip - 2 \jot

\beginAbschnittsebene

\medskip

\Abschnitt Molek\"ulrechnungen in der Quantenchemie

\smallskip

\aktTag = 0

Nehmen wir an, da{\ss} man quantenchemische {\it ab-initio}-Rechnungen an
einem Molek\"ul, das aus $N$ Elektronen und $\Omega$ Atomkernen besteht,
im Rahmen der Born-Oppenheimer-N\"aherung [Born und Oppenheimer 1927]
durchf\"uhren will. Dann mu{\ss} man Eigenfunktionen und Eigenwerte von
Hamiltonoperatoren des folgenden Typs
berechnen{\footnote[\dagger]{Quantenmechanische Rechnungen an Atomen und
Molek\"ulen werden fast ausschlie{\ss}lich in der {\it Ortsdarstellung\/}
durchgef\"uhrt. Rechnungen in der {\it Impulsdarstellung\/} sind auf
kleine Systeme beschr\"ankt. Einen \"Uberblick \"uber Atom- und
Molek\"ulrechnungen in der Impulsdarstellung einschlie{\ss}lich weiterer
Referenzen findet man beispielsweise in einem Artikel von Fischer,
Defranceschi und Delhalle [1992]}}:
$$
\hat{H} \; = \; \sum_{j=1}^{N} \,
\Biggl\{ - \frac {1} {2} \vec{\nabla}_j^2 \, - \,
\sum_{\alpha=1}^{\Omega} \, \frac
{Z_{\alpha}} {\vert \vec{r}_j - \vec{R}_{\alpha} \vert} \, + \,
\frac {1} {2} \, \sum^{N} \Sb{k=1 \\ k \ne j} \,
\frac {1} {\vert \vec{r}_j - \vec{r}_k \vert}
\Biggr\} \, + \, \frac {1} {2} \,
\sum^{\Omega} \Sb{ \alpha, \beta = 1 \\ \alpha \ne \beta} \,
\frac {Z_{\alpha} Z_{\beta}}
{\vert \vec{R}_{\alpha} - \vec{R}_{\beta} \vert} \, .
\tag
$$
Dabei werden die Ortsvektoren der $N$ Elektronen mit $\vec{r}_1$,
$\vec{r}_2$, $\ldots$ , $\vec{r}_N$ und die Ortsvektoren der $\Omega$
Atomkerne mit $\vec{R}_1$, $\vec{R}_2$, $\ldots$ , $\vec{R}_{\Omega}$
bezeichnet. Die Laplaceoperatoren $\vec{\nabla}_1^2$,
$\vec{\nabla}_2^2$, $\ldots$ , $\vec{\nabla}_N^2$ wirken nur auf die
Koordinaten der $N$ Elektronen.

Kato [1951a] konnte beweisen, da{\ss} ein solcher Hamiltonoperator {\it
selbstadjungiert\/} ist. Das bedeutet, da{\ss} ein {\it vollst\"andiger} Satz
von Eigenfunktionen existiert, die zum diskreten oder kontinuierlichen
Teil des Spektrums geh\"oren. Au{\ss}erdem konnte Kato [1951b] explizit
beweisen, da{\ss} der Hamiltonoperator des Heliumatoms, der ein Spezialfall
des allgemeinen molekularen Hamiltonoperators (8.1-1) ist, eine
Grundzustandswellenfunktion mit minimaler Energie besitzt.

Allerdings gibt es bei der Berechnung der Eigenwerte und Eigenfunktionen
solcher Hamilton\-operatoren das Problem, da{\ss} die entsprechende
zeitunabh\"angige Schr\"odingergleichung
$$
\hat{H} \, \Psi \; = \; E \, \Psi
\tag
$$
aufgrund der Elektron-Elektron-Wechselwirkung nicht {\it separierbar}
ist. Das impliziert, da{\ss} eine exakte L\"osung $\Psi$ der
Schr\"odingergleichung (8.1-2) auf solche Weise von den Koordinaten
$\vec{r}_n$ und Spins $\sigma_n$ der beteiligten $N$ Elektronen abh\"angt,
da{\ss} sie nicht durch eine endliche Linearkombination von Produkten aus
Funktionen dargestellt werden kann, die jeweils nur von Koordinaten und
dem Spin eines {\it einzelnen} Elektrons abh\"angen.

Ungl\"ucklicherweise sind keine praktikablen mathematischen Techniken
bekannt, mit deren Hilfe man derartige h\"oherdimensionale
nichtseparierbare partielle Differentialgleichungen exakt l\"osen kann. In
der Praxis ist es deswegen unumg\"anglich, zus\"atzliche N\"aherungen
einzuf\"uhren.

Die Mehrheit aller quantenchemischen Rechnungen an Molek\"ulen basiert auf
der sogenannten {\it Orbitaln\"aherung}. Im Rahmen dieser N\"aherung wird
eine von den Koordinaten $\vec{r}_1$, $\vec{r}_2$, $\ldots~$,
$\vec{r}_N$ und Spins $\sigma_1$, $\sigma_2$, $\ldots$ , $\sigma_N$ der
$N$ Elektronen abh\"angende L\"osung $\Psi$ der zeitunabh\"angigen
Schr\"odingergleichung (8.1-2) durch eine endliche Linearkombination von
antisymmetrisierten Produkten aus sogenannten {\it Spinorbitalen} oder
{\it Einteilchenfunktionen} $\psi (\vec{r}, \sigma)$, die von den
Koordinaten $\vec{r}$ und dem Spin $\sigma$ eines einzigen Elektrons
abh\"angen, dargestellt.

Die elektronische Schr\"odingergleichung (8.1-2) kann im Rahmen der
Orbitaln\"aherung als gel\"ost betrachtet werden, sobald alle Spinorbitale
$\psi_j (\vec{r}, \sigma)$, aus denen die $N$-Elektronenfunktion $\Psi$
aufgebaut wird, mit Hilfe des von Hartree [1928] und Fock [1930]
eingef\"uhrten SCF{\footnote[\ddagger]{Self-Consistent Field}}-Verfahrens
berechnet sind.

Es gibt zwei grunds\"atzlich verschiedene Varianten des SCF-Verfahrens:

\beginEingezogeneBeschreibung \zu \Laenge{(ii):}

\item{(i):} Beim {\it numerischen} Hartree-Fock-Verfahren f\"uhrt das
Ritzsche Variationsverfahren zu nichtlinearen
Integrodifferentialgleichungen f\"ur die Ortsanteile $\varphi_j (\vec{r})$
der Spinorbitale $\psi_j (\vec{r}, \sigma)$, die dann punktweise
selbstkonsistent gel\"ost werden k\"onnen [Hartree 1957; Froese Fischer
1977].

\item{(ii):} Bei der anderen Variante des SCF-Verfahrens, die manchmal
auch als {\it analytisches} SCF-Verfahren bezeichnet wird, werden die
Ortsanteile $\varphi_j (\vec{r})$ der zu bestimmenden Spinorbitale
$\psi_j (\vec{r}, \sigma)$ durch endliche Linearkombinationen
sogenannter Basisfunktionen $\{ \xi_k (\vec{r}) \}_{k=1}^{K}$, die in
geschlossener Form angegeben werden, approximiert:
$$
\varphi_j (\vec{r}) \; = \;
\sum_{k=1}^{K} \, C_k^{(j)} \, \xi_k (\vec{r}) \, .
\tag
$$

\endEingezogeneBeschreibung

Da bei der analytischen Variante des SCF-Verfahrens nur die unbestimmten
Koeffizienten $C_k^{(j)}$ variiert werden, erfolgt die Variation nicht
mehr im gesamten Hilbert-Raum der Einteilchenfunktionen, sondern nur
noch in dem durch die $K$ Basisfunktionen $\xi_k (\vec{r})$
aufgespannten endlich\-dimensionalen Teilraum{\footnote[\dagger]{Die
Basisfunktionen $\{ \xi_k (\vec{r}) \}_{k=1}^{K}$ k\"onnen noch weitere
Variationsparameter enthalten, die man selbstverst\"and\-lich ebenfalls
optimieren kann, W\"ahrend der Variation der Koeffizienten $C_k^{(j)}$
werden diese zus\"atzlichen Variationsparameter aber festgehalten.}}. Das
Ritzsche Variationsverfahren ergibt dann ein System gekoppelter
nichtlinearer Gleichungen f\"ur die unbestimmten Koeffizienten
$C_k^{(j)}$, die sogenannten Roothaanschen Gleichungen [Roothaan 1951],
das iterativ gel\"ost werden kann [Weissbluth 1978, S. 561 - 565].
Voraussetzung daf\"ur ist, da{\ss} die numerischen Werte aller im Rahmen eines
solchen Ansatzes vorkommenden Ein- und Zweielektronenintegrale bekannt
sind. Im Rahmen des N\"aherungsansatzes (8.1-3) treten die folgenden
Integraltypen auf [Weissbluth 1978, S. 561]:

\beginEingezogeneBeschreibung \zu \Laenge{(3):}

\item{(1):} \"Uberlappungsintegrale:
$$
S_{\mu \nu} \; = \;
\int \, \bigl[ \xi_{\mu} (\vec{r}) \bigr]^{\ast}
\, \xi_{\nu} (\vec{r}) \, \d^3 \vec{r} \, .
\tag
$$

\item{(2):} Integrale des Einelektronanteils $\hat{H}_0$ des
vollen atomaren oder molekularen Hamiltonoperators (8.1-1):
$$
H_{\mu \nu} \; = \;
\int \, \bigl[ \xi_{\mu} (\vec{r}) \bigr]^{\ast}
\, \hat{H}_0 \, \xi_{\nu} (\vec{r}) \, \d^3 \vec{r} \, .
\tag
$$

\item{(3):} Elektronenwechselwirkungsintegrale:
$$
\langle \mu \nu \vert \lambda \sigma \rangle \; = \; \int \,
\bigl[ \xi_{\mu} (\vec{r}_1) \bigr]^{\ast} \,
\bigl[ \xi_{\nu} (\vec{r}_2) \bigr]^{\ast} \,
\frac {1} {\vert \vec{r}_1 - \vec{r}_2 \vert} \,
\xi_{\lambda} (\vec{r}_1) \, \xi_{\sigma} (\vec{r}_2) \,
\d^3 \vec{r}_1 \, \d^3 \vec{r}_2 \, .
\tag
$$

\endEingezogeneBeschreibung

Ein unbestreitbarer Nachteil des Ansatzes (8.1-3) besteht darin, da{\ss} der
Erfolg eines solchen analytischen SCF-Verfahrens entscheidend sowohl von
der Zahl $K$ als auch von der Art der verwendeten Basisfunktionen $\{
\xi_k (\vec{r}) \}_{k=1}^{K}$ abh\"angt. Bei Verwendung einer endlichen
Basis kann man sich demzufolge im Prinzip nie v\"ollig sicher sein, da{\ss}
der Basissatz gro{\ss} genug ist und da{\ss} die verwendeten Basisfunktionen dem
Problem gut genug angepa{\ss}t sind, um Ergebnisse von ausreichender
Genauigkeit zu liefern. Dagegen hat das numerische
Hartree-Fock-Verfahren den unbestreitbaren Vorteil, da{\ss} es keine
st\"orende Basissatzabh\"angigkeit gibt und da{\ss} man auf diese Weise
automatisch die {\it bestm\"ogliche} Approximation erh\"alt, die bei
Verwendung der Orbitaln\"aherung \"uberhaupt m\"oglich ist.

Das numerische Hartree-Fock-Verfahren ist besonders f\"ur Rechnungen an
Atomen geeignet. Bei Atomen ist es aufgrund ihrer Kugelsymmetrie
vorteilhaft, den Ortsanteil eines Spinorbitals als Produkt aus einer
Radialfunktion und einer Kugelfl\"achenfunktion $Y_{\ell}^m (\theta,
\phi)$ anzusetzen:
$$
\varphi_{n, \ell}^{m} (\vec{r}) \; = \;
R_{n, \ell} (r) \, Y_{\ell}^m (\theta, \phi) \, .
\tag
$$
Dabei sind $\ell$ und $m$ die \"ublichen Drehimpulsquantenzahlen, und $n$
ist als eine Verallgemeinerung der Hauptquantenzahl einer
Wasserstoffeigenfunktion [Bethe und Salpeter 1977, S. 15] zu betrachten.

Wenn man die Phasenkonvention von Condon und Shortley [1970, S. 48]
verwendet, sind die Kugelfl\"achenfunktionen folgenderma{\ss}en definiert
[Weissbluth 1978, S. 3]:
$$
Y_{\ell}^m (\theta, \phi) \; = \;
i^{m + \abs{m}} \, \left\{
\frac {(2 \ell + 1) (\ell - \abs{m})!} {4 \pi (\ell + \abs{m})!}
\right\}^{1/2} \, P_{\ell}^{\abs{m}} (\cos \theta) \,
\e^{i m \phi} \, .
\tag
$$
$P_{\ell}^{\abs{m}} (\cos \theta)$ ist ein zugeordnetes Legendrepolynom,
das folgenderma{\ss}en definiert ist [Condon und Odaba\c{s}i 1980, S. 155]:
$$
\beginAligntags
" P_{\ell}^{m} (z) " \; = \; " (1 - z^2)^{m/2} \,
\frac {\d^{\ell + m}} {\d z^{\ell + m}} \,
\left\{ \frac {(z^2 - 1)^{\ell}} {2^{\ell} {\ell}!} \right\} \\
" " \; = \; " (1 - z^2)^{m/2} \, \frac {\d^m} {\d z^m} \,
P_{\ell} (z) \, , \qquad m \ge 0 \, .
\\ \tag
\endAligntags
$$

Da das Coulombpotential ein Zentralpotential ist, kann man erreichen,
da{\ss} die gekoppelten Integrodifferentialgleichungen f\"ur die Ortsanteile
$\varphi_{n, \ell}^{m} (\vec{r})$ in Gl. (8.1-7) nicht explizit von den
Winkelvariablen $\theta$ und $\phi$ abh\"angen. Man mu{\ss} also bei Atomen
nur den Radialanteil $R_{n, \ell} (r)$ aus Gl. (8.1-7) mit Hilfe des
SCF-Verfahrens bestimmen, was zu einer betr\"achtlichen Verringerung des
Rechenaufwandes f\"uhrt. Ein weiterer Vorteil des Ansatzes (8.1-7) ist,
da{\ss} man auf diese Weise die hochentwickelte Theorie quantenmechanischer
Drehimpulse verwenden kann.

Molek\"ule besitzen normalerweise \"uberhaupt keine Symmetrie, und wenn
doch, dann besitzen sie eine geringere Symmetrie als Kugelsymmetrie. Bei
Molek\"ulen erreicht man demzufolge durch den Ansatz (8.1-7), der die
Kugelsymmetrie der Atome ausn\"utzt, keine Vereinfachung der gekoppelten
Integrodifferentialgleichungen f\"ur die Ortsanteile $\varphi_j
(\vec{r})$. Eine rein numerische punktweise L\"osung der gekoppelten
Integrodifferentialgleichungen ist demzufolge bei Molek\"ulen \"au{\ss}erst
schwierig. Dem Autor sind keine Molek\"ulprogramme bekannt, die auf dem
numerischen Hartree-Fock-Verfahren basieren, und die routinem\"a{\ss}ig
angewendet werden.

Deswegen ist man bei Molek\"ulen mehr oder weniger gezwungen,
SCF-Rechnungen auf der Basis der oben erw\"ahnten Roothaanschen
Gleichungen [Roothaan 1951] durchzuf\"uhren. Die heute in der
Quantenchemie \"ubliche Vorgehensweise besteht darin, da{\ss} man die
Ortsanteile $\varphi_j (\vec{r})$, die \"uber das gesamte Molek\"ul
delokalisiert sein k\"onnen und deswegen auch MO's oder Molek\"ulorbitale
genannt werden, durch eine endliche Basis gem\"a{\ss} Gl. (8.1-3) darstellt.
Eine solche Darstellung durch Basisfunktionen l\"a{\ss}t aber noch zahlreiche
M\"oglichkeiten offen.

Eine wesentliche Forderung ist, da{\ss} ein Basissatz {\it vollst\"andig} sein
mu{\ss}. Nur dann ist im Prinzip garantiert, da{\ss} eine SCF-Rechnung
konvergiert. Praktisch bedeutet die Vollst\"andigkeit der Basis, da{\ss} jedes
MO $\varphi_j (\vec{r})$ im Prinzip mit beliebiger Genauigkeit
approximiert werden kann, wenn man die Basis $\{ \xi_k (\vec{r})
\}_{k=1}^{K}$ nur ausreichend gro{\ss} werden l\"a{\ss}t. Die
Vollst\"andigkeitseigenschaften quantenchemischer Basiss\"atze in
verschiedenen Funktionenr\"aumen und ihre Implikationen f\"ur die Konvergenz
von Molek\"ulrechnungen wurden von Klahn [1981; 1984; 1985a; 1985b], von
Klahn und Bingel [1977a; 1977b; 1977c] und von Klahn und Morgan [1984]
eingehend untersucht.

Beispielsweise k\"onnte man als Basissatz ein vollst\"andiges System von
Einteilchenfunktionen verwenden, dessen Funktionen alle an einem
beliebigen Punkt des Raumes zentriert sind. Aufgrund der Vollst\"andigkeit
des Basissatzes kann der Ortsanteil $\varphi_j (\vec{r})$ eines
Spinorbitals $\psi_j (\vec{r}, \sigma)$ auf diese Weise im Prinzip mit
beliebiger Genauigkeit approximiert werden, unabh\"angig davon, welches
vollst\"andige System verwendet wird und wo die Funktionen zentriert sind.
Der Nachteil solcher {\it single-center}-Ans\"atze besteht darin, da{\ss} sie
in der Regel nur relativ schlecht konvergieren. Laut Bishop [1967, S.
28] sind {\it single-center}-Ans\"atze h\"ochstens bei Rechnungen an
hochsymmetrischen Systemen konkurrenzf\"ahig.

Der heute in der Quantenchemie fast ausschlie{\ss}lich verwendete Ansatz zur
Darstellung der Ortsanteile $\varphi_j (\vec{r})$ der zu bestimmenden
Spinorbitale $\psi_j (\vec{r}, \sigma)$ besteht darin, einen Basissatz
zu verwenden, dessen Funktionen an den verschiedenen Atomkernen des
Molek\"uls zentriert sind. Durch die Verwendung eines solchen
LCAO{\footnote[\dagger]{Linear Combination of Atomic Orbitals}}-Ansatzes
kann man in vielen F\"allen eine ausreichend schnelle Konvergenz des
SCF-Verfahrens erreichen. Nur dadurch sind Rechnungen an gr\"o{\ss}eren
Molek\"ulen \"uberhaupt m\"oglich geworden. Trotzdem besitzt die LCAO-N\"aherung
-- wie jede andere N\"aherung auch -- ihre spezifischen Beschr\"ankungen und
Probleme, die ihre Anwendbarkeit begrenzen.

Wie schon fr\"uher erw\"ahnt, k\"onnen die Roothaanschen Gleichungen [Roothaan
1951] zur Bestimmung der Koeffizienten $C_k^{(j)}$ in Gl. (8.1-3)
iterativ gel\"ost werden, wenn die numerischen Werte der Ein- und
Zweielektronenintegrale (8.1-4) - (8.1-7) bekannt sind. Bei Atomen ist
die Berechnung dieser Integrale normalerweise unproblematisch. Selbst
wenn man keine geschlossene Darstellung finden kann, k\"onnen atomare
Integrale in der Regel sowohl genau genug als auch schnell genug mit
Hilfe von Quadraturverfahren berechnet werden. Bei Molek\"ulen ist dagegen
die Berechnung dieser Integrale, bedingt durch die Verwendung einer
Mehrzentrenbasis, in der Regel der schwierigste und zeitaufwendigste
Teil einer molekularen SCF-Rechnung. Eine Klassifikation der im Rahmen
einer LCAO-MO-Rechnung auftretenden Integrale findet man beispielsweise
in einem \"Ubersichtsartikel von Steinborn [1983, S. 45 - 48]. Wenn man
\"uberhaupt geschlossene Ausdr\"ucke f\"ur Mehrzentrenintegrale finden kann,
sind sie in der Regel \"au{\ss}erst kompliziert und enthalten oft mehrfache
unendliche Reihen. Eine ausschlie{\ss}lich numerische Berechnung der drei-
oder sechsdimensionalen Mehrzentrenintegrale mit Hilfe von
Quadraturverfahren d\"urfte beim heutigen Stand des Wissens \"uber
numerische Integration [Bra{\ss} 1977; Davis und Rabinowitz 1984; Piessens,
de Doncker-Kapenga, \"Uberhuber und Kahaner 1983] ebenfalls nicht
praktikabel sein [Boerrigter, te Velde und Baerends 1988, te Velde und
Baerends 1992].

Als weitere Komplikation kommt hinzu, da{\ss} die Berechnung molekularer
Mehrzentrenintegrale in der Regel nicht nur \"au{\ss}erst schwierig ist,
sondern da{\ss} man auch noch eine sehr gro{\ss}e Anzahl von Integralen
berechnen mu{\ss} (siehe \v{C}\'{a}rsky und Urban [1980, Table 2.3]).
Verwendet man beispielsweise in einer molekularen SCF-Rechnung einen
LCAO-Basissatz, bei dem jedes MO durch $M$ {\it reelle} Basisfunktionen
dargestellt wird, so m\"ussen $M(M+1)/2$ Einelektronenintegrale und
$[M(M+1)/2] [M(M+1)/2+1]/2$ Zweielektronenintegrale berechnet werden
[Daudel, Leroy, Peeters und Sana 1983, Gl. (5.87)].

Die st\"orende Rolle der Mehrzentrenintegrale in der Entwicklung der
Quantenchemie wird durch das folgende Zitat von Huzinaga [1967, S. 52]
treffend beschrieben:

\medskip

\beginSchmaeler
\noindent {\sl The difficulty of evaluating so-called molecular
integrals is essentially of mathematical nature and has never occupied a
position of primal importance in the conceptual development of quantum
chemistry but it has been notoriously persistent and extremely annoying.}
\endSchmaeler

\medskip

Au{\ss}erdem sei noch erw\"ahnt, da{\ss} bereits das \"alteste Lehrbuch der
Quantenchemie [Hellmann 1937] auf S. 331 - 345 einen mathematischen
Anhang enth\"alt, in dem Mehrzentrenintegrale relativ ausf\"uhrlich
behandelt werden.

\medskip

\Abschnitt Exponentialartige Basisfunktionen f\"ur Molek\"ulrechnungen

\smallskip

\aktTag = 0

Wie im letzten Unterabschnitt erw\"ahnt, basieren praktisch alle
Molek\"ulrechnungen auf den LCAO-MO-SCF-Verfahren. Dabei wird ein MO
$\varphi_j (\vec{r})$ gem\"a{\ss} Gl. (8.1-3) als Linearkombination von
Basisfunktionen, die an den verschiedenen Atomkernen des Molek\"uls
zentriert sind, dargestellt. Aus historischen Gr\"unden werden die
Basisfunktionen AO's oder Atomorbitale genannt, was f\"alschlicherweise
suggeriert, da{\ss} es sich bei einer Basisfunktion $\xi_k (\vec{r})$ um
eine L\"osung einer atomaren Schr\"odingergleichung handelt. Im modernen
Sprachgebrauch sind AO's einfach Funktionen $\xi_k : \R^3 \to \C$, die
\"ublicherweise, aber nicht notwendigerweise an den Atomkernen des
Molek\"uls zentriert sind, und die aufgrund ihrer mathematischen
Eigenschaften geeignet sind, molekulare Wellenfunktionen zu
approximieren [Steiner 1976, S. 86]. Das bedeutet beispielsweise, da{\ss}
Basisfunktionen bez\"uglich einer Integration \"uber den gesamten
dreidimensionalen Raum $\R^3$ {\it quadratintegrabel} sein m\"ussen,
oder -- was gleichbedeutend ist -- da{\ss} sie Elemente des Hilbertraumes
$L^2 (\R^3)$ sein m\"ussen. Au{\ss}erdem m\"ussen die Basisfunktionen
ausreichend oft differenzierbar sein.

Die Wahl des Basissatzes ist ein ganz entscheidender Schritt f\"ur jede
Molek\"ulrechnung, da sowohl die G\"ute der Approximation als auch der
numerische Aufwand durch den Basissatz determiniert werden. Ein
Basissatz $\{ \xi_k (\vec{r}) \}_{k=1}^{K}$ mu{\ss} deswegen noch bestimmte
weitere mathematische Eigenschaften besitzen, um sinnvoll und effizient
in Molek\"ulrechnungen verwendet werden zu k\"onnen.

Wie schon erw\"ahnt, ist die Konvergenz einer SCF-Rechnung nur dann
gew\"ahrleistet, wenn der Basissatz vollst\"andig ist. Allerdings ist die
Vollst\"andigkeit eines Basissatzes zwar eine notwendige, aber keine
hinreichende Bedingung f\"ur die praktische Verwendbarkeit eines
Basissatzes. Wenn die Konvergenzgeschwindigkeit zu langsam ist, werden
auch die gr\"o{\ss}ten Rechner durch die Dimensionalit\"at der Roothaanschen
Gleichungen [Roothaan 1951] und die gro{\ss}e Zahl der zu berechnenden
Mehrzentrenintegrale \"uberfordert. Ein guter Basissatz sollte also in der
Lage sein, die physikalische Realit\"at so gut zu beschreiben, da{\ss} man
schon mit einer relativ kleinen Basis aussagekr\"aftige Ergebnisse erh\"alt.
Das ist besonders wichtig, wenn man sich nicht mit dem sogenannten {\it
Hartree-Fock-Limit}{\footnote[\dagger]{Optimale Energie und
Wellenfunktion, die f\"ur das betreffende System im Rahmen der
Orbitaln\"aherung erreichbar ist.}} [\v{C}\'{a}rsky und Urban 1980, S. 8],
zufrieden geben kann oder will. Die heute \"ublichen Verfahren zur
expliziten Ber\"ucksichtigung der Elektronenkorrelation (siehe
beispielsweise Ahlrichs und Scharf [1987], Bartlett [1981], Bartlett,
Dykstra und Paldus [1984], Bartlett und Stanton [1994], {\v C}{\' \i}{\v
z}ek [1966; 1969], Freed [1971], Fulde [1991], Geertsen, Eriksen und
Oddershede [1991], Harris, Monkhorst und Freeman [1992], Harter und
Patterson [1976], Hinze [1981], Hoffmann und Schaefer [1986], Hurley
[1976], Linderberg und \"Ohrn [1973], Lindgren und Morrison [1982],
Jankowski [1987; 1992], J{\o}rgensen und Simons [1981], Karwowski
[1992], Kucharski und Bartlett [1986], K\"ummel, L\"uhrmann und Zabolitzky
[1978], Kutzelnigg [1977], Kvasni\v{c}ka, Laurinc und Biskupi\v{c}
[1982], Matsen und Pauncz [1986], McWeeny [1992], Mukherjee und Pal
[1989], Oddershede [1992], Oddershede, J{\o}rgensen und Yeager [1984],
Paldus [1974; 1976; 1981; 1988; 1992], Paldus und {\v C}{\' \i}{\v z}ek
[1975], Pauncz [1979], Roos [1987], Roos und Siegbahn [1977], Sharma
[1976], Shavitt [1977], Surj\'{a}n [1989], Sutcliffe [1983], Urban,
\v{C}ernu\v{s}\'{a}k, Kell\"{o} und Noga [1987], Werner [1987], Wilson
[1981; 1984; 1985; 1992a; 1992b; 1992c; 1992d; 1992e]) sind so
aufwendig, da{\ss} sie bei gr\"o{\ss}eren Basiss\"atzen nicht mehr durchf\"uhrbar
werden.

Wenn man Basisfunktionen $\xi_k (\vec{r})$ verwenden will, die eine
ausreichende \"Ahnlichkeit mit den exakten Eigenfunktionen der Atome des
Molek\"uls besitzen, mu{\ss} man sich analog zu Gl. (8.1-7) auf
Basisfunktionen beschr\"anken, die irreduzible sph\"arische Tensoren
[Biedenharn und Louck 1981, Gl. (3.206)] sind. Auf diese Weise ist
gew\"ahrleistet, da{\ss} die Basisfunktionen Drehimpulseigenfunktionen sind
und da{\ss} man bei der Berechnung der Mehrzentrenintegrale die
hochentwickelte mathematische Theorie quantenmechanischer Drehimpulse
verwenden kann.

Die L\"osungen einer atomaren Einteilchen-Schr\"odingergleichung mit einem
Zentralpotential, das nicht singul\"arer ist als das Coulombpotential,
verhalten sich f\"ur kleine Abst\"ande vom Kraftzentrum asymptotisch wie die
L\"osungen der homogenen dreidimensionalen Laplacegleichung
$$
\vec{\nabla}^2 \, f (\vec{r}) \; = \; 0 \, .
\qquad \vec{r} \in \R^3 \, .
\tag
$$
Bekanntlich sind die {\it regul\"aren r\"aumlichen Kugelfunktionen}
$$
{\cal Y}_{\ell}^{m} (\vec{r}) \; = \;
r^{\ell} \, Y_{\ell}^{m} (\theta, \phi) \, ,
\tag
$$
die Polynome $\ell$-ten Grades in den cartesischen Komponenten $x$, $y$
und $z$ des Vektors $\vec{r}$ sind [Biedenharn und Louck 1981, S. 71],
die am Ursprung endlichen L\"osungen der homogenen dreidimensionalen
Laplacegleichung (8.2-1), wogegen die {\it irregul\"aren r\"aumlichen
Kugelfunktionen}
$$
{\cal Z}_{\ell}^{m} (\vec{r}) \; = \;
r^{- \ell - 1} \, Y_{\ell}^{m} (\theta, \phi)
\tag
$$
die am Ursprung singul\"aren L\"osungen der homogenen dreidimensionalen
Laplacegleichung (8.2-1) sind. Da Basisfunktionen und ihre zweiten
Ableitungen quadratintegrabel sein m\"ussen, folgt, da{\ss} Basisfunktionen
sich f\"ur kleine Abst\"ande $r$ asymptotisch wie regul\"are r\"aumliche
Kugelfunktionen verhalten m\"ussen.

Exakte L\"osungen atomarer und molekularer Schr\"odingergleichungen besitzen
an jedem Atomkern einen {\it Cusp} [Kato 1957]. Au{\ss}erdem fallen exakte
L\"osungen atomarer und molekularer Schr\"odingergleichungen f\"ur gro{\ss}e
Abst\"ande {\it exponentiell\/} (siehe beispielsweise [Agmon 1982; 1985;
Ahlrichs 1989; Cycon, Froese, Kirsch und Simon 1987, Abschnitt 4; Herbst
1993]). Diese beiden asymptotischen Bedingungen k\"onnen problemlos
erf\"ullt werden, wenn man als Basisfunktionen {\it Slaterfunktionen}
verwendet:
$$
\chi_{n, \ell}^{m} (\alpha \vec{r}) \; = \;
(\alpha r)^{n - 1} \, \e^{- \alpha r} \, Y_{\ell}^{m} (\theta, \phi)
\, , \qquad \alpha \in \R_{+} \, .
\tag
$$
Diese Funktionen wurden von Slater [1930] in Verbindung mit den
sogenannten Slaterschen Regeln eingef\"uhrt, um die Radialanteile
numerischer Hartree-Fock-Funktionen durch analytische Funktionen zu
approximieren. Wie Slater [1932, S. 42] selbst schrieb, war er dabei
nicht an einer m\"oglichst genauen Approximation interessiert. Sein
Hauptziel war eine m\"oglichst einfache und kompakte Approximation, die
nat\"urlich auch eine befriedigende Genauigkeit liefern sollte. Die
Betonung der analytischen Einfachheit wird verst\"andlich, wenn man
bedenkt, da{\ss} damals nur Handrechenmaschinen zur Verf\"ugung standen.

Da Slaterfunktionen das asymptotische Verhalten exakter atomarer und
molekularer Wellenfunktionen f\"ur kleine und gro{\ss}e Abst\"ande reproduzieren
k\"onnen, sollten schon wenige Funktionen f\"ur eine sehr genaue
Beschreibung atomarer und molekularer Wellenfunktionen ausreichen. Es
ist also keineswegs \"uberraschend, da{\ss} Slaterfunktionen mit gro{\ss}em Erfolg
als Basisfunktionen in atomaren SCF-Rechnungen verwendet wurden. Die
numerischen und algorithmischen Probleme, die bei solchen atomaren
SCF-Rechnungen mit einer Slater-Basis auftreten, wurden von Roothaan und
Bagus [1963] ausf\"uhrlich behandelt.

Da Slaterfunktionen bei Atomen \"au{\ss}erst erfolgreich waren, lag es nahe,
sie auch als LCAO-Basissatz in Molek\"ulrechnungen zu verwenden. Hier
tritt allerdings das Problem auf, da{\ss} die im Rahmen eines LCAO-Ansatzes
unvermeidlich und in gro{\ss}er Zahl auftretenden Mehrzentrenintegrale bei
Slaterfunktionen \"uberaus kompliziert sind. So sind trotz zahlreicher
Versuche bis heute keine Rechenverfahren bekannt, die eine ausreichend
schnelle und genaue Berechnung der schwierigen Mehrzentrenintegrale von
Slaterfunktionen erm\"oglichen. Diese mathematischen Probleme waren zum
Teil daf\"ur verantwortlich, da{\ss} semiempirische Verfahren entwickelt
wurden, bei denen die Berechnung der Mehrzentrenintegrale wenigstens
teilweise vermieden werden kann. Einen \"Uberblick \"uber die historische
Entwicklung und \"uber die insgesamt nicht befriedigende Situation der
Integralberechnung bei Slaterfunktionen findet man in Arbeiten von
Barnett [1963], Barnett und Coulson [1951], Browne [1971], Corbat\'{o}
und Switendick [1963], Coulson 1937, Harris und Michels [1967] und
Huzinaga [1967] und in dem von Weatherford und Jones [1982]
herausgegebenen Tagungsband. In diesem Zusammenhang sei auch erw\"ahnt,
da{\ss} schon relativ bald versucht wurde, verallgemeinerte
Summationsprozesse bei der Berechnung von Mehrzentrenintegralen von
Slaterfunktionen zu verwenden [Petersson und McKoy 1967].

Da aber die Verwendung von Slaterfunktionen oder anderen
exponentialartigen Basisfunktionen in Molek\"ulrechnungen \"au{\ss}erst
w\"unschenswert w\"are, wird auch weiterhin intensiv \"uber die Berechnung der
Mehrzentrenintegrale von Slaterfunktionen gearbeitet, beispielsweise in
Canada [Talman 1984; 1986; 1989; 1993], in Frankreich [Bouferguene und
Rinaldi 1994], in Spanien [Carb\'{o} und Besal\'{u} 1992; Fern\'andez Rico
1993; Fern\'andez Rico, L\'opez, Paniagua und Fern\'andez-Alonso 1986;
Fern\'andez Rico, L\'opez und Ram\'{\i}rez 1989a; 1989b; 1991; 1992a; 1992b;
Fern\'andez Rico, L\'opez, Paniagua und Ram\'{\i}rez 1991; Fern\'andez Rico,
L\'opez, Ram\'{\i}rez und Fern\'andez-Alonso 1993; L\'opez und Ram\'{\i}rez
1994; Mart\'{\i}n Pend\'{a}s und Francisco 1991], in den Vereinigten
Staaten [Jones 1986a; 1986b; 1987; 1988; 1991; 1992a; 1992b; 1993; Jones
und Etemadi 1993; Jones, Etemadi und Brown 1992], oder auch in
Deutschland [Hierse und Oppeneer 1993].

Am hiesigen Institut f\"ur Physikalische und Theoretische Chemie der
Universit\"at Regensburg wurde ebenfalls sehr intensiv \"uber
Mehrzentrenmolek\"ulintegrale exponentialartiger Basisfunktionen
gearbeitet. Das Hauptgewicht lag dabei nicht auf den Slaterfunktionen,
sondern auf einer anderen Funktionenklasse, den sogenannten
$B$-Funktionen [Filter und Steinborn 1978b, S. 2],
$$
B^m_{n,\ell} (\alpha, \vec r) \; = \;
[2^{n+\ell}(n+\ell)!]^{-1} \, \widehat{k}_{n-1/2} (\alpha r) \,
{\cal Y}^m_{\ell} (\alpha \vec{r}) \, ,
\tag
$$
die anisotrope Verallgemeinerungen der reduzierten Besselfunktionen sind
[Steinborn und Filter 1975, S. 275]:
$$
\widehat k_{\nu} (z) \; = \; (2/\pi)^{1/2} z^{\nu} K_{\nu} (z) \, .
\tag
$$
$K_{\nu} (z)$ ist eine modifizierte Besselfunktion der zweiten Art, die
in Gl. (7.1-5) definiert ist.

Wenn die Ordnung $\nu$ einer reduzierten Besselfunktion $\widehat
k_{\nu} (z)$ halbzahlig und positiv ist, $\nu = n + 1/2$ und $n \in
\N_0$, kann die reduzierte Besselfunktion geschrieben werden als
Exponentialfunktion multipliziert mit einer abbrechenden
hypergeometrischen Reihe ${}_1 F_1$ [Weniger und Steinborn 1983b, S.
2028]:
$$
{\widehat k}_{n+1/2} (z) \; = \; 2^n \, (1/2)_n \, \e^{- z} \,
{}_1 F_1 (- n; - 2 n; 2 z) \, , \qquad n \in \N_0 \, .
\tag
$$
Der Polynomanteil auf der rechten Seite von Gl. (8.2-7) wurde unabh\"angig
von den in Regensburg durchgef\"uhrten Arbeiten \"uber
Mehrzentrenmolek\"ulintegrale auch in der mathematischen Literatur
eingehend untersucht [Grosswald 1978]. Dort wird die folgende Notation
verwendet [Grosswald 1978, S. 34]:
$$
\theta_n (z) \; = \; \e^z \, {\widehat k}_{n+1/2} (z) \; = \;
2^n \, (1/2)_n \, {}_1 F_1 (- n; - 2 n; 2 z) \, , \qquad n \in \N_0 \, .
\tag
$$
Zusammen mit einigen anderen, eng verwandten Polynomen werden die
$\theta_n (z)$ als {\it Besselpolynome\/} bezeichnet. In Abschnitt 14
des Buches von Grosswald [1978] wird gezeigt, da{\ss} diese Polynome in so
verschiedenen Gebieten wie der Zahlentheorie, der Statistik oder der
Analyse komplexer elektrischer Netzwerke vorkommen.

Interessanterweise kommen Besselpolynome auch in der Theorie der
Pad\'e-Approximation vor. In Band \Roemisch{1} der Monographie von Baker
und Graves-Morris \"uber Pad\'e-Approximation wird gezeigt, da{\ss} die
Pad\'e-Approximationen $[ \ell / m ]$ der Exponentialfunktion $\exp (z)$
die folgende Darstellung besitzen [Baker und Graves-Morris [1981a, Gl.
(2.12)]:
$$
[ \, \ell \, / \, m \, ] \; = \; \frac
{ {}_1 F_1 (- \ell; - \ell - m; z)}
{ {}_1 F_1 (- m; - \ell - m;- z)} \, ,
\qquad \ell, m \in \N_0 \, .
\tag
$$
Ein Vergleich der Gln. (8.2-8) und (8.2-9) ergibt, da{\ss} die
Diagonalelemente $[ \, n \, / \, n \, ]$ der Pad\'e-Tafel der
Exponentialfunktion $\exp (z)$ als Quotient zweier Besselpolynome
geschrieben werden k\"onnen [Weniger 1989, Gl. (14.3-15)]:
$$
[ \, n \, / \, n \, ] \; = \; \frac
{ \theta_n (z/2) } { \theta_n (- z/2) } \, ,
\qquad n \in \N_0 \, .
\tag
$$

Es mag befremdlich erscheinen, warum man gerade $B$-Funktionen anstelle
von Slaterfunktionen als LCAO-Basisfunktionen verwenden will. So folgt
aus Gln. (8.2-5) und (8.2-7), da{\ss} eine $B$-Funktion ein relativ
kompliziertes mathematisches Objekt ist. Ein Vergleich der Gln. (8.2-4),
(8.2-5) und (8.2-7) ergibt au{\ss}erdem, da{\ss} eine $B$-Funktion als
Linearkombination von Slaterfunktionen ausgedr\"uckt werden kann. Da die
Mehrzentrenintegrale von Slaterfunktionen bis heute nicht auf
befriedigende Weise berechnet werden k\"onnen, ist es {\it a priori\/}
keineswegs einleuchtend, da{\ss} die doch relativ komplizierten
$B$-Funktionen hier irgendwelche Vorteile aufweisen k\"onnten. Es konnte
aber gezeigt werden, da{\ss} $B$-Funktionen mathematische Eigenschaften
besitzen, die ihnen eine ganz besondere Position unter allen in der
Quantenchemie verwendeten exponentiell fallenden Basisfunktionen geben.

Eine der wichtigsten Methoden zur Berechnung von
Mehrzentrenmolek\"ulintegralen ist die auf Geller [1962; 1963a; 1963b;
1964a; 1964b], Prosser und Blanchard [1962], und Bonham, Peacher und Cox
[1964] zur\"uckgehende {\it Fouriertransformationsmethode}. Mit Hilfe
dieser Methode k\"onnen Mehrzentrenmolek\"ulintegrale unter Verwendung des
{\it Fourier-Konvolutionstheorems\/} [Bochner 1948, Satz 13] in inverse
Fourierintegrale transformiert werden. Ob und wie leicht die inversen
Fourierintegrale berechnet werden k\"onnen, h\"angt in entscheidender Weise
von den Fouriertransformierten der verwendeten Basisfunktionen ab. Es
ist also sicherlich sinnvoll, nach Basisfunktionen zu suchen, deren
Fouriertransformierte besonders g\"unstige Eigenschaften be\-sitzen.
Ausf\"uhrliche Beschreibungen der Fouriertransformationsmethode im
Zusammenhang mit $B$-Funktionen findet man in Artikeln von Grotendorst
und Steinborn [1988], Trivedi und Steinborn [1983], Weniger, Grotendorst
und Steinborn [1986b] und Weniger und Steinborn [1983a], sowie in den
Dissertationen von Grotendorst [1985] und Homeier [1990].

Die im Zusammenhang mit Mehrzentrenmolek\"ulintegralen {\it
fundamentale\/} mathematische Eigenschaft der $B$-Funktionen ist
deswegen ihre au{\ss}ergew\"ohnlich einfache Fouriertransformierte [Weni\-ger
und Steinborn 1983a, Gl. (3.7)]:
$$
\beginAligntags
" \bar{B}_{n, \ell}^m (\alpha, {\vec p}) " \; = \;
" (2 \pi)^{- 3/2} \, \int \e^{- i {\vec p} \cdot {\vec r}} \,
B_{n, \ell}^{m} (\alpha, {\vec r}) \, \d^3 {\vec r} \\
" " \; = \; " (2/\pi)^{1/2} \,
\frac {\alpha^{2 n + \ell - 1}} {[\alpha^2 + p^2]^{n + \ell + 1}} \,
{\cal Y}_{\ell}^m (- i {\vec p}) \, .
\\ \tag
\endAligntags
$$
Interessanterweise ist diese Beziehung nicht auf den dreidimensionalen
Raum $\R^3$ beschr\"ankt, sondern sie gilt im Fall {\it skalarer\/}
$B$-Funktionen ($\ell = 0$) ganz analog auch im $n$-dimensionalen Raum
$\R^n$ [Homeier 1990, Gl. (3.2-47); Nikol'ski\v{\i} 1975, S. 289]. Die
Fouriertransformierte einer skalaren $B$-Funktion spielt eine wichtige
Rolle in der Theorie der {\it Besselpotentiale\/} und den damit
verkn\"upften Funktionenr\"aumen [Blanchard und Br\"uning 1992, Appendix E.1;
Aronszajn und Smith 1961; Kufner, John und Fu\v{c}ik 1977, Abschnitt
8.6].

Die Fouriertransformierte einer $B$-Funktion ist wesentlich einfacher
als die Fouriertransformierte einer Slaterfunktion (vergleiche
beispielsweise Homeier [1990, S. 58] oder Kaijser und Smith [1977, S. 47
- 48]) oder als die Fouriertransformierten solcher exponentialartigen
Basisfunktionen, die wie die Wasserstoffeigenfunktionen auf den
verallgemeinerten Laguerrepolynomen basieren [Weniger 1985, S. 282 -
283]. Laut Niukkanen [1984, S. 953], der die Fouriertransformierten
exponentialartiger quantenchemischer Basisfunktionen auf sehr allgemeine
Weise untersuchte, besitzt die $B$-Funktion von allen exponentialartigen
Funktionen die einfachste Fouriertransformierte.

Au{\ss}erdem k\"onnen die Fouriertransformierten der Slaterfunktionen als auch
der anderen, in der Quantenchemie verwendeten exponentialartigen
Basisfunktionen ausnahmslos als Linearkombinationen von
Fouriertransformierten von $B$-Funktionen geschrieben werden. Da die
Fouriertransformation eine {\it lineare} Operation ist, folgt daraus,
da{\ss} alle in der Quantenchemie \"ublicherweise verwendeten
exponentialartigen Basisfunktionen wie etwa Slaterfunktionen oder
Wasserstoffeigenfunktionen durch einfache Linearkombinationen von
$B$-Funktionen dargestellt werden k\"onnen (siehe beispielsweise Filter
und Steinborn [1978a, S. 83], Homeier [1990, S. 56 - 60] und Weniger
[1985, S. 282 - 283]). Es ist also im Prinzip ausreichend,
ausschlie{\ss}lich Mehrzentrenmolek\"ulintegrale von $B$-Funktionen zu
untersuchen, da die Mehrzentrenmolek\"ulintegrale anderer
exponentialartiger Basisfunktionen durch einfache Linearkombinationen
der entsprechenden Integrale \"uber $B$-Funktionen dargestellt werden
k\"onnen.

Neben der au{\ss}ergew\"ohnlich einfachen Fouriertransformierten (8.2-11)
besitzen die $B$-Funktio\-nen noch einige andere mathematische
Eigenschaften, die bei der Berechnung von Mehrzentrenmolek\"ulintegralen
sehr hilfreich sind. Die zur Zeit aktuellste und vollst\"andigste
Behandlung der mathematischen Eigenschaften der $B$-Funktionen und ihre
Implikationen f\"ur die Berechnung von Mehrzentrenmolek\"ulintegralen findet
man in der Dissertation von Homeier [1990]. In diesem Abschnitt sollen
nur einige Differentiationseigenschaften der $B$-Funktionen beschrieben
werden, die f\"ur die Ableitung expliziter Darstellungen von {\it
\"Uberlappungs-\/} und {\it Coulombintegralen\/} von $B$-Funktionen
[Weniger und Steinborn 1983a; Weniger, Grotendorst und Steinborn 1986b]
und f\"ur die Konstruktion von {\it Additionstheoremen} [Weniger und
Steinborn 1985; 1989b] eine wesentliche Rolle gespielt haben.

Bestimmte Teilmengen der Polynome in den cartesischen Komponenten eines
Vektors $\vec{r} = (x, y, z)$ k\"onnen durch Transformationseigenschaften
oder Symmetrien charakterisiert werden. Die {\it homogenen} Polynome
$P_{\ell} (x, y, z)$ von Grade $\ell$ sind beispielsweise durch die
folgende Beziehung definiert:
$$
P_{\ell} (\eta x, \eta y, \eta z) \; = \;
\eta^{\ell} \, P_{\ell} (x, y, z) \, .
\tag
$$
Eine spezielle Teilmenge der homogenen Polynome vom Grade $\ell$ sind
die sogenannten {\it harmonischen} Polynome $H_{\ell} (x, y, z)$ vom
Grade $\ell$, die L\"osungen der homogenen dreidimensionalen
La\-place\-gleichung (8.2-1) sind:
$$
\vec{\nabla}^2 \, H_{\ell} (x, y, z) \; = \;
\left( \frac {\partial^2} {\partial x^2} \, + \,
\frac {\partial^2} {\partial y^2} \, + \,
\frac {\partial^2} {\partial z^2}
\right) \, H_{\ell} (x, y, z) \; = \; 0 \, .
\tag
$$
F\"ur jedes $\ell \in \N_0$ gibt es genau $2 \ell + 1$ linear unabh\"angige
harmonische Polynome $H_{\ell} (x, y, z)$ [Hobson 1965, S. 123; Normand
1980, Appendix H.3.]. Die in Gl. (8.2-2) definierten regul\"aren
r\"aumlichen Kugelfunktionen ${\cal Y}_{\ell}^{m} (\vec{r})$, die als
homogene Polynome $\ell$-ten Grades in $x$, $y$ und $z$ L\"osungen der
homogenen dreidimensionalen Laplacegleichung sind und die sich wie
sph\"arische Tensoren $\ell$-ter Stufe transformieren, k\"onnen also die
Menge der harmonischen Polynome vom Grade $\ell$ aufspannen:
$$
H_{\ell} (x, y, z) \; = \; \sum_{m = - \ell}^{\ell} \,
C_{\ell}^{m} \, {\cal Y}_{\ell}^{m} (\vec{r}) \, .
\tag
$$

Man kann dieses Klassifikationsschema ebenfalls verwenden, wenn man
keine Polynome in den cartesischen Komponenten eines Vektors $\vec{r} =
(x, y, z)$ betrachtet, sondern Differentialoperatoren, die Polynome in
den cartesischen Komponenten des Nablaoperators $\vec{\nabla} =
(\partial / \partial x, \partial / \partial y, \partial / \partial z)$
sind. Anstelle der Kugelfunktion ${\cal Y}_{\ell}^{m} (\vec{r})$ mu{\ss} man
dann den Differentialoperator ${\cal Y}_{\ell}^{m} (\vec{\nabla})$
betrachten, der sich wie ein sph\"arischer Tensor $\ell$-ter Stufe
transformiert [Biedenharn und Louck 1981, S. 312], und den man erh\"alt,
indem man in dem expliziten Ausdruck f\"ur ${\cal Y}_{\ell}^{m} (\vec{r})$
[Biedenharn und Louck 1981, S. 71] die cartesischen Komponenten von
$\vec{r}$ durch die cartesischen Komponenten von $\vec{\nabla}$ ersetzt.

Der Differentialoperators ${\cal Y}_{\ell}^{m} (\vec{\nabla})$ wurde im
Prinzip bereits von Hobson [1892] verwendet, um die in Gl. (8.2-3)
definierte irregul\"are r\"aumliche Kugelfunktion ${\cal Z}_{\ell}^{m}$
durch Differentiation des Coulombpotentials zu erhalten:
$$
{\cal Z}_{\ell}^{m} (\vec{r}) \; = \;
\frac {(- 1)^{\ell}} {(2 \ell - 1)!!} \,
{\cal Y}_{\ell}^{m} (\vec{\nabla}) \, \frac {1} {r} \, .
\tag
$$
Sp\"ater wurde der Differentialoperator ${\cal Y}_{\ell}^{m}
(\vec{\nabla})$ von Bayman [1978], Fieck [1980], Fortunelli und
Carrravetta [1992], Fortunelli und Salvetti [1993], Fujimura und
Matsuoka [1992], Grotendorst und Steinborn [1985], Homeier [1990],
Matsuoka [1992a; 1992b], Niukkanen [1983], Rashid [1986], Rowe [1978],
Santos [1973], Stuart [1981], und Weniger und Steinborn [1983a; 1983c;
1985; 1989a] verwendet.

Wenn man diesen Differentialoperator, der sich wie ein sph\"arischer
Tensor $\ell$-ter Stufe transformiert [Biedenharn und Louck 1981, S.
312], auf eine abstandsabh\"angige Funktion $\phi (r)$, die ein
sph\"arischer Tensor nullter Stufe ist, anwendet, erh\"alt man eine
Funktion, die in \"Ubereinstimmung mit den Kopplungsregeln
quantenmechanischer Drehimpulse ein sph\"arischer Tensor $\ell$-ter Stufe
ist:
$$
{\cal Y}_{\ell}^{m} (\vec{\nabla}) \, \phi (r) \; = \;
\left[ \left( \frac {1} {r} \frac {\d} {\d r} \right)^{\ell}
\phi (r) \right] \, {\cal Y}_{\ell}^{m} (\vec{r}) \, .
\tag
$$
Diese Beziehung kann leicht mit Hilfe eines Theorems von Hobson [1892,
S. 65; 1965, S. 127] bewiesen werden (siehe Weniger und Steinborn
[1983a, S. 6126]).

Wenn man den Differentialoperator ${\cal Y}_{\ell_1}^{m_1}
(\vec{\nabla})$ auf einen sph\"arischen Tensor
$$
F_{\ell_2}^{m_2} (\vec{r}) \; = \;
f_{\ell_2} (r) Y_{\ell_2}^{m_2} (\vec{r}/r)
\tag
$$
anwendet, erh\"alt man ein Ergebnis, das bis auf eine noch zu bestimmende
Funktion $\gamma_{\ell_1 \ell_2}^{\ell} (r)$ durch die Kopplungsregeln
quantenmechanischer Drehimpuls determiniert ist [Weniger und Steinborn
1983c, S. 2555]:
$$
{\cal Y}_{\ell_1}^{m_1} (\vec{\nabla}) \, F_{\ell_2}^{m_2} (\vec{r})
\; = \; \gntsm \, \gntcf{}{m_1 + m_2}{1}{m_1}{2}{m_2} \,
\gamma_{\ell_1 \ell_2}^{\ell} (r) \,
Y_{\ell}^{m_1 + m_2} (\vec{r}/r) \, .
\tag
$$
Dabei ist $\langle \ell \, m_1 + m_2 \vert \ell_1 \, m_1 \vert \ell_2 \,
m_2 \rangle$ ein Gaunt-Koeffizient, der allgemein als Integral von drei
Kugel\-fl\"achen\-funktio\-nen \"uber die Oberfl\"ache der dreidimensionalen
Einheitskugel definiert ist [Gaunt 1929, Appendix]:
$$
\gntcf{3}{m_3}{2}{m_2}{1}{m_1} \; = \;
\int \bigl[ Y_{\ell_3}^{m_3} (\Omega) \bigr]^{\ast} \,
Y_{\ell_2}^{m_2} (\Omega) \,
Y_{\ell_1}^{m_1} (\Omega) \, \d \Omega \, .
\tag
$$
Gaunt-Koeffizienten k\"onnen auf folgende Weise durch die Wignerschen
$3jm$-Symbole [Wigner 1959, Gl. (24.9)] dargestellt werden [Weissbluth
1978; Gl. (1.2-29)]:
$$
\beginAligntags
" \gntcf{3}{m_3}{2}{m_2}{1}{m_1} \\
" \quad \; = \; (-1)^{m_3}
\left[
  \frac{(2\ell_1+1)(2\ell_2+1)(2\ell_3+1)}{4\pi}
\right]^{1/2} \pmatrix{\ell_1 " \ell_2 " \ell_3 \\
                          0   " 0      " 0}
                 \pmatrix{\ell_1 " \ell_2 " \ell_3 \\
                          m_1    " m_2    " - m_3} \, . \qquad
\\ \tag
\endAligntags
$$
In allen Arbeiten des Autors wurden Gaunt-Koeffizienten mit Hilfe dieser
Beziehung berechnet. In der Literatur sind zahlreiche
Rekursionsbeziehungen f\"ur $3jm$-Symbole bekannt [Rotenberg, Bivins,
Metropolis und Wooten 1959, S. 9 - 11; Varshalovich, Moskalev und
Khersonskii 1988, S. 252 - 257]. Besonders geeignet f\"ur die Berechnung
der in Mehrzentrenintegralen vorkommenden Gaunt-Koeffizienten ist die
folgende Dreitermrekursion der $3jm$-Symbole [Schulten und Gordon 1975,
Gl. (6)]:
$$
\beginAligntags
" j_1 \, A (j_1 + 1) \pmatrix{j_1 + 1 " j_2 " j_3 \\ m_1 " m_2 " m_3}
\, + \,
B (j_1) \pmatrix{j_1 " j_2 " j_3 \\ m_1 " m_2 " m_3} \\
" \qquad \, + \,
(j_1 + 1) \, A (j_1) \pmatrix{j_1 - 1 " j_2 " j_3 \\ m_1 " m_2 " m_3}
\; = \; 0 \, ,
\erhoehe\aktTag \\ \tag*{\tagnr a}
" A (j_1) \; = \; \bigl\{ [j_1^2 - (j_2 - j_3)^2] \,
[(j_2 + j_3 + 1)^2 - j_1^2] \, [j_1^2 - m_1^2] \bigr\}^{1/2} \, ,
\\ \tag*{\tagform\aktTagnr b}
" B (j_1) \; = \; - (2 j_1 + 1) \, \bigl\{ [j_2 (j_2 + 1) -
j_3 (j_3 + 1)] m_1 - j_1 (j_1 + 1) (m_3 - m_2) \bigr\} \, .
\\ \tag*{\tagform\aktTagnr c}
\endAligntags
$$
Obwohl diese Rekursionsformel weder aufw\"arts- noch abw\"artsstabil ist,
konnten Schulten und Gordon einen eleganten Algorithmus entwickeln, mit
dessen Hilfe man $3jm$-Symbole selbst f\"ur sehr gro{\ss}e $j$-Werte schnell
und genau berechnen kann, ohne die Startwerte der Rekursion (8.2-21)
kennen zu m\"ussen [Schulten und Gordon 1975; 1976]. Wenn man diesen
rekursiven Algorithmus geeignet modifiziert, kann man auf diese Weise
alle Gaunt-Koeffizienten, die bei der Linearisierung des Produktes
zweier Kugelfl\"achenfunktionen auftreten,
$$
Y_{\ell_1}^{m_1} (\Omega) \, Y_{\ell_2}^{m_2} (\Omega) \; = \;
\gntsm \, \gntcf{}{m_1 + m_2}{1}{m_1}{}{m_2} \,
Y_{\ell}^{m_1 + m_2} (\Omega) \, ,
\tag
$$
selbst f\"ur extrem gro{\ss}e Drehimpulsquantenzahlen ($\ell_1, \ell_2 > 100$)
schnell und genau berechnen [Weniger und Steinborn 1982].

Das Symbol $\sum {}^{(2)}$ in Gl. (8.2-22) bedeutet, da{\ss} die Summe von
$\ell_{min}$ bis $\ell_{max}$ in Zweierschritten verl\"auft. Die
Summationsgrenzen in Gl. (8.2-22) sind direkte Konsequenzen der
Auswahlregeln, die von den Gaunt-Koeffizienten $\langle \ell \, m_1 +
m_2 \vert \ell_1 \, m_1 \vert \ell_2 \, m_2 \rangle$ erf\"ullt werden
[Weniger und Steinborn 1982, S. 151]:
$$
\beginAligntags
" \ell_{\max} " \; = \; " \ell_1 + \ell_2 \, ,
\erhoehe\aktTag \\ \tag*{\tagnr a}
" \ell_{\min} " \; = \; "
\cases{
  \max(\abs{\ell_1-\ell_2},\abs{m_1+m_2}) @
  wenn $\max(\abs{\ell_1-\ell_2},\abs{m_1+m_2})+\ell_{\max}$\\
  \quad @ gerade ist,
\\
\\
  \max(\abs{\ell_1-\ell_2},\abs{m_1+m_2}) +1 @
  wenn $\max(\abs{\ell_1-\ell_2},\abs{m_1+m_2})+\ell_{\max}$\\
  \quad @ ungerade ist.
}
\\ \tag*{\tagform\aktTagnr b}
\endAligntags
$$

F\"ur die Funktion $\gamma_{\ell_1 \ell_2}^{\ell} (r)$ in Gl. (8.2-18)
konnten zahlreiche verschiedene Darstellungen abgeleitet werden [Bayman
1978, Gl. (11); Niukkanen 1983, Gln. (3) - (6); Rashid 1986, Gln. (5)
und (6); Santos 1973, S. 360; Stuart 1981, S. 373; Weniger und Steinborn
1983c, S. 2555 - 2559]. Ein Beispiel [Weniger und Steinborn 1983c, Gl.
(3.29)]:
$$
\gamma_{\ell_1 \ell_2}^{\ell} (r) \; = \; \sum_{q = 0}^{\Delta \ell} \,
\frac {( - \Delta \ell)_q (- \sigma (\ell) - 1/2)_q} {q!} \, 2^q \,
r^{\ell_1 + \ell_2 - 2 q} \,
\left( \frac {1} {r} \frac {\d} {\d r} \right)^{\ell_1 - q} \,
\frac {f_{\ell_2} (r)} {r^{\ell_2}} \, .
\tag
$$
Aus den Summengrenzen (8.2-23) der Gauntkoeffizienten folgt sofort, da{\ss}
die in Gl. (8.2-24) vorkommenden Gr\"o{\ss}en
$$
\Delta \ell \; = \; (\ell_1 + \ell_2 - \ell)/2
\tag
$$
und
$$
\sigma (\ell) \; = \; (\ell_1 + \ell_2 + \ell)/2
\tag
$$
entweder positive ganze Zahlen oder Null sind.

Bei $B$-Funktionen liefert der Differentialoperator ${\cal Y}_{\ell}^{m}
(\vec{\nabla})$ besonders einfache Ergebnisse. So gilt im Fall skalarer
$B$-Funktionen [Weniger und Steinborn 1983a, Gl. (4.12)]:
$$
B_{n, \ell}^{m} (\alpha, \vec{r}) \; = \; (- \alpha)^{- \ell} \,
(4 \pi)^{1/2} \, {\cal Y}_{\ell}^{m} (\vec{\nabla}) \,
B_{n + \ell, 0}^{m} (\alpha, \vec{r}) \, .
\tag
$$
Die Anwendung des Differentialoperators ${\cal Y}_{\ell}^{m}
(\vec{\nabla})$ auf eine skalare Slaterfunktion w\"urde zu einem
wesentlich komplizierteren Ergebnis f\"uhren.

Auch im Fall nichtskalarer $B$-Funktionen ergibt die Anwendung dieses
Differentialoperators ein bemerkenswert einfaches Ergebnis [Weniger und
Steinborn 1983c, Gl. (6.25)]:
$$
\beginAligntags
" {\cal Y}_{\ell_1}^{m_1} (\vec{\nabla}) \,
B_{n_2, \ell_2}^{m_2} (\alpha, \vec{r}) " \; = \; "
(- \alpha)^{\ell_1} \, \gntsm \, \gntcf{}{m_1 + m_2}{1}{m_1}{2}{m_2} \\
" " " \times \,
\sum_{t = 0}^{\Delta \ell} \, (- 1)^t \, \binom {\Delta \ell} {t} \,
B_{n_2 + \ell_2 - \ell - t, \ell}^{m_1 + m_2} (\alpha, \vec{r}) \, .
\\ \tag
\endAligntags
$$

$B$-Funktionen sind eine relativ gro{\ss}e Funktionenklasse, die Funktionen
mit v\"ollig verschiedenen Eigenschaften enth\"alt. So folgt aus Gln.
(8.2-5) - (8.2-7), da{\ss} eine $B$-Funktion nur dann als LCOA-Basisfunktion
geeignet ist, wenn $n \ge 1$ gilt. Wenn $- \ell \le n \le 1$ gilt,
verh\"alt sich eine $B$-Funktion asymptotisch f\"ur kleine Abst\"ande $r$ vom
Ursprung nicht mehr wie eine regul\"are r\"aumliche Kugelfunktion und kann
sogar singul\"ar sein. F\"ur $n = - \ell$ gilt beispielsweise [Weniger und
Steinborn 1989b, Gl. (3.9)]:
$$
B_{- \ell, \ell}^{m} (\alpha, \vec{r}) \; = \;
\widehat{k}_{\ell + 1/2} (\alpha r) \,
{\cal Z}_{\ell}^{m} (\alpha \vec{r}) \, .
\tag
$$
Aufgrund der Fakult\"at $(n + \ell)!$, die auf der rechten Seite von Gl.
(8.2-5) vorkommt, sind $B$-Funktionen nur dann im Sinne der klassischen
Analysis definiert, wenn $n + \ell \ge 0$ gilt. Es konnte aber gezeigt
werden, da{\ss} die Definition einer $B$-Funktionen gem\"a{\ss} Gl. (8.2-5)
sinnvoll bleibt, wenn die Ordnung $n$ eine negative ganze Zahl mit $n +
\ell < 0$ ist [Weniger und Steinborn 1983c]. Allerdings ben\"otigt man
dazu die Theorie der verallgemeinerten Funktionen oder Distributionen im
Sinne von Laurent Schwartz [1978].

Die dreidimensionale Diracsche Deltafunktion kann bekanntlich \"uber die
Poissonsche Gleichung
$$
\vec{\nabla}^2 \, \frac {1} {r} \; = \; - 4 \pi \, \delta (\vec{r})
\tag
$$
einer Einheitspunktladung [Jackson 1975, Gl. (1.31)] eingef\"uhrt werden.
Ganz analog kann eine {\it sph\"arische Deltafunktion\/} durch Anwendung
des Laplaceoperators auf eine irregul\"are r\"aumliche Kugelfunktion
definiert werden [Rowe 1978, Gl. (29)]:
$$
\vec{\nabla}^2 \, {\cal Z}_{\ell}^{m} (\vec{r}) \; = \;
- 4 \pi \, \delta_{\ell}^{m} (\vec{r}) \, .
\tag
$$
Eine alternative Definition der sph\"arischen Deltafunktion lautet [Rowe
1978, Gl. (30); Weniger und Steinborn 1983c, Gl. (6.1)]:
$$
\delta_{\ell}^{m} (\vec{r}) \; = \;
\frac {(- 1)^{\ell}} {(2 \ell - 1)!!} \,
{\cal Y}_{\ell}^{m} (\vec{\nabla}) \, \delta (\vec{r}) \, .
\tag
$$
Mit Hilfe von Gl. (8.2-27) und der Differentiationsbeziehung [Weniger
und Steinborn 1983c, Gl. (5.6)]
$$
\bigl[ 1 - \alpha^{- 2} \vec{\nabla}^2 \bigr] \,
B_{n, \ell}^{m} (\alpha, \vec{r}) \; = \;
B_{n - 1, \ell}^{m} (\alpha, \vec{r})
\tag
$$
konnte gezeigt werden, da{\ss} $B$-Funktionen des Typs $B_{- \ell - \nu,
\ell}^{m} (\alpha, \vec{r})$ mit $\nu \ge 1$ Ableitungen der sph\"arischen
Deltafunktion sind [Weniger und Steinborn 1983c, Gl. (6.20)]:
$$
B_{- \ell - \nu, \ell}^{m} (\alpha, \vec{r}) \; = \; (2 \ell - 1)!! \,
\bigl(4 \pi / \alpha^{\ell + 3} \bigr) \,
\bigl[ 1 - \alpha^{- 2} \vec{\nabla}^2 \bigr]^{\nu - 1} \,
\delta_{\ell}^{m} (\vec{r}) \, , \qquad \nu \in \N \, .
\tag
$$

Es ist naheliegend, aber falsch zu glauben, da{\ss} distributive
$B$-Funktionen im Zusammenhang mit Mehrzentrenmolek\"ulintegralen, die im
Rahmen eines LCAO-Ansatzes auftreten, keine Rolle spielen.
Selbstverst\"andlich k\"onnen distributive $B$-Funktionen keinen Beitrag zum
{\it numerischen Wert\/} eines Mehrzentrenintegrals leisten. Bei {\it
analytischen Operationen\/} d\"urfen distributive Anteile aber keinesfalls
vernachl\"assigt werden. Betrachten wir als Beispiel das
\"Uberlappungsintegral zweier $B$-Funktionen mit verschiedenen
Exponentialparametern $\alpha$ und $\beta$,
$$
S_{n_1 \ell_1 m_1}^{n_2 \ell_2 m_2}(\alpha, \beta, \vec{R}) \; = \;
\int \, \bigl\[ B_{n_1, \ell_1}^{m_1}(\alpha, \vec{r}) \bigr\]^{*} \,
B_{n_2, \ell_2}^{m_2}(\beta, \vec{r} - \vec{R}) \, \d^3 \vec{r} \, ,
\tag
$$
das schon in zahlreichen Artikel behandelt wurde [Antolovi\'{c} und
Delhalle 1980; Bhattacharya und Dhabal 1986; Filter und Steinborn 1978a;
1978b; Grotendorst 1985; Grotendorst, Weniger und Steinborn 1986;
Homeier 1990; Homeier und Steinborn 1992; Homeier, Weniger und Steinborn
1992b; Steinborn und Weniger 1992; Weniger, Grotendorst und Steinborn
1986b; Weniger und Steinborn 1983a; 1983b; 1984; 1987; 1988; Yamaguchi
1983].

Filter und Steinborn [1978b, Gln. (4.6) und (4.7)] leiteten f\"ur das
\"Uberlappungsintegral (8.2-35) die folgende Darstellung ab, die nur
endlich viele Terme enth\"alt:
$$
\beginAligntags
S_{n_1 \ell_1 m_1}^{n_2 \ell_2 m_2} " (\alpha, \beta, \vec{R}) \; = \;
(-1)^{\ell_2} \, 4 \pi \,
\gntsm \, \gntcf{2}{m_2}{1}{m_1}{}{m_2 - m_1} \\
\times \; \biggl\[
" \frac {(- 1)^{n_1 + \ell_1} (\alpha / \beta)^{\ell_2}}
{\beta^3 [1 - (\alpha / \beta)^2]^{n_2 + \ell_2 + 1}} \\
\times \, " \sum_{s=0}^{n_1 + \ell_1} \, (- 1)^s \,
P_{n_1 + \ell_1 - s}^{(s - n_1 - \Delta \ell_2, n_2 + \Delta \ell_1)} \,
\Bigl( \frac {\beta^2 + \alpha^2} {\beta^2 - \alpha^2} \Bigr) \,
B_{s - \ell, \ell}^{m_2 - m_1} (\alpha, \vec{R} ) \\
+ \, " \frac {(- 1)^{n_2 + \ell_2} (\beta / \alpha)^{\ell_1}}
{\alpha^3 [1 - (\beta / \alpha)^2]^{n_1 + \ell_1 + 1}} \\
\times \, " \sum_{s=0}^{n_2 + \ell_2} \, (- 1)^s \,
P_{n_2 + \ell_2 - s}^{(s - n_2 - \Delta \ell_1, n_1 + \Delta \ell_2)} \,
\Bigl( \frac {\alpha^2 + \beta^2} {\alpha^2 - \beta^2} \Bigr) \,
B_{s - \ell, \ell}^{m_2 - m_1} (\beta, \vec{R} )
\biggr\] \, .
\\ \tag
\endAligntags
$$
$P_n^{(\alpha, \beta)} (x)$ ist ein Jacobipolynom [Magnus, Oberhettinger
und Soni 1966, Abschnitt 5.2]. F\"ur die speziellen Jacobipolynome in Gl.
(8.2-36) konnte eine lineare Dreitermrekursion abgeleitet werden, die
eine sehr bequeme und effiziente Berechnung der Polynome erm\"oglicht
[Weniger und Steinborn 1983b, Gl. (6.6)].

Die Summengrenzen $\ell_{\min}$ und $\ell_{\max}$ in Gl. (8.2-36) sind
direkte Konsequenzen der Auswahlregeln, die von den Gaunt-Koeffizienten
$\langle \ell_2 \, m_2 \vert \ell_1 \, m_1 \vert \ell \, m_2 - m_1
\rangle$ erf\"ullt werden [Weniger und Steinborn 1982, S. 151]:
\neueSeite
$$
\beginAligntags
" \ell_{\max} " \; = \; " \ell_1 + \ell_2 \, ,
\erhoehe\aktTag \\ \tag*{\tagnr a}
" \ell_{\min} " \; = \; "
\cases{
  \max(\abs{\ell_1-\ell_2},\abs{m_2-m_1}) @
  wenn $\max(\abs{\ell_1-\ell_2},\abs{m_2-m_1})+\ell_{\max}$\\
  \quad @ gerade ist,
\\
\\
  \max(\abs{\ell_1-\ell_2},\abs{m_2-m_1}) +1 @
  wenn $\max(\abs{\ell_1-\ell_2},\abs{m_2-m_1})+\ell_{\max}$\\
  \quad @ ungerade ist.
}
\\ \tag*{\tagform\aktTagnr b}
\endAligntags
$$
Aus den in Gl. (8.2-37) definierten Summengrenzen $\ell_{\min}$ und
$\ell_{\max}$ folgt sofort, da{\ss} die in Gl. (8.2-36) vorkommenden Gr\"o{\ss}en
$$
\Delta \ell_1 \; = \; (\ell - \ell_1 + \ell_2)/2
\tag
$$
und
$$
\Delta \ell_2 \; = \; (\ell + \ell_1 - \ell_2)/2
\tag
$$
entweder positive ganze Zahlen oder Null sind.

Die von Filter und Steinborn [1978b, Gln. (4.6) und (4.7)] abgeleitete
Jacobipolynomdarstellung (8.2-36) kann aufgrund numerischer
Instabilit\"aten nicht verwendet werden, wenn die Exponentialparameter
$\alpha$ und $\beta$ sich nur wenig unterscheiden, da f\"ur $\alpha \to
\beta$ sich gegenseitig kompensierende singul\"are Terme auftreten.
Au{\ss}erdem kann Gl. (8.2-26) nicht f\"ur kleine Abst\"ande $R$ verwendet
werden, da $B$-Funktionen mit negativen Ordnungen vorkommen, die f\"ur $R
\to 0$ singul\"ar werden. Wenn aber $R$ nicht klein ist und wenn die
beiden Exponentialparameter $\alpha$ und $\beta$ sich ausreichend
unterscheiden, kann man mit Hilfe von Gl. (8.2-36) \"Uberlappungsintegrale
von $B$-Funktionen sehr bequem und \"au{\ss}erst schnell berechnen [Weniger
und Steinborn 1983b; Homeier 1990; Homeier und Steinborn 1992; Homeier,
Weniger und Steinborn 1992b].

Trotzdem ist die Jacobipolynomdarstellung (8.2-36) nicht v\"ollig korrekt.
Bei der Ableitung wurden n\"amlich alle distributiven $B$-Funktionen, die
gem\"a{\ss} Gl. (8.2-34) Ableitungen der sph\"arischen Deltafunktion sind,
vernachl\"assigt [Weniger, Grotendorst und Steinborn 1986b, S. 3696]. Das
bedeutet, da{\ss} die Jacobipolynomdarstellung (8.2-36) f\"ur $R > 0$
abgesehen von m\"oglichen Rundungsfehlern korrekte numerische Ergebnisse
liefert. Wenn man aber die Jacobipolynomdarstellung (8.2-36) in
Integralen verwenden will, wird sie normalerweise falsche Ergebnisse
liefern.

Distributive Anteile von Mehrzentrenintegralen spielen eine
entscheidende Rolle, wenn man beispielsweise eine auf O-Ohata und
Ruedenberg [1966] und auf Silver und Ruedenberg [1968] zur\"uckgehende
Methode zur Berechnung von Kernanziehungs- und
Elektronenwechselwirkungsintegralen verwendet, bei der die
Mehrzentrenintegrale als L\"osungen inhomogener dreidimensionaler
Laplacegleichungen berechnet werden.

Diese Vorgehensweise kann anhand des zweizentrigen
Kernanziehungsintegrals
$$
{\cal A} (f; {\vec r}) \; = \;
\int \frac {1}
{\vert \vec{r} - \vec{\rho} \vert} \, f ({\vec \rho}) \,
\d^3 {\vec \rho}
\tag
$$
verdeutlicht werden. Die Anwendung des Laplaceoperators auf diese
Beziehung ergibt mit Hilfe von Gl. (8.2-30) die folgende inhomogene
Laplacegleichung,
$$
{\vec \nabla}^2 {\cal A} (f; {\vec r}) \; = \;
- \, 4 \pi \, f ({\vec r}) \, ,
\tag
$$
mit deren Hilfe man das Kernanziehungsintegral (8.2-40) berechnen kann.

Diese Technik funktioniert nicht nur bei Kernanziehungsintegralen,
sondern auch bei Mehrzentrenintegralen des Typs
$$
{\cal C} (f, g; {\vec r}) \; = \;
\int \! \! \int \bigl[ f ({\vec \rho}) \bigr]^{\ast} \,
\frac {1} {\vert {\vec r} - {\vec \rho} + {\vec \xi} \vert} \,
g ({\vec \xi}) \, \d^3 {\vec \rho} \, \d^3 {\vec \xi} \, ,
\tag
$$
welche die Wechselwirkung zweier Ladungsdichten $f$ und $g$ beschreiben,
die an verschiedenen, durch einen Vektor ${\vec r}$ getrennten Zentren
lokalisiert sind. Mit Hilfe von Gl. (8.2-30) erhalten wir dann:
$$
\beginAligntags
" {\vec \nabla}^2 {\cal C} (f, g; {\vec r}) " \; = \; "
- \, 4 \pi \, \int \! \! \int \bigl[ f ({\vec \rho}) \bigr]^{\ast} \,
\delta ({\vec r} - {\vec \rho} + {\vec \xi}) \,
g ({\vec \xi}) \, \d^3 {\vec \rho} \, \d^3 {\vec \xi} \\ \tag
" " \; = \; " - \, 4 \pi \,
\int \, \bigl[ f ({\vec \rho}) \bigr]^{\ast} \,
g ({\vec \rho} - {\vec r}) \, \d^3 {\vec \rho} \, .
\\ \tag
\endAligntags
$$
Wenn wir f\"ur das \"Uberlappungsintegral in Gl. (8.2-44) die Schreibweise
$$
{\cal S} (f, g; {\vec r}) \; = \;
\int \, \bigl[ f ({\vec \rho}) \bigr]^{\ast}
\, g ({\vec \rho}-{\vec r}) \,
\d^3 {\vec \rho}
\tag
$$
einf\"uhren, sehen wir, da{\ss} das Elektronenwechselwirkungsintegral ${\cal
C} (f, g; \vec{r})$ die L\"osung einer Laplacegleichung ist, deren
Inhomogenit\"at ein \"Uberlappungsintegral ist:
$$
{\vec \nabla}^2 {\cal C} (f, g; {\vec r}) \; = \;
- \, 4 \pi \, {\cal S} (f, g; {\vec r}) \, .
\tag
$$

Man kann mit Hilfe der Laplacegleichung (8.2-46) beispielsweise
explizite Darstellungen f\"ur Coulombintegrale des Typs
$$
C_{n_1 \ell_1 m_1}^{n_2 \ell_2 m_2}(\alpha, \beta, \vec{R})
\; = \; \int \! \! \int \,
\bigl\[ B_{n_1, \ell_1}^{m_1}(\alpha, \vec{r}) \bigr\]^{*} \,
\frac{1}{\abs{\vec{R} - \vec{r} + \vec{\rho}}} \,
B_{n_2, \ell_2}^{m_2}(\beta, \vec{\rho}) \,
\d^3 \vec{r} \, \d^3 \vec{\rho}
\tag
$$
berechnen [Steinborn und Weniger 1992, Abschnitt 4]. In diesem Fall darf
man aber keinesfalls die Jacobipolynomdarstellung (8.2-36) als
Inhomogenit\"at in der Laplacegleichung (8.2-46) verwenden, da man dann
aufgrund der unzul\"assigen Vernachl\"assigung der distributiven Anteile in
Gl. (8.2-36) falsche Ergebnisse erhalten w\"urde [Steinborn und Weniger
1992, Abschnitt 5].

Die korrekte Form der Jacobipolynomdarstellung des \"Uberlappungsintegrals
zweier $B$-Funktio\-nen mit verschiedenen Exponentialparametern $\alpha$
und $\beta$ lautet [Steinborn und Weniger 1992, Gl. (5.8)]:
$$
\beginAligntags
S_{n_1 \ell_1 m_1}^{n_2 \ell_2 m_2} " (\alpha, \beta, \vec{R})
\; = \; (-1)^{\ell_2} \, 4 \pi \,
\gntsm \, \gntcf{2}{m_2}{1}{m_1}{}{m_2 - m_1} \\
\times \; \Biggl\{
" \frac {(- 1)^{n_1 + \ell_1} (\alpha / \beta)^{\ell_2}}
{\beta^3 [1 - (\alpha / \beta)^2]^{n_2 + \ell_2 + 1}} \\
\times \, " \Biggl\[ \sum_{s=0}^{n_1 + \ell_1} \, (- 1)^s \,
P_{n_1 + \ell_1 - s}^{(s - n_1 - \Delta \ell_2, n_2 + \Delta \ell_1)} \,
\Bigl( \frac {\beta^2 + \alpha^2} {\beta^2 - \alpha^2} \Bigr) \,
B_{s - \ell, \ell}^{m_2 - m_1} (\alpha, \vec{R} ) \\
+ \, " \frac
{(\alpha/\beta)^{2 (n_1 + \ell_1)}}
{[1 - (\alpha/\beta)^2]^{n_1 + \ell_1}} \,
\frac
{(n_1 + n_2 + \ell_1 + \ell_2)!}
{(n_1 + \ell_1)! \, (n_2 + \ell_2)!}
\, \sum_{q=0}^{\Delta \ell - 1} \, (- 1)^{q+1}
\binom {\Delta \ell} {q + 1} \,
B_{- q - \ell - 1, \ell}^{m_2 - m_1} (\alpha, \vec{R} ) \\
\times \,
" {}_3 F_2 \left( q - \Delta \ell + 1, - n_1 - \ell_1, 1; q + 2,
- n_1 - n_2 - \ell_1 - \ell_2;
\frac {\alpha^2 - \beta^2}{\alpha^2} \right)
\Biggr\] \\
+ \, " \frac {(- 1)^{n_2 + \ell_2} (\beta / \alpha)^{\ell_1}}
{\alpha^3 [1 - (\beta / \alpha)^2]^{n_1 + \ell_1 + 1}} \\
\times \, " \Biggl\[ \sum_{s=0}^{n_2 + \ell_2} \, (- 1)^s \,
P_{n_2 + \ell_2 - s}^{(s - n_2 - \Delta \ell_1, n_1 + \Delta \ell_2)} \,
\Bigl( \frac {\alpha^2 + \beta^2} {\alpha^2 - \beta^2} \Bigr) \,
B_{s - \ell, \ell}^{m_2 - m_1} (\beta, \vec{R} ) \\
+ \, " \frac
{(\beta/\alpha)^{2 (n_2 + \ell_2)}}
{[1 - (\beta/\alpha)^2]^{n_2 + \ell_2}} \,
\frac
{(n_1 + n_2 + \ell_1 + \ell_2)!}
{(n_1 + \ell_1)! \, (n_2 + \ell_2)!}
\, \sum_{q=0}^{\Delta \ell - 1} \, (- 1)^{q+1}
\binom {\Delta \ell} {q + 1} \,
B_{- q - \ell - 1, \ell}^{m_2 - m_1} (\beta, \vec{R} ) \\
\times \,
" {}_3 F_2 \left( q - \Delta \ell + 1, - n_2 - \ell_2, 1; q + 2,
- n_1 - n_2 - \ell_1 - \ell_2;
\frac {\beta^2 - \alpha^2}{\beta^2} \right)
\Biggr\] \Biggr\} \, .
\\ \tag
\endAligntags
$$
$\Delta \ell$ ist in Gl. (8.2-25) definiert. Die distributiven
$B$-Funktionen in Gl. (8.2-48) verschwinden f\"ur $\Delta \ell = 0$, da
die Summen von $q = 0$ bis $q = \Delta \ell - 1$ dann leere Summen sind,
die definitionsgem\"a{\ss} gleich Null sind. Da $\ell_1 = \ell_2 = 0$ immer
$\Delta \ell = 0$ impliziert, kann die Jacobipolynomdarstellung des
\"Uberlappungsintegrals skalarer $B$-Funktionen mit Ordnungen $n_1, n_2
\ge 1$ keine distributiven Anteile enthalten. Bei nicht\-skalaren
$B$-Funktionen ist aber normalerweise mit distributiven Anteilen zu
rechnen.

Wenn man die korrekte Jacobipolynomdarstellung (8.2-48) in der
inhomogenen Laplacegleichung (8.2-46) verwendet, erh\"alt man die folgende
Darstellung f\"ur das Coulombintegral (8.2-47) [Steinborn und Weniger
1992, Gl. (5.9)]:
$$
\beginAligntags
C_{n_1 \ell_1 m_1}^{n_2 \ell_2 m_2} " (\alpha, \beta, \vec{R})
\; = \; (-1)^{\ell_2} \, 4 \pi \,
\gntsm \, \gntcf{2}{m_2}{1}{m_1}{}{m_2 - m_1} \\
\times \; \Biggl\{
" \frac {(- 1)^{n_1 + \ell_1} (\alpha / \beta)^{\ell_2}}
{\beta^3 [1 - (\alpha / \beta)^2]^{n_2 + \ell_2 + 1}} \\
\times \, " \Biggl\[ \sum_{s=0}^{n_1 + \ell_1} \, (- 1)^s \,
P_{n_1 + \ell_1 - s}^{(s - n_1 - \Delta \ell_2, n_2 + \Delta \ell_1)} \,
\Bigl( \frac {\beta^2 + \alpha^2} {\beta^2 - \alpha^2} \Bigr) \,
A_{s - \ell, \ell}^{m_2 - m_1} (\alpha, \vec{R} ) \\
+ \, " \frac
{(\alpha/\beta)^{2 (n_1 + \ell_1)}}
{[1 - (\alpha/\beta)^2]^{n_1 + \ell_1}} \,
\frac
{(n_1 + n_2 + \ell_1 + \ell_2)!}
{(n_1 + \ell_1)! \, (n_2 + \ell_2)!}
\, \sum_{q=0}^{\Delta \ell - 1} \, (- 1)^{q+1}
\binom {\Delta \ell} {q + 1} \,
A_{- q - \ell - 1, \ell}^{m_2 - m_1} (\alpha, \vec{R} ) \\
\times \,
" {}_3 F_2 \left( q - \Delta \ell + 1, - n_1 - \ell_1, 1; q + 2,
- n_1 - n_2 - \ell_1 - \ell_2;
\frac {\alpha^2 - \beta^2}{\alpha^2} \right)
\Biggr\] \\
+ \, " \frac {(- 1)^{n_2 + \ell_2} (\beta / \alpha)^{\ell_1}}
{\alpha^3 [1 - (\beta / \alpha)^2]^{n_1 + \ell_1 + 1}} \\
\times \, " \Biggl\[ \sum_{s=0}^{n_2 + \ell_2} \, (- 1)^s \,
P_{n_2 + \ell_2 - s}^{(s - n_2 - \Delta \ell_1, n_1 + \Delta \ell_2)} \,
\Bigl( \frac {\alpha^2 + \beta^2} {\alpha^2 - \beta^2} \Bigr) \,
A_{s - \ell, \ell}^{m_2 - m_1} (\beta, \vec{R} ) \\
+ \, " \frac
{(\beta/\alpha)^{2 (n_2 + \ell_2)}}
{[1 - (\beta/\alpha)^2]^{n_2 + \ell_2}} \,
\frac
{(n_1 + n_2 + \ell_1 + \ell_2)!}
{(n_1 + \ell_1)! \, (n_2 + \ell_2)!}
\, \sum_{q=0}^{\Delta \ell - 1} \, (- 1)^{q+1}
\binom {\Delta \ell} {q + 1} \,
A_{- q - \ell - 1, \ell}^{m_2 - m_1} (\beta, \vec{R} ) \\
\times \,
" {}_3 F_2 \left( q - \Delta \ell + 1, - n_2 - \ell_2, 1; q + 2,
- n_1 - n_2 - \ell_1 - \ell_2;
\frac {\beta^2 - \alpha^2}{\beta^2} \right)
\Biggr\] \Biggr\} \, .
\\ \tag
\endAligntags
$$
Die in Gl. (8.2-49) vorkommenden Kernanziehungsintegrale $A_{n,
\ell}^{m}$ sind folgenderma{\ss}en definiert [Steinborn und Weniger 1992,
Gl. (3.1)]:
$$
A_{n, \ell}^m (\alpha, {\vec R}) \; = \;
\int \frac{1}{\abs{{\vec R}-{\vec r}}} \,
B_{n,\ell}^m (\alpha,{\vec r}) \, \d^3 {\vec r} \, .
\tag
$$
Diese Kernanziehungsintegrale k\"onnen auf folgende Weise berechnet werden
[Filter und Steinborn 1978b, Gln. (6.4) und (6.5)]:
$$
\beginAligntags
" A_{n, \ell}^m (\alpha, {\vec R}) " \; = \; "
\frac {4 \pi} {\alpha^2} \,
\Bigl\{ (2 \ell - 1)!! \, {\cal Z}_{\ell}^m (\alpha {\vec R}) \, - \,
\sum_{\nu=0}^{n + \ell} \,
B_{\nu - \ell, \ell}^m (\alpha, {\vec R}) \Bigr\}
\\ \tag
" " \; = \; "
\frac {4 \pi} {\alpha^2} \,
\sum_{\nu=0}^{\infty} B_{n + \nu + 1, \ell}^m (\alpha, {\vec R}) \, .
\\ \tag
\endAligntags
$$
Wenn die $B$-Funktion in Gl. (8.2-50) eine Ableitung der sph\"arischen
Deltafunktion gem\"a{\ss} Gl. (8.2-34) ist, d.~h., wenn $n + \ell < 0$ gilt,
enth\"alt auch das Kernanziehungsintegral distributive Anteile [Steinborn
und Weniger 1992, Gl. (5.13)]:
$$
A_{- q - \ell - 1, \ell}^m (\alpha, {\vec R}) \; = \;
\frac {4 \pi} {\alpha^2} \,
\Bigl\{ (2 \ell -1)!! {\cal Z}_{\ell}^{m} (\alpha {\vec r}) \, + \,
\sum_{\sigma=0}^{q - 1} \,
B_{\sigma - q - \ell, \ell}^{m} (\alpha, {\vec R}) \Bigl\} \, .
\tag
$$
Die $B$-Funktionen mit Ordnungen $\sigma - q - \ell$ sind Ableitungen
der sph\"arischen Deltafunktion gem\"a{\ss} Gl. (8.2-34) und tragen f\"ur $R > 0$
nicht zum numerischen Wert des Kernanziehungsintegrals bei. F\"ur $q = 0$
verschwinden alle distributiven Anteile auf der rechten Seite von Gl.
(8.2-53), da die Summe dann eine leere Summe ist, und das
Kernanziehungsintegral $A_{- \ell - 1, \ell}^m$ ist bis auf einen
Vorfaktor identisch mit einer irregul\"aren Kugelfunktion.

Die bemerkenswert einfache Fouriertransformierte (8.2-11) der
$B$-Funktionen ist nicht nur daf\"ur verantwortlich, da{\ss} man f\"ur
Mehrzentrenintegrale von $B$-Funktionen in den meisten F\"allen kompaktere
Darstellungen finden kann als f\"ur die analogen Integrale anderer
exponentialartiger Funktionen, sondern sie erkl\"art auch [Ruedenberg
1967; Silverstone 1967], warum es bei reduzierten Besselfunktionen und
$B$-Funktionen vergleichsweise einfach ist, sowohl punktweise
konvergente Additionstheoreme [Steinborn und Filter 1975; Weniger und
Steinborn 1985; 1989b] als auch normkonvergente Aditionstheoreme [Filter
und Steinborn 1980; Homeier, Weniger und Steinborn 1992a; Weniger 1985]
abzuleiten. Ein Beispiel f\"ur ein punktweise konvergentes
Additionstheorem folgt [Weniger und Steinborn 1989b, Gl. (4.18)]:
$$
\beginAligntags
" B_{n,\ell}^m \bigl( \alpha, \vec{r}_{<} + \vec{r}_{>} \bigr)
\; = \; \frac {4 \pi} {(- 2)^{n + \ell} (n + \ell)!} \,
\sum_{\ell_1 = 0}^{\infty} \;
\sum_{m_1=-\ell_1}^{\ell_1} \, (-1)^{\ell_1} \,
\bigl[ Y_{\ell_1}^{m_1} (\vec{r}_{<} / r_{<}) \bigr]^{*} \\
" \quad \times
{{}\,\Stapel{{}_{\ell_2^{max}} \\ \sum{}^{(2)} \\
{}^{\ell_2 = \ell_2^{min}}} \, {}}
\, \gntcf{2}{m + m_1}{}{m}{1}{m_1} \,
Y_{\ell_2}^{m+m_1} (\vec r_{>} / r_{>}) \\
" \qquad \times
\sum_{q=0}^{n+\ell} \, (-1)^q \, \binom {n + \ell} {q} \,
(\alpha r_{<})^{n + \ell - q - 1/2} \,
I_{\ell_1 - n - \ell + q + 1/2} (\alpha r_{<}) \\
" \qquad \quad \times
\sum^{q}_{s=0} \, 2^s \, \binom {q} {s} \, (\Delta \ell + 1/2)_{s} \,
(\alpha r_{>})^{q - s - 1/2} \, K_{\ell_2-q+s+1/2} (\alpha r_>) \, .
\\ \tag
\endAligntags
$$
Dabei ist $\vec{r}_{<}$ der betragsm\"a{\ss}ig kleinere und $\vec{r}_{>}$ der
betragsm\"a{\ss}ig gr\"o{\ss}ere Vektor, und $\Delta \ell$ ist in Gl. (8.2-25)
definiert.

\medskip

\Abschnitt Linear konvergente Reihenentwicklungen f\"ur
Zweizentrenintegrale von $B$-Funktionen

\smallskip

\aktTag = 0

Wie schon erw\"ahnt, wird die Jacobipolynomdarstellung (8.2-36) f\"ur das
\"Uberlappungsintegral zweier $B$-Funktionen mit verschiedenen
Exponentialparametern $\alpha$ und $\beta$ numerisch instabil, wenn
$\alpha$ und $\beta$ sich nur wenig unterscheiden oder wenn der Abstand
$R$ klein ist. Mit Hilfe der Jacobi\-polynomdarstellung (8.2-36) allein
kann man das \"Uberlappungsintegral (8.2-35) also nicht f\"ur alle praktisch
relevanten Bereiche der Exponentialparameter und Verschiebungsvektoren
verl\"a{\ss}lich berechnen. Man ben\"otigt noch alternative Darstellungen.

Das \"Uberlappungsintegral zweier $B$-Funktionen mit gleichen
Exponentialparametern hat eine bemerkenswert kompakte Struktur, da es
eine einfache Linearkombination von Gauntkoeffizienten und
$B$-Funktionen ist [Filter und Steinborn 1978b, Gl. (4.3)]:
$$
\beginAligntags
" S_{n_1 \ell_1 m_1}^{n_2 \ell_2 m_2} (\alpha, \alpha, \vec{R})
" \; = \; " (-1)^{\ell_2} \frac {4 \pi} {\alpha^3} \,
\gntsm \, \gntcf{2}{m_2}{1}{m_1}{}{m_2 - m_1} \\
" " " \times \, \sum_{t=0}^{\Delta \ell} \,
(-1)^t \binom{\Delta \ell}{t} \,
B_{n_1 + n_2 + 2 \Delta \ell - t + 1, \ell}^{m_2 - m_1}(\alpha, \vec{R})
\, .
\\ \tag
\endAligntags
$$
Die Einfachheit dieses \"Uberlappungsintegrals ist eine direkte Konsequenz
der Einfachheit der Fouriertransformierten (8.2-11) einer $B$-Funktion
[Weniger und Steinborn 1983a, Abschnitt 5].

Es ist also relativ naheliegend, das \"Uberlappungsintegral zweier
$B$-Funktionen mit {\it verschiedenen\/} Exponentialparametern durch
eine unendliche Reihe von \"Uberlappungsintegralen mit {\it gleichen\/}
Exponentialparametern darzustellen. Dies kann leicht mit Hilfe des {\it
Multiplikationstheorems\/} der $B$-Funktionen geschehen [Filter und
Steinborn 1978b, Gl. (4.6)]:
$$
B_{n, \ell}^{m} (\gamma, {\vec r}) \; = \;
(\gamma/\delta)^{2 n + \ell - 1} \, \sum_{\nu=0}^{\infty} \,
\frac {(n + \ell + 1)_{\nu}} {{\nu}!} \,
\left\{\frac {\delta^2 - \gamma^2} {\delta^2} \right\}^{\nu} \,
B_{n + \nu, \ell}^{m} (\delta, {\vec r}) \, .
\tag
$$
Das Multiplikationstheorem konvergiert, wenn $\abs{1 -
(\gamma/\delta)^2} < 1$ gilt. Wenn man dieses Multiplikationstheorem in
Gl. (8.2-35) verwendet, erh\"alt man die beiden folgenden unendlichen
Reihen f\"ur das \"Uberlappungsintegral zweier $B$-Funktionen mit
verschiedenen Exponentialparametern $\alpha$ und $\beta$ [Filter und
Steinborn 1978b, Gl. (4.9)]:
$$
\beginAligntags
" S_{n_1 \ell_1 m_1}^{n_2 \ell_2 m_2} (\alpha, \beta, \vec{R}) \\
" \qquad \; = \; (\alpha/\beta)^{2 n_1 + \ell_1 - 1} \,
\sum_{\nu=0}^{\infty} \, \frac {(n_1 + \ell_1 + 1)_{\nu}} {{\nu}!} \,
\left\{\frac {\beta^2 - \alpha^2} {\beta^2} \right\}^{\nu} \,
S_{n_1 + \nu \ell_1 m_1}^{n_2 \ell_2 m_2} (\beta, \beta, \vec{R})
\\ \tag
" \qquad \; = \; (\beta/\alpha)^{2 n_2 + \ell_2 - 1} \,
\sum_{\nu=0}^{\infty} \, \frac {(n_2 + \ell_2 + 1)_{\nu}} {{\nu}!} \,
\left\{\frac {\alpha^2 - \beta^2} {\alpha^2} \right\}^{\nu} \,
S_{n_1 \ell_1 m_1}^{n_2 + \nu \ell_2 m_2} (\alpha, \alpha, \vec{R}) \, .
\\ \tag
\endAligntags
$$
Um die Konvergenzeigenschaften der unendlichen Reihen (8.3-3) und
(8.3-4) theoretisch analysie\-ren zu k\"onnen, ben\"otigt man das
asymptotische Verhalten einer $B$-Funktion $B_{n, \ell}^{m}$ f\"ur gro{\ss}e
Werte der Ordnung $n$.

Mit Hilfe bekannter Monotonieeigenschaften der modifizierten
Besselfunktion $K_{\nu} (z)$ [Magnus, Oberhettinger und 1966, S. 151]
kann man leicht zeigen, da{\ss} die reduzierten Besselfunktionen ${\widehat
k}_{\nu} (z)$ f\"ur $\nu > 0$ und $z \ge 0$ positiv und beschr\"ankt sind
[Weniger und Steinborn 1983b, Gl. (3.1)]. Bei halbzahligen Ordnungen
$\nu = n + 1/2$ mit $n \in \N_0$ folgt damit aufgrund von Gl. (8.2-7):
$$
0 \; < \; {\widehat k}_{n+1/2} (z) \; \le \;
{\widehat k}_{n+1/2} (0) \; = \; 2^n \, (1/2)_n \, ,
\qquad 0 \le z < \infty \, , \quad n \in \N_0 \, .
\tag
$$

Das Buch von Grosswald [1978] enth\"alt einen Abschnitt \"uber die
asymptotischen Eigenschaften der Besselpolynome. Dort wird gezeigt, da{\ss}
das Besselpolynom $\theta_n (z)$ f\"ur festes Argument $z$ die folgende
asymptotische Beziehung erf\"ullt [Grosswald 1978, S. 125]:
$$
\theta_n (z) \; \sim \; \frac {(2 n)!} {2^n n!} \, \e^z \; = \;
2^n \, (1/2)_n \, \e^z \, , \qquad n \to \infty \, .
\tag
$$
Aus Gln. (8.2-8) und (8.3-6) folgt dann, da{\ss} der dominante Term der
asymptotischen Entwicklung einer reduzierten Besselfunktion ${\widehat
k}_{n+1/2} (z)$ f\"ur festes Argument $z \ge 0$ durch ihren Wert am
Nullpunkt gegeben ist [Weniger und Steinborn 1983b, Gl. (3.9)]:
$$
{\widehat k}_{n+1/2} (z) \; = \;
2^n \, (1/2)_n \, [1 + O (n^{- 1})] \; = \;
{\widehat k}_{n+1/2} (0) \, [1 + O (n^{- 1})]
\, , \qquad n \to \infty \, .
\tag
$$
H\"ohere Terme der asymptotischen Entwicklung einer reduzierten
Besselfunktion ${\widehat k}_{n+1/2} (z)$ in Potenzen von $1/n$ kann man
aus einer analogen asymptotischen Entwicklung f\"ur das Besselpolynom $y_n
(z)$ erhalten. Auf S. 130 des Buches von Grosswald [1978] kann man die
Koeffizienten bis zur Ordnung $O (n^{- 3})$ und in einem Artikel von
Salzer [1983] die Koeffizienten bis zur Ordnung $O (n^{- 4})$ finden.

Aus Gl. (8.3-7) erh\"alt man mit Hilfe von Gl. (7.2-1) den f\"uhrenden Term
der asymptotischen Entwicklung einer $B$-Funktion [Grotendorst, Weniger
und Steinborn 1986, Gl. (3.8)]:
$$
B_{n, \ell}^{m} (\alpha, \vec{r}) \; \sim \;
(2 /\pi)^{1/2} \, {\cal Y}_{\ell}^{m} (\alpha \vec{r}) \,
(2 n)^{- \ell - 3/2} \, , \qquad n \to \infty \, .
\tag
$$
Diese Beziehung gilt f\"ur konstante, aber ansonsten beliebige $\alpha >
0$ und $\vec{r} \in \R^3$.

Die beiden unendlichen Reihen in Gln. (8.3-3) und (8.3-4) k\"onnen auf
folgende Weise umgeschrieben werden [Grotendorst, Weniger und Steinborn
1986, Gln. (5.1) und (5.2)]:
$$
\beginAligntags
" S_{n_1 \ell_1 m_1}^{n_2 \ell_2 m_2} (\alpha, \beta, \vec{R}) \\
" \qquad \; = \; (- 1)^{\ell_2} \, \frac {4 \pi} {\beta^3} \,
(\alpha/\beta)^{2 n_1 + \ell_1 - 1} \,
\sum_{\nu=0}^{\infty} \, \frac {(n_1 + \ell_1 + 1)_{\nu}} {{\nu}!} \,
\left\{\frac {\beta^2 - \alpha^2} {\beta^2} \right\}^{\nu} \\
" \qquad \quad \times \,
\gntsm \, \gntcf{2}{m_2}{1}{m_1}{}{m_2 - m_1} \\
" \qquad \qquad \times \,
\sum_{t=0}^{\Delta \ell} \, (-1)^t \binom{\Delta \ell}{t} \,
B_{n_1 + n_2 + \nu + 2 \Delta \ell - t + 1, \ell}^{m_2 - m_1}
(\beta, \vec{R})
\\ \tag
" \qquad \; = \; (- 1)^{\ell_2} \, \frac {4 \pi} {\alpha^3}
(\beta/\alpha)^{2 n_2 + \ell_2 - 1} \,
\sum_{\nu=0}^{\infty} \, \frac {(n_2 + \ell_2 + 1)_{\nu}} {{\nu}!} \,
\left\{\frac {\alpha^2 - \beta^2} {\alpha^2} \right\}^{\nu} \\
" \qquad \quad \times \,
\gntsm \, \gntcf{2}{m_2}{1}{m_1}{}{m_2 - m_1} \\
" \qquad \qquad \times \,
\sum_{t=0}^{\Delta \ell} \, (-1)^t \binom{\Delta \ell}{t} \,
B_{n_1 + n_2 + \nu + 2 \Delta \ell - t + 1, \ell}^{m_2 - m_1}
(\alpha, \vec{R}) \, . \quad
\\ \tag
\endAligntags
$$

Im n\"achsten Schritt werden alle Beitr\"age in Gln. (8.3-9) und (8.3-10),
die nicht vom Summationsindex $\nu$ abh\"angen, vernachl\"assigt. Au{\ss}erdem
werden die Quotienten $(n_1 + \ell_1 + 1)_{\nu} / {\nu}!$ und $(n_2 +
\ell_2 + 1)_{\nu} / {\nu}!$ mit Hilfe von Gl. (7.2-1) vereinfacht. Mit
Hilfe von Gl. (8.3-8) erh\"alt man dann die dominanten Beitr\"age der
asymptotischen Entwicklungen der Terme der unendlichen Reihen in Gln.
(8.3-9) und (8.3-10) f\"ur $\nu \to \infty$ [Grotendorst, Weniger und
Steinborn 1986, Gln. (5.4) und (5.5)]:
$$
\beginAligntags
" \frac {(n_1 + \ell_1 + 1)_{\nu}} {{\nu}!} \,
\left\{\frac {\beta^2 - \alpha^2} {\beta^2} \right\}^{\nu} \,
B_{n_1 + n_2 + \nu + 2 \Delta \ell - t + 1, \ell}^{m_2 - m_1}
(\beta, \vec{R}) \\
" \qquad \; \sim \;
\frac {(2 / \pi)^{1/2}} {2^{n_1 + \ell_1} (n_1 + \ell_1)!} \,
\left\{\frac {\beta^2 - \alpha^2} {\beta^2} \right\}^{\nu} \,
(2 \nu)^{n_1 + \ell_1 - \ell - 3/2} \,
{\cal Y}_{\ell}^{m_2 - m_1} (\beta \vec{R}) \, ,
\\ \tag
" \frac {(n_2 + \ell_2 + 1)_{\nu}} {{\nu}!} \,
\left\{ \frac {\alpha^2 - \beta^2} {\alpha^2} \right\}^{\nu} \,
B_{n_1 + n_2 + \nu + 2 \Delta \ell - t + 1, \ell}^{m_2 - m_1}
(\alpha, \vec{R}) \\
" \qquad \; \sim \;
\frac {(2 / \pi)^{1/2}} {2^{n_2 + \ell_2} (n_2 + \ell_2)!} \,
\left\{\frac {\alpha^2 - \beta^2} {\alpha^2} \right\}^{\nu} \,
(2 \nu)^{n_2 + \ell_2 - \ell - 3/2} \,
{\cal Y}_{\ell}^{m_2 - m_1} (\alpha \vec{R}) \, .
\\ \tag
\endAligntags
$$
Diese asymptotischen Absch\"atzungen zeigen, da{\ss} die unendlichen Reihen
(8.3-3) und (8.3-4) offensichtlich {\it linear\/} konvergieren. Au{\ss}erdem
folgt aus Gl. (8.3-11), da{\ss} die unendliche Reihe in Gl. (8.3-3) f\"ur
$\abs{1 - (\alpha/\beta)^2} < 1$ konvergiert und da{\ss} die Konvergenz f\"ur
wachsende Werte von $n_1 + \ell_1$ schlechter werden sollte. Analog
folgt aus Gl. (8.3-12), da{\ss} die unendliche Reihe in Gl. (8.3-4) f\"ur
$\abs{1 - (\beta/\alpha)^2} < 1$ konvergiert und da{\ss} die Konvergenz f\"ur
wachsende Werte von $n_2 + \ell_2$ schlechter werden sollte. Numerische
Ergebnisse best\"atigen diese theoretischen Vorhersagen [Weniger und
Steinborn 1983b, Table \Roemisch{2}].

Au{\ss}erdem sollten beide Reihenentwicklungen (8.3-3) und (8.3-4)
schneller konvergieren, wenn $\ell_{\min}$ gr\"o{\ss}er wird. Daraus folgt,
da{\ss} die Konvergenzgeschwindigkeit der beiden Reihenentwicklungen (8.3-3)
und (8.3-4) im Umweg \"uber die Summengrenzen $\ell_{\min}$ und
$\ell_{\max}$, die in Gl. (8.2-37) definiert sind, von den magnetischen
Quantenzahlen $m_1$ und $m_2$ abh\"angen sollte. Auch diese theoretische
Voraussage wird durch numerische Ergebnisse best\"atigt [Weniger und
Steinborn 1983b, Table \Roemisch{2}].

Ungl\"ucklicherweise zeigen diese numerischen Ergebnisse aber auch, da{\ss}
die beiden unendlichen Reihen (8.3-3) und (8.3-4) f\"ur gr\"o{\ss}ere
Unterschiede der Exponentialparameter $\alpha$ und $\beta$ so langsam
konvergieren, da{\ss} eine effiziente Berechnung des \"Uberlappungsintegrals
(8.2-35) auf diese Weise nicht m\"oglich ist [Weniger und Steinborn 1983b,
Table \Roemisch{2}]. Es war also n\"otig, nach alternativen
Reihendarstellungen mit besseren Konvergenzeigenschaften zu suchen.

Auf den ersten Blick scheint die folgende Reihendarstellung des
\"Uberlappungsintegrals (8.2-35) mit verschiedenen Exponentialparametern
$\alpha$ und $\beta$ [Weniger, Grotendorst und Steinborn 1986b, Gl.
(5.9)] eine wesentlich kompliziertere Struktur zu besitzen als die
beiden Reihendarstellungen (8.3-3) und (8.3-4):
$$
\beginAligntags
S_{n_1 \ell_1 m_1}^{n_2 \ell_2 m_2} " (\alpha, \beta, \vec{R}) \; = \;
\frac
{\alpha^{2 n_1 + \ell_1 - 1} \beta^{2 n_2 + \ell_2 - 1}}
{[(\alpha^2 + \beta^2)/2]^{n_1 + n_2 + (\ell_1 + \ell_2)/2 - 1}} \\
" \quad \times \, \sum_{\nu=0}^{\infty} \,
{}_2 F_1 \bigl( - \nu, n_1 + \ell_1 + 1;
n_1 + n_2 + \ell_1 + \ell_2 + 2; 2 \bigr) \, \\
" \qquad \times \, \frac
{(n_1 + n_2 + \ell_1 + \ell_2 + 2)_{\nu}} {{\nu}!} \,
\left\{\frac {\alpha^2 - \beta^2} {\alpha^2 + \beta^2} \right\}^{\nu} \\
" \qquad \quad \times \, S_{n_1 + \nu \ell_1 m_1}^{n_2 \ell_2 m_2}
\bigl([(\alpha^2 + \beta^2)/2]^{1/2}, [(\alpha^2 + \beta^2)/2]^{1/2},
\vec{R} \bigr) \, .
\\ \tag
\endAligntags
$$
Man kann aber die abbrechende hypergeometrische Reihe in Gl. (8.3-13)
auf einfache Weise mit Hilfe einer linearen Dreitermrekursion berechnen
[Weniger, Grotendorst und Steinborn 1986b, Gl. (4.22)]. Au{\ss}erdem
vereinfacht sich die unendliche Reihe in Gl. (8.3-14) erheblich, wenn
$n_1 + \ell_1 = n_2 + \ell_2$ gilt, da die abbrechende hypergeometrische
Reihe ${}_2 F_1$ dann in geschlossener Form ausgedr\"uckt werden kann
[Weniger, Grotendorst und Steinborn 1986b, Gl. (5.10)]:
$$
\beginAligntags
S_{n_1 \ell_1 m_1}^{n_2 \ell_2 m_2} (\alpha, \beta, \vec{R})
\; = \; " \frac
{\alpha^{2 n_1 + \ell_1 - 1} \beta^{2 n_2 + \ell_2 - 1}}
{[(\alpha^2 + \beta^2)/2]^{n_1 + n_2 + (\ell_1 + \ell_2)/2 - 1}} \,
\sum_{\nu=0}^{\infty} \, \frac
{(n_1 + \ell_1 + 1)_{\nu}} {{\nu}!} \, \left\{
\frac {\alpha^2 - \beta^2} {\alpha^2 + \beta^2} \right\}^{2 \nu} \\
" \quad \times \,
S_{n_1 + 2 \nu \, \ell_1 m_1}^{n_2 \ell_2 m_2}
\bigl([(\alpha^2 + \beta^2)/2]^{1/2}, [(\alpha^2 + \beta^2)/2]^{1/2},
\vec{R} \bigr) \, .
\\ \tag
\endAligntags
$$

Die unendliche Reihe in Gln. (8.3-13) kann auf folgende Weise
umgeschrieben werden [Grotendorst, Weniger und Steinborn 1986, Gl.
(5.3)]:
$$
\beginAligntags
" S_{n_1 \ell_1 m_1}^{n_2 \ell_2 m_2} (\alpha, \beta, \vec{R})
\; = \;
(- 1)^{\ell_2} \, 4 \pi \, \frac
{\alpha^{2 n_1 + \ell_1 - 1} \beta^{2 n_2 + \ell_2 - 1}}
{[(\alpha^2 + \beta^2)/2]^{n_1 + n_2 + (\ell_1 + \ell_2)/2 - 1}} \\
" \qquad \times
\sum_{\nu=0}^{\infty} \, {}_2 F_1 \bigl( - \nu, n_1 + \ell_1 + 1;
n_1 + n_2 + \ell_1 + \ell_2 + 2; 2 \bigr) \\
" \qquad \quad \times
\frac {(n_1 + n_2 + \ell_1 + \ell_2 + 2)_{\nu}} {{\nu}!} \, \left\{
\frac {\alpha^2 - \beta^2} {\alpha^2 + \beta^2} \right\}^{\nu} \,
\gntsm \, \gntcf{2}{m_2}{1}{m_1}{}{m_2 - m_1} \\
" \qquad \qquad \times
\sum_{t=0}^{\Delta \ell} \, (-1)^t \binom{\Delta \ell}{t} \,
B_{n_1 + n_2 + \nu + 2 \Delta \ell - t + 1, \ell}^{m_2 - m_1}
\bigl( [(\alpha^2 + \beta^2)/2]^{1/2}, \vec{R} \bigr) \, .
\\ \tag
\endAligntags
$$

Um die Konvergenzeigenschaften der unendlichen Reihe (8.3-13)
theoretisch analysieren zu k\"onnen, werden im n\"achsten Schritt alle
Beitr\"age in Gl. (8.3-15), die nicht vom Summationsindex $\nu$ abh\"angen,
vernachl\"assigt. Au{\ss}erdem wird der Quotient $(n_1 + n_2 + \ell_1 + \ell_2
+ 2)_{\nu} / {\nu}!$ mit Hilfe von Gl. (7.2-1) vereinfacht, und f\"ur die
hypergeometrische Reihe wird die folgende asymptotische Absch\"atzung
verwendet [Grotendorst, Weniger und Steinborn 1986, Gl. (A7)]:
$$
\beginAligntags
" {}_2 F_1 (- \nu, m + 1; m + n + 2; 2) \\
" \quad \; \sim \;
\frac {(m + n + 1)!} {m! n!} \, \left[
\frac {m!} {(2 \nu)^{m + 1}} \, + \, \cdots \, + \,
(- 1)^{- \nu} \, \frac {n!} {(2 \nu)^{n + 1}} \, + \, \cdots \,
\right] \, , \qquad \nu \to \infty \, .
\\ \tag
\endAligntags
$$
Durch Kombination der Gln. (8.3-8), (8.3-15) und (8.3-16) erh\"alt man
dann den dominanten Term der asymptotischen Entwicklung der Terme der
unendlichen Reihe in Gl. (8.3-15) f\"ur $\nu \to \infty$ [Grotendorst,
Weniger und Steinborn 1986, Gl. (5.6)]:
$$
\beginAligntags
" {}_2 F_1 \bigl( - \nu, n_1 + \ell_1 + 1;
n_1 + n_2 + \ell_1 + \ell_2 + 2; 2 \bigr) \,
\frac {(n_1 + n_2 + \ell_1 + \ell_2 + 2)_{\nu}} {{\nu}!} \\
" \quad \times \left\{
\frac {\alpha^2 - \beta^2} {\alpha^2 + \beta^2} \right\}^{\nu} \,
B_{n_1 + n_2 + \nu + 2 \Delta \ell - t + 1, \ell}^{m_2 - m_1}
\bigl( [(\alpha^2 + \beta^2)/2]^{1/2}, \vec{R} \bigr) \\
" \qquad \sim \;
\frac {(2 / \pi)^{1/2}} {2^{n_1 + n_2 + \ell_1 + \ell_2 + 1}} \,
{\cal Y}_{\ell}^{m_2 - m_1} \,
\bigl( [(\alpha^2 + \beta^2)/2]^{1/2} \vec{R} \bigr) \,
\left\{\frac {\alpha^2 - \beta^2} {\alpha^2 + \beta^2} \right\}^{\nu} \\
" \qquad \quad \times
\left[
\frac {(2 \nu)^{n_2 + \ell_2 - \ell - 3/2}} {(n_2 + \ell_2)!} \, + \,
\cdots \, + (- 1)^{\nu} \,
\frac {(2 \nu)^{n_1 + \ell_1 - \ell - 3/2}} {(n_1 + \ell_1)!}
\, + \, \cdots \, \right] \, .
\\ \tag
\endAligntags
$$
Da immer $(\alpha^2 - \beta^2)/(\alpha^2 + \beta^2) < 1$ gilt, folgt aus
dieser asymptotischen Absch\"atzung, da{\ss} die unendliche Reihe (8.3-13) f\"ur
alle $\alpha, \beta > 0$ {\it linear\/} konvergiert. Im Gegensatz zu den
unendlichen Reihen (8.3-3) und (8.3-4) ist die Rolle, die die Wertepaare
$n_1 + \ell_1$ und $n_2 + \ell_2$ in Gl. (8.3-13) spielen, {\it
symmetrisch}, da eine Erh\"ohung von entweder $n_1 + \ell_1$ oder $n_2 +
\ell_2$ immer zu einer Verschlechterung der Konvergenz f\"uhren sollte.
Ebenso wie die Reihenentwicklungen (8.3-3) und (8.3-4) sollte die
unendliche Reihe in Gl. (8.3-13) schneller konvergieren, wenn
$\ell_{\min}$ gr\"o{\ss}er wird. Die Konvergenzgeschwindigkeit h\"angt also
wieder im Umweg \"uber die Summengrenzen $\ell_{\min}$ und $\ell_{\max}$,
die in Gl. (8.2-37) definiert sind, von den magnetischen Quantenzahlen
$m_1$ und $m_2$ ab.

Aus den Ungleichungen
$$
\abs{(\alpha^2 - \beta^2)/\alpha^2} \; < \;
\abs{(\alpha^2 - \beta^2)/(\alpha^2 + \beta^2)}
\tag
$$
und
$$
\abs{(\beta^2 - \alpha^2)/\beta^2} \; < \;
\abs{(\alpha^2 - \beta^2)/(\alpha^2 + \beta^2)}
\tag
$$
und den asymptotischen Absch\"atzungen (8.3-11), (8.3-12) und (8.3-17)
folgt dann noch, da{\ss} die unendliche Reihe (8.3-13) nur dann nicht
schneller konvergieren sollte als entweder die unendliche Reihe (8.3-3)
oder (8.3-4), wenn sowohl die Paare $n_1 + \ell_1$ und $n_2 + \ell_2$
als auch die Exponentialparameter $\alpha$ und $\beta$ sich betr\"achtlich
unterscheiden. Numerische Ergebnisse best\"atigen diese Vorhersagen
[Grotendorst, Weniger und Steinborn 1986, Table \Roemisch{3}].

Obwohl die unendliche Reihe in Gl. (8.3-13) f\"ur das \"Uberlappungsintegral
(8.2-35) deutlich bessere Konvergenzeigenschaften besitzt als entweder
die unendliche Reihe in Gl. (8.3-3) oder (8.3-4) [Grotendorst, Weniger
und Steinborn 1986, Table \Roemisch{3}], war es trotzdem w\"unschenswert,
die Konvergenzgeschwindigkeit noch weiter zu verbessern. Aus diesem
Grund wurde dann auch versucht, die Konvergenz der unendlichen Reihe in
Gl. (8.3-13) mit Hilfe von verallgemeinerten Summationsprozessen wie dem
Wynnschen $\epsilon$-Algorithmus, Gl. (2.4-10), dem $\theta$-Algorithmus
(4.4-13) oder der Levinschen $u$-Transformation, Gl. (5.2-13), zu
verbessern. Dabei zeigte es sich, da{\ss} der Wynnsche
$\epsilon$-Algorithmus die Konvergenz der unendlichen Reihe deutlich
verbessern konnte [Grotendorst, Weniger und Steinborn 1986, Table
\Roemisch{4}]. Die $u$-Transformation und der $\theta$-Algorithmus waren
etwas weniger effizient als der $\epsilon$-Algorithmus und au{\ss}erdem
numerisch weniger stabil [Grotendorst, Weniger und Steinborn 1986, S.
3716; Weniger und Steinborn 1988, S. 328].

Trotz des unbestreitbaren Erfolges verallgemeinerter Summationsprozesse
im Zusammenhang mit Reihendarstellungen f\"ur das \"Uberlappungsintegral
zweier $B$-Funktionen mit verschiedenen Exponentialparametern sollen
diese hier nicht weiter behandelt werden. Der Grund ist, da{\ss} inzwischen
noch ein effizienteres Verfahren zur Berechnung solcher
\"Uberlappungsintegrale bekannt ist. Der Ausgangspunkt ist die sogenannte
{\it Feynmanidentit\"at\/} [Feynman 1949, Appendix, Gl. (14a)]
$$
\frac {1} {a b} \; = \; \int\nolimits_{0}^{1} \,
\frac {\d t} {[a t + b (1 - t)]^2} \, ,
\tag
$$
die leicht auf folgende Weise verallgemeinert werden kann [Joachain
1975, Gl. (D.3)]:
$$
\frac {1} {a^m b^n} \; = \;
\frac {(m + n - 1)!} {(m - 1)! (n - 1)!} \,
\int\nolimits_{0}^{1} \, \frac
{t^{m - 1} (1 - t)^{n - 1}} {[a t + b (1 - t)]^{m + n}} \, \d t \, .
\tag
$$
Wenn man diese Beziehung in der Fourierdarstellung des
\"Uberlappungsintegrals zweier $B$-Funktionen [Weniger und Steinborn
1983a, Gl. (5.3)],
$$
\beginAligntags
" S_{n_1 \ell_1 m_1}^{n_2 \ell_2 m_2}(\alpha, \beta, \vec{R})
\; = \; (- 1)^{\ell_2} \, 4 \pi \,
\alpha^{2 n_1 + \ell_1 - 1} \, \beta^{2 n_2 + \ell_2 - 1} \\
" \qquad \times
\gntsm \, \gntcf{2}{m_2}{1}{m_1}{}{m_2 - m_1} \,
Y_{\ell}^{m_2 - m_1} (\vec{R} / R) \\
" \qquad \quad \times
\left[ \frac {2} {R \pi} \right]^{1/2} \,
\int\nolimits_{0}^{\infty} \, \frac
{p^{\ell_1 + \ell_2 + 3/2} J_{\ell + 1/2} (R p)}
{[\alpha^2 + p^2]^{n_1 + \ell_1 + 1}
[\beta^2 + p^2]^{n_2 + \ell_2 + 1}} \, \d p \, ,
\\ \tag
\endAligntags
$$
einsetzt, erh\"alt man die folgende eindimensionale Integraldarstellung
f\"ur das \"Uberlappungsintegral zweier $B$-Funktionen mit verschiedenen
Exponentialparametern $\alpha$ und $\beta$ [Trivedi und Steinborn 1983,
Gl. (3.1); Weniger und Steinborn 1988, Gl. (2.14)]:
$$
\beginAligntags
" S_{n_1 \ell_1 m_1}^{n_2 \ell_2 m_2}(\alpha, \beta, \vec{R})
\; = \; \alpha^{2 n_1 + \ell_1 - 1} \, \beta^{2 n_2 + \ell_2 - 1} \,
\frac {(n_1 + n_2 + \ell_1 + \ell_2 + 1)!}
{(n_1 + \ell_1)! (n_2 + \ell_2)!} \\
" \qquad \times
\int\nolimits_{0}^{1} \, \frac
{t^{n_1 + \ell_1} (1 - t)^{n_2 + \ell_2}} {\bigl[ \gamma
(\alpha, \beta, t) \bigr]^{n_1 + n_2 + \ell_1 + \ell_2 + 2}} \,
S_{n_1 \ell_1 m_1}^{n_2 \ell_2 m_2}
(\gamma (\alpha, \beta, t), \gamma (\alpha, \beta, t), \vec{R}) \, \d t
\, ,
\erhoehe\aktTag \\ \tag*{\tagnr a}
" \qquad \qquad \gamma (\alpha, \beta, t) \; = \;
\bigl[ \alpha^2 t + \beta^2 (1 - t) \bigr]^{1/2} \, .
\\ \tag*{\tagform\aktTagnr b}
\endAligntags
$$
F\"ur das Integral auf der rechten Seite in Gl. (8.3-23a) ist kein
geschlossener Ausdruck bekannt. Man kann diese Integraldarstellung aber
verwenden, um den Wert des \"Uberlappungsintegrals mit Hilfe geeigneter
Quadraturverfahren zu berechnen.

Trivedi und Steinborn [1983] berechneten das Integral mit Hilfe eines
{\it adaptiven Quadraturverfahrens\/} unter Verwendung des Programmes
DCADRE der IMSL Library, das auch in dem Buch von Rice [1983]
beschrieben ist. Der Vorteil dieser Vorgehensweise ist, da{\ss} man die
gew\"unschte Genauigkeit vorgeben kann und da{\ss} die Konvergenzanalyse vom
Programm durchgef\"uhrt wird. Der Nachteil einer adaptiven Quadratur ist,
da{\ss} sie nie so effizient sein kann wie ein Quadratur\-ver\-fahren mit
einer festen Anzahl von Abszissen. Ein guter adaptiver Algorithmus mu{\ss}
bei der Bewertung der bisher erreichten Genauigkeit sehr vorsichtig sein
und wird deswegen immer versuchen, die bisherigen Schlu{\ss}folgerungen
bez\"uglich der bisher erreichten Genauigkeit durch zus\"atzliche
Berechnungen des Integranden zu best\"atigen, was zu einer Verringerung
der Effizienz des adaptiven Quadraturverfahrens f\"uhren mu{\ss}.

Wesentlich effizienter als adaptive Verfahren sind die nach Gau{\ss}
benannten Quadraturverfahren [Davis und Rabinowitz 1984, Abschnitte 2.7
und 3.7], bei denen ein Integral, das eine Gewichtsfunktion $w (x)$
enth\"alt, durch eine Quadratursumme und einen Quadraturfehler $E_n (f)$
dargestellt wird:
$$
\int\nolimits_{a}^{b} \, w (x) \, f (x) \, \d x \; = \;
\sum_{k = 1}^{n} \, w_k \, f (x_k) \, + \, E_n (f) \, .
\tag
$$
Bei der Gau{\ss}-Quadratur werden die $n$ Gewichte $w_k$ und die $n$
Abszissen $x_k$ so gew\"ahlt, da{\ss} der Quadraturfehler $E_n (f)$ {\it
minimal\/} wird. Eine Gau{\ss}-Regel wird nach den Polynomen benannt, die
auf dem Intervall $[a, b]$ bez\"uglich der Gewichtsfunktion $w (x)$
orthogonal sind.

Bhattacharya und Dhabal [1986] verwendeten eine Gau{\ss}-Legendre-Regel zur
Berechnung des Integrals in Gl. (8.3-23). Weniger und Steinborn [1988,
Table \Roemisch{1} - \Roemisch{3}] zeigten, da{\ss} eine Gau{\ss}-Legendre-Regel
in etwa gleich gute Ergebnisse liefert wie die unendlichen Reihe in Gl.
(8.3-13), wenn ihre Konvergenz durch den Wynnschen
$\epsilon$-Algorithmus, Gl. (2.4-10), beschleunigt wird.

Nach der allgemeinen Theorie der Gau{\ss}-Quadratur sollte aber eine
Gau{\ss}-Legendre-Regel, die immer eine Gewichtsfunktion $w (x) = 1$ im
Integral (8.3-24) voraussetzt, im Falle der Integraldarstellung (8.3-23)
nicht optimal sein. Aufgrund der Potenzen $t^{n_1 + \ell_1} (1 - t)^{n_2
+ \ell_2}$ im Integranden der Integraldarstellung (8.3-23) sollte eine
geeignete Gau{\ss}-Jacobi-Regel besonders f\"ur gr\"o{\ss}ere Werte von $n_1 +
\ell_1$ und $n_2 + \ell_2$ deutlich bessere Ergebnisse
liefern{\footnote[\dagger]{Die Jacobipolynome $P_n^{(\alpha, \beta)}
(x)$ sind orthogonal auf dem Intervall $[- 1, 1]$ bez\"uglich der
Gewichtsfunktion $w (x) = (1 - x)^{\alpha} (1 + x)^{\beta}$ [Magnus,
Oberhettinger und Soni 1966, S. 212]. Die Variablensubstitution $x = 2 t
- 1$ transformiert das Integrationsintervall $[0, 1]$ in Gl. (8.3-23a)
in das passende Integrationsintervall $[- 1, 1]$ und die
Gewichtsfunktion $t^{n_1 + \ell_1} (1 - t)^{n_2 + \ell_2}$ in die
passende Gewichtsfunktion $[(1 - x)/2]^{n_2 + \ell_2} [(1 + x)/2]^{n_1 +
\ell_1}$.}}. Numerische Ergebnisse best\"atigen, da{\ss} diese Vermutung in
den meisten F\"allen zutrifft [Weniger und Steinborn 1988, Table
\Roemisch{1} - \Roemisch{3}; Homeier 1990, Tabelle 6.4.\Roemisch{3}].
Man mu{\ss} aber ber\"ucksichtigen, da{\ss} der Effizienzgewinn einer
Gau{\ss}-Jacobi-Regel zum Teil dadurch zunichte gemacht wird, da{\ss} man bei
einer Gau{\ss}-Jacobi-Quadratur f\"ur jedes Paar $n_1 + \ell_1$ und $n_2 +
\ell_2$ die Gewichte und Abszissen neu berechnen mu{\ss}. In dieser
Beziehung ist eine Gau{\ss}-Legendre-Quadratur g\"unstiger: Solange die {\it
Zahl\/} der verwendeten Abszissen sich nicht \"andert, kann man immer die
gleichen Gewichte und Abszissen verwenden. Wie vorteilhaft die
Verwendung einer Gau{\ss}-Jacobi-Regel verglichen mit einer
Gau{\ss}-Legendre-Regel letztlich ist, h\"angt demzufolge sehr stark sowohl
von der Implementierung als auch von den \"Uberlappungsintegralen ab, die
man zu berechnen hat.

Die Effizienz von Gau{\ss}-Quadraturen kann im Falle der Integraldarstellung
(8.3-23) noch deutlich verbessert werden, wenn man modifizierte
Gau{\ss}-Regeln verwendet, die auf einer M\"obius\-trans\-for\-mation basieren
[Homeier und Steinborn 1990]. Mit Hilfe dieser M\"obiustransfor\-mation
ist es oft m\"oglich, Integranden, die Spitzen oder andere Pathologien
enthalten, in Integranden mit numerisch g\"unstigeren Eigenschaften zu
transformieren.

Homeier und Steinborn [1992, Table \Roemisch{2}] erhielten im Falle der
Integraldarstellung (8.3-23) wesentlich bessere Ergebnisse bei
Verwendung sogenannter M\"obius-Legendre- und M\"obius-Jacobi-Regeln als bei
Verwendung von Gau{\ss}-Legendre- oder Gau{\ss}-Jacobi-Regeln. Ein vor kurzer
Zeit ver\"offentlichtes Programmpaket f\"ur Uberlappungsintegrale von
$B$-Funktionen [Homeier, Weniger und Steinborn 1992b] enth\"alt deswegen
keine Programme f\"ur \"Uberlappungsintegrale mit verschiedenen
Exponentialparametern, die auf den unendlichen Reihen (8.3-3), (8.3-4)
oder (8.3-13) basieren, sondern nur Programme, die entweder die \"au{\ss}erst
effiziente, aber numerisch h\"aufig instabile Jacobipolynomdarstellung
(8.2-36) verwenden oder die Integraldarstellung (8.3-23) mit Hilfe einer
M\"obius-Regel numerisch auswerten.

Die auf einer M\"obiustransformation basierenden Quadraturverfahren
[Homeier und Steinborn 1990] haben sich auch bei anderen
Mehrzentrenintegralen von $B$-Funktionen als sehr leistungsf\"ahig
erwiesen [Homeier 1990; Homeier und Steinborn 1990; 1991; 1993;
Steinborn und Homeier 1990; Steinborn, Homeier und Weniger 1992].

Bei den Coulombintegralen (8.2-47) zweier $B$-Funktionen mit
verschiedenen Exponentialparametern $\alpha$ und $\beta$ gibt es
\"ahnliche Probleme wie bei den \"Uberlappungsintegralen. Ebenso wie die
Jacobipolynomdarstellung (8.2-36) wird auch die daraus abgeleitete
Jacobipolynomdarstellung (8.2-49) des Coulombintegrals numerisch
instabil f\"ur $\alpha \to \beta$ und f\"ur $R \to 0$. Man ben\"otigt also
noch alternative Darstellungen.

Das Coulombintegral zweier $B$-Funktionen mit gleichen
Exponentialparametern hat eine bemerkenswert kompakte Struktur, da es
eine einfache Linearkombination von Gauntkoeffizienten und den in Gl.
(8.2-50) definierten Kernanziehungsintegralen $A_{n, \ell}^{m}$ ist
[Steinborn und Weniger 1992, Gl. (4.9); Weniger, Grotendorst und
Steinborn 1986b, Gln. (6.16) und (7.1]:
$$
\beginAligntags
" C_{n_1 \ell_1 m_1}^{n_2 \ell_2 m_2} (\alpha, \alpha, \vec{R})
" \; = \; " (-1)^{\ell_2} \frac {4 \pi} {\alpha^3} \,
\gntsm \, \gntcf{2}{m_2}{1}{m_1}{}{m_2 - m_1} \\
" " " \times \, \sum_{t=0}^{\Delta \ell} \,
(-1)^t \binom{\Delta \ell}{t} \,
A_{n_1 + n_2 + 2 \Delta \ell - t + 1, \ell}^{m_2 - m_1}(\alpha, \vec{R})
\, .
\\ \tag
\endAligntags
$$
Diese Beziehung ist formal fast v\"ollig identisch mit dem
\"Uberlappungsintegral zweier $B$-Funktio\-nen mit gleichen
Exponentialparametern, Gl. (8.3-1). Der einzige Unterschied ist, da{\ss} die
$B$-Funktionen in Gl. (8.3-1) durch die in Gl. (8.2-50) definierten
Kernanziehungsintegrale ersetzt werden m\"ussen.

Aufgrund der Einfachheit des Coulombintegrals (8.3-25) ist es
naheliegend, das Coulombintegral (8.2-47) zweier $B$-Funktionen mit {\it
verschiedenen\/} Exponentialparametern durch eine unendliche Reihe von
Coulombintegralen mit {\it gleichen\/} Exponentialparametern
darzustellen. Dies kann leicht mit Hilfe des Multiplikationstheorems
(8.3-2) der $B$-Funktionen geschehen. Wenn man dieses
Multiplikationstheorem in Gl. (8.2-47) verwendet, erh\"alt man die beiden
folgenden unendlichen Reihen f\"ur das Coulombintegral zweier
$B$-Funktionen mit verschiedenen Exponentialparametern $\alpha$ und
$\beta$ [Weniger, Grotendorst und Steinborn 1986b, Gln. (7.14) und
(7.15)]:
$$
\beginAligntags
" C_{n_1 \ell_1 m_1}^{n_2 \ell_2 m_2} (\alpha, \beta, \vec{R}) \\
" \; = \; (\alpha/\beta)^{2 n_1 + \ell_1 - 1} \,
\sum_{\nu=0}^{\infty} \, \frac {(n_1 + \ell_1 + 1)_{\nu}} {{\nu}!} \,
\left\{\frac {\beta^2 - \alpha^2} {\beta^2} \right\}^{\nu} \,
C_{n_1 + \nu \ell_1 m_1}^{n_2 \ell_2 m_2} (\beta, \beta, \vec{R})
\\ \tag
" \; = \; (\beta/\alpha)^{2 n_2 + \ell_2 - 1} \,
\sum_{\nu=0}^{\infty} \, \frac {(n_2 + \ell_2 + 1)_{\nu}} {{\nu}!} \,
\left\{\frac {\alpha^2 - \beta^2} {\alpha^2} \right\}^{\nu} \,
C_{n_1 \ell_1 m_1}^{n_2 + \nu \ell_2 m_2} (\alpha, \alpha, \vec{R}) \, .
\\ \tag
\endAligntags
$$

Um die Konvergenzeigenschaften der unendlichen Reihen in Gln. (8.3-26)
und (8.3-27) theoretisch analysieren zu k\"onnen, ben\"otigen wir die
folgende asymptotische Absch\"atzung [Grotendorst, Weniger und Steinborn
1986, Gl. (7.18)]:
$$
A_{n, \ell}^{m} (\alpha, \vec{R}) \; \sim \;
\frac {(32 \pi)^{1/2}} {\alpha^2 (2 \ell + 1)} \,
{\cal Y}_{\ell}^{m} (\alpha \vec{R}) \,
(2 n)^{- \ell - 1/2} \, , \qquad n \to \infty \, .
\tag
$$

Die beiden unendlichen Reihen in Gln. (8.3-26) und (8.3-27) k\"onnen auf
folgende Weise umgeschrieben werden [Grotendorst, Weniger und Steinborn
1986, Gln. (7.14) und (7.15)]:
$$
\beginAligntags
" C_{n_1 \ell_1 m_1}^{n_2 \ell_2 m_2} (\alpha, \beta, \vec{R}) \\
" \qquad \; = \; (- 1)^{\ell_2} \, \frac {4 \pi} {\beta^3} \,
(\alpha/\beta)^{2 n_1 + \ell_1 - 1} \,
\sum_{\nu=0}^{\infty} \, \frac {(n_1 + \ell_1 + 1)_{\nu}} {{\nu}!} \,
\left\{\frac {\beta^2 - \alpha^2} {\beta^2} \right\}^{\nu} \\
" \qquad \quad \times \,
\gntsm \, \gntcf{2}{m_2}{1}{m_1}{}{m_2 - m_1} \\
" \qquad \qquad \times \,
\sum_{t=0}^{\Delta \ell} \, (-1)^t \binom{\Delta \ell}{t} \,
A_{n_1 + n_2 + \nu + 2 \Delta \ell - t + 1, \ell}^{m_2 - m_1}
(\beta, \vec{R})
\\ \tag
" \qquad \; = \; (- 1)^{\ell_2} \, \frac {4 \pi} {\alpha^3}
(\beta/\alpha)^{2 n_2 + \ell_2 - 1} \,
\sum_{\nu=0}^{\infty} \, \frac {(n_2 + \ell_2 + 1)_{\nu}} {{\nu}!} \,
\left\{\frac {\alpha^2 - \beta^2} {\alpha^2} \right\}^{\nu} \\
" \qquad \quad \times \,
\gntsm \, \gntcf{2}{m_2}{1}{m_1}{}{m_2 - m_1} \\
" \qquad \qquad \times \,
\sum_{t=0}^{\Delta \ell} \, (-1)^t \binom{\Delta \ell}{t} \,
A_{n_1 + n_2 + \nu + 2 \Delta \ell - t + 1, \ell}^{m_2 - m_1}
(\alpha, \vec{R}) \, . \quad
\\ \tag
\endAligntags
$$

Im n\"achsten Schritt werden alle Beitr\"age in Gln. (8.3-29) und (8.3-30),
die nicht vom Summationsindex $\nu$ abh\"angen, vernachl\"assigt. Au{\ss}erdem
werden die Quotienten $(n_1 + \ell_1 + 1)_{\nu} / {\nu}!$ und $(n_2 +
\ell_2 + 1)_{\nu} / {\nu}!$ mit Hilfe von Gl. (7.2-1) vereinfacht. Mit
Hilfe von Gl. (8.3-28) erh\"alt man dann die dominanten Beitr\"age der
asymptotischen Entwicklungen der Terme der unendlichen Reihen in Gln.
(8.3-29) und (8.3-30) f\"ur $\nu \to \infty$ [Grotendorst, Weniger und
Steinborn 1986, Gln. (7.19) und (7.20)]:
\neueSeite
$$
\beginAligntags
" \frac {(n_1 + \ell_1 + 1)_{\nu}} {{\nu}!} \,
\left\{\frac {\beta^2 - \alpha^2} {\beta^2} \right\}^{\nu} \,
A_{n_1 + n_2 + \nu + 2 \Delta \ell - t + 1, \ell}^{m_2 - m_1}
(\beta, \vec{R}) \\
" \qquad \; \sim \;
\frac {(32 \pi)^{1/2}}
{\beta^2 (2 \ell + 1) 2^{n_1 + \ell_1} (n_1 + \ell_1)!}
\, \left\{\frac {\beta^2 - \alpha^2} {\beta^2} \right\}^{\nu} \,
(2 \nu)^{n_1 + \ell_1 - \ell - 1/2} \,
{\cal Y}_{\ell}^{m_2 - m_1} (\beta \vec{R}) \, ,
\\ \tag
" \frac {(n_2 + \ell_2 + 1)_{\nu}} {{\nu}!} \,
\left\{ \frac {\alpha^2 - \beta^2} {\alpha^2} \right\}^{\nu} \,
A_{n_1 + n_2 + \nu + 2 \Delta \ell - t + 1, \ell}^{m_2 - m_1}
(\alpha, \vec{R}) \\
" \qquad \; \sim \;
\frac {(32 \pi))^{1/2}}
{\alpha^2 (2 \ell + 1) 2^{n_2 + \ell_2} (n_2 + \ell_2)!}
\, \left\{\frac {\alpha^2 - \beta^2} {\alpha^2} \right\}^{\nu} \,
(2 \nu)^{n_2 + \ell_2 - \ell - 1/2} \,
{\cal Y}_{\ell}^{m_2 - m_1} (\alpha \vec{R}) \, .
\\ \tag
\endAligntags
$$
Ein Vergleich der Potenzen des Laufindex $\nu$ in Gln. (8.3-11) und
(3.3-12) und in Gln. (8.3-31) und (8.3-32) zeigt, da{\ss} die unendlichen
Reihen (8.3-26) und (8.3-27) geringf\"ugig langsamer konvergieren sollten
als die unendlichen Reihen (8.3-3) und (8.3-4) f\"ur das
\"Uberlappungsintegral. Ansonsten sollte die Konvergenzgeschwindigkeit der
unendlichen Reihen (8.3-26) und (8.3-27) auf gleiche Weise von den
beteiligten Quantenzahlen und Exponentialparametern abh\"angen wie die
Konvergenzgeschwindigkeit der unendlichen Reihen (8.3-3) und (8.3-4).

Analog zur Reihenentwicklung (8.3-13) f\"ur das \"Uberlappungsintegral kann
man auch die folgende Reihenentwicklung f\"ur das Coulombintegral zweier
$B$-Funktionen mit verschiedenen Exponentialparametern $\alpha$ und
$\beta$ ableiten [Weniger, Grotendorst und Steinborn 1986b, Gl. (7.16)]:
$$
\beginAligntags
" C_{n_1 \ell_1 m_1}^{n_2 \ell_2 m_2} (\alpha, \beta, \vec{R}) \; = \;
\frac
{\alpha^{2 n_1 + \ell_1 - 1} \beta^{2 n_2 + \ell_2 - 1}}
{[(\alpha^2 + \beta^2)/2]^{n_1 + n_2 + (\ell_1 + \ell_2)/2 - 1}} \\
" \qquad \times
\sum_{\nu=0}^{\infty} \,
{}_2 F_1 \bigl( - \nu, n_1 + \ell_1 + 1;
n_1 + n_2 + \ell_1 + \ell_2 + 2; 2 \bigr) \, \\
" \qquad \quad \times
\frac {(n_1 + n_2 + \ell_1 + \ell_2 + 2)_{\nu}} {{\nu}!} \,
\left\{\frac {\alpha^2 - \beta^2} {\alpha^2 + \beta^2} \right\}^{\nu} \\
" \qquad \qquad \times
C_{n_1 + \nu \ell_1 m_1}^{n_2 \ell_2 m_2}
\bigl([(\alpha^2 + \beta^2)/2]^{1/2}, [(\alpha^2 + \beta^2)/2]^{1/2},
\vec{R} \bigr) \, .
\\ \tag
\endAligntags
$$
Diese unendliche Reihe kann auf folgende Weise umgeschrieben werden
[Grotendorst, Weniger und Steinborn 1986, Gl. (7.16)]:
$$
\beginAligntags
" C_{n_1 \ell_1 m_1}^{n_2 \ell_2 m_2} (\alpha, \beta, \vec{R})
\; = \;
(- 1)^{\ell_2} \, 4 \pi \, \frac
{\alpha^{2 n_1 + \ell_1 - 1} \beta^{2 n_2 + \ell_2 - 1}}
{[(\alpha^2 + \beta^2)/2]^{n_1 + n_2 + (\ell_1 + \ell_2)/2 - 1}} \\
" \qquad \times
\sum_{\nu=0}^{\infty} \, {}_2 F_1 \bigl( - \nu, n_1 + \ell_1 + 1;
n_1 + n_2 + \ell_1 + \ell_2 + 2; 2 \bigr) \\
" \qquad \quad \times
\frac {(n_1 + n_2 + \ell_1 + \ell_2 + 2)_{\nu}} {{\nu}!} \, \left\{
\frac {\alpha^2 - \beta^2} {\alpha^2 + \beta^2} \right\}^{\nu} \,
\gntsm \, \gntcf{2}{m_2}{1}{m_1}{}{m_2 - m_1} \\
" \qquad \qquad \times
\sum_{t=0}^{\Delta \ell} \, (-1)^t \binom{\Delta \ell}{t} \,
A_{n_1 + n_2 + \nu + 2 \Delta \ell - t + 1, \ell}^{m_2 - m_1}
\bigl( [(\alpha^2 + \beta^2)/2]^{1/2}, \vec{R} \bigr) \, .
\\ \tag
\endAligntags
$$

Durch Kombination der Gln. (8.3-16), (8.3-28) und (8.3-34) erh\"alt man
dann den dominanten Term der asymptotischen Entwicklung der Terme der
unendlichen Reihe in Gl. (8.3-33) f\"ur $\nu \to \infty$ [Grotendorst,
Weniger und Steinborn 1986, Gl. (7.21)]:
$$
\beginAligntags
" {}_2 F_1 \bigl( - \nu, n_1 + \ell_1 + 1;
n_1 + n_2 + \ell_1 + \ell_2 + 2; 2 \bigr) \,
\frac {(n_1 + n_2 + \ell_1 + \ell_2 + 2)_{\nu}} {{\nu}!} \\
" \quad \times \left\{
\frac {\alpha^2 - \beta^2} {\alpha^2 + \beta^2} \right\}^{\nu} \,
A_{n_1 + n_2 + \nu + 2 \Delta \ell - t + 1, \ell}^{m_2 - m_1}
\bigl( [(\alpha^2 + \beta^2)/2]^{1/2}, \vec{R} \bigr) \\
" \qquad \sim \;
\frac {(2 \pi)^{1/2}}
{(\alpha^2 + \beta^2) (2 \ell + 1) 2^{n_1 + n_2 + \ell_1 + \ell_2 + 1}}
\, {\cal Y}_{\ell}^{m_2 - m_1} \,
\bigl( [(\alpha^2 + \beta^2)/2]^{1/2} \vec{R} \bigr) \,
\left\{\frac {\alpha^2 - \beta^2} {\alpha^2 + \beta^2} \right\}^{\nu} \\
" \qquad \quad \times
\left[
\frac {(2 \nu)^{n_2 + \ell_2 - \ell - 1/2}} {(n_2 + \ell_2)!} \, + \,
\cdots \, + (- 1)^{\nu} \,
\frac {(2 \nu)^{n_1 + \ell_1 - \ell - 1/2}} {(n_1 + \ell_1)!}
\, + \, \cdots \, \right] \, .
\\ \tag
\endAligntags
$$
Ein Vergleich der Potenzen des Laufindex $\nu$ in Gl. (8.3-17) und in
Gl. (8.3-35) zeigt, da{\ss} die unendliche Reihe (8.3-34) geringf\"ugig
langsamer konvergieren sollte als die unendliche Reihe (8.3-13) f\"ur das
\"Uberlappungsintegral. Ansonsten sollte die Konvergenzgeschwindigkeit der
unendlichen Reihe (8.3-35) auf gleiche Weise von den beteiligten
Quantenzahlen und Exponentialparametern abh\"angen wie die
Konvergenzgeschwindigkeit der unendlichen Reihe (8.3-13). Numerische
Tests best\"atigen diese Vermutung [Grotendorst, Weniger und Steinborn
1986, Table \Roemisch{8}]. Diese Tests zeigten auch, da{\ss} die Konvergenz
der unendlichen Reihe in Gl. (8.3-34) durch Anwendung des Wynnschen
$\epsilon$-Algorithmus, Gl. (2.4-10), erheblich verbessert werden kann.

In Analogie zu den \"Uberlappungsintegralen zweier $B$-Funktionen mit
verschiedenen Exponentialparametern ist aber auch bei den
Coulombintegralen zu vermuten, da{\ss} man bessere Ergebnisse erh\"alt, wenn
man die Integraldarstellung [Homeier 1990, Gl. (8.1-19a); Steinborn,
Homeier und Weniger 1992, Gl. (45a)]
$$
\beginAligntags
" C_{n_1 \ell_1 m_1}^{n_2 \ell_2 m_2}(\alpha, \beta, \vec{R})
\; = \; \alpha^{2 n_1 + \ell_1 - 1} \, \beta^{2 n_2 + \ell_2 - 1} \,
\frac {(n_1 + n_2 + \ell_1 + \ell_2 + 1)!}
{(n_1 + \ell_1)! (n_2 + \ell_2)!} \\
" \qquad \times
\int\nolimits_{0}^{1} \, \frac
{t^{n_1 + \ell_1} (1 - t)^{n_2 + \ell_2}} {\bigl[ \gamma
(\alpha, \beta, t) \bigr]^{n_1 + n_2 + \ell_1 + \ell_2 + 2}} \,
C_{n_1 \ell_1 m_1}^{n_2 \ell_2 m_2}
(\gamma (\alpha, \beta, t), \gamma (\alpha, \beta, t), \vec{R}) \, \d t
\, ,
\erhoehe\aktTag \\ \tag*{\tagnr a}
" \qquad \qquad \gamma (\alpha, \beta, t) \; = \;
\bigl[ \alpha^2 t + \beta^2 (1 - t) \bigr]^{1/2} \, ,
\\ \tag*{\tagform\aktTagnr b}
\endAligntags
$$
unter Verwendung einer entsprechenden M\"obius-Regel [Homeier und
Steinborn 1990] auswertet. Allerdings sind bisher noch keine numerischen
Tests durchgef\"uhrt worden.

\medskip

\Abschnitt Logarithmisch konvergente Reihenentwicklungen f\"ur
Zweizentrenintegrale von $B$-Funktionen

\smallskip

\aktTag = 0

Im letzten Unterabschnitt wurde gezeigt, da{\ss} man zur Berechnung der
\"Uberlappungs- und Coulombintegrale zweier $B$-Funktionen mit
verschiedenen Exponentialparametern $\alpha$ und $\beta$ nicht mit den
Jacobipolynomdarstellungen (8.2-36) und (8.2-49) auskommt, da diese
Darstellungen, die nur endlich viele Terme enthalten, f\"ur $\alpha \to
\beta$ und f\"ur $R \to 0$ numerisch instabil werden. Man kann aber mit
Hilfe des Multiplikationstheorems (8.3-2) der $B$-Funktionen die linear
konvergenten Reihenentwicklungen (8.3-3), (8.3-4) und (8.3-13) f\"ur das
\"Uberlappungsintegral und (8.3-26), (8.3-27) und (8.3-33) f\"ur das
Coulombintegral ableiten{\footnote[\dagger]{Die Reihenentwicklungen
(8.3-13) und (8.3-33) k\"onnen ebenfalls mit Hilfe des
Multiplikationstheorems (8.3-2) abgeleitet werden. Man mu{\ss} allerdings
{\it beide\/} $B$-Funktionen im \"Uberlappungs- und Coulombintegral mit
Hilfe des Multiplikationstheorems nach $B$-Funktionen mit dem
Exponentialparameter $[(\alpha^2 + \beta^2)/2]^{1/2}$ entwickeln und die
Summationsreihenfolge der resultierenden Doppelsumme umkehren.}}. Aus dem
asymptotischen Verhalten ihrer Terme kann man schlie{\ss}en, da{\ss} diese
Reihenentwicklungen relativ gut konvergieren, wenn die beiden
Exponentialparameter sich nur wenig unterscheiden, und da{\ss} sie schlecht
konvergieren, wenn die beiden Exponentialparameter sich stark
unterscheiden. Die Konvergenz dieser Reihen kann aber mit Hilfe des
Wynnschen $\epsilon$-Algorithmus, Gl. (2.4-10), betr\"achtlich verbessert
werden [Grotendorst, Weniger und Steinborn 1986].

Es gibt zahlreiche formale Analogien zwischen \"Uberlappungs- und
Coulombintegralen. Dabei spielt das in Gl. (8.2-50) definierte
Zweizentrenkernanziehungsintegral $A_{n, \ell}^{m} (\alpha, \vec{R})$
eine besondere Rolle. Wenn man beispielsweise in den Darstellungen
(8.2-48) und (8.3-1) des \"Uberlappungsinte\-grals die $B$-Funktionen
durch Kernanziehungsintegrale mit identischen Indizes ersetzt, erh\"alt
man die analogen Darstellungen (8.2-49) und (8.3-25) f\"ur
Coulombintegrale.

In diesem Zusammenhang stellt sich die Frage nach der effizienten und
verl\"a{\ss}lichen Berechnung dieser Kernanziehungsintegrale. Mit Hilfe der
Darstellung (8.2-51), die nur aus einer irregul\"aren r\"aumlichen
Kugelfunktion und einer einfachen endlichen Summe von $B$-Funktionen
besteht, ist ohne Zweifel eine sehr effiziente Berechnung m\"oglich.
Ungl\"ucklicherweise wird die Darstellung (8.2-51) numerisch instabil f\"ur
$R \to 0$. Bei LCAO-Rechnungen an realen Molek\"ulen k\"onnen sehr kleine
Kernabst\"ande $R$ nicht vorkommen, und die Darstellung (8.2-51) ist f\"ur
diese Zwecke v\"ollig ausreichend. Wenn man aber Coulombintegrale mit
verschiedenen Exponentialparametern unter Verwendung der von Homeier und
Steinborn [1990] eingef\"uhrten M\"obius-Quadratur durch Auswertung der
Integraldarstellung (8.3-36) berechnen will, was wahrscheinlich das
effizienteste Verfahren ist, dann mu{\ss} man in der Lage sein,
Kernanziehungsintegrale $A_{n, \ell}^{m} (\alpha, \vec{R})$ auch f\"ur
sehr kleine Werte von $\alpha R$ schnell und zuverl\"assig zu berechnen.

Wenn die $B$-Funktion im Kernanziehungsintegral $A_{n, \ell}^{m}
(\alpha, \vec{R})$ eine Ordnung $n \ge 1$ besitzt, kann man sie im
Prinzip als Basisfunktion in einer LCAO-MO-Rechnung verwenden. Au{\ss}erdem
ist die Reihendarstellung (8.2-52) dann f\"ur alle Drehimpulsquantenzahlen
$\ell \ge 0$ frei von Beitr\"agen, die f\"ur $R \to 0$ singul\"ar werden.

Die unendliche Reihe (8.2-52) ist ein Spezialfall der folgenden
Reihenentwicklung einer irregul\"aren r\"aumlichen Kugelfunktion nach
$B$-Funktionen [Filter und Steinborn 1978a, Gl. (6.7)]:
$$
{\cal Z}_{\ell}^m (\alpha \vec R) \; = \;
\frac {1} {(2 \ell - 1)!!} \,
\sum_{\nu=0}^{\infty} \, B_{\nu - \ell, \ell}^m (\alpha, \vec R) \, .
\tag
$$

Man kann leicht zeigen, da{\ss} die unendliche Reihe (8.4-1) konvergiert.
Wenn man in der unendlichen Reihe f\"ur den Abbruchfehler
$$
{\cal Z}_{\ell}^m (\vec R) \, - \,
\frac {1} {(2 \ell - 1)!!} \,
\sum_{\nu=0}^{n + \ell} \, B_{\nu - \ell, \ell}^m (\alpha, \vec R)
\; = \;
\frac {1} {(2 \ell - 1)!!} \,
\sum_{\nu=0}^{\infty} \, B_{n + \nu + 1, \ell}^m (\alpha, \vec R)
\, ,
\tag
$$
die bis auf einen konstanten Vorfaktor identisch ist mit der unendlichen
Reihe in Gl. (8.2-52), die reduzierten Besselfunktionen gem\"a{\ss} Gl.
(8.3-5) durch ihren Wert am Nullpunkt ersetzt, erh\"alt man eine
hypergeometrische Reihe ${}_2 F_1$ mit Einheitsargument. Diese Reihe
kann mit Hilfe eines Summationstheorems von Gau{\ss} [Slater 1966, Gl.
(\Roemisch{3}.1)] in geschlossener Form aufsummiert werden, und man
erh\"alt die folgende obere Schranke:
$$
\biggl\vert \frac {1} {(2 \ell - 1)!!} \,
\sum_{\nu=0}^{\infty} \, B_{n + \nu + 1, \ell}^m (\alpha, \vec R)
\biggr\vert
\; \le \; \frac {(1/2)_n} {2^{\ell} (2 \ell + 1)!! (n + \ell)!} \,
\Bigl\vert {\cal Y}_{\ell}^{m} (\alpha \vec{R}) \Bigr\vert \, \, .
\tag
$$

Auf analoge Weise kann man auch den dominanten Term der asymptotischen
Entwicklung des Abbruchfehlers f\"ur $n \to \infty$ bestimmen
[Grotendorst, Weniger und Steinborn 1986, Gln. (7.17) und (7.18)]:
$$
\frac {1} {(2 \ell - 1)!!} \, \sum_{\nu=0}^{\infty} \,
B_{n + \nu + 1, \ell}^m (\alpha, \vec R)
\; \sim \; \frac {(2/\pi)^{1/2}} {(2 \ell + 1)!!} \,
{\cal Y}_{\ell}^{m} (\alpha \vec{R}) \,
(2 n)^{- \ell - 1/2} \, .
\tag
$$
Ein Vergleich dieser Beziehung mit Gl. (2.1-8) zeigt, da{\ss} die unendliche
Reihe (8.4-1) und damit auch die Reihendarstellung (8.2-52) f\"ur das
Kernanziehungsintegral $A_{n, \ell}^{m} (\alpha \vec{R})$ {\it
logarithmisch\/} konvergiert. Man kann ein \"ahnliches Konvergenzverhalten
erwarten wie bei der Reihenentwicklung (2.1-4) f\"ur die Riemannsche
Zetafunktion $\zeta (\ell + 3/2)$, die f\"ur ihre schlechte Konvergenz
bekannt ist.

Da der Abbruchfehler von der Ordnung $O (n^{- \ell - 1/2})$ f\"ur $n \to
\infty$ ist, wird die Konvergenz der unendlichen Reihe (8.4-1) mit
wachsender Drehimpulsquantenzahl $\ell$ besser. Das ist ein ganz
ungew\"ohnliches Verhalten. Bei Mehrzentrenmolek\"ulintegralen h\"angt
normalerweise sowohl die analytische Komplexit\"at als auch der numerische
Aufwand sehr stark von der Gr\"o{\ss}e der beteiligten Drehimpulsquantenzahlen
ab. Mehrzentrenintegrale {\it skalarer\/} Funktionen $(\ell = 0)$ kann
man oft selbst dann noch vergleichsweise leicht berechnen, wenn die
analogen Integrale {\it anisotroper\/} Funktionen $(\ell > 0)$ nicht
mehr mit vertretbarem Aufwand berechnet werden
k\"onnen{\footnote[\dagger]{Diese Erfahrung erkl\"art, warum
Mehrzentrenintegrale skalarer Funktionen bei der Pr\"asentation
sogenannter {\it benchmarks\/} so beliebt sind}}.

Die unendliche Reihe (8.2-52) f\"ur das Kernanziehungsintegral $A_{n,
\ell}^{m} (\alpha \vec{R})$ wird also immer sehr schlecht konvergieren,
und die schlechteste Konvergenz ist f\"ur $\ell = 0$ zu erwarten. In
diesem Fall entspricht die Reihenentwicklung (8.4-1) der folgenden
Reihenentwicklung des Coulombpotentials nach reduzierten
Besselfunktionen [Filter und Steinborn 1978a, Gl. (6.5)]:
$$
1 / z \; = \; \sum_{m=0}^{\infty} \,
{\widehat k}_{m-1/2} (z) \, / \, [2^m \, m!] \, , \qquad z > 0 \, .
\tag
$$

Aus der asymptotischen Absch\"atzung (8.4-4) folgt, da{\ss} der Abbruchfehler
dieser unendlichen Reihe von der Ordnung $O (n^{- 1/2})$ f\"ur $n \to
\infty$ ist. Das bedeutet, da{\ss} man nur eine einzige Dezimalstelle
gewinnt, wenn man die Summengrenze $n$ der Partialsumme
$$
s_n (z) \; = \; \sum_{m=0}^{n} \,
{\widehat k}_{m-1/2} (z) \, / \, [2^m \, m!]
\tag
$$
um eine Faktor 100 vergr\"o{\ss}ert. Die unendliche Reihe (8.4-5) ist also f\"ur
numerische Zwecke nicht verwendbar. Beispielsweise ergaben 1~000~000
Terme der unendlichen Reihe (8.4-5) f\"ur $z = 1$ nur eine Genauigkeit von
3 Dezimalstellen [Grotendorst, Weniger und Steinborn 1986, Table
\Roemisch{1}].

Gl\"ucklicherweise kann man die Konvergenz der unendlichen Reihe (8.4-5)
mit Hilfe von verallgemeinerten Summationsprozessen ganz erheblich
verbessern. In Table \Roemisch{2} von Grotendorst, Weniger und Steinborn
[1986] wurde gezeigt, da{\ss} die Levinsche $u$-Transformation, Gl.
(5.2-13), mit $\zeta = 1$ die Konvergenz so sehr beschleunigt, da{\ss} die
unendliche Reihe (8.4-5) tats\"achlich f\"ur numerische Zwecke verwendbar
wird.

Die oben erw\"ahnten Ergebnisse sind aber noch nicht das Optimum, das man
mit Hilfe der Levinschen Transformation ${\cal L}_{k}^{(n)} (\zeta ,
s_n, \omega_n)$, Gl. (5.2-5), und ihren Varianten erreichen kann
[Weniger 1989, Abschnitt 14.4; Steinborn und Weniger 1990]. Die
Levinschen Transformation ${\cal L}_{k}^{(n)} (\zeta , s_n, \omega_n)$
enth\"alt den Parameter $\zeta$, der als positiv vorausgesetzt wird,
ansonsten aber im Prinzip frei w\"ahlbar ist. In den meisten F\"allen
verwendet man quasi automatisch den Wert $\zeta = 1$. Manchmal erh\"alt
man aber bessere Ergebnisse, wenn man einen anderen Wert f\"ur $\zeta$
w\"ahlt. Numerische Tests ergaben beispielsweise, da{\ss} $\zeta = 1/2$ im
Falle der unendlichen Reihe (8.4-5) etwas bessere Ergebnisse liefert als
der \"ublichere Wert $\zeta = 1$ [Weniger 1989, S. 359].

Wenn man die Levinsche $u$-Transformation, Gl. (5.2-13), zur
Beschleunigung der Konvergenz der unendlichen Reihe (8.4-5) verwendet,
geht man aufgrund von Gl. (5.2-12) implizit davon aus, da{\ss} das Produkt
$(\zeta + n) \, {\widehat k}_{n-1/2} (z)$ in der Lage ist, den
Abbruchfehler der unendlichen Reihe (8.4-5) mit ausreichender
Genauigkeit zu approximieren. F\"ur kleinere Werte von $n$ und f\"ur gr\"o{\ss}ere
Werte von $z$ ist das aber nicht unbedingt gew\"ahrleistet. Bei der
Ableitung sowohl der Ungleichung (8.4-3) als auch der asymptotischen
Absch\"atzung (8.4-4) wurden die reduzierten Besselfunktionen in der
unendlichen Reihe (8.4-1) durch ihren Wert am Nullpunkt ersetzt. Das ist
eine zul\"assige N\"aherung, da eine reduzierte Besselfunktion ${\widehat
k}_{n+1/2} (z)$ mit $z > 0$ und $n \in \N_0$ gem\"a{\ss} Gl. (8.3-5) positiv
und durch ihren Wert am Nullpunkt beschr\"ankt ist. Da aber reduzierte
Besselfunktionen exponentiell fallen, ist ihr Wert am Nullpunkt nur dann
eine gute Approximation f\"ur Funktionen mit von Null verschiedenem
Argument, wenn $n$ gro{\ss} ist. Numerische Tests ergaben beispielsweise,
da{\ss} man $n \ge 1400$ ben\"otigt, um eine reduzierte Besselfunktion
${\widehat k}_{n+1/2} (z)$ mit dem Argument $z = 8$ mit einem Fehler von
h\"ochstens einem Prozent durch ihren Wert am Nullpunkt zu approximieren.
F\"ur $z = 4$ ben\"otigt man immerhin noch $n \ge 400$.

Es ist also zu vermuten, da{\ss} die Partialsummen (8.4-6) der unendlichen
Reihe (8.4-5) f\"ur kleinere Werte von $n$ sich im wesentlichen wie
Linearkombinationen exponentiell fallender Gr\"o{\ss}en und damit irregul\"ar
verhalten. Man wird also erst f\"ur relativ gro{\ss}e Werte von $n$ ein
regul\"ares Verhalten beobachten, das in Einklang mit der asymptotischen
Absch\"atzung (8.4-4) des Abbruchfehlers ist.

Man sollte also die Effizienz der Levinschen Transformation im Falle der
unendlichen Reihe (8.4-5) noch verbessern k\"onnen, wenn man anstelle der
Restsummenabsch\"atzung
$$
\omega_n \; = \; (\zeta + n) \, {\widehat k}_{n-1/2} (z) \, ,
\qquad n \in \N_0 \, ,
\tag
$$
die f\"ur die $u$-Transformation, Gl. (5.2-13), charakteristisch ist, in
Gl. (5.2-5) bessere Restsummenabsch\"atzungen verwendet. Beispielsweise
legt die Ungleichung (8.4-3) die folgende Restsummenabsch\"atzung nahe:
$$
\omega_n \; = \; \frac {(1/2)_n} {n!} \; = \;
\frac {(2 n - 1)!!} {(2 n)!!} \, , \qquad n \in \N_0 \, .
\tag
$$
Als Alternative k\"onnte man auch einfach den $n$-abh\"angigen Anteil des
dominanten Terms (8.4-4) der asymptotischen Entwicklung des
Abbruchfehlers als Restsummenabsch\"atzung verwenden:
$$
\omega_n \; = \; (n + 1)^{- 1/2} \, ,
\qquad n \in \N_0 \, .
\tag
$$

Wenn man diese beiden Restsummenabsch\"atzungen in Gl. (5.2-5), verwendet,
ist ${\cal L}_{k}^{(n)} (\zeta , s_n, \omega_n)$ eine {\it lineare\/}
Transformation. Trotzdem erh\"alt man auf diese Weise bessere Ergebnisse
als mit Hilfe der {\it nichtlinearen\/} $u$-Transformation.

In Tabelle 8-1 wird die Konvergenz der Partialsummen (8.4-6) f\"ur $z = 1
/ 2$ durch $u_k^{(n)} (\zeta, s_n)$, Gl. (5.2-13), und durch ${\cal
L}_k^{(n)} (\zeta, s_n, \omega_n)$, Gl. (5.2-6), mit entweder $\omega_n
= (n + 1)^{- 1/2}$ oder $\omega_n = (1/2)_n / n!$ beschleunigt. In allen
F\"allen wurde $\zeta = 1/2$ verwendet.

\beginFloat

\medskip

\beginTabelle [to \kolumnenbreite]
\beginFormat \rechts " \mitte " \mitte " \mitte " \mitte
\endFormat
\+ " \links {\bf Tabelle 8-1} \@ \@ \@ \@ " \\
\+ " \links {Beschleunigung der Konvergenz der Reihenentwicklung (8.4-5)
f\"ur $z = 1/2$} \@ \@ \@ \@ " \\
\- " \- " \- " \- " \- " \- " \\ \sstrut {} {1.5 \jot} {1.5 \jot}
\+ " \rechts {$n$} " \mitte {Partialsumme $s_n (z)$}
" $u_n^{(0)} (1/2, s_0)$
" ${\cal L}_n^{(0)} (1/2, s_0, \omega_0)$
" ${\cal L}_n^{(0)} (1/2, s_0, \omega_0)$
" \\ \sstrut {} {1 \jot} {1.5 \jot}
\+ " " \mitte {Gl. (8.4-6)} " Gl. (5.2-13) " Gl. (5.2-6) " Gl. (5.2-6)
" \\ \sstrut {} {1.5 \jot} {1 \jot}
\+  " " " " $\omega_n = (n+1)^{- 1/2}$ "
$\omega_n = (1/2)_n / n!$ " \\ \sstrut {} {1 \jot} {1 \jot}
\- " \- " \- " \- " \- " \- " \\ \sstrut {} {1 \jot} {1 \jot}
\+ "  7 " 1.79119644919306  "  2.00215775399122 "  1.99994898411663 "
1.99996077568249 " \\
\+ "  8 " 1.80416331715800  "  2.00043327863322 "  2.00000078796349 "
1.99999884090829 " \\
\+ "  9 " 1.81498286983696  "  1.99962798058347 "  2.00000243519952 "
2.00000228871071 " \\
\+ " 10 " 1.82418849438350  "  2.00008193424340 "  1.99999946697387 "
1.99999957527198 " \\
\+ " 11 " 1.83214495396498  "  2.00000542454101 "  1.99999999411426 "
1.99999997936909 " \\
\+ " 12 " 1.83911121482253  "  1.99999184482571 "  2.00000002628991 "
2.00000002482268 " \\
\+ " 13 " 1.84527686403448  "  2.00000213245876 "  1.99999999452425 "
1.99999999541264 " \\
\+ " 14 " 1.85078430023364  "  1.99999990763032 "  2.00000000003907 "
1.99999999991230 " \\
\+ " 15 " 1.85574282782753  "  1.99999989059952 "  2.00000000023090 "
2.00000000022303 " \\
\+ " 16 " 1.86023792768766  "  2.00000003775919 "  1.99999999994587 "
1.99999999995245 " \\
\+ " 17 " 1.86433754045711  "  1.99999999543626 "  2.00000000000271 "
2.00000000000158 " \\
\+ " 18 " 1.86809643694590  "  1.99999999919590 "  2.00000000000160 "
2.00000000000160 " \\
\+ " 19 " 1.87155932764697  "  2.00000000047323 "  1.99999999999952 "
1.99999999999956 " \\
\+ " 20 " 1.87476311978502  "  1.99999999990910 "  2.00000000000005 "
2.00000000000004 " \\
\+ " 21 " 1.87773858493861  "  2.00000000000144 "  2.00000000000001 "
2.00000000000001 " \\
\+ " 22 " 1.88051161088152  "  2.00000000000405 "  2.00000000000000 "
2.00000000000000 " \\
\+ " 23 " 1.88310415482930  "  1.99999999999882 "  2.00000000000000 "
2.00000000000000 " \\
\- " \- " \- " \- " \- " \- " \\ \sstrut {} {1 \jot} {1 \jot}
\+ " \links {Exakt} \@      "  2.00000000000000 "  2.00000000000000 "
2.00000000000000 " \\
\- " \- " \- " \- " \- " \- " \\ \sstrut {} {1 \jot} {1 \jot}

\endTabelle

\medskip

\endFloat

Die Ergebnisse in Tabelle 8-1 zeigen, da{\ss} die Restsummenabsch\"atzungen
(8.4-8) und (8.4-9), die aus der Levinschen Transformation einen {\it
linearen\/} verallgemeinerten Summationsproze{\ss} machen, deutlich bessere
Ergebnisse produzieren als die Restsummenabsch\"atzung (8.4-7), die die
Basis der {\it nichtlinearen\/} $u$-Transformation, Gl. (5.2-13), ist.

Die Partialsummen und die Transformationen in Tabelle 8-1 wurden in
FORTRAN in QUA\-DRUPLE PRECISION (31 - 32 Dezimalstellen) berechnet. In
DOUBLE PRECISION (15 - 16 Dezimalstellen) produzierte die
$u$-Transformation f\"ur $n = 15$ eine relative Genauigkeit von 8
Dezimalstellen. Die beiden anderen Transformationen produzierten
ebenfalls f\"ur $n = 15$ eine relative Genauigkeit von 10 Dezimalstellen.
F\"ur gr\"o{\ss}ere Werte von $n$ wurden die Ergebnisse wieder schlechter.
Dieses Beispiel zeigt deutlich, da{\ss} Rundungsfehler und daraus
resultierende numerische Instabilit\"aten bei der Beschleunigung
logarithmischer Konvergenz ein ernstzunehmendes Problem darstellen.

Die Konvergenz der unendlichen Reihe (8.4-1) und ganz besonders des
skalaren Spezialfalles (8.4-5) ist extrem langsam. Die effiziente und
genaue Berechnung der Kernanziehungsintegrale $A_{n, \ell}^{m} (\alpha,
\vec{R})$ auch f\"ur sehr kleine Werte von $\alpha R$ und f\"ur kleine Werte
der Drehimpulsquantenzahl $\ell$ ist aber sehr wichtig, wenn man
Coulombintegrale mit verschiedenen Exponentialparametern durch
Auswertung der Integraldarstellung (8.3-36) unter Verwendung der von
Homeier und Steinborn [1990] eingef\"uhrten M\"obius-Quadraturen berechnen
will. Demzufolge ist es sicherlich sinnvoll, nach anderen
verallgemeinerten Summationsprozessen zu suchen, die m\"oglicherweise noch
bessere Ergebnisse liefern als die in Tabelle 8-1 verwendeten Varianten
der Levinsche Transformation.

Aufgrund der asymptotischen Absch\"atzung (8.4-4) des Abbruchfehlers der
unendlichen Reihe (8.4-1) sollten die Elemente von Modellfolgen des Typs
$$
s_n \; = \; s \, + \,
\sum_{j=0}^{\infty} \, c_j / (n+1)^{\alpha + j} \, ,
\qquad n \in \N_0 \, ,
\tag
$$
die durch einen konstanten Abklingparameter $\alpha > 0$ charakterisiert
sind, zumindest f\"ur gr\"o{\ss}ere Werte von $n$ in der Lage sein, die
Partialsummen der logarithmisch konvergenten unendlichen Reihe (8.4-1)
mit ausreichender Genauigkeit zu approximieren, wenn man in Gl. (8.4-10)
den Abklingparameter $\alpha = \ell + 1/2$ verwendet. Verallgemeinerte
Summationsprozesse, die die Konvergenz der Modellfolge (8.4-10) auf
effiziente Weise beschleunigen, sollten also im Prinzip auch zur
Beschleunigung der Konvergenz der Partialsummen der unendlichen Reihe
(8.4-1) geeignet sein.

Wenn der Wert des Abklingparameters $\alpha$ einer Folge vom Typ von Gl.
(8.4-10) bekannt ist, kann die Konvergenz einer solchen Folge laut Osada
[1990a, Gl. (3.1)] mit Hilfe der folgenden Verallgemeinerung der
Standardform des Wynnschen $\rho$-Algorithmus, Gl. (3.3-15),
beschleunigt werden:
$$
\beginAligntags
" {\bar \rho}_{-1}^{(n)} " \; = \; " 0 \, ,
\hfill {\bar \rho}_0^{(n)} \; = \; s_n \, ,
\erhoehe\aktTag \\ \tag*{\tagnr a}
" {\bar \rho}_{k+1}^{(n)} " \; = \; " {\bar \rho}_{k-1}^{(n+1)} \, + \,
\frac {k+\alpha} {{\bar \rho}_{k}^{(n+1)} - {\bar \rho}_{k}^{(n)} }
\, , \qquad \alpha > 0 \, , \quad k,n \in \N_0 \, .
\\ \tag*{\tagform\aktTagnr b}
\endAligntags
$$
F\"ur $\alpha = 1$ stimmt Osada's Variante des $\rho$-Algorithmus mit der
Standardform des $\rho$-Algorithmus, Gl. (3.3-15), \"uberein.

Osada konnte zeigen, da{\ss} seine Variante des $\rho$-Algorithmus die
Konvergenz einer Folge vom Typ von Gl. (8.4-10) beschleunigt, und da{\ss}
der Transformationsfehler die folgende asymptotische Absch\"atzung erf\"ullt
[Osada 1990a, Theorem 4]:
$$
{\bar \rho}_{2 k}^{(n)} \, - \, s \; = \;
O (n^{- \alpha - 2 k}) \, , \qquad n \to \infty \, .
\tag
$$

Osada's Variante des $\rho$-Algorithmus kann analog zu der in Abschnitt
3.3 beschriebenen Vorgehensweise iteriert werden. Dabei mu{\ss} man ${\bar
\rho}_2^{(n)}$ durch die Folgenelemente $s_n$, $s_{n+1}$ und $s_{n+2}$
ausdr\"ucken. Aus Gl. (8.4-11) folgt [Weniger 1991, Gl. (2.28)]:
$$
{\bar \rho}_2^{(n)} \; = \;
s_{n+1} \, - \, \frac {(\alpha + 1)} {\alpha} \,
\frac {[\Delta s_n] [\Delta s_{n+1}]} {[\Delta^2 s_n]} \, ,
\qquad n \in \N_0 \, .
\tag
$$
Wenn man diesen Ausdruck so iteriert, da{\ss} $\alpha$ nach jedem
Rekursionsschritt durch $\alpha + 2$ ersetzt wird, erh\"alt man die
folgende Transformation [Weniger 1991, Gl. (2.29)]:
$$
\beginAligntags
" {\overline {\cal W}}_{0}^{(n)} \; " = \; " s_n \, ,
\qquad n \in \N_0 \, , \quad
\erhoehe\aktTag \\ \tag*{\tagnr a}
" {\overline {\cal W}}_{k+1}^{(n)} \; " = \;
" {\overline {\cal W}}_{k}^{(n+1)} \, - \,
\frac {(2 k + \alpha + 1)} {(2 k + \alpha)} \,
\frac
{\bigl[ \Delta {\overline {\cal W}}_k^{(n+1)} \bigr]
\bigl[ \Delta {\overline {\cal W}}_k^{(n)} \bigr] }
{\Delta^2 {\overline {\cal W}}_k^{(n)} }
\, , \qquad \alpha > 0 \, , \quad k,n \in \N_0 \, . \qquad
\\ \tag*{\tagform\aktTagnr b}
\endAligntags
$$
Diese Iteration von ${\bar \rho}_2^{(n)}$ ist aber identisch mit einer
Modifikation des Aitkenschen iterierten $\Delta^2$-Algorithmus, Gl.
(3.3-8), die von Bj{\o}rstad, Dahlquist und Grosse [1981, Gl. (2.4)]
eingef\"uhrt wurde. F\"ur $\alpha = 1$ stimmt die Transformation ${\overline
{\cal W}}_k^{(n)}$ mit der Standardform (3.3-18) der durch Iteration des
Wynnschen $\rho$-Algorithmus abgeleiteten Transformation ${\cal
W}_k^{(n)}$, Gl. (3.3-17), \"uberein.

Die Transformation ${\overline {\cal W}}_k^{(n)}$ beschleunigt die
Konvergenz einer Folge vom Typ von Gl. (8.4-10), und der
Transformationsfehler erf\"ullt die folgende asymptotische Absch\"atzung
[Bj{\o}rstad, Dahlquist und Grosse 1981, Gl. (3.1)]:
$$
{\overline {\cal W}}_k^{(n)} - s \; = \;
O (n^{- \alpha - 2 k}) \, , \qquad n \to \infty \, .
\tag
$$

Ein Vergleich der asymptotischen Fehlerabsch\"atzungen (8.4-12) und
(8.4-15) zeigt, da{\ss} ${\bar \rho}_{2 k}^{(n)}$ und ${\overline {\cal
W}}_k^{(n)}$, die zu ihrer Berechnung den gleichen String $s_n$,
$s_{n+1}$, $\ldots$ , $s_{n + 2 k}$ von Folgenelementen ben\"otigen, im
Falle einer Folge vom Typ von Gl. (8.4-10) asymptotisch \"aquivalente
Ergebnisse liefern.

Die Fehlerabsch\"atzungen (8.4-12) und (8.4-15) implizieren, da{\ss} durch
zweimalige Anwendungen der Rekursion (8.4-11b) beziehungsweise durch
einmalige Anwendung der Rekursion (8.4-14b) die beiden f\"uhrenden Terme
der Summe auf der rechten Seite von Gl. (8.4.10) eliminiert werden. Da
man zur Elimination der beiden f\"uhrenden Terme die numerischen Werte
zweier Elemente der Folge ben\"otigt, kann man folgern, da{\ss} die
Transformationen (8.4-11) und (8.4-14) zumindest asymptotisch {\it
optimal\/} sind. Kein verallgemeinerter Summationsproze{\ss}, der nur die
numerischen Werte von $2 k + 1$ Elementen $s_n$, $s_{n+1}$, $\ldots$ ,
$s_{n + 2 k}$ einer Folge vom Typ von Gl. (8.4-10) und den Wert des
Abklingparameters $\alpha$ als Eingabedaten verwendet, kann also eine bessere
asymptotische Absch\"atzung des Transformationsfehlers als $O (n^{- \alpha
- 2 k})$ erreichen.

Bei Osada's Variante des $\rho$-Algorithmus, Gl. (8.4-11), werden die
Approximationen zum Grenzwert $s$ der zu transformierenden Folge $\Seqn
s$ wie bei der Standardform des $\rho$-Algorithmus, Gl. (3.3-15),
gew\"ahlt:
$$
\left\{ s_{m - 2 \Ent {m/2}}, s_{m - 2 \Ent {m/2} + 1}, \ldots ,
s_m \right\} \; \to \;
{\bar \rho}_{2 \Ent {m/2}}^{(m - 2 \Ent {m/2})} \, .
\tag
$$

Bei der von Bj{\o}rstad, Dahlquist und Grosse [1981, Gl. (2.4)]
eingef\"uhrten Transformation (8.4-14) werden die Approximationen zum
Grenzwert $s$ der zu transformierenden Folge $\Seqn s$ folgenderma{\ss}en
gew\"ahlt:
$$
\left\{ s_{m - 2 \Ent {m/2}}, s_{m - 2 \Ent {m/2} + 1}, \ldots ,
s_m \right\} \; \to \;
{\overline {\cal W}}_{\Ent {m/2}}^{(m - 2 \Ent {m/2})} \, .
\tag
$$

In Gln. (8.4-16) und (8.4-17) wurde die Notation $\Ent x$ f\"ur den
ganzzahligen Anteil von $x$ verwendet, der die gr\"o{\ss}te ganze Zahl $\nu$
ist, welche die Beziehung $\nu \le x$ erf\"ullt.

Ein anderer verallgemeinerter Summationsproze{\ss}, der die Konvergenz von
Folgen vom Typ von Gl. (8.4-10) beschleunigen kann, ist die folgende
Iteration des von Brezinski [1971] eingef\"uhrten $\theta$-Algorithmus,
Gl. (4.4-13) [Weniger 1989, Abschnitt 10.3]:
$$
\beginAligntags
" {\cal J}_0^{(n)} \; " = \; " s_n \, , \qquad n \in \N_0 \, ,
\erhoehe\aktTag \\ \tag*{\tagnr a}
" {\cal J}_{k+1}^{(n)} \; " = \;
" {\cal J}_k^{(n+1)} \, - \, \frac
{[\Delta {\cal J}_k^{(n)}] \, [\Delta {\cal J}_k^{(n+1)}]
\, [\Delta^2 {\cal J}_k^{(n+1)}]}
{[\Delta {\cal J}_k^{(n+2)}] \, [\Delta^2 {\cal J}_k^{(n)}]
\, - \,
[\Delta {\cal J}_k^{(n)}] \, [\Delta^2 {\cal J}_k^{(n+1)}]}
\, , \qquad k,n \in \N_0 \, .
\\ \tag*{\tagform\aktTagnr b}
\endAligntags
$$
Bei dieser Transformation werden die Approximationen zum Grenzwert $s$
der zu transformierenden Folge $\Seqn s$ folgenderma{\ss}en gew\"ahlt [Weniger
1989, Gl. (10.4-7)]:
$$
\left\{ s_{m - 3 \Ent {m/3}}, s_{m - 3 \Ent {m/3} + 1}, \ldots , s_m
\right\} \; \to \;
{\cal J}_{\Ent {m/3}}^{(m - 3 \Ent {m/3})} \, .
\tag
$$

Bei praktischen Problemen kommt es relativ h\"aufig vor, da{\ss} man vermutet,
da{\ss} die Elemente einer Folge $\Seqn s$ vom Typ von Gl. (8.4-10) sind,
ohne aber den Wert des Abklingparameters $\alpha$ explizit zu kennen.
Eine Approximation von $\alpha$ erh\"alt man mit Hilfe der Transformation
$$
T_n \; = \; \frac
{ [\Delta^2 s_n ] \, [\Delta^2 s_{n+1} ]}
{[\Delta s_{n+1} ] \, [\Delta^2 s_{n+1} ] \, - \,
[\Delta s_{n+2} ] \, [\Delta^2 s_{n}]} \; - \; 1 \, ,
\qquad n \in \N_0 \, ,
\tag
$$
die zuerst in einer etwas versteckten Form von Drummond [1976, S. 419]
und sp\"ater von Bj{\o}rstad, Dahlquist und Grosse [1981] erneut
abgeleitet wurde. Bj{\o}rstad, Dahlquist und Grosse [1981, Gl. (4.1)]
bewiesen ebenfalls die folgende Absch\"atzung:
$$
\alpha \; = \; T_n \, + \, O (n^{- 2}) \, , \qquad n \to \infty \, .
\tag
$$

Wenn man den Abklingparameter $\alpha$ einer Folge vom Typ von Gl.
(8.4-10) durch die Approximation (8.4-20) ersetzt und diese Beziehung in
Gl. (8.4-13) verwendet, erh\"alt man [Drummond 1976, S. 419; Weniger
1991, S. 38]:
$$
\theta_2^{(n)} \; = \; s_{n+1} \, - \, \frac
{[\Delta s_n] \, [\Delta s_{n+1}] \, [\Delta^2 s_{n+1}]}
{[\Delta s_{n+2}] \, [\Delta^2 s_n] -
[\Delta s_n] \, [\Delta^2 s_{n+1}]}
\, , \qquad n \in \N_0 \, .
\tag
$$
Diese Beziehung kann iteriert werden und ergibt Gl. (8.4-18) [Weniger
1989, Abschnitt 10.3].

Die Anwendung von ${\cal J}_k^{(n)}$ auf eine Folge vom Typ von Gl.
(8.4-10) kann also folgenderma{\ss}en interpretiert werden: Zuerst wird eine
{\it lokale\/} Approximation des Abklingparameters $\alpha$ gem\"a{\ss} Gl.
(8.4-20) berechnet, die in Gl. (8.4-13) eingesetzt wird. In n\"achsten
Schritt wird der modifizierte $\Delta^2$-Proze{\ss} ${\bar \rho}_2^{(n)}$,
Gl. (8.4-13), auf die Elemente der zu transformierenden Folge
angewendet. Danach wird aus den Elementen ${\cal J}_1^{(n)}$ eine neue
lokale Approximation des Abklingparameters $\alpha$ berechnet, die dann
wieder in Gl. (8.4-13) eingesetzt wird.

Da man zur Approximation des lokalen Abklingparameters gem\"a{\ss} Gl.
(8.4-20) ${\cal J}_{k-1}^{(n)}$, ${\cal J}_{k-1}^{(n+1)}$, ${\cal
J}_{k-1}^{(n+2)}$ und ${\cal J}_{k-1}^{(n+3)}$ ben\"otigt, zur
eigentlichen Transformation gem\"a{\ss} Gl. (8.4-13) aber nur ${\cal
J}_{k-1}^{(n)}$, ${\cal J}_{k-1}^{(n+1)}$ und ${\cal J}_{k-1}^{(n+2)}$,
sollte ${\cal J}_k^{(n)}$ die Konvergenz einer Folge vom Typ von Gl.
(8.4-10) etwas weniger gut beschleunigen als ${\bar \rho}_k^{(n)}$, Gl.
(8.4-11), oder ${\overline {\cal W}}_k^{(n)}$, Gl. (8.4-11), die beide
von einem festen und bekanntem Abklingparameter $\alpha$ ausgehen.

In Tabelle 8-2 wird die Konvergenz der Partialsummen (8.4-6) f\"ur $z = 1
/ 2$ durch ${\cal J}_k^{(n)}$, Gl. (8.4-18), ${\bar \rho}_k^{(n)}$, Gl.
(8.4-11), und ${\overline {\cal W}}_k^{(n)}$, Gl. (8.4-11), beschleunigt
[ Weniger 1991, Table 3].

\beginFloat

\medskip

\beginTabelle [to \kolumnenbreite]
\beginFormat \rechts " \mitte " \mitte " \mitte " \mitte
\endFormat
\+ " \links {\bf Tabelle 8-2} \@ \@ \@ \@ " \\
\+ " \links {Beschleunigung der Konvergenz der Reihenentwicklung (8.4-5)
f\"ur $z = 1/2$} \@ \@ \@ \@ " \\
\- " \- " \- " \- " \- " \- " \\ \sstrut {} {1.5 \jot} {1.5 \jot}
\+ " \rechts {$n$} " \mitte {Partialsumme $s_n (z)$}
" ${\cal J}_{\Ent {n/3}}^{(n - 3 \Ent {n/3})}$
" ${\bar \rho}_{2 \Ent {n/2}}^{(n - 2 \Ent {n/2})}$
" ${\overline {\cal W}}_{\Ent {n/2}}^{(n - 2 \Ent {n/2})}$ "
\\ \sstrut {} {1 \jot} {1.5 \jot}
\+ " " \mitte {Gl. (8.4-6)} " Gl. (8.4-18) " Gl. (8.4-11) " Gl. (8.4-14)
" \\ \sstrut {} {1.5 \jot} {1 \jot}
\- " \- " \- " \- " \- " \- " \\ \sstrut {} {1 \jot} {1 \jot}
\+ "  5 " 1.75503950544225 " 1.98841710819656 " 1.99972008107657 "
1.99973616846417 " \\
\+ "  6 " 1.77526500936798 " 1.99726498861717 " 2.00007282508438 "
2.00005026956598 " \\
\+ "  7 " 1.79119644919306 " 1.99939364880553 " 2.00000572616163 "
2.00000474873864 " \\
\+ "  8 " 1.80416331715800 " 1.99986117743342 " 1.99999840648539 "
1.99999930118309 " \\
\+ "  9 " 1.81498286983696 " 1.99999473527258 " 1.99999990208882 "
1.99999994421466 " \\
\+ " 10 " 1.82418849438350 " 1.99999668298955 " 2.00000002833230 "
2.00000000720034 " \\
\+ " 11 " 1.83214495396498 " 1.99999923108310 " 2.00000000143007 "
2.00000000065046 " \\
\+ " 12 " 1.83911121482253 " 1.99999900242050 " 1.99999999957556 "
1.99999999995198 " \\
\+ " 13 " 1.84527686403448 " 1.99999999108074 " 1.99999999998181 "
2.00000000000660 " \\
\+ " 14 " 1.85078430023364 " 1.99999999865937 " 2.00000000000550 "
1.99999999997799 " \\
\+ " 15 " 1.85574282782753 " 2.00000000033829 " 2.00000000000020 "
1.99999999998031 " \\
\+ " 16 " 1.86023792768766 " 2.00000000000460 " 1.99999999999994 "
1.99999999997750 " \\
\+ " 17 " 1.86433754045711 " 2.00000000000018 " 2.00000000000000 "
1.99999999999999 " \\
\+ " 18 " 1.86809643694590 " 1.99999999999931 " 2.00000000000000 "
2.00000000000000 " \\
\+ " 19 " 1.87155932764697 " 1.99999999999959 " 2.00000000000000 "
2.00000000000000 " \\
\+ " 20 " 1.87476311978502 " 1.99999999999958 " 2.00000000000000 "
2.00000000000000 " \\
\- " \- " \- " \- " \- " \- " \\ \sstrut {} {1 \jot} {1 \jot}
\+ " \links {Exakt} \@ " 2.00000000000000 " 2.00000000000000 "
2.00000000000000 " \\
\- " \- " \- " \- " \- " \- " \\ \sstrut {} {1 \jot} {1 \jot}

\endTabelle

\medskip

\endFloat

Die Partialsummen und die Transformationen in Tabelle 8-1 wurden in
FORTRAN in QUA\-DRUPLE PRECISION (31 - 32 Dezimalstellen) berechnet. Die
stabilsten Ergebnisse in DOUBLE PRECISION (15 - 16 Dezimalstellen)
wurden von ${\bar \rho}_k^{(n)}$, Gl. (8.4-11), erzieht. Diese
Transformation verlor f\"ur $n = 17$ nur drei dezimale Stellen durch
Rundungsfehler.

Die Ergebnisse in Tabelle 8-2 zeigen, da{\ss} wie vermutet sowohl ${\bar
\rho}_k^{(n)}$ als auch ${\overline {\cal W}}_k^{(n)}$ bessere
Ergebnisse liefern als ${\cal J}_k^{(n)}$. Ein Vergleich der Tabellen
8-1 und 8-2 zeigt au{\ss}erdem, da{\ss} die Transformationen ${\bar
\rho}_k^{(n)}$ und ${\overline {\cal W}}_k^{(n)}$ deutlich bessere
Ergebnisse liefern als die Varianten der Levinschen Transformation, die in
Tabelle 8-1 verwendet wurden.

Wenn man das Coulombintegral zweier $B$-Funktionen mit verschiedenen
Exponentialparametern $\alpha$ und $\beta$ durch Auswertung der
Integraldarstellung (8.3-36) berechnen will, wobei das Coulombintegral
mit gleichen Exponentialparametern im Integral in Gl. (8.3-36) unter
Verwendung der Darstellung (8.3-25) berechnet werden soll, mu{\ss} man f\"ur
jeden $\ell$-Wert in Gl. (8.3-25) h\"ochstens das Kernanziehungsintegral
$A_{n_1 + n_2 + \ell_1 + \ell_2 - \ell + 1, \ell}^{m_2 - m_1}$ durch
Auswertung der unendlichen Reihe in Gl. (8.2-52) berechnen. Die anderen
Kernanziehungsintegrale erh\"alt man mit Hilfe der Rekursionsformel
[Weniger, Grotendorst und Steinborn 1986b, Gl. (6.18)]
$$
A_{n, \ell}^{m} (\alpha, \vec{R}) \; = \;
A_{n + 1, \ell}^{m} (\alpha, \vec{R}) \, + \,
\frac {4 \pi} {\alpha^2} \, B_{n, \ell}^{m} (\alpha, \vec{R}) \, ,
\tag
$$
die abw\"artsstabil ist.

Eine Alternative zur \"ublichen Darstellung von
Mehrelektronenwellenfunktionen durch Linearkombinationen von
Slaterdeterminanten ist die Verwendung von explizit korrelierten
Funktionen. Solche Basisfunktionen enthalten nicht nur Produkte von
Einteilchenfunktionen, sondern h\"angen auch von den Relativkoordinaten
$\vert \vec{r}_1 - \vec{r}_2 \vert$ zweier Elektronen ab. Mit Hilfe
solcher Funktionen kann man die Wechselwirkung der Elektronen wesentlich
besser beschreiben als durch die \"ublichen
Mehrelektronenwellenfunktionen, die Linearkombinationen von
Slaterdeterminanten sind. Demzufolge konvergieren Rechnungen mit
explizit korrelierten Basisfunktionen wesentlich schneller als etwa
CI-Verfahren [Kutzelnigg 1988, S. 47; Rychlewski 1994]. Allerdings
treten bei einem solchen Ansatz wesentlich kompliziertere Integrale auf
als bei SCF-Rechnungen auf der Basis der Roothaanschen Gleichungen
[Roothaan 1951]. Schon die Berechnung atomarer Drei- und
Vierelektronenintegrale bereitet gro{\ss}e Probleme [Abbott und Maslen 1987;
Fromm und Hill 1987; Gottschalk, Abbott und Maslen 1987; Gottschalk und
Maslen 1987; King 1991; 1993; King, Dykema und Lund 1992]. L\"uchow [1993]
und L\"uchow und Kleindienst [1993] konnten zeigen, da{\ss} verallgemeinerte
Summationsprozesse auch bei solchen Integralen sehr n\"utzlich sind.

\endAbschnittsebene

\endAbschnittsebene

\keinTitelblatt\neueSeite

\beginAbschnittsebene
\aktAbschnitt = 8

\Abschnitt Rationale Approximationen f\"ur Hilfsfunktionen in der
Quantenchemie

\vskip - 2 \jot

\beginAbschnittsebene

\medskip

\Abschnitt Gau{\ss}funktionen als quantenchemische Basisfunktionen

\smallskip

\aktTag = 0

Wie schon in Abschnitt 8.2 erw\"ahnt wurde, besitzen exakte L\"osungen
atomarer und molekularer Schr\"odingergleichungen an den Kernorten einen
{\it Cusp\/} [Kato 1957], und f\"ur gro{\ss}e Abst\"ande fallen sie {\it
exponentiell\/} (siehe beispielsweise [Agmon 1982; 1985; Ahlrichs 1989;
Cycon, Froese, Kirsch und Simon 1987, Abschnitt 4; Herbst 1993]). Wenn
der verwendete Basissatz bestimmte Vollst\"andigkeitskriterien wie
beispielsweise die Kriterien von Michlin oder Kato [Klahn und Bingel
1977a, Abschnitte 4 und 5] erf\"ullt, dann konvergiert eine
Variationsrechnung f\"ur die Energie{\footnote[\dagger]{Das bedeutet
nicht, da{\ss} man damit auch alle physikalischen Eigenschaften des
betrachteten Systems berechnen kann. Klahn und Morgan [1984] zeigten,
da{\ss} die Konvergenz eines Variationsansatzes gegen die korrekte Energie
{\it nicht\/} die Konvergenz der Erwartungswerte physikalischer
Eigenschaften impliziert.}} auch dann, wenn eine endliche Anzahl von
Basisfunktionen weder die Cusps noch den exponentiellen Abfall
reproduzieren k\"onnen. Die Konvergenzgeschwindigkeit eines
Variationsansatzes wird aber betr\"achtlich erh\"oht, wenn der verwendete
Basissatz die Cusps und den exponentiellen Abfall exakter L\"osungen gut
beschreiben kann [Hill 1985; Klopper und Kutzelnigg 1986; Kutzelnigg und
Morgan 1992; Morgan 1989]. Dabei d\"urfte der exponentielle Abfall
zumindest bei Berechnungen der Energie weniger wichtig sein als eine
ausreichend genaue Beschreibung der Cusps.

Wenn man in LCAO-MO-Rechnungen exponentialartige Basisfunktionen
verwendet, deren Radialteile die allgemeine Struktur {\it
Exponentialfunktion in $r$ $\times$ Polynom in $r$\/} besitzen, kann man
sowohl die Cusps als auch den exponentiellen Abfall exakter L\"osungen
reproduzieren, und man erh\"alt schon mit relativ kleinen Basiss\"atzen gute
Ergebnisse. Dieser unbestreitbare Vorteil exponentialartiger
Basisfunktionen wird aber zumindest bei Molek\"ulrechnungen dadurch
zunichte gemacht, da{\ss} man die komplizierten Mehrzentrenmolek\"ulintegrale,
die im Rahmen eines LCAO-MO-Ansatzes zwangsl\"aufig auftreten, immer noch
nicht auf befriedigende Weise berechnen kann.

Offensichtlich determinieren die analytischen Eigenschaften der
verwendeten Basisfunktionen weitgehend die Ergebnisse, die man in
atomaren oder molekularen SCF-Rechnungen auf der Basis der
Root\-haanschen Gleichungen [Roothaan 1951] erhalten kann. Vielfach wird
aber \"ubersehen, da{\ss} die Probleme, die man mit der Berechnung der
entsprechenden Mehrzentrenmolek\"ulintegrale hat, ebenfalls sehr stark von
den analytischen Eigenschaften der Basisfunktionen beeinflu{\ss}t werden.
Allerdings besteht hier eine {\it inverse\/} Beziehung: Je besser ein
Basissatz die Singularit\"aten exakter L\"osungen atomarer und molekularer
Schr\"odingergleichungen und ihren exponentiellen Abfall reproduzieren
kann, desto komplizierter scheinen die entsprechenden
Mehrzentrenmolek\"ulintegrale zu sein.

Um Mehrzentrenmolek\"ulintegrale auf effiziente Weise berechnen zu k\"onnen,
mu{\ss} man \"ahnlich wie bei partiellen Differentialgleichungen eine {\it
Separation\/} der Variablen durchf\"uhren. Auf diese Weise kann man die im
Rahmen eines LCAO-MO-Ansatzes auftretenden drei- und sechsdimensionalen
Mehrzentrenintegrale als Produkte von Integralen niederer
Dimensionalit\"at darstellen, die dann entweder analytisch oder numerisch
berechnet werden. Bei exponentialartigen Basisfunktionen wird eine
solche Separation der Variablen aber gerade durch diejenigen
Eigenschaften erschwert, die eine schnelle Konvergenz atomarer oder
molekularer SCF-Rechnungen erm\"oglichen. Demzufolge sind die Hauptvorz\"uge
exponentialartiger Basisfunktionen in LCAO-MO-Rechnungen gleichzeitig
verantwortlich f\"ur ihre schwerwiegendsten Nachteile.

Eine Separation der Integrationsvariablen in
Mehrzentrenmolek\"ulintegralen wird erleichtert, wenn die Basisfunktionen
{\it analytische\/} Funktionen der cartesischen Komponenten des
Ortsvektors $\vec{r} = (x, y, z)$ im Sinne der Funktionentheorie sind,
da dann konvergente Reihenentwicklungen in $x$, $y$ und $z$ existieren.
Die Exponentialfunktion $\exp (- z)$ ist zwar ein Lehrbuchbeispiel einer
analytischen Funktion in der Variablen $z \in \C$, aber $\exp ( - \alpha
r)$ mit $\alpha > 0$ ist {\it keine\/} analytische Funktion der
cartesischen Komponenten des Ortsvektors $\vec{r} = (x, y, z)$, da der
Abstand $r = [x^2 + y^2 + z^2]^{1/2}$ keine analytische Funktion von
$x$, $y$ und $z$ ist. Dagegen ist $r^2 = x^2 + y^2 + z^2$ oder jedes
Polynom in $x$, $y$ und $z$ offensichtlich eine analytische Funktion von
$x$, $y$ und $z$.

Wenn man fordert, da{\ss} Basisfunktionen in einem LCAO-MO-Ansatz
irreduzible sph\"arische Tensoren sind,
$$
F_{\ell}^{m} (\vec{r}) \; = \; f_{\ell} (r)
{\cal Y}_{\ell}^{m} (\vec{r}) \, ,
\tag
$$
wobei ${\cal Y}_{\ell}^{m}$ eine in Gl. (8.2-2) definierte regul\"are
r\"aumliche Kugelfunktion ist, dann kann eine solche Basisfunktion nur
dann analytisch in $x$, $y$ und $z$ sein, wenn die Reihenentwicklung der
Radialfunktion $f_{\ell} (r)$ um $r = 0$ nur gerade Potenzen enth\"alt
[Weniger 1985, S. 284]:
$$
f_{\ell} (r) \; = \; \sum_{m=0}^{\infty} \,
\left\{ \left( \frac {\d} {\d r} \right)^{2 m} \, f_{\ell} (r)
\right\}_{r = 0} \, \frac {r^{2 m}} {(2 m)!} \, .
\tag
$$
Daraus folgt sofort, da{\ss} alle sph\"arische Tensoren, deren Radialteile die
allgemeine Form {\it Exponentialfunktion in $r$ $\times$ Polynom in
$r$\/} besitzen, nicht analytisch in $x$, $y$ und $z$ sein k\"onnen.

Dagegen sind die von Boys [1955] eingef\"uhrten cartesischen
Gau{\ss}funktionen
$$
{\cal C} (\lambda, \mu, \nu; \alpha, \vec{r}) \; = \;
x^{\lambda} y^{\mu} z^{\nu} \, \exp (- \alpha r^2) \, ,
\qquad \lambda, \mu, \nu \in \N_0 \, , \quad \alpha > 0 \, ,
\tag
$$
offensichtlich analytische Funktionen in $x$, $y$ und $z$. Aus Gln.
(9.1-1) und (9.1-2) folgt, da{\ss} {\it sph\"arische\/} Gau{\ss}funktionen des
Typs
$$
G_{n, \ell}^{m} (\alpha, \vec{r}) \; = \;
r^{2 n} \, \exp (- \alpha r^2) \, {\cal Y}_{\ell}^{m} (\vec{r}) \, ,
\qquad \alpha > 0 \, ,
\tag
$$
nur dann analytische Funktionen in $x$, $y$ und $z$ sind, wenn die
Ordnung $n$ eine nichtnegative ganze Zahl ist.

Aufgrund der Vollst\"andigkeitseigenschaften der Gau{\ss}funktionen [Klahn und
Bingel 1977b, Abschnitt 5] ist die Konvergenz von LCAO-MO-Rechnungen
gew\"ahrleistet, obwohl endliche Linearkombinationen von Gau{\ss}funktionen
weder einen Cusp noch den exponentiellen Abfall exakter L\"osungen
atomarer und molekularer Schr\"odingergleichungen reproduzieren k\"onnen.
Allerdings konvergieren LCAO-MO-Rechnungen mit einer Gau{\ss}basis relativ
schlecht, und man ben\"otigt deutlich mehr analytische Gau{\ss}funktionen vom
Typ von Gl. (9.1-3) oder (9.1-4) als Slaterfunktionen, um Ergebnisse
vergleichbarer Genauigkeit zu erzielen{\footnote[\dagger]{Eine
ausreichend genaue Darstellung der Cusps an den Kernorten ist
offensichtlich sehr wichtig f\"ur eine schnelle Konvergenz einer
Variationsrechnung. Klopper und Kutzelnigg [1986] konnten zeigen, da{\ss}
man die Cusps relativ genau reproduzieren und damit die Konvergenz
deutlich verbessern kann, wenn man in einer Variationsrechnung nicht nur
{\it analytische\/} Gau{\ss}funktionen, sondern auch {\it
nichtanalytische\/} Gau{\ss}funktionen des Typs $r^{2 n + 1} \exp (- \alpha
r^2) {\cal Y}_{\ell}^{m} (\vec{r})$ mit $n \in \N_0$ verwendet.}}. Der
einzige und allerdings entscheidende Vorteil von Gau{\ss}funktionen im
Vergleich zu exponentialartigen Funktionen ist die relative
Leichtigkeit, mit der Mehrzentrenmolek\"ulintegrale berechnet werden
k\"onnen.

Ein prinzipielles Verfahren zur Separation von Variablen in
Mehrzentrenmolek\"ulintegralen besteht in der Verwendung sogenannter
Additionstheoreme. Ein Additionstheorem ist eine Reihenentwicklung eines
sph\"arischen Tensors $F_{\ell}^{m}$, der von der Summe zweier Vektoren
$\vec{r}_1$ und $\vec{r}_2$ abh\"angt, nach irreduziblen sph\"arischen
Tensoren $U_{n_1, \ell_1}^{m_1}$ und $V_{n_2, \ell_2}^{m_2}$, die
entweder von $\vec{r}_1$ oder von $\vec{r}_2$ abh\"angen:
$$
F_{\ell}^{m} (\vec{r}_1 + \vec{r}_2) \; = \;
\sum_{n_1 \, \ell_1 \, m_1} \, \sum_{n_2 \, \ell_2 \, m_2} \,
U_{n_1, \ell_1}^{m_1} (\vec{r}_1) \,
V_{n_2, \ell_2}^{m_2} (\vec{r}_2) \, .
\tag
$$
Wenn ein solches Additionstheorem {\it punktweise\/} konvergiert, dann
ist es eine Umordnung einer dreidimensionalen Taylorentwicklung von
$F_{\ell}^{m} (\vec{r}_1 + \vec{r}_2)$ um entweder $\vec{r}_1$ oder
$\vec{r}_2$. Es gibt auch Additionstheoreme sph\"arischer Tensoren
$F_{\ell}^{m}$, die {\it im Mittel\/}, d.~h., bez\"uglich der Norm eines
geeigneten Hilbert- oder Sobolevraumes{\footnote[\dagger]{Sobolevr\"aume
und ihre mathematischen Eigenschaften werden beispielsweise in B\"uchern
von Adams [1975], Blanchard und Br\"uning [1992, Appendix D], Maz'ja
[1985], Michlin [1978, Abschnitt 4], Sobolev [1963], Weidmann [1980,
Abschnitt 10.2], Wloka [1982, Abschnitt \Roemisch{1}], und Ziemer [1989]
behandelt. Anwendungen von Sobolevr\"aumen in der Quantenmechanik findet
man beispielsweise in dem Buch von Albeverio, Gesztesy, H{\o}egh-Krohn
und Holden [1988]}} konvergieren [Filter und Steinborn 1980; Homeier,
Weniger und Steinborn 1992a; Novosadov 1983; Pauli und Alder 1976;
Sawaguri und Tobocman 1967; Weniger 1985]. Solche Additionstheoreme
sollen hier aber nicht behandelt werden, da nicht bekannt ist, wie die
analytischen Eigenschaften des sph\"arischen Tensors $F_{\ell}^{m}$ die
Eigenschaften eines normkonvergenten Additionstheorems beeinflussen.

Die Vorteile von Gau{\ss}funktionen gegen\"uber exponentialartigen Funktionen
werden augenf\"allig, wenn man die entsprechenden punktweise konvergenten
Additionstheoreme vergleicht. So besitzt beispielsweise eine
$1s$-Gau{\ss}funktion das folgende punktweise konvergente Additionstheorem
[Kaufmann und Baumeister 1989, Gl. (9); Seeger 1982, Gl. (19)]:
$$
\beginAligntags
" \exp \bigl( - \alpha [\vec{r}_1 + \vec{r}_2]^2 \bigr) \; = \;
4 \pi \, \exp \bigl( - \alpha [r_1^2 + r_2^2] \bigr) \\
" \qquad \times \,
\sum_{\ell = 0}^{\infty} \, \sum_{m = - \ell}^{\ell} \,
(- \i)^{\ell} \, j_{\ell} (\i \alpha r_1 r_2) \,
\bigl[ Y_{\ell}^{m} (\vec{r}_1 / r_1) \bigr]^{*} \,
Y_{\ell}^{m} (\vec{r}_2 / r_2) \, ,
\\ \tag
\endAligntags
$$
wobei $j_{\ell}$ eine sph\"arische Besselfunktion ist [Abramowitz und
Stegun 1972, S. 437]:
$$
j_{\ell} (z) \; = \; \bigl[ \pi / (2 z) \bigr]^{1/2} \,
J_{\ell + 1/2} (z) \, , \qquad \ell \in \N_0 \, .
\tag
$$

Gau{\ss}funktionen sind analytische Funktionen der cartesischen Komponenten
des Ortsvektors $\vec{r} = (x, y, z)$. Demzufolge kann man eine
$1s$-Gau{\ss}funktion mit dem Argument $\vec{r}_1 + \vec{r}_2$ sowohl um den
Punkt $\vec{r}_1$ als auch um den Punkt $\vec{r}_2$ in eine
dreidimensionale Taylorreihe entwickeln. Das Additionstheorem (9.1-6),
das eine Umordnung einer dreidimensionalen Taylorreihe ist, konvergiert
also f\"ur alle Werte der Ortsvektoren $\vec{r}_1$ und $\vec{r}_2$, und
die Rolle der Variablen $\vec{r}_1$ und $\vec{r}_2$ im Additionstheorem
ist symmetrisch.

Die Situation ist v\"ollig anders bei exponentialartigen Funktionen, die
am Ursprung nicht analytisch sind. Eine dreidimensionale Taylorreihe in
$x$, $y$ und $z$ um den Nullpunkt existiert nicht. Demzufolge kann man
eine exponentialartige Funktion mit dem Argument $\vec{r}_1 + \vec{r}_2$
nicht sowohl um $\vec{r}_1$ als auch um $\vec{r}_2$ in eine
dreidimensionale Taylorreihe entwickeln, sondern nur um den betragsm\"a{\ss}ig
gr\"o{\ss}eren der beiden Vektoren $\vec{r}_1$ und $\vec{r}_2$, und der
Verschiebungsvektor mu{\ss} der betragsm\"a{\ss}ig kleinere der beiden Vektoren
$\vec{r}_1$ und $\vec{r}_2$ sein. Ein punktweise konvergentes
Additionstheorem einer exponentialartigen Funktion besitzt also einen
{\it endlichen\/} Konvergenzradius, und die nat\"urlichen Variablen im
Additionstheorem sind nicht die Vektoren $\vec{r}_1$ und $\vec{r}_2$,
sondern die Vektoren $\vec{r}_{<}$ und $\vec{r}_{>}$, die laut
Voraussetzung die Beziehung $\vert \vec{r}_{<} \vert < \vert \vec{r}_{>}
\vert$ erf\"ullen.

Bekanntlich hat das Yukawapotential $\exp (- \alpha r)/r$ [Yukawa 1935]
von allen skalaren exponentialartigen Funktionen das einfachste
punktweise konvergente Additionstheorem. Aufgrund der Beziehung [Weniger
und Steinborn 1985, Gl. (6.10)]
$$
\exp (- \alpha r)/r \; = \; (4 \pi)^{1/2} \, \alpha \,
B_{0, 0}^{0} (\alpha, \vec{r})
\tag
$$
kann dieses Additionstheorem unter Verwendung der Variablen
$\vec{r}_{<}$ und $\vec{r}_{>}$ auch auf folgende Weise geschrieben
werden [Weniger und Steinborn 1985, Gl. (6.11)]:
$$
\beginAligntags
" B_{0, 0}^{0} (\alpha, \vec{r}_{<} + \vec{r}_{>}) \\
" \qquad \; = \; (2 \pi^2)^{1/2} \,
\sum_{\ell = 0}^{\infty} \, \sum_{m = - \ell}^{\ell} \, (- 1)^{\ell} \,
(\alpha r_{<})^{- \ell - 1/2} I_{\ell} (\alpha r_{<}) \,
\bigl[ {\cal Y}_{\ell}^{m} (\alpha \vec{r}_{<}) \bigr]^{*} \,
B_{-\ell, \ell}^{m} (\alpha \vec{r}_{>}) \, .
\\ \tag
\endAligntags
$$
Dieses Additionstheorem besitzt aufgrund seiner Abh\"angigkeit von den
Variablen $\vec{r}_{<}$ und $\vec{r}_{>}$ eine \"ahnliche Struktur wie die
bekannte Laplaceentwicklung des Coulompotentials, die auch auf folgende
Weise geschrieben werden kann [Weniger und Steinborn 1985, Gl. (4.2)]:
$$
\frac {1} {\vert \vec{r}_{<} + \vec{r}_{>} \vert} \; = \;
4 \pi \, \sum_{\ell = 0}^{\infty} \, \sum_{m = - \ell}^{\ell} \,
\frac {(- 1)^{\ell}} {2 \ell + 1} \,
\bigl[ {\cal Y}_{\ell}^{m} (\vec{r}_{<}) \bigr]^{*} \,
{\cal Z}_{\ell}^{m} (\alpha \vec{r}_{>}) \, .
\tag
$$

Die strukturellen Unterschiede der beiden Additionstheoreme (9.1-6) und
(9.1-9) sind direkte Konsequenzen der unterschiedlichen Eigenschaften
der $1s$-Gau{\ss}funktion und des Yukawapotentials. Die $1s$-Gau{\ss}funktion
ist f\"ur alle $\vec{r} \in \R^3$ analytisch. Demzufolge konvergiert eine
dreidimensionale Taylorreihe in $x$, $y$ und $z$ um den Nullpunkt, und
das Additionstheorem (9.1-6) besitzt aufgrund seiner Symmetrie in
$\vec{r}_1$ und $\vec{r}_2$ eine {\it Einbereichsform\/}. Dagegen ist
das Yukawapotential am Ursprung nicht analytisch. Demzufolge besitzt das
Additionstheorem (9.1-9) einen endlichen Konvergenzradius, der zu einer
charakteristischen {\it Zweibereichsform\/} f\"uhrt.

Additionstheoreme k\"onnen zur Separation der Variablen in
Mehrzentrenmolek\"ulintegralen verwendet werden. Dabei kann das
Additionstheorem (9.1-6) aufgrund seiner Einbereichsform relativ leicht
angewendet werden, da die beiden Variablen $\vec{r}_1$ und $\vec{r}_2$
v\"ollig separiert sind und im Additionstheorem auf symmetrische Weise
auftreten. Nach Ausf\"uhrung der Winkelintegrationen mu{\ss} man in einem
Mehrzentrenintegral dann nur noch vergleichsweise einfache
Radialintegrale mit den Integrationsgrenzen Null und Unendlich berechnen.

Dagegen ist die Verwendung des Additionstheorems (9.1-9) in
Mehrzentrenmolek\"ulintegralen aufgrund seiner Zweibereichsform keineswegs
einfach. Da die nat\"urlichen Variablen dieses Additionstheorems die
Vektoren $\vec{r}_{<}$ und $\vec{r}_{>}$ sind, mu{\ss} man beim Integrieren
zwischen $r_1 < r_2$ und $r_1 > r_2$ unterscheiden, was ein
schwerwiegender Nachteil ist. Nach Ausf\"uhrung der Winkelintegrationen
ist man deswegen mit Radialintegralen konfrontiert, deren funktionale
Form sich in Abh\"angigkeit von der Gr\"o{\ss}e der Integrationsvariablen
\"andert. Zur Berechnung solcher Radialintegrale ben\"otigt man {\it
unbestimmte\/} Integrale spezieller Funktionen. Ein Blick in eine
Integraltafel wie beispielsweise Gradshteyn und Ryzhik [1980] lehrt
aber, da{\ss} nur relativ wenige unbestimmte Integrale spezieller Funktionen
bekannt sind, und wenn sie bekannt sind, dann sind sie in der Regel
wesentlich komplizierter als die analogen bestimmten Integrale mit den
Integrationsgrenzen Null und Unendlich.

Die unterschiedlichen Eigenschaften der beiden Additionstheoreme (9.1-6)
und (9.1-9) sind nicht spezifisch f\"ur die $1s$-Gau{\ss}funktion und das
Yukawapotential. Beim Additionstheorem anisotroper Gau{\ss}funktionen des
Typs $\exp(- \alpha [\vec{r}_1 + \vec{r}_2]^2) {\cal Y}_{\ell}^{m}
(\vec{r}_1 + \vec{r}_2)$ ist die Rolle der beiden Vektoren $\vec{r}_1$
und $\vec{r}_2$ ebenfalls v\"ollig symmetrisch [Kaufmann und Baumeister
1989, Gln. (24) und (25); Seeger 1982, Gl. (20)]. Dagegen weist das
punktweise konvergente Additionstheorem (8.2-54) einer $B$-Funktion die
charakteristische Zweibereichsform auf, die bei allen am Ursprung
nichtanalytischen Funktionen auftreten mu{\ss}. Da alle in der Quantenchemie
\"ublicherweise verwendeten exponentialartigen Basisfunktionen wie etwa
Slaterfunktionen oder Wasserstoffeigenfunktionen durch einfache
Linearkombinationen von $B$-Funktionen dargestellt werden k\"onnen (siehe
beispielsweise Filter und Steinborn [1978a, S. 83], Homeier [1990, S. 56
- 60] und Weniger [1985, S. 282 - 283]), weisen ihre punktweise
konvergenten Additionstheoreme ebenfalls die typische Zweibereichsform
auf.

Wahrscheinlich beeinflussen die analytischen Eigenschaften der Funktion
auch die Konvergenz\-eigenschaften des betreffenden Additionstheorems.
Wenn eine Potenzreihe
$$
f (z) \; = \; \sum_{\nu=0}^{\infty} \, \gamma_{\nu} z^{\nu}
\tag
$$
nur innerhalb eines Konvergenzkreises mit Radius $R$ konvergiert, k\"onnen
die Koeffizienten $\gamma_{\nu}$ f\"ur $\nu \to \infty$ bestenfalls
relativ langsam gegen Null gehen. Demzufolge kann eine Potenzreihe mit
endlichem Konvergenzradius nur dann sehr schnell konvergieren, wenn das
Argument $z$ betragsm\"a{\ss}ig sehr klein ist. Wenn dagegen eine solche
Potenzreihe einen unendlichen Konvergenzradius besitzt, m\"ussen die
Koeffizienten $\gamma_{\nu}$ f\"ur $\nu \to \infty$ sehr schnell gegen
Null gehen. Potenzreihen mit unendlichem Konvergenzradius konvergieren
also nur dann schlecht, wenn das Argument $z$ betragsm\"a{\ss}ig sehr gro{\ss}
ist.

Punktweise konvergente Additionstheoreme sind Umordnungen
dreidimensionaler Potenzreihen. Man m\"u{\ss}te noch genauer untersuchen, wie
Umordnungen sich auf das Konvergenzverhalten einer dreidimensionalen
Potenzreihe auswirken. Trotzdem ist es eine plausible Annahme, da{\ss}
Additionstheoreme nichtanalytischer Funktionen, die einen endlichen
Konvergenzradius besitzen, normalerweise schlechter konvergieren als
Additionstheoreme analytischer Funktionen, die einen unendlichen
Konvergenzradius besitzen.

\medskip

\Abschnitt Mehrzentrenmolek\"ulintegrale von Gau{\ss}funktionen

\smallskip

\aktTag = 0

Im letzten Unterabschnitt wurde am Beispiel des Additionstheorems
(9.1-9) des Yukawapotentials gezeigt, da{\ss} eine Separation der Variablen
in Mehrzentrenmolek\"ulintegralen bei exponentialartigen Funktionen mit
Hilfe von Additionstheoremen aufgrund ihrer Nichtanalytizit\"at am
Ursprung \"au{\ss}erst schwierig ist. Man kann nat\"urlich trotzdem versuchen,
Mehrzentrenintegrale exponentialartiger Funktionen unter Verwendung
punktweise konvergenter Additionstheoreme zu berechnen. Da die
\"ublicherweise verwendeten exponentialartigen Basisfunktionen irreduzible
sph\"arische Tensoren sind, ist man im allgemeinen Fall eines
vierzentrigen Elektronenwechselwirkungsintegrals mit mehrfach
unendlichen Reihenentwicklungen nach Kugelfl\"achenfunktionen
konfrontiert, deren allgemeine Struktur laut Steinborn und Filter [1979]
eine direkte Konsequenz des Verhaltens der verwendeten Basisfunktionen
gegen\"uber Drehungen ist. Die effiziente und genaue Berechnung solcher
mehrfach unendlichen Reihen ist ohne Zweifel ein schwieriges numerisches
Problem. Inwieweit verallgemeinerte Summationsprozesse dabei helfen
k\"onnen, m\"u{\ss}te noch untersucht werden. Das Hauptproblem dieser
Reihenentwicklungen d\"urften aber die nach Ausf\"uhrung der
Winkelintegrationen verbleibenden Radialintegrale sein, die aufgrund der
charakteristischen Zweibereichsform der Additionstheoreme
nichtanalytischer Funktionen \"au{\ss}erst kompliziert sind.

Dagegen ist bei den Gau{\ss}funktionen, die am Ursprung analytisch sind,
eine Separation der Variablen in Mehrzentrenmolek\"ulintegralen durch
Additionstheoreme wesentlich einfacher. Zwar w\"are man auch bei
Gau{\ss}funktionen mit mehrfach unendlichen Reihen nach
Kugelfl\"achenfunktio\-nen konfrontiert. Die Radialintegrale, die in
diesem Zusammenhang auftreten w\"urden, k\"onnten aber aufgrund der
Einbereichsform des Additionstheorems einer Gau{\ss}funktion wesentlich
leichter berechnet werden. Es sollte also im Prinzip m\"oglich sein, die
bei einem LCAO-MO-Ansatz auftretenden Mehrzentrenmolek\"ulintegrale von
Gau{\ss}funktionen mit Hilfe von Additionstheoremen zu berechnen. In der
Praxis geht man nicht so vor, aber nur deswegen, weil es einen
wesentlich einfacheren Weg gibt.

Der Ausgangspunkt ist das bekannte Produkttheorem zweier
$1s$-Gau{\ss}funktionen (siehe beispielsweise Daudel, Leroy, Peeters und
Sana [1983, Gl. (5.45)]),
$$
\beginAligntags
" \exp \bigl(- \alpha [\vec{r} - \vec{A}]^2 \bigr) \,
\exp \bigl(- \beta [\vec{r} - \vec{B}]^2 \bigr) \\
" \qquad \; = \; \exp \bigl(- \, \alpha \beta \,
[\vec{A} - \vec{B}]^2 / [\alpha + \beta] \bigr) \,
\exp \bigl(- [\alpha + \beta] [\vec{r} - \vec{P}]^2 \bigr) \, ,
\\ \tag
\endAligntags
$$
mit dessen Hilfe eine Zweizentrendichte aus $1s$-Gau{\ss}funktionen in eine
Einzentrendichte transformiert werden kann, die aus einer
$1s$-Gau{\ss}funktion an einem neuen Zentrum $\vec{P}$ multipliziert mit
einer $\vec{r}$-unabh\"angigen $1s$-Gau{\ss}funktion besteht. Das Zentrum
$\vec{P}$ der Einzentrendichte ist ein gewichteter Mittelwert der beiden
urspr\"unglichen Zentren $\vec{A}$ und $\vec{B}$:
$$
\vec{P} \; = \;
\frac {\alpha \vec{A} + \beta \vec{B}} {\alpha + \beta} \, .
\tag
$$

Es sind auch explizite Darstellungen f\"ur Zweizentrenprodukte
exponentialartiger Funktionen bekannt. Beispielsweise konnte Homeier
[1990, Abschnitt 3.5.2] f\"ur das Zweizentrenprodukt zweier
$B$-Funktionen, die in Gl. (8.2-5) definiert sind, eine Darstellung
ableiten, deren Komplexit\"at stark von den beteiligten
Drehimpulsquantenzahlen abh\"angt. Aber auch im Falle zweier
$1s$-Funktionen ist diese Darstellung wesentlich komplizierter als die
analoge Beziehung (9.2-1) bei Gau{\ss}funktionen, da sie ein
eindimensionales Integral enth\"alt, das mit Hilfe von Quadraturverfahren
ausgewertet werden mu{\ss}.

In diesem Abschnitt sollen die Mehrzentrenmolek\"ulintegrale von
Gau{\ss}funktionen nur anhand einiger charakteristischer Beispiele behandelt
werden, um die prinzipiell unterschiedlichen Eigenschaften {\it
analytischer\/} Gau{\ss}funktionen und {\it nichtanalytischer\/}
exponentialartiger Funktionen zu verdeutlichen. Eine ausf\"uhrliche
Behandlung der im Rahmen eines LCAO-MO-Ansatzes auftretenden
Mehrzentrenmolek\"ulintegrale von Gau{\ss}funktionen kann man beispielsweise
in Artikeln von Clementi und Corongiu [1989, S. 250 - 269], Hegarty und
van der Velde [1983], Saunders [1975; 1983] und Shavitt [1963, Abschnitt
\Roemisch{4}] oder auch in den B\"uchern von Daudel, Leroy, Peeters und
Sana [1983, Abschnitt 5.2] und Glaeske, Reinhold und Volkmer [1987,
Abschnitt 11.8] finden.

Mit Hilfe des Produkttheorems (9.2-1) kann man das \"Uberlappungsintegral
von $1s$-Gau{\ss}funktio\-nen auf ein Einzentrenintegral reduzieren, dessen
Berechnung trivial ist [Shavitt 1963, Gl. (85)]:
$$
\beginAligntags
" \int \exp \bigl(- \alpha [\vec{r} - \vec{A}]^2 \bigr) \,
\exp \bigl(- \beta [\vec{r} - \vec{B}]^2 \bigr) \, \d^3 \vec{r} \\
" \qquad \; = \; \exp \bigl(- \, \alpha \beta \,
[\vec{A} - \vec{B}]^2 / [\alpha + \beta] \bigr) \,
\int \exp \bigl(- [\alpha + \beta] [\vec{r} - \vec{P}]^2 \bigr) \,
\d^3 \vec{r} \\ \tag
" \qquad \; = \; \bigl( \pi / [\alpha + \beta] \bigr)^{3/2} \,
\exp \bigl(- \, \alpha \beta \,
[\vec{A} - \vec{B}]^2 / [\alpha + \beta] \bigr) \, .
\\ \tag
\endAligntags
$$

Das Produkttheorem der $1s$-Gau{\ss}funktionen kann auch auf anisotrope
Gau{\ss}funktionen verallgemeinert werden, beispielsweise durch
Differentiation nach den beiden Exponentialparametern $\alpha$ und
$\beta$ oder nach den cartesischen Komponenten der Vektoren $\vec{A}$
und $\vec{B}$ [Clementi und Corongiu 1989, Appendix 6B; Saunders 1983,
Abschnitt 3]. Demzufolge k\"onnen auch Zweizentrendichten anisotroper
Gau{\ss}funktionen auf einfache Weise durch Linearkombinationen von
Einzentrendichten ausgedr\"uckt werden. Bei Gau{\ss}funktionen mu{\ss} man
deswegen auch keine Vierzentrenelektronenwechselwirkungsintegrale oder
Dreizentrenkernanziehungsintegrale berechnen, deren effiziente und
verl\"a{\ss}liche Berechnung bei exponentialartigen Funktionen so schwierig
ist, sondern nur Linearkombinationen der wesentlich einfacheren
Zweizentrenelektronenwechselwirkungsintegrale oder
Zweizentrenkernanziehungsintegrale.

Mit Hilfe des Produkttheorems (9.2-1) kann man beispielsweise ein
Vierzentrenintegral, das die Wechselwirkung einer Ladungsdichte aus
$1s$-Gau{\ss}funktionen an den Zentren $\vec{A}$ und $\vec{B}$ mit einer
Ladungsdichte aus $1s$-Gau{\ss}funktionen an den Zentren $\vec{C}$ und
$\vec{D}$ beschreibt, in ein Zweizentren\-integral transformieren, das
die Wechselwirkung der in Gl. (9.2-1) definierten Einzentrendichten mit
einer anderen, am Zentrum
$$
\vec{Q} \; = \;
\frac {\gamma \vec{C} + \delta \vec{D}} {\gamma + \delta}
\tag
$$
lokalisierten Einzentrendichte beschreibt:
$$
\beginAligntags
" \int \! \int
\frac {\exp \bigl(
- \alpha [\vec{r}_1 - \vec{A}]^2 - \beta [\vec{r}_1 - \vec{B}]^2
- \gamma [\vec{r}_2 - \vec{C}]^2 - \delta [\vec{r}_2 - \vec{D}]^2
\bigr)} {\vert \vec{r}_1 - \vec{r}_2 \vert}
\, \d^3 \vec{r}_1 \, \d^3 \vec{r}_2 \\ \sstrut {} {1.5 \jot} {1.5 \jot}
" \qquad \; = \;
\exp \left( - \frac {\alpha \beta \bigl[ \vec{A} - \vec{B} \bigr]^2}
{\alpha + \beta}
- \frac {\gamma \delta \bigl[ \vec{C} - \vec{D} \bigr]^2}
{\gamma + \delta} \right) \\ \sstrut {} {1.5 \jot} {1.5 \jot}
" \qquad \quad \times \, \int \! \int
\frac {\exp \bigl(
- [\alpha + \beta] [\vec{r}_1 - \vec{P}]^2
- [\gamma + \delta] [\vec{r}_2 - \vec{Q}]^2
\bigr)} {\vert \vec{r}_1 - \vec{r}_2 \vert}
\, \d^3 \vec{r}_1 \, \d^3 \vec{r}_2 \, .
\\ \tag
\endAligntags
$$
Mit Hilfe der Integraldarstellung [Gradshteyn und Ryzhik 1980, S. 307]
$$
\frac {1} {\vert \vec{r}_1 - \vec{r}_2 \vert} \; = \;
\frac {2} {\pi^{1/2}} \, \int\nolimits_{0}^{\infty}
\exp \bigl( - \vert \vec{r}_1 - \vec{r}_2 \vert^2 u^2 \bigr) \d u
\tag
$$
kann man das Integral auf der rechten Seite von Gl. (9.2-6) ausf\"uhren
und man erh\"alt f\"ur das Vierzentrenintegral die folgende, \"au{\ss}erst
kompakte geschlossene Darstellung [Shavitt 1963, Gl. (88)]:
$$
\beginAligntags
" \int \! \int
\frac {\exp \bigl(
- \alpha [\vec{r}_1 - \vec{A}]^2 - \beta [\vec{r}_1 - \vec{B}]^2
- \gamma [\vec{r}_2 - \vec{C}]^2 - \delta [\vec{r}_2 - \vec{D}]^2
\bigr)} {\vert \vec{r}_1 - \vec{r}_2 \vert}
\, \d^3 \vec{r}_1 \, \d^3 \vec{r}_2 \\ \sstrut {} {1 \jot} {1 \jot}
" \qquad \; = \; \frac {2 \pi^{5/2}}
{(\alpha + \beta) (\gamma + \delta)
\bigl[\alpha + \beta + \gamma + \delta \bigr]^{1/2}} \\
\sstrut {} {1 \jot} {1 \jot}
" \qquad \quad \times \,
\exp \left( - \frac {\alpha \beta \bigl[ \vec{A} - \vec{B} \bigr]^2}
{\alpha + \beta}
- \frac {\gamma \delta \bigl[ \vec{C} - \vec{D} \bigr]^2}
{\gamma + \delta} \right) \,
F_0 \left( \frac {(\alpha + \beta) (\gamma + \delta)}
{\alpha + \beta + \gamma + \delta} \bigl( \vec{P} - \vec{Q} \bigr)^2
\right) \, . \quad
\\ \tag
\endAligntags
$$

Die in Gl. (9.2-8) vorkommende Funktion $F_0$ ist ein Spezialfall der
folgenden Klasse von Hilfsfunktionen [Shavitt 1963, Gl. (22)]:
$$
F_{m} (z) \; = \;
\int\nolimits_{0}^{1} \, u^{2 m} \, \exp \bigl(- z u^2 \bigr) \, \d u
\, , \qquad m \in \N_0, \quad z > 0 \, .
\tag
$$
Au{\ss}erdem ist die Funktion $F_0$ gem\"a{\ss} [Shavitt 1963, S. 7]
$$
F_{0} (z) \; = \; \frac {1} {z^{1/2}} \, \Funk {erf} (z^{1/2}) \, ,
\tag
$$
ein Spezialfall der Fehlerfunktion $\Funk {erf}$, die folgenderma{\ss}en
definiert ist [Abramowitz und Stegun 1972, Gl. (7.1.1)]:
$$
\Funk {erf} (z) \; = \; \frac {2} {\pi^{1/2}} \,
\int\nolimits_{0}^{z} \, \exp \bigl(- t^2 \bigr) \, \d t \, .
\tag
$$

Auch f\"ur das Dreizentrenkernanziehungsintegral von $1s$-Gau{\ss}funktionen
kann man mit Hilfe des Produkttheorems (9.2-1) und der
Integraldarstellung (9.2-7) einen kompakten geschlossenen Ausdruck
ableiten [Shavitt 1963, Gl. (87)]:
$$
\beginAligntags
" \int
\frac {\exp \bigl(
- \alpha [\vec{r} - \vec{A}]^2 - \beta [\vec{r} - \vec{B}]^2
\bigr)} {\vert \vec{r} - \vec{C} \vert}
\, \d^3 \vec{r} \\ \sstrut {} {1 \jot} {1 \jot}
" \qquad \; = \; \frac {2 \pi}
{\alpha + \beta} \,
\exp \Bigl( - \alpha \beta \bigl[ \vec{A} - \vec{B} \bigr]^2 /
[\alpha + \beta] \Bigr) \,
F_0 \Bigl( (\alpha + \beta) \bigl( \vec{C} - \vec{P} \bigr)^2 \Bigr)
\, .
\\ \tag
\endAligntags
$$

F\"ur Vierzentrenelektronenwechselwirkungsintegrale oder
Dreizentrenkernanziehungsintegrale cartesischer Gau{\ss}funktionen kann man
auf analoge Weise geschlossener Ausdr\"ucke ableiten (siehe beispielsweise
Clementi und Corongiu [1989, Gln. (30) und (47)]). Die resultierenden
Formeln sind nat\"urlich komplizierter als die extrem kompakten
Darstellungen (9.2-8) und (9.2-12) f\"ur $1s$-Gau{\ss}funktionen, enthalten
aber im Prinzip die gleichen Strukturelemente, da sie endliche
Linearkombinationen der in Gl. (9.2-9) definierten Hilfsfunktionen $F_m$
sind. Das Hauptproblem bei der Berechnung der
Elektronenwechselwirkungsintegrale und Kernanziehungsintegralen ist also
die effiziente Berechnung der $F_m$-Funktionen.

Bei exponentialartigen Funktionen wie etwa den in Gl. (8.2-5)
definierten $B$-Funktionen sind keine auch nur ann\"ahernd so kompakte
Darstellungen f\"ur Mehrzentrenmolek\"ulintegrale bekannt. So wurden
beispielsweise f\"ur das Dreizentrenkernanziehungsintegral
$$
\int \, \frac
{\bigl[ B_{n_1, \ell_1}^{m_1} (\alpha, \vec{r}) \bigr]^{*}
B_{n_2, \ell_2}^{m_2} (\beta, \vec{r} - \vec{B})}
{\vert \vec{r} - \vec{C} \vert} \, \d^3 \vec{r}
\tag
$$
verschiedene \"aquivalente ein- und zweidimensionale Integraldarstellungen
abgeleitet, die mit Hilfe von Quadraturverfahren ausgewertet werden
k\"onnen [Grotendorst und Steinborn 1988, Gl. (3.24); Homeier 1990, Gln.
(7.1-34), (7.1-35), (7.1-46), (7.1-57), (7.1-62) und (7.1-67); Homeier
und Steinborn 1991, Gl. (29); Trivedi und Steinborn 1983, Gl. (4.3)].
F\"ur das Vierzentrenintegral
$$
\int \! \! \int \, \frac
{\bigl[ B_{n_1, \ell_1}^{m_1} (\alpha, \vec{r}_1 - \vec{A}) \bigr]^{*}
B_{n_2, \ell_2}^{m_2} (\beta, \vec{r}_1 - \vec{B})
\bigl[ B_{n_3, \ell_3}^{m_3} (\gamma, \vec{r}_2 - \vec{C}) \bigr]^{*}
B_{n_4, \ell_4}^{m_4} (\delta, \vec{r}_2 - \vec{D})}
{\vert \vec{r}_1 - \vec{r}_2 \vert}
\, \d^3 \vec{r}_1 \d^3 \vec{r}_2
\tag
$$
konnten ebenfalls verschiedene dreidimensionale Integraldarstellungen
abgeleitet werden [Grotendorst und Steinborn 1988, Gl. (3.28); Homeier
1990, Gln. (8.1-25) und (8.1-43); Steinborn und Homeier 1990, Gl. (19);
Trivedi und Steinborn 1983, Gl. (5.1)]. Auch in diesem Fall m\"ussen die
dreidimensionalen Integraldarstellungen mit Hilfe von Quadraturverfahren
ausgewertet werden.

Aufgrund der Nichtanalytizit\"at exponentialartiger Funktionen am Ursprung
ist zu bef\"urchten, da{\ss} man f\"ur Mehrzentrenmolek\"ulintegrale
exponentialartiger Funktionen keine Darstellungen finden kann, die auch
nur ann\"ahernd so einfach sind wie die extrem kompakten geschlossenen
Darstellungen der Mehrzentrenmolek\"ulintegrale von Gau{\ss}funktionen.

\medskip

\Abschnitt Eigenschaften der $F_m$-Funktionen

\smallskip

\aktTag = 0

Wie schon im letzten Unterabschnitt erw\"ahnt wurde, k\"onnen die
Elektronenwechselwirkungsintegrale und
Dreizentrenkernanziehungsintegrale von Gau{\ss}funktionen durch endliche
Linearkombinationen der in Gl. (9.2-9) definierten Hilfsfunktionen $F_m
(z)$ dargestellt werden. Wenn man die in Gl. (9.1-4) definierten
cartesischen Gau{\ss}funktionen ${\cal C} (\lambda, \mu, \nu; \alpha,
\vec{r})$ als Basisfunktionen verwendet, h\"angen die Indizes $m$ der in
solchen Linearkombinationen auftretenden Hilfsfunktionen $F_m (z)$ nur
von den Exponenten $\lambda$, $\mu$ und $\nu$ der im Integral
vorkommenden cartesischen Gau{\ss}funktionen ab, und die Argumente $z$ der
Hilfsfunktionen h\"angen nur von den Exponentialparametern der
Gau{\ss}funktionen und den Koordinaten der beteiligten Zentren ab (siehe
beispielsweise Rys, Dupuis und King [1983, S. 155]).

Bei den heute \"ublicherweise verwendeten Molek\"ulprogrammen werden die
Ortsanteile $\varphi_j (\vec{r})$ der zu bestimmenden Spinorbitale
$\psi_j (\vec{r}, \sigma)$ durch endliche Linearkombinationen von
Gau{\ss}funktionen mit verschiedenen Exponentialparametern dargestellt.
Aufgrund der relativ schlechten Konvergenz von SCF-Rechnungen bei
Verwendung einer Basis aus {\it analytischen\/} Gau{\ss}funktionen ben\"otigt
man eine gro{\ss}e Zahl von Basisfunktionen, um eine befriedigende
Genauigkeit zu erzielen. Das f\"uhrt letztlich dazu, da{\ss} man bei einer
Molek\"ulrechnung eine sehr gro{\ss}e Anzahl der in Gl. (9.2-9) definierten
Hilfsfunktionen $F_m (z)$ mit verschiedenen Indizes $m$ und
verschiedenen Argumenten $z$ berechnen mu{\ss}. Demzufolge wird ein
betr\"achtlicher Anteil der Rechenzeit, die zur Berechnung der Integrale
ben\"otigt wird, f\"ur die Berechnung der Hilfsfunktionen $F_m (z)$
verbraucht. Es ist also sicherlich lohnend, die Berechnung der
Hilfsfunktionen $F_m (z)$ soweit wie m\"oglich zu optimieren, da der
akkumulative Effekt selbst kleiner Zeitgewinne zu einer betr\"achtlichen
Reduktion der Gesamtrechenzeit f\"uhren kann.

Dementsprechend umfangreich ist die Literatur \"uber die $F_m$-Funktionen.
Die f\"ur die Berechnung der $F_m$-Funktionen wichtigsten mathematischen
Eigenschaften werden beispielsweise in Artikeln von Clementi und
Corongiu [1989, Appendix 6D], Saunders [1975, Abschnitt 5; 1983,
Abschnitt 5.6] und Shavitt [1963, Abschnitt \Roemisch{2}.C.1] oder in
den B\"uchern von Daudel, Leroy, Peeters und Sana [1983, Abschnitt
5.2.2.7] und Glaeske, Reinhold und Volkmer [1987, Abschnitt 11.8.7]
behandelt.

Einen einfachen Zugang zu den mathematischen Eigenschaften der
$F_m$-Funktionen erm\"oglicht die folgende Beziehung [Shavitt 1963, Gl.
(22)],
$$
F_m (z) \; = \; \frac {1} {2 z^{m+1/2}} \, \gamma (m + 1/2, z) \, ,
\tag
$$
die zeigt, da{\ss} die $F_m$-Funktion im wesentlichen ein Spezialfall der
unvollst\"andigen Gammafunktion ist, die folgenderma{\ss}en definiert ist
[Magnus, Oberhettinger und Soni 1966, S. 337]:
$$
\gamma (a, x) \; = \;
\int\nolimits_{0}^{x} \, t^{a - 1} \, \e^{- t} \, \d t \, ,
\qquad Re (a) > 0 \, .
\tag
$$
Man mu{\ss} also nur bekannte Beziehungen f\"ur die unvollst\"andigen
Gammafunktion $\gamma (a, x)$ gem\"a{\ss} Gl. (9.3-1) \"ubersetzen, um die
entspechenden Beziehungen f\"ur die $F_m$-Funktion zu erhalten.

Aus den Reihendarstellungen [Magnus, Oberhettinger und Soni 1966, S. 337]
$$
\beginAligntags
" \gamma (a, x) " \; = \; " \frac {x^a} {a} \e^{- x} \,
{}_1 F_1 (1; a + 1; x)
\\ \tag
" " \; = \; " \frac {x^a} {a} \, {}_1 F_1 (a; a + 1; - x)
\\ \tag
\endAligntags
$$
der unvollst\"andigen Gammafunktion folgt sofort, da{\ss} die $F_m$-Funktion
die folgenden Reihendarstellungen besitzt:
$$
\beginAligntags
" F_m (z) " \; = \; " \frac {\e^{- z}} {2 m + 1} \,
{}_1 F_1 (1; m + 3/2; z) \; = \;
\frac {\e^{- z}} {2 m + 1} \,
\sum_{\nu = 0}^{\infty} \, \frac {z^{\nu}} {(m + 3/2)_{\nu}}
\\ \tag
" " \; = \; "
\frac {1} {2 m + 1} \, {}_1 F_1 (m + 1/2; m + 3/2; - z)
\; = \; \sum_{\nu = 0}^{\infty} \,
\frac {(- z)^{\nu}} {{\nu}! (2 m + 2 \nu + 1)}
\, .
\\ \tag
\endAligntags
$$
Diese beiden Reihenentwicklungen konvergieren f\"ur alle $z \in \C$.
Trotzdem kann man nur dann eine ausreichend schnelle Konvergenz
erwarten, wenn $z$ klein oder h\"ochstens mittelgro{\ss} ist. Wenn $z$ sehr
gro{\ss} ist, ist eine \"okonomische Berechnung der $F_m$-Funktion mit Hilfe
dieser Reihendarstellungen nicht m\"oglich. Man kann aber den
Anwendungsbereich der beiden Reihenentwicklungen (9.3-5) und (9.3-6)
deutlich erweitern, wenn man verallgemeinerte Summationsprozesse
verwendet [Grotendorst und Steinborn 1986; Harris 1983; Weniger und
Steinborn 1989a]. Au{\ss}erdem sind bei den beiden Reihendarstellungen
(9.3-5) und (9.3-6) deutlich unterschiedliche numerische Eigenschaften
zu erwarten. Bei der Reihendarstellung (9.3-5), die auch als
Fakult\"atenreihe in der Variablen $m + 1/2$ interpretiert werden kann,
ist zu erwarten, da{\ss} die Konvergenzgeschwindigkeit mit zunehmender
Ordnung $m$ der $F_m$-Funktion deutlich besser wird, wogegen die
Konvergenzgeschwindigkeit der Reihendarstellung (9.3-6) nur relativ
schwach von $m$ abh\"angen d\"urfte. F\"ur die Reihendarstellung (9.3-6)
spricht dagegen, da{\ss} ihre Terme strikt alternieren. Normalerweise k\"onnen
verallgemeinerte Summationsprozesse die Konvergenz alternierender Reihen
besser beschleunigen als die Konvergenz von Reihen, deren Terme das
gleiche Vorzeichen besitzen (siehe beispielsweise Smith und Ford [1979;
1982]). Selbst wenn die Reihendarstellung (9.3-5) f\"ur gr\"o{\ss}ere Werte von
$m$ schneller konvergieren sollte als die Reihendarstellung (9.3-6), ist
damit nicht ausgeschlossen, da{\ss} die Reihendarstellung (9.3-6) in
Verbindung mit geeigneten verallgemeinerten Summationsprozessen bessere
Ergebnisse liefern k\"onnte.

Aus den Beziehungen [Magnus, Oberhettinger und Soni 1966, S. 337]
$$
\beginAligntags
" \Gamma (a, x) " \; = \; "
\int\nolimits_{x}^{\infty} \, t^{a - 1} \, \e^{- t} \, \d t \, ,
\\ \tag
" \Gamma (a) " \; = \; " \gamma (a, x) \, + \, \Gamma (a, x) \, ,
\\ \tag
\endAligntags
$$
und [Magnus, Oberhettinger und Soni 1966, S. 341]
$$
\Gamma (a, x) \; \sim \; \e^{- x} \, x^{a - 1} \,
{}_2 F_0 (1, 1 - a; - 1/x) \, \qquad x \to \infty
\tag
$$
folgt, da{\ss} die $F_m$-Funktion die folgende asymptotische Reihe besitzt:
$$
F_m (z) \; \sim \;
\frac {\Gamma (m + 1/2)} {2 z^{m + 1/2}} \, - \,
\frac {e^{- z}} {2 z } \, {}_2 F_0 (1, 1/2 - m; - 1/z) \, ,
\qquad z \to \infty \, .
\tag
$$
Die nichtabbrechende verallgemeinerte hypergeometrische Reihe ${}_2 F_0$
in Gl. (9.3-10) divergiert f\"ur jedes endliche
Argument{\footnote[\dagger]{Glaeske, Reinhold und Volkmer [1987]
behaupten f\"alschlicherweise auf S. 605 ihres Buches, da{\ss} die Terme ihrer
Reihenentwicklung (11.8.62) rasch kleine Betr\"age annehmen. Wenn man aber
die Beziehung [Magnus, Oberhettinger und Soni 1966, S. 2]
$$
\Gamma (z-n) \; = \;
(-1)^n \Gamma (z) \frac {\Gamma (1-z)} {\Gamma (n + 1 - z)}
$$
in Gl. (11.8.62) verwendet, sieht man sofort, da{\ss} diese
Reihenentwicklung identisch mit der divergenten asymptotischen Reihe
(9.3-10) ist.}}. F\"ur sehr gro{\ss}e Werte von $z$ kann Gl. (9.3-10) aber
relativ genaue Approximationen liefern, wenn man die divergente Reihe
${}_2 F_0$ auf geeignete Weise abbricht. Au{\ss}erdem kann man die
divergente asymptotische Reihe (9.3-10) mit Hilfe von verallgemeinerten
Summationsprozessen summieren. Auf diese Weise kann man die
$F_m$-Funktion f\"ur nicht zu kleine Argumente sehr effizient berechnen.
Dabei tritt allerdings das Problem auf, da{\ss} die hypergeometrische Reihe
${}_2 F_0$ in Gl. (9.3-10) f\"ur gr\"o{\ss}ere Werte des Index $m$ deutlich
st\"arker divergiert als f\"ur kleine Werte von $m$. Au{\ss}erdem besitzen die
ersten $m$ Terme der hypergeometrischen Reihe ${}_2 F_0$ in Gl. (9.3-10)
das gleiche Vorzeichen. Dementsprechend ist es dann auch deutlich
schwieriger, die asymptotische Reihe (9.3-10) f\"ur gr\"o{\ss}ere Werte von $m$
zu summieren als f\"ur kleinere.

Die $F_m$-Funktion erf\"ullt die folgende inhomogene Zweitermrekursion
[Shavitt 1963, Gl. (24)]:
$$
(2 m + 1) F_{m} (z) \; = \; 2 z F_{m+1} (z) \, + \, e^{- z} \, .
\tag
$$
Diese Rekursionsformel ist f\"ur alle $z \ge 0$ {\it abw\"artsstabil\/}. Nur
wenn $z$ sehr gro{\ss} ist, kann diese Rekursionsformel auch in {\it
Aufw\"artsrichtung\/} verwendet werden, da die Exponentialfunktion dann
vernachl\"assigbar klein wird. Au{\ss}erdem erf\"ullt die $F_m$-Funktion noch
die folgende homogene Dreitermrekursion [Harris 1983, Gl. (7)], die
leicht aus Gl. (9.3-11) abgeleitet werden kann:
$$
2 z F_{m+1} (z) \; = \; (2 m + 2 z + 1) F_m (z) \, - \,
(2 m - 1) F_{m-1} (z) \, .
\tag
$$

Aus der Differentiationsbeziehung
$$
\frac {\d} {\d z} F_m (z) \; = \; - \int\nolimits_{0}^{1} \,
u^{2 m + 2} \, \exp \bigl(- z u^2 \bigr) \, \d u \; = \;
- F_{m+1} (z)
\tag
$$
folgt [Truesdell 1948, Theorem 14.1; 1950, Gl. (2)], da{\ss} die
$F_m$-Funktion das folgende Additionstheorem erf\"ullt:
$$
F_m (x + y) \; = \;
\sum_{k=0}^{\infty } \, \frac {( - x)^k} {k!} \, F_{m+k} (y) \, .
\tag
$$
Wenn man in Gl. (9.3-14) $y = 0$ setzt und die Beziehung [Saunders 1975,
Gl. (36)]
$$
F_m (0) \; = \; \frac {1} {2 m + 1}
\tag
$$
verwendet, erh\"alt man die Reihenentwicklung (9.3-6).

In der Literatur kann man noch zahlreiche andere Verfahren zur
Berechnung der $F_m$-Funktionen finden. So wurden von Gill, Johnson und
Pople [1991] und von Yahiro und Gondo [1992] Algorithmen beschrieben,
die auf Entwicklungen nach Tschebyscheffpolynomen basieren und die
besonders f\"ur Vektorrechner geeignet sind.

Ein Programm, das eine rationale Approximation f\"ur die $F_m$-Funktionen
mit $m = 0, \ldots , 5$ berechnet, die auf Tschebyscheffapproximationen
in Teilintervallen basiert, wurde von Spellucci und Pulay [1975]
ver\"offentlicht.

Harris [1983] verglich die numerischen Eigenschaften der folgenden
Verfahren zur Berechnung der $F_m$-Funktionen: Entwicklungen nach
sph\"arischen modifizierten Besselfunktionen erster Art [Harris 1983, Gln.
(14) und (15)], Darstellung durch Kettenbr\"uche [Harris 1983, Gln. (21) -
(26), und au{\ss}erdem die Darstellung durch die unendlichen Reihen (9.3-5)
und (9.3-6).

Eine rationale Approximation des Typs
$$
F_m (z) \; \approx \; \left[
\frac
{a_0^{(m)} + a_1^{(m)} z + a_2^{(m)} z^2 + \cdots + a_n^{(m)} z^n}
{1 + b_1^{(m)} z + b_2^{(m)} z^2 + \cdots + b_n^{(m)} z^n}
\right]^{m + 1/2}
\tag
$$
mit tabellierten Werten der Koeffizienten $a_0^{(m)}$, $a_1^{(m)}$,
$\ldots$ , $a_n^{(m)}$ und $b_1^{(m)}$, $b_2^{(m)}$, $\ldots$ ,
$b_n^{(m)}$ wurde von Schaad und Morrell [1971] konstruiert.

Die Berechnung der $F_m$-Funktionen mit Hilfe einer
Gau{\ss}-Legendre-Quadratur wurde von Jakab [1979] vorgeschlagen.

Wenn man St\"o{\ss}e von Atomen mit Hilfe quantenchemischer Methoden unter
Verwendung einer Gau{\ss}basis behandelt [Mac{\' \i}as und Riera 1982], ist
man letztlich mit $F_m$-Funktionen mit komplexen Argumenten
konfrontiert. Errea, M\'endez und Riera [1984] und Jakab [1984]
ver\"offentlichten Programme zur Berechnung solcher $F_m$-Funktionen.
Jones und Thron [1985] verglichen verschiedene Algorithmen zur
Berechnung unvollst\"andiger Gammafunktionen mit komplexen Argumenten.

\medskip

\Abschnitt Numerische Beispiele

\smallskip

\aktTag = 0

Wie schon im letzten Unterabschnitt erw\"ahnt wurde, h\"angen die Argumente
der bei einer LCAO-MO-Rechnung auftretenden $F_m$-Funktionen nur von den
Exponentialparametern und Zentren der im Basissatz vorkommenden
Gau{\ss}funktionen ab. Wenn man die in Gl. (9.1-3) definierten cartesischen
Gau{\ss}funktionen verwendet, h\"angen die Indizes der $F_m$-Funktionen
aus\-schlie{\ss}\-lich von den im Basissatz vorkommenden Exponenten
$\lambda$, $\mu$ und $\nu$ ab, und wenn man die in Gl. (9.1-4)
definierten sph\"arischen Gau{\ss}funktionen verwendet, h\"angen die Indizes der
$F_m$-Funktionen aus\-schlie{\ss}\-lich von den Quantenzahlen $n$, $\ell$
und $m$ der im Basissatz vorkommenden Gau{\ss}funktionen ab. Das bedeutet,
da{\ss} man in einer Molek\"ulrechnung in der Lage sein mu{\ss}, $F_m$-Funktionen
f\"ur alle $0 \le m \le m_{\Text{max}}$ und $0 \le z \le z_{\Text{max}}$
auf effiziente Weise mit ausreichender Genauigkeit zu berechnen, wobei
die Werte von $m_{\Text{max}}$ und $z_{\Text{max}}$ von den im Basissatz
enthaltenen Gau{\ss}funktionen und den im Molek\"ul vorkommenden Kernabst\"anden
festgelegt werden.

In diesem Abschnitt soll diskutiert werden, wie man -- ausgehend von den
Potenzreihen (9.3-5) und (9.3-6) und der asymptotischen Reihe (9.3-9)
-- alle im Rahmen einer LCAO-MO-Rechnung auftretenden $F_m$-Funktionen
auf effiziente Weise mit Hilfe von verallgemeinerten Summationsprozessen
berechnen kann. Dabei ist man mit dem Problem konfrontiert, da{\ss} der
Index $m$ und das Argument $z$ einer $F_m$-Funktion die numerischen
Eigenschaften der Potenzreihen (9.3-5) und (9.3-6) und der
asymptotischen Reihe (9.3-9) auf unterschiedliche Weise beeinflussen.

Die in der Literatur beschriebenen Algorithmen gehen von der Pr\"amisse
aus, da{\ss} es ineffizient w\"are, die Werte der im Rahmen einer
LCAO-MO-Rechnung auftretenden Hilfsfunktionen $F_m (z)$ mit $0 \le m \le
m_{\Text{max}}$ und $0 \le z \le z_{\Text{max}}$ direkt zu berechnen. Es
ist wesentlich effizienter, wenn man unbekannte Funktionswerte soweit
wie m\"oglich unter Verwendung bereits bekannter Funktionswerte berechnet,
beispielsweise mit Hilfe des Addtionstheorems (9.3-14). So folgt aus der
Integraldarstellung (9.2-9), da{\ss} $F_m (z)$ f\"ur alle $z \ge 0$ positiv
ist und eine monoton fallende Funktion des Index $m$ ist. Demzufolge
konvergiert das Additionstheorem (9.3-14) f\"ur $F_m (x + y)$ mindestens
so schnell wie die Potenzreihe f\"ur $\exp(- x)$. Wenn $x$ klein genug
ist, kann man also unter Verwendung der Funktionswerte $F_{m} (y)$,
$F_{m+1} (y)$, $\ldots$ , $F_{m+n} (y)$ den Funktionswert von
$F_m (x + y)$ mit Hilfe des Additionstheorems (9.3-14) ausreichend genau
berechnen.

Das legt die folgende Vorgehensweise nahe: Man w\"ahlt einen Index $M >
m_{\Text{max}}$ und eine endliche Zahlenfolge $\{ z_0, z_1, \ldots , z_k
\}$, deren Elemente $z_j$ alle in dem Intervall $0 \le z \le
z_{\Text{max}}$ enthalten sind. Die Abst\"ande $z_{j+1} - z_j$ zwischen
den Elementen der Zahlenfolge $\{ z_0, z_1, \ldots , z_k \}$ m\"ussen
dabei so klein sein, da{\ss} man alle Funktionswerte $F_m (z)$ mit $0 \le z
\le z_{\Text{max}}$ und $0 \le m \le m_{\Text{max}}$ unter Verwendung
des Additionstheorems (9.3-14) mit ausreichender Genauigkeit berechnen
kann, wenn die Funktionswerte $F_m (z_j)$ mit $0 \le m \le M$ und $z_j
\in \{ z_0, z_1, \ldots , z_k \}$ bekannt sind.

Bei dieser Vorgehensweise mu{\ss} man nur noch die Funktionswerte der
Hilfsfunktionen $F_m (z)$ auf dem Gitter $0 \le m \le M$ und $\{ z_0,
z_1, \ldots , z_k \}$ schnell und genau berechnen und abspeichern. Da
die $F_m$-Funktion die inhomogene Zweitermrekursion (9.3-11) erf\"ullt,
mu{\ss} man f\"ur jedes Argument $z_j \in \{ z_0, z_1, \ldots , z_k \}$ nur
einen einzigen Funktionswert $F_{\mu} (z_j)$ mit $\mu = 0$ oder $\mu =
M$ berechnen.

Die inhomogene Zweitermrekursion (9.3-11) kann f\"ur kleine oder
mittelgro{\ss}e Argumente der $F_m$-Funktion nur in Abw\"artsrichtung
verwendet werden. F\"ur kleine oder mittelgro{\ss}e $z_j$ mu{\ss} man also die
Funktionswerte $F_M (z_j)$ berechnen. Wenn $M$ gro{\ss} ist, sind bei
Verwendung der Reihendarstellung (9.3-5) deutlich bessere Ergebnisse zu
erwarten als bei Verwendung der Reihendarstellung (9.3-6). Numerische
Tests ergaben, da{\ss} man die Reihendarstellung (9.3-5) noch f\"ur
erstaunlich gro{\ss}e Argumente $z$ verwenden kann, wenn man ihre Konvergenz
mit Hilfe der in Abschnitt 5 besprochenen verallgemeinerten
Summationsprozessen beschleunigt.

\beginFloat

\medskip

\beginTabelle [to \kolumnenbreite]
\beginFormat \rechts " \rechts " \mitte " \mitte " \mitte
\endFormat
\+ " \links {\bf Tabelle 9-1} \@ \@ \@ \@ " \\
\+ " \links {Beschleunigung der Konvergenz der Reihendarstellung}
\@ \@ \@ \@ " \\
\+ " \links {$F_m (z) \; = \; \frac {\e^{- z}} {2 m + 1} \,
{}_1 F_1 (1; m + 3/2; z)$
f\"ur $m \; = \; 16$ und $z \; = \; 8$} \@ \@ \@ \@ " \\
\+ " \links {Alle Zahlen der Tabelle sind mit einem Faktor $10^{- 4}$
zu multiplizieren} \@ \@ \@ \@ " \\
\- " \- " \- " \- " \- " \- " \\ \sstrut {} {1.5 \jot} {1.5 \jot}
\+ " \rechts {$n$} " \mitte {Partialsumme $s_n (16, 8)$} "
$u_n^{(0)} (1, s_0)$ " $y_n^{(0)} (1, s_0)$ "
$Y_n^{(0)} (12, s_0)$ " \\
\+ " " \mitte {Gl. (9.4-1)} " Gl. (5.2-13) " Gl. (5.4-11) "
Gl. (5.5-13) " \\
\- " \- " \- " \- " \- " \- " \\ \sstrut {} {1 \jot} {1 \jot}
\+ " $ 0$ " $0.10165534178864$ " $0.10165534178864$ " $0.10165534178864$
" $0.10165534178864$ " \\
\+ " $ 1$ " $0.14812635517773$ " $0.64381716466139$ " $0.64381716466139$
" $0.19372055699345$ " \\
\+ " $ 2$ " $0.16822192853518$ " $0.18090405371667$ " $0.18090405371667$
" $0.18090405371667$ " \\
\+ " $ 3$ " $0.17646626632285$ " $0.18137817409500$ " $0.18141029847891$
" $0.18151711086320$ " \\
\+ " $ 4$ " $0.17968356887413$ " $0.18149435383639$ " $0.18150723273573$
" $0.18151983734319$ " \\
\+ " $ 5$ " $0.18088070470717$ " $0.18151602088745$ " $0.18151874245424$
" $0.18151970615451$ " \\
\+ " $ 6$ " $0.18130635300336$ " $0.18151925399816$ " $0.18151964902945$
" $0.18151969823835$ " \\
\+ " $ 7$ " $0.18145125455100$ " $0.18151965250174$ " $0.18151969638909$
" $0.18151969791552$ " \\
\+ " $ 8$ " $0.18149856934207$ " $0.18151969392321$ " $0.18151969788841$
" $0.18151969790483$ " \\
\+ " $ 9$ " $0.18151341319809$ " $0.18151969760313$ " $0.18151969790531$
" $0.18151969790461$ " \\
\+ " $10$ " $0.18151789436217$ " $0.18151969788485$ " $0.18151969790464$
" $0.18151969790461$ " \\
\+ " $11$ " $0.18151919797354$ " $0.18151969790349$ " $0.18151969790461$
" $0.18151969790461$ " \\
\+ " $12$ " $0.18151956389954$ " $0.18151969790456$ " $0.18151969790461$
" $0.18151969790461$ " \\
\+ " $13$ " $0.18151966313371$ " $0.18151969790461$ " $0.18151969790461$
" $0.18151969790461$ " \\
\- " \- " \- " \- " \- " \- " \\ \sstrut {} {1 \jot} {1 \jot}
\+ " \links {Exakt} \@ " $0.18151969790461$  " $0.18151969790461$ "
$0.18151969790461$ " \\
\- " \- " \- " \- " \- " \- " \\ \sstrut {} {1 \jot} {1 \jot}
\endTabelle

\medskip

\endFloat

In Tabelle 9-1 werden die verallgemeinerten Summationsprozesse
$u_n^{(0)} (\zeta, s_0)$, Gl. (5.2-13), und $y_n^{(0)} (\zeta, s_0)$,
Gl. (5.4-11), mit $\zeta = 1$ und $Y_n^{(0)} (\xi, s_0)$, Gl. (5.5-13),
mit $\xi = 12$ auf die Partialsummen
$$
s_n (m, z) \; = \; \frac {\e^{- z}} {2 m + 1} \,
\sum_{\nu = 0}^{n} \, \frac {z^{\nu}} {(m + 3/2)_{\nu}}
\tag
$$
der Reihendarstellung (9.3-5) mit $m = 16$ und $z = 8$ angewendet.

Die Partialsummen und die Transformationen in Tabelle 9-1 wurden in
QUADRUPLE PRECISION (31 - 32 Dezimalstellen) berechnet. Um die
numerische Stabilit\"at dieser Summationen \"uberpr\"ufen zu k\"onnen, wurden
die Rechnungen in DOUBLE PRECISION (15 - 16 Dezimalstellen) wiederholt.
Dabei ergab sich eine \"Ubereinstimmung aller ausgegebenen 14
Dezimalstellen.

Der klare Gewinner in Tabelle 9-1 ist $Y_k^{(n)} (\xi, s_n)$, Gl.
(5.5-13), gefolgt von $y_k^{(n)} (\zeta, s_n)$, Gl. (5.4-11), und das
schlechteste Ergebnis erzielte die Levinsche $u$-Transformation, Gl.
(5.2-13). Im Prinzip identische, wenn auch geringf\"ugig schlechtere
Ergebnisse ergaben auch die anderen Varianten der verallgemeinerten
Summationsprozesse ${\cal L}_k^{(n)} (\zeta, s_n, \omega_n)$, Gl.
(5.2-6), ${\cal S}_k^{(n)} (\zeta, s_n, \omega_n)$, Gl. (5.4-6), und
${\cal M}_k^{(n)} (\xi, s_n, \omega_n)$, Gl. (5.5-8), die auf den
Restsummenabsch\"atzungen (5.2-14), (5.2-16) und (5.2-19) basieren. In
allen F\"allen war die Variante des verallgemeinerten Summationsprozesses
${\cal M}_k^{(n)} (\xi, s_n, \omega_n)$ etwas leistungsf\"ahiger als die
analoge Variante von ${\cal S}_k^{(n)} (\zeta, s_n, \omega_n)$, und die
entsprechende Levinsche Transformation ${\cal L}_k^{(n)} (\zeta, s_n,
\omega_n)$ war immer das Schlu{\ss}licht.

\beginFloat

\medskip

\beginTabelle [to \kolumnenbreite]
\beginFormat \rechts " \rechts " \mitte " \mitte " \mitte
\endFormat
\+ " \links {\bf Tabelle 9-2} \@ \@ \@ \@ " \\
\+ " \links {Beschleunigung der Konvergenz der Reihendarstellung}
\@ \@ \@ \@ " \\
\+ " \links {$F_m (z) \; = \; \frac {\e^{- z}} {2 m + 1} \,
{}_1 F_1 (1; m + 3/2; z)$
f\"ur $m \; = \; 16$ und $z \; = \; 8$} \@ \@ \@ \@ " \\
\+ " \links {Alle Zahlen der Tabelle sind mit einem Faktor $10^{- 4}$ zu
multiplizieren} \@ \@ \@ \@ " \\
\- " \- " \- " \- " \- " \- " \\ \sstrut {} {1.5 \jot} {1.5 \jot}
\+ " \rechts {$n$} " \mitte {Partialsumme $s_n (16, 8)$}
" $\epsilon_{2 \Ent {n/2}}^{(n - 2 \Ent {n/2})}$
" ${\cal A}_{\Ent {n/2}}^{(n - 2 \Ent {n/2})}$
" $\theta_{2 \Ent {n/3}}^{(n - 3 \Ent {n/3})}$ " \\
\+ " " \mitte {Gl. (9.4-1)} " Gl. (2.4-10) " Gl. (3.3-8)
" Gl. (4.4-13) " \\
\- " \- " \- " \- " \- " \- " \\ \sstrut {} {1 \jot} {1 \jot}
\+ " $ 6$ " $0.18130635300336$ " $0.18152111252918$ " $0.18151982531105$
" $0.18151970554062$ " \\
\+ " $ 7$ " $0.18145125455100$ " $0.18152005232207$ " $0.18151973026652$
" $0.18151969785579$ " \\
\+ " $ 8$ " $0.18149856934207$ " $0.18151966539056$ " $0.18151969768681$
" $0.18151969752398$ " \\
\+ " $ 9$ " $0.18151341319809$ " $0.18151969067471$ " $0.18151969783430$
" $0.18151969761596$ " \\
\+ " $10$ " $0.18151789436217$ " $0.18151969856749$ " $0.18151969791312$
" $0.18151969716825$ " \\
\+ " $11$ " $0.18151919797354$ " $0.18151969803722$ " $0.18151969790546$
" $0.18151969789129$ " \\
\+ " $12$ " $0.18151956389954$ " $0.18151969789264$ " $0.18151969790465$
" $0.18151969800406$ " \\
\+ " $13$ " $0.18151966313371$ " $0.18151969790244$ " $0.18151969790462$
" $0.18151969790465$ " \\
\+ " $14$ " $0.18151968916234$ " $0.18151969790481$ " $0.18151969790461$
" $0.18151969790462$ " \\
\+ " $15$ " $0.18151969577279$ " $0.18151969790465$ " $0.18151969790461$
" $0.18151969790461$ " \\
\+ " $16$ " $0.18151969739997$ " $0.18151969790461$ " $0.18151969790461$
" $0.18151969790461$ " \\
\- " \- " \- " \- " \- " \- " \\ \sstrut {} {1 \jot} {1 \jot}
\+ " \links {Exakt} \@ " $0.18151969790461$  " $0.18151969790461$ "
$0.18151969790461$ " \\
\- " \- " \- " \- " \- " \- " \\ \sstrut {} {1 \jot} {1 \jot}
\endTabelle

\medskip

\endFloat

Trotzdem sind aber selbst die Varianten der Levinschen Transformation im
Falle der Reihendarstellung (9.3-5) offensichtlich wesentlich
effizienter als verallgemeinerte Summationsprozesse, die keine
expliziten Restsummenabsch\"atzungen verwenden. In Tabelle 9-2 wird der
Wynnsche $\epsilon$-Algorithmus, Gl. (2.4-10), der iterierte Aitkensche
$\Delta^2$-Proze{\ss}, Gl. (3.3-8), und der Brezinskische
$\theta$-Algorithmus, Gl. (4.4-13), auf die Partialsummen (9.4-1)
angewendet.

Die Partialsummen und Transformationen in Tabelle 9-2 wurden
in QUADRUPLE PRECISION berechnet. Eine Wiederholung dieser Rechnungen
in DOUBLE PRECISION ergab v\"ollige \"Ubereinstimmung aller ausgegebenen
14 Dezimalstellen.

Ein Vergleich der beiden Reihendarstellungen (9.3-5) und (9.3-6) ergab,
da{\ss} die Reihendarstellung (9.3-5) im Falle gro{\ss}er Indizes $m$ deutlich
bessere Ergebnisse liefert. So w\"urden $y_n^{(0)} (\zeta, s_0)$, Gl.
(5.4-11), und $Y_n^{(0)} (\xi, s_0)$, Gl. (5.5-13), die Partialsummen
$s_0$, $s_1$, $\ldots$ , $s_{21}$ ben\"otigen, um die $F_m$-Funktion mit
$m = 16$ und $z = 8$ wie in Tabelle 9-1 mit einer relativen Genauigkeit
von 14 Dezimalstellen zu berechnen.

Die Ergebnisse in den Tabellen 9-1 und 9-2 dokumentieren, da{\ss} man die
Reihendarstellung (9.3-5) in Verbindung mit geeigneten verallgemeinerten
Summationsprozessen auch f\"ur bemerkenswert gro{\ss}e Argumente $z$ zur
Berechnung der $F_m$-Funktionen verwenden kann. Trotzdem ist es aber im
Falle gr\"o{\ss}erer Argumente effizienter, die divergente asymptotische Reihe
(9.3-10) mit Hilfe von geeigneten verallgemeinerten Summationsprozessen
zu summieren. Da man die inhomogene Zweitermrekursion (9.3-11) f\"ur gro{\ss}e
Argumente $z$ in Aufw\"artsrichtung verwenden kann, sollte man in diesem
Fall immer die divergente Reihe (9.3-10) f\"ur die Funktion $F_0 (z)$
summieren, und die restlichen Hilfsfunktionen $F_{\mu} (z)$ mit $1 \le
\mu \le M$ rekursiv berechnen.

In Tabelle 9-3 werden die verallgemeinerten Summationsprozesse
$d_n^{(0)} (\zeta, s_0)$, Gl. (5.2-18), und ${\delta}_n^{(0)} (\zeta,
s_0)$, Gl. (5.4-13), mit $\zeta = 1$ und ${\Delta}_n^{(0)} (\xi, s_0)$,
Gl. (5.5-15), mit $\xi = 8$ auf die Partialsummen
$$
s_n (m, z) \; = \; \frac {\Gamma (m + 1/2)} {2 z^{m + 1/2}} \, - \,
\frac {e^{- z}} {2 z } \,
\sum_{\nu=0}^{n} \, \frac {(1/2 - m)_{\nu}} {(- z)^{\nu}}
\tag
$$
der asymptotischen Reihe (9.3-10) mit $m = 0$ und $z = 8$ angewendet.

\beginFloat

\medskip

\beginTabelle [to \kolumnenbreite]
\beginFormat \rechts " \rechts " \mitte " \mitte " \mitte
\endFormat
\+ " \links {\bf Tabelle 9-3} \@ \@ \@ \@ " \\
\+ " \links {Summation der divergenten asymptotischen Reihe}
\@ \@ \@ \@ " \\
\+ " \links {$F_m (z) \; \sim \;
\frac {\Gamma (m + 1/2)} {2 z^{m + 1/2}} \, - \,
\frac {e^{- z}} {2 z } \; {}_2 F_0 (1, 1/2 - m; - 1/z)$
f\"ur $m \; = \; 0$ und $z \; = \; 8$} \@ \@ \@ \@ " \\
\- " \- " \- " \- " \- " \- " \\ \sstrut {} {1.5 \jot} {1.5 \jot}
\+ " \rechts {$n$} " \mitte {Partialsumme $s_n (0, 8)$}
" $d_n^{(0)} (1, s_0)$
" ${\delta}_n^{(0)} (1, s_0)$
" ${\Delta}_n^{(0)} (8, s_0)$ " \\
\+ " " \mitte {Gl. (9.4-2)} " Gl. (5.2-18) "
Gl. (5.4-13) " Gl. (5.5-15) " \\
\- " \- " \- " \- " \- " \- " \\ \sstrut {} {1 \jot} {1 \jot}
\+ " $ 0 $ " $0.31330756791463$ " $0.31330756791463$ "
$0.31330756791463$ " $0.31330756791463$ " \\
\+ " $ 1 $ " $0.31330887831552$ " $0.31330867141012$ "
$0.31330867141012$ " $0.31330867141012$ " \\
\+ " $ 2 $ " $0.31330863261535$ " $0.31330868941820$ "
$0.31330868941820$ " $0.31330868748044$ " \\
\+ " $ 3 $ " $0.31330870939666$ " $0.31330868703791$ "
$0.31330868712586$ " $0.31330868731932$ " \\
\+ " $ 4 $ " $0.31330867580484$ " $0.31330868736050$ "
$0.31330868733607$ " $0.31330868732129$ " \\
\+ " $ 5 $ " $0.31330869470024$ " $0.31330868731583$ "
$0.31330868732043$ " $0.31330868732131$ " \\
\+ " $ 6 $ " $0.31330868170965$ " $0.31330868732207$ "
$0.31330868732134$ " $0.31330868732131$ " \\
\+ " $ 7 $ " $0.31330869226450$ " $0.31330868732120$ "
$0.31330868732131$ " $0.31330868732131$ " \\
\+ " $ 8 $ " $0.31330868236933$ " $0.31330868732132$ "
$0.31330868732131$ " $0.31330868732131$ " \\
\+ " $ 9 $ " $0.31330869288295$ " $0.31330868732131$ "
$0.31330868732131$ " $0.31330868732131$ " \\
\- " \- " \- " \- " \- " \- " \\ \sstrut {} {1 \jot} {1 \jot}
\+ " \links {Exakt} \@ " $0.31330868732131$ " $0.31330868732131$ "
$0.31330868732131$ " \\
\- " \- " \- " \- " \- " \- " \\ \sstrut {} {1 \jot} {1 \jot}
\endTabelle

\medskip

\endFloat

Die Partialsummen und Transformationen in Tabelle 9-3 wurden in
QUADRUPLE PRECISION berechnet. Eine Wiederholung dieser Rechnungen in
DOUBLE PRECISION ergab v\"ollige \"Ubereinstimmung aller ausgegebenen 14
Dezimalstellen.

Der klare Gewinner in Tabelle 9-3 ist $\Delta_k^{(n)} (\xi, s_n)$, Gl.
(5.5-15), gefolgt von $\delta_k^{(n)} (\zeta, s_n)$, Gl. (5.4-13), und
die Levinsche Transformation $d_k^{(n)} (\zeta, s_n)$, Gl. (5.2-18), ist
wieder das Schlu{\ss}licht.

Mit Hilfe der Transformation $\Delta_k^{(n)} (\xi, s_n)$, die im Falle
der Hilfsfunktion $F_m (z)$ erstaunlich leistungsf\"ahig ist, kann man
$F_0 (z)$ auch f\"ur relativ kleine Argumente durch Summation der
divergenten asymptotischen Reihe (9.3-10) effizient berechnen. So
ben\"otigt $\Delta_k^{(n)} (\xi, s_n)$ nur die Partialsummen $s_0$, $s_1$,
$\ldots$ , $s_{13}$ der divergenten Reihe in Gl. (9.3-10), um $F_0 (z)$
f\"ur $z = 2$ mit einer relativen Genauigkeit von 14 Dezimalstellen zu
berechnen.

Dagegen ist es relativ schwierig, die divergente asymptotische Reihe in
Gl. (9.3-10) f\"ur gro{\ss}e Werte von $m$ zu summieren. So ben\"otigen
$\delta_k^{(n)} (\zeta, s_n)$, Gl. (5.4-13), und $\Delta_k^{(n)} (\xi,
s_n)$, Gl. (5.5-15), immerhin die Partialsummen $s_0$, $s_1$, $\ldots$ ,
$s_{17}$ der asymptotischen Reihe (9.3-10), um $F_m (z)$ f\"ur $m = 16$
und $z = 8$ mit einer relativen Genauigkeit von 14 Dezimalstellen zu
berechnen.

Die Ergebnisse in Tabelle 9-4 zeigen, da{\ss} die in Tabelle 9-2 verwendeten
verallgemeinerte Summationsprozesse -- der Wynnsche
$\epsilon$-Algorithmus, Gl. (2.4-10), der iterierte Aitkensche
$\Delta^2$-Proze{\ss}, Gl. (3.3-8), und der Brezinskische
$\theta$-Algorithmus, Gl. (4.4-13), -- die divergente Reihe in Gl.
(9.3-10) weniger effizient summieren als verallgemeinerte
Summationsprozesse, die explizite Restsummenabsch\"atzungen $\Seqn \omega$
verwenden. Bemerkenswert ist allerdings das Ergebnis des
$\theta$-Algorithmus, der ebenso leistungsf\"ahig war wie die Levinsche
Transformation $d_k^{(n)} (\zeta, s_n)$, Gl. (5.2-18), in Tabelle 9-3.

\beginFloat

\medskip

\beginTabelle [to \kolumnenbreite]
\beginFormat \rechts " \rechts " \mitte " \mitte " \mitte
\endFormat
\+ " \links {\bf Tabelle 9-4} \@ \@ \@ \@ " \\
\+ " \links {Summation der divergenten asymptotischen Reihe}
\@ \@ \@ \@ " \\
\+ " \links {$F_m (z) \; \sim \;
\frac {\Gamma (m + 1/2)} {2 z^{m + 1/2}} \, - \,
\frac {e^{- z}} {2 z } \; {}_2 F_0 (1, 1/2 - m; - 1/z)$
f\"ur $m \; = \; 0$ und $z \; = \; 8$} \@ \@ \@ \@ " \\
\- " \- " \- " \- " \- " \- " \\ \sstrut {} {1.5 \jot} {1.5 \jot}
\+ " \rechts {$n$} " \mitte {Partialsumme $s_n (0, 8)$}
" $\epsilon_{2 \Ent {n/2}}^{(n - 2 \Ent {n/2})}$
" ${\cal A}_{\Ent {n/2}}^{(n - 2 \Ent {n/2})}$
" $\theta_{2 \Ent {n/3}}^{(n - 3 \Ent {n/3})}$ " \\
\+ " " \mitte {Gl. (9.4-2)} " Gl. (2.4-10) " Gl. (3.3-8)
" Gl. (4.4-13) " \\
\- " \- " \- " \- " \- " \- " \\ \sstrut {} {1 \jot} {1 \jot}
\+ " $ 6$ " $0.31330868170965$ " $0.31330868729057$ " $0.31330868731727$
" $0.31330868732107$ " \\
\+ " $ 7$ " $0.31330869226450$ " $0.31330868733146$ " $0.31330868732273$
" $0.31330868732141$ " \\
\+ " $ 8$ " $0.31330868236933$ " $0.31330868731894$ " $0.31330868732124$
" $0.31330868732126$ " \\
\+ " $ 9$ " $0.31330869288295$ " $0.31330868732217$ " $0.31330868732133$
" $0.31330868732131$ " \\
\+ " $10$ " $0.31330868039802$ " $0.31330868732108$ " $0.31330868732131$
" $0.31330868732131$ " \\
\+ " $11$ " $0.31330869678449$ " $0.31330868732140$ " $0.31330868732131$
" $0.31330868732131$ " \\
\+ " $12$ " $0.31330867322894$ " $0.31330868732128$ " $0.31330868732131$
" $0.31330868732131$ " \\
\+ " $13$ " $0.31330871003448$ " $0.31330868732132$ " $0.31330868732131$
" $0.31330868732131$ " \\
\+ " $14$ " $0.31330864792514$ " $0.31330868732130$ " $0.31330868732131$
" $0.31330868732131$ " \\
\+ " $15$ " $0.31330876049832$ " $0.31330868732131$ " $0.31330868732131$
" $0.31330868732131$ " \\
\- " \- " \- " \- " \- " \- " \\ \sstrut {} {1 \jot} {1 \jot}
\+ " \links {Exakt} \@ " $0.31330868732131$ " $0.31330868732131$ "
$0.31330868732131$ " \\
\- " \- " \- " \- " \- " \- " \\ \sstrut {} {1 \jot} {1 \jot}
\endTabelle

\medskip

\endFloat

Die Partialsummen und Transformationen in Tabelle 9-4 wurden in
QUADRUPLE PRECISION berechnet. Eine Wiederholung dieser Rechnungen in
DOUBLE PRECISION ergab v\"ollige \"Ubereinstimmung aller ausgegebenen 14
Dezimalstellen.

\endAbschnittsebene

\endAbschnittsebene

\keinTitelblatt\neueSeite

\beginAbschnittsebene
\aktAbschnitt = 9

\define\BeweisEnde{\ \Rahmen .7 em mal .7 em{\ }}

\Abschnitt Die Summation der St\"orungsreihe f\"ur die Grundzustandsenergie
anharmonischer Oszillatoren

\vskip - 2 \jot

\beginAbschnittsebene

\medskip

\Abschnitt Konvergente und divergente St\"orungsreihen

\smallskip

\aktTag = 0

Die Quantenmechanik ist im wesentlichen eine {\it Eigenwerttheorie\/}.
Ungl\"ucklicherweise k\"onnen die meisten Eigenwertgleichungen, die in
diesem Zusammenhang auftreten, nicht exakt gel\"ost werden. Man ist also
auf N\"aherungsverfahren angewiesen.

Die St\"orungstheorie ist neben dem Variationsverfahren das ohne Zweifel
wichtigste systematische Verfahren, um Eigenwertprobleme aus dem Bereich
der Quantenmechanik n\"aherungsweise zu l\"osen. Laut Simon [1991, S. 303]
basiert die St\"orungstheorie f\"ur Eigenwerte, die auch zur L\"osung von
Matrixeigenwertproblemen verwendet wird [Baumg\"artel 1985], einerseits
auf Arbeiten Lord Rayleigh's auf dem Gebiet der Akustik um die
Jahrhundertwende [Rayleigh 1945a; 1945b] und andererseits auf einer der
bahnbrechenden Arbeiten Schr\"odingers \"uber die Grundlagen der
Quantenmechanik [Schr\"odinger 1926].

Die Literatur \"uber quantenmechanische St\"orungstheorie ist sehr
umfangreich. Abgesehen davon, da{\ss} St\"orungstheorie in nahezu jedem
Lehrbuch der Quantenmechanik, der Quantenfeldtheorie oder auch der
Quantenchemie mehr oder weniger ausf\"uhrlich behandelt wird, gibt es noch
zahlreiche Monographien und \"Ubersichtsartikel, die sich speziell mit
St\"orungstheorie besch\"aftigen. Die mathematischen Grundlagen der
St\"orungstheorie werden in Monographien von Friedrichs [1965], Kato
[1976], Maslov [1972] und Rellich [1969] behandelt. Eine gute Behandlung
der mathematischen Grundlagen findet man auch in Abschnitt \Roemisch{12}
des Buches von Reed und Simon [1978]. Zahlreiche Anwendungen der
St\"orungstheorie in der Atom- und Molek\"ulphysik werden in B\"uchern von
Arteca, Fern\'{a}ndez und Castro [1990], Nicolaides, Clark und Nayfeh
[1990] und Wilcox [1966] als auch in \"Ubersichtsartikeln von Dalgarno
[1961], Fern\'{a}ndez und Castro [1992], Hirschfelder, Byers Brown und
Epstein [1964], Killingbeck [1977] und Kutzelnigg [1992; 1993]
beschrieben. Die St\"orungstheorie ist unverzichtbar f\"ur eine theoretische
Analyse der Atomspektren [Condon und Odaba\c{s}i 1980; Condon und
Shortley 1970; Cowan 1981; Herzberg 1944; Mizushima 1970]. Au{\ss}erdem
spielt die St\"orungstheorie eine wesentliche Rolle bei der Behandlung von
Vielteilchenproblemen der Atom- und Molek\"ulphysik [Avery 1976; Davidson
1976; Harris, Monkhorst und Freeman 1992; Lefebvre und Moser 1969;
Lindgren und Morrison 1982; Paldus und {\v C}{\' \i}{\v z}ek 1975;
Wilson 1981; 1984; 1985; 1992a; 1992b; 1992c; 1992d; 1992e], der
Festk\"orperphysik [Brown 1972; Fetter und Walecka 1971; Gross und Runge
1986; Inkson 1984; Mahan 1981; Mattuck 1976; Raimes 1972] und der
Kernphysik [Kumar 1962].

Die Rayleigh-Schr\"odingersche St\"orungstheorie ist f\"ur Systeme geeignet,
deren Hamilton\-opera\-to\-ren $\hat{H} (\beta)$ in einen l\"osbaren
Anteil $\hat{H}_0$, dessen Eigenwerte und Eigenfunktionen vollst\"andig
bekannt sind, und einen St\"orterm $\beta \hat{V}$ aufgespalten werden
k\"onnen:
$$
\hat{H} (\beta) \; = \; \hat{H}_0 \, + \, \beta \hat{V} \, .
\tag
$$
Der Formalismus der Rayleigh-Schr\"odingerschen St\"orungstheorie liefert
f\"ur einen Eigenwert $E (\beta)$ von $\hat{H} (\beta)$ eine {\it formale
Potenzreihe\/} in der Kopplungskonstante $\beta$:
$$
E (\beta) \; = \; \sum_{m=0}^{\infty} \, c_m \beta^m \, .
\tag
$$
Hier stellt sich nat\"urlich die Frage, ob diese Potenzreihe in $\beta$
konvergiert oder divergiert. Die Konvergenz dieser Reihe ist nur
garantiert, wenn der Energieeigenwert $E (\beta)$ in einer Umgebung von
$\beta = 0$ eine {\it analytische\/} Funktion der Kopplungskonstante
$\beta$ im Sinne der Funktionentheorie ist. Das ist eine sehr
restriktive Bedingung, die in vielen F\"allen nicht erf\"ullt ist.

Die heuristische Begr\"undung der Rayleigh-Schr\"odingerschen
St\"orungstheorie basiert auf der Annahme, da{\ss} die Kleinheit der
Kopplungskonstante $\beta$ -- von der man \"ublicherweise stillschweigend
ausgeht -- dazu f\"uhrt, da{\ss} der St\"orterm $\beta \hat{V}$ ebenfalls eine
{\it kleine St\"orung\/} ist. Wenn diese Bedingung erf\"ullt ist, sollten
das ungest\"orte und das gest\"orte System, die von den Hamiltonoperatoren
$\hat{H}_0$ beziehungsweise $\hat{H} (\beta)$ beschrieben werden, sich
nur geringf\"ugig unterscheiden. Man kann dann erwarten, da{\ss} schon wenige
Terme der St\"orungsreihe (10.1-2) eine numerisch n\"utzliche Approximation
des Energieeigenwertes $E (\beta)$ ergeben, wenn $\beta$ nur klein genug
ist. In diesem Geist wird vornehmlich in \"alteren Lehrb\"uchern der
Quantenmechanik behauptet, da{\ss} St\"orungsreihen vom Typ von Gl. (10.1-2)
f\"ur kleine Werte der Kopplungskonstante {\it konvergieren\/} (siehe
beispielsweise Byron und Fuller [1970, S. 539], Condon und Shortley
[1970, S. 35], Powell und Crasemann [1961, S. 382], Sakurai [1985, S.
293]). Auf S. 687 von Messiah [1970] wird sogar behauptet, es sei
vern\"unftig anzunehmen, da{\ss} eine St\"orungsreihe vom Type von Gl. (10.1-2)
f\"ur kleine Werte der Kopplungskonstante {\it schnell\/} konvergiert.

Leider ist die Situation wesentlich komplizierter. Der St\"orterm $\beta
\hat{V}$ ist ein Operator, und quan\-ten\-mechanische Operatoren sind in
der Regel {\it unbeschr\"ankt\/}{\footnote[\dagger]{Die Unbeschr\"anktheit
quantenmechanischer Operatoren ist trotz der damit verbundenen
mathematischen Komplikationen eine {\it unverzichtbare\/} Voraussetzung
beim formalen Aufbau der Quantenmechanik. Wenn $\hat{A}$ und $\hat{B}$
{\it beschr\"ankte\/} Operatoren sind und $\hat{I}$ der Einheitsoperator,
kann beispielsweise die Heisenbergsche Vertauschungsrelation
$$
\hat{A} \hat{B} \, - \, \hat{B} \hat{A}
\; = \; \frac {\hbar} {i} \hat{I}
$$
nicht erf\"ullt werden [Hellwig 1967, S. 108, Theorem 4].}}. Das bedeutet,
da{\ss} nicht alle Funktionen $\psi$ aus dem zugrundeliegenden Hilbertraum
der quadratintegrablen Funktionen, die die Normierbarkeitsbedingung
$\Vert \psi \Vert < \infty$ erf\"ullen, notwendigerweise auch die
Bedingungen $\Vert \hat{H}_0 \psi \Vert < \infty$ beziehungsweise $\Vert
\hat{V} \psi \Vert < \infty$ erf\"ullen. Die {\it Kleinheit\/} der St\"orung
$\beta \hat{V}$ wird also normalerweise nicht nur vom numerischen Wert
der Kopplungskonstante $\beta$ abh\"angen, sondern auch von den
Eigenschaften und den Definitionsbereichen der Operatoren $\hat{H}_0$
und $\hat{V}$. In der Quantenmechanik und in der Quantenfeldtheorie sind
zahlreiche St\"orungen $\beta \hat{V}$ bekannt, die {\it gro{\ss}e
St\"orungen\/} sind, sobald die Kopplungskonstante $\beta$ {\it ungleich
Null\/} ist{\footnote[\ddagger]{Beispielsweise tr\"agt \S 1 der Einleitung
des Buches von Rellich [1969] den bezeichnenden Titel {\it A small
pertur\-bation parameter does not mean a small perturbation\/}.}}. In
einem solchen Fall f\"uhrt die St\"orung $\beta \hat{V}$ zu einer {\it
fundamentalen\/} \"Anderung des ungest\"orten Systems, das durch den
Hamiltonoperator $\hat{H}_0$ beschrieben wird. Der Energieeigenwert $E
(\beta)$ ist dann keine analytische Funktion der Kopplungskonstante, was
impliziert, da{\ss} die St\"orungsreihe (10.1-2) {\it divergiert\/}.

In der Mathematik wurde die Konvergenz der St\"orungsreihe (10.1-2) schon
relativ bald intensiv untersucht [Kato 1949; 1950a; 1950b; Titchmarsh
1949; 1950]. Kato [1949, Gl. (30)] konnte zeigen, da{\ss} der
Energieeigenwert $E (\beta)$ in einer Umgebung von $\beta = 0$ {\it
analytisch\/} in $\beta$ ist, wenn der St\"oroperator $\hat{V}$ relativ
zum ungest\"orten Hamiltonoperator $\hat{H}_0$ {\it beschr\"ankt\/} ist:
Wenn ${\cal D} (\hat{H}_0)$ und ${\cal D} (\hat{V})$ die
Definitionsbereiche des ungest\"orten Hamiltonoperators $\hat{H}_0$
beziehungsweise des St\"oroperators $\hat{V}$ sind, dann konvergiert die
St\"orungsreihe (10.1-2) f\"ur ausreichend kleine Kopplungskonstanten
$\beta$, wenn der Definitionsbereich des ungest\"orten Hamiltonoperators
$\hat{H}_0$ im Definitionsbereich des St\"oroperators $\hat{V}$ enthalten
ist,
$$
{\cal D} (\hat{H}_0) \; \subset \; {\cal D} (\hat{V}) \, ,
\tag
$$
und wenn die Ungleichung
$$
\Vert \hat{V} \psi \Vert \; \le \;
a \, \Vert \hat{H}_0 \psi \Vert \, + \, b \, \Vert \psi \Vert
\tag
$$
f\"ur alle $\psi \in {\cal D} (\hat{H}_0)$ erf\"ullt ist, wobei $a$ und $b$
geeignete positive Konstanten sind [Reed und Simon 1978, S. 16, Lemma,
S. 17, Theorem \Roemisch{12}.9, und S. 21, Theorem \Roemisch{12}.11]. In
der Literatur werden St\"oroperatoren, die relativ zum ungest\"orten
Operator beschr\"ankt sind und deswegen zu konvergenten St\"orungsreihen
f\"uhren, {\it ana\-lytische\/} St\"orungen genannt{\footnote[\dagger]{Auch
bei konvergenten St\"orungsreihen k\"onnen Verfahren zur
Konvergenzverbesserung sehr hilfreich sein. So erschienen vor kurzem
einige Arbeiten \"uber die Verbesserung der Konvergenz von
Vielteilchenst\"orungsreihen [Dietz, Schmidt, Warken und He{\ss} 1993a; 1993b;
1993c].}}.

Bei unbeschr\"ankten Operatoren $\hat{H}_0$ und $\hat{V}$ kann man aber
nicht ausschlie{\ss}en, da{\ss} der Definitionsbereich ${\cal D} (\hat{V})$ des
St\"oroperators kleiner ist als der Definitionsbereich ${\cal D}
(\hat{H}_0)$ des ungest\"orten Hamiltonoperators. Das bedeutet, da{\ss} es
Wellenfunktionen aus dem Definitionsbereich von $\hat{H}_0$ geben kann,
die nicht die Bedingung $\Vert \hat{V} \psi \Vert < \infty$ erf\"ullen und
die deswegen auch nicht zum Definitionsbereich des St\"oroperators
$\hat{V}$ geh\"oren k\"onnen. Solche Wellenfunktionen $\psi \in {\cal D}
(\hat{H}_0)$ k\"onnen die Beschr\"anktheitsbedingung (10.1-4) offensichtlich
nicht erf\"ullen.

Bei unbeschr\"ankten St\"oroperatoren ist es also nicht garantiert, da{\ss} eine
St\"orungsreihe vom Typ von Gl. (10.1-2) in einer Umgebung von $\beta = 0$
konvergiert, und man mu{\ss} damit rechnen, da{\ss} die St\"orungsreihe f\"ur alle
Argumente $\beta \ne 0$ {\it divergiert\/}. St\"oroperatoren, die zu
divergenten St\"orungsreihen f\"uhren, werden in der Literatur {\it
singul\"are\/} St\"orungen genannt.

Wenn man die Freiheit h\"atte, den ungest\"orten Hamiltonoperator
$\hat{H}_0$ und den St\"oroperator $\hat{V}$ in Gl. (10.1-1) beliebig zu
w\"ahlen, k\"onnte man immer erreichen, da{\ss} die entsprechende St\"orungsreihe
f\"ur ausreichend kleine Kopplungskonstanten konvergiert. Aus den
Absch\"atzungen (10.1-3) und (10.1-4) folgt, da{\ss} die St\"orungsreihe
(10.1-2) dann in einer Umgebung von $\beta = 0$ konvergiert, wenn $\beta
\hat{V}$ relativ zu $\hat{H}_0$ eine kleine St\"orung ist. Man m\"u{\ss}te also
$\hat{H}_0$ und $\hat{V}$ so w\"ahlen, da{\ss} $\hat{H}_0$ die singul\"areren
Anteile des Hamiltonoperators $\hat{H} (\beta)$ enth\"alt und $\hat{V}$
die weniger singul\"aren Anteile. Da der Formalismus der
Rayleigh-Schr\"odingerschen St\"orungstheorie aber davon ausgeht, da{\ss} die
Eigenwerte und Eigenfunktionen von $\hat{H}_0$ bekannt sind, kann man
$\hat{H}_0$ und $\hat{V}$ normalerweise nicht frei w\"ahlen. Deswegen kann
man auch nicht ausschlie{\ss}en, da{\ss} der St\"oroperator $\hat{V}$ singul\"arer
ist als der ungest\"orte Hamiltonoperator $\hat{H}_0$. Dann wird es aber
Wellenfunktionen aus dem Definitionsbereich des ungest\"orten
Hamiltonoperators $\hat{H}_0$ geben, die {\it nicht\/} zum
Definitionsbereich des St\"oroperators $\hat{V}$ geh\"oren, und die
Beschr\"anktheitsbedingung (10.1-4) kann nicht f\"ur alle $\psi \in {\cal D}
(\hat{H}_0)$ erf\"ullt sein.

Wenn eine St\"orungsreihe divergiert, kann man nur noch hoffen, da{\ss} sie
wenigstens eine asymp\-totische Entwicklung des Energieeigenwertes $E
(\beta)$ f\"ur $\beta \to 0$ ist. Titchmarsh [1949; 1950] und Kato [1950a;
1950b] konnte allgemeine Bedingungen formulieren, die dies garantieren.
Eine ausf\"uhrliche und mathematisch sehr anspruchsvolle Diskussion
asymptotischer St\"orungen findet man in Abschnitt 8 des Buches von Kato
[1976] oder in einem Artikel von Hunziker [1988]. Die asymptotische
Natur von St\"orungsreihen vom Typ von Gl. (10.1-2) wurde auch in einem
Artikel von Krieger [1968] diskutiert. Die asymptotische Natur
divergenter St\"orungsreihen ist daf\"ur verantwortlich, da{\ss} die ersten
Terme divergenter St\"orungsreihen in vielen F\"allen relativ gute
Ergebnisse liefern, wenn die Kopplungskonstante nur klein genug ist.

Die mathematischen Erkenntnisse \"uber die Natur quantenmechanischer
St\"orungsreihen wurden in der physikalischen Literatur bis in die j\"ungere
Vergangenheit weitgehend ignoriert. Da die Rechenkapazit\"at vor der
allgemeinen Verf\"ugbarkeit leistungsf\"ahiger Computer in den meisten
F\"allen nicht ausreichte, um mehr als nur einige wenige Terme einer
St\"orungsreihe berechnen zu k\"onnen, erschien die Frage, ob der exakte
Energieeigenwert durch die ersten Terme einer konvergenten oder einer
divergenten asymptotischen St\"orungsreihe approximiert wird, eine
mathematische Spitzfindigkeit ohne besondere praktische Relevanz zu
sein. Immerhin waren ja gen\"ugend Beispiele aus der Quantenmechanik
bekannt, bei denen die St\"orungstheorie offensichtlich physikalisch
sinnvolle Ergebnisse lieferte. Da st\"orte auch ein Artikel von Dyson
[1952] nicht, in dem argumentiert wurde, da{\ss} bestimmte St\"orungsreihen in
der Quantenelektrodynamik aus physikalischen Gr\"unden divergieren
sollten.

Die Situation \"anderte sich schlagartig, als es Bender und Wu [1969]
gelang, die ersten 75 Koeffizienten der St\"orungsreihe f\"ur die
Grundzustandsenergie des anharmonischen Oszillators mit einer
$\hat{x}^4$-Anharmonizit\"at zu berechnen und zu zeigen, da{\ss} die
Koeffizienten mit wachsendem Index $n$ in etwa wie $n!/n^{1/2}$ wachsen
(siehe Gl. (2.2-5a)). Aus dieser asymptotischen Absch\"atzung folgt, da{\ss}
diese St\"orungsreihe f\"ur alle von Null verschiedenen Kopplungskonstanten
divergiert, und da{\ss} man Summationsverfahren verwenden mu{\ss}, um die in der
divergenten St\"orungsreihe enthaltene Information extrahieren zu k\"onnen.
Trotzdem gibt es erstaunlicherweise auch heute noch Artikel, in denen
versucht wird, den Konvergenzradius der St\"orungsreihe anharmonischer
Oszillatoren abzusch\"atzen [Popescu und Popescu 1987].

Die Arbeiten von Bender und Wu [1969; 1971; 1973] \"uber die Divergenz der
St\"orungsreihen anharmonischer Oszillatoren und die durch die
Fortschritte der Computertechnologie stark erh\"ohte Rechenkapazit\"at
f\"uhrten dazu, da{\ss} man das asymptotische Verhalten der Koeffizienten
quantenmechanischer St\"orungsreihen auch bei anderen elementaren Systemen
genauer untersuchte. In vielen F\"allen fand man sehr \"ahnliche
Divergenzprobleme wie bei den anharmonischen Oszillatoren. Auf diese
Weise entstand in den letzten Jahren eine neue Forschungsrichtung, die
in der englischsprachigen Literatur {\it large order perturbation
theory\/} genannt wird.

Die intensive Forschung auf dem Gebiet der {\it large order perturbation
theory\/} wird beispielsweise durch die {\it Proceedings of the Sanibel
workshop on perturbation theory at large order\/} [L\"owdin 1982a], durch
B\"ucher von Arteca, Fern\'{a}ndez und Castro [1990] und Le Guillou und
Zinn-Justin [1990], durch Dissertationen von Breen [1982] und Vrscay
[1983], oder durch Artikel von Adams, Avron, {\v C}{\' \i}{\v z}ek,
Otto, Paldus, Moats und Silverstone [1980], Alvarez [1988], Avron [1981;
1982], Avron, Adams, {\v C}{\' \i}{\v z}ek, Clay, Glasser, Otto, Paldus
und Vrscay [1979], Baker und Pirner [1983], Bender [1970; 1982], Breen
[1983; 1987], Br\'ezin, Le Guillou und Zinn-Justin [1977], Br\'ezin und
Parisi [1978], Br\'ezin, Parisi und Zinn-Justin [1977], {\v C}{\'\i}{\v
z}ek, Clay und Paldus [1980], {\v C}{\' \i}{\v z}ek, Damburg, Graffi,
Grecchi, Harrell, Harris, Nakai, Paldus, Propin und Silverstone [1986],
{\v C}{\' \i}{\v z}ek und Vrscay [1982; 1984], Collins und Soper [1978],
\'{C}wiok, Jeziorski, Ko{\l}os, Moszynski, Rychlewski und Szalewicz
[1992], Damburg und Propin [1983], Damburg, Propin, Graffi, Grecchi,
Harrell, {\v C}{\' \i}{\v z}ek, Paldus und Silverstone [1984], Damburg,
Propin und Martyshenko [1984], Fern\'{a}ndez [1992a], Graffi, Grecchi,
Harrell und Silverstone [1985], Janke [1990a; 1990b; 1990c], Janke und
Kleinert [1990], Kleinert [1993a], McRae und Vrscay [1992],
M\"uller-Kirsten [1986; 1987], Patil [1982], Popov, Weinberg und Mur
[1986], Reinhardt [1982], Silverman [1983a; 1983b], Silverman und Bishop
[1986], Silverman, Bishop und Pipin [1986], Silverman, Bonchev und
Polansky [1986], Silverman und Hinze [1986], Silverman und Nicolaides
[1991], Silverstone [1978; 1990], Silverstone, Adams, {\v C}{\' \i}{\v
z}ek und Otto [1979], Silverstone, Harrell und Grot [1981], Silverstone,
Harris, {\v C}{\' \i}{\v z}ek und Paldus [1985], Simon [1982],
Verbaarschot und West [1991], Verbaarschot, West und Wu [1990a; 1990b],
Vrscay [1984; 1985; 1986; 1987], Vrscay und {\v C}{\' \i}{\v z}ek
[1986], Vrscay und Handy [1989], Wu [1982] und Zinn-Justin [1979; 1981a;
1981b] dokumentiert{\footnote[\dagger]{Zahlreiche weitere Artikel aus
dem Bereich der {\it large order perturbation theory\/} sind in dem Buch
von Le Guillou und Zinn-Justin [1990] abgedruckt.}}.

\medskip

\Abschnitt Anharmonische Oszillatoren

\smallskip

\aktTag = 0

Die anharmonischen Oszillatoren, die durch den Hamiltonoperator
$$
\hat{H}^{(m)} (\beta) \; = \; \hat{p}^2 + \hat{x}^2 +
\beta \hat{x}^{2m} \, , \qquad m = 2, 3, 4, \ldots \, ,
\tag
$$
beschrieben werden, wobei $\hat{p} = - \i \d / \! \d x$ der
Impulsoperator ist, geh\"oren ohne Zweifel zu den wichtigsten
Modellsystemen der theoretischen Physik und der physikalischen und
theoretischen Chemie. Beispielsweise verwendet man anharmonische
Oszillatoren, um Schwingungen von Molek\"ulen oder Kristallgittern zu
beschreiben (siehe beispielsweise Graybeal [1988, Abschnitte 13.1 -
13.3] oder Steinfeld [1985, S. 128 - 132]). Weitere physikalische und
chemische Anwendungen anharmonischer Oszillatoren findet man in Anhang B
des Buches von Arteca, Fern\'{a}ndez und Castro [1990]. Von besonderem
Interesse sind anharmonische Oszillatoren auch als Modellsysteme in der
Quantenfeldtheorie. Deswegen wird die Rayleigh-Schr\"odingersche
St\"orungstheorie anharmonischer Oszillatoren ausf\"uhrlich in vielen
Lehrb\"uchern der Quantenfeldtheorie und verwandter Gebiete behandelt
(siehe beispielsweise Itzykson und Zuber [1980, S. 467 - 473], Kleinert
[1993b, Abschnitt 17.9], Negele und Orland [1987, S. 376 - 382], Parisi
[1988, S. 311 - 313], Schulman [1981, S. 92 - 107], und Zinn-Justin
[1989, S. 835 - 847]).

Die anharmonischen Oszillatoren geh\"oren ohne Zweifel zu den einfachsten
nicht geschlossen l\"osbaren quantenmechanischen Systemen. Trotzdem treten
bei der approximativen L\"osung der zeitunabh\"angigen Schr\"odingergleichung
der anharmonischen Oszillatoren hochinteressante und auch sehr
schwierige mathematische und konzeptionelle Probleme auf.
Dementsprechend intensiv wurde und wird \"uber anharmonische Oszillatoren
gearbeitet. Es gibt so viele Artikel \"uber anharmonische Oszillatoren,
da{\ss} ein Versuch einer halbwegs vollst\"andigen Bibliographie den Rahmen
dieser Arbeit sprengen w\"urde. Deswegen sollen in dieser Arbeit
haupts\"achlich Referenzen erw\"ahnt werden, die sich mit der Summation der
divergenten St\"orungsreihen der anharmonischen Oszillatoren und mit der
genauen Berechnung ihrer Energieeigenwerte besch\"aftigen. Zahlreiche
\"altere Arbeiten, in denen die Berechnung der Energieeigenwerte
anharmonischer Oszillatoren behandelt wird, werden beispielsweise in
Artikeln von Killingbeck [1980] und von Chaudhuri und Mukherjee [1985]
erw\"ahnt.

Der Anharmonizit\"atsterm $x^{2m}$ mit $m = 2, 3, 4$ w\"achst f\"ur $\vert x
\vert \to \pm \infty$ wesentlich st\"arker als das Parabelpotential $x^2$.
Man kann also problemlos Wellenfunktionen $\psi (x)$ aus dem
Definitionsbereich
$$
{\cal D} (\hat{H}_0) \; = \; {\cal D} (\hat{p}^2) \; \cap \;
{\cal D} (\hat{x}^2)
\tag
$$
des Hamiltonoperators des harmonischen Oszillators finden, die die
Beschr\"anktheitsbedingung
$$
\Vert \hat{x}^{2 m} \psi \Vert \; = \;
\int\nolimits_{- \infty}^{+ \infty}
\, \vert x^{2 m} \, \psi (x) \vert^2 \, \d x
\; < \; \infty \, ,
\qquad m = 2, 3, 4 \, ,
\tag
$$
nicht erf\"ullen. Die Ungleichung (10.1-4) kann also im Falle der
anharmonischen Oszillatoren (10.2-1) offensichtlich nicht f\"ur alle $\psi
\in {\cal D} (\hat{H}_0)$ erf\"ullt sein, was impliziert, da{\ss} der
St\"oroperator
$$
\hat{V}^{(m)} \; = \; \hat{x}^{2 m} \, , \qquad m = 2, 3, 4 \, ,
$$
keine analytische St\"orung sein kann. Dementsprechend ist die
St\"orungsreihe
$$
E^{(m)} (\beta) \; = \;
\sum_{n=0}^{\infty} \, b_{n}^{(m)} \, \beta^{n}
\tag
$$
f\"ur die Grundzustandsenergie des Hamiltonoperators (10.2-1) eine
Potenzreihe, die f\"ur jede Kopplungskonstante $\beta \ne 0$ divergiert.

Man kann auch physikalische Gr\"unde anf\"uhren, warum die St\"orungsreihe
(10.2-4) divergieren sollte. Bekanntlich kann eine komplexwertige
Funktion $f$ nur dann {\it analytisch\/} in einem Punkt $z_0 \in \C$
sein, wenn man eine {\it offene\/} Umgebung von $z_0$ finden kann, in
der die Potenzreihe
$$
f (z) \; = \; \sum_{\nu=0}^{\infty} \, \gamma_{\nu} \, (z - z_0)^{\nu}
\tag
$$
konvergiert. Das bedeutet, da{\ss} die Grundzustandsenergie $E^{(m)}
(\beta)$ des Hamiltonoperators (10.2-1) nur dann eine {\it
analytische\/} Funktion der Kopplungskonstante $\beta$ sein kann, wenn
die St\"orungsreihe (10.2-4) in einer {\it offenen\/} Umgebung von $\beta
= 0$ konvergiert. Aus physikalischen Gr\"unden ist man an sich nur an
positiven Kopplungskonstanten $\beta > 0$ interessiert. Eine offene
Umgebung von $\beta = 0$ mu{\ss} aber immer einen Teil der negativen reellen
Achse enthalten. Das bedeutet, da{\ss} die Grundzustandsenergie $E^{(m)}
(\beta)$ nur dann analytisch in einer Umgebung von $\beta = 0$ sein
kann, wenn der \"Ubergang von einer {\it positiven\/} zu einer {\it
negativen\/} Kopplungskonstante die Natur des betrachteten Systems nicht
fundamental \"andert, und wenn die St\"orungsreihe (10.2-4) auch f\"ur $\beta
< 0$ konvergiert.

F\"ur $\beta > 0$ unterscheidet sich die potentielle Energie
$\hat{U}^{(m)} (\beta) = \hat{x}^2 + \beta \hat{x}^{2 m}$ des
anharmonischen Oszillators qualitativ nur wenig von dem Parabelpotential
$\hat{U}_0 = \hat{x}^2$ des ungest\"orten Hamilton\-operators $\hat{H}_0$.
Der Hauptunterschied ist, da{\ss} die Potentialkurve f\"ur $\beta > 0$ eine
gr\"o{\ss}ere Steilheit aufweist als f\"ur $\beta = 0$, und da{\ss} $\hat{U}^{(m)}
(\beta)$ demzufolge f\"ur $\vert x \vert \to \pm \infty$ rascher gegen $+
\infty$ w\"achst als $\hat{U}_0$. Wichtig ist, da{\ss} $\hat{U}^{(m)} (\beta)$
f\"ur alle $\beta > 0$ einen gebundenen Zustand ergibt.

Wenn die Kopplungskonstante dagegen negativ ist ($\beta < 0$), ist die
potentielle Energie $\hat{U}^{(m)} (\beta) = \hat{x}^2 - \vert \beta
\vert \hat{x}^{2 m}$ ein Doppelmaximumpotential, das f\"ur $\vert x \vert
\to \pm \infty$ gegen $- \infty$ geht. Ein solches Potential ergibt aber
keine gebundenen Zust\"ande. Wenn man also im Hamiltonoperator (10.2-1)
von einer betragsm\"a{\ss}ig infinitesimal kleinen {\it positiven\/}
Kopplungskonstante zu einer betragsm\"a{\ss}ig infinitesimal kleinen {\it
negativen\/} Kopplungskonstante \"ubergeht, \"andert sich die Natur des
betrachteten Systems auf \"uberaus drastische Weise, was impliziert, da{\ss}
$E^{(m)} (\beta)$ in einer Umgebung von $\beta = 0$ nicht analytisch
sein kann. Demzufolge kann man auch nicht erwarten, da{\ss} die
St\"orungsreihe (10.2-4) konvergiert.

Bender und Wu [1969; 1971; 1973] konnten zeigen, da{\ss} die Koeffizienten
$b_{n}^{(m)}$ der St\"orungsreihe (10.2-4) f\"ur gro{\ss}e Indizes $n$ das
folgende asymptotische Verhalten besitzen{\footnote[\dagger]{Bender und
Wu [1969; 1971; 1973] und auch zahlreiche andere Autoren verwendeten in
ihren Artikeln nicht die Hamiltonoperatoren (10.2-1), die
charakteristisch sind f\"ur Arbeiten von Simon [1970; 1972; 1982; 1991],
sondern die Hamiltonoperatoren
$$
{\hat{H}}^{(m)}_{\Text{BW}} (\lambda) \; = \; -
\frac {\d^2} {\d x^2} + \frac {x^2} {4} +
\lambda \frac {x^{2 m}} {2^m} \, , \qquad m = 2, 3, 4 \, .
$$
Die Skalentransformation $x \mapsto 2^{1/2} \xi$ ergibt
$$
{\hat{H}}^{(m)}_{\Text{BW}} (\lambda) \; = \; - \frac {1} {2}
\frac {\d^2} {\d \xi^2} + \frac {\xi^2} {2} +
\lambda \xi^{2 m} \, , \qquad m = 2, 3, 4 \, .
$$
Wenn $E^{(m)}_{\Text{BW}} (\lambda)$ der Grundzustandseigenwert des
Bender-Wu-Hamiltonoperators ${\hat{H}}^{(m)}_{\Text{BW}} (\lambda)$ ist
und wenn man $\lambda = \beta / 2$ w\"ahlt, dann gilt offensichtlich
$$
E^{(m)} (\beta) \; = \; 2 E^{(m)}_{\Text{BW}} (\beta / 2) \, .
$$
Man kann $E^{(m)}_{\Text{BW}} (\lambda)$ analog zu Gl. (10.2-4) durch
eine formale Potenzreihe in $\lambda$ darstellen,
$$
E^{(m)}_{\Text{BW}} (\lambda) \; = \;
\sum_{n=0}^{\infty} \, a_{n}^{(m)} \, \lambda^{n} \, .
$$
Durch Vergleich mit Gl. (10.2-4) erh\"alt man f\"ur $\lambda = \beta / 2$:
$$
b_{n}^{(m)} \; = \; a_{n}^{(m)} / 2^{n-1} \, , \qquad n \in \N_0 \, .
$$
}}:
$$
\beginAligntags
" b_{n}^{(2)} \, " \sim \, " (-1)^{n+1} \,
\frac{(24)^{1/2}}{\pi^{3/2}} \, \Gamma(n+1/2) \, (3/2)^{n} \, ,
\erhoehe\aktTag \\ \tag*{\tagnr a}
" b_{n}^{(3)} \, " \sim \, " (-1)^{n+1} \,
\frac{(128)^{1/2}}{\pi^{2}} \, \Gamma(2 n + 1/2) \,
(16/\pi^{2})^{n} \, ,
\\ \tag*{\tagform\aktTagnr b}
" b_{n}^{(4)} \, " \sim \, " (-1)^{n+1} \,
\left\{
\frac{270 \, [ \Gamma(2/3)]^{3}}{\pi^{5}} \right\}^{1/2} \,
\Gamma (3 n + 1/2) \, (250)^{n} \,
\left\{ \frac{3 \, [ \Gamma (2/3)]^{3}}{4 \pi ^{2}} \right\}^{3n} \, .
\\ \tag*{\tagform\aktTagnr c}
\endAligntags
$$

Wenn man die Grundzustandsenergie $E^{(m)} (\beta)$ eines anharmonischen
Oszillators mit einer $\hat{x}^{2 m}$-Anharmonizit\"at durch Summation der
hochgradig divergenten St\"orungsreihe (10.2-4) bestimmen will, mu{\ss} man
zuerst kl\"aren, ob man der St\"orungsreihe \"uberhaupt einen eindeutig
bestimmten Energieeigenwert zuordnen kann.

Aus der asymptotischen Absch\"atzung (10.2-6a) folgt, da{\ss} die
Koeffizienten $b_{n}^{(2)}$ der St\"orungs\-reihe (10.2-4) des
anharmonischen Oszillators mit einer $\hat{x}^4$-Anharmonizit\"at eine
Ungleichung des Typs von Gl.~(6.5-13) erf\"ullen. Die St\"orungsreihe
(10.2-4) mit $m = 2$ ist also eine starke asymptotische Reihe gem\"a{\ss}
Gl. (6.5-12), der man die Grundzustandsenergie $E^{(2)} (\beta)$ auf
eindeutige Weise zuordnen kann [Reed und Simon 1978, S. 41, Example 1].

Analog folgt aus den asymptotischen Absch\"atzungen (10.2-6b) und
(10.2-6c), da{\ss} die Koeffizienten $b_{n}^{(3)}$ und $b_{n}^{(4)}$ der
St\"orungsreihe (10.2-4) eines anharmonischen Oszillators mit einer
$\hat{x}^6$- beziehungsweise $\hat{x}^8$-Anharmonizit\"at eine Ungleichung
des Typs von Gl. (6.5-15) mit $k = 2$ beziehungsweise $k = 3$ erf\"ullen.
Die St\"orungsreihe (10.2-4) mit $m = 3, 4$ ist also eine starke
asymptotische Reihe der Ordnung $k = 2$ beziehungsweise $k = 3$ gem\"a{\ss}
Gl.(6.5-14), der man die Grundzustandsenergie $E^{(3)} (\beta)$
beziehungsweise $E^{(4)} (\beta)$ auf eindeutige Weise zuordnen kann
[Reed und Simon 1978, S. 43, Example 3].

Aus den asymptotischen Absch\"atzungen (10.2-6) f\"ur die Koeffizienten
$b_n^{(m)}$ kann man hypergeometrische Modellreihen konstruieren, die
ebenso stark divergieren wie die St\"orungsreihe (10.2-4) f\"ur $m = 2, 3,
4$. So folgt aus Gl. (10.2-6a), da{\ss} die St\"orungsreihe (10.2-4) f\"ur $m =
2$ ebenso stark divergiert wie die hypergeometrische Reihe [Weniger
1990, Gl. (1.10)]
$$
{}_2 F_0 (1/2, 1; - 3 \beta / 2) \; = \;
\sum_{m=0}^{\infty} \, (1/2)_m \, (- 3 \beta / 2)^m \, .
\tag
$$
Eine nichtabbrechende hypergeometrische Reihe ${}_2 F_0$ dieses Typs
kommt auch in der divergenten asymptotischen Reihe der komplement\"aren
Fehlerfunktion vor [Abramowitz und Stegun 1972, Gln. (7.1.2) und
(7.1.23)]:
$$
\beginAligntags
" \Funk {erfc} (z) \; " = \; " \frac {2} {\pi^{1/2}}
\int\nolimits_{z}^{\infty} \, \e^{- t^2} \, \d t \, ,
\erhoehe\aktTag \\ \tag*{\tagnr a}
" \pi^{1/2} \, z \, \exp (z^2) \, \Funk {erfc} (z) \; " \sim \;
" {}_2 F_0 \bigl(1/2, 1; - 1 / z^2 \bigr) \, ,
\qquad \vert z \vert \to \infty
\quad \vert \arg (z) \vert < 3 \pi / 4 \, .
\\ \tag*{\tagform\aktTagnr b}
\endAligntags
$$
Daraus folgt, da{\ss} die divergente Reihe ${}_2 F_0$ in Gl. (10.2-7) im
wesentlichen eine komplement\"are Fehlerfunktion mit einem relativ
komplizierten Argument darstellt [Weniger 1990, Gl. (3.2)]:
$$
[2 \pi / (3 \beta)]^{1/2} \,
\exp \bigl( 2/(3 \beta) \bigr) \,
\Funk {erfc} \bigl([2/(3 \beta)]^{1/2}\bigr) \; \sim \;
{}_2 F_0 \bigl( 1/2, 1; - 3 \beta/ 2 \bigr) \, ,
\qquad \vert \beta \vert \to 0 \, .
\tag
$$
Aus dieser Beziehung folgt, da{\ss} die St\"orungsreihe (10.2-4) f\"ur die
Grundzu\-stands\-energie des anharmonischen Oszillators mit einer
${\hat{x}}^4$-Anharmonizit\"at in etwa so stark divergiert wie die
asymptotische Reihe (9.3-10) f\"ur die $F_m$-Funktion mit $m = 0$ und $z =
2/(3 \beta)$.

Wenn man das Multiplikationstheorem der Gammafunktion [Luke 1969a, S.
11, Gl. (1)],
$$
\Gamma (n z) \; = \; (2 \pi)^{(1-n)/2} \, n^{n z - 1/2} \,
\prod_{\nu=0}^{n-1} \, \Gamma(z + \nu / n) \, ,
\qquad n \in \N_0 \, ,
\tag
$$
f\"ur $n = 2$ mit der asymptotischen Absch\"atzung (10.2-6b) kombiniert,
kann man zeigen, da{\ss} die St\"orungsreihe (10.2-4) f\"ur $m = 3$ ebenso stark
divergiert wie die divergente hypergeometrische Reihe [Weniger, {\v
C}{\'\i}{\v z}ek und Vinette 1993, Gl. (3.9)]
$$
{}_3 F_0 (1/4, 3/4, 1; - 64 \beta / \pi^2) \; \sim \;
\sum_{\nu=0}^{\infty} \, (1/4)_{\nu} \, (3/4)_{\nu} \,
(- 64 \beta / \pi^2)^{\nu} \, .
\tag
$$
Nichtabbrechende hypergeometrische Reihen ${}_3 F_0$ kommen in der
Theorie der speziellen Funktionen nur sehr selten vor. Beispiele sind
die asymptotischen Entwicklungen der Lommelfunktionen $S_{\mu, \nu} (z)$
[Luke 1969a, S. 219, Gl. (22)] oder der Struvefunktionen ${\bf H}_{\nu}
(z)$ und ${\bf L}_{\nu} (z)$ [Luke 1969a, S. 219, Gln. (23) und (24)].
Dem Autor ist aber keine Referenz bekannt, in der eine nichtabbrechende
hypergeometrische Reihe ${}_3 F_0$ zur Berechnung der zugeh\"origen
Funktion verwendet wurde.

Wenn man das Multiplikationstheorem (10.2-10) f\"ur $n = 3$ mit der
asymptotischen Absch\"atzung (10.2-6c) kombiniert, kann man zeigen, da{\ss}
die St\"orungsreihe (10.2-4) f\"ur $m = 4$ ebenso stark divergiert wie die
hypergeometrische Reihe [Weniger, {\v C}{\'\i}{\v z}ek und Vinette 1993,
Gl. (3.10)]
$$
{}_4 F_0 (1/6, 1/2, 5/6, 1; - z) \; \sim \;
\sum_{\nu=0}^{\infty} \, (1/6)_{\nu} \, (1/2)_{\nu} \, (5/6)_{\nu}
\, (- z)^{\nu}
\tag
$$
mit $z = 91125 [\Gamma (2/3)]^9 \beta / (23 \pi^6)$. Dem Autor sind
keine asymptotische Entwicklungen spezieller Funktionen bekannt, die
nichtabbrechende hypergeometrische Reihen ${}_4 F_0$ enthalten.

Aufgrund der asymptotischen Absch\"atzung (10.2-6a) f\"ur die
St\"orungstheoriekoeffizienten $b_n^{(2)}$ oder der hypergeometrischen
Modellreihe (10.2-7) kann man vermuten, da{\ss} die St\"orungsreihe (10.2-4)
f\"ur die Grundzustandsenergie eines anharmonischen Oszillators mit einer
$\hat{x}^4$-An\-har\-mo\-ni\-zi\-t\"at sich in Summationsprozessen in etwa
so verhalten sollte wie divergente asymptotische Reihen f\"ur spezielle
Funktionen, die eine nichtabbrechende hypergeometrische Reihe ${}_2 F_0$
enthalten. Divergente hypergeometrische Reihen dieses Typs k\"onnen aber,
wie anhand der asymptotischen Reihe (6.7-12) f\"ur das Exponentialintegral
$E_1 (z)$, der asymptotischen Reihe (7.1-11) f\"ur die modifizierte
Besselfunktion $K_{\nu} (z)$ oder der asymptotischen Reihe (9.3-10) f\"ur
die $F_m$-Funktion in dieser Arbeit demonstriert wurde, f\"ur einen
relativ gro{\ss}en Bereich von Argumenten erfolgreich summiert werden.
Demzufolge kann man erwarten, da{\ss} man die Grundzustandsenergie $E^{(2)}
(\beta)$ eines anharmonischen Oszillators mit einer
$\hat{x}^4$-Anharmonizit\"at f\"ur nicht zu gro{\ss}e Kopplungskonstanten
$\beta$ mit ausreichender Genauigkeit berechnen kann, wenn man die
St\"orungsreihe (10.2-4) mit Hilfe geeigneter Varianten der in Abschnitt 7
dieser Arbeit beschriebenen verallgemeinerten Summationsprozesse
summiert.

Bei den anharmonischen Oszillatoren mit einer $\hat{x}^6$- und besonders
mit einer $\hat{x}^8$-Anharmonizit\"at folgt dagegen aus den
asymptotischen Absch\"atzungen (10.2-6b) und (10.2-6c) f\"ur die
St\"orungs\-theo\-rie\-koeffizienten $b_n^{(3)}$ und $b_n^{(4)}$ oder aus
den hypergeometrischen Modellreihen (10.2-11) und (10.2-12), da{\ss} man mit
\"au{\ss}erst schwierigen Summationsproblemen konfrontiert ist, die wesentlich
anspruchsvoller sind als alle Summationsprobleme aus der Theorie der
speziellen Funktionen, die in fr\"uheren Abschnitten dieser Arbeit
behandelt wurden.

Die unterschiedlich starke Divergenz der St\"orungstheoriekoeffizienten
$b_{n}^{(2)}$, $b_{n}^{(3)}$ und $b_{n}^{(4)}$ f\"ur $n \to \infty$ gem\"a{\ss}
Gl. (10.2-6) spielt eine erhebliche Rolle, wenn man versucht, die
St\"orungsreihe (10.2-4) mit Hilfe von Pad\'e-Approximationen zu summieren.

Simon [1970, Theorem \Roemisch{4}.2.1] konnte explizit zeigen, da{\ss} die
St\"orungstheoriekoeffizienten $b_{n}^{(2)}$ des anharmonischen
Oszillators mit einer $\hat{x}^4$-Anharmonizit\"at -- abgesehen vom ersten
Koeffizienten $b_{0}^{(2)}$ -- Momente eines Stieltjesma{\ss}es mit
negativem Vorzeichen gem\"a{\ss} Gl. (4.3-2) sind. Da die $b_{n}^{(2)}$ gem\"a{\ss}
Gl. (10.2-6a) im wesentlichen wie $n!/n^{1/2}$ wachsen, ist die
Carlemanbedingung (4.3-5) erf\"ullt, was impliziert, da{\ss} die
Pad\'e-Approximationen $[n + j / n]$ f\"ur festes $j \ge - 1$ und f\"ur $n \to
\infty$ gegen $E^{(2)} (\beta)$ konvergieren.

Im Falle des anharmonischen Oszillators mit einer
$\hat{x}^6$-Anharmonizit\"at sind die St\"orungstheorie\-koeffizienten
$b_{n}^{(3)}$ -- abgesehen vom ersten Koeffizienten $b_{0}^{(3)}$ --
ebenfalls Momente eines Stieltjesma{\ss}es mit negativem Vorzeichen gem\"a{\ss}
Gl. (4.3-2) [Simon 1972, S. 403; Graffi und Grecchi 1978, Abschnitt
\Roemisch{4}]. Da die $b_{n}^{(3)}$ gem\"a{\ss} Gl. (10.2-6b) im wesentlichen
wie $(2 n)!/n^{1/2}$ wachsen, ist die Carlemanbedingung (4.3-5) gerade
noch erf\"ullt, und die Pad\'e-Approximationen $[n + j / n]$ konvergieren
f\"ur festes $j \ge - 1$ und f\"ur $n \to \infty$ gegen $E^{(3)} (\beta)$.
Der anharmonische Oszillator mit einer $\hat{x}^6$-Anharmonizit\"at ist
ein Grenzfall: Numerische Tests ergaben n\"amlich, da{\ss}
Pad\'e-Approximationen so langsam gegen $E^{(3)} (\beta)$ konvergieren,
da{\ss} sie f\"ur praktische Zwecke ungeeignet sind.

Da die St\"orungstheoriekoeffizienten $b_{n}^{(4)}$ des anharmonischen
Oszillators mit einer $\hat{x}^8$-Anharmo\-nizit\"at gem\"a{\ss} Gl. (10.2-6c)
im wesentlichen wie $(3 n)!/n^{1/2}$ wachsen, ist die Carlemanbedingung
(4.3-5), die eine hinreichende, aber keine notwendige Bedingung ist,
nicht erf\"ullt. Es ist also nicht klar, ob die Pad\'e-Approximationen $[n +
j / n]$ f\"ur festes $j \ge - 1$ und f\"ur $n \to \infty$ gegen $E^{(4)}
(\beta)$ konvergieren. Auf der Basis numerischer Ergebnisse vermuteten
Graffi, Grecchi und Turchetti [1971, S. 314], da{\ss} Pad\'e-Approximationen
nicht konvergieren. Graffi und Grecchi [1978, Theorem 4.1] konnten
sp\"ater explizit zeigen, da{\ss} das entsprechende Momentenproblem in Falle
des anharmonischen Oszillators mit einer $\hat{x}^8$-Anharmo\-nizit\"at
keine eindeutige L\"osung besitzt. Damit ist es immer noch m\"oglich, da{\ss}
die Pad\'e-Approximationen $[n + j / n]$ f\"ur festes $j \ge - 1$ und f\"ur $n
\to \infty$ gegen einen wohldefinierten Grenzwert konvergieren. Da das
Momentenproblem unbestimmt ist, ist nicht gew\"ahrleistet, da{\ss} die
Pad\'e-Approximationen $[n + j / n]$ f\"ur {\it jedes\/} $j \ge - 1$ gegen
den {\it gleichen\/} Grenzwert konvergieren. Bender, Mead und
Papanicolaou [1987, S. 1016] bezeichneten deswegen $E^{(4)} (\beta)$
auch als {\it verallgemeinerte Stieltjesfunktion\/}.

Trotzdem ist es auch im diesem Fall m\"oglich, unter Verwendung der
St\"orungstheoriekoeffizienten $b_{n}^{(4)}$ Approximationen f\"ur das
betreffende Stieltjesma{\ss} und damit f\"ur $E^{(4)} (\beta)$ zu
konstruieren, wenn man die sogenannte {\it Methode der maximalen
Entropie\/} verwendet. Diese Methode, die urspr\"unglich aus der
Thermodynamik stammt, wird beispielsweise in den B\"uchern von Bevensee
[1993] und Kapur [1989] oder in Artikeln von Mead und Papanicolaou
[1984] und Tagliani [1993] beschrieben. Bender, Mead und Papanicolaou
[1987, Table \Roemisch{4}] und Bhattacharyya [1989b, Table 1]
verwendeten diese Methode der maximalen Entropie, um die
Grundzustandsenergie $E^{(4)} (\beta)$ aus den ersten Termen der
divergenten St\"orungsreihe (10.2-4) mit $m = 4$ zu bestimmen. Allerdings
waren ihre Ergebnisse nicht besonders gut.

\medskip

\Abschnitt Symanzik-Scaling

\smallskip

\aktTag = 0

Eine geschlossene L\"osung der zeitunabh\"angigen Schr\"odingergleichung
$$
\hat{H}^{(m)} (\beta) \, \psi (x) \; = \;
\bigl\{ \hat{p}^2 + \hat{x}^2 + \beta \hat{x}^{2m} \bigr\} \, \psi (x)
\; = \; E^{(m)} (\beta) \, \psi (x)
\tag
$$
der anharmonischen Oszillatoren mit $m = 2, 3, 4$ f\"ur beliebiges $\beta
> 0$ ist nicht bekannt. Trotzdem ist es m\"oglich, bestimmte Aussagen \"uber
das Verhalten der Energieeigenwerte $E^{(m)} (\beta)$ als Funktion der
Kopplungskonstante $\beta$ zu machen, wenn man Invarianzeigenschaften
des Hamilton\-operators $\hat{H}^{(m)} (\beta)$ und seiner Eigenwerte
gegen\"uber {\it Skalentransformationen\/} ausn\"utzt.

Das in diesem Unterabschnitt beschriebene Verfahren basiert laut Simon
[1970, S. 85] auf einer pers\"onlichen Mitteilung von K. Symanzik an A.S.
Wightman. Dementsprechend wird es in der Literatur \"ublicherweise {\it
Symanzik-Scaling\/} genannt.

Zur Ableitung der oben erw\"ahnten Invarianzeigenschaften gegen\"uber
Skalentransformationen betrachten wir die folgende Klasse von
Hamiltonoperatoren,
$$
\hat{H}^{(m)} (\alpha, \beta) \; = \; \hat{p}^2 +
\alpha \hat{x}^2 + \beta \hat{x}^{2m} \, ,
\qquad m = 2, 3, 4, \ldots \, ,
\tag
$$
die Verallgemeinerungen der in Gl. (10.2-1) definierten
Hamiltonoperatoren $\hat{H}^{(m)} (\beta)$ sind, da sie von zwei reellen
Parametern $\alpha$ und $\beta > 0$ abh\"angen. Au{\ss}erdem sei $E^{(m)}
(\alpha, \beta)$ ein Eigenwert dieses Hamiltonoperators zur
Eigenfunktion $\psi (x)$:
$$
\hat{H}^{(m)} (\alpha, \beta) \, \psi (x) \; = \;
E^{(m)} (\alpha, \beta) \, \psi (x) \, .
\tag
$$

Wir definieren nun einen unit\"aren Operator $\hat{U} (\tau)$ mit $\tau >
0$, der bei den kanonisch konjugierten Variablen $\hat{x}$ und $\hat{p}$
die folgenden Skalentransformationen bewirkt [Simon 1970, S. 85]:
$$
\beginAligntags
" \hat{U} (\tau) \, \hat{x} \, \hat{U}^{- 1} (\tau) " \; = \;
" \tau^{1/2} \hat{X} \, ,
\\ \tag
" \hat{U} (\tau) \, \hat{p} \, \hat{U}^{- 1} (\tau) " \; = \;
" \tau^{- 1/2} \hat{P} \, .
\\ \tag
\endAligntags
$$
Die Anwendung des unit\"aren Operator $\hat{U} (\tau)$ auf eine
Funktion $f (x)$ ergibt [Simon 1970, S. 85]:
$$
\hat{U} (\tau) \, f (x) \; = \; \tau^{1/4} \,
f \bigl( \tau^{1/2} X \bigl) \, .
\tag
$$
Auf diese Weise ist gew\"ahrleistet, da{\ss} die Normierung einer
Wellenfunktion $\psi (x)$ durch Skalierung nicht ver\"andert wird. Aus Gl.
(10.3-6) folgt n\"amlich:
$$
\int\nolimits_{- \infty}^{+ \infty} \,
\bigl\vert \psi (x) \bigl\vert^2 \d x \; = \;
\tau^{1/2} \, \int\nolimits_{- \infty}^{+ \infty} \,
\bigl\vert \psi (\tau^{1/2} X) \bigl\vert^2 \d X
\; = \; \int\nolimits_{- \infty}^{+ \infty} \,
\bigl\vert \psi (y) \bigl\vert^2 \d y \, .
\tag
$$

Durch Anwendung des unit\"aren Operators $\hat{U} (\tau)$ auf den in Gl.
(10.3-2) definierten Hamilton\-operator $\hat{H}^{(m)} (\alpha, \beta)$
kann man diesen durch die neuen Variablen $\hat{P}$ und $\hat{X}$
ausdr\"ucken:
$$
\beginAligntags
" \hat{H}^{(m)} (\alpha, \beta; \tau) " \; = \; " \hat{U} (\tau)
\, \bigl\{ \hat{p}^2 + \hat{x}^2 + \beta \hat{x}^{2m} \bigr\}
\, \hat{U}^{- 1} (\tau) \\
" " \; = \; " \tau^{- 1} \hat{P}^2 + \alpha \tau \hat{X}^2 +
\beta \tau^m \hat{X}^{2m} \, .
\\ \tag
\endAligntags
$$
Diese Beziehung kann auf folgende Weise umgeschrieben werden:
$$
\hat{H}^{(m)} (\alpha, \beta; \tau) \; = \;
\tau^{- 1} \, \left\{ \hat{P}^2 + \alpha \tau^2 \hat{X}^2
+ \beta \tau^{m+1} \hat{X}^{2m} \right\} \, .
\tag
$$
Ein Vergleich der Gln. (10.3-2), (10.3-3) und (10.3-9) zeigt, da{\ss} der
transformierte Hamiltonoperator $\hat{H}^{(m)} (\alpha, \beta;
\tau)$ die Eigenwertgleichung [Simon 1970, S. 112, Gl. (\Roemisch{3}.1)]
$$
\hat{H}^{(m)} (\alpha, \beta; \tau) \, \Psi (X) \; = \;
\tau^{- 1} \, E^{(m)} (\alpha \tau^2, \beta \tau^{m+1}) \, \Psi (X)
\tag
$$
erf\"ullt, wobei $\Psi (X)$ eine gem\"a{\ss} Gl. (10.3-6) transformierte
Eigenfunktion ist.

Die Eigenwerte hermitischer Operatoren, die durch eine {\it
\"Ahnlichkeits\-trans\-formation\/} miteinander verkn\"upft sind, sind
gleich. Das bedeutet, da{\ss} die Eigenwerte des Hamiltonoperators
$\hat{H}^{(m)} (\alpha, \beta)$ nicht davon abh\"angen k\"onnen, ob die
urspr\"unglichen Variablen $\hat{p}$ und $\hat{x}$ verwendet werden oder
ob man ihn durch die skalierten Variablen $\hat{P}$ und $\hat{X}$ gem\"a{\ss}
Gl. (10.3-9) ausdr\"uckt. F\"ur alle zul\"assigen $\alpha$, $\beta$ und $\tau$
mu{\ss} also die folgende Beziehung gelten:
$$
E^{(m)} (\alpha, \beta) \; = \;
\tau^{- 1} \, E^{(m)} (\alpha \tau^2, \beta \tau^{m+1}) \, .
\tag
$$
Wir w\"ahlen jetzt $\alpha = 1$ und $\beta \tau^{m+1} = 1$. Damit erhalten
wir:
$$
\beginAligntags
" \tau^{- 1} " \; = \; " \beta^{1/(m+1)} \, ,
\erhoehe\aktTag \\ \tag*{\tagnr a}
" \tau^2 " \; = \; " \beta^{- 2/(m+1)} \, .
\\ \tag*{\tagform\aktTagnr b}
\endAligntags
$$
Wenn wir diese Beziehungen in Gln. (10.3-9) und (10.3-11) einsetzen,
erhalten wir:
$$
\beginAligntags
" \hat{H}^{(m)} (1, \beta; \beta^{- 1/(m+1)}) " \; = \; "
\beta^{1/(m+1)} \, \left\{ \hat{P}^2 \, + \, \beta^{- 2/(m+1)} \hat{X}^2
\, + \, \hat{X}^{2m} \right\} \, ,
\\ \tag
" E^{(m)} (1, \beta) " \; = \; " \beta^{1/(m+1)} \,
E^{(m)} (\beta^{- 2/(m+1)}, 1) \, .
\\ \tag
\endAligntags
$$
Diese Beziehungen zeigen, da{\ss} die Parameter $\alpha$ und $\beta$ in dem
in Gl. (10.3-2) definierten Hamiltonoperator $\hat{H}^{(m)} (\alpha,
\beta)$ nicht v\"ollig unabh\"angig sind, da man durch eine geeignete
Skalentransformation entweder $\alpha = 1$ oder $\beta = 1$ erzwingen
kann. Aus Gl. (10.3-14) folgt au{\ss}erdem, da{\ss} ein Eigenwert $E^{(m)} (1,
\beta)$, der mit dem entsprechenden Eigenwert $E^{(m)} (\beta)$ des in
Gl. (10.2-1) definierten Hamiltonoperators $\hat{H}^{(m)} (\beta)$
identisch ist, das folgende asymptotische Verhalten besitzt [Simon 1970,
S. 112, Gl. (\Roemisch{3}.2)]:
$$
E^{(m)} (1, \beta) \; \sim \; \beta^{1/(m+1)} \, ,
\qquad \beta \to \infty \, .
\tag
$$

Im n\"achsten Schritt f\"uhren wir den transformierten Hamiltonoperator
$$
\beginAligntags
" \hat{\bf H}^{(m)} (\beta) " \; = \; " \beta^{- 1/(m+1)} \,
\hat{H}^{(m)} (1, \beta; \beta^{- 1/(m+1)}) \\
" " \; = \; "
\hat{P}^2 \, + \, \beta^{- 2/(m+1)} \hat{X}^2 \, + \, \hat{X}^{2m}
\\ \tag
\endAligntags
$$
ein, der die folgende Eigenwertgleichung erf\"ullt:
$$
\hat{\bf H}^{(m)} (\beta) \, \Psi (X) \; = \;
{\bf E}^{(m)} (\beta) \, \Psi (X) \, .
\tag
$$
Ein Vergleich mit Gl. (10.3-15) zeigt, da{\ss} der Eigenwert ${\bf E}^{(m)}
(\beta)$ in Gl. (10.3-17) die folgende Beziehung erf\"ullt:
$$
{\bf E}^{(m)} (\beta) \; = \; \beta^{- 1/(m+1)} \, E^{(m)} (1, \beta)
\; = \; E^{(m)} (\beta^{- 2/(m+1)}, 1) \, .
\tag
$$

Der in Gl. (10.3-16) definierte Hamiltonoperator $\hat{\bf H}^{(m)}
(\beta)$ kann auch folgenderma{\ss}en geschrie\-ben werden:
$$
\hat{\bf H}^{(m)} (\beta) \; = \;
\hat{\bf H}^{(m)}_0 \, + \, \beta^{- 2/(m+1)} \, \hat{\bf V}^{(m)} \, .
\tag
$$
Dabei ist
$$
\hat{\bf H}^{(m)}_0 \; = \; \hat{P}^2 \, + \, \hat{X}^{2 m}
\tag
$$
der ungest\"orte Hamiltonoperator, und
$$
\hat{\bf V}^{(m)} \; = \; \hat{X}^2
\tag
$$
der St\"oroperator mit der Kopplungskonstanten $\beta^{- 2/(m+1)}$. Das
impliziert, da{\ss} der Energieeigenwert ${\bf E}^{(m)} (\beta)$ in Gl.
(10.3-17) durch eine St\"orungsreihe dargestellt werden kann, die eine
formale Potenzreihe in der Variablen $\beta^{- 2/(m+1)}$ ist:
$$
{\bf E}^{(m)} (\beta) \; = \;
\sum_{n=0}^{\infty} \, K_n^{(m)} \, \beta^{- 2 n/(m+1)} \, .
\tag
$$
Analog folgt aus Gln. (10.3-18) und (10.3-22), da{\ss} der Eigenwert
$E^{(m)} (1, \beta)$ des Hamiltonoperators $\hat{H}^{(m)} (1, \beta)$,
der mit dem Energieeigenwert $E^{(m)} (\beta)$ in Gl. (10.2-1)
definiertem Hamiltonoperator $\hat{H}^{(m)} (\beta)$ \"ubereinstimmt,
durch die folgende St\"orungsreihe dargestellt werden kann,
$$
E^{(m)} (1, \beta) \; = \; \beta^{1/(m+1)} \,
\sum_{n=0}^{\infty} \, K_n^{(m)} \, \beta^{- 2 n/(m+1)} \, ,
\tag
$$
die eine Potenzreihe in der Variablen $\beta^{- 2/(m+1)}$ ist. In der
englischsprachigen Literatur wird diese St\"orungsreihe \"ublicherweise als
{\it strong coupling expansion\/} bezeichnet.

Die St\"orungsreihe (10.2-4) divergiert, weil der St\"oroperator
$\hat{V}^{(m)} = \hat{x}^{2 m}$ mit $m = 2, 3, 4$ relativ zum
ungest\"orten Hamiltonoperator $\hat{H}_0 = \hat{p}^2 + \hat{x}^2$
unbeschr\"ankt ist. Im Falle des in Gl. (10.3-16) definierten
Hamiltonoperators $\hat{\bf H}^{(m)} (\beta)$ ist der singul\"are Term
$\hat{X}^{2 m}$ gem\"a{\ss} Gl. (10.3-20) im ungest\"orten Operator $\hat{\bf
H}^{(m)}_0$ enthalten. Man kann also erwarten, da{\ss} die Absch\"atzung
$$
\Vert \hat{X}^2 \Psi \Vert \; \le \;
a \, \Vert (\hat{P}^2 + \hat{X}^{2 m}) \Psi \Vert \, + \,
b \, \Vert \Psi \Vert
\tag
$$
f\"ur geeignete positive Konstanten $a$ und $b$ g\"ultig ist, was aufgrund
der Absch\"atzung (10.1-4) die Konvergenz der St\"orungsreihe (10.3-23) f\"ur
ausreichend gro{\ss}e Werte von $\beta$ beweisen w\"urde. F\"ur $m = 2$ wurde
die Absch\"atzung (10.3-24) schon von Simon [1970, S. 81, Lemma
\Roemisch{2}.1 und S. 82, Theorem \Roemisch{2}.1.2] bewiesen. Die
Erweiterung auf beliebiges $m = 2, 3, 4, \ldots $ ist leicht m\"oglich.
Der Beweis verl\"auft in zwei Schritten.

\medskip

\beginEinzug \sl \parindent = 0 pt

\Auszug {\bf Satz 10-1:} Alle Wellenfunktionen $\Psi$ aus dem
Definitionsbereich des Hamiltonoperators
$$
\hat{\bf H}^{(m)}_0 \; = \; \hat{P}^2 \, + \, \hat{X}^{2m}
\tag
$$
erf\"ullen die Absch\"atzung
$$
\Vert \hat{P}^2 \Psi \Vert^2 \, + \, \Vert \hat{X}^{2m} \Psi \Vert^2
\; \le \;
2 \, \Vert (\hat{P}^2 + \hat{X}^{2m}) \Psi \Vert^2 \, + \,
b_0 \, \Vert \Psi \Vert^2 \, ,
\tag
$$
wobei $b_0$ eine geeignete positive Konstante ist.

\endEinzug

\medskip

\noindent {\it Beweis:\/} Unter Verwendung der Kommutatorbeziehungen
$$
\bigl\[ \hat{P}, \hat{X} \bigr\] \; = \; \hat{P} \, \hat{X}
\, - \, \hat{X} \, \hat{P} \; = \; - i
\tag
$$
und
$$
\bigl\[ \hat{P}, \hat{X}^{\alpha} \bigr\] \; = \;
\hat{P} \, \hat{X}^{\alpha} \, - \, \hat{X}^{\alpha} \, \hat{P}
\; = \; - i \alpha \, \hat{X}^{\alpha - 1}
\tag
$$
sch\"atzen wir den folgenden Ausdruck ab:
$$
\beginAligntags
" \bigl\{ \hat{P}^2 + \hat{X}^{2 m} \bigr\}^2 " \; = \; "
\hat{P}^4 \, + \, \hat{X}^{4 m} \, + \, \hat{P}^2 \hat{X}^{2 m}
\, + \, \hat{X}^{2 m} \hat{P}^2 \\ \tag
" " \; = \; " \hat{P}^4 \, + \, \hat{X}^{4 m} \, + \,
\bigl\[ \hat{P}, \bigl\[ \hat{P}, \hat{X}^{2 m} \bigr\] \bigr\]
\, + \, 2 \, \big( \hat{P} \, \hat{X}^{m} \bigr)
\big( \hat{X}^{m} \, \hat{P} \bigr) \quad
\\ \tag
" " \; = \; " \hat{P}^4 \, + \, \hat{X}^{4 m} \, - \, (2 m) \, i \,
\bigl\[ \hat{P}, \hat{X}^{2 m - 1} \bigr\] \, + \,
2 \, \big( \hat{P} \, \hat{X}^{m} \bigr)
\big( \hat{X}^{m} \, \hat{P} \bigr)
\\ \tag
" " \; = \; " \hat{P}^4 \, + \, \hat{X}^{4 m} \, - \,
(2 m - 1) (2 m) \, \hat{X}^{2 m - 2} \, + \,
2 \, \big( \hat{P} \, \hat{X}^{m} \bigr)
\big( \hat{X}^{m} \, \hat{P} \bigr) \, .
\\ \tag
\endAligntags
$$
Zur Absch\"atzung der rechten Seite von Gl. (10.3-32) w\"ahlen wir eine
Konstante $b_0 > 0$ auf solche Weise, da{\ss}
$$
\frac {1} {2} x^{4 m} \, - \, (2 m - 1) (2 m) \, x^{2 m - 2}
\, + \, \frac {1} {2} b_0 \; \ge \; 0
\tag
$$
f\"ur alle $x \in \R$ gilt. Damit erhalten wir:
$$
\beginAligntags
\bigl\{ \hat{P}^2 + \hat{X}^{2 m} \bigr\}^2 \, + \, \frac {1} {2} b_0
\; = \; " \frac {1} {2}
\bigl\{ \hat{P}^4 + \hat{X}^{4 m} \bigr\} \, + \,
2 \, \big( \hat{P} \, \hat{X}^{m} \bigr)
\big( \hat{X}^{m} \, \hat{P} \bigr) \, + \,
\frac {1} {2} \hat{P}^4 \\
" \, + \, \frac {1} {2} \hat{X}^4 \, - \,
(2 m - 1) (2 m) \, \hat{X}^{2 m - 2} \, + \, \frac {1} {2} b_0 \, .
\\ \tag
\endAligntags
$$
Aufgrund von Gl. (10.3-33) gilt aber
$$
2 \, \big( \hat{P} \, \hat{X}^{m} \bigr)
\big( \hat{X}^{m} \, \hat{P} \bigr) \, + \,
\frac {1} {2} \hat{P}^4 \, + \, \frac {1} {2} \hat{X}^4 \, - \,
(2 m - 1) (2 m) \, \hat{X}^{2 m - 2} \, + \, \frac {1} {2} b_0
\; \ge \; 0
\tag
$$
im Sinne eines Erwartungswertes, was impliziert, da{\ss} auch
$$
\hat{P}^4 + \hat{X}^{4 m} \; \le \;
2 \, \bigl\{ \hat{P}^2 + \hat{X}^{2 m} \bigr\}^2 \, + \, b_0
\tag
$$
im Sinne eines Erwartungswertes gilt. Wenn man jetzt den Erwartungswert
der Ungleichung (10.3-36) mit einer Funktion $\Psi \in {\cal D}
(\hat{P}^2) \cap {\cal D} (\hat{X}^{2 m})$ bildet, erh\"alt man Gl.
(10.3-26). \BeweisEnde

Mit Hilfe von Satz 10-1 kann man beweisen, da{\ss} der St\"oroperator
$\hat{X}^2$ relativ zum Operator $\hat{P}^2 + \hat{X}^{2 m}$ beschr\"ankt
ist.

\medskip

\beginEinzug \sl \parindent = 0 pt

\Auszug {\bf Satz 10-2:} Der Operator $\hat{X}^2$ ist relativ zu dem in
Gl. (10.3-25) definierten Operator $\hat{\bf H}^{(m)}_0$ beschr\"ankt. F\"ur
jedes $a > 0$ gibt es n\"amlich ein $b > 0$ mit
$$
\Vert \hat{X}^2 \Psi \Vert \; \le \; a \,
\Vert \hat{\bf H}^{(m)}_0 \Psi \Vert \, + \, b \, \Vert \Psi \Vert
\; = \; a \, \Vert (\hat{P}^2 + \hat{X}^{2m}) \Psi \Vert \, + \,
b \, \Vert \Psi \Vert \, .
\tag
$$

\endEinzug

\medskip

\noindent {\it Beweis:\/} F\"ur jedes $a > 0$ kann man eine Konstante $c >
0$ so w\"ahlen, da{\ss}
$$
x^2 \; \le \; \frac {a} {2^{1/2}} \, x^{2 m} \, + \, c
\tag
$$
f\"ur alle $x \in \R$ gilt. Wenn man diese Beziehung als
Operatorungleichung interpretiert und auf eine Funktion $\Psi \in {\cal
D} (\hat{P}^2) \cap {\cal D} (\hat{X}^{2 m})$ anwendet, erh\"alt man mit
Hilfe der Dreiecksungleichung f\"ur Normen [Mitrinovi\'{c},
Pe\v{c}ari\'{c} und Fink 1993, Abschnitt \Roemisch{17}] die folgende
Absch\"atzung:
$$
\Vert \hat{X}^2 \Psi \Vert \; \le \;
\frac {a} {2^{1/2}} \, \Vert \hat{X}^{2 m} \Psi \Vert \, + \,
c \, \Vert \Psi \Vert \, .
\tag
$$
Wir definieren jetzt:
$$
\beginAligntags
" A " \; = \; " \Vert \hat{X}^{2 m} \Psi \Vert \, ,
\\ \tag
" B " \; = \; " \Bigl\{ \Vert \hat{P}^2 \Psi \Vert^2 \, + \,
\Vert \hat{X}^{2 m} \Psi \Vert^2 \Bigr\}^{1/2} \, ,
\\ \tag
" C " \; = \; "
\Bigl\{ 2 \, \Vert (\hat{P}^2 + \hat{X}^{2 m}) \Psi \Vert^2
\, + \, b_0 \, \Vert \Psi \Vert^2 \Bigr\}^{1/2} \, ,
\\ \tag
" D " \; = \; " 2^{1/2} \, \Vert (\hat{P}^2 + \hat{X}^{2 m}) \Psi \Vert
\, + \, b_0^{1/2} \, \Vert \Psi \Vert \, .
\\ \tag
\endAligntags
$$
Aus den Gln. (10.3-26) und (10.3-40) - (10.3-43) erhalten wir die
Absch\"atzungen
$$
A^2 \; \le \; B^2 \; \le \; C^2 \; \le \; D^2 \, .
\tag
$$
Die Ungleichung $D^2 - A^2 = (D-A)(D+A) \ge 0$ impliziert aber auch $D
\ge A$ oder
$$
\Vert \hat{X}^{2 m} \Psi \Vert \; \le \;
2^{1/2} \, \Vert (\hat{P}^2 + \hat{X}^{2 m}) \Psi \Vert
\, + \, b_0^{1/2} \, \Vert \Psi \Vert \, .
\tag
$$
Wenn wir diese Beziehung in Gl. (10.3-39) verwenden, erhalten wir:
$$
\Vert \hat{X}^2 \Psi \Vert \; \le \;
a \, \Vert (\hat{P}^2 + \hat{X}^{2 m}) \Psi \Vert \, + \,
a \, (b_{0}/2)^{1/2} \, \Vert \Psi \Vert \, + \,
c \, \Vert \Psi \Vert \, .
\tag
$$
Wenn wir jetzt $b = a (b_0/2)^{1/2} + c$ setzen, erhalten wir die
Absch\"atzung (10.3-37). \BeweisEnde

Aus den S\"atzen 10-1 und 10-2 folgt, da{\ss} die {\it strong coupling
expansion\/} (10.3-23) f\"ur ausreichend gro{\ss}e Werte der
Kopplungskonstante $\beta$ konvergiert. Sie besitzt also im Prinzip
wesentlich g\"unstigere Eigenschaften als die komplement\"are St\"orungsreihe
(10.2-4), die eine divergente asymp\-totische Reihe f\"ur $\beta \to 0$
ist, und die mit Hilfe geeigneter verallgemeinerter Summationsprozesse
summiert werden mu{\ss}, wenn man sie f\"ur numerische Zwecke verwenden will.

Der Hauptnachteil der St\"orungsreihe (10.3-23) ist, da{\ss} die Eigenwerte
und Eigenfunktionen des ungest\"orten Hamiltonoperators (10.3-20) nicht in
geschlossener Form bekannt sind. Der \"ubliche Formalismus der
Rayleigh-Schr\"odingerschen St\"orungstheorie zur Berechnung der
St\"orungstheoriekoeffizienten kann also im Fall der {\it strong coupling
expansion\/} (10.3-23) nicht angewendet werden. Hioe und Montroll [1975,
Gl.~(\Roemisch{3}.6)] gelang es mit Hilfe einer Fitprozedur, die
Koeffizienten $K_0^{(2)}$, $K_1^{(2)}$ und $K_2^{(2)}$ der St\"orungsreihe
(10.3-23) des anharmonischen Oszillators mit einer
$\hat{x}^4$-Anharmonizit\"at approximativ zu bestimmen (siehe auch Hioe,
MacMillen und Montroll [1978, Gl.~(3.3)]). Auf analoge Weise bestimmten
Hioe, MacMillen und Montroll [1976, Gl.~(\Roemisch{3}.6)]
Approximationen f\"ur die Koeffizienten $K_0^{(3)}$, $K_1^{(3)}$ und
$K_2^{(3)}$ der St\"orungsreihe (10.3-23) des Oszillators mit einer
$\hat{x}^6$-Anharmonizit\"at. Allerdings lieferte die Fitprozedur nur
relativ ungenaue Ergebnisse. Turbiner und Ushveridze [1988,
Table~\Roemisch{2}] gelang es im Falle der $\hat{x}^4$-Anharmonizit\"at,
die Koeffizienten $K_0^{(2)}$, $K_1^{(2)}$ und $K_2^{(2)}$ der
St\"orungsreihe (10.3-23) mit Hilfe einer Pfadintegralmethode zu
berechnen. Guardiola, Sol\'{\i}s und Ros [1992, Table~\Roemisch{4}]
verwendeten eine Skalentransformation, um den Hamiltonoperator (10.3-16)
so zu transformieren, da{\ss} sie die St\"orungstheoriekoeffizienten
$K_0^{(2)}$, $\ldots$ , $K_{11}^{(2)}$, $K_0^{(3)}$, $\ldots$ ,
$K_{11}^{(3)}$, $K_0^{(4)}$, $\ldots$ , $K_7^{(3)}$, und $K_0^{(5)}$,
$\ldots$ , $K_7^{(5)}$ f\"ur die Grundzustandsenergie eines anharmonischen
Oszillators mit einer $\hat{x}^4$-, $\hat{x}^6$-, $\hat{x}^8$- und sogar
$\hat{x}^{10}$-Anharmonizit\"at berechnen konnten. Allerdings gibt es bei
dem von Guardiola, Sol\'{\i}s und Ros [1992] verwendeten Verfahren noch
einen freien Parameter, der auf geeignete Weise gew\"ahlt werden mu{\ss}.
Fern\'{a}ndez [1992b, Table~1] gelang es, die Koeffizienten $K_0^{(2)}$,
$\ldots$ , $K_4^{(2)}$ der St\"orungsreihe (10.3-23) des anharmonischen
Oszillators mit einer $\hat{x}^4$-Anharmonizit\"at zu berechnen, indem er
die Schr\"odingergleichung durch eine entsprechende Riccatigleichung
ersetzte und die resultierende Reihenentwicklung in Pad\'e-Approximationen
transformierte. Weitere Arbeiten \"uber die Berechnung der Koeffizienten
$K_n^{(m)}$ der St\"orungsreihe (10.3-23) scheint es bisher nicht zu
geben.

\medskip

\Abschnitt Renormierung

\smallskip

\aktTag = 0

Die Tatsache, da{\ss} ein Eigenwert $E^{(m)} (\beta)$ des Hamiltonoperators
(10.2-1) f\"ur $\beta \to \infty$ gem\"a{\ss} Gl. (10.3-15) wie
$\beta^{1/(m+1)}$ w\"achst, hat weitreichende Konsequenzen, wenn man
versucht, die divergente St\"orungsreihe (10.2-4) mit Hilfe von
verallgemeinerten Summationsprozessen zu summieren, die aus den
Partialsummen der St\"orungsreihe rationale Approximationen konstruieren.
Man kann davon ausgehen, da{\ss} ein Summationsverfahren die divergente
St\"orungsreihe (10.2-4) nur dann f\"ur gr\"o{\ss}ere Kopplungskonstanten
effizient summieren kann, wenn es in der Lage ist, das asymptotische
Verhalten eines Energieeigenwertes gem\"a{\ss} Gl. (10.3-15) zu reproduzieren.

In dieser Arbeit werden ausschlie{\ss}lich verallgemeinerte
Summationsprozesse behandelt, bei denen die Partialsummen einer formalen
Potenzreihe
$$
f (z) \; = \; \sum_{\nu=0}^{\infty} \, \gamma_{\nu} \, z^{\nu}
\tag
$$
in eine Folge rationaler Funktionen
$$
{\cal R}_{m, n} (z) \; = \;
\frac {\displaystyle A_m (z)} {\displaystyle B_n (z)}
\; = \;
\frac
{\displaystyle \sum_{\mu=0}^{m} \, a_{\mu} \, z^{\mu}}
{\displaystyle \sum_{\nu=0}^{n} \, b_{\nu} \, z^{\nu}}
\tag
$$
transformiert werden. Man ist also mit dem Problem konfrontiert, das
bekannte asymptotische Verhalten einer Funktion $f (z)$ f\"ur $z \to
\infty$ bei der Konstruktion rationaler Approximationen zu
ber\"uck\-sich\-ti\-gen.

Wenn man die Folge der Partialsummen einer langsam konvergierenden oder
divergenten Potenzreihe in eine Folge von zweifach indizierten
rationalen Funktionen ${\cal R}_{m, n} (z)$ transformiert, mu{\ss} man sich
zuerst entscheiden, wie man die Z\"ahler- und Nennerpolynome $A_m (z)$
und $B_n (z)$ in Gl. (10.4-2) w\"ahlen will. Normalerweise versucht man
immer, die Z\"ahler- und Nennerpolynome $A_m (z)$ und $B_n (z)$ so zu
w\"ahlen, da{\ss} ihre Polynomgrade $m$ und $n$ entweder gleich sind oder da{\ss}
sie sich so wenig wie m\"oglich unterscheiden. Im Rahmen dieser Arbeit
wurde bisher immer so vorgegangen. Beispielsweise wurden beim Wynnschen
$\epsilon$-Algorithmus, Gl. (2.4-10), die Approximationen zum
(verallgemeinerten) Grenzwert einer Folge $\Seqn s$ immer gem\"a{\ss} Gl.
(4.4-8) gew\"ahlt. Wenn die Eingabedaten die Partialsummen
$\bigSeq {f_n (z)} {n=0}$ einer Potenzreihe vom Typ von Gl. (10.4-1)
sind, erh\"alt man auf diese Weise die treppenartige Folge (4.4-9) in der
Pad\'e-Tafel, die aus den Diagonalelementen $[n / n]$ und den
Nebendiagonalelementen $[n + 1 / n]$ besteht.

In manchen F\"allen kann man explizit zeigen, da{\ss} man auf diese Weise die
besten Ergebnisse in Konvergenzbeschleunigungs- oder Summationsverfahren
erh\"alt. Wenn man beispielsweise die Partialsummen $\bigSeq {f_n (z)}
{n=0}$ einer Stieltjesreihe vom Typ von Gl. (4.3-4) in
Pad\'e-Approximationen transformiert, dann liefern die diagonalen
Approximationen $[n / n]$ die besten Ergebnisse, wenn man die
Eingabedaten $f_0 (z)$, $f_1 (z)$, $\cdots$, $f_{2 n} (z)$ verwendet,
und wenn man die Eingabedaten $f_0 (z)$, $f_1 (z)$, $\cdots$, $f_{2 n +
1} (z)$ verwendet, liefern entweder $[n + 1 / n]$ oder $[n / n + 1]$ die
besten Ergebnisse [Wynn 1968, Theorem 5].

Auch bei anderen verallgemeinerten Summationsprozessen wurden in dieser
Arbeit bevorzugt diagonale Approximationen verwendet. Wenn man
beispielsweise die verallgemeinerten Summationsprozesse $d_k^{(n)}
(\zeta, s_n)$, Gl. (5.2-18), und ${\delta}_k^{(n)} (\zeta, s_n)$, Gl.
(5.4-13), auf die Partialsummen $\bigSeq {f_n (z)} {n=0}$ der formalen
Potenzreihe (10.4-1) anwendet, erh\"alt man gem\"a{\ss} Gln. (5.7-3) und (5.7-4)
aus\-schlie{\ss}\-lich diagonale Approximationen, wenn man die Folgen
$$
d_0^{(0)} \bigl(\zeta, f_0 (z) \bigr),
d_1^{(0)} \bigl(\zeta, f_0 (z) \bigr),
d_2^{(0)} \bigl(\zeta, f_0 (z) \bigr),
\ldots, d_n^{(0)} \bigl(\zeta, f_0 (z) \bigr), \ldots
\tag
$$
und
$$
\delta_0^{(0)} \bigl(\zeta, f_0 (z) \bigr),
\delta_1^{(0)} \bigl(\zeta, f_0 (z) \bigr),
\delta_2^{(0)} \bigl(\zeta, f_0 (z) \bigr),
\ldots, \delta_n^{(0)} \bigl(\zeta, f_0 (z) \bigr), \ldots
\tag
$$
zur Approximation von $f (z)$ verwendet.

Nehmen wir jetzt an, da{\ss} die durch die formale Potenzreihe (10.4-1)
definierte Funktion $f (z)$ das folgende asymptotische Verhalten
aufweist:
$$
f (z) \; \sim \; z^{\alpha} \, , \qquad z \to \infty \, ,
\quad \alpha \in \R \, .
\tag
$$
Wenn $z$ klein ist, sollte man eine Funktion dieses Typs normalerweise
ausreichend genau durch rationale Funktionen ${\cal R}_{m, n} (z)$
approximieren k\"onnen. Wenn $z$ aber gro{\ss} ist, spielt das asymptotische
Verhalten von $f (z)$ eine zentrale Rolle, und der Parameter $\alpha$ in
Gl. (10.4-5) entscheidet letztlich, ob und wie gut rationale Funktionen
zur Approximation von $f (z)$ geeignet sind.

Eine diagonale rationale Funktion ${\cal R}_{n, n} (z)$ mit $a_n \ne 0$
und $b_n \ne 0$ geht f\"ur $z \to \infty$ gegen die Konstante $a_n/b_n$.
Dagegen w\"achst eine nichtdiagonale rationale Funktion ${\cal R}_{m, n}
(z)$ f\"ur $z \to \infty$ wie eine positive ganzzahlige Potenz von $z$,
wenn das Z\"ahlerpolynom $A_m (z)$ einen h\"oheren Grad hat als das
Nennerpolynom $B_n (z)$, und sie f\"allt wie eine negative ganzzahlige
Potenz von $z$, wenn das Z\"ahlerpolynom einen niedrigeren Grad hat als das
Nennerpolynom. Das bedeutet, da{\ss} $f (z)$ nur dann gut durch diagonale
rationale Funktionen approximiert werden kann, wenn $f (z)$ f\"ur $z \to
\infty$ gegen eine von Null verschiedene Konstante geht, d.~h., wenn
$\alpha = 0$ gilt. Wenn $\alpha$ dagegen eine positive ganze Zahl ist,
sollte man $f (z)$ durch rationale Funktionen approximieren, deren
Z\"ahlerpolynome einen entsprechend h\"oheren Grad haben als die
Nennerpolynome, und wenn $\alpha$ eine negative ganze Zahl ist, sollten
die Nennerpolynome einen entsprechend h\"oheren Grad aufweisen als die
Z\"ahlerpolynome.

Besonders schwierig ist die Situation, wenn $\alpha$ einen
nichtganzzahligen Wert hat, da dann {\it keine\/} rationale Funktion
${\cal R}_{m, n} (z)$ das asymptotische Verhalten von $f (z)$ f\"ur $z \to
\infty$ reproduzieren kann. Das bedeutet, da{\ss} rationale Funktionen
dieses Typs prinzipiell nicht zur Approximation von $f (z)$ geeignet
sind, wenn $z$ gro{\ss} ist.

Die Eigenwerte $E^{(m)} (\beta)$ des Hamiltonoperators (10.2-1) wachsen
gem\"a{\ss} Gl. (10.3-15) wie $\beta^{1/(m+1)}$ f\"ur $\beta \to \infty$.
Rationale Approximationen k\"onnen die St\"orungsreihe (10.2-4) also nur
dann einigerma{\ss}en effizient summieren, wenn $\beta$ klein oder h\"ochstens
mittelgro{\ss} ist. Wenn die Kopplungskonstante $\beta$ gro{\ss} ist, sind
rationale Funktionen des Typs von Gl. (10.4-2) aufgrund ihres falschen
asymptotischen Verhaltens f\"ur $\beta \to \infty$ prinzipiell ungeeignet
f\"ur eine genaue Approximation eines Eigenwertes $E^{(m)} (\beta)$. Eine
effiziente Summation der St\"orungsreihe (10.2-4) durch rationale
Funktionen in $\beta$ ist dann nicht m\"oglich.

Rationale Funktionen besitzen Eigenschaften, die f\"ur numerische Zwecke
\"au{\ss}erst vorteilhaft sind. Es w\"are deswegen sicherlich w\"unschenswert, wenn
man rationale Funktionen auch f\"ur gro{\ss}e Kopplungskonstanten $\beta$ zur
Summation der St\"orungsreihe (10.2-4) verwenden k\"onnte. Das setzt aber
voraus, da{\ss} man Mittel und Wege findet, das asymptotische Verhalten eines
Energieeigenwertes $E^{(m)} (\beta)$ f\"ur $\beta \to \infty$ explizit zu
ber\"ucksichtigen. Allerdings kann dies offensichtlich nicht auf der Ebene
der rationalen Approximationen geschehen. Es ist vielmehr notwendig, die
St\"orungsreihe (10.2-4) auf geeignete Weise zu transformieren.

In den letzten Jahren wurden Methoden zur Transformation von
St\"orungsreihen entwickelt, die in der Literatur \"ublicherweise {\it
Renormierung\/} genannt werden (siehe beispielsweise Austin [1984],
Austin und Killingbeck [1982], Banerjee [1979], Bonnier [1978], Caswell
[1979], Cohen und Kais [1984; 1986], Dmitrieva und Plindov [1980a],
[1980b], Fern\'{a}ndez und Castro [1983], Hsue und Chern [1984],
Killingbeck [1981], Makarewicz [1984], Maluendes, Fern\'{a}ndez und
Castro [1985], Parisi [1977], Seznec und Zinn-Justin [1979], Vrscay
[1988], Yamazaki [1984], Zinn-Justin [1981a]). Mit Hilfe solcher
Renormierungsverfahren kann das bekannte asymptotische Verhalten eines
Energieeigenwertes $E^{(m)} (\beta)$ f\"ur $\beta \to \infty$ bei der
Konstruktion rationaler Approximationen explizit ber\"ucksichtigt werden,
wodurch die Summation einer St\"orungsreihe auch f\"ur gro{\ss}e
Kopplungskonstanten praktikabel wird. Eine vergleichende \"Ubersicht \"uber
verschiedene Renormierungstechniken findet man auf S. 126 - 135 des
Buches von Arteca, Fern\'{a}ndez und Castro [1990].

Mathematisch gesehen sind Renormierungsverfahren
Variablensubstitutionen, bei denen die Kopplungskonstante $\beta$ einer
St\"orungsreihe vom Typ von Gl. (10.1-2) in eine neue Kopplungskonstante
transformiert wird. \"Ublicherweise erreicht man dies durch eine geeignete
Transformation des Hamiltonoperators. Im zweiten Schritt wird dann mit
Hilfe des Formalismus der Rayleigh-Schr\"odingerschen St\"orungstheorie eine
formale Potenzreihe f\"ur den Energieeigenwert $E (\beta)$ in der
renormierten Kopplungskonstante konstruiert. Ein Renormierungsverfahren
ist dann erfolgreich, wenn die neue St\"orungsreihe g\"unstigere numerische
Eigenschaften besitzt als die urspr\"ungliche St\"orungsreihe in $\beta$.

In dieser Arbeit wird die St\"orungsreihe (10.2-4) f\"ur die
Grundzustandsenergie anharmonischer Oszillatoren mit Hilfe eines
Verfahrens renormiert, das auf der in Abschnitt 10.3 beschriebenen
Skalentransformation basiert. Dieses Renormierungsverfahren wurde zuerst
von {\v C}{\' \i}{\v z}ek und Vrscay [1986] verwendet, um die
Grundzustandsenergie anharmonischer Oszillatoren mit einer $\hat{x}^4$-
und $\hat{x}^6$-Anharmonizit\"at unter Verwendung der von L\"owdin [1966;
1982b] eingef\"uhrten Methode der inneren Projektion zu berechnen. Sp\"ater
wurde diese Technik von Vinette, {\v C}{\' \i}{\v z}ek und Vrscay
[1987a; 1987b] auch bei anharmonischen Oszillatoren mit einer
$\hat{x}^8$-Anharmonizit\"at angewendet. In Abschnitt 2 der Dissertation
von Vinette [1989] und in den Abschnitten \Roemisch{2} und \Roemisch{3}
von Vinette und {\v C}{\'\i}{\v z}ek [1991] findet man eine detaillierte
Beschreibung dieses Renormierungsverfahrens. Die Verwendung dieses
Renormierungsverfahrens bei der Summation der St\"orungsreihe (10.2-4) f\"ur
die Grundzustandsenergie anharmonischer Oszillatoren mit einer
$\hat{x}^4$-, $\hat{x}^6$- und $\hat{x}^8$-Anharmonizit\"at wird in
Abschnitt \Roemisch{3} von Weniger, {\v C}{\'\i}{\v z}ek und Vinette
[1993] ausf\"uhrlich behandelt. Weitere Anwendungen dieses
Renormierungsverfahrens findet man in Artikeln von {\v C}{\'\i}{\v z}ek
und Vinette [1988], {\v C}{\'\i}{\v z}ek, Vinette und Vrscay [1987], {\v
C}{\'\i}{\v z}ek, Vinette und Weniger [1991; 1993], Vinette und {\v
C}{\'\i}{\v z}ek [1989], Weniger [1992] und Weniger, {\v C}{\'\i}{\v
z}ek und Vinette [1991].

Das oben erw\"ahnte Renormierungsverfahren basiert darauf, da{\ss} man den im
vorherigen Unterabschnitt definierten unit\"aren Operator $\hat{U} (\tau)$
auf den Hamiltonoperator (10.2-1) anwendet. Unter Verwendung der Gln.
(10.3-4) und (10.3-5) erh\"alt man dann:
$$
\beginAligntags
" \hat{H}^{(m)} (\beta; \tau) " \; = \; "
\tau^{-1} \hat{P}^{2} \, + \, \tau \hat{X}^{2}
\, + \, \beta \tau^{m} \hat{X}^{2m} \\
" " \; = \; " \tau^{-1} \bigl\{ \hat{P}^{2} \, + \, \hat{X}^{2}
\, + \, \beta \tau^{m+1} \hat{X}^{2m}
\, - \, (1 - \tau^2) \hat{X}^{2} \bigl\} \, .
\\ \tag
\endAligntags
$$
Durch Multiplikation dieser Beziehung mit dem Skalenparameter $\tau$
erh\"alt man einen neuen, {\ renormierten\/} Hamiltonoperator:
$$
\beginAligntags
" \hat{\cal H}^{(m)} (\beta; \tau) " \; = \; "
\tau \hat{H}^{(m)} (\beta; \tau) \\
" " \; = \; "
\hat{P}^{2} \, + \, \hat{X}^{2} \, + \,
\beta \tau^{m+1} \hat{X}^{2m} \, - \, (1 - \tau^2) \hat{X}^{2} \, .
\\ \tag
\endAligntags
$$

Der Hamiltonoperator (10.4-6) enth\"alt noch den bisher unspezifizierten
positiven Skalenparameter $\tau$. Wir w\"ahlen $\tau$ so, da{\ss} der
Erwartungswert des transformierten Hamiltonoperators (10.4-6) bez\"uglich
der Grundzustandseigenfunktion $\Phi_{0}$ des ungest\"orten renormierten
Hamiltonoperators $\hat{P}^{2} + \hat{X}^{2}$ minimal wird [Vinette und
{\v C}{\'\i}{\v z}ek 1991, Gl. (10)]:
$$
\frac{{\rm d}}{{\rm d} \tau } \,
\left\{
\frac { \langle \Phi_{0}
\mid \hat{H}^{(m)} (\beta; \tau) \mid \Phi_{0} \rangle }
{ \langle \Phi_{0} \mid \Phi_{0} \rangle } \right\}
\; = \; 0 \, .
\tag
$$
Diese physikalisch motivierbare Extremalbedingung ergibt f\"ur beliebiges
$m = 2, 3, 4, \ldots$ eine einfache explizite Beziehung, die die
Kopplungskonstante $\beta$ mit dem Skalenparameter $\tau$ verkn\"upft
[Vinette und {\v C}{\'\i}{\v z}ek 1991, Gl. (18)]:
$$
B_m \tau^{m + 1} \beta \; = \; 1 - \tau^{2} \, .
\tag
$$
Dabei sind die Konstanten $B_m$ folgenderma{\ss}en definiert [Vinette und {\v
C}{\'\i}{\v z}ek 1991, Gl. (19)]:
$$
B_m \; = \; \frac {m \, (2 m - 1)!!} {2^{m - 1}} \, ,
\qquad m \, = \, 2,3,4, \ldots \, .
\tag
$$
F\"ur $m = 2, 3, 4$ ergibt dies $B_2 = 3$, $B_3 = 45/4$ und $B_4 = 105/2$.

Anstelle der einfachen Extremalbedingung (10.4-8) kann man auch
aufwendigere Verfahren zur Bestimmung des Skalenparameters $\tau$
verwenden, beispielsweise die Verfahren von Austin [1984], Austin und
Killingbeck [1982], Caswell [1979] und Killingbeck [1981]. Es besteht
kein Zweifel, da{\ss} es dadurch im Prinzip m\"oglich sein sollte, bessere
Resultate zu erhalten. Allerdings sind diese Verfahren wesentlich
komplizierter als das hier beschriebene Renormierungsverfahren, das
nicht nur auf einfache Weise angewendet werden kann, sondern auch sehr
transparente analytische Ergebnisse liefert.

Wenn man jetzt eine {\it renormierte\/} Kopplungskonstante
$$
\kappa \; = \; 1 - \tau^{2}
\tag
$$
einf\"uhrt, dann folgt aus Gln. (10.4-9) und (10.4-11), da{\ss} man die
Kopplungskonstante $\beta$ auf folgende Weise durch die renormierte
Kopplungskonstante $\kappa$ ausdr\"ucken kann [Weniger, {\v C}{\'\i}{\v
z}ek und Vinette 1993, Gl. (3.19)]:
$$
\beta \; = \; \frac {1} { B_m } \,
\frac {\kappa} {(1 - \kappa)^{(m + 1)/2}} \, ,
\qquad m = 2,3,4 \, .
\tag
$$
F\"ur $m = 2, 3, 4$ ergibt das die folgenden Beziehungen zwischen $\beta$
und $\kappa$:
$$
\beginAligntags
" \beta \; " = \; " \frac{\kappa}{3(1- \kappa )^{3/2}} \, ,
" \qquad m \; = \; 2 \, ,
\erhoehe\aktTag \\ \tag*{\tagnr a}
" \beta \; " = \; " \frac{4 \kappa}{45(1- \kappa)^{2}} \, ,
" \qquad m \; = \; 3 \, ,
\\ \tag*{\tagform\aktTagnr b}
" \beta \; " = \; " \frac{2 \kappa}{105(1- \kappa)^{5/2}} \, ,
" \qquad m \; = \; 4 \, .
\\ \tag*{\tagform\aktTagnr c}
\endAligntags
$$

Der physikalisch relevante Bereich der Kopplungskonstante $\beta$ ist
die positive reelle Halbachse $[0, \infty)$. Aus Gl. (10.4-12) folgt,
da{\ss} das halbunendliche Intervall $[0, \infty)$ f\"ur $\beta$ auf das
kompakte Intervall $[0, 1)$ f\"ur $\kappa$ abgebildet wird. Diese
Kontraktion l\"a{\ss}t vermuten, da{\ss} man durch Renormierung die Summation der
divergenten St\"orungsreihe (10.2-4) tats\"achlich erleichtern kann.

Aus Gln. (10.4-7), (10.4-9) und (10.4-11) folgt, da{\ss} der renormierte
Hamiltonoperator $\hat{\cal H}^{(m)}$ auf folgende Weise geschrieben
werden kann,
$$
\hat{\cal H}^{(m)} (\kappa) \; = \; \hat{\cal H}_{0} +
\kappa \, \hat{\cal V}^{(m)} \, ,
\tag
$$
wobei
$$
\hat{\cal H}_{0} \; = \; \hat{P}^{2} + \hat{X}^{2}
\tag
$$
der ungest\"orte renormierte Hamiltonoperator ist. Der renormierte
St\"oroperator $\hat{\cal V}^{(m)}$ erf\"ullt die folgende Beziehung [Vinette
und {\v C}{\'\i}{\v z}ek 1991, Gl. (31)]:
$$
\hat{\cal V}^{(m)} \; = \; \frac {1} {\, B_m} \,
\bigl\{ \hat{X}^{2 m} \, - \, B_m \, \hat{X}^{2} \bigr\} \, .
\tag
$$
F\"ur $m = 2, 3, 4$ ergibt das die folgenden St\"oroperatoren:
$$
\beginAligntags
" \hat{\cal V}^{(2)} \; & = \; & \frac{1}{3} \,
\bigl[ \hat{X}^{4} - 3 \hat{X}^{2} \bigr] \, ,
\erhoehe\aktTag \\ \tag*{\tagnr a}
" \hat{\cal V}^{(3)} \; & = \; & \frac{4}{45} \,
\bigl[ \hat{X}^{6} - \frac{45}{4} \hat{X}^{2} \bigr] \, ,
\\ \tag*{\tagform\aktTagnr b}
" \hat{\cal V}^{(4)} \; & = \; & \frac{2}{105} \,
\bigl[ \hat{X}^{8} - \frac{105}{2} \hat{X}^{2}\bigr] \, .
\\ \tag*{\tagform\aktTagnr c}
\endAligntags
$$
Diese Beziehungen zeigen, da{\ss} das hier beschriebene
Renormierungsverfahren den parabelartigen St\"oroperator $\hat{x}^{2 m}$
im urspr\"unglichen Hamiltonoperator (10.2-1) in ein
Doppelminimumpotential transformiert.

Der renormierte Hamiltonoperator $\hat{\cal H}^{(m)} (\kappa)$ erf\"ullt
die zeitunabh\"angige Schr\"odingergleichung
$$
\hat{\cal H}^{(m)} (\kappa) \, \Psi^{(m)} (X) \; = \;
E_{R}^{(m)} (\kappa) \, \Psi^{(m)} (X) \, ,
\tag
$$
wobei der Index $R$ am Grundzustandsenergieeigenwert $E_{R}^{(m)}
(\kappa)$ betonen soll, da{\ss} es sich um eine renormierte Energie handelt.

Der Formalismus der Rayleigh-Schr\"odingerschen St\"orungstheorie liefert
f\"ur die renormierte Grundzustandsenergie
$$
E_{R}^{(m)}(\kappa) \; = \; (1 - \kappa )^{1/2} \, E^{(m)} (\beta)
\tag
$$
eine formale Potenzreihe in $\kappa$ [Weniger, {\v C}{\'\i}{\v z}ek und
Vinette 1993, Gl. (3.31)]:
$$
E_{R}^{(m)} (\kappa) \; = \;
\sum_{n=0}^{\infty} \, c_{n}^{(m)} \, \kappa^{n} \, .
\tag
$$
Die Berechnung der Koeffizienten $b_{n}^{(m)}$ der St\"orungsreihe
(10.2-4) und $c_{n}^{(m)}$ der St\"orungsreihe (10.4-20) mit Hilfe
nichtlinearer zweidimensionaler Differenzenschemata wird im n\"achsten
Unterabschnitt beschrieben. Man kann die Koeffizienten $c_{n}^{(m)}$
aber auch direkt aus den $b_{n}^{(m)}$ berechnen. Aus Gln. (10.2-4) und
(10.4-19) folgt, da{\ss} man $E_{R}^{(m)}(\kappa)$ auch durch die folgende
formale Potenzreihe in $\beta$ darstellen kann:
$$
E_{R}^{(m)}(\kappa) \; = \;
(1-\kappa)^{1/2} \, \sum_{n=0}^{\infty} \, b_{n}^{(m)} \, \beta^{n} \, .
\tag
$$
Wenn man jetzt $\beta$ gem\"a{\ss} Gl. (10.4-12) durch $\kappa$ ersetzt, mu{\ss}
man nur noch eine Taylorentwicklung der rechten Seite von Gl. (10.4-21)
durchf\"uhren. In MAPLE kann dies leicht mit Hilfe der Kommandos {\it
subs\/}, {\it taylor\/} und {\it coeff\/} geschehen [Char, Geddes,
Gonnet, Leong, Monagan und Watt 1991b, S. 19, 206 und 213].

Ein Renormierungsverfahren ist dann erfolgreich, wenn die transformierte
St\"orungsreihe g\"unsti\-gere numerische Eigenschaften besitzt als die
urspr\"ungliche St\"orungsreihe. Da der St\"oroperator (10.4-16) den Term
$\hat{X}^{2 m}$ enth\"alt, kann er nicht relativ zum ungest\"orten
renormierten Hamilton\-operator (10.4-15) beschr\"ankt sein. Das
impliziert, da{\ss} die Katosche Ungleichung (10.1-4) nicht erf\"ullt sein
kann. Der St\"oroperator (10.4-16) ist also keine analytische St\"orung, und
man hat keinen Grund zu der Annahme, da{\ss} die renormierte St\"orungsreihe
(10.4-20) konvergieren k\"onnte.

Trotzdem f\"uhrt das hier beschriebene Renormierungsverfahren -- wie man
leicht zeigen kann -- zu einer signifikanten Verbesserung der
numerischen Eigenschaften. Banks und Bender [1972] analysierten das
asymptotische Verhalten der St\"orungstheoriekoeffizienten von
Hamiltonoperatoren des Typs von Gl. (10.2-1) und von transformierten
Hamiltonoperatoren, die den renormierten Hamiltonoperator (10.4-14) als
Spezialfall enthalten. Mit Hilfe einer asymptotischen Absch\"atzung von
Banks und Bender [1972, Gl. (4)] kann man zeigen, da{\ss} die folgenden
Grenzwerte von den St\"orungstheoriekoeffizienten $b_{n}^{(m)}$ und den
renormierten St\"orungstheoriekoeffizienten $c_{n}^{(m)}$ erf\"ullt werden
[Weniger, {\v C}{\'\i}{\v z}ek und Vinette 1993, Gln. (3.33) - (3.35)]:
$$
\beginAligntags
" \lim_{n \rightarrow \infty} \,
\frac {b_{n}^{(2)}} {3^{n} \, c_{n}^{(2)}} " \; = \; " {\rm e}^{3} \, ,
\erhoehe\aktTag \\ \tag*{\tagnr a}
" \lim_{n \rightarrow \infty } \,
\frac {b_{n}^{(3)}} {(45/4)^{n} \, c_{n}^{(3)}}
" \; = \; " 1 \, ,
\\ \tag*{\tagform\aktTagnr b}
" \lim_{n \rightarrow \infty } \,
\frac {b_{n}^{(4)}} {(105/2)^{n} \, c_{n}^{(4)}}
" \; = \; " 1 \, .
\\ \tag*{\tagform\aktTagnr c}
\endAligntags
$$
Numerische Tests ergaben, da{\ss} die Quotienten in Gl. (10.4-22) ihre
Grenzwerte relativ schnell erreichen. Wenn man diese Grenzwerte mit Gl.
(10.2-6) kombiniert, erh\"alt man die f\"uhrenden Terme der asymptotischen
Entwicklungen der renormierten St\"orungstheoriekoeffizienten $c_{n}^{(m)}$
f\"ur $n \to \infty$ [Weniger, {\v C}{\'\i}{\v z}ek und Vinette 1993, Gln.
(3.36) - (3.38)]:
$$
\beginAligntags
" c_{n}^{(2)} \, " \sim \, " (-1)^{n+1} \,
\frac{(24)^{1/2}}{e^3 \, \pi^{3/2}} \,
\Gamma(n+1/2) \, (1/2)^{n} \, ,
\erhoehe\aktTag \\ \tag*{\tagnr a}
" c_{n}^{(3)} \, " \sim \, " (-1)^{n+1} \,
\frac{(128)^{1/2}}{\pi^{2}} \, \Gamma(2 n + 1/2) \,
\bigl( 64 / [45 \pi^{2}] \bigr)^{n} \, ,
\\ \tag*{\tagform\aktTagnr b}
" c_{n}^{(4)} \, " \sim \, " (-1)^{n+1} \,
\left\{
\frac
{270 [ \Gamma(2/3)]^{3}} {\pi^{5}} \right\}^{1/2}
\, \Gamma (3 n + 1/2) \, \left\{
\frac{225 [ \Gamma (2/3)]^{9}} {112 \pi^{6}} \right\}^{n} \, .
\\ \tag*{\tagform\aktTagnr c}
\endAligntags
$$
Offensichtlich kann das hier beschriebene Renormierungsverfahren nichts
daran \"andern, da{\ss} die St\"orungstheoriekoeffizienten eines anharmonischen
Oszillators mit einer $\hat{x}^4$-, $\hat{x}^6$- und
$\hat{x}^8$-Anharmonizit\"at f\"ur $n \to \infty$ im wesentlichen wie
$n!/n^{1/2}$, $(2 n)!/n^{1/2}$, beziehungsweise $(3 n)!/n^{1/2}$
wachsen. Trotzdem folgt aus Gl. (10.4-22), da{\ss} die
St\"orungstheoriekoeffizienten $b_{n}^{(m)}$ wesentlich schneller wachsen
als die renormierten St\"orungstheoriekoeffizienten $c_{n}^{(m)}$.

Um den Effekt des Renormierungsverfahrens genauer analysieren zu k\"onnen,
bezeichnen wir $\bigl\[ B_m \bigr\]^n c_{n}^{}$ als effektive
renormierte St\"orungstheoriekoeffizienten und $\kappa/B_m$ als eine
effektive renormierte Kopplungskonstante. Aus Gl. (10.4-22) folgt
$$
[ B_m ]^n \, c_n^{(m)} \; = \; O \bigl( b_n^{(m)} \bigr) \, ,
\qquad n \to \infty \, .
\tag
$$
Au{\ss}erdem impliziert Gl. (10.4-12)
$$
\beta \; > \; \kappa / B_m \, , \qquad m = 2,3,4 \, ,
\tag
$$
wenn $\beta > 0$ gilt. Das hier beschriebene Renormierungsverfahren
transformiert also die urspr\"ungliche St\"orungsreihe (10.2-4) mit der
unbeschr\"ankten Kopplungskonstante $\beta \in [0, \infty)$ in eine neue
St\"orungsreihe mit effektiven St\"orungstheoriekoeffizienten $[ B_m ]^n
c_n^{(m)}$, die gem\"a{\ss} Gl. (10.4-24) f\"ur gro{\ss}e Indizes $n$ ebenso rasch
wachsen wie die urspr\"unglichen St\"orungstheoriekoeffizienten
$b_{n}^{(m)}$. Im Gegensatz zur urspr\"unglichen Kopplungskonstante
$\beta$ ist die effektive Kopplungskonstante $\kappa/B_m$ aber
beschr\"ankt und au{\ss}erdem gem\"a{\ss} Gl. (10.4-12) immer kleiner als $\beta$.
Aufgrund der Gln. (10.4-24) und (10.4-25) k\"onnen wir also folgern, da{\ss}
die renormierte St\"orungsreihe (10.4-20) f\"ur alle $\beta > 0$ weniger
stark divergiert als die urspr\"ungliche St\"orungsreihe (10.2-4).

Die unterschiedlichen Eigenschaften der urspr\"unglichen St\"orungsreihe
(10.2-4) und der renor\-mier\-ten St\"orungsreihe (10.4-20) werden besonders
evident, wenn $\beta$ gro{\ss} ist. \"Uberzeugende Beispiele f\"ur die
prinzipielle \"Uberlegenheit der renormierten St\"orungsreihe (10.4-20) sind
die {\it Grenzf\"alle unendlicher Kopplung\/}
$$
k_m \; = \; \lim_{\beta \to \infty} \,
\frac {E^{(m)} (\beta)} {\beta^{1/(m+1)}} \, ,
\qquad m = 2, 3, 4 \, .
\tag
$$
Aus Gln. (10.3-13), (10.3-14) und (10.3-16) - (10.3-19) folgt, da{\ss} $k_m$
identisch ist mit dem Grundzustandsenergieeigenwert des Hamiltonoperators
$$
\hat{\bf H}^{(m)}_0 \; = \; \hat{P}^2 \, + \, \hat{X}^{2 m} \, .
\tag
$$
Eine Berechnung der $k_m$ durch Summation der St\"orungsreihe (10.2-4)
w\"urde voraussetzen, da{\ss} wir die St\"orungsreihe (10.2-4) auch f\"ur $\beta
\to \infty$ effizient summieren k\"onnen. Es ist fraglich, ob dies
\"uberhaupt m\"oglich ist, und wenn es tats\"achlich m\"oglich sein sollte, wird
es sicherlich nicht einfach sein. Wenn man aber die renormierte
St\"orungsreihe (10.4-20) verwendet, kann man die Grenzf\"alle unendlicher
Kopplung relativ leicht durch Summation berechnen.

Aus Gln. (10.4-19) und (10.4-20) folgt, da{\ss} man den Energieeigenwert
$E^{(m)} (\beta)$ folgenderma{\ss}en darstellen kann:
$$
E^{(m)} (\beta) \; = \; (1 - \kappa)^{- 1/2} \,
\sum_{n=0}^{\infty} \, c_{n}^{(m)} \, \kappa^{n} \, .
\tag
$$
Weiterhin folgt aus Gl. (10.4-12)
$$
\beta^{1/(m+1)} \; \sim \; (1 - \kappa)^{- 1/2} \, ,
\qquad \beta \to \infty \, .
\tag
$$
Man sieht also, da{\ss} der Energieeigenwert $E^{(m)} (\beta)$ durch das hier
verwendete Renormierungsverfahren aufgespalten wird in den renormierten
Energieeigenwert $E_{R}^{(m)} (\kappa)$, der f\"ur $\kappa \to 1$ endlich
bleibt, und in einen Anteil $(1 - \kappa)^{- 1/2}$, der f\"ur $\kappa \to
1$ so divergiert wie $\beta^{1/(m+1)}$ f\"ur $\beta \to \infty$.

Zur Berechnung der Grenzf\"alle unendlicher Kopplung mu{\ss} man also nur die
divergente Reihe
$$
E_{R}^{(m)} (1) \; = \; \sum_{n=0}^{\infty} \, c_{n}^{(m)}
\tag
$$
summieren, da aus Gln. (10.4-12), (10.2-26) und (10.4-29)
$$
k_m \; = \; [B_m]^{1/(m+1)} \, E_{R}^{(m)} (1)
\tag
$$
folgt. F\"ur $m = 2, 3, 4$ ergibt das die folgenden Beziehungen:
$$
\beginAligntags
" k_{2} \; " = \; " 3^{1/3} E_{R}^{(2)} (1) \, ,
\erhoehe\aktTag \\ \tag*{\tagnr a}
" k_{3} \; " = \; " (45/4)^{1/4} E_{R}^{(3)} (1) \, ,
\\ \tag*{\tagform\aktTagnr b}
" k_{4} \; " = \; " (105/2)^{1/5} E_{R}^{(4)} (1) \, .
\\ \tag*{\tagform\aktTagnr c}
\endAligntags
$$

Diese Reihenentwicklungen f\"ur die Grenzf\"alle unendlicher Kopplung sind
die schwierigsten Summationsprobleme, die es im Zusammenhang mit der
renormierten St\"orungsreihe (10.4-22) geben kann. Aus diesem Grund sind
diese divergenten Reihen auch sehr gut zum Testen des Leistungsverm\"ogens
eines verallgemeinerten Summationsprozesses geeignet.

Die genauesten bisher bekannten numerischen Werte (62, 33
beziehungsweise 21 Dezimalstellen) f\"ur die Grenzf\"alle unendlicher
Kopplung $k_2$, $k_3$ und $k_4$ wurden durch eine Kombination der von
L\"owdin [1966; 1982b] eingef\"uhrten Methode der inneren Projektion mit dem
in diesem Abschnitt beschriebenen Renormierungsverfahren berechnet
[Vinette und {\v C}{\'\i}{\v z}ek 1991, Gln. (66), (69), und (71)]:
$$
\beginAligntags
" k_2 " \; = \; 1. " 060 \, 362 \, 090 \, 484 \, 182 \, 899 \, 647
\, 046 \, 016 \, 692 \, 663 \\
" "  " 545 \, 515 \, 208 \, 728 \, 528 \, 977 \, 933 \, 216 \, 245
\, 24 \, ,
\erhoehe\aktTag \\ \tag*{\tagnr a}
" k_3 " \; = \; 1. " 144 \, 802 \, 453 \, 797 \, 052 \, 763 \, 765 \,
457 \, 534 \, 149 \, 549 \, ,
\\ \tag*{\tagform\aktTagnr b}
" k_4 " \; = \; 1. " 225 \, 820 \, 113 \, 800 \, 492 \, 191 \, 591 \, .
\\ \tag*{\tagform\aktTagnr c}
\endAligntags
$$
Andere genaue Rechnungen f\"ur $k_2$ wurden von Banerjee, Bhatnagar,
Choudry und Kanwal [1978, Table 1], von Fern\'{a}ndez, Mes\'{o}n und
Castro [1985, Table 4], von Killingbeck, Jones und Thompson [1985, S.
800], und von Richardson und Blankenbecler [1979, S. 502, Ref.~13]
durchgef\"uhrt. Ihre Ergebnisse reproduzierten 14, 15, 19 beziehungsweise
26 Dezimalstellen des extrem genauen Wertes (10.4-33a) von Vinette und
{\v C}{\'\i}{\v z}ek [1991, Gl. (66)]. Im Falle von $k_3$ und $k_4$ gibt
es genaue Rechnungen von Guardiola, Sol\'{\i}s und Ros [1992, Gl. (29)]
und von Fern\'{a}ndez, Mes\'{o}n und Castro [1985, Table 5 und 6], die
jeweils 11 beziehungsweise 15 Dezimalstellen der extrem genauen Werte in
Gln. (10.4-33b) und (10.4-33c) reproduzieren.

\medskip

\Abschnitt Die Berechnung der St\"orungstheoriekoeffizienten

\smallskip

\aktTag = 0

Der \"ubliche Formalismus der Rayleigh-Schr\"odingerschen St\"orungstheorie
zur Berechnung der St\"orungstheoriekoeffizienten, der beispielsweise in
\S 7 des Buches von Arteca, Fern\'{a}ndez und Castro [1990] beschrieben
wird, f\"uhrt zu \"au{\ss}erst komplizierten Ausdr\"ucken mit den f\"ur diesen
Ansatz charakteristischen Energienennern. Man kann auf diese Weise nur
St\"o\-rungs\-theo\-rie\-ko\-effizienten niedriger Ordnung mit
vertretbarem Aufwand berechnen. Es gibt aber alternative Verfahren, die
eine effiziente Berechnung von St\"orungstheoriekoeffizienten auch im
Falle sehr gro{\ss}er Indizes erm\"oglichen (siehe beispielsweise Arteca,
Fern\'{a}ndez und Castro [1990, \S 8 und \S 9]).

Besonders geeignet zur Berechnung der St\"orungstheoriekoeffizienten
$b_{n}^{(m)}$ und $c_{n}^{(m)}$ ist ein Verfahren, das auf Bender und Wu
[1969, S. 1233 und Appendix D; 1973, S. 1629] zur\"uckgeht. Dabei wird die
zeitunabh\"angige Schr\"odingergleichung in jeder Ordnung von $\beta$
explizit als Differentialgleichung gel\"ost, was letztlich ein
nichtlineares Differenzenschema f\"ur die St\"orungstheoriekoeffizien\-ten
ergibt.

In diesem Unterabschnitt werden zweidimensionale nichtlineare
Differenzenschemata f\"ur die St\"orungstheoriekoeffizienten $b_n^{(m)}$ und
$c_n^{(m)}$ abgeleitet und ihre numerischen Eigenschaften analy\-siert.

F\"ur die Grundzustandseigenfunktion $\psi^{(m)} (x)$ des
Hamiltonoperators (10.2-1) machen wir in Analogie zu Bender und Wu
[1969, Gl. (2.5); 1973, Gl. (6.2)] den folgenden Ansatz:
$$
\psi^{(m)} (x) \; = \; \e^{ - x^2 / 2} \,
\sum_{n=0}^{\infty} \, (- \beta)^n \, \phi_n (x) \, .
\tag
$$
Die Funktionen $\phi_n (x)$ sind Polynome in $x$, die bestimmt werden
m\"ussen. In diesem Unterabschnitt wird stillschweigend davon ausgegangen,
da{\ss} die Polynome $\phi_n (x)$ au{\ss}erdem noch explizit von $m$ abh\"angen.

Wenn man den Ansatz (10.5-1) zusammen mit der formalen Potenzreihe
(10.2-4) f\"ur $E^{(m)} (\beta)$ in der Schr\"odingergleichung
$$
\left\{ - \frac {{\rm d}^2} {{\rm d} x^2} + x^2 + \beta x^{2 m} -
 E^{(m)} (\beta) \right\} \, \psi^{(m)} (x) \; = \; 0
\tag
$$
verwendet, erh\"alt man das folgende Gleichungssystem, das die Polynome
$\phi_n (x)$ erf\"ullen m\"ussen:
$$
2 x \phi'_n (x) \, - \, \phi''_n (x) \, - \,
x^{2 m} \phi_{n - 1} (x) \; = \;
\sum_{k=1}^{n} \, (- 1)^k \, b_k^{(m)} \, \phi_{n - k} (x) \, .
\tag
$$
Bei der Ableitung dieser Beziehung wurde die Tatsache verwendet, da{\ss} der
Term $b_0^{(m)}$ in der St\"orungsreihe (10.2-4) identisch ist mit der
Grundzustandsenergie des ungest\"orten harmonischen Oszillators, was
$b_0^{(m)} = 1$ impliziert. Au{\ss}erdem wurde die Konvention verwendet, da{\ss}
f\"ur jedes $l \in \N$
$$
\phi_{- l} (x) \; = \; 0
\tag
$$
gelten soll.

F\"ur die Polynome $\phi_n (x)$ machen wir jetzt den folgenden Ansatz:
$$
\phi_n (x) \; = \; \sum_{j=0}^{m n} \, g_j^{(n)} \, x^{2 j} \, .
\tag
$$
Aufgrund der in der St\"orungstheorie \"ublicherweise verwendeten
intermedi\"aren Normierung m\"ussen die Koeffizienten $g_j^{(n)}$ in Gl.
(10.5-5) f\"ur alle $j, n \in \N_0$ die folgenden Beziehungen erf\"ullen:
$$
\beginAligntags
" g_j^{(0)} " \; = \; " \delta_{j,0} \, ,
\\ \tag
" g_0^{(n)} " \; = \; " \delta_{n,0} \, .
\\ \tag
\endAligntags
$$
Au{\ss}erdem verwenden wir die Konvention, da{\ss} f\"ur alle $n \in \N_0$ und f\"ur
alle $k \in \N$
$$
g_{- k}^{(n)} \; = \; 0
\tag
$$
gelten soll. Wenn wir diese Beziehungen im Gleichungssystem (10.5-3)
verwenden, erhalten wir:
$$
\beginAligntags
4 \, \sum_{j=0}^{m n} \, j g_j^{(n)} x^{2 j} \; = \; "
\sum_{j=0}^{m n} \, g_{j-m}^{(n-1)} x^{2 j} \, + \,
2 \sum_{j=0}^{m n - 1} \, (j + 1) (2 j + 1) g_{j+1}^{(n)} x^{2 j} \\
" \qquad - \, \sum_{k=1}^{n} \, (- 1)^k b_k^{(m)} \,
\sum_{j=0}^{m (n - k)} \, g_j^{(n - k)} x^{2 j} \, .
\\ \tag
\endAligntags
$$
Wenn wir jetzt in Gl. (10.5-9) $x = 0$ setzen, erhalten wir
$$
2 \, g_1^{(n)} \, + \,
\sum_{k=1}^{n} \, (- 1)^k b_k^{(m)} \, g_0^{(n-k)} \; = \; 0 \, .
\tag
$$
Aufgrund von Gl. (10.5-7) ist aber nur der Term mit $n = k$ von Null
verschieden. Wir erhalten also f\"ur alle $n \in \N$:
$$
\frac {1} {2} b_n^{(m)} \; = \; (- 1)^{n + 1} \, g_{1}^{(n)} \, .
\tag
$$
Wenn wir diese Beziehung in Gl.~(10.5-9) einsetzen, erhalten wir das
folgende System von nicht\-linearen Differenzengleichungen f\"ur die
Koeffizienten $g_j^{(n)}$ mit $n \ge 1$ und $1 \le j \le m n$:
$$
4 j \, g_j^{(n)} \; = \; 2 (j + 1) (2 j + 1) g_{j+1}^{(n)}
\, + \, g_{j - m}^{(n-1)} \, - \,
2 \, \sum_{k=1}^{n-1} \, g_1^{(k)} \, g_j^{(n-k)} \, .
\tag
$$

F\"ur die Koeffizienten $c_n^{(m)}$ der renormierten St\"orungsreihe
(10.4-20) kann man auf ganz analoge Weise ein nichtlineares
Differenzenschema ableiten.

F\"ur die Grundzustandseigenfunktionen $\Psi^{(m)} (X)$ des renormierten
Hamiltonoperators (10.4-14) machen wir dazu den folgenden Ansatz:
$$
\Psi^{(m)} (X) \; = \; \e^{ - X^2 / 2} \,
\sum_{n=0}^{\infty} \, ( - \kappa)^n \, \Phi_n (X) \, .
\tag
$$
Analog zu den Funktionen $\phi_n (x)$ in Gl. (10.5-1) sind die
Funktionen $\Phi_n (X)$ Polynome in $X$, die bestimmt werden m\"ussen und
die au{\ss}erdem noch explizit von $m$ abh\"angen.

Wenn wir den Ansatz (10.5-13) zusammen mit der formalen Potenzreihe
(10.4-20) f\"ur $E_R^{(m)} (\kappa)$ in der renormierten
Schr\"odingergleichung
$$
\left\{ - \frac {{\rm d}^2} {{\rm d} X^2} + X^2 +
\frac {\kappa} {B_m} X^{2 m} - \kappa X^2 \right\} \,
\Psi^{(m)} (X) \; = \;
E_R^{(m)} (\kappa) \Psi^{(m)} (X)
\tag
$$
verwenden, erhalten wir das folgende Gleichungssystem, das die Polynome
$\Phi_n (X)$ erf\"ullen m\"ussen:
$$
2 X \Phi'_n (X) \, - \, \Phi''_n (X) \, + \,
\left\{ X^2 - \frac {X^{2 m}} {B_m} \right\} \Phi_{n - 1} (X)
\; = \; \sum_{k=1}^{n} \, (- 1)^k \, c_k^{(m)} \,
\Phi_{n - k} (X) \, .
\tag
$$
Bei der Ableitung dieser Beziehung wurde die Tatsache verwendet, da{\ss} der
Term $c_0^{(m)}$ in der St\"orungsreihe (10.4-20) identisch ist mit der
Grundzustandsenergie des ungest\"orten harmonischen Oszillators, was
$c_0^{(m)} = 1$ impliziert. Au{\ss}erdem wurde die Konvention verwendet, da{\ss}
f\"ur jedes $l \in \N$
$$
\Phi_{- l} (X) \; = \; 0
\tag
$$
gelten soll. F\"ur die Polynome $\Phi_n (X)$ machen wir jetzt den
folgenden Ansatz:
$$
\Phi_n (X) \; = \; \sum_{j=0}^{m n} \, G_j^{(n)} \, X^{2 j} \, .
\tag
$$
Au{\ss}erdem sollen die Koeffizienten $G_j^{(n)}$ in Gl. (10.5-13) die
Konventionen
$$
\beginAligntags
" G_j^{(0)} " \; = \; " \delta_{j,0} \, , \\ \tag
" G_0^{(n)} " \; = \; " \delta_{n,0} \, , \\ \tag
" G_{-k}^{(n)} " \; = \; " 0 \, , \\ \tag
\endAligntags
$$
f\"ur alle $j, n \in \N_0$ und f\"ur alle $k \in \N$ erf\"ullen.

Wenn wir jetzt so vorgehen wie im Falle der Koeffizienten $b_n^{(m)}$,
erhalten wir
$$
\frac {1} {2} c_n^{(m)} \; = \; (- 1)^{n + 1} \, G_{1}^{(n)} \, .
\tag
$$
Mit Hilfe dieser Beziehung erhalten wir das folgende System von
nichtlinearen Differenzengleichungen f\"ur die Koeffizienten $G_j^{(n)}$
mit $n \ge 1$ und $1 \le j \le m n$:
$$
4 j \, G_j^{(n)} \; = \; 2 (j + 1) (2 j + 1) \, G_{j+1}^{(n)} \, + \,
\frac {1} {B_m} \, G_{j - m}^{(n-1)}
\, - \, G_{j - 1}^{(n-1)} \, - \,
2 \, \sum_{k=1}^{n-1} \, G_1^{(k)} \, G_j^{(n-k)} \, .
\tag
$$

Mit Hilfe der nichtlinearen Differenzenschemata (10.5-12) und (10.5-22)
k\"onnen die St\"orungs\-theoriekoeffizienten $b_{n}^{(m)}$ und
$c_{n}^{(m)}$ gem\"a{\ss} Gl.~(10.5-11) beziehungsweise (10.5-21) berechnet
werden. Allerdings gibt es dabei numerische Probleme, die die Berechnung
von St\"o\-rungs\-theoriekoeffizienten mit gro{\ss}en Indizes schwierig macht,
wenn man eine konventionelle Pro\-grammier\-sprache wie FORTRAN 77
verwendet.

Bender und Wu [1969] berechneten die ersten 75
St\"orungstheoriekoeffizienten
$$
a_{n}^{(2)} \; = \; 2^{n - 1} b_{n}^{(2)}
\tag
$$
des anharmonischen Oszillators mit einer $\hat{x}^4$-Anharmonizit\"at mit
Hilfe eines nichtlinearen Differenzenschemas [Bender und Wu 1969, Gln.
(D2) und (D3)], das sich von dem Differenzenschema (10.5-12) mit $m = 2$
im wesentlichen durch eine andere Skalierung unterscheidet, ansonsten
aber \"aquivalent ist. Die numerischen Probleme von Bender und Wu [1969]
sind demzufolge auch f\"ur uns von Interesse.

Laut Bender und Wu [1969, Appendix D] war OVERFLOW bei der Berechnung
der St\"orungs\-theoriekoeffizienten $a_{n}^{(2)}$ bereits f\"ur $n \ge 20$
ein erhebliches Problem. Mit Hilfe einer geeigneten Skalierung kann
OVERFLOW aber f\"ur $n \le 100$ vermieden werden [Bender und Wu 1969, Gl.
D4]. Da die St\"orungstheoriekoeffizienten $b_{n}^{(m)}$ und $c_{n}^{(m)}$
mit wachsendem $n$ betragsm\"a{\ss}ig wesentlich schneller f\"ur $m = 3$ und $m
= 4$ wachsen als f\"ur $m = 2$, d\"urfte OVERFLOW in Falle einer
$\hat{x}^6$- oder $\hat{x}^8$-Anharmonizit\"at ein noch wesentlich
gr\"o{\ss}eres Problem darstellen.

Ein weiteres Problem, das bei dem Differenzenschema von Bender und Wu
[1969, Gln. (D2) und (D3)] auftritt, sind Rundungsfehler. Auf der Basis
ihrer numerischen Ergebnisse schlossen Bender und Wu [1969, Appendix D],
die die St\"orungstheoriekoeffizienten $a_{n}^{(2)}$ in FORTRAN vermutlich
auf einer IBM in DOUBLE PRECISION (15 bis 16 Dezimalstellen Genauigkeit)
berechneten, da{\ss} die dabei auftretenden Rundungsfehler weitgehend
statistischer und nicht systematischer Natur sind. Demzufolge sollte der
kummulative Fehler proportional zur Quadratwurzel der Anzahl der
arithmetischen Operationen sein. Bender und Wu [1969, S. 1251] sch\"atzten
ab, da{\ss} zur Berechnung des St\"orungstheoriekoeffizienten $a_{75}^{(2)}$
mit dem h\"ochsten Index $n$ in etwa $10^8$ arithmetische Operationen
ben\"otigt werden, was einen Genauigkeitsverlust von 4 der urspr\"unglich 16
Dezimalstellen erwarten l\"a{\ss}t. Die Richtigkeit dieser
Rundungsfehleranalyse wird indirekt dadurch best\"atigt, da{\ss} alle 12
Stellen der St\"orungstheoriekoeffizienten $a_{\nu}^{(2)}$ mit $1 \le \nu
\le 75$ korrekt sind, die von Bender und Wu [1969, Table \Roemisch{1}]
angegeben wurden.

Aus den nichtlinearen Differenzengleichungen (10.5-12) und (10.5-22)
folgt aber, da{\ss} die Anzahl der arithmetischen Operationen, die zur
Berechnung der St\"orungstheoriekoeffizienten $b_{n}^{(m)}$ und
$c_{n}^{(m)}$ ben\"otigt werden, mit wachsendem $m$ rasch zunimmt.
Numerische Instabilit\"aten d\"urften also im Falle einer $\hat{x}^6$- oder
$\hat{x}^8$-Anharmonizit\"at noch deutlich gr\"o{\ss}ere Problem aufwerfen,
selbst wenn die Rundungsfehler nur statistischer und nicht
systematischer Natur sind.

Wenn man eine konventionelle Programmiersprache wie FORTRAN 77
verwendet, ist es demzufolge auch bei Verwendung von DOUBLE PRECISION
auf einer Cyber (ungef\"ahr 29 Dezimalstellen) oder QUADRUPLE PRECISION
auf einer IBM (31 - 32 Dezimalstellen) nicht m\"oglich, eine sehr gro{\ss}e
Zahl von St\"orungstheoriekoeffizienten $b_{n}^{(m)}$ und $c_{n}^{(m)}$
mit einer so gro{\ss}en Genauigkeit zu berechnen, da{\ss} man aufwendige
Summationen mit hohen Transformationsordnungen durchf\"uhren k\"onnte.

Die Anfangsbedingungen $b_{0}^{(m)} = 1$ und $c_{0}^{(m)} = 1$ der
Differenzenschemata (10.5-12) und (10.5-22) implizieren aber, da{\ss} die
St\"orungstheoriekoeffizienten $b_{n}^{(m)}$ und $c_{n}^{(m)}$ {\it
rationale\/} Zahlen sind. Es ist also eine relativ naheliegende Idee,
die St\"orungstheoriekoeffizienten $b_{n}^{(m)}$ und $c_{n}^{(m)}$ mit
Hilfe der im Prinzip exakten {\it rationalen Arithmetik\/} von MAPLE zu
berechnen.

Mit Hilfe der folgenden, sehr kompakten MAPLE-Prozedur kann man die in
Gl. (10.2-4) vorkommenden Rayleigh-Schr\"odingerschen
St\"orungstheoriekoeffizienten $b_{\nu}^{(m)}$ mit $0 \le \nu \le n$ unter
Verwendung des zweidimensionalen nichtlinearen Differenzenschemas
(10.5-12) {\it rundungsfehlerfrei\/} berechnen:

\medskip

\listenvon{cf_ns.mpl}

Die Option {\it sparse\/} bei der Definition des zweidimensionalen
Feldes $g$ bewirkt, da{\ss} alle Feldelemente $g [i,j]$ automatisch mit Null
vorbesetzt werden [Char, Geddes, Gonnet, Leong, Monagan und Watt 1992,
S. 89]. Das erleichtert die Programmierung des Differenzenschemas
erheblich.

Die {\it rundungsfehlerfreie\/} Berechnung der in Gl. (10.4-20)
vorkommenden renormierten St\"orungs\-theoriekoeffizienten $c_{\nu}^{(m)}$
mit $0 \le \nu \le n$ unter Verwendung des zweidimensionalen
nichtlinearen Differenzenschemas (10.5-22) erfolgt mit Hilfe der
folgenden, ebenfalls sehr kompakten MAPLE-Prozedur:

\medskip

\listenvon{cf_rs.mpl}

Die Berechnung der St\"orungstheoriekoeffizienten $b_{n}^{(m)}$ und
$c_{n}^{(m)}$ unter Verwendung dieser Prozeduren hat den unbestreitbaren
Vorteil, da{\ss} die St\"orungstheoriekoeffizienten als exakt rationale
Zahlen berechnet werden. Rundungsfehler und OVERFLOW sind dabei kein
Thema.

Allerdings hat diese Vorgehensweise auch einen gravierenden Nachteil.
Schon bei der Berechnung der 75 St\"orungstheoriekoeffizienten
$a_{n}^{(2)}$ in FORTRAN DOUBLE PRECISION war der Spei\-cherbedarf des
Differenzenschemas der komplexit\"atsbestimmende Faktor [Bender und Wu
1969, S. 184]. Die Verwendung der rationalen Arithmetik von MAPLE
versch\"arft dieses Problem erheblich. Eine doppeltgenaue Gleitkommazahl
hat beispielsweise auf einem IBM-Gro{\ss}rechner nur einen
Speicherbedarf von 64 Bit. Dagegen kann der Speicherbedarf einer im
Prinzip un\-be\-schr\"ank\-ten rationalen Zahl beliebig gro{\ss}
werden. Seit den Rechnungen von Bender und Wu [1969] hat es einen
enormen Fortschritt auf dem Gebiet der Computertechnologie gegeben,
wobei besonders Speicherchips um Gr\"o{\ss}enordnungen billiger und
leistungsf\"ahiger geworden sind. Trotzdem ist auch bei modernen
Computern der verf\"ugbare Speicher der komplexit\"atsbestimmende Faktor
bei der Berechnung der St\"orungstheoriekoeffizienten $b_{n}^{(m)}$ und
$c_{n}^{(m)}$.

Dagegen ist der Rechenzeit- und Speicherbedarf der in den n\"achsten
Unterabschnitten beschrie\-benen Summationen unter Verwendung der
Gleitkommaarithmetik von MAPLE selbst dann unproblematisch, wenn man die
Rechnungen mit einer Genauigkeit von 1000 Dezimalstellen durchf\"uhrt. Ein
gro{\ss}er Teil der in den n\"achsten Unterabschnitten pr\"asentierten
Summationsergebnisse wurde mit Hilfe der DOS-Version von MAPLE auf einem
486/33 PC mit 8 MB RAM erzeugt.

Da nur Computer mit maximal 64 MB RAM zur Verf\"ugung standen (eine
Silicon Graphics 4D-340S in Waterloo und eine SUN IPX in Regensburg),
konnten im Falle des anharmonischen Oszillators mit einer
$\hat{x}^4$-Anharmonizit\"at nur die Koeffizienten $b_{n}^{(2)}$ und
$c_{n}^{(2)}$ mit $n \le 200$ exakt rational berechnet werden. Im Falle
der $\hat{x}^6$-Anharmonizit\"at konnten die Koeffizienten $b_{n}^{(3)}$
und $c_{n}^{(3)}$ mit $n \le 165$ exakt rational berechnet werden, und
im Falle der $\hat{x}^8$-Anharmonizit\"at lag die Grenze bei den
Koeffizienten $b_{n}^{(4)}$ und $c_{n}^{(4)}$ mit $n \le 139$.

Ein weiteres Problem der rationalen Arithmetik von MAPLE ist, da{\ss} sie
nat\"urlich um Gr\"o{\ss}enordnungen langsamer ist als die doppelt- oder
vierfachgenaue Gleitkommaarithmetik von FORTRAN. Beispielsweise war die
SUN IPX jeweils mehrere Tage lang mit der Berechnung eines der oben
genannten Strings von St\"orungstheoriekoeffizienten besch\"aftigt. Aufgrund
des gro{\ss}en Speicherbedarfes der rationalen Arithmetik von MAPLE bei der
Berechnung der St\"orungs\-theo\-rie\-ko\-effizi\-en\-ten sollte in dieser
Zeit auch kein anderer Benutzer die SUN IPX verwenden, da sonst ein
st\"orungsfreier Betrieb nicht gew\"ahrleistet ist.

Einige Beispiel f\"ur exakt rational berechnete
St\"orungstheoriekoeffizienten folgen:
$$
\beginAligntags
" b_{10}^{(2)}" \; = \; "
- \frac {6417007431590595} {134217728} \, ,
\\ \tag [1\jot]
" c_{10}^{(2)}" \; = \; "
- \frac {123340039323181} {2641807540224} \, ,
\\ \tag [1\jot]
" b_{10}^{(3)}" \; = \; "
- \frac {21160722559334931139552067094465} {274877906944} \, ,
\\ \tag [1\jot]
" c_{10}^{(3)}" \; = \; "
- \frac {505349738352007529610640051753} {297538935552000000000} \, ,
\\ \tag [1\jot]
" b_{10}^{(4)}" \; = \; "
- \frac {1730368698886327054172173366535535848834046705} {137438953472}
 \, ,
\\ \tag [1\jot]
" c_{10}^{(4)}" \; = \; " - \frac
{2285486899421981602701856519846256674169803}
{29745106932989952000000000} \, .
\\ \tag
\endAligntags
$$

In den Tabellen 10-1, 10-2 und 10-3 werden die exakt rational
berechneten St\"orungs\-theorie\-koeffizienten $b_{\nu}^{(m)}$ und
$c_{\nu}^{(m)}$ f\"ur $m = 2, 3, 4$ und f\"ur $0 \le \nu \le 40$ mit einer
relativen Genauigkeit von 28 Dezimalstellen aufgelistet. Diese Tabellen
zeigen, da{\ss} die Rayleigh-Schr\"odingerschen St\"orungs\-theoriekoeffizienten
$b_{n}^{(m)}$ in \"Ubereinstimmung mit den asymptotischen Absch\"atzungen
(10.2-6) und (10.4-22) tats\"achlich betragsm\"a{\ss}ig wesentlich schneller
wachsen als die renormierten St\"orungstheoriekoeffizienten
$c_{\nu}^{(m)}$. Das ist ein weiteres Indiz, da{\ss} das im letzten
Unterabschnitt beschriebene Renormierungsverfahren tats\"achlich in der
Lage sein sollte, die Summation der divergenten St\"orungsreihe f\"ur die
Grundzustandsenergie eines anharmonischen Oszillators zu erleichtern.

\neueSeite

\beginTabelle [to \kolumnenbreite]
\beginFormat \rechts " \rechts " \rechts " \rechts \endFormat
\+ " \links {\bf Tabelle 10-1} \@ \@ " \\
\+ " \links {Grundzustandsenergie $E^{(2)} (\beta)$ des anharmonischen
Oszillators mit einer $\hat{x}^4$-Anharmonizit\"at} \@ \@ " \\
\+ " \links {Rayleigh-Schr\"odingersche und renormierte
St\"orungstheoriekoeffizienten} \@ \@ " \\
\+ " \links {Relative Genauigkeit 28 Dezimalstellen} \@ \@ " \\
\- " \- " \- " \- " \\ \sstrut {} {1.5 \jot} {1.5 \jot}
\+ " \rechts {$n$} " \mitte {$b_n^{(2)}$} " \mitte {$c_n^{(2)}$} \\
\- " \- " \- " \- " \\ \sstrut {} {1 \jot} {1 \jot}
\+ "  0 " $ 0.1000000000000000000000000000 \times 10^{+01}$
" $ 0.1000000000000000000000000000 \times 10^{+01}$ " \\
\+ "  1 " $ 0.7500000000000000000000000000 \times 10^{+00}$
" $-0.2500000000000000000000000000 \times 10^{+00}$ " \\
\+ "  2 " $-0.1312500000000000000000000000 \times 10^{+01}$
" $-0.2083333333333333333333333333 \times 10^{-01}$ " \\
\+ "  3 " $ 0.5203125000000000000000000000 \times 10^{+01}$
" $ 0.1562500000000000000000000000 \times 10^{-01}$ " \\
\+ "  4 " $-0.3016113281250000000000000000 \times 10^{+02}$
" $-0.2860966435185185185185185185 \times 10^{-01}$ " \\
\+ "  5 " $ 0.2238112792968750000000000000 \times 10^{+03}$
" $ 0.6576425057870370370370370370 \times 10^{-01}$ " \\
\+ "  6 " $-0.1999462921142578125000000000 \times 10^{+04}$
" $-0.1836971078880529835390946502 \times 10^{+00}$ " \\
\+ "  7 " $ 0.2077708948516845703125000000 \times 10^{+05}$
" $ 0.6040323830435796039094650206 \times 10^{+00}$ " \\
\+ "  8 " $-0.2456891772873401641845703125 \times 10^{+06}$
" $-0.2285197581882939035618855738 \times 10^{+01}$ " \\
\+ "  9 " $ 0.3256021887746751308441162109 \times 10^{+07}$
" $ 0.9777776663767784547740376372 \times 10^{+01}$ " \\
\+ " 10 " $-0.4781043106012489646673202515 \times 10^{+08}$
" $-0.4668774596378165568567985430 \times 10^{+02}$ " \\
\+ " 11 " $ 0.7708333164092826973646879196 \times 10^{+09}$
" $ 0.2461225127523891491139679428 \times 10^{+03}$ " \\
\+ " 12 " $-0.1354432468922861661587376148 \times 10^{+11}$
" $-0.1419892831505181806046516902 \times 10^{+04}$ " \\
\+ " 13 " $ 0.2577262349393415855157363694 \times 10^{+12}$
" $ 0.8898353658197510201075563986 \times 10^{+04}$ " \\
\+ " 14 " $-0.5281751322678385961494150251 \times 10^{+13}$
" $-0.6019903263267595986917389427 \times 10^{+05}$ " \\
\+ " 15 " $ 0.1160166746583067834466055947 \times 10^{+15}$
" $ 0.4372719262409609670191132705 \times 10^{+06}$ " \\
\+ " 16 " $-0.2719757615246875955788085614 \times 10^{+16}$
" $-0.3394365729204465505490556399 \times 10^{+07}$ " \\
\+ " 17 " $ 0.6778794692977178488378423041 \times 10^{+17}$
" $ 0.2804294100493724809188467068 \times 10^{+08}$ " \\
\+ " 18 " $-0.1790210195015489075756957099 \times 10^{+19}$
" $-0.2456780647171045555469970280 \times 10^{+09}$ " \\
\+ " 19 " $ 0.4994011921119654708440035423 \times 10^{+20}$
" $ 0.2274994897166960194815507415 \times 10^{+10}$ " \\
\+ " 20 " $-0.1467514010204401499367828849 \times 10^{+22}$
" $-0.2220268390998453123923548567 \times 10^{+11}$ " \\
\+ " 21 " $ 0.4531136296684817382367060002 \times 10^{+23}$
" $ 0.2277750616809672779626695114 \times 10^{+12}$ " \\
\+ " 22 " $-0.1466652370037317717988127036 \times 10^{+25}$
" $-0.2450498476816026530633628568 \times 10^{+13}$ " \\
\+ " 23 " $ 0.4966283069462673713596194460 \times 10^{+26}$
" $ 0.2758766955970795125808880544 \times 10^{+14}$ " \\
\+ " 24 " $-0.1755839492534921737512895028 \times 10^{+28}$
" $-0.3243646392632414667688996382 \times 10^{+15}$ " \\
\+ " 25 " $ 0.6470221042946596755711674695 \times 10^{+29}$
" $ 0.3975816471140009754721618702 \times 10^{+16}$ " \\
\+ " 26 " $-0.2480994545016985113346802602 \times 10^{+31}$
" $-0.5071919119002102672589343609 \times 10^{+17}$ " \\
\+ " 27 " $ 0.9884377883559940386628660317 \times 10^{+32}$
" $ 0.6723657044660422034919183881 \times 10^{+18}$ " \\
\+ " 28 " $-0.4085842008364257515453173876 \times 10^{+34}$
" $-0.9249308298148636222936018197 \times 10^{+19}$ " \\
\+ " 29 " $ 0.1750075894533248498982509849 \times 10^{+36}$
" $ 0.1318592381939034953862653752 \times 10^{+21}$ " \\
\+ " 30 " $-0.7757967334354601595192872285 \times 10^{+37}$
" $-0.1945700535996911601773475188 \times 10^{+22}$ " \\
\+ " 31 " $ 0.3555183256998040970346514133 \times 10^{+39}$
" $ 0.2968298394125291210879904734 \times 10^{+23}$ " \\
\+ " 32 " $-0.1682432154270258570153329268 \times 10^{+41}$
" $-0.4676697500161363867361514596 \times 10^{+24}$ " \\
\+ " 33 " $ 0.8213752926846649649678781627 \times 10^{+42}$
" $ 0.7602111978952624779577099733 \times 10^{+25}$ " \\
\+ " 34 " $-0.4133016264153364753814592029 \times 10^{+44}$
" $-0.1273743627372844316885629424 \times 10^{+27}$ " \\
\+ " 35 " $ 0.2141561571497207326689983107 \times 10^{+46}$
" $ 0.2197840547215710427449652636 \times 10^{+28}$ " \\
\+ " 36 " $-0.1141747926177188161195498195 \times 10^{+48}$
" $-0.3902225729034262624911774400 \times 10^{+29}$ " \\
\+ " 37 " $ 0.6258125322225467287382748361 \times 10^{+49}$
" $ 0.7123386442335327910654531765 \times 10^{+30}$ " \\
\+ " 38 " $-0.3523944748523644918671660939 \times 10^{+51}$
" $-0.1335958494723964559115194976 \times 10^{+32}$ " \\
\+ " 39 " $ 0.2037126840152550363058543882 \times 10^{+53}$
" $ 0.2572310028622833893270814111 \times 10^{+33}$ " \\
\+ " 40 " $-0.1208147982683254411720135688 \times 10^{+55}$
" $-0.5081417489889012827342881937 \times 10^{+34}$ " \\
\- " \- " \- " \- " \\ \sstrut {} {1 \jot} {1 \jot}
\endTabelle

\neueSeite

\beginTabelle [to \kolumnenbreite]
\beginFormat \rechts " \rechts " \rechts " \rechts \endFormat
\+ " \links {\bf Tabelle 10-2} \@ \@ " \\
\+ " \links {Grundzustandsenergie $E^{(3)} (\beta)$ des anharmonischen
Oszillators mit einer $\hat{x}^6$-Anharmonizit\"at} \@ \@ " \\
\+ " \links {Rayleigh-Schr\"odingersche und renormierte
St\"orungstheoriekoeffizienten} \@ \@ " \\
\+ " \links {Relative Genauigkeit 28 Dezimalstellen} \@ \@ " \\
\- " \- " \- " \- " \\ \sstrut {} {1.5 \jot} {1.5 \jot}
\+ " \rechts {$n$} " \mitte {$b_n^{(3)}$} " \mitte {$c_n^{(3)}$} \\
\- " \- " \- " \- " \\ \sstrut {} {1 \jot} {1 \jot}
\+ "  0 " $ 0.1000000000000000000000000000 \times 10^{+001}$
" $ 0.1000000000000000000000000000 \times 10^{+001}$ " \\
\+ "  1 " $ 0.1875000000000000000000000000 \times 10^{+001}$
" $-0.3333333333333333333333333333 \times 10^{+000}$ " \\
\+ "  2 " $-0.2730468750000000000000000000 \times 10^{+002}$
" $-0.9074074074074074074074074074 \times 10^{-001}$ " \\
\+ "  3 " $ 0.1210620117187500000000000000 \times 10^{+004}$
" $ 0.3451646090534979423868312757 \times 10^{+000}$ " \\
\+ "  4 " $-0.1019996141052246093750000000 \times 10^{+006}$
" $-0.3064808584819387288523090992 \times 10^{+001}$ " \\
\+ "  5 " $ 0.1382973447847366333007812500 \times 10^{+008}$
" $ 0.4145321167060407458212670833 \times 10^{+002}$ " \\
\+ "  6 " $-0.2758405889345072507858276367 \times 10^{+010}$
" $-0.8011680848747756100865382987 \times 10^{+003}$ " \\
\+ "  7 " $ 0.7624090761643007420003414154 \times 10^{+012}$
" $ 0.2103995758599676692709068363 \times 10^{+005}$ " \\
\+ "  8 " $-0.2794538968079133336657076143 \times 10^{+015}$
" $-0.7225346394287017811898386232 \times 10^{+006}$ " \\
\+ "  9 " $ 0.1313042915039690044448743938 \times 10^{+018}$
" $ 0.3148105305746268747625734095 \times 10^{+008}$ " \\
\+ " 10 " $-0.7698226021360799182567304195 \times 10^{+020}$
" $-0.1698432299001382459616174045 \times 10^{+010}$ " \\
\+ " 11 " $ 0.5510195542199472404222847408 \times 10^{+023}$
" $ 0.1112192278350429700410653539 \times 10^{+012}$ " \\
\+ " 12 " $-0.4729313989051410578294591182 \times 10^{+026}$
" $-0.8693791326282078233903279177 \times 10^{+013}$ " \\
\+ " 13 " $ 0.4794568649016500869708791279 \times 10^{+029}$
" $ 0.7998709458256125310465977701 \times 10^{+015}$ " \\
\+ " 14 " $-0.5668638845105089889893469760 \times 10^{+032}$
" $-0.8558133511622717537064529071 \times 10^{+017}$ " \\
\+ " 15 " $ 0.7730868708392225116585659118 \times 10^{+035}$
" $ 0.1053809184929675849938435395 \times 10^{+020}$ " \\
\+ " 16 " $-0.1204677876887977082686406782 \times 10^{+039}$
" $-0.1479857266619485112172349713 \times 10^{+022}$ " \\
\+ " 17 " $ 0.2127125807659420801080170541 \times 10^{+042}$
" $ 0.2351135259457285370074949664 \times 10^{+024}$ " \\
\+ " 18 " $-0.4224790967656847410984775785 \times 10^{+045}$
" $-0.4196185697357360024328831693 \times 10^{+026}$ " \\
\+ " 19 " $ 0.9377149163648241043543969958 \times 10^{+048}$
" $ 0.8359916593280975065227197941 \times 10^{+028}$ " \\
\+ " 20 " $-0.2312340008724852652612445042 \times 10^{+052}$
" $-0.1848636007437448746319484353 \times 10^{+031}$ " \\
\+ " 21 " $ 0.6301776563411529926112680544 \times 10^{+055}$
" $ 0.4514146424160909327171625697 \times 10^{+033}$ " \\
\+ " 22 " $-0.1889023053176330615906768360 \times 10^{+059}$
" $-0.1211586508289993573862394094 \times 10^{+036}$ " \\
\+ " 23 " $ 0.6201471510847780850187133912 \times 10^{+062}$
" $ 0.3559158544311975012093842606 \times 10^{+038}$ " \\
\+ " 24 " $-0.2220849365882392672661166959 \times 10^{+066}$
" $-0.1139910826991979199070027621 \times 10^{+041}$ " \\
\+ " 25 " $ 0.8644440087852245564670979801 \times 10^{+069}$
" $ 0.3966243637285127789424151214 \times 10^{+043}$ " \\
\+ " 26 " $-0.3645024287481314864469944715 \times 10^{+073}$
" $-0.1494337957478330322610461426 \times 10^{+046}$ " \\
\+ " 27 " $ 0.1659867657272409519616032382 \times 10^{+077}$
" $ 0.6078032985359067768960067735 \times 10^{+048}$ " \\
\+ " 28 " $-0.8139891556691932823257041773 \times 10^{+080}$
" $-0.2661350333660720674204398633 \times 10^{+051}$ " \\
\+ " 29 " $ 0.4287329844714913047493868238 \times 10^{+084}$
" $ 0.1251214623248615367928351238 \times 10^{+054}$ " \\
\+ " 30 " $-0.2419404721267758975615865337 \times 10^{+088}$
" $-0.6300798666272061547826135013 \times 10^{+056}$ " \\
\+ " 31 " $ 0.1459435925329802383474525164 \times 10^{+092}$
" $ 0.3390832468952766295496582425 \times 10^{+059}$ " \\
\+ " 32 " $-0.9390362679921760987027239489 \times 10^{+095}$
" $-0.1945986954576207938163805083 \times 10^{+062}$ " \\
\+ " 33 " $ 0.6431686094898852043394318269 \times 10^{+099}$
" $ 0.1188582905603568060699598285 \times 10^{+065}$ " \\
\+ " 34 " $-0.4680474008411081675049791233 \times 10^{+103}$
" $-0.7711878010767234185411720094 \times 10^{+067}$ " \\
\+ " 35 " $ 0.3612460689256218583590889294 \times 10^{+107}$
" $ 0.5305967693868868722457673167 \times 10^{+070}$ " \\
\+ " 36 " $-0.2952124889344106855771325009 \times 10^{+111}$
" $-0.3864728396039321633672770434 \times 10^{+073}$ " \\
\+ " 37 " $ 0.2550323026091156629476319641 \times 10^{+115}$
" $ 0.2975358682722591106809092907 \times 10^{+076}$ " \\
\+ " 38 " $-0.2325585752026781909585439424 \times 10^{+119}$
" $-0.2417566302591801229336823865 \times 10^{+079}$ " \\
\+ " 39 " $ 0.2235261717348247217617087662 \times 10^{+123}$
" $ 0.2070249769815577569885667706 \times 10^{+082}$ " \\
\+ " 40 " $-0.2261502639532406528609923432 \times 10^{+127}$
" $-0.1865907885953646428428074125 \times 10^{+085}$ " \\
\- " \- " \- " \- " \\ \sstrut {} {1 \jot} {1 \jot}
\endTabelle

\neueSeite

\beginTabelle [to \kolumnenbreite]
\beginFormat \rechts " \rechts " \rechts " \rechts \endFormat
\+ " \links {\bf Tabelle 10-3} \@ \@ " \\
\+ " \links {Grundzustandsenergie $E^{(4)} (\beta)$ des anharmonischen
Oszillators mit einer $\hat{x}^8$-Anharmonizit\"at} \@ \@ " \\
\+ " \links {Rayleigh-Schr\"odingersche und renormierte
St\"orungstheoriekoeffizienten} \@ \@ " \\
\+ " \links {Relative Genauigkeit 28 Dezimalstellen} \@ \@ " \\
\- " \- " \- " \- " \\ \sstrut {} {1.5 \jot} {1.5 \jot}
\+ " \rechts {$n$} " \mitte {$b_n^{(4)}$} " \mitte {$c_n^{(4)}$} \\
\- " \- " \- " \- " \\ \sstrut {} {1 \jot} {1 \jot}
\+ "  0 " $ 0.1000000000000000000000000000 \times 10^{+001}$
" $ 0.1000000000000000000000000000 \times 10^{+001}$ " \\
\+ "  1 " $ 0.6562500000000000000000000000 \times 10^{+001}$
" $-0.3750000000000000000000000000 \times 10^{+000}$ " \\
\+ "  2 " $-0.1054921875000000000000000000 \times 10^{+004}$
" $-0.2577380952380952380952380952 \times 10^{+000}$ " \\
\+ "  3 " $ 0.7842751464843750000000000000 \times 10^{+006}$
" $ 0.4010069444444444444444444444 \times 10^{+001}$ " \\
\+ "  4 " $-0.1551763724836120605468750000 \times 10^{+010}$
" $-0.1705985873918515819026023108 \times 10^{+003}$ " \\
\+ "  5 " $ 0.6449194871343744850158691406 \times 10^{+013}$
" $ 0.1437156124323559202120001440 \times 10^{+005}$ " \\
\+ "  6 " $-0.4875280106462811879916906357 \times 10^{+017}$
" $-0.2144037100761293781798518052 \times 10^{+007}$ " \\
\+ "  7 " $ 0.6075883460179283253360016331 \times 10^{+021}$
" $ 0.5201687548704109399079088804 \times 10^{+009}$ " \\
\+ "  8 " $-0.1162693617615391939064463060 \times 10^{+026}$
" $-0.1923208169065830465390043883 \times 10^{+012}$ " \\
\+ "  9 " $ 0.3238248310327800948866547042 \times 10^{+030}$
" $ 0.1030297295315763364591802181 \times 10^{+015}$ " \\
\+ " 10 " $-0.1259008931000663909196863363 \times 10^{+035}$
" $-0.7683572644639494219972552204 \times 10^{+017}$ " \\
\+ " 11 " $ 0.6608813150574499539973494683 \times 10^{+039}$
" $ 0.7722310693438535576086953358 \times 10^{+020}$ " \\
\+ " 12 " $-0.4557555541698597279045592497 \times 10^{+044}$
" $-0.1018368546772651985018708094 \times 10^{+024}$ " \\
\+ " 13 " $ 0.4036175702989948211936892173 \times 10^{+049}$
" $ 0.1723105152706815772832516685 \times 10^{+027}$ " \\
\+ " 14 " $-0.4502722752726406643033065360 \times 10^{+054}$
" $-0.3670371714010763820444015058 \times 10^{+030}$ " \\
\+ " 15 " $ 0.6224155093583807327174540505 \times 10^{+059}$
" $ 0.9682866830479761205912370800 \times 10^{+033}$ " \\
\+ " 16 " $-0.1050932916972165018533205831 \times 10^{+065}$
" $-0.3119122832454623005458438450 \times 10^{+037}$ " \\
\+ " 17 " $ 0.2140549197427382484102110749 \times 10^{+070}$
" $ 0.1211706118235308218863653480 \times 10^{+041}$ " \\
\+ " 18 " $-0.5201600731241721250063550958 \times 10^{+075}$
" $-0.5614747700461266967511103754 \times 10^{+044}$ " \\
\+ " 19 " $ 0.1493307901840972910870610336 \times 10^{+081}$
" $ 0.3073190699125756090442242265 \times 10^{+048}$ " \\
\+ " 20 " $-0.5020535421991564271301839675 \times 10^{+086}$
" $-0.1969597092214332040649531052 \times 10^{+052}$ " \\
\+ " 21 " $ 0.1961138382030910719954145312 \times 10^{+092}$
" $ 0.1466473608314397434043080126 \times 10^{+056}$ " \\
\+ " 22 " $-0.8837310139672377641259822149 \times 10^{+097}$
" $-0.1259461408396291518493433010 \times 10^{+060}$ " \\
\+ " 23 " $ 0.4564193368578255788127856410 \times 10^{+103}$
" $ 0.1239636847577184018876908397 \times 10^{+064}$ " \\
\+ " 24 " $-0.2685739891523011783114746739 \times 10^{+109}$
" $-0.1390056334988531530617148620 \times 10^{+068}$ " \\
\+ " 25 " $ 0.1790847055896280467174271816 \times 10^{+115}$
" $ 0.1766207197726964878558025097 \times 10^{+072}$ " \\
\+ " 26 " $-0.1346403763380976333237766628 \times 10^{+121}$
" $-0.2530191504390037751762720629 \times 10^{+076}$ " \\
\+ " 27 " $ 0.1136080780112106037549778200 \times 10^{+127}$
" $ 0.4067853658325892651811070793 \times 10^{+080}$ " \\
\+ " 28 " $-0.1071281417696968264564796886 \times 10^{+133}$
" $-0.7308414977965929785301028104 \times 10^{+084}$ " \\
\+ " 29 " $ 0.1124435215572944829487571340 \times 10^{+139}$
" $ 0.1461520955658283761146730027 \times 10^{+089}$ " \\
\+ " 30 " $-0.1308870821264095846760956068 \times 10^{+145}$
" $-0.3241212363807525573104203604 \times 10^{+093}$ " \\
\+ " 31 " $ 0.1683808194298817425271504127 \times 10^{+151}$
" $ 0.7943902607396363612524392935 \times 10^{+097}$ " \\
\+ " 32 " $-0.2386275657753780247074260718 \times 10^{+157}$
" $-0.2144786992069770817573194280 \times 10^{+102}$ " \\
\+ " 33 " $ 0.3714215021304125281315027942 \times 10^{+163}$
" $ 0.6359830356276262390280745601 \times 10^{+106}$ " \\
\+ " 34 " $-0.6331371960660631764442476245 \times 10^{+169}$
" $-0.2065307743480187063920509288 \times 10^{+111}$ " \\
\+ " 35 " $ 0.1178830859607103006220941649 \times 10^{+176}$
" $ 0.7325565419640661184224880519 \times 10^{+115}$ " \\
\+ " 36 " $-0.2391293894141623727506028335 \times 10^{+182}$
" $-0.2830871822667374211336129230 \times 10^{+120}$ " \\
\+ " 37 " $ 0.5272395298414059636402686908 \times 10^{+188}$
" $ 0.1189017492629281666371195184 \times 10^{+125}$ " \\
\+ " 38 " $-0.1260660655056164563596955258 \times 10^{+195}$
" $-0.5415859844649942520349644991 \times 10^{+129}$ " \\
\+ " 39 " $ 0.3261946501529033075736678567 \times 10^{+201}$
" $ 0.2669508541398678477745494011 \times 10^{+134}$ " \\
\+ " 40 " $-0.9115165995589978247727728346 \times 10^{+207}$
" $-0.1421022695098991431015589663 \times 10^{+139}$ " \\
\- " \- " \- " \- " \\ \sstrut {} {1 \jot} {1 \jot}
\endTabelle

\neueSeite

Abgesehen von der unterschiedlich starken Divergenz der
Rayleigh-Schr\"odingerschen und der renormierten
St\"orungstheoriekoeffizienten $b_{n}^{(m)}$ und $c_{n}^{(m)}$ gibt es
auch sonst noch einige auff\"allige Unterschiede. Mit Ausnahme des ersten
Koeffizienten $b_{0}^{(m)} = 1$ besitzen die $b_{n}^{(m)}$ strikt
alternierende Vorzeichen. Das ist eine notwendige Voraussetzung daf\"ur,
da{\ss} die $b_{n}^{(2)}$ und $b_{n}^{(3)}$ f\"ur $n \ge 1$ Momente eines
Stieltjesma{\ss}es mit negativem Vorzeichen gem\"a{\ss} Gl. (4.3-2) sein k\"onnen.
Au{\ss}erdem wachsen die $b_{n}^{(m)}$ betragsm\"a{\ss}ig monoton mit wachsendem
$n$.

Die renormierten St\"orungstheoriekoeffizienten $c_{n}^{(m)}$ zeigen
dagegen ein deutlich weniger regul\"ares Verhalten. Die $c_{n}^{(m)}$
besitzen erst f\"ur $n \ge 2$ ein strikt alternierendes Vorzeichen. Das
impliziert, da{\ss} die renormierten St\"orungstheoriekoeffizienten
$c_{n}^{(m)}$ -- wenn \"uberhaupt -- nur f\"ur $n \ge 2$ Momente eines
Stieltjesma{\ss}es mit negativem Vorzeichen gem\"a{\ss} Gl. (4.3-2) sein k\"onnen.

Au{\ss}erdem wachsen die $c_{n}^{(m)}$ betragsm\"a{\ss}ig nicht monoton mit
wachsendem $n$. Dieser Effekt ist besonders ausgepr\"agt im Falle des
anharmonischen Oszillators mit einer $\hat{x}^4$-Anharmonizit\"at. Die
renormierten St\"orungstheoriekoeffizienten $c_{n}^{(2)}$ werden zuerst
betragsm\"a{\ss}ig kleiner und wachsen erst f\"ur $n \ge 6$ so, wie man es
aufgrund von Gl. (10.4-22a) erwarten w\"urde. Die
St\"orungstheoriekoeffizienten $c_{n}^{(3)}$ und $c_{n}^{(4)}$ zeigen im
Prinzip ein \"ahnlich irregul\"ares Verhalten, das allerdings weniger stark
ausgepr\"agt ist als im $\hat{x}^4$-Fall.

\medskip

\Abschnitt Die Divergenz der Levinschen Transformation

\smallskip

\aktTag = 0

Es gibt nur wenige theoretische Probleme aus dem Bereich der
Naturwissenschaften, die mit Hilfe mathematischer Methoden {\it exakt\/}
und in {\it geschlossener Form\/} l\"osbar sind. Der Versuch einer
mathematischen Behandlung naturwissenschaftlicher Probleme impliziert
deswegen quasi automatisch, da{\ss} man numerische und approximative
Verfahren verwenden mu{\ss}. Bedingt durch die gro{\ss}en Fortschritte sowohl
der Computertechnologie als auch der numerischen Mathematik ist es
inzwischen m\"oglich, \"au{\ss}erst komplexe Probleme {\it theoretisch\/} mit
Hilfe numerischer Methoden zu behandeln. Die Quantenchemie, die in den
letzten Jahren enorme Fortschritte gemacht hat, ist meiner Meinung nach
ein sehr gutes Beispiel f\"ur die Leistungsf\"ahigkeit einer ausschlie{\ss}lich
numerisch orientierten Vorgehensweise{\footnote[\dagger]{Wilson, der
1982 f\"ur seine Arbeiten \"uber die Renormierungsgruppe [Wilson 1975;
Wilson und Kogut 1974] den Nobelpreis in Physik erhalten hatte, vertritt
beispielsweise den Standpunkt, da{\ss} die computer\-orientierte
Quantenchemie in methodischer Hinsicht wesentlich weiter entwickelt ist
als die Rechentechniken, die \"ublicherweise in der statistischen Physik
und in der Hochenergiephysik verwendet werden, und da{\ss} man der
Quantenchemie deswegen eine gewisse Vorbildfunktion zubilligen sollte
[Wilson 1990].}}.

Bei allen numerischen und approximativen Verfahren ist man aber mit der
keineswegs trivialen Fragestellung konfrontiert, unter welchen Umst\"anden
ein solches Verfahren erfolgreich ist, oder -- anders formuliert --
inwieweit ausschlie{\ss}lich numerische Ergebnisse \"uberhaupt beweiskr\"aftig
sind, wenn man die Konvergenz des verwendeten numerischen Verfahrens
nicht durch explizite mathematische Fehlerabsch\"atzungen beweisen kann.

Fragen dieser Art spielen eine wichtige Rolle, wenn man divergente
asymptotische St\"orungs\-reihen summieren will. In Abschnitt 6.5 wurde
beispielsweise gezeigt, da{\ss} man einer asymptotischen divergenten
Potenzreihe
$$
f (z) \; \sim \; \sum_{\nu=0}^{\infty} \, \gamma_{\nu} \, z^{\nu} \, ,
\qquad z \to \infty \, ,
\tag
$$
nur dann auf eindeutige Weise eine Funktion $f (z)$ zuordnen kann, wenn
die Abbruchfehler
$$
R_n (z) \; = \;
f (z) \, - \, \sum_{\nu=0}^{n} \, \gamma_{\nu} \, z^{\nu}
\tag
$$
f\"ur alle hinreichend gro{\ss}en Werte von $n$ die Absch\"atzungen (6.5-12)
beziehungsweise (6.5-14) erf\"ullen. Bei quantenmechanischen
St\"orungsreihen sind normalerweise aber nur die numerischen Werte einer
endlichen und oft relativ kleinen Zahl von St\"orungstheoriekoeffizienten
bekannt. Die G\"ultigkeit der Absch\"atzungen (6.5-12) beziehungsweise
(6.5-14) f\"ur die Abbruchfehler $R_n (z)$ k\"onnen unter diesen Umst\"anden
prinzipiell nicht bewiesen werden, und man ist gezwungen, die Existenz
einer durch die divergente asymptotische St\"orungs\-reihe eindeutig
bestimmten Funktion $f (z)$ zu postulieren.

Meiner Meinung nach wird die in diesem Zusammenhang auftretende
Problematik durch das folgende Zitat [Olver 1974, S. 519] treffend
beschrieben:

\medskip

\beginSchmaeler
\noindent {\sl In consequence, the numerical use of an asymptotic
approximation (or expansion), without rigorous investigation of the
error term, has to be regarded as being in the nature of a hypothesis.
Nevertheless, it would be extravagant to reject the use of an
approximation for this reason alone. The essence of progress in the
physical sciences is the development and application of hypotheses that
have a high probability of being correct: it would be artificial to
exclude those of a purely mathematical character. Instead, what we need
to do is examine the approximations by carefully devised tests.}
\endSchmaeler

\medskip

Man kann nur zustimmen, da{\ss} ein ausschlie{\ss}lich numerisches Ergebnis nur
dann als \anf{Beweis} betrachtet werden sollte, wenn die dabei
verwendeten numerischen Verfahren sehr sorgf\"altig ge\-testet wurden.
Aber auch dann kann man die M\"oglichkeit einer Fehlinterpretation nie
v\"ollig ausschlie{\ss}en, wie ich selbst feststellen mu{\ss}te.

In meiner ersten Arbeit \"uber die Summation der divergenten St\"orungsreihe
anharmonischer Oszillatoren [Weniger 1990] wurde der Wynnsche
$\epsilon$-Algorithmus, Gl. (2.4-10), und die beiden eng verwandten
verallgemeinerten Summationsprozesse $d_k^{(n)} (\zeta, s_n)$, Gl.
(5.2-18), und ${\delta}_k^{(n)} (\zeta, s_n)$, Gl. (5.4-13), mit $\zeta
= 1$ zur Summation sowohl der divergenten St\"orungsreihe (10.2-4) f\"ur die
Grundzustandsenergie des anharmonischen Oszillators mit einer
$\hat{x}^4$-Anharmonizit\"at als auch der divergenten hypergeometrischen
Modellreihe (10.2-7) verwendet. Die divergente hypergeometrische
Modellreihe (10.2-7) wurde deswegen zum Vergleich gew\"ahlt, weil ihre
Terme per Konstruktion ebenso schnell wachsen wie die Terme der
divergenten St\"orungsreihe (10.2-4) mit $m = 2$. Die beiden Reihen
divergieren also gleich stark.

Tabelle 10-4 [Weniger 1990, Table \Roemisch{1}] zeigt den Effekt der
oben genannten verallgemeinerten Summationsprozesse auf die Partialsummen
$$
s_n \; = \; \sum_{\nu=0}^{n} \, (1/2)_{\nu} (- 3 \beta/ 2)^{\nu}
\tag
$$
der hypergeometrischen Modellreihe (10.2-7).

\beginFloat

\medskip

\beginTabelle [to \kolumnenbreite]
\beginFormat \rechts " \rechts " \mitte " \mitte " \mitte
\endFormat
\+ " \links {\bf Tabelle 10-4} \@ \@ \@ \@ " \\
\+ " \links {Summation der divergenten asymptotischen Reihe}
\@ \@ \@ \@ " \\
\+ " \links {${}_2 F_0 \bigl( 1/2, 1; - 3 \beta/ 2 \bigr)
\; = \; [2 \pi / (3 \beta)]^{1/2} \,
\exp \bigl( 2/(3 \beta) \bigr) \,
\Funk {erfc} \bigl([2/(3 \beta)]^{1/2}\bigr)$
f\"ur $\beta \, = \, 3 / 10$} \@ \@ \@ \@ " \\
\- " \- " \- " \- " \- " \- " \\ \sstrut {} {1.5 \jot} {1.5 \jot}
\+ " \rechts {$n$} " \mitte {Partialsumme $s_n$}
" $\epsilon_{2 \Ent {n/2}}^{(n - 2 \Ent {n/2})}$
" $d_n^{(0)} (1, s_0)$ " ${\delta}_n^{(0)} (1, s_0)$ " \\
\+ " " \mitte {Gl. (10.6-3)} " Gl. (2.4-10) " Gl. (5.2-18)
" Gl. (5.4-13) " \\
\- " \- " \- " \- " \- " \- " \\ \sstrut {} {1 \jot} {1 \jot}
\+ " $  6$ " $   0.1828897704 \times 10^{+01}$ " $  0.85408449265462$ "
$  0.85373155284794$ " $  0.85373166671109$ " \\
\+ " $  7$ " $  -0.2116088355 \times 10^{+01}$ " $  0.85349016639700$ "
$  0.85373141552735$ " $  0.85373175595228$ " \\
\+ " $  8$ " $   0.1119823959 \times 10^{+02}$ " $  0.85381735606149$ "
$  0.85373181185485$ " $  0.85373173450373$ " \\
\+ " $  9$ " $  -0.3972906481 \times 10^{+02}$ " $  0.85367102805092$ "
$  0.85373172488413$ " $  0.85373173103853$ " \\
\+ " $ 10$ " $   0.1779851615 \times 10^{+03}$ " $  0.85375556509951$ "
$  0.85373172909698$ " $  0.85373173105887$ " \\
\+ " $ 11$ " $  -0.8507145578 \times 10^{+03}$ " $  0.85371437764065$ "
$  0.85373173190273$ " $  0.85373173114150$ " \\
\+ " $ 12$ " $   0.4472806489 \times 10^{+04}$ " $  0.85373908949052$ "
$  0.85373173109439$ " $  0.85373173115425$ " \\
\+ " $ 13$ " $  -0.2547199940 \times 10^{+05}$ " $  0.85372625810352$ "
$  0.85373173112605$ " $  0.85373173115413$ " \\
\+ " $ 14$ " $   0.1564426964 \times 10^{+06}$ " $  0.85373419680300$ "
$  0.85373173116267$ " $  0.85373173115370$ " \\
\+ " $ 15$ " $  -0.1030550694 \times 10^{+07}$ " $  0.85372986477609$ "
$  0.85373173115342$ " $  0.85373173115361$ " \\
\+ " $ 16$ " $   0.7248728202 \times 10^{+07}$ " $  0.85373261447538$ "
$  0.85373173115307$ " $  0.85373173115360$ " \\
\+ " $ 17$ " $  -0.5422491760 \times 10^{+08}$ " $  0.85373105257911$ "
$  0.85373173115372$ " $  0.85373173115360$ " \\
\+ " $ 18$ " $   0.4298800431 \times 10^{+09}$ " $  0.85373206580547$ "
$  0.85373173115362$ " $  0.85373173115360$ " \\
\+ " $ 19$ " $  -0.3600293754 \times 10^{+10}$ " $  0.85373147080071$ "
$  0.85373173115359$ " $  0.85373173115360$ " \\
\+ " $ 20$ " $   0.3176448132 \times 10^{+11}$ " $  0.85373186412957$ "
$  0.85373173115361$ " $  0.85373173115360$ " \\
\+ " $ 21$ " $  -0.2944755687 \times 10^{+12}$ " $  0.85373162655800$ "
$  0.85373173115361$ " $  0.85373173115360$ " \\
\- " \- " \- " \- " \- " \- " \\ \sstrut {} {1 \jot} {1 \jot}
\+ " \links {NAG FUNCTION S15ADF} \@ " $  0.85373173115360$ " $
0.85373173115360$ " $  0.85373173115360$ " \\
\- " \- " \- " \- " \- " \- " \\ \sstrut {} {1 \jot} {1 \jot}
\endTabelle

\medskip

\endFloat

Die Ergebnisse in Tabelle 10-4 ergeben ein inzwischen gewohntes Bild:
Der verallgemeinerte Summationsproze{\ss} ${\delta}_k^{(n)} (\zeta, s_n)$
liefert ganz vorz\"ugliche Ergebnisse, die deutlich besser sind als die
Resultate der Levinschen Transformation $d_k^{(n)} (\zeta, s_n)$,
w\"ahrend der Wynnsche $\epsilon$-Algorithmus klar abgeschlagen das
Schlu{\ss}licht ist. Aufgrund der in Tabelle 10-4 pr\"asentierten Ergebnisse
sollten sowohl der Wynnsche $\epsilon$-Algorithmus als auch die beiden
verallgemeinerten Summationsprozesse $d_k^{(n)} (\zeta, s_n)$ und
${\delta}_k^{(n)} (\zeta, s_n)$ die divergente hypergeometrische Reihe
(10.2-7) zumindest f\"ur nicht zu gro{\ss}e Werte von $\beta$ problemlos
summieren k\"onnen.

Tabelle 10-5 [Weniger 1990, Table \Roemisch{2}] zeigt den Effekt der
oben genannten verallgemeinerten Summationsprozesse auf die Partialsummen
$$
s_n \; = \; \sum_{\nu=0}^{n} \, b_{\nu}^{(2)} \, \beta^{\nu} \, ,
\qquad 0 \le n \le 21 \, ,
\tag
$$
der St\"orungsreihe (10.2-4) f\"ur die Grundzustandsenergie des
anharmonischen Oszillators mit einer $\hat{x}^4$-Anharmonizit\"at. Die
\anf{exakte} Energie in Tabelle 10-5 stammt aus einem Artikel von
Marziani [1984, Table \Roemisch{3}].

\beginFloat

\medskip

\beginTabelle [to \kolumnenbreite]
\beginFormat \rechts " \rechts " \mitte " \mitte " \mitte
\endFormat
\+ " \links {\bf Tabelle 10-5} \@ \@ \@ \@ " \\
\+ " \links {Grundzustandsenergie $E^{(2)} (\beta)$ des anharmonischen
Oszillators mit einer $\hat{x}^4$-Anharmonizit\"at} \@ \@ \@ \@ " \\
\+ " \links {Summation der divergenten St\"orungsreihe (10.2-4) f\"ur
$\beta \, = \, 3 / 10$} \@ \@ \@ \@ " \\
\- " \- " \- " \- " \- " \- " \\ \sstrut {} {1.5 \jot} {1.5 \jot}
\+ " \rechts {$n$} " \mitte {Partialsumme $s_n$}
" $\epsilon_{2 \Ent {n/2}}^{(n - 2 \Ent {n/2})}$
" $d_n^{(0)} (1, s_0)$ " ${\delta}_n^{(0)} (1, s_0)$ " \\
\+ " " \mitte {Gl. (10.6-4)} " Gl. (2.4-10) " Gl. (5.2-18)
" Gl. (5.4-13) " \\
\- " \- " \- " \- " \- " \- " \\ \sstrut {} {1 \jot} {1 \jot}
\+ " $  6$ " $   0.8930713840 \times 10^{-01}$ " $  1.16352355861439$ "
$  1.16406000134219$ " $  1.16404717346312$ " \\
\+ " $  7$ " $   0.4633256609 \times 10^{+01}$ " $  1.16432005116001$ "
$  1.16404418517836$ " $  1.16404770164600$ " \\
\+ " $  8$ " $  -0.1148641031 \times 10^{+02}$ " $  1.16391853160463$ "
$  1.16404614223662$ " $  1.16404689014490$ " \\
\+ " $  9$ " $   0.5260186850 \times 10^{+02}$ " $  1.16411849704021$ "
$  1.16404880640477$ " $  1.16404713867428$ " \\
\+ " $ 10$ " $  -0.2297139459 \times 10^{+03}$ " $  1.16401107091330$ "
$  1.16404633642795$ " $  1.16404722008755$ " \\
\+ " $ 11$ " $   0.1135794149 \times 10^{+04}$ " $  1.16406814453682$ "
$  1.16404704603055$ " $  1.16404714738494$ " \\
\+ " $ 12$ " $  -0.6062215308 \times 10^{+04}$ " $  1.16403595361348$ "
$  1.16404760987659$ " $  1.16404714186144$ " \\
\+ " $ 13$ " $   0.3502767110 \times 10^{+05}$ " $  1.16405392327546$ "
$  1.16404692357883$ " $  1.16404716184730$ " \\
\+ " $ 14$ " $  -0.2175968573 \times 10^{+06}$ " $  1.16404338730170$ "
$  1.16404706293922$ " $  1.16404716201218$ " \\
\+ " $ 15$ " $   0.1447115618 \times 10^{+07}$ " $  1.16404950527943$ "
$  1.16404733576627$ " $  1.16404715605033$ " \\
\+ " $ 16$ " $  -0.1026054911 \times 10^{+08}$ " $  1.16404580226005$ "
$  1.16404711573105$ " $  1.16404715568364$ " \\
\+ " $ 17$ " $   0.7728091605 \times 10^{+08}$ " $  1.16404802332121$ "
$  1.16404707849489$ " $  1.16404715753428$ " \\
\+ " $ 18$ " $  -0.6162831931 \times 10^{+09}$ " $  1.16404664259960$ "
$  1.16404721674024$ " $  1.16404715796080$ " \\
\+ " $ 19$ " $   0.5188064429 \times 10^{+10}$ " $  1.16404749361828$ "
$  1.16404717588561$ " $  1.16404715745285$ " \\
\+ " $ 20$ " $  -0.4598098516 \times 10^{+11}$ " $  1.16404695236331$ "
$  1.16404711676123$ " $  1.16404715717850$ " \\
\+ " $ 21$ " $   0.4279918756 \times 10^{+12}$ " $  1.16404729383850$ "
$  1.16404716286765$ " $  1.16404715725758$ " \\
\- " \- " \- " \- " \- " \- " \\ \sstrut {} {1 \jot} {1 \jot}
\+ " \links {Exakte Energie} \@ " $  1.16404715735384$
" $1.16404715735384$ " $  1.16404715735384$ " \\
\- " \- " \- " \- " \- " \- " \\ \sstrut {} {1 \jot} {1 \jot}
\endTabelle

\medskip

\endFloat

Auf den ersten Blick bietet Tabelle 10-5 das inzwischen gewohnte Bild:
Die besten Ergebnisse werden von dem verallgemeinerten Summationsproze{\ss}
${\delta}_k^{(n)} (\zeta, s_n)$ erzielt, und der Wynnsche
$\epsilon$-Algorithmus ist das Schlu{\ss}licht. Wenn man aber die
Summationsergebnisse in den Tabellen 10-4 und 10-5 genauer analysiert,
dann f\"allt auf, da{\ss} die Levinsche Transformation $d_k^{(n)} (\zeta,
s_n)$ in Tabelle 10-4 fast so leistungsf\"ahig war wie ${\delta}_k^{(n)}
(\zeta, s_n)$, wogegen sie in Tabelle 10-5 nur geringf\"ugig
leistungsf\"ahiger war als der Wynnsche $\epsilon$-Algorithmus. Au{\ss}erdem
f\"allt auf, da{\ss} der $\epsilon$-Algorithmus in beiden Tabellen in etwa
gleich leistungsf\"ahig war, wogegen sowohl ${\delta}_k^{(n)} (\zeta,
s_n)$ als auch $d_k^{(n)} (\zeta, s_n)$ in Tabelle 10-4 deutlich bessere
Ergebnisse lieferten als in Tabelle 10-5.

Die St\"orungsreihe (10.2-4) des anharmonischen Oszillators mit einer
$\hat{x}^4$-Anharmonizit\"at ist, abgesehen vom ersten Term $b_{0}^{(2)}$,
eine Stieltjesreihe [Simon 1970, Theorem \Roemisch{4}.2.1]. Aus der
asymptotischen Absch\"atzung (10.2-6a) folgt au{\ss}erdem noch, da{\ss} die
Carlemanbedingung (4.3-5) erf\"ullt ist. Damit ist garantiert, da{\ss} die
durch den Wynnschen $\epsilon$-Algorithmus gem\"a{\ss} Gl. (4.4-8) berechneten
Pad\'e-Approximationen $\epsilon_{2 n}^{(0)} = [n / n]$ und $\epsilon_{2
n}^{(1)} = [n + 1 / n]$ f\"ur $n \to \infty$ gegen $E^{(2)} (\beta)$
konvergieren. In Tabelle 10-5 konvergieren sowohl ${\delta}_k^{(n)}
(\zeta, s_n)$ als auch $d_k^{(n)} (\zeta, s_n)$ schneller als die
Pad\'e-Approximationen. Obwohl in Tabelle 10-5 nur die
St\"orungstheoriekoeffizienten $b_{n}^{(2)}$ mit $n \le 22$ verwendet
wurden, scheinen die Ergebnisse demzufolge zu \anf{beweisen}, da{\ss} die
divergente St\"orungsreihe (10.2-4) f\"ur die Grundzustandsenergie $E^{(2)}
(\beta)$ des anharmonischen Oszillators mit einer
$\hat{x}^4$-Anharmonizit\"at zumindest f\"ur kleinere Kopplungskonstanten
auch durch die verallgemeinerten Summationsprozesse $d_k^{(n)} (\zeta,
s_n)$ und ${\delta}_k^{(n)} (\zeta, s_n)$ summiert werden kann.
Aufwendigere Rechnungen ergaben aber, da{\ss} diese naheliegende
Schlu{\ss}folgerung im Falle der Levinschen Transformation {\it falsch\/}
ist.

In sp\"ateren Rechnungen wurden sowohl die St\"orungsreihe (10.2-4) als auch
die renormierte St\"orungsreihe (10.4-20) mit Hilfe des Wynnschen
$\epsilon$-Algorithmus, Gl. (2.4-19), und der verallgemeinerten
Summationsprozesse $d_k^{(n)} (\zeta, s_n)$, Gl. (5.2-18), und
${\delta}_k^{(n)} (\zeta, s_n)$, Gl. (5.4-13), mit $\zeta = 1$ f\"ur $m =
2, 3, 4$ summiert. Dabei wurde entweder die Grundzustandsenergie
$E^{(m)} (\beta)$ gem\"a{\ss} Gln. (10.2-4) und (10.4-28) oder der in Gl.
(10.4-26) definierte Grenzfall unendlicher Kopplung $k_m$ gem\"a{\ss} Gl.
(10.4-31) berechnet.

Die Rechnungen wurden zuerst in FORTRAN auf einer Cyber 180-995 E in
DOUBLE PRECISION (ungef\"ahr 29 Dezimalstellen) durchgef\"uhrt. Sowohl im
Falle der $\hat{x}^4$-, der $\hat{x}^6$- als auch der
$\hat{x}^8$-Anharmonizit\"at wurden jeweils die
St\"orungstheoriekoeffizienten $b_{n}^{(m)}$ und $c_{n}^{(m)}$ mit $n \le
50$ verwendet. Allerdings lieferten auch diese Rechnungen in FORTRAN
keine v\"ollig eindeutigen Ergebnisse. In allen F\"allen wurde aber
beobachtet, da{\ss} die Transformationen $\{ {\delta}_{\nu}^{(0)} \}$
vergleichsweise schnell konvergierten, wogegen die eng verwandten
Transformationen $\{ d_{\nu}^{(0)} \}$ deutlich weniger leistungsf\"ahig
waren. Au{\ss}erdem schien es, da{\ss} die Transformationen $\{ d_{\nu}^{(0)}
\}$ st\"arker unter Rundungsfehlern litten als die anderen hier
verwendeten Transformationen, da die erreichte Genauigkeit mit
wachsender Transformationsordnung $\nu$ zuerst langsam zunahm, bei
weiterer Vergr\"o{\ss}erung von $\nu$ dann aber wieder abnahm.

Um herauszufinden, inwieweit die entt\"auschenden Ergebnisse der
Levinschen Transformation tats\"achlich auf Rundungsfehler zur\"uckzuf\"uhren
waren, wurden ein Teil der oben beschriebenen Rechnungen auch in MAPLE
mit h\"oherer Genauigkeit wiederholt, wobei immer die maximal m\"ogliche
Zahl der St\"orungstheoriekoeffizienten $b_{n}^{(m)}$ und $c_{n}^{(m)}$
verwendet wurde ($m = 2$: $n \le 200$; $m = 3$: $n \le 165$; $m = 4$: $n
\le 139$). Dabei wurden immer die folgenden Beobachtungen gemacht: Die
Transformationen $\{ d_{\nu}^{(0)} \}$ schienen f\"ur kleinere Werte von
$\nu$ zuerst relativ langsam zu {\it konvergieren\/}. F\"ur gr\"o{\ss}ere Werte
der Transformationsordnung $\nu$ {\it divergierten\/} die Levinschen
Transformationen aber ausnahmslos. In Anbetracht der sehr guten
Reputation der Levinschen Transformation ${\cal L}_k^{(n)} (\zeta, s_n,
\omega_n)$, Gl. (5.2-6), und seiner Varianten ist diese Divergenz
sicherlich ein sehr \"uberraschendes Ergebnis.

Es mu{\ss} hier betont werden, da{\ss} die Divergenz der Folge $\{ d_{\nu}^{(0)}
\}$ nicht die Konsequenz numerischer Instabilit\"aten ist. Normalerweise
wurden die MAPLE-Rechnungen mit einer Genauigkeit von 300 - 500
Dezimalstellen durchgef\"uhrt. Im Falle der Levinschen Transformation
wurden aber auch einige Rechnungen mit einer Genauigkeit von 1000
Dezimalstellen wiederholt, ohne da{\ss} eine \"Anderung der Divergenzen
beobachtet wurde.

Die Divergenz der Transformationen $\{ d_{\nu}^{(0)} \}$ wurde
ausnahmslos beobachtet, d.~h., sowohl bei der St\"orungsreihe (10.2-4) als
auch bei der renormierten St\"orungsreihe (10.4-20) mit $m = 2, 3, 4$.
Allerdings hing sowohl die Schnelligkeit als auch die St\"arke, mit der
die Transformationen $\{ d_{\nu}^{(0)} \}$ divergierten, signifikant von
der St\"arke der Divergenz der zu summierenden St\"orungsreihe ab. Das
bedeutet, da{\ss} die Folge der Transformationen $\{ d_{\nu}^{(0)} \}$ f\"ur
gr\"o{\ss}ere Kopplungskonstanten und im Falle der St\"orungsreihe (10.2-4)
st\"arker divergierten als f\"ur kleinere Kopplungskonstanten und im Falle
der renormierten St\"orungsreihe (10.4-20). Au{\ss}erdem divergierten die
Transformationen nur relativ schwach im Falle der
$\hat{x}^4$-Anharmonizit\"at, wogegen sie im Falle der
$\hat{x}^8$-Anharmonizit\"at sehr stark divergierten.

Die erste Frage, die sich in diesem Zusammenhang aufdr\"angte, war, ob die
Levinsche Transformation \"uberhaupt prinzipiell in der Lage ist,
divergente Reihen zu summieren, deren Koeffizienten in etwa wie $(2 n)!
/ n^{1/2}$ oder gar $(3 n)! / n^{1/2}$ wachsen. Geeignete Modellprobleme
zur Beantwortung dieser Frage sind die hochgradig divergenten
asymptotischen Reihen f\"ur die beiden folgenden Integrale:
$$
\beginAligntags
" {\cal I}_3 (\kappa) " \; = \; " \int\nolimits_{0}^{\infty} \,
\frac
{t^{- 1/2} \e^{- t} \d t}
{1 + \{64 \kappa t^2 / [45 \pi^2]\}} \\
" " \sim \; " \pi^{1/2} \,
\sum_{\nu = 0}^{\infty} \, (
1/2)_{2 \nu} \, \bigl\{- 64 \kappa / [45 \pi^2] \bigr\}^{\nu} \, ,
\qquad \kappa \to 0 \, ,
\\ \tag
" {\cal I}_4 (\kappa) " \; = \; " \int\nolimits_{0}^{\infty} \,
\frac
{t^{- 1/2} \e^{- t} \d t}
{1 + (225 [\Gamma (2/3)]^9 \kappa t^3 / [112 \pi^6])} \\
" " \sim \; " \pi^{1/2} \, \sum_{\nu = 0}^{\infty} \, (1/2)_{3 \nu} \,
\bigl\{- 225 [\Gamma (2/3)]^9 \kappa / [112 \pi^6] \bigr\}^{\nu} \, ,
\qquad \kappa \to 0 \, .
\\ \tag
\endAligntags
$$
Die Koeffizienten dieser Modellreihen sind identisch mit den f\"uhrenden
Termen der asymptotischen Absch\"atzungen (10.4-23b) und (10.4-23c) f\"ur
die renormierten St\"orungstheorie\-koeffizienten $c_{n}^{(3)}$
beziehungsweise $c_{n}^{(4)}$. Diese beiden Reihen divergieren
demzufolge ebenso schnell wie die renormierte St\"orungsreihe (10.4-20)
f\"ur $m = 3$ beziehungsweise $m = 4$. F\"ur $\kappa = 1$ divergieren die
Modellreihen (10.6-5) und (10.6-6) also ebenso stark wie die
Reihenentwicklung (10.4-31) f\"ur $k_3$ beziehungsweise $k_4$.

Die Integrale ${\cal I}_3 (1)$ und ${\cal I}_4 (1)$ wurden in MAPLE
unter Verwendung einer Clenshaw-Curtis-Quadratur berechnet [Char,
Geddes, Gonnet, Leong, Monagan und Watt 1991b, S. 111], und mit den
Summationsergebnissen verglichen, die durch Anwendung der
verallgemeinerten Summationsprozesse $d_k^{(n)} (\zeta, s_n)$, Gl.
(5.2-18), und ${\delta}_k^{(n)} (\zeta, s_n)$, Gl. (5.4-13), mit $\zeta
= 1$ und des Wynnschen $\epsilon$-Algorithmus, Gl. (2.4-19), auf die
Partialsummen der divergenten Reihen in Gln. (10.6-5) und (10.6-6)
erhalten wurden.

Im Falle der Modellreihe in Gl. (10.6-5) f\"ur $k_3$ gab es die folgenden
Resultate:
$$
\beginAligntags
" {\cal I}_3 (1) " \; = \; " 1.660~177~206~668~046~735~308~9 \, ,
\erhoehe\aktTag \\ \tag*{\tagnr a}
" d_{100}^{(0)} (1, s_0) " \; = \;
" 1.660~177~206~668~046~735~310~4 \, ,
\\ \tag*{\tagform\aktTagnr b}
" {\delta}_{100}^{(0)} (1, s_0) " \; = \; "
1.660~177~206~668~046~735~234~6 \, ,
\\ \tag*{\tagform\aktTagnr c}
" {[ 50 / 49 ]} " \; = \; " 1.660~060 \, ,
\\ \tag*{\tagform\aktTagnr d}
" {[ 50 / 50 ]} " \; = \; " 1.660~270 \, .
\\ \tag*{\tagform\aktTagnr e}
\endAligntags
$$
Im Falle der Modellreihe in Gl. (10.6-6) f\"ur $k_4$ gab es die folgenden
Resultate:
$$
\beginAligntags
" {\cal I}_4 (1) " \; = \; " 1.718~397~518~229~0 \, ,
\erhoehe\aktTag \\ \tag*{\tagnr a}
" d_{100}^{(0)} (1, s_0) " \; = \; " 1.718~397~518~230~0 \, ,
\\ \tag*{\tagform\aktTagnr b}
" {\delta}_{100}^{(0)} (1, s_0) " \; = \; " 1.718~397~518~433~3 \, ,
\\ \tag*{\tagform\aktTagnr c}
" {[ 50 / 49 ]} " \; = \; " 1.707~4 \, ,
\\ \tag*{\tagform\aktTagnr d}
" {[ 50 / 50 ]} " \; = \; " 1.726~6 \, .
\\ \tag*{\tagform\aktTagnr e}
\endAligntags
$$
Diese Ergebnisse zeigen, da{\ss} die Levinsche Transformation $d_k^{(n)}
(\zeta, s_n)$, Gl. (5.2-18), die Modellreihen (10.6-5) und (10.6-6)
sogar noch etwas effizienter summiert als ${\delta}_k^{(n)} (\zeta,
s_n)$, Gl. (5.4-13). Au{\ss}erdem sieht man, da{\ss} Pad\'e-Approximationen in
Falle der Reihe (10.6-5) langsam konvergieren, wogegen die beiden Folgen
$[n / n]$ und $[n + 1/ n ]$ im Falle der divergenten Reihe (10.6-6),
deren Koeffizienten die Carlemanbedingung (4.3-5) nicht erf\"ullen,
anscheinend gegen verschiedene Grenzwerte konvergieren.

Es gibt bisher keine v\"ollig befriedigende Erkl\"arung, warum die Levinsche
Transformation $d_k^{(n)} (\zeta, s_n)$ die Modellreihen (10.6-5) und
(10.6-6) sehr effizient summiert, aber im Falle der eng verwandten Reihe
(10.4-31) f\"ur die Grenzwerte der unendlichen Kopplung $k_m$ als auch der
Reihen (10.2-4) und (10.4-28) f\"ur die Grundzustandsenergie $E^{(m)}
(\beta)$ divergente Ergebnisse liefert. Eine m\"ogliche Erkl\"arung w\"are,
da{\ss} {\it subdominante\/} Beitr\"age, die in den asymptotischen
Entwicklungen der St\"orungs\-theorie\-koeffizienten $b_{n}^{(m)}$ und
$c_{n}^{(m)}$ vorkommen, bei der rekursiven Berechnung der Levinschen
Transformationen allm\"ahlich vergr\"o{\ss}ert werden und damit die anf\"angliche
Konvergenz zerst\"oren.

Im Gegensatz dazu produzierte die eng verwandte Transformation
${\delta}_k^{(n)} (\zeta, s_n)$, Gl. (5.4-13), immer sehr gute
Ergebnisse. Divergenzen wie bei der Levinschen Transformationen wurden
nie beobachtet. Man kann zwar nicht mit letzter Sicherheit ausschlie{\ss}en,
da{\ss} eine \"ahnliche destruktive Vergr\"o{\ss}erung subdominanter Terme auch hier
vorkommen k\"onnte. In Anbetracht der vorhandenen numerischen Ergebnisse
w\"aren solche Divergenzen -- wenn \"uberhaupt -- aber nur bei wesentlich
gr\"o{\ss}eren Transformationsordnungen vorstellbar. Da der
komplexit\"atsbestimmende Faktor der Speicherbedarf bei der Berechnung der
St\"orungs\-theorie\-koeffizienten $b_{n}^{(m)}$ und $c_{n}^{(m)}$ unter
Verwendung der im Prinzip exakten rationalen Arithmetik von MAPLE ist,
und weil nur Rechner mit maximal 64 MB RAM zur Verf\"ugung stehen, ist es
zur Zeit leider unm\"oglich, umfangreichere Summationen durchzuf\"uhren als
die, die in diesem und im n\"achsten Unterabschnitt beschrieben werden.

\medskip

\Abschnitt Summationsergebnisse f\"ur die Grundzustandsenergien und die
Grenzf\"alle unendlicher Kopplung

\smallskip

\aktTag = 0

In diesem Unterabschnitt soll untersucht werden, wie gut man die
Grundzustandsenergie $E^{(m)} (\beta)$ oder den Grenzfall unendlicher
Kopplung $k_m$ durch Summation der St\"orungsreihe (10.2-4)
beziehungsweise der renormierten St\"orungsreihe (10.4-20) bestimmen kann.
Dabei werden ausschlie{\ss}lich der verallgemeinerte Summationsproze{\ss}
${\delta}_k^{(n)} (\zeta, s_n)$, Gl. (5.4-13), und der Wynnsche
$\epsilon$-Algorithmus, Gl. (2.4-10), zur Summation der divergenten
St\"orungsreihen verwendet.

Bei Pad\'e-Approximationen von Stieltjesreihen gibt es bekanntlich eine
hochentwickelte Dar\-stellungs- und Konvergenztheorie [Baker 1975; 1990;
Baker und Graves-Morris 1981a; Borel 1928; Bowman und Shenton 1989;
Perron 1957; Wall 1973]. Da die Koeffizienten $b_{n}^{(m)}$ der
St\"orungsreihe (10.2-4) des anharmonischen Oszillators mit einer
$\hat{x}^4$- oder $\hat{x}^6$-Anharmonizit\"at mit Ausnahme von
$b_{0}^{(m)}$ Momente eines Stieltjesma{\ss}es mit negativem Vorzeichen
gem\"a{\ss} Gl.~(4.3-2) sind [Simon 1970, Theorem \Roemisch{4}.2.1; Simon
1972, S. 403; Graffi und Grecchi 1978, Abschnitt \Roemisch{4}] und
au{\ss}erdem die Carlemanbedingung~(4.3-5) erf\"ullen, ist garantiert, da{\ss} die
durch den Wynnschen $\epsilon$-Algorithmus gem\"a{\ss} Gl.~(4.4-12)
berechneten Pad\'e-Approximationen $[ n / n ]$ und $[ n + 1 / n ]$ f\"ur $n
\to \infty$ gegen die Grundzustandsenergie $E^{(2)} (\beta)$
beziehungsweise $E^{(3)} (\beta)$ konvergieren. Wenn man dagegen die
renormierte St\"orungsreihe (10.4-28) f\"ur $E^{(m)} (\beta)$, die {\it
keine\/} Stieltjesreihe ist, mit Hilfe des $\epsilon$-Algorithmus
summiert, ist es nicht {\it a priori\/} klar, ob die
Pad\'e-Approximationen $[ n / n ]$ und $[ n + 1 / n ]$ konvergieren, und
man kann diese Frage beim augenblicklichen Stand des Wissens nur durch
numerische Untersuchungen beantworten.

Verglichen mit der hochentwickelten Konvergenztheorie der
Pad\'e-Approximationen ist die Konvergenztheorie sowohl des
verallgemeinerten Summationsprozesses ${\delta}_k^{(n)} (\zeta, s_n)$,
Gl. (5.4-13), als auch der meisten anderen verallgemeinerten
Summationsprozesse noch v\"ollig unterentwickelt. Man ist also fast
ausschlie{\ss}lich auf numerische Ergebnisse angewiesen, die aber -- wie im
letzten Unterabschnitt gezeigt wurde -- aufgrund ihrer Unvollst\"andigkeit
irref\"uhrend sein k\"onnen. Deswegen wird in diesem Unterabschnitt
versucht, sowohl die Grundzustandsenergie $E^{(m)} (\beta)$ als auch den
Grenzfall unendlicher Kopplung $k_m$ immer so genau wie nur m\"oglich zu
berechnen. In einigen F\"allen werden dabei mit Hilfe des
verallgemeinerten Summationsprozesses ${\delta}_k^{(n)} (\zeta, s_n)$
extrem hohe Genauigkeiten erreicht, die jenseits dessen liegen, was
physikalisch sinnvoll ist. Trotzdem sind derartig hohe Genauigkeiten
wertvoll und informativ, da sie es \"au{\ss}erst wahrscheinlich machen, da{\ss}
das Summationsverfahren tats\"achlich konvergiert und da{\ss} keine
Divergenzen wie bei der Levinschen Transformation $d_k^{(n)} (\zeta,
s_n)$, Gl. (5.2-18), auftreten.

Vergleichswerte f\"ur die in diesem Unterabschnitt angegebenen
Summationsergebnisse kann man beispielsweise in Artikeln von Banerjee,
Bhatnagar, Choudry und Kanwal [1978], Biswas, Datta, Saxena, Srivastava
und Varma [1973], Chaudhuri und Mukherjee [1985], Chhajlany, Letov und
Malnev [1991], Hioe, MacMillen und Montroll [1976; 1978], Schiffrer und
Stanzial [1985], Ta\c{s}eli und Demiralp [1988] und Vinette und {\v
C}{\'\i}{\v z}ek [1991] finden.

Sowohl der erste Term $b_0^{(m)}$ der St\"orungsreihe (10.2-4) als auch
der erste Term $c_0^{(m)}$ der renormierten St\"orungsreihe (10.4-20)
entspricht der Energie des ungest\"orten harmonischen Oszillators. Es gilt
also aufgrund der Konvention (10.2-1) immer $b_0^{(m)} = c_0^{(m)} = 1$.
Deswegen k\"onnen die St\"orungsreihen (10.2-4) und (10.4-20) auf folgende
Weise umgeschrieben werden:
$$
\beginAligntags
" E^{(m)} (\beta) " \; = \; " 1 \, + \, \beta \Delta E^{(m)} (\beta) \, ,
\\ \tag
" E_R^{(m)} (\kappa) " \; = \; "
1 \, + \, \kappa \Delta E_R^{(m)} (\kappa) \, .
\\ \tag
\endAligntags
$$
Die {\it Energieverschiebungen\/} $\Delta E^{(m)} (\beta)$ und $\Delta
E_R^{(m)} (\kappa)$, die den Effekt des St\"orterms beschreiben, sind
dabei folgenderma{\ss}en definiert:
$$
\beginAligntags
" \Delta E^{(m)} (\beta) " \; = \; "
\sum_{n=0}^{\infty} \, b_{n+1}^{(m)} \, \beta^{n} \, ,
\\ \tag
" \Delta E_R^{(m)} (\kappa) " \; = \; "
\sum_{n=0}^{\infty} \, c_{n+1}^{(m)} \, \kappa^{n} \, .
\\ \tag
\endAligntags
$$
Wenn man die Grundzustandsenergie $E^{(m)} (\beta)$ oder die renormierte
Energie $E_R^{(m)} (\kappa)$ berechnen will, kann man also auch so
vorgehen, da{\ss} man nur die St\"orungsreihen (10.7-3) und (10.7-4) f\"ur die
Energieverschiebungen $\Delta E^{(m)} (\beta)$ und $\Delta E_R^{(m)}
(\kappa)$ summiert. F\"ur diese Vorgehensweise spricht, da{\ss} die
St\"orungsreihe (10.7-3) f\"ur die Energieverschiebung $\Delta E^{(m)}
(\beta)$ des anharmonischen Oszillators mit einer $\hat{x}^4$- und einer
$\hat{x}^6$-Anharmonizit\"at eine Stieltjesreihe ist [Simon 1970, Theorem
\Roemisch{4}.2.1; Simon 1972, S. 403; Graffi und Grecchi 1978, Abschnitt
\Roemisch{4}]. Damit ist garantiert, da{\ss} die durch den Wynnschen
$\epsilon$-Algorithmus gem\"a{\ss} Gl. (4.4-12) berechneten
Pad\'e-Approximationen $[ n / n ]$ und $[ n + 1 / n ]$ f\"ur $m = 2$ und $m
= 3$ gegen die Grundzustandsenergie $E^{(m)} (\beta)$ konvergieren.

Die Grundzustandsenergie $E^{(m)} (\beta)$ kann gem\"a{\ss} Gl.~(10.4-28) auch
durch Summation der re\-normierten St\"orungsreihe (10.4-20) berechnet
werden. Wenn man die Gln.~(10.4-19), (10.4-28), (10.7-2) und (10.7-4)
kombiniert, erh\"alt man:
$$
\beginAligntags
" E^{(m)} (\beta) " \; = \; " (1 - \kappa)^{-1/2} \,
\bigl\{ 1 \, + \, \kappa \, \Delta E_R^{(m)} (\kappa) \bigr\}
\\ \tag
" " \; = \; " (1 - \kappa)^{-1/2} \,
\bigl\{ 1 \, + \, \kappa \,
\sum_{n=0}^{\infty} \, c_{n+1}^{(m)} \, \kappa^{n} \bigr\} \, .
\\ \tag
\endAligntags
$$
Aus Gl. (10.4-29) folgt, da{\ss} die Terme und Partialsummen der
St\"orungsreihe (10.7-6) das richtige asymptotische Verhalten f\"ur $\beta
\to \infty$ besitzen. Demzufolge sollte es auf diese Weise m\"oglich sein,
die Grundzustandsenergie $E^{(m)} (\beta)$ auch f\"ur gr\"o{\ss}ere Werte von
$\beta$ durch rationale Funktionen zu summieren{\footnote[\dagger]{Wenn
man die Partialsummen der St\"orungsreihe (10.7-6) als Eingabedaten f\"ur
den Wynnschen $\epsilon$-Algorithmus, Gl. (2.4-10), oder den
verallgemeinerten Summationsproze{\ss} ${\delta}_k^{(n)} (\zeta, s_n)$, Gl.
(5.4-13), verwendet, sind die resultierenden Approximationen f\"ur
$E^{(m)} (\beta)$ aufgrund des Termes $(1 - \kappa)^{-1/2}$ keine {\it
rationale\/} Funktionen in der renormierten Kopplungskonstante $\kappa$.
Wenn man dagegen die Partialsummen der St\"orungsreihe (10.7-4) f\"ur
$\Delta E_R^{(m)} (\kappa)$, die den Term $(1 - \kappa)^{-1/2}$ nicht
enthalten, als Eingabedaten verwendet, erh\"alt man Approximationen f\"ur
die renormierte Energieverschiebung $\Delta E_R^{(m)} (\kappa)$, die
{\it rational\/} in $\kappa$ sind. Aus diesen Ausdr\"ucken kann man dann
gem\"a{\ss} Gln. (10.4-28) und (10.7-2) {\it nichtrationale\/} Approximationen
f\"ur $E^{(m)} (\beta)$ berechnen.}}. Eine effiziente Summation der
St\"orungsreihe (10.2-4) durch rationale Funktionen d\"urfte dagegen f\"ur
gr\"o{\ss}ere Kopplungskonstanten $\beta$ kaum m\"oglich sein.

In den Tabellen 10-6, 10-7 und 10-8 [Weniger, {\v C}{\'\i}{\v z}ek und
Vinette 1993, Tables \Roemisch{4} - \Roemisch{6}] wird verglichen, wie
gut die Grundzustandsenergie $E^{(m)} (\beta)$ f\"ur $m = 2, 3, 4$ durch
Summation der St\"orungsreihe (10.7-3) f\"ur die Energieverschiebung $\Delta
E^{(m)} (\beta)$ gem\"a{\ss} Gl. (10.7-1) beziehungsweise der St\"orungsreihe
(10.7-4) f\"ur die renormierte Energieverschiebung $\Delta E_R^{(m)}
(\kappa)$ gem\"a{\ss} Gl. (10.7-6) berechnet werden kann. Dabei wurden die
Partialsummen
$$
s_n^{(m)} (\beta) \; = \;
\sum_{j=0}^{n} \, b_{j+1}^{(m)} \, \beta^j
\tag
$$
und
$$
\sigma_n^{(m)} (\kappa) \; = \; (1 - \kappa)^{- 1/2} \,
\sum_{j=0}^{n} \, c_{j+1}^{(m)} \, \kappa^j
\tag
$$
der St\"orungsreihen f\"ur $\Delta E^{(m)} (\beta)$ beziehungsweise $(1 -
\kappa)^{-1/2} \Delta E_R^{(m)} (\kappa)$ als Eingabedaten verwendet.
Alle Rechnungen wurden in FORTRAN auf einer Cyber 180-995 E in DOUBLE
PRECISION (ungef\"ahr 29 Dezimalstellen) durchgef\"uhrt. Au{\ss}erdem wurden
immer nur die St\"orungs\-theorie\-ko\-effizi\-enten $b_{n}^{(m)}$ und
$c_{n}^{(m)}$ mit $0 \le n \le 50$ verwendet. F\"ur einen gegebenen Wert
der Kopplungskonstante $\beta$ wurde die entsprechende renormierte
Kopplungskonstante $\kappa$ gem\"a{\ss} Gl.~(10.4-12) f\"ur $m = 2, 3, 4$
numerisch berechnet. In FORTRAN geschah dies mit Hilfe des Newtonschen
Iterationsverfahrens, und bei den sp\"ater beschriebenen MAPLE-Rechnungen
wurde dazu das Kommando {\it fsolve\/} verwendet [Char, Geddes, Gonnet,
Leong, Monagan und Watt 1991b, S. 97].

\beginFloat

\medskip

\beginTabelle [to \kolumnenbreite]
\beginFormat \rechts " \mitte " \mitte " \mitte " \mitte
\endFormat
\+ " \links {\bf Tabelle 10-6} \@ \@ \@ \@ " \\
\+ " \links {Berechnung der Grundzustandsenergie $E^{(2)} (\beta)$ f\"ur
$\beta = 2$} \@ \@ \@ \@ " \\
\+ " \links {Summation der St\"orungsreihen (10.7-3) und (10.7-4) f\"ur
$\Delta E^{(2)} (\beta)$ und $\Delta E_R^{(2)} (\kappa)$}
\@ \@ \@ \@ " \\ \sstrut {} {1.5 \jot} {1.5 \jot}
\= " \= " \= " \= " \= " \= " \\ \sstrut {} {1 \jot} {1 \jot}
\+ " " \links {St\"orungsreihe (10.7-3)}
\@ " \links {Renormierte St\"orungsreihe (10.7-4)} \@ " \\
\- " \- " \- " \- " \- " \- " \\ \sstrut {} {1 \jot} {1 \jot}
\+ " $n$ " \links {Pad\'e}
" \mitte {${\delta}_n^{(0)} (1, s_0^{(2)} (\beta))$}
" \links {Pad\'e}
" \mitte {${\delta}_n^{(0)} (1, \sigma_0^{(2)} (\kappa))$} " \\
\- " \- " \- " \- " \- " \- " \\ \sstrut {} {1 \jot} {1 \jot}
\+ " 15 " 1.58719872130224 " 1.60741239541032 " 1.60753998246438
" 1.60754130148468 " \\
\+ " 16 " 1.61490803497579 " 1.60788606258563 " 1.60754134289198
" 1.60754130184499 " \\
\+ " 17 " 1.59437892786761 " 1.60774000912502 " 1.60754115010134
" 1.60754130210150 " \\
\+ " 18 " 1.61261134950870 " 1.60732301758607 " 1.60754130865941
" 1.60754130226049 " \\
\+ " 19 " 1.59878408544814 " 1.60728921713028 " 1.60754126443087
" 1.60754130235364 " \\
\+ " 20 " 1.61109538687340 " 1.60759433576965 " 1.60754130049978
" 1.60754130240642 " \\
\+ " 21 " 1.60157813198874 " 1.60776516006506 " 1.60754129054942
" 1.60754130243549 " \\
\+ " 22 " 1.61007253342611 " 1.60763364063421 " 1.60754129905132
" 1.60754130245117 " \\
\+ " 23 " 1.60339995400794 " 1.60743740132082 " 1.60754129819839
" 1.60754130245950 " \\
\+ " 24 " 1.60936925356861 " 1.60740600172367 " 1.60754129901702
" 1.60754130246389 " \\
\+ " 25 " 1.60461594952095 " 1.60751428256417 " 1.60754130077405
" 1.60754130246617 " \\
\+ " 26 " 1.60887769423078 " 1.60760771878698 " 1.60754129851286
" 1.60754130246735 " \\
\+ " 27 " 1.60544404826871 " 1.60760879629622 " 1.60754130173572
" 1.60754130246796 " \\
\+ " 28 " 1.60852910836743 " 1.60755743028065 " 1.60754119852643
" 1.60754130246826 " \\
\+ " 29 " 1.60601793374047 " 1.60752016064055 " 1.60754130212779
" 1.60754130246841 " \\
\+ " 30 " 1.60827870607604 " 1.60751703195156 " 1.60754130366700
" 1.60754130246848 " \\
\+ " 31 " 1.60642181112352 " 1.60752799543141 " 1.60754130230074
" 1.60754130246852 " \\
\+ " 32 " 1.60809674086083 " 1.60753439095606 " 1.60754130278104
" 1.60754130246854 " \\
\+ " 33 " 1.60670995434205 " 1.60753737125348 " 1.60754130238248
" 1.60754130246854 " \\
\+ " 34 " 1.60796311871296 " 1.60754480727720 " 1.60754130257785
" 1.60754130246855 " \\
\+ " 35 " 1.60691805906270 " 1.60755502116582 " 1.60754130242332
" 1.60754130246855 " \\
\- " \- " \- " \- " \- " \- " \\ \sstrut {} {1 \jot} {1 \jot}

\endTabelle

\medskip

\endFloat

In Table \Roemisch{1} von Vinette und {\v C}{\'\i}{\v z}ek [1991], wo
obere und untere Schranken f\"ur $E^{(2)} (\beta)$ angegeben sind, findet
man $E^{(2)} (2) = 1.607~541~302~468~548$, das nach Rundung der letzten
Stelle mit dem besten Summationsergebnis in der letzten Spalte von
Tabelle 10-6 \"ubereinstimmt.

In einer analogen Rechnung f\"ur $k_2$ ergab die Transformation
$\delta_{35}^{(0)} \bigl( 1, \Sigma_{0}^{(2)}\bigr)$ den Wert $k_2 =
1.060~362~090~484~2$. Alle angegebenen Stellen stimmen mit dem extrem
genauen Wert (10.4-33a) von Vinette und {\v C}{\'\i}{\v z}ek [1991, Gl.
(66)] \"uberein. Pad\'e-Approximationen konnten unter den gleichen
Bedingungen nur 8 Dezimalstellen reproduzieren. Bei der Berechnung der
Grenzwerte unendlicher Kopplung $k_m$ durch Summation der renormierten
St\"orungsreihe (10.4-31) wurden die Partialsummen
$$
\Sigma_n^{(m)} \; = \; \sum_{j=0}^{n} \, c_{j+1}^{(m)}
\tag
$$
als Eingabedaten verwendet.

Die Ergebnisse aus Tabelle 10-6 und anderer Rechnungen zeigen, da{\ss} man
die Grundzustands\-energie $E^{(2)} (\beta)$ des anharmonischen
Oszillators mit einer $\hat{x}^4$-Anharmonizit\"at selbst f\"ur kleine
Kopplungskonstanten $\beta$ wesentlich leichter durch Summation der
renormierten St\"orungsreihe (10.7-4) f\"ur $\Delta E_R^{(2)} (\kappa)$
berechnen kann als durch Summation der St\"orungsreihe (10.7-3) f\"ur
$\Delta E^{(2)} (\beta)$. F\"ur gr\"o{\ss}ere Werte der Kopplungskonstanten
$\beta$ wird die \"Uberlegenheit der renormierten St\"orungsreihe noch
wesentlich auff\"alliger. Diese Beobachtung ist nicht typisch f\"ur den
Oszillator mit einer $\hat{x}^4$-Anharmonizit\"at, sondern sie gilt ganz
analog auch im Falle einer $\hat{x}^6$- und $\hat{x}^8$-Anharmonizit\"at.

Ein Vergleich der Ergebnisse von Rechnungen in FORTRAN auf einer Cyber
180-995 E in DOUBLE PRECISION und in MAPLE mit h\"oherer Genauigkeit
ergab, da{\ss} numerische Instabilit\"aten bei der Summation der renormierten
St\"orungsreihe (10.4-20) im Falle des anharmonischen Oszillators mit
einer $\hat{x}^4$-Anharmonizit\"at keine ernsthaften Probleme aufwerfen.
Der Grenzfall unendlicher Kopplung $k_2$ ist das schwierigste
Summationsproblem, das im Zusammenhang mit der renormierten
St\"orungsreihe (10.4-20) mit $m = 2$ vorkommen kann. Da $k_2$ -- wie oben
gezeigt -- in FORTRAN auf einer Cyber 180-995 E in DOUBLE PRECISION mit
ausreichender Genauigkeit aus den St\"orungstheoriekoeffizienten
$c_{n}^{(2)}$ mit $n \le 50$ berechnet werden kann, folgt, da{\ss} die
Summation der St\"orungsreihe (10.7-6) f\"ur $E^{(2)} (\beta)$ durch die
Transformationen $\bigl\{ \delta_{n}^{(0)} \bigr\}$ f\"ur den gesamten
physikalisch relevanten Bereich $0 \le \beta < \infty$ Ergebnisse von
v\"ollig befriedigender Genauigkeit liefert, wenn man die Rechnungen in
FORTRAN mit einer Genauigkeit von 29 Dezimalstellen durchf\"uhrt.

Tabelle 10-6 zeigt, da{\ss} die Pad\'e-Approximationen $\epsilon_{2 n}^{(0)} =
[n / n]$ und $\epsilon_{2 n}^{(1)} = [n + 1 / n]$ deutlich weniger
effizient sind als die Transformationen $\bigl\{ \delta_{n}^{(0)}
\bigr\}$. Trotzdem sind die Ergebnisse, die man durch Summation der
renormierten St\"orungsreihe (10.7-6) durch Pad\'e-Approximationen erh\"alt,
besser als die Ergebnisse, die Graffi, Grecchi, und Simon [1970, Tables
2 und 3] oder Hirsbrunner [1982, Tables 1 und 2] durch eine
Borel-Pad\'e-Summation der St\"orungsreihe (10.2-4) mit $m = 2$ erzielen
konnten. Das ist ein weiteres Indiz daf\"ur, da{\ss} die renormierte
St\"orungsreihe (10.4-20) in numerischer Hinsicht deutlich g\"unstigere
Eigenschaften besitzt als die St\"orungsreihe (10.2-4).

\beginFloat

\medskip

\beginTabelle [to \kolumnenbreite]
\beginFormat \rechts " \mitte " \mitte " \mitte " \mitte
\endFormat
\+ " \links {\bf Tabelle 10-7} \@ \@ \@ \@ " \\
\+ " \links {Berechnung der Grundzustandsenergie $E^{(3)} (\beta)$ f\"ur
$\beta = 2/10$}
\@ \@ \@ \@ " \\
\+ " \links {Summation der St\"orungsreihen (10.7-3) und (10.7-4) f\"ur
$\Delta E^{(3)} (\beta)$ und $\Delta E_R^{(3)} (\kappa)$}
\@ \@ \@ \@ " \\ \sstrut {} {1.5 \jot} {1.5 \jot}
\= " \= " \= " \= " \= " \= " \\ \sstrut {} {1 \jot} {1 \jot}
\+ " " \links {St\"orungsreihe (10.7-3)}
\@ " \links {Renormierte St\"orungsreihe (10.7-4)} \@ " \\
\- " \- " \- " \- " \- " \- " \\ \sstrut {} {1 \jot} {1 \jot}
\+ " $n$ " \links {Pad\'e}
" \mitte {${\delta}_n^{(0)} (1, s_0^{(3)} (\beta))$}
" \links {Pad\'e}
" \mitte {${\delta}_n^{(0)} (1, \sigma_0^{(3)} (\kappa))$} " \\
\- " \- " \- " \- " \- " \- " \\ \sstrut {} {1 \jot} {1 \jot}
\+ " 25 " 1.14639468647148 " 1.17390088173818 " 1.17387739089120 "
1.17388935983299 " \\
\+ " 26 " 1.18357215348621 " 1.17388873060312 " 1.17401741295442 "
1.17388935168880 " \\
\+ " 27 " 1.14895126786169 " 1.17388086741464 " 1.17388294261549 "
1.17388934686938 " \\
\+ " 28 " 1.18284073309867 " 1.17387841206729 " 1.17399706557154 "
1.17388934405220 " \\
\+ " 29 " 1.15111151972269 " 1.17388038694433 " 1.17388703297933 "
1.17388934247616 " \\
\+ " 30 " 1.18220691231111 " 1.17388474304821 " 1.17398088281009 "
1.17388934168882 " \\
\+ " 31 " 1.15295800654519 " 1.17388931750900 " 1.17389008380525 "
1.17388934140238 " \\
\+ " 32 " 1.18165269027090 " 1.17389253201858 " 1.17396784296236 "
1.17388934142601 " \\
\+ " 33 " 1.15455224235697 " 1.17389372029792 " 1.17389238195332 "
1.17388934163324 " \\
\+ " 34 " 1.18116420136612 " 1.17389307644954 " 1.17395721201425 "
1.17388934194092 " \\
\+ " 35 " 1.15594092629916 " 1.17389133573861 " 1.17389412676387 "
1.17388934229375 " \\
\+ " 36 " 1.18073059778939 " 1.17388937002385 " 1.17394845326765 "
1.17388934265437 " \\
\+ " 37 " 1.15716010150736 " 1.17388786414682 " 1.17389545941309 "
1.17388934299825 " \\
\+ " 38 " 1.18034327477069 " 1.17388715905137 " 1.17394116801666 "
1.17388934331114 " \\
\+ " 39 " 1.15823799762542 " 1.17388725715669 " 1.17389648157606 "
1.17388934358709 " \\
\+ " 40 " 1.17999532210504 " 1.17388793008340 " 1.17393505577502 "
1.17388934382587 " \\
\+ " 41 " 1.15919701616766 " 1.17388886098168 " 1.17389726758032 "
1.17388934403039 " \\
\+ " 42 " 1.17968112860856 " 1.17388978397298 " 1.17392988709495 "
1.17388934420412 " \\
\+ " 43 " 1.16005514426458 " 1.17389064733678 " 1.17389787249174 "
1.17388934434760 " \\
\+ " 44 " 1.17939609207394 " 1.17389194935022 " 1.17392548466396 "
1.17388934444794 " \\
\+ " 45 " 1.16082697898754 " 1.17389557438344 " 1.17389833759864 "
1.17388934443612 " \\
\- " \- " \- " \- " \- " \- " \\ \sstrut {} {1 \jot} {1 \jot}

\endTabelle

\medskip

\endFloat

In Tabelle \Roemisch{2} von Vinette und {\v C}{\'\i}{\v z}ek [1991], wo
obere und untere Schranken f\"ur $E^{(3)} (\beta)$ angegeben sind, findet
man nach Rundung $E^{(3)} (2/10) = 1.173~889~345$, was abgesehen von der
letzten Stelle mit dem besten Summationsergebnis in der letzten Spalte
von Tabelle 10-7 \"ubereinstimmt. In einer analogen Rechnung f\"ur $k_3$
ergab die Transformation $\delta_{41}^{(0)} \bigl( 1,
\Sigma_{0}^{(3)}\bigr)$ den Wert $k_3 = 1.144~802$. Alle hier
angegebenen Stellen stimmen mit dem extrem genauen Wert (10.4-33b) von
Vinette und {\v C}{\'\i}{\v z}ek [1991, Gl. (69)] \"uberein.
Pad\'e-Approximationen konnten unter den gleichen Bedingungen nur 3
Dezimalstellen reproduzieren. Dieses Ergebnis zeigt, da{\ss} im Falle des
anharmonischen Oszillators mit einer $\hat{x}^6$-Anharmonizit\"at die
praktische N\"utzlichkeit von Pad\'e-Approximationen auch durch Renormierung
nicht wesentlich vergr\"o{\ss}ert werden kann.

Ein Vergleich der Ergebnisse von Rechnungen in FORTRAN auf einer Cyber
180-995 E in DOUBLE PRECISION und in MAPLE mit h\"oherer Genauigkeit
ergab, da{\ss} numerische Instabilit\"aten bei der Summation der renormierten
St\"orungsreihe (10.4-20) im Falle des anharmonischen Oszillators mit
einer $\hat{x}^6$-Anharmonizit\"at ein nicht mehr vernachl\"assigbares
Problem darstellen. So wurde bei der Summation der St\"orungsreihe
(10.4-31) f\"ur $k_3$ in FORTRAN beobachtet, da{\ss} die Genauigkeit der
Transformationen $\bigl\{ {\delta}_{n}^{(0)} \bigr\}$ f\"ur $n \ge 41$
aufgrund von Rundungsfehlern rasch abnimmt. Extrem genaue
Summationsergebnisse f\"ur $E^{(3)} (\beta)$ oder $k_3$, wie sie sp\"ater
noch pr\"asentiert werden, sind also bei Verwendung einer konventionellen
Programmiersprache wie FORTRAN aufgrund numerischer Instabilit\"aten nicht
m\"oglich.

\beginFloat

\medskip

\beginTabelle [to \kolumnenbreite]
\beginFormat \rechts " \mitte " \mitte " \mitte " \mitte
\endFormat
\+ " \links {\bf Tabelle 10-8} \@ \@ \@ \@ " \\
\+ " \links {Berechnung der Grundzustandsenergie $E^{(4)} (\beta)$ f\"ur
$\beta = 2/100$}
\@ \@ \@ \@ " \\
\+ " \links {Summation der St\"orungsreihen (10.7-3) und (10.7-4) f\"ur
$\Delta E^{(4)} (\beta)$ und $\Delta E_R^{(4)} (\kappa)$}
\@ \@ \@ \@ " \\ \sstrut {} {1.5 \jot} {1.5 \jot}
\= " \= " \= " \= " \= " \= " \\ \sstrut {} {1 \jot} {1 \jot}
\+ " " \links {St\"orungsreihe (10.7-3)}
\@ " \links {Renormierte St\"orungsreihe (10.7-4)} \@ " \\
\- " \- " \- " \- " \- " \- " \\ \sstrut {} {1 \jot} {1 \jot}
\+ " $n$ " \links {Pad\'e}
" \mitte {${\delta}_n^{(0)} (1, s_0^{(3)} (\beta))$}
" \links {Pad\'e}
" \mitte {${\delta}_n^{(0)} (1, \sigma_0^{(3)} (\kappa))$} " \\
\- " \- " \- " \- " \- " \- " \\ \sstrut {} {1 \jot} {1 \jot}
\+ " 25 " 0.99092116316557 " 1.06421035678877 " 1.06155163294552 "
1.06421371850029 " \\
\+ " 26 " 1.08212594990847 " 1.06421318766816 " 1.06748583611372 "
1.06421276806101 " \\
\+ " 27 " 0.99273913256429 " 1.06421455351208 " 1.06165363799713 "
1.06421195422154 " \\
\+ " 28 " 1.08182406934377 " 1.06421473220647 " 1.06741798520934 "
1.06421126615365 " \\
\+ " 29 " 0.99433529238189 " 1.06421406048243 " 1.06174222286937 "
1.06421069158403 " \\
\+ " 30 " 1.08155534138883 " 1.06421285612716 " 1.06735759957236 "
1.06421021654500 " \\
\+ " 31 " 0.99574979872390 " 1.06421138785051 " 1.06181994983029 "
1.06420982638728 " \\
\+ " 32 " 1.08131426249317 " 1.06420987065378 " 1.06730343954841 "
1.06420950695803 " \\
\+ " 33 " 0.99701350352910 " 1.06420846929170 " 1.06188876253958 "
1.06420924544745 " \\
\+ " 34 " 1.08109651160186 " 1.06420730111644 " 1.06725453179744 "
1.06420903080149 " \\
\+ " 35 " 0.99815052269549 " 1.06420643697412 " 1.06195016260197 "
1.06420885379222 " \\
\+ " 36 " 1.08089864508760 " 1.06420590258545 " 1.06721010060402 "
1.06420870688867 " \\
\+ " 37 " 0.99917999458812 " 1.06420568313976 " 1.06200532887796 "
1.06420858404518 " \\
\+ " 38 " 1.08071788211592 " 1.06420573216720 " 1.06716951954598 "
1.06420848047644 " \\
\+ " 39 " 1.00011731254261 " 1.06420598396757 " 1.06205520009218 "
1.06420839244816 " \\
\+ " 40 " 1.08055195051214 " 1.06420636891641 " 1.06713227678385 "
1.06420831709103 " \\
\+ " 41 " 1.00097500668322 " 1.06420683673568 " 1.06210053327082 "
1.06420825224212 " \\
\+ " 42 " 1.08039897403989 " 1.06420741845962 " 1.06709794967396 "
1.06420819629714 " \\
\+ " 43 " 1.00176338652181 " 1.06420845912081 " 1.06214194587826 "
1.06420814776269 " \\
\+ " 44 " 1.08025738860810 " 1.06421154217121 " 1.06706618589512 "
1.06420810196258 " \\
\+ " 45 " 1.00249101700116 " 1.06422308340851 " 1.06217994672486 "
1.06420802867976 " \\
\- " \- " \- " \- " \- " \- " \\ \sstrut {} {1 \jot} {1 \jot}

\endTabelle

\medskip

\endFloat

Zur \"Uberpr\"ufung der Summationsergebnisse in Tabelle 10-8 wurde $E^{(4)}
(2/100)$ auch mit Hilfe der von Vinette und {\v C}{\'\i}{\v z}ek [1991]
beschriebenen Methode der inneren Projektion berechnet. Diese Rechnung
ergab $E^{(4)} (2/100) = 1.064~207~854~737$ [Weniger, {\v C}{\'\i}{\v
z}ek und Vinette 1993, S. 599]. Ein Vergleich mit den besten
Summationsergebnissen in der letzten Spalte von Tabelle 10-8 ergibt nach
Rundung eine \"Ubereinstimmung von 6 Dezimalstellen. Dieses Ergebnis und
auch andere Rechnungen, die sp\"ater besprochen werden, zeigen, da{\ss}
${\delta}_k^{(n)} (\zeta, s_n)$, Gl. (5.4-13), anscheinend in der Lage
ist, die renormierte St\"orungsreihe (10.4-20) f\"ur den anharmonischen
Oszillator mit einer $\hat{x}^8$-Anharmonizit\"at zu summieren. Auch im
Falle der Partialsummen (10.7-7) scheinen die Transformationen $\bigl\{
\delta_{n}^{(0)} \bigr\}$ -- wenn auch deutlich langsamer -- zu
konvergieren.

Die Pad\'e-Ergebnisse in der zweiten und der vierten Spalte von Tabelle
10-8 zeigen dagegen, da{\ss} die Transformationen $\epsilon_{2 n}^{(0)} = [n
/ n]$ und $\epsilon_{2 n}^{(1)} = [n + 1 / n]$ sowohl im Falle der
St\"orungsreihe (10.7-3) als auch der renormierten St\"orungsreihe (10.7-4)
anscheinend gegen verschiedene Grenzwerte konvergieren. Das hier
verwendete Renormierungsverfahren kann also offensichtlich nichts daran
\"andern, da{\ss} Pad\'e-Approximationen starke asymptotische Reihen der Ordnung
$k > 2$ nicht summieren k\"onnen.

In einer analogen Rechnung f\"ur $k_4$ ergab die Transformation
$\delta_{47}^{(0)} \bigl( 1, \Sigma_{0}^{(4)}\bigr)$ den Wert $k_4 =
1.226$. Nach Rundung stimmen alle angegebenen Stellen mit dem extrem
genauen Wert (10.4-33c) von Vinette und {\v C}{\'\i}{\v z}ek [1991, Gl.
(71)] \"uberein. Pad\'e-Approximationen f\"ur $k_4$ konvergieren nicht.

Aufgrund der \"au{\ss}erst starken Divergenz der St\"orungsreihe (10.2-4) als
auch der renormierten St\"orungsreihe (10.4-20) sind im Falle des
anharmonischen Oszillators mit einer $\hat{x}^8$-Anharmonizit\"at
Rundungsfehler ein noch gr\"o{\ss}eres Problem als bei einer
$\hat{x}^6$-Anharmonizit\"at. In Tabelle 10-8 macht sich Stellenverlust
durch Rundungsfehler schon sehr stark bemerkbar, und es ist sicherlich
eine gute Idee, aufwendigere Rechnungen ausschlie{\ss}lich in MAPLE -- oder
einer anderen Sprache f\"ur symbolische Manipulationen -- mit einer
deutlich h\"oheren Genauigkeit und nicht in FORTRAN durchzuf\"uhren.

Die Ergebnisse in Tabellen 10-6, 10-7 und 10-8 zeigen einerseits, da{\ss}
Pad\'e-Approximationen deutlich weniger effizient sind als der
verallgemeinerte Summationsproze{\ss} ${\delta}_k^{(n)} (\zeta, s_n)$, und
andererseits, da{\ss} die Summation der renormierten St\"orungsreihe
wesentlich bessere Ergebnisse liefert als die Summation der
St\"orungsreihe (10.2-4). Die Tabellen 10-6, 10-7 und 10-8 wurden auf
einer Cyber 180-995 E in DOUBLE PRECISION berechnet, wobei immer nur die
St\"orungs\-theorie\-ko\-effizi\-enten $b_{n}^{(m)}$ und $c_{n}^{(m)}$ mit
$0 \le n \le 50$ verwendet wurde. Es w\"are sicherlich von Interesse,
herauszufinden, wie gut man die Grundzustandsenergie $E^{(m)} (\beta)$
mit Hilfe des verallgemeinerten Summationsprozesses ${\delta}_k^{(n)}
(\zeta, s_n)$ durch Summation der renormierten St\"orungsreihe (10.7-6)
berechnen kann, wenn man die Summation in MAPLE mit einer deutlich
h\"oheren Genauigkeit durchf\"uhrt und dabei alle verf\"ugbaren renormierten
St\"o\-rungs\-theorie\-ko\-effizi\-enten $c_{n}^{(m)}$ verwendet.

\beginTabelle % [to \kolumnenbreite]
\beginFormat \links " \links " \endFormat
\+ " \links {\bf Tabelle 10-9} \@ " \\
\+ " \links {Anharmonischer Oszillator mit einer $\hat{x}^4$-Anharmonizit\"at}
\@ " \\
\+ " \links {Summation der St\"orungsreihe (10.7-6) f\"ur die
Grundzustandsenergie $E^{(2)} (\beta)$} \@ " \\
\sstrut {} {1.5 \jot} {1.5 \jot}
\= " \= " \=  " \\ \sstrut {} {1 \jot} {1 \jot}
\+ " \links{$\beta$}
" \links{$\qquad \qquad {\cal E}_{n}^{(2)} (\beta)$}
" \\ \sstrut {} {1.5 \jot} {1.5 \jot}
\= " \= " \=  " \\ \sstrut {} {1 \jot} {1 \jot}
\+ " 0.2 "
${\cal E}_{196}^{(2)} =
1.11829265436703915343081315383965718542276478699214171151312824$
" \\ \sstrut {} {1.25 \jot} {1.25 \jot}
\+ " "
${\cal E}_{197}^{(2)} =
1.11829265436703915343081315383965718542276478699214171151312820$
" \\ \sstrut {} {1.25 \jot} {1.25 \jot}
\+ " "
${\cal E}_{198}^{(2)} =
1.11829265436703915343081315383965718542276478699214171151312817$
" \\ \sstrut {} {1.25 \jot} {1.25 \jot}
\- " \- " \-  " \\ \sstrut {} {1.25 \jot} {1.25 \jot}
\+ " 0.6 "
${\cal E}_{196}^{(2)} =
1.27598356634255705890504639597965876233802059727461$
" \\ \sstrut {} {1.25 \jot} {1.25 \jot}
\+ " "
${\cal E}_{197}^{(2)} =
1.27598356634255705890504639597965876233802059727449$ " \\
\+ " "
${\cal E}_{198}^{(2)} =
1.27598356634255705890504639597965876233802059727440$
" \\ \sstrut {} {1.25 \jot} {1.25 \jot}
\- " \- " \-  " \\ \sstrut {} {1.25 \jot} {1.25 \jot}
\+ " 1.0 "
${\cal E}_{196}^{(2)} =
1.3923516415302918556575078766099341846000667126$
" \\ \sstrut {} {1.25 \jot} {1.25 \jot}
\+ " "
${\cal E}_{197}^{(2)} =
1.3923516415302918556575078766099341846000667122$
" \\ \sstrut {} {1.25 \jot} {1.25 \jot}
\+ " "
${\cal E}_{198}^{(2)} =
1.3923516415302918556575078766099341846000667119$
" \\ \sstrut {} {1.25 \jot} {1.25 \jot}
\- " \- " \-  " \\ \sstrut {} {1.25 \jot} {1.25 \jot}
\+ " 2.0 "
${\cal E}_{196}^{(2)} =
1.6075413024685475387081719294732483820810503$
" \\ \sstrut {} {1.25 \jot} {1.25 \jot}
\+ " "
${\cal E}_{197}^{(2)} =
1.6075413024685475387081719294732483820810494$
" \\ \sstrut {} {1.25 \jot} {1.25 \jot}
\+ " "
${\cal E}_{198}^{(2)} =
1.6075413024685475387081719294732483820810487$
" \\ \sstrut {} {1.25 \jot} {1.25 \jot}
\- " \- " \-  " \\ \sstrut {} {1.25 \jot} {1.25 \jot}
\+ " 4.0 "
${\cal E}_{196}^{(2)} =
1.90313694545900002229385072220102393181777$
" \\ \sstrut {} {1.25 \jot} {1.25 \jot}
\+ " "
${\cal E}_{197}^{(2)} =
1.90313694545900002229385072220102393181764$
" \\ \sstrut {} {1.25 \jot} {1.25 \jot}
\+ " "
${\cal E}_{198}^{(2)} =
1.90313694545900002229385072220102393181755$
" \\ \sstrut {} {1.25 \jot} {1.25 \jot}
\- " \- " \-  " \\ \sstrut {} {1.25 \jot} {1.25 \jot}
\+ " 100.0 "
${\cal E}_{196}^{(2)} =
4.9994175451375878292946320373496527204$
" \\ \sstrut {} {1.25 \jot} {1.25 \jot}
\+ " "
${\cal E}_{197}^{(2)} =
4.9994175451375878292946320373496527199$
" \\ \sstrut {} {1.25 \jot} {1.25 \jot}
\+ " "
${\cal E}_{198}^{(2)} =
4.9994175451375878292946320373496527196$
" \\ \sstrut {} {1.25 \jot} {1.25 \jot}
\- " \- " \-  " \\ \sstrut {} {1.25 \jot} {1.25 \jot}
\+ " 400.0 "
${\cal E}_{196}^{(2)} =
7.8618626782758914111858091254519727848$
" \\ \sstrut {} {1.25 \jot} {1.25 \jot}
\+ " "
${\cal E}_{197}^{(2)} =
7.8618626782758914111858091254519727835$
" \\ \sstrut {} {1.25 \jot} {1.25 \jot}
\+ " "
${\cal E}_{198}^{(2)} =
7.8618626782758914111858091254519727826$
" \\ \sstrut {} {1.25 \jot} {1.25 \jot}
\- " \- " \-  " \\ \sstrut {} {1.25 \jot} {1.25 \jot}
\+ " 2000.0 "
${\cal E}_{196}^{(2)} =
13.388441701008061939006176902807286534$
" \\ \sstrut {} {1.25 \jot} {1.25 \jot}
\+ " "
${\cal E}_{197}^{(2)} =
13.388441701008061939006176902807286531$
" \\ \sstrut {} {1.25 \jot} {1.25 \jot}
\+ " "
${\cal E}_{198}^{(2)} =
13.388441701008061939006176902807286529$
" \\ \sstrut {} {1.25 \jot} {1.25 \jot}
\- " \- " \-  " \\ \sstrut {} {1.25 \jot} {1.25 \jot}
\+ " 16000.0 "
${\cal E}_{196}^{(2)} =
26.733815088189002423231037392721473642$
" \\ \sstrut {} {1.25 \jot} {1.25 \jot}
\+ " "
${\cal E}_{197}^{(2)} =
26.733815088189002423231037392721473635$
" \\ \sstrut {} {1.25 \jot} {1.25 \jot}
\+ " "
${\cal E}_{198}^{(2)} =
26.733815088189002423231037392721473631$
" \\ \sstrut {} {1.25 \jot} {1.25 \jot}
\- " \- " \-  " \\ \sstrut {} {1.25 \jot} {1.25 \jot}
\+ " 40000.0 "
${\cal E}_{196}^{(2)} =
36.274458133736835470376382678474479672$
" \\ \sstrut {} {1.25 \jot} {1.25 \jot}
\+ " "
${\cal E}_{197}^{(2)} =
36.274458133736835470376382678474479664$
" \\ \sstrut {} {1.25 \jot} {1.25 \jot}
\+ " "
${\cal E}_{198}^{(2)} =
36.274458133736835470376382678474479657$
" \\ \sstrut {} {1.25 \jot} {1.25 \jot}
\= " \= " \= " \\ \sstrut {} {1 \jot} {1 \jot}

\endTabelle

\neueSeite

\beginTabelle % [to \kolumnenbreite]
\beginFormat & \links \endFormat
\+ " \links {\bf Tabelle 10-10} \@ \@ \@ " \\
\+ " \links {Anharmonischer Oszillator mit einer
$\hat{x}^6$-Anharmonizit\"at} \@ \@ \@ " \\
\+ " \links {Summation der St\"orungsreihe (10.7-6) f\"ur die
Grundzustandsenergie $E^{(3)} (\beta)$} \@ \@ \@ " \\
\sstrut {} {1.5 \jot} {1.5 \jot}
\= " \= " \= " \= " \=  " \\ \sstrut {} {1 \jot} {1 \jot}
\+ " $\beta$ " {$\quad E_{<}^{(3)} (\beta)$}
" {$\quad E_{>}^{(3)} (\beta)$}
" {$\qquad \qquad {\cal E}_{n}^{(3)} (\beta)$}
" \\ \sstrut {} {1.5 \jot} {1.5 \jot}
\= " \= " \= " \= " \= " \\ \sstrut {} {1 \jot} {1 \jot}
\+ " 0.2   " $1.173~889~345~117$  " $1.173~889~345~130$ "
${\cal E}_{161}^{(3)} \; = \;  1.173~889~345~125~433~175$
" \\ \sstrut {} {1.25 \jot} {1.25 \jot}
\+ "       "                      "
" ${\cal E}_{162}^{(3)} \; = \;  1.173~889~345~125~433~172$
" \\ \sstrut {} {1.25 \jot} {1.25 \jot}
\+ "       "                      "
" ${\cal E}_{163}^{(3)} \; = \;  1.173~889~345~125~433~169$ "
\\ \sstrut {} {1.25 \jot} {1.25 \jot}
\- " \- " \- " \- " \- " \\ \sstrut {} {1.25 \jot} {1.25 \jot}
\+ " 0.6   " $1.332~895~943~33$   " $1.332~895~943~43$
" ${\cal E}_{161}^{(3)} \; = \;  1.332~895~943~373~395~84$
" \\ \sstrut {} {1.25 \jot} {1.25 \jot}
\+ "       "                      "
" ${\cal E}_{162}^{(3)} \; = \;  1.332~895~943~373~395~49$
" \\ \sstrut {} {1.25 \jot} {1.25 \jot}
\+ "       "                      "
" ${\cal E}_{163}^{(3)} \; = \;  1.332~895~943~373~395~18$
" \\ \sstrut {} {1.25 \jot} {1.25 \jot}
\- " \- " \- " \- " \- " \\ \sstrut {} {1.25 \jot} {1.25 \jot}
\+ " 1.0   " $1.435~624~618~9$    " $1.435~624~619~1$
" ${\cal E}_{161}^{(3)} \; = \;  1.435~624~619~003~406$
" \\ \sstrut {} {1.25 \jot} {1.25 \jot}
\+ "       "                      "
" ${\cal E}_{162}^{(3)} \; = \;  1.435~624~619~003~405$
" \\ \sstrut {} {1.25 \jot} {1.25 \jot}
\+ "       "                      "
" ${\cal E}_{163}^{(3)} \; = \;  1.435~624~619~003~403$
" \\ \sstrut {} {1.25 \jot} {1.25 \jot}
\- " \- " \- " \- " \- " \\ \sstrut {} {1.25 \jot} {1.25 \jot}
\+ " 2.0   " $1.609~931~951~9$    " $1.609~931~952~4$
" ${\cal E}_{161}^{(3)} \; = \;  1.609~931~952~023~15$
" \\ \sstrut {} {1.25 \jot} {1.25 \jot}
\+ "       "                      "
" ${\cal E}_{162}^{(3)} \; = \;  1.609~931~952~023~14$
" \\ \sstrut {} {1.25 \jot} {1.25 \jot}
\+ "       "                      "
" ${\cal E}_{163}^{(3)} \; = \;  1.609~931~952~023~14$ "
 \\ \sstrut {} {1.25 \jot} {1.25 \jot}
\- " \- " \- " \- " \- " \\ \sstrut {} {1.25 \jot} {1.25 \jot}
\+ " 4.0   " $1.830~437~343~6$    " $1.830~437~344~4$
" ${\cal E}_{161}^{(3)} \; = \;  1.830~437~343~750~32$
" \\ \sstrut {} {1.25 \jot} {1.25 \jot}
\+ "       "                      "
" ${\cal E}_{162}^{(3)} \; = \;  1.830~437~343~750~29$ "
 \\ \sstrut {} {1.25 \jot} {1.25 \jot}
\+ "       "                      "
" ${\cal E}_{163}^{(3)} \; = \;  1.830~437~343~750~27$
" \\ \sstrut {} {1.25 \jot} {1.25 \jot}
\- " \- " \- " \- " \- " \\ \sstrut {} {1.25 \jot} {1.25 \jot}
\+ " 100   " $3.716~974~728$      " $3.716~974~732$
" ${\cal E}_{161}^{(3)} \; = \;  3.716~974~729~211~6$ "
\\ \sstrut {} {1.25 \jot} {1.25 \jot}
\+ "       "                      "
" ${\cal E}_{162}^{(3)} \; = \;  3.716~974~729~211~3$ "
\\ \sstrut {} {1.25 \jot} {1.25 \jot}
\+ "       "                      "
" ${\cal E}_{163}^{(3)} \; = \;  3.716~974~729~211~0$ "
\\ \sstrut {} {1.25 \jot} {1.25 \jot}
\- " \- " \- " \- " \- " \\ \sstrut {} {1.25 \jot} {1.25 \jot}
\+ " 400   " $5.188~358~853$      " $5.188~358~859$
" ${\cal E}_{161}^{(3)} \; = \;  5.188~358~854~443~9$ "
\\ \sstrut {} {1.25 \jot} {1.25 \jot}
\+ "       "                      "
" ${\cal E}_{162}^{(3)} \; = \;  5.188~358~854~443~4$
" \\ \sstrut {} {1.25 \jot} {1.25 \jot}
\+ "       "                      "
" ${\cal E}_{163}^{(3)} \; = \;  5.188~358~854~442~9$ "
\\ \sstrut {} {1.25 \jot} {1.25 \jot}
\- " \- " \- " \- " \- " \\ \sstrut {} {1.25 \jot} {1.25 \jot}
\+ " 2~000 " $7.701~738~363$      " $7.701~738~372$
" ${\cal E}_{161}^{(3)} \; = \;  7.701~738~364~619$
" \\  \sstrut {} {1.25 \jot} {1.25 \jot}
\+ "       "                      "
" ${\cal E}_{162}^{(3)} \; = \;  7.701~738~364~618$
" \\  \sstrut {} {1.25 \jot} {1.25 \jot}
\+ "       "                      "
" ${\cal E}_{163}^{(3)} \; = \;  7.701~738~364~617$
" \\  \sstrut {} {1.25 \jot} {1.25 \jot}
\- " \- " \- " \- " \- " \\ \sstrut {} {1.25 \jot} {1.25 \jot}
\+ " 16~000" $12.902~759~968$     " $12.902~759~984$
" ${\cal E}_{161}^{(3)} \; = \;  12.902~759~971~040$
" \\ \sstrut {} {1.25 \jot} {1.25 \jot}
\+ "       "                      "
" ${\cal E}_{162}^{(3)} \; = \;  12.902~759~971~037$
" \\ \sstrut {} {1.25 \jot} {1.25 \jot}
\+ "       "                      "
" ${\cal E}_{163}^{(3)} \; = \;  12.902~759~971~037$
" \\ \sstrut {} {1.25 \jot} {1.25 \jot}
\- " \- " \- " \- " \- " \\ \sstrut {} {1.25 \jot} {1.25 \jot}
\+ " 40~000" $16.211~718~261$     " $16.211~718~281$
" ${\cal E}_{161}^{(3)} \; = \;  16.211~718~264~770$ "
\\ \sstrut {} {1.25 \jot} {1.25 \jot}
\+ "       "                      "
" ${\cal E}_{162}^{(3)} \; = \;  16.211~718~264~768$ "
\\ \sstrut {} {1.25 \jot} {1.25 \jot}
\+ "       "                      "
" ${\cal E}_{163}^{(3)} \; = \;  16.211~718~264~766$
" \\ \sstrut {} {1.25 \jot} {1.25 \jot}
\= " \= " \= " \= " \= " \\ \sstrut {} {1 \jot} {1 \jot}

\endTabelle

\neueSeite

\beginTabelle % [to \kolumnenbreite]
\beginFormat & \links \endFormat
\+ " \links {\bf Tabelle 10-11} \@ \@ \@ " \\
\+ " \links {Anharmonischer Oszillator mit einer
$\hat{x}^8$-Anharmonizit\"at} \@ \@ \@ " \\
\+ " \links {Summation der St\"orungsreihe (10.7-6) f\"ur die
Grundzustandsenergie $E^{(4)} (\beta)$} \@ \@ \@ " \\
\sstrut {} {1.5 \jot} {1.5 \jot}
\= " \= " \= " \= " \=  " \\ \sstrut {} {1 \jot} {1 \jot}
\+ " $\beta$ " {$\quad E_{<}^{(4)} (\beta)$}
" {$\quad E_{>}^{(4)} (\beta)$}
" {$\qquad \qquad {\cal E}_{n}^{(4)} (\beta)$}
" \\ \sstrut {} {1.5 \jot} {1.5 \jot}
\= " \= " \= " \= " \= " \\ \sstrut {} {1 \jot} {1 \jot}
\+ " 0.2    " 1.241~027~88  " 1.241~027~94
" ${\cal E}_{135}^{(4)} \; = \; 1.241~027~811$
" \\ \sstrut {} {1.25 \jot} {1.25 \jot}
\+ "        "               "
" ${\cal E}_{136}^{(4)} \; = \; 1.241~027~814$
" \\ \sstrut {} {1.25 \jot} {1.25 \jot}
\+ "        "               "
" ${\cal E}_{137}^{(4)} \; = \; 1.241~027~818$
" \\ \sstrut {} {1.25 \jot} {1.25 \jot}
\- " \- " \- " \- " \- " \\ \sstrut {} {1.25 \jot} {1.25 \jot}
\+ " 0.6    " 1.397~708~69  " 1.397~708~86
" ${\cal E}_{135}^{(4)} \; = \; 1.397~708~36$
" \\ \sstrut {} {1.25 \jot} {1.25 \jot}
\+ "        "               "
" ${\cal E}_{136}^{(4)} \; = \; 1.397~708~37$
" \\ \sstrut {} {1.25 \jot} {1.25 \jot}
\+ "        "               "
" ${\cal E}_{137}^{(4)} \; = \; 1.397~708~38$
" \\ \sstrut {} {1.25 \jot} {1.25 \jot}
\- " \- " \- " \- " \- " \\ \sstrut {} {1.25 \jot} {1.25 \jot}
\+ " 1.0    " 1.491~019~79  " 1.491~020~04
" ${\cal E}_{135}^{(4)} \; = \; 1.491~019~23$
" \\ \sstrut {} {1.25 \jot} {1.25 \jot}
\+ "        "               "
" ${\cal E}_{136}^{(4)} \; = \; 1.491~019~25$
" \\ \sstrut {} {1.25 \jot} {1.25 \jot}
\+ "        "               "
" ${\cal E}_{137}^{(4)} \; = \; 1.491~019~27$
" \\ \sstrut {} {1.25 \jot} {1.25 \jot}
\- " \- " \- " \- " \- " \\ \sstrut {} {1.25 \jot} {1.25 \jot}
\+ " 2.0    " 1.641~370~19  " 1.641~370~57
" ${\cal E}_{135}^{(4)} \; = \; 1.641~369~18$
" \\ \sstrut {} {1.25 \jot} {1.25 \jot}
\+ "        "               "
" ${\cal E}_{136}^{(4)} \; = \; 1.641~369~22$
" \\ \sstrut {} {1.25 \jot} {1.25 \jot}
\+ "        "               "
" ${\cal E}_{137}^{(4)} \; = \; 1.641~369~25$
" \\ \sstrut {} {1.25 \jot} {1.25 \jot}
\- " \- " \- " \- " \- " \\ \sstrut {} {1.25 \jot} {1.25 \jot}
\+ " 4.0    " 1.822~179~6   " 1.822~180~2
" ${\cal E}_{135}^{(4)} \; = \; 1.822~178~02$
" \\ \sstrut {} {1.25 \jot} {1.25 \jot}
\+ "        "               "
" ${\cal E}_{136}^{(4)} \; = \; 1.822~178~07$
" \\ \sstrut {} {1.25 \jot} {1.25 \jot}
\+ "        "               "
" ${\cal E}_{137}^{(4)} \; = \; 1.822~178~13$
" \\ \sstrut {} {1.25 \jot} {1.25 \jot}
\- " \- " \- " \- " \- " \\ \sstrut {} {1.25 \jot} {1.25 \jot}
\+ " 100    " 3.188~653~6   " 3.188~655~1
" ${\cal E}_{135}^{(4)} \; = \; 3.188~647~25$
" \\ \sstrut {} {1.25 \jot} {1.25 \jot}
\+ "        "               "
" ${\cal E}_{136}^{(4)} \; = \; 3.188~647~45$
" \\ \sstrut {} {1.25 \jot} {1.25 \jot}
\+ "        "               "
" ${\cal E}_{137}^{(4)} \; = \; 3.188~647~64$
" \\ \sstrut {} {1.25 \jot} {1.25 \jot}
\- " \- " \- " \- " \- " \\ \sstrut {} {1.25 \jot} {1.25 \jot}
\+ " 400    " 4.146~187~5   " 4.146~189~6
" ${\cal E}_{135}^{(4)} \; = \; 4.146~178~1$
" \\ \sstrut {} {1.25 \jot} {1.25 \jot}
\+ "        "               "
" ${\cal E}_{136}^{(4)} \; = \; 4.146~178~4$
" \\ \sstrut {} {1.25 \jot} {1.25 \jot}
\+ "        "               "
" ${\cal E}_{137}^{(4)} \; = \; 4.146~178~7$
" \\ \sstrut {} {1.25 \jot} {1.25 \jot}
\- " \- " \- " \- " \- " \\ \sstrut {} {1.25 \jot} {1.25 \jot}
\+ " 2~000  " 5.666~202~3   " 5.666~205~4
" ${\cal E}_{135}^{(4)} \; = \; 5.666~188~4$
" \\ \sstrut {} {1.25 \jot} {1.25 \jot}
\+ "        "               "
" ${\cal E}_{136}^{(4)} \; = \; 5.666~188~8$
" \\ \sstrut {} {1.25 \jot} {1.25 \jot}
\+ "        "               "
" ${\cal E}_{137}^{(4)} \; = \; 5.666~189~2$
" \\ \sstrut {} {1.25 \jot} {1.25 \jot}
\- " \- " \- " \- " \- " \\ \sstrut {} {1.25 \jot} {1.25 \jot}
\+ " 16~000 " 8.536~649     " 8.536~653
" ${\cal E}_{135}^{(4)} \; = \; 8.536~626$
" \\ \sstrut {} {1.25 \jot} {1.25 \jot}
\+ "        "               "
" ${\cal E}_{136}^{(4)} \; = \; 8.536~627$
" \\ \sstrut {} {1.25 \jot} {1.25 \jot}
\+ "        "               "
" ${\cal E}_{137}^{(4)} \; = \; 8.536~628$
" \\ \sstrut {} {1.25 \jot} {1.25 \jot}
\- " \- " \- " \- " \- " \\ \sstrut {} {1.25 \jot} {1.25 \jot}
\+ " 40~000 " 10.238~865    " 10.238~871
" ${\cal E}_{135}^{(4)} \; = \; 10.238~838$
" \\ \sstrut {} {1.25 \jot} {1.25 \jot}
\+ "        "               "
" ${\cal E}_{136}^{(4)} \; = \; 10.238~839$ "
\\ \sstrut {} {1.25 \jot} {1.25 \jot}
\+ "        "               "
" ${\cal E}_{137}^{(4)} \; = \; 10.238~840$ "
\\ \sstrut {} {1.25 \jot} {1.25 \jot}
\= " \= " \= " \= " \= " \\ \sstrut {} {1 \jot} {1 \jot}

\endTabelle

\neueSeite

In den Tabellen 10-9, 10-10 und 10-11 werden die Approximationen
$$
{\cal E}_n^{(m)} (\beta) \; = \; (1 - \kappa)^{- 1/2} \, + \,
\kappa \, {\delta}_n^{(0)} (1, \sigma_0^{(m)} (\kappa)) \, ,
\tag
$$
die durch Anwendung des verallgemeinerten Summationsprozesses
${\delta}_k^{(n)} (\zeta, s_n)$, Gl. (5.4-13), auf die Partialsummen
(10.7-8) der St\"orungsreihe (10.7-6) entstehen, f\"ur $m = 2, 3, 4$ als
Funktion der Kopplungskonstante $\beta$ aufgelistet. Alle Rechnungen
wurden in MAPLE mit einer Genauigkeit von 300 Dezimalstellen
durchgef\"uhrt. In allen F\"allen wurde die maximal m\"ogliche Zahl von
renormierten St\"orungs\-theorie\-ko\-effizi\-enten $c_{n}^{(m)}$
verwendet ($m = 2$: $n \le 200$; $m = 3$: $n \le 165$; $m = 4$: $n \le
139$). In Tabellen 10-9, 10-10 und 10-11 wurden die gleichen
$\beta$-Werte verwendet wie in Tables \Roemisch{1} - \Roemisch{3} von
Vinette und {\v C}{\'\i}{\v z}ek [1991], wo obere und untere Schranken
$E_{>}^{(m)} (\beta)$ und $E_{<}^{(m)} (\beta)$ f\"ur die
Grundzustandsenergie $E^{(m)} (\beta)$ mit $m = 2, 3, 4$ angegeben
werden.

F\"ur einen gegebenen Wert von $\beta$ wurde die entsprechende renormierte
Kopplungskonstante $\kappa$ immer gem\"a{\ss} Gl.~(10.4-12) numerisch mit
Hilfe des MAPLE-Kommandos {\it fsolve\/} [Char, Geddes, Gonnet, Leong,
Monagan und Watt 1991b, S. 97] berechnet.

Die Ergebnisse in Tabelle 10-9 f\"ur die Grundzustandsenergie des
anharmonischen Oszillators mit einer $\hat{x}^4$-Anharmonizit\"at sind
deutlich besser als die oberen und unteren Schranken $E_{>}^{(2)}
(\beta)$ und $E_{<}^{(3)} (\beta)$ in Table \Roemisch{1} von Vinette und
{\v C}{\'\i}{\v z}ek [1991]. Weitere Rechnungen ergaben, da{\ss} die
Ergebnisse, die man unter Verwendung der renormierten
St\"orungs\-theorie\-ko\-effizi\-enten $c_{n}^{(2)}$ mit $n \le 200$ gem\"a{\ss}
Gl. (10.7-10) erhalten kann, auch deutlich genauer sind als die
ebenfalls sehr genauen Werte von Schiffrer und Stanzial [1985, Table
\Roemisch{1}].

Es besteht kein Zweifel, da{\ss} es mit der von Vinette und {\v C}{\'\i}{\v
z}ek [1991] verwendeten Methode der inneren Projektion m\"oglich sein
sollte, die sehr genauen Werte f\"ur die Grundzustandsenergie $E^{(2)}
(\beta)$ eines anharmonischen Oszillators mit einer
$\hat{x}^4$-Anharmonizit\"at in Tabelle 10-9 noch zu \"ubertreffen.
Abgesehen von dem extrem genauen Wert (10.4-33a) f\"ur $k_2$ von Vinette
und {\v C}{\'\i}{\v z}ek [1991, Gl. (66)] d\"urften die
Summationsergebnisse in Tabelle 10-9 aber die genauesten Werte sein, die
zur Zeit f\"ur $E^{(2)} (\beta)$ bekannt sind.

In den Tabellen 10-10 und 10-11 werden zus\"atzlich zu den in Gl.
(10.7-10) definierten Approximationen ${\cal E}_{n}^{(m)} (\beta)$ auch
noch die von Vinette und {\v C}{\'\i}{\v z}ek [1991, Tables \Roemisch{2}
und \Roemisch{3}] berechneten oberen und unteren Schranken $E_{>}^{(m)}
(\beta)$ und $E_{<}^{(m)} (\beta)$ f\"ur $m = 3, 4$ angegeben.

Die Summationsergebnisse in Tabelle 10-10 f\"ur die Grundzustandsenergie
des anharmonischen Oszillators mit einer $\hat{x}^6$-Anharmonizit\"at, die
man unter Verwendung der renormierten
St\"o\-rungs\-theorie\-ko\-effizi\-enten $c_{n}^{(3)}$ mit $n \le 165$
gem\"a{\ss} Gl. (10.7-10) erh\"alt, sind ebenfalls deutlich besser als die
oberen und unteren Schranken $E_{>}^{(3)} (\beta)$ und $E_{<}^{(4)}
(\beta)$ in Table \Roemisch{2} von Vinette und {\v C}{\'\i}{\v z}ek
[1991], und sie d\"urften mit Ausnahme des extrem genauen Wertes
(10.4-33b) f\"ur $k_3$ von Vinette und {\v C}{\'\i}{\v z}ek [1991, Gl.
(69)] die genauesten Werte sein, die zur Zeit f\"ur $E^{(3)} (\beta)$
bekannt sind.

Die Ergebnisse in Tabelle 10-11 f\"ur die Grundzustandsenergie des
anharmonischen Oszillators mit einer $\hat{x}^8$-Anharmonizit\"at, die man
unter Verwendung der renormierten St\"orungs\-theorie\-ko\-effizi\-enten
$c_{n}^{(4)}$ mit $n \le 139$ gem\"a{\ss} Gl. (10.7-10) erh\"alt, sind
schlechter als die oberen und unteren Schranken $E_{>}^{(4)} (\beta)$
und $E_{<}^{(4)} (\beta)$ in Table \Roemisch{3} von Vinette und {\v
C}{\'\i}{\v z}ek [1991]. In Anbetracht der extrem starken Divergenz der
St\"orungsreihe (10.7-6) f\"ur die Grundzustandsenergie $E^{(4)} (\beta)$
sind die Summationsergebnisse aber trotzdem vergleichsweise gut.

Die Ergebnisse in Tabelle 10-11 sind kein definitiver Beweis, da{\ss}
${\delta}_k^{(n)} (\zeta, s_n)$ tats\"achlich in der Lage ist, die
renormierte St\"orungsreihe (10.7-6) f\"ur die Grundzustandsenergie $E^{(4)}
(\beta)$, die eine starke asymptotische Reihe der Ordnung 3 ist, zu
summieren. Es w\"are deswegen sicherlich interessant, diese Rechnungen zu
wiederholen, sobald mehr als nur die
St\"o\-rungs\-theo\-rie\-ko\-effi\-zi\-enten $c_{n}^{(4)}$ mit $n \le 139$
zur Verf\"ugung stehen. Dazu w\"urde man allerdings einen Rechner mit
deutlich mehr als 64 MB RAM ben\"otigen.

In Anbetracht der Tatsache, da{\ss} die Folge $\bigl\{ \delta_{\nu}^{(0)}
\bigr\}$ mit $\nu \le 137$ in Tabelle 10-11 f\"ur alle betrachteten Werte
der Kopplungskonstante $\beta$ Ergebnisse produziert, die nicht sehr
stark von den oberen und unteren Schranken $E_{>}^{(4)} (\beta)$ und
$E_{<}^{(4)} (\beta)$ in Table \Roemisch{3} von Vinette und {\v
C}{\'\i}{\v z}ek [1991] abweichen, ist es relativ wahrscheinlich, da{\ss}
${\delta}_k^{(n)} (\zeta, s_n)$, Gl. (5.4-13), tats\"achlich die
renormierte St\"orungsreihe (10.7-6) auch f\"ur $m = 4$ summieren kann.

Wie schon fr\"uher erw\"ahnt, k\"onnen Pad\'e-Approximationen nur starke
asymptotische Reihen der Ordnung $k \le 2$ summieren. Da
Pad\'e-Approximationen rationale Funktionen sind, ist es ein relativ
naheliegender Analogieschlu{\ss}, da{\ss} auch andere rationale Funktionen nur
starke asymptotische Reihen der Ordnung $k \le 2$ summieren k\"onnen. Die
Ergebnisse in Tabelle 10-11 zeigen aber, da{\ss} dieser Analogieschlu{\ss}
wahrscheinlich falsch ist, und da{\ss} die Beschr\"ankung auf starke
asymptotische Reihen der Ordnung $k \le 2$ nur f\"ur Pad\'e-Approximationen
typisch ist und nicht allgemein f\"ur rationale Approximationen gilt. Eine
definitive Kl\"arung dieser Frage durch einen expliziten Beweis, da{\ss} der
verallgemeinerte Summationsproze{\ss} ${\delta}_k^{(n)} (\zeta, s_n)$, Gl.
(5.4-13), der die Partialsummen einer formalen Potenzreihe gem\"a{\ss} Gl.
(5.7-4) in rationale Funktionen transformiert, in der Lage ist, starke
asymptotische Reihen der Ordnung $k > 2$ zu summieren, w\"are deswegen
\"au{\ss}erst w\"unschenswert.

Aus Gl.~(10.4-12) folgt, da{\ss} f\"ur jedes $\beta \in [0, \infty)$ die
entsprechende renormierte Kopplungskonstante $\kappa$ die Ungleichung $0
\le \kappa < 1$ erf\"ullt. Demzufolge ist -- wie schon mehrfach erw\"ahnt --
die St\"orungsreihe (10.4-31) f\"ur die Grenzwerte unendlicher Kopplung
$k_m$ das schwierigste Summationsproblem, das im Zusammenhang mit der
renormierten St\"orungsreihe (10.4-20) auftreten kann. Es ist deswegen
sicherlich von Interesse, herauszufinden, wie genau man die Grenzwerte
unendlicher Kopplung $k_m$ mit Hilfe des verallgemeinerten
Summationsprozesses ${\delta}_k^{(n)} (\zeta, s_n)$ durch Summation
berechnen kann, wenn man die Summation in MAPLE mit einer deutlich
h\"oheren Genauigkeit durchf\"uhrt und dabei alle verf\"ugbaren renormierten
St\"o\-rungs\-theorie\-ko\-effizi\-enten $c_{n}^{(m)}$ verwendet.

In den Tabellen 10-12, 10-13 und 10-14 [Weniger, {\v C}{\'\i}{\v z}ek
und Vinette 1991, Tables \Roemisch{1} - \Roemisch{3}] wird der
verallgemeinerte Summationsproze{\ss} ${\delta}_k^{(n)} (\zeta, s_n)$ zur
Summation der St\"orungsreihe
$$
\beginAligntags
" k_m " \; = \; " [B_m]^{1/(m+1)} \,
\Bigl\{ 1 \, + \, \Delta E_R^{(m)} (1) \Bigr\} \\ \tag
" " \; = \; " [B_m]^{1/(m+1)} \,
\Bigl\{ 1 \, + \, \sum_{\nu=0}^{\infty} \, c_{\nu + 1}^{(m)} \Bigr\}
\\ \tag
\endAligntags
$$
f\"ur die Grenzf\"alle unendlicher Kopplung verwendet. Dabei werden die
Approximationen
$$
k_m^{(n)} \; = \; [B_m]^{1/(m+1)} \,
\Bigl\{ 1 \, + \, \delta_n^{(0)} \bigl(\Sigma_0^{(m)} \bigr) \Bigr\}
\tag
$$
f\"ur verschiedene Werte von $n$ und f\"ur $m = 2, 3, 4$ aufgelistet. Als
Eingabedaten wurden die Partialsummen (10.7-9) verwendet. Alle
Rechnungen wurden in MAPLE mit einer Genauigkeit von 300 Dezimalstellen
durchgef\"uhrt. Au{\ss}erdem wurde immer die maximal m\"ogliche Zahl von
renormierten St\"orungs\-theorie\-ko\-effizi\-enten $c_{n}^{(m)}$
verwendet ($m = 2$: $n \le 200$; $m = 3$: $n \le 165$; $m = 4$: $n \le
139$).

\beginFloat

\medskip

\beginTabelle % [to \kolumnenbreite]
\beginFormat \rechts " \rechts " \endFormat
\+ " \links {\bf Tabelle 10-12} \@ " \\
\+ " \links {Anharmonischer Oszillator mit einer $\hat{x}^4$-Anharmonizit\"at}
\@ " \\
\+ " \links {Summation der St\"orungsreihe (10.7-12) f\"ur den Grenzfall
unendlicher Kopplung $k_2$} \@ " \\
\sstrut {} {1.5 \jot} {1.5 \jot}
\- " \- " \- " \\ \sstrut {} {1 \jot} {1 \jot}
\+ " $n$ " \mitte{$k_{2}^{(n)}$}
" \\ \sstrut {} {1.5 \jot} {1.5 \jot}
\- " \- " \-  " \\ \sstrut {} {1 \jot} {1 \jot}
\+ " 10  " 1.060~362~388~851~781~684~290~023~000~982~889~433~242 " \\
\+ " 30  " 1.060~362~090~481~907~462~133~112~895~036~642~497~314 " \\
\+ " 50  " 1.060~362~090~484~183~251~949~681~617~798~800~941~942 " \\
\+ " 75  " 1.060~362~090~484~182~899~629~961~495~541~518~481~673 " \\
\+ " 100 " 1.060~362~090~484~182~899~647~047~435~874~856~320~511 " \\
\+ " 125 " 1.060~362~090~484~182~899~647~046~016~970~889~226~728 " \\
\+ " 150 " 1.060~362~090~484~182~899~647~046~016~692~256~571~685 " \\
\+ " 175 " 1.060~362~090~484~182~899~647~046~016~692~663~692~532 " \\
\+ " 194 " 1.060~362~090~484~182~899~647~046~016~692~663~547~291 " \\
\+ " 195 " 1.060~362~090~484~182~899~647~046~016~692~663~546~840 " \\
\+ " 196 " 1.060~362~090~484~182~899~647~046~016~692~663~546~498 " \\
\+ " 197 " 1.060~362~090~484~182~899~647~046~016~692~663~546~240 " \\
\+ " 198 " 1.060~362~090~484~182~899~647~046~016~692~663~546~047 " \\
\- " \- " \- " \\ \sstrut {} {1 \jot} {1 \jot}

\endTabelle

\medskip

\endFloat

Ein Vergleich der Summationsergebnisse in Tabelle 10-12 mit dem extrem
genauen Ergebnis (10.4-33a) von Vinette und {\v C}{\'\i}{\v z}ek [1991,
Gl. (66)] f\"ur $k_2$ ergibt, da{\ss} die Transformationen $\bigl\{
\delta_{n}^{(0)} \bigr\}$ eine f\"ur Summationsverfahren erstaunliche
Genauigkeit von 35 Dezimalstellen erreichen. Unter den gleichen
Bedingungen reproduzieren Pad\'e-Approximationen nur 21 Dezimalstellen.

Ein Vergleich der Summationsergebnisse in Tabelle 10-13 mit dem extrem
genauen Ergebnis (10.4-33b) von Vinette und {\v C}{\'\i}{\v z}ek [1991,
Gl. (69)] f\"ur $k_3$ ergibt, da{\ss} die Transformationen $\bigl\{
\delta_{n}^{(0)} \bigr\}$ eine Genauigkeit von 11 Dezimalstellen
erreichen. Die Summationsergebnisse f\"ur $k_3$ sind deutlich schlechter
als die analogen Ergebnisse f\"ur $k_2$ in Tabelle 10-12. Man darf hier
aber nicht vergessen, da{\ss} die St\"orungsreihe (10.7-12) f\"ur $k_3$ ein
wesentlich anspruchsvolleres Summationsproblem darstellt als f\"ur $k_2$,
da die Koeffizienten $c_{n}^{(3)}$ wesentlich schneller divergieren als
die Koeffizienten $c_{n}^{(2)}$. Hinzu kommt, da{\ss} f\"ur $m = 2$ die
Koeffizienten mit $n \le 200$ zur Verf\"ugung standen, f\"ur $m = 3$ dagegen
nur die Koeffizienten mit $n \le 165$. In Anbetracht der starken
Divergenz der St\"o\-rungs\-rei\-he (10.7-12) f\"ur $m = 3$ ist eine
Genauigkeit von 11 Stellen sicherlich ein sehr gutes Ergebnis. Unter den
gleichen Bedingungen reproduzieren Pad\'e-Approximationen nur 3
Dezimalstellen.

\beginFloat

\medskip

\beginTabelle % [to \kolumnenbreite]
\beginFormat \rechts " \rechts " \endFormat
\+ " \links {\bf Tabelle 10-13} \@ " \\
\+ " \links {Anharmonischer Oszillator mit einer $\hat{x}^6$-Anharmonizit\"at}
\@ " \\
\+ " \links {Summation der St\"orungsreihe (10.7-12) f\"ur den Grenzfall
unendlicher Kopplung $k_3$} \@ " \\
\sstrut {} {1.5 \jot} {1.5 \jot}
\- " \- " \-  " \\ \sstrut {} {1 \jot} {1 \jot}
\+ " $n$ " \rechts{$k_{3}^{(n)}$ \qquad \qquad \qquad \quad}
" \\ \sstrut {} {1.5 \jot} {1.5 \jot}
\- " \- " \-  " \\ \sstrut {} {1 \jot} {1 \jot}
\+ " 10  " 1.145~406~742~444~731 " \\
\+ " 30  " 1.144~801~650~628~644 " \\
\+ " 50  " 1.144~802~299~173~531 " \\
\+ " 75  " 1.144~802~450~126~817 " \\
\+ " 100 " 1.144~802~453~942~561 " \\
\+ " 125 " 1.144~802~453~834~731 " \\
\+ " 150 " 1.144~802~453~801~448 " \\
\+ " 159 " 1.144~802~453~798~862 " \\
\+ " 160 " 1.144~802~453~798~684 " \\
\+ " 161 " 1.144~802~453~798~521 " \\
\+ " 162 " 1.144~802~453~798~373 " \\
\+ " 163 " 1.144~802~453~798~239 " \\
\- " \- " \- " \\ \sstrut {} {1 \jot} {1 \jot}

\endTabelle

\medskip

\endFloat

Ein Vergleich der Summationsergebnisse in Tabelle 10-14 mit dem extrem
genauen Ergebnis (10.4-33c) von Vinette und {\v C}{\'\i}{\v z}ek [1991,
Gl. (71)] f\"ur $k_4$ ergibt, da{\ss} die Transformationen $\bigl\{
\delta_{n}^{(0)} \bigr\}$ eine Genauigkeit von 5 Dezimalstellen
erreichen. Pad\'e-Approximationen f\"ur $k_4$ konvergieren nicht.

Die Summationsergebnisse f\"ur $k_4$ sind nicht so gut wie die f\"ur $k_2$
und $k_3$. Man sollte hier aber nicht vergessen, da{\ss} die Summation einer
divergenten Reihe, deren Terme im wesentlichen wie $(3 n)!/n^{1/2}$
divergieren, ein extrem schwieriges Problem ist. Au{\ss}erdem sind f\"ur $m =
4$ nur die St\"orungs\-theorie\-ko\-effizi\-enten mit $n \le 139$
verf\"ugbar.

\beginFloat

\medskip

\beginTabelle % [to \kolumnenbreite]
\beginFormat \rechts " \rechts " \endFormat
\+ " \links {\bf Tabelle 10-14} \@ " \\
\+ " \links {Anharmonischer Oszillator mit einer $\hat{x}^8$-Anharmonizit\"at}
\@ " \\
\+ " \links {Summation der St\"orungsreihe (10.7-12) f\"ur den Grenzfall
unendlicher Kopplung $k_4$} \@ " \\ \sstrut {} {1.5 \jot} {1.5 \jot}
\- " \- " \-  " \\ \sstrut {} {1 \jot} {1 \jot}
\+ " $n$ " \rechts{$k_{4}^{(n)}$ \qquad \quad}
" \\ \sstrut {} {1.5 \jot} {1.5 \jot}
\- " \- " \-  " \\ \sstrut {} {1 \jot} {1 \jot}
\+ " 10  " 1.239~331~145 " \\
\+ " 30  " 1.226~787~840 " \\
\+ " 50  " 1.225~925~653 " \\
\+ " 75  " 1.225~815~816 " \\
\+ " 100 " 1.225~811~764 " \\
\+ " 125 " 1.225~815~298 " \\
\+ " 132 " 1.225~816~126 " \\
\+ " 133 " 1.225~816~234 " \\
\+ " 134 " 1.225~816~340 " \\
\+ " 135 " 1.225~816~447 " \\
\+ " 136 " 1.225~816~545 " \\
\+ " 137 " 1.225~816~643 " \\
\- " \- " \- " \\ \sstrut {} {1 \jot} {1 \jot}

\endTabelle

\medskip

\endFloat

Sowohl die Ergebnisse in den Tabellen 10-6, 10-7 und 10-8 als auch
andere Rechnungen zeigen ganz deutlich, da{\ss} der verallgemeinerte
Summationsproze{\ss} ${\delta}_k^{(n)} (\zeta, s_n)$ die in diesem Abschnitt
behandelten St\"orungsreihen anharmonischer Oszillatoren wesentlich
effizienter summiert als die durch den Wynnschen $\epsilon$-Algorithmus
gem\"a{\ss} Gl. (4.4-8) berechneten Pad\'e-Approximationen $\epsilon_{2 n}^{(0)}
= [n / n]$ und $\epsilon_{2 n}^{(1)} = [n + 1 / n]$. Meiner Meinung nach
ist die \"Uberlegenheit des verallgemeinerten Summationsprozesses ${\cal
S}_k^{(n)} (\zeta, s_n, \omega_n)$, Gl. (5.4-6), und seiner Varianten
bei der Summation hochgradig divergenter Reihen eine Konsequenz der
Tatsache, da{\ss} als Eingabedaten nicht nur die Partialsummen $\Seqn s$
einer unendlichen Reihe verwendet werden, sondern auch explizite
Restsummenabsch\"atzungen $\Seqn \omega$.

Man sollte sich aber vor der irrigen Vorstellung h\"uten, da{\ss} die
Verwendung expliziter Restsummenabsch\"atzungen $\Seqn \omega$ {\it
automatisch\/} zu besseren Summationsergebnissen f\"uhrt. Der
verallgemeinerte Summationsproze{\ss} ${\delta}_k^{(n)} (\zeta, s_n)$
verwendet die Partialsummen $s_n$, $s_{n+1}$, $\ldots$, $s_{n+k}$, und
die Restsummenabsch\"atzungen $\omega_n = \Delta s_n$, $\omega_{n+1} =
\Delta s_{n+1}$, $\ldots$, $\omega_{n+k} = \Delta s_{n+k}$, um eine
Approximation des tats\"achlichen Abbruchfehlers $r_n = s_n - s$ zu
konstruieren, die dann aus den Partialsummen $s_n$ eliminiert werden
kann. Dadurch erh\"alt man eine Approximation des (verallgemeinerten)
Grenzwertes $s$ der Folge $\Seqn s$.

Wenn die Eingabedaten $s_n$ die Partialsummen
$$
f_n (z) \; = \; \sum_{\nu=0}^{n} \, \gamma_{\nu} \, z^{\nu}
\tag
$$
einer formalen Potenzreihe $f (z)$ sind, wird implizit bei der
Berechnung der Transformationen ${\delta}_k^{(n)} \bigl( \zeta, f_n (z)
\bigr)$ angenommen, da{\ss} der Abbruchfehler $f_n (z) - f (z)$ der formalen
Potenzreihe approximiert werden kann durch den ersten Term, der nicht in
der Partialsumme $f_n (z)$ enthalten ist, multipliziert mit eine
endlichen Summe, die Pochhammersymbole und unspezifizierte Koeffizienten
$c_0$, $c_1$, $\ldots$, $c_{k-1}$ enth\"alt:
$$
f_n (z) \, - \, f (z) \; \approx \;
\gamma_{n+1} z^{n+1} \, \sum_{j=0}^{k-1} \, c_j / (\zeta + n)_j \, .
\tag
$$
Wenn der Term $\gamma_{n+1} z^{n+1}$ schon f\"ur kleine oder mittelgro{\ss}e
Werte von $n$ eine halbwegs gute Absch\"atzung des tats\"achlichen
Abbruchfehlers $f_n (z) - f (z)$ liefert, dann sollte die rechte Seite
von Gl. (10.7-15) den Abbruchfehler sehr genau approximieren k\"onnen, und
${\delta}_k^{(n)} \bigl( \zeta, f_n (z) \bigr)$ sollte sehr gute
Summationsergebnisse liefern.

Ungl\"ucklicherweise kann man aber nicht davon ausgehen, da{\ss} die
Koeffizienten $\gamma_n$ einer formalen Potenzreihe $f (z)$ schon f\"ur
kleine oder mittelgro{\ss}e Indizes $n$ ein regul\"ares Verhalten zeigen,
selbst wenn die $\gamma_n$ f\"ur $n \to \infty$ ein wohldefiniertes
asymptotisches Verhalten besitzen. In einem solchen Fall wird der Term
$\gamma_{n+1} z^{n+1}$ erst f\"ur sehr gro{\ss}e Werte von $n$ eine halbwegs
gute Absch\"atzung des Abbruchfehlers $f_n (z) - f (z)$ liefern. Deswegen
mu{\ss} man davon ausgehen, da{\ss} die Effizienz des verallgemeinerten
Summationsprozesses ${\delta}_k^{(n)} \bigl( \zeta, f_n (z) \bigr)$ sehr
stark von dem Verhalten der Koeffizienten $\gamma_n$ abh\"angt. Wenn die
Koeffizienten $\gamma_n$ sich schon f\"ur kleine Werte von $n$ so
verhalten, wie es ihrem asymptotischen Verhalten f\"ur $n \to \infty$
entspricht, kann man erwarten, da{\ss} ${\delta}_k^{(n)} \bigl( \zeta, f_n
(z) \bigr)$ sehr gute Ergebnisse liefert. Wenn dagegen die Koeffizienten
$\gamma_n$ f\"ur kleine und mittelgro{\ss}e Werte von $n$ ein irregul\"ares
Verhalten zeigen, werden die Restsummenabsch\"atzungen $\omega_{n+j} =
\gamma_{n+j+1} z^{n+j+1}$ mit $0 \le j \le k$ keine guten Absch\"atzungen
der tats\"achlichen Abbruchfehler liefern. Man kann dann nicht erwarten,
da{\ss} ein verallgemeinerter Summationsproze{\ss} wie ${\delta}_k^{(n)} \bigl(
\zeta, f_n (z) \bigr)$, dessen Effizienz sehr stark von den verwendeten
Restsummenabsch\"atzungen abh\"angt, besonders gute Ergebnisse liefert.

In Tabelle 10-1 sieht man, da{\ss} die Koeffizienten $c_{n}^{(2)}$ der
renormierten St\"orungsreihe (10.4-20) f\"ur den anharmonischen Oszillator
mit einer $\hat{x}^4$-Anharmonizit\"at erst f\"ur $n \ge 10$ betragsm\"a{\ss}ig so
wachsen, wie man es aufgrund der asymptotischen Absch\"atzung (10.4-23a)
vermuten w\"urde. In Tabellen 10-2 und 10-3 sieht man, da{\ss} die
Koeffizienten $c_n^{(3)}$ und $c_n^{(4)}$ ein \"ahnlich irregul\"ares
Verhalten zeigen, da{\ss} allerdings weniger stark ausgepr\"agt ist.

Da diese Irregularit\"aten der Koeffizienten $c_n^{(m)}$ f\"ur kleinere
Werte von $n$ zu einem irregul\"aren Verhalten der Restsummenabsch\"atzungen
f\"uhren, ist es sicherlich von Interesse, den Einflu{\ss} dieser
Irregularit\"aten auf die Konvergenz der Transformationen $\bigl\{
\delta_{\nu}^{(0)} \bigr\}$ am Beispiel der Grenzf\"alle unendlicher
Kopplung $k_m$ genauer zu untersuchen. F\"ur $\ell = 0, 1, 2, \ldots$
werden deswegen die Partialsummen
$$
\Sigma_{n, \ell}^{(m)} \; = \; \sum_{\nu=0}^{n} \, c_{\nu+\ell}^{(m)}
\tag
$$
als Eingabedaten verwendet. Das bedeutet, da{\ss} die ersten $\ell$ Terme
$c_0^{(m)}$, $c_{1}^{(m)}$, $\ldots$, $c_{\ell-1}^{(m)}$ beim
Summationsproze{\ss} \"ubersprungen werden, und da{\ss} nur die verbleibende
renormierte St\"orungsreihe
$$
k_m \, - \, [B_m]^{1/(m+1)} \,
\Bigl\{ 1 \, + \sum_{\lambda=0}^{\ell-1} \, c_{\lambda}^{(m)} \Bigr\}
\; = \; [B_m]^{1/(m+1)} \, \sum_{\nu=0}^{\infty} \, c_{\nu+\ell}^{(m)}
\tag
$$
transformiert wird. In den folgenden Tabellen 10-15, 10-16 und 10-17
werden f\"ur $m = 2, 3, 4$ die Approximationen
$$
k_m^{(n, \ell)} \; = \; [B_m]^{1/(m+1)} \,
\biggl\{ \sum_{\lambda=0}^{\ell-1} \, c_{\lambda}^{(m)} \, + \,
\delta_{n-\ell}^{(0)} \bigl(1, \Sigma_{0, \ell}^{(m)} \bigr)
\biggr\}
\tag
$$
aufgelistet. Dabei wird $n$ immer so gro{\ss} wie m\"oglich gew\"ahlt. F\"ur $m
=2$ gilt also $n=199$, f\"ur $m = 3$ gilt $n = 164$, und f\"ur $m = 4$ gilt
$n = 138$.

\beginFloat

\medskip

\beginTabelle % [to \kolumnenbreite]
\beginFormat \rechts " \rechts " \endFormat
\+ " \links {\bf Tabelle 10-15} \@ " \\
\+ " \links {Anharmonischer Oszillator mit einer $\hat{x}^4$-Anharmonizit\"at}
\@ " \\
\+ " \links {Approximation des Grenzfalles
unendlicher Kopplung $k_2$ gem\"a{\ss} Gl. (10.7-18)} \@
" \\ \sstrut {} {1.5 \jot} {1.5 \jot}
\- " \- " \-  " \\ \sstrut {} {1 \jot} {1 \jot}
\+ " $\ell$ " \mitte{$k_{2}^{(199, \ell)}$ \qquad \quad}
" \\ \sstrut {} {1.5 \jot} {1.5 \jot}
\- " \- " \-  " \\ \sstrut {} {1 \jot} {1 \jot}
\+ "  0 " 1.060~362~090~484~182~899~647~046~016~692~663~548~348~056~14 " \\
\+ "  1 " 1.060~362~090~484~182~899~647~046~016~692~663~546~046~551~12 " \\
\+ "  2 " 1.060~362~090~484~182~899~647~046~016~692~663~545~543~206~44 " \\
\+ "  3 " 1.060~362~090~484~182~899~647~046~016~692~663~545~498~561~59 " \\
\+ "  4 " 1.060~362~090~484~182~899~647~046~016~692~663~545~508~994~13 " \\
\+ "  5 " 1.060~362~090~484~182~899~647~046~016~692~663~545~514~158~93 " \\
\+ "  6 " 1.060~362~090~484~182~899~647~046~016~692~663~545~515~182~08 " \\
\+ "  7 " 1.060~362~090~484~182~899~647~046~016~692~663~545~515~245~16 " \\
\+ "  8 " 1.060~362~090~484~182~899~647~046~016~692~663~545~515~219~44 " \\
\+ "  9 " 1.060~362~090~484~182~899~647~046~016~692~663~545~515~210~06 " \\
\+ " 10 " 1.060~362~090~484~182~899~647~046~016~692~663~545~515~208~65 " \\
\+ " 11 " 1.060~362~090~484~182~899~647~046~016~692~663~545~515~208~54 " \\
\+ " 12 " 1.060~362~090~484~182~899~647~046~016~692~663~545~515~208~70 " \\
\+ " 13 " 1.060~362~090~484~182~899~647~046~016~692~663~545~515~252~63 " \\
\- " \- " \- " \\ \sstrut {} {1 \jot} {1 \jot}

\endTabelle

\medskip

\endFloat

Ein Vergleich von Tabelle 10-15 mit dem extrem genauen Ergebnis
(10.4-33a) von Vinette und {\v C}{\'\i}{\v z}ek [1991, Gl. (66)] zeigt,
da{\ss} man 8 Stellen gewinnt, wenn man $\ell$ von 0 auf 12 erh\"oht. In
Anbetracht der Tatsache, da{\ss} die Transformationsordnung von
$\delta_{199-\ell}^{(0)} \bigl(1, \Sigma_{0, \ell}^{(2)} \bigr)$
abnimmt, wenn $\ell$ zunimmt, ist das eine bemerkenswerte Verbesserung,
und die beste Approximation $k_2^{(199, 12)}$ in Tabelle 10-15 hat eine
Genauigkeit von 43 Dezimalstellen. Die Approximation $k_2^{(199, 1)}$ in
Tabelle 10-15 entspricht dem besten Ergebnis $k_{2}^{(198)}$ in Tabelle
10-12.

\beginFloat

\medskip

\beginTabelle % [to \kolumnenbreite]
\beginFormat \rechts " \rechts " \endFormat
\+ " \links {\bf Tabelle 10-16} \@ " \\
\+ " \links {Anharmonischer Oszillator mit einer $\hat{x}^6$-Anharmonizit\"at}
\@ " \\
\+ " \links {Approximation des Grenzfalles
unendlicher Kopplung $k_3$ gem\"a{\ss} Gl. (10.7-18)} \@
" \\ \sstrut {} {1.5 \jot} {1.5 \jot}
\- " \- " \-  " \\ \sstrut {} {1 \jot} {1 \jot}
\+ " $\ell$ " \rechts{$k_{3}^{(164, \ell)}$ \qquad \qquad \qquad}
" \\ \sstrut {} {1.5 \jot} {1.5 \jot}
\- " \- " \-  " \\ \sstrut {} {1 \jot} {1 \jot}
\+ " 0 " 1.144~802~453~801~733 " \\
\+ " 1 " 1.144~802~453~798~239 " \\
\+ " 2 " 1.144~802~453~797~147 " \\
\+ " 3 " 1.144~802~453~796~963 " \\
\+ " 4 " 1.144~802~453~796~994 " \\
\+ " 5 " 1.144~802~453~797~034 " \\
\+ " 6 " 1.144~802~453~797~053 " \\
\+ " 7 " 1.144~802~453~796~937 " \\
\+ " 8 " 1.144~802~453~797~471 " \\
\- " \- " \- " \\ \sstrut {} {1 \jot} {1 \jot}

\endTabelle

\medskip

\endFloat

Ein Vergleich von Tabelle 10-16 mit dem extrem genauen Ergebnis
(10.4-33b) von Vinette und {\v C}{\'\i}{\v z}ek [1991, Gl. (69)] zeigt,
da{\ss} man 4 Stellen gewinnt, wenn man $\ell$ von 0 auf 6 erh\"oht. Das ist
wiederum eine bemerkenswerte Verbesserung der Genauigkeit, und die beste
Approximation $k_3^{(164, 6)}$ in Tabelle 10-16 hat eine Genauigkeit von
15 Dezimalstellen. Die Approximation $k_3^{(164, 1)}$ in Tabelle 10-16
entspricht dem besten Ergebnis $k_{3}^{(163)}$ in Tabelle 10-13.

\beginFloat

\medskip

\beginTabelle % [to \kolumnenbreite]
\beginFormat \rechts " \rechts " \endFormat
\+ " \links {\bf Tabelle 10-17} \@ " \\
\+ " \links {Anharmonischer Oszillator mit einer $\hat{x}^8$-Anharmonizit\"at}
\@ " \\
\+ " \links {Approximation des Grenzfalles
unendlicher Kopplung $k_4$ gem\"a{\ss} Gl. (10.7-18)} \@
" \\ \sstrut {} {1.5 \jot} {1.5 \jot}
\- " \- " \-  " \\ \sstrut {} {1 \jot} {1 \jot}
\+ " $\ell$ " \rechts{$k_{4}^{(138, \ell)}$ \qquad}
" \\ \sstrut {} {1.5 \jot} {1.5 \jot}
\- " \- " \-  " \\ \sstrut {} {1 \jot} {1 \jot}
\+ " 0 " 1.225~814~645 " \\
\+ " 1 " 1.225~816~643 " \\
\+ " 2 " 1.225~818~504 " \\
\+ " 3 " 1.225~819~614 " \\
\+ " 4 " 1.225~819~895 " \\
\+ " 5 " 1.225~820~791 " \\
\+ " 6 " 1.225~798~264 " \\
\- " \- " \- " \\ \sstrut {} {1 \jot} {1 \jot}

\endTabelle

\medskip

\endFloat

Ein Vergleich von Tabelle 10-17 mit dem extrem genauen Ergebnis
(10.4-33c) von Vinette und {\v C}{\'\i}{\v z}ek [1991, Gl. (71)] zeigt,
da{\ss} man hier nur 1 - 2 Stellen gewinnt, wenn man $\ell$ von 0 auf 4 oder
5 erh\"oht. Die beste Approximation $k_4^{(138, 4)}$ in Tabelle 10-17 hat
nach Rundung eine Genauigkeit von 6 Dezimalstellen. Die Approximation
$k_4^{(138, 1)}$ in Tabelle 10-17 entspricht dem besten Ergebnis
$k_{4}^{(137)}$ in Tabelle 10-14.

Wenn man den Wynnschen $\epsilon$-Algorithmus, Gl. (2.4-10), auf die
Partialsummen (10.7-16) anwendet, beobachtet man bei einer Vergr\"o{\ss}erung
von $\ell$ in Gl. (10.7-16) keine Verbesserung der Konvergenz.
Stattdessen beobachtet man immer, da{\ss} die Ergebnisse mit wachsendem
$\ell$ schlechter werden, da eine Vergr\"o{\ss}erung von $\ell$ nur zu einer
Verkleinerung der maximal m\"oglichen Transformationsordnung f\"uhrt. Dieses
Beispiel zeigt, da{\ss} es ganz fundamentale Unterschiede gibt zwischen
verallgemeinerten Summationsprozessen, die wie der Wynnsche
$\epsilon$-Algorithmus nur die Elemente einer Folge $\Seqn s$ von
Partialsummen verwenden, und verallgemeinerten Summationsprozessen wie
${\cal S}_k^{(n)} (\zeta, s_n, \omega_n)$, Gl. (5.4-6), die au{\ss}erdem
noch explizite Restsummenabsch\"atzungen $\Seqn \omega$ ben\"otigen.

Ein Vergleich der Tabellen 10-12, 10-13 und 10-14 einerseits und der
Tabellen 10-15, 10-16 und 10-17 andererseits zeigt aber auch, da{\ss} es bei
verallgemeinerten Summationsprozessen, die explizite
Restsummenabsch\"atzungen verwenden, zus\"atzlich zu allen anderen m\"oglichen
Fehlerquellen auch noch den {\it Absch\"atzungsfehler\/} gibt. Diese
zus\"atzliche Fehlerquelle ist sicherlich eine unwillkommene Komplikation,
die die praktische Anwendung von verallgemeinerten Summationsprozessen
mit expliziten Restsummenabsch\"atzungen zum Teil erheblich erschwert.
Wenn nur die numerischen Werte einer kleinen Zahl von Termen einer
schlecht konvergierenden oder divergenten Reihe bekannt sind, ist es
normalerweise nicht {\it a priori\/} klar, ob die Terme dieser Reihe
sich schon f\"ur kleine Indizes so regul\"ar verhalten, da{\ss} sie halbwegs
vern\"unftige Absch\"atzungen der tats\"achlichen Abbruchfehler liefern
k\"onnen. Weiterhin ist nicht {\it a priori\/} klar, welche der in
Abschnitt 5.2 beschriebenen einfachen Restsummenabsch\"atzungen die besten
Ergebnisse liefert.

Diese Probleme sind aber ein unvermeidlicher Preis daf\"ur, da{\ss}
verallgemeinerte Summationsprozesse, die wie ${\cal S}_k^{(n)} (\zeta,
s_n, \omega_n)$ auch explizite Restsummenabsch\"atzungen verwenden, in
vielen F\"allen signifikant bessere Ergebnisse liefern als
Summationsprozesse, die wie der Wynnsche $\epsilon$-Algorithmus
ausschlie{\ss}lich die Elemente einer Folge von Partialsummen als
Eingabedaten verwenden.

Sowohl die Levinsche Transformation ${\cal L}_{k}^{(n)} (\zeta, s_n,
\omega_n)$, Gl. (5.2-6), als auch ${\cal S}_{k}^{(n)} (\zeta, s_n,
\omega_n)$, Gl. (5.4-6), sind gewichtete Mittelwerte der Quotienten
$s_n/\omega_n$, $s_{n+1}/\omega_{n+1}$, $\ldots$ ,
$s_{n+k}/\omega_{n+k}$ vom Typ von Gl. (5.6-1), die sich nur bez\"uglich
der Gewichtsfaktoren unterscheiden. Da die Pochhammersymbole $(n + j +
\zeta)_{k - 1}$ in Gl. (5.4-6) f\"ur zunehmendes $n$ und festes $k$
schneller wachsen als Potenzen $(n + j + \zeta)^{k - 1}$ in Gl. (5.2-6),
erhalten die Quotienten $s_{n+j}/\omega_{n+j}$ mit gr\"o{\ss}eren Werten von
$j$ in ${\cal S}_{k}^{(n)} (\zeta, s_n, \omega_n)$ ein gr\"o{\ss}eres Gewicht
als in der Levinschen Transformation ${\cal L}_{k}^{(n)} (\zeta, s_n,
\omega_n)$.

Da die Levinsche Transformation ${\cal L}_{k}^{(n)} (\zeta, s_n,
\omega_n)$ und der eng verwandte verallgemeinerte Summationsproze{\ss}
${\cal S}_{k}^{(n)} (\zeta, s_n, \omega_n)$ deutlich unterschiedliche
Eigenschaften bei der Summation der divergenten St\"orungsreihen
anharmonischer Oszillatoren aufweisen, kann man versuchen, das
unterschiedliche Verhalten dieser beiden verallgemeinerten
Summationsprozesse durch die unterschiedlichen Gewichte der Quotienten
$s_n/\omega_n$, $s_{n+1}/\omega_{n+1}$, $\ldots$ ,
$s_{n+k}/\omega_{n+k}$ zu erkl\"aren.

In Abschnitt 5.6 wurde der verallgemeinerte Summationsproze{\ss} ${\cal
C}_{k}^{(n)} (\alpha, \zeta, s_n, \omega_n)$, Gl. (5.6-7), beschrieben,
der in Abh\"angigkeit von einem kontinuierlichen Parameter $\alpha$
zwischen der Levinschen Transformation ${\cal L}_{k}^{(n)} (\zeta, s_n,
\omega_n)$ und ${\cal S}_{k}^{(n)} (\zeta, s_n, \omega_n)$ interpolieren
kann [Weniger 1992]. F\"ur $\alpha = 1$ ist ${\cal C}_{k}^{(n)} (\alpha,
\zeta, s_n, \omega_n)$ identisch mit ${\cal S}_{k}^{(n)} (\zeta, s_n,
\omega_n)$, und f\"ur $\alpha \to \infty$ geht ${\cal C}_{k}^{(n)}
(\alpha, \zeta, s_n, \omega_n)$ aufgrund von Gl. (5.6-2) in die
Levinsche Transformation ${\cal L}_{k}^{(n)} (\zeta, s_n, \omega_n)$
\"uber.

Es ist also eine interessante Frage, wie die Summationsergebnisse sich
in Abh\"angigkeit von $\alpha$ \"andern, wenn man nicht ${\delta}_k^{(n)}
(\zeta, s_n)$, Gl. (5.4-13), sondern ${\cal C}_{k}^{(n)} (\alpha, \zeta,
s_n, \omega_n)$ mit $\omega_n = \Delta s_n$ auf die Partialsummen
$$
{\bf S}_{n}^{(2)} \; = \; 3^{1/3} \, \sum_{\nu=0}^{n} \, c_{\nu}^{(2)}
\tag
$$
der St\"orungsreihe (10.4-31) f\"ur $k_2$ anwendet.

In Tabelle 10-18 [Weniger 1992, Table 2] werden die Approximationen
$$
k_2^{(199)} (\alpha) \; = \;
{\cal C}_{199}^{(0)} (\alpha, 1, {\bf S}_0^{(2)}, \omega_0)
\tag
$$
f\"ur verschiedene Werte von $\alpha$ aufgelistet. Als Eingabedaten wurden
die Partialsummen (10.7-19) und die Restsummenabsch\"atzungen $\omega_n =
c_{n+1}^{(2)}$ verwendet. ${\cal C}_{k}^{(n)} (\alpha, \zeta, s_n,
\omega_n)$ wird also in Verbindung mit der Restsummenabsch\"atzung
$\omega_n = \Delta s_n$, Gl. (5.2-16) verwendet. Das bedeutet, da{\ss}
${\cal C}_{k}^{(n)} (\alpha, \zeta, s_n, \Delta s_n)$ f\"ur $\alpha = 1$
identisch ist mit ${\delta}_k^{(n)} (\zeta, s_n)$, Gl. (5.4-13), und f\"ur
$\alpha \to \infty$ in die Levinsche Transformation $d_k^{(n)} (\zeta,
s_n)$, Gl. (5.2-18), \"ubergeht.

\beginFloat

\medskip

\beginTabelle
\beginFormat \rechts " \rechts \endFormat
\+ " \links {\bf Tabelle 10-18} \@ " \\
\+ " \links {Anharmonischer Oszillator mit einer $\hat{x}^4$-Anharmonizit\"at}
\@ " \\
\+ " \links {Approximation des Grenzfalles
unendlicher Kopplung $k_2$ gem\"a{\ss} Gl. (10.7-20)} \@ " \\
\- " \- " \- " \\ \sstrut {} {1.5 \jot} {1.5 \jot}
\+ " \rechts {$\alpha$} "
\mitte {$k_2^{(199)} (\alpha)$} " \\
\- " \- " \- " \\ \sstrut {} {1 \jot} {1 \jot}
\+ " 0.1   "  1.060~362~090~484~182~900~369~547~959~430~562~403~080~971~787 " \\
\+ " 0.3   "  1.060~362~090~484~182~899~647~041~385~115~655~285~042~236~782 " \\
\+ " 0.5   "  1.060~362~090~484~182~899~647~046~016~924~259~468~431~397~969 " \\
\+ " 0.7   "  1.060~362~090~484~182~899~647~046~016~692~615~203~171~647~728 " \\
\+ " 0.9   "  1.060~362~090~484~182~899~647~046~016~692~663~559~227~360~603 " \\
\+ " 1.0   "  1.060~362~090~484~182~899~647~046~016~692~663~548~348~056~144 " \\
\+ " 1.1   "  1.060~362~090~484~182~899~647~046~016~692~663~545~615~616~218 " \\
\+ " 1.3   "  1.060~362~090~484~182~899~647~046~016~692~663~545~514~195~837 " \\
\+ " 1.5   "  1.060~362~090~484~182~899~647~046~016~692~663~545~515~209~234 " \\
\+ " 1.7   "  1.060~362~090~484~182~899~647~046~016~692~663~545~514~151~069 " \\
\+ " 2.0   "  1.060~362~090~484~182~899~647~046~016~692~663~582~622~309~201 " \\
\+ " 3.0   "  1.060~362~090~484~182~899~647~047~772~179~412~541~330~299~183 " \\
\+ " 4.0   "  1.060~362~090~484~182~897~421~907~913~740~122~912~198~363~562 " \\
\+ " 5.0   "  1.060~362~090~484~193~604~224~009~929~803~842~171~266~754~798 " \\
\+ " 10.0  "  1.060~363~842~860~957~816~375~091~493~473~477~564~455~856~113 " \\
\+ " 30.0  "  14.920~402~244~013~914~000~996~666~181~030~744~566~630~690~993 " \\
\+ " 50.0  "  191.867~622~455~971~044~784~451~769~928~222~993~465~919~717~363 " \\
\+ " 500.0 "  -22896.205~013~871~809~816~130~347~915~214~288~474~796~377~994~633 " \\
\+ " $\infty$ " -37427.488~079~987~781~721~423~775~948~893~388~944~626~750~668~209 " \\
\- " \- " \- " \\ \sstrut {} {1 \jot} {1 \jot}

\endTabelle

\medskip

\endFloat

Die Ergebnisse in Tabelle 10-18 zeigen ganz deutlich, da{\ss} ${\cal
C}_{k}^{(n)} (\alpha, \zeta, s_n, \Delta s_n)$ nicht konvergiert, wenn
$\alpha$ gro{\ss} ist. Das ist im Einklang mit der Beobachtung, da{\ss} die
Levinsche Transformation ${\cal L}_{k}^{(n)} (\zeta, s_n, \omega_n)$,
Gl. (5.2-6), nicht in der Lage ist, die in dieser Arbeit behandelten
St\"orungsreihen f\"ur anharmonische Oszillatoren zu summieren.

Interessant ist, da{\ss} die besten Summationsergebnisse f\"ur $k_2^{(199)}
(\alpha)$ nicht f\"ur $\alpha = 1$ erzielt wurden, was den
verallgemeinerten Summationsproze{\ss} ${\delta}_k^{(n)} (\zeta, s_n)$, Gl.
(5.4-13), entspr\"ache, sondern f\"ur $\alpha = 1.49$ [Weniger 1992, Gl.
(4.2)],
$$
k_2^{(199)} (1.49) \; = \;
1.060~362~090~484~182~899~647~046~016~692~663~545~515~208~066 \, ,
\tag
$$
was mehr als 40 Stellen des extrem genauen Ergebnisses (10.4-33a) von
Vinette und {\v C}{\'\i}{\v z}ek [1991, Gl. (66)] reproduziert.

Die Gewichte der Quotienten $s_n/\omega_n$, $s_{n+1}/\omega_{n+1}$,
$\ldots$ , $s_{n+k}/\omega_{n+k}$, die f\"ur den verallgemeinerten
Summationsproze{\ss} ${\cal S}_{k}^{(n)} (\zeta, s_n, \omega_n)$ typisch
sind, ergeben im Falle der St\"orungsreihe (10.4-31) f\"ur $k_2$ also noch
nicht die optimalen Ergebnisse. Die Tatsache, da{\ss} der optimale Wert
$\alpha = 1.49$ ist, bedeutet, da{\ss} die Quotienten $s_{n+j}/\omega_{n+j}$
mit gr\"o{\ss}eren Werten von $j$ in ${\cal C}_{k}^{(n)} (\alpha, \zeta, s_n,
\omega_n)$ ein etwas geringeres Gewicht haben als in ${\cal S}_{k}^{(n)}
(\zeta, s_n, \omega_n)$, aber immer noch ein deutlich gr\"o{\ss}eres als in
der Levinschen Transformation ${\cal L}_{k}^{(n)} (\zeta, s_n,
\omega_n)$.

Diese Ergebnisse zeigen einmal mehr, da{\ss} nicht nur die
Restsummenabsch\"atzungen $\Seqn \omega$, sondern auch die in Gl. (5.1-3)
definierten Korrekturterme $\Seqn z$ das Leistungsverm\"ogen eines
verallgemeinerten Summationsprozesses vom Typ von Gl. (5.1-7) ganz
erheblich beeinflussen. Leider ist die Rolle der Korrekturterme $\Seqn
z$ in Konvergenzbeschleunigungs- und Summa\-tionsverfahren bisher nur
unzureichend verstanden.

\medskip

\Abschnitt Eine konvergente renormierte St\"orungsreihe f\"ur die
Grundzustandsenergie eines anharmonischen Oszillators

\smallskip

\aktTag = 0

In Unterabschnitt 10.3 wurde gezeigt, da{\ss} die Grundzustandsenergie
$E^{(m)} (\beta)$ eines anharmonischen Oszillators mit einer $\hat{x}^{2
m}$-Anharmonizit\"at f\"ur $\beta \to \infty$ wie $\beta^{1/(m+1)}$ w\"achst.
In Unterabschnitt 10.4 wurde au{\ss}erdem gezeigt, da{\ss} rationale
Approximationen nicht in der Lage sind, eine divergente Potenzreihe f\"ur
gr\"o{\ss}ere Argumente effizient zu summieren, wenn die entsprechende
Funktion sich f\"ur gro{\ss}e Argumente wie eine nichtganzzahlige Potenz
verh\"alt. Aus dem asymptotischen Verhalten von $E^{(m)} (\beta)$ folgt
also, da{\ss} die Rayleigh-Schr\"odingersche St\"orungsreihe (10.2-4), die eine
formale Potenzreihe in der Kopplungskonstante $\beta$ ist, nur f\"ur
kleinere Werte von $\beta$ effizient durch rationale Funktionen in
$\beta$ summiert werden kann. Im Falle gr\"o{\ss}erer Kopplungskonstanten
$\beta$ ist eine Summation der St\"orungsreihe (10.2-4) durch rationale
Funktionen nicht praktikabel, und man mu{\ss} alternative Verfahren zur
Berechnung von $E^{(m)} (\beta)$ verwenden.

Eine prinzipielle Alternative ist die sogenannte {\it strong coupling
expansion\/} (10.3-23), die eine Potenzreihe in der Variablen $\beta^{-
2/(m+1)}$ ist. Aufgrund des Vorfaktors $\beta^{1/(m+1)}$ besitzen sowohl
die Terme als auch die Partialsummen dieser Reihe automatisch das
korrekte asymptotische Verhalten f\"ur $\beta \to \infty$. Deswegen ist
diese St\"orungsreihe besonders gut geeignet f\"ur die Berechnung der
Grundzustandsenergie $E^{(m)} (\beta)$ im Falle gro{\ss}er
Kopplungskonstanten. Au{\ss}erdem folgt aus den S\"atzen 10-1 und 10-2, da{\ss}
die {\it strong coupling expansion\/} (10.3-23) die Katosche Ungleichung
(10.1-4) erf\"ullt und demzufolge f\"ur ausreichend gro{\ss}e Werte von $\beta$
konvergiert. Wenn $\beta$ sehr gro{\ss} ist, sollten also schon einige
wenige Terme dieser Reihe gen\"ugen, um $E^{(m)} (\beta)$ mit
ausreichender Genauigkeit zu berechnen. Da die Potenzen $\beta^{- 2 n /
(m+1)}$ mit $n \in \N_0$ f\"ur $\beta \to 0$ singul\"ar werden, kann die
St\"orungsreihe (10.3-23) aber nicht in einer Umgebung von $\beta = 0$
konvergieren, sondern nur f\"ur ausreichend gro{\ss}e Werte von $\beta$. Der
Konvergenzradius der St\"orungsreihe (10.3-23) scheint aber nicht bekannt
zu sein. Trotzdem w\"are die Verwendung der {\it strong coupling
expansion\/} (10.3-23) an sich \"au{\ss}erst w\"unschenswert. Das
Haupthindernis, das bisher eine Verwendung dieser St\"orungsreihe
verhindert hat, sind die in Unterabschnitt 10.3 beschriebenen
Schwierigkeiten, die bei der Berechnung der St\"orungstheoriekoeffizienten
$K_{n}^{(m)}$ auftreten.

In Unterabschnitt 10.4 wurde ein Renormierungsverfahren beschrieben, das
als eine Variablensubstitution interpretiert werden kann. Der
physikalisch relevante Bereich der Kopplungs\-konstante $\beta$ ist die
positive reelle Halbachse $[0, \infty)$, die durch das
Renormierungsverfahren gem\"a{\ss} Gl. (10.4-12) auf das kompakte Intervall
$[0, 1)$ f\"ur die renormierte Kopplungskonstante $\kappa$ abgebildet
wird. Wenn man die Grundzustandsenergie $E^{(m)} (\beta)$ durch
Summation der renormierten St\"orungsreihe gem\"a{\ss} Gl. (10.4-28) berechnet,
hat man den Vorteil, da{\ss} die Terme und Partialsummen der St\"orungsreihe
(10.4-28) aufgrund von Gl. (10.4-29) automatisch das richtige
asymptotische Verhalten f\"ur $\beta \to \infty$ besitzen. Au{\ss}erdem ist
die Berechnung der renormierten St\"orungstheoriekoeffizienten
$c_{n}^{(m)}$ -- wie in Unterabschnitt 10.5 gezeigt wurde -- nicht
schwieriger als die Berechnung der Rayleigh-Schr\"odingerschen
St\"orungstheoriekoeffizienten $b_{n}^{(m)}$ in Gl. (10.2-4).

Die St\"orungsreihe (10.4-28) vereinigt also einige der Vorteile der
Rayleigh-Schr\"odingerschen St\"orungsreihe (10.2-4) einerseits und der {\it
strong coupling expansion\/} (10.3-23) andererseits. Verglichen mit der
{\it strong coupling expansion\/} (10.3-23) hat die St\"orungsreihe
(10.4-28) aber den unbestreitbaren Nachteil, da{\ss} sie f\"ur alle $\beta >
0$ eine hochgradig divergente Reihe ist. Au{\ss}erdem wird die renormierte
St\"orungsreihe (10.4-28) im kopplungsfreien Fall $\beta = 0$
beziehungsweise $\kappa = 0$ trivial, wogegen die {\it strong coupling
expansion\/} f\"ur $\beta \to \infty$ trivial wird. Da $\beta \to \infty$
gem\"a{\ss} Gl. (10.4-12) $\kappa \to 1$ impliziert, ist man bei der
St\"orungsreihe (10.4-28) dann mit dem schwierigsten Summationsproblem
konfrontiert, das in diesem Zusammenhang \"uberhaupt auftreten kann. Trotz
des korrekten asymp\-totischen Verhaltens der Terme und Partialsummen
f\"ur $\beta \to \infty$ ist die renormierte St\"orungsreihe (10.4-28) also
eine {\it weak coupling expansion\/}, die allerdings -- wie im letzten
Unterabschnitt gezeigt wurde -- auch f\"ur sehr gro{\ss}e Kopplungskonstanten
$\beta$ erfolgreich zur Berechnung von $E^{(m)} (\beta)$ verwendet
werden kann.

Man kann aber beweisen, da{\ss} eine renormierte St\"orungsreihe f\"ur die
Grundzustandsenergie $E^{(m)} (\beta)$ existiert, die im Grenzfall
unendlicher Kopplung $\kappa = 1$ trivial wird und die man deswegen als
eine {\it renormalized strong coupling expansion\/} bezeichnen kann.

Ebenso wie bei dem formalen Beweis der Existenz der {\it strong coupling
expansion\/} (10.3-23) geht man auch beim Beweis der Existenz einer {\it
renormalized strong coupling expansion\/} von Gl. (10.3-9) aus. Wenn man
in dieser Beziehung $\beta$ gem\"a{\ss} Gl. (10.4-9) durch $(1 - \tau^{2}) /
[B_m \tau^{m+1}]$ ersetzt, erh\"alt man den transformierten
Hamiltonoperator
$$
\hat{H}^{(m)} (\alpha, (1 - \tau^2)/[B_m \tau^{m+1}]; \tau) \; = \;
\tau^{- 1} \, \left\{ \hat{P}^2 + \alpha \tau^2 \hat{X}^2
+ \frac {1 - \tau^{2}} {B_m} \, \hat{X}^{2m} \right\} \, .
\tag
$$
Aus Gl. (10.3-10) folgt, da{\ss} dieser Hamiltonoperator die
Eigenwertgleichung
$$
\hat{H}^{(m)}
\bigl( \alpha, (1 - \tau^2)/[B_m \tau^{m+1}]; \tau \bigr) \,
\Psi (X) \; = \; \tau^{- 1} \,
E^{(m)} \bigl( \alpha \tau^2, (1 - \tau^{2}) / B_m \bigr) \, \Psi (X)
\tag
$$
erf\"ullt, wobei $\Psi (X)$ eine gem\"a{\ss} Gl. (10.3-6) transformierte
Eigenfunktion des Hamiltonoperators (10.3-2) ist.

Wenn man in Gl. (10.8-1) $\alpha = 1$ setzt, erh\"alt man
$$
\hat{H}^{(m)} \bigl( 1, (1 - \tau^2)/[B_m \tau^{m+1}]; \tau \bigr)
\; = \;
\tau^{- 1} \, \left\{ \hat{P}^2 \, + \,
\frac {1} {B_m} \, \hat{X}^{2m} \, + \, \tau^2 \bigl\[
\hat{X}^2 \, - \, \frac {1} {B_m} \, \hat{X}^{2m} \bigr\]
\right\} \, .
\tag
$$
Diese Beziehung legt die Definition eines {\it renormierten\/}
Hamiltonoperators
$$
\beginAligntags
" {\bf \hat{H}}^{(m)} (\tau) " \; = \; " \tau \,
\hat{H}^{(m)} \bigl( 1, (1 - \tau^2)/[B_m \tau^{m+1}]; \tau \bigr) \\
" " \; = \; " \hat{P}^2 \, + \,
\frac {1} {B_m} \, \hat{X}^{2m} \, + \, \tau^2 \bigl\[
\hat{X}^2 \, - \, \frac {1} {B_m} \, \hat{X}^{2m} \bigr\]
\\ \tag
\endAligntags
$$
\noindent nahe, der auf folgende Weise in einen ungest\"orten renormierten
Hamiltonoperator ${\bf \hat{H}}_0^{(m)}$ und in einen renormierten
St\"oroperator ${\bf \hat{V}}^{(m)}$ aufgespalten werden kann:
$$
\beginAligntags
" {\bf \hat{H}}^{(m)} (\tau) " \; = \; " {\bf \hat{H}}_0^{(m)} \, + \,
\tau^2 \, {\bf \hat{V}}^{(m)} \, , \\ \tag
" {\bf \hat{H}}_0^{(m)} " \; = \; " \hat{P}^2 \, + \,
\frac {1} {B_m} \, \hat{X}^{2m} \, , \\ \tag
" {\bf \hat{V}}^{(m)} " \; = \; "
\hat{X}^2 \, - \, \frac {1} {B_m} \, \hat{X}^{2m} \, . \\ \tag
\endAligntags
$$
Der Grundzustandsenergieeigenwert ${\cal E}_{R}^{(m)} (\tau)$ des
renormierten Hamiltonoperators ${\bf \hat{H}}^{(m)} (\tau)$ kann also
durch die folgende St\"orungsreihe dargestellt werden, die eine formale
Potenzreihe in $\tau^2$ ist:
$$
{\cal E}_{R}^{(m)} (\tau) \; = \; \sum_{n=0}^{\infty} \,
\Gamma_{n}^{(m)} \, \tau^{2 n} \, .
\tag
$$
Weiterhin folgt aus Gln. (10.8-2) und (10.8-4), da{\ss} der
Grundzustandsenergieeigenwert $E^{(m)} (1, \beta)$ des Hamiltonoperators
(10.3-2) mit $\alpha = 1$ auf folgende Weise durch den
Grundzustandsenergieeigenwert ${\cal E}_{R}^{(m)} (\tau)$ des in Gl.
(10.8-4) definierten renormierten Hamiltonoperators ${\bf \hat{H}}^{(m)}
(\tau)$ dargestellt werden kann:
$$
E^{(m)} (1, \beta) \; = \;
E^{(m)} \bigl( \tau^2, (1 - \tau^{2}) / B_m \bigr) \; = \;
\tau^{- 1} \,{\cal E}_{R}^{(m)} (\tau) \, .
\tag
$$
Das bedeutet, da{\ss} die Grundzustandsenergie $E^{(m)} (1, \beta)$, die mit
dem entsprechenden Eigenwert $E^{(m)} (\beta)$ des in Gl. (10.2-1)
definierten Hamiltonoperators $\hat{H}^{(m)} (\beta)$ identisch ist,
auch eine Darstellung durch die folgende Potenzreihe in $\tau^2$
besitzt:
$$
E^{(m)} (1, \beta) \; = \; \tau^{- 1} \, \sum_{n=0}^{\infty} \,
\Gamma_{n}^{(m)} \, \tau^{2 n} \, .
\tag
$$
Aus Gl. (10.4-9) folgt, da{\ss} $\beta = (1 - \tau^2) / [B_m \tau^{m+1}]$
gilt. Der kopplungsfreie Fall $\beta = 0$ impliziert also $\tau = 1$,
und der Grenzfall unendlicher Kopplung $\beta \to \infty$ impliziert
$\tau \to 0$. Au{\ss}erdem folgt aus Gl. (10.4-9) die asymptotische
Beziehung
$$
\beta^{1/(m+1)} \; \sim \; \tau^{- 1} \, , \qquad \beta \to \infty \, .
\tag
$$
Die St\"orungsreihe (10.8-10) ist also tats\"achlich eine {\it renormalized
strong coupling expansion\/} der Grundzustandsenergie $E^{(m)} (1,
\beta)$, da sie f\"ur $\tau \to 0$ trivial wird.

Trotz dieser Analogie zwischen den St\"orungsreihen (10.3-23) und
(10.8-10) unterscheidet sich der Hamiltonoperator (10.8-4) ganz
erheblich vom Hamiltonoperator (10.3-19), der die Basis der {\it strong
coupling expansion\/} (10.3-23) ist. Der Hamiltonoperator (10.8-4) ist
f\"ur alle $\tau \in [0, 1]$ ein mathematisch sinnvolles Objekt. Im
kopplungsfreien Fall $\tau = 1$ gilt ${\bf \hat{H}}^{(m)} (1) =
\hat{P}^2 + \hat{X}^2$, und im Grenzfall unendlicher Kopplung $\tau = 0$
gilt ${\bf \hat{H}}^{(m)} (0) = \hat{P}^2 + [\hat{X}^{2 m} / B_m]$.
Dagegen divergiert die effektive Kopplungskonstante $\beta^{- 2 /
(m+1)}$ des Hamiltonoperators (10.3-19) f\"ur $\beta \to 0$. Der
Hamiltonoperator (10.3-19) ist also nur f\"ur $\beta \in [\varepsilon,
\infty)$ mit $\varepsilon > 0$ ein mathematisch sinnvolles Objekt. Man
kann also hoffen, da{\ss} die {\it renormalized strong coupling expansion\/}
(10.8-10) im Gegensatz zur {\it strong coupling expansion\/} (10.3-23)
nicht nur in einer Umgebung des Entwicklungspunktes $\tau = 0$
konvergiert, sondern f\"ur alle physikalisch relevanten renormierten
Kopplungskonstanten $\tau \in [0, 1]$.

Bekanntlich konvergiert eine St\"orungsreihe nur dann in einer Umgebung
des Entwicklungs\-punktes, wenn der Definitionsbereich des ungest\"orten
Hamiltonoperators im Definitionsbereich des St\"oroperators enthalten ist,
und wenn die Katosche Ungleichung (10.1-4) f\"ur alle Wellenfunktionen
$\Psi$ aus dem Definitionsbereich des ungest\"orten Hamiltonoperators
erf\"ullt ist [Reed und Simon 1978, S. 16, Lemma, S. 17, Theorem
\Roemisch{12}.9, und S. 21, Theorem \Roemisch{12}.11].

Der in Gl. (10.4-16) definierte St\"or\-ope\-ra\-tor $\hat{\cal V}^{(m)} =
\{ \hat{X}^{2 m} - B_m \hat{X}^2 \} / B_m$ ist aber nicht relativ zum
ungest\"orten renormierten Hamiltonoperator $\hat{\cal H}_0 = \hat{P}^2 +
\hat{X}^2$ beschr\"ankt. Die Katosche Ungleichung (10.1-4) kann also nicht
erf\"ullt sein, und die renormierte St\"orungsreihe (10.4-20) divergiert
folglich f\"ur alle $\kappa \ne 0$. Im Falle des in Gl. (10.8-4)
definierten Hamiltonoperators $\hat{\bf H}^{(m)} (\tau)$ kommt der
singul\"are Term $\hat{X}^{2 m}$ dagegen nicht nur im St\"oroperator ${\bf
\hat{V}}^{(m)}$, sondern auch im ungest\"orten Operator $\hat{\bf
H}^{(m)}_0$ vor. Man kann also hoffen, die G\"ultigkeit einer
Absch\"atzung des Typs
$$
\Vert (\hat{X}^2 - [\hat{X}^{2 m} / B_m]) \Psi \Vert \; \le \;
a \, \Vert (\hat{P}^2 + [\hat{X}^{2 m} / B_m]) \Psi \Vert \, + \,
b \, \Vert \Psi \Vert \, , \qquad a, b > 0 \, ,
\tag
$$
die der Katoschen Ungleichung (10.1-4) entspricht, f\"ur alle $\Psi \in
{\cal D} (\hat{P}^2) \cap {\cal D} (\hat{X}^{2 m})$ beweisen zu k\"onnen.
Auf diese Weise w\"are dann bewiesen, da{\ss} die St\"orungsreihe (10.8-8) in
einer Umgebung von $\tau = 0$ konvergiert.

Der Beweis der Konvergenz der St\"orungsreihe (10.8-8) verl\"auft analog zum
Beweis der Konvergenz der {\it strong coupling expansion\/} (10.3-23) in
zwei Schritten.

\medskip

\beginEinzug \sl \parindent = 0 pt

\Auszug {\bf Satz 10-3:} Alle Wellenfunktionen $\Psi$ aus dem
Definitionsbereich des Hamiltonoperators
$$
\hat{\bf H}^{(m)}_0 \; = \; \hat{P}^2 \, + \,
\frac {1} {B_m} \, \hat{X}^{2m}
\tag
$$
erf\"ullen die Absch\"atzung
$$
\Vert \hat{P}^2 \Psi \Vert^2 \, + \, \frac {1} {(B_{m})^{2}} \,
\Vert \hat{X}^{2m} \Psi \Vert^2 \; \le \;
2 \, \Vert (\hat{P}^2 + [\hat{X}^{2m} / B_m]) \Psi \Vert^2 \, + \,
b_0 \, \Vert \Psi \Vert^2 \, ,
\tag
$$
wobei $b_0$ eine geeignete positive Konstante ist.

\endEinzug

\medskip

\noindent {\it Beweis:\/} Unter Verwendung der Kommutatorbeziehung
(10.3-28) sch\"atzen wir den folgenden Ausdruck ab:
$$
\beginAligntags
" \Bigl\{ \hat{P}^2 + \frac {\hat{X}^{2 m}} {B_m} \Bigr\}^2 \; = \;
\hat{P}^4 \, + \, \frac {\hat{X}^{4 m}} {(B_m)^2} \, + \,
\frac {1} {B_m} \bigl\{ \hat{P}^2 \hat{X}^{2 m}
\, + \, \hat{X}^{2 m} \hat{P}^2 \bigr\} \\ \tag
" \qquad \quad \; = \;
\hat{P}^4 \, + \, \frac {\hat{X}^{4 m}} {(B_m)^2} \, + \,
\frac {1} {B_m} \, \Bigl\{ 2 \, \big( \hat{P} \, \hat{X}^{m} \bigr)
\big( \hat{X}^{m} \, \hat{P} \bigr) \, + \,
\bigl\[ \hat{P}, \bigl\[ \hat{P}, \hat{X}^{2 m} \bigr\] \bigr\]
\Bigr\}
\\ \tag
" \qquad \quad \; = \; \hat{P}^4 \, + \, \frac {\hat{X}^{4 m}} {(B_m)^2}
\, + \, \frac {1} {B_m} \, \Bigl\{
2 \, \big( \hat{P} \, \hat{X}^{m} \bigr)
\big( \hat{X}^{m} \, \hat{P} \bigr)
\, - \, (2 m) \, i \, \bigl\[ \hat{P}, \hat{X}^{2 m - 1} \bigr\] \Bigr\}
\\ \tag
" \qquad \quad \; = \; \hat{P}^4 \, + \, \frac {\hat{X}^{4 m}} {(B_m)^2}
\, + \,
\frac {1} {B_m} \, \Bigl\{ 2 \, \big( \hat{P} \, \hat{X}^{m} \bigr)
\big( \hat{X}^{m} \, \hat{P} \bigr)
\, - \, (2 m - 1) (2 m) \, \hat{X}^{2 m - 2} \Bigr\} \, . \qquad
\\ \tag
\endAligntags
$$
Zur Absch\"atzung der rechten Seite von Gl. (10.8-18) w\"ahlen wir eine
Konstante $b_0 > 0$ auf solche Weise, da{\ss}
$$
\frac {1} {2} \, \frac {x^{4 m}} {(B_m)^2} \, - \,
\frac {1} {B_m} \, (2 m - 1) (2 m) \, x^{2 m - 2}
\, + \, \frac {1} {2} \, b_0 \; \ge \; 0
\tag
$$
f\"ur alle $x \in \R$ gilt. Damit erhalten wir:
$$
\beginAligntags
\Bigl\{ \hat{P}^2 + \frac {\hat{X}^{2 m}} {B_m} \Bigr\}^2 " \, + \,
\frac {1} {2} \, b_0 \; = \; \frac {1} {2} \,
\Bigl\{ \hat{P}^4 + \frac {\hat{X}^{4 m}} {(B_m)^2} \Bigr\}
\, + \, \frac {1} {2} \, \hat{P}^4\, + \, \frac {1} {2} \,
\frac {\hat{X}^{4 m}} {(B_m)^2} \\
" \, + \, \frac {1} {B_m} \, \Bigl\{
2 \, \big( \hat{P} \, \hat{X}^{m} \bigr)
\big( \hat{X}^{m} \, \hat{P} \bigr)
\, - \,
(2 m - 1) (2 m) \, \hat{X}^{2 m - 2} \Bigr\} \, + \,
\frac {1} {2} b_0 \, . \qquad
\\ \tag
\endAligntags
$$
Aufgrund von Gl. (10.8-19) gilt aber
$$
\frac {1} {2} \, \hat{P}^4\, + \, \frac {1} {2} \,
\frac {\hat{X}^{4 m}} {(B_m)^2}
\, + \, \frac {1} {B_m} \, \Bigl\{
2 \, \big( \hat{P} \, \hat{X}^{m} \bigr)
\big( \hat{X}^{m} \, \hat{P} \bigr)
\, - \,
(2 m - 1) (2 m) \, \hat{X}^{2 m - 2} \Bigr\} \, + \,
\frac {1} {2} b_0 \; \ge \; 0
\tag
$$
im Sinne eines Erwartungswertes, was impliziert, da{\ss} auch
$$
\hat{P}^4 + \frac {\hat{X}^{4 m}} {(B_m)^2} \; \le \;
2 \, \Bigl\{ \hat{P}^2 + \frac {\hat{X}^{2 m}} {B_m} \Bigr\}^2
\, + \, b_0
\tag
$$
im Sinne eines Erwartungswertes gilt. Wenn man jetzt den Erwartungswert
der Ungleichung (10.8-22) mit einer Funktion $\Psi \in {\cal D}
(\hat{P}^2) \cap {\cal D} (\hat{X}^{2 m})$ bildet, erh\"alt man Gl.
(10.8-14). \BeweisEnde

Mit Hilfe von Satz 10-3 kann man beweisen, da{\ss} der St\"oroperator
$\hat{\bf V}^{(m)} = \hat{X}^2 - [\hat{X}^{2 m} / B_m]$ relativ zum
ungest\"orten Operator $\hat{\bf H}^{(m)}_0 = \hat{P}^2 + [\hat{X}^{2 m}/
B_m]$ beschr\"ankt ist.

\medskip

\beginEinzug \sl \parindent = 0 pt

\Auszug {\bf Satz 10-4:} Der Operator $\hat{\bf V}^{(m)} = \hat{X}^2 -
[\hat{X}^{2 m} / B_m]$ ist relativ zu dem in Gl. (10.8-6) definierten
Operator $\hat{\bf H}^{(m)}_0$ beschr\"ankt. Es gibt n\"amlich ein $b > 0$
mit
$$
\beginAligntags
" \Vert (\hat{X}^2 - [\hat{X}^{2 m} / B_m]) \Psi \Vert
" \; \le \; " 2^{1/2} \,
\Vert \hat{\bf H}^{(m)}_0 \Psi \Vert \, + \, b \, \Vert \Psi \Vert \\
" " \; = \; " 2^{1/2} \,
\Vert (\hat{P}^2 + [\hat{X}^{2m} / B_m]) \Psi \Vert
\, + \, b \, \Vert \Psi \Vert \, .
\\ \tag
\endAligntags
$$

\endEinzug

\medskip

\noindent {\it Beweis:\/} Man kann eine Konstante $c > 0$ so w\"ahlen, da{\ss}
$$
\left\vert \frac {x^{2 m}} {B_m} \, - \, x^2 \right\vert \; \le \;
\frac {x^{2 m}} {B_m} \, + \, c
\tag
$$
f\"ur alle $x \in \R$ gilt. Wenn man diese Beziehung als
Operatorungleichung interpretiert und auf eine Funktion $\Psi \in {\cal
D} (\hat{P}^2) \cap {\cal D} (\hat{X}^{2 m})$ anwendet, erh\"alt man mit
Hilfe der Dreiecksungleichung f\"ur Normen [Mitrinovi\'{c},
Pe\v{c}ari\'{c} und Fink 1993, Abschnitt \Roemisch{17}] die folgende
Absch\"atzung:
$$
\Vert (\hat{X}^2 - [X^{2 m} / B_m]) \Psi \Vert \; \le \;
\frac {1} {B_m} \, \Vert \hat{X}^{2 m} \Psi \Vert
\, + \, c \, \Vert \Psi \Vert \, .
\tag
$$
Wir definieren jetzt:
$$
\beginAligntags
" {\bf A} " \; = \; "
\frac {1} {B_m} \, \Vert \hat{X}^{2 m} \Psi \Vert \, ,
\\ \tag
" {\bf B} " \; = \; " \Bigl\{ \Vert \hat{P}^2 \Psi \Vert^2 \, + \,
\frac {1} {(B_m)^2} \, \Vert \hat{X}^{2 m} \Psi \Vert^2
\Bigr\}^{1/2} \, ,
\\ \tag
" {\bf C} " \; = \; "
\Bigl\{ 2 \, \Vert (\hat{P}^2 + [\hat{X}^{2 m} / B_m]) \Psi \Vert^2
\, + \, b_0 \, \Vert \Psi \Vert^2 \Bigr\}^{1/2} \, ,
\\ \tag
" {\bf D} " \; = \; " 2^{1/2} \,
\Vert (\hat{P}^2 + [\hat{X}^{2 m} / B_m]) \Psi \Vert
\, + \, b_0^{1/2} \, \Vert \Psi \Vert \, .
\\ \tag
\endAligntags
$$
Aus den Gln. (10.8-14) und (10.8-26) - (10.8-29) erhalten wir die
Absch\"atzungen
$$
{\bf A}^2  \; \le \; {\bf B}^2 \; \le \; {\bf C}^2
\; \le \; {\bf D}^2 \, .
\tag
$$
Die Ungleichung ${\bf D}^2 - {\bf A}^2 = ({\bf D}-{\bf A})({\bf
D}+{\bf A}) \ge 0$ impliziert aber auch ${\bf D} \ge {\bf A}$ oder
$$
\frac {1} {B_m} \, \Vert \hat{X}^{2 m} \Psi \Vert \; \le \;
2^{1/2} \, \Vert (\hat{P}^2 + [\hat{X}^{2 m} / B_m]) \Psi \Vert
\, + \, b_0^{1/2} \, \Vert \Psi \Vert \, .
\tag
$$
Wenn wir diese Beziehung in Gl. (10.8-25) verwenden, erhalten wir:
$$
\Vert (\hat{X}^2 - [\hat{X}^{2 m} / B_m] \Psi \Vert \; \le \;
2^{1/2} \, \Vert (\hat{P}^2 + [\hat{X}^{2 m} / B_m]) \Psi \Vert
\, + \, b_{0}^{1/2} \, \Vert \Psi \Vert \, + \,
c \, \Vert \Psi \Vert \, .
\tag
$$
Wir m\"ussen jetzt nur noch $b = b_0^{1/2} + c$ setzen, um die Absch\"atzung
(10.8-23) zu erhalten. \BeweisEnde

Aus den S\"atzen 10-3 und 10-4 folgt, da{\ss} die {\it renormalized strong
coupling expansion\/} (10.8-8) f\"ur ${\cal E}_{R}^{(m)} (\tau)$ in einer
Umgebung von $\tau = 0$ konvergiert. F\"ur gro{\ss}e Werte von $\beta$ oder
kleine Werte von $\tau$ sollten demzufolge schon einige Terme der
St\"orungsreihe (10.8-10) ausreichen, um die Grundzustandsenergie $E^{(m)}
(\beta)$ ausreichend genau berechnen zu k\"onnen. Die St\"orungsreihe
(10.8-10) f\"ur $E^{(m)} (\beta)$ besitzt also im Prinzip wesentlich
g\"unstigere Eigenschaften als die komplement\"are renormierte St\"orungsreihe
(10.4-28), die f\"ur alle $\beta > 0$ und damit auch f\"ur alle $\kappa > 0$
eine hochgradig divergente Reihe ist, die mit Hilfe geeigneter
verallgemeinerter Summationsprozesse summiert werden mu{\ss}, wenn man sie
f\"ur numerische Zwecke verwenden will.

Allerdings ist die {\it renormalized strong coupling expansion\/}
(10.8-10) ebenso wie die {\it strong coupling expansion\/} (10.3-23)
praktisch nutzlos, solange es nicht gelingt, eine gen\"ugend gro{\ss}e Anzahl
von St\"orungstheoriekoeffizienten mit ausreichender Genauigkeit zu
berechnen. Im Falle der St\"orungsreihe (10.3-23) ist die Berechnung der
St\"orungstheoriekoeffizienten $K_{n}^{(m)}$ trotz einiger Versuche
[Fern\'{a}ndez 1992b; Guardiola, Sol\'{\i}s und Ros 1992; Hioe,
MacMillen und Montroll 1978; Hioe und Montroll 1975; Turbiner und
Ushveridze 1988] ein bis heute nicht befriedigend gel\"ostes Problem.
Meiner Meinung nach besteht das Hauptproblem darin, da{\ss} die
Grundzustandsenergie $E^{(m)} (\beta)$ eines anharmonischen Oszillators
f\"ur $\beta \to \infty$ gem\"a{\ss} Gl. (10.3-15) wie $\beta^{1/(m+1)}$
divergiert. Demzufolge sind analytische Manipulationen der
Grundzustandsenergie $E^{(m)} (\beta)$ extrem schwierig, wenn $\beta$
gegen Unendlich geht. Deswegen kann man die St\"orungsreihe (10.2-4) auch
nicht unmittelbar zur Berechnung der St\"orungstheorieokeffizienten
$K_{n}^{(m)}$ in Gl. (10.3-23) verwenden.

Das in Unterabschnitt 10.4 beschriebene Renormierungsverfahren f\"uhrt zu
einer Aufspaltung der Grundzustandsenergie $E^{(m)} (\beta)$ gem\"a{\ss} Gl.
(10.4-19) in den renormierten Energieeigenwert $E_{R}^{(m)} (\kappa)$,
der f\"ur $\kappa \to 1$ endlich bleibt und analytischen Manipulationen
zug\"anglich ist, und in einen Anteil $(1 - \kappa)^{- 1/2}$, der f\"ur
$\kappa \to 1$ gem\"a{\ss} Gl. (10.4-29) so w\"achst wie $\beta^{1/(m+1)}$ f\"ur
$\beta \to \infty$. Aufgrund dieser Aufspaltung k\"onnen die Koeffizienten
$\Gamma_{n}^{(m)}$ in Gl. (10.8-8) auf vergleichsweise einfache Weise
aus den Koeffizienten $c_{n}^{(m)}$ der renormierten St\"orungsreihe
(10.4-20) f\"ur $E_{R}^{(m)} (\kappa)$ berechnet werden.

Aus Gl. (10.4-11) folgt
$$
\tau^2 \; = \; 1 \, - \, \kappa \, .
\tag
$$
Die St\"orungsreihen (10.8-8) und (10.8-10) k\"onnen also auch
folgenderma{\ss}en geschrieben werden:
$$
\beginAligntags
" {\cal E}_{R}^{(m)} \bigl( (1 - \kappa)^{1/2} \bigr) " \; = \;
" \sum_{n=0}^{\infty} \, \Gamma_{n}^{(m)} \, (1 - \kappa)^n \, ,
\\ \tag
" E^{(m)} (\beta) " \; = \; " (1 - \kappa)^{- 1/2} \,
\sum_{n=0}^{\infty} \, \Gamma_{n}^{(m)} \, (1 - \kappa)^n \, .
\\ \tag
\endAligntags
$$
Ein Vergleich der Gln. (10.4-19) und (10.8-35) ergibt aber
$$
{\cal E}_{R}^{(m)} \bigl( (1 - \kappa)^{1/2} \bigr) \; = \;
E_{R}^{(m)} (\kappa)
\tag
$$
oder
$$
E_{R}^{(m)} (\kappa) \; = \;
\sum_{n=0}^{\infty} \, \Gamma_{n}^{(m)} \, (1 - \kappa)^n \, .
\tag
$$

Die Koeffizienten $\Gamma_{n}^{(m)}$ der formalen Potenzreihe (10.8-37)
k\"onnen durch Koeffizientenvergleich bestimmt werden. Dazu dr\"uckt man
$\kappa^n$ mit Hilfe des Binomialtheorems [Abramowitz und Stegun 1972,
Gl. (3.1.1)] auf folgende Weise aus:
$$
\kappa^n \; = \; \bigl[ 1 - (1 - \kappa) \bigr]^n \; = \;
\sum_{j=0}^{n} \, (- 1)^j \, \binom {n} {j} \,(1 - \kappa)^j \, .
\tag
$$
Im n\"achsten Schritt wird diese Beziehung auf der rechten Seite von Gl.
(10.4-20) verwendet. Man erh\"alt dann:
$$
\beginAligntags
" \sum_{n=0}^{\infty} \, c_{n}^{(m)} \, \kappa^n
" \; = \; "
\sum_{n=0}^{\infty} \, c_{n}^{(m)} \,
\sum_{j=0}^{n} \, (- 1)^j \, \binom {n} {j} \,(1 - \kappa)^j
\\ \tag
" " \; = \; "
\sum_{j=0}^{\infty} \, \frac {(- 1)^j (1 - \kappa)^j} {j!} \,
\sum_{n=j}^{\infty} \, \frac {n!} {(n-j)!} \, c_{n}^{(m)}
\\ \tag
" " \; = \; " \sum_{\mu=0}^{\infty} \,
\frac {(- 1)^{\mu} (1 - \kappa)^{\mu}} {{\mu}!} \,
\sum_{\nu=0}^{\infty} \, \frac {(\mu + \nu)!} {{\nu}!}
\, c_{\mu + \nu}^{(m)}
\\ \tag
" " \; = \; " \sum_{\mu=0}^{\infty} \,
\frac {(- 1)^{\mu} (1 - \kappa)^{\mu}} {{\mu}!} \,
\sum_{\nu=0}^{\infty} \, (\nu + 1)_{\mu} \, c_{\mu + \nu}^{(m)} \, .
\\ \tag
\endAligntags
$$
Wenn man dieses Ergebnis in Gl. (10.4-20) verwendet, folgt:
$$
E_{R}^{(m)} (\kappa) \; = \; \sum_{\mu=0}^{\infty} \,
\frac {(- 1)^{\mu} (1 - \kappa)^{\mu}} {{\mu}!} \,
\sum_{\nu=0}^{\infty} \, (\nu+1)_{\mu} \, c_{\mu + \nu}^{(m)} \, .
\tag
$$
Durch Vergleich der unendlichen Reihen (10.8-37) und (10.8-43) erh\"alt
man die folgende Dar\-stellung der St\"orungs\-theorie\-koeffizienten
$\Gamma_{n}^{(m)}$ durch eine divergente Reihe:
$$
\Gamma_{n}^{(m)} \; = \; \frac {(- 1)^n} {n!} \,
\sum_{\nu=0}^{\infty} \, (\nu+1)_{n} \, c_{n + \nu}^{(m)} \, .
\tag
$$

Durch $n$-fache Differentiation der St\"orungsreihe (10.4-20) f\"ur
$E_{R}^{(m)} (\kappa)$ nach $\kappa$ an der Stelle $\kappa = 1$ findet
man, da{\ss} $\Gamma_{n}^{(m)}$ proportional zur $n$-ten Ableitung der
renormierten Energie $E_{R}^{(m)} (\kappa)$ f\"ur den Grenzfall
unendlicher Kopplung $\kappa = 1$ ist:
$$
\Gamma_{n}^{(m)} \; = \; \frac {(- 1)^n} {n!} \,
\frac {\d^n} {\d \kappa^n} \, E_{R}^{(m)} (\kappa) \,
\Big\vert_{\kappa=1} \, .
\tag
$$
Wenn man diese Beziehung in Gl. (10.8-37) einsetzt, folgt:
$$
E_{R}^{(m)} (\kappa) \; = \; \sum_{n=0}^{\infty} \,
\frac {(\kappa - 1)^n} {n!} \,
\frac {\d^n} {\d \kappa^n} \, E_{R}^{(m)} (\kappa) \,
\Big\vert_{\kappa=1} \, .
\tag
$$
Diese Beziehung zeigt, da{\ss} die {\it renormalized strong coupling
expansion\/} (10.8-37) eine formale Taylorentwicklung der renormierten
Energie $E_{R}^{(m)} (\kappa)$ um $\kappa = 1$ ist, was dem Grenzfall
unendlicher Kopplung entspricht. Im Gegensatz dazu kann die renormierte
St\"orungsreihe (10.4-20) als {\it weak coupling expansion\/}
interpretiert werden kann, da sie eine formale Taylorentwicklung der
renormierten Energie $E_{R}^{(m)} (\kappa)$ um $\kappa = 0$ ist, was dem
Grenzfall verschwindender Kopplung entspricht. Analog kann die
Rayleigh-Schr\"odingersche St\"orungsreihe (10.2-4) als eine formale
Taylorentwicklung von $E^{(m)} (\beta)$ um den Grenzfall verschwindender
Kopplung $\beta = 0$ interpretiert werden, und die {\it strong coupling
expansion\/} (10.3-23) als eine formale Taylorentwicklung des Quotienten
$E^{(m)} (\beta) / \beta^{1/(m+1)}$ um $1/\beta = 0$.

Es gibt aber einen wesentlichen Unterschied zwischen der
Rayleigh-Schr\"odingerschen St\"orungs\-reihe (10.2-4) und der renormierten
St\"orungsreihe (10.4-20): Sowohl $E_{R}^{(m)} (\kappa)$ als auch h\"ohere
Ableitungen von $E_{R}^{(m)} (\kappa)$ k\"onnen auch im Grenzfall
unendlicher Kopplung $\kappa = 1$ durch Summation berechnet werden. Eine
analoge Vorgehensweise ist bei der St\"orungsreihe (10.2-4) nicht m\"oglich,
da man eine divergente Potenzreihe in $\beta$ nicht f\"ur $\beta \to
\infty$ summieren kann.

Bei der Ableitung der formalen Taylorentwicklung (10.8-46) wurde die
Frage, ob und unter welchen Voraussetzungen eine divergente, aber
summierbare Potenzreihe gliedweise differenziert werden kann, v\"ollig
ignoriert. Eine Diskussion dieser Problematik findet man beispielsweise
in \S 9 eines Artikels von Hardy [1904].

Die praktische N\"utzlichkeit der {\it renormalized strong coupling
expansion\/} (10.8-35) h\"angt ganz entscheidend davon ab, wie viele
Koeffizienten $\Gamma_{n}^{(m)}$ man durch Summation der divergenten
Reihe (10.8-44) mit ausreichender Genauigkeit berechnen kann.
Offensichtlich divergiert die Reihe (10.8-44) f\"ur $\Gamma_{n}^{(m)}$ f\"ur
alle $n \in \N_0$ mindestens so stark wie die renormierte St\"orungsreihe
(10.4-20) f\"ur $E_{R}^{(m)} (1)$. Man kann also erwarten, da{\ss} die
Berechnung der St\"orungstheoriekoeffizienten $\Gamma_{n}^{(m)}$ durch
Summation der divergenten Reihe (10.8-44) mindestens so schwierig ist
wie die Berechnung der Grenzf\"alle unendlicher Kopplung $k_m$ gem\"a{\ss} Gln.
(10.4-30) und (10.4-31).

Wie im letzten Unterabschnitt gezeigt wurde, erh\"alt man f\"ur die
Grenzf\"alle unendlicher Kopplung $k_m$ die besten Summationsergebnisse,
wenn man die renormierte St\"orungsreihe (10.2-40) f\"ur $\kappa = 1$ mit
Hilfe des verallgemeinerten Summationsprozesses ${\delta}_k^{(n)}
(\zeta, s_n)$, Gl.(5.4-13), summiert. Es ist also eine relativ
naheliegende Idee, die Folge der Partialsummen
$$
s_{n}^{(m, \nu)} \; = \; \frac {(- 1)^\nu} {\nu!} \,
\sum_{j=0}^{n} \, (j+1)_{\nu} \, c_{\nu + j}^{(m)} \, .
\tag
$$
der divergenten Reihe (10.8-44) f\"ur $\Gamma_{n}^{(m)}$ ebenfalls durch
diesen verallgemeinerten Summationsproze{\ss} zu summieren.

In Tabelle 10-19 werden numerische Werte f\"ur die
St\"orungstheoriekoeffizienten $\Gamma_{\nu}^{(2)}$ mit $0 \le \nu \le 20$
aufgelistet, die durch Summation der Partialsummen $s_{n}^{(2, \nu)}$
mit Hilfe des verallgemeinerten Summationsprozesses ${\delta}_k^{(n)}
\bigl(\zeta, s_n^{(2, \nu)} \bigr)$, Gl. (5.4-13), mit $\zeta = 1$
berechnet wurden. Dabei wurden immer alle verf\"ugbaren
St\"orungstheoriekoeffizienten $c_{\nu}^{(2)}$ mit $0 \le \nu \le 200$
verwendet. Zur Berechnung des St\"orungs\-theorie\-koeffizienten
$\Gamma_{n}^{(2)}$ standen demzufolge alle Partialsummen $s_{\nu}^{(2,
n)}$ mit $0 \le \nu \le 200 - n - 1$ zur Verf\"ugung. Alle Rechnungen
wurden in MAPLE mit einer Genauigkeit von 400 Dezimalstellen
durchgef\"uhrt.

\beginFloat

\medskip

\beginTabelle
\beginFormat \rechts " \rechts \endFormat
\+ " \links {\bf Tabelle 10-19} \@ " \\
\+ " \links {Anharmonischer Oszillator mit einer
$\hat{x}^4$-Anharmonizit\"at} \@ " \\
\+ " \links {Renormierte St\"orungstheoriekoeffizienten
$\Gamma_{n}^{(2)}$} \@ " \\
\+ " \links {Berechnung durch Summation der divergenten Reihe (10.8-44)}
\@ " \\
\- " \- " \- " \\ \sstrut {} {1.5 \jot} {1.5 \jot}
\+ " \rechts {$n$} " \mitte {$\Gamma_{n}^{(2)}$} " \\
\- " \- " \- " \\ \sstrut {} {1 \jot} {1 \jot}
\+ "  0 " $  0.735~214~010~331~216~080~772~291~445~276~89$ \\
\+ "  1 " $  0.277~055~672~879~946~971~403~937~539~329~61$ \\
\+ "  2 " $- 0.011~178~897~209~645~025~734~252~454~055~43$ \\
\+ "  3 " $- 0.000~466~149~311~582~119~933~320~771~099~17$ \\
\+ "  4 " $- 0.000~293~444~235~328~683~814~525~579~234~90$ \\
\+ "  5 " $- 0.000~148~065~256~807~374~148~499~884~645~99$ \\
\+ "  6 " $- 0.000~076~915~442~996~337~825~207~489~433~94$ \\
\+ "  7 " $- 0.000~042~320~618~648~808~367~469~055~867~96$ \\
\+ "  8 " $- 0.000~024~285~794~062~144~242~438~164~268~23$ \\
\+ "  9 " $- 0.000~014~425~057~950~710~159~752~050~632~78$ \\
\+ " 10 " $- 0.000~008~820~357~960~483~973~108~096~772~82$ \\
\+ " 11 " $- 0.000~005~528~678~496~533~919~643~979~804~30$ \\
\+ " 12 " $- 0.000~003~540~659~029~089~413~300~771~355~~~$ \\
\+ " 13 " $- 0.000~002~310~600~583~349~358~753~376~00~~~~$ \\
\+ " 14 " $- 0.000~001~533~235~811~195~624~243~65~~~~~~~~$ \\
\+ " 15 " $- 0.000~001~032~678~241~419~866~930~9~~~~~~~~~$ \\
\+ " 16 " $- 0.000~000~704~933~716~265~410~645~~~~~~~~~~~$ \\
\+ " 17 " $- 0.000~000~487~094~742~781~166~8~~~~~~~~~~~~~$ \\
\+ " 18 " $- 0.000~000~340~325~648~653~58~~~~~~~~~~~~~~~~$ \\
\+ " 19 " $- 0.000~000~240~209~641~371~4~~~~~~~~~~~~~~~~~$ \\
\+ " 20 " $- 0.000~000~171~140~073~87~~~~~~~~~~~~~~~~~~~~$ \\
\- " \- " \- " \\ \sstrut {} {1 \jot} {1 \jot}

\endTabelle

\medskip

\endFloat

Bei den Rechnungen f\"ur Tabelle 10-19 wurde versucht, m\"oglichst viele
Koeffizienten $\Gamma_{n}^{(2)}$ mit einer {\it absoluten\/} Genauigkeit
von 32 Dezimalstellen{\footnote[\dagger]{Dies entspricht der Genauigkeit
von FORTRAN QUADRUPLE PRECISION auf einem IBM-Gro{\ss}recher mit dem
Betriebssystem VM/CMS.}} zu berechnen. Dies gelang aber nur f\"ur $n \le
11$. F\"ur $n \ge 12$ nimmt die Genauigkeit der
St\"orungstheoriekoeffizienten $\Gamma_{n}^{(2)}$ mit wachsendem $n$ rasch
ab. Hier macht sich bemerkbar, da{\ss} die Reihe (10.8-44) f\"ur h\"ohere Werte
von $n$ nicht nur st\"arker divergiert, sondern da{\ss} zur Berechnung von
$\Gamma_{n}^{(2)}$ nur noch die renormierten
St\"orungstheorie\-koeffi\-zi\-enten $c_{n + \nu}^{(2)}$ mit $0 \le \nu
\le 200 - n$ zur Verf\"ugung stehen.

Am Anfang dieses Unterabschnittes wurde spekuliert, da{\ss} die {\it
renormalized strong coupling expansion\/} (10.8-8) f\"ur den
Grundzustandseigenwert ${\cal E}_{R}^{(m)} (\tau)$ des in Gl. (10.8-4)
definierten Hamiltonoperators ${\bf H}^{(m)} (\tau)$ im Gegensatz zur
{\it strong coupling expansion\/} (10.3-23) f\"ur alle physika\-lisch
relevanten renormierten Kopplungskonstanten $\tau \in [0, 1]$
konvergieren k\"onnte, weil ${\bf H}^{(m)} (\tau)$ sowohl im
kopplungsfreien Fall $\tau = 1$ als auch im Grenzfall unendlicher
Kopplung $\tau = 0$ ein mathematisch sinnvolles Objekt ist. Aus den Gln.
(10.8-2) - (10.8-4) folgt, da{\ss} die Grundzustandsenergie ${\cal
E}_{R}^{(m)} (\tau)$ des Hamiltonoperators ${\bf H}^{(m)} (\tau)$ im
kopplungsfreien Fall $\tau = 1$ mit der Grundzustandsenergie eines
ungest\"orten harmonischen Oszillators identisch ist. Es gilt also
$$
{\cal E}_{R}^{(m)} (1) \; = \; 1
\tag
$$
oder
$$
\sum_{\nu=0}^{\infty} \, \Gamma_{\nu}^{(m)} \; = \; 1 \, .
\tag
$$

Wenn die St\"orungsreihe (10.8-8) tats\"achlich f\"ur alle physikalisch
relevanten renormierten Kopplungskonstanten $\tau = [0, 1]$ konvergiert,
dann mu{\ss} die Normierungsrelation (10.8-49) ebenfalls erf\"ullt sein. In
Tabelle 10-20 wird die Konvergenz der Partialsummen
$$
\Sigma_{n}^{(m)} \; = \; \sum_{\nu=0}^{n} \, \Gamma_{\nu}^{(m)}
\tag
$$
der unendlichen Reihe (10.8-49) f\"ur $m = 2$ numerisch \"uberpr\"uft.
Au{\ss}erdem werden der Wynnsche $\epsilon$-Algorithmus, Gl. (2.4-10), der
$\theta$-Algorithmus von Brezinski, Gl. (4.4-13), und seine Iteration
${\cal J}_{k}^{(n)}$, Gl. (8.4-18), zur Beschleunigung der Konvergenz
verwendet.

\beginFloat

\medskip

\beginTabelle [to \kolumnenbreite]
\beginFormat \rechts " \rechts " \rechts " \rechts " \rechts
\endFormat
\+ " \links {\bf Tabelle 10-20} \@ \@ \@ \@ " \\
\+ " \links {Anharmonischer Oszillator mit einer
$\hat{x}^4$-Anharmonizit\"at} \@ \@ \@ \@ " \\
\+ " \links {Renormalized Strong Coupling Expansion (10.8-35)}
\@ \@ \@ \@ " \\
\+ " \links {Konvergenz der Normierungsrelation (10.8-49)}
\@ \@ \@ \@ " \\
\- " \- " \- " \- " \- " \- " \\ \sstrut {} {1 \jot} {1 \jot}
\+ " $n$ " \mitte{$\Sigma_n^{(m)}$}
" \mitte{$\epsilon_{2 \Ent {n/2}}^{(n - 2 \Ent {n/2})}$}
" \mitte{$\theta_{2 \Ent {n/3}}^{(n - 3 \Ent {n/3})}$}
" \mitte{${\cal J}_{\Ent {n/3}}^{(n - 3 \Ent {n/3})}$} " \\
\+ " " \mitte{Gl. (10.8-50)} " \mitte{Gl. (2.4-10)}
" \mitte{Gl. (4.4-13)} " \mitte{Gl. (8.4-18)} " \\
\- " \- " \- " \- " \- " \- " \\ \sstrut {} {1 \jot} {1 \jot}
\+ "  0 " 0.73521401033122 " 0.73521401033122 " 0.73521401033122
" 0.73521401033122 \\
\+ "  1 " 1.01226968321116 " 1.01226968321116 " 1.01226968321116
" 1.01226968321116 \\
\+ "  2 " 1.00109078600152 " 1.00152434864734 " 1.00109078600152
" 1.00109078600152 \\
\+ "  3 " 1.00062463668994 " 1.00060435289509 " 1.00056073064240
" 1.00056073064240 \\
\+ "  4 " 1.00033119245461 " 1.00016276804237 " 1.00183276077985
" 1.00183276077985 \\
\+ "  5 " 1.00018312719780 " 1.00003101102118 " 1.00015336227328
" 1.00015336227328 \\
\+ "  6 " 1.00010621175480 " 1.00002299099165 " 0.99881799035120
" 0.99969984724536 \\
\+ "  7 " 1.00006389113615 " 1.00005556027141 " 0.99999113722276
" 0.99999384366168 \\
\+ "  8 " 1.00003960534209 " 1.00000211248569 " 0.99999826317426
" 0.99999815529995 \\
\+ "  9 " 1.00002518028414 " 1.00000154792974 " 1.00000204700707
" 1.00001017672140 \\
\+ " 10 " 1.00001635992618 " 1.00001609537184 " 1.00000062338577
" 1.00000082018019 \\
\+ " 11 " 1.00001083124768 " 1.00000023405850 " 1.00000045841046
" 1.00000053638914 \\
\+ " 12 " 1.00000729058866 " 1.00000013303662 " 1.00000048549501
" 1.00000057943773 \\
\+ " 13 " 1.00000497998807 " 1.00000005099980 " 1.00000048506886
" 1.00000056673334 \\
\+ " 14 " 1.00000344675226 " 1.00000002135071 " 1.00000024644100
" 1.00000021627657 \\
\+ " 15 " 1.00000241407402 " 1.00000001025699 " 1.00000048277847
" 1.00000055870953 \\
\+ " 16 " 1.00000170914030 " 1.00000000488650 " 1.00000000654683
" 1.00000000043488 \\
\+ " 17 " 1.00000122204556 " 1.00000000145736 " 1.00000000115994
" 1.00000000049190 \\
\+ " 18 " 1.00000088171991 " 1.00000000106639 " 1.00000000426423
" 1.00000000046535 \\
\+ " 19 " 1.00000064151027 " 1.00000000120513 " 1.00000000003117
" 1.00000000020586 \\
\+ " 20 " 1.00000047037020 " 1.00000000109679 " 1.00000000008496
" 1.00000000012512 \\
\- " \- " \- " \- " \- " \- " \\ \sstrut {} {1 \jot} {1 \jot}

\endTabelle

\medskip

\endFloat

Die Ergebnisse in Tabelle 10-20 zeigen, da{\ss} die in Gl. (10.8-50)
definierten Partialsummen $\Sigma_{n}^{(2)}$ anscheinend gegen Eins
konvergieren. Die in Tabelle 10-20 verwendeten verallgemeinerten
Summationsprozesse beschleunigen offensichtlich die Konvergenz der
Partialsummen gegen Eins, obwohl die Verbesserung der Konvergenz nicht
spektakul\"ar ist. Nichtsdestoweniger ist es bemerkenswert, da{\ss} {\it
verschiedene\/} verallgemeinerte Summationsprozesse zu einer {\it
stabilen\/} Verbesserung der Konvergenz f\"uhren. Die Ergebnisse in
Tabelle 10-20 sind sicherlich kein {\it definitiver\/} Beweis daf\"ur, da{\ss}
die Koeffizienten $\Gamma_{n}^{(2)}$ tats\"achlich die Normierungsrelation
(10.8-49) erf\"ullen. Trotzdem machen diese Ergebnisse es \"au{\ss}erst
wahrscheinlich, da{\ss} die {\it renormalized strong coupling expansion\/}
(10.8-8) im Falle des anharmonischen Oszillators mit einer
$\hat{x}^4$-Anharmonizit\"at tats\"achlich f\"ur alle physikalisch relevanten
renormierten Kopplungskonstanten $\tau \in [0, 1]$ konvergiert.

Die {\it renormalized strong coupling expansion\/} (10.8-8) f\"ur ${\cal
E}_{R}^{(m)} (\tau)$ ist eine Potenzreihe in der Variablen $\tau^2$ mit
$\tau \in [0, 1]$. Diese Reihe konvergiert am schlechtesten im
kopplungsfreien Fall $\tau = 1$, und sie konvergiert am besten im
Grenzfall unendlicher Kopplung $\tau = 0$, da dann nur der konstante
Term $\Gamma_{0}^{(m)}$ von Null verschieden ist. In Tabelle 10-21 wird
gezeigt, wie gut die Partialsummen
$$
\Sigma_n^{(m)} (\tau) \; = \; \tau^{- 1} \,
\sum_{\nu = 0}^{n} \, \Gamma_{\nu}^{(m)} \, \tau^{2 \nu} \, .
\tag
$$
der St\"orungsreihe (10.8-10) f\"ur $E^{(m)} (\beta)$ f\"ur $m = 2$ und f\"ur
verschiedene Werte von $\beta$ konvergieren. Alle Rechnungen wurden in
MAPLE mit einer Genauigkeit von 32 Dezimalstellen durchgef\"uhrt. F\"ur
jeden Wert von $\beta$ wurde die entsprechende renormierte
Kopplungskonstante $\tau$ gem\"a{\ss} Gl. (10.4-9) mit Hilfe des Kommandos
{\it fsolve\/} [Char, Geddes, Gonnet, Leong, Monagan und Watt 1991b, S.
97] berechnet. Die \anf{exakten} Energien in Tabelle 10-21 wurden aus
Tabelle 10-9 entnommen.

\beginFloat

\medskip

\beginTabelle [to \kolumnenbreite]
\beginFormat \links " \rechts " \rechts " \rechts " \rechts
\endFormat
\+ " \links {\bf Tabelle 10-21} \@ \@ \@ \@ " \\
\+ " \links {Anharmonischer Oszillator mit einer
$\hat{x}^4$-Anharmonizit\"at} \@ \@ \@ \@ " \\
\+ " \links {Renormalized Strong Coupling Expansion (10.8-10)}
\@ \@ \@ \@ " \\
\+ " \links {Berechnung der Grundzustandsenergie $E^{(2)} (\beta)$}
\@ \@ \@ \@ " \\
\+ " \links {Konvergenz der Partialsummen $\Sigma^{(2)}_{n} (\tau)$,
Gl. (10.8-51), f\"ur verschiedene Werte von $\beta$} \@ \@ \@ \@ " \\
\- " \- " \- " \- " \- " \- " \\ \sstrut {} {1 \jot} {1 \jot}
\+ " " \mitte{$\beta = 2/10$}
" \mitte{$\beta = 1$}
" \mitte{$\beta = 4$}
" \mitte{$\beta = 100$} " \\
\- " \- " \- " \- " \- " \- " \\ \sstrut {} {1 \jot} {1 \jot}
\+ " $\Sigma^{(2)}_{0} (\tau)$ " 0.89784091305038 " 1.22905717406338
" 1.79012948684709 " 4.95837296372249 \\
\+ " $\Sigma^{(2)}_{1} (\tau)$ " 1.12471319006773 " 1.39479007619489
" 1.90391746354918 " 4.99945402273789 \\
\+ " $\Sigma^{(2)}_{2} (\tau)$ " 1.11857497485217 " 1.39239717907118
" 1.90314302502752 " 4.99941757901686 \\
\+ " $\Sigma^{(2)}_{3} (\tau)$ " 1.11840334336663 " 1.39236147370523
" 1.90313757784800 " 4.99941754560521 \\
\+ " $\Sigma^{(2)}_{4} (\tau)$ " 1.11833089538973 " 1.39235343071761
" 1.90313699944455 " 4.99941754514277 \\
\+ " $\Sigma^{(2)}_{5} (\tau)$ " 1.11830638317175 " 1.39235197851117
" 1.90313695021599 " 4.99941754513765 \\
\+ " $\Sigma^{(2)}_{6} (\tau)$ " 1.11829784486493 " 1.39235170856814
" 1.90313694590243 " 4.99941754513759 \\
\+ " $\Sigma^{(2)}_{7} (\tau)$ " 1.11829469465870 " 1.39235165541928
" 1.90313694550208 " 4.99941754513759 \\
\+ " $\Sigma^{(2)}_{8} (\tau)$ " 1.11829348247609 " 1.39235164450544
" 1.90313694546333 " 4.99941754513759 \\
\+ " $\Sigma^{(2)}_{9} (\tau)$ " 1.11829299968185 " 1.39235164218577
" 1.90313694545945 " 4.99941754513759 \\
\+ " $\Sigma^{(2)}_{10} (\tau)$ " 1.11829280173004 " 1.39235164167822
" 1.90313694545905 " 4.99941754513759 \\
\+ " $\Sigma^{(2)}_{11} (\tau)$ " 1.11829271853003 " 1.39235164156438
" 1.90313694545900 " 4.99941754513759 \\
\+ " $\Sigma^{(2)}_{12} (\tau)$ " 1.11829268280152 " 1.39235164153829
" 1.90313694545900 " 4.99941754513759 \\
\+ " $\Sigma^{(2)}_{13} (\tau)$ " 1.11829266716700 " 1.39235164153220
" 1.90313694545900 " 4.99941754513759 \\
\+ " $\Sigma^{(2)}_{14} (\tau)$ " 1.11829266021040 " 1.39235164153075
" 1.90313694545900 " 4.99941754513759 \\
\+ " $\Sigma^{(2)}_{15} (\tau)$ " 1.11829265706857 " 1.39235164153041
" 1.90313694545900 " 4.99941754513759 \\
\+ " $\Sigma^{(2)}_{16} (\tau)$ " 1.11829265563046 " 1.39235164153032
" 1.90313694545900 " 4.99941754513759 \\
\+ " $\Sigma^{(2)}_{17} (\tau)$ " 1.11829265496413 " 1.39235164153030
" 1.90313694545900 " 4.99941754513759 \\
\+ " $\Sigma^{(2)}_{18} (\tau)$ " 1.11829265465195 " 1.39235164153029
" 1.90313694545900 " 4.99941754513759 \\
\+ " $\Sigma^{(2)}_{19} (\tau)$ " 1.11829265450421 " 1.39235164153029
" 1.90313694545900 " 4.99941754513759 \\
\+ " $\Sigma^{(2)}_{20} (\tau)$ " 1.11829265443362 " 1.39235164153029
" 1.90313694545900 " 4.99941754513759 \\
\- " \- " \- " \- " \- " \- " \\ \sstrut {} {1 \jot} {1 \jot}
\+ " Exakt                 "  1.11829265436704 " 1.39235164153029
" 1.90313694545900 " 4.99941754513759 \\
\- " \- " \- " \- " \- " \- " \\ \sstrut {} {1 \jot} {1 \jot}

\endTabelle

\medskip

\endFloat

Die Ergebnisse in Tabelle 10-21 zeigen, da{\ss} die {\it renormalized strong
coupling expansion\/} (10.8-10) selbst f\"ur so kleine Kopplungskonstanten
wie beispielsweise $\beta = 2/10$ schon relativ schnell gegen die
Grundzustandsenergie $E^{(2)} (\beta)$ konvergiert. F\"ur gr\"o{\ss}ere Werte
von $\beta$ wird die Konvergenz extrem schnell. Bemerkenswert ist, da{\ss}
man die Grundzustandsenergie $E^{(2)} (\beta)$ eines anharmonischen
Oszillators mit einer $\hat{x}^4$-Anharmonizit\"at mit Hilfe der {\it
renormalized strong coupling expansion\/} (10.8-10) auf extrem einfache
Weise berechnen kann. Der schwierigste Schritt ist die numerische L\"osung
der nichtlinearen Gleichung (10.4-9) zur Bestimmung der renormierten
Kopplungskonstante $\tau$. Ansonsten mu{\ss} man nur noch den Wert eines
Polynoms in $\tau^2$ mit tabellierten Koeffizienten
bestimmen{\footnote[\dagger]{Mein Taschenrechner vom Typ HP-28C besitzt
eine {\it solve\/}-Funktion, die nichtlineare Gleichungen l\"osen kann.
Man w\"urde also nur einen etwas leistungsf\"ahigeren Taschenrechner
ben\"otigen, um die Grundzustandsenergie $E^{(2)} (\beta)$ f\"ur alle $\beta
\in [0, \infty)$ unter Verwendung der {\it renormalized strong coupling
expansion\/} (10.8-10) berechnen zu k\"onnen. Die einzige wesentliche
Einschr\"ankung best\"unde darin, da{\ss} Taschenrechner normalerweise mit einer
festen Stellenzahl von 10 Dezimalstellen rechnen.}}. Mir ist zur Zeit
kein anderes Verfahren bekannt, das $E^{(2)} (\beta)$ auf \"ahnlich
einfache Weise auch f\"ur extrem gro{\ss}e Kopplungskonstanten berechnen kann.

In Tabelle 10-22 werden numerische Werte f\"ur die
St\"orungstheoriekoeffizienten $\Gamma_{\nu}^{(3)}$ mit $0 \le \nu \le 9$
aufgelistet, die durch Summation der Partialsummen $s_{n}^{(3, \nu)}$,
Gl. (10.8-51), mit Hilfe des verallgemeinerten Summationsprozesses
${\delta}_k^{(n)} \bigl(\zeta, s_n^{(3, \nu)}\bigr)$, Gl. (5.4-13), mit
$\zeta = 1$ berechnet wurden. Dabei wurden immer alle verf\"ugbaren
St\"orungstheoriekoeffizienten $c_{\nu}^{(3)}$ mit $0 \le \nu \le 165$
verwendet. Zur Berechnung des St\"orungs\-theorie\-koeffizienten
$\Gamma_{n}^{(3)}$ standen demzufolge alle Partialsummen $s_{\nu}^{(3,
n)}$ mit $0 \le \nu \le 165 - n - 1$ zur Verf\"ugung. Alle Rechnungen
wurden in MAPLE mit einer Genauigkeit von 500 Dezimalstellen
durchgef\"uhrt.

\beginFloat

\medskip

\beginTabelle
\beginFormat \rechts " \rechts \endFormat
\+ " \links {\bf Tabelle 10-22} \@ " \\
\+ " \links {Anharmonischer Oszillator mit einer
$\hat{x}^6$-Anharmonizit\"at} \@ " \\
\+ " \links {Renormierte St\"orungstheoriekoeffizienten
$\Gamma_{n}^{(3)}$} \@ " \\
\+ " \links {Berechnung durch Summation der divergenten Reihe (10.8-44)}
\@ " \\
\- " \- " \- " \\ \sstrut {} {1.5 \jot} {1.5 \jot}
\+ " \rechts {$n$}
" \rechts{$\Gamma_{n}^{(3)} \qquad \qquad \;$} " \\
\- " \- " \- " \\ \sstrut {} {1 \jot} {1 \jot}
\+ " 0 " $ 0.625~089~812~561$ \\
\+ " 1 " $ 0.407~659~180~68~$ \\
\+ " 2 " $-0.031~516~476~79~$ \\
\+ " 3 " $ 0.000~642~929~71~$ \\
\+ " 4 " $-0.000~553~405~4~~$ \\
\+ " 5 " $-0.000~392~921~7~~$ \\
\+ " 6 " $-0.000~244~097~4~~$ \\
\+ " 7 " $-0.000~167~765~~~~$ \\
\+ " 8 " $-0.000~118~7~~~~~~$ \\
\+ " 9 " $-0.000~10~~~~~~~~~$ \\
\- " \- " \- " \\ \sstrut {} {1 \jot} {1 \jot}

\endTabelle

\medskip

\endFloat

Bei den Rechnungen f\"ur Tabelle 10-22 wurde versucht, m\"oglichst viele
Koeffizienten $\Gamma_{n}^{(3)}$ mit einer {\it absoluten\/} Genauigkeit
von 12 Dezimalstellen zu berechnen. Dies gelang aber nur f\"ur $n = 0$.
F\"ur $n \ge 1$ nimmt die Genauigkeit der St\"orungstheoriekoeffizienten
$\Gamma_{n}^{(3)}$ mit wachsendem $n$ rasch ab. F\"ur $n \ge 10$
erh\"alt man keine aussagekr\"aftige Summationsergebnisse.

Ein Vergleich der Tabellen 10-19 und 10-22 zeigt sofort, da{\ss} die {\it
renormalized strong coupling expansion\/} (10.8-10) im Falle eines
Oszillators mit einer $\hat{x}^6$-Anharmonizit\"at auch nicht ann\"ahernd so
gute Ergebnisse liefern kann wie im Falle eines Oszillators mit einer
$\hat{x}^4$-Anharmonizit\"at. Hier macht sich einerseits bemerkbar, da{\ss}
die renormierten St\"orungstheoriekoeffizienten $c_{n}^{(3)}$ gem\"a{\ss} Gl.
(10.4-23b) mit wachsendem $n$ betragsm\"a{\ss}ig in etwa wie $(2 n)!/n^{1/2}$
wachsen, wogegen die $c_{n}^{(2)}$ gem\"a{\ss} Gl. (10.4-23a) betragsm\"a{\ss}ig nur
in etwa wie $n!/n^{1/2}$ wachsen, und andererseits, da{\ss} trotz der
wesentlich st\"arkeren Divergenz f\"ur $m = 3$ nur die Koeffizienten
$c_{\nu}^{(3)}$ mit $0 \le \nu \le 165$ bekannt sind, wogegen f\"ur $m =
2$ die Koeffizienten $c_{\nu}^{(2)}$ mit $0 \le \nu \le 200$ zur
Verf\"ugung stehen.

Trotz ihrer geringeren Genauigkeit sind die St\"orungstheoriekoeffizienten
$\Gamma_{n}^{(3)}$ aber keineswegs nutzlos. In Tabelle 10-23 wird
gezeigt, wie gut die Partialsummen (10.8-51) der {\it renormalized
strong coupling expansion\/} (10.8-10) f\"ur $E^{(3)} (\beta)$ f\"ur
verschiedene Werte von $\beta$ konvergieren. Alle Rechnungen wurden in
MAPLE mit einer Genauigkeit von 12 Dezimalstellen durchgef\"uhrt. F\"ur
jeden Wert von $\beta$ wurde die entsprechende renormierte
Kopplungskonstante $\tau$ gem\"a{\ss} Gl. (10.4-9) mit Hilfe des Kommandos
{\it fsolve\/} [Char, Geddes, Gonnet, Leong, Monagan und Watt 1991b, S.
97] berechnet. Die \anf{exakten} Energien in Tabelle 10-23 wurden aus
Tabelle 10-10 entnommen.

\beginFloat

\medskip

\beginTabelle [to \kolumnenbreite]
\beginFormat \links " \rechts " \rechts " \rechts " \rechts " \rechts
\endFormat
\+ " \links {\bf Tabelle 10-23} \@ \@ \@ \@ \@ " \\
\+ " \links {Anharmonischer Oszillator mit einer
$\hat{x}^6$-Anharmonizit\"at} \@ \@ \@ \@ \@ " \\
\+ " \links {Berechnung der Grundzustandsenergie $E^{(3)} (\beta)$}
\@ \@ \@ \@ \@ " \\
\+ " \links {Konvergenz der Partialsummen $\Sigma^{(3)}_{n} (\tau)$,
Gl. (10.8-51),} \@ \@ \@ \@ \@ " \\
\+ " \links {f\"ur verschiedene Werte von $\beta$} \@ \@ \@ \@ \@ " \\
\- " \- " \- " \- " \- " \- " \- " \\ \sstrut {} {1 \jot} {1 \jot}
\+ " " \mitte{$\beta = 0$} " \mitte{$\beta = 2/10$}
" \mitte{$\beta = 1$}
" \mitte{$\beta = 4$}
" \mitte{$\beta = 100$} " \\
\- " \- " \- " \- " \- " \- " \- " \\ \sstrut {} {1 \jot} {1 \jot}
\+ " $\Sigma^{(3)}_{0} (\tau)$ " 0.625089813 " 0.901764112 " 1.233054419
" 1.680412101 " 3.647266282 \\
\+ " $\Sigma^{(3)}_{1} (\tau)$ " 1.032748993 " 1.184347547 " 1.439714885
" 1.832055618 " 3.717133295 \\
\+ " $\Sigma^{(3)}_{2} (\tau)$ " 1.001232516 " 1.173850042 " 1.435608892
" 1.830433369 " 3.716974636 \\
\+ " $\Sigma^{(3)}_{3} (\tau)$ " 1.001875446 " 1.173952941 " 1.435630418
" 1.830437948 " 3.716974731 \\
\+ " $\Sigma^{(3)}_{4} (\tau)$ " 1.001322041 " 1.173910382 " 1.435625657
" 1.830437403 " 3.716974729 \\
\+ " $\Sigma^{(3)}_{5} (\tau)$ " 1.000929119 " 1.173895863 " 1.435624788
" 1.830437349 " 3.716974729 \\
\+ " $\Sigma^{(3)}_{6} (\tau)$ " 1.000685022 " 1.173891529 " 1.435624649
" 1.830437344 " 3.716974729 \\
\+ " $\Sigma^{(3)}_{7} (\tau)$ " 1.000517257 " 1.173890097 " 1.435624624
" 1.830437344 " 3.716974729 \\
\+ " $\Sigma^{(3)}_{8} (\tau)$ " 1.000398557 " 1.173889611 " 1.435624620
" 1.830437344 " 3.716974729 \\
\+ " $\Sigma^{(3)}_{9} (\tau)$ " 1.000298557 " 1.173889414 " 1.435624619
" 1.830437344 " 3.716974729 \\
\- " \- " \- " \- " \- " \- " \- " \\ \sstrut {} {1 \jot} {1 \jot}
\+ " Exakt                " 1.000000000 " 1.173889345 " 1.435624619
" 1.830437344 " 3.716974729 \\
\- " \- " \- " \- " \- " \- " \- " \\ \sstrut {} {1 \jot} {1 \jot}

\endTabelle

\medskip

\endFloat

Bemerkenswert an den Ergebnissen in Tabelle 10-23 ist, da{\ss} die
Partialsummen $\Sigma^{(3)}_{n} (\tau)$ im kopplungsfreien Fall $\beta =
0$ beziehungsweise $\tau = 1$ anscheinend gegen Eins konvergieren. Das
bedeutet, da{\ss} die Normierungsrelation (10.8-49) anscheinend auch f\"ur $m
= 3$ erf\"ullt wird. Die Erf\"ullung der Normierungsrelation impliziert, da{\ss}
die {\it renormalized strong coupling expansion\/} (10.8-8) auch f\"ur $m
= 3$ f\"ur alle physikalisch relevanten renormierten Kopplungskonstanten
$\tau \in [0, 1]$ konvergiert. Weiterhin zeigen die Ergebnisse in
Tabelle 10-23, da{\ss} $E^{(3)} (\beta)$ selbst f\"ur eine Kopplungskonstante,
die so klein ist wie $\beta = 2/10$, mit einer bemerkenswert hohen
Genauigkeit berechnet werden kann. Solange die Kopplungskonstante
$\beta$ nicht extrem klein ist, d\"urfte das in Tabelle 10-23 verwendete
Verfahren zur Berechnung von $E^{(3)} (\beta)$ zur Zeit konkurrenzlos
sein.

In Tabelle 10-24 werden numerische Werte f\"ur die
St\"orungstheoriekoeffizienten $\Gamma_{\nu}^{(4)}$ mit $0 \le \nu \le 5$
aufgelistet, die durch Summation der Partialsummen $s_{n}^{(4, \nu)}$
mit Hilfe des verallgemeinerten Summationsprozesses ${\delta}_k^{(n)}
\bigl(\zeta, s_n^{(4, \nu)}\bigr)$, Gl. (5.4-13), mit $\zeta = 1$
berechnet wurden. Dabei wurden immer alle verf\"ugbaren
St\"orungstheoriekoeffizienten $c_{\nu}^{(4)}$ mit $0 \le \nu \le 139$
verwendet. Zur Berechnung des St\"orungs\-theorie\-koeffizienten
$\Gamma_{n}^{(4)}$ standen demzufolge alle Partialsummen $s_{\nu}^{(4,
n)}$ mit $0 \le \nu \le 139 - n - 1$ zur Verf\"ugung. Alle Rechnungen
wurden in MAPLE mit einer Genauigkeit von 600 Dezimalstellen
durchgef\"uhrt.

Bei den Rechnungen f\"ur Tabelle 10-24 wurde versucht, m\"oglichst viele
Koeffizienten $\Gamma_{n}^{(4)}$ mit einer {\it absoluten\/} Genauigkeit
von 8 Dezimalstellen zu berechnen. Dies gelang aber nur f\"ur $n = 0$. F\"ur
$n \ge 1$ nimmt die Genauigkeit der St\"orungstheoriekoeffizienten
$\Gamma_{n}^{(4)}$ mit wachsendem $n$ rasch ab. F\"ur $n \ge 6$ erh\"alt man
keine aussagekr\"aftige Summationsergebnisse.

Auf den ersten Blick sind die in Tabelle 10-24 pr\"asentierten
Summationsergebnisse f\"ur die St\"orungstheoriekoeffizienten
$\Gamma_{\nu}^{(4)}$ mit $0 \le \nu \le 5$ nicht sehr beeindruckend.
Man sollte aber bedenken, da{\ss} die renormierten
St\"orungstheoriekoeffizienten $c_{n}^{(4)}$ gem\"a{\ss} Gl. (10.4-23c) mit
wachsendem $n$ betragsm\"a{\ss}ig in etwa wie $(3 n)!/n^{1/2}$ wachsen, und
da{\ss} demzufolge die Summation der divergenten Reihe (10.8-44) f\"ur $m = 4$
\"au{\ss}erst schwierig ist.

\beginFloat

\medskip

\beginTabelle
\beginFormat \rechts " \rechts \endFormat
\+ " \links {\bf Tabelle 10-24} \@ " \\
\+ " \links {Anharmonischer Oszillator mit einer
$\hat{x}^8$-Anharmonizit\"at} \@ " \\
\+ " \links {Renormierte St\"orungstheoriekoeffizienten
$\Gamma_{n}^{(4)}$} \@ " \\
\+ " \links {Berechnung durch Summation der divergenten Reihe (10.8-44)}
\@ " \\
\- " \- " \- " \\ \sstrut {} {1.5 \jot} {1.5 \jot}
\+ " \rechts {$n$}
" \rechts{$\Gamma_{n}^{(4)} \qquad \qquad \;$} " \\
\- " \- " \- " \\ \sstrut {} {1 \jot} {1 \jot}
\+ " 0 " $ 0.555~127~76$ \\
\+ " 1 " $ 0.500~911~5~$ \\
\+ " 2 " $-0.058~060~2~$ \\
\+ " 3 " $ 0.004~592~5~$ \\
\+ " 4 " $-0.000~58~~~~$ \\
\+ " 5 " $-0.000~5~~~~~$ \\

\- " \- " \- " \\ \sstrut {} {1 \jot} {1 \jot}

\endTabelle

\medskip

\endFloat

Au{\ss}erdem kann man zeigen, da{\ss} die St\"orungstheoriekoeffizienten
$\Gamma_{n}^{(4)}$ trotz ihrer geringen Genauigkeit keineswegs nutzlos
sind. In Tabelle 10-25 wird gezeigt, wie gut die Partialsummen (10.8-51)
der {\it renormalized strong coupling expansion\/} (10.8-10) f\"ur
$E^{(4)} (\beta)$ f\"ur verschiedene Werte von $\beta$ konvergieren. Alle
Rechnungen wurden in MAPLE mit einer Genauigkeit von 12 Dezimalstellen
durchgef\"uhrt. F\"ur jeden Wert von $\beta$ wurde die entsprechende
renormierte Kopplungskonstante $\tau$ gem\"a{\ss} Gl. (10.4-9) mit Hilfe des
Kommandos {\it fsolve\/} [Char, Geddes, Gonnet, Leong, Monagan und Watt
1991b, S. 97] berechnet. Die \anf{exakten} Energien in Tabelle 10-25
wurden aus Tabelle 10-11 entnommen.

\beginFloat

\medskip

\beginTabelle
\beginFormat \links " \rechts " \rechts " \rechts " \rechts " \rechts
\endFormat
\+ " \links {\bf Tabelle 10-25} \@ \@ \@ \@ \@ " \\
\+ " \links {Anharmonischer Oszillator mit einer
$\hat{x}^8$-Anharmonizit\"at} \@ \@ \@ \@ \@ " \\
\+ " \links {Berechnung der Grundzustandsenergie $E^{(4)} (\beta)$}
\@ \@ \@ \@ \@ " \\
\+ " \links {Konvergenz der Partialsummen $\Sigma^{(4)}_{n} (\tau)$,
Gl. (10.8-51),} \@ \@ \@ \@ \@ " \\
\+ " \links {f\"ur verschiedene Werte von $\beta$} \@ \@ \@ \@ \@ " \\
\- " \- " \- " \- " \- " \- " \- " \\ \sstrut {} {1 \jot} {1 \jot}
\+ " " \mitte{$\beta = 0$} " \mitte{$\beta = 2/10$}
" \mitte{$\beta = 1$}
" \mitte{$\beta = 4$}
" \mitte{$\beta = 100$} " \\
\- " \- " \- " \- " \- " \- " \- " \\ \sstrut {} {1 \jot} {1 \jot}
\+ " $\Sigma^{(4)}_{0} (\tau)$ " 0.55513 " 0.96316 " 1.27815 " 1.65647
" 3.09926 \\
\+ " $\Sigma^{(4)}_{1} (\tau)$ " 1.05604 " 1.25187 " 1.49571 " 1.82434
" 3.18897 \\
\+ " $\Sigma^{(4)}_{2} (\tau)$ " 0.99798 " 1.24075 " 1.49095 " 1.82216
" 3.18864 \\
\+ " $\Sigma^{(4)}_{3} (\tau)$ " 1.00257 " 1.24104 " 1.49102 " 1.82218
" 3.18864 \\
\+ " $\Sigma^{(4)}_{4} (\tau)$ " 1.00199 " 1.24103 " 1.49102 " 1.82218
" 3.18864 \\
\+ " $\Sigma^{(4)}_{5} (\tau)$ " 1.00149 " 1.24103 " 1.49102 " 1.82218
" 3.18864 \\
\- " \- " \- " \- " \- " \- " \- " \\ \sstrut {} {1 \jot} {1 \jot}
\+ " Exakt                " 1.00000 " 1.24103 " 1.49102 " 1.82218
" 3.18864 \\
\- " \- " \- " \- " \- " \- " \- " \\ \sstrut {} {1 \jot} {1 \jot}

\endTabelle

\medskip

\endFloat

Obwohl die Ergebnisse in Tabelle 10-25 absolut gesehen nicht besonders
genau sind -- es wurden nur 6 Dezimalstellen angegeben -- zeigen sie
doch, da{\ss} die Partialsummen (10.8-51) die Grundzustandsenergie $E^{(4)}
(\beta)$ eines anharmonischen Oszillators mit einer
$\hat{x}^8$-Anharmonizit\"at schon f\"ur relativ kleine Werte von $\beta$
mit bemerkenswerter Genauigkeit approximieren k\"onnen. Weiterhin zeigen
die Ergebnisse f\"ur den kopplungsfreien Fall $\beta = 0$ beziehungsweise
$\tau = 1$, da{\ss} die Normierungsrelation (10.8-49) anscheinend auch f\"ur
$m = 4$ erf\"ullt ist. Es scheint also, da{\ss} die {\it renormalized strong
coupling expansion\/} (10.8-8) auch f\"ur $m = 4$ f\"ur alle physikalisch
relevanten renormierten Kopplungskonstanten $\tau \in [0, 1]$
konvergiert.

Der in diesem Unterabschnitt beschriebene Formalismus macht es also
m\"oglich, ausgehend von der hochgradig divergenten
Rayleigh-Schr\"odingerschen St\"orungsreihe (10.2-4) eine Reihenentwicklung
f\"ur die Grundzustandsenergie $E^{(m)} (\beta)$ zu konstruieren, die
anscheinend f\"ur jede physikalisch relevante Kopplungskonstante $\beta
\in [0, \infty)$ konvergiert.

\endAbschnittsebene

\endAbschnittsebene

\keinTitelblatt\neueSeite

\beginAbschnittsebene
\aktAbschnitt = 10

\Abschnitt Konvergenzverbesserung von quantenmechanischen Rechnungen an
Polyacetylen mit Hilfe von Extrapolationsverfahren

\vskip - 2 \jot

\beginAbschnittsebene

\medskip

\Abschnitt Quantenmechanische Rechenverfahren f\"ur quasi-eindimensionale
stereo\-regul\"are Polymere

\smallskip

\aktTag = 0

Die heute \"ublichen quantenchemischen {\it ab-initio\/}-Rechnungen in der
Ortsdarstellung basieren auf dem sogenannten analytischen SCF-Verfahren,
bei dem die Ortsanteile $\varphi_j (\vec{r})$ der zu bestimmenden
Spinorbitale $\psi_j (\vec{r}, \sigma)$ durch endliche
Linearkombinationen sogenannter Basisfunktionen $\{ \xi_m (\vec{r})
\}_{m=1}^{M}$ dargestellt werden:
$$
\varphi_j (\vec{r}) \; = \;
\sum_{m=1}^{M} \, C_m^{(j)} \, \xi_m (\vec{r}) \, .
\tag
$$
Da die Basisfunktionen $\{ \xi_m (\vec{r}) \}_{m=1}^{M}$ als bekannt
vorausgesetzt werden, m\"ussen in einer LCAO-MO-SCF-Rechnung nur noch die
unbekannten Koeffizienten $C_m^{(j)}$ variativ bestimmt werden. Dazu mu{\ss}
man die Roothaanschen Gleichungen [Roothaan 1951] l\"osen, die in Form
eines verallgemeinerten Matrixeigenwertproblems geschrieben werden
k\"onnen:
$$
{\bf F \, C} \; = \; {\cal E} \, {\bf S \, C} \, .
\tag
$$

\noindent Dabei ist ${\bf F}$ die Fockmatrix, ${\bf C}$ die
Koeffizientenmatrix, ${\bf S}$ die \"Uberlappungsmatrix und ${\cal E}$ die
Diagonalmatrix der Orbitalenergien. Da die Elemente der Fockmatrix ${\bf
F}$ nicht nur von den Ein- und Zweielektronenintegralen der
Basisfunktionen (den ber\"uchtigten Mehrzentrenmolek\"ulintegralen)
abh\"angen, sondern auch von den unbekannten Koeffizienten $C_m^{(j)}$ in
Gl.~(11.1-1), m\"ussen die Root\-haanschen Gleichungen iterativ gel\"ost
werden, bis {\it Selbstkonsistenz} erreicht ist.

Die Dimensionalit\"at der Roothaanschen Gleichungen entspricht der Zahl
der Basisfunktionen. Allgemein gilt, da{\ss} die Zahl der Basisfunktionen
mindestens so gro{\ss} sein mu{\ss} wie die Zahl der Elektronen des Molek\"uls.
Laut Cioslowski [1993b, S. 3] wurden inzwischen schon SCF-Rechnungen mit
bis zu 1800 Basisfunktionen
durchgef\"uhrt{\footnote[\dagger]{Bemerkenswert an dieser zur Zeit
g\"ultigen oberen Schranke von 1800 Basisfunktionen f\"ur {\it
ab-initio\/}-Rechnungen ist, da{\ss} man mit Hilfe semiempirischer Verfahren
anscheinend nur geringf\"ugig gr\"o{\ss}ere Systeme rechnen kann. Laut
Cioslowski [1993b, S. 3] wurden bei den bisher umfangreichsten
MNDO-Rechnungen ungef\"ahr 2000 Basisfunktionen verwendeten.}}. Da die
Computertechnologie aber immer noch rasche Fortschritte macht, und weil
man auch bei Molek\"ulprogrammen noch weitere signifikante
algo\-rith\-mi\-sche Verbesserungen erwarten kann, ist damit zu rechnen,
da{\ss} diese Schranke von 1800 Basisfunktionen bald weiter nach oben
geschraubt werden wird.

Zusammenfassend kann man sagen, da{\ss} sowohl die zur Verf\"ugung stehenden
Gro{\ss}rechner als auch die Molek\"ulprogramme f\"ur {\it
ab-initio\/}-Rechnungen ein sehr hohes Niveau erreicht haben. So kann
man inzwischen sowohl die Elektronenstruktur als auch Eigenschaften
kleiner und mittelgro{\ss}er Molek\"ule sehr genau berechnen. Au{\ss}erdem wird in
zahlreichen Arbeitsgruppen hart daran gearbeitet, dem folgenden
Idealzustand noch n\"aher zu kommen [Schaefer 1984, S. 31]:

\medskip

\beginSchmaeler
\noindent {\sl Visionaries have for some years imagined a futuristic
``black box'' computer program, to which the bench chemist specifies a
desired molecule and a series of properties of interest. After a few
moments of cogitation, the computer politely returns the answers,
reliable to the specified tolerances.}
\endSchmaeler

\medskip

Wie schon oben erw\"ahnt, h\"angt die Komplexit\"at einer quantenchemischen
{\it ab-initio\/}-Rechnung ganz entscheidend von der Dimensionalit\"at der
Roothaanschen Gleichungen (11.1-2) und damit von der Zahl der
verwendeten Basisfunktionen ab. Das impliziert, da{\ss} man auf diese Weise
immer nur Rechnungen an Molek\"ulen {\it endlicher\/} Gr\"o{\ss}e durchf\"uhren
kann. Quantenchemische {\it ab-initio\/}-Rechnungen an im Prinzip
unendlich ausgedehnten Polymeren auf der Basis der Roothaanschen
Gleichungen sind also nicht m\"oglich.

Allerdings gibt es zahlreiche {\it stereoregul\"are} quasi-eindimensionale
Polymere wie beispielsweise Polyethylen, Polyacetylen oder kettenf\"ormige
Biopolymere, bei denen {\it ab-initio\/}-Rechnungen doch wieder m\"oglich
sind, wenn man Rechentechniken der Quantenchemie und der
Festk\"orperphysik auf geeignete Weise kombiniert{\footnote[\dagger]{Eine
\"Ubersicht \"uber festk\"orperphysikalische Rechenverfahren zur Bestimmung
der Gesamtenergie findet man beispielsweise in einem \"Ubersichtsartikel
von Ihm [1988].}}. Dazu mu{\ss} man ausnutzen, da{\ss} ein stereoregul\"arer
quasi-eindimensionaler Polymer formal durch Translation einer
Elementarzelle erzeugt werden kann. Diese Translationsinvarianz f\"uhrt zu
einer erheblichen Reduktion der Komplexit\"at der zugrundeliegenden
Gleichungen, wodurch {\it ab-initio\/}-Rechnungen an unendlich
ausgedehnten Systemen dann doch praktisch m\"oglich werden.

In Artikeln von Andr\'e, Gouverneur und Leroy [1967a; 1967b] und Del Re,
Ladik und Bicz\'{o} [1967] wurden die Roothaanschen Gleichungen (11.1-2)
f\"ur endliche Molek\"ule so modifiziert, da{\ss} man unter Verwendung
zyklischer Randbedingungen SCF-Rechnungen an stereoregul\"aren
quasi-eindimensionalen Polymeren in der
Ortsdarstellung{\footnote[\ddagger]{SCF-Rechnungen an stereoregul\"aren
Polymeren k\"onnen auch in der Impulsdarstellung durchgef\"uhrt werden
[Defranceschi und Delhalle 1986; Delhalle und Harris 1985; Harris 1972;
1975a; 1975b; 1978].}} durchf\"uhren kann. Das den Roothaanschen
Gleichungen (11.1-2) entsprechende Gleichungssystem kann in Matrixform
folgenderma{\ss}en geschrieben werden (siehe beispielsweise Kert\'esz [1982,
S. 168]):
$$
{\bf F} ({\vec k}) \, {\bf C} ({\vec k}) \; = \;
{\cal E} ({\vec k}) \, {\bf S} ({\vec k}) \, {\bf C} ({\vec k}) \, .
\tag
$$
Im Gegensatz zu den formal scheinbar identischen Roothaanschen
Gleichungen (11.1-2) f\"ur endliche Molek\"ule h\"angt dieses Gleichungssystem
explizit vom reziproken Gittervektor ${\vec k}$ ab. Es mu{\ss} also f\"ur
jeden Wert von ${\vec k}$ separat gel\"ost werden. Trotzdem ist das
Gleichungssystem (11.1-3) ein gro{\ss}er Fortschritt, der
Hartree-Fock-Rechnungen an Polymeren m\"oglich macht: Anstelle eines
Matrixproblems mit einer im Prinzip unendlichen Dimensionalit\"at mu{\ss} man
nur noch viele, aber vergleichsweise niedrigdimensionale ${\vec
k}$-abh\"angige Matrixprobleme l\"osen.

Das hier skizzierte Verfahren der sogenannten {\it Kristallorbitale\/}
wurde mit gro{\ss}em Erfolg f\"ur Rechnungen an quasi-eindimensionalen
stereoregul\"aren Polymeren verwendet, was beispielsweise belegt wird
durch Artikel von Andr\'e, Bodart, Br\'edas, Delhalle und Fripiat [1984],
Andr\'e, Fripiat, Demanet, Br\'edas und Delhalle [1978], Andr\'e, Vercauteren,
Bodart und Fripiat [1984], Andr\'e, Vercauteren und Fripiat [1984], Calais
[1986], Champagne, Mosley und Andr\'e [1993], Delhalle und Calais [1986;
1987], Delhalle, Delvaux, Fripiat, Andr\'e und Calais [1988], Delhalle,
Fripiat und Harris [1984], Delhalle, Fripiat und Piela [1980], Delhalle,
Piela, Br\'edas und Andr\'e [1980], Dovesi [1984], Fripiat, Andr\'e, Delhalle
und Calais [1991], Fripiat und Delhalle [1979], Fripiat, Delhalle, Andr\'e
und Calais [1989], Karpfen [1978], I'Haya, Narita, Fujita und Ujino
[1984], Liegener [1985; 1988], Liegener, Beleznay und Ladik [1987],
Mintmire und White [1987], Mosley, Fripiat, Champagne und Andr\'e [1993],
Piela und Delhalle [1978], Roszak und Kaufman [1991], Suhai [1983;
1992], Teramae [1986], Vra\v{c}ko, Liegener und Ladik [1988] und Weniger
und Liegener [1990], durch \"Ubersichtsartikel von Andr\'e [1980] und
Kert\'esz [1982] und durch die Monographien von Andr\'e, Delhalle und Br\'edas
[1991], Ladik [1988] und Pisani, Dovesi und Roetti [1988].

Trotz des unbestreitbaren Erfolges der Kristallorbitalrechnungen gibt es
aber immer noch erhebliche technische Probleme, die nicht v\"ollig
befriedigend gel\"ost sind. Bei Kristallorbitalrechnungen an
quasi-eindimensionalen Polymeren wird das im Prinzip unendlich
ausgedehnte System unter Verwendung zyklischer
Randbedingungen durch $2 N + 1$ Elementarzellen approximiert. Es ist im
Prinzip erstrebenswert, $N$ so gro{\ss} wie m\"oglich werden zu lassen, da
dann die Wechselwirkung der Elektronen und der Atomkerne aus der
Referenzzelle mit den Elektronen und Atomkernen aus den anderen
Elementarzellen des Polymers gut beschrieben werden kann. Allerdings
nimmt der numerische Aufwand einer Kristallorbitalrechnung sehr stark
mit wachsendem $N$ zu.

Kristallorbitalrechnungen sind ebenso wie Hartree-Fock-Rechnungen an
Molek\"ulen effektive Einteilchenrechnungen. In Molek\"ulrechnungen werden
die Ortsanteile der Spinorbitale gem\"a{\ss} Gl.~(11.1-1) als
Linearkombinationen von Basisfunktionen dargestellt, die normalerweise
an den verschiedenen Atomkernen des Molek\"uls zentriert sind. Auch bei
Kristallorbitalrechnungen macht man f\"ur die Ortsanteile der Spinorbitale
einen LCAO-Ansatz. Allerdings kann eine solche Linearkombination nur
dann translationsinvariant sein, wenn die Basisfunktionen nicht nur \"uber
eine {\it einzige\/} Elementarzelle delokalisiert sind, sondern \"uber
{\it alle\/} $2 N + 1$ Elementarzellen, durch die der im Prinzip
unendlich ausgedehnte quasi-eindimensionale Polymer approximiert wird.
Wenn ${\vec a}$ ein Gittervektor ist, und wenn innerhalb einer
Elementarzelle $M$ an den verschiedenen Atomkernen zentrierte
Basisfunktionen $\{ \xi_{\mu} \}_{\mu=1}^{M}$ verwendet werden, ist ein
translationsinvariantes Kristallorbital $\Phi_n ({\vec k}, {\vec r})$
durch die folgende Linearkombination von Basisfunktionen definiert
(siehe beispielsweise Andr\'e [1980, Gl. (1)]):
$$
\Phi_n ({\vec k}, {\vec r}) \; = \; (2 N + 1)^{- 1/2}
\sum_{\nu = - N}^{N} \, \sum_{\mu = 1}^{M} \, C_{\mu}^{(n)} ({\vec k})
\, \exp (\i \nu {\vec k} \cdot {\vec a}) \,
\xi_{\mu} ({\vec r} - {\vec R}_{\mu} - \nu {\vec a}) \, .
\tag
$$

Zum Aufbau der Fockmatrix ${\bf F} ({\vec k})$ in Gl. (11.1-3) m\"ussen
Ein- und Zweielektronenmatrixelemente von Kristallorbitalen berechnet
werden. Aus Gl. (11.1-4) folgt, da{\ss} diese Matrixelemente durch Summen
analoger Matrixelemente von Basisfunktionen dargestellt werden k\"onnen,
die an Atomkernen aus allen $2 N + 1$ Elementarzellen zentriert sind.
Die Berechnung solcher Summen von Matrixelementen wird \"au{\ss}erst
aufwendig, wenn sowohl $M$ und $N$ gro{\ss} ist. Um den sowieso schon gro{\ss}en
Rechenaufwand einigerma{\ss}en im Rahmen zu halten, mu{\ss} man versuchen, diese
Summen so effizient wie m\"oglich zu berechnen. Eine M\"oglichkeit zur
Verbesserung der Effi\-zienz besteht darin, da{\ss} man alle Matrixelemente
vernachl\"assigt, die betragsm\"a{\ss}ig unterhalb einer gewissen Schranke
liegen. Allerdings besteht in der Literatur keine Einigkeit dar\"uber,
welche Abbruchkriterien optimal sind, und es ist auch nur unzureichend
bekannt, wie der Abbruch solcher Wechselwirkungssummen die Konvergenz
von Kristallorbitalrechnungen beeinflu{\ss}t. Die verschiedenen
Polymerprogramme{\footnote[\dagger]{Eine kurze Beschreibung
verschiedener Polymerprogramme findet man in Abschnitt \Roemisch{1}.3a
des Buches von Pisani, Dovesi und Roetti [1988].}}, die bisher
entwickelt worden sind, unterscheiden sich haupts\"achlich in der
Behandlung solcher Summen von Wechselwirkungsmatrixelementen. Aufgrund
der unterschiedlichen Strategien bei der Vernachl\"assigung von
Matrixelementen liefern verschiedene Polymerprogramme auch nicht exakt
die gleichen Resultate, selbst wenn identische Basiss\"atze und Geometrien
verwendet werden [Andr\'e, Bodart, Br\'edas, Delhalle und Fripiat 1984, S.
1].

Aufgrund der gro{\ss}en Reichweite des Coulombpotentials gibt es bei solchen
Summen von Matrixelementen von Basisfunktionen au{\ss}erdem noch
Konvergenzprobleme (siehe beispielsweise Delhalle und Calais [1986;
1987]), die man schon anhand des einfachsten Beispiels eines
hypothetischen eindimensionalen Ionenkristalls verstehen kann.
Betrachten wir als Modell eine lineare Kette gegens\"atzlich geladener
Einheitspunktladungen, die voneinander um einen festen Abstand $R$
entfernt sind. Die elektrostatische Energie $E_{pot}$ dieser Kette ist
gegeben durch die Coulombwechselwirkung eines Referenzions mit allen
Nachbarn, die relativ zum Referenzion an den Positionen $j R$ mit $j =
\pm 1$, $\pm 2$, $\ldots$ sitzen. Offensichtlich gilt [Levy 1968, S. 74]
$$
E_{pot} \; = \; - \frac {2} {R} \, + \, \frac {2} {2 R}
\, - \, \frac {2} {3 R} \, + \, \ldots
\; = \; - \frac {2} {R} \, \Bigl\{ 1 \, - \, \frac {1} {2}
\, + \, \frac {1} {3} \, - \, \ldots \Bigr\}
\; = \; - \, \frac {2} {R} \,
\sum_{k=0}^{\infty} \, \frac {(- 1)^k} {k + 1} \, .
\tag
$$
Die {\it Madelungkonstante} $\alpha$ dieses eindimensionalen Kristalls
ist der numerische Faktor, mit dem $1 / R$ im Ausdruck f\"ur die
elektrostatische Energie multipliziert werden mu{\ss} [Levy 1968, S. 75]:
$$
\alpha \; = \;
- 2 \, \sum_{k=0}^{\infty} \, \frac {(- 1)^k} {k + 1} \, .
\tag
$$
In diesem einfachen Spezialfall bereitet die Berechnung der
Madelungkonstante keine Schwierigkeiten. Wegen [Magnus, Oberhettinger
und Soni 1966, S. 38]
$$
\ln (1+z) \; = \;
z \, \sum_{m=0}^{\infty} \, \frac {(-1)^m z^m} {m+1}
\tag
$$
folgt sofort
$$
\alpha \; = \; - 2 \ln (2) \, .
\tag
$$
Obwohl die Berechnung der Madelungkonstante einer linearen Kette
offensichtlich trivial ist, kann man an diesem einfachen Beispiel die
typischen Probleme erkennen, mit denen man trotz erheblicher
methodischer und technologischer Fortschritte in den letzten Jahren bei
der Berechnung der Madelungkonstanten realer Ionenkristalle immer noch
konfrontiert ist [Adams und Duby 1987; Borwein, Borwein und Shail 1989;
Borwein, Borwein und Taylor 1985; Dias und Chaba 1986; Glasser 1988;
Harris 1975a; 1975b; 1977; 1978; Hautot 1974; Hosoya 1982; Kukhtin und
Shramko 1991; Lowery und House 1984; Taylor 1987; Sarkar und
Bhattacharyya 1992a; 1992b; 1993; Singh und Pathria 1989; Zucker und
Robertson 1984]. Die alternierende Reihe (11.1-6) konvergiert nur {\it
bedingt\/}, d.~h., die positiven und negativen Beitr\"age f\"ur sich allein
genommen {\it divergieren\/}. Au{\ss}erdem konvergiert diese Reihe sehr
langsam, was eine direkte Konsequenz der gro{\ss}en Reichweite der
Coulombwechselwirkung ist. Man ben\"otigt demzufolge verallgemeinerte
Summationsprozesse, wenn man diese Reihe f\"ur numerische Zwecke verwenden
will [Bender und Orszag 1978, S. 372 - 373, Example 2; Weniger 1989,
Table 13-5].

Beim Aufbau der Fockmatrix ${\bf F} ({\vec k})$ in Gl. (11.1-3) durch
Summen von Matrixelementen von Basisfunktionen, die an Atomkernen aus
allen $2 N + 1$ Elementarzellen zentriert sind, gibt es ganz analoge
Konvergenzprobleme wie bei der unendlichen Reihe (11.1-6) f\"ur die
Madelungkonstante der linearen Kette. Wenn $N$ gro{\ss} wird, divergieren
beispielsweise die Summen von Matrixelementen, die die Kern-Elektron-
beziehungsweise die Elektron-Elektron-Wechselwirkung in einem
quasi-eindimensionalen Polymer beschreiben. Nur wenn man die Terme
dieser beiden Summen so anordnet, da{\ss} die divergierenden Beitr\"age sich
gegenseitig kompensieren, erh\"alt man Wechselwirkungssummen, die f\"ur
gro{\ss}e $N$ konvergieren. Allerdings konvergieren diese Summen aufgrund
der gro{\ss}en Reichweite der Coulombwechselwirkung nur relativ
langsam{\footnote[\dagger]{Aufgrund der schlechten Konvergenz solcher
Wechselwirkungssummen ist es an sich eine naheliegende Idee,
verallgemeinerte Summationsprozesse auch bei der Berechnung solcher
Wechselwirkungssummen zu verwenden. Es scheint aber bisher keine
Arbeiten zu geben, in denen diese Fragestellung systematisch untersucht
worden ist.}} (siehe beispielsweise Andr\'e, Fripiat, Demanet, Br\'edas und
Delhalle [1978], Delhalle, Fripiat und Piela [1980] oder Piela und
Delhalle [1978]).

Aufgrund der Schwierigkeiten, mit denen man bei der genauen Berechnung
der Elemente der Fockmatrix ${\bf F} ({\vec k})$ konfrontiert sein kann,
ist es nicht verwunderlich, da{\ss} es besonders bei guten Basiss\"atzen, die
diffuse Polarisationsfunktionen enthalten, manchmal Probleme mit der
Konvergenz der SCF-Iterationen gibt. In dieser Beziehung haben {\it
ab-initio\/}-Programme f\"ur Molek\"ule deutlich g\"unstigere Eigenschaften.
Zwar k\"onnen auch bei diesen Programmen Konvergenzprobleme im SCF-Teil
nie v\"ollig ausgeschlossen werden, und es wird auch immer noch \"uber
Probleme dieser Art gearbeitet (siehe beispielsweise Badziag und Solms
[1988], Cioslowski [1988d], Hamilton und Pulay [1986], Sarkar,
Bhattacharyya und Bhattacharyya [1989], Schlegel und McDouall [1991],
Sellers [1991; 1993], Srivastava [1984], oder Stanton [1981]). Trotzdem
sind Konvergenzprobleme im SCF-Teil bei Molek\"ulrechnungen deutlich
weniger wahrscheinlich als bei Kristallorbitalrechnungen.

Ein anderes Problem bei Kristallorbitalrechnungen ist die explizite
Behandlung der Elektronenkorrelation. Es gibt zwar
Kristallorbitalrechnungen, bei denen Korrelationseffekte ber\"uck\-sichtigt
werden [Ladik 1988, Abschnitt 5; Ladik und Otto 1993; Liegener 1985;
1988; Suhai 1983; 1992; 1993; Vra\v{c}ko, Liegener und Ladik 1988; Ye,
F\"orner und Ladik 1993]. Trotzdem scheint eine Behandlung der
Elektronenkorrelation in endlichen Molek\"ulen deutlich einfacher zu sein
als in quasi-eindimensionalen Polymeren (siehe beispielsweise Pisani,
Dovesi und Roetti [1988, Abschnitt \Roemisch{1}.4c] oder Cioslowski und
Lepetit [1991, S. 3536]).

Aus diesem Grund verwendet man in letzter Zeit vermehrt einen
alternativen Ansatz f\"ur Rechnungen an stereoregul\"aren
quasi-eindimensionalen Polymeren: Man approximiert Polymere durch
endliche Molek\"ule oder Cluster, wobei freie Valenzen an den Enden des
Clusters abges\"attigt werden m\"ussen [Andr\'e, Delhalle, Fripiat, Hennico
und Piela 1988; Augspurger und Dykstra 1992; Champagne, Mosley und Andr\'e
1993; Cioslowski 1988c; 1990a; 1990b; 1993a; Cioslowski und Bessis 1988;
Cioslowski und Lepetit 1991; Cioslowski und Weniger 1993; Cui und
Kertesz 1990; Cui, Kertesz und Jiang 1990; Grimes, Catlow und Shluger
1992; Delhalle, Bodart, Dory, Andr\'e und Zyss 1986; Distefano, Jones,
Guerra, Favaretto, Modelli und Mengoli 1991; Evangelisti 1990; Hurst,
Dupuis und Clementi 1988; Karpfen und Kertesz 1991; Kirtman 1992;
Kirtman, Hasan und Chipman 1991; Kirtman, Nilson und Palke 1983; Lee und
Kertesz 1987; Liegener, Beleznay und Ladik 1987; Mazumdar, Guo und
Dixit 1992; Roszak und Kaufman 1991; Seel 1988; Shuai, Beljonne und
Br\'edas 1992; Villar, Dupuis, Watts, Hurst und Clementi 1988; Weniger und
Liegener 1990].

Diese Vorgehensweise hat den Vorteil, da{\ss} man die hochentwickelte
Technologie quantenchemischer {\it ab-initio\/}-Rechnungen an endlichen
Molek\"ulen verwenden kann. Allerdings haben solche Clusterrechnungen auch
einen sehr schwerwiegenden Nachteil: In vielen F\"allen konvergieren
Clusterrechnungen deutlich langsamer als analoge
Kristallorbitalrechnungen{\footnote[\dagger]{Selbstverst\"andlich h\"angen
die Konvergenzgeschwindigkeiten von Kristallorbital- beziehungsweise
Clusterrechnungen nicht nur vom betrachteten System ab, sondern auch von
der Eigenschaft, bez\"uglich der die Konvergenz beurteilt wird.
Beispielsweise ergaben {\it ab-initio\/}-Rechnungen mit dem gleichen
Basissatz eine ganz ausgezeichnete \"Ubereinstimmung der Bindungsl\"angen
und Bindungswinkel von Polyethylen und $n$-Pentan [Kert\'esz 1982, Table
\Roemisch{1}]. Allerdings sollte man bedenken, da{\ss} Bindungsl\"angen und
Bindungswinkel aliphatischer Kohlenwasserstoffketten normalerweise nur
wenig von der L\"ange der Kohlenstoffkette abh\"angen. Demzufolge ist es in
diesem Fall nicht \"uberraschend, da{\ss} bereits kleine Cluster (nur 5 $\rm
CH_2$-Einheiten) schon sehr gute Ergebnisse liefern. Leider kann man
nicht erwarten, da{\ss} auch andere Rechnungen \"ahnlich gute Ergebnisse
liefern werden.}}. Diese schlechtere Konvergenz ist eine direkte
Konsequenz der Tatsache, da{\ss} bei Molek\"ulen endlicher Gr\"o{\ss}e {\it
End\-effekte\/} wie beispielsweise die unvollst\"andige Delokalisation
konjugierter $\pi$-Bindungen oder die notwendige Abs\"attigung freier
Valenzen an den Enden eines Clusters eine erhebliche Rolle spielen,
wogegen diese Effekte bei einem unendlich ausgedehnten System nicht mehr
auftreten. Wirklich gute Ergebnisse kann man bei Clusterrechnungen also
nur dann erwarten, wenn man die st\"orenden Einfl\"usse der {\it
Endlichkeit\/} des Clusters soweit wie m\"oglich eliminieren kann.

Eine M\"oglichkeit, solche Endeffekte weitgehend zu eliminieren, ist die
Verwendung von {\it Extrapolationsverfahren\/}. Dabei wird der unendlich
ausgedehnte quasi-eindimensionale Polymer nicht durch einen einzigen
Cluster endlicher Gr\"o{\ss}e approximiert{\footnote[\ddagger]{Die
Approximation eines Polymers durch einen {\it einzigen\/} Cluster hat
den zus\"atzlichen Nachteil, da{\ss} man keine Aussage \"uber die Konvergenz des
Verfahrens machen kann. Konvergenzaussagen sind nur m\"oglich, wenn man
den Polymer durch eine Folge von Clustern wachsender Gr\"o{\ss}e
approximiert.}}. Stattdessen wird eine Folge von endlichen Clustern
wachsender Gr\"o{\ss}e berechnet und die so erhaltenen Ergebnisse mit Hilfe
geeigneter Extrapolationsverfahren auf unendliche Kettenl\"ange
extrapoliert. Auf diese Weise sollte es im Prinzip m\"oglich sein, die
Konvergenz von Clusterrechnungen signifikant zu verbessern.

Die Idee, Rechnungen an Clustern endlicher Gr\"o{\ss}e auf Cluster unendlicher
Gr\"o{\ss}e zu extrapolieren, ist keineswegs neu (siehe beispielsweise
Champagne, Mosley und Andr\'e [1993], Cioslowski [1988c; 1990a; 1990b;
1993a], Cioslowski und Bessis [1988], Cioslowski und Lepetit [1991],
Cioslowski und Weniger [1993], Cui, Kertesz und Jiang [1990], Hurst,
Dupuis und Clementi [1988], Kirtman [1992], Kirtman, Hasan und Chipman
[1991], Kirtman, Nilson und Palke [1983], Liegener, Beleznay und Ladik
[1987], Mazumdar, Guo und Dixit [1992], Villar, Dupuis, Watts, Hurst und
Clementi [1988], Weniger und Liegener [1990]). Allerdings ist in der
Mehrheit dieser Arbeiten das mathematische Niveau der verwendeten
Extrapolationsverfahren vergleichsweise niedrig. In den meisten F\"allen
wurden nur einige wenige und oft auch relativ primitive
Extrapolationsverfahren verwendet, die auch nicht immer zur
Extrapolation von Clusterrechnungen geeignet waren. Die in den oben
genannten Arbeiten erhaltenen Schlu{\ss}folgerungen sind folglich nur
bedingt aussagekr\"aftig. Hier macht sich sehr st\"orend bemerkbar, da{\ss} in
der mathematischen Ausbildung an Universit\"aten Extrapolationsverfahren
und verwandte numerische Techniken stiefm\"utterlich behandelt werden. Die
gro{\ss}en Fortschritte auf dem Gebiet der Extrapolation in den letzten
Jahren, die beispielsweise in der Monographie von Brezinski und Redivo
Zaglia [1991] ausf\"uhrlich beschrieben sind, wurden bisher von der
Mehrheit der numerisch arbeitenden Naturwissenschaftler weitgehend
ignoriert.

Viele der in dieser Arbeit beschriebenen verallgemeinerten
Summationsprozesse k\"onnen auch zur Extrapolation verwendet werden. Die
Eingabedaten $\Seqn s$ sind dann nicht wie sonst die Partialsummen
$$
s_n \; = \; \sum_{k=0}^{n} \, a_k
\tag
$$
einer unendlichen Reihe, sondern die Elemente einer Folge von Zahlen,
die man beispielsweise durch {\it ab-initio\/}-Rechnungen an einer Folge
von endlichen Clustern wachsender Kettenl\"ange erh\"alt.

Extrapolationsverfahren sind aber nicht nur bei Clusterrechnungen
n\"utzlich. Man kann auch versuchen, die Konvergenz von
Kristallorbitalrechnungen durch Extrapolation zu verbessern. Wie schon
fr\"uher erw\"ahnt, wird in Kristallorbitalrechnungen der unendlich
ausgedehnte quasi-eindimensionale Festk\"orper unter Verwendung zyklischer
Randbedingungen durch $2 N + 1$ wechselwirkende Elementarzellen
approximiert. Man kann f\"ur $N = 1, 2, 3 \ldots$ eine Folge von
Kristallorbitalrechnungen mit wachsender Komplexit\"at durchf\"uhren. Die so
erhaltenen Ergebnisse werden dann auf ein System extrapoliert, bei dem
die Wechselwirkungen sich \"uber unendlich viele Elementarzellen
erstreckt.

In dieser Arbeit soll am Beispiel des {\it trans}-Polyacetylens
untersucht werden, welche der zahlreichen bekannten
Extrapolationsverfahren besonders gut zur Verbesserung der Konvergenz
von Kristallorbital- beziehungsweise Clusterrechnungen geeignet sind.
Dies geschah bereits in einer fr\"uheren Arbeit [Weniger und Liegener
1990]. In der Zwischenzeit sind aber die bei der Extrapolation der
Kristallorbital- und Clusterenergien des {\it trans}-Polyacetylens
auftretenden Probleme wesentlich besser verstanden, was auch eine
partielle Modifikation fr\"uherer Schlu{\ss}folgerungen [Weniger und Liegener
1990] erzwang.

\medskip

\Abschnitt Polyacetylen als Modellsystem

\smallskip

\aktTag = 0

Das einfachste realistische Modellsystem, an dem man die Verbesserung
der Konvergenz von Kristallorbital- und Clusterrechnungen durch
Extrapolationsverfahren testen kann, ist das Polyethylen $\rm
(CH_2)_{\infty}$. Dieses System hat aber den Nachteil, da{\ss}
ausschlie{\ss}lich $\rm (C \! - \! C)$- und
$\rm (C \! - \! H)$-Einfachbindungen vorkommen, deren Energien, Abst\"ande
und Winkel bekanntlich nur wenig von Art und Gr\"o{\ss}e aliphatischer
Nachbargruppen abh\"angen. Ein in dieser Beziehung wesentlich
anspruchsvolleres System ist das Polyacetylen $\rm (HC \! = \!
CH)_{\infty}$, dessen konjugierte $\pi$-Bindungen wesentlich sensibler
auf eine \"Anderung der Nachbargruppen reagieren sollten als die
weitgehend inerten $\sigma$-Bindungen des Polyethylens.

Aber auch sonst ist Polyacetylen ein \"au{\ss}erst interessantes
System{\footnote[\dagger]{Das gro{\ss}e Interesse am Polyacetylen wird unter
anderem dadurch dokumentiert, da{\ss} es eine Monographie mit dem Titel {\it
Polyacetylene\/} gibt [Chien 1984].}}. Bedingt durch die konjugierten
$\pi$-Bingungen und die damit verbundene hochgradige Anisotropie besitzt
Polyacetylen sowohl f\"ur Theoretiker als auch f\"ur Experimentatoren
\"au{\ss}erst interessante Eigenschaften.

Von besonderer Bedeutung war die Entdeckung, da{\ss} die Leitf\"ahigkeit von
{\it trans}-Polyacetylen durch Dotierung mit Elektronenakzeptoren wie
$\rm Br_2$, $\rm I_2$ oder $\rm AsF_5$ oder Elektronendonatoren wie Li
oder Na sehr stark ver\"andert werden kann. So kann die Leitf\"ahigkeit von
Filmen von {\it trans}-Polyacetylen durch Dotierung in einem Bereich von
12 Gr\"o{\ss}enordnun\-gen variiert werden (zwischen $\sigma \approx 10^{- 9}
\, \Omega^{- 1} cm^{- 1}$ und $\sigma \approx 10^3 \, \Omega^{- 1} cm^{-
1}$ [Ladik 1988, S. 64]). Zum Vergleich: Die spezifische Leitf\"ahigkeit
von Ag bei $273 K$ liegt bei ungef\"ahr $6.6 \times 10^5 \Omega^{- 1}
cm^{- 1}$ [Hellwege 1976, S. 442, Tabelle 44.1]. Dotiertes Polyacetylen
wird beispielsweise in einem Artikel von Heeger und MacDiarmid [1980]
ausf\"uhrlich behandelt. Der Mechanismus der Abspeicherung negativer
Ladung in Polyenen nach Dotierung durch $\rm Na$ wird in einem Artikel
von Stafstr\"om, Br\'edas, L\"ogdlund und Saleneck [1993] behandelt. In diesem
Zusammenhang sind neben der Monographie von Chien [1984] auch eine von
Seymour [1981] herausgegebene Artikelsammlung und ein Buch von Lu
[1988], in dem unter anderem zahlreiche \"altere Orginalarbeiten \"uber
Polyacetylen abgedruckt sind, von Interesse.

Polyacetylen ist auch ein \"au{\ss}erst wichtiges Modellsystem f\"ur
Theoretiker. Br\'edas, Chance, Silbey, Nicolas und Durand [1981]
verwendeten effektive Hamiltonoperatoren, um Rechnungen an verschiedenen
Konformeren von Polyacetylen durchzuf\"uhren. Einen \"Uberblick \"uber
Hartree-Fock-Rechnungen an {\it cis}- und {\it trans}-Polyacetylen
findet man in Abschnitt 2.2.1 des Buches von Ladik [1988]. Dovesi [1984]
berechnete die Gesamtenergie pro Monomereinheit von verschiedenen
Polyacetylenisomeren mit Hilfe von {\it ab-initio\/}-Verfahren. I'Haya,
Narita, Fujita und Ujino [1984] f\"uhrten Kristallorbitalrechnungen an
{\it trans}-Polyacetylen mit umfangreicheren Basiss\"atzen durch. Die
Energetik der {\it cis-trans}-Isomerisierung von Polyacetylen wurde von
Teramae [1986] untersucht. L\"u, Tachibana und Yamabe [1992] untersuchten
die Dimerisierung von Polyacetylen. Angeregte Zust\"ande des {\it
trans}-Polyacetylen wurden von Vra\v{c}ko und Zaider [1993] unter
Verwendung verschiedener N\"aherungsverfahren untersucht.
Korrelationseffekte in {\it ab-initio\/}-Rechnungen an Polyacetylen
wurden von Ladik [1988, Abschnitt 5], Liegener [1988] und von Fulde [1991,
Abschnitt 8.5] diskutiert. Ye, F\"orner und Ladik [1993] behandelten
Korrelationseffekte in Poly\-acetylen und in anderen
quasi-eindimensionalen Systemen mit Hilfe der Coupled-Cluster-Methode.
Der Einflu{\ss} der Elektronenkorrelation auf die Bindungsalternanz und
andere strukturelle und elektronische Eigenschaften von Polyacetylen
wurde von Suhai [1983; 1992] untersucht. Suhai analysierte au{\ss}erdem noch
UV-Spektren von Polyacetylen [Suhai 1986], und verglich Polyacetylen mit
analogen Polymeren, die auf Silizium basieren [Suhai 1993]. Dann gibt es
Arbeiten von Paldus und Chin [1983], Paldus, Chin und Grey [1983],
Paldus und Takahashi [1984], Paldus, Takahashi und Cho [1984], Pauncz
und Paldus [1983], Takahashi und Paldus [1984; 1985a; 1985b] und
Takahashi, Paldus und {\v C}{\'\i}{\v z}ek [1983], in denen
Korrelationseffekte und Bindungsalternanz im Polyacetylen sowohl mit
Hilfe der Coupled-Cluster-Methode als auch st\"orungstheoretisch behandelt
wurden. Optische Eigenschaften von Polyacetylen wurden von Champagne,
Mosley und Andr\'e [1993], von Mosley, Fripiat, Champagne und Andr\'e
[1993], von Sekino und Bartlett [1992] sowie von Soos, McWilliams und Hayden
[1992] behandelt. Mazumdar, Guo und Dixit [1992] untersuchten
hochenergetische Zweiphotonenzust\"ande im Polyacetylen. Yang, You-Liang
und Bo [1990] berechneten die Bandstruktur von Li-dotiertem
Polyacetylen. Die Frequenzabh\"angigkeit linearer und nichtlinearer
optischer Eigenschaften von Polyacetylen wurden von Karna, Talapatra,
Wijekoon und Prasad [1992] studiert. Solitonen in Polyacetylen wurden
von Craw, Reimers, Bacskay, Wong und Hush [1992a; 1992b], Kirtman, Hasan
und Chipman [1991], Nakahara, Waxman und Williams [1990] und von
Tachibana, Ishikawa, Asai, Katagiri und Yamabe [1992] behandelt. Ein
hypothetischer antiferromagnetischer Zustand des {\it
trans}-Polyacetylens wurde von Yoshizawa, Ito, Tanaka und Yamabe [1993]
diskutiert. Elektronische Phasen\"uberg\"ange im Polyacetylen und in
dotiertem Polyacetylen wurden von Tanaka, Kobayashi, Okada, Kobashi und
Yamabe [1992] beziehungsweise von Tanaka, Takata, Okada und Yamabe
[1993] diskutiert. Maekawa und Imamura [1993] behandelten den Einflu{\ss}
von Fehlstellen auf die Elektronenstruktur von {\it trans}-Polyacetylen.

\medskip

\Abschnitt Kristallorbital- und Clusterrechnungen an Polyacetylen

\smallskip

\aktTag = 0

Ein Vergleich von zwei so unterschiedlichen Verfahren wie Cluster- und
Kristallorbitalrechnungen ist nur dann sinnvoll, wenn man in beiden
F\"allen soweit wie m\"oglich ein konsistentes System von N\"aherungen
verwendet. Das bedeutet nat\"urlich, da{\ss} man sowohl bei Kristallorbital-
als auch bei Clusterrechnungen den gleichen Basissatz und die gleiche
Geometrie verwenden mu{\ss}.

Da eine $\rm (HC \! = \! CH)$-Einheit 14 Elektronen enth\"alt, und da man
zur Extrapolation eine ausreichend gro{\ss}e Anzahl von Folgenelementen
ben\"otigt, war es aus Kapazit\"atsgr\"unden leider nicht m\"oglich, gro{\ss}e
Basiss\"atze zu verwenden. Im Rahmen der hier beschriebenen Rechnungen
wurden die Energien der Cluster $\rm C_{2 N} H_{2 N + 2}$ mit $1 \le N
\le 7$ berechnet [Weniger und Liegener 1990]. Im Falle von 7
Acetyleneinheiten hat man dann ein Molek\"ul mit 100 Elektronen, was zu
einer nicht ganz kleinen {\it ab-initio}-Rechnung f\"uhrt. Demzufolge
wurde sowohl bei den Clusterrechnungen als auch bei den
Kristallorbitalrechnungen nur eine relativ kleine Clementi-Basis
[Gianolio, Pavani und Clementi 1978] verwendet, die f\"ur die Energien
etwas bessere Ergebnisse liefert als eine STO-3G Basis von Hehre,
Stewart und Pople [1969].

Um den Rechenaufwand einigerma{\ss}en klein zu halten, wurde bei allen hier
beschriebenen Rechnungen [Weniger und Liegener 1990] nie die Geometrie
optimiert{\footnote[\dagger]{Es gibt noch einen weiteren Grund, warum
man auf eine Geometrieoptimierung verzichten sollte: Eine Optimierung
der Geometrie jedes einzelnen endlichen Clusters $\rm C_{2 N} H_{2 N +
2}$ mit $1 \le N \le 7$ f\"uhrt zu einer St\"orung im Verhalten der
Folgenelemente, deren Gr\"o{\ss}e man nicht kontrollieren kann. Deswegen kann
man auch nicht vorhersagen, ob und wie eine Geometrieoptimierung die
Konvergenz von Extrapolationsverfahren beeinflu{\ss}t. Mit \"ahnlichen
Problemen w\"are man konfrontiert, wenn man bei Kristallorbitalrechnungen,
die die Wechselwirkungen auf $2 N + 1$ Elementarzellen mit $1 \le N \le
7$ beschr\"anken, f\"ur jedes $N$ eine separate Geometrieoptimierung
durchf\"uhren w\"urde.}}. Stattdessen wurde immer eine feste Geometrie mit
den folgenden Bindungswinkeln und Abst\"anden verwendet ($R_{C-C}$ = 1.450
{\AA}, $R_{C=C}$ = 1.366 {\AA}, ${\angle}_{CCC}$ = 123.9{\Grad},
$R_{C-H}$ = 1.085 {\AA}), die von Suhai [1983] unter Ber\"ucksichtigung
von Korrelationseffekten bestimmt wurden.

Bei den Kristallorbitalrechnungen wurden die Integrationen \"uber den
irreduziblen Teil der Brillouinzone mit Hilfe einer Simpsonregel mit 17
$k$-Punkten durchgef\"uhrt. Bei den Clusterrechnungen wurden die beiden
endst\"andigen freien Valenzen durch Wasserstoffatome abges\"attigt, die
sich an den Positionen der hypothetischen n\"achsten Kohlenstoffatome der
unendlichen Kette befanden.

Alle Rechnungen wurde mit Hilfe eines von Prof.~P.~Otto, Universit\"at
Erlangen, modifizierten IBMOL-Programmes durchgef\"uhrt, bei dem alle
Wechselwirkungssummen strikt nach Elementarzellen abgeschnitten werden.

\beginFloat

\medskip

\beginTabelle
\beginFormat & \mitte \endFormat
\+ " \links {\bf Tabelle 11-1} \@ \@ " \\
\+ " \links {SCF-Rechnungen am {\it trans}-Polyacetylen} \@ \@ " \\
\+ " \links {Energien pro Acetyleneinheit in atomaren Einheiten}
\@ \@ " \\
\- " \- " \- " \- " \\ \sstrut {} {1.5 \jot} {1.5 \jot}
\+ " N " {Kristallorbitalrechnung} " {Clusterrechnung} " \\
\- " \- " \- " \- " \\ \sstrut {} {1 \jot} {1 \jot}
\+ "  1 " $-76.5813406536$ " $-77.6380003705$ " \\
\+ "  2 " $-76.5921411052$ " $-77.1124557954$ " \\
\+ "  3 " $-76.5918638429$ " $-76.9386177465$ " \\
\+ "  4 " $-76.5917242457$ " $-76.8517965289$ " \\
\+ "  5 " $-76.5916424980$ " $-76.7997219706$ " \\
\+ "  6 " $-76.5915903132$ " $-76.7650095717$ " \\
\+ "  7 " $-76.5915555082$ " $-76.7402157547$ " \\
\- " \- " \- " \- " \\ \sstrut {} {1 \jot} {1 \jot}
\endTabelle

\medskip

\endFloat

In Tabelle 11-1 werden die Folgen $\bigl\{ E_N^{(ko)} \bigr\}_{N=1}^{7}$
und $\bigl\{ E_N^{(cl)} \bigr\}_{N=1}^{7}$ der Kristallorbital-
beziehungs\-weise Clusterenergien pro Acetyleneinheit in atomaren
Einheiten aufgelistet [Weniger und Liege\-ner 1990, Table \Roemisch{1}].
Dabei darf man nicht vergessen, da{\ss} $N$ bei Cluster- und
Kristallorbitalrechnungen eine unterschiedliche Bedeutung hat: Bei
Clusterrechnungen entspricht $N$ der Zahl der Acetyleneinheiten des
Clusters $\rm C_{2 N} H_{2 N + 2}$, und ist somit ein Ma{\ss} f\"ur die Gr\"o{\ss}e
des molekularen Systems. Dagegen bezeichnet $N$ bei
Kristallorbitalrechnungen die Anzahl der wechselwirkenden
Acetyleneinheiten auf jeder Seite der Referenzzelle und ist somit ein
Ma{\ss} f\"ur die Reichweite der Wechselwirkungen, die bei der Polymerrechnung
ber\"ucksichtigt werden.

Die Ergebnisse in Tabelle 11-1 zeigen, da{\ss} die Folge der
Kristallorbitalrechnungen f\"ur die Energie pro Acetyleneinheit
vergleichsweise gut konvergiert. Die Folge der Clusterrechnungen
konvergiert dagegen so langsam, da{\ss} Clusterrechnungen an sich \"uberhaupt
nicht konkurrenzf\"ahig sind. So scheint eine Kristallorbitalrechnung mit
einer einzigen wechselwirkenden Elementarzelle auf jeder Seite der
Referenzzelle eine bessere Energie zu liefern als eine Clusterrechnung
am $\rm C_{14} H_{16}$ mit 7 Acetyleneinheiten.

\medskip

\Abschnitt Abbruchfehler bei Kristallorbital- und Clusterrechnungen

\smallskip

\aktTag = 0

Alle im Rahmen dieser Arbeit behandelten verallgemeinerten
Summationsprozesse gehen entweder implizit oder explizit davon aus, da{\ss}
die Elemente einer konvergenten Folge $\Seqn s$ f\"ur alle $n \in \N_0$
gem\"a{\ss}
$$
s_n \; = \; s \, + \, r_n
\tag
$$
in den Grenzwert $s$ und den Rest $r_n$ aufgespalten werden k\"onnen.

Ein verallgemeinerter Summationsproze{\ss} versucht, durch Elimination der
Reste $r_n$ aus den Folgenelementen $s_n$ die Ausgangsfolge $\Seqn s$ in
eine neue Folge $\Seqn {s'}$ mit g\"unstigeren numerischen Eigenschaften
zu transformieren. Normalerweise ist ein verallgemeinerter
Summationsproze{\ss} nicht v\"ollig erfolgreich, d.~h., er ist nicht in der
Lage, mit einer {\it endlichen Anzahl\/} von Rechenoperationen den Rest
$r_n$ {\it vollst\"andig\/} aus dem Folgenelement $s_n$ zu elimieren. Man
kann also die Elemente der transformierten Folge $\Seqn {s'}$ ebenfalls
in den Grenzwert $s$ und die transformierten, aber im Normalfall von
Null verschiedenen Reste $\Seqn {r'}$ gem\"a{\ss}
$$
s'_n \; = \; s \, + \, r'_n
\tag
$$
aufspalten. Ein verallgemeinerter Summationsproze{\ss} verbessert aber
offensichtlich die Konvergenz der Ausgangsfolge $\Seqn s$, wenn die
transformierten Reste $\Seqn {r'}$ f\"ur $n \to \infty$ schneller gegen
Null gehen als die urspr\"unglichen Reste $\Seqn r$.

Bei der Elimination der Reste $\Seqn r$ aus den Elementen der Folge
$\Seqn s$ gibt es aber einige prinzipielle Probleme. Abgesehen von
einigen f\"ur die Praxis wenig relevanten Modellproblemen sind die Reste
einer Folge normalerweise entweder unbekannt oder in einer numerisch
wenig zug\"anglichen Form. Die Bestimmung der Reste einer Folge ist also
in der Regel nicht einfacher als die Bestimmung des Grenzwertes $s$.
Weiterhin ist man hier mit dem Problem konfrontiert, da{\ss} es eine im
Prinzip beliebig gro{\ss}e Anzahl von m\"oglichen $n$-Abh\"angigkeiten der Reste
gibt. Man kann also nicht erwarten, da{\ss} ein {\it einziger\/}
verallgemeinerter Summationsproze{\ss} die Konvergenz {\it aller\/} langsam
konvergierender Folgen beschleunigen kann{\footnote[\dagger]{Delahaye
und Germain-Bonne [1980; 1982] konnten dies auch explizit beweisen.}}.

In vielen F\"allen sind aber gewisse {\it strukturelle\/} Informationen
\"uber das Verhalten der Reste $r_n$ als Funktion des Index $n$ bekannt.
Auf der Basis solcher strukturellen Informationen, die nicht sehr
detailliert sein m\"ussen, gelingt es oft, leistungsf\"ahige
verallgemeinerte Summationsprozesse zu konstruieren.

Hier gibt es aber das Problem, da{\ss} ein verallgemeinerter
Summationsproze{\ss}, der auf der Basis solcher struktureller Annahmen
konstruiert wurde, nur dann die Konvergenz einer Folge $\Seqn s$
erfolgreich beschleunigen wird, wenn das Verhalten der Reste $\Seqn r$
sich nicht zu sehr von den Annahmen unterscheidet, die bei der
Konstruktion des verallgemeinerten Summationsprozesses gemacht wurden.
Wenn diese grundlegenden Annahmen \"uber das Verhalten der Reste $\Seqn r$
als Funktion des Index $n$ nicht erf\"ullt sind, wird man entweder keine
Verbesserung der Konvergenz erreichen k\"onnen oder die transformierte
Folge $\Seqn {s'}$ wird m\"oglicherweise sogar divergieren.

Man wird also nur dann verallgemeinerte Summationsprozesse finden
k\"onnen, die zu einer deutlichen Verbesserung der Konvergenz der Folgen
$\bigl\{ E_N^{(ko)} \bigr\}_{N=1}^{7}$ und $\bigl\{ E_N^{(cl)}
\bigr\}_{N=1}^{7}$ f\"uhren, wenn wenigstens gewisse strukturelle
Informationen \"uber die Abh\"angigkeiten der Abbruchfehler $E_N^{(ko)} -
E_{\infty}^{(ko)}$ und $E_N^{(cl)} - E_{\infty}^{(cl)}$ von $N$ bekannt
sind.

Im Falle der Clusterrechnungen kann man relativ leicht quantitative
Aussagen \"uber die Abh\"angig\-keit des Abbruchfehlers $E_N^{(cl)} -
E_{\infty}^{(cl)}$ von der Zahl $N$ der Acetyleneinheiten machen. Wie
schon fr\"uher erw\"ahnt, m\"ussen die beiden endst\"andigen freien Valenzen der
$\rm (CH\! = \! CH)_N$-Kette durch H-Atome abges\"attigt werden. Die
Bindungsenergien einer $\rm C\! - \! C$- und einer $\rm C\! - \!
H$-Bindung sind aber verschieden. Da Bindungsenergien wenigstens in
nullter N\"aherung additiv und damit unabh\"angig von der L\"ange der
restlichen Kette sind, folgt daraus, da{\ss} die Gesamtenergie eines $\rm
H\! - \! (HC\! = \!CH)_N \! - \! H$-Molek\"uls sich um einen ann\"ahernd
konstanten Betrag von der Gesamtenergie eines hypothetischen $\rm (HC\!
= \! CH)_N$-Ringes unterscheiden sollte{\footnote[\dagger]{Dabei m\"ussen
wir nat\"urlich noch annehmen, da{\ss} $N$ so gro{\ss} ist, da{\ss} beispielsweise
Spannungsenergien oder andere Effekte, die auf den Ringschlu{\ss}
zur\"uckzuf\"uhren sind, vernachl\"assigbar sind.}}. Da wir aber nicht die
Gesamtenergie des Clusters, sondern die Energie pro Acetyleneinheit
betrachten, folgt daraus, da{\ss} der Abbruchfehler bei einem Cluster aus
$N$ Acetyleneinheiten proportional zu $1 / N$ sein sollte.

Im Prinzip ist eine $1 / N$-Abh\"angigkeit des Abbruchfehlers von der Zahl
$N$ der Monomere eine Art Todesurteil. Ein Abbruchfehler, der
proportional zu $1 / N$ ist, impliziert, da{\ss} man nur eine einzige
Dezimalstelle gewinnt, wenn man $N$ um einen Faktor 10 erh\"oht. Man m\"u{\ss}te
also Clusterrechnungen mit mehreren tausend Acetyleneinheiten
durchf\"uhren, um eine \"ahnliche Genauigkeit zu erreichen wie die
Kristallorbitalrechnungen in Tabelle 11-1. Auch der gr\"o{\ss}te Optimist kann
nicht erwarten, da{\ss} derartig umfangreiche {\it ab-initio\/}-Rechnungen
in absehbarer Zeit m\"oglich sein werden.

Es gibt weitere Argumente, die f\"ur eine $1 / N$-Abh\"angigkeit des
Abbruchfehlers von Clusterrechnungen sprechen. Cioslowski und Lepetit
[1991] konnten unter Verwendung st\"orungstheoretischer Argumente zeigen,
da{\ss} die Energie pro Monomereinheit $\varepsilon_N$ eines Clusters $\rm X
\! - \! (A)_N\!\! -\! Y$ einen Abbruchfehler $\varepsilon_N -
\varepsilon_{\infty}$ hat, der auch dann proportional zu $1 / N$ ist,
wenn es keine Abs\"attigung der freien Valenzen durch die funktionellen
Gruppen X und Y gibt oder wenn die Abs\"attigung zu keiner \"Anderung der
Bindungsenergie f\"uhrt. Diese Art von Abbruchfehler ist laut Cioslowski
und Lepetit [1991, S. 3545] ein ausschlie{\ss}lich topologischer Effekt, der
darauf zur\"uckzuf\"uhren ist, das es in einem Cluster $\rm X \! - \!
(A)_N\!\! -\! Y$ weniger langreichweitige als kurzreichweitige
Wechselwirkungen zwischen Monomereinheiten A gibt als in einem
quasi-eindimensionalen Polymer $(A)_{\infty}$.

Es ist also eine relativ naheliegende Idee, die Energie pro
Monomereinheit $\varepsilon_N$ eines Clusters $\rm X \! - \! (A)_N\!\!
- \! Y$ durch eine Potenzreihe in $1/N$ darzustellen [Cioslowski und
Lepetit 1991, S. 3537]:
$$
\varepsilon_N \; = \; \varepsilon_{\infty} \, + \,
\sum_{j=0}^{\infty} \, c_j / N^j \, .
\tag
$$
Allerdings gibt es im Zusammenhang mit dieser Reihenentwicklung noch
offene Fragen. Laut Cioslowski und Lepetit [1991, S. 3544] kann eine
Reihenentwicklung der Energie pro Monomereinheit $\varepsilon_N$
zus\"atzlich zu den Potenzen von $1/N$ noch nichtanalytische exponentielle
Terme des Typs $\exp(- \gamma N)$ mit $\gamma > 0$ enthalten, die auf
Austauschwechselwirkungen zur\"uckzuf\"uhren sind und die nur f\"ur relativ
kleine Werte von $N$ eine wesentliche Rolle spielen. Trotzdem sind
solche nichtanalytische Terme \"au{\ss}erst st\"orend: Die Potenzreihe (11.4-3)
w\"are dann nur noch eine asymptotische Reihe, die {\it nicht\/} gegen
$\varepsilon_N$ konvergiert, und die auch nicht summiert werden kann, da
das Carlemansche Theorem [Carleman 1926, Abschnitt \Roemisch{5}; Reed
und Simon 1978, Theoreme XII.17 und XII.18; Baker 1990, S. 223] nicht
erf\"ullt w\"are. Au{\ss}erdem k\"onnen exponentielle Beitr\"age bei Extrapolationen
eine \"au{\ss}erst st\"orende Rolle spielen: Extrapolationsverfahren, die
Potenzen von $1/N$ aus der unendlichen Reihe (11.4-3) eliminieren
k\"onnen, sind normalerweise nicht zur Elimination exponentieller Terme
geeignet.

Im Falle von Kristallorbitalrechnungen ist es dagegen \"au{\ss}erst schwierig,
quantitative Aussagen \"uber die Abh\"angigkeit der Abbruchfehler
$E_N^{(ko)} - E_{\infty}^{(ko)}$ von der Zahl $N$ der wechselwirkenden
Acetyleneinheiten auf jeder Seite der Referenzzelle zu machen. Mir ist
keine Arbeit bekannt, in der die Abh\"angigkeit der Konvergenz einer
Kristallorbitalrechnung vom Wechselwirkungsradius behandelt wurde.
Aufgrund der Ergebnisse in Tabelle 11-1 ist es aber sicher, da{\ss}
Kristallorbitalrechnungen deutlich schneller konvergieren als
Clusterrechnungen, deren Abbruchfehler proportional zu $1 / N$ ist.

Man kann auch auf rein numerische Weise versuchen, Aussagen \"uber die
$N$-Abh\"angigkeit der Abbruchfehler $E_N^{(ko)} - E_{\infty}^{(ko)}$
beziehungsweise $E_N^{(cl)} - E_{\infty}^{(cl)}$ zu machen. Die
Transformation $T_n$, Gl. (8.4-20), die zuerst in einer etwas
versteckten Form von Drummond [1976, S. 419] und sp\"ater von Bj{\o}rstad,
Dahlquist und Grosse [1981] erneut abgeleitet wurde, liefert eine
Approximation des Abklingparameters $\alpha > 0$ von logarithmisch
konvergenten Folgen des Typs
$$
s_n \; = \; s \, + \, (n+1)^{- \alpha}\,
\sum_{j=0}^{\infty} \, c_j / (n+1)^{j} \, ,
\qquad n \in \N_0 \, ,
\tag
$$
wobei die $c_j$ unspezifizierte Koeffizienten sind. Bj{\o}rstad,
Dahlquist und Grosse [1981, Gl. (4.1)] konnten au{\ss}erdem zeigen, da{\ss} die
asymptotische Absch\"atzung
$$
\alpha \; = \; T_n \, + \, O (n^{- 2}) \, , \qquad n \to \infty \, ,
\tag
$$
erf\"ullt ist, wenn die Eingabedaten $\Seqn s$ von der Form von Gl.
(11.4-4) sind.

Tabelle 11-2 zeigt den Effekt der Transformation $T_n$, Gl. (8.4-20),
auf die in Tabelle 11-1 aufgelisteten Kristallorbital- und
Clusterenergien $\bigl\{ E_N^{(ko)} \bigr\}_{N=1}^{7}$ und $\bigl\{
E_N^{(cl)} \bigr\}_{N=1}^{7}$.

\beginFloat

\medskip

\beginTabelle
\beginFormat  \mitte " \rechts " \mitte \endFormat
\+ " \links {\bf Tabelle 11-2} \@ \@ " \\
\+ " \links {Approximative Bestimmung des Abklingparameters $\alpha$
der} \@ \@ " \\
\+ " \links {Kristallorbital- und Clusterenergien gem\"a{\ss} Gl. (8.4-20)}
\@ \@ " \\
\- " \- " \- " \- " \\ \sstrut {} {1.5 \jot} {1.5 \jot}
\+ " n " \mitte {Kristallorbitalrechnung} " {Clusterrechnung} " \\
\- " \- " \- " \- " \\ \sstrut {} {1 \jot} {1 \jot}
\+ " 0 " $-0.0375950240296 \, \quad$ " $0.9861690376284$ " \\
\+ " 1 " $ 1.5057940501080 \, \quad$ " $0.9962449996809$ " \\
\+ " 2 " $ 1.8400812808299 \, \quad$ " $0.9975193297047$ " \\
\+ " 3 " $ 3.2123043560583 \, \quad$ " $0.9983189958783$ " \\
\- " \- " \- " \- " \\ \sstrut {} {1 \jot} {1 \jot}
\endTabelle

\medskip

\endFloat

Im Falle der Kristallorbitalenergien wurden die Eingabedaten f\"ur $T_n$,
Gl. (8.4-20), gem\"a{\ss}
$$
s_n \; = \; E_{n+1}^{(ko)}
\tag
$$
mit $0 \le n \le 6$ gew\"ahlt, im Falle der Clusterenergien gem\"a{\ss}
$$
s_n \; = \; E_{n+1}^{(cl)} \, .
\tag
$$

Die Ergebnisse in Tabelle 11-2 zeigen, da{\ss} die Clusterenergien $\bigl\{
E_N^{(cl)} \bigr\}_{N=1}^{7}$ tats\"achlich in guter N\"aherung durch den
Abklingparameter $\alpha = 1$ charakterisiert werden k\"onnen. Die
\"Uberlegung, da{\ss} die Abs\"attigung der endst\"andigen freien Valenzen durch
Wasserstoffatome einen zu $1/N$ proportionalen Abbruchfehler ergeben
mu{\ss}, scheint also richtig zu sein.

Dagegen zeigen die Transformationen $T_n$ in Tabelle 11-2 im Falle der
Kristallorbitalenergien $\bigl\{ E_N^{(ko)} \bigr\}_{N=1}^{7}$ ein
v\"ollig irregul\"ares Verhalten. Auf der Basis dieser Ergebnisse kann man
den Kristallorbitalenergien keinen definierten Abklingparameter $\alpha$
zuordnen.

Es gibt zwei einfache Erkl\"arungsversuche f\"ur das erratische Verhalten
der Approximationen des Abklingparameters der Kristallorbitalenergien in
Tabelle 11-2:

\item {(1):} Die Kristallorbitalenergien $E_{n+1}^{(ko)}$ mit $n \in
\N_0$ k\"onnen durch Reihen des Typs von Gl. (11.4-4) mit einem
wohldefinierten Abklingparameter $\alpha > 1$ dargestellt werden. Die
Transformation $T_n$, Gl. (8.4-20), die zur Approximation des
Abklingparameters verwendet wird, ist im Prinzip ein gewichteter
$\Delta^3$-Algorithmus und damit sehr anf\"allig f\"ur Rundungsfehler. Da
die Kristallorbitalenergien in Tabelle 11-1 relativ gut konvergieren,
unterscheiden sie sich nur bez\"uglich der hinteren Stellen. Es ist aber
unklar, wie gro{\ss} die {\it relative\/} Genauigkeit der
Kristallorbital\-energien tats\"achlich ist und ob alle in Tabelle 11-1
angegebenen Stellen auch wirklich korrekt sind. Das erratische Verhalten
der Approximationen des Abklingparameters der Kristallorbitalenergien in
Tabelle 11-2 w\"are dann auf den akkumulativen Effekt von Rundungsfehlern
zur\"uckzuf\"uhren.

\item {(2):} Die Kristallorbitalenergien $E_{n+1}^{(ko)}$ mit $n \in
\N_0$ k\"onnen durch Reihen des Typs von Gl. (11.4-4) mit einem
wohldefinierten Abklingparameter $\alpha > 1$ dargestellt werden. Diese
Reihenentwicklungen konvergieren aber nicht, sondern sind nur
asymptotisch f\"ur $n \to \infty$. Das bedeutet, da{\ss} eine {\it
vollst\"andige\/} Reihenentwicklung von $E_{n+1}^{(ko)}$ au{\ss}erdem noch
nichtanalytische exponentielle Beitr\"age enth\"alt, die f\"ur gr\"o{\ss}ere Werte
von $n$ rasch verschwinden. Im Rahmen dieser Arbeit konnten aber nur die
Kristallorbitalenergien $\bigl\{ E_{n+1}^{(ko)} \bigr\}_{n=0}^{6}$
berechnet werden. F\"ur so kleine Werte von $n$ k\"onnen nichtanalytische
Beitr\"age des Typs $\exp (- \gamma n)$ mit $\gamma > 0$ aber eine ganz
erhebliche Rolle spielen. Man kann unter diesen Umst\"anden nicht
ausschlie{\ss}en, da{\ss} die f\"uhrenden Terme der vollst\"andigen
Reihenentwicklung von $E_{n+1}^{(ko)}$ f\"ur niedrige Werte von $n$ nicht
die Potenzen $(n+1)^{- \alpha}$ und $(n+1)^{- \alpha - 1}$ sind, sondern
exponentielle Terme. In einem solchen Fall w\"are es durchaus vorstellbar,
da{\ss} $T_n$, Gl. (8.4-20), keine stabilen Approximationen f\"ur den
Abklingparameter $\alpha$ produzieren kann.

Ein naheliegender Einwand gegen den Interpretationsversuch (1) ist, da{\ss}
man den Cluster\-energien $\bigl\{ E_{n+1}^{(cl)} \bigr\}_{n=0}^{6}$
gem\"a{\ss} Tabelle 11-2 den Abklingparameter $\alpha = 1$ zuordnen kann. Wenn
numerische Instabilit\"aten die Ursache f\"ur das erratische Verhalten der
Approximationen des Abklingparameters der Kristallorbitalenergien in
Tabelle 11-2 sein sollten, m\"u{\ss}te man \"ahnliche numerische Instabilit\"aten
eigentlich auch bei den Clusterenergien beobachten. Ein solcher
Analogieschlu{\ss} klingt zwar plausibel, mu{\ss} aber nicht unbedingt richtig
sein. Die ver\-glichen mit den Kristallorbitalenergien wesentlich
schlechtere Konvergenz der Clusterenergien ist bei der approximativen
Bestimmung des Abklingparameters ein gro{\ss}er Vorteil, da wesentlich mehr
signifikante Stellen zur Verf\"ugung stehen{\footnote[\dagger]{Hier wird
stillschweigend angenommen, da{\ss} Kristallorbital- und Clusterenergien in
etwa die gleiche relative Genauigkeit besitzen.}}. Numerische
Instabilit\"aten sind also im Falle der Clusterenergien deutlich weniger
wahrscheinlich.

Ein naheliegender Einwand gegen den Interpretationsversuch (2) ist, da{\ss}
es nicht {\it a priori\/} klar ist, ob Potenzreihenentwicklungen in
$1/(n+1)$ des Typs von Gl. (11.4-4) mit dem Abklingparameter $\alpha =
1$ bei den Clusterenergien zur vollst\"andigen Beschreibung ausreichen.
Eine solche Reihe ist m\"oglicherweise nur asymptotisch f\"ur $n \to \infty$
und konvergiert nicht f\"ur endliche Werte von $n$. Nichtanalytische
exponentielle Beitr\"age w\"urden dann f\"ur kleinere Werte von $n$ einen
wesentlichen Beitrag zur Clusterenergie $E_{n+1}^{(cl)}$ liefern. Unter
diesen Umst\"anden sollten die Approximationen des Abklingparameters der
Clusterenergien eigentlich auch ein irregul\"ares Verhalten aufweisen. Es
gibt aber einen wesentlichen Unterschied zwischen den Kristallorbital-
und den Clusterenergien: Da der Abbruchfehler der Clusterenergien
proportional zu $1/(n+1)$ und damit relativ gro{\ss} ist, d\"urften sich
nichtanalytische exponentielle Terme bei den Clusterenergien deutlich
weniger stark bemerkbar machen als bei den Kristallorbitalenergien. Es
ist also durchaus denkbar, da{\ss} nichtanalytische exponentielle Terme die
Bestimmung des Abklingparameters der Kristallorbitalenergien mit Hilfe
der Transformation $T_n$, Gl. (8.4-20), verhindern k\"onnen, ohne bei der
Bestimmung des Abklingparameters der Clusterenergien besonders zu
st\"oren.

Diese \"Uberlegungen sind aber ausschlie{\ss}lich spekulativer Natur. Auf der
Basis der verf\"ugbaren Daten -- den Kristallorbitalenergien $\bigl\{
E_{n+1}^{(ko)} \bigr\}_{n=0}^{6}$ beziehungsweise den Clusterenergien
$\bigl\{ E_{n+1}^{(cl)} \bigr\}_{n=0}^{6}$ -- ist es nicht m\"oglich,
definitive Aussagen zu machen \"uber das asymptotische Verhalten dieser
Folgen f\"ur gro{\ss}e Indizes $n$. Man w\"urde dazu wesentlich l\"angere Folgen
von Kristallorbital- und Cluster\-energien ben\"otigen. Eine wesentliche
Voraussetzung f\"ur eine detaillierte numerische Analyse w\"aren au{\ss}erdem
verl\"a{\ss}liche Aussagen \"uber die relative Genauigkeit der Kristallorbital-
und Clusterenergien. Das ist aber \"au{\ss}erst schwierig, da die Energien
Resultate extrem aufwendiger Rechnungen sind. Aufgrund der Komplexit\"at
von {\it ab-initio\/}-Programme k\"onnen die in der numerischen Mathematik
\"ublichen Techniken zur Untersuchung des Einflusses von Rundungsfehlern
[Wilkinson 1969] nicht angewendet werden.

\medskip

\Abschnitt Extrapolationsverfahren

\smallskip

\aktTag = 0

Die im letzten Unterabschnitt durchgef\"uhrte Untersuchung der
$n$-Abh\"angigkeit der Abbruchfehler $E_{n+1}^{(ko)} - E_{\infty}^{(ko)}$
beziehungsweise $E_{n+1}^{(cl)} - E_{\infty}^{(cl)}$ ergab kein v\"ollig
eindeutiges Bild: Einerseits gibt es Anzeichen daf\"ur, da{\ss} die
Clusterenergien und m\"oglicherweise auch die Kristallorbitalenergien
durch logarithmisch konvergente Reihenentwicklungen des Typs von Gl.
(11.4-4) mit einem definierten Abklingparameter $\alpha$ dargestellt
werden k\"onnen. Andererseits kann man auf der Basis der verf\"ugbaren Daten
nicht ausschlie{\ss}en, da{\ss} diese Reihenentwicklungen nur asymptotisch f\"ur
$n \to \infty$ sind, und da{\ss} f\"ur kleinere Werte von $n$ nichtanalytische
Beitr\"age des Typs $\exp (- \gamma n)$ mit $\gamma > 0$ eine erhebliche
Rolle spielen.

Bei der Extrapolation der Kristallorbital- und Clusterenergien ist man
also m\"oglicherweise mit dem Problem konfrontiert, da{\ss} man nicht nur
Potenzen des Typs $(n+1)^{- \alpha - j}$ mit $\alpha \ge 1$ und $j \in
\N_0$ elimieren mu{\ss}, sondern zus\"atzlich noch exponentielle Terme.
Ungl\"ucklicherweise k\"onnen die meisten verallgemeinerten
Summationsprozesse entweder nur Potenzen oder nur exponentielle Terme
effizient elimieren. Durch das gleichzeitige Auftreten von Potenzen und
exponentiellen Termen in den Abbruchfehlern $E_{n+1}^{(ko)} -
E_{\infty}^{(ko)}$ und $E_{n+1}^{(cl)} - E_{\infty}^{(cl)}$ w\"are man
dann mit einem \"au{\ss}erst schwierigen Eliminationsproblem und damit auch
Extrapolationsproblem konfrontiert.

Im folgenden Text werden die verallgemeinerten Summationsprozesse
beschrieben, die zur Extrapolation der Kristallorbital- und
Clusterenergien in Tabelle 11-1 verwendet wurden. Au{\ss}erdem werden
diejenigen Eigenschaften der verallgemeinerten Summationsprozesse
besprochen, die bei der Extrapolation der Kristallorbital- und
Clusterenergien von besonderer Bedeutung sind. Man darf in diesem
Zusammenhang nicht vergessen, da{\ss} bei der Extrapolation von
Kristallorbital- und Clusterenergien zum Teil ganz andere Probleme
auftreten als bei den in fr\"uheren Abschnitten behandelten
Anwendungsbeispielen.

Bei den in fr\"uheren Abschnitten behandelten Anwendungsbeispielen war die
relative Genauigkeit der Eingabedaten nie ein Problem. Bei den
Rechnungen in FORTRAN konnte man praktisch immer davon ausgehen, da{\ss} die
Eingabedaten f\"ur die verallgemeinerten Summationsprozesse ann\"ahernd
maschinengenau waren, und bei den in Abschnitt 10 behandelten
Summationen hochgradig divergenter St\"orungsreihen anharmonischer
Oszillatoren wurden einige Rechnungen sogar mit einer relativen
Genauigkeit von 1000 Dezimalstellen in MAPLE [Char, Geddes, Gonnet,
Leong, Monagan und Watt 1991a] durchgef\"uhrt. Bei allen im Rahmen dieser
Arbeit durchgef\"uhrten Rechnungen stand also immer eine ausreichend gro{\ss}e
Anzahl signifikanter Stellen zur Verf\"ugung, und es gab demzufolge keine
un\"uberwindlichen Probleme mit numerischen Instabilit\"aten, die bei
verallgemeinerten Summationsprozessen nie ausgeschlossen werden k\"onnen.

Die Kristallorbital- und Clusterenergien in Tabelle 11-1 werden mit
einer relativen Genauigkeit von 12 Dezimalstellen angegeben, was
deutlich kleiner ist als Maschinengenauigkeit (15 - 16 Dezimalstellen
Genauigkeit in FORTRAN DOUBLE PRECISION). Es ist aber durchaus m\"oglich,
da{\ss} nicht alle der in Tabelle 11-1 angegebenen 12 Dezimalstellen
tats\"achlich korrekt sind. Die im Rahmen dieser Arbeit behandelten
verallgemeinerten Summationsprozesse erreichen eine Beschleunigung der
Konvergenz, indem sie gewichtete Differenzen der Eingabedaten bilden.
Besonders bei der Beschleunigung logarithmischer Konvergenz f\"uhrt diese
Differenzenbildung relativ rasch zu einer Akkumulation von
Rundungsfehlern und zu einem erheblichen Verlust signifikanter Stellen,
wenn man verallgemeinerte Summationsprozesse mit h\"oheren
Transformationsordnungen verwendet. Die Anzahl der signifikanten Stellen
der Eingabedaten determiniert also letztlich, welche
Transformationsordnungen der verallgemeinerten Summationsprozesse noch
sinnvoll sind.

In dieser Arbeit wurde bisher immer versucht, m\"oglichst effiziente
verallgemeinerte Summationsprozesse zu verwenden, und die Robustheit der
verallgemeinerten Summationsprozesse war nur von sekund\"arer Bedeutung.
Bei der Extrapolation der Kristallorbital- und Clusterenergien ist die
Effizienz eines verallgemeinerten Summationsprozesses aber weniger
wichtig als seine F\"ahigkeit, selbst bei relativ ungenauen Eingabedaten
noch verl\"a{\ss}liche Ergebnisse produzieren k\"onnen.

Zur Extrapolation der Kristallorbital- und Clusterenergien in Tabelle
11-1 wurden drei prinzipiell verschiedene Typen von verallgemeinerten
Summationsprozesse verwendet:

\item{(1):} Verallgemeinerte Summationsprozesse, die in der Lage sind,
die Konvergenz einer logarithmisch konvergenten Folge vom Typ von Gl.
(11.4-4) zu beschleunigen, wenn der Wert des Abklingparameters $\alpha$
bekannt ist. Summationsprozesse dieses Typs sind unf\"ahig, lineare
Konvergenz zu beschleunigen.

\item{(2):} Verallgemeinerte Summationsprozesse, die in der Lage sind,
die Konvergenz einer logarithmisch konvergenten Folge vom Typ von Gl.
(11.4-4) zu beschleunigen, auch wenn der Wert des Abklingparameters
$\alpha$ nicht bekannt ist. Alle Transformationen dieses Typs sind
au{\ss}erdem in der Lage, lineare Konvergenz zu beschleunigen.

\item{(3):} Verallgemeinerte Summationsprozesse, die in der Lage sind,
lineare Konvergenz zu beschleunigen, nicht aber logarithmische
Konvergenz.

Einige der leistungsf\"ahigsten Beschleuniger f\"ur logarithmische
Konvergenz basieren auf Interpolationsverfahren. Dabei wird die Existenz
einer Funktion ${\cal S}$ einer kontinuierlichen Variablen $x$
postuliert, die auf einer diskreten Menge $\Seqn x$ mit den Elementen
einer Folge $\Seqn s$ \"ubereinstimmt:
$$
{\cal S} (x_n) \; = \; s_n \, , \qquad n \in \N_0 \, .
\tag
$$
Wenn $k+1$ Folgenelemente $s_m, s_{m+1}, \ldots , s_{m+k}$ bekannt sind,
ist es im Prinzip m\"oglich, eine Approximation ${\cal S}_k (x)$ von
${\cal S} (x)$ zu konstruieren, die die $k+1$ Interpolationsbedingungen
$$
{\cal S}_k (x_{m+j}) \; = \; s_{m+j} \, ,
\qquad 0 \le j \le k \, ,
\tag
$$
erf\"ullt. Mit Hilfe der Approximation ${\cal S}_k (x)$ kann man
versuchen, den Funktionswert ${\cal S} (x)$ auch f\"ur den Grenzwert
$x_{\infty}$ der Folge $\Seqn x$ der Interpolationspunkte zu
approximieren. Wenn man den Wert der Approximation ${\cal S}_k (x)$ f\"ur
$x = x_{\infty}$ bestimmen kann, und wenn die interpolierende Funktion
die unbekannte Funktion ${\cal S} (x)$ wenigstens in einer Umgebung des
Extrapolationspunktes $x_{\infty}$ ausreichend genau approximieren kann,
sollte ${\cal S}_k (x_{\infty})$ eine bessere Approximation des
Grenzwertes $s$ der Folge $\Seqn s$ liefern als das letzte Folgenelement
$s_{n+k}$, das zur Konstruktion der N\"aherung ${\cal S}_k (x)$ verwendet
wurde.

Die am h\"aufigsten verwendeten Interpolationsfunktionen sind entweder
Polynome oder rationale Funktionen. Im Falle einer Polynominterpolation
wird \"ublicherweise vorausgesetzt, da{\ss} die Approximation ${\cal S}_k (x)$
ein Polynom $k$-ten Grades in $x$ ist,
$$
{\cal S}_k (x) \; = \;
c_0 \, + \, c_1 x \, + \, \cdots \, + \, c_k x^k \, ,
\tag
$$
und da{\ss} $x_{\infty} = 0$, der Grenzwert der Folge $\Seqn x$, der
Extrapolationspunkt ist. Das bedeutet, da{\ss} man den Grenzwert $s$ der zu
transformierenden Folge $\Seqn s$ durch den konstanten Term $c_0$ des
Interpolationspolynoms ${\cal S}_k (x)$ approximiert.

Nehmen wir jetzt an, da{\ss} die Interpolationspunkte $\Seqn x$ die
folgenden Bedingungen erf\"ullen:
$$
\beginMultiline
x_0 > x_1 > x_2 > \cdots > x_m > x_{m+1} > \cdots > 0 \, ,
\erhoehe\aktTag \\ \tag*{\tagnr a}
\lim_{n \to \infty} \; x_n \; = \; 0 \, .
\\ \tag*{\tagform\aktTagnr b}
\endMultiline
$$
Mit Hilfe der folgenden Variante des Nevilleschen Rekursionsschemas
[Neville 1934], die in der Literatur \"ublicherweise als das
Richardsonsche Extrapolationsverfahren bezeichnet wird [Richardson
1927], kann {\it sukzessive\/} ein ganzer String ${\cal S}_0 (0),
{\cal S}_1 (0), {\cal S}_2 (0), \ldots $ von interpolierenden
Polynom\-approximationen rekursiv berechnet werden (siehe beispielsweise
Brezinski und Redivo Zaglia [1991, S. 73] oder Weniger [1989; Abschnitt
6.1]):
$$
\beginAligntags
" {\cal N}_0^{(n)} (s_n , x_n ) \, " \; = \; " \, s_n
\, , \qquad n \in \N_0 \, ,
\erhoehe\aktTag \\ \tag*{\tagnr a}
" {\cal N}_{k+1}^{(n)} (s_n , x_n ) \, " \; = \; " \,
\frac
{x_n {\cal N}_{k}^{(n+1)} (s_{n+1} , x_{n+1} ) \, - \,
x_{n+k+1} {\cal N}_{k}^{(n)} (s_n , x_n ) }
{x_n \, - \, x_{n+k+1}} \; ,
\qquad k,n \in \N_0 \, . \quad
\\ \tag*{\tagform\aktTagnr b}
\endAligntags
$$
Aus diesem Rekursionsschema folgt, da{\ss} zur Berechnung von ${\cal
N}_k^{(n)} (s_n, x_n)$ mit $k,n \in \N_0$ die numerischen Werte der
$k+1$ Folgenelemente $s_n, s_{n+1}, \ldots, s_{n+k}$ und der $k+1$
Interpolationspunkte $x_n, x_{n+1}, \ldots, x_{n+k}$ bekannt sein
m\"ussen.

Wenn nur relativ wenige Elemente der Folge $\Seqn s$ zur Verf\"ugung
stehen, ist es normalerweise empfehlenswert, die Transformationen mit
maximalem Subskript $k$ und minimalem Superskript $n$ als
Approximationen des Grenzwertes $s$ zu verwenden. Im Falle des
Richardsonschen Extrapolationsverfahren werden in dieser Arbeit immer
die folgenden Approximationen zum Grenzwert $s$ der zu transformierenden
Folge $\Seqn s$ verwendet:
$$
\{s_0, x_0; s_1, x_1; \ldots ; s_m, x_m \} \,
\to \; {\cal N}_m^{(0)} (s_0, x_0) \, , \qquad m \in \N_0 \, .
\tag
$$

Das lineare Extrapolationsverfahren ${\cal N}_k^{(n)} (s_n , x_n)$ ist
offensichtlich exakt f\"ur Modellfolgen, die Polynome $k$-ten Grades in
den Interpolationspunkten $\Seqn x$ sind:
$$
s_n \; = \; s \, + \, \sum_{j=0}^{k-1} \, c_j \, x_n^{j+1} \, ,
\qquad k,n \in \N_0 \, .
\tag
$$
Es sollte demzufolge nicht \"uberraschen, da{\ss} das Richardsonsche
Extrapolationsverfahren die Konvergenz von Folgen, deren Elemente durch
Potenzreihen des folgenden Typs dargestellt werden k\"onnen,
$$
s_n \; = \; s \, + \, \sum_{j=0}^{\infty} \, c_j \, x_n^{j+1} \, ,
\qquad n \in \N_0 \, ,
\tag
$$
effizient beschleunigen kann. Theoretische Aussagen \"uber die Konvergenz
des Richardsonschen Extrapolationsverfahrens findet man beispielsweise
in dem Buch von Brezinski und Redivo Zaglia [1991, S. 73, Theorems 2.15
- 2.17].

Trotz aller unbestreitbaren Erfolge des Richardsonschen
Extrapolationsverfahrens gibt es aber ein prinzipielles Problem:
Polynome sind nicht in allen F\"allen zur Approximation unbekannter
Funktionen geeignet. Weiterhin ist bekannt, da{\ss} rationale Funktionen in
Approximationsverfahren h\"aufig bessere und stabilere Ergebnisse liefern
als Polynome (siehe beispielsweise Powell [1981, S. 28 und S. 111 -
112]). In unserem Fall bedeutet dies, da{\ss} man versuchen sollte,
Extrapolationsverfahren zu konstruieren, die auf rationalen
Interpolationsfunktionen des Typs
$$
{\cal S}_{2 k} (x) \; = \; \frac
{a_0 + a_1 x + a_2 x^2 + \cdots + a_k x^k}
{b_0 + b_1 x + b_2 x^2 + \cdots + b_k x^k} \, ,
\qquad k \in \N_0 \, ,
\tag
$$
basieren. Eine solche rationale Funktion ${\cal S}_{2 k} (x)$ enth\"alt $2
k + 2$ Polynomkoeffizienten $a_0, \ldots, a_k$ und $b_0, \ldots, b_k$.
Allerdings sind nur $2 k + 1$ der $2 k + 2$ Koeffizienten unabh\"angig, da
sie nur bis auf einen gemeinsamen, von Null verschiedenen Faktor
bestimmt sind. Man ben\"otigt also nur $2 k + 1$ Interpolationsbedingungen
vom Typ von Gl. (11.5-2) zur Bestimmung der rationalen Funktion ${\cal
S}_{2 k} (x)$.

Auch bei rationalen Interpolationsfunktionen ${\cal S}_{2 k} (x)$ kann
man den Nullpunkt als Extrapolationspunkt w\"ahlen. Das bedeutet, da{\ss} der
Grenzwert $s$ der zu transformierenden Folge durch den Quotienten $a_0 /
b_0$ approximiert wird. Bei rationalen Interpolationsfunktionen ist es
aber aus algorithmischen und numerischen Gr\"unden g\"unstiger, gegen
Unendlich zu extrapolieren, was impliziert, da{\ss} der Grenzwert $s$ durch
den Quotienten $a_k / b_k$ approximiert wird.

Die rationale Funktion ${\cal S}_{2 k} (x)$ kann auch als N\"aherungsbruch
eines Thieleschen interpolierenden Kettenbruchs [Thiele 1909, 3. Teil]
interpretiert werden. In Abschnitt \Roemisch{4}.1.4 des Buches von Cuyt
und Wuytack [1987] wird gezeigt, da{\ss} der von Wynn [1956b] eingef\"uhrte
$\rho$-Algorithmus, Gl. (3.3-12), die Berechnung des N\"aherungsbruches
${\cal S}_{2 k} (x)$ und die Extrapolation nach Unendlich auf
effizientere Weise durchf\"uhrt als andere rekursive Algorithmen. Dabei
mu{\ss} man die Interpolationspunkte $\Seqn x$ so w\"ahlen, da{\ss} sie Gl.
(3.3-13) erf\"ullen. Ein weiterer Vorzug des Wynnschen $\rho$-Algorithmus
ist, da{\ss} er anscheinend nicht besonders anf\"allig ist f\"ur Rundungsfehler
und auch bei vergleichsweise ungenauen Eingabedaten noch einigerma{\ss}en
gut funktioniert [Smith und Ford 1979; 1982; Weniger 1989; Abschnitt
14.4]. Das ist im Zusammenhang mit der Extrapolation von
Kristallorbitall- und Clusterenergien in Tabelle 11-1 von besonderer
Wichtigkeit, da die relative Genauigkeit der Eingabedaten mit an
Sicherheit grenzender Wahrscheinlichkeit deutlich kleiner ist als
Maschinengenauigkeit.

Der Wynnsche $\rho$-Algorithmus, Gl. (3.3-12), ist exakt f\"ur
Modellfolgen, deren Elemente durch rationale Funktionen des folgenden
Typs in den Interpolationspunkten $\Seqn x$ dargestellt werden k\"onnen
[Brezinski und Redivo Zaglia 1991, S. 102]:
$$
s_n \; = \, \frac
{ s x_n^k \, + \, a_1 x_n^{k-1} \, + \, \cdots \, + \, a_k }
{ x_n^k \, + \, b_1 x_n^{k-1} \, + \, \cdots \, + \, b_k }
\tag
$$
Eine Determinantendarstellung f\"ur den Wynnschen $\rho$-Algorithmus ist
ebenfalls bekannt [Brezinski und Redivo Zaglia 1991, S. 101].

Der verallgemeinerte Summationsproze{\ss} ${\cal W}_{k}^{(n)}$, Gl.
(3.3-17), der durch Iteration der Transformation $\rho_{2}^{(n)}$, Gl.
(3.3-16), abgeleitet wurde [Weniger 1989, Abschnitt 6.3], besitzt
\"ahnliche Eigenschaften wie der $\rho$-Algorithmus, Gl. (3.3-12). F\"ur die
Berechnung der Transformationen ${\cal W}_{k}^{(n)}$ und $\rho_{2
k}^{(n)}$ ben\"otigt man sowohl die numerischen Werte der $2 k + 1$
Folgenelemente $s_{n}$, $s_{n+1}$, $\ldots$ , $s_{n+2k}$ als auch der $2
k + 1$ Interpolationspunkte $x_{n}$, $x_{n+1}$, $\ldots$ , $x_{n+2k}$.
In beiden F\"allen m\"ussen die Interpolationspunkte $\Seqn x$ so gew\"ahlt
werden, da{\ss} sie Gl. (3.3-13) erf\"ullen.

Bei allen Extrapolationsalgorithmen, die auf Interpolationsverfahren
basieren, und die als Eingabedaten nicht nur die Elemente einer Folge
$\Seqn s$ ben\"otigen, sondern auch eine Folge $\Seqn x$ von
Interpolationspunkten, gibt es ein prinzipielles Problem: Wie mu{\ss} man
f\"ur eine gegebene Folge $\Seqn s$ von Eingabedaten die
Interpolationspunkte $\Seqn x$ w\"ahlen, um eine m\"oglichst schnelle
Konvergenz der Transformationen zu erreichen.

Das ist keineswegs eine triviale Frage. Beispielsweise wird das
Extrapolationsverfahren von Richardson, Gl. (11.5-5), am h\"aufigsten in
Verbindung mit den Interpolationspunkten
$$
x_n \; = \; 1/(n + \beta) \, , \qquad \beta > 0 \, ,
\quad n \in \N_0 \, ,
\tag
$$
verwendet. In Theorem 14-4 von Weniger [1989] wurde gezeigt, da{\ss} diese
Variante des Richardsonschen Extrapolationsverfahrens die Konvergenz von
Folgen des Typs von Gl. (11.4-4) deutlich verbessern kann, wenn der
Abklingparameter $\alpha$ eine positive ganze Zahl ist. Wenn $\alpha$
dagegen keine positive ganze Zahl ist, ist diese Variante des
Richardsonschen Extrapolationsverfahrens wirkungslos und f\"uhrt zu keiner
Konvergenzbeschleunigung.

Die Ergebnisse in Tabelle 14-3 von Weniger [1989] zeigen aber, da{\ss}
${\cal N}_{k}^{(n)} (s_n , x_n)$, Gl. (11.5-5), anscheinend die
Konvergenz einer Folge vom Typ von Gl. (11.4-4) mit einem {\it
nichtganzzahligen\/} Abklingparameter $\alpha > 0$ verbessern kann, wenn
man die Interpolationspunkte gem\"a{\ss}
$$
x_n \; = \; (n + \beta)^{- \alpha} \, , \qquad \beta > 0 \, ,
\quad n \in \N_0 \, ,
\tag
$$
w\"ahlt.

Im Falle des Wynnschen $\rho$-Algorithmus, Gl. (3.3-12), und seiner
Iteration ${\cal W}_{k}^{(n)}$, Gl. (3.3-17), ist die Situation ganz
\"ahnlich. Es scheint [Weniger 1989, Tabelle 14-3], da{\ss} die Konvergenz
einer Folge vom Typ von Gl. (11.4-4) mit dem festen Abklingparameter
$\alpha > 0$ nur dann verbessert werden kann, wenn die
Interpolationspunkte $\Seqn {x}$ gem\"a{\ss}
$$
x_n \; = \; (n + \beta)^{\alpha} \, , \qquad \beta > 0 \, ,
\quad n \in \N_0 \, ,
\tag
$$
gew\"ahlt werden.

Sowohl im Falle des Richardsonschen Extrapolationsprozesses, Gl.
(11.5-5), als auch im Falle des Wynnschen $\rho$-Algorithmus, Gl.
(3.3-12), und seiner Iteration ${\cal W}_{k}^{(n)}$, Gl. (3.3-17), ist
es also \"au{\ss}erst wichtig, den Abklingparameter $\alpha$ zu kennen. Nur
dann kann man erwarten, die Konvergenz einer Folge vom Typ von Gl.
(11.4-4) wirksam beschleunigen zu k\"onnen.

Zur Beschleunigung der Konvergenz von Folgen vom Typ von Gl. (11.4-4)
mit einem festen Abklingparameter $\alpha > 0$ wurden au{\ss}erdem noch die
von Osada [1990a, Gl. (3.1)] eingef\"uhrte Verallgemeinerung ${\bar
\rho}_{k}^{(n)}$, Gl. (8.4-11), des Wynnschen $\rho$-Algorithmus, Gl.
(3.3-12), und eine von Bj{\o}rstad, Dahlquist und Grosse [1981, Gl.
(2.4)] eingef\"uhrte Modifikation ${\overline {\cal W}}_k^{(n)}$, Gl.
(8.4-14), des Aitkenschen iterierten $\Delta^2$-Algorithmus, Gl.
(3.3-8), verwendet. In beiden F\"allen mu{\ss} der Wert des Abklingparameters
$\alpha > 0$ explizit bekannt sein, da er in den Rekursionen (8.4-11)
und (8.4-14), mit denen ${\bar \rho}_{k}^{(n)}$ beziehungsweise
${\overline {\cal W}}_k^{(n)}$ berechnet wird, vorkommt. F\"ur $\alpha =
1$ ist ${\bar \rho}_{k}^{(n)}$ identisch mit der Standardform des
Wynnschen $\rho$-Algorithmus, Gl. (3.3-15), die man erh\"alt, wenn man in
dem Rekursionsschema (3.3-12) $x_n = n + \beta$ mit $\beta > 0$ setzt.
Analog ist ${\overline {\cal W}}_k^{(n)}$ f\"ur $\alpha = 1$ identisch mit
der Standardform (3.3-18) von ${\cal W}_{k}^{(n)}$, die man erh\"alt,
indem man im Rekursionsschema (3.3-17) $x_n = n + \beta$ mit $\beta > 0$
setzt.

Aus den asymptotischen Fehlerabsch\"atzungen (8.4-12) beziehungsweise
(8.4-15) folgt, da{\ss} ${\bar \rho}_{k}^{(n)}$ und ${\overline {\cal
W}}_k^{(n)}$ im Falle von Folgen des Typs von Gl. (11.4-4) asymptotisch
\"aquivalent sind. Wie schon in Abschnitt 8.4 ausgef\"uhrt wurde, sind die
asymptotischen Absch\"atzungen (8.4-12) und (8.4-15) {\it optimal\/}. Das
bedeutet, da{\ss} kein verallgemeinerter Summationsproze{\ss}, der nur die
numerischen Werte von $2 k + 1$ Elementen $s_n$, $s_{n+1}$, $\ldots$ ,
$s_{n + 2 k}$ einer Folge vom Typ von Gl. (11.4-4) und den Wert des
Abklingparameters $\alpha$ als Eingabedaten verwendet, eine bessere
asymptotische Absch\"atzung des Transformationsfehlers erreichen kann.

Der von Brezinski [1971] eingef\"uhrte $\theta$-Algorithmus, Gl. (4.4-13),
und der eng verwandte verallgemeinerte Summationsproze{\ss} ${\cal
J}_{k}^{(n)}$, Gl. (8.4-18), der durch Iteration des expliziten
Ausdrucks (8.4-22) f\"ur $\theta_{2}^{(n)}$ abgeleitet werden kann
[Weniger 1989, Abschnitt 10.3], sind \"au{\ss}erst vielseitig: Sie sind nicht
nur in der Lage, alternierende divergente Reihen zu summieren, sondern
auch die Konvergenz sowohl von linear konvergenten Folgen als auch von
logarithmisch konvergenten Folgen des Typs von Gl. (11.4-4) mit einem
definierten Abklingparameter $\alpha$ zu beschleunigen. Bei Folgen des
Typs von Gl. (11.4-4) sind $\theta_{k}^{(n)}$ und ${\cal J}_{k}^{(n)}$
allerdings etwas weniger effizient als beispielsweise ${\bar
\rho}_{k}^{(n)}$, Gl. (8.4-11), oder ${\overline {\cal W}}_k^{(n)}$, Gl.
(8.4-14), da sie die Information, da{\ss} die Eingabedaten durch einen
festen Abklingparameter $\alpha$ charakterisiert werden, nicht
nutzbringend verwenden k\"onnen.

\"Ahnlich vielseitig wie $\theta_{k}^{(n)}$ und ${\cal J}_{k}^{(n)}$ ist
auch die von Levin [1973] eingef\"uhrte $u$-Transformation, Gl. (5.2-13).
Ein weiterer verallgemeinerter Summationsproze{\ss}, der divergente
alternierende Reihen summieren kann und der sowohl lineare als auch
logarithmische Konvergenz beschleunigen kann ist die von Levin [1973]
eingef\"uhrte $v$-Transformation, Gl. (5.2-20). Verglichen mit der
$u$-Transformation hat die $v$-Transformation aber den Nachteil, da{\ss} man
zur Berechnung von $v_{m}^{(0)} (\zeta, s_0)$ die Folgenelemente $s_0$,
$s_1$, $\ldots$, $s_{m+1}$ ben\"otigt, wogegen man zur Berechnung von
$u_{m}^{(0)} (\zeta, s_0)$ nur die Folgenelemente $s_0$, $s_1$,
$\ldots$, $s_m$ ben\"otigt. Da nur die Kristallorbital- und
Clusterenergien $E_N^{(ko)}$ und $E_N^{(cl)}$ mit $1 \le N \le 7$ als
Eingabedaten zur Verf\"ugung stehen, wurde ausschlie{\ss}lich die Levinsche
$u$-Transformation, Gl. (5.2-13), zur Extrapolation verwendet.

Verallgemeinerte Summationsprozesse des dritten Typus sind der von Wynn
[1956a] eingef\"uhrte $\epsilon$-Algorithmus, Gl. (2.4-10), und der eng
verwandte nach Aitken [1926] benannte iterierte $\Delta^2$-Algorithmus,
Gl. (3.3-8). Weiterhin wurde noch die von Levin [1973] eingef\"uhrte
$t$-Transformation, Gl. (5.2-15), zur Extrapolation verwendet. Diese
drei Transformationen sind in der Lage, lineare Konvergenz zu
beschleunigen, versagen aber v\"ollig im Falle logarithmischer Konvergenz.
Das bedeutet, da{\ss} diese Transformationen aus den Kristallorbital- und
Clusterenergien in Tabelle 11-1 zwar exponentielle Terme des Typs $\exp
(- \gamma n)$ mit $\gamma > 0$ eliminieren k\"onnen, nicht aber Potenzen
des Typs $(n+1)^{- \alpha - j}$ mit $\alpha > 0$ und $j \in \N_0$.

\medskip

\Abschnitt Extrapolationsergebnisse

\smallskip

\aktTag = 0

In Tabelle 11-3 wird versucht, die Konvergenz der Clusterenergien
$\bigl\{ E_{n+1}^{(cl)} \bigr\}_{n=0}^{6}$ aus Tabelle 11-1 mit Hilfe
der im letzten Abschnitt beschriebenen verallgemeinerten
Summationsprozesse zu beschleunigen.

Aufgrund der Ergebnisse in Tabelle 11-2 ist es eine relativ naheliegende
Annahme, da{\ss} die Clusterenergien in Tabelle 11-1 durch
Reihenentwicklungen vom Typ von Gl. (11.4-4) mit dem festen
Abklingparameter $\alpha = 1$ dargestellt werden k\"onnen. Auf der Basis
dieser Annahme kann man versuchen, Vorhersagen \"uber die Effizienz der in
Tabelle 11-3 verwendeten verallgemeinerten Summationsprozesse zu machen.

Der Richardsonsche Extrapolationsproze{\ss} ${\cal N}_{k}^{(n)} (s_n ,
x_n)$, Gl. (11.5-5), wurde in Tabelle 11-3 sowohl mit den
Interpolationspunkten $x_n = (n+1)^{- 1}$ als auch mit den
Interpolationspunkten $x_n = (n+1)^{- 2}$ verwendet. Analog wurde der
Wynnsche $\rho$-Algorithmus, Gl. (3.3-12), und seine Iteration ${\cal
W}_{k}^{(n)}$, Gl. (3.3-17), sowohl mit den Interpolationspunkten $x_n =
n+1$ als auch mit den Interpolationspunkten $x_n = (n+1)^2$ verwendet.
Auf der Basis der oben genannten Annahme ist zu erwarten, da{\ss} der
Richardsonsche Extrapolationsproze{\ss} gute Extrapolationsergebnisse
liefern sollte, wenn man die Interpolationspunkte $x_n = (n+1)^{- 1}$
verwendet, und er sollte relativ schlechte Extrapolationsergebnisse
liefern, wenn man die Interpolationspunkte $x_n = (n+1)^{- 2}$
verwendet. Analog sollte der Wynnsche $\rho$-Algorithmus und seine
Iteration ${\cal W}_{k}^{(n)}$ gute Ergebnisse liefern, wenn man die
Interpolationspunkte $x_n = n+1$ verwendet, und schlechte Ergebnisse,
wenn man die Interpolationspunkte $x_n = (n+1)^2$ verwendet.

Wie schon im letzten Unterabschnitt erw\"ahnt wurde, wird die von Osada
[1990a, Gl. (3.1)] eingef\"uhrte Modifikation ${\bar \rho}_{k}^{(n)}$, Gl.
(8.4-11), des Wynnschen $\rho$-Algorithmus f\"ur $\alpha = 1$ identisch
mit der Standardform des Wynnschen $\rho$-Algorithmus, Gl. (3.3-15).
Analog wird die von Bj{\o}rstad, Dahlquist und Grosse [1981, Gl. (2.4)]
eingef\"uhrte Modifikation ${\overline {\cal W}}_k^{(n)}$, Gl. (8.4-14),
des Aitkenschen iterierten $\Delta^2$-Algorithmus f\"ur $\alpha = 1$
identisch mit der Standardform (3.3-18) von ${\cal W}_{k}^{(n)}$. In
Tabelle 11-3 werden diese beiden Transformationen deswegen nur in
Verbindung mit dem Abklingparameter $\alpha = 2$ verwendet. Auf der
Basis der oben gemachten Annahme sind f\"ur ${\bar \rho}_{k}^{(n)}$ und
${\overline {\cal W}}_k^{(n)}$ nur relativ schlechte
Extrapolationsergebnisse zu erwarten.

Der $\theta$-Algorithmus, Gl. (4.4-13), und seine Iteration ${\cal
J}_{k}^{(n)}$, Gl. (8.4-18), sowie die Levinsche $u$-Transformation, Gl.
(5.2-13), sollten die Konvergenz einer Folge vom Typ von Gl. (11.4-4)
unabh\"angig vom Wert des Abklingparameters $\alpha > 0$ verbessern
k\"onnen. Allerdings sollten sie etwas weniger effizient sein als
diejenigen Verfahren des ersten Typus, die optimal sind f\"ur Folgen des
Typs von Gl. (11.4-4) mit dem Abklingparameter $\alpha = 1$.

Auf der Basis der obigen Annahme ist zu erwarten, da{\ss} die
verallgemeinerten Summationsprozesse des dritten Typus, die
ausschlie{\ss}lich lineare Konvergenz beschleunigen k\"onnen, die Konvergenz
der Clusterenergien nicht verbessern k\"onnen.

\beginFloat

\medskip

\beginTabelle % [to \kolumnenbreite]
\beginFormat
\mitte " \mitte " \mitte " \mitte " \mitte " \mitte
\endFormat
\+ " \links {\bf Tabelle 11-3} \@ \@ \@ \@ \@ " \\
\+ " \links {Extrapolation der Clusterenergien aus Tabelle 11-1}
\@ \@ \@ \@ \@ " \\
\+ " \links {Eingabedaten $s_n \, = \; E_{n+1}^{(cl)}$ mit $0
\le n \le 6$}  \@ \@ \@ \@ \@ " \\
\sstrut {$E_{n+1}^{(cl)}$} {0.2 \jot} {1.2 \jot}
\= \\ \sstrut {} {1.5 \jot} {1.5 \jot}
\+ " \rechts {$n$} " $u_{n}^{(0)} (1, s_0)$ " $t_{n}^{(0)} (1, s_0)$
" $\epsilon_{2 \Ent {n/2}}^{(n - 2 \Ent {n/2})}$
" ${\cal A}_{\Ent {n/2}}^{(n - 2 \Ent {n/2})}$
" $\theta_{2 \Ent {n/3}}^{(n - 3 \Ent {n/3})}$ \\
\sstrut {$\theta_{2 \Ent {n/3}}^{(n - 3 \Ent {n/3})}$} {0.8 \jot} {1.2 \jot}
\+ " " Gl. (5.2-13) " Gl. (5.2-15) " Gl. (2.4-10) " Gl. (3.3-8)
" Gl. (4.4-13) \\ \sstrut {} {1.5 \jot} {1.5 \jot}
\= \\ \sstrut {} {1 \jot} {1 \jot}
\+ " 0 " $-77.638000$ " $-77.638000$ " $-77.638000$ " $-77.638000$ " $-77.638000$ \\
\+ " 1 " $-77.119476$ " $-77.115989$ " $-77.112456$ " $-77.112456$ " $-77.112456$ \\
\+ " 2 " $-76.591899$ " $-76.800337$ " $-76.852695$ " $-76.852695$ " $-76.938618$ \\
\+ " 3 " $-76.588410$ " $-76.706611$ " $-76.765170$ " $-76.765170$ " $-76.589291$ \\
\+ " 4 " $-76.593203$ " $-76.666017$ " $-76.707476$ " $-76.678719$ " $-76.591069$ \\
\+ " 5 " $-76.590531$ " $-76.643416$ " $-76.678297$ " $-76.656600$ " $-76.591236$ \\
\+ " 6 " $-76.591827$ " $-76.630034$ " $-76.656685$ " $-76.624767$ " $-76.592929$ \\
\sstrut {} {1.5 \jot} {1.5 \jot}
\= \\ \sstrut {} {1 \jot} {1 \jot}
\+ " \rechts {$n$}
" ${\cal J}_{\Ent {n/3}}^{(n - 3 \Ent {n/3})}$
" ${\cal N}_n^{(0)} (s_0, x_0)$ " ${\cal N}_n^{(0)} (s_0, x_0)$
" ${\bar \rho}_{2 \Ent {n/2}}^{(n - 2 \Ent {n/2})}$
" ${\overline {\cal W}}_{\Ent {n/2}}^{(n - 2 \Ent {n/2})}$ \\ \sstrut
{${\overline {\cal W}}_{\Ent {n/2}}^{(n - 2 \Ent {n/2})}$} {0.8 \jot} {1.2 \jot}
\+ " " Gl. (8.4-18) " Gl. (11.5-5) " Gl. (11.5-5) " Gl. (8.4-11) " Gl.
(8.4-14) \\ \sstrut {} {1.5 \jot} {1.5 \jot}
\+ " " " $x_n = (n + 1)^{-1}$ " $x_n = (n + 1)^{-2}$ " $\alpha = 2$
" $\alpha = 2$ \\ \sstrut {} {1.5 \jot} {1.5 \jot}
\= \\ \sstrut {} {1 \jot} {1 \jot}
\+ " 0 " $-77.638000$ " $-77.638000$ " $-77.638000$ " $-77.638000$ " $-77.638000$ \\
\+ " 1 " $-77.112456$ " $-76.586911$ " $-76.937274$ " $-77.112456$ " $-77.112456$ \\
\+ " 2 " $-76.938618$ " $-76.592957$ " $-76.782331$ " $-76.722814$ " $-76.722814$ \\
\+ " 3 " $-76.589291$ " $-76.591313$ " $-76.716246$ " $-76.678447$ " $-76.678447$ \\
\+ " 4 " $-76.591069$ " $-76.591485$ " $-76.681105$ " $-76.630658$ " $-76.624739$ \\
\+ " 5 " $-76.591236$ " $-76.591407$ " $-76.659826$ " $-76.620447$ " $-76.615922$ \\
\+ " 6 " $-76.592925$ " $-76.591429$ " $-76.645806$ " $-76.608414$ " $-76.602997$ \\
\sstrut {} {1.5 \jot} {1.5 \jot}
\= \\ \sstrut {} {1 \jot} {1 \jot}
\+ " \rechts {$n$} " $\rho_{2 \Ent {n/2}}^{(n - 2 \Ent {n/2})}$
" ${\cal W}_{\Ent {n/2}}^{(n - 2 \Ent {n/2})}$
" $\rho_{2 \Ent {n/2}}^{(n - 2 \Ent {n/2})}$
" ${\cal W}_{\Ent {n/2}}^{(n - 2 \Ent {n/2})}$ \\ \sstrut
{${\cal W}_{\Ent {n/2}}^{(n - 2 \Ent {n/2})}$} {0.8 \jot} {1.2 \jot}
\+ " " Gl. (3.3-12) " Gl. (3.3-17) " Gl. (3.3-12) " Gl. (3.3-17) \\
\+ " " $x_n = n + 1$ " $x_n = n + 1$
" $x_n = (n + 1)^2$ " $x_n = (n + 1)^2$ "
\\ \sstrut {} {1.5 \jot} {1.5 \jot}
\= \\ \sstrut {} {1 \jot} {1 \jot}
\+ " 0 " $-77.638000$ " $-77.638000$ " $-77.638000$ " $-77.638000$ " \\
\+ " 1 " $-77.112456$ " $-77.112456$ " $-77.112456$ " $-77.112456$ " \\
\+ " 2 " $-76.592934$ " $-76.592934$ " $-76.765445$ " $-76.765445$ " \\
\+ " 3 " $-76.591723$ " $-76.591723$ " $-76.707239$ " $-76.707239$ " \\
\+ " 4 " $-76.591472$ " $-76.591472$ " $-76.660997$ " $-76.655863$ " \\
\+ " 5 " $-76.591417$ " $-76.591417$ " $-76.643530$ " $-76.639667$ " \\
\+ " 6 " $-76.591425$ " $-76.591425$ " $-76.628693$ " $-76.620491$ " \\
\sstrut {} {1.5 \jot} {1.5 \jot}
\= \\ \sstrut {} {1 \jot} {1 \jot}

\endTabelle

\medskip

\endFloat

Die Ergebnisse in Tabelle 11-3 sind weitgehend in \"Ubereinstimmung mit
der Annahme, da{\ss} die Clusterenergien durch Reihenentwicklungen des Typs
von Gl.~(11.4-4) mit dem Abklingparameter $\alpha = 1$ dargestellt
werden k\"onnen. Sowohl der Richardsonsche Extrapolationsproze{\ss} ${\cal
N}_{k}^{(n)} (s_n , x_n)$ in Verbindung mit den Interpolationspunkten
$x_n = (n+1)^{- 1}$ als auch der Wynnsche $\rho$-Algorithmus und seine
Iteration ${\cal W}_{k}^{(n)}$ in Verbindung mit den
Interpolationspunkten $x_n = n + 1$ f\"uhrten in Tabelle 11-3 zu einer
bemerkenswerten Verbesserung der Konvergenz. Die
Extrapolationsergebnisse dieser drei Transformationen unterschieden sich
untereinander erst in der sechsten Stelle nach dem Komma, und man erh\"alt
$E_{\infty}^{(cl)} = - 76.591425 \pm 0.000005$ als
Extrapolationsergebnis. Man kann also durch Extrapolation der
Clusterenergien $\bigl\{ E_{n+1}^{(cl)} \bigr\}_{n=0}^{6}$ mindestens
vier oder m\"oglicherweise sogar f\"unf Dezimalstellen gewinnen, was in
Anbetracht der geringen Zahl der zur Verf\"ugung stehenden Eingabedaten
eine bemerkenswerte Verbesserung der Konvergenz ist. Um diese
Genauigkeit aus unextrapolierten Clusterrechnungen zu erhalten, m\"u{\ss}te
man Rechnungen an Molek\"ulen aus $10^4$ oder vielleicht sogar $10^5$
Acetyleneinheiten durchf\"uhren.

Der $\theta$-Algorithmus und seine Iteration ${\cal J}_{k}^{(n)}$
verhielten sich wie erwartet, da sie ebenfalls eine Verbesserung der
Konvergenz bewirkten. Sie waren aber wie erwartet deutlich weniger
wirksam als diejenigen verallgemeinerten Summationsprozesse, die von der
Information profitieren k\"onnen, da{\ss} die Clusterenergien in Tabelle 11-1
durch Reihenentwicklungen des Typs von Gl. (11.4-4) mit dem festen
Abklingparameter $\alpha = 1$ dargestellt werden k\"onnen. \"Uberraschend
ist das schlechte Ergebnis der Levinschen $u$-Transformation, die
normalerweise zu den besten Beschleunigern f\"ur logarithmisch konvergente
Reihen des Typs von Gl. (11.4-4) geh\"ort.

Der Richardsonsche Extrapolationsproze{\ss} ${\cal N}_{k}^{(n)} (s_n , x_n)$
in Verbindung mit den Interpolationspunkten $x_n = (n+1)^{- 2}$, der
Wynnsche $\rho$-Algorithmus und seine Iteration ${\cal W}_{k}^{(n)}$ in
Verbindung mit den Interpolationspunkten $x_n = (n + 1)^2$, als auch
${\bar \rho}_{k}^{(n)}$ und ${\overline {\cal W}}_k^{(n)}$, die den
Abklingparameter $\alpha = 2$ verwenden, bewirkten wie erwartet keine
Verbesserung der Konvergenz. Die verallgemeinerten Summationsprozesse
des dritten Typus verhielten sich ebenfalls wie erwartet, da sie die
Konvergenz der Clusterenergien nicht verbessern konnten.

In Tabelle 11-4 wird versucht, die Konvergenz der
Kristallorbitalenergien $\bigl\{ E_{n+1}^{(ko)} \bigr\}_{n=0}^{6}$ aus
Tabelle 11-1 mit Hilfe derjenigen verallgemeinerten Summationsprozesse
zu verbessern, die in Tabelle 11-3 zur Extrapolation der Clusterenergien
verwendet wurden.

\beginFloat

\medskip

\beginTabelle % [to \kolumnenbreite]
\beginFormat
\mitte " \mitte " \mitte " \mitte " \mitte " \mitte
\endFormat
\+ " \links {\bf Tabelle 11-4} \@ \@ \@ \@ \@ " \\
\+ " \links {Extrapolation der Kristallorbitalenergien aus Tabelle 11-1}
\@ \@ \@ \@ \@ " \\
\+ " \links {Eingabedaten $s_n \, = \; E_{n+1}^{(ko)}$ mit $0
\le n \le 6$}  \@ \@ \@ \@ \@ " \\
\sstrut {$E_{n+1}^{(cl)}$} {0.2 \jot} {1.2 \jot}
\= \\ \sstrut {} {1.5 \jot} {1.5 \jot}
\+ " \rechts {$n$} " $u_{n}^{(0)} (1, s_0)$ " $t_{n}^{(0)} (1, s_0)$
" $\epsilon_{2 \Ent {n/2}}^{(n - 2 \Ent {n/2})}$
" ${\cal A}_{\Ent {n/2}}^{(n - 2 \Ent {n/2})}$
" $\theta_{2 \Ent {n/3}}^{(n - 3 \Ent {n/3})}$ \\
\sstrut {$\theta_{2 \Ent {n/3}}^{(n - 3 \Ent {n/3})}$} {0.8 \jot} {1.2 \jot}
\+ " " Gl. (5.2-13) " Gl. (5.2-15) " Gl. (2.4-10) " Gl. (3.3-8)
" Gl. (4.4-13) \\ \sstrut {} {1.5 \jot} {1.5 \jot}
\= \\ \sstrut {} {1 \jot} {1 \jot}
\+ " 0 " $-76.581341$ " $-76.581341$ " $-76.581341$ " $-76.581341$ " $-76.581341$ \\
\+ " 1 " $-76.592144$ " $-76.592143$ " $-76.592141$ " $-76.592141$ " $-76.592141$ \\
\+ " 2 " $-76.591877$ " $-76.591873$ " $-76.591871$ " $-76.591871$ " $-76.591864$ \\
\+ " 3 " $-76.592816$ " $-76.590852$ " $-76.591583$ " $-76.591583$ " $-76.599061$ \\
\+ " 4 " $-76.591789$ " $-76.591575$ " $-76.591527$ " $-76.591514$ " $-76.591396$ \\
\+ " 5 " $-76.591061$ " $-76.591421$ " $-76.591481$ " $-76.591467$ " $-76.591420$ \\
\+ " 6 " $-76.591538$ " $-76.591488$ " $-76.591479$ " $-76.591475$ " $-76.591403$ \\
\sstrut {} {1.5 \jot} {1.5 \jot}
\= \\ \sstrut {} {1 \jot} {1 \jot}
\+ " \rechts {$n$}
" ${\cal J}_{\Ent {n/3}}^{(n - 3 \Ent {n/3})}$
" ${\cal N}_n^{(0)} (s_0, x_0)$ " ${\cal N}_n^{(0)} (s_0, x_0)$
" ${\bar \rho}_{2 \Ent {n/2}}^{(n - 2 \Ent {n/2})}$
" ${\overline {\cal W}}_{\Ent {n/2}}^{(n - 2 \Ent {n/2})}$ \\ \sstrut
{${\overline {\cal W}}_{\Ent {n/2}}^{(n - 2 \Ent {n/2})}$} {0.8 \jot} {1.2 \jot}
\+ " " Gl. (8.4-18) " Gl. (11.5-5) " Gl. (11.5-5) " Gl. (8.4-11) " Gl.
(8.4-14) \\ \sstrut {} {1.5 \jot} {1.5 \jot}
\+ " " " $x_n = (n + 1)^{-1}$ " $x_n = (n + 1)^{-2}$ " $\alpha = 2$
" $\alpha = 2$ \\ \sstrut {} {1.5 \jot} {1.5 \jot}
\= \\ \sstrut {} {1 \jot} {1 \jot}
\+ " 0 " $-76.581341$ " $-76.581341$ " $-76.581341$ " $-76.581341$ " $-76.581341$ \\
\+ " 1 " $-76.592141$ " $-76.602942$ " $-76.595741$ " $-76.592141$ " $-76.592141$ \\
\+ " 2 " $-76.591864$ " $-76.585493$ " $-76.591130$ " $-76.591736$ " $-76.591736$ \\
\+ " 3 " $-76.599061$ " $-76.593238$ " $-76.591538$ " $-76.591442$ " $-76.591442$ \\
\+ " 4 " $-76.591396$ " $-76.590878$ " $-76.591459$ " $-76.591425$ " $-76.591424$ \\
\+ " 5 " $-76.591420$ " $-76.591505$ " $-76.591442$ " $-76.591425$ " $-76.591425$ \\
\+ " 6 " $-76.591401$ " $-76.591498$ " $-76.591443$ " $-76.591424$ " $-76.591424$ \\
\sstrut {} {1.5 \jot} {1.5 \jot}
\= \\ \sstrut {} {1 \jot} {1 \jot}
\+ " \rechts {$n$} " $\rho_{2 \Ent {n/2}}^{(n - 2 \Ent {n/2})}$
" ${\cal W}_{\Ent {n/2}}^{(n - 2 \Ent {n/2})}$
" $\rho_{2 \Ent {n/2}}^{(n - 2 \Ent {n/2})}$
" ${\cal W}_{\Ent {n/2}}^{(n - 2 \Ent {n/2})}$ \\ \sstrut
{${\cal W}_{\Ent {n/2}}^{(n - 2 \Ent {n/2})}$} {0.8 \jot} {1.2 \jot}
\+ " " Gl. (3.3-12) " Gl. (3.3-17) " Gl. (3.3-12) " Gl. (3.3-17)
\\ \sstrut {} {1.5 \jot} {1.5 \jot}
\+ " " $x_n = n + 1$ " $x_n = n + 1$
" $x_n = (n + 1)^2$ " $x_n = (n + 1)^2$ "
\\ \sstrut {} {1.5 \jot} {1.5 \jot}
\= \\ \sstrut {} {1 \jot} {1 \jot}
\+ " 0 " $-76.581341$ " $-76.581341$ " $-76.581341$ " $-76.581341$ " \\
\+ " 1 " $-76.592141$ " $-76.592141$ " $-76.592141$ " $-76.592141$ " \\
\+ " 2 " $-76.591600$ " $-76.591600$ " $-76.591704$ " $-76.591704$ " \\
\+ " 3 " $-76.591302$ " $-76.591302$ " $-76.591490$ " $-76.591490$ " \\
\+ " 4 " $-76.591337$ " $-76.591336$ " $-76.591450$ " $-76.591448$ " \\
\+ " 5 " $-76.591649$ " $-76.591557$ " $-76.591435$ " $-76.591434$ " \\
\+ " 6 " $-76.591217$ " $-76.591376$ " $-76.591442$ " $-76.591440$ " \\
\sstrut {} {1.5 \jot} {1.5 \jot}
\= \\ \sstrut {} {1 \jot} {1 \jot}

\endTabelle

\medskip

\endFloat

Eine Vorhersage des Leistungsverm\"ogens der verallgemeinerten
Summationsprozesse ist im Falle der Kristallorbitalenergien aber
wesentlich schwieriger als im Falle der Clusterenergien. Aufgrund der
Ergebnisse in Tabelle 11-2 ist es nicht klar, ob die
Kristallorbitalenergien \"uberhaupt durch Reihenentwicklungen vom Typ von
Gl. (11-4-4) mit einem festen Abklingparameter $\alpha > 0$ dargestellt
werden k\"onnen, und wenn ja, welchen Wert dieser Abklingparameter hat.
Aufgrund der relativ guten Konvergenz der Kristallorbitalenergien mu{\ss}
der Abklingparameter der Kristallorbitalenergien -- wenn er existiert --
aber deutlich gr\"o{\ss}er sein als der Abklingparameter der Clusterenergien.

Wenn die Kristallorbitalenergien tats\"achlich durch Reihenentwicklungen
des Typs von Gl. (11.4-4) mit einem festen Abklingparameter $\alpha$
dargestellt werden k\"onnen, sollte der Richardsonsche
Extrapolationsproze{\ss} ${\cal N}_{k}^{(n)} (s_n , x_n)$ in Verbindung mit
den Interpolationspunkten $x_n = (n+1)^{- 1}$ schlechtere
Extrapolationsergebnisse liefern als in Verbindung mit den
Interpolationspunkten $x_n = (n+1)^{- 2}$. Analog sollte der Wynnsche
$\rho$-Algorithmus und seine Iteration ${\cal W}_{k}^{(n)}$ in
Verbindung mit den Interpolationspunkten $x_n = n + 1$ schlechtere
Ergebnisse liefern als in Verbindung mit den Interpolationspunkten $x_n
= (n+1)^2$. Allerdings ist es keineswegs klar, ob die
Interpolationspunkte $x_n = (n+1)^{- 2}$ f\"ur ${\cal N}_{k}^{(n)} (s_n ,
x_n)$ beziehungsweise $x_n = (n+1)^2$ f\"ur $\rho_{k}^{(n)}$ und ${\cal
W}_{k}^{(n)}$, die optimal w\"aren f\"ur die Extrapolation einer Folge vom
Typ von Gl. (11.4-4) mit dem festen Abklingparameter $\alpha = 2$, im
Falle der Kristallorbitalenergien tats\"achlich die besten Ergebnisse
liefern, oder ob Interpolationspunkte, die f\"ur einen Abklingparameter
$\alpha \ne 2$ optimal w\"aren, besser zur Extrapolation der
Kristallorbitalenergien geeignet w\"aren. Ebensowenig ist klar, ob ${\bar
\rho}_{k}^{(n)}$, Gl. (8.4-11), oder ${\overline {\cal W}}_k^{(n)}$, Gl.
(8.4-14), in Verbindung mit dem Abklingparameter $\alpha = 2$ oder mit
einem anderen Abklingparameter die besten Ergebnisse liefern w\"urde.

Beim augenblicklichen Stand des Wissens \"uber das asymptotische Verhalten
der Kristallorbitalenergien $E_{N}^{(ko)}$ f\"ur gro{\ss}e Werte von $N$ kann
man auch keine Vorhersagen machen, ob -- und wenn ja wie gut -- die
verallgemeinerten Summationsprozesse des dritten Typus, die nur lineare
Konvergenz beschleunigen k\"onnen, zur Extrapolation der
Kristallorbitalenergien in Tabelle 11-1 geeignet sind.

Die Kristallorbitalenergien in Tabelle 11-1 konvergieren wesentlich
schneller als die Clusterenergien. Man kann also nicht unbedingt
erwarten, da{\ss} man in Tabelle 11-4 \"ahnlich spektakul\"are Verbesserungen
der Konvergenz beobachten wird wie in Tabelle 11-3.

Die Extrapolationsergebnisse in Tabelle 11-4 sind widerspr\"uchlich und
keineswegs einfach zu interpretieren. Wenn die Kristallorbitalenergien
tats\"achlich durch Reihenentwicklungen des Typs von Gl. (11.4-4) mit
einem festen Abklingparameter $\alpha > 0$ dargestellt werden k\"onnen,
sollten die verallgemeinerten Summationsprozesse des dritten Typus, die
aus\-schlie{\ss}lich lineare Konvergenz beschleunigen k\"onnen, wirkungslos
sein. Die Levinsche $t$-Transformation, der Wynnsche
$\epsilon$-Algorithmus und der iterierte $\Delta^2$-Proze{\ss} von Aitken
waren aber in der Lage, mit $E_{\infty}^{(ko)} = - 76.59148 \pm 0.00001$
ein Extrapolationsergebnis zu produzieren, das sich nur wenig von dem
besten Resultat in Tabelle 11-3 ($E_{\infty}^{(cl)} = - 76.591425 \pm
0.000005$) unterscheidet. Die erfolgreiche Beschleunigung der Konvergenz
der Kristallorbitalenergien durch die Transformationen des dritten Typus
steht im Widerspruch zu der Annahme, da{\ss} die Kristallorbitalenergien in
Tabelle 11-1 durch logarithmisch konvergente Reihenentwicklungen vom Typ
von Gl. (11.4-4) mit einem festen Abklingparameter $\alpha >0$
dargestellt werden k\"onnen. Stattdessen sprechen diese Ergebnisse daf\"ur,
da{\ss} nichtanalytische exponentielle Terme bei den Abbruchfehlern
$E_{N}^{(ko)} - E_{\infty}^{(ko)}$ der Kristallorbitalenergien zumindest
f\"ur kleinere Werte von $N$ eine wesentliche Rolle spielen.

Bei den verallgemeinerten Summationsprozessen, die logarithmisch
konvergente Folgen des Typs von Gl. (11.4-4) beschleunigen k\"onnen, war
die Levinsche $u$-Transformation \"ahnlich wie in Tabelle 11-3 relativ
wenig wirksam. Eine plausible Erkl\"arung f\"ur das Versagen der
$u$-Transformation ist nicht bekannt. Ansonsten war das Verhalten der
verallgemeinerten Summationsprozesse des ersten Typus in den Tabellen
11-3 und 11-4 weitgehend komplement\"ar. In Tabelle 11-3 ergaben der
Richardsonsche Extrapolationsproze{\ss} ${\cal N}_{k}^{(n)} (s_n , x_n)$ in
Verbindung mit den Interpolationspunkten $x_n = (n+1)^{- 1}$, sowie der
Wynnschen $\rho$-Algorithmus und seine Iteration ${\cal W}_{k}^{(n)}$ in
Verbindung mit den Interpolationspunkten $x_n = n + 1$, die optimal sind
f\"ur die Beschleunigung der Konvergenz von Folgen des Typs von Gl.
(11.4-4) mit dem Abklingparameter $\alpha = 1$, die besten
Extrapolationsergebnisse ($E_{\infty}^{(cl)} = - 76.591425 \pm
0.000005$). In Tabelle 11-4 waren diese Transformationen nur wenig
wirksam. Dagegen produzierten ${\cal N}_{k}^{(n)} (s_n , x_n)$ in
Verbindung mit den Interpolationspunkten $x_n = (n+1)^{- 2}$,
$\rho_{k}^{(n)}$ und seine Iteration ${\cal W}_{k}^{(n)}$ in Verbindung
mit den Interpolationspunkten $x_n = (n + 1)^2$, sowie ${\bar
\rho}_{k}^{(n)}$ und ${\overline {\cal W}}_k^{(n)}$ mit den
Abklingparameter $\alpha = 2$, die alle optimal sind f\"ur die
Beschleunigung der Konvergenz von Folgen des Typs von Gl. (11.4-4) mit
dem Abklingparameter $\alpha = 2$, die besten Extrapolationsergebnisse
in Tabelle 11-4. Eine {\it konservative\/} Sch\"atzung ergibt
$E_{\infty}^{(ko)} = - 76.59145 \pm 0.00005$ als Extrapolationsergebnis.
Hervorzuheben ist noch die \"au{\ss}erst schnelle Konvergenz von ${\bar
\rho}_{k}^{(n)}$ und ${\overline {\cal W}}_k^{(n)}$, deren
Extrapolationsergebnisse bis auf die sechste Stelle nach dem Komma mit
den besten Extrapolationsergebnissen aus Tabelle 11-3 \"ubereinstimmen.

Die besten Extrapolationsergebnisse in den Tabellen 11-3 und 11-4 sind
also in etwa gleich gut, und sie unterscheiden sich erst in der f\"unften
Stelle nach dem Komma. In Anbetracht der sehr unterschiedlichen
Konvergenz der Kristallorbital- und Clusterenergien in Tabelle 11-1, die
au{\ss}erdem auf v\"ollig unterschiedliche Weise berechnet wurden, ist das
sicherlich eine bemerkenswerte \"Ubereinstimmung.

Das wirft nat\"urlich die Frage auf, ob es nicht m\"oglich ist, die
Extrapolationsergebnisse in Tabelle 11-4 noch zu verbessern. In Tabelle
11-4 wurde der Richardsonsche Extrapolationsproze{\ss} ${\cal N}_{k}^{(n)}
(s_n , x_n)$, der Wynnsche $\rho$-Agorithmus und seine Iteration ${\cal
W}_{k}^{(n)}$ in Verbindung mit Interpolationspunkten verwendet, die f\"ur
die Extrapolation von Folgen des Typs von Gl. (11.4-4) mit dem festen
Abklingparameter $\alpha = 2$ optimal sind. Analog wurden ${\bar
\rho}_{k}^{(n)}$ und ${\overline {\cal W}}_k^{(n)}$ mit den
Abklingparameter $\alpha = 2$ verwendet. Wie schon erw\"ahnt, ist die Wahl
$\alpha = 2$ aber mehr oder weniger willk\"urlich, und es ist es
keineswegs klar, ob man nicht mit einem Abklingparameter $\alpha \ne 2$
bessere Extrapolationsergebnisse erhalten k\"onnte.

Die Frage, f\"ur welchen Wert des Abklingparameters man die besten
Extrapolationsergebnisse erh\"alt, kann untersucht werden, indem man die
Kristallorbitalenergien in Tabelle 11-1 unter Verwendung von ${\cal
N}_{k}^{(n)} (s_n , x_n)$ in Verbindung mit den Interpolationspunkten
$x_n = (n + 1)^{- \alpha}$, $\rho_{k}^{(n)}$ und seine Iteration ${\cal
W}_{k}^{(n)}$ in Verbindung mit Interpolationspunkten $x_n = (n +
1)^{\alpha}$, sowie ${\bar \rho}_{k}^{(n)}$ und ${\overline {\cal
W}}_k^{(n)}$ f\"ur verschiedene Werte von $\alpha$ extrapoliert. Es zeigt
sich aber, da{\ss} die so erzielten Extrapolationsergebnisse in einem
weiten Bereich nur ganz schwach vom Abklingparameter $\alpha$ abh\"angen.

\beginFloat

\medskip

\beginTabelle % [to \kolumnenbreite]
\beginFormat
\mitte " \links " \mitte " \mitte " \mitte " \mitte " \mitte
\endFormat
\+ " \links {\bf Tabelle 11-5} \@ \@ \@ \@ \@ \@ " \\
\+ " \links {Extrapolation der Kristallorbitalenergien f\"ur verschiedene
Werte von $\alpha$}
\@ \@ \@ \@ \@ \@ " \\
\+ " \links {Eingabedaten $s_n \, = \; E_{n+1}^{(ko)}$ mit $0
\le n \le 6$}  \@ \@ \@ \@ \@ \@ " \\
\sstrut {$E_{n+1}^{(ko)}$} {0.2 \jot} {1.2 \jot}
\= \\
\+ " \links {$\alpha$} \| \rechts {$n$}
" ${\cal N}_n^{(0)} (s_0, x_0)$
" $\rho_{2 \Ent {n/2}}^{(n - 2 \Ent {n/2})}$
" ${\cal W}_{\Ent {n/2}}^{(n - 2 \Ent {n/2})}$
" ${\bar \rho}_{2 \Ent {n/2}}^{(n - 2 \Ent {n/2})}$
" ${\overline {\cal W}}_{\Ent {n/2}}^{(n - 2 \Ent {n/2})}$ \\ \sstrut
{${\overline {\cal W}}_{\Ent {n/2}}^{(n - 2 \Ent {n/2})}$} {0.8 \jot} {1.2 \jot}
\+ " \| " Gl. (11.5-5) " Gl. (3.3-12) " Gl. (3.3-17)
" Gl. (8.4-11) " Gl. (8.4-14) \\ \sstrut {} {1 \jot} {1 \jot}
\+ " \= \| \= " \= " \= " \= " \= " \= " \\ \sstrut {} {1 \jot} {1 \jot}
\+ " 1.8 \| 0 " $-76.581341$ " $-76.581341$ " $-76.581341$ " $-76.581341$ " $-76.581341$ \\
\+ "     \| 1 " $-76.596492$ " $-76.592141$ " $-76.592141$ " $-76.592141$ " $-76.592141$ \\
\+ "     \| 2 " $-76.590821$ " $-76.591688$ " $-76.591688$ " $-76.591721$ " $-76.591721$ \\
\+ "     \| 3 " $-76.591543$ " $-76.591464$ " $-76.591464$ " $-76.591426$ " $-76.591426$ \\
\+ "     \| 4 " $-76.591436$ " $-76.591433$ " $-76.591432$ " $-76.591415$ " $-76.591415$ \\
\+ "     \| 5 " $-76.591429$ " $-76.591426$ " $-76.591425$ " $-76.591418$ " $-76.591418$ \\
\+ "     \| 6 " $-76.591437$ " $-76.591432$ " $-76.591430$ " $-76.591413$ " $-76.591417$ \\
\sstrut {} {1 \jot} {1 \jot}
\+ " \= \| \= " \= " \= " \= " \= " \= " \\ \sstrut {} {1 \jot} {1 \jot}
\+ " 3.2 \| 0 " $-76.581341$ " $-76.581341$ " $-76.581341$ " $-76.581341$ " $-76.581341$ \\
\+ "     \| 1 " $-76.593460$ " $-76.592141$ " $-76.592141$ " $-76.592141$ " $-76.592141$ \\
\+ "     \| 2 " $-76.591708$ " $-76.591774$ " $-76.591774$ " $-76.591786$ " $-76.591786$ \\
\+ "     \| 3 " $-76.591615$ " $-76.591591$ " $-76.591591$ " $-76.591495$ " $-76.591495$ \\
\+ "     \| 4 " $-76.591543$ " $-76.591522$ " $-76.591513$ " $-76.591458$ " $-76.591456$ \\
\+ "     \| 5 " $-76.591504$ " $-76.591483$ " $-76.591478$ " $-76.591442$ " $-76.591440$ \\
\+ "     \| 6 " $-76.591483$ " $-76.591472$ " $-76.591469$ " $-76.591452$ " $-76.591448$ \\
\= \\

\endTabelle

\medskip

\endFloat

In Tabelle 11-5 wurden die oben erw\"ahnten verallgemeinerten
Summationsprozesse f\"ur $\alpha = 1.8$ und f\"ur $\alpha = 3.2$ auf die
Kristallorbitalenergien aus Tabelle 11-1 angewendet. Die
Extrapolationsergebnisse in Tabelle 11-5 f\"ur $\alpha = 1.8$ und $\alpha
= 3.2$ unterscheiden sich nur unwesentlich voneinander und von den
analogen Ergebnissen in Tabelle 11-4, die von einem Abklingparameter
$\alpha = 2$ ausgingen. In dieser Beziehung unterscheiden sich
Kristallorbitalenergien und Cluster\-energien ganz erheblich. Wenn man
die verallgemeinerten Summationsergebnisse aus Tabelle 11-5 auf die
Clusterenergien anwendet und nur geringf\"ugig vom optimalen
Abklingparameter $\alpha = 1$ abweicht, erh\"alt man sofort deutlich
schlechtere Extrapolationsergebnisse.

Die Ergebnisse in Tabelle 11-5 zeigen, da{\ss} es offensichtlich nicht
m\"oglich ist, die Extrapolationsergebnisse in Tabelle 11-4 durch
Variation des Abklingparameters $\alpha$ zu verbessern. Au{\ss}erdem kann
man in Anbetracht dieser Ergebnisse nicht l\"anger davon ausgehen, da{\ss} die
Kristallorbitalenergien durch Reihenentwicklungen des Typs von Gl.
(11.4-4) mit einem festen Abklingparameter $\alpha > 0$ dargestellt
werden k\"onnen. Stattdessen kann man nur noch annehmen, da{\ss} die
Kristallorbitalenergien in Tabelle 11-1 durch Ausdr\"ucke des Typs
$$
E_{N}^{(ko)} \; = \; E_{\infty}^{(ko)} \, + \, \Psi (N)
\tag
$$
dargestellt werden k\"onnen, wobei $\Psi (N)$ eine {\it unbekannte\/}
Funktion von $N$ ist, die f\"ur gro{\ss}e Werte von $N$ deutlich schneller
verschwindet als $1/N$.

Es ist keineswegs unwahrscheinlich, da{\ss} die Funktion $\Psi (N)$ in Gl.
(11.6-1) im wesentlichen nichtanalytische exponentielle Terme enth\"alt,
die f\"ur gro{\ss}e Werte von $N$ schnell verschwinden. Beim augenblicklichen
Stand des Wissens kann man aber keine definitiven Aussagen \"uber das
Verhalten von $\Psi (N)$ als Funktion von $N$ machen.

Eine genauere Kenntnis der unbekannten Funktion $\Psi (N)$ in
Gl.~(11.6-1) und ihrer $N$-Abh\"angig\-keit w\"are an sich \"au{\ss}erst
w\"unschenswert. Wie in Unterabschnitt 11.4 diskutiert wurde, kann man die
Konvergenz einer Folge $\Seqn s$ normalerweise nur dann {\it
effizient\/} durch verallgemeinerte Summationsprozesse beschleunigen,
wenn wenigstens gewisse strukturelle Informationen \"uber die Abh\"angigkeit
der Reste $\Seqn r$ vom Index $n$ bekannt sind. Leider ist mir keine
Arbeit bekannt, in der die Abh\"angigkeit der Konvergenz einer
Kristallorbitalrechnung von der Zahl $N$ der wechselwirkenden
Elementarzellen rechts und links von der Referenzzelle theoretisch
behandelt wird{\footnote[\dagger]{Es ist zu bef\"urchten, da{\ss} eine
theoretische Analyse der $N$-Abh\"angigkeit von Kristallorbitalenergien
\"au{\ss}erst schwierig ist. Die von einem Kristallorbitalprogramm mit einer
festen Basis und einer festen Geometrie berechneten Energiewerte h\"angen
ja nicht nur von der Zahl $N$ der wechselwirkenden Elementarzellen ab,
sondern auch davon, welche N\"aherungen man bei der Auswertung der
Wechselwirkungssummen macht. Wie schon erw\"ahnt, liefern verschiedene
Kristallorbitalprogramme aufgrund der unterschiedlichen N\"aherungen nicht
exakt die gleichen Resultate, selbst wenn identische Basiss\"atze und
Geometrien verwendet werden [Andr\'e, Bodart, Br\'edas, Delhalle und Fripiat
1984, S. 1].}}. Eine rein numerische Analyse des Verhaltens der
unbekannten Funktion $\Psi (N)$ auf der Basis der verf\"ugbaren
Kristallorbitalenergien $\bigl\{ E_N^{(ko)} \bigr\}_{N=1}^{7}$ scheint
jedenfalls nicht m\"oglich zu sein.

Die $N$-Abh\"angigkeit der Clusterenergien aus Tabelle 11-1 kann ebenfalls
noch genauer untersucht werden. Die Ergebnisse in Tabellen 11-2 und 11-3
zeigen, da{\ss} die Clusterenergien in Tabelle 11-1 durch Ausdr\"ucke des Typs
$$
E_{N}^{(cl)} \; = \; E_{\infty}^{(cl)} \, + \, C / N \, + \,  \Phi (N)
\tag
$$
dargestellt werden k\"onnen, wobei $C \ne 0$ eine Konstante ist und $\Phi
(N)$ eine unbekannte Funktion von $N$, die f\"ur gro{\ss}e Werte von $N$
deutlich schneller verschwindet als $1/N$. Damit ist aber noch nicht
bewiesen, da{\ss} die Clusterenergien tats\"achlich durch Reihenentwicklungen
des Typs von Gl. (11.4-4) mit einem Abklingparameter $\alpha = 1$
dargestellt werden k\"onnen. Dazu mu{\ss} noch gezeigt werden, da{\ss} $\Phi (N)$
von der Ordnung $O (1/N^2)$ f\"ur $N \to \infty$ ist und durch eine
Potenzreihe in $1/N$ dargestellt werden kann:
$$
\Phi (N) \; = \; \frac {1} {N^2} \,
\sum_{j=0}^{\infty} \, \frac {c'_j} {N^j} \, .
\tag
$$
Offensichtlich kann man den zu $1/N$ proportionalen Term in Gl. (11.6-2)
eliminieren, wenn man die folgenden gewichteten Differenzen der
Clusterenergien aus Tabelle 11-1 bildet:
$$
\beginAligntags
" \delta E_{1}^{(cl)} " \; = \; " E_{1}^{(cl)} \, ,
\erhoehe\aktTag \\ \tag*{\tagnr a}
" \delta E_{N}^{(cl)} " \; = \;
" \Delta \bigl\[(N-1) \, E_{N-1}^{(cl)} \bigr\] \; = \;
N E_{N}^{(cl)} \, - \, (N - 1) E_{N-1}^{(cl)} \, ,
\qquad N \ge 2 \, .
\\ \tag*{\tagform\aktTagnr b}
\endAligntags
$$
Aus Gln. (11.6-2) und (11.6-4) folgt dann:
$$
\delta E_{N}^{(cl)} \; = \; E_{\infty}^{(cl)} \, + \,
\Delta \bigl\[(N-1) \, \Phi (N-1) \bigr\] \, ,
\qquad N \ge 1 \, .
\tag
$$
Wenn die unbekannte Funktion $\Phi (N)$ tats\"achlich gem\"a{\ss} Gl. (11.6-3)
durch eine Potenzreihe in $1/N$ dargestellt werden kann, dann gilt
offensichtlich
$$
\delta E_{N}^{(cl)} \, - \, E_{_{}^{}}^{(cl)} \; = \;
\Delta \bigl\[(N-1) \Phi (N-1) \bigr\] \; = \;
c'_0/N^2 \, + \, O (1/N^3) \, , \qquad N \to \infty \, .
\tag
$$

Cui, Kertesz und Jiang [1990] schlugen vor, bei Clusterrechnungen an
quasi-eindimensionalen stereoregul\"aren Polymeren die Energie pro
Monomereinheit nicht \"uber die mittlere Energie pro Monomereinheit zu
berechnen, sondern \"uber die in Gl. (11.6-4) definierten gewichteten
Energiedifferenzen. Ihr Argument war, da{\ss} die st\"orenden Endeffekte sich
bei Clustern aus $N+1$ beziehungsweise $N$ Monomereinheiten nur relativ
wenig unterscheiden sollten, wenn $N$ gro{\ss} genug ist, und da{\ss} sich die
Gesamtenergien der beiden Cluster im wesentlichen um die Energie einer
{\it inneren\/} Monomereinheit unterscheiden. F\"ur ausreichend gro{\ss}e
Werte von $N$ sollte eine solche innere Monomereinheit sich nur relativ
wenig vom sogenannten {\it bulk limit\/} unterscheiden. Wenn diese
Annahme richtig ist, dann sollten die gewichteten Energiedifferenzen
wesentlich schneller konvergieren als die mittleren Energien pro
Monomereinheit, deren Abbruchfehler proportional zu $1/N$ ist.

Die in Gl. (11.6-4) definierten gewichteten Energiedifferenzen k\"onnen
auch als Spezialf\"alle des Richardsonschen Extrapolationsprozesses
interpretiert werden. Wenn man in Gl. (11.5-5) die Interpolationspunkte
$x_n = 1/(n+1)$ verwendet und $k = 1$ setzt, erh\"alt man:
$$
{\cal N}_1^{(n)} \bigl(s_n , 1/(n+1) \bigr)
\; = \; (n + 2) s_{n+1} \, - \, (n+1) s_n \, ,
\qquad n = 0, 1, 2, \ldots \, .
\tag
$$
Wenn man die Eingabedaten $\Seqn s$ f\"ur den Richardsonschen
Extrapolationsproze{\ss} gem\"a{\ss} Gl. (11.4-7) w\"ahlt, erh\"alt man:
$$
{\cal N}_1^{(n)} \bigl(E_{n+1}^{(cl)}, 1/(n+1) \bigr)
\; = \; (n + 2) E_{n+2}^{(cl)} \, - \, (n+1) E_{n+1}^{(cl)}
\; = \; \delta \, E_{n+2}^{(cl)} \, , \qquad 0 \le n \le 5 \, .
\tag
$$
Die in Gl. (11.6-4) definierten gewichteten Energiedifferenzen $\delta
E_{N}^{(cl)}$ sind also bis auf eine Indexverschiebung identisch mit den
Elementen aus der ersten Spalte ${\cal N}_1^{(n)} \bigl(E_{n+1}^{(cl)},
1/(n+1) \bigr)$ des in Gl. (11.5-5) definierten Richardsonschen
Extrapolationsprozesses.

\beginFloat

\medskip

\beginTabelle
\beginFormat & \mitte \endFormat
\+ " \links {\bf Tabelle 11-6} \@ " \\
\+ " \links {Gewichtete Differenzen der Clusterenergien} \@ " \\
\+ " \links {aus Tabelle 11-1 gem\"a{\ss} Gl. (11.6-4)} \@ " \\
\- \\ \sstrut {} {1 \jot} {1 \jot}
\+ " N " $\delta E_{N}^{(cl)}$ \\
\sstrut {$\delta E_{N}^{(cl)}$} {0.2 \jot} {1.2 \jot}
\- \\ \sstrut {} {1 \jot} {1 \jot}
\+ "  1 " $-77.6380003705$ \\
\+ "  2 " $-76.5869112203$ \\
\+ "  3 " $-76.5909416487$ \\
\+ "  4 " $-76.5913328761$ \\
\+ "  5 " $-76.5914237374$ \\
\+ "  6 " $-76.5914475772$ \\
\+ "  7 " $-76.5914528527$ \\
\- \\ \sstrut {} {1 \jot} {1 \jot}
\endTabelle

\medskip

\endFloat

In Tabelle 11-6 werden die in Gl. (11.6-4) definierten gewichteten
Clusterenergiedifferenzen $\delta E_{N}^{(cl)}$ aufgelistet. Die
Ergebnisse zeigen ganz deutlich, da{\ss} die Elimination des zu $1/N$
proportionalen Anteils der Clusterenergien eine deutliche Verbesserung
der Konvergenz bewirkt, da $\delta E_{5}^{(cl)}$, $\delta E_{6}^{(cl)}$
und $\delta E_{7}^{(cl)}$ sich erst in der f\"unften Stelle nach dem Komma
von den besten Extrapolationsergebnissen in den Tabellen 11-3 und 11-4
unterscheiden. Bemerkenswert ist, da{\ss} die gewichteten
Clusterenergiedifferenzen $\delta E_{N}^{(cl)}$ in Tabelle 11-6 deutlich
schneller konvergieren als die Kristallorbitalenergien $E_{N}^{(ko)}$ in
Tabelle 11-1.

Mit Hilfe der in Gl. (8.4-20) definierten Transformation $T_n$ kann man
untersuchen, ob die unbekannte Funktion $\Phi (N)$ in Gl. (11.6-2)
tats\"achlich von der Form von Gl. (11.6-3) ist, d.~h., ob sie durch eine
Reihenentwicklung des Typs von Gl. (11.4-4) mit dem Abklingparameter
$\alpha = 2$ dargestellt werden kann.

Tabelle 11-7 zeigt den Effekt der in Gl. (8.4-20) definierten
Transformation $T_n$ auf die gewichteten Clusterenergiedifferenzen aus
Tabelle 11-6. Die Eingabedaten $\Seqn s$ f\"ur $T_n$ wurden dabei
folgenderma{\ss}en gew\"ahlt:
$$
s_n \; = \; \delta E_{n+1}^{(cl)} \; = \;
\Delta \bigl\[n \, E_{n}^{(cl)} \bigr\] \, ,
\qquad 0 \le n \le 6 \, .
\tag
$$

\beginFloat

\medskip

\beginTabelle
\beginFormat & \mitte \endFormat
\+ " \links {\bf Tabelle 11-7} \@ \\
\+ " \links {Approximative Bestimmung des Abklingparameters $\alpha$} \@ \\
\+ " \links {der gewichteten Differenzen der Clusterenergien
$\delta E_{n+1}^{(cl)}$} \@ \\
\+ " \links {gem\"a{\ss} Gl. (8.4-20)} \@ \\
\- \\ \sstrut {} {1 \jot} {1 \jot}
\+ " n " $\alpha$ \\
\- \\ \sstrut {} {1 \jot} {1 \jot}
\+ "  0 " $~~7.98282682854$ \\
\+ "  1 " $~~4.12825261666$ \\
\+ "  2 " $~17.79631367181$ \\
\+ "  3 " $-14.98028361646$ \\
\- \\ \sstrut {} {1 \jot} {1 \jot}
\endTabelle

\medskip

\endFloat

Die v\"ollig erratischen Ergebnisse in Tabelle 11-7 zeigen, da{\ss} man den
gewichteten Energiedifferenzen $\delta E_{N}^{(cl)}$ offensichtlich
keinen definierten Abklingparameter $\alpha$ zuordnen kann, und da{\ss} Gl.
(11.6-3) offensichtlich nicht erf\"ullt ist.

Wie schon in Unterabschnitt 11.4 erw\"ahnt, schlossen Cioslowski und
Lepetit [1991] auf der Basis st\"orungstheoretischer Argumente, da{\ss} die
Energie pro Monomereinheit $\varepsilon_N$ eines Clusters $\rm X \! - \!
(A)_N\!\! -\! Y$ gem\"a{\ss} Gl. (11.4-3) durch eine Potenzreihe in $1/N$
dargestellt werden kann. Die numerischen Ergebnisse in Tabelle 11-7 sind
nicht v\"ollig im Widerspruch zu den Schlu{\ss}folgerungen von Cioslowski und
Lepetit [1991]. Allerdings ist die \"Ubereinstimmung minimal, da in den
hier beschriebenen Rechnungen nur die Existenz des f\"uhrenden, zu $1/N$
proportionalen Termes der postulierten Potenzreihendarstellung f\"ur die
Clusterenergie $E_{N}^{(cl)}$ nachgewiesen werden konnte.

Die Ergebnisse in Tabelle 11-7 machen es wahrscheinlich, da{\ss} die
unbekannte Funktion $\Phi (N)$ in Gl. (11.6-2) im wesentlichen
nichtanalytische Terme enth\"alt, die f\"ur gr\"o{\ss}ere Werte von $N$ rasch
verschwinden.

Man kann allerdings nicht v\"ollig ausschlie{\ss}en, da{\ss} die Ergebnisse in
Tabelle 11-7 auf numerische Instabilit\"aten zur\"uckzuf\"uhren sind.
Beispielsweise unterscheiden sich die gewichteten Energiedifferenzen
$\delta E_{5}^{(cl)}$, $\delta E_{6}^{(cl)}$ und $\delta E_{7}^{(cl)}$
erst in der f\"unften Stelle nach dem Komma, und es ist nicht klar, ob die
weiteren, in Tabelle 11-6 ausgegebenen Stellen tats\"achlich korrekt sind.
Die Transformation $T_n$, Gl. (8.4-20), ist im Prinzip ein gewichteter
$\Delta^3$-Algorithmus und damit sehr anf\"allig f\"ur Rundungsfehler. Eine
katastrophale Akkumulation numerischer Instabilit\"aten kann deswegen nie
ausgeschlossen werden. Auf der Basis der verf\"ugbaren Daten sind
definitive Aussagen aber nicht m\"oglich.

\beginFloat

\medskip

\beginTabelle % [to \kolumnenbreite]
\beginFormat
\mitte " \mitte " \mitte " \mitte " \mitte " \mitte
\endFormat
\+ " \links {\bf Tabelle 11-8} \@ \@ \@ \@ \@ " \\
\+ " \links {Extrapolation der gewichteten Clusterenergiedifferenzenn
aus Tabelle 11-6} \@ \@ \@ \@ \@ " \\
\+ " \links {Eingabedaten $s_n \, = \; \delta E_{n+1}^{(cl)}$ mit $0
\le n \le 6$}  \@ \@ \@ \@ \@ " \\
\sstrut {$E_{n+1}^{(cl)}$} {0.2 \jot} {1.2 \jot}
\= \\ \sstrut {} {1.5 \jot} {1.5 \jot}
\+ " \rechts {$n$} " $u_{n}^{(0)} (1, s_0)$ " $t_{n}^{(0)} (1, s_0)$
" $\epsilon_{2 \Ent {n/2}}^{(n - 2 \Ent {n/2})}$
" ${\cal A}_{\Ent {n/2}}^{(n - 2 \Ent {n/2})}$
" $\theta_{2 \Ent {n/3}}^{(n - 3 \Ent {n/3})}$ \\
\sstrut {$\theta_{2 \Ent {n/3}}^{(n - 3 \Ent {n/3})}$} {0.8 \jot} {1.2 \jot}
\+ " " Gl. (5.2-13) " Gl. (5.2-15) " Gl. (2.4-10) " Gl. (3.3-8)
" Gl. (4.4-13) \\ \sstrut {} {1.5 \jot} {1.5 \jot}
\= \\ \sstrut {} {1 \jot} {1 \jot}
\+ " 0 " $-77.638000$ " $-77.638000$ " $-77.638000$ " $-77.638000$ " $-77.638000$ \\
\+ " 1 " $-76.614621$ " $-76.600951$ " $-76.586911$ " $-76.586911$ " $-76.586911$ \\
\+ " 2 " $-76.590965$ " $-76.590939$ " $-76.590926$ " $-76.590926$ " $-76.590942$ \\
\+ " 3 " $-76.591444$ " $-76.591411$ " $-76.591375$ " $-76.591375$ " $-76.591429$ \\
\+ " 4 " $-76.591491$ " $-76.591476$ " $-76.591451$ " $-76.591467$ " $-76.591480$ \\
\+ " 5 " $-76.591450$ " $-76.591454$ " $-76.591456$ " $-76.591456$ " $-76.591458$ \\
\+ " 6 " $-76.591453$ " $-76.591454$ " $-76.591455$ " $-76.591455$ " $-76.591448$ \\
\sstrut {} {1.5 \jot} {1.5 \jot}
\= \\ \sstrut {} {1 \jot} {1 \jot}
\+ " \rechts {$n$}
" ${\cal J}_{\Ent {n/3}}^{(n - 3 \Ent {n/3})}$
" ${\cal N}_n^{(0)} (s_0, x_0)$ " ${\cal N}_n^{(0)} (s_0, x_0)$
" ${\bar \rho}_{2 \Ent {n/2}}^{(n - 2 \Ent {n/2})}$
" ${\overline {\cal W}}_{\Ent {n/2}}^{(n - 2 \Ent {n/2})}$ \\ \sstrut
{${\overline {\cal W}}_{\Ent {n/2}}^{(n - 2 \Ent {n/2})}$} {0.8 \jot} {1.2 \jot}
\+ " " Gl. (8.4-18) " Gl. (11.5-5) " Gl. (11.5-5) " Gl. (8.4-11) " Gl.
(8.4-14) \\ \sstrut {} {1.5 \jot} {1.5 \jot}
\+ " " " $x_n = (n + 1)^{-1}$ " $x_n = (n + 1)^{-2}$ " $\alpha = 2$
" $\alpha = 2$ \\ \sstrut {} {1.5 \jot} {1.5 \jot}
\= \\ \sstrut {} {1 \jot} {1 \jot}
\+ " 0 " $-77.638000$ " $-77.638000$ " $-77.638000$ " $-77.638000$ " $-77.638000$ \\
\+ " 1 " $-76.586911$ " $-75.535822$ " $-76.236548$ " $-76.586911$ " $-76.586911$ \\
\+ " 2 " $-76.590942$ " $-77.130593$ " $-76.638868$ " $-76.592934$ " $-76.592934$ \\
\+ " 3 " $-76.591429$ " $-76.404483$ " $-76.587872$ " $-76.591592$ " $-76.591592$ \\
\+ " 4 " $-76.591480$ " $-76.641179$ " $-76.591670$ " $-76.591472$ " $-76.591484$ \\
\+ " 5 " $-76.591458$ " $-76.580239$ " $-76.591414$ " $-76.591381$ " $-76.591420$ \\
\+ " 6 " $-76.591448$ " $-76.593789$ " $-76.591431$ " $-76.591425$ " $-76.591440$ \\
\sstrut {} {1.5 \jot} {1.5 \jot}
\= \\ \sstrut {} {1 \jot} {1 \jot}
\+ " \rechts {$n$} " $\rho_{2 \Ent {n/2}}^{(n - 2 \Ent {n/2})}$
" ${\cal W}_{\Ent {n/2}}^{(n - 2 \Ent {n/2})}$
" $\rho_{2 \Ent {n/2}}^{(n - 2 \Ent {n/2})}$
" ${\cal W}_{\Ent {n/2}}^{(n - 2 \Ent {n/2})}$ \\ \sstrut
{${\cal W}_{\Ent {n/2}}^{(n - 2 \Ent {n/2})}$} {0.8 \jot} {1.2 \jot}
\+ " " Gl. (3.3-12) " Gl. (3.3-17) " Gl. (3.3-12) " Gl. (3.3-17)
\\ \sstrut {} {1.5 \jot} {1.5 \jot}
\+ " " $x_n = n + 1$ " $x_n = n + 1$
" $x_n = (n + 1)^2$ " $x_n = (n + 1)^2$
\\ \sstrut {} {1.5 \jot} {1.5 \jot}
\= \\ \sstrut {} {1 \jot} {1 \jot}
\+ " 0 " $-77.638000$ " $-77.638000$ " $-77.638000$ " $-77.638000$ \\
\+ " 1 " $-76.586911$ " $-76.586911$ " $-76.586911$ " $-76.586911$ \\
\+ " 2 " $-76.594941$ " $-76.594941$ " $-76.593345$ " $-76.593345$ \\
\+ " 3 " $-76.591808$ " $-76.591808$ " $-76.591662$ " $-76.591662$ \\
\+ " 4 " $-76.591378$ " $-76.591464$ " $-76.591482$ " $-76.591502$ \\
\+ " 5 " $-76.591318$ " $-76.591405$ " $-76.591419$ " $-76.591443$ \\
\+ " 6 " $-76.591363$ " $-76.591428$ " $-76.591430$ " $-76.591445$ \\
\sstrut {} {1.5 \jot} {1.5 \jot}
\= \\

\endTabelle

\medskip

\endFloat

Obwohl die gewichteten Energiedifferenzen in Tabelle 11-6 anscheinend
nicht durch eine Potenzreihe in $1/N$ dargestellt werden k\"onnen, kann
man trotzdem versuchen, ihre Konvergenz durch verallgemeinerte
Summationsprozesse zu verbessern.

In Tabelle 11-8 werden diejenigen verallgemeinerten Summationsprozesse,
die in den Tabellen 11-3 und 11-4 zur Beschleunigung der Konvergenz der
Cluster- und Kristallorbitalenergien aus Tabelle 11-1 verwendet
wurden, auf die gewichteten Clusterenergiedifferenzen $\bigl\{ \delta
E_{n+1}^{(cl)} \bigr\}_{n=0}^{6}$ aus Tabelle 11-6 angewendet.

Mit Ausnahme des Richardsonschen Extrapolationsprozesses ${\cal
N}_{k}^{(n)} (s_n , x_n)$ in Verbindung mit den Interpolationspunkten
$x_n = (n+1)^{- 1}$ und des Wynnschen $\rho$-Algorithmus und seiner
Iteration ${\cal W}_{k}^{(n)}$ in Verbindung mit den
Interpolationspunkten $x_n = n + 1$ scheinen alle verallgemeinerten
Summationsprozesse in Tabelle 11-8 sehr gute Extrapolationsergebnisse zu
produzieren, die v\"ollig im Einklang mit den konservativen Sch\"atzungen
$E_{\infty}^{(cl)} = E_{\infty}^{(ko)} = - 76.59145 \pm 0.00005$ sind.

Trotzdem kann man daraus nicht folgern, da{\ss} die in Tabelle 11-8
verwendeten verallgemeinerten Summationsprozesse in der Lage sind, die
Konvergenz der gewichteten Energiedifferenzen aus Tabelle 11-6 zu
verbessern. Wenn ein verallgemeinerter Summationsproze{\ss} tats\"achlich in
der Lage ist, die Konvergenz einer Folge $\Seqn s$ gegen ihren Grenzwert
$s$ zu beschleunigen, dann m\"ussen die transformierten Gr\"o{\ss}en $s'_n$
wesentlich schneller gegen $s$ konvergieren als die Eingabedaten.

Beispielsweise f\"uhrten in Tabelle 11-3 sowohl ${\cal N}_{k}^{(n)} (s_n ,
x_n)$ in Verbindung mit den Interpolationspunkten $x_n = (n+1)^{- 1}$
als auch $\rho_{k}^{(n)}$ und seine Iteration ${\cal W}_{k}^{(n)}$ in
Verbindung mit den Interpolationspunkten $x_n = n + 1$ zu einer
betr\"achtlichen Verbesserung der Konvergenz der Clusterenergien, und in
Tabelle 11-4 fiel die \"au{\ss}erst schnelle Konvergenz von ${\bar
\rho}_{k}^{(n)}$ und ${\overline {\cal W}}_k^{(n)}$ auf. Derartige
Verbesserungen der Konvergenz werden in Tabelle 11-8 aber bei keinem
einzigen verallgemeinerten Summationsproze{\ss} beobachtet: Die
transformierten Gr\"o{\ss}en konvergierten in keinem einzigen Fall signifikant
schneller gegen $E_{\infty}^{(cl)} = - 76.59145 \pm 0.00005$ als die
Eingabedaten.

Auf der Basis dieser Ergebnisse mu{\ss} man also folgern, da{\ss} die Konvergenz
der gewichteten Clusterenergiedifferenzen aus Tabelle 11-6 durch die in
Tabelle 11-8 verwendeten verallgemeinerten Summationsprozesse nicht noch
weiter verbessert werden kann, und da{\ss} man sich mit der konservativen
Sch\"atzung $E_{\infty}^{(cl)} = E_{\infty}^{(ko)} = - 76.59145 \pm
0.00005$ f\"ur die Grundzustandsenergie des {\it trans}-Polyacetylens pro
Acetyleneinheit zufrieden geben mu{\ss}.

Es gibt auch andere Extrapolationsverfahren, die auf der Annahme
basieren, da{\ss} die Eingabedaten $\Seqn s$ dargestellt werden k\"onnen durch
Reihenentwicklungen des Typs von Gl. (11.4-4) mit einem festen
Abklingparameter $\alpha > 0$, dessen Wert aber nicht unbekannt sein
mu{\ss}.

Beleznay [1986] verwendete das Richardsonsche Extrapolationsverfahren
${\cal N}_{k}^{(n)} (s_n , x_n)$, Gl. (11.5-5), in Verbindung mit den
Interpolationspunkten
$$
x_n = (n + 1)^{- \gamma} \, \qquad \gamma > 0 \, .
\tag
$$
Der Exponent $\gamma$ ist ein freier Parameter, der noch bestimmt werden
mu{\ss}.

Nehmem wir an, da{\ss} die Folgenelemente $s_0, s_1, \ldots, s_m$ als
Inputdaten zur Verf\"ugung stehen. Dann kann man mit Hilfe des
Rekursionsschemas (11.5-5) alle Elemente ${\cal N}_{j}^{(\mu-j)}
(s_{\mu-j}, (\mu-j+1)^{- \gamma})$ mit $0 \le j \le \mu$ und $0 \le \mu
\le m$ berechnen. Beleznay [1986, S. 553] schlug vor, den unbestimmten
Parameter $\gamma$ in Gl. (11.6-10) so zu w\"ahlen, da{\ss} der Fehlerterm
$\vert {\cal N}_{m-1}^{(0)} - {\cal N}_{m-1}^{(1)} \vert$ minimal wird.
Dieses Extrapolationsverfahren wurde von Liegener, Beleznay und Ladik
[1987] zur Extrapolation von Hatree-Fock-Rechnungen an Ringen und Ketten
aus $\rm H_2$-Molek\"ulen verwendet, und von Weniger und Liegener [1990]
zur Extrapolation von Cluster- und Kristallorbitalrechnungen an {\it
trans}-Polyacetylen.

Neben der von Beleznay [1986] zur Optimierung von $\gamma$ verwendeten
Fehlerbedingung gibt es aber noch andere sinnvolle Fehlerbedingungen.
Weniger und Liegener [1990, S. 66] w\"ahlten den Parameter $\gamma$ in Gl.
(11.6-10) auch so, da{\ss} der Fehlerterm $\vert {\cal N}_{m-1}^{(0)} -
{\cal N}_m^{(0)} \vert$ minimal wird.

Die hier skizzierten Verfahren, die auf einer Optimierung des Exponenten
$\gamma$ in Gl. (11.6-10) basieren, sind nicht auf das Richardsonsche
Extrapolationsverfahren, Gl. (11.5-5), beschr\"ankt. Man kann das von
Beleznay [1986] eingef\"uhrte Konzept der Optimierung eines freien
Parameters auch bei nichtlinearen Extrapolationsverfahren wie
beispielsweise dem Wynnschen $\rho$-Algorithmus, Gl. (3.3-12), und
seiner Iteration ${\cal W}_{k}^{(n)}$, Gl. (3.3-17), verwenden [Weniger
und Liegener [1990, S. 66]. Dabei verwendet man die Transformationen
$\rho_{k}^{(n)}$ und ${\cal W}_{k}^{(n)}$ in Verbindung mit den
Interpolationspunkten
$$
x_n \; = \; (n + 1)^{\gamma} \, , \qquad \gamma > 0 \, .
\tag
$$
Nehmen wir wieder an, da{\ss} die Inputdaten $s_0, s_1,
\ldots, s_m$ zur Verf\"ugung stehen. Wenn man bei der Optimierung von
$\gamma$ so vorgeht wie Beleznay [1986], mu{\ss} man im Falle des
$\rho$-Algorithmus den Parameter $\gamma$ in Gl. (11.6-11) dann so
w\"ahlen, da{\ss} der Fehlerterm $\vert \rho_{2 \mu - 2}^{(1)} - \rho_{2 \mu -
2}^{(2)} \vert$ minimal wird, wenn $m = 2 \mu$ gerade ist, und $\vert
\rho_{2 \mu}^{(0)} - \rho_{2 \mu}^{(1)} \vert$, wenn $m = 2 \mu + 1$
ungerade ist [Weniger und Liegener 1990, S. 66]. Im Falle von ${\cal
W}_{k}^{(n)}$ mu{\ss} man dann den Fehlerterm $\vert {\cal W}_{\mu-1}^{(1)}
- {\cal W}_{\mu-1}^{(2)} \vert$ minimisieren, wenn $m = 2 \mu$ gerade
ist, und den Fehlerterm $\vert {\cal W}_{\mu}^{(0)} - {\cal
W}_{\mu}^{(1)} \vert$, wenn $m = 2 \mu + 1$ ungerade ist [Weniger und
Liegener 1990, S. 66]. Alternativ kann man auch die Fehlerterme $\vert
\rho_{2 \mu}^{(0)} - \rho_{2 \mu - 2}^{(1)} \vert$ beziehungsweise
$\vert {\cal W}_{\mu}^{(0)} - {\cal W}_{\mu - 1}^{(1)} \vert$
minimisieren, wenn $m = 2 \mu$ gerade ist, und die Fehlerterme $\vert
\rho_{2 \mu}^{(1)} - \rho_{2 \mu} ^{(0)} \vert$ beziehungsweise $\vert
{\cal W}_{\mu}^{(1)} - {\cal W}_{\mu}^{(0)} \vert$, wenn $m = 2 \mu + 1$
ungerade ist [Weniger und Liegener 1990, S. 66].

Konzeptionell eng verwandte lineare und nichtlineare
Extrapolationsalgorithmen wurden von Henkel und Sch\"utz [1988]
beschrieben und sp\"ater haupts\"achlich im Zusammenhang mit Problemen aus
der statistischen Physik verwendet [Frachebourg und Henkel 1991; Henkel
und Herrmann 1990; Henkel 1990].

Die Tatsache, da{\ss} sowohl die Cluster- als auch die
Kristallorbitalenergien in Tabelle 11-1 mit gro{\ss}er Wahrscheinlichkeit
nicht durch Reihenentwicklungen vom Typ von Gl. (11.4-4) mit einem
festen Abklingparameter $\alpha$ dargestellt werden k\"onnen, ist von gro{\ss}er
Wichtigkeit, wenn man die oben beschriebenen Extrapolationsalgorithmen
zur Verbesserung der Konvergenz der Cluster- und Kristallorbitalenergien
verwenden will. Es ist zwar m\"oglich, mit Hilfe dieser Algorithmen
durchaus vern\"unftige Extrapolationsergebnisse zu erhalten [Weniger und
Liegener 1990, S. 70 und 72]. Trotzdem ist diese Vorgehensweise, die
eine gewisse \"Ahnlichkeit mit dem Ritzschen Variationsverfahren hat, hier
nicht ausreichend begr\"undet, da man den Cluster- und
Kristallorbitalenergien keine festen Abklingparameter zuordnen kann, und
da au{\ss}erdem nicht bekannt ist, ob die Cluster- und
Kristallorbitalenergien \"uberhaupt irgendwelche Extremalbedingungen
dieses Typs erf\"ullen.

Wenn man die Interpolationspunkte f\"ur den Richardsonschen
Extrapolationsprozess gem\"a{\ss} Gl. (11.6-10) oder f\"ur den Wynnschen
$\rho$-Algorithmus und seine Iteration ${\cal W}_{k}^{(n)}$ gem\"a{\ss} Gl.
(11.6-11) w\"ahlt und den Parameter $\gamma$ entsprechend optimiert,
erh\"alt man Extrapolationsergebnisse, die von der Zahl der verwendeten
Eingabedaten abh\"angen. Man hat aber keinen Grund zu der Annahme, da{\ss} die
oben beschriebenen Optimierungskriterien f\"ur $\gamma$ eine Konvergenz
der ordnungsabh\"angigen Extrapolationsergebnisse gegen den richtigen Wert
gew\"ahrleisten, wenn die Zahl der verwendeten Eingabedaten gro{\ss} wird. Die
Konvergenz solcher Verfahren gegen einen falschen Wert kann also nie
ausgeschlossen werden.

Zusammenfassend kann man sagen, da{\ss} man die Konvergenz der Cluster- und
Kristallorbitalenergien aus Tabelle 11-1 durch Extrapolation
offensichtlich deutlich verbessern kann, auch wenn in theoretischer
Hinsicht noch viele Fragen offen bleiben. Hervorzuheben ist auch die
bemerkenswert gute \"Ubereinstimmung der besten Extrapolationsergebnisse
aus den Tabellen 11-3 und 11-4. Konservative Sch\"atzungen ergaben
$E_{\infty}^{(cl)} = E_{\infty}^{(ko)} = - 76.59145 \pm 0.00005$.

Eine derartig gute \"Ubereinstimmung der Extrapolationsergebnisse war an
sich nicht zu erwarten, da die Kristallorbitalenergien deutlich
schneller konvergieren als die Clusterenergien. Es scheint aber, da{\ss} die
Folgen $\bigl\{ E_N^{(ko)} \bigr\}_{N=1}^{7}$ und $\bigl\{ E_N^{(cl)}
\bigr\}_{N=1}^{7}$ -- dem Augenschein zum Trotz -- in etwa den gleichen
Informationsgehalt besitzen. Die deutlich schlechtere Konvergenz der
Clusterenergien wird dadurch ausgeglichen, da{\ss} man sie offensichtlich
wesentlich effizienter extrapolieren kann.

\endAbschnittsebene

\endAbschnittsebene

\keinTitelblatt\neueSeite

\beginAbschnittsebene
\aktAbschnitt = 11

\Abschnitt Zusammenfassung und Ausblicke

\medskip

\aktTag = 0

In Abschnitt 2 dieser Arbeit wurden am Beispiel einfacher Modellreihen
typische Konvergenzprobleme beschrieben, die bei unendlichen Reihen
auftreten. In Abschnitt 3 wurde eine allgemeine Einf\"uhrung in die
Theorie verallgemeinerter Summationsprozesse gegeben. In Abschnitt 4
wurden Pad\'e-Approximationen behandelt, die ohne Zweifel die am besten
verstandenen und auch die am h\"aufigsten angewendeten verallgemeinerten
Summationsprozesse sind. In Abschnitt 5 wurden einige vom Autor
entwickelte verallgemeinerte Summationsprozesse [Weniger 1989; 1992]
genauer beschrieben. In Abschnitt 6 wurden die Konvergenzeigenschaften
dieser Transformationen theoretisch analysiert. In den Abschnitten 7 -
11 wurde die praktische N\"utzlichkeit dieser Verfahren anhand einiger
Beispiele aus der Quantenmechanik und der theoretischen Chemie
demonstriert.

In den Abschnitten 7 und 9 wurde am Beispiel der modifizierten
Besselfunktion der zweiten Art sowie der Hilfsfunktion $F_m (z)$, die
eine zentrale Rolle bei der Berechnung der Mehrzentrenmolek\"ulintegrale
von Gau{\ss}funktionen spielt, gezeigt, wie man spezielle Funktionen und
Hilfsfunktionen unter Verwendung verallgemeinerter Summationsprozesse
auf effiziente Weise berechnen kann. In Abschnitt 8 wurden
verallgemeinerte Summationsprozesse zur Auswertung komplizierter,
logarithmisch konvergenter Reihendarstellungen f\"ur
Mehrzentrenmolek\"ulintegrale exponential\-artiger Basisfunktionen
verwendet.

Neben dem Variationsverfahren ist die St\"orungstheorie das ohne Zweifel
wichtigste systematische N\"aherungsverfahren zur L\"osung
quantenmechanischer Eigenwertprobleme. Der Formalismus der
Rayleigh-Schr\"odingerschen St\"orungstheorie ergibt f\"ur die
Energieeigenwerte Reihenentwicklungen in der Kopplungskonstanten. Leider
sind bei St\"orungsreihen Konvergenzprobleme eher die Regel als die
Ausnahme.

Besonders gut untersuchte Modellsysteme sind die anharmonischen
Oszillatoren mit einer $\hat{x}^{2 m}$-Anharmonizit\"at ($m = 2, 3, 4$).
Die Koeffizienten $c_{n}^{(m)}$ der Rayleigh-Schr\"odingerschen
St\"orungsreihe f\"ur die Grundzustandsenergie eines solchen anharmonischen
Oszillators wachsen betragsm\"a{\ss}ig im wesentlichen wie
$([m-1]n)!/n^{1/2}$. In Abschnitt 10 wurde gezeigt, da{\ss} verallgemeinerte
Summationsprozesse diese St\"orungsreihen trotz ihrer extrem starken
Divergenz effizient summieren k\"onnen. Mit Hilfe eines
Renormierungsverfahrens gelang es au{\ss}erdem, aus den hochgradig
divergenten Rayleigh-Schr\"odingerschen St\"orungsreihen {\it
transformierte\/} St\"orungsreihen zu konstruieren, die anscheinend f\"ur
alle physikalisch relevanten Werte der Kopplungskonstante konvergieren.
Die neuen St\"orungsreihen sind so einfach, da{\ss} man die
Grundzustandsenergie eines anharmonischen Oszillators mit einer
$\hat{x}^{2 m}$-Anharmonizit\"at ($m = 2, 3, 4$) f\"ur alle
Kopplungskonstanten $0 \le \beta < \infty$ problemlos mit einem
Taschenrechner, der eine {\it solve}-Funktion besitzt, berechnen k\"onnte.

In Abschnitt 11 wurde am Beispiel des {\it trans}-Polyacetylens gezeigt,
wie man die Konvergenz von Kristallorbital- und Clusterrechnungen an
quasi-eindimensionalen Polymeren mit Hilfe von verallgemeinerten
Summationsprozessen deutlich verbessern kann.

In mathematischer Hinsicht war man in den Abschnitten 7, 9 und 10 immer
mit langsam konvergenten oder divergenten Potenzreihen konfrontiert, in
den Abschnitten 8 und 11 dagegen mit der Beschleunigung logarithmischer
Konvergenz.

In methodischer Hinsicht lag das Schwergewicht dieser Arbeit auf den in
Abschnitt 5 beschriebenen verallgemeinerten Summationsprozessen mit
expliziten Restsummenabsch\"atzungen. Wenn man diese Transformationen auf
die Folge der Partialsummen einer Potenzreihe anwendet, erh\"alt man
\"ahnlich wie bei Pad\'e-Approximationen eine Folge zweifach indizierter
rationaler Funktionen. Trotz dieser \"Ahnlichkeit unterscheiden sich
Pad\'e-Approximationen und die verallgemeinerten Summationsprozesse aus
Abschnitt 5 in numerischer Hinsicht ganz erheblich. In den
Anwendungsbeispielen aus den Abschnitten 7, 9 und 10 waren die
verallgemeinerten Summationsprozesse mit expliziten
Restsummenabsch\"atzungen immer deutlich leistungsf\"ahiger als
Pad\'e-Approximationen. Besonders bemerkenswert waren die Ergebnisse aus
Abschnitt 10: Mit Hilfe einer vom Autor entwickelten Transformation
[Weniger 1989, Abschnitt 8] konnten selbst die extrem stark divergenten
quantenmechanischen St\"orungsreihen f\"ur die Grundzustandsenergie eines
anharmonischen Oszillators mit einer $\hat{x}^{6}$- beziehungsweise
$\hat{x}^{8}$-Anharmonizit\"at summiert werden. Im Falle der
$\hat{x}^{6}$-Anharmonizit\"at konvergieren Pad\'e-Approximationen so
langsam, da{\ss} sie praktisch nutzlos sind, und im Falle der
$\hat{x}^{8}$-Anharmonizit\"at k\"onnen Pad\'e-Approximationen zur Summation
nicht verwendet werden, da sie divergieren.

Die Anwendungsbeispiele aus Abschnitt 11 unterschieden sich in
methodischer Hinsicht deutlich von den Anwendungsbeispielen der fr\"uheren
Abschnitte. In den Abschnitten 7 - 9 waren die Ein\-gabedaten der
verallgemeinerten Summationsprozesse in Form expliziter mathematischer
Ausdr\"ucke gegeben, und die Eingabedaten konnten immer mit ausreichender
Genauigkeit berechnet werden. Au{\ss}erdem war der Konvergenztyp der
unendlichen Reihen nie fraglich, da die Abh\"angigkeit des Abbruchfehlers
vom Index $n$ bekannt war.

In Abschnitt 10 war die Situation etwas ung\"unstiger, da keine
geschlossenen Ausdr\"ucke f\"ur die Koeffizienten der quantenmechanischen
St\"orungsreihen der anharmonischen Oszillatoren bekannt sind. Trotzdem
war die Situation in Abschnitt 10 immer noch vergleichsweise g\"unstig, da
eine ausreichend gro{\ss}e Zahl dieser Koeffizienten {\it rundefehlerfrei\/}
unter Verwendung der exakten rationalen Arithmetik der
Programmiersprache MAPLE [Char, Geddes, Gonnet, Leong, Monagan und Watt
1991a] berechnet werden konnte. Au{\ss}erdem stand der Konvergenztyp der
St\"orungsreihen -- oder besser gesagt ihr Divergenzverhalten -- fest, da
asymptotische N\"aherungen f\"ur die St\"orungstheoriekoeffizienten im Falle
gro{\ss}er Indizes $n$ bekannt sind.

In Abschnitt 11 waren die Eingabedaten f\"ur die verallgemeinerten
Summationsprozesse Ergebnisse aufwendiger Kristallorbital- und
Clusterrechnungen. Aufgrund des gro{\ss}en Aufwandes solcher Rechnungen
stand nur eine relativ kleine Anzahl von Eingabedaten zur Verf\"ugung, und
die relative Genauigkeit der Eingabedaten konnte bestenfalls abgesch\"atzt
werden. Au{\ss}erdem ist der Konvergenztyp sowohl der Kristallorbital- als
auch der Clusterenergien nicht mit Sicherheit bekannt. Aber auch hier
konnte durch Extrapolation eine deutliche Verbesserung der Konvergenz
bewirkt werden. In Abschnitt 11 war es aber n\"otig, andere
verallgemeinerte Summationsprozesse zu verwenden als in den fr\"uheren
Abschnitten.

In dieser Arbeit wurden die mathematischen Eigenschaften einiger
verallgemeinerter Summationsprozesse beschrieben, wobei das
Schwergewicht auf Transformationen lag, die vom Autor entwickelt wurden
[Weniger 1989; 1992]. Einige wichtige Verfahren zur
Konvergenzverbesserung und zur Summation wurden \"uberhaupt nicht
behandelt. Diese Arbeit ist also {\it kein\/} Versuch einer halbwegs
vollst\"andigen Beschreibung der mathematischen Eigenschaften
verallgemeinerter Summationsprozesse. Hierzu sei auf die B\"ucher von
Baker und Graves-Morris [1981a; 1981b] \"uber Pad\'e-Approximationen, auf
das Buch von Wimp [1981] und vor allem auf die Monographie von Brezinski
und Redivo Zaglia [1991] verwiesen, welche die zur Zeit aktuellste und
vollst\"andigste Monographie \"uber Extrapolationsverfahren ist.

Konvergenzprobleme gibt es nicht nur bei Potenzreihen oder bei
logarithmisch konvergenten Folgen und Reihen. Deswegen erlauben die in
dieser Arbeit pr\"asentierten Anwendungsbeispiele nat\"urlich keine {\it
definitiven\/} Aussagen \"uber die N\"utzlichkeit von verallgemeinerten
Summationsprozessen im allgemeinen oder \"uber das Leistungsverm\"ogen und
die Grenzen der in dieser Arbeit behandelten verallgemeinerten
Summationsprozesse im speziellen. Dazu w\"urde man noch wesentlich mehr
praktische Erfahrungen ben\"otigen.

Ich bin aber \"uberzeugt, da{\ss} verallgemeinerte Summationsprozesse auch in
anderen Bereichen \"ahnlich leistungsf\"ahig und hilfreich sein k\"onnen wie
in dieser Arbeit. Man mu{\ss} nur nach weiteren Anwendungen suchen. In
dieser Hinsicht gibt es aber das Problem, da{\ss} die {\it moderne\/}
Theorie der verallgemeinerten Summationsprozesse eine sehr junge
mathematische Disziplin ist, die auf zwei Arbeiten von Shanks [1955]
beziehungsweise Wynn [1956a] zur\"uckgeht. Au{\ss}erdem hat es in den letzten
Jahren enorme Fortschritte auf diesem Gebiet gegeben. Die modernen
nichtlinearen Verfahren zur Konvergenzverbesserung und zur Summation
geh\"oren deswegen noch nicht zum Allgemeinwissen der numerischen
Mathematiker und der mathematisch orientierten Naturwissenschaftler. Aus
diesem Grund m\"ochte ich noch einige weitere Techniken zur
Konvergenzverbesserung und zur Summation skizzieren, die nicht nur f\"ur
quantenmechanische und quantenchemische Rechnungen von Interesse sein
d\"urften.

Fourierreihen sind in den Naturwissenschaften oder der Technik von
gr\"o{\ss}ter Bedeutung. Bekanntlich konvergieren Fourierreihen ebenso wie
andere Orthogonalentwicklungen im allgemeinen nicht punktweise, sondern
nur im Mittel. Deswegen kann es vorkommen, da{\ss} eine im Mittel
konvergente Fourierreihe f\"ur einige Argumente divergiert. Derartige
Divergenzen von Fourierreihen werden in Abschnitt 18 der B\"ucher von
K\"orner [1988; 1993] behandelt. Die Summation divergenter Fourier- und
Orthogonalreihen wird in einem Buch von Okuyama [1984] diskutiert. Die
Anwendung verallgemeinerter Summationsprozesse auf Fourierreihen wird
beispielsweise in Abschnitt 6.1.3 des Buches von Brezinski und Redivo
Zaglia [1991], in Abschnitt 2.4 des Buches von Wimp [1981], oder in
Artikeln von Kiefer und Weiss [1981], Longman [1985; 1986; 1987] und
Tasche [1991] behandelt. Auch am hiesigen Institut f\"ur Physikalische und
Theoretische Chemie der Universit\"at Regensburg wurde schon erfolgreich
\"uber die Beschleunigung der Konvergenz von Fourierreihen gearbeitet
[Homeier 1992; 1993].

Orthogonalentwicklungen sind in allen Bereichen der Mathematik, der
Naturwissenschaften und der Technik von gr\"o{\ss}ter Bedeutung.
Selbstverst\"andlich gibt es auch bei Orthogonalentwicklungen
Konvergenzprobleme. Die Beschleunigung der Konvergenz von
Orthogonalentwicklungen wird beispielsweise in Abschnitt 1.6 des Buches
von Baker und Graves-Morris [1981b] oder in Artikeln von Baker und
Gubernatis [1981], Fleischer [1972; 1973a; 1973b], Gabutti und
Sacripante [1991], Garibotti und Grinstein [1978a; 1978b; 1979],
Garibotti, Grinstein und Miraglia [1980], Grinstein [1980], Holdeman
[1969], Holvorcem [1992], Lepora und Gabutti [1987] und Longman [1987]
behandelt.

In dieser Arbeit wurden verallgemeinerte Summationsprozesse
ausschlie{\ss}lich auf Folgen $\Seqn s$ von {\it Zahlen\/} angewendet. In
der Literatur werden Verfahren zur Konvergenzverbesserung oder zur
Summation von Zahlenfolgen h\"aufig {\it skalare Algorithmen\/} genannt.
Die zugrundeliegenden Konzepte sind aber auch im Falle wesentlich
allgemeinerer mathematischer Objekte anwendbar. Extrapolationsverfahren
f\"ur {\it Vektorfolgen\/} werden in Teil 4 und Teil 6 des Buches von
Brezinski und Redivo Zaglia [1991] ausf\"uhrlich behandelt. Au{\ss}erdem ist
in den letzten Jahren eine sehr gro{\ss}e Zahl von Artikeln \"uber die
Extrapolation von Vektorfolgen erschienen [Brezinski und Sadok 1992;
Cuyt 1989/90; Graves-Morris 1992; Jbilou und Sadok 1991; Le Ferrand
1992; MacLeod 1986; Matos 1992; Midy 1992; Nievergelt 1991; Osada 1991;
1992; Sidi 1986a; 1988b; 1989/90; 1991; Sidi und Bridger 1988; Sidi und
Ford 1991; Sidi, Ford und Smith 1986; Smith, Ford und Sidi 1987; 1988].
In einem Buch von Cuyt [1984] werden Pad\'e-Approximationen von {\it
Operatoren\/} behandelt, und Graves-Morris [1990] und Graves-Morris und
Thukral [1992] verwendeten Pad\'e-Approximationen, um die Konvergenz von
{\it Funktionenfolgen\/} zu beschleunigen, die bei der L\"osung von
Integralgleichungen auftreten.

Schlie{\ss}lich sei noch erw\"ahnt, da{\ss} Konvergenzbeschleunigung auch im
Zusammenhang mit {\it Monte-Carlo}-Verfahren verwendet wird [Swendsen
1991].

Ich m\"ochte diese Arbeit abschlie{\ss}en mit der Schlu{\ss}bemerkung des
Artikels, in dem Wynn [1956a, S. 96] seinen ber\"uhmten
$\epsilon$-Algorithmus einf\"uhrte:

\medskip

\beginSchmaeler
\noindent {\sl It is the author's hope that by demonstrating the ease
with which the various transformations may be effected, their field of
application might be widened, and deeper insight thereby obtained into
the problems for whose solution the transformations have been used.}
\endSchmaeler

\medskip

\endAbschnittsebene

\keinTitelblatt\neueSeite

\beginAbschnittsebene
\aktAbschnitt = 12

\Abschnitt Literaturverzeichnis

\parindent = 0 pt

\medskip

\aktTag = 0

{\sc Abbott, P.C., Maslen, E.N.} [1987], {\it Coordinate systems and
analytic expansions for three-body atomic wavefunctions: \Roemisch{1}.
Partial summation for the Fock expansion in hyperspherical coordinates},
J. Phys. A {\bf 20}, 2043 - 2075.
% Abschnitt 8.4

{\sc Abramowitz, M., Stegun, I.A.} [1972], {\it Handbook of
mathematical functions} (National Bureau of Standards, Washington,
D.~C.).
% Abschnitt 7.1, 9.1, 9.2, 10.2, 10.8

{\sc Adams, B.G.} [1988], {\it Application of 2-point Pad\'e approximants
to the ground state of the 2-dimensional hydrogen atom in an external
magnetic field}, Theor. Chim. Acta {\bf 73}, 459 - 465.
% Abschnitt 4.5

{\sc Adams, B.G., Avron, J.E., {\v C}{\' \i}{\v z}ek, J., Otto, P.,
Paldus, J., Moats, R.K., Silverstone, H.J.} [1980], {\it Bender-Wu
formulas for degenerate eigenvalues}, Phys. Rev. A {\bf 21}, 1914 -
1916.
% Abschnitt 10.1

{\sc Adams, B.G., {\v C}{\' \i}{\v z}ek, J., Paldus, J.} [1988], {\it
Lie algebraic methods and their applications to simple quantum
systems}, Adv. Quantum Chem. {\bf 19} 1 - 85.
% Abschnitt 1.2

{\sc Adams, D.J., Duby, G.S.} [1987], {\it Taming the Ewald sum in
computer simulations of charged systems}, J. Comput. Phys. {\bf 72}, 156
- 176.
% Abschnitt 11.1

{\sc Adams, R.A.} [1975], {\it Sobolev spaces} (Academic Press, New
York).
% Abschnitt 9.1

{\sc Agmon, S.} [1982], {\it Lectures on exponential decay of solutions
of second-order elliptic equations: Bounds on eigenfunctions of $N$-body
Schr\"odinger operators} (Princeton U.~P., Princeton).
% Abschnitt 8.2, 9.1

{\sc Agmon, S.} [1985], {\it Bounds on exponential decay of
eigenfunctions of Schr\"odinger operators}, in Graffi, S. (Herausgeber),
{\it Schr\"odinger operators} (Springer-Verlag, Berlin), 1 - 38.
% Abschnitt 8.2, 9.1

{\sc Ahlrichs, R.} [1989], {\it Basic mathematical properties of
electronic wave functions in momentum space}, in Defranceschi, M.,
Delhalle, J. (Herausgeber), {\it Numerical determination of the
electronic structure of atoms, diatomic and polyatomic molecules}
(Kluwer, Dordrecht), 1 - 15.
% Abschnitt 8.2, 9.1

{\sc Ahlrichs, R., Scharf, P.} [1987], {\it The coupled pair
approximation}, in Lawley, K.P. (Herausgeber), {\it Ab initio methods
in quantum chemistry \Roemisch{1}\/} (Wiley, Chichester), 501 - 537.
% Abschnitt 8.2

{\sc Airey, J.R.} [1937], {\it The ``converging factor'' in asymptotic
series and the calculation of Bessel, Laguerre and other functions},
Philos. Mag. {\bf 24}, 521 - 552.
% Abschnitt 5.7

{\sc Aitken, A.C.} [1926], {\it On Bernoulli's numerical solution of
algebraic equations}, Proc. Roy. Soc. Edinburgh {\bf 46}, 289 - 305.
% Abschnitt 2.4, 3.3, 11.5

{\sc Albeverio, S., Gesztesy, F., H{\o}egh-Krohn, R., Holden, H.}
[1988], {\it Solvable models in quantum mechanics} (Springer-Verlag, New
York).
% Abschnitt 9.1

{\sc Albeverio, S., Tirrozzi, B., Zegarlinski, B.} [1992], {\it Rigorous
results for the free energy in the Hopfield model}, Commun. Math. Phys.
{\bf 150}, 337 - 373.
% Abschnitt 1.3

{\sc Allen, G.D., Chui, C.K., Madych, W.R., Narcowich, F.J., Smith,
P.W.} [1975], {\it Pad\'e approximation of Stieltjes series}, J. Approx.
Theor. {\bf 14}, 302 - 316.
% Abschnitt 6.6, 6.7

{\sc Alvarez, G.} [1988], {\it Coupling-constant behavior of the cubic
anharmonic oscillator}, Phys. Rev. A {\bf 37}, 4079 - 4083.
% Abschnitt 2.3, 5.5, 10.1

{\sc Amos, A.T.} [1978], {\it Pad\'e approximants and
Rayleigh-Schr\"odinger perturbation theory}, J. Phys. B {\bf 11}, 2053 -
2060.
% Abschnitt 4.5

{\sc Andr\'e, J.-M.} [1980], {\it Quantum mechanical methods for regular
polymers}, Adv. Quantum Chem. {\bf 12}, 65 - 102.
% Abschnitt 11.1

{\sc Andr\'e, J.-M., Bodart, V.P., Br\'edas, J.-L. Delhalle, J., Fripiat,
J.G.} [1984], {\it Towards specific ab initio programs for polymer
calculations}, in Ladik, J., Andr\'e, J.-M., Seel, M. (Herausgeber), {\it
Quantum chemistry of polymers -- solid state aspects} (Reidel,
Dordrecht), 1 - 21.
% Abschnitt 11.1, 11-6

{\sc Andr\'e, J.-M., Delhalle, J., Br\'edas, J.-L.} [1991], {\it Quantum
chemistry aided design of organic polymers} (World Scientific,
Singapore).
% Abschnitt 11.1

{\sc Andr\'e, J.-M., Delhalle, J., Fripiat, J.G., Hennico, G., Piela, L.}
[1988], {\it Calculations of molecular polarizabilities from
electric-field-variant atomic orbitals: An analysis of the problem and
its application to the hydrogen molecule and to the alkane series}, Int.
J. Quantum Chem. Symp. {\bf 22}, 665 - 678.
% Abschnitt 11.1

{\sc Andr\'e, J.-M., Fripiat, J.G., Demanet, C., Br\'edas, J.-L. Delhalle,
J.} [1978], {\it Long-range Coulombic interactions in the theory of
polymers: A statement of the problem and a method for calculation by the
Fourier transformation technique}, Int. J. Quantum Chem. Symp. {\bf 12},
233 - 247.
% Abschnitt 11.1

{\sc Andr\'e, J.-M., Gouverneur, L., Leroy, G.} [1967a], {\it L'etude
th\'eorique des syst\`emes p\'erio\-diques. \Roemisch{1}. La m\'ethode
LCAO-HCO}, Int. J. Quantum Chem. {\bf 1}, 427 - 450.
% Abschnitt 11.1

{\sc Andr\'e, J.-M., Gouverneur, L., Leroy, G.} [1967b], {\it L'etude
th\'eorique des syst\`emes p\'erio\-diques. \Roemisch{2}. La m\'ethode
LCAO-SCF-CO}, Int. J. Quantum Chem. {\bf 1}, 451 - 461.
% Abschnitt 11.1

{\sc Andr\'e, J.-M., Vercauteren, D.P., Bodart, V.P. Fripiat, J.G.}
[1984], {\it {\rm Ab initio} calculations of the electronic structure of
helical polymers}, J. Comput. Chem. {\bf 5}, 535 - 537.
% Abschnitt 11.1

{\sc Andr\'e, J.-M., Vercauteren, D.P., Fripiat, J.G.} [1984], {\it
Electron density and related properties in stereoregular polymers and
bioplymers}, J. Comput. Chem. {\bf 5}, 349 - 352.
% Abschnitt 11.1

{\sc Antolovi\'{c}, D., Delhalle, J.} [1980], {\it Multipole and overlap
integrals over reduced Bessel functions in molecular quantum mechanics},
Phys. Rev. A {\bf 21}, 1815 - 1828.
% Abschnitt 8.2

{\sc Aronszajn, N., Smith, K.T.} [1961], {\it Theory of Bessel
potentials. Part \Roemisch{1}}, Ann. Inst. Fourier (Grenoble) {\bf 11},
385 - 475.
% Abschnitt 8.2

{\sc Arteca, G.A., Fern\'{a}ndez, F.M., Castro, E.A.} [1990], {\it
Large order perturbation theory and summation methods in quantum
mechanics} (Springer-Verlag, Berlin).
% Abschnitt 2.2, 2.3, 4.5, 10.1, 10.2, 10.4, 10.5

{\sc Augspurger, J.D., Dykstra, C.E.} [1992], {\it Evolution of
polarizabilities and hyperpolarizabilities with molecular aggregation: A
model study of acetylene clusters}, Int. J. Quantum Chem. {\bf 43}, 135 -
146.
% Abschnitt 11.1

{\sc Austin, E.J.} [1984], {\it Further applications of the renormalised
series technique}, J. Phys. A {\bf 17}, 367 - 373.
% Abschnitt 10.4

{\sc Austin, E.J., Killingbeck, J.} [1982], {\it Some extensions of the
renormalised series approach}, J. Phys. A {\bf 15}, L443 - L445.
% Abschnitt 10.4

{\sc Avery, J.} [1976], {\it Creation and annihilation operators}
(McGraw-Hill, New York).
% Abschnitt 10.1

{\sc Avron, J.E.} [1981], {\it Bender-Wu formulas for the Zeeman effect
in hydrogen}, Ann. Phys. (NY) {\bf 131}, 73 - 94.
% Abschnitt 2.3, 10.1

{\sc Avron, J.E.} [1982], {\it Bender-Wu formulas and classical
trajectories: Higher dimensions and degeneracies}, Int. J. Quantum Chem.
{\bf 21}, 119 - 124.
% Abschnitt 10.1

{\sc Avron, J.E., Adams, B.G., {\v C}{\' \i}{\v z}ek, J., Clay, M.,
Glasser, M.L., Otto, P., Paldus, J., Vrscay, E.} [1979], {\it Bender-Wu
formula, the $SO (4,2)$ dynamical group, and the Zeeman effect in
hydrogen}, Phys. Rev. Lett. {\bf 43}, 691 - 693.
% Abschnitt 10.1

{\sc Avron, J., Herbst, I., Simon, B.} [1977], {\it The Zeeman effect
revisited}, Phys. Lett. A {\bf 62}, 214 - 216.
% Abschnitt 2.3

{\sc Badziag, P., Solms, F.} [1988], {\it An improved SCF iteration
scheme}, Comput. Chem. {\bf 12}, 233 - 236.
% Abschnitt 11.1

{\sc Baker, Jr., G.A.} [1965], {\it The theory and application of the
Pad\'e approximant method}, Adv. Theor. Phys. {\bf 1}, 1 - 58.
% Abschnitt 2.4, 4.1, 4.3

{\sc Baker, Jr., G.A.} [1969], {\it Best error bounds for Pad\'e
approximants to convergent series of Stieltjes}, J. Math. Phys. {\bf
10}, 814 - 820.
% Abschnitt 6.6, 6.7

{\sc Baker, Jr., G.A.} [1972], {\it The Pad\'e approximant and related
material}, in Bessis, D. (Herausgeber), {\it Carg\`ese lectures in
physics} (Gordon and Breach, New York), Vol. 5, 349 - 383.
% Abschnitt 2.4, 4.1

{\sc Baker, Jr., G.A.} [1975], {\it Essentials of Pad\'e approximants}
(Academic Press, New York).
% Abschnitt 2.4, 4.1, 4.2, 4.3, 4.5, 6.1, 6.4, 10.7

{\sc Baker, Jr., G.A.} [1990], {\it Quantitative theory of critical
phenomena} (Academic Press, San Diego).
% Abschnitt 4.1, 4.3, 4.5, 6.5, 10.7, 11.4

{\sc Baker, Jr., G.A., Chisholm, R.} [1966], {\it The validity of
perturbation series with zero radius of convergence}, J. Math. Phys.
{\bf 7}, 1900 - 1902.
% Abschnitt 4.5

{\sc Baker, Jr., G.A., Gammel, J.L.} (Herausgeber) [1970], {\it The
Pad\'e approximant in theoretical physics} (Academic Press, New York).
% Abschnitt 2.4, 4.1

{\sc Baker, Jr., G.A., Graves-Morris, P.} [1981a], {\it Pad\'e
approximants. Part \Roemisch{1}: Basic theory} (Addison-Wesley, Reading,
Mass.).
% Abschnitt 2.4, 4.1, 4.3, 4.5, 6.1, 6.5, 7.2, 8.2, 10.7, 13

{\sc Baker, Jr., G.A., Graves-Morris, P.} [1981b], {\it Pad\'e
approximants. Part \Roemisch{2}: Extensions and applications}
(Addison-Wesley, Reading, Mass.).
% Abschnitt 2.4, 4.1, 4.5, 6.1, 13

{\sc Baker, Jr., G.A., Gubernatis, J.E.} [1981], {\it An asymptotic,
Pad\'e approximant method for Legendre series}, in Butzer, P.L., Feh\'er, F.
(Herausgeber), {\it E.B. Christoffel -- The influence of his work on
mathematics and the physical sciences} (Birkh\"auser, Basel), 232 - 242.
% Abschnitt 13

{\sc Baker, Jr., G.A., Pirner, H.J.} [1983], {\it Asymptotic estimate
of large orders in perturbation theory for the many-fermion ground state
energy}, Ann. Phys. (NY) {\bf 148}, 168 - 191.
% Abschnitt 2.3, 10.1

{\sc Balser, W.} [1992], {\it A different characterization of
multi-summable power series}, Analysis {\bf 12}, 57 - 65.
% Abschnitt 2.2

{\sc Balser, W., Braaksma, B.L.J., Ramis, J.-P., Sibuya, Y.} [1991],
{\it Multisummability of formal power series solutions of linear
ordinary differential equations}, Asympt. Anal. {\bf 5}, 27 - 45.
% Abschnitt 2.2

{\sc Banerjee, K.} [1979], {\it Rescaling the perturbation series},
Proc. Roy. Soc. London A {\bf 368}, 155 -162.
% Abschnitt 10.4

{\sc Banerjee, K., Bhatnagar, S.P., Choudry, V., Kanwal, S.S.} [1978],
{\it The anharmonic oscillator}, Proc. Roy. Soc. London A {\bf 360}, 575
- 586.
% Abschnitt 10.4, 10.7

{\sc Banks, T.I., Bender, C.M.} [1972], {\it Anharmonic oscillator with
polynomial self-interaction}, J. Math. Phys. {\bf 13}, 1320 - 1324.
% Abschnitt 10.4

{\sc Bansal, M., Srivastava, S., Vishwamittar} [1991], {\it Energy
eigenvalues of double-well oscillator with mixed quartic and sextic
anharmonicities}, Phys. Rev. A {\bf 44}, 8012 - 8019.
% Abschnitt 4.5

{\sc Bansal, M., Srivastava, S., Vishwamittar} [1992], {\it Energy
eigenvalues for double-well oscillator with mixed cubic-quartic
anharmonicities}, Chem. Phys. Lett. {\bf 195}, 505 - 508.
% Abschnitt 4.5

{\sc Baltus, C., Jones, W.B.} [1989], {\it Truncation error bounds for
modified continued fractions with applications to special functions},
Numer. Math. {\bf 55}, 281 - 307.
% Abschnitt 4.1

{\sc Baram, A., Luban, M.} [1979], {\it Divergence of the virial series
for hard discs and hard spheres at closest packing}, J. Phys. C {\bf
12}, L659 - L664.
% Abschnitt 2.1

{\sc Barnett, A.R.} [1982], {\it Continued-fraction evaluation of
Coulomb functions $F_{\lambda} (\eta, x)$, $G_{\lambda} (\eta, x)$ and
their derivatives}, J. Comput. Phys. {\bf 46}, 171 - 188.
% Abschnitt 4.1

{\sc Barnett, M.P.} [1963], {\it The evaluation of molecular integrals
by the zeta-function expansion}, in Alder, B., Fernbach, S., Rotenberg,
M. (Herausgeber), {\it Methods in computational physics {\bf 2}. Quantum
mechanics} (Academic Press, New York), 95 - 153.
% Abschnitt 8.2

{\sc Barnett, M.P., Coulson, C.A.} [1951], {\it The evaluation of
integrals occurring in the theory of molecular structure. Part
\Roemisch{1} \& \Roemisch{2}}, Phil. Trans. Roy. Soc. London A {\bf
243}, 221 - 249.
% Abschnitt 8.2

{\sc Bartlett, R.J.} [1981], {\it Many-body perturbation theory and
coupled cluster theory for electron correlation in molecules}, Annu.
Rev. Phys. Chem. {\bf 32}, 359 - 401.
% Abschnitt 1.2, 8.2

{\sc Bartlett, R.J., Dykstra, C.E., Paldus, J.} [1984], {\it Coupled
cluster methods for molecular calculations}, in Dykstra, C.E.
(Herausgeber), {\it Advanced theories and computational approaches to
the electronic structure of molecules} (Reidel, Dordrecht), 127 - 159.
% Abschnitt 1.2, 8.2

{\sc Bartlett, R.J., Stanton, J.F.} [1994], {\it Applications of
post-Hartree-Fock methods: A tutorial}, in Lipkowitz, K.B., Boyd, D.B.
(Herausgeber), {\it Reviews in computational chemistry \Roemisch{5}}
(VCH Verlagsgesellschaft, Weinheim), 65 - 169.
% Abschnitt 1.2, 8.2

{\sc Basdevant, J.L.} [1972], {\it The Pad\'e approximation and its
physical applications}, Fortschr. Physik {\bf 20}, 283 - 331.
% Abschnitt 4.1, 4.5

{\sc Bates, D.R., Ledsham, K., Stewart, A.L.} [1953], {\it Wave
functions of the hydrogen molecular ion}, Phil. Trans. Roy. Soc. London
A {\bf 246}, 215 - 240.
% Abschnitt 1.1

{\sc Bates, D.R., Reid, R.G.H.} [1968], {\it Electronic eigenenergies of
the hydrogen molecular ion}, Adv. At. Mol. Phys. {\bf 4}, 13 - 35.
% Abschnitt 1.1

{\sc Baumg\"artel, H.} [1985], {\it Analytic perturbation theory for
matrices and operators} (Birkh\"auser, Basel).
% Abschnitt 10.1

{\sc Bayman, B.F.} [1978], {\it A generalization of the spherical
harmonic gradient formula}, J. Math. Phys. {\bf 19}, 2558 - 2562.
% Abschnitt 8.2

{\sc Beleznay, F.} [1986], {\it Estimations for asymptotic series using
a modified Romberg algorithm: \Roemisch{1}. Finite-size scaling
calculations}, J. Phys. A {\bf 19}, 551 - 562.
% Abschnitt 11.6

{\sc Belki\'{c}, D.} [1989], {\it New hybrid non-linear transformations
of divergent perturbation series for quadratic Zeeman effects}, J. Phys.
A {\bf 22}, 3003 - 3010.
% Abschnitt 4.5

{\sc Bender, C.M.} [1970], {\it Generalized anharmonic oscillator}, J.
Math. Phys. {\bf 11}, 796 - 804.
% Abschnitt 10.1

{\sc Bender, C.M.} [1982], {\it Perturbation theory in large order for
some elementary systems}, Int. J. Quantum Chem. {\bf 21}, 93 - 104.
% Abschnitt 10.1

{\sc Bender, C.M., Mead, L.R., Papanicolaou, N.} [1987], {\it Maximum
entropy summation of divergent perturbation series}, J. Math. Phys. {\bf
28}, 1016 - 1018.
% Abschnitt 10.2

{\sc Bender, C.M., Orszag, S.A.} [1978], {\it Advanced mathematical
methods for scientists and engineers} (McGraw-Hill, New York).
% Abschnitt 2.1, 11.1

{\sc Bender, C.M., Wu, T.T.} [1969], {\it Anharmonic oscillator},
Phys. Rev. {\bf 184}, 1231 - 1260.
% Abschnitt 2.2, 4.5, 10.1, 10.2, 10.5

{\sc Bender, C.M., Wu, T.T.} [1971], {\it Large-order behavior of
perturbation theory}, Phys. Rev. Lett. {\bf 27}, 461 - 465.
% Abschnitt 2.2, 4.5, 10.1, 10.2

{\sc Bender, C.M., Wu, T.T.} [1973], {\it Anharmonic oscillator.
\Roemisch {2}. A study in perturbation theory in large order}, Phys.
Rev. D {\bf 7}, 1620 - 1636.
% Abschnitt 2.2, 4.5, 10.1, 10.2, 10.5

{\sc Berry, M.V.} [1989], {\it Uniform asymptotic smoothing of Stokes'
discontinuities}, Proc. Roy. Soc. London A {\bf 422}, 7 - 21.
% Abschnitt 2.2

{\sc Berry, M.V.} [1991], {\it Asymptotics, superasymptotics,
hyperasymptotics}, in Segur, H., Tanveer, S. (Herausgeber), {\it
Asymptotics beyond all orders} (Plenum Press, New York), 1 - 14.
% Abschnitt 2.2

{\sc Berry, M.V., Howls, C.J.} [1990], {\it Hyperasymptotics}, Proc.
Roy. Soc. London A {\bf 430}, 653 - 668.
% Abschnitt 2.2

{\sc Berry, M.V., Howls, C.J.} [1991], {\it Hyperasymptotics for
integrals with saddles}, Proc. Roy. Soc. London A {\bf 434}, 657 - 675. 
% Abschnitt 2.2

{\sc Bethe, H.A.} [1947], {\it The electromagnetic shift of energy
levels}, Phys. Rev. {\bf 72}, 339 - 341.
% Abschnitt 2.1

{\sc Bethe, H.A., Salpeter, E.A.} [1977], {\it Quantum mechanics of one-
and two-electron atoms} (Plenum Press, New York).
% Abschnitt 8.1

{\sc Bevensee, R.M.} [1993], {\it Maximum entropy solution to scientific
problems} (Prentice Hall, Englewood Cliffs).
% Abschnitt 10.2

{\sc Bhattacharya, A.K., Dhabal, S.C.} [1986], {\it Molecular overlap
integrals with exponential-type orbitals}, J. Chem. Phys. {\bf 84},
1598 - 1605.
% Abschnitt 8.2, 8.3

{\sc Bhattacharya, R., Roy, D., Bhowmick, S.} [1989], {\it On the
regularity of the Levin $u$-transform}, Comput. Phys. Commun. {\bf 55},
297 - 301.
% Abschnitt 4.5

{\sc Bhattacharyya, K.} [1982], {\it Generalized Euler transformation
in extracting useful information from divergent (asymptotic)
perturbation series and the construction of Pad\'e approximants}, Int. J.
Quantum Chem. {\bf 22}, 307 - 330.
% Abschnitt 4.5

{\sc Bhattacharyya, K.} [1987], {\it On the use of generalized Euler
transformation in handling divergent series}, Proc. Indian Acad. Sci
(Chem. Sci.) {\bf 99}, 9 - 20.
% Abschnitt 4.5

{\sc Bhattacharyya, K.} [1989a], {\it Asymptotic response of observables
from divergent power series expansions}, Phys. Rev. A {\bf 39}, 6124 -
6128.
% Abschnitt 4.5

{\sc Bhattacharyya, K.} [1989b], {\it On the maximum entropy method of
handling divergent perturbation series}, Chem. Phys. Lett. {\bf 161},
259 - 265.
% Abschnitt 10.2

{\sc Bhowmick, S., Bhattacharya, R., Roy, D.} [1989], {\it Iterations
of convergence accelerating nonlinear transforms}, Comput. Phys. Commun.
{\bf 54}, 31 - 36.
% Abschnitt 3.3, 4.5

{\sc Bj{\o}rstad, P., Dahlquist, G., Grosse, E.} [1981], {\it
Extrapolations of asymptotic expansions by a modified Aitken
$\delta^2$-formula}, BIT {\bf 21}, 56 - 65.
% Abschnitt 4.5, 8.4, 11.4, 11.5, 11.6

{\sc Biedenharn, L.C., Louck, J.D.} [1981], {\it Angular momentum in
quantum physics} (Addison-Wesley, Reading, Mass.).
% Abschnitt 8.1, 8.2

{\sc Bishop, D.M.} [1967], {\it Single-center molecular wave
functions}, Adv. Quantum Chem. {\bf 3}, 25 - 59.
% Abschnitt 8.1

{\sc Bishop, R.F.} [1991], {\it An overview of coupled cluster theory
and its applications in physics}, Theor. Chim. Acta {\bf 80}, 95 - 148.
% Abschnitt 1.3

{\sc Biswas, S.N., Datta, K., Saxena, R.P., Srivastava, P.K., Varma,
V.S.} [1973], {\it Eigenvalues of $\lambda x^{2 m}$ anharmonic
oscillators}, J. Math. Phys. {\bf 14}, 1190 - 1195.
% Abschnitt 10.7

{\sc Blanchard, P., Br\"uning, E.} [1992], {\it Variational methods in
mathematical physics} (Springer-Verlag, Berlin).
% Abschnitt 8.2, 9.1

{\sc Bleistein, N., Handelsman, R.A.} [1986], {\it Asymptotic
expansions of integrals} (Dover, New York).
% Abschnitt 2.2

{\sc Bochner, S.} [1948], {\it Vorlesungen \"uber Fouriersche Integrale}
(Chelsea, New York).
% Abschnitt 8.2

{\sc Boerrigter, P.M., te Velde, G., Baerends, E.J.} [1988], {\it
Three-dimensional numerical integration for electronic structure
calculations}, Int. J. Quantum Chem. {\bf 33}, 87 - 113.
% Abschnitt 8.1

{\sc Bolcer, J.D., Hermann, R.B.} [1994], {\it The development of
computational chemistry in the United States}, in Lipkowitz, K.B., Boyd,
D.B. (Herausgeber), {\it Reviews in computational chemistry
\Roemisch{5}} (VCH Verlagsgesellschaft, Weinheim), 1 - 63.
% Abschnitt 1.1

{\sc Bonham, R.A., Peacher, J.L., Cox, Jr., H.L.} [1964], {\it On the
calculation of multicenter two-electron repulsion integrals involving
Slater functions}, J. Chem. Phys. {\bf 40}, 3083 - 3086.
% Abschnitt 8.2

{\sc Bonnier, B.} [1978], {\it Strong coupling regime of the anharmonic
oscillator from perturbation theory and application to the
$\Phi_{4}^{4}$ model}, Phys. Lett. B {\bf 78}, 107 - 109.
% Abschnitt 10.4

{\sc Borel E.} [1899], {\it M\'emoires sur les s\'eries divergentes}, Ann.
sci. Ec. norm. sup. Paris {\bf 16}, 9 - 136.
% Abschnitt 2.3

{\sc Borel E.} [1928], {\it Le{\c c}ons sur les s\'eries divergentes}
(Gautier-Villars, Paris). Nachgedruckt durch \'Editions Jacques Gabay,
Paris, 1988.
% Abschnitt 2.2, 4.3, 5.3, 6.5, 10.7

{\sc Born, M., Oppenheimer, R.} [1927], {\it Zur Quantentheorie von
Molekeln}, Ann. Physik {\bf 84}, 457 - 484.
% Abschnitt 1.1, 8.1

{\sc Borwein, D., Borwein, J.M., Shail, R.} [1989], {\it Analysis of
certain lattice sums}, J. Math. Anal. Appl. {\bf 143}, 126 - 137.
% Abschnitt 11.1

{\sc Borwein, D., Borwein, J.M., Taylor, K.F.} [1985], {\it Convergence
of lattice sums and Made\-lung's constant}, J. Math. Phys. {\bf 26},
2999 - 3009.
% Abschnitt 11.1

{\sc Bouferguene, A., Rinaldi, D.} [1994], {\it A new single-center
method to compute molecular integrals of quantum chemistry in
Slater-type orbital basis of functions}, Int. J. Quantum Chem. {\bf 50},
21 - 42.
% Abschnitt 8.2

{\sc Bowman, F.} [1958], {\it Introduction to Bessel functions} (Dover,
New York).
% Abschnitt 7.1

{\sc Bowman, K.O., Shenton, L.R.} [1989], {\it Continued fractions in
statistical applications} (Marcel Dekker, New York).
% Abschnitt 2.2, 4.1, 4.3, 6.5, 10.6

{\sc Boyd, W.G.C.} [1990a], {\it Stieltjes transforms and the Stokes
phenomenon}, Proc. Roy. Soc. London A {\bf 429}, 227 - 246.
% Abschnitt 2.2

{\sc Boyd, W.G.C.} [1990b], {\it Asymptotic expansions for the
coefficient functions associated with linear second-order differential
equations: The simple pole case}, in Wong, R. (Herausgeber), {\it
Asymptotics and computational analysis} (Marcel Dekker, New York), 53 -
73.
% Abschnitt 2.2

{\sc Boyle, A., Caviness, B.F.} (Herausgeber) [1990], {\it Future
directions for research in symbolic computation} (Society of Industrial
and Applied Mathematics, Philadelphia).
% Abschnitt 4.4

{\sc Boys, S.F.} [1955], {\it Electronic wave functions. \Roemisch{1}. A
general method of calculation for the stationary states of any molecular
system}, Proc. Roy. Soc. London A {\bf 200}, 542 - 554.
% Abschnitt 9.1

{\sc Braaksma, B.L.J.} [1992], {\it Multisummability of formal power
series solutions of nonlinear meromorphic differential equations}, Ann.
Inst. Fourier (Grenoble) {\bf 42}, 517 - 540. % Abschnitt 2.2

{\sc Br\"andas, E., Goscinski, O.} [1970], {\it Variation-perturbation
expansions and Pad\'e approximants to the energy}, Phys. Rev. A {\bf 1},
552 - 560.
% Abschnitt 4.5

{\sc Braess, D.} [1986], {\it Nonlinear approximation theory}
(Springer-Verlag, Berlin).
% Abschnitt 2.4

{\sc Bra{\ss}, H.} [1977], {\it Quadraturverfahren} (Vandenhoek \&
Rupprecht, G\"ottingen). % Abschnitt 8.2

{\sc Br\'edas, J.L., Chance, R.R., Silbey, R., Nicolas, G., Durand, Ph.}
[1981], {\it A nonempirical effective Hamiltonian technique for
polymers: Application to polyacetylene and polydiacetylene}, J. Chem.
Phys. {\bf 75}, 255 - 267. % Abschnitt 11.2

{\sc Breen, S.A.} [1982], {\it Large order perturbation theory for the
anharmonic oscillator} (Ph.D. Thesis, Graduate Program in Mathematics,
Rutgers University, New Brunswick, New Jersey).
% Abschnitt 10.1

{\sc Breen, S.} [1983], {\it Leading large order asymptotics for
$(\varphi^4)_2$ perturbation theory}, Commun. Math. Phys. {\bf 92}, 179
- 194.
% Abschnitt 10.1

{\sc Breen, S.} [1987], {\it Large-order estimates for ground-state
energy perturbation series}, J. Stat. Phys. {\bf 46}, 1233 - 1280.
% Abschnitt 10.1

{\sc Br\'ezin, E., Le Guillou, J.C., Zinn-Justin, J.} [1977], {\it
Perturbation theory at large order: \Roemisch{1}. The $\varphi^{2 N}$
interaction}, Phys. Rev. D {\bf 15}, 1544 - 1557.
% Abschnitt 2.3, 10.1

{\sc Br\'ezin, E., Parisi, G.} [1978], {\it Critical exponents and
large-order behavior of perturbation theory}, J. Stat. Phys. {\bf 19},
269 - 292. % Abschnitt 10.1

{\sc Br\'ezin, E., Parisi, G., Zinn-Justin, J.} [1977], {\it Perturbation
theory at large orders for a potential with degenerate minima}, Phys.
Rev. D {\bf 16}, 408 - 412. % Abschnitt 10.1

{\sc Brezinski, C.} [1971], {\it Acc\'el\'eration de suites \`a convergence
logarithmique}, C. R. Acad. Sc. Paris {\bf 273}, 727 - 730.
% Abschnitt 4.4, 4.5, 6.5, 6.6, 8.4, 11.5

{\sc Brezinski, C.} [1976], {\it A bibliography on Pad\'e approximation
and related matters}, in Cabannes, H. (Herausgeber), {\it Pad\'e
approximation method and its application to mechanics} (Springer-Verlag,
Berlin), 245 -267.
% Abschnitt 4.1, 4.5

{\sc Brezinski, C.} [1977], {\it Acc\'el\'eration de la convergence en
analyse num\'erique} (Springer-Verlag, Berlin).
% Abschnitt 2.4, 4.1, 4.5, 6.1

{\sc Brezinski, C.} [1978], {\it Algorithmes d'acc\'el\'eration de la
convergence -- \'Etude num\'erique} (\'Editions Technip, Paris).
% Abschnitt 2.4, 4.1, 4.4, 4.5

{\sc Brezinski, C.} [1980a], {\it Pad\'e-type approximation and general
orthogonal polynomials} (Birk\-h\"au\-ser, Basel).
% Abschnitt 2.4, 4.1

{\sc Brezinski, C.} [1980b], {\it A general extrapolation algorithm},
Numer. Math. {\bf 35}, 175 - 180.
% Abschnitt 3.2

{\sc Brezinski, C.} [1982], {\it Algorithm 585: A subroutine for the
general interpolation and extrapolation problems}, ACM Trans. Math.
Software {\bf 8}, 290 - 301.
% Abschnitt 3.2

{\sc Brezinski, C.} [1985a], {\it Convergence acceleration methods: The
past decade}, J. Comput. Appl. Math. {\bf 12} \& {\bf 13}, 19 - 36.
% Abschnitt 2.4

{\sc Brezinski, C.} [1985b], {\it Prediction properties of some
extrapolation methods}, Appl. Numer. Math. {\bf 1}, 457 - 462.
% Abschnitt 4.2

{\sc Brezinski, C.} [1988], {\it Introduction and historical survey},
in Delahaye, J. P., {\it Sequence transformations} (Springer-Verlag,
Berlin), \Roemisch{9} - \Roemisch{15} (Vorwort).
% Abschnitt 2.4

{\sc Brezinski, C.} [1989], {\it A survey of iterative interpolation by
the $E$-algorithm}, in H{\aa}vie, T. (Herausgeber), {\it Romberg
seminar on quadrature/interpolation/extrapolation and rational
approximations (Pad\'e-continued fractions)} (Det Kongelige Norske
Videnskabers Selskab Skrifters 2, Tapir Forlag, Trondheim), 1 - 26.
% Abschnitt 2.4

{\sc Brezinski, C.} [1991a], {\it History of continued fractions and
Pad\'e approximants} (Springer-Verlag, Berlin).
% Abschnitt 2.4, 4.1

{\sc Brezinski, C.} (Herausgeber) [1991b], {\it Continued fractions and
Pad\'e approximants} (North-Holland, Amsterdam).
% Abschnitt 2.4, 4.1

{\sc Brezinski, C.} [1991c], {\it A bibliography on continued fractions,
Pad\'e approximation, sequence transformation and related subjects}
(Prensas Universitarias de Zaragoza, Zaragoza).
% Abschnitt 4.1, 4.5

{\sc Brezinski, C., Delahaye, J.P., Germain-Bonne, B.} [1983], {\it
Convergence acceleration by extraction of linear subsequences}, SIAM J.
Numer. Anal. {\bf 20}, 1099 - 1105.
% Abschnitt 4.5

{\sc Brezinski, C., Draux, A., Magnus, A.P., Maroni, P., Ronveaux, A.}
(Herausgeber) [1985], {\it Polyn\^{o}mes orthogonaux et applications}
(Springer-Verlag, Berlin).
% Abschnitt 2.4, 4.1

{\sc Brezinski, C., Lembarki, A.} [1986], {\it Acceleration of extended
Fibonacci sequences}, Appl. Numer. Math. {\bf 2}, 1 - 8.
% Abschnitt 3.3

{\sc Brezinski, C., Matos, A.C.} [1993], {\it A derivation of
extrapolation algorithms based on error estimates} (Publication ANO-319,
Laboratoire d'Analyse Num\'erique et d'Optimisation, Universit\'e des
Sciences et Technolgie de Lille).
% Abschnitt 5.1

{\sc Brezinski, C., Redivo Zaglia, M.} [1991], {\it Extrapolation
methods} (North-Holland, Amsterdam).
% Abschnitt 2.4, 4.1, 4.5, 5.2, 5.3, 5.4, 5.5, 6.1, 11.1, 11.5, 13

{\sc Brezinski, C., Redivo Zaglia, M.} [1993a], {\it A general
extrapolation algorithm revisited} (Publication ANO-306, Laboratoire
d'Analyse Num\'erique et d'Optimisation, Universit\'e des Sciences et
Technolgie de Lille).
% Abschnitt 5.1

{\sc Brezinski, C., Redivo Zaglia, M.} [1993b], {\it On the kernel of
sequence transformations} (Publication ANO-309, Laboratoire d'Analyse
Num\'erique et d'Optimisation, Universit\'e des Sciences et Technolgie de
Lille).
% Abschnitt 5.1

{\sc Brezinski, C., Sadok, H.} [1992], {\it Some vector sequence
transformations with applications to systems of equations}, Numer.
Algor. {\bf 3}, 75 - 80.
% Abschnitt 13

{\sc Brezinski, C., van Iseghem J.} [1991], {\it Pad\'e approximations},
in Ciarlet, P.G., Lions, J.L. (Herausgeber), {\it Handbook of
numerical analysis vol. \Roemisch{3}} (North-Holland, Amsterdam), im
Druck.
% Abschnitt 2.4, 4.1, 6.1

{\sc Brezinski, C., Walz, G.} [1991], {\it Sequences of transformations
and triangular recursion schemes}, J. Comput. Appl. Math. {\bf 34}, 361
- 383.
% Abschnitt 6.6

{\sc Bromwich, T.J.I'A.} [1926], {\it An introduction to the theory
of infinite series} (Macmillan, New York).
% Abschnitt 2.1

{\sc Brown, G.E.} [1972], {\it Many-body problems} (North-Holland,
Amsterdam).
% Abschnitt 10.1

{\sc Browne, J.C.} [1971], {\it Molecular wave functions: Calculation
and use in atomic and molecular processes}, Adv. At. Mol. Phys. {\bf 7},
47 - 95.
% Abschnitt 8.2

{\sc Buchholz, H.} [1969], {\it The confluent hypergeometric function}
(Springer-Verlag, Berlin).
% Abschnitt 7.2

{\sc Bultheel, A.} [1987], {\it Laurent series and their Pad\'e
approximations} (Birkh\"auser, Basel).
% Abschnitt 2.4, 4.1

{\sc Byron, Jr., F.W., Fuller, R.W.} [1970], {\it Mathematics of
classical and quantum physics \Roemisch{2}} (Addison-Wesley, Reading,
Mass.).
% Abschnitt 10.1

{\sc Cabannes, H.} (Herausgeber) [1976], {\it Pad\'e approximation
method and its application to mechanics} (Springer-Verlag, Berlin).
% Abschnitt 2.4, 4.1, 4.5

{\sc Cabay, S., Kossowski, P.} [1990], {\it Power series remainder
sequences and Pad\'e fractions over an integral domain}, J. Symb. Comput.
{\bf 10}, 139 - 163.
% Abschnitt 4.4

{\sc Calais, J.-L.} [1986], {\it Orthonormalization and symmetry
adaptation of crystal orbitals}, Int. J. Quantum Chem. Symp. {\bf 19},
655 - 667.
% Abschnitt 11.1

{\sc Campbell, J.B.} [1980], {\it On Temme's algorithm for the modified
Bessel function of the third kind}, ACM Trans. Math. Software {\bf 6},
581 - 586.
% Abschnitt 7.1

{\sc Campbell, J.B.} [1981], {\it Bessel functions $I_{\nu} (z)$ and
$K_{\nu} (z)$ of real order and complex argument}, Comput. Phys. Commun.
{\bf 24}, 97 - 105.
% Abschnitt 7.1

{\sc Campbell, J.B.} [1982], {\it Erratum note: Bessel functions
$I_{\nu} (z)$ and $K_{\nu} (z)$ of real order and complex argument},
Comput. Phys. Commun. {\bf 25}, 207.
% Abschnitt 7.1

{\sc Carb\'{o}, R., Besal\'{u}, E.} [1992], {\it Many center AO integral
evaluation using Cartesian exponential type orbitals (CETOS)}, Adv.
Quantum Chem. {\bf 24}, 115 - 237.
% Abschnitt 8.2

{\sc Carleman, T.} [1926], {\it Les fonctions quasi-analytiques},
(Gautiers Villars, Paris).
% Abschnitt 2.2, 6.5, 11.4

{\sc Carlson, B.C.} [1977], {\it Special functions of applied
mathematics} (Academic Press, New York).
% Abschnitt 3.1, 5.2

{\sc \v{C}\'{a}rsky, P., Urban, M.} [1980], {\it Ab initio calculations}
(Springer-Verlag, Berlin).
% Abschnitt 8.1, 8.2

{\sc Carstensen, C.} [1990], {\it On a general $\rho$-algorithm}, J.
Comput. Appl. Math. {\bf 33}, 61 - 71.
% Abschnitt 2.4, 4.5

{\sc Caswell, W.E.} [1979], {\it Accurate energy levels for the
anharmonic oscillator and a summable series for the double-well
potential in perturbation theory}, Ann. Phys. (NY) {\bf 123}, 153 - 184.
% Abschnitt 2.3, 10.4

{\sc Cedillo, A., Robles, J., G\'{a}zquez, J.L.} [1988], {\it New
nonlocal exchange-energy functional from a kinetic-energy-density
Pad\'e-approximant model}, Phys. Rev. A {\bf 38}, 1697 - 1701.
% Abschnitt 4.5

{\sc Champagne, B., Mosley, D.H., Andr\'e, J.-M.} [1993], {\it {\rm Ab
initio} coupled and uncoupled Hartree-Fock calculations of the
polarizabilities of finite and infinite polyacetylene chains}, Int. J.
Quantum Chem. Symp. {\bf 27}, 667 - 685.
% Abschnitt 11.1, 11.2

{\sc Chandra, A.K., Bhattacharyya, K.} [1993], {\it Perturbation
expansions, Symanzik scaling, and Pad\'e-type approximants: The anharmonic
oscillator problem}, Int. J. Quantum Chem. {\bf 45}, 251 - 262.
% Abschnitt 4.5

{\sc Char, B.W., Geddes, K.O., Gonnet, G.H., Leong, B.L., Monagan,
M.B., Watt, S.M.} [1991a], {\it Maple \Roemisch{5} language reference
manual} (Springer-Verlag, Berlin).
% Abschnitt 3.2, 4.4, 11.5, 13

{\sc Char, B.W., Geddes, K.O., Gonnet, G.H., Leong, B.L., Monagan, M.B.,
Watt, S.M.} [1991b], {\it Maple \Roemisch{5} library reference manual}
(Springer-Verlag, Berlin).
% Abschnitt 10.4, 10.6, 10.7, 10.8

{\sc Char, B.W., Geddes, K.O., Gonnet, G.H., Leong, B.L., Monagan,
M.B., Watt, S.M.} [1992], {\it First leaves: A tutorial introduction to
Maple \Roemisch{5}} (Springer-Verlag, Berlin).
% Abschnitt 10.5

{\sc Chaudhuri, R.N., Mukherjee, B.} [1985], {\it Eigenvalues of the
$\mu x^2 + \lambda x^{2 m}$ interaction}, Pramana {\bf 24}, 685 - 693.
% Abschnitt 10.2, 10.7

{\sc Chhajlany, S.C., Letov, D.A., Malnev, V.N.} [1991], {\it Energy
spectrum of the potential $V = a x^2 + x^4$}, J. Phys. A. {\bf 24}, 2731
- 2741.
% Abschnitt 10.7

{\sc Chien, J.C.W.} [1984], {\it Polyacetylene} (Academic Press, Orlando).
% Abschnitt 11.2

{\sc Cioslowski, J.} [1988a], {\it Hartree-Fock exchange energy from the
application of generalized Pad\'e approximants to the first-order density
matrix}, Phys. Rev. A {\bf 37}, 4023 - 4025.
% Abschnitt 4.5

{\sc Cioslowski, J.} [1988b], {\it The generalized Pad\'e approximant and
chemical graph theory}, Int. J. Quantum Chem. {\bf 34}, 217 - 224.
% Abschnitt 4.5

{\sc Cioslowski, J.} [1988c], {\it On extracting the bulk properties
from results on small cluster calculations}, Chem. Phys. Lett. {\bf
153}, 446 - 450.
% Abschnitt 4.5, 11.1

{\sc Cioslowski, J.} [1988d], {\it Why does Aitken extrapolation often
help to attain convergence in self-consistent field calculations}, J.
Chem. Phys. {\bf 89}, 2126 - 2127.
% Abschnitt 11.1

{\sc Cioslowski, J.} [1990a], {\it Bulk properties from finite-cluster
calculations: \Roemisch{4}. Linear chains of hydrogen fluoride}, Theor.
Chim. Acta {\bf 77}, 253 - 261.
% Abschnitt 4.5, 11.1

{\sc Cioslowski, J.} [1990b], {\it Bulk properties from finite-cluster
calculations: \Roemisch{5}. Pseudo-Wannier orbitals from molecular orbital
calculations on finite clusters}, J. Chem. Phys. {\bf 92}, 1236 - 1239.
% Abschnitt 11.1

{\sc Cioslowski, J.} [1993a], {\it \it Bulk properties from
finite-cluster calculations: \Roemisch{7}. Accurate\/} ab initio {\it
calculations on the Peierls distortion in polyacene}, J. Chem. Phys.
{\bf 98}, 473 - 477.
% Abschnitt 4.5, 11.1

{\sc Cioslowski, J.} [1993b], {\it Ab initio calculations on large
molecules: Methodology and applications}, in Lipkowitz, K.B., Boyd, D.B.
(Herausgeber), {\it Reviews in computational chemistry \Roemisch{4}}
(VCH Verlagsgesellschaft, Weinheim), 1 - 33.
% Abschnitt 11.1

{\sc Cioslowski, J., Bessis, D.} [1988], {\it Extrapolation of the total
energy of polymers to the bulk limit using generalized Pad\'e
approximants}, Int. J. Quantum Chem. {\bf 34}, 225 - 229.
% Abschnitt 4.5, 11.1

{\sc Cioslowski, J., Lepetit, M.P.} [1991], {\it Bulk properties from
finite cluster calculations. \Roemisch{6}. A finite-size perturbation
theory for the Hartree-Fock energy of linear oligomers}, J. Chem. Phys.
{\bf 95}, 3536 - 3548.
% Abschnitt 11.1, 11.4, 11.6

{\sc Cioslowski, J., Weniger, E.J.} [1993], {\it Bulk properties from
finite cluster calculations. \Roemisch{8}. Benchmark calculations on the
efficiency of extrapolation methods for the HF and MP2 energies of
polyacenes}, J. Comput. Chem. {\bf 14}, 1468 - 1481.
% Abschnitt 4.5, 11.1

{\sc {\v C}{\' \i}{\v z}ek, J.} [1966], {\it On the correlation problem
in atomic and molecular systems. Calculation of wavefunction components
in Ursell-type expansion using quantum-field theoretical methods}, J.
Chem. Phys. {\bf 45}, 4256 - 4266.
% Abschnitt 1.2 8.2

{\sc {\v C}{\' \i}{\v z}ek, J.} [1969], {\it On the use of the cluster
expansion and the technique of diagrams in calculations of correlation
effects in atoms and molecules}, Adv. Chem. Phys. {\bf 14}, 35 - 89.
% Abschnitt 8.2

{\sc {\v C}{\'\i}{\v z}ek, J., Clay, M.R., Paldus, J.} [1980], {\it
Asymptotic behavior of the ground-state-energy expansion for $H_2^+$ in
terms of internuclear separation}, Phys. Rev. A {\bf 22}, 793 - 796.
% Abschnitt 10.1

{\sc {\v C}{\' \i}{\v z}ek, J., Damburg, R. J., Graffi, S., Grecchi, V.,
Harrell \Roemisch{2}, E.M., Harris, J.G., Nakai, S., Paldus, J.,
Propin, R.Kh., Silverstone, H.J.} [1986], {\it $1 / R$ expansion for
$H_2^+$: Calculation of exponentially small terms and asymptotics},
Phys. Rev. A {\bf 33}, 12 - 54.
% Abschnitt 2.3, 5.5, 10.1

{\sc {\v C}{\'\i}{\v z}ek, J., Paldus, J., Ramgulam, U.W., Vinette,
F.} [1987], {\it Two-point Pad\'e approximants in electrochemical kinetic
currents}, Prog. Surface Sci. {\bf 25}, 17 - 39.
% Abschnitt 4.5

{\sc {\v C}{\'\i}{\v z}ek, J., {\v S}pirko, V., Bludsk\'{y}, O.} [1993],
{\it On the use of divergent series in vibrational spectroscopy. Two-
and three-dimensional oscillators}, J. Chem. Phys. {\bf 99}, 7331 -
7336.
% Abschnitt 4.5

{\sc {\v C}{\'\i}{\v z}ek, J., Vinette, F} [1988], {\it Symbolic
computation in quantum mechanics. Several simple examples}, Int. J.
Quantum Chem. Symp. {\bf 22}, 537 - 548.
% Abschnitt 10.4

{\sc {\v C}{\'\i}{\v z}ek, J., Vinette, F., Vrscay, E.R.} [1987], {\it
Renormalized inner projection, symbolic computation, and L\"owdin rational
approximants in explicit form}, Int. J. Quantum Chem. Symp. {\bf 21},
757 - 758.
% Abschnitt 10.4

{\sc {\v C}{\'\i}{\v z}ek, J., Vinette, F., Weniger, E.J.} [1991],
{\it Examples on the use of symbolic computation in physics and
chemistry: Applications of the inner projection technique and of a new
summation method for divergent series}, Int. J. Quantum Chem. Symp.
{\bf 25}, 209 - 223.
% Abschnitt 4.4, 4.5, 5.1, 5.4, 6.6, 6.7, 10.4

{\sc {\v C}{\'\i}{\v z}ek, J., Vinette, F., Weniger, E.J.} [1993], {\it
On the use of the symbolic language Maple in physics and chemistry:
Several examples}, in de Groot, R.A., Nadrchal, J. (Herausgeber), {\it
Pro\-ceed\-ings of the Fourth International Conference on Computational
Physics PHYSICS COMPUTING '92}, (World Scientific, Singapore), 31 - 44.
% Abschnitt 4.5, 5.1, 5.4, 6.6, 6.7, 10.4

{\sc {\v C}{\' \i}{\v z}ek, J., Vrscay, E.R.} [1982], {\it Large order
perturbation theory in the context of atomic and molecular physics --
Interdisciplinary aspects}, Int. J. Quantum Chem. {\bf 21}, 27 - 68.
% Abschnitt 2.2, 2.3, 4.1, 4.5, 10.1

{\sc {\v C}{\' \i}{\v z}ek, J., Vrscay, E.R.} [1984], {\it Continued
fractions and quantum-mechanical large-order perturbation theory: The
anharmonic operator revisited}, Phys. Rev. A {\bf 30}, 1550 - 1553.
% Abschnitt 4.1, 10.1

{\sc {\v C}{\' \i}{\v z}ek, J., Vrscay, E.R.} [1985], {\it Inner
projection theory with and without perturbation theory: The anharmonic
oscillator revisited and the quadratic Zeeman effect in ground-state
hydrogen}, Int. J. Quantum Chem. {\bf 28}, 665 - 686.
% Abschnitt 4.1

{\sc {\v C}{\' \i}{\v z}ek, J., Vrscay, E.R.} [1986], {\it Lower bounds
to ground state eigenvalues of the Schr\"{o}dinger equation via
optimized inner projection: Application to quartic and sextic anharmonic
oscillators}, Int. J. Quantum Chem. Symp. {\bf 20}, 65 - 72.
% Abschnitt 10.4

{\sc Clementi, E.} [1992], {\it Chemistry and computers: On research
aims from my preparatory period until early 1991}, Int. J. Quantum
Chem. {\bf 42}, 547 - 580.
% Abschnitt 1.1

{\sc Clementi, E., Corongiu, G.} [1989], {\it KGNMOL: A program for
molecular interactions}, in Clementi, E. (Herausgeber), {\it Modern
techniques in computational chemistry: MO\-TECC{\small${}^{\, TM}$}-89}
(Escom, Leiden), 243 - 306.
% Abschnitt 9.2, 9.3

{\sc Coester, F.} [1958], {\it Bound states of a many-particle system},
Nucl. Phys. {\bf 7}, 421 - 424.
% Abschnitt 1.2

{\sc Coester, F., K\"ummel, H.} [1960], {\it Short range correlations in
nuclear wave functions}, Nucl. Phys. {\bf 17}, 477 - 485.
% Abschnitt 1.2

{\sc Cohen, A.} [1991], {\it A Pad\'e approximant to the inverse Langevin
function}, Rheol. Acta {\bf 30}, 207 - 273.
% Abschnitt 4.5

{\sc Cohen, M., Kais, S.} [1984], {\it Scaling, renormalization and
accuracy of perturbation calculations}, Chem. Phys. Lett. {\bf 105}, 295
- 298.
% Abschnitt 10.4

{\sc Cohen, M., Kais, S.} [1986], {\it Rayleigh-Schr\"odinger perturbation
theory with a strong perturbation: Anharmonic oscillators}, J. Phys. A
{\bf 19}, 683 - 690.
% Abschnitt 10.4

{\sc Coleman, J.P.} [1976], {\it Iteration-variation methods and the
epsilon algorithm}, J. Phys. B {\bf 9}, 1079 - 1093.
% Abschnitt 4.5

{\sc Collins, J.C., Soper, D.E.} [1978], {\it Large order expansion in
perturbation theory}, Ann. Phys. (NY) {\bf 112}, 209 - 234.
% Abschnitt 10.1

{\sc Common, A.K.} [1968], {\it Pad\'e approximants and bounds to series
of Stieltjes}, J. Math. Phys. {\bf 9}, 32 - 38.
% Abschnitt 6.6, 6.7

{\sc Common, A.K.} [1969], {\it Properties of generalizations to Pad\'e
approximants}, J. Math. Phys. {\bf 10}, 1875 - 1880.
% Abschnitt 6.6

{\sc Condon, E.U., Odaba\c{s}i, H.} [1980], {\it Atomic Structure}
(Cambridge U. P., Cambridge).
% Abschnitt 8.1, 10.1

{\sc Condon, E.U., Shortley, G.H.} [1970], {\it The theory of atomic
spectra} (Cambridge U. P., Cambridge).
% Abschnitt 8.1, 10.1

{\sc Cook, D.M., Dubisch, R., Sowell, G., Tam, P., Donnelly, D.}
[1992a], {\it A comparison of several symbol-manipulating programs: Part
\Roemisch{1}}, Computers in Physics {\bf 6}, 411 - 420.
% Abschnitt 4.4

{\sc Cook, D.M., Dubisch, R., Sowell, G., Tam, P., Donnelly, D.}
[1992b], {\it A comparison of several symbol-manipulating programs: Part
\Roemisch{2}}, Computers in Physics {\bf 6}, 530 - 540.
% Abschnitt 4.4

{\sc Copson, E.T.} [1965], {\it Asymptotic expansions} (Cambridge U.
P., Cambridge).
% Abschnitt 2.2

{\sc Corbat\'{o}, F.J., Switendick, A.C.} [1963], {\it Integrals for
diatomic molecular calculations}, in Alder, B., Fernbach, S., Rotenberg,
M. (Herausgeber), {\it Methods in computational physics {\bf 2}. Quantum
mechanics} (Academic Press, New York), 155 - 179.
% Abschnitt 8.2

{\sc Cordellier, F.} [1979], {\it Sur la r\'egularit\'e des proc\'ed\'es
$\delta^2$ d'Aitken et W de Lubkin}, in Wuytack, L. (Herausgeber), {\it
Pad\'e approximation and its applications} (Springer-Verlag, Berlin),
20 - 35.
% Abschnitt 4.5

{\sc Coulson, C.A.} [1937], {\it The evaluation of certain integrals
occurring in studies of molecular structure}, Proc. Cambridge Phil. Soc.
{\bf 33}, 104 - 110.
% Abschnitt 8.2

{\sc Coulson, C.A.} [1960], {\it Present state of molecular structure
calculations}, Rev. Mod. Phys. {\bf 32}, 170 - 177.
% Abschnitt 1.2

{\sc Cowan, R.D.} [1981], {\it The theory of atomic structure and
spectra} (University of California Press, Berkeley).
% Abschnitt 10.1

{\sc Craw, J.S., Reimers, J.R., Bacskay, G.B., Wong, A.T., Hush, N.S.}
[1992a], {\it Solitons in finite- and infinite-length negative-defect
trans-polyacetylene and the corresponding Brooker (polymethinecyanine)
cations. \Roemisch{1}. Geometry}, Chem. Phys. {\bf 167}, 77 - 99.
% Abschnitt 11.2

{\sc Craw, J.S., Reimers, J.R., Bacskay, G.B., Wong, A.T., Hush, N.S.}
[1992b], {\it Solitons in finite- and infinite-length negative-defect
trans-polyacetylene and the corresponding Brooker (polymethinecyanine)
cations. \Roemisch{2}. Charge density waves}, Chem. Phys. {\bf 167}, 101
- 109.
% Abschnitt 11.2

{\sc Cui, C.X., Kertesz, M.} [1990], {\it Quantum-mechanical oligomer
approach for the calculation of vibrational spectra of polymers}, J.
Chem. Phys. {\bf 93}, 5257 - 5266.
% Abschnitt 11.1

{\sc Cui, C.X., Kertesz, M., Jiang, Y.} [1990], {\it Extraction of
polymer properties from oligomer calculations}, J. Phys. Chem. {\bf
94}, 5172 - 5179.
% Abschnitt 11.1, 11.6

{\sc Cullum, J.K., Willoughby, R.A.} [1985a], {\it Lanczos algorithms
for large symmetric eigenvalue computations \Roemisch{1}. Theory}
(Birkh\"auser, Basel).
% Abschnitt 1.2

{\sc Cullum, J.K., Willoughby, R.A.} [1985b], {\it Lanczos algorithms
for large symmetric eigenvalue computations \Roemisch{2}. Programs}
(Birkh\"auser, Basel).
% Abschnitt 1.2

{\sc Cullum, J.K., Willoughby, R.A.} [1985c], {\it A survey of Lanczos
procedures for very large real `symmetric' eigenvalue problems}, J.
Comput. Appl. Math. {\bf 12} \& {\bf 13}, 37 - 60.
% Abschnitt 1.2

{\sc Cullum, J.K., Willoughby, R.A.} (Herausgeber) [1986], {\it Large
scale eigenvalue problems} (North-Holland, Amsterdam).
% Abschnitt 1.2

{\sc Cuyt, A.} [1984], {\it Pad\'e approximants for operators: Theory and
applications} (Springer-Verlag, Berlin).
% Abschnitt 4.1, 13

{\sc Cuyt, A.} (Herausgeberin) [1988], {\it Nonlinear numerical
methods and rational approximation} (Reidel, Dordrecht).
% Abschnitt 2.4, 4.1

{\sc Cuyt, A.} [1989/90], {\it Old and new multidimensional convergence
accelerators}, Appl. Numer. Math. {\bf 6}, 169 - 185.
% Abschnitt 13

{\sc Cuyt, A., Wuytack, L.} [1987], {\it Nonlinear methods in numerical
analysis} (North-Holland, Amsterdam).
% Abschnitt 2.4, 4.1, 6.1, 11.5

{\sc \'{C}wiok, T., Jeziorski, B., Ko{\l}os, W., Moszynski, R.,
Rychlewski, J., Szalewicz, K.} [1992], {\it Convergence properties and
large-order behavior of the polarization expansion for the interaction
energy of hydrogen atoms}, Chem. Phys. Lett. {\bf 195}, 67 - 76.
% Abschnitt 10.1

{\sc Cycon, H.L., Froese, R.G., Kirsch, W., Simon, B.} [1987], {\it
Schr\"odinger operators} (Springer-Verlag, Berlin).
% Abschnitt 8.2, 9.1

{\sc Dalgarno, A.} [1961], {\it Stationary perturbation theory}, in
Bates, D.R. (Herausgeber), {\it Quantum theory \Roemisch{1}. Elements}
(Academic Press, New York), 170 - 209.
% Abschnitt 10.1

{\sc Damburg, R.J., Propin, R.Kh.} [1983], {\it Large-order expansion of
the imaginary part of the Stark effect in hydrogen and of the anharmonic
oscillator with negative anharmonicity}, J. Phys. B {\bf 16}, 3741 -
3745.
% Abschnitt 10.1

{\sc Damburg, R.J., Propin, R.Kh., Graffi, S., Grecchi, V., Harrell
\Roemisch{2}, E.M., {\v C}{\' \i}{\v z}ek, J., Paldus, J., Silverstone,
H.J.} [1984], {\it $1/R$-Expansion for $H_2^{+}$: Analyticity,
summability, asymptotics, and calculations of exponentially small
terms}, Phys. Rev. Lett. {\bf 52}, 1112 - 1115.
% Abschnitt 2.3, 10.1

{\sc Damburg, R.J., Propin, R.Kh., Martyshenko, V.} [1984], {\it Large
order perturbation theory for the $O (2)$ anharmonic oscillator with
negative anharmonicity and for the double-well potential}, J. Phys. A
{\bf 17}, 3493 - 3503.
% Abschnitt 10.1

{\sc Davidson, E.R.} [1976], {\it Reduced density matrices in quantum
chemistry} (Academic Press, New York).
% Abschnitt 10.1

{\sc Davidson, E.R.} [1983], {\it Matrix eigenvector methods}, in
Diercksen, G.H.F., Wilson, S. (Herausgeber), {\it Methods in
computational molecular physics} (Reidel, Dordrecht), 95 - 113.
% Abschnitt 1.2

{\sc Davis, H.T.} [1962], {\it The summation of series} (The Principia
Press of Trinity University, San Antonio, Texas). % Abschnitt 2.1

{\sc Davis, P.J., Rabinowitz, P.} [1984], {\it Methods of numerical
integration} (Academic Press, Orlando).
% Abschnitt 2.4, 8.2, 8.3

{\sc Daudel, R., Leroy, G., Peeters, D., Sana, M.} [1983], {\it Quantum
chemistry} (Wiley, Chichester).
% Abschnitt 8.1, 9.2, 9.3

{\sc de Bruijn, N.G.} [1981], {\it Asymptotic methods in analysis}
(Dover, New York).
% Abschnitt 2.2

{\sc de Bruin, M.G., Van Rossum, H.} (Herausgeber) [1981], {\it Pad\'e
approximation and its applications, Amsterdam 1980} (Springer-Verlag,
Berlin).
% Abschnitt 2.4, 4.1

{\sc Defranceschi, M., Delhalle, J.} [1986], {\it Numerical solution of
the Hartree-Fock equations for quasi one-dimensional systems:
Prototypical calculations on the $(-\!H\!-)_x$ chain}, Phys. Rev. D {\bf
34}, 5862 - 5873.
% Abschnitt 11.1

{\sc Delahaye, J.P.} [1988], {\it Sequence transformations}
(Springer-Verlag, Berlin).
% Abschnitt 2.2, 2.4, 6.1

{\sc Delahaye, J.P., Germain-Bonne, B.} [1980], {\it R\'esultats
n\'egatifs en acc\'el\'eration de la convergence}, Numer. Math. {\bf 35},
443 - 457.
% Abschnitt 4.5, 11.4

{\sc Delahaye, J.P., Germain-Bonne, B.} [1982], {\it The set of
logarithmically convergent sequences cannot be accelerated}, SIAM J.
Numer. Anal. {\bf 19}, 840 - 844.
% Abschnitt 4.5, 11.4

{\sc Delhalle, J., Bodart, V.P., Dory, M., Andr\'e, J.M., Zyss, J.}
[1986], {\it Equilibrium geometry and longitudinal electric
polarizability of allene, diallene, and triallene: An {\rm ab initio}
study}, Int. J. Quantum Chem. Symp. {\bf 19}, 313 - 321.
% Abschnitt 11.1

{\sc Delhalle, J., Calais, J.-L.} [1986], {\it Convergence of direct
space exchange lattice sums in polymer band structure calculations}, J.
Chem. Phys. {\bf 85}, 5286 - 5298.
% Abschnitt 11.1

{\sc Delhalle, J., Calais, J.-L.} [1987], {\it Pathological aspects of
restricted Hartree-Fock band calculations for metallic chains}, Int. J.
Quantum Chem. Symp. {\bf 21}, 115 - 129.
% Abschnitt 11.1

{\sc Delhalle, J., Delvaux, M.H., Fripiat, J.G., Andr\'e, J.M., Calais,
J.-L.} [1988], {\it RHF energy band shapes for metallic chains:
Dependence on the summation of exchange contributions. An illustration
on the linear chain of hydrogen atoms}, J. Chem. Phys. {\bf 88}, 3141 -
3146.
% Abschnitt 11.1

{\sc Delhalle, J., Fripiat, J.G., Harris, F.E.} [1984], {\it Evaluation
of computational aspects of a modified CS-LCAO-SCF-CO strategy for
electronic structure calculations of extended model chains}, Int. J.
Quantum Chem. Symp. {\bf 18}, 141 - 152.
% Abschnitt 11.1

{\sc Delhalle, J., Fripiat, J.G., Piela, L.} [1980], {\it On the use of
Laplace transform to evaluate one-dimensional lattice summations in
quantum calculations of model polymers}, Int. J. Quantum Chem. Symp.
{\bf 14}, 431 - 442.
% Abschnitt 11.1

{\sc Delhalle, J., Harris, F.E.} [1985], {\it Fourier-representation
method for electronic structure of chainlike systems: Restricted
Hartree-Fock equations and applications to the $(H)_x$ chain in a basis
of Gaussian functions}, Phys. Rev. B {\bf 31}, 6755 - 6765.
% Abschnitt 11.1

{\sc Delhalle, J., Piela, L., Br\'edas, J.-L., Andr\'e, J.-M.} [1980], {\it
Multipole expansion in tight-binding Hartree-Fock calculations for
infinite model polymers}, Phys. Rev. B {\bf 22}, 6254 - 6267.
% Abschnitt 11.1

{\sc Del Re, G., Ladik, J., Bicz\'{o}} [1967], {\it Self-consistent
field tight-binding treatment of polymers. \Roemisch{1}. Infinite
three-dimensional case}, Phys. Rev. {\bf 155}, 997 - 1003.
% Abschnitt 11.1

{\sc DePristo A.E., Kress, J.D.} [1987], {\it Kinetic-energy
functionals via Pad\'e approximations}, Phys. Rev. A {\bf 35}, 438 - 441.
% Abschnitt 4.5

{\sc Dias, M., Chaba, A.N.} [1986], {\it Equivalence between the methods
involving Fourier series and Poisson's summation formula, and the class
of lattice sums in arbitrary dimensions}, Hadronic J. {\bf 9}, 141 - 146.
% Abschnitt 11.1

{\sc Dietz, K., Schmidt, C., Warken, M., He{\ss}, B.A.} [1993a], {\it On the
acceleration of convergence of many-body perturbation theory:
\Roemisch{1}. General theory}, J. Phys. B {\bf 26}, 1885 - 1896.
% Abschnitt 8.1

{\sc Dietz, K., Schmidt, C., Warken, M., He{\ss}, B.A.} [1993b], {\it On the
acceleration of convergence of many-body perturbation theory:
\Roemisch{2}. Benchmark checks for small systems}, J. Phys. B {\bf 26},
1897 - 1914.
% Abschnitt 8.1

{\sc Dietz, K., Schmidt, C., Warken, M., He{\ss}, B.A.} [1993c], {\it The
acceleration of convergence of many-body perturbation theory:
Unlinked-graph shift in M{\o}ller-Plesset perturbation theory}, Chem.
Phys. Lett. {\bf 207}, 281 - 286.
% Abschnitt 8.1

{\sc Dingle, R.B.} [1973], {\it Asymptotic expansions: Their
derivation and interpretation} (Academic Press, London).
% Abschnitt 2.2, 5.7

{\sc Dirac, P.A.M.} [1929], {\it Quantum mechanics of many-electron
systems}, Proc. Roy. Soc. London A {\bf 123}, 714 - 733.
% Abschnitt 1.1

{\sc Distefano, G., Jones, D., Guerra, M., Favaretto, L., Modelli, A.,
Mengoli, G.} [1991], {\it Determination of the electronic structure of
oligofurans and extrapolation to polyfuran}, J. Phys. Chem. {\bf 95},
9746 - 9753.
% Abschnitt 11.1

{\sc Dmitrieva, I.K., Plindov, G.I.} [1980a], {\it Perturbation theory
and hypervirial theorems for the anharmonic oscillator}, Phys. Lett. A
{\bf 79}, 47 - 50.
% Abschnitt 10.4

{\sc Dmitrieva, I.K., Plindov, G.I.} [1980b], {\it Rearranged
perturbation theory for the Schr\"odinger equation}, Phys. Scr. {\bf 22},
386 - 388.
% Abschnitt 10.4

{\sc D\"orrie, H.} [1951], {\it Unendliche Reihen} (Oldenbourg,
M\"unchen).
% Abschnitt 2.1

{\sc Doetsch, G.} [1955], {\it Handbuch der Laplace-Transformation. Band
\Roemisch{2}: Anwendungen der Laplace-Transformation} (Birkh\"auser,
Basel).
% Abschnitt 5.3

{\sc Dolgov, A.D., Popov, V.S.} [1978], {\it Modified perturbation
theories for an anharmonic oscillator}, Phys. Lett. B {\bf 79}, 403 -
405.
% Abschnitt 2.3

{\sc Dovesi, R.} [1984], {\it Ab initio Hartree-Fock approach to the
study of polymers: Applications to polyacetylene}, Int. J. Quantum Chem.
{\bf 24}, 197 - 212.
% Abschnitt 11.1, 11.2

{\sc Draux, A., van Ingelandt, P.} [1987], {\it Polyn\^omes orthogonaux
et approximants de Pad\'e} (\'Editions Technip, Paris).
% Abschnitt 2.4, 4.1, 4.4

{\sc Drummond, J.E.} [1972], {\it A formula for accelerating the
convergence of a general series}, Bull. Austral. Math. Soc. {\bf 6}, 69
- 74.
% Abschnitt 6.7

{\sc Drummond, J.E.} [1976], {\it Summing a common type of slowly
convergent series of positive terms}, J. Austral. Math. Soc. {\bf B 19},
416 - 421.
% Abschnitt 4.5, 8.4, 11.4

{\sc Drummond, J.E.} [1981], {\it The anharmonic oscillator:
perturbation series for cubic and quartic energy distortion}, J. Phys. A
{\bf 14}, 1651 - 1661.
% Abschnitt 3.3

{\sc Drummond, J.E.} [1984], {\it Convergence speeding, convergence,
and summability}, J. Comput. Appl. Math. {\bf 11}, 145 - 159.
% Abschnitt 4.5

{\sc Dunster, T.M., Lutz, D.A.} [1991], {\it Convergent factorial
series expansion for Bessel functions}, SIAM J. Math. Anal. {\bf 22},
1156 - 1172.
% Abschnitt 5.3, 7.3

{\sc Dyson, D.J.} [1952], {\it Divergence of perturbation theory in
quantum electrodynamics}, Phys. Rev. {\bf 85}, 32- 33.
% Abschnitt 10.1

{\sc Elbaz, E.} [1985], {\it Algebre de Racah et analyse vectorielle
graphiques} (Ellipses, Paris).
% Abschnitt 1.2

{\sc El Baz, E., Castel, B.} [1972], {\it Graphical methods of spin
algebras} (Marcel Dekker, New York).
% Abschnitt 1.2

{\sc Englefield, M.J.} [1972], {\it Group theory and the Coulomb
problem} (Wiley, New York).
% Abschnitt 1.2

{\sc Erd\'elyi, A.} [1956], {\it Asymptotic expansions} (Dover, New
York).
% Abschnitt 2.1

{\sc Erd\'elyi, A., Magnus, W., Oberhettinger, F., Tricomi, F.G.}
[1953], {\it Higher transcendental functions \Roemisch{2}}
(McGraw-Hill, New York).
% Abschnitt 5.2

{\sc Errea, L.F., M\'endez, L., Riera, A.} [1984], {\it Calculation of
the auxiliary function $F_m (z)$}, Comput. Phys. Commun. {\bf 31}, 47 -
52.
% Abschnitt 9.3

{\sc Euler, L.} [1755], {\it Institutiones calculi differentialis cum
eius usu in analysi finitorum ac doctrina serium. Pars II.1. De
transformatione serium} (Academia Imperialis Scientiarum Petropolitana).
Dieses Buch wurde 1913 als Band \Roemisch {10} von {\it Leonardi Euleri
Opera Omnia, Seria Prima} (Teubner, Leipzig und Berlin) nachgedruckt.
% Abschnitt 2.2

{\sc Evangelisti, S.} [1990], {\it Extrapolation to the infinite length
limit of polymer energies. \Roemisch{1}. H\"uckel approximations}, J.
Chem. Phys. {\bf 92}, 4383 - 4386.
% Abschnitt 11.1

{\sc Feenberg, E.} [1958], {\it Analysis of the Schr\"odinger energy
series}, Ann. Phys. (NY) {\bf 3}, 292 - 303.
% Abschnitt 4.1

{\sc Fern\'{a}ndez, F.M.} [1992a], {\it Large-order perturbation theory
without a wavefunction for the LoSurdo-Stark effect in hydrogenic
atoms}, J. Phys. A {\bf 25}, 495 - 501.
% Abschnitt 10.1

{\sc Fern\'{a}ndez, F.M.} [1992b], {\it Strong coupling expansion for
anharmonic oscillators and perturbed Coulomb potentials}, Phys. Lett. A
{\bf 166}, 173- 176.
% Abschnitt 4.5, 10.3, 10.8

{\sc Fern\'{a}ndez, F.M., Castro, E.A.} [1983], {\it Scaling-variational
treatment of anharmonic oscillators}, Phys. Rev. A {\bf 27}, 663 - 669.
% Abschnitt 10.4

{\sc Fern\'{a}ndez, F.M., Castro, E.A.} [1992], {\it Variation principle
and perturbation theory}, in Fraga, S. (Herausgeber), {\it Computational
chemistry: Structure, interactions and reactivity. Part A} (Elsevier,
Amsterdam), 3 - 25.
% Abschnitt 10.1

{\sc Fern\'{a}ndez, F.M., Ma, Q., Tipping, R.H.} [1989], {\it
Eigenvalues of the Schr\"odinger equation via the Riccati-Pad\'e method},
Phys. Rev. A {\bf 40}, 6149 - 6153.
% Abschnitt 4.5

{\sc Fern\'{a}ndez, F.M., Mes\'{o}n, A.M., Castro, E.A.} [1985], {\it A
simple iterative solution of the Schr\"odinger equation in matrix
representation form}, J. Phys. A {\bf 18}, 1389 - 1398.
% Abschnitt 10.4

{\sc Fern\'andez Rico, J.,} [1993], {\it Long-range multicenter integrals
with Slater functions: Gauss transform-based methods}, J. Comput. Chem.
{\bf 14}, 1203 - 1211.
% Abschnitt 8.2

{\sc Fern\'andez Rico, J., L\'opez, R., Paniagua, M., Fern\'andez-Alonso,
J.I.} [1986], {\it Atomic partitioning of two-center potentials for
Slater basis}, Int. J. Quantum Chem. {\bf 29}, 1155 - 1164.
% Abschnitt 8.2

{\sc Fern\'andez Rico, J., L\'opez, R., Paniagua, M., Ram\'{\i}rez, G.}
[1991], {\it Calculation of two-center one-electron molecular integrals
with STOs}, Comput. Phys. Commun. {\bf 64}, 329 - 342.
% Abschnitt 8.2

{\sc Fern\'andez Rico, J., L\'opez, R., Ram\'{\i}rez, G.} [1989a], {\it
Molecular integrals with Slater basis. \Roemisch{1}. General approach},
J. Chem. Phys. {\bf 91}, 4204 - 4212.
% Abschnitt 8.2

{\sc Fern\'andez Rico, J., L\'opez, R., Ram\'{\i}rez, G.} [1989b], {\it
Molecular integrals with Slater basis. \Roemisch{2}. Fast computational
algorithms}, J. Chem. Phys. {\bf 91}, 4213 - 4222.
% Abschnitt 8.2

{\sc Fern\'andez Rico, J., L\'opez, R., Ram\'{\i}rez, G.} [1991], {\it
Molecular integrals with Slater basis. \Roemisch{3}. Three center
nuclear attraction integrals}, J. Chem. Phys. {\bf 94}, 5032 - 5039.
% Abschnitt 8.2

{\sc Fern\'andez Rico, J., L\'opez, R., Ram\'{\i}rez, G.} [1992a], {\it
Molecular integrals with Slater basis. \Roemisch{4}. Ellipsoidal
coordinate methods for three-center nuclear attraction integrals}, J.
Chem. Phys. {\bf 97}, 7613 - 7622.
% Abschnitt 8.2

{\sc Fern\'andez Rico, J., L\'opez, R., Ram\'{\i}rez, G.} [1992b], {\it
Molecular integrals with Slater func\-tions: One-center expansion
methods}, in Fraga, S. (Herausgeber), {\it Computational chemistry:
Structure, interactions and reactivity. Part A} (Elsevier, Amsterdam),
241 - 272.
% Abschnitt 8.2

{\sc Fern\'andez Rico, J., L\'opez, R., Ram\'{\i}rez, G., Fern\'andez-Alonso,
J.I.} [1993], {\it Auxiliary functions for Slater molecular integrals},
Theor. Chim. Acta {\bf 85}, 101 - 107.
% Abschnitt 8.2

{\sc Ferrar, W.L.} [1938], {\it A text-book on convergence} (Oxford U.
P., Oxford).
% Abschnitt 2.1

{\sc Fessler, T., Ford, W.F., Smith, D.A.} [1983a], {\it HURRY: An
acceleration algorithm for scalar sequences and series}, ACM Trans.
Math. Software {\bf 9 }, 346 - 354.
% Abschnitt 5.2,

{\sc Fessler, T., Ford, W.F., Smith, D.A.} [1983b], {\it ALGORITHM
602 HURRY: An acceleration algorithm for scalar sequences and series},
ACM Trans. Math. Software {\bf 9 }, 355 - 357.
% Abschnitt 5.2

{\sc Fetter, A., Walecka, J.D.} [1971], {\it Quantum theory of
many-particle systems} (McGraw-Hill, New York).
% Abschnitt 10.1

{\sc Feynman, R.P.} [1949], {\it Space-time approach to quantum
electrodynamics}, Phys. Rev. {\bf 76}, 769 - 789.
% Feynman-Graphen
% Abschnitt 1.2, 8.3

{\sc Fichtenholz, G.M.} [1970a], {\it Infinite series: Rudiments}
(Gordon and Breach, New York).
% Abschnitt 2.1

{\sc Fichtenholz, G.M.} [1970b], {\it Infinite series: Ramifications}
(Gordon and Breach, New York).
% Abschnitt 2.1

{\sc Fichtenholz, G.M.} [1970c], {\it Functional series}
(Gordon and Breach, New York).
% Abschnitt 2.1

{\sc Fieck, G.} [1980], {\it The multi-centre integrals of derivative,
spherical GTOs}, Theor. Chim. Acta {\bf 54}, 323 - 332.
% Abschnitt 8.2

{\sc Filter, E., Steinborn, E.O.} [1978a], {\it The three-dimensional
convolution of reduced Bessel functions and other functions of physical
interest}, J. Math. Phys. {\bf 19}, 79 - 84.
% Abschnitt 8.2, 8.4, 9.1

{\sc Filter, E., Steinborn, E.O.} [1978b], {\it Extremely compact
formulas for molecular one-electron integrals and Coulomb integrals over
Slater-type atomic orbitals}, Phys. Rev. A {\bf 18}, 1 - 11.
% Abschnitt 8.2, 8.3,

{\sc Filter, E., Steinborn, E.O.} [1980], {\it A matrix representation
of the translation operator with respect to a basis of exponentially
declining functions}, J. Math. Phys. {\bf 21}, 2725 - 2736.
% Abschnitt 8.2, 9.1

{\sc Fischer, P., Defranceschi, M., Delhalle, J.} [1992], {\it Molecular
Hartree-Fock equations for iteration-variation calculations in momentum
space}, Numer. Math. {\bf 63}, 67 - 82.
% Abschnitt 8.1

{\sc Fleischer, J.} [1972], {\it Analytic continuation of scattering
amplitudes and Pad\'e approximants}, Nucl. Phys. B {\bf 37}, 59 - 62.
% Abschnitt 12

{\sc Fleischer, J.} [1973a], {\it Generalizations of Pad\'e approximants},
in Graves-Morris, P.R. (Herausgeber) [1973], {\it Pad\'e approximants}
(The Institute of Physics, London), 126 - 131.
% Abschnitt 12

{\sc Fleischer, J.} [1973b], {\it Nonlinear Pad\'e approximants for
Legendre series}, J. Math. Phys. {\bf 14}, 246 - 248.
% Abschnitt 12

{\sc Fock, V.} [1930], {\it N\"aherungsmethoden zur L\"osung des
quantenmechanischen Mehrk\"orperproblems}, Z. Physik {\bf 61}, 126 - 148.
% Abschnitt 1.1, 8.1

{\sc Ford, W.B.} [1960], {\it Studies on divergent series and
summability and the asymptotic developments of functions defined by
Maclaurin series} (Chelsea, New York).
% Abschnitt 2.2, 2.4

{\sc Ford, W.F., Sidi, A.} [1987], {\it An algorithm for a
generalization of the Richardson extrapolation process}, SIAM J. Numer.
Anal. {\bf 24}, 1212 - 1232.
% Abschnitt 3.2

{\sc Fortunelli, A., Carrravetta, V.} [1992], {\it Alternative basis
functions for the molecular continuum. \Roemisch{2}. Integrals with
higher-order functions}, Phys. Rev. A {\bf 45}, 4438 - 4451.
% Abschnitt 8.2

{\sc Fortunelli, A., Salvetti, O.} [1993], {\it Recurrence relations for
the evaluation of electron repulsion integrals over spherical Gaussian
functions}, Int. J. Quantum Chem. {\bf 48}, 257 - 265.
% Abschnitt 8.2

{\sc Frachebourg, L., Henkel, M.} [1991], {\it Transfer matrix spectrum
of two-dimensional layered Ising models}, J. Phys. A {\bf 24}, 5121 -
5136.
% Abschnitt 11.6

{\sc Franceschini, V., Grecchi, V., Silverstone, H.J.} [1985], {\it
Complex energies from real perturbation series for the LoSurdo-Stark
effect in hydrogen by Borel-Pad\'e approximants}, Phys. Rev. A {\bf 32},
1338 - 1340.
% Abschnitt 2.3

{\sc Freed, K.F.} [1971], {\it Many-body theories of the electronic
structure of atoms and molecules}, Annu. Rev. Phys. Chem. {\bf 22}, 313
- 346.
% Abschnitt 8.2

{\sc Fried, L.E., Ezra, G.S.} [1989], {\it Avoided crossings and
resummation of nearly resonant molecular vibrations: Reconstruction of
an effective secular equation}, J. Chem. Phys. {\bf 90}, 6378 - 6390.
% Abschnitt 4.5

{\sc Friedrichs, K.O.} [1965], {\it Perturbation of spectra in Hilbert
space} (American Mathematical Society, Providence, Rhode Island).
% Abschnitt 10.1

{\sc Fripiat, J.G., Andr\'e, J.M., Delhalle, J., Calais, J.L.} [1991],
{\it An {\rm ab initio} computational scheme for polymeric chains with
fully converged Coulomb and exchange lattice sums}, Int. J. Quantum
Chem. Symp. {\bf 25}, 603 - 618.
% Abschnitt 11.1

{\sc Fripiat, J.G., Delhalle, J.} [1979], {\it Fourier representations
of the Coulombic contributions to polymer chains}, J. Comput. Phys. {\bf
33}, 425 - 431.
% Abschnitt 1.1

{\sc Fripiat, J.G., Delhalle, J., Andr\'e, J.M., Calais, J.L.} [1989],
{\it Potentials of an alternative scheme for ab initio polymer band
structure calculations. Illustration on the chain of hydrogen atoms},
Int. J. Quantum. Chem. {\bf 34}, 341 - 351.
% Abschnitt 11.1

{\sc Froese Fischer, C.} [1977], {\it The Hartree-Fock method for atoms}
(Wiley, New York).
% Abschnitt 8.1

{\sc Fromm, D.M., Hill, R.N.} [1987], {\it Analytic evaluation of
three-electron integrals}, Phys. Rev. A {\bf 36}, 1013 - 1044.
% Abschnitt 8.4

{\sc Fujimura, N., Matsuoka, O.} [1992], {\it Molecular integrals of
Breit interaction over Laguerre Gaussian-type functions}, Int. J.
Quantum Chem. {\bf 42}, 751 - 759.
% Abschnitt 8.2

{\sc Fulde, P.} [1991], {\it Electron correlation in atoms and
molecules} (Springer-Verlag, Berlin).
% Abschnitt 8.2, 11.2

{\sc Gabutti, B., Sacripante, L.} [1991], {\it Numerical inversion of
the Mellin transform by accelerated series of Laguerre polynomials}, J.
Comput. Appl. Math. {\bf 34}, 191 - 200.
% Abschnitt 13

{\sc Gailitis, M., Silverstone, H.J.} [1988], {\it On the use of
asymptotic expansions}, Theor. Chim. Acta {\bf 73}, 105 - 114.
% Abschnitt 2.3, 7.2

{\sc Garbow, B.S., Boyle, J.M., Dongarra, J.J., Moler, C.B.} [1977],
{\it Matrix eigensystem routines -- EISPACK guide extension}
(Springer-Verlag, Berlin).
% Abschnitt 1.2

{\sc Gargantini, I., Henrici, P.} [1967], {\it A continued fraction
algorithm for the computation of higher transcendental functions in the
complex plane}, Math. Comput. {\bf 21}, 18 - 29.
% Abschnitt 7.2

{\sc Garibotti, C.R., Grinstein, F.F.} [1978a], {\it Summation of
partial wave expansions in the scattering by long range potentials.
\Roemisch{1}}, J. Math. Phys. {\bf 19}, 821 - 829.
% Abschnitt 13

{\sc Garibotti, C.R., Grinstein, F.F.} [1978b], {\it Punctual Pad\'e
approximants as a regularization procedure for divergent and oscillatory
partial wave expansions of the scattering amplitude}, J. Math. Phys.
{\bf 19}, 2405 - 2409.
% Abschnitt 13

{\sc Garibotti, C.R., Grinstein, F.F.} [1979], {\it Summation of partial
wave expansions in the scattering by long range potentials.
\Roemisch{2}. Numerical applications}, J. Math. Phys. {\bf 20}, 141 -
147.
% Abschnitt 13

{\sc Garibotti, C.R., Grinstein, F.F., Miraglia, J.E.} [1980], {\it
Calculation of the non forward scattering amplitude for long range
potentials}, Z. Physik A {\bf 297}, 189 - 197.
% Abschnitt 13

{\sc Gaunt, D.S., Guttmann, A.J.} [1974], {\it Asymptotic analysis of
coefficients}, in Domb, C., Green, M.S. (Herausgeber), {\it Phase
transitions and critical phenomena\/} {\bf 3} (Academic Press,
London), 181 - 243.
% Abschnitt 2.4

{\sc Gaunt, J.A.} [1929], {\it The triplets of helium}, Phil. Trans.
Roy. Soc. A {\bf 228}, 151 - 196.
% Abschnitt 8.2

{\sc Gautschi, W.} [1967], {\it Computational aspects of three-term
recurrence relations}, SIAM Rev. {\bf 9}, 24 - 82.
% Abschnitt 7.1

{\sc Gautschi, W.} [1975], {\it Computational methods in special
functions -- A survey}, in Askey, R.A. (Herausgeber), {\it Theory and
application of special functions} (Academic Press, New York), 1 - 98.
% Abschnitt 7.1

{\sc Geddes, K.} [1979], {\it Symbolic computation of Pad\'e
approximants}, ACM Trans. Math. Software {\bf 5}, 218 - 233.
% Abschnitt 4.4

{\sc Geertsen, J., Eriksen, S., Oddershede, J.} [1991], {\it Some
aspects of the coupled cluster based polarization propagator method},
Adv. Quantum Chem. {\bf 22}, 167 - 209.
% Abschnitt 8.2

{\sc Geller, M.} [1962], {\it Two-center, nonintegral, Slater-orbital
calculations: Integral formulation and application to the Hydrogen
molecule-ion}, J. Chem. Phys {\bf 36}, 2424 - 2428.
% Abschnitt 8.2

{\sc Geller, M.} [1963a], {\it Two-center integrals over solid spherical
harmonics}, J. Chem. Phys {\bf 39}, 84 - 89.
% Abschnitt 8.2

{\sc Geller, M.} [1963b], {\it Two-electron, one- and two-center
integrals}, J. Chem. Phys {\bf 39}, 853 - 854.
% Abschnitt 8.2

{\sc Geller, M.} [1964a], {\it Zero-field splitting, one- and two-center
Coulomb-type integrals}, J. Chem. Phys {\bf 40}, 2309 - 3225.
% Abschnitt 8.2

{\sc Geller, M.} [1964b], {\it Two-center Coulomb integrals}, J. Chem.
Phys {\bf 41}, 4006 - 4007.
% Abschnitt 8.2

{\sc Georgi, H.} [1982], {\it Lie algebras in particle physics}
(Benjamin/Cummings, Reading, Mass.).
% Abschnitt 1.2

{\sc Gerck, E., d'Olivera, A.B.} [1980], {\it Continued fraction
calculation of the eigenvalues of tridiagonal matrices arising from the
Schr\"odinger equation}, J. Comput. Appl. Math. {\bf 6}, 81 - 82.
% Abschnitt 4.1

{\sc Germain-Bonne, B.} [1973], {\it Transformations de suites}, Rev.
Fran\c{c}aise Automat. Informat. Rech. Operat. {\bf 7} (R-1), 84 - 90.
% Abschnitt 4.5, 6.2, 6.3, 6.4

{\sc Gianolio, L., Pavani, R., Clementi, E.} [1978], {\it A new
algorithm for obtaining contracted basis set from Gaussian type
function}, Gazz. Chim. Italiana {\bf 108}, 181 - 205.
% Abschnitt 11.3

{\sc Gibson, W.M., Pollard, B.R.} [1976], {\it Symmetry principles in
elementary particle physics} (Cambridge  U. P., Cambridge).
% Abschnitt 1.2

{\sc Gilewicz, J.} [1978], {\it Approximants de Pad\'e} (Springer-Verlag,
Berlin).
% Abschnitt 2.4, 4.1

{\sc Gilewicz, J., Pindor, M., Siemaszko, W.} (Herausgeber) [1985],
{\it Rational approximation and its applications in mathematics and
physics} (Springer-Verlag, Berlin).
% Abschnitt 2.4, 4.1

{\sc Gill, P.W.M., Johnson, B.G., Pople, J.A.} [1991], {\it
Two-electron repulsion integrals over Gaussian $s$ functions}, Int. J.
Quantum Chem. {\bf 40}, 745 - 752.
% Abschnitt 9.3

{\sc Glaeske, H.-J., Reinhold, J., Volkmer, P.} [1987], {\it
Quantenchemie Band 5. Ausgew\"ahlte mathematische Methoden der Chemie}
(VEB Deutscher Verlag der Wissenschaften).
% Abschnitt 9.2, 9.3

{\sc Glasser, M.L.} [1988], {\it Electrostatic sums for polymer chains},
Theor. Chim. Acta {\bf 73}, 229 - 232.
% Abschnitt 11.1

{\sc Goldstein, M., Thaler, R. M.} [1959], {\it Recurrence techniques
for the calculation of Bessel functions}, Math. Tables Aids Comput. {\bf
13}, 102 - 108.
% Abschnitt 7.1

{\sc Golub, G.H., Van Loan, C.F.} [1983], {\it Matrix computations} (The
Johns Hopkins University Press, Baltimore).
% Abschnitt 1.2

{\sc Goodson, D.Z., Herschbach, D.R.} [1992], {\it Summation methods
for dimensional perturbation theory}, Phys. Rev. A {\bf 46}, 5428 -
5436.
% Abschnitt 1.3, 4.5

{\sc Goodson, D.Z., L\'{o}pez-Cabrera, M.} [1993], {\it Singularity
analysis and summation of $1/D$ expansions}, in Herschbach, D.R., Avery,
J., Goscinski, O. (Herausgeber), {\it Dimensional scaling in chemical
physics} (Kluwer, Dordrecht), 275 - 314.
% Abschnitt 1.3, 4.5

{\sc Goscinski, O.} [1967], {\it Continued fractions and upper and lower
bounds in the Brillouin-Wigner perturbation scheme}, Int. J. Quantum
Chem. {\bf 1}, 769 - 780.
% Abschnitt 4.1

{\sc Gottschalk, J.E., Abbott, P.C., Maslen, E.N.} [1987], {\it
Coordinate systems and analytic expansions for three-body atomic
wavefunctions: \Roemisch{2}. Closed form wavefunction to second order in
$r$}, J. Phys. A {\bf 20}, 2077 - 2104.
% Abschnitt 8.4

{\sc Gottschalk, J.E., Maslen, E.N.} [1987], {\it Coordinate systems and
analytic expansions for three-body atomic wavefunctions: \Roemisch{3}.
Derivative continuity via solutions to Laplace's equation}, J. Phys. A
{\bf 20}, 2781 - 2803.
% Abschnitt 8.4

{\sc Gradshteyn, I.S., Ryzhik, I.M.} [1980], {\it Table of integrals,
series, and products} (Academic Press, New York).
% Abschnitt 2.1, 7.2, 9.1, 9.2

{\sc Graffi, S., Grecchi, V.} [1978], {\it Borel summability and
indeterminacy of the Stieltjes moment problem: Application to the
anharmonic oscillators}, J. Math. Phys. {\bf 19}, 1002 - 1006.
% Abschnitt 2.3, 4.5, 10.2, 10.7

{\sc Graffi, S., Grecchi, V., Harrell \Roemisch{2}, E.M., Silverstone,
H.J.} [1985], {\it The $1/R$-expansion for $H_2^{+}$: Analyticity,
summability, and asymptotics}, Ann. Phys. (NY) {\bf 165}, 441 - 483.
% Abschnitt 2.3, 10.1

{\sc Graffi, S., Grecchi, V., Simon, B.} [1970], {\it Borel
summability: Application to the anharmonic oscillator}, Phys. Lett. B
{\bf 32}, 631 - 634.
% Abschnitt 2.3, 4.5, 10.7

{\sc Graffi, S., Grecchi, V., Turchetti, G.} [1971], {\it Summation
methods for the perturbation series of the generalized anharmonic
oscillators}, Il Nuovo Cimento B {\bf 4}, 313 - 340.
% Abschnitt 2.3, 4.5, 10.2

{\sc Gragg, W.B.} [1972], {\it The Pad\'e table and its relation to
certain algorithms of numerical analysis}, SIAM Rev. {\bf 14}, 1 - 62.
% Abschnitt 2.4, 4.1

{\sc Graves-Morris, P.R.} (Herausgeber) [1973a], {\it Pad\'e
approximants} (The Institute of Physics, London).
% Abschnitt 2.4, 4.1

{\sc Graves-Morris, P.R.} (Herausgeber) [1973b], {\it Pad\'e
approximants and their applications} (Academic Press, London).
% Abschnitt 2.4, 4.1

{\sc Graves-Morris, P.R.} [1981], {\it The convergence of ray
sequences of Pad\'e approximants of Stieltjes functions}, J. Comput.
Appl. Math. {\bf 7}, 191 - 201.
% Abschnitt 6.6, 6.7

{\sc Graves-Morris, P.R.} [1990], {\it Solution of integral equations
using generalised inverse, function-valued Pad\'e approximants,
\Roemisch{1}}, J. Comput. Appl. Math. {\bf 32}, 117 - 124.
% Abschnitt 13

{\sc Graves-Morris, P.R.} [1992], {\it Extrapolation methods for vector
sequences}, Numer. Math. {\bf 61}, 475 - 487.
% Abschnitt 13

{\sc Graves-Morris, P.R., Saff, E.B., Varga, R.S.} (Herausgeber)
[1984], {\it Rational approximation and interpolation} (Springer-Verlag,
Berlin).
% Abschnitt 2.4, 4.1

{\sc Graves-Morris, P.R., Thukral, R.} [1992], {\it Solution of integral
equations using function-valued Pad\'e approximants, \Roemisch{2}}, Numer.
Algor. {\bf 3}, 223 - 234.
% Abschnitt 13

{\sc Gray, C.G., Gubbins, K.E.} [1984], {\it Theory of molecular fluids.
Volume 1: Fundamentals} (Clarendon Press, Oxford).
% Abschnitt 4.5

{\sc Graybeal, J.D.} [1988], {\it Molecular spectroscopy} (McGraw-Hill,
New York).
% Abschnitt 10.2

{\sc Grecchi, V., Maioli, M.} [1984a], {\it Borel summability beyond the
factorial growth}, Ann. Inst. Henri Poincar\'e A {\bf 41}, 37 - 47.
% Abschnitt 2.3

{\sc Grecchi, V., Maioli, M.} [1984b], {\it Generalized logarithmic
Borel summability}, J. Math. Phys. {\bf 25}, 3439 - 3443.
% Abschnitt 2.3

{\sc Grimes, R.W., Catlow, C.R.A., Shluger, A.L.} (Herausgeber) [1992],
{\it Quantum mechanical cluster calculations in solid state studies}
(World Scientific, Singapore).
% Abschnitt 11.1

{\sc Grinstein, F.F.} [1980], {\it Summation of partial wave expansions
in the scattering by short-range potentials}, J. Math. Phys. {\bf 21},
112 -  119.
% Abschnitt 13

{\sc Gross, E.K.U., Runge, E.} [1986], {\it Vielteilchentheorie}
(Teubner, Stuttgart).
% Abschnitt 10.1

{\sc Grossman, R.} (Herausgeber) [1989], {\it Symbolic computation:
Applications to scientific computing} (SIAM, Philadelphia).
% Abschnitt 4.4

{\sc Grosso, G., Pastori Parravicini, G.} [1985a], {\it Continued
fractions in the theory of relaxation}, Adv. Chem. Phys. {\bf 62}, 81 -
132.
% Abschnitt 4.1

{\sc Grosso, G., Pastori Parravicini, G.} [1985b], {\it Memory function
method in solid state physics}, Adv. Chem. Phys. {\bf 62}, 133 - 181.
% Abschnitt 4.1

{\sc Grosswald, E.} [1978], {\it Bessel polynomials} (Springer-Verlag,
Berlin).
% Abschnitt 8.2, 8.3

{\sc Grotendorst, J.} [1985], {\it Berechnung der
Mehrzentrenmolek\"ulintegrale mit exponentialartigen Basisfunktionen durch
systematische Anwendung der Fourier-Transformationsmethode}
(Dissertation, Universit\"at Regensburg).
% Abschnitt 8.2

{\sc Grotendorst, J.} [1989], {\it A Maple programs for converting
series expansions to rational functions using the Levin
transformation}, Comput. Phys. Commun. {\bf 55}, 325 - 335.
% Abschnitt 5.2

{\sc Grotendorst, J.} [1990], {\it Approximating functions by means of
symbolic computation and a general extrapolation method}, Comput. Phys.
Commun. {\bf 59}, 289 - 301.
% Abschnitt 3.2, 4.1

{\sc Grotendorst, J.} [1991], {\it A Maple package for transforming
series, sequences and functions}, Comput. Phys. Commun. {\bf 67}, 325 -
342.
% Abschnitt 4.5, 5.1, 5.2, 5.4, 5.6, 6.6, 6.7

{\sc Grotendorst, J., Steinborn, E. O.} [1985], {\it The Fourier
transform of a two-center product of exponential-type functions and its
efficient evaluation}, J. Comput. Phys. {\bf 61}, 195 - 217.
% Abschnitt 8.2

{\sc Grotendorst, J., Steinborn, E. O.} [1986], {\it Use of nonlinear
convergence accelerators for the efficient evaluation of GTO molecular
integrals}, J. Chem. Phys. {\bf 84}, 5617 - 5623.
% Abschnitt 9.3

{\sc Grotendorst, J., Steinborn, E.O.} [1988], {\it Numerical
evaluation of molecular one- and two-center multicenter integrals with
exponential-type orbitals via the Fourier-transform method}, Phys. Rev.
A {\bf 38}, 3857 - 3876.
% Abschnitt 8.2, 9.2

{\sc Grotendorst, J., Weniger, E.J., Steinborn, E.O.} [1986], {\it
Efficient evaluation of infinite-series representations for overlap,
two-center nuclear attraction, and Coulomb integrals using nonlinear
convergence accelerators}, Phys. Rev. A {\bf 33}, 3706 - 3726.
% Abschnitt 4.5, 8.2, 8.3, 8.4

{\sc Guardiola, R., Ros, J.} [1992], {\it The anharmonic oscillator $-
{\d}^2 / \d x^2 + x^2 + b / x^4 + c / x^6$ for extreme values of the
anharmonicity constant}, J. Phys. A {\bf 25}, 1351 - 1372.
% Abschnitt 4.5

{\sc Guardiola, R., Sol\'{\i}s, M.A., Ros, J.} [1992], {\it
Strong-coupling expansion for the anharmonic oscillators $- {\d}^2 / \d
x^2 + x^2 + \lambda x^{2 N}$}, Il Nuovo Cimento B {\bf 107}, 713 - 724.
% Abschnitt 4.5, 10.3, 10.4, 10.8

{\sc Gurlay, A.R., Watson, G.A.} [1973], {\it Computational methods for
matrix eigenproblems} (Wiley, Chichester).
% Abschnitt 1.2

{\sc Guttmann, A.J.} [1989], {\it Asymptotic analysis of power series
expansions}, in Domb, C., Lebowitz, J.L. (Herausgeber), {\it Phase
transitions and critical phenomena\/} {\bf 13} (Academic Press, London),
3 - 234.
% Abschnitt 2.4

{\sc H\"anggi, P., Roesel, F., Trautmann, D.} [1980], {\it Evaluation of
infinite series by use of continued fraction expansions: A numerical
study}, J. Comput. Phys. {\bf 37}, 242 - 258.
% Abschnitt 4.1

{\sc Hamilton, T.P., Pulay, P.} [1986], {\it Direct inversion in the
iterative subspace (DIIS) optimization of open-shell, excited-state, and
small multiconfiguration SCF wave functions}, J. Chem. Phys. {\bf 84},
5728 - 5734.
% Abschnitt 11.1

{\sc Hansen, E.R.} [1975], {\it A table of series and products}
(Prentice-Hall, Englewood-Cliffs).
% Abschnitt 2.1

{\sc Hansen, J.-P., McDonald, I.R.} [1986], {\it Theory of simple
liquids} (Academic Press, London).
% Abschnitt 4.5

{\sc Hardy, G.H.} [1904], {\it On differentiation and integration of
divergent series}, Trans. Cambridge Phil. Soc. {\bf 19}, 297 - 321.
% Abschnitt 10.8

{\sc Hardy, G.H.} [1949], {\it Divergent series} (Clarendon Press,
Oxford).
% Abschnitt 2.2, 2.3, 2.4, 6.1

{\sc Harper, D., Wooff, C., and Hodgkinson, D.} [1991], {\it A guide to
computer algebra systems} (Wiley, Chichester).
% Abschnitt 4.4

{\sc Harris, F.E.} [1972], {\it Fourier representation methods for
electronic structures of linear polymers}, J. Chem. Phys. {\bf 56}, 4422
- 4425.
% Abschnitt 11.1

{\sc Harris, F.E.} [1975a], {\it Hartree-Fock studies of electronic
structures of crystalline solids}, in Eyring, H., Henderson, D.
(Herausgeber), {\it Theoretical Chemistry\/} {\bf 1} (Academic Press,
New York), 147 - 218.
% Abschnitt 11.1

{\sc Harris, F.E.} [1975b], {\it Electronic structure calculations on
crystals and polymers}, in Andr\'e, J.-M., Ladik, J., Delhalle, J.
(Herausgeber), {\it Electronic structure of polymers and molecular
crystals} (Plenum Press, New York), 453 - 477.
% Abschnitt 11.1

{\sc Harris, F.E.} [1977], {\it Convergence acceleration technique for
lattice sums arising in electronic-structure studies of crystalline
solids}, J. Math. Phys. {\bf 18}, 2377 - 2381.
% Abschnitt 11.1

{\sc Harris, F.E.} [1978], {\it Fourier representation methods in
electronic structure studies of crystals and polymers}, in Andr\'e, J.-M.,
Delhalle, J., Ladik, J. (Herausgeber), {\it Quantum theory of polymers}
(Reidel, Dordrecht), 117 - 135.
% Abschnitt 11.1

{\sc Harris, F.E.} [1983], {\it Evaluation of GTO molecular integrals},
Int. J. Quantum Chem. {\bf 23}, 1469 - 1478.
% Abschnitt 9.3

{\sc Harris, F.E., Michels, H.H.} [1967], {\it The evaluation of
molecular integrals for Slater-type orbitals}, Adv. Chem. Phys. {\bf 8},
205 - 266.
% Abschnitt 8.2

{\sc Harris, F.E., Monkhorst, H.J., Freeman, D.L.} [1992], {\it
Algebraic and diagrammatic methods in many-fermion theory} (Oxford U.
P., Oxford).
% Abschnitt 1.2, 8.2, 10.1

{\sc Hart, J.F., Cheney, E.W., Lawson, C.L., Maehly, H.J.,
Mesztenyi, C.M., Rice, J.R., Thacher, Jr., H.G., Witzgall, C.},
[1968], {\it Computer approximations} (Wiley, New York)
% Abschnitt 3.2

{\sc Harter, W.G.} [1993], {\it Principles of symmetry, dynamics, and
spectroscopy} (Wiley, New York). % Abschnitt 1.2

{\sc Harter, W.G., Patterson, C.W.} [1976], {\it A unitary calculus for
electronic orbitals} (Springer-Verlag, Berlin).
% Abschnitt 1.2, 8.2

{\sc Hartree, D.R.} [1928], {\it The wave mechanics of an atom with a
non-coulomb central field}, Proc. Cambridge Phil. Soc. {\bf 24}, 89 -
132.
% Abschnitt 1.1, 8.1

{\sc Hartree, D.R.} [1957], {\it The calculation of atomic structures}
(Wiley, New York).
% Abschnitt 1.1, 8.1

{\sc Hautot, A.} [1974], {\it A new method for the evaluation of slowly
convergent series}, J. Math. Phys. {\bf 15}, 1722 - 1727.
% Abschnitt 11.1

{\sc Hautot, A.} [1982], {\it Application of generalized Pad\'e
approximants to the special function evaluation problem}, J. Comput.
Phys. {\bf 47}, 477 - 488.
% Abschnitt 7.1

{\sc H{\aa}vie, T.} [1979], {\it Generalized Neville type extrapolation
schemes}, BIT {\bf 19}, 204 - 213.
% Abschnitt 3.2

{\sc H{\aa}vie, T.} (Herausgeber) [1989], {\it Romberg seminar on
quadrature/interpolation/extrapolation and rational approximations
(Pad\'e-continued fractions)} (Det Kongelige Norske Videnskabers Selskab
Skrifters 2, Tapir Forlag, Trondheim).
% Abschnitt 2.4

{\sc Haywood, S.E., Morgan \Roemisch{3}, J.D.} [1985], {\it Discrete
basis-set approach for calculating Bethe logarithms}, Phys. Rev. A {\bf
32}, 3179 - 3186.
% Abschnitt 2.1

{\sc Heeger, A.J., MacDiarmid, A.G.} [1980], {\it Conducting organic
polymers: Doped polyacetylene}, in Alc\'acer, L. (Herausgeber), {\it The
physics and chemistry of low dimensional solids} (Reidel, Dordrecht),
353 - 392.
% Abschnitt 11.2

{\sc Hegarty, D., van der Velde, G.} [1983], {\it Integral evaluation
algorithms and their implementation}, Int. J. Quantum Chem. {\bf 23},
1135 - 1153.
% Abschnitt 9.2

{\sc Hehre, W.J., Stewart, R.F., Pople, J.A.} [1969], {\it
Self-consistent molecular-orbital method. \Roemisch{1}. Use of Gaussian
expansions of Slater-type molecular orbitals}, J. Chem. Phys. {\bf 51},
2657 - 2664.
% Abschnitt 11.3

{\sc Hellmann, H.} [1937], {\it Einf\"uhrung in die Quantenchemie}
(Deuticke, Leipzig).
% Abschnitt 1.1, 8.1

{\sc Hellwege, K.-H.} [1976], {\it Einf\"uhrung in die Festk\"orperphysik}
(Springer-Verlag, Berlin).
% Abschnitt 11.2

{\sc Hellwig, G.} [1967], {\it Differential operators of mathematical
physics} (Addison Wesley, Reading, Mass.).
% Abschnitt 10.1

{\sc Henkel, M.} [1990], {\it Applications of the Hamiltonian limit to
critical phenomena: Numerical methods and applications}, in Privman, V.
(Herausgeber), {\it Finite size scaling and numerical simulation of
statistical systems} (World Scientific, Singapore), 353 - 433.
% Abschnitt 11.6

{\sc Henkel, M., Herrmann, H.J.} [1990], {\it The Hamiltonian spectrum
of directed percolation}, J. Phys. A {\bf 23}, 3719 - 3727.
% Abschnitt 11.6

{\sc Henkel, M., Sch\"utz, G.} [1988], {\it Finite-lattice extrapolation
algorithms}, J. Phys. A {\bf 21}, 2617 - 2633.
% Abschnitt 11.6

{\sc Henrici, P.} [1974], {\it Applied and computational complex
analysis \Roemisch{1}} \/ (Wiley, New York).
% Abschnitt 2.2

{\sc Henrici, P.} [1977], {\it Applied and computational complex
analysis \Roemisch{2}} \/ (Wiley, New York).
% Abschnitt 4.4, 7.2

{\sc Herbst, I.} [1993], {\it Perturbation theory for the decay rate of
eigenfunctions in the generalized $N$-body problem}, Commun. Math. Phys.
{\bf 158}, 517 - 536.
% Abschnitt 8.2, 9.1

{\sc Herbst, I., Simon, B.} [1978], {\it Stark effect revisited}, Phys.
Rev. Lett. {\bf 41}, 67 - 69.
% Abschnitt 2.3

{\sc Herrick, D.} [1983], {\it New symmetry properties of atoms and
molecules}, Adv. Chem. Phys. {\bf 52}, 1 - 115.
% Abschnitt 1.2

{\sc Herschbach, D.R.} [1992], {\it Chemical reaction dynamics and
electronic structure}, in Zewail, A. (Herausgeber), {\it The chemical
bond} (Academic Press, San Diego), 175 - 222.
% Abschnitt 1.2

{\sc Herschbach, D.R., Avery, J., Goscinski, O.} (Herausgeber) [1993],
{\it Dimensional scaling in chemical physics} (Kluwer, Dordrecht).
% Abschnitt 1.2

{\sc Herzberg, G.} [1944], {\it Atomic spectra and atomic structure}
(Dover, New York).
% Abschnitt 10.1

{\sc Heuser, H.} [1981], {\it Lehrbuch der Analysis Teil \Roemisch{2}\/}
(Teubner, Stuttgart).
% Abschnitt 2.2

{\sc Hierse, W., Oppeneer, P.M.} [1993], {\it Unified kernel function
approach to two-center integrations in quantum-chemical calculations},
J. Chem. Phys. {\bf 99}, 1278 - 1287.
% Abschnitt 8.2

{\sc Hill, R.N.} [1985], {\it Rates of convergence and error estimation
formulas for the Rayleigh-Ritz variational method}, J. Chem. Phys. {\bf
83}, 1173 - 1196.
% Abschnitt 9.1

{\sc Hinze, J.} (Herausgeber) [1981], {\it The unitary group}
(Springer-Verlag, Berlin).
% Abschnitt 1.2, 8.2

{\sc Hioe, F.T., MacMillen, D., Montroll, E.W.} [1976], {\it Quantum
theory of anharmonic oscillators. \Roemisch{2}. Energy levels of
oscillators with $x^{2 \alpha}$ anharmonicity}, J. Math. Phys. {\bf 17},
1320 - 1337.
% Abschnitt 10.3, 10.7

{\sc Hioe, F.T., MacMillen, D., Montroll, E.W.} [1978], {\it Quantum
theory of anharmonic oscillators: Energy levels of a single and a pair
of coupled oscillators with quartic coupling}, Phys. Rep. {\bf 43}, 305
- 335.
% Abschnitt 10.3, 10.7, 10.8

{\sc Hioe, F.T., Montroll, E.W.} [1975], {\it Quantum theory of
anharmonic oscillators. \Roemisch{1}. Energy levels of oscillators with
positive quartic anharmonicity}, J. Math. Phys. {\bf 16}, 1945 - 1955.
% Abschnitt 10.3, 10.7, 10.8

{\sc Hirsbrunner, B.} [1982], {\it Approximants de Borel}, Helv. Phys.
Acta {\bf 55}, 295 - 329.
% Abschnitt 2.3, 10.7

{\sc Hirschfelder, J.O., Byers Brown, W., Epstein, S.T.} [1964], {\it
Recent developments in perturbation theory}, Adv. Quantum Chem. {\bf 1},
255 - 374.
% Abschnitt 10.1

{\sc Hirschman, Jr., I.I.} [1962], {\it Infinite series} (Holt,
Rinehart, and Winston, New York).
% Abschnitt 2.1

{\sc Hitotumatu, S.} [1967], {\it On the numerical computation of Bessel
functions through continued fractions}, Comment. Math. Univ. St. Paul
{\bf 16}, 90 - 113.
% Abschnitt 7.1

{\sc Hobson, E.W.} [1892], {\it On a theorem in differentiation, and its
application to spherical harmonics}, Proc. London Math. Soc. {\bf 24},
54 - 67.
% Abschnitt 8.2

{\sc Hobson, E.W.} [1965], {\it The theory of spherical and ellipsoidal
harmonics} (Chelsea, New York).
% Abschnitt 8.2

{\sc Hoffmann, M.R., Schaefer \Roemisch{3}, H.F.} [1986], {\it A full
coupled-cluster singles, doubles and triples model for the description
of electron correlation}, Adv. Quantum Chem. {\bf 18}, 207 - 279.
% Abschnitt 8.2

{\sc Holdeman, Jr., J.T.} [1969], {\it A method for the approximation of
functions defined by formal series expansions in orthogonal
polynomials}, Math. Comput. {\bf 23}, 275 - 287.
% Abschnitt 13

{\sc Holmes, M.H., Bell, J.} [1991], {\it The application of symbolic
computing to chemical kinetic reaction schemes}, J. Comput. Chem. {\bf
10}, 1223 - 1231.
% Abschnitt 4.4

{\sc Holvorcem, P.R.} [1992], {\it Asymptotic summation of Hermite
series}, J. Phys. A {\bf 25}, 909 - 924.
% Abschnitt 13

{\sc Homeier, H.H.H.} [1990], {\it Integraltransformationsmethode und
Quadraturverfahren f\"ur Molek\"ul\-integrale mit $B$-Funktionen}
(Dissertation, Universit\"at Regensburg, S. Roderer Verlag, Regensburg).
% Abschnitt 8.2, 8.3, 9.1, 9.2

{\sc Homeier, H.H.H.} [1992], {\it A Levin-type algorithm for
accelerating the convergence of Fourier series}, Numer. Algor. {\bf 3},
245 - 254.
% Abschnitt 13

{\sc Homeier, H.H.H.} [1993], {\it Some applications of nonlinear
convergence accelerators}, Int. J. Quantum Chem. {\bf 45}, 545 - 562.
% Abschnitt 3.3, 13

{\sc Homeier, H.H.H., Steinborn, E.O.} [1990], {\it Numerical
integration of a function with a sharp peak at or near one boundary
using M\"obius transformations}, J. Comput. Phys. {\bf 87}, 61 - 72.
% Abschnitt 8.3, 8.4

{\sc Homeier, H.H.H., Steinborn, E.O.} [1991], {\it Improved
quadrature methods for three-center nuclear attraction integrals with
exponential-type basis functions}, Int. J. Quantum Chem. {\bf 39}, 625 -
645.
% Abschnitt 8.3

{\sc Homeier, H.H.H., Steinborn, E.O.} [1992], {\it On the evaluation
of overlap integrals with exponential-type basis function}, Int. J.
Quantum Chem. {\bf 42}, 761 - 778.
% Abschnitt 8.2

{\sc Homeier, H.H.H., Steinborn, E.O.} [1993], {\it Programs for the
evaluation of nuclear attraction integrals with $B$ functions}, Comput.
Phys. Commun. {\bf 77}, 135 - 151. % Abschnitt 8.3

{\sc Homeier, H.H.H., Weniger, E.J., Steinborn, E.O.} [1992a], {\it
Simplified derivation of a one-range addition theorem of the Yukawa
potential}, Int. J. Quantum Chem. {\bf 44}, 405 - 411.
% Abschnitt 8.2, 9.1

{\sc Homeier, H.H.H., Weniger, E.J., Steinborn, E.O.} [1992b], {\it
Programs for the evaluation of overlap integrals with $B$ functions},
Comput. Phys. Commun. {\bf 72}, 269 - 287.
% Abschnitt 8.2, 8.3

{\sc Hor\'{a}\v{c}ek, J., Sasakawa, T.} [1983], {\it Method of
continued fractions with applications to atomic physics}, Phys. Rev. A
{\bf 28}, 2151 - 2156.
% Abschnitt 4.1

{\sc Hosoya, H.} [1982], {\it Damping factors for the calculation of the
Madelung constants of ionic crystals -- Mathematical relations between
Evjen's method and Euler's transformation of series}, Physica B {\bf
113}, 175 - 188.
% Abschnitt 11.1

{\sc Houghton, A., Reeve, J.S., Wallace, D.J.} [1978], {\it High-order
behavior in $\Phi^3$ field theories and the percolation problem}, Phys.
Rev. A {\bf 17}, 2956 - 2964.
% Abschnitt 2.3

{\sc Hsue, C.-S., Chern, J.L.} [1984], {\it Two-step approach to
one-dimensional anharmonic oscillators}, Phys. Rev. D {\bf 29}, 643 -
647.
% Abschnitt 10.4

{\sc Hunziker, W.} [1988], {\it Notes on asymptotic perturbation theory
for Schr\"odinger eigenvalue problems}, Helv. Phys. Acta {\bf 61}, 257 -
304.
% Abschnitt 10.1

{\sc Hurley, A.C.} [1976], {\it Electron correlation in small
molecules} (Academic Press, London).
% Abschnitt 8.2

{\sc Hurst, J.B., Dupuis, M., Clementi, E.} [1988], {\it {\rm Ab initio}
analytic polarizability, first and second hyperpolarizabilities of large
conjugated organic molecules: Application to polyenes $C_4 H_6$ to
$C_{22} H_{24}$}, J. Chem. Phys. {\bf 89}, 385 - 395.
% Abschnitt 11.1

{\sc Huzinaga, S.} [1967], {\it Molecular integrals}, Prog. Theor. Phys.
Suppl. {\bf 40}, 52 - 77.
% Abschnitt 8.1, 8.2

{\sc Iafrate, G.J., Mendelsohn, L.B.} [1970], {\it High-order
perturbation theory and Pad\'e approximants for a one-electron atom in a
generalized central-field potential}, Phys. Rev. A {\bf 2}, 561 - 565.
% Abschnitt 4.5

{\sc I'Haya, Y.J., Narita, S., Fujita, Y., Ujino, H.} [1984], {\it Ab
initio crystal orbital calculations on $({\rm CH})_n$ and $({\rm HF})_n$
with extended basis sets}, Int. J. Quantum Chem. Symp. {\bf 18}, 153 -
159.
% Abschnitt 11.1, 11.2

{\sc Ihm, J.} [1988], {\it Total energy calculations in solid state
physics}, Rep. Prog. Phys. {\bf 51}, 105 - 142.
% Abschnitt 11.1

{\sc Inkson, J.C.} [1984], {\it Many-body theory of solids} (Plenum
Press, New York).
% Abschnitt 10.1

{\sc Iseki, S., Iseki, Y.} [1980], {\it Asymptotic expansion for the
remainder of a factorial series}, Mem. Natl. Defense Acad. Japan {\bf
20}, 1 - 6.
% Abschnitt 5.3

{\sc Itzykson, C., Zuber, J.-B.} [1980], {\it Quantum field theory}
(McGraw-Hill, New York).
% Abschnitt 2.3, 10.2

{\sc Jackson, J.D.} [1975], {\it Classical electrodynamics} (Wiley, New
York).
% Abschnitt 8.2

{\sc Jackson, R.I., Swain, S.} [1981], {\it Comparison between the
projection operator and continued fraction approaches to perturbation
theory}, J. Phys. A {\bf 14}, 3169 - 3179.
% Abschnitt 4.1

{\sc Jakab, L.} [1979], {\it An improved approach for computing the $F_m
(y)$ in molecular quantum mechanics}, J. Chem. Phys. {\bf 70}, 4421 -
4422.
% Abschnitt 9.3

{\sc Jakab, L.} [1984], {\it A simple procedure for numerical
approximation of the $F_m (z)$ functions with complex arguments},
Comput. Phys. Commun. {\bf 31}, 89 - 95.
% Abschnitt 9.3

{\sc Janke, W.} [1990a], {\it Path-integral derivation of large-order
behaviour of perturbation theory for anisotropic anharmonic
oscillators}, Phys. Lett. A {\bf 143}, 107 - 113.
% Abschnitt 10.1

{\sc Janke, W.} [1990b], {\it Zeeman effect from four-dimensional
anisotropic anharmonic oscillator}, Phys. Lett. A {\bf 144}, 116 - 122.
% Abschnitt 10.1

{\sc Janke, W.} [1990c], {\it Large-order perturbation theory of the
Zeeman effect in hydrogen from a four-dimensional anharmonic
oscillator}, Phys. Rev. A {\bf 41}, 6071 - 6084.
% Abschnitt 10.1

{\sc Janke, W., Kleinert, H.} [1990], {\it Large-order perturbation
expansion of three-dimensional Coulomb systems from four-dimensional
anisotropic anharmonic oscillators}, Phys. Rev. A {\bf 42}, 2792 - 2805.
% Abschnitt 10.1

{\sc Jankowski, K.} [1987], {\it Electron correlation in atoms}, in
Wilson, S. (Herausgeber), {\it Methods in computational chemistry {\bf
1}. Electron correlation in atoms and molecules} (Plenum Press, New
York), 1 - 116.
% Abschnitt 8.2

{\sc Jankowski, K.} [1992], {\it Electron correlation effects on atomic
properties}, in Wilson, S. (Herausgeber), {\it Methods in computational
chemistry {\bf 5}. Atomic and molecular properties} (Plenum Press, New
York), 1 - 98.
% Abschnitt 8.2

{\sc Janse van Rensburg, E.J.} [1993], {\it Virial coefficients for hard
discs and hard spheres}, J. Phys. A {\bf 26}, 4805 - 4818.
% Abschnitt 4.5

{\sc Jbilou, K., Sadok, H.} [1991], {\it Some results about vector
extrapolation methods and related fixed-point iterations}, J. Comput.
Appl. Math. {\bf 36}, 385 - 398.
% Abschnitt 13

{\sc Joachain, C.J.} [1975], {\it Quantum collision theory}
(North-Holland, Amsterdam).
% Abschnitt 8.3

{\sc J{\o}rgensen, P., Simons, J.} [1981], {\it Second
quantization-based methods in quantum chemistry} (Academic Press, New
York). % Abschnitt 8.2

{\sc Johnson, B.R., Reinhardt, W.P.} [1984], {\it Observation of Pad\'e
summability in divergent $L^2$ complex coordinate calculations of
$t$-matrix amplitudes in the presence of long-range forces}, Phys. Rev.
A {\bf 29}, 2933 - 2935.
% Abschnitt 4.5

{\sc Jones, D.S.} [1990], {\it Uniform asymptotic remainders}, in Wong,
R. (Herausgeber), {\it Asymptotics and computational analysis} (Marcel
Dekker, New York), 241 - 264.
% Abschnitt 2.2

{\sc Jones, H.W.} [1986a], {\it Computer-generated formulas for
four-center integrals over Slater-type orbitals}, Int. J. Quantum Chem.
{\bf 29}, 177 - 183. % Abschnitt 8.2

{\sc Jones, H.W.} [1986b], {\it Exact formulas for multipole moments
using Slater-type molecular orbitals}, Phys. Rev. A {\bf 33}, 2081 -
2083.
% Abschnitt 8.2

{\sc Jones, H.W.} [1987], {\it Exact formulas and their evaluation for
Slater-type orbital overlap integrals with large quantum numbers}, Phys.
Rev. A {\bf 35}, 1923 - 1926.
% Abschnitt 8.2

{\sc Jones, H.W.} [1988], {\it Analytical evaluation of multicenter
molecular integrals over Slater-type orbitals using expanded L\"owdin
alpha function}, Phys. Rev. A {\bf 38}, 1065 - 1068. % Abschnitt 8.2

{\sc Jones, H.W.} [1991], {\it Analytic L\"owdin alpha-function method for
two-center electron-repulsion integrals over Slater-type orbitals}, J.
Comput. Chem. {\bf 12}, 1217 - 1222.
% Abschnitt 8.2

{\sc Jones, H.W.} [1992a], {\it L\"owdin $\alpha$-function, overlap
integral, and computer algebra}, Int. J. Quantum Chem. {\bf 41}, 749 -
754.
% Abschnitt 8.2

{\sc Jones, H.W.} [1992b], {\it Semianalytical methods for four-center
molecular integrals over Slater-type orbitals}, Int. J. Quantum Chem.
{\bf 42}, 779 - 784.
% Abschnitt 8.2

{\sc Jones, H.W.} [1993], {\it Benchmark values for two-center Coulomb
integrals over Slater-type orbitals}, Int. J. Quantum Chem. {\bf 45},
21 - 30.
% Abschnitt 8.2

{\sc Jones, H.W., Etemadi, B.} [1993], {\it Accurate ground state
calculations of\/ ${\rm H}_{2}^{+}$ using basis sets of atom-centered
Slater-type orbitals}, Phys. Rev. A {\bf 47}, 3430 - 3432.
% Abschnitt 8.2

{\sc Jones, H.W., Etemadi, B., Brown, F.B.} [1992], {\it Restricted
basis functions for ${\rm H}_{2}^{+}$ with use of overlap integrals of
Slater-type orbitals}, Int. J. Quantum Chem. Symp. {\bf 26}, 265 - 270.
% Abschnitt 8.2

{\sc Jones, W.B., Thron, W.T.} [1980], {\it Continued fractions}
(Addison Wesley, Reading, Mass.).
% Abschnitt 4.1

{\sc Jones, W.B., Thron, W.T.} [1988], {\it Continued fractions in
numerical analysis}, Appl. Numer. Anal. {\bf 4}, 143 - 230.
% Abschnitt 4.1

{\sc Jones, W.B., Thron, W.T.} [1985], {\it On the computation of the
incomplete gamma function in the complex domain}, J. Comput. Appl.
Math. {\bf 12} \& {\bf 13}, 401 - 417.
% Abschnitt 4.1, 9.3

{\sc Judd, B.R.} [1967], {\it Second quantization and atomic
spectroscopy} (The Johns Hopkins Press, Baltimore).
% Abschnitt 1.2

{\sc Judd, B.R.} [1968], {\it Group theory in atomic spectroscopy}, in
Loebl, E.M. (Herausgeber), {\it Group theory and its applications
\Roemisch{1}} (Academic Press, New York), 183 - 220.
% Abschnitt 1.2

{\sc Judd, B.R.} [1975], {\it Angular momentum theory for diatomic
molecules} (Academic Press, New York).
% Abschnitt 1.2

{\sc Judd, B.R.} [1985], {\it Complex atomic spectra}, Rep. Prog. Phys.
{\bf 48}, 907 - 954.
% Abschnitt 1.2

{\sc Kaijser, P., Smith, Jr., V.H.} [1977], {\it Evaluation of momentum
distributions and Compton profiles for atomic and molecular systems},
Adv. Quantum Chem. {\bf 10}, 37 - 76.
% Abschnitt 8.2

{\sc Kaplan, I.G.} [1975], {\it Symmetry of many-electron systems}
(Academic Press, New York).
% Abschnitt 1.2

{\sc Kapur, J.N.} [1989], {\it Maximum-entropy models in science and
engineering} (Wiley, New York).
% Abschnitt 10.2

{\sc Karlsson, J., von Sydow, B.} [1976], {\it The convergence of Pad\'e
approximants to series of Stieltjes}, Ark. Matem. {\bf 14}, 43 - 53.
% Abschnitt 6.6, 6.7

{\sc Karna, S.P., Talapatra, G.B., Wijekoon, W.M.K.P., Prasad, P.N.}
[1992], {\it Frequency dependence of linear and nonlinear optical
properties of conjugated polyenes: An {\rm ab initio} time-dependent
coupled Hartree-Fock study}, Phys. Rev. A {\bf 45}, 2765 - 2770.
% Abschnitt 11.2

{\sc Karpfen, A.} [1978], {\it Ab initio calculations on model chains},
Theor. Chim. Acta {\bf 50}, 49 - 65.
% Abschnitt 11.1

{\sc Karpfen, A., Kertesz, M.} [1991], {\it Energetics and geometry of
conducting polymers from oligomers}, J. Phys. Chem. {\bf 95}, 7680 -
7681.
% Abschnitt 11.1

{\sc Karwowski, J.} [1992], {\it The configuration interaction approach
to electron correlation}, in Wilson, S., Diercksen, G.H.F.
(Herausgeber), {\it Methods in computational molecular physics} (Plenum
Press, New York), 65 - 98.
% Abschnitt 8.2

{\sc Kaufmann, K., Baumeister, W.} [1989], {\it Single-centre expansion
of Gaussian basis functions and the angular decomposition of their
overlap integrals}, J. Phys. B {\bf 22}, 1 - 12.
% Abschnitt 9.1

{\sc Kato, T.} [1949], {\it On the convergence of the perturbation
method. \Roemisch{1}.}, Prog. Theor. Phys. {\bf 4}, 514 - 523.
% Abschnitt 10.1

{\sc Kato, T.} [1950a], {\it On the convergence of the perturbation
method. \Roemisch{2}. 1.}, Prog. Theor. Phys. {\bf 5}, 95 - 101.
% Abschnitt 10.1

{\sc Kato, T.} [1950b], {\it On the convergence of the perturbation
method. \Roemisch{2}. 2.}, Prog. Theor. Phys. {\bf 5}, 207 - 212.
% Abschnitt 10.1

{\sc Kato, T.} [1951a], {\it Fundamental properties of Hamiltonian
operators of Schr\"odinger type}, Trans. Amer. Math. Soc. {\bf 70}, 195 -
211.
% Abschnitt 8.1

{\sc Kato, T.} [1951b], {\it On the existence of solutions of the Helium
wave equation}, Trans. Amer. Math. Soc. {\bf 70}, 212 - 218.
% Abschnitt 8.1

{\sc Kato, T.} [1957], {\it On the eigenfunctions of many-particle
systems in quantum mechanics}, Commun. Pure Appl. Math. {\bf 10}, 151 -
177.
% Abschnitt 8.2, 9.1

{\sc Kato, T.} [1976], {\it Perturbation theory for linear operators}
(Springer-Verlag, Berlin).
% Abschnitt 10.1

{\sc Kazakov, D.I., Shirkov, D.V.} [1980], {\it Asymptotic series of
quantum field theory and their summation}, Fortschr. Phys. {\bf 28},
465 - 499.
% Abschnitt 2.2, 2.3, 4.5

{\sc Kert\'esz, M.} [1982], {\it Electronic structure of polymers}, Adv.
Quantum Chem. {\bf 15}, 161 - 214.
% Abschnitt 11.1

{\sc Khalil, A.B.} [1992], {\it A relation between two perturbation
methods generating Pad\'e upper and lower bounds to ground state
eigenvalues}, J. Math. Phys. {\bf 33}, 639 - 642.
% Abschnitt 4.5

{\sc Khovanskii, A.N.} [1963], {\it The application of continued
fractions and their generalizations to problems in approximation
theory} (Noordhoff, Groningen).
% Abschnitt 4.1

{\sc Kiefer, J.E., Weiss, G.H.} [1981], {\it A comparison of two methods
for accelerating the convergence of Fourier series}, Comput. Math. Appl.
{\bf 7}, 527 - 535.
% Abschnitt 13

{\sc Killingbeck, J.} [1977], {\it Quantum-mechanical perturbation
theory}, Rep. Prog. Phys. {\bf 40}, 963 - 1031.
% Abschnitt 2.2, 2.3, 4.5, 10.1

{\sc Killingbeck, J.} [1980], {\it The harmonic oscillator with $\lambda
x^M$ perturbation}, J. Phys. A {\bf 13}, 49 - 56.
% Abschnitt 10.2

{\sc Killingbeck, J.} [1981], {\it Renormalised perturbation series}, J.
Phys. A {\bf 14}, 1005 - 1008.
% Abschnitt 10.4

{\sc Killingbeck, J., Jones, M.N., Thompson, M.J.} [1985], {\it Inner
product methods for eigenvalue calculations}, J. Phys. A {\bf 18}, 793 -
801.
% Abschnitt 10.4

{\sc King, F.W.} [1991], {\it Analysis of some integrals arising in
the atomic three-electron problem}, Phys. Rev. A {\bf 44}, 7108 - 7133.
% Abschnitt 8.4

{\sc King, F.W.} [1993], {\it Analysis of some integrals arising in
the atomic four-electron problem}, J. Chem. Phys. {\bf 99}, 3622 - 3628.
% Abschnitt 8.4

{\sc King, F.W., Dykema, K.J., Lund, A.D.} [1992], {\it Calculation of
some integrals for the atomic three-electron problem}, Phys. Rev. A {\bf
46}, 5406 - 5416.
% Abschnitt 8.4

{\sc Kirtman, B.} [1992], {\it Nonlinear optical properties of
conjugated polymers from ab initio finite oligomer calculations}, Int.
J. Quantum Chem. {\bf 43}, 147 - 158.
% Abschnitt 11.1

{\sc Kirtman, B., Hasan, M., Chipman, D.M.} [1991], {\it Solitons in
polyacetylene: Magnetic hyperfine constants from {\rm ab initio}
calculations}, J. Chem. Phys. {\bf 95}, 7698 - 7716.
% Abschnitt 11.1, 11.2

{\sc Kirtman, B., Nilson, W.B., Palke, W.E.} [1983], {\it Small chain
approximation for the electronic structure of polyacetylene}, Solid
State Commun. {\bf 46}, 791 - 794.
% Abschnitt 11.1

{\sc Klahn, B.} [1981], {\it Review of linear independence properties of
infinite sets of functions used in quantum chemistry}, Adv. Quantum
Chem. {\bf 13}, 155 - 209.
% Abschnitt 8.1

{\sc Klahn, B.} [1984], {\it Rates of convergence of variational
calculations and of expectation values}, J. Chem. Phys. {\bf 81}, 410 -
433.
% Abschnitt 8.1

{\sc Klahn, B.} [1985a], {\it A generalization of the M\"untz-Sz\'{a}sz
theorem to floating exponents with applications to Gauss- and
Slater-type functions}, J. Chem. Phys. {\bf 83}, 5749 - 5753.
% Abschnitt 8.1

{\sc Klahn, B.} [1985b], {\it The convergence of CI calculations for
atomic and molecular electronic bound states in a basis of floating
Gauss orbitals}, J. Chem. Phys. {\bf 83}, 5754 - 5755.
% Abschnitt 8.1

{\sc Klahn, B., Bingel, W.A.} [1977a], {\it The convergence of the
Rayleigh-Ritz method in quantum chemistry. \Roemisch{1}. The criteria
for convergence}, Theor. Chim. Acta {\bf 44}, 9 - 26.
% Abschnitt 8.2, 9.1

{\sc Klahn, B., Bingel, W.A.} [1977b], {\it The convergence of the
Rayleigh-Ritz method in quantum chemistry. \Roemisch{2}. Investigation
of the convergence for special systems of Slater, Gauss and Two-electron
functions}, Theor. Chim. Acta {\bf 44}, 27 - 43.
% Abschnitt 8.1, 9.1

{\sc Klahn, B., Bingel, W.A.} [1977c], {\it Completeness and linear
independence of basis sets used in quantum chemistry}, Int. J. Quantum
Chem. {\bf 11}, 943 - 957.
% Abschnitt 8.1

{\sc Klahn, B., Morgan \Roemisch{3}, J.D.} [1984], {\it Rates of
convergence of variational calculations and of expectation values}, J.
Chem. Phys. {\bf 81}, 410 - 433.
% Abschnitt 8.1, 9.1

{\sc Kleinert, H.} [1993a], {\it Variational approach to tunneling.
Beyond the semiclassical approximation of Langer and Lipatov --
perturbation coefficients to all order}, Phys. Lett. B {\bf 300}, 261 -
270.
% Abschnitt 10.1

{\sc Kleinert, H.} [1993b], {\it Pfadintegrale in Quantenmechanik,
Statistik und Polymerphysik} (B.I. Wissenschaftsverlag, Mannheim).
% Abschnitt 10.2

{\sc Kline, M.} [1972], {\it Mathematical thought from ancient to
modern times} (Oxford U. P., Oxford).
% Abschnitt 2.1, 2.2

{\sc Klopper, W., Kutzelnigg, W.} [1986], {\it Gaussian basis sets and
the nuclear cusp problem}, J. Mol. Struct. (Theochem) {\bf 135}, 339 -
356.
% Abschnitt 9.1

{\sc Knopp, K.} [1964], {\it Theorie und Anwendung der unendlichen
Reihen} (Springer-Verlag, Berlin).
% Abschnitt 2.1, 2.4, 5.2, 6.1

{\sc K\"orner, T.W.} [1988], {\it Fourier analysis} (Cambridge U. P.,
Cambridge).
% Abschnitt 2.2, 13

{\sc K\"orner, T.W.} [1993], {\it Exercises for Fourier analysis}
(Cambridge U. P., Cambridge).
% Abschnitt 13

{\sc Kostroun, V.O.} [1980], {\it Simple numerical evaluation of
modified Bessel functions $K_{\nu} (x)$ of fractional order and the
integral $\int\nolimits_{x}^{\infty} K_{\nu} (\eta) \d \eta$}, Nucl.
Inst. Meth. {\bf 172}, 371 - 374.
% Abschnitt 7.1

{\sc Kowalewski, C.} [1981], {\it Acc\'el\'eration de la convergence pour
certaines suites a convergence logarithmique}, in de Bruin, M.G., Van
Rossum, H. (Herausgeber), {\it Pad\'e approximation and its applications,
Amsterdam 1980} (Springer-Verlag, Berlin), 263 - 272.
% Abschnitt 4.5

{\sc Krieger, J.B.} [1968], {\it Asymptotic properties of perturbation
theory}, J. Math. Phys. {\bf 9}, 432 - 435.
% Abschnitt 10.1

{\sc Kucharski, S.A., Bartlett, R.J.} [1986], {\it Fifth-order
many-body perturbation theory and its relationship to various
coupled-cluster approaches}, Adv. Quantum Chem. {\bf 18}, 281 - 344.
% Abschnitt 8.2

{\sc K\"ummel, H., L\"uhrmann, K.H., Zabolitzky, J.G.} [1978], {\it
Many-fermion theory in $\exp S$- (or coupled cluster) form}, Phys. Rep.
{\bf 36}, 1 - 63.
% Abschnitt 8.2

{\sc Kufner, A., John, O., Fu\v{c}ik, S.} [1977], {\it Function spaces}
(Noordhoff, Leyden).
% Abschnitt 8.2

{\sc Kukhtin, V.V., Shramko, O.V.} [1991], {\it Lattice sums within the
Euler-MacLaurin approach}, Phys. Lett. A {\bf 156}, 257 - 259.
% Abschnitt 11.1

{\sc Kukulin, V.I., Krasnopol'sky, V.M., Hor\'a\v{c}ek, J.} [1989],
{\it Theory of resonances} (Kluwer, Dordrecht).
% Abschnitt 2.1, 4.5

{\sc Kumar, K.} [1962], {\it Perturbation theory and the nuclear many
body problem} (North-Holland, Amsterdam).
% Abschnitt 10.1

{\sc Kummer, E.E.} [1837], {\it Eine neue Methode, die numerischen
Summen langsam convergirender Reihen zu berechnen}, J. Reine Angew.
Math. {\bf 16}, 206 - 214.
% Abschnitt 2.4

{\sc Kuntzmann, J.} [1971], {\it Unendliche Reihen} (Vieweg,
Braunschweig).
% Abschnitt 2.1

{\sc Kutzelnigg, W.} [1977], {\it Pair correlation theories}, in
Schaefer \Roemisch{3}, H.F. (Herausgeber), {\it Methods of electronic
structure theory} (Plenum Press, New York), 129 - 188.
% Abschnitt 8.2

{\sc Kutzelnigg, W.} [1988], {\it Present and future trends in quantum
chemical calculations}, J. Mol. Struct. (Theochem) {\bf 181}, 33 - 54.
% Abschnitt 8.4

{\sc Kutzelnigg, W.} [1992], {\it Stationary perturbation theory.
\Roemisch{1}. Survey of basic concepts}, Theor. Chim. Acta {\bf 83}, 263
- 312.
% Abschnitt 10.1

{\sc Kutzelnigg, W.} [1993], {\it Stationary perturbation theory.
\Roemisch{2}. Electron correlation and its effect on properties}, Theor.
Chim. Acta {\bf 86}, 41 - 81.
% Abschnitt 10.1

{\sc Kutzelnigg, W., Morgan \Roemisch{3}, J.D.} [1992], {\it Rates of
convergence of the partial-wave expansions of atomic correlation
energies}, J. Chem. Phys. {\bf 96}, 4484 - 4508.
% Abschnitt 9.1

{\sc Kvasni\v{c}ka, V., Laurinc, V., Biskupi\v{c}, S.} [1982], {\it
Coupled-cluster approach in electronic structure theory of molecules},
Phys. Rep. {\bf 90}, 159 - 202.
% Abschnitt 1.2, 8.2

{\sc Lackner, T.} [1987], {\it Matrix continued-fraction
representation of the dynamical self-structure factor $S_s (q,
\omega)$}, Phys. Rev. A {\bf 35}, 987 - 998.
% Abschnitt 4.1

{\sc Ladik, J.L.} [1988], {\it Quantum theory of polymers as solids}
(Plenum Press, New York).
% Abschnitt 11.1, 11.2

{\sc Ladik, J., Otto, P.} [1993], {\it Correlation corrected band
structures of quasi 1d and 2d periodic systems and level distributions
of disordered chains; New method with correlation for dynamic nonlinear
properties of periodic polymers}, Int. J. Quantum Chem. Symp. {\bf 27},
111 - 129.
% Abschnitt 11.1

{\sc Landau, E.} [1906], {\it \"Uber die Grundlagen der Theorie der
Fakult\"atenreihen}, Sitzungsb. K\"onigl. Bay. Akad. Wissensch. M\"unchen,
math.-phys. Kl. {\bf 36}, 151 - 218.
% Abschnitt 5.3

{\sc Langhoff, P.W., Karplus, M.} [1970], {\it Application of Pad\'e
approximants to dispersion force and optical polarizability
computations}, in Baker, Jr., G.A., Gammel, J.L. (Herausgeber), {\it
The Pad\'e approximant in theoretical physics} (Academic Press, New
York), 41 - 97.
% Abschnitt 4.5

{\sc Lebedev, N.N.} [1972], {\it Special functions and their
applications} (Dover, New York).
% Abschnitt 7.1

{\sc Lee, Y.S., Kertesz, M.} [1987], {\it The effect of additional
fused rings on the stabilities and the band gaps of heteroconjugated
polymers}, Int. J. Quantum Chem. {\bf 21}, 163 - 170.
% Abschnitt 11.1

{\sc Lefebvre, R., Moser, C.} (Herausgeber) [1969], {\it Correlation
effects in atoms and molecules}, Adv. Chem. Phys. {\bf 14}.
% Abschnitt 10.1

{\sc Le Ferrand, H.} [1992], {\it The quadratic convergence of the
topological epsilon algorithm for systems of nonlinear equations},
Numer. Algor. {\bf 3}, 273 - 284.
% Abschnitt 13

{\sc Le Guillou, J.C., Zinn-Justin, J.} [1977], {\it Critical exponents
for the $n$-vector model in three dimensions from field theory}, Phys.
Rev. Lett. {\bf 39}, 95 - 98.
% Abschnitt 2.3

{\sc Le Guillou, J.C., Zinn-Justin, J.} [1983], {\it The hydrogen atom
in strong magnetic fields: Summation of the weak field series
expansion}, Ann. Phys. (NY) {\bf 147}, 57 - 84.
% Abschnitt 2.3

{\sc Le Guillou, J.C., Zinn-Justin, J.} [1984], {\it The $H_2^{+}$ ion
in an intense magnetic field: Improved adiabatic approximations}, Ann.
Phys. (NY) {\bf 154}, 440 - 455.
% Abschnitt 2.3

{\sc Le Guillou, J.C., Zinn-Justin, J.} (Herausgeber) [1990], {\it
Large-order behaviour of perturbation theory} (North-Holland,
Amsterdam).
% Abschnitt 2.2, 2.3, 4.5, 10.1

{\sc Leibbrandt, G.} [1975], {\it Introduction to the technique of
dimensional regularization}, Rev. Mod. Phys. {\bf 47}, 849 -876.
% Abschnitt 1.2

{\sc Lense, J.} [1953], {\it Reihenentwicklungen in der mathematischen
Physik} (Walter de Gruyter, Berlin).
% Abschnitt 2.1

{\sc Lepora, P., Gabutti, B.} [1987], {\it An algorithm for the
summation of series}, Appl. Numer. Math. {\bf 3}, 523 - 528.
% Abschnitt 13

{\sc Levin, D.} [1973], {\it Development of non-linear transformations
for improving convergence of sequences}, Int. J. Comput. Math. B {\bf
3}, 371 - 388.
% Abschnitt 3.1, 5.1, 5.2, 5.5, 11.5

{\sc Levy, R.A.} [1968], {\it Principles of solid state physics}
(Academic Press, New York).
% Abschnitt 11.1

{\sc Lichtenberg, D.B.} [1978], {\it Unitary symmetry and elementary
particles} (Academic Press, New York).
% Abschnitt 1.2

{\sc Liegener, C.-M.} [1985], {\it Third-order many-body perturbation
theory in the M{\o}ller-Plesset partitioning applied to an infinite
alternating hydrogen chain}, J. Phys. C {\bf 18}, 6011 - 6022.
% Abschnitt 11.1

{\sc Liegener, C.-M.} [1988], {\it {\rm Ab initio} calculations of
correlation effects in trans-polyacetylene}, J. Chem. Phys. {\bf 88},
6999 - 7004.
% Abschnitt 11.1, 11.2

{\sc Liegener, C.-M., Beleznay, F., Ladik, J.} [1987], {\it Application
of a modified Romberg algorithm to Hartree-Fock calculations on periodic
chains}, Phys. Lett. A {\bf 123}, 399 - 401.
% Abschnitt 11.1, 11.6

{\sc Linderberg, J., \"Ohrn, Y.} [1973], {\it Propagators in quantum
chemistry} (Academic Press, London).
% Abschnitt 8.2

{\sc Lindgren, I., Morrison, J.} [1982], {\it Atomic many-body theory}
(Springer-Verlag, Berlin).
% Abschnitt 1.2, 8.2, 10.1

{\sc Lipkin, H.J.} [1967], {\it Anwendungen von Lieschen Gruppen in der
Physik} (Bibliographisches Institut, Mannheim).
% Abschnitt 1.2

{\sc Loeffel, J.J., Martin, A.} [1972], {\it Propri\'et\'es analytiques
des niveaux de l'oscillateur anharmonique et convergence des
approximants de Pad\'e}, in Bessis, D. (Herausgeber), {\it Carg\`ese
lectures in physics} (Gordon and Breach, New York), Vol. 5, 415 - 429.
% Abschnitt 4.5

{\sc Loeffel, J.J., Martin, A., Simon, B., Wightman, A.S.} [1969],
{\it Pad\'e approximants and the anharmonic oscillator}, Phys. Lett. B
{\bf 30}, 656 - 658.
% Abschnitt 4.5

{\sc L\"osch, F., Schoblik, F.} [1951], {\it Die Fakult\"at} (Teubner,
Leipzig). % Abschnitt

{\sc L\"owdin, P.O.} [1966], {\it The calculation of upper and lower
bounds of energy eigenvalues in perturbation theory by means of
partitioning techniques}, in Wilcox, C.H. (Herausgeber), {\it
Perturbation theory and its application in quantum mechanics} (Wiley,
New York), 255 - 294.
% Abschnitt 10.4

{\sc L\"owdin, P.O.} [1982a], {\it Proceedings of the Sanibel workshop on
perturbation theory at large order, Sanibel Conference, Florida, 1981},
Int. J. Quantum Chem. {\bf 21}, 1 - 209.
% Abschnitt 10.1

{\sc L\"owdin, P.O.} [1982b], {\it Partitioning technique, perturbation
theory, and rational approximants}, Int. J. Quantum Chem. {\bf 21}, 69 -
92.
% Abschnitt 10.4

{\sc Longman, I.M.} [1981], {\it Difficulties of convergence
acceleration}, in de Bruin, M.G., van Rossum, H. (Herausgeber), {\it
Pad\'e approximation and its applications Amsterdam 1980}
(Springer-Verlag, Berlin), 273 - 289.
% Abschnitt 5.2

{\sc Longman, I.M.} [1985], {\it The summation of power series and
Fourier series}, J. Comput. Appl. Math. {\bf 12} \& {\bf 13}, 447 - 457.
% Abschnitt 13

{\sc Longman, I.M.} [1986], {\it The summation of series}, Appl. Numer.
Math. {\bf 2}, 135 - 141.
% Abschnitt 13

{\sc Longman, I.M.} [1987], {\it The summation of Fourier, Chebyshev,
and Legendre series}, Appl. Math. Comput. {\bf 23}, 61 - 70.
% Abschnitt 13

{\sc L\'opez-Cabrera, M., Goodson, D.Z., Herschbach, D.R., Morgan
\Roemisch{3}, J.D.} [1992], {\it Large-order dimensional perturbation
theory for $H_2^{+}$}, Phys. Rev. Lett. {\bf 68}, 1992 - 1995.
% Abschnitt 1.3

{\sc L\'opez, R., Ram\'{\i}rez, G.} [1994], {\it Calculation of two-center
exchange integrals with STOs using M\"obius transformations}, Int. J.
Quantum Chem. {\bf 49}, 11 - 19.
% Abschnitt 8.2

{\sc Lorentzen, L.} [1992], {\it Computation of hypergeometric functions
by means of continued fractions}, Comput. Appl. Math. {\bf 1}, 305 -
314.
% Abschnitt 4.1

{\sc Lorentzen, L., Waadeland, H.} [1992], {\it Continued fractions with
applications} (North-Holland, Amsterdam).
% Abschnitt 4.1

{\sc Lowery, M.D., House, Jr., J.E.} [1984], {\it An algorithm for
computing the Madelung constant for the sodium chloride lattice},
Comput. Chem. {\bf 8}, 249 - 254.
% Abschnitt 11.1

{\sc Lu, Y.} [1988], {\it Solitons and Polarons in conducting polymers}
(World Scientific).
% Abschnitt 11.2

{\sc Lubkin, S.} [1952], {\it A method of summing infinite series}, J.
Res. Natl. Bur. Stand. {\bf 48}, 228 - 254.
% Abschnitt 4.5

{\sc L\"u, T., Tachibana, A., Yamabe, T.} [1992], {\it Molecular dynamic
structure and dimerization of polyacetylene}, Int. J. Quantum Chem. {\bf
41}, 475 - 488.
% Abschnitt 11.2

{\sc L\"uchow, A.} [1993], {\it Eigenwertschranken f\"ur das atomare
Vier-Teilchen-System} (Dissertation, Heinrich-Heine-Universit\"at
D\"usseldorf).
% Abschnitt 8.4

{\sc L\"uchow, A., Kleindienst, H.} [1993], {\it Atomic integrals
containing $r_{2 3}^{\lambda} r_{13}^{\mu} r_{12}^{\nu}$ with $\lambda,
\mu, \nu \ge - 2$}, Int. J. Quantum Chem. {\bf 45}, 445 - 470.
% Abschnitt 8.4

{\sc Luke, Y.L.} [1969a], {\it The special functions and their
approximations \Roemisch{1}} (Academic Press, New York).
% Abschnitt 7.1, 10.2

{\sc Luke, Y.L.} [1969b], {\it The special functions and their
approximations \Roemisch{2}} (Academic Press, New York).
% Abschnitt 7.1

{\sc Luke, Y.L.} [1971], {\it Miniaturized tables of Bessel functions},
Math. Comput. {\bf 25}, 323 - 330.
% Abschnitt 7.1

{\sc Luke, Y.L.} [1975], {\it Mathematical functions and their
approximations} (Academic Press, New York).
% Abschnitt 7.1

{\sc Luke, Y.L.} [1977], {\it Algorithms for the computation of
mathematical functions} (Academic Press, New York).
% Abschnitt 7.1

{\sc Mac\'{\i}as, A., Riera, A.} [1982], {\it Ab initio quantum
chemistry in the molecular model of atomic collisions}, Phys. Rep. {\bf
90}, 299 - 386.
% Abschnitt 9.3

{\sc MacLeod, A. J.} [1986], {\it Acceleration of vector sequences by
multi-dimensional $\Delta^2$ methods}, Commun. Appl. Numer. Meth. {\bf
2}, 385 - 392.
% Abschnitt 13

{\sc Maekawa, K., Imamura, A.} [1993], {\it Electronic structures around
local defects in all-{\rm trans}-polyacetylene: An analysis by the
cluster series-model}, Int. J. Quantum Chem. {\bf 47}, 449 - 467.
% Abschnitt 11.2

{\sc Magnus, W., Oberhettinger, F., Soni, R.P.} [1966], {\it Formulas
and theorems for the special functions of mathematical physics},
(Springer-Verlag, New York).
% Abschnitt 2.2, 4.5, 5.3, 6.7, 7.1, 7.2, 8.2, 8.3, 9.3, 11.3

{\sc Mahan, G.D.} [1981], {\it Many-particle physics} (Plenum Press, New
York).
% Abschnitt 10.1

{\sc Maioli, M.} [1981], {\it Exponential perturbations of the
anharmonic oscillator}, J. Math. Phys. {\bf 22}, 1952 - 1958.
% Abschnitt 2.3

{\sc Makarewicz, J.} [1984], {\it Renormalised perturbation theory for a
general $D$-dimensional anharmonic oscillator}, J. Phys. A {\bf 17},
1449 - 1460.
% Abschnitt 10.4

{\sc Malgrange, B., Ramis, J.-P.} [1992], {\it Fonctions
multisommables}, Ann. Inst. Fourier (Grenoble) {\bf 42}, 1 - 16.
% Abschnitt 2.2

{\sc Malmqvist, P.-\AA.} [1992], {\it Mathematical tools in quantum
chemistry}, in Roos, B. (Herausgeber), {\it Lecture notes in quantum
chemistry} (Springer-Verlag, Berlin), 1 - 35.
% Abschnitt 1.2, 1.3

{\sc Maluendes, S.A., Fern\'{a}ndez, F.M., Castro, E.A.} [1985], {\it
Convergent renormalized perturbation series for the Stark rotational
energies of diatomic molecules}, J. Chem. Phys. {\bf 83}, 4599 - 4603.
% Abschnitt 10.4

{\sc Markushevich, A.I.} [1977], {\it Theory of functions of a complex
variable} (Chelsea, New York).
% Abschnitt 2.2

{\sc Mart\'{\i}n Pend\'{a}s, A., Francisco, E.} [1991], {\it Overlap,
effective-potential, and projection-operator integrals over complex
Slater-type orbitals}, Phys. Rev. A {\bf 43}, 3384 - 3391.
% Abschnitt 8.2

{\sc Martinet, J., Ramis, J.-P.} [1991], {\it Elementary acceleration
and multisummability \Roemisch{1}}, Ann. Inst. Henri Poincar\'e Phys.
Th\'eor. A {\bf 54}, 331 - 401.
% Abschnitt 2.2

{\sc Marziani, M.F.} [1984], {\it Perturbative solutions for the
generalised anharmonic oscillators}, J. Phys. A {\bf 17}, 547 - 557.
% Abschnitt 2.3, 4.5, 10.6

{\sc Marziani, M.F.} [1987], {\it Convergence of a class of Borel-Pad\'e
approximants}, Il Nuovo Cimento B {\bf 99}, 145 - 154.
% Abschnitt 2.3

{\sc Maslov, V.P.} [1972], {\it Th\'eorie des perturbations et m\'ethodes
asymptotiques} (Dunod, Paris).
% Abschnitt 10.1

{\sc Masson, D.} [1983], {\it The rotating harmonic oscillator
eigenvalue problem. \Roemisch{1}. Continued fractions and analytic
continuation}, J. Math. Phys. {\bf 24}, 2074 - 2088.
% Abschnitt 4.1

{\sc Masson, D.} [1986], {\it Exact Bogoliubov limits for the
Bassichis-Foldy model and continued fractions}, J. Math. Phys. {\bf
27}, 1093 - 1098.
% Abschnitt 4.1

{\sc Matos, A.} [1989], {\it Construction de methodes d'extrapolation \`a
partir de developpements asymptotiques} (Dissertation, Universit\'e
des Sciences et Techniques de Lille de Flandres-Artois).
% Abschnitt 4.5

{\sc Matos, A.} [1990a], {\it Acceleration methods based upon
convergence tests}, Numer. Math. {\bf 58}, 329 - 340.
% Abschnitt 4.5

{\sc Matos, A.} [1990b], {\it A convergence acceleration method based
upon a good estimation of the absolute value of the error}, IMA J.
Numer. Anal. {\bf 10}, 243 - 251.
% Abschnitt 4.5

{\sc Matos, A.} [1992], {\it Convergence and acceleration properties for
the vector $\epsilon$-algorithm}, Numer. Algor. {\bf 3}, 313 - 320.
% Abschnitt 13

{\sc Matsen, F.A., Pauncz, R.} [1986], {\it The unitary group in
quantum chemistry} (Elsevier, Amsterdam).
% Abschnitt 1.2, 8.2

{\sc Matsuoka, O.} [1992a], {\it Molecular integrals over spherical
Laguerre Gaussian-type functions}, J. Chem. Phys. {\bf 92}, 4364 - 4371.
% Abschnitt 8.2

{\sc Matsuoka, O.} [1992b], {\it Molecular integrals over Laguerre
Gaussian-type functions of real spherical harmonics}, Can. J. Chem. {\bf
70}, 388 - 392.
% Abschnitt 8.2

{\sc Mattuck, R.D.} [1976], {\it A guide to Feynman diagrams in the
many-body problem} (McGraw-Hill, New York).
% Abschnitt 1.2, 10.1

{\sc Maz'ja, V.G.} [1985], {\it Sobolev spaces} (Springer-Verlag,
Berlin).
% Abschnitt 9.1

{\sc Mazumdar, S., Guo, D., Dixit, S.N.} [1992], {\it High energy
two-photon states in finite versus infinite polyenes}, J. Chem. Phys.
{\bf 96}, 6862 - 6867.
% Abschnitt 11.1, 11.2

{\sc McRae, S.M., Vrscay, E.R.} [1992], {\it Canonical perturbation
expansions to large order from classical hypervirial and
Hellman\/-Feynman theorems}, J. Math. Phys. {\bf 33}, 3004 - 3024.
% Abschnitt 10.1

{\sc McWeeny, R.} [1992], {\it The electron correlation problem}, in
Wilson, S., Diercksen, G.H.F. (Herausgeber), {\it Methods in
computational molecular physics} (Plenum Press, New York), 47 - 56.
% Abschnitt 8.2

{\sc Mead, L.R., Papanicolaou, N.} [1984], {\it Maximum entropy method
in the problem of moments}, J. Math. Phys. {\bf 25}, 2404 - 2417.
% Abschnitt 10.2

{\sc Meschkowski, H.} [1959], {\it Differenzengleichungen} (Vandenhoek
\& Rupprecht, G\"ottingen).
% Abschnitt 5.3

{\sc Meschkowski, H.} [1962], {\it Unendliche Reihen}
(Bibliographisches Institut, Mannheim).
% Abschnitt 2.1

{\sc Meschkowski, H.} [1963], {\it Reihenentwicklungen in der
mathematischen Physik} (Bibliographisches Institut, Mannheim).
% Abschnitt 2.1

{\sc Messiah, A.} [1970], {\it Quantum mechanics \Roemisch{2}}
(North-Holland, Amsterdam).
% Abschnitt 10.1

{\sc Meyer, B.} [1985], {\it On continued fractions corresponding to
asymptotic series}, Rocky Mountain J. Math. {\bf 15}, 167 - 172.
% Abschnitt 4.1

{\sc Meyer, B.} [1986], {\it Real continued fractions and asymptotic
expansions}, SIAM J. Math. Anal. {\bf 17}, 1218 - 1221.
% Abschnitt 4.1

{\sc Meyer, W.} [1977], {\it Configuration expansions by means of
pseudonatural orbitals}, in Schaefer \Roemisch{3}, H.F. (Herausgeber),
{\it Methods of electronic structure theory} (Plenum Press, New York),
413 - 446.
% Abschnitt 8.2

{\sc Michlin, S.G.} [1978], {\it Partielle Differentialgleichungen in
der Mathematischen Physik} (Verlag Harri Deutsch, Thun).
% Abschnitt 9.1

{\sc Midy, P.} [1992], {\it Scaling transformations and extrapolation
algorithms for vector sequences}, Comput. Phys. Commun. {\bf 70}, 285 -
291.
% Abschnitt 13

{\sc Miller, J.C.P.} [1952], {\it A method for the determination of
converging factors, applied to the asymptotic expansions for the
parabolic cylinder function}, Proc. Cambridge Phil. Soc. {\bf 48}, 243
- 254.
% Abschnitt 5.7

{\sc Milne-Thomson, L.M.} [1981], {\it The calculus of finite
differences} (Chelsea, New York).
% Abschnitt 3.1, 5.1, 5.2, 5.3, 5.4, 5.5, 5.6, 5.7, 6.6, 6.7

{\sc Mintmire, J.W., White, C.T.} [1987], {\it Brillouin zone treatment
in total energy calculations of Peierls distorted chains}, Int. J.
Quantum Chem. Symp. {\bf 21}, 131 - 136.
% Abschnitt 11.1

{\sc Mitrinovi\'{c}, D.S., Pe\v{c}ari\'{c}, J.E., Fink, A.M.} [1993],
{\it Classical and new inequalities in analysis} (Kluwer, Dordrecht).
% Abschnitt 10.3, 10.8

{\sc Mizushima, M.} [1970], {\it Quantum mechanics of atomic spectra and
atomic structure} (Benjamin, New York).
% Abschnitt 10.1

{\sc Mlodinow, L.D., Shatz, M.P.} [1984], {\it Solving the Schr\"odinger
equation with use of $1/N$ perturbation theory}, J. Math. Phys. {\bf
25}, 943 - 950.
% Abschnitt 1.3

{\sc Morgan \Roemisch{3}, J.D.} [1989], {\it The analytic structure of
atomic and molecular wavefunctions and its impact on the rate of
convergence of molecular calculations} in Defranceschi, M., Delhalle,~J.
(Herausgeber), {\it Numerical determination of the electronic structure
of atoms, diatomic and polyatomic molecules} (Kluwer, Dordrecht), 49 -
84.
% Abschnitt 9.1

{\sc Mosley, D.H., Fripiat, J.G., Champagne, B., Andr\'e, J.-M.} [1993],
{\it Efficient computation of electron-repulsion integrals in {\rm ab
initio} studies of polymeric systems}, Int. J. Quantum Chem. Symp. {\bf
27}, 793 - 806.
% Abschnitt 11.1, 11.2

{\sc M\"uller-Kirsten, H.J.W.} [1986], {\it Perturbation theory,
level-splitting and large-order behaviour}, Fortschr. Phys. {\bf 34},
775 - 790.
% Abschnitt 10.1

{\sc M\"uller-Kirsten, H.J.W.} [1987], {\it Relationship between the level
splitting and the large order behaviour of perturbation theory}, Arab.
Gulf. J. Scient. Res. Math. Phys. Sci. {\bf A5}, 283 - 287.
% Abschnitt 10.1

{\sc Mukherjee, D., Pal, S.} [1989], {\it Use of cluster expansions in
the open-shell correlation problem}, Adv. Quantum Chem. {\bf 20}, 291 -
373.
% Abschnitt 8.2

{\sc Murray, J.D.} [1984], {\it Asymptotic analysis} (Springer-Verlag,
New York).
% Abschnitt 2.2

{\sc Nakahara, M., Waxman, D., Williams, G.} [1990], {\it Functional
treatment of solitons in polyacetylene}, J. Phys. A {\bf 23}, 5017 -
5027.
% Abschnitt 11.2

{\sc Narison, S.} [1982], {\it Techniques of dimensional regularization
and the two-point functions of QCD and QED}, Phys. Rep. {\bf 84}, 263 -
399.
% Abschnitt 1.2

{\sc Navarro, R.} [1990], {\it Application of high- and low-temperature
series expansions to two-dimensio\-nal systems}, in de Jongh, L.J.
(Herausgeber), {\it Magnetic properties of layered transition metal
compounds} (Kluwer, Dordrecht), 105 - 190.
% Abschnitt 4.5

{\sc Ne'eman, Y.} [1967], {\it Algebraic theory of particle physics}
(Benjamin, New York).
% Abschnitt 1.2

{\sc Negele, J.W., Orland, H.} [1987], {\it Quantum many-particle
systems} (Addison-Wesley, Reading, Mass.).
% Abschnitt 2.3, 10.2

{\sc Neuhaus, W., Schottlaender, S.} [1975], {\it Die Weiterentwicklung
der Aireyschen Konvergenzfaktoren f\"ur das Exponentialintegral zu einer
Darstellung mit Restglied}, Computing {\bf 15}, 41- 52.
% Abschnitt 5.7

{\sc Nevanlinna, F.} [1919], {\it Zur Theorie asymptotischer
Potenzreihen}, Ann. Acad. Sci. Fenn. Ser. A {\bf 12}, 1 - 81.
% Abschnitt 2.3

{\sc Neville, E.H.} [1934], {\it Iterative interpolation}, J. Indian
Math. Soc. {\bf 20}, 87 - 120.
% Abschnitt 11.5

{\sc Newton, R.G.} [1982], {\it Scattering theory of waves and
particles} (Springer-Verlag, New York).
% Abschnitt 4.5

{\sc Nicolaides, C.A., Clark, C.W., Nayfeh, M.H.} (Herausgeber)
[1990], {\it Atoms in strong fields} (Plenum Press, New York).
% Abschnitt 10.1

{\sc Nielsen, N.} [1965], {\it Die Gammafunktion} (Chelsea, New York).
% Abschnitt 5.3, 6.6

{\sc Nievergelt, Y.} [1991], {\it Aitken's and Steffensen's
accelerations in several variables}, Numer. Math. {\bf 59}, 295 - 310.
% Abschnitt 13

{\sc Nikishin, E.M., Sorokin, V.N.} [1991], {\it Rational
approximations and orthogonality} (American Mathematical Society,
Providence, Rhode Island).
% Abschnitt 2.3

{\sc Nikol'ski\v{\i}, S.M.} [1975], {\it Approximation of functions of
several variables and imbedding theorems} (Springer-Verlag, Berlin).
% Abschnitt 8.2

{\sc Niukkanen, A.W.} [1983], {\it A new class of gradient formulas in
the angular momentum theory}, J. Math. Phys. {\bf 24}, 1989 - 1991.
% Abschnitt 8.2

{\sc Niukkanen, A.W.} [1984], {\it Fourier transforms of atomic
orbitals. \Roemisch{1}. Reduction to four-dimensio\-nal harmonics and
quadratic transformations}, Int. J. Quantum. Chem. {\bf 25}, 941 - 955.
% Abschnitt 8.2

{\sc N\"orlund, N.E.} [1926], {\it Le{\c c}ons sur les s\'eries
d'interpolation} (Gautier-Villars, Paris).
% Abschnitt 5.3

{\sc N\"orlund, N.E.} [1954], {\it Vorlesungen \"uber
Differenzenrechnung} (Chelsea, New York).
% Abschnitt 5.1, 5.3

{\sc Normand, J.-M.} [1980], {\it A Lie group: Rotations in quantum
mechanics} (North-Holland, Amsterdam).
% Abschnitt 1.2, 8.2

{\sc Novosadov, B.K.} [1983], {\it Hydrogen-like atomic orbitals:
Addition and expansion theorems, integrals}, Int. J. Quantum Chem. {\bf
24}, 1 - 18.
% Abschnitt 9.1

{\sc Oddershede, J.} [1992], {\it Response and propagator methods}, in
Wilson, S., Diercksen, G.H.F. (Herausgeber), {\it Methods in
computational molecular physics} (Plenum Press, New York), 303 - 324.
% Abschnitt 8.2

{\sc Oddershede, J., J{\o}rgensen, P., Yeager, D.L.} [1984], {\it
Polarization propagator methods in atomic and molecular calculations},
Comput. Phys. Rep. {\bf 2}, 33 - 92.
% Abschnitt 8.2

{\sc Okuyama, Y.} [1984], {\it Absolute summability of Fourier series
and orthogonal series} (Springer-Verlag, Berlin).
% Abschnitt 2.4, 12

{\sc Olde Daalhuis, A.B.} [1992], {\it Hyperasymptotic expansions of
confluent hypergeometric functions}, IMA J. Appl. Math. {\bf 49}, 203 -
216.
% Abschnitt 2.2

{\sc Olde Daalhuis, A.B.} [1993], {\it Uniform, hyper-, and
$q$-asymptotics} (Dissertation, Universiteit van Amsterdam).
% Abschnitt 2.2

{\sc Olver, F.W.J.} [1974], {\it Asymptotics and special functions}
(Academic Press, New York).
% Abschnitt 2.2, 4.5, 7.1, 10.6

{\sc Olver, F.W.J.} [1990], {\it On Stokes' phenomenon and converging
factors}, in Wong, R. (Herausgeber), {\it Asymptotics and
computational analysis} (Marcel Dekker, New York), 329 - 355.
% Abschnitt 2.2

{\sc Olver, F.W.J.} [1991a], {\it Uniform, exponentially improved,
asymptotic expansions for the generalized exponential integral}, SIAM J.
Math. Anal. {\bf 22}, 1460 - 1474.
% Abschnitt 2.2

{\sc Olver, F.W.J.} [1991b], {\it Uniform, exponentially improved,
asymptotic expansions for the confluent hypergeometric function and
other integral transforms}, SIAM J. Math. Anal. {\bf 22},
1475 - 1489.
% Abschnitt 2.2

{\sc Onody, R.N., Neves, U.P.C.} [1992], {\it Series expansion of the
direct percolation probability}, J. Phys. A {\bf 25}, 6609 - 6615.
% Abschnitt 4.5

{\sc O-Ohata, K., Ruedenberg, K.} [1966], {\it Two-center Coulomb
integrals between atomic orbitals}, J. Math. Phys. {\bf 7}, 547 - 559.
% Abschnitt 8.2

{\sc Osada, N.} [1990a], {\it A convergence acceleration method for
some logarithmically convergent sequences}, SIAM J. Numer. Anal. {\bf
27}, 178 - 189.
% Abschnitt 2.4, 4.5, 8.4, 11.5, 11.6

{\sc Osada, N.} [1990b], {\it Accelerable subsets of logarithmic
sequences}, J. Comput. Appl. Math. {\bf 32}, 217 - 227.
% Abschnitt 4.5

{\sc Osada, N.} [1991], {\it Acceleration methods for vector sequences},
J. Comput. Appl. Math. {\bf 38}, 361 - 371.
% Abschnitt 13

{\sc Osada, N.} [1992], {\it Extensions of Levin's transformation to
vector sequences}, Numer. Algor. {\bf 2}, 121 - 132.
% Abschnitt 13

{\sc Paldus, J.} [1974], {\it Group theoretical approach to the
configuration interaction and perturbation theory calculations for
atomic and molecular systems}, J. Chem. Phys. {\bf 61}, 5321 -  5330.
% Abschnitt 1.2, 8.2

{\sc Paldus, J.} [1976], {\it Many-electron correlation problem. A group
theoretical approach}, in Eyring, H., Henderson, D. (Herausgeber), {\it
Theoretical Chemistry\/} {\bf 2} (Academic Press, New York), 131 - 290.
% Abschnitt 1.2, 8.2

{\sc Paldus, J.} [1981], {\it Diagrammatical methods for many-fermion
systems} (Vorlesungsmitschrift, Institute of Theoretical Chemistry,
Catholic University of Nijmegen).
% Abschnitt 8.2

{\sc Paldus, J.} [1988], {\it Lie algebraic approach to the
many-electron correlation problem}, in Truhlar, D.G. (Herausgeber),
{\it Mathematical frontiers in computational chemical physics}
(Springer-Verlag, Berlin), 262 - 299.
% Abschnitt 1.2, 8.2, 10.1

{\sc Paldus, J.} [1992], {\it Coupled cluster theory}, in Wilson, S.,
Diercksen, G.H.F. (Herausgeber), {\it Methods in computational molecular
physics} (Plenum Press, New York), 99 - 194.
% Abschnitt 1.2, 8.2

{\sc Paldus, J., Chin, E.} [1983], {\it Bond length alternation in
cyclic polyenes. \Roemisch{1}. Restricted Hartree-Fock method}, Int.
J. Quantum Chem. {\bf 24}, 373 - 394.
% Abschnitt 11.2

{\sc Paldus, J., Chin, E., Grey, M.G.} [1983], {\it Bond length
alternation in cyclic polyenes. \Roemisch{2}. Unrestricted Hartree-Fock
method}, Int. J. Quantum Chem. {\bf 24}, 395 - 409.
% Abschnitt 11.2

{\sc Paldus, J., {\v C}{\' \i}{\v z}ek, J.} [1975], {\it
Time-independent diagrammatic approach to perturbation theory of
fermion systems}, Adv. Quantum Chem. {\bf 9}, 105 - 197.
% Abschnitt 1.2, 8.2

{\sc Paldus, J., Takahashi, M.} [1984], {\it Bond length alternation in
cyclic polyenes. \Roemisch{4}. Finite-order perturbation theory}, Int.
J. Quantum Chem. {\bf 25}, 423 - 443.
% Abschnitt 11.2

{\sc Paldus, J., Takahashi, M., Cho, R.W.H.} [1984], {\it
Coupled-cluster approach to electron correlation in one dimension:
Cyclic polyene model in delocalized basis}, Phys. Rev. B {\bf 30}, 4267
- 4291.
% Abschnitt 11.2

{\sc Parisi, G.} [1977], {\it The perturbative expansion and the
infinite coupling limit}, Phys. Lett. B {\bf 69}, 329 - 331.
% Abschnitt 2.3, 10.4

{\sc Parisi, G.} [1988], {\it Statistical field theory} (Addison-Wesley,
Reading, Mass.).
% Abschnitt 2.3, 10.1

{\sc Parlett, B.N.} [1980], {\it The symmetric eigenvalue problem}
(Prentice-Hall, Englewood Cliffs).
% Abschnitt 1.2

{\sc Patil, S.H.} [1982], {\it Zeeman effect in the hydrogen atom}, J.
Phys. B {\bf 15}, 1161 - 1173.
% Abschnitt 10.1

{\sc Pauli, M., Alder, K.} [1976], {\it An addition theorem for the
Coulomb function}, J. Phys. A {\bf 9}, 905 - 929.
% Abschnitt 9.1

{\sc Pauling, L., Wilson, E.B.} [1935], {\it Introduction to quantum
mechanics} (McGraw-Hill, New York). % Abschnitt 1.1

{\sc Pauncz, R.} [1979], {\it Spin eigenfunctions} (Plenum Press, New
York).
% Abschnitt 1.2, 8.2

{\sc Pauncz, R., Paldus, J.} [1983], {\it Bond length alternation in
cyclic polyenes. \Roemisch{3}. Alternant molecular orbital method}, Int.
J. Quantum Chem. {\bf 24}, 411 - 423. % Abschnitt 11.2

{\sc Peratt, A.L.} [1984], {\it Continued fraction expansions for the
complete, incomplete, and relativistic dispersion function}, J. Math.
Phys. {\bf 25}, 466 - 468.
% Abschnitt 4.1

{\sc Perron, O.} [1957], {\it Die Lehre von den Kettenbr\"uchen, Band
\Roemisch{2}: Analytisch-funktionentheore\-ti\-sche Kettenbr\"uche}
(Teubner, Stuttgart).
% Abschnitt 4.1, 4.3, 6.1, 6.5, 10.7

{\sc Petersen, G.M.} [1966], {\it Regular matrix transformations}
(McGraw-Hill, London).
% Abschnitt 2.4, 6.1

{\sc Petersson G.A., McKoy, V.} [1967], {\it Application of nonlinear
transformations to the evaluation of multicenter integrals}, J. Chem.
Phys. {\bf 46}, 4362 - 4368.
% Abschnitt 8.2

{\sc Petrushev, P.P., Popov, V.A.} [1987], {\it Rational
approximation of real functions} (Cambridge U. P., Cambridge).
% Abschnitt 2.4

{\sc Pettifor, D.G., Weaire, D.L.} (Herausgeber) [1985], {\it The
recursion method and its applications} (Springer-Verlag, Berlin).
% Abschnitt 4.1

{\sc Peyerimhoff, A.} [1969], {\it Lectures on summability}
(Springer-Verlag, Berlin).
% Abschnitt 2.2, 2.4, 6.1

{\sc Piela, L., Delhalle, J.} [1978], {\it An efficient procedure to
evaluate long-range Coulombic interactions within the framework of the
LCAO-CO method for infinite polymers}, Int. J. Quantum Chem. {\bf 13},
605 - 617.
% Abschnitt 11.1

{\sc Piessens, R., de Doncker-Kapenga, E., \"Uberhuber, C.W., Kahaner, D.
K.} [1983], {\it QUADPACK} (Springer-Verlag, Berlin).
% Abschnitt 2.4, 8.2

{\sc Pisani, C., Dovesi, R., Roetti, C.} [1988], {\it Hartree-Fock ab
initio treatment of crystalline system} (Springer-Verlag, Berlin).
% Abschnitt 11.1

{\sc Poincar\'e, H.} [1886], {\it Sur les int\'egrales irr\'eguli\`eres des
\'equations lin\'eaires}, Acta Math. {\bf 8}, 295 - 344.
% Abschnitt 2.2

{\sc Popescu, V.A., Popescu, I.M.} [1987], {\it An estimate of the
radius of convergence of the perturbation series for the anharmonic
oscillator}, Rev. Roum. Phys. (Bucarest) {\bf 32}, 279 - 281.
% Abschnitt 10.1

{\sc Popov, V.S., Weinberg, V.M.} [1982], {\it On summation of
perturbation series in quantum mechanics}, Phys. Lett. A {\bf 90}, 107 -
109.
% Abschnitt 2.3

{\sc Popov, V.S., Weinberg, V.M., Mur, V.D.} [1986], {\it High orders
of perturbation theory, classical mechanics, and the $1/n$ expansion},
Sov. J. Nucl. Phys. {\bf 44}, 714 - 720.
% Abschnitt 10.1

{\sc Powell, J.L., Crasemann, B.} [1961], {\it Quantum mechanics}
(Addison-Wesley, Reading, Mass.).
% Abschnitt 10.1

{\sc Powell, M.J.D.} [1981], {\it Approximation theory and methods}
(Cambridge U. P., Cambridge).
% Abschnitt 5.1, 11.5

{\sc Powell, R.E., Shah, S.M.} [1988], {\it Summability theory and
its applications} (Prentice-Hall of India, New Delhi).
% Abschnitt 2.2, 2.4, 6.1

{\sc Pre\v{s}najder, P., Kubinec, P.} [1991], {\it On the Borel
summation of perturbative series}, Acta Phys. Slov. {\bf 41}, 3 - 13.
% Abschnitt 2.3

{\sc Prosser, F.P., Blanchard, C.H.} [1962], {\it On the evaluation of
two-center integrals}, J. Chem. Phys. {\bf 36}, 1112.
% Abschnitt 8.2

{\sc Raimes, S.} [1972], {\it Many-electron theory} (North-Holland,
Amsterdam).
% Abschnitt 10.1

{\sc Rainville, E.D.} [1967], {\it Infinite series} (Macmillan, New
York).
% Abschnitt 2.1

{\sc Ramis, J.P., Thomann, J.} [1981], {\it Some comments about the
numerical utilization of factorial series}, in Della Dora, J.,
Demongeot, J., Lacolle, B. (Herausgeber), {\it Numerical methods in the
study of critical phenomena} (Springer-Verlag, Berlin), 12 - 25.
% Abschnitt 5.3

{\sc Rashid, M.A.} [1986], {\it Simple expressions for radial functions
appearing in the expansions of {\rm ${\cal Y}_{\ell_1}^{m_1} (\nabla)
F_{\ell_2}^{m_2} ({\bf r})$} and {\rm $\nabla^{2 n} {\cal
Y}_{\ell_1}^{m_1} (\nabla) F_{\ell_2}^{m_2} ({\bf r})$}}, J. Math. Phys.
{\bf 27}, 549 - 551.
% Abschnitt 8.2

{\sc Rayleigh, J.W.S.} [1945a], {\it The theory of sound \Roemisch{1}}
(Dover, New York).
% Abschnitt 10.1

{\sc Rayleigh, J.W.S.} [1945b], {\it The theory of sound \Roemisch{2}}
(Dover, New York).
% Abschnitt 10.1

{\sc Reed, M., Simon, B.} [1978], {\it Methods of modern mathematical
physics \Roemisch{4}: Analysis of operators} (Academic Press, New York).
% Abschnitt 2.2, 2.3, 6.5, 10.1, 10.2, 10.8, 11.4

{\sc Reid, C.E.} [1967], {\it Transformation of perturbation series
into continued fractions, with applications to an anharmonic
oscillator}, Int. J. Quantum Chem. {\bf 1}, 521 - 534.
% Abschnitt 4.1

{\sc Reinhardt, W.P.} [1982], {\it Pad\'e summations for the real and
imaginary parts of atomic Stark eigenvalues}, Int. J. Quantum Chem.
{\bf 21}, 133 - 146.
% Abschnitt 4.5, 10.1

{\sc Rellich, F.} [1969], {\it Perturbation theory of eigenvalue
problems} (Gordon and Breach, New York).
% Abschnitt

{\sc Rice, J.R.} [1983], {\it Numerical methods, software, and analysis}
(McGraw-Hill, New York).
% Abschnitt 8.3

{\sc Richardson, J.L., Blankenbecler, R.} [1979], {\it Moment recursions
and the Schr\"odinger prob\-lem}, Phys. Rev. D {\bf 19}, 496 - 502.
% Abschnitt 10.4

{\sc Richardson, L.F.} [1927], {\it The deferred approach to the
limit. I. Single lattice}, Phil. Trans. Roy. Soc. London A {\bf 226},
229 - 349.
% Abschnitt 2.4, 5.1, 11.5

{\sc Rivasseau, V.} [1991], {\it From perturbative to constructive
renormalization} (Princeton U.~P., Princeton).
% Abschnitt 2.3

{\sc Rogers, C., Ames, W.F.} [1989], {\it Nonlinear boundary value
problems in science and engineering} (Academic Press, San Diego).
% Abschnitt 4.5

{\sc Roos, B.O.} [1987], {\it The complete active space self-consistent
field method and its application in electronic structure calculations},
in Lawley, K.P. (Herausgeber), {\it Ab initio methods in quantum
chemistry \Roemisch{2}\/} (Wiley, Chichester), 399 - 445.
% Abschnitt 8.2

{\sc Roos, B.} (Herausgeber) [1992], {\it Lecture notes in quantum
chemistry} (Springer-Verlag, Berlin).
% Abschnitt 1.3

{\sc Roos, B.O., Siegbahn, P.E.M.} [1977], {\it The direct configuration
interaction method from molecular integrals}, in Schaefer \Roemisch{3},
H.F. (Herausgeber), {\it Methods of electronic structure theory} (Plenum
Press, New York), 277 - 318.
% Abschnitt 8.2

{\sc Roothaan, C.C.J.} [1951], {\it New developments in molecular
orbital theory}, Rev. Mod. Phys. {\bf 23}, 69 - 89.
% Abschnitt 8.1, 8.2, 8.4, 9.1, 11.1

{\sc Roothaan, C.C.J., Bagus, P.S.} [1963], {\it Atomic
self-consistent field calculations by the expansion method}, in Alder,
B., Fernbach, S., Rotenberg, M. (Herausgeber), {\it Methods in
computational physics {\bf 2}. Quantum mechanics} (Academic Press, New
York), 47 - 94.
% Abschnitt 8.2

{\sc Ross, B.} [1987], {\it Methods of summation} (Descartes Press,
Koriyama).
% Abschnitt 2.1

{\sc Roszak, S., Kaufman, J.J.} [1991], {\it The {\rm ab-initio} hybrid
crystal orbital/molecular cluster approach to study the electronic
structure of molecular crystals and reactions in the solid environment},
Int. J. Quantum Chem. {\bf 25}, 619 - 628.
% Abschnitt 11.1

{\sc Rotenberg, M., Bivins, R., Metropolis, M., Wooten, Jr., J.K.}
[1959], {\it The 3-$j$ and 6-$j$ symbols} (The Technology Press,
Massachusetts Institute of Technology, Cambridge, Mass.).
% Abschnitt 8.2

{\sc Rowe, E.G.P.} [1978], {\it Spherical delta functions and multipole
expansions}, J. Math. Phys. {\bf 19}, 1962 - 1968.
% Abschnitt 8.2

{\sc Ruedenberg, K.} [1967], {\it Bipolare Entwicklungen,
Fouriertransformation und Molekulare Mehr\-zentren-Integrale}, Theor.
Chim. Acta {\bf 7}, 359 - 366.
% Abschnitt 8.2

{\sc Ruiz, J.M.} [1993], {\it The basic theory of power series} (Vieweg,
Braunschweig).
% Abschnitt 2.1

{\sc Rychlewski, J.} [1994], {\it On the use of explicitly correlated
functions in variational computations for small molecules}, Int. J.
Quantum Chem. {\bf 49}, 477 - 494.
% Abschnitt 8.4

{\sc Rys, J., Dupuis, M., King, H.F.} [1983], {\it Computation of
electron repulsion integrals using the Rys quadrature method}, J.
Comput. Chem. {\bf 4}, 154 - 157.
% Abschnitt 9.3

{\sc Sablonniere, P.} [1991], {\it Comparison of four algorithms
accelerating the convergence of a subset of logarithmic fixed point
sequences}, Numer. Algor. {\bf 1}, 177 - 197.
% Abschnitt 4.5

{\sc Sablonniere, P.} [1992], {\it Asymptotic behaviour of iterated
modified $\Delta^2$ and $\theta_2$ transforms on some slowly convergent
sequences}, Numer. Algor. {\bf 3}, 401 - 409.
% Abschnitt 4.5

{\sc Saff, E.B., Varga, R.S.} (Herausgeber) [1977], {\it Pad\'e and
rational approximation} (Academic Press, New York).
% Abschnitt 2.4, 4.1

{\sc Sakurai, J.J.} [1985], {\it Modern quantum mechanics},
(Benjamin/Cummings, Menlo Park, Cal.).
% Abschnitt 10.1

{\sc Salzer, H.E.} [1954], {\it A simple method for summing certain
slowly convergent series}, J. Math. and Phys. (Cambridge, Mass.) {\bf
33}, 356 - 359.
% Abschnitt 4.5

{\sc Salzer, H.E.} [1956], {\it Formulas for the partial summation of
series}, Math. Tables Aids Comput. {\bf 10}, 149 - 156.
% Abschnitt 4.5

{\sc Salzer, H.E.} [1983], {\it Note on the Do{\v c}ev-Grosswald
asymptotic series for generalized Bessel polynomials}, J. Comput. Appl.
Math. {\bf 9}, 131 - 135.
% Abschnitt 8.3

{\sc Salzer, H.E., Kimbro, G.M.} [1961], {\it Improved formulas for
complete and partial summation of certain series}, Math. Comput. {\bf
15}, 23 - 39.
% Abschnitt 4.5

{\sc Santos, F.D.} [1973], {\it Finite range approximation in direct
transfer reactions}, Nucl. Phys. A {\bf 212}, 341 - 364.
% Abschnitt 8.2

{\sc Sarkar, B., Bhattacharyya, K.} [1992a], {\it Solution of convergence
difficulties in the Madelung-sum problem: An extrapolation scheme for
sawtooth sequences}, J. Math. Phys. {\bf 33}, 349 - 357.
% Abschnitt 11.1

{\sc Sarkar, B., Bhattacharyya, K.} [1992b], {\it Accurate evaluation of
lattice constants using multipoint Pad\'e-approximants}, Phys. Rev. B
{\bf 45}, 4594 - 4599. % Abschnitt 11.1

{\sc Sarkar, B., Bhattacharyya, K.} [1993], {\it Accurate calculation of
Coulomb sums. Efficacy of Pad\'e-like methods}, Phys. Rev. B {\bf 48},
6913 - 6918.
% Abschnitt 11.1

{\sc Sarkar, B., Bhattacharyya, K., Bhattacharyya, S.P.} [1989], {\it
Forcing the convergence in pathological self-consistent-field
calculations: A Pad\'e-(MC)SCF strategy}, Chem. Phys. Lett. {\bf 162}, 61
- 66.
% Abschnitt 4.5, 11.1

{\sc Saunders, V.R.} [1975], {\it An introduction to molecular integral
evaluation}, in Diercksen, G.H.F., Sutcliffe, B.T., Veillard, A.
(Herausgeber), {\it Computational techniques in quantum chemistry and
molecular physics} (Reidel, Dordrecht), 347 - 424.
% Abschnitt 9.2, 9.3

{\sc Saunders, V.R.} [1983], {\it Molecular integrals for Gaussian type
functions}, in Diercksen, G.H.F., Wilson, S. (Herausgeber), {\it Methods
in computational molecular physics} (Reidel, Dordrecht), 1 - 36.
% Abschnitt 9.2, 9.3

{\sc Sawaguri, T., Tobocman, W.} [1967], {\it Finite-range effects in
distorted-wave Born-approxima\-tion calculations of nucleon transfer
reactions}, J. Math. Phys. {\bf 8}, 2223 - 2230.
% Abschnitt 9.1

{\sc Schaad, L.J., Morrell, G.O.} [1971], {\it Approximations for the
functions $F_m (z)$ occurring in molecular calculations with a Gaussian
basis}, J. Chem. Phys. {\bf 54}, 1965 - 1967.
% Abschnitt 9.3

{\sc Schaefer {\Roemisch 3}, H.F.} [1984], {\it Quantum chemistry -- The
development of {\rm ab initio} methods in molecular structure theory}
(Clarendon Press, Oxford).
% Abschnitt 1.2, 11.1

{\sc Schlegel, H.B., McDouall, J.J.W.} [1991], {\it Do you have SCF
stability and convergence problems?}, in \"Ogretir, C., Csizmadia, I.G.,
Lang, E.A. (Herausgeber), {\it Computational advances in organic
chemistry: Molecular structure and reactivity} (Kluwer, Dordrecht), 167
- 185.
% Abschnitt 11.1

{\sc Schiffrer, G., Stanzial, D.} [1985], {\it Improved calculations for
anharmonic oscillators using the gradient method}, Il Nuovo Cimento B
{\bf 90}, 74 - 83.
% Abschnitt 10.7

{\sc Schneider, C.} [1975], {\it Vereinfachte Rekursionen zur
Richardson-Extrapolation in Speziallf\"allen}, Numer. Math. {\bf 24}, 177
- 184.
% Abschnitt 3.2

{\sc Schr\"odinger, E.} [1926], {\it Quantisierung als Eigenwertproblem.
\Roemisch{3}. St\"orungstheorie, mit Anwendung auf den Starkeffekt der
Balmerlinien}, Ann. Physik {\bf 80}, 437 - 490.
% Abschnitt 10.1

{\sc Schulman, L.S.} [1981], {\it Techniques and applications of path
integration} (Wiley, New York).
% Abschnitt 10.1

{\sc Schulten, K., Gordon, R.G.} [1975], {\it Exact recursive evaluation
of $3j$- and $6j$-coefficients for quantum-mechanical coupling of
angular momenta}, J. Math. Phys. {\bf 16}, 1961 - 1970.
% Abschnitt 8.2

{\sc Schulten, K., Gordon, R.G.} [1976], {\it Recursive evaluation of
$3j$ and $6j$ coefficients}, Comput. Phys. Commun. {\bf 11}, 269 - 278.
% Abschnitt 8.2

{\sc Schwartz, L.} [1978], {\it Th\'{e}orie des distributions} (Hermann,
Paris).
% Abschnitt 8.2

{\sc Scott, T.C., Moore, R.A., Fee, G.J., Monagan, M.B., Labahn, G.,
Geddes, K.O.} [1990], {\it Perturbative solutions of quantum mechanical
problems by symbolic computation: A review}, Int. J. Mod. Phys. C {\bf
1}, 53 - 76.
% Abschnitt 4.4

{\sc Sedogbo, G.A.} [1990], {\it Convergence acceleration of some
logarithmic sequences}, J. Comput. Appl. Math. {\bf 32}, 253 - 260.
% Abschnitt 4.5

{\sc Seeger, R.} [1982], {\it Integrals of Gaussian and continuum
functions for polyatomic molecules. An addition theorem for solid
harmonic Gaussians}, Chem. Phys. Lett. {\bf 92}, 493 - 497.
% Abschnitt 9.1

{\sc Seel, M.} [1988], {\it Atomic clusters and cluster models in solid
state physics}, Int. J. Quantum Chem. Symp. {\bf 22}, 265 - 274.
% Abschnitt 11.1

{\sc Sekino, H., Bartlett, R.J.} [1992], {\it New algorithm for
high-order time-dependent Hartree-Fock theory for nonlinear optical
properties}, Int. J. Quantum Chem. {\bf 43}, 119 - 134.
% Abschnitt 11.2

{\sc Sellers, H.} [1991], {\it ADEM-DIOS: An SCF convergence algorithm
for difficult cases}, Chem. Phys. Lett. {\bf 180}, 461 - 465.
% Abschnitt 11.1

{\sc Sellers, H.} [1993], {\it The ${\rm C}^2$-{\rm DIIS} convergence
acceleration algorithm}, Int. J. Quantum Chem. {\bf 45}, 31 - 41.
% Abschnitt 11.1

{\sc Seymour, R.S.} (Herausgeber) [1981], {\it Conductive polymers}
(Plenum Press, New York).
% Abschnitt 11.2

{\sc Seznec, R., Zinn-Justin, J.} [1979], {\it Summation of divergent
series by order dependent mappings: Application to the anharmonic
oscillator and critical exponents in field theory}, J. Math. Phys. {\bf
20}, 1398 - 1408.
% Abschnitt 2.3, 10.4

{\sc Shanks, D.} [1955], {\it Non-linear transformations of divergent
and slowly convergent sequences}, J. Math. and Phys. (Cambridge, Mass.)
{\bf 34}, 1 - 42.
% Abschnitt 2.4, 4.5, 13

{\sc Sharma, C.S.} [1976], {\it Correlation energies in atoms}, Phys.
Rep. {\bf 26}, 1 - 67.
% Abschnitt 8.2

{\sc Shavitt, I.} [1963], {\it The Gaussian function in calculations of
statistical mechanics and quantum mechanics}, in Alder, B., Fernbach,
S., Rotenberg, M. (Herausgeber), {\it Methods in computational physics
{\bf 2}. Quantum mechanics} (Academic Press, New York), 1 - 45.
% Abschnitt 9.2, 9.3

{\sc Shavitt, I.} [1977], {\it The method of configuration interaction},
in Schaefer \Roemisch{3}, H.F. (Herausgeber), {\it Methods of
electronic structure theory} (Plenum Press, New York), 189 - 275.
% Abschnitt 1.2, 8.2

{\sc Shelef, R.} [1987], {\it New numerical quadrature formulas for
Laplace transform inversion by Bromwich's integral} (auf Hebr\"aisch)
(Master Thesis, Technion, Israel Institute of Technology, Haifa).
% Abschnitt 5.4

{\sc Sherman, A.V.} [1987], {\it A modified Lanczos algorithm and the
continued-fraction representation of correlation functions. An example:
A correlation function of the exciton-phonon system}, J. Phys. A {\bf
20}, 569 - 576.
% Abschnitt 4.1

{\sc Shohat, J.A., Tamarkin, J.D.} [1950], {\it The problem of
moments} (American Mathematical Society, Providence, Rhode Island).
% Abschnitt 4.3

{\sc Shuai, Z., Beljonne, D., Br\'edas, J.L.} [1992], {\it Nonlinear
optical processes in short polyenes: Configuration interaction
description of two-photon absorption and third harmonic generation}, J.
Chem. Phys. {\bf 97}, 1132 - 1137.
% Abschnitt 11.1

{\sc Sidi, A.} [1979], {\it Convergence properties of some nonlinear
sequence transformations}, Math. Comput. {\bf 33}, 315 - 326.
% Abschnitt 5.2, 6.6

{\sc Sidi, A.} [1980], {\it Analysis of convergence of the
$T$-transformation for power series}, Math. Comput. {\bf 35}, 833 - 850.
% Abschnitt 6.6

{\sc Sidi, A.} [1981], {\it A new method for deriving Pad\'e approximants
for some hypergeometric functions}, J. Comput. Appl. Math. {\bf 7}, 37 -
40.
% Abschnitt 5.4, 6.7

{\sc Sidi, A.} [1982], {\it An algorithm for a special case of a
generalization of the Richardson extrapolation process}, Numer. Math.
{\bf 38}, 299 - 307.
% Abschnitt 3.2, 5.1, 5.2

{\sc Sidi, A.} [1986a], {\it Convergence and stability properties of
minimal polynomial and reduced rank extrapolation algorithms}, SIAM J.
Numer. Anal. {\bf 23}, 197 - 209.
% Abschnitt 13

{\sc Sidi, A.} [1986b], {\it Borel summability and converging factors
for some everywhere divergent series}, SIAM J. Math. Anal. {\bf 17},
1222 - 1231.
% Abschnitt 6.6

{\sc Sidi, A.} [1988a], {\it Generalization of Richardson extrapolation
with application to numerical integration}, in Bra{\ss}, H., H\"ammerlin, G.
(Herausgeber), {\it Numerical integration \Roemisch 3} (Birkh\"auser,
Basel), 237 - 250.
% Abschnitt 3.2

{\sc Sidi, A.} [1988b], {\it Extrapolation vs. projection methods for
linear systems of equations}, J. Comput. Appl. Math. {\bf 22}, 71 - 88.
% Abschnitt 13

{\sc Sidi, A.} [1989/90], {\it Application of vector extrapolation
methods to consistent singular linear systems}, Appl. Numer. Math. {\bf
6}, 487 - 500.
% Abschnitt 13

{\sc Sidi, A.} [1990], {\it On a generalization of the Richardson
extrapolation process}, Numer. Math. {\bf 47}, 365 - 377.
% Abschnitt 6.6

{\sc Sidi, A.} [1991], {\it Efficient implementation of minimal
polynomial and reduced rank extrapolation methods}, J. Comput. Appl.
Math. {\bf 36}, 305 - 337.
% Abschnitt 13

{\sc Sidi, A., Bridger, J.} [1988], {\it Convergence and stability
analyses for some vector extrapolation methods in the presence of
defective iteration matrices}, J. Comput. Appl. Math. {\bf 22}, 35 - 61.
% Abschnitt 13

{\sc Sidi, A., Ford, W.F.} [1991], {\it Quotient-difference type
generalizations of the power method and their analysis}, J. Comput.
Appl. Math. {\bf 32}, 261 - 272.
% Abschnitt 13

{\sc Sidi, A., Ford, W.F., Smith, D.A.} [1986], {\it Acceleration of
convergence of vector sequences}, SIAM J. Numer. Anal. {\bf 23}, 178 -
196.
% Abschnitt 13

{\sc Sidi, A., Levin, D.} [1983], {\it Prediction properties of the
$t$-transformation}, SIAM J. Numer. Anal. {\bf 20}, 589 - 598.
% Abschnitt 4.2

{\sc Silver, D.M., Ruedenberg, K.} [1968], {\it Coulomb integrals over
Slater-type atomic orbitals}, J. Chem. Phys. {\bf 49}, 4306 - 4311.
% Abschnitt 8.2

{\sc Silverman, J.N.} [1983a], {\it Generalized Euler transformation for
summing strongly divergent Rayleigh-Schr\"odinger perturbation series: The
Zeeman effect}, Phys. Rev. A {\bf 28}, 498 - 501.
% Abschnitt 10.1

{\sc Silverman, J.N.} [1983b], {\it Extension of the
perturbational-variational Rayleigh-Ritz formalism to large order}, J.
Phys. A {\bf 16}, 3471 - 3483.
% Abschnitt 10.1

{\sc Silverman, J.N., Bishop, D.M.} [1986], {\it A 30-th order
perturbational study of the Born-Oppenheimer electronic polarizabilities
for $\rm H_{2}^{+}$ via the perturbational-variational Rayleigh-Ritz
formalism}, Chem. Phys. Lett. {\bf 130}, 132 - 138.
% Abschnitt 10.1

{\sc Silverman, J.N., Bishop, D.M., Pipin, J.} [1986], {\it
Twentieth-order perturbation study of the nonadiabatic electric
polarizabilities for $\rm H_{2}^{+}$ via the perturbational-variational
Rayleigh-Ritz formalism}, Phys. Rev. Lett. {\bf 56}, 1358 - 1361.
% Abschnitt 10.1

{\sc Silverman, J.N., Bonchev, D., Polansky, O.E.} [1986], {\it
Information-theoretic approach to the convergence of perturbation
expansions}, Phys. Rev. A {\bf 34}, 1736 - 1747.
% Abschnitt 10.1

{\sc Silverman, J.N., Hinze, J.} [1986], {\it High-order Stark effect
perturbation series for hydrogenic ions via the
perturbational-variational Rayleigh-Ritz formalism}, Chem. Phys. Lett.
{\bf 128}, 466 - 473.
% Abschnitt 10.1

{\sc Silverman, J.N., Nicolaides, C.A.} [1991], {\it Complex Stark
eigenvalues for excited states of hydrogenic ions from analytic
continuation of real variationally based large-order perturbation
theory}, Chem. Phys. Lett. {\bf 184}, 321 - 329.
% Abschnitt 10.1

{\sc Silverstone, H.J.} [1967], {\it Expansion about an arbitrary point
of three-dimensional functions by the Fourier-transform convolution
theorem}, J. Chem. Phys. {\bf 47}, 537 - 540.
% Abschnitt 8.2

{\sc Silverstone, H.J.} [1978], {\it Perturbation theory of the Stark
effect in hydrogen to arbitrarily high order}, Phys. Rev. A {\bf 18},
1853 - 1864.
% Abschnitt 10.1

{\sc Silverstone, H.J.} [1985], {\it JWKB connection-formula problem
revisited via Borel summation}, Phys. Rev. Lett. {\bf 55}, 2523 - 2526.
% Abschnitt 2.3

{\sc Silverstone, H.J.} [1986], {\it Reality and complexity in
asymptotic expansions for eigenvalues and eigenfunctions, with
application to the JWKB connection-formula problem}, Int. J. Quantum
Chem. {\bf 29}, 261 - 272.
% Abschnitt 2.3

{\sc Silverstone, H.J.} [1990], {\it High-order perturbation theory
and its application to atoms in strong fields}, in Nicolaides, C.A.,
Clark, C.W., Nayfeh, M.H. (Herausgeber), {\it Atoms in strong fields}
(Plenum Press, New York), 295 - 307.
% Abschnitt 5.5, 10.1

{\sc Silverstone, H.J., Adams, B.G., {\v C}{\' \i}{\v z}ek, J., Otto,
P.} [1979], {\it Stark effect in hydrogen: Dispersion relation,
asymptotic formula, and calculation of the ionization rate via
high-order perturbation theory}, Phys. Rev. Lett. {\bf 43}, 1498 - 1501.
% Abschnitt 5.5, 10.1

{\sc Silverstone, H.J., Harrell, E., Grot, C.} [1981], {\it High-order
perturbation theory of the imaginary part of the resonance eigenvalues
of the Stark effect in hydrogen and of the anharmonic oscillator with
negative anharmonicity}, Phys. Rev. A {\bf 24}, 1925 - 1934.
% Abschnitt 10.1

{\sc Silverstone, H.J., Harris, J.G., {\v C}{\' \i}{\v z}ek, J.,
Paldus, J.} [1985], {\it Asymptotics of high-order perturbation theory
for the one-dimensional anharmonic oscillator by quasisemiclassical
methods}, Phys. Rev. A {\bf 32}, 1965 - 1980.
% Abschnitt 5.5, 10.1

{\sc Silverstone, H.J., Nakai, S., Harris, J.G.} [1985], {\it
Observations on the summability of confluent hypergeometric functions
and on semiclassical quantum mechanics}, Phys. Rev. A {\bf 32}, 1341 -
1345.
% Abschnitt 2.3, 7.2

{\sc Simon, B.} [1970], {\it Coupling constant analyticity for the
anharmonic oscillator}, Ann. Phys. (NY) {\bf 58}, 76 - 136.
% Abschnitt 4.5, 10.2, 10.3, 10.6, 10.7

{\sc Simon, B.} [1972], {\it The anharmonic oscillator: A singular
perturbation theory}, in Bessis, D. (Herausgeber), {\it Carg\`ese lectures
in physics} (Gordon and Breach, New York), Vol. 5, 383 - 414.
% Abschnitt 2.2, 6.5, 10.2, 10.7

{\sc Simon, B.} [1982], {\it Large orders and summability of eigenvalue
perturbation theory: A mathematical overview}, Int. J. Quantum Chem.
{\bf 21}, 3 - 25.
% Abschnitt 2.2, 2.3, 4.3, 4.5, 6.5, 10.1, 10.2

{\sc Simon, B.} [1991], {\it Fifty years of eigenvalue perturbation
theory}, Bull. Amer. Math. Soc. {\bf 24}, 303 - 319.
% Abschnitt 2.2, 2.3, 4.5, 10.1, 10.2

{\sc Singh, S., Pathria, R.K.} [1989], {\it Analytical evaluation of a
class of lattice sums in arbitrary dimensions}, J. Phys. A {\bf 22},
1883 - 1897.
% Abschnitt 11.1

{\sc Slater, J.C.} [1930], {\it Atomic shielding constants}, Phys. Rev.
{\bf 36}, 57 - 64.
% Abschnitt 8.2

{\sc Slater, J.C.} [1932], {\it Analytic atomic wave functions}, Phys.
Rev. {\bf 42}, 33 - 43.
% Abschnitt 8.2

{\sc Slater, L.J.} [1966], {\it Generalized hypergeometric functions}
(Cambridge University Press, Cambridge).
% Abschnitt 7.2, 8.4

{\sc Smith, B.T., Boyle, J.M., Dongarra, J.J., Garbow, B.S., Ikebe, Y.,
Klema, V.C., Moler, C.B.} [1976], {\it Matrix eigensystem routines --
EISPACK guide} (Springer-Verlag, Berlin).
% Abschnitt 1.2

{\sc Smith, D.A., Ford, W.F.} [1979], {\it Acceleration of linear and
logarithmic convergence}, SIAM J. Numer. Anal. {\bf 16}, 223 - 240.
% Abschnitt 3.1, 4.5, 5.2, 5.4, 5.5, 5.7, 6.5, 6.6, 9.3, 11.5

{\sc Smith, D.A., Ford, W.F.} [1982], {\it Numerical comparisons of
nonlinear convergence accelerators}, Math. Comput. {\bf 38}, 481 - 499.
% Abschnitt 3.1, 4.5, 5.2, 6.6, 9.3, 11.5

{\sc Smith, D.A., Ford, W.F., Sidi, A.} [1987], {\it Extrapolation
methods for vector sequences}, SIAM Rev. {\bf 29}, 199 - 233.
% Abschnitt 13

{\sc Smith, D.A., Ford, W.F., Sidi, A.} [1988], {\it Correction to
``Extrapolation methods for vector sequences''}, SIAM Rev. {\bf 30}, 623
- 624.
% Abschnitt 13

{\sc Smith, K.T.} [1987], {\it Power series from a computational point
of view} (Springer-Verlag, Berlin).
% Abschnitt 2.1

{\sc Sobolev, S.L.} [1963], {\it Applications of functional analysis in
mathematical physics} (American Mathematical Society, Providence, Rhode
Island).
% Abschnitt 9.1

{\sc Sobelman, G.E.} [1979], {\it Asymptotic estimates and Borel
resummation for a doubly anharmonic oscillator}, Phys. Rev. D {\bf 19},
3754 - 3767.
% Abschnitt 2.3

{\sc Sokal, A.D.} [1980], {\it An improvement of Watson's theorem on
Borel summability}, J. Math. Phys. {\bf 21}, 261 - 263.
% Abschnitt 2.3

{\sc Soos, Z.G., McWilliams, P.C.M., Hayden, G.W.} [1992], {\it Exact
nonlinear optical coefficients of quantum cell models with interacting
electrons}, Int. J. Quantum Chem. {\bf 43}, 37 - 60.
% Abschnitt 11.2

{\sc Spellucci, P., Pulay, P.} [1975], {\it Effective calculation of the
incomplete gamma function for parameter values $\alpha = (2 n + 1)/2$,
$n = 0, \ldots, 5$}, Angew. Informatik {\bf 17}, 30 - 32.
% Abschnitt 9.3

{\sc Srivastava, G.P.} [1984], {\it Broyden's method for self-consistent
field convergence acceleration}, J. Phys. A {\bf 17}, L317 - L321.
% Abschnitt 11.1

{\sc Srivastava, S., Vishwamittar} [1991], {\it Renormalized
hypervirial-Pad\'e calculations of energies for $V (x) = m (\frac {1}{2}
\omega^2 x^2 + \frac {1}{3} \alpha x^3 + \frac {1}{4} \beta x^4)$},
Chem. Phys. Lett. {\bf 176}, 266 - 272.
% Abschnitt 4.5

{\sc Stafstr\"om, S., Br\'edas, J.L., L\"ogdlund, M., Saleneck, W.R.} [1993],
{\it Charge storage states in polyenes}, J. Chem. Phys. {\bf 99}, 7938 -
7945.
% Abschnitt 11.2

{\sc Stanton, R.E.} [1981], {\it Intrinsic convergence in closed-shell
SCF calculations. A general criterion}, J. Chem. Phys. {\bf 75}, 5416 -
5422. % Abschnitt 11.1

{\sc Stauffer, D., Aharony, A.} [1992], {\it Introduction to percolation
theory} (Taylor \& Francis, London).
% Abschnitt 4.5

{\sc Stedman, G.E.} [1990], {\it Diagram techniques in group theory}
(Cambridge U. P., Cambridge).
% Abschnitt 1.2

{\sc Steinborn, E.O.} [1983], {\it On the evaluation of exponential
(Slater) type integrals}, in Diercksen, G.H.F., Wilson, S.
(Herausgeber), {\it Methods in computational molecular physics} (Reidel,
Dordrecht), 37 - 69.
% Abschnitt 8.1

{\sc Steinborn, E.O., Filter, E.} [1975], {\it Translations of fields
represented by spherical-harmonic expansions for molecular calculations.
\Roemisch{3}. Translations of reduced Bessel functions, Slater-type
$s$-orbitals, and other functions}, Theor. Chim. Acta {\bf 38}, 273 -
281.
% Abschnitt 8.2

{\sc Steinborn, E.O., Filter, E.} [1979], {\it Symmetrie und analytische
Struktur der Additionstheoreme r\"aumlicher Funktionen und der
Mehrzentren-Molek\"ulintegrale \"uber beliebige Atomfunktionen}, Theor.
Chim. Acta {\bf 52}, 189 - 208.
% Abschnitt 9.2

{\sc Steinborn, E.O., Homeier, H.H.H.} [1990], {\it M\"obius-type
quadrature of electron repulsion integrals with $B$ functions}, Int. J.
Quantum Chem. Symp. {\bf 24}, 349 - 363.
% Abschnitt 8.3, 9.2

{\sc Steinborn, E.O., Homeier, H.H.H., Weniger, E.J.} [1992], {\it
Recent progress on representa\-tions for Coulomb integrals of
exponential-type orbitals}, J. Mol. Struct. (THEOCHEM) {\bf 260}, 207 -
221.
% Abschnitt 8.3

{\sc Steinborn, E.O., Weniger, E.J.} [1990], {\it Sequence
transformations for the efficient evaluation of infinite series
representations of some molecular integrals with exponentially decaying
basis functions}, J. Mol. Struct. (Theochem) {\bf 210}, 71 - 78.
% Abschnitt 4.5, 5.2, 8.4

{\sc Steinborn, E.O., Weniger, E.J.} [1992], {\it Nuclear attraction
and electron interaction integrals of exponentially decaying functions
and the Poisson equation}, Theor. Chim. Acta {\bf 83}, 105 - 121.
% Abschnitt 8.2, 8.3

{\sc Steiner, E.} [1976], {\it The determination and interpretation of
molecular wave functions} (Cambridge U. P., Cambridge).
% Abschnitt 8.2

{\sc Steinfeld, J.I.} [1985], {\it Molecules and radiation} (MIT Press,
Cambridge, Mass.).
% Abschnitt 10.2

{\sc Stevens, R.E., Kinsey, J.L., Johnson, B.R.} [1992], {\it
Rational approximants as analytic poly\-atomic potential surfaces}, J.
Chem. Phys. {\bf 96}, 7619 - 7624.
% Abschnitt 4.5

{\sc Stieltjes, T.J.} [1894], {\it Recherches sur les fractions
continues}, Ann. Fac. Sci. Toulouse {\bf 8}, 1 - 122.
% Abschnitt 4.3

{\sc Stirling, J.} [1730], {\it Methodus differentialis sive tractatus
de summatione et interpolatione serium infinitarum} (London). Englische
\"Ubersetzung: Holliday, F. [1749], {\it The differential method, or, a
treatise concerning the summation and interpolation of infinite series}
(London).
% Abschnitt 2.2, 5.3

{\sc Stuart, S.N.} [1981], {\it Non-classical integrals of Bessel
functions}, J. Austral. Math. Soc. B {\bf 22}, 368 - 378.
% Abschnitt 8.2

{\sc Suhai, S.} [1983], {\it Bond alteration in infinite polyene:
Peierls distortion reduced by electron correlation}, Chem. Phys. Lett.
{\bf 96}, 619 - 625.
% Abschnitt 11.1, 11.2, 11.3

{\sc Suhai, S.} [1986], {\it On the excitonic nature of the first UV
absorption peak in polyene}, Int. J. Quantum Chem. {\bf 29}, 469 - 476.
% Abschnitt 11.2

{\sc Suhai, S.} [1992], {\it Structural and electronic properties of
infinite {\rm cis} and {\rm trans} polyenes: Perturbation theory of
electron correlation effects}, Int. J. Quantum Chem. {\bf 42}, 193 - 216.
% Abschnitt 11.1, 11.2

{\sc Suhai, S.} [1993], {\it Third-order M{\o}ller-Plesset perturbation
theory of electron correlation in infinite systems: A comparison of
carbon- and silicon-based polymers}, Int. J. Quantum Chem. {\bf 27}, 131
- 146.
% Abschnitt 11.1, 11.2

{\sc Surj\'{a}n, P.} [1989], {\it Second quantized approach to quantum
chemistry} (Springer-Verlag, Berlin).
% Abschnitt 8.2

{\sc Sutcliffe, B.T.} [1983], {\it Group theory applied to CI methods},
in Diercksen, G.H.F., Wilson, S. (Herausgeber), {\it Methods in
computational molecular physics} (Reidel, Dordrecht), 115 - 159.
% Abschnitt 1.2, 8.2

{\sc Swain, S.} [1976], {\it Continued fraction solutions to systems of
linear equations}, J. Phys. A {\bf 9}, 1811 - 1821.
% Abschnitt 4.1

{\sc Swain, S.} [1986], {\it Continued-fraction methods in atomic
physics}, Adv. Atom. Mol. Phys. {\bf 22}, 387 - 431.
% Abschnitt 4.1

{\sc Swendsen, R.H.} [1991], {\it Acceleration methods for Monte Carlo
computer simulations}, Comput. Phys. Commun. {\bf 65}, 281 - 288.
% Abschnitt 13

{\sc Sz\'{a}sz, O.} [1952], {\it Introduction to the theory of divergent
series} (Stechert-Hafner, New York).
% Abschnitt 2.2

{\sc Tachibana, A., Ishikawa, S., Asai, Y., Katagiri, H., Yamabe, T.}
[1992], {\it A quantum chemical study of interchain hopping model of
negatively charged solitons in polyacetylene}, Int. J. Quantum Chem.
{\bf 41}, 461 - 474. % Abschnitt 11.2

{\sc Tagliani, A.} [1993], {\it On the application of maximum entropy to
the moments problem}, J. Math. Phys. {\bf 34}, 326 - 337.
% Abschnitt 10.2

{\sc Takahashi, M., Paldus, J.} [1984], {\it Bond length alternation in
cyclic polyenes. \Roemisch{5}. Local finite-order perturbation theory
approach}, Int. J. Quantum Chem. {\bf 26}, 349 - 371.
% Abschnitt 11.2

{\sc Takahashi, M., Paldus, J.} [1985a], {\it Coupled-cluster approach
to electron correlation in one dimension. \Roemisch{2}. Cyclic polyene
model in localized basis}, Phys. Rev. B {\bf 31}, 5121 - 5142.
% Abschnitt 11.2

{\sc Takahashi, M., Paldus, J.} [1985b], {\it Bond length alternation in
cyclic polyenes. \Roemisch{6}. Coupled cluster approach with Wannier
orbital basis}, Int. J. Quantum Chem. {\bf 28}, 459 - 479.
% Abschnitt 11.2

{\sc Takahashi, M., Paldus, J., {\v C}{\'\i}{\v z}ek, J.} [1983], {\it
Perturbation theory and electron correlation in extended systems: Cyclic
polyene model}, Int. J. Quantum Chem. {\bf 24}, 707 - 727.
% Abschnitt 11.2

{\sc Talman, J.D.} [1984], {\it Numerical calculation of four-center
Coulomb integrals}, J. Chem. Phys. {\bf 80}, 2000 - 2008.
% Abschnitt 8.2

{\sc Talman, J.D.} [1986], {\it Numerical calculation of nuclear
attraction three-center integrals for arbitrary orbitals}, J. Chem.
Phys. {\bf 84}, 6879 - 6885.
% Abschnitt 8.2

{\sc Talman, J.D.} [1989], {\it Numerical methods for calculating
multicenter integrals for arbitrary orbitals} in Defranceschi, M.,
Delhalle, J. (Herausgeber), {\it Numerical determination of the
electronic structure of atoms, diatomic and polyatomic molecules}
(Kluwer, Dordrecht), 335 - 339.
% Abschnitt 8.2

{\sc Talman, J.D.} [1993], {\it Expressions for overlap integrals of
Slater orbitals}, Phys. Rev. A {\bf 48}, 243 - 249.
% Abschnitt 8.2

{\sc Tanaka, K., Kobayashi, H., Okada, M., Kobashi, M., Yamabe, T.}
[1992], {\it Electronic-phase transition associated with instability of
Hartree-Fock solution of infinite one-dimensional system}, Int. J.
Quantum Chem. {\bf 42}, 45 - 54.
% Abschnitt 11.2

{\sc Tanaka, K., Takata, A., Okada, M., Yamabe, T.} [1993], {\it
Electronic-phase transition of the Hartree-Fock solution of the infinite
one-dimensional system: Structural change in an arbitrarily doped
polyacetylene chain}, Int. J. Quantum Chem. {\bf 47}, 447 - 467.
% Abschnitt 11.2

{\sc Tasche, M.} [1991], {\it Accelerating convergence of univariate and
bivariate Fourier approximations}, Zeitschr. Anal. Anwend. {\bf 10}, 239
- 250.
% Abschnitt 13

{\sc Ta\c{s}eli, H., Demiralp, M.} [1988], {\it Studies on algebraic
methods to solve linear eigenvalue problems: generalised anharmonic
oscillators}, J. Phys. A {\bf 21}, 3903 - 3919.
% Abschnitt 10.7

{\sc Taylor, K.F.} [1987], {\it On Madelung's constant}, J. Comput.
Chem. {\bf 8}, 291 - 295.
% Abschnitt 11.1

{\sc Temme, N.M.} [1975], {\it On the numerical evaluation of modified
Bessel function of the third kind}, J. Comput. Phys. {\bf 19}, 324 -
337.
% Abschnitt 7.1

{\sc Teramae, H.} [1986], {\it {\rm Ab initio} study of the {\rm
cis-trans} energetics of polyacetylene}, J. Chem. Phys. {\bf 85}, 990 -
996.
% Abschnitt 11.1, 11.2

{\sc te Velde, G., Baerends, E.J.} [1992], {\it Numerical integration
for polyatomic systems}, J. Comput. Phys. {\bf 99}, 84 - 98.
% Abschnitt 8.1

{\sc Thiele, T.N.} [1909], {\it Interpolationsrechnung} (Teubner,
Leipzig).
% Abschnitt 11.5

{\sc 't Hooft, G., Veltman, M.V.} [1972], {\it Regularization and
renormalization of gauge fields}, Nucl. Phys. B {\bf 44}, 189 - 213.
% Abschnitt 1.2

{\sc 't Hooft, G.} [1973], {\it Dimensional regularization and the
renormalization group}, Nucl. Phys. B {\bf 61}, 455 - 468.
% Abschnitt 1.2

{\sc Thomann, J.} [1990], {\it Resommation des series formelles},
Numer. Math. {\bf 58}, 503 - 535.
% Abschnitt 2.2, 5.3

{\sc Thompson, I.J., Barnett, A.R.} [1987], {\it Modified Bessel
functions $I_{\nu} (z)$ and $K_{\nu} (z)$ of real order and complex
argument, to selected accuracy}, Comput. Phys. Commun. {\bf 47}, 245 -
257.
% Abschnitt 7.1

{\sc Titchmarsh, E.C.} [1949], {\it Some theorems on perturbation theory},
Proc. Roy. Soc. A {\bf 200}, 34 - 46.
% Abschnitt 10.1

{\sc Titchmarsh, E.C.} [1950], {\it Some theorems on perturbation theory
\Roemisch{2}}, Proc. Roy. Soc. A {\bf 201}, 473 - 479.
% Abschnitt 10.1

{\sc Todd, J.} [1962], {\it Motivation for working in numerical
analysis}, in Todd, J. (Herausgeber), {\it Survey of numerical
analysis} (McGraw-Hill, New York), 1 - 26.
% Abschnitt 2.4

{\sc Trivedi, H.P., Steinborn, E.O.} [1983], {\it Fourier transform of a
two-center product of expo\-nential-type orbitals. Application to one-
and two-electron multicenter integrals.}, Phys. Rev. A {\bf 27}, 670 -
679.
% Abschnitt 8.2, 8.3, 9.2

{\sc Truesdell, C.} [1948], {\it An essay toward a unified theory of
special functions based upon the functional equation $\frac {\partial}
{\partial z} F (z, \alpha) = F (z, \alpha + 1)$} (Princeton U.~P.,
Princeton).
% Abschnitt 9.3

{\sc Truesdell, C.} [1950], {\it On the addition and multiplication
theorems of special functions}, Proc. Natl. Acad. Sci. (USA) {\bf 36},
752 - 755.
% Abschnitt 9.3

{\sc Turbiner, A.V., Ushveridze, A.G.} [1988], {\it Anharmonic
oscillator: Constructing the strong coupling expansions}, J. Math. Phys.
{\bf 29}, 2053 - 2063.
% Abschnitt 10.3

{\sc Urban, M., \v{C}ernu\v{s}\'{a}k, I., Kell\"{o}, V., Noga, J.}
[1987], {\it Electron correlation in molecules}, in Wilson, S.
(Herausgeber), {\it Methods in computational chemistry {\bf 1}. Electron
correlation in atoms and molecules} (Plenum Press, New York), 117 - 250.
% Abschnitt 8.2

{\sc Ursell, H.D.} [1927], {\it The evaluation of Gibb's phase integral
for imperfect gases}, Proc. Cambridge Phil. Soc. {\bf 23}, 685 - 697.
% Abschnitt 1.2

{\sc van den Berg, I.} [1987], {\it Nonstandard asymptotic analysis}
(Springer-Verlag, Berlin).
% Abschnitt 2.2

{\sc van der Laan, C.G., Temme, N.M.} [1980], {\it Calculation of
special functions: The gamma function, the exponential integrals and
error-like functions} (Centrum voor Wiskunde en Informatica, Amsterdam).
% Abschnitt 4.1, 7.1

{\sc Van Vleck, E.B.} [1905], {\it Selected topics in the theory of
divergent series and of continued fractions}, in Van Vleck, E.B.,
White, H.S., Woods, F.S. (Herausgeber), {\it The Boston Colloquium
Lectures} (Macmillan, London), 75 - 187.
% Abschnitt 2.2

{\sc Varshalovich, D.A., Moskalev, A.N., Khersonskii, V.K.} [1988], {\it
Quantum theory of angular momentum} (World Scientific, Singapore).
% Abschnitt 1.2, 8.2

{\sc Verbaarschot, J.J.M., West, P., Wu, T.T.} [1990a], {\it $N = 4$
supersymmetric quantum mechanics and its large order behaviour}, Phys.
Lett. B {\bf 240}, 401 - 403.
% Abschnitt 10.1

{\sc Verbaarschot, J.J.M., West, P., Wu, T.T.} [1990b], {\it Large-order
behavior of the supersymmetric anharmonic oscillator}, Phys. Rev. D {\bf
42}, 1276 - 1284.
% Abschnitt 10.1

{\sc Verbaarschot, J.J.M., West, P.} [1991], {\it Instantons and Borel
resummability for the perturbed supersymmetric anharmonic oscillator},
Phys. Rev. D {\bf 43}, 2718 - 2725.
% Abschnitt 10.1

{\sc Villar, H.O., Dupuis, M., Watts, J.D., Hurst, J.B., Clementi, E.}
[1988], {\it Structure, vibrational spectra, and IR intensities of
polyenes from {\rm ab initio} SCF calculations}, J. Chem. Phys. {\bf
88}, 1003 - 1009.
% Abschnitt 11.1

{\sc Vinette, F.} [1989], {\it Symbolic computation in quantum
mechanics: Determination of lower bounds using the inner projection
technique and some related problems} (Ph.D. Thesis, Department of
Applied Mathematics, University of Waterloo).
% Abschnitt 10.4

{\sc Vinette, F., {\v C}{\'\i}{\v z}ek, J.} [1989], {\it The use of
symbolic computation in solving some non-relativistic quantum mechanical
problems}, in Gianni, P. (Herausgeber), {\it Symbolic and Algebraic
Computation. International Symposium ISSAC '88 -- Rome, Italy}
(Springer-Verlag, Berlin), 85 - 95.
% Abschnitt 10.4

{\sc Vinette, F., {\v C}{\'\i}{\v z}ek, J.} [1991], {\it Upper and
lower bounds of the ground state energy of anharmonic oscillators using
renormalized inner projection}, J. Math. Phys. {\bf 32}, 3392 - 3404.
% Abschnitt 10.4, 10.7

{\sc Vinette, F., {\v C}{\' \i}{\v z}ek, J., Vrscay, E.R.} [1987a],
{\it Renormalized inner projection -- a case study: The octic
oscillator}, Int. J. Quantum Chem. {\bf 32}, 663 - 667.
% Abschnitt 10.4

{\sc Vinette, F., {\v C}{\' \i}{\v z}ek, J., Vrscay, E.R.} [1987b],
{\it Projection interne renormalis\'{e}. \'{E}tude d'une cas:
l'oscillateur octique}, C.R. Acad. Sci. Paris {\bf 306}, 21 - 25.
% Abschnitt 10.4

{\sc Voros, A.} [1983], {\it The return of the quartic oscillator. The
complex WKB method}, Ann. Inst. Henri Poincar\'e Phys. Th\'eor. A {\bf 39},
211 - 338.
% Abschnitt 5.5

{\sc Vra\v{c}ko, M., Liegener, C.-M., Ladik, J.} [1988], {\it
Quasi-particle band structure and exciton spectrum of a two-dimensional
system of hydrogen atoms using M{\o}ller-Plesset many-body perturbation
theory up to second order}, Chem. Phys. {\bf 126}, 255 - 261.
% Abschnitt 11.1

{\sc Vra\v{c}ko, M., Zaider, M.} [1993], {\it A study of excited states
in {\rm trans}-polyacetylene in the Hartree-Fock, Tamm-Dancoff, and
random phase approximation}, Int. J. Quantum Chem. {\bf 47}, 119 - 127.
% Abschnitt 11.2

{\sc Vrscay, E.R.} [1983], {\it Continued fractions, algebraic methods
and quantum mechanical large order perturbation theory} (Ph.D. Thesis,
Department of Applied Mathematics, University of Waterloo).
% Abschnitt 10.1

{\sc Vrscay, E.R.} [1984], {\it Perturbation theory in high order for
generalized charmonium potentials}, Phys. Rev. Lett. {\bf 53}, 2521 -
2524.
% Abschnitt 10.1

{\sc Vrscay, E.R.} [1985], {\it Algebraic methods, Bender-Wu formulas,
and continued fractions at large order for charmonium}, Phys. Rev. A
{\bf 31}, 2054 - 2069.
% Abschnitt 4.1, 10.1

{\sc Vrscay, E.R.} [1986], {\it Hydrogen atom with a Yukawa potential:
Perturbation theory and con\-tinued\/-fractions\/-Pad\'e approximants at
large order}, Phys. Rev. A {\bf 33}, 1433 - 1436.
% Abschnitt 4.1, 10.1

{\sc Vrscay, E.R.} [1987], {\it Rayleigh-Schr\"odinger perturbation
expansions for a hydrogen atom in a polynomial perturbation}, Int. J.
Quantum Chem. {\bf 32}, 613 - 620.
% Abschnitt 10.1

{\sc Vrscay, E.R.} [1988], {\it Renormalized Rayleigh-Schr\"odinger
perturbation theory}, Theor. Chim. Acta {\bf 73}, 365 - 382.
% Abschnitt 10.4

{\sc Vrscay, E.R., {\v C}{\' \i}{\v z}ek, J.} [1986], {\it Continued
fractions and Rayleigh-Schr\"odinger perturbation theory at large order},
J. Math. Phys. {\bf 27}, 185 - 201.
% Abschnitt 4.1, 10.1

{\sc Vrscay, E.R., Handy, C.R.} [1989], {\it The perturbed
two-dimensional oscillator: eigenvalues and infinite field limits via
continued fractions, renormalized perturbation theory and moment
methods}, J. Phys. A {\bf 22}, 823 - 834.
% Abschnitt 4.1, 10.1

{\sc Wall, H.S.} [1973], {\it Analytic theory of continued fractions}
(Chelsea, New York).
% Abschnitt 4.1, 4.3, 6.1, 6.5, 10.7

{\sc Wasow, W.} [1987], {\it Asymptotic expansions for ordinary
differential equations} (Dover, New York).
% Abschnitt 2.2, 5.3

{\sc Watson, G.N.} [1912a], {\it A theory of asymptotic series}, Phil.
Trans. Roy. Soc. London {\bf 211}, 279 - 313.
% Abschnitt 2.2, 2.3

{\sc Watson, G.N.} [1912b], {\it The transformation of an asymptotic
series into a convergent series of inverse factorials}, Rend. Circ.
Mat. Palermo {\bf 34}, 41 - 88.
% Abschnitt 5.3

{\sc Watson, G.N.} [1966], {\it A treatise on the theory of Bessel
functions} (Cambridge U. P., Cambridge).
% Abschnitt 7.1

{\sc Weatherford, C.A., Jones, H.W.} (Herausgeber) [1982], {\it ETO
multicenter molecular integrals} (Reidel, Dordrecht).
% Abschnitt 8.2

{\sc Weidmann, J.} [1980], {\it Linear operators in Hilbert space}
(Springer-Verlag, New York). % Abschnitt 9.1

{\sc Weissbluth, M.} [1978], {\it Atoms and molecules} (Academic Press).
% Abschnitt 8.1

{\sc Weniger, E.J.} [1985], {\it Weakly convergent expansions of a
plane wave and their use in Fourier integrals}, J. Math. Phys. {\bf 26},
276 - 291.
% Abschnitt 8.2, 9.1

{\sc Weniger, E.J.} [1989], {\it Nonlinear sequence transformations
for the acceleration of convergence and the summation of divergent
series}, Comput. Phys. Rep. {\bf 10}, 189 - 371.
% Abschnitt 1.3, 2.4, 3.1, 3.2, 3.3, 4.4, 4.5, 5.1, 5.2, 5.3, 5.4, 5.5,
% 5.6, 6.1, 6.2, 6.3, 6.5, 6.6, 6.7, 7.3, 8.2, 8.4, 11.1, 11.5, 13

{\sc Weniger, E.J.} [1990], {\it On the summation of some divergent
hypergeometric series and related perturbation expansions}, J. Comput.
Appl. Math. {\bf 32}, 291 - 300.
% Abschnitt 4.5, 5.1, 5.4, 5.6, 6.6, 6.7, 10.2, 10.6

{\sc Weniger, E.J.} [1991], {\it On the derivation of iterated
sequence transformations for the acceleration of convergence and the
summation of divergent series}, Comput. Phys. Commun. {\bf 64}, 19 - 45.
% Abschnitt 1.3, 3.3, 4.5, 6.2, 6.7, 8.4

{\sc Weniger, E.J.} [1992], {\it Interpolation between sequence
transformations}, Numer. Algor. {\bf 3}, 477 - 486.
% Abschnitt 1.3, 4.5, 5.1, 5.4, 5.6, 6.6, 6.7, 10.4, 13

{\sc Weniger, E.J., {\v C}{\'\i}{\v z}ek, J.} [1990], {\it Rational
approximations for the modified Bessel function of the second kind},
Comput. Phys. Commun. {\bf 59}, 471 - 493.
% Abschnitt 3.3, 4.5, 5.1, 5.4, 5.5, 5.6, 6.6, 6.7, 7.1, 7.2

{\sc Weniger, E.J., {\v C}{\'\i}{\v z}ek, J., Vinette, F.} [1991],
{\it Very accurate summation for the infinite coupling limit of the
perturbation series expansions of anharmonic oscillators}, Phys. Lett.
A {\bf 156}, 169 - 174.
% Abschnitt 4.4, 4.5, 5.1, 5.4, 5.6, 6.6, 6.7, 10.4, 10.7

{\sc Weniger, E.J., {\v C}{\'\i}{\v z}ek, J., Vinette, F.} [1993],
{\it The summation of the ordinary and renormalized perturbation series
for the ground state energy of the quartic, sextic, and octic
anharmonic oscillators using nonlinear sequence transformations}, J.
Math. Phys. {\bf 34}, 571 - 609.
% Abschnitt 4.4, 4.5, 5.1, 5.4, 5.6, 6.6, 6.7, 10.2, 10.4, 10.7

{\sc Weniger, E.J., Grotendorst, J., Steinborn, E.O.} [1986a], {\it
Some applications of nonlinear convergence accelerators}, Int. J.
Quantum Chem. Symp. {\bf 19}, 181 - 191.
% Abschnitt 4.5

{\sc Weniger, E.J., Grotendorst, J., Steinborn, E.O.} [1986b], {\it
Unified analytical treatment of overlap, two-center nuclear attraction,
and Coulomb integrals of $B$ functions via the Fourier transform
method}, Phys. Rev. A {\bf 33}, 3688 - 3705.
% Abschnitt 8.2, 8.3, 8.4

{\sc Weniger, E.J., Liegener, C.-M.} [1990], {\it Extrapolation of
finite cluster and crystal-orbital calculations on trans-polyacetylene},
Int. J. Quantum Chem. {\bf 38}, 55 - 74.
% Abschnitt 4.5, 11.1, 11.3, 11.6

{\sc Weniger, E.J., Steinborn, E.O.} [1982], {\it Programs for the
coupling of spherical harmonics}, Comput. Phys. Commun. {\bf 25}, 149 -
157.
% Abschnitt 8.2

{\sc Weniger, E.J., Steinborn, E.O.} [1983a], {\it The Fourier
transforms of some exponential-type functions and their relevance to
multicenter problems}, J. Chem. Phys. {\bf 78}, 6121 - 6132.
% Abschnitt 8.2, 8.3

{\sc Weniger, E.J., Steinborn, E.O.} [1983b], {\it Numerical
properties of the convolution theorems of $B$ functions}, Phys. Rev. A
{\bf 28}, 2026 - 2041.
% Abschnitt 8.2, 8.3

{\sc Weniger, E.J., Steinborn, E.O.} [1983c], {\it New representations
for the spherical tensor gradient and the spherical delta function}, J.
Math. Phys. {\bf 24}, 2553 - 2563.
% Abschnitt 8.2

{\sc Weniger, E.J., Steinborn, E.O.} [1984], {\it Comment on
``Hydrogenlike orbitals referred to another coordinate''}, Phys. Rev. A
{\bf 29}, 2268 - 2271.
% Abschnitt 8.2

{\sc Weniger, E.J., Steinborn, E.O.} [1985], {\it A simple derivation
of the addition theorems of the irregular solid harmonics, the Helmholtz
harmonics, and the modified Helmholtz harmonics}, J. Math. Phys. {\bf
26}, 664 - 670.
% Abschnitt 8.2, 9.1

{\sc Weniger, E.J., Steinborn, E.O.} [1987], {\it Comment on ``Molecular
overlap integrals with exponential-type integrals''}, J. Chem. Phys.
{\bf 87}, 3709 - 3711.
% Abschnitt 8.2

{\sc Weniger, E.J., Steinborn, E.O.} [1988], {\it Overlap integrals of
B functions. A numerical study of infinite series representations and
integral representations}, Theor. Chim. Acta {\bf 73}, 323 - 336.
% Abschnitt 8.2, 8.3

{\sc Weniger, E.J., Steinborn, E.O.} [1989a], {\it Nonlinear sequence
transformations for the efficient evaluation of auxiliary functions for
GTO molecular integrals}, in Defranceschi, M., Delhalle, J.
(Herausgeber), {\it Numerical determination of the electronic structure
of atoms, diatomic and polyatomic molecules} (Kluwer, Dordrecht), 341 -
346.
% Abschnitt 4.5, 5.1, 5.4, 5.5, 5.6, 6.6, 6.7, 9.3

{\sc Weniger, E.J., Steinborn, E.O.} [1989b], {\it Addition theorems
for $B$ functions and other exponentially declining functions}, J. Math.
Phys. {\bf 30}, 774 - 784.
% Abschnitt 8.2

{\sc Werner, H., B\"unger, H.J.} (Herausgeber) [1984], {\it Pad\'e
approximations and its applications, Bad Honnef 1983} (Springer-Verlag,
Berlin).
% Abschnitt 2.4, 4.1

{\sc Werner, H.-J.} [1987], {\it Matrix-formulated direct
multi-configuration self-consistent field and multi-configuration
reference configuration-interaction methods}, in Lawley, K.P.
(Herausgeber), {\it Ab initio methods in quantum chemistry
\Roemisch{2}\/} (Wiley, Chichester), 1 - 62.
% Abschnitt 8.2

{\sc Wheeler, J.C.} [1984], {\it Modified moments and continued
fraction coefficients for the diatomic linear chain}, J. Chem. Phys.
{\bf 80}, 472 - 476.
% Abschnitt 4.1

{\sc Wheelon, A.D.} [1968], {\it Tables of summable series and integrals
involving Bessel functions} (Holden-Day, San Francisco).
% Abschnitt 2.1

{\sc Whittaker, E.T., Watson, G.N.} [1927], {\it A course of modern
analysis} (Cambridge U. P., Cambridge).
% Abschnitt 2.3

{\sc Widder, D.V.} [1938], {\it The Stieltjes transform}, Trans. Amer.
Math. Soc. {\bf 43}, 5 - 60.
% Abschnitt 4.3

{\sc Wigner, E. P.} [1959], {\it Group theory and its application to the
quantum mechanics of atomic spectra} (Academic Press, New York).
% Abschnitt 8.2

{\sc Wilansky, A.} [1984], {\it Summability through functional analysis}
(North-Holland, Amsterdam).
% Abschnitt 2.4

{\sc Wilcox, C.H.} (Herausgeber) [1966], {\it Perturbation theory and
its application in quantum mechanics} (Wiley, New York).
% Abschnitt 2.2, 2.3, 4.5, 10.1

{\sc Wilkinson, J.H.} [1965], {\it The algebraic eigenvalue problem}
(Clarendon Press, Oxford).
% Abschnitt 1.2

{\sc Wilkinson, J.H.} [1969], {\it Rundungsfehler} (Springer-Verlag,
Berlin).
% Abschnitt 11.4

{\sc Wilkinson, J.H., Reinsch, C.} (Herausgeber) [1971], {\it Handbook
for automatic computation \Roemisch{2}. Linear algebra}
(Springer-Verlag, Berlin).
% Abschnitt 1.2

{\sc Wilson, K.G.} [1975], {\it The renormalization group: Critical
phenomena and the Kondo problem}, Rev. Mod. Phys. {\bf 47}, 773 - 840.
% Abschnitt 10.6

{\sc Wilson, K.G.} [1990], {\it Ab initio quantum chemistry: A source
of ideas for lattice gauge theorists}, Nucl. Phys. B (Proc. Suppl.) {\bf
17}, 82 - 92.
% Abschnitt 10.6

{\sc Wilson, K.G., Kogut, J.} [1974], {\it The renormalization group
and the $\varepsilon$ expansion}, Phys. Rep. {\bf 12}, 75 - 200.
% Abschnitt 10.6

{\sc Wilson, S.} [1981], {\it Many-body perturbation theory of
molecules}, in Thomson C. (Herausgeber), {\it Theoretical Chemistry}
{\bf 4} (Royal Society of Chemistry, London), 1 - 48.
% Abschnitt 1.2, 8.2, 10.1

{\sc Wilson, S.} [1984], {\it Electron correlation in molecules}
(Clarendon Press, Oxford).
% Abschnitt 1.2, 8.2, 10.1

{\sc Wilson, S.} [1985], {\it Diagrammatic many-body perturbation
theory of atomic and molecular electronic structure}, Comput. Phys.
Rep. {\bf 2}, 389 - 480.
% Abschnitt 1.2, 8.2, 10.1

{\sc Wilson, S.} [1992a], {\it Perturbation theory for atomic and
molecular properties}, in Wilson, S. (Herausgeber), {\it Methods in
computational chemistry {\bf 5}. Atomic and molecular properties}
(Plenum Press, New York), 267 - 333.
% Abschnitt 1.2, 8.2, 10.1

{\sc Wilson, S.} [1992b], {\it The perturbation theory of electron
correlation. \Roemisch{1}. Basic Rayleigh-Schr\"odin\-ger perturbation
theory}, in Wilson, S., Diercksen, G.H.F. (Herausgeber), {\it Methods in
computational molecular physics} (Plenum Press, New York), 195 - 223.
% Abschnitt 1.2, 8.2

{\sc Wilson, S.} [1992c], {\it The perturbation theory of electron
correlation. \Roemisch{2}. Many-body perturbation theory and the
algebraic approximation}, in Wilson, S., Diercksen, G.H.F.
(Herausgeber), {\it Methods in computational molecular physics} (Plenum
Press, New York), 225 - 252.
% Abschnitt 1.2, 8.2

{\sc Wilson, S.} [1992d], {\it The perturbation theory of electron
correlation. \Roemisch{3}. Relativistic many-body perturbation theory},
in Wilson, S., Diercksen, G.H.F. (Herausgeber), {\it Methods in
computational molecular physics} (Plenum Press, New York), 253 - 271.
% Abschnitt 1.2, 8.2

{\sc Wilson, S.} [1992e], {\it The perturbation theory of electron
correlation. \Roemisch{4}. {\rm Concurrent computation} many-body
perturbation theory}, in Wilson, S., Diercksen, G.H.F. (Herausgeber),
{\it Methods in computational molecular physics} (Plenum Press, New
York), 272 - 301.
% Abschnitt 1.2, 8.2

{\sc Wilson, S., Silver, D.M., Farrel, R.A.} [1977], {\it Special
invariance properties of the $[N+1/N]$ Pad\'e approximants in
Rayleigh-Schr\"odinger perturbation theory}, Proc. Roy. Soc. London A
{\bf 356}, 363 - 374.
% Abschnitt 4.5

{\sc Wimp, J.} [1972], {\it Some transformations of monotone
sequences}, Math. Comput. {\bf 26}, 251 - 254.
% Abschnitt 4.5

{\sc Wimp, J.} [1981], {\it Sequence transformations and their
applications} (Academic Press, New York).
% Abschnitt 2.1, 2.4, 4.1, 4.5, 5.3, 6.1, 6.2, 13

{\sc Wimp, J.} [1984], {\it Computation with recurrence relations}
(Pitman, Boston).
% Abschnitt 7.1

{\sc Wimp. J.} [1991], {\it Current trends in asymptotics: Some
problems and some solutions}, J. Comput. Appl. Math. {\bf 35}, 53 - 79.
% Abschnitt 2.2

{\sc Witwit, M.R.M., Killingbeck, J.P.} [1992], {\it Energy levels of
the Schr\"odinger equation for some rational potentials using the
hypervirial method}, Can. J. Phys. {\bf 70}, 1261 - 1266.
% Abschnitt 4.5

{\sc Wloka, J.} [1982], {\it Partielle Differentialgleichungen}
(Teubner, Stuttgart).
% Abschnitt 9.1

{\sc Wong, R.} [1989], {\it Asymptotic approximations of integrals},
(Academic Press, San Diego).
% Abschnitt 2.2

{\sc Wu, T.T.} [1982], {\it Large-order perturbation theory}, Int. J.
Quantum Chem. {\bf 21}, 105 - 118.
% Abschnitt 10.1

{\sc Wulfman, C.} [1971], {\it Dynamical groups in atomic and molecular
physics}, in Loebl, E.M. (Herausgeber), {\it Group theory and its
applications \Roemisch{2}} (Academic Press, New York), 145 - 197.
% Abschnitt 1.2

{\sc Wulfman, C.} [1979], {\it Dynamical groups in atomic and molecular
physics}, in Donini, J.C. (Herausgeber), {\it Recent advances in group
theory and their application to spectroscopy} (Plenum Press, New York),
329 - 403.
% Abschnitt 1.2

{\sc Wuytack, L.} (Herausgeber) [1979a], {\it Pad\'e approximations and
its applications} (Springer-Verlag, Berlin).
% Abschnitt 2.4, 4.1

{\sc Wuytack, L.} [1979b], {\it Commented bibliography on techniques for
computing Pad\'e approximants}, in Wuytack, L. (Herausgeber), {\it Pad\'e
approximations and its applications} (Springer-Verlag, Berlin), 375 -
392.
% Abschnitt 4.4

{\sc Wybourne, B.G.} [1970], {\it Symmetry principles and atomic
spectroscopy} (Wiley, New York). % Abschnitt 1.2

{\sc Wynn, P.} [1956a], {\it On a device for computing the $e_m (S_n)$
transformation}, Math. Tables Aids Comput. {\bf 10}, 91 - 96.
% Abschnitt 2.4, 4.5, 11.5, 13

{\sc Wynn, P.} [1956b], {\it On a Procrustean technique for the
numerical transformation of slowly convergent sequences and series},
Proc. Cambridge Phil. Soc. {\bf 52}, 663 - 671.
% Abschnitt 2.4, 3.3, 4.5, 11.5

{\sc Wynn, P.} [1963], {\it Singular rules for certain non-linear
algorithms}, BIT {\bf 3}, 175 - 195.
% Abschnitt 4.4

{\sc Wynn, P.} [1965], {\it A note on programming repeated applications
of the $\epsilon$-algorithm}, R.F.T.I. - Chiffres {\bf 8}, 23 - 62.
% Abschnitt 4.4

{\sc Wynn, P.} [1966], {\it On the convergence and the stability of the
epsilon algorithm}, SIAM J. Numer. Anal. {\bf 3}, 91 - 122.
% Abschnitt 4.5

{\sc Wynn, P.} [1968], {\it Upon the Pad\'e table derived from a Stieltjes
series}, SIAM J. Numer. Anal. {\bf 5}, 805 - 834.
% Abschnitt 4.4, 6.6, 6.7, 10.4

{\sc Yahiro, S., Gondo, Y.} [1992], {\it A vector processing algorithm
of auxiliary integral evaluation for two-electron Gaussian integrals},
J. Comput. Chem. {\bf 13}, 12 - 16.
% Abschnitt 9.3

{\sc Yamaguchi, T.} [1983], {\it Hydrogenlike orbitals of bound state
referred to another coordinate}, Phys. Rev. A {\bf 27}, 646 - 656.
% Abschnitt 8.2

{\sc Yamazaki, K.} [1984], {\it Quantum theory of anharmonic oscillators
-- a variational and systematic general approximation method}, J. Phys.
A {\bf 17}, 345 - 365.
% Abschnitt 10.4

{\sc Yang, C., You-Liang, W., Bo, C.} [1990], {\it Calculation of band
structure of doped polyacetylene}, Int. J. Quantum Chem. {\bf 37}, 679 -
683.
% Abschnitt 11.2

{\sc Ye, Y.-J., F\"orner, W., Ladik, J.} [1993], {\it Numerical
application of the coupled cluster theory with localized orbitals to
polymers. \Roemisch{1}. Total correlation energy per unit cell}, Chem.
Phys. {\bf 178}, 1 - 23.
% Abschnitt 11.1, 11-2

{\sc Yoshino, S.} [1988], {\it Analytic expressions for asymptotic forms
of continued-fraction coefficients in the presence of a spectral gap},
J. Phys. A {\bf 21}, 1533 - 1547.
% Abschnitt 4.1

{\sc Yoshizawa, K., Ito, A., Tanaka, K., Yamabe, T.} [1993], {\it
Unrestricted Hartree-Fock method for infinite systems with
antiferromagnetic array: Analysis of antiferromagnetic state of {\rm
trans}-polyacetylene}, Int. J. Quantum Chem. {\bf 45}, 391 - 400.
% Abschnitt 11.2

{\sc Yukawa, H.} [1935], {\it On the interaction of elementary
particles. \Roemisch{1}.}, Proc. Phys.-Math. Soc. Japan {\bf 17}, 48 -
57.
% Abschnitt 9.1

{\sc Yutsis, A.P., Levinson, I.B., Vanagas, V.V.} [1962], {\it The
theory of angular momentum} (Israel Program for Scientific Translations,
Jerusalem).
% Abschnitt 1.2

{\sc Zeller, K., Beekmann, W.} [1970], {\it Theorie der
Limitierungsverfahren} (Springer-Verlag, Ber\-lin).
% Abschnitt 2.4, 6.1

{\sc Ziemer, W.P.} [1989], {\it Weakly differentiable functions}
(Springer-Verlag, New York).
% Abschnitt 9.1

{\sc Zinn-Justin, J.} [1971], {\it Strong interaction dynamics with
Pad\'e approximants}, Phys. Rep. {\bf 1}, 55 - 102.
% Abschnitt 4.1, 4.5

{\sc Zinn-Justin, J.} [1979], {\it Large order estimates in perturbation
theory}, Phys. Rep. {\bf 49}, 205 - 213.
% Abschnitt 10.1

{\sc Zinn-Justin, J.} [1981a], {\it Perturbation series at large orders
in quantum mechanics and field theories: Application to the problem of
resummation}, Phys. Rep. {\bf 70}, 109 - 167.
% Abschnitt 2.2, 2.3, 4.5, 10.1

{\sc Zinn-Justin, J.} [1981b], {\it Expansions around instantons in
quantum mechanics}, J. Math. Phys. {\bf 22}, 511 - 520.
% Abschnitt 10.1

{\sc Zinn-Justin, J.} [1984], {\it The principles of instanton calculus:
A few applications}, in Zuber, J.-B., Stora, R. (Herausgeber), {\it
Recent advances in field theory and statistical mechanics}
(North-Holland, Amsterdam), 39 - 172.
% Abschnitt 2.3

{\sc Zinn-Justin, J.} [1989], {\it Quantum field theory and critical
phenomena} (Clarendon Press, Oxford).
% Abschnitt 2.3, 10.2

{\sc Znojil, M.} [1988], {\it Pad\'e oscillators and a new formulation of
perturbation theory}, J. Math. Phys. {\bf 29}, 2611 - 2617.
% Abschnitt 4.5

{\sc Znojil, M.} [1991], {\it The exact bound-state ansaetze as
zero-order approximations in perturbation theory: \Roemisch{1}. The
formalism and Pad\'e oscillators}, Czechoslovak J. Phys. {\bf 41}, 397 -
408.
% Abschnitt 4.5

{\sc Znojil, M.} [1993], {\it The three-point Pad\'e resummation of
perturbation series for anharmonic oscillators}, Phys. Lett. A {\bf
177}, 111 - 120.
% Abschnitt 4.5

{\sc Zucker, I.J., Robertson, M.M.} [1984], {\it Further aspects of the
evaluation of\/} $\sum_{(m, n \ne 0, 0)} (a m^2 + b n m + c n^2)^{- s}$,
Math. Proc. Cambridge Phil. Soc. {\bf 95}, 5 - 13.
% Abschnitt 11.1

\endAbschnittsebene

\keinTitelblatt\neueSeite
\Ueberschrift \large Danksagungen

\Inhaltspez {\protect\bigbreak}

\Inhaltszeile {\large Danksagungen}

\medskip

Ich m\"ochte Herrn Professor E.O. Steinborn danken f\"ur stimulierende und
kritische Diskussionen, f\"ur seine Unterst\"utzung und Ermutigung, und f\"ur
die vorz\"uglichen Arbeitsbedingungen am Institut f\"ur Physikalische und
Theoretische Chemie der Universit\"at Regensburg.

Weiterhin m\"ochte ich mich noch bei Herrn Professor B. Dick und bei Herrn
Professor H. Yersin f\"ur ihre Unterst\"utzung bedanken.

Den Mitarbeitern des Rechenzentrums der Universit\"at Regensburg danke ich
f\"ur ihre Hilfe bei der Durchf\"uhrung von numerischen Rechnungen und beim
Satz dieser Arbeit in \RzTeX.

F\"ur stimulierende und kritische Diskussionen sowie ihre Geduld beim
Korrekturlesen zahlreicher vorl\"aufiger Versionen dieser Arbeit m\"ochte
ich mich bei Herrn Dr.~H.~Homeier und bei Herrn Dipl.~Phys.~H.~Mei{\ss}ner
bedanken.

F\"ur vertrauensvolle und produktive Zusammenarbeit m\"ochte ich mich bei
Frau Professor F. Vinette, York University, Herrn Professor J.
Cioslowski, Florida State University, und Herrn PD C.-M. Liegener,
Universit\"at Erlangen, bedanken.

Die Forschungen, die letztlich zu dieser Arbeit f\"uhrten, wurden durch
einen Aufenthalt als Postdoctoral Fellow am Department of Applied
Mathematics der University of Waterloo, Ontario, Canada, initiiert. Ich
m\"ochte mich deswegen bei Herrn Professor J. {\v C}{\' \i}{\v z}ek und
bei Herrn Professor J. Paldus f\"ur ihre Einladung bedanken, in der
Quantum Theory Group des Department of Applied Mathematics zu arbeiten.
Ich habe von der inspirierenden Atmosph\"are dieser multinationalen
Arbeitsgruppe sowohl in wissenschaftlicher als auch in pers\"onlicher
Hinsicht in hohem Ma{\ss}e profitiert. Besonderer Dank geb\"uhrt Professor J.
{\v C}{\' \i}{\v z}ek, der mein latentes Interesse an asymptotischen
Techniken, an divergenten Reihen und an rationalen Approximationen
erfolg\-reich geweckt hat, der mir eine folgenreiche erste Einf\"uhrung in
das symbolische Rechnen mit Hilfe der in Waterloo entwickelten
Programmiersprache MAPLE gab, und der auch einen weiteren
Forschungsaufenthalt am Department of Applied Mathematics der University
of Waterloo als Visiting Assistant Professor erm\"oglichte.

\bye